\pdfoutput=1
\documentclass[12pt]{book}

\usepackage[english]{babel}
\usepackage[letterpaper,top=1in,bottom=1in,left=1in,right=1in]{geometry}
\usepackage{amsmath}
\usepackage{amssymb}
\allowdisplaybreaks[1]
\usepackage{graphicx}
\usepackage[colorlinks=true, allcolors=blue]{hyperref}
\usepackage{xcolor}
\usepackage{setspace}
\usepackage{parskip}
\usepackage{indentfirst}
\usepackage{titlesec}
\usepackage{fancyhdr}
\usepackage{natbib}
\usepackage{makeidx}
\makeindex

\hypersetup{pdftitle={Statistical Machine Learning for Astronomy}, pdfauthor={Yuan-Sen Ting}}
\usepackage{fvextra}
\usepackage[breakable]{tcolorbox}
\usepackage{newunicodechar}
\newunicodechar{±}{\ensuremath{\pm}}
\newunicodechar{°}{\ensuremath{^\circ}}
\newunicodechar{×}{\ensuremath{\times}}
\newunicodechar{·}{\ensuremath{\cdot}}
\newunicodechar{→}{\ensuremath{\rightarrow}}
\newunicodechar{√}{\ensuremath{\surd}}
\newunicodechar{☉}{\ensuremath{\odot}}
\newunicodechar{²}{\ensuremath{^2}}
\newunicodechar{³}{\ensuremath{^3}}
\newunicodechar{©}{\textcopyright{}}
\newunicodechar{Å}{\AA{}}
\newunicodechar{α}{\ensuremath{\alpha}}
\newunicodechar{γ}{\ensuremath{\gamma}}
\newunicodechar{Δ}{\ensuremath{\Delta}}
\newunicodechar{η}{\ensuremath{\eta}}
\newunicodechar{κ}{\ensuremath{\kappa}}
\newunicodechar{λ}{\ensuremath{\lambda}}
\newunicodechar{μ}{\ensuremath{\mu}}
\newunicodechar{ν}{\ensuremath{\nu}}
\newunicodechar{π}{\ensuremath{\pi}}
\newunicodechar{σ}{\ensuremath{\sigma}}
\newunicodechar{Σ}{\ensuremath{\Sigma}}
\newunicodechar{φ}{\ensuremath{\phi}}
\newunicodechar{Φ}{\ensuremath{\Phi}}

\makeatletter
\def\PYG@reset{\let\PYG@it=\relax \let\PYG@bf=\relax%
    \let\PYG@ul=\relax \let\PYG@tc=\relax%
    \let\PYG@bc=\relax \let\PYG@ff=\relax}
\def\PYG@tok#1{\csname PYG@tok@#1\endcsname}
\def\PYG@toks#1+{\ifx\relax#1\empty\else%
    \PYG@tok{#1}\expandafter\PYG@toks\fi}
\def\PYG@do#1{\PYG@bc{\PYG@tc{\PYG@ul{%
    \PYG@it{\PYG@bf{\PYG@ff{#1}}}}}}}
\def\PYG#1#2{\PYG@reset\PYG@toks#1+\relax+\PYG@do{#2}}

\@namedef{PYG@tok@w}{\def\PYG@tc##1{\textcolor[rgb]{0.73,0.73,0.73}{##1}}}
\@namedef{PYG@tok@c}{\let\PYG@it=\textit\def\PYG@tc##1{\textcolor[rgb]{0.24,0.48,0.48}{##1}}}
\@namedef{PYG@tok@cp}{\def\PYG@tc##1{\textcolor[rgb]{0.61,0.40,0.00}{##1}}}
\@namedef{PYG@tok@k}{\let\PYG@bf=\textbf\def\PYG@tc##1{\textcolor[rgb]{0.00,0.50,0.00}{##1}}}
\@namedef{PYG@tok@kp}{\def\PYG@tc##1{\textcolor[rgb]{0.00,0.50,0.00}{##1}}}
\@namedef{PYG@tok@kt}{\def\PYG@tc##1{\textcolor[rgb]{0.69,0.00,0.25}{##1}}}
\@namedef{PYG@tok@o}{\def\PYG@tc##1{\textcolor[rgb]{0.40,0.40,0.40}{##1}}}
\@namedef{PYG@tok@ow}{\let\PYG@bf=\textbf\def\PYG@tc##1{\textcolor[rgb]{0.67,0.13,1.00}{##1}}}
\@namedef{PYG@tok@nb}{\def\PYG@tc##1{\textcolor[rgb]{0.00,0.50,0.00}{##1}}}
\@namedef{PYG@tok@nf}{\def\PYG@tc##1{\textcolor[rgb]{0.00,0.00,1.00}{##1}}}
\@namedef{PYG@tok@nc}{\let\PYG@bf=\textbf\def\PYG@tc##1{\textcolor[rgb]{0.00,0.00,1.00}{##1}}}
\@namedef{PYG@tok@nn}{\let\PYG@bf=\textbf\def\PYG@tc##1{\textcolor[rgb]{0.00,0.00,1.00}{##1}}}
\@namedef{PYG@tok@ne}{\let\PYG@bf=\textbf\def\PYG@tc##1{\textcolor[rgb]{0.80,0.25,0.22}{##1}}}
\@namedef{PYG@tok@nv}{\def\PYG@tc##1{\textcolor[rgb]{0.10,0.09,0.49}{##1}}}
\@namedef{PYG@tok@no}{\def\PYG@tc##1{\textcolor[rgb]{0.53,0.00,0.00}{##1}}}
\@namedef{PYG@tok@nl}{\def\PYG@tc##1{\textcolor[rgb]{0.46,0.46,0.00}{##1}}}
\@namedef{PYG@tok@ni}{\let\PYG@bf=\textbf\def\PYG@tc##1{\textcolor[rgb]{0.44,0.44,0.44}{##1}}}
\@namedef{PYG@tok@na}{\def\PYG@tc##1{\textcolor[rgb]{0.41,0.47,0.13}{##1}}}
\@namedef{PYG@tok@nt}{\let\PYG@bf=\textbf\def\PYG@tc##1{\textcolor[rgb]{0.00,0.50,0.00}{##1}}}
\@namedef{PYG@tok@nd}{\def\PYG@tc##1{\textcolor[rgb]{0.67,0.13,1.00}{##1}}}
\@namedef{PYG@tok@s}{\def\PYG@tc##1{\textcolor[rgb]{0.73,0.13,0.13}{##1}}}
\@namedef{PYG@tok@sd}{\let\PYG@it=\textit\def\PYG@tc##1{\textcolor[rgb]{0.73,0.13,0.13}{##1}}}
\@namedef{PYG@tok@si}{\let\PYG@bf=\textbf\def\PYG@tc##1{\textcolor[rgb]{0.64,0.35,0.47}{##1}}}
\@namedef{PYG@tok@se}{\let\PYG@bf=\textbf\def\PYG@tc##1{\textcolor[rgb]{0.67,0.36,0.12}{##1}}}
\@namedef{PYG@tok@sr}{\def\PYG@tc##1{\textcolor[rgb]{0.64,0.35,0.47}{##1}}}
\@namedef{PYG@tok@ss}{\def\PYG@tc##1{\textcolor[rgb]{0.10,0.09,0.49}{##1}}}
\@namedef{PYG@tok@sx}{\def\PYG@tc##1{\textcolor[rgb]{0.00,0.50,0.00}{##1}}}
\@namedef{PYG@tok@m}{\def\PYG@tc##1{\textcolor[rgb]{0.40,0.40,0.40}{##1}}}
\@namedef{PYG@tok@gh}{\let\PYG@bf=\textbf\def\PYG@tc##1{\textcolor[rgb]{0.00,0.00,0.50}{##1}}}
\@namedef{PYG@tok@gu}{\let\PYG@bf=\textbf\def\PYG@tc##1{\textcolor[rgb]{0.50,0.00,0.50}{##1}}}
\@namedef{PYG@tok@gd}{\def\PYG@tc##1{\textcolor[rgb]{0.63,0.00,0.00}{##1}}}
\@namedef{PYG@tok@gi}{\def\PYG@tc##1{\textcolor[rgb]{0.00,0.52,0.00}{##1}}}
\@namedef{PYG@tok@gr}{\def\PYG@tc##1{\textcolor[rgb]{0.89,0.00,0.00}{##1}}}
\@namedef{PYG@tok@ge}{\let\PYG@it=\textit}
\@namedef{PYG@tok@gs}{\let\PYG@bf=\textbf}
\@namedef{PYG@tok@ges}{\let\PYG@bf=\textbf\let\PYG@it=\textit}
\@namedef{PYG@tok@gp}{\let\PYG@bf=\textbf\def\PYG@tc##1{\textcolor[rgb]{0.00,0.00,0.50}{##1}}}
\@namedef{PYG@tok@go}{\def\PYG@tc##1{\textcolor[rgb]{0.44,0.44,0.44}{##1}}}
\@namedef{PYG@tok@gt}{\def\PYG@tc##1{\textcolor[rgb]{0.00,0.27,0.87}{##1}}}
\@namedef{PYG@tok@err}{\def\PYG@bc##1{{\setlength{\fboxsep}{\string -\fboxrule}\fcolorbox[rgb]{1.00,0.00,0.00}{1,1,1}{\strut ##1}}}}
\@namedef{PYG@tok@kc}{\let\PYG@bf=\textbf\def\PYG@tc##1{\textcolor[rgb]{0.00,0.50,0.00}{##1}}}
\@namedef{PYG@tok@kd}{\let\PYG@bf=\textbf\def\PYG@tc##1{\textcolor[rgb]{0.00,0.50,0.00}{##1}}}
\@namedef{PYG@tok@kn}{\let\PYG@bf=\textbf\def\PYG@tc##1{\textcolor[rgb]{0.00,0.50,0.00}{##1}}}
\@namedef{PYG@tok@kr}{\let\PYG@bf=\textbf\def\PYG@tc##1{\textcolor[rgb]{0.00,0.50,0.00}{##1}}}
\@namedef{PYG@tok@bp}{\def\PYG@tc##1{\textcolor[rgb]{0.00,0.50,0.00}{##1}}}
\@namedef{PYG@tok@fm}{\def\PYG@tc##1{\textcolor[rgb]{0.00,0.00,1.00}{##1}}}
\@namedef{PYG@tok@vc}{\def\PYG@tc##1{\textcolor[rgb]{0.10,0.09,0.49}{##1}}}
\@namedef{PYG@tok@vg}{\def\PYG@tc##1{\textcolor[rgb]{0.10,0.09,0.49}{##1}}}
\@namedef{PYG@tok@vi}{\def\PYG@tc##1{\textcolor[rgb]{0.10,0.09,0.49}{##1}}}
\@namedef{PYG@tok@vm}{\def\PYG@tc##1{\textcolor[rgb]{0.10,0.09,0.49}{##1}}}
\@namedef{PYG@tok@sa}{\def\PYG@tc##1{\textcolor[rgb]{0.73,0.13,0.13}{##1}}}
\@namedef{PYG@tok@sb}{\def\PYG@tc##1{\textcolor[rgb]{0.73,0.13,0.13}{##1}}}
\@namedef{PYG@tok@sc}{\def\PYG@tc##1{\textcolor[rgb]{0.73,0.13,0.13}{##1}}}
\@namedef{PYG@tok@dl}{\def\PYG@tc##1{\textcolor[rgb]{0.73,0.13,0.13}{##1}}}
\@namedef{PYG@tok@s2}{\def\PYG@tc##1{\textcolor[rgb]{0.73,0.13,0.13}{##1}}}
\@namedef{PYG@tok@sh}{\def\PYG@tc##1{\textcolor[rgb]{0.73,0.13,0.13}{##1}}}
\@namedef{PYG@tok@s1}{\def\PYG@tc##1{\textcolor[rgb]{0.73,0.13,0.13}{##1}}}
\@namedef{PYG@tok@mb}{\def\PYG@tc##1{\textcolor[rgb]{0.40,0.40,0.40}{##1}}}
\@namedef{PYG@tok@mf}{\def\PYG@tc##1{\textcolor[rgb]{0.40,0.40,0.40}{##1}}}
\@namedef{PYG@tok@mh}{\def\PYG@tc##1{\textcolor[rgb]{0.40,0.40,0.40}{##1}}}
\@namedef{PYG@tok@mi}{\def\PYG@tc##1{\textcolor[rgb]{0.40,0.40,0.40}{##1}}}
\@namedef{PYG@tok@il}{\def\PYG@tc##1{\textcolor[rgb]{0.40,0.40,0.40}{##1}}}
\@namedef{PYG@tok@mo}{\def\PYG@tc##1{\textcolor[rgb]{0.40,0.40,0.40}{##1}}}
\@namedef{PYG@tok@ch}{\let\PYG@it=\textit\def\PYG@tc##1{\textcolor[rgb]{0.24,0.48,0.48}{##1}}}
\@namedef{PYG@tok@cm}{\let\PYG@it=\textit\def\PYG@tc##1{\textcolor[rgb]{0.24,0.48,0.48}{##1}}}
\@namedef{PYG@tok@cpf}{\let\PYG@it=\textit\def\PYG@tc##1{\textcolor[rgb]{0.24,0.48,0.48}{##1}}}
\@namedef{PYG@tok@c1}{\let\PYG@it=\textit\def\PYG@tc##1{\textcolor[rgb]{0.24,0.48,0.48}{##1}}}
\@namedef{PYG@tok@cs}{\let\PYG@it=\textit\def\PYG@tc##1{\textcolor[rgb]{0.24,0.48,0.48}{##1}}}

\def\PYGZbs{\char`\\}
\def\PYGZus{\char`\_}
\def\PYGZob{\char`\{}
\def\PYGZcb{\char`\}}
\def\PYGZca{\char`\^}
\def\PYGZam{\char`\&}
\def\PYGZlt{\char`\<}
\def\PYGZgt{\char`\>}
\def\PYGZsh{\char`\#}
\def\PYGZpc{\char`\%}
\def\PYGZdl{\char`\$}
\def\PYGZhy{\char`\-}
\def\PYGZsq{\char`\'}
\def\PYGZdq{\char`\"}
\def\PYGZti{\char`\~}

\makeatother

\def\PYGZsq{\textquotesingle}
\providecommand{\tightlist}{\setlength{\itemsep}{0pt}\setlength{\parskip}{0pt}}
\fvset{breaklines, breakafter={,(=+]\}}, baselinestretch=1}
\definecolor{codeshade}{gray}{0.95}
\newtcolorbox{codecell}{breakable, colback=codeshade, colframe=codeshade, boxrule=0pt, arc=0pt, outer arc=0pt,
  left=4pt, right=4pt, top=2pt, bottom=2pt, before skip=6pt, after skip=6pt}
\makeatletter
\let\tutorial@addtocurcol\@addtocurcol
\usepackage{flafter}
\let\tutorial@flafter\@addtocurcol
\let\@addtocurcol\tutorial@addtocurcol
\newsavebox{\tutorialbox}
\newcommand{\tutorialfloats}{\let\@addtocurcol\tutorial@flafter
  \renewcommand{\topfraction}{0.95}\renewcommand{\textfraction}{0.05}%
  \setcounter{topnumber}{3}\setcounter{totalnumber}{5}}
\newcommand{\tutorialplot}[2]{\par\penalty\z@
  \sbox{\tutorialbox}{#1}%
  \ifdim\pagegoal<\maxdimen
    \dimen@\pagegoal \advance\dimen@-\pagetotal \advance\dimen@-\pagedepth \advance\dimen@-\intextsep
    \advance\dimen@-3pt
    \ifdim\ht\tutorialbox>\dimen@ \ifdim0.93\ht\tutorialbox<\dimen@
      \edef\tutorialheight{\the\dimen@}%
      \sbox{\tutorialbox}{#2}%
    \fi\fi
  \fi
  \begin{figure}[!ht]\centering\usebox{\tutorialbox}\end{figure}}
\newcommand{\tutorialbarrier}{\par\ifx\@deferlist\@empty\else\newpage\fi}
\newcommand{\tutorialdisplay}[2]{\sbox{\tutorialbox}{$\displaystyle #1\qquad #2$}%
  \ifdim\wd\tutorialbox>\linewidth\[\begin{gathered}#1\\#2\end{gathered}\]\else\[#1\qquad #2\]\fi}
\makeatother
\newcommand{\tutorialchapter}[2]{\cleardoublepage\phantomsection
  \addcontentsline{toc}{chapter}{Tutorial #1: #2}\thispagestyle{plain}\vspace*{50pt}
  {\raggedright\normalfont\huge\bfseries Tutorial #1\par\nobreak\vspace{20pt}\Huge\hyphenpenalty=10000 #2\par}
  \nobreak\vspace{40pt}}

\titleformat{\chapter}[display]
{\normalfont\huge\bfseries}{\chaptertitlename\ \thechapter}{20pt}{\Huge}

\title{Statistical Machine Learning for Astronomy}
\author{Yuan-Sen Ting}
\date{\today}

\begin{document}

\begin{titlepage}
    \centering
    \vspace*{2cm}
    
    {\Huge\bfseries Statistical Machine Learning\par}
    {\Huge\bfseries for Astronomy\par}
    \vspace{1.5cm}
    {\large Yuan-Sen Ting\par}
    \vspace{1cm}
    {\large The Ohio State University\par}
    \vspace{2cm}    
    
    \vfill
    {\large \today\par}
    {\large Version 1.0\par}
\end{titlepage}

\newpage
\thispagestyle{empty}
\vspace*{8cm}
\begin{center}
    \textit{To my beloved Popo and Baba, and my family\\
    who supported my journey with remarkable patience.}
\end{center}
\vspace{3cm}
\begin{flushright}
    \vfill
    \includegraphics[height=7cm]{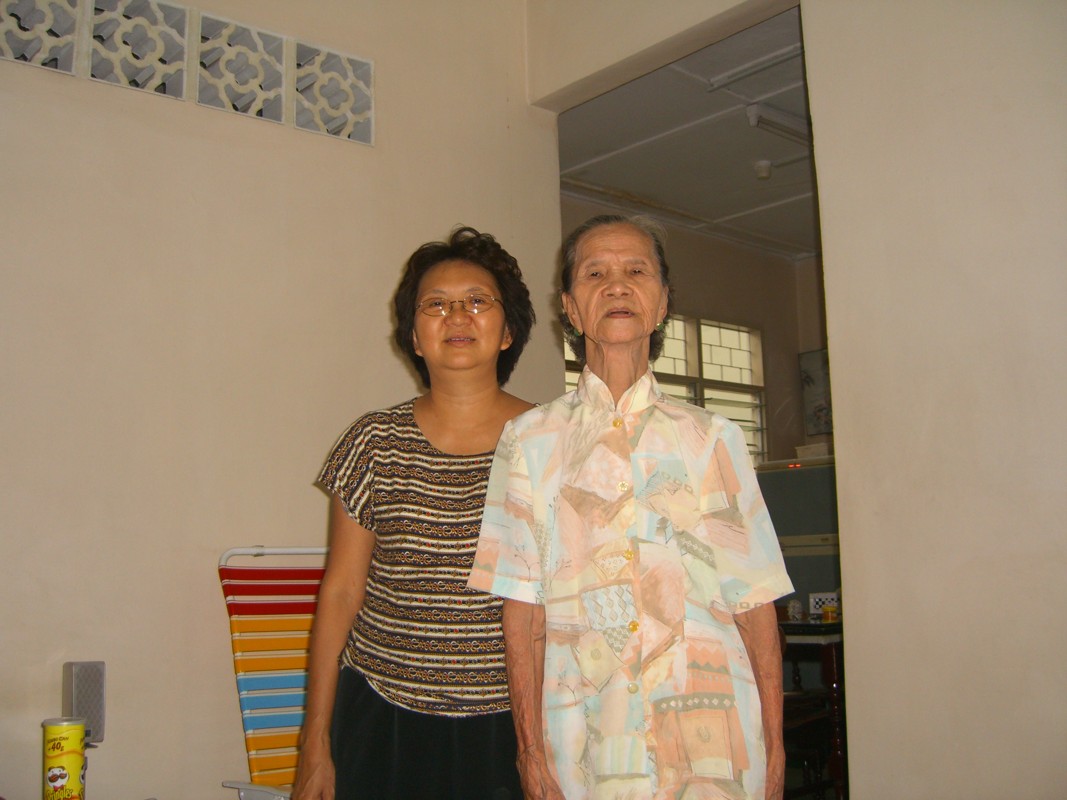}\hspace{0.5cm}\includegraphics[height=7cm]{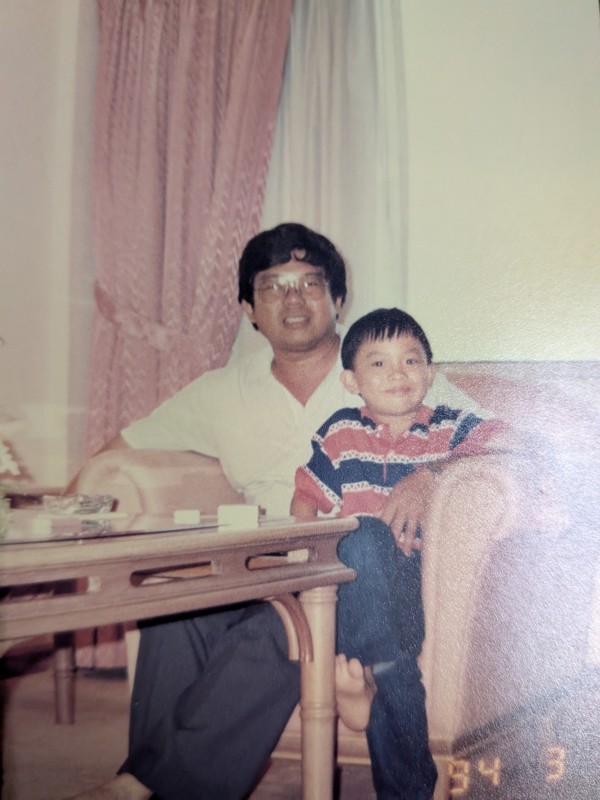}
\end{flushright}
\newpage

\vspace{1cm}
\noindent Columbus, Ohio\\
\today
\newpage

\tableofcontents
\newpage

\chapter{Preface and Overview}

Machine learning is now common in astronomical research because modern data are so large and complex. Astronomers now routinely work with datasets of billions of objects and petabytes of information. This holds from the Sloan Digital Sky Survey to the Vera Rubin Observatory and Nancy Grace Roman Space Telescope. At this scale, automated analysis is necessary.

Discussions about machine learning in astronomy are often confused. The term ``machine learning'' has become nearly synonymous with neural networks and deep learning, which are recent and popular. But machine learning is a much broader set of techniques with a long history in statistical analysis. Treating the two as the same makes productive discussion harder than it needs to be.

Machine learning has also become a polarizing issue in the research community. Much of this comes from misconceptions about what machine learning is and how it connects to the statistical foundations that astronomers have long used. Some presentations treat the methods as ``black boxes'' that avoid the need for rigorous statistical thinking. This has divided researchers into those who embrace the methods and those who distrust them.

The root of the problem is how we usually present the techniques. Existing resources fall into three groups. Classical statistics textbooks are rigorous but often stop before modern methods like neural networks. Machine learning textbooks focus on implementation and give less space to the statistical foundations that make the methods valid for science. Astronomy-focused books are rare. Those that exist tend to focus on practical applications, which is a different goal from teaching the theory.

This gap matters most in astronomy, where the need for interpretability and uncertainty quantification is greater than in most other fields. Computer scientists often view statistics and machine learning differently from physical scientists. In commercial applications, predictive power often justifies the means. In astronomy, we need to know what our models predict, why they predict it, and how confident we should be in the results. Sometimes we use these techniques to question existing physical models or to claim new discoveries. Then we must be clear about what counts as statistical evidence of discovery.

The result is a fragmented approach. Researchers often learn one technique for one problem. They use clustering for galaxy classification, regression for stellar parameter estimation, and neural networks for image analysis. They may not see the statistical principles that connect these techniques. This piecemeal approach limits how well researchers use the techniques. It also feeds the perception that machine learning is only a collection of tools, without the theoretical depth of physics.

This textbook closes that gap by showing that machine learning techniques are natural extensions of the statistical methods astronomers have long used. We do not present them as revolutionary departures from older approaches. We show how they follow from the same probability theory and Bayesian inference that underlie scientific data analysis. We build this foundation step by step and give a helicopter view of the field. We hope this demystifies these techniques and helps astronomers apply them with more thought.

\paragraph{What is Machine Learning}

Machine\index{machine learning}\index{machine learning|(} learning has the same goal as scientific modeling. Both try to build a coherent worldview from observed data. Newton developed the law of gravity to predict planetary positions without explicit lookup tables. In the same way, machine learning tries to learn general rules that apply to new situations. In this view, ``pure learning'' and ``pure modeling'' are the two ends of a spectrum. Real approaches lie somewhere between them.

Machine learning methods rely on inductive bias, the set of assumptions that guide learning when the data alone cannot determine a unique solution. In linear regression, we assume relationships follow hyperplanes, flat surfaces that generalize straight lines. When we choose a kernel function for a Gaussian Process, we encode beliefs about the smoothness of a function. When we design a convolutional neural network, we assume translation invariance in images. These biases are not weaknesses. Without them, learning from finite data is impossible.

Stronger assumptions let us learn efficiently from limited data, but they can reduce flexibility. This trade-off between modeling (adding assumptions) and learning (finding patterns in data) appears throughout science. Even physical theories rest on implicit assumptions about mathematical form, symmetry, and which phenomena are relevant. Machine learning makes these assumptions explicit through the choice of algorithm. It gives a principled way to balance prior knowledge against evidence from data.

This view shows why machine learning is not so different from the astronomical analysis we already do. Choosing between a parametric stellar evolution model and a neural network to predict stellar properties is a choice of position on the modeling-learning spectrum. It is not a choice between science and black-box algorithms. Both combine theory with data-driven discovery. Machine learning is most powerful when it augments scientific understanding and does not replace it.\index{machine learning|)}

\paragraph{What This Book Is and What It Is Not}

This textbook is about statistical machine learning for astronomy. Its main aim is to build the theoretical foundations needed to understand these methods and apply them correctly. The book has a substantial chapter on neural networks and deep learning, but it is not mainly a deep learning textbook. We see deep learning as the natural end point of a progression that starts with basic probability theory and moves through more sophisticated statistical techniques. The theory of deep learning is still catching up with its success, much as thermodynamics came after decades of steam engines.

Our approach is deliberately ``classical-centric'' and puts first the statistical principles that make machine learning methods work. We have three reasons. First, classical techniques often give analytical solutions and well-understood uncertainties, which remain useful in many astronomical applications. Second, knowing these foundations shows why and when more complex methods like neural networks become necessary. Third, many supposedly ``modern'' techniques are natural extensions of classical ideas. This connection is easiest to see when we place both in the same statistical framework.

We use a musical analogy to frame this choice. Classical machine learning methods are like classical music. They need more initial investment to appreciate, but this foundation helps us understand more complex forms. Neural networks and deep learning are more like contemporary popular music, or K-pop. They are immediately accessible and often quite powerful, but you appreciate them more when you know some of the principles behind the music.

This book is not a survey of machine learning techniques or a handbook of implementations. We aim to build conceptual understanding and mathematical intuition, which we believe lead to better application in the end. Our goal is to develop researchers who can think critically about which methods suit which problems, and why.

We derive methods from first principles instead of presenting them as given algorithms. This makes the connections between techniques that look unrelated clear. We are unrelenting in our mathematical approach, from the uniqueness proof for MCMC to the induction proof for Principal Component Analysis. This might seem like too much, but we believe this rigor is critical for the motivation of this book. Without the theoretical foundations, we cannot judge when methods are appropriate, how they might fail, or how they fit into statistical inference. We balance this rigor with analogies, geometric interpretations, and physical intuition that show what the mathematics means.

The astronomical focus also sets this book apart from general machine learning texts. Astronomical data bring particular difficulties. These include irregular sampling, heteroscedastic uncertainties that differ from point to point, physical constraints, and the need for proper error propagation. We use astronomical examples as the central problems that motivate the techniques.

This textbook does not claim new theoretical results or new algorithms. Its mathematical foundations, statistical principles, and machine learning techniques are well established in the literature. The individual parts sit in specialist texts, research papers, and technical resources. They do not appear to be available elsewhere in one compact form. Our role is curation and pedagogical organization. We collect, connect, and present established results in one coherent framework for astronomical applications, so that researchers can build a systematic understanding.

\paragraph{Online Tutorials and Further Reading}

Understanding comes first, but practice matters too. The companion tutorials are runnable and work through astronomical examples, many with real data. Readers who want to implement the methods should work through them as they read the matching chapters. The tutorials, and further reading for each chapter, are at \url{https://statml.org/}.

\paragraph{The Structure of This Textbook}

The book has a planned progression that keeps a clear link to astronomical applications. The structure follows from our belief that machine learning is best understood as an extension of statistical inference, not as a separate discipline.

We begin with probability theory and, above all, Bayesian inference (Chapters 2-3). The philosophy of Bayesian inference as a framework for uncertainty quantification runs through the book. This investment in foundations pays off throughout, because it lets us show how methods that look different follow from the same principles. We then look at summary statistics, showing that finite data limit what we can learn about probability distributions. This creates the need for more sophisticated approaches.

Supervised\index{supervised learning} learning (Chapters 4-9) follows a logical sequence in which complexity grows out of the limits of simpler methods. Supervised learning maps inputs to known outputs. We learn the relationship between variables from examples of both. We start with linear regression. Maximum likelihood leads to the method of least squares that readers may know from physics, and a Bayesian treatment gives uncertainty quantification. Bayesian linear regression then shows how to include prior knowledge formally, and how regularization connects to Bayesian priors.

Chapter 6 treats input uncertainties. This matters particularly in astronomy, where measurement errors in both dependent and independent variables are common. The chapter leads to the iterative optimization methods needed for logistic regression (Chapters 7-8). These in turn motivate the Bayesian treatment of classification in Chapter 9.

Unsupervised\index{unsupervised learning} learning (Chapters 10-11) follows a similar philosophy. It looks for patterns without known target outputs and finds hidden structure rather than input-output relationships. We show how Principal Component Analysis follows from constrained optimization. We also show how K-means and Gaussian Mixture Models give related ways to find structure in data. Both rely on the expectation-maximization algorithm, a general framework for hidden variables. This development makes clear how the methods depend on their assumptions about the data.

Computational methods (Chapters 12-13) come next. As models grow more complex, we often cannot solve them analytically. Then we need sampling. Our treatment of Monte Carlo and Markov Chain Monte Carlo methods gives the numerical tools for complex probabilistic models. It stays tied to the Bayesian framework from the earlier chapters. It also prepares us for simulation-based inference at the end of the book.

Gaussian Processes (Chapter 14) join linear algebra, kernel methods, and Bayesian inference. We can view them as kernelized linear models or as distributions over functions. The same mathematics describes both. The chapter explains how a kernel defines a prior over functions and how a Gaussian Process makes predictions with uncertainties. It also shows how Gaussian Processes connect to neural networks.

Finally, neural networks (Chapter 15) appear as a natural extension of the earlier methods. Autoencoders extend Principal Component Analysis, and mixture density networks extend Gaussian Mixture Models. Neural networks give up some of the mathematical tractability and uncertainty quantification of classical methods, and gain scalability and flexibility. By this point, readers have the background to see both what neural networks can do and what limits they inherit. We show several approaches to uncertainty quantification, from Monte Carlo dropout to the Bayesian treatments available for simpler models. The chapter closes with neural density estimators and simulation-based inference, where the probability models, sampling methods, and neural networks of the earlier chapters come together.

\paragraph{What You Will Learn}

By the end of this textbook, you will have a systematic understanding of statistical machine learning. You will be able to apply these methods to astronomical research problems with confidence. This understanding works at several levels.

At the conceptual level, you will understand how machine learning techniques follow from probability theory and Bayesian inference. This will let you judge new methods critically and understand their assumptions, limits, and proper domains of use.

At the mathematical level, you will be comfortable with the derivations and analysis of key algorithms. This includes how maximum likelihood estimation leads to specific loss functions. Mathematical understanding is the basis for confident application and troubleshooting.

At the practical level, especially with the accompanying tutorials, you will know how to match statistical techniques to astronomical problems. This includes knowing when linear models are enough and when you need more complex approaches. It includes knowing how to build physical knowledge into statistical models. It also includes knowing how to evaluate and compare different approaches to the same problem. You will also understand the computing trade-offs between methods, and how to balance statistical rigor against practical constraints.

Most important, you will learn to think statistically about astronomical problems. This means understanding how uncertainty propagates through analysis pipelines and how to tell different sources of uncertainty apart. Techniques that first seem complicated often turn out to be simple in the right framework. We hope this book gives you that framework, and that you can use it throughout your research career, whatever direction the tools take.

\chapter{Bayesian Inference}

The core of scientific data analysis lies in measuring, quantifying, and reasoning about uncertainty. Throughout this textbook, we develop machine learning methods based on Bayesian principles. These methods help us understand astronomical phenomena under uncertainty. Before we apply them, we must establish the probability theory behind the Bayesian approach.

Inference means drawing conclusions about unknown quantities, such as model parameters, from measured data. Bayesian inference treats both observations and model parameters as random variables. Treating observations as random seems intuitive, because our measurements have uncertainty. Treating model parameters as random may seem odd. Physical constants like the gravitational constant, or the mass of a galaxy, should have definite values. Yet our knowledge of these values is uncertain, and Bayesian inference gives a consistent way to update that knowledge as we collect more observations.

This view is especially useful in astronomy, where we cannot repeat many phenomena under controlled conditions. Astronomers often study unique events and have limited data. Treating model parameters as random variables lets us make statements about uncertainty even when we have limited data.

In this chapter, we start with deterministic and random variables, then cover the probability distributions common in astronomy and the rules for manipulating probabilities. We then compare frequentist and Bayesian approaches and show how Bayesian inference works in practice, even with minimal data. Many methods, from linear regression to neural networks, extend one Bayesian idea. This idea is to update our beliefs about model parameters using observed data.

\section{Deterministic vs. Random Variables}\label{sec:ch2_deterministic_random}

To\index{random variable|(} see why scientific inference needs a probabilistic approach, we first distinguish between deterministic\index{deterministic variable} and random\index{random variable} variables.

\textbf{Deterministic Variables} are quantities that yield the same value under the same conditions. For example, the orbital period of a planet around a star of known mass at a known distance follows deterministic physical laws. By ``the same conditions'' we mean that we specify and fix all relevant physical parameters. Kepler's Third Law for a planet orbiting a star is
\begin{equation}
P^2 = \frac{4\pi^2}{GM}a^3.
\end{equation}
\noindent
Here $P$ is the orbital period, $G$ is the gravitational constant, and $a$ is the semi-major axis of the orbit. $M$ is the mass of the central star, which we assume is much larger than the mass of the planet ($M \gg m_\text{planet}$).

The variables $P$, $M$, and $a$ are deterministic variables. If we specify $M$ and $a$, they completely determine the period $P$. If we use the same values for $M$ and $a$ again, we obtain exactly the same period $P$. Most quantities in basic physics and mathematics courses are deterministic in this way. Consider solving the equation
\begin{equation}
x^2 + 2x - 3 = 0.
\end{equation}

The solutions $x = 1$ and $x = -3$ are deterministic. They are the exact values that satisfy the equation, and they do not change however many times we solve it.

\textbf{Random Variables}, in contrast, are quantities whose values are subject to uncertainty or variation. Even when we specify all known conditions, we can only predict probabilities of different outcomes, not the exact outcome. In astronomy, the true orbital period of a planet is deterministic, but our measurement of it is a random variable. To measure the period $P$, we make many observations of the star's radial velocity or the planet's transit. Each measurement has some uncertainty from instrumental noise and atmospheric effects. Our final estimate of $P$ comes with an uncertainty, often written as $P \pm \sigma_P$.

This uncertainty does not arise because the true period is random. The randomness comes from our imperfect measurement process.

Random variables come in two main forms.

\paragraph{Random by Nature} Some quantities appear to be probabilistic at the most basic level, such as quantum phenomena. Examples include the exact time of decay of a radioactive atom and the spin measurement of an electron in a superposition state.

\paragraph{Incomplete Knowledge or Measurement Uncertainty} Many astronomical quantities appear random because of practical limits, not because they are probabilistic. Consider again the orbital period, which Kepler's Third Law fixes exactly for given values of mass and semi-major axis. Our measurements of these quantities still have uncertainties. The star's mass may have error bars from spectroscopic analysis, and observational limits may affect the semi-major axis.

\begin{figure}[ht!]
    \centering
    \includegraphics[width=0.6\textwidth]{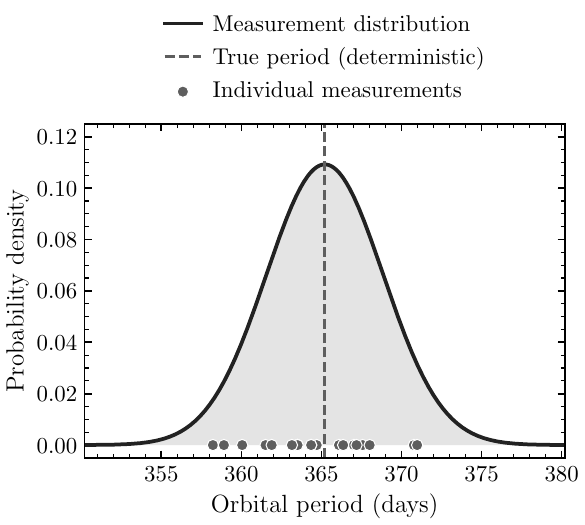}
    \caption{Deterministic versus random variables. The dashed vertical line is the true (deterministic) value of some quantity. The solid curve is the probability distribution of the measured values, including their uncertainties. Individual dots are specific measurements, which scatter around the true value because of various sources of uncertainty. A deterministic quantity becomes a random variable when we include measurement or other uncertainties.}
    \label{fig:random_vs_deterministic}
\end{figure}

The true period is a fixed value, but because our inputs and our period measurements are uncertain, we can only make probabilistic statements about it. This turning of deterministic quantities into random variables is common in astronomical analysis. Figure~\ref{fig:random_vs_deterministic} contrasts a true value with the distribution of its measurements.

This distinction matters when we analyze physical systems. We often know the physical laws, such as Kepler's Third Law, but our measurements of the parameters are uncertain. This creates an asymmetry in scientific inference. We can easily predict the observations that a given set of parameters should produce, which is the forward\index{forward problem} problem. Finding the parameters from the observations, which is the inverse\index{inverse problem} problem, requires us to include the sources of uncertainty.

For instance, the mass of the black hole at the center of a galaxy has a definite value. Our estimate of this mass, based on stellar velocities, gas dynamics, or other techniques, has uncertainty. We represent this uncertainty with a probability distribution that describes our current knowledge. As we gather more data, this distribution becomes narrower around the true value, but some uncertainty remains.

Describing our knowledge of physical quantities with probability distributions gives a consistent mathematical language for scientific inference. We start with prior knowledge, add new observations, and update our understanding step by step.\index{random variable|)}

\section{Probability Distributions}

We\index{random variable|(} now need a mathematical way to describe the uncertainty of these random variables. Probability distributions do this. They describe how probability is spread over the possible values of a random variable.

A probability distribution describes our state of knowledge about a parameter such as the mass of a black hole. It does not mean that the mass itself fluctuates. Its peak marks our best estimate, and its width shows our uncertainty. For observational data, it describes both the expected value and the uncertainty of our measurements.

An event $A$ is a set of possible values of a random variable $X$, such as a measured period between 365.2 and 365.3 days. We write its probability as $P(A)$. This probability must satisfy three axioms.

\begin{enumerate}
    \item Non-negativity means $P(A) \geq 0$ for any event $A$
    \item Normalization means $P(\Omega) = 1$, where $\Omega$ is the sample space, which contains all possible outcomes
    \item Additivity means that for mutually exclusive events $A$ and $B$, which cannot both happen, $P(A \cup B) = P(A) + P(B)$, where $A \cup B$ means that $A$ or $B$ happens
\end{enumerate}

\looseness=1 A probability gives one number for a specific event. A probability distribution covers all possible values of a random variable at once. This concept is central to statistical inference because it gives a complete description of uncertainty rather than a single-point estimate.

Astronomy has both discrete and continuous random variables, which we treat differently.

\paragraph{Discrete Case} Some astronomical quantities take only specific, countable values. Examples are photon counts in a given time interval and galaxy morphological types, such as spiral, elliptical, or irregular. For a discrete random variable $X$, we define a probability\index{probability mass function (PMF)} mass function (PMF)\index{PMF|see{probability mass function}} $p(x)$ where
\begin{itemize}
    \item $p(x) = P(X = x)$ is the probability of $X$ taking the specific value $x$
    \item $\sum_x p(x) = 1$ (normalization)
\end{itemize}

\paragraph{Continuous Case} Most astronomical measurements are continuous. For these random variables, we define a probability\index{probability density function (PDF)} density function (PDF)\index{PDF|see{probability density function}} $p(x)$ where
\begin{itemize}
    \item $P(a \leq X \leq b) = \int_a^b p(x)\mathrm{d}x$ is the probability of $X$ falling in the interval $[a,b]$
    \item $\int_{-\infty}^{\infty} p(x)\mathrm{d}x = 1$ (normalization)
\end{itemize}

A random variable $X$ does not take a fixed value. It follows a probability distribution, and we write this as
\begin{equation}
    X \sim P(x).
\end{equation}
This notation says concisely that the random variable follows that distribution.

A subtle point about continuous distributions is that the probability of measuring any exact value is zero. If we measured an orbital period with infinite precision, what would be the probability of getting exactly 365.256363... days? There are infinitely many possible values, so the probability of any specific one is zero. We can only discuss the probability that a measurement falls in an interval, such as between 365.2 and 365.3 days. We find it by integrating the PDF. The value of $p(x)$ itself gives the relative density of measurements near $x$.

This has an implication for inference with continuous parameters. When a probability distribution peaks at a certain value, it marks the most likely small range of values around that point. The shape of the distribution around the peak shows the uncertainty. A narrow peak means high confidence, and a broad distribution means greater uncertainty.\index{random variable|)}

\section{Common Distributions in Astronomy}\label{sec:ch2_common_distributions}

Several probability distributions appear often in astronomy. We can derive some of them, such as the Gaussian, from the principle of maximum\index{maximum entropy} entropy. Entropy measures how uncertain a distribution is (Section~8.6). The principle picks the most uncertain distribution that still matches the constraints we know, such as a given mean and variance. We can also understand where these distributions come from through physical intuition, which shows when and why they arise.

\paragraph{Gaussian Distribution}

\looseness=1 The\index{Gaussian distribution|(} Gaussian or normal\index{normal distribution|see{Gaussian distribution}} distribution arises when we only know, or constrain, the mean and variance of a quantity. Physically, it appears when many small, independent effects add together (Section~2.5.1 defines independence). The Central\index{central limit theorem} Limit Theorem explains this. Consider measuring a star's brightness, where the measurement has countless sources of uncertainty.

For a single random variable $X$, we write
\begin{equation}
    X \sim \mathcal{N}(\mu,\sigma^2)
\end{equation}
to denote that it follows a Gaussian\index{Gaussian distribution} distribution with probability density
\begin{equation}
p(x|\mu,\sigma) = \frac{1}{\sigma\sqrt{2\pi}} \exp\left(-\frac{(x-\mu)^2}{2\sigma^2}\right).
\end{equation}Here, $p(x|\mu,\sigma)$ is the probability density at value $x$ given parameters $\mu$ and $\sigma$, where
\begin{itemize}
    \item $x$ is the random variable (e.g., the measured brightness)
    \item $\mu$ is the mean or expected value
    \item $\sigma$ is the standard deviation (Chapter~3), measuring the spread of the distribution
    \item The factor $1/(\sigma\sqrt{2\pi})$ normalizes the total probability to 1
\end{itemize}

To see why Gaussian distributions are so common in astronomical measurements, consider how we measure stellar brightness. Many small air pockets of varying density and temperature disturb each photon's path through the atmosphere. The telescope's optics add small aberrations from surface imperfections and thermal variations. The detector adds electronic noise from millions of independent electron movements in the CCD, and the analog-to-digital conversion adds quantization noise.

\begin{figure}[ht!]
    \centering
    \includegraphics[width=0.6\textwidth]{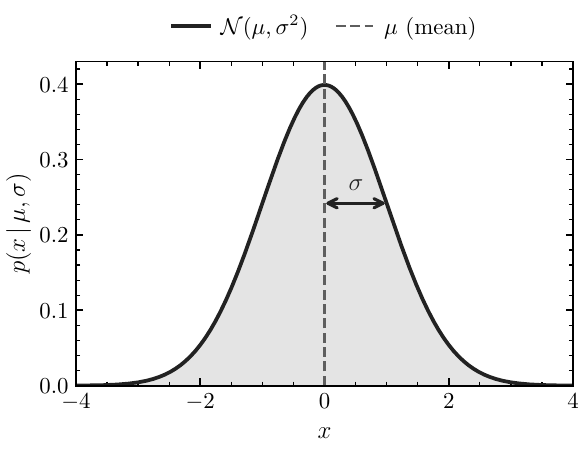}
    \caption{A standard normal (Gaussian) distribution with mean $\mu=0$ and standard deviation $\sigma=1$. The dashed vertical line marks the mean ($\mu$), which is the peak of the distribution. The arrow shows one standard deviation ($\sigma$) from the mean, a main measure of the spread. The distribution is symmetric around its mean, and about 68\% of the probability mass lies within one standard deviation of the mean. Gaussian distributions arise when many small, independent effects add together, which is why they are common in astronomical measurements, as the Central Limit Theorem explains.}
    \label{fig:gaussian_distribution}
\end{figure}

Each effect is small and independent, but they add up. By the Central Limit Theorem, a sum of many small, independent effects approaches a Gaussian distribution, whatever their individual distributions. Figure~\ref{fig:gaussian_distribution} shows the standard Gaussian.

This is why the Gaussian distribution is so common in nature. It is the choice that adds the fewest assumptions when we know only the mean and variance. It is also the result when many small, independent effects combine and nothing else constrains them.

For multivariate\index{Gaussian distribution!multivariate} data, meaning many different random variables, we write
\begin{equation}
    \mathbf{X} \sim \mathcal{N}(\boldsymbol{\mu},\boldsymbol{\Sigma})
\end{equation}
to denote a multivariate Gaussian distribution with probability density
\begin{equation}
p(\mathbf{x}|\boldsymbol{\mu},\boldsymbol{\Sigma}) = \frac{1}{(2\pi)^{D/2}\det(\boldsymbol{\Sigma})^{1/2}} \exp\left(-\frac{1}{2}(\mathbf{x}-\boldsymbol{\mu})^T\boldsymbol{\Sigma}^{-1}(\mathbf{x}-\boldsymbol{\mu})\right).
\end{equation}In this expression
\begin{itemize}
    \item $\mathbf{x}$ is a $D$-dimensional vector of random variables
    \item $\boldsymbol{\mu}$ is the mean vector
    \item $\boldsymbol{\Sigma}$ is the $D \times D$ covariance matrix
    \item $\det(\boldsymbol{\Sigma})$ is the determinant of the covariance matrix
    \item $\boldsymbol{\Sigma}^{-1}$ is the inverse of the covariance matrix
\end{itemize}

The covariance matrix $\boldsymbol{\Sigma}$ holds the variance of each variable (diagonal elements) and also how pairs of variables are correlated (off-diagonal elements). This matters for multivariate astronomical data, such as the correlation between different spectral lines or between mass, age, and metallicity in stellar populations.\index{Gaussian distribution|)}

\paragraph{Poisson Distribution}

The\index{Poisson distribution|(} Poisson\index{Poisson distribution} distribution arises when we count independent events that occur at a constant average rate and we know only the mean rate. It is central to photon detection and many other counting processes in astronomy. Its probability mass function for $k$ events at the mean rate $\lambda$ is
\begin{equation}
P(X=k|\lambda) = \frac{\lambda^k e^{-\lambda}}{k!}.
\end{equation}

To see how the Poisson distribution arises from physical principles, consider detecting photons from a distant star. We divide the observation time into short intervals, so that the chance of two or more photons in one interval is negligible. Three principles then apply, namely independence, uniformity, and randomness.

\begin{figure}[ht!]
    \centering
    \includegraphics[width=0.6\textwidth]{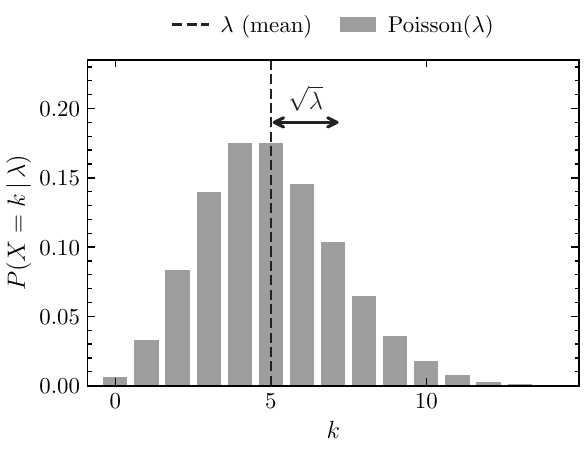}
    \caption{A Poisson distribution with mean rate $\lambda=5$. The bars show the discrete probability mass function, and the dashed vertical line marks the mean. The standard deviation equals $\sqrt{\lambda}$ (arrow), so the distribution is relatively more spread out for small $\lambda$ and more concentrated for large $\lambda$. It describes counting processes in astronomy, such as photon arrivals, cosmic ray hits, or galaxy counts in small volumes, where events occur independently at a constant average rate.}
    \label{fig:poisson_distribution}
\end{figure}

Photon arrivals in different intervals are independent, so one arrival has no effect on another. The chance of detecting a photon in an interval is proportional to its length, so doubling an interval's length doubles its chance. Finally, the arrival times are random, since no two photons arrive at exactly the same instant and no times are preferred. Figure~\ref{fig:poisson_distribution} shows the Poisson distribution.

These simple assumptions lead to the Poisson distribution, which describes random, independent counting events. This is why photon counts, cosmic ray hits, and galaxy counts in small volumes tend to follow Poisson statistics.

\begin{figure}[ht!]
    \centering
    \includegraphics[width=0.9\textwidth]{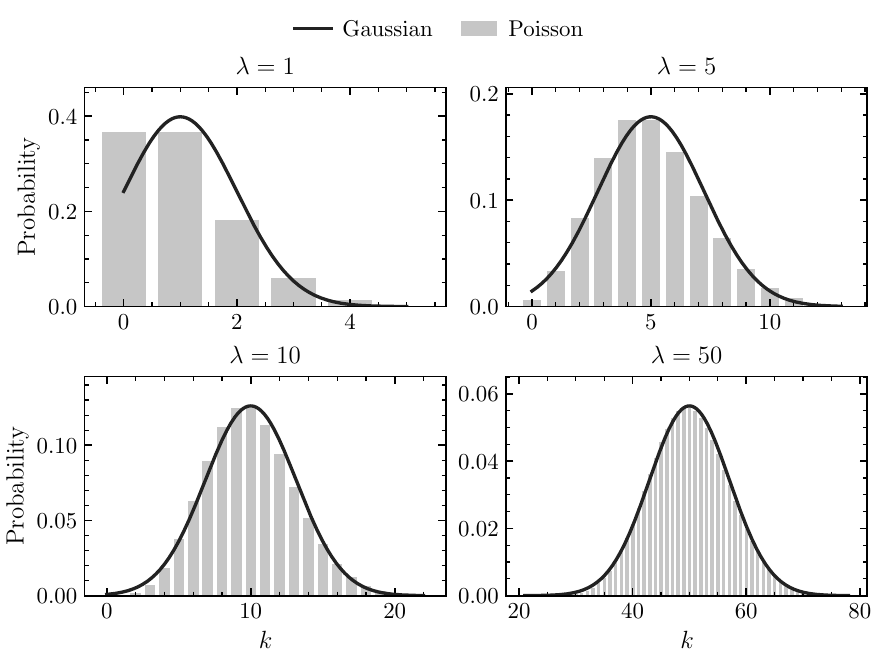}
    \caption{How the Poisson distribution approaches a Gaussian distribution as the mean rate $\lambda$ increases. For small $\lambda$, the distribution is discrete and skewed. As $\lambda$ increases, it becomes more symmetric and less discrete. For large $\lambda$, it is nearly indistinguishable from a Gaussian distribution with the same mean and variance ($\sigma^2=\lambda$). This convergence, a consequence of the Central Limit Theorem, is why we can often approximate photon counts with Gaussian distributions when count rates are high.}
    \label{fig:poisson_gaussian_convergence}
\end{figure}

The Poisson distribution is closely related to the Gaussian distribution. When the mean rate $\lambda$ is large, the Poisson distribution approaches a Gaussian with mean $\mu=\lambda$ and variance $\sigma^2=\lambda$. This is a special case of the Central Limit Theorem. In astronomy, photon counting follows Poisson statistics, but for bright sources with high count rates we can often approximate the counts with a Gaussian. Figure~\ref{fig:poisson_gaussian_convergence} shows how the Poisson distribution approaches the Gaussian.\index{Poisson distribution|)}

\paragraph{Power Law Distribution}

Power\index{power law distribution}\index{power law distribution|(} laws appear when a process has no characteristic scale, so the physics works the same way regardless of size. An example is the mass distribution of stars, the Initial\index{initial mass function (IMF)} Mass Function, for a stellar mass $x$ and a power-law index $\alpha$
\begin{equation}
p(x) \propto x^{-\alpha}.
\end{equation}

When physical limits restrict $x$ to a range between $x_{\min}$ and $x_{\max}$, the normalized power law is
\begin{equation}
p(x) = \frac{1-\alpha}{x_{\max}^{1-\alpha} - x_{\min}^{1-\alpha}} x^{-\alpha}.
\end{equation}
The prefactor makes the integral of $p(x)$ from $x_{\min}$ to $x_{\max}$ equal to 1.

Why do power laws arise in star formation? Consider the fragmentation of a molecular cloud. As a massive cloud collapses under its own gravity, it becomes unstable and fragments into smaller clouds. Each of these can fragment again through the same physical processes. The physics of this fragmentation stays similar at different size scales.

\begin{figure}[ht!]
    \centering
    \includegraphics[width=0.6\textwidth]{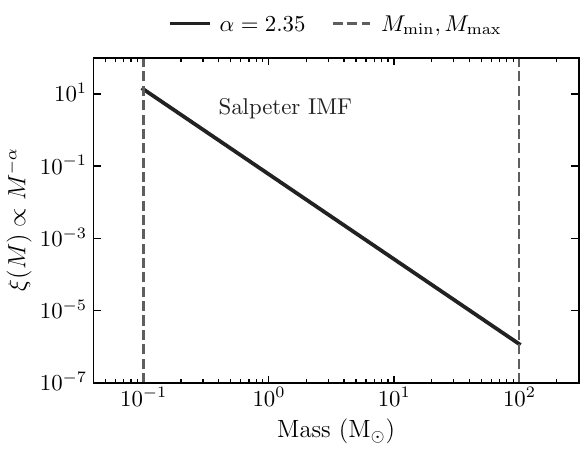}
    \caption{A power law distribution, using the Salpeter Initial Mass Function (IMF) as an example. The IMF describes the distribution of stellar masses at birth, that is, how many stars of each mass form in a population. The distribution follows $\xi(M) \propto M^{-\alpha}$ with $\alpha = 2.35$ (Salpeter's value), on logarithmic scales. The vertical dashed lines mark the physical limits of stellar masses ($M_{\mathrm{min}}$ and $M_{\mathrm{max}}$). This power law arises from the scale-invariant nature of star formation, where the same physical mechanisms operate over many orders of magnitude in mass. The negative slope shows that lower-mass stars are much more numerous than high-mass stars. The frequency drops by a factor of $10^{2.35} \approx 224$ for each factor of 10 increase in mass. The distribution is normalized over the finite mass range, but it appears as a straight line in log-log space.}
    \label{fig:power_law_imf}
\end{figure}

The same processes drive the fragmentation in a massive cloud complex of hundreds of parsecs and in a small protostellar core of 0.01 parsecs. Gravitational collapse competes with turbulent pressure, magnetic fields add support, and thermal pressure from heating and cooling affects the dynamics. The cloud does not ``know'' whether it is breaking into pieces of 1 parsec or 0.1 parsec. The process looks nearly the same at different scales, within the limits set by the interstellar medium. Figure~\ref{fig:power_law_imf} shows the Initial Mass Function.

Power laws also appear in many other astronomical settings. Galaxy luminosity functions, the distribution of cosmic structures, and the energy distributions of particles in astrophysical plasmas often follow power laws. This shows the scale-invariant nature of many astronomical processes, where similar mechanisms operate over many orders of magnitude.\index{power law distribution|)}

\section{Joint and Conditional Probability}\label{sec:ch2_joint_conditional}

Real astronomical problems involve several variables at once. Even with one parameter and one observable, we have a pair of related random variables that we must analyze together. So we need to understand how random variables relate to each other.

In practice, astronomical problems have many more dimensions. To characterize a stellar spectrum, we need several parameters for the physical properties of the star. These include its effective temperature, surface gravity, metallicity, rotational velocity, and possibly magnetic field strength. Our data are flux measurements at thousands of wavelengths, each with its own uncertainty. This creates a web of relationships between many variables that we must untangle statistically.

We therefore manipulate these variables in different ways. Sometimes we want the observations we expect for given model parameters, which is the forward problem. At other times we want to infer model parameters from observations, which is the inverse problem. To handle both, we need joint and conditional probabilities.

\paragraph{Joint Probability} For\index{joint probability|(} two random variables $x$ and $y$, the joint\index{joint probability} probability distribution $p(x,y)$ gives the probability that both variables take specific values at the same time. We look at both the discrete and the continuous case in astronomy.

For discrete variables, joint probability is simple. Consider classifying galaxies by morphology (spiral or elliptical) and size (large or small). Here $p(\text{spiral}, \text{large})$ is the probability of finding a galaxy that is both spiral and large. It measures how often this combination occurs in the galaxy population.

The concept extends to continuous variables. Consider stars described by temperature and luminosity. The joint distribution $p(T,L)$ gives the probability density of finding stars in the temperature-luminosity space, which produces the Hertzsprung-Russell\index{Hertzsprung-Russell diagram} diagram. Not all combinations are equally likely. Stars cluster along the main sequence, which follows from the physics of stellar evolution.

As for single continuous variables, the probability of any exact pair of values is zero. To get meaningful probabilities, we integrate over finite regions. For example, the probability of finding a star with temperature between $T_1$ and $T_2$ and luminosity between $L_1$ and $L_2$ is\index{joint probability|)}
\begin{equation}
P(T_1 \leq T \leq T_2, L_1 \leq L \leq L_2) = \int_{L_1}^{L_2} \int_{T_1}^{T_2} p(T,L) \, \mathrm{d}T \, \mathrm{d}L.
\end{equation}

\paragraph{Conditional Probability} Joint\index{conditional probability|(} probability describes two events that occur together. Conditional\index{conditional distribution|see{conditional probability; Gaussian distribution, conditional}}\index{conditional probability} probability asks a different question. Given that we know one event has occurred, what is the probability of another? We write this as $p(x|y)$, read as ``the probability of $x$ given $y$.''

\begin{figure}[ht!]
    \centering
    \includegraphics[width=0.9\textwidth]{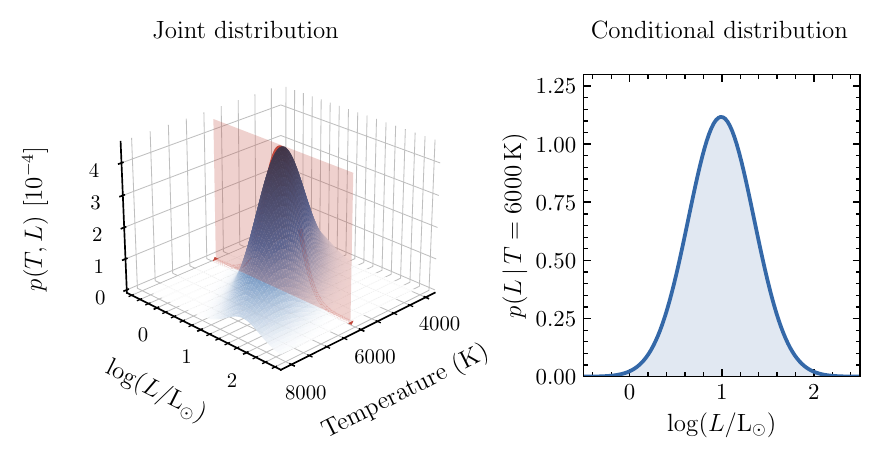}
    \caption{Joint and conditional probability distributions of stellar temperature and luminosity. The left panel shows the joint distribution $p(T,L)$, the probability density of finding stars with different combinations of temperature and luminosity. The red plane marks a slice through the joint distribution at a fixed temperature $T=6000$K. The right panel shows the conditional distribution $p(L|T=6000\text{K})$, the probability distribution of luminosities for stars at that temperature.}
    \label{fig:joint_conditional}
\end{figure}

Consider again the stars. The joint distribution $p(T,L)$ describes the overall distribution of stars in temperature-luminosity space. Now suppose we observe a star and measure its temperature precisely. The conditional probability $p(L|T)$ gives the distribution of possible luminosities for stars at that temperature. This ``slice'' through the joint distribution at fixed temperature usually covers a much narrower range of luminosities than the range for all stars. Figure~\ref{fig:joint_conditional} shows the joint distribution and such a slice.

Conditional probability is central to statistical inference. Two conditional probabilities describe the relationship between the forward and inverse problems.

\begin{itemize}
    \item In the forward problem, we know the model parameters and ask what observations to expect. This is $p(\text{data}|\text{model})$.

    \item In the inverse problem, we have the observations and ask what they tell us about the model parameters. This is $p(\text{model}|\text{data})$.
\end{itemize}\index{conditional probability|)}

\section{Probability Rules and Independence}\label{sec:ch2_probability_rules}

We now look at the rules that connect these probabilities. They are the basis of statistical inference and lead to one of the most important results in probability theory, Bayes' theorem.

\paragraph{The Sum Rule}

We\index{sum rule|(} often want information about a single variable from a joint distribution of several variables. Suppose we have the joint distribution $p(T,L)$ of stellar temperature and luminosity, but we only want the distribution of luminosities, whatever the temperature. How do we get it?

The answer is the marginal\index{marginal distribution} distribution and the sum\index{sum rule} rule. A marginal distribution describes one variable while we include all possible values of the other variables. The sum rule tells us how to compute it. Summing or integrating over the variables we do not need is called marginalization.\index{marginalization}

For discrete variables, the probability that $X$ takes a specific value $x$ is the sum over all possible values of $Y$
\begin{equation}
p(x) = \sum_y p(x,y).
\end{equation}

\begin{figure}[ht!]
    \centering
    \includegraphics[width=0.7\textwidth]{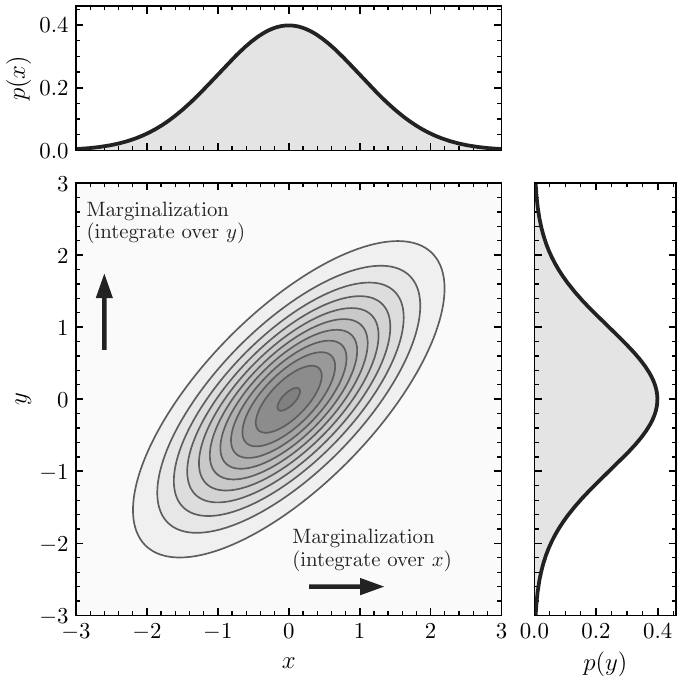}
    \caption{Marginalization for a bivariate Gaussian distribution. The central panel shows contours of a joint distribution $p(x,y)$ with positive correlation ($\rho=0.7$, the correlation coefficient of Section~3.5). The top and right panels show the marginal distributions $p(x)$ and $p(y)$, obtained by integrating the joint distribution over the other variable. The arrows show this process. Integrating over $y$ (vertical arrow) gives the marginal distribution of $x$ (top panel), and integrating over $x$ (horizontal arrow) gives the marginal distribution of $y$ (right panel). Marginalization reduces a multivariate distribution to its component distributions by including all possible values of the marginalized variables.}
    \label{fig:marginalization}
\end{figure}

For continuous variables, as in our stellar example, we integrate over all possible values of the other variable
\begin{equation}
p(L) = \int p(T,L) \, \mathrm{d}T.
\end{equation}Here $p(L)$ is the probability of finding a star with luminosity $L$, whatever its temperature. We can picture this as ``collapsing'' the two-dimensional joint distribution along the temperature axis. Figure~\ref{fig:marginalization} shows this for a bivariate Gaussian.

Marginalization is important for multivariate data. A study of galaxies may have a joint distribution of several properties, such as mass, star formation rate, and metallicity. The sum rule gives the distribution of any one property by integrating over the others. This helps when we want to compare our results with earlier studies that used different subsets of properties.\index{sum rule|)}

\paragraph{The Product Rule}

The\index{product rule|(} product\index{product rule} rule relates joint probabilities to conditional probabilities
\begin{equation}
p(x,y) = p(x|y)p(y) = p(y|x)p(x).
\end{equation}

To see this, consider our stellar example. We can split the joint probability $p(T,L)$ of finding a star with temperature $T$ and luminosity $L$ in two equivalent ways.

1. First take the probability $p(T)$ of a star with temperature $T$, then multiply by the probability $p(L|T)$ of luminosity $L$ at that temperature
\begin{equation}
p(T,L) = p(L|T)p(T).
\end{equation}

2. Or first take the probability $p(L)$ of a star with luminosity $L$, then multiply by the probability $p(T|L)$ of temperature $T$ at that luminosity
\begin{equation}
p(T,L) = p(T|L)p(L).
\end{equation}

\looseness=1 The equality of these two expressions, $p(L|T)p(T) = p(T|L)p(L)$, matters more than it first appears. Stellar evolution gives the probability of a luminosity for a given temperature, $p(L|T)$. But we may want to infer temperature from an observed luminosity, $p(T|L)$. The product rule shows that these viewpoints are connected. We can convert between them if we know the overall distributions of temperature and luminosity in the stellar population.

The conditional probabilities $p(L|T)$ and $p(T|L)$ are generally not equal. Knowing the probability of luminosity given temperature does not directly give the probability of temperature given luminosity. For example, a star with temperature 6000K might have an 80\% chance of a luminosity between 0.5 and 2 solar luminosities. But a star with 1 solar luminosity might have only a 30\% chance of a temperature between 5800 and 6200K.

This asymmetry is at the heart of scientific inference. Our theories predict observables from physical parameters (forward direction), but we observe phenomena and want their physical properties (inverse direction). The product rule and the sum rule together give the mathematical basis for connecting these two perspectives.\index{product rule|)}

\subsection{Independence}

Before\index{independence|(} we connect the product rule to statistical inference, we look at one more useful concept, independence, which we will meet often in this book. When we have many variables, it helps to know whether we can handle them separately without changing the conclusion of the inference.

Two random variables $X$ and $Y$ are independent\index{independence} if knowing the value of one gives no information about the other. In other words, observing $Y$ does not change our probability for $X$
\begin{equation}
    P(X|Y) = P(X) \quad \text{for all values of } Y.
\end{equation}This definition matches our intuition. If knowing $Y$ tells us nothing new about $X$, the probability of $X$ stays the same for any value of $Y$ that we observe.

\begin{figure}[ht!]
    \centering
    \includegraphics[width=0.9\textwidth]{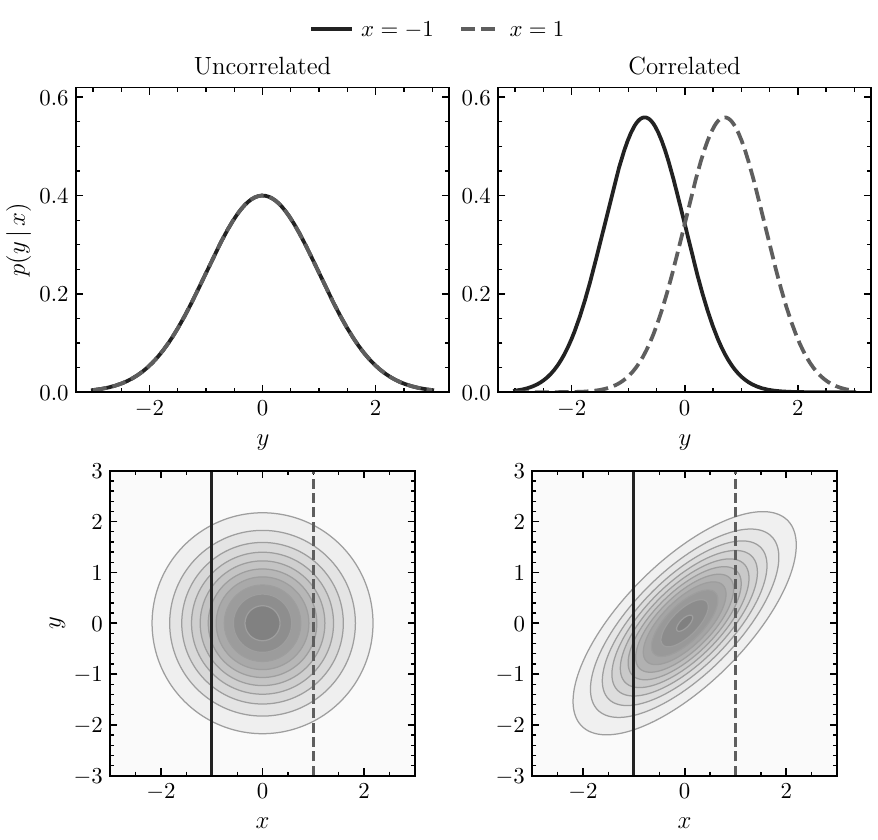}
    \caption{Dependent versus independent variables. Left panels show independent variables $x$ and $y$. The conditional distributions $p(y|x)$ stay the same whatever the value of $x$, so knowing $x$ gives no information about $y$. Right panels show dependent variables, where $p(y|x)$ changes with $x$, so knowing $x$ changes what we expect for $y$. Top panels show the conditional distributions for two values of $x$. Bottom panels show the corresponding joint distributions, with vertical lines marking where we took the conditional slices.}
    \label{fig:independence_dependence}
\end{figure}

From this definition and the product rule, we can derive another property of independent variables. Recall that the product rule gives
\begin{equation}
    P(X,Y) = P(X|Y)P(Y).
\end{equation}
If $X$ and $Y$ are independent, then $P(X|Y) = P(X)$, so
\begin{equation}
    P(X,Y) = P(X)P(Y).
\end{equation}So for independent variables, the joint probability factorizes into a product of individual probabilities. Many texts use this factorization as the definition of independence. Here it follows from the more intuitive definition through conditional probabilities. Figure~\ref{fig:independence_dependence} contrasts independent and dependent variables.

For continuous random variables, the same equations hold with probability densities in place of probabilities. Independence still means that knowing one variable tells us nothing about the other
\begin{equation}
    p(x|y) = p(x) \quad \text{leads to} \quad p(x,y) = p(x)p(y).
\end{equation}

Consider photons from two widely separated stars. The arrival time of a photon from one star is independent of the arrivals from the other. Detecting a photon from the first star tells us nothing about when we detect one from the second. Similarly, star counts in non-overlapping regions of the sky that are far enough apart are independent. So we can analyze different patches of the sky separately when we study stellar populations. In contrast, the temperature and luminosity of a star are not independent, because a star's temperature tells us much about its likely luminosity.

Independence simplifies analysis. When random variables are independent, we can analyze them separately. For example, for multiple independent measurements of the same quantity, the joint probability of all measurements is the product of the individual probabilities. This helps with large datasets, because we can break a complex problem into simpler pieces.

We must not assume independence without justification, because many astronomical properties are correlated in subtle ways. In galaxy surveys, for example, neighboring galaxies often have correlated properties because they share an environment and formation history. Knowing when we can assume independence is important for sound statistical analysis in astronomy.

The sum rule, the product rule, and independence are the tools we need for inference with astronomical data. The product rule leads directly to Bayes' theorem, which we derive below.\index{independence|)}

\section{Frequentist and Bayesian Approaches}\label{sec:ch2_frequentist_bayesian}

We can now compare two approaches to statistical inference, frequentist and Bayesian. The choice affects how we approach machine learning in astronomy, and astronomical problems often favor a Bayesian perspective.

\paragraph{The Frequentist Approach}

The\index{frequentist inference|(} frequentist\index{frequentist inference} approach, as its name suggests, defines probability through repeated experiments, that is, through frequency. Consider whether a die is fair. A frequentist rolls the die many times and plots the relative frequency of each outcome. For a fair die, these frequencies converge to 1/6 for each face as the number of rolls grows. This convergence is the law\index{law of large numbers} of large numbers. It is the basis of frequentist probability, because with enough trials the sample average approaches the true population parameter.

This approach is intuitive as well as mathematically sound. In many areas of science, especially laboratory physics, we can repeat experiments many times under controlled conditions. Each trial is independent, and our statistical uncertainties shrink in proportion to $1/\sqrt{N}$, where $N$ is the number of measurements. Section~3.6 derives this scaling. The Central Limit Theorem adds that the average of many measurements follows a nearly Gaussian distribution.

The confidence\index{confidence interval} interval is a good example of the frequentist approach. A 95\% confidence interval means that if we repeated our experiment many times, about 95\% of similarly constructed intervals would contain the true parameter value. This statement describes the long-run behavior of the procedure. It does not give our degree of belief in the parameter value itself.

In astronomy, we often cannot repeat an observation. For cosmological parameters such as the densities of dark matter and dark energy, we have only one observable universe. Many questions concern unique objects or events, such as the Milky Way's central black hole, a particular supernova, or the Solar System's formation. In these cases, the repeated experiment is only hypothetical, which makes the frequentist interpretation hard to apply.

Frequentist methods remain useful in astronomy, for example for large surveys with many similar objects and for random measurement errors. However, these limits explain why astronomers often need tools beyond frequentist methods, particularly for unique systems or complex systematic uncertainties. This leads us to the Bayesian approach.\index{frequentist inference|)}

\paragraph{The Bayesian Approach}

Bayesian\index{Bayesian inference}\index{Bayesian inference|(} inference asks what we can infer about the physical system from our current knowledge and observations. It also asks how we should update this inference when new data arrive. In the frequentist approach, parameters are fixed unknown numbers and only the data are random. The central insight of Thomas Bayes was to treat the observations and the model parameters as random variables on equal footing. Both play similar roles, like two sides of the same coin.

Treating model parameters as uncertain may seem unsettling. In physics and astronomy, we believe that true physical laws describe how the universe works. Examples are the equation of state of dark energy, the mass of the Milky Way's central black hole, and the mechanism that drives galaxy evolution. The Bayesian view does not deny that these truths exist. It says that our knowledge of them is uncertain, because our observations and theory are incomplete. With finite samples, as with our one observable universe, or with systematic uncertainties, we cannot be certain about our models.

Our ``belief'' in a theory is therefore itself a random variable. The probability distribution that represents our state of knowledge becomes more narrowly peaked as evidence accumulates, but some uncertainty remains. This matches how science progresses, by continued refinement of our understanding as new evidence narrows our uncertainty.

\paragraph{Mathematical Formalism} Bayes'\index{Bayes' theorem}\index{Bayes' theorem|(} theorem follows directly from the probability rules above. For two random variables $\mathbf{x}$ and $\mathbf{y}$, the product rule gives two equivalent ways to write their joint probability
\begin{equation}
p(\mathbf{x},\mathbf{y}) = p(\mathbf{x}|\mathbf{y})p(\mathbf{y}) = p(\mathbf{y}|\mathbf{x})p(\mathbf{x}).
\end{equation}This symmetry in the product rule leads directly to Bayes' theorem. Since both expressions equal the joint probability, we can write
\begin{equation}
p(\mathbf{x}|\mathbf{y})p(\mathbf{y}) = p(\mathbf{y}|\mathbf{x})p(\mathbf{x}).
\end{equation}
Dividing both sides by $p(\mathbf{y})$ gives
\begin{equation}
p(\mathbf{x}|\mathbf{y}) = \frac{p(\mathbf{y}|\mathbf{x})p(\mathbf{x})}{p(\mathbf{y})}.
\end{equation}

Because $\mathbf{x}$ and $\mathbf{y}$ are both random variables, neither plays a privileged role, and we can reverse the direction of conditioning. This is useful because it lets us infer model parameters from data. We apply Bayes' theorem to astronomical research by writing $\mathcal{D}$ for the observational data and $\boldsymbol{\theta}$ for the model parameters. Both can be complex. $\mathcal{D}$ might contain millions of data points from a large survey, and $\boldsymbol{\theta}$ might contain hundreds of parameters that describe detailed physical processes. Bayes' theorem then becomes
\begin{equation}
p(\boldsymbol{\theta} | \mathcal{D}) = \frac{p(\mathcal{D} | \boldsymbol{\theta}) p(\boldsymbol{\theta})}{p(\mathcal{D})}.
\end{equation}

\paragraph{Terms of Bayes' Theorem}

This equation may look like simple algebra, but each term has a specific meaning in scientific inference.

\begin{enumerate}
    \item $p(\boldsymbol{\theta} | \mathcal{D})$ is the \textbf{posterior distribution}.\index{posterior distribution} It represents our updated knowledge about the model parameters after seeing the data. It tells us which combinations of parameters are most likely to explain our observations. The posterior distribution gives a complete picture of our uncertainty about the parameters, in addition to the best-fit values.

    \item $p(\mathcal{D} | \boldsymbol{\theta})$ is the \textbf{likelihood function}.\index{likelihood} It measures how well our model explains the observed data. For a specific choice of model parameters, how likely are we to observe the data we measured? The likelihood function connects model predictions to reality. We read it as a function of $\boldsymbol{\theta}$ with the data held fixed. It need not integrate to one over $\boldsymbol{\theta}$.

    \item $p(\boldsymbol{\theta})$ is our \textbf{prior distribution}.\index{prior} It encodes what we know about the model parameters before we consider the data. This may include knowing that some parameter values are physically unreasonable, or that previous studies suggest a range. The prior lets us add existing scientific knowledge formally.

    \item $p(\mathcal{D})$ is the \textbf{evidence}.\index{evidence|see{marginal likelihood}} It is the probability of the data after we average over the parameter values, weighted by the prior. For continuous parameters, the sum rule and the product rule give $p(\mathcal{D}) = \int p(\mathcal{D} | \boldsymbol{\theta}) p(\boldsymbol{\theta}) \, \mathrm{d}\boldsymbol{\theta}$. This quantity is also called the marginal\index{marginal likelihood} likelihood. It matters for model comparison, and it acts as a normalization\index{normalization constant} constant when we estimate parameters.
\end{enumerate}

We return to the evidence $p(\mathcal{D})$ when we discuss information criteria and model selection in Section~11.8. Because it does not depend on $\boldsymbol{\theta}$, it is a constant in parameter estimation. In this book and in many practical applications, we therefore often use the proportional form
\begin{equation}
p(\boldsymbol{\theta} | \mathcal{D}) \propto p(\mathcal{D} | \boldsymbol{\theta}) p(\boldsymbol{\theta}).
\end{equation}

To see how Bayes' theorem guides scientific inference, imagine we propose a set of model parameters $\boldsymbol{\theta}$. We can ask how likely we are to observe our actual data $\mathcal{D}$ if these were the true parameter values. Models with high likelihood $p(\mathcal{D} | \boldsymbol{\theta})$ become more probable in the posterior distribution $p(\boldsymbol{\theta} | \mathcal{D})$. This mirrors how we evaluate scientific theories in general. For example, we accept that the Earth is round because this model better explains ships that disappear hull-first and the circular shadow in lunar eclipses.

The Bayesian approach has several advantages for astronomical inference. It gives the full posterior distribution of the parameters, in addition to point estimates. It adds prior knowledge from previous studies or physical constraints in a direct way. The same approach works for a single observation or a large dataset. We can compare models through the evidence term $p(\mathcal{D})$. It also allows hierarchical modeling, where parameters depend on higher-level parameters (Section~13.15).

These advantages suit astronomical data analysis, where we often have limited observations of unique phenomena and need to combine theoretical constraints with empirical evidence.\index{Bayes' theorem|)}

\section{Bayesian Inference in Practice}\label{sec:ch2_bayesian_practice}

Treating model parameters as random variables lets Bayesian inference handle limited data, even a single observation. We explore this with two examples that connect the theory to practice.

\subsection{Inference with One Data Point}

\paragraph{The Die Example} Consider four candidate models for the behavior of a die, which we call Fair, Bad, Worse, and Worst. Each model gives a specific probability for each outcome.

\begin{table}[ht!]
\centering
\begin{tabular}{|c|cccc|}
\hline
Outcome & Fair & Bad & Worse & Worst \\
\hline
6 & 1/6 & 1/8 & 1/12 & 1/20 \\
5 & 1/6 & 1/8 & 1/12 & 1/20 \\
4 & 1/6 & 1/6 & 1/8 & 1/10 \\
3 & 1/6 & 1/6 & 1/6 & 1/10 \\
2 & 1/6 & 5/24 & 5/24 & 1/5 \\
1 & 1/6 & 5/24 & 1/3 & 1/2 \\
\hline
\end{tabular}
\caption{Probability distributions for four die models. Each column is a hypothesis about the die, with Fair an unbiased die and each later model more biased toward lower numbers.}
\label{tab:die_models}
\end{table}

Under the Fair model, each outcome has equal probability ($p(\mathcal{D}=i|\boldsymbol{\theta}=\text{Fair})=1/6$ for all $i$). The Bad model is slightly biased toward lower numbers, and the Worse and Worst models are progressively more biased. For each model $\boldsymbol{\theta}$, the probabilities $p(\mathcal{D}|\boldsymbol{\theta})$ must sum to one, so each column of the table sums to one.

Now suppose we roll the die once and observe a 6. A frequentist approach can say little about a single roll, but Bayesian inference still gives a direct answer, a probability for each model. If we start with equal prior probabilities for each model ($p(\boldsymbol{\theta}=\text{Fair}) = p(\boldsymbol{\theta}=\text{Bad}) = p(\boldsymbol{\theta}=\text{Worse}) = p(\boldsymbol{\theta}=\text{Worst}) = 1/4$), Bayes' theorem gives
\begin{equation}
p(\boldsymbol{\theta}|\{6\}) \propto p(\{6\}|\boldsymbol{\theta})p(\boldsymbol{\theta}).
\end{equation}

Braces list the observed values, so $\{6\}$ is one roll that gave a 6. Since we have equal priors $p(\boldsymbol{\theta}) = \text{constant}$, the posterior $p(\boldsymbol{\theta}|\mathcal{D}=\{6\})$ is proportional to the likelihood $p(\mathcal{D}=\{6\}|\boldsymbol{\theta})$. We read this likelihood directly from the first row of the table, which gives the probability of rolling a 6 under each model. The evidence $p(\mathcal{D})$ does not change our relative beliefs about the models, but it normalizes the posterior. Each column of the table, which is $p(\mathcal{D}|\boldsymbol{\theta})$ for a fixed $\boldsymbol{\theta}$, sums to one, but the rows do not. To get proper posterior probabilities, we divide the likelihoods by their sum over all models.

For our single roll of 6, the likelihoods are
\begin{itemize}
    \item The Fair model gives 1/6 = 0.167
    \item The Bad model gives 1/8 = 0.125
    \item The Worse model gives 1/12 = 0.083
    \item The Worst model gives 1/20 = 0.050
\end{itemize}

The sum of these likelihoods is 0.425, so the posterior probabilities are
\begin{itemize}
    \item $p(\text{Fair}|\{6\}) = 0.167/0.425 = 0.393$
    \item $p(\text{Bad}|\{6\}) = 0.125/0.425 = 0.294$
    \item $p(\text{Worse}|\{6\}) = 0.083/0.425 = 0.195$
    \item $p(\text{Worst}|\{6\}) = 0.050/0.425 = 0.118$
\end{itemize}

Our beliefs have shifted. We started with equal probability (0.25) for each model. After observing a single 6, we favor the Fair model (0.393) and consider the Worst model least likely (0.118). This makes sense, because a 6 is most consistent with the Fair model and least consistent with the Worst model. The Worst model gives much more probability to lower numbers.

Now suppose we observe two more outcomes, 1 and 4, so the likelihood becomes $p(\{6,1,4\}|\boldsymbol{\theta})$. Die rolls are independent events, by the definition of independence we saw earlier, so this joint likelihood factorizes
\begin{equation}
p(\{6,1,4\}|\boldsymbol{\theta}) = p(\{6\}|\boldsymbol{\theta})p(\{1\}|\boldsymbol{\theta})p(\{4\}|\boldsymbol{\theta}).
\end{equation}

We calculate this product for each model
\begin{itemize}
    \item For the Fair model, $1/6 \times 1/6 \times 1/6 = 1/216 \approx 0.0046$
    \item For the Bad model, $1/8 \times 5/24 \times 1/6 \approx 0.0043$
    \item For the Worse model, $1/12 \times 1/3 \times 1/8 \approx 0.0035$
    \item For the Worst model, $1/20 \times 1/2 \times 1/10 = 1/400 = 0.0025$
\end{itemize}

After normalizing, the posterior probabilities are
\begin{itemize}
    \item $p(\text{Fair}|\{6,1,4\}) \approx 0.31$
    \item $p(\text{Bad}|\{6,1,4\}) \approx 0.29$
    \item $p(\text{Worse}|\{6,1,4\}) \approx 0.23$
    \item $p(\text{Worst}|\{6,1,4\}) \approx 0.17$
\end{itemize}

The results are now more balanced between the models. The Fair model still has the highest probability, but the differences between models are smaller. This makes sense, because we observed a mixture of high (6) and low (1) values, which is somewhat consistent with all the models.

This example shows a main strength of Bayesian inference. If we can specify the probability of the observations under each model, which is the likelihood, we can update our beliefs even with minimal data. The same approach works for a single observation or for thousands of data points.

\paragraph{The Gaussian Mean Example} The same idea extends to continuous parameters. Consider measuring an astronomical quantity that follows a Gaussian distribution with known standard deviation $\sigma$ and unknown mean $\mu$. In the frequentist approach, we would take many measurements and use their average as our best estimate. But what if we only have one or two measurements?

Here the model parameter is the unknown mean $\mu$. For simplicity, we fix $\sigma=1$ to get a Gaussian with unit variance. Suppose we make a single measurement $x = 6$. As in the die example, we apply Bayes' theorem
\begin{equation}
p(\mu|\mathcal{D}=\{6\}) \propto p(\mathcal{D}=\{6\}|\mu)p(\mu).
\end{equation}

We take a flat\index{flat prior|see{uninformative prior}}\index{uninformative prior} prior, so $p(\mu)$ is a constant that does not depend on $\mu$. Instead of a ``look-up'' table as for the die, the likelihood function $p(\mathcal{D}=\{6\}|\mu)$ is
\begin{equation}
p(x=6|\mu) = \frac{1}{\sqrt{2\pi\sigma^2}} \exp\left(-\frac{(6-\mu)^2}{2\sigma^2}\right).
\end{equation}If we ``propose'' a value for $\mu$, the probability density at the value 6 is the Gaussian PDF with this $\mu$ evaluated at $x=6$.

\begin{figure}[ht!]
    \centering
    \includegraphics[width=0.7\textwidth]{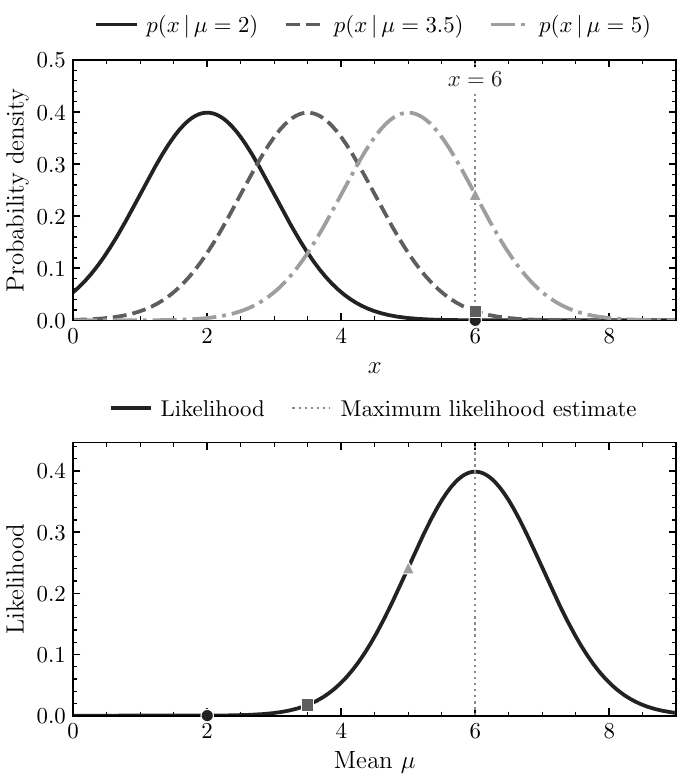}
    \caption{The top panel shows the probability density $p(x|\mu)$ of the measurement for three proposed values of $\mu$. The bottom panel shows the likelihood $p(\{6\}|\mu)$ as a function of $\mu$ for our single observation.}
    \label{fig:gaussian_one_point}
\end{figure}

In Figure~\ref{fig:gaussian_one_point}, different proposed values of $\mu$ lead to different probability densities at $x = 6$. When $\mu = 2$, the density at $x = 6$ is low, as the small value of the solid curve at $x=6$ shows. When $\mu = 5$, it is much higher, as the larger value of the dash-dot curve at $x=6$ shows.

The bottom panel shows the likelihood for each proposed value of $\mu$ given our single observation. With this flat prior, also called an uninformative prior, the likelihood is proportional to the posterior distribution. The maximum\index{maximum likelihood estimation (MLE)} likelihood estimate, which is the value of $\mu$ with the highest likelihood, is $\mu = 6$. This makes sense, because with a single measurement and no informative prior, our best estimate of the mean is the measured value.

We fixed $\sigma$ here on purpose. With a single data point, the maximum likelihood estimate of $\sigma$ tends toward zero. This gives an infinitely narrow Gaussian centered at 6, unless we assume a reasonable prior that prevents this degenerate solution.

The problem becomes more interesting with more measurements. Suppose we make a second measurement $x = 4$. We assume that the measurement errors are independent of each other, so we can multiply the individual likelihoods
\begin{equation}
p(x_1,x_2|\mu) = p(x_1|\mu)p(x_2|\mu).
\end{equation}

\begin{figure}[ht!]
    \centering
    \includegraphics[width=0.7\textwidth]{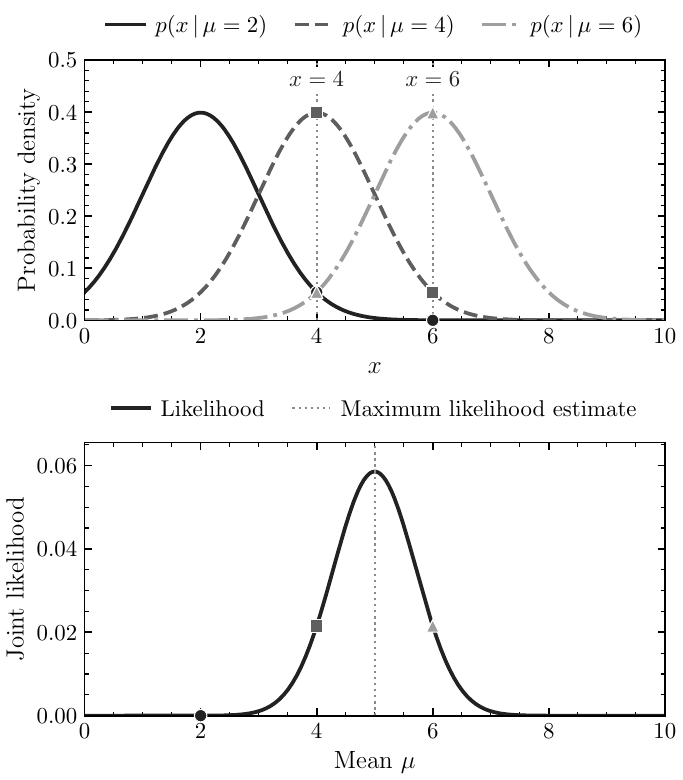}
    \caption{The top panel shows the densities $p(x|\mu)$ for three values of $\mu$, with vertical lines at our two observations $x=4$ and $x=6$. The bottom panel shows the joint likelihood as a function of $\mu$.}
    \label{fig:gaussian_two_points}
\end{figure}

In Figure~\ref{fig:gaussian_two_points}, the bottom panel shows the joint likelihood, which is the probability of observing both measurements for each proposed value of $\mu$. The maximum likelihood estimate has moved to $\mu = 5$, the average of our two measurements. Each new measurement multiplies in another likelihood factor, so the posterior uses all the available information and becomes more concentrated.

If we add a third measurement, say $x = 5$, the likelihood becomes
\begin{equation}
p(x_1,x_2,x_3|\mu) = p(x_1|\mu)p(x_2|\mu)p(x_3|\mu).
\end{equation}The maximum likelihood estimate is still $\mu = 5$, the average of our three measurements (4, 5, and 6). The likelihood curve becomes narrower still, which shows increased confidence in our estimate. Uncertainty decreases as we collect more data.

\paragraph{Implications for Scientific Inference}

Treating model parameters as random variables lets us work with small or large amounts of evidence, whether or not we can repeat the experiment.

In the die example, Bayesian inference let us update our beliefs about different models as we gathered evidence. Even a single data point let us quantify the relative plausibility of each model. In the Gaussian example, our uncertainty about the mean decreased as we gathered more measurements, because the likelihood function became more concentrated.

Bayesian inference can also use prior knowledge. We used uninformative priors in our examples, but we could use more informative priors from previous studies or theory. For instance, if previous surveys suggest that the Gaussian mean is close to 3, we might use a prior $p(\mu) = \mathcal{N}(\mu|3,1)$ (mean 3, variance 1). This pulls the posterior toward 3, especially when we have limited data. As we gather more data, the likelihood typically dominates the prior and reduces its influence on our conclusions.

Some see the need to specify priors as a weakness, but it can also be a strength. The balance between the prior $p(\boldsymbol{\theta})$ and the likelihood $p(\mathcal{D} | \boldsymbol{\theta})$ matters for extraordinary claims in science. As Carl Sagan noted, ``extraordinary claims require extraordinary evidence.'' In Bayesian terms, we need a correspondingly strong likelihood $p(\mathcal{D} | \boldsymbol{\theta})$ to overcome a strong skeptical prior, such as our prior against faster-than-light travel. This balance expresses both the open-mindedness and the skepticism that scientific work needs.

The die and Gaussian examples are deliberately simple, to show the core principles of Bayesian inference. The same principles apply to complex models with many parameters. There the posterior is a high-dimensional distribution over parameter space. The logic of updating our beliefs from observed data stays the same in the more sophisticated machine learning methods that follow.\index{Bayesian inference|)}

\section{Tutorials: Cluster Detection and Binary Orbits}\label{sec:ch2_tutorials}

Two tutorials apply the Poisson distribution, the likelihood, Bayes' theorem and marginalization to astronomy. The first searches for intermediate-mass black holes through their star clusters. The second reads the shape of a binary orbit from one snapshot. The full tutorials can be found at \url{https://statml.org/} as 2a and 2b.

\paragraph{Finding Star Clusters} Astronomers\index{black holes!intermediate-mass|(}\index{star clusters|(} search the Milky Way halo for intermediate-mass\index{black holes!intermediate-mass} black holes, which lie between stellar-mass and supermassive ones. Theory predicts that each one keeps a cluster\index{star clusters} of stars around it, an excess over the smooth halo background. A search for such excesses can find candidate black holes.

To find a cluster, we first need a model of the stars without one. Stars at random positions fall independently of each other (Section~\ref{sec:ch2_probability_rules}). The count in a small window therefore follows the Poisson distribution (Section~\ref{sec:ch2_common_distributions}), whose mean is the density times the area. This gives the likelihood of the count at each position under the background alone (Section~\ref{sec:ch2_bayesian_practice}).

\begin{figure}[ht!]
    \centering
    \includegraphics[width=0.95\textwidth]{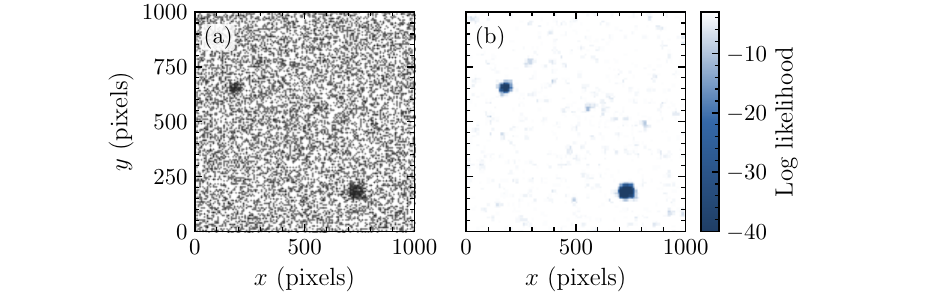}
    \caption{The first tutorial searches a simulated star field for clusters. Panel~(a) shows the stars, which include two clusters on a uniform background. Panel~(b) shows the logarithm of the likelihood of the star count in a small window around each position under the background alone. Dark blue marks counts far above the average for the background.}
    \label{fig:tutorial_clusters}
\end{figure}

With this model, a cluster is a place where the count is far above the average, so the likelihood of the count is very small. In a simulated field the two clusters stand out to the eye (Figure~\ref{fig:tutorial_clusters}a). The likelihood of the count at each position replaces this judgment with a number (Figure~\ref{fig:tutorial_clusters}b). A detection is therefore a statement about probability, and it depends on the background model. A search that finds no cluster also limits the number of such black holes, because each one would leave a very unlikely count.\index{black holes!intermediate-mass|)}\index{star clusters|)}

\begin{figure}[ht!]
    \centering
    \includegraphics[width=0.95\textwidth]{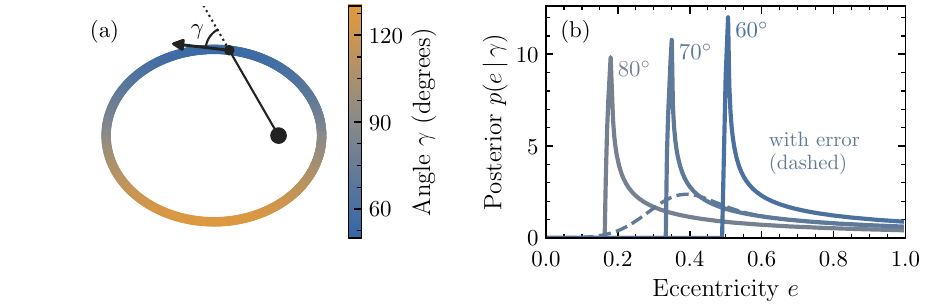}
    \caption{The second tutorial infers the shape of a binary orbit from one measured angle. Panel~(a) shows a face-on orbit colored by the angle $\gamma$. The dots mark the two stars, the line and arrow their relative position and velocity, and the arc the angle $\gamma$. Panel~(b) shows the posterior of the eccentricity $e$ for three observed angles, colored as in panel~(a), with a uniform prior. The dashed curve shows the $70^\circ$ case with a measurement error.}
    \label{fig:tutorial_binary}
\end{figure}

\paragraph{Reading a Binary Orbit from One Snapshot} Binary\index{binary stars}\index{binary stars|(} stars give us stellar masses through Kepler's Third Law (Section~\ref{sec:ch2_deterministic_random}), and their orbits record how they formed. Measuring an orbit needs observations over multiple orbital periods, which is too slow for large samples. A snapshot gives one angle $\gamma$, between the relative position and velocity of the two stars. Figure~\ref{fig:tutorial_binary}a colors an orbit by this angle. On a circular orbit it is $90^\circ$ at each point. On an eccentric orbit it changes, and the eccentricity $e$ measures how elongated it is.

Physics gives the angle from the eccentricity and the orbital phase. The phase tells where the stars are along the orbit, and we do not know it when we observe. We want the reverse, $e$ from the angle. Bayes' theorem (Section~\ref{sec:ch2_frequentist_bayesian}) does this, and marginalization (Section~\ref{sec:ch2_probability_rules}) removes the unknown phase by averaging over phases that we treat as equally likely. With a uniform prior (Section~\ref{sec:ch2_bayesian_practice}), the posterior of $e$ is proportional to the likelihood (Figure~\ref{fig:tutorial_binary}b).

The likelihood has a simple geometry. The larger the eccentricity, the farther the angle can swing from $90^\circ$ (Figure~\ref{fig:tutorial_binary}a). An angle far from $90^\circ$ therefore needs a large eccentricity, so one angle sets a lower limit on $e$. In Figure~\ref{fig:tutorial_binary}b each posterior starts at this limit, which is higher for angles farther from $90^\circ$, and has a long tail above it. The angle has a measurement error, and marginalizing over the true angle blurs the posterior (dashed).

Many binaries reveal the eccentricity distribution of a population, which depends on how the binaries formed and evolved. Physics predicts the data, an unknown such as the phase is marginalized, and Bayes' theorem turns the data into what we want to know.\index{binary stars|)}

\paragraph{Further Reading} Bayes (1763) wrote a posthumous essay on inverse probability. Jaynes (2003) treats probability theory as an extension of logic and links Bayesian methods to science. Box and Tiao (1973) treat Bayesian methods systematically. Gelman et~al.\ (2013) give worked examples. Sivia and Skilling (2006) give an introduction for physical scientists. Loredo (1992) argued for Bayesian methods in astronomy with examples, including supernova neutrinos. Gregory (2005b) compares frequentist and Bayesian approaches. Trotta (2008) reviews Bayesian inference and model selection in cosmology. Eadie et~al.\ (2023) discuss Bayesian methods in astronomy. Thrane and Talbot (2019) cover parameter estimation, model selection and hierarchical models in gravitational wave astronomy. Greene et~al.\ (2021) searched for clusters around wandering black holes (first tutorial). Hwang et~al.\ (2022) measured wide binary eccentricities from single observations (second tutorial).

\chapter{Statistical Foundations and Summary Statistics}

A probability distribution describes uncertainty, but we do not observe it directly. We see only finite, noisy samples drawn from it. When we study an astronomical population, we have a limited set of measurements, each with its own uncertainty. Because our observations are limited and uncertain, our models and inferences carry uncertainty too. How do we extract reliable information about the distribution from such data?

The methods in later chapters depend on this. Linear regression (Chapters 4 and 5) estimates means, variances, and covariances from samples to determine the best model parameters and to quantify their uncertainties. In the M-$\sigma$ relation\index{M-sigma relation@$M$--$\sigma$ relation} between black hole mass and stellar velocity dispersion, we need the best-fit line, its confidence, and predictions for new observations.

Bayesian inference (Chapter 5) needs complete probability distributions over the parameters. This requires methods that separate measurement noise, the irreducible randomness in the data, from model uncertainty, our incomplete knowledge of the parameters. Section~3.8 calls these aleatoric and epistemic uncertainty. Model comparison depends on this distinction too.

The solution is to use summary\index{summary statistics} statistics. Limited data rarely determine the full distribution, but summary statistics capture its main features, and we can estimate them from finite samples. Expectations give central tendencies, variances give scatter and measurement precision, and covariances give the relationships between observed quantities, which lets us build predictive models. We build these tools from the expectation operator and the moments of a distribution. We then study how they behave under transformations and how uncertainties combine when we average measurements, combine datasets, or propagate errors through calculations.

\section{Moments of Distributions}\label{sec:ch3_moments}

Moments\index{moments of a distribution}\index{moments of a distribution|(} give a systematic way to describe a probability distribution. They support the parameter estimation and uncertainty\index{uncertainty quantification} quantification that machine learning needs.

For a random variable $X$ with probability distribution $p(x)$, we define the $k$th moment as
\begin{align}
\text{Discrete case:} \quad \mu_k &= \sum_x x^k p(x), \\*
\text{Continuous case:} \quad \mu_k &= \int x^k p(x) \mathrm{d}x.
\end{align}

We often write these expressions with the expectation operator
\begin{equation}
\mu_k = \mathbb{E}[x^k],
\end{equation}
where $\mathbb{E}[\cdot]$ denotes the expectation, or average, of a quantity weighted by its probability distribution. We use this notation throughout the chapter.

For discrete variables, we sum over the possible values (finite or countably infinite), with $p(x)$ the probability mass function. For continuous variables, we integrate over the range of possible values, with $p(x)$ the probability density function. We need this distinction because a continuous variable has zero probability of taking any exact value, so we work with probability densities over intervals.

Each moment $\mu_k$ weights the distribution by a higher power of $x$, so each moment emphasizes different features of its shape.

\begin{figure}[ht!]
    \centering
    \includegraphics[width=0.7\textwidth]{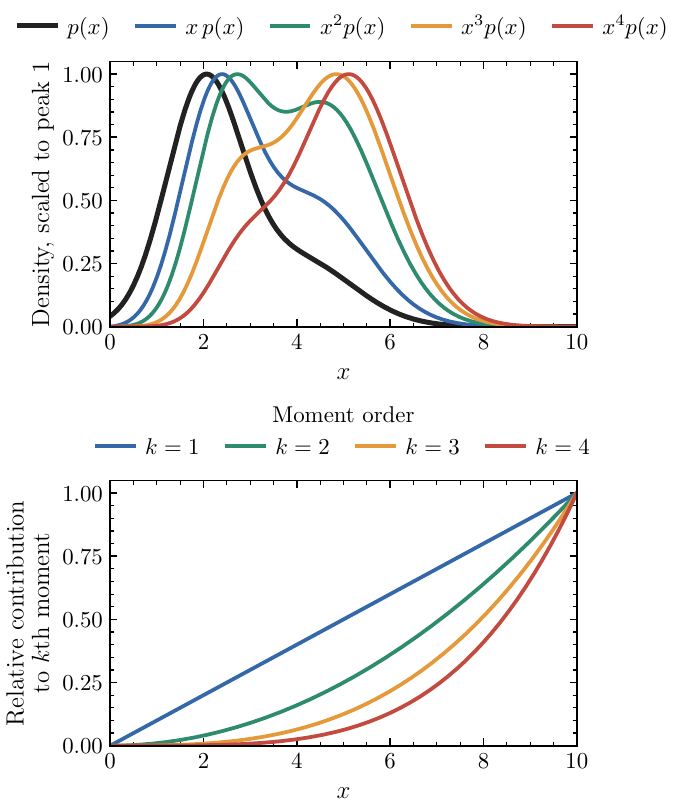}
    \caption{Moments weight a probability distribution differently. The top panel shows $p(x)$ in black and $x^k p(x)$ for several values of $k$ in color. As $k$ increases, the peak of $x^k p(x)$ moves to larger $x$, so higher moments are more sensitive to the tails. The bottom panel shows the relative contribution of $x^k$ to each moment. Powers with $k > 1$ give progressively more weight to large values. This matters in astronomy, where rare, extreme objects often carry important physical information.}
    \label{fig:moments_weights}
\end{figure}

The first moment ($k=1$) is the mean or expected value
\begin{equation}
\mu_1 = \mathbb{E}[x] = \int x p(x) \mathrm{d}x.
\end{equation}
It gives the center or typical value of the distribution, which is the basis for parameter estimation in linear regression. Recall that $p(x)$ is the probability density, which tells us how likely each value $x$ is. Multiplying $x$ by $p(x)$ and integrating gives a probability-weighted average of all possible values, which is why the first moment measures central tendency.

The second moment ($k=2$) measures the mean squared value
\begin{equation}
\mu_2 = \mathbb{E}[x^2] = \int x^2 p(x) \mathrm{d}x.
\end{equation}
Combined with the first moment, it gives the variance, the main quantity for uncertainty in machine learning. When the mean is zero, squaring makes positive and negative deviations both contribute positively, so the second moment measures spread. The spread of a distribution relates directly to the precision of our parameter estimates.

Higher moments ($k > 2$) are increasingly sensitive to the ``tails'' of the distribution, the regions far from the center where extreme values occur. As Figure~\ref{fig:moments_weights} shows, raising values to higher powers ($k$) amplifies large values much more than small ones. If a value is twice the typical scale, its contribution to the $k$th moment is $2^k$ times that of a typical value. With $k=3$ it contributes 8 times more, and with $k=4$ it contributes 16 times more. Because the amplification is exponential in $k$, even rare extreme values can dominate the higher moments. This makes higher moments useful for studying rare events in astronomy and for testing how sensitive our machine learning models are to outliers.

In the next sections, we study how moments transform, how they relate to marginal and conditional distributions, and how we estimate them from data.\index{moments of a distribution|)}

\section{Transformation of Random Variables}

Computing\index{transformation of random variables|(} the $k$th moment transforms\index{transformation of random variables} the random variable through $x \rightarrow x^k$. This is an example of the transformation of random variables, which machine learning uses throughout data preprocessing. Preprocessing means preparing the input variables of a model before fitting it. Machine learning calls these variables features.

As we saw in Chapter 2, $x$ is an outcome of the random variable $X$, not a fixed number. A function of a quantity that varies must also vary. So $x^k$ is an outcome of a new random variable $X^k$. A function of a random variable is itself a random variable with its own distribution. Preprocessing steps such as taking logarithms or ratios create new random variables whose statistics we need to understand. Standardizing a feature, which means subtracting its mean and dividing by its standard deviation, is another example.

In astronomy we often transform random variables. Examples are converting units (parsecs to light-years), applying calibration corrections, and computing derived quantities such as absolute magnitude from apparent magnitude and distance. To see how a transformation affects a distribution, consider a random variable $X$ with probability density $p_X(x)$ and a function $g$ that defines $Y = g(X)$. The new variable $Y$ has its own density $p_Y(y)$. Given $p_X(x)$ and $g$, how do we find $p_Y(y)$?

\begin{figure}[ht!]
    \centering
    \includegraphics[width=0.7\textwidth]{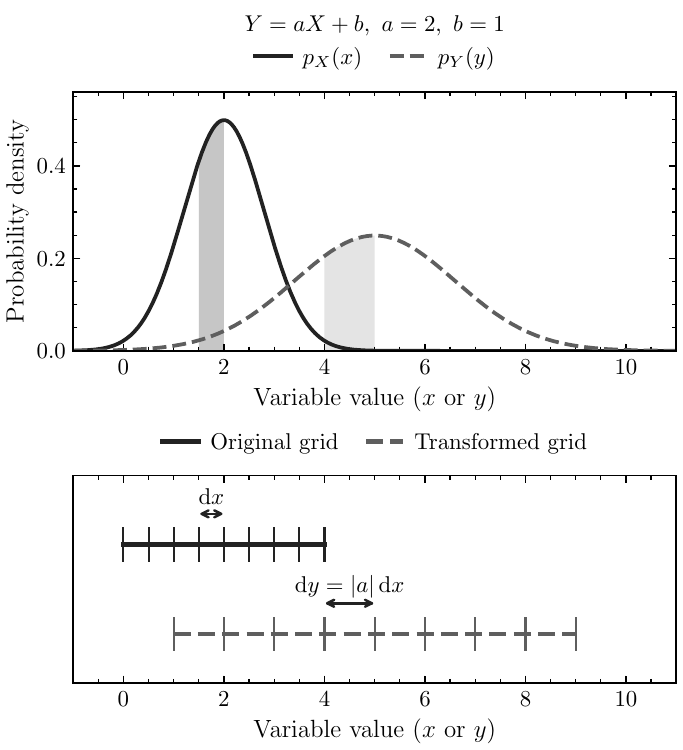}
    \caption{A probability distribution under the linear transformation $Y = 2X + 1$. The top panel shows the original distribution $p_X(x)$ as a solid curve and the transformed distribution $p_Y(y)$ as a dashed curve. The shaded regions have equal areas, so probability is conserved. The bottom panel shows how a uniform grid transforms. The grid spacing grows by a factor $|a|$, so the density must fall by a factor $1/|a|$ to keep the total probability fixed. This is the origin of the Jacobian factor $|\mathrm{d}x/\mathrm{d}y|$ in the transformation formula.}
    \label{fig:prob_transform}
\end{figure}

We start with the simplest case, a linear transformation of the form
\begin{equation}
Y = aX + b,
\end{equation}
where $a$ and $b$ are constants. The probability densities are then related by
\begin{equation}
p_Y(y) = p_X\left(\frac{y-b}{a}\right)\frac{1}{|a|}.
\end{equation}

The factor $1/|a|$ comes from probability conservation. A probability density gives the chance of an outcome per ``unit'' of the outcome space. As Figure~\ref{fig:prob_transform} illustrates, stretching the distribution by a factor $a$ (here $a=2$) changes what one unit means. Each unit is now $a$ times larger, so the density per unit must fall by $1/|a|$ for the total probability to stay equal to one. This factor, which measures how the ``unit'' changes size, is the Jacobian.\index{Jacobian} It appears whenever we transform between variables, for example when we standardize or rescale features.

For a general transformation $Y = g(X)$, where $g$ is differentiable and monotonic (strictly increasing or strictly decreasing), the density transforms as
\begin{equation}
p_Y(y) = p_X(x)\left|\frac{\mathrm{d}x}{\mathrm{d}y}\right| = p_X(g^{-1}(y))\left|\frac{\mathrm{d}}{\mathrm{d}y}g^{-1}(y)\right|.
\end{equation}The term $|\mathrm{d}x/\mathrm{d}y|$ is the Jacobian of the transformation. It measures how the transformation stretches or compresses intervals, and it keeps the total probability equal to one. The absolute value keeps the density non-negative when $g$ is decreasing, because the transformation then reverses the orientation. Probability conservation holds for each small interval, because matching intervals $\mathrm{d}x$ and $\mathrm{d}y$ contain the same probability, $p_X(x)|\mathrm{d}x| = p_Y(y)|\mathrm{d}y|$. Dividing by $|\mathrm{d}y|$ turns the ratio of the two interval lengths into the derivative $|\mathrm{d}x/\mathrm{d}y|$.

For a multivariate transformation, the Jacobian becomes a matrix determinant. For a vector of random variables $\mathbf{X} = (X_1, ..., X_D)^T$ that transforms to $\mathbf{Y} = \mathbf{g}(\mathbf{X})$, we have
\begin{equation}
p_Y(\mathbf{y}) = p_X(\mathbf{g}^{-1}(\mathbf{y}))\left|\det\left(\frac{\partial\mathbf{g}^{-1}}{\partial\mathbf{y}}\right)\right|.
\end{equation}The matrix of partial derivatives $\partial\mathbf{g}^{-1}/\partial\mathbf{y}$ has element $(i,j)$ equal to $\partial(\mathbf{g}^{-1})_i/\partial y_j$.

The multivariate case applies when we change coordinate systems (such as equatorial to galactic coordinates) or compute derived quantities from several observables. The choice of coordinates can change how linear a relationship looks, and so it can affect the performance of linear regression.

Another class of transformations combines several random variables, for example in a sum. The sum $Z = X + Y$ of two random variables is a new random variable. Its distribution depends on the distributions of $X$ and $Y$ and on how they relate to each other. This matters in linear regression, where we model the output we want to predict as a linear combination of the inputs. Next, we look at how transformations affect the moments, starting with the expectation.\index{transformation of random variables|)}

\section{Expectation}

In\index{expectation|(} the discussion of moments, we introduced the first moment as the mean or expected value of a random variable. This concept is so important that it deserves its own treatment. We now call it the expectation\index{expectation} and extend it to functions of $X$. It involves a weighting, in which each outcome is weighted by its probability, and, for higher moments, a transformation of the variable ($X \rightarrow X^k$).

We use probability-weighted averages throughout machine learning, from parameter estimation to loss function optimization (Chapter~4), so we give them their own notation and name. We write the expectation of a random variable $X$ as $\mathbb{E}[X]$, defined as
\begin{align}
\text{Discrete case:} \quad \mathbb{E}[X] &= \sum_x x \, p_X(x), \\*
\text{Continuous case:} \quad \mathbb{E}[X] &= \int x \, p_X(x) \, \mathrm{d}x.
\end{align}The argument inside $\mathbb{E}[\cdot]$ is the random variable whose expectation we compute. In machine learning we need to distinguish between different averages. In linear regression, for example, we meet the expectation over data points, over the random noise in repeated measurements, and over parameter distributions. Each plays a different role in fitting and uncertainty quantification.

\looseness=1 The name ``expectation'' suggests what we expect to measure, but this can mislead. Consider the Initial Mass Function (IMF) of stars. $\mathbb{E}[M]$ gives the average stellar mass, but most stars have masses below this value. Stellar masses roughly follow a power law over three orders of magnitude, from the hydrogen-burning limit at 0.08 solar masses to above 100 solar masses. For a Salpeter\index{initial mass function (IMF)!Salpeter} IMF ($p(M) \propto M^{-2.35}$), the density of 0.5 solar mass stars is about 50,000 times that of 50 solar mass stars. The few massive stars still pull the average up, because their large values contribute substantially even though they are rare.

We now see how expectation behaves under the transformations of random variables from the previous section. These properties show how preprocessing steps affect our machine learning models and how to propagate uncertainties through calculations.

For a transformation $Y = g(X)$, we can compute the expectation in two equivalent ways.
\begin{enumerate}
\item Using the original variable $X$ and its distribution
\begin{equation}
\mathbb{E}[g(X)] = \int g(x) \, p_X(x) \, \mathrm{d}x.
\end{equation}

\item Using the transformed variable $Y$ and its distribution
\begin{equation}
\mathbb{E}[Y] = \int y \, p_Y(y) \, \mathrm{d}y = \int g(x) \, p_X(x) \, \mathrm{d}x.
\end{equation}
\end{enumerate}

These expressions are equivalent when we account for the transformation of densities with $p_Y(y) = p_X(g^{-1}(y))|\mathrm{d}g^{-1}/\mathrm{d}y|$. This equivalence is useful in machine learning, where we may transform features before fitting a model. We can compute expected values with either the original or the transformed representation, as long as we handle the probability transformation correctly.

This equivalence is why we can write our definition of moments either way
\begin{equation}
\mu_k = \mathbb{E}[X^k] = \mathbb{E}[Y] \text{ where } Y = X^k.
\end{equation}

\paragraph{Linearity and Additivity}

We now look at the main properties of expectation. For any constants $a$ and $b$, expectation is linear
\begin{equation}
\mathbb{E}[aX + b] = a\mathbb{E}[X] + b.
\end{equation}

\begin{figure}[ht!]
\centering
\includegraphics[width=0.9\textwidth]{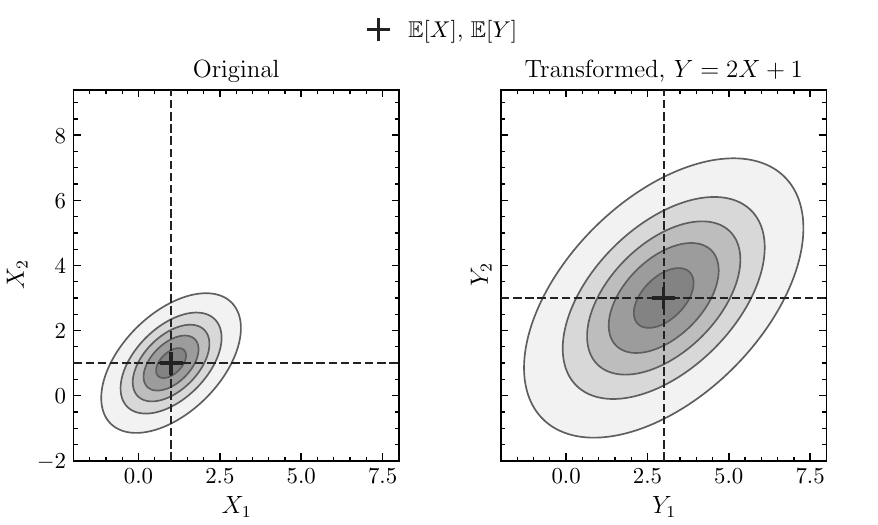}
    \caption{Linearity of expectation for a bivariate Gaussian distribution. The left panel shows the original distribution, with its expectation $\mathbb{E}[X]$ marked by a cross. The right panel shows the distribution under $Y = 2X + 1$. Contours mark levels of probability density. The distribution stretches and shifts, and the expectation follows the same rule, $\mathbb{E}[Y] = 2\mathbb{E}[X] + 1$ (dashed lines). This holds whatever the shape or spread of the distribution.}
    \label{fig:expectation_transform}
\end{figure}

Multiplying a random variable by $a$ stretches or compresses its values by that factor, so the center stretches by $a$ too. Adding $b$ shifts the values by that amount, so the center shifts by $b$, as Figure~\ref{fig:expectation_transform} shows. This holds whatever the shape of the distribution, because expectation represents the center of mass of the distribution. In machine learning, linearity tells us how data preprocessing affects model parameters. For example, when we standardize features by subtracting the mean and dividing by the standard deviation, linearity shows that the standardized features have zero mean.

Formally, we prove this with the definition of expectation
\begin{align}
\mathbb{E}[aX + b] &= \int (ax + b)p(x)\mathrm{d}x \\*
&= a\int xp(x)\mathrm{d}x + b\int p(x)\mathrm{d}x \\*
&= a\mathbb{E}[X] + b,
\end{align}
where we used $\int p(x)\mathrm{d}x = 1$ for any probability distribution.

\begin{figure}[ht!]
    \centering
    \includegraphics[width=0.9\textwidth]{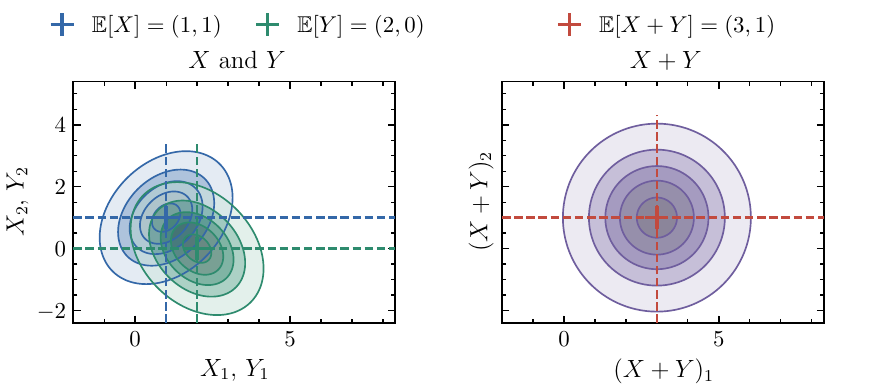}
    \caption{Additivity of expectation for bivariate Gaussian distributions. The left panel shows the distributions of $X$ (blue contours) and $Y$ (green contours). Colored crosses mark their expectations $\mathbb{E}[X]$ and $\mathbb{E}[Y]$, and dashed lines mark the coordinates of the expectations. The right panel shows the distribution of the sum $Z = X + Y$ (purple contours), with $\mathbb{E}[Z]$ marked by a red cross. The expectations add component-wise, $\mathbb{E}[X + Y] = \mathbb{E}[X] + \mathbb{E}[Y]$, whatever the correlation structure of the original distributions.}
    \label{fig:expectation_additivity}
\end{figure}

A subscript names the variables we average over, so $\mathbb{E}_{X,Y}$ averages over $X$ and $Y$ together and $\mathbb{E}_X$ over $X$ alone (Section~3.4). For any two random variables $X$ and $Y$, expectation is additive
\begin{equation}
\mathbb{E}_{X,Y}[X + Y] = \mathbb{E}_X[X] + \mathbb{E}_Y[Y].
\end{equation}The expectation of a sum is the sum of the expectations, whether or not $X$ and $Y$ are independent (Figure~\ref{fig:expectation_additivity}). Linear regression relies on this, because it models the output as a sum of contributions from different inputs. It follows from
\begin{align}
\mathbb{E}_{X,Y}[X + Y] &= \iint (x + y)p(x,y)\mathrm{d}x\mathrm{d}y \\*
&= \iint xp(x,y)\mathrm{d}x\mathrm{d}y + \iint yp(x,y)\mathrm{d}x\mathrm{d}y \\
&= \int x\left(\int p(x,y)\mathrm{d}y\right)\mathrm{d}x + \int y\left(\int p(x,y)\mathrm{d}x\right)\mathrm{d}y \\
&= \int xp(x)\mathrm{d}x + \int yp(y)\mathrm{d}y \\*
&= \mathbb{E}_X[X] + \mathbb{E}_Y[Y],
\end{align}
where we used that integrating the joint distribution $p(x,y)$ over one variable gives the marginal distribution of the other. For example, if we measure the total light from two stars, the average total luminosity is the sum of their average luminosities.

We now extend these ideas to the multivariate setting of most machine learning applications. We represent several random variables as a random vector $\mathbf{X} = (X_1, \ldots, X_D)^T$. Its expectation $\mathbb{E}[\mathbf{X}]$ is the vector of the expectations of its components
\begin{equation}
\mathbb{E}[\mathbf{X}] \equiv \begin{pmatrix} \mathbb{E}[X_1] \\ \vdots \\ \mathbb{E}[X_D] \end{pmatrix}.
\end{equation}

When we apply a linear transformation with an $m \times D$ matrix $\mathbf{A}$ to $\mathbf{X}$, we get a new random vector $\mathbf{Y} = \mathbf{A}\mathbf{X} + \mathbf{b}$ of dimension $m$. Expanding the matrix product and using the linearity of expectation gives
\begin{align}
Y_i &= \sum_{j=1}^D A_{ij}X_j + b_i, \\*
\mathbb{E}[Y_i] &= \sum_{j=1}^D A_{ij}\mathbb{E}[X_j] + b_i.
\end{align}Hence, in the multivariate case
\begin{equation}
\mathbb{E}[\mathbf{Y}] = \mathbf{A}\mathbb{E}[\mathbf{X}] + \mathbf{b},
\end{equation}
where $\mathbf{A}$ is an $m \times D$ transformation matrix and $\mathbf{b}$ is an $m$-dimensional offset vector.

Linearity holds for general linear transformations, in which matrix multiplication mixes the components of $\mathbf{X}$. Such transformations rotate the coordinate system or create new variables that are linear combinations of the old ones. They are at the heart of linear regression and principal component analysis (Chapter\index{expectation|)}~10).

\section{Law of Total Expectation}

Next,\index{law of total expectation|(} we ask how expectation relates to the marginal and conditional distributions of Chapter 2. This applies when we have only partial information or a hierarchical structure, where the distribution of one variable depends on another.

\looseness=1 Suppose we want to estimate a star's mass, but we can only measure its luminosity directly. We then combine the luminosity measurement with theory about the mass distribution of stars with a given luminosity. This leads to the law\index{law of total expectation} of total expectation.

To state the law, we must say which variables we average over. For two random variables $X$ and $Y$, we use subscripts for this
\begin{align}
\mathbb{E}_{X,Y}[g(X,Y)] &= \iint g(x,y) \, p(x,y) \, \mathrm{d}x \, \mathrm{d}y, \\*
\mathbb{E}_X[g(X,Y)\,|\,Y=y] &= \int g(x,y) \, p(x|y) \, \mathrm{d}x,
\end{align}
where the second equation averages only over $X$ and treats $Y$ as fixed. It uses the conditional probability $p(x|y)$, the distribution of $X$ for a specific value of $Y$.

This average is the conditional\index{conditional expectation} expectation of $g(X,Y)$ given $Y=y$. It changes with $y$, so with $Y$ left unfixed, $\mathbb{E}_X[g(X,Y)|Y]$ is itself a random variable.

The law of total expectation says that the overall expectation of $X$ equals the average, over $Y$, of its conditional expectations
\begin{equation}
\mathbb{E}_X[X] = \mathbb{E}_Y[\mathbb{E}_X[X|Y]].
\end{equation}

To prove this, we write out the right-hand side
\begin{align}
\mathbb{E}_Y[\mathbb{E}_X[X|Y]] &= \int \mathbb{E}_X[X|Y=y] \, p(y) \, \mathrm{d}y \\*
&= \int \left(\int x \, p(x|y) \, \mathrm{d}x\right) p(y) \, \mathrm{d}y \\
&= \int\int x \, p(x|y) \, p(y) \, \mathrm{d}x \, \mathrm{d}y \\
&= \int\int x \, p(x,y) \, \mathrm{d}x \, \mathrm{d}y \\
&= \int x \left(\int p(x,y) \, \mathrm{d}y\right) \mathrm{d}x \\
&= \int x \, p(x) \, \mathrm{d}x \\*
&= \mathbb{E}_X[X],
\end{align}
where we used the product rule $p(x,y) = p(x|y)p(y)$ and the marginal distribution $p(x) = \int p(x,y) \, \mathrm{d}y$.

\begin{figure}[ht!]
    \centering
    \includegraphics[width=0.9\textwidth]{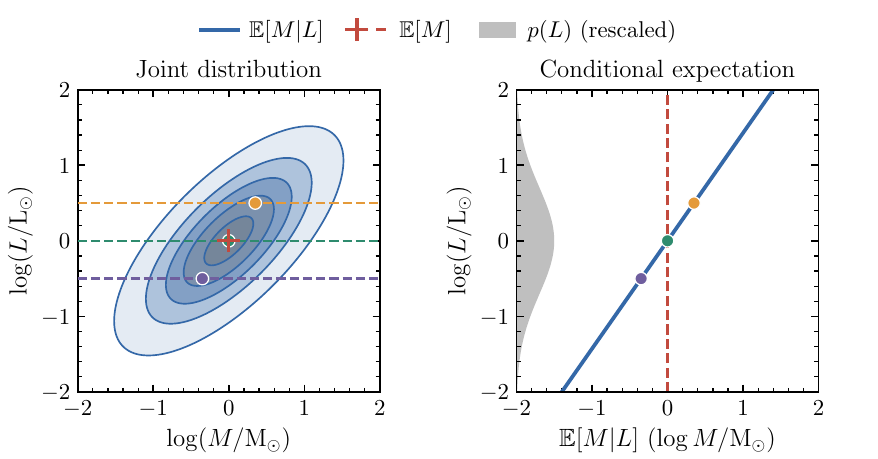}
    \caption{The law of total expectation for the stellar mass-luminosity relation. The left panel shows the joint distribution of mass and luminosity (solar units, log scale) as blue contours. The red cross marks the overall expectation $\mathbb{E}[M]$, and colored circles mark the conditional expectations $\mathbb{E}[M|L]$ at three luminosities (dashed lines). The right panel shows $\mathbb{E}[M|L]$ as a function of luminosity (blue line), which lets us estimate stellar masses from luminosities. The gray curve is the marginal distribution $p(L)$. Weighting $\mathbb{E}[M|L]$ by $p(L)$ recovers the overall expectation $\mathbb{E}[M]$ (red dashed line), even when direct mass measurements are difficult.}
    \label{fig:conditional_expectation}
\end{figure}

The law of total expectation helps when direct measurements are hard or biased. Consider the problem of estimating the average stellar mass, shown in Figure~\ref{fig:conditional_expectation}. The most direct approach averages all stellar masses
\begin{equation}
\mathbb{E}_M[M] = \int M \, p(M) \, \mathrm{d}M.
\end{equation}

However, we rarely know the true distribution $p(M)$ in advance, and observational effects such as selection bias make it hard to estimate. In a magnitude-limited survey, for example, we preferentially observe brighter, typically more massive stars. Measuring $\mathbb{E}_M[M]$ directly requires an accurate sample of the full mass distribution, which observational limits often make impossible.

A better approach uses conditional expectations. We first find the average mass of stars at a given luminosity, then average over all luminosities
\begin{equation}
\mathbb{E}_M[M] = \mathbb{E}_L[\mathbb{E}_M[M|L]] = \int \left(\int M \, p(M|L) \, \mathrm{d}M\right) p(L) \, \mathrm{d}L.
\end{equation}

As the figure shows, this hierarchical decomposition has practical advantages. For each luminosity (the horizontal dashed lines in the left panel), we compute the conditional expectation $\mathbb{E}_M[M|L]$ (the colored circles). The blue contours show the correlation between mass and luminosity. The blue line in the right panel traces how the conditional expectation varies with luminosity. This line is a regression function (Chapter~4) that predicts mass from luminosity.

We then weight these expectations by the luminosity distribution $p(L)$ (the gray curve in the right panel). We can measure $p(L)$ directly in our magnitude-limited survey.

This approach handles observational constraints well. We can model the luminosity distribution $p(L)$ with the selection\index{selection effects} effects of the survey included, because luminosity relates directly to the survey's limits. Horizontal slices through the contours in the left panel show the mass distribution at fixed luminosity, $p(M|L)$. Stellar evolution theory describes it well, and it typically has much smaller observational biases.

As in the univariate case, the law of total expectation extends to random vectors $\mathbf{X}$ and $\mathbf{Y}$
\begin{equation}
\mathbb{E}_{\mathbf{X}}[\mathbf{X}] = \mathbb{E}_{\mathbf{Y}}[\mathbb{E}_{\mathbf{X}}[\mathbf{X}|\mathbf{Y}]].
\end{equation}For each component of the random vector, we find the conditional expectation given $\mathbf{Y}$ and then take the expectation over all values of $\mathbf{Y}$. The proof follows the univariate steps with extra integration variables, exchanging the order of the multiple integrals over the components of $\mathbf{X}$ and $\mathbf{Y}$.

This principle underlies the posterior predictive distribution, the distribution of a new observation given the data, which Section~5.7 develops. In Bayesian analysis, model parameters such as regression coefficients are random variables with uncertainty, and predictions must include that uncertainty. Let $\theta$ be the model parameters and $y_*$ a new observation to predict. The law of total expectation gives
\begin{equation}
\mathbb{E}[y_*] = \mathbb{E}_{\theta}[\mathbb{E}[y_*|\theta]].
\end{equation}We compute the expected prediction for each parameter value $\theta$ (the inner expectation), then average over the posterior distribution of $\theta$ (the outer expectation). This process, called marginalization, includes the parameter uncertainty in our predictions.\index{law of total expectation|)}

\section{Variance, Covariance, and Correlation}\label{sec:ch3_variance_covariance}

The\index{variance|(} first moment $\mathbb{E}[X]$ gives the center of a distribution, but the center alone is not enough. We also need to quantify uncertainty and the relationships between variables. The second moment $\mathbb{E}[X^2]$ seems a reasonable measure of spread, because it weights values far from zero more. But $\mathbb{E}[X^2]$ measures spread only when the distribution is centered at zero. Otherwise it mixes the spread with the location of the center, which makes it unsuitable for uncertainty quantification in most practical cases.

To isolate the spread, we measure deviations from the center $\mathbb{E}[X]$, not from zero. This defines the variance\index{variance} of a random variable $X$, written $\text{Var}[X]$ or $\sigma^2_X$
\begin{equation}
\text{Var}[X] \equiv \mathbb{E}[(X - \mathbb{E}[X])^2].
\end{equation}We first center the distribution by subtracting $\mathbb{E}[X]$, then average the squared deviation from this center. The variance is the main quantity for prediction uncertainty and for the reliability of parameter estimates. It is the second central moment, meaning a moment about the mean, whereas the moments $\mu_k$ of Section~3.1 are taken about zero.

We can expand the variance into a more convenient form using the properties of expectation. While $X$ is a random variable, $\mathbb{E}[X]$ is a constant (just a number), so linearity applies
\begin{align}
\text{Var}[X] &= \mathbb{E}[(X - \mathbb{E}[X])^2] \\*
&= \mathbb{E}[X^2 - 2X\mathbb{E}[X] + (\mathbb{E}[X])^2] \\
&= \mathbb{E}[X^2] - 2\mathbb{E}[X]\mathbb{E}[X] + \mathbb{E}[(\mathbb{E}[X])^2] \\
&= \mathbb{E}[X^2] - 2(\mathbb{E}[X])^2 + (\mathbb{E}[X])^2 \\*
&= \mathbb{E}[X^2] - (\mathbb{E}[X])^2.
\end{align}So we can compute the variance from the first and second moments, without computing deviations from the mean directly.

The standard\index{standard deviation} deviation $\sigma_X$ is the square root of the variance
\begin{equation}
\sigma_X = \sqrt{\text{Var}[X]}.
\end{equation}

\looseness=-1 We use the standard deviation because it has the same units as the original variable. If we measure stellar masses in solar masses ($\mathrm{M}_\odot$), the variance is in $\mathrm{M}_\odot^2$ and the standard deviation is in $\mathrm{M}_\odot$. This makes the standard deviation easier to interpret. If the mean stellar mass is $1\mathrm{M}_\odot$ with a standard deviation of $0.5\mathrm{M}_\odot$, we immediately see the scale of the variations.

\begin{figure}[ht!]
    \centering
    \includegraphics[width=0.9\textwidth]{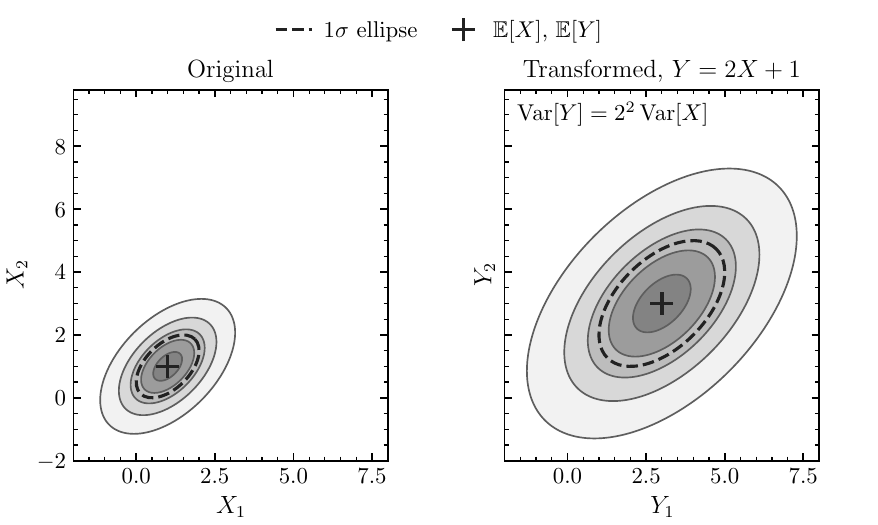}
    \caption{Variance and standard deviation under the linear transformation $Y = 2X + 1$. The left panel shows a bivariate distribution (contours) with its expectation (cross) and its $1\sigma$ ellipse (dashed). The ellipse is the contour where the density has fallen to $e^{-1/2}$ of its peak, and its shape and orientation follow the covariance matrix. The right panel shows the transformed distribution. The mean shifts as $\mathbb{E}[Y] = 2\mathbb{E}[X] + 1$. The standard deviation scales linearly, $\sigma_Y = 2\sigma_X$, and the variance scales quadratically, $\text{Var}[Y] = 4\text{Var}[X]$, so the dashed ellipse is larger.}
    \label{fig:variance_transform}
\end{figure}

Next, we look at how variance behaves under transformations, starting with a linear transformation that scales and shifts a random variable. Unlike expectation, which is linear, variance scales differently (Figure~\ref{fig:variance_transform}). For any constants $a$ and $b$
\begin{equation}
\text{Var}[aX + b] = a^2\text{Var}[X].
\end{equation}

We prove this with the definition of variance and the properties of expectation
\begin{align}
\text{Var}[aX + b] &= \mathbb{E}[(aX + b - \mathbb{E}[aX + b])^2] \\*
&= \mathbb{E}[(aX + b - (a\mathbb{E}[X] + b))^2] \quad \text{(using linearity of expectation)} \\
&= \mathbb{E}[(aX - a\mathbb{E}[X])^2] \\
&= \mathbb{E}[a^2(X - \mathbb{E}[X])^2] \\
&= a^2\mathbb{E}[(X - \mathbb{E}[X])^2] \\*
&= a^2\text{Var}[X].
\end{align}

Multiplying a random variable by $a$ stretches or compresses the deviations from the mean by $a$. Variance measures squared deviations, so it scales by $a^2$. Adding a constant $b$ shifts the distribution without changing its spread, so it does not change the variance. Scaling variables therefore changes their variance, which affects both interpretation and computation.

\paragraph{Covariance and Correlation} Variance\index{covariance|(} describes the spread of a single variable. Covariance\index{covariance} and correlation describe how two variables vary together.

The covariance of two random variables $X$ and $Y$, written $\text{Cov}(X,Y)$, measures their joint variability
\begin{equation}
\text{Cov}(X,Y) = \mathbb{E}[(X - \mathbb{E}[X])(Y - \mathbb{E}[Y])].
\end{equation}

Covariance measures whether two variables tend to deviate from their means together. When both are above or below their means at the same time, their deviations multiply to positive values, and the covariance is positive. When one is above its mean while the other is below, the products are negative, and so is the covariance. With no linear relationship, positive and negative products cancel, and the covariance is zero.

\looseness=1 In astronomical measurements, covariance has two distinct sources. It can come from a physical relationship. For example, more massive stars tend to be more luminous than average, which gives a positive covariance between mass and luminosity. It can also come from the measurement process. Systematic instrument errors that affect several measurements at once create covariance even when the underlying quantities are independent.

Using the properties of expectation, we can write covariance in a more convenient form
\begin{equation}
\text{Cov}(X,Y) = \mathbb{E}[XY] - \mathbb{E}[X]\mathbb{E}[Y].
\end{equation}

To prove this, we expand the product
\begin{align}
\text{Cov}(X,Y) &= \mathbb{E}[(X - \mathbb{E}[X])(Y - \mathbb{E}[Y])] \\*
&= \mathbb{E}[XY - X\mathbb{E}[Y] - Y\mathbb{E}[X] + \mathbb{E}[X]\mathbb{E}[Y]] \\
&= \mathbb{E}[XY] - \mathbb{E}[X]\mathbb{E}[Y] - \mathbb{E}[X]\mathbb{E}[Y] + \mathbb{E}[X]\mathbb{E}[Y] \\*
&= \mathbb{E}[XY] - \mathbb{E}[X]\mathbb{E}[Y].
\end{align}

Covariance depends on the scale of the variables, which makes it a poor way to compare relationships. To see this, we rescale the variables by constants $a$ and $b$
\begin{align}
\text{Cov}(aX, bY) &= \mathbb{E}[(aX - \mathbb{E}[aX])(bY - \mathbb{E}[bY])] \\*
&= \mathbb{E}[(aX - a\mathbb{E}[X])(bY - b\mathbb{E}[Y])] \\
&= \mathbb{E}[ab(X - \mathbb{E}[X])(Y - \mathbb{E}[Y])] \\*
&= ab\text{Cov}(X,Y).
\end{align}If we change measurement units, for example from solar masses to kilograms (multiplying by $2 \times 10^{30}$), the covariance changes in proportion. This makes it hard to compare covariances between different pairs of variables.

\begin{figure}[ht!]
    \centering
    \includegraphics[width=0.9\textwidth]{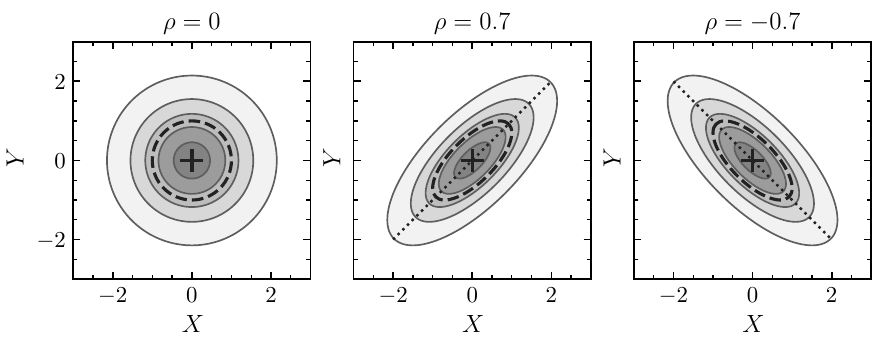}
    \caption{Correlation in bivariate normal distributions with identical marginal variances. The left panel shows uncorrelated variables, which give circular contours. The middle panel shows positive correlation, which stretches the distribution along the diagonal (dotted line). High values of one variable go with high values of the other. The right panel shows negative correlation, which tilts the distribution the other way. High values of one variable go with low values of the other. The dashed ellipses are contours of constant probability. They rotate and deform with the correlation, while the individual scales stay the same.}
    \label{fig:correlation_structure}
\end{figure}

To remove this scale dependence, we define the correlation\index{correlation coefficient} coefficient
\begin{equation}
\rho_{XY} = \frac{\text{Cov}(X,Y)}{\sigma_X\sigma_Y},
\end{equation}
which is dimensionless and independent of scale. It lets us compare the strength of relationships between different pairs of variables and see which relationships are strongest in multivariate data. Figure~\ref{fig:correlation_structure} shows examples.

We can verify that correlation is scale invariant. For a constant $a$, variance scales quadratically, with $\text{Var}[aX] = a^2\text{Var}[X]$. The standard deviation therefore scales as $\sigma_{aX} = |a|\sigma_X$. Combining this with the scaling of covariance gives
\begin{align}
\rho_{aX,bY} &= \frac{\text{Cov}(aX,bY)}{\sigma_{aX}\sigma_{bY}} \\*
&= \frac{ab\text{Cov}(X,Y)}{|a|\sigma_X|b|\sigma_Y} \\*
&= \text{sign}(ab)\rho_{XY}.
\end{align}Correlation keeps its magnitude under scaling, but its sign depends on the transformation. This is sensible, because multiplying one variable by $-1$ reverses the direction of the relationship.\index{covariance|)}

\paragraph{Bounds on Correlation}

The correlation coefficient lies between $-1$ and $+1$. A correlation of $+1$ means perfect positive correlation, so the variables are linearly related with a positive slope. A correlation of $-1$ means perfect negative correlation, so one decreases linearly as the other increases. A correlation of zero means no linear relationship. Independent variables have zero correlation, but zero correlation does not make variables independent, because a nonlinear relationship can leave it at zero.

The bound follows from the Cauchy-Schwarz\index{Cauchy-Schwarz inequality} inequality. For vectors $\mathbf{u}$ and $\mathbf{v}$ in an inner product space, it states
\begin{equation}
|\mathbf{u} \cdot \mathbf{v}| \leq \|\mathbf{u}\| \|\mathbf{v}\|,
\end{equation}
where $\|\cdot\|$ denotes the vector norm.

For random variables, the expectation $\mathbb{E}[UV]$ plays the role of the inner product of $U$ and $V$. For centered random variables, which have their means subtracted, we can show
\begin{align}
|\mathbb{E}[(X - \mathbb{E}[X])(Y - \mathbb{E}[Y])]| \leq \sqrt{\mathbb{E}[(X - \mathbb{E}[X])^2]\mathbb{E}[(Y - \mathbb{E}[Y])^2]}.
\end{align}The left side is the absolute value of the covariance, $|\text{Cov}(X,Y)|$, and the right side is $\sqrt{\text{Var}(X)\text{Var}(Y)} = \sigma_X\sigma_Y$. Dividing both sides by $\sigma_X\sigma_Y$ gives the bound on correlation
\begin{equation}
|\rho_{XY}| = \left|\frac{\text{Cov}(X,Y)}{\sigma_X\sigma_Y}\right| \leq 1.
\end{equation}The magnitude of the covariance cannot exceed the product of the standard deviations, which is why correlation lies between $-1$ and $1$.

This bound matters in astronomy. A measured correlation tells us how close a linear relationship is to the strongest possible one. The Tully-Fisher\index{Tully-Fisher relation} relation, for example, shows a strong positive correlation between galaxy luminosity and rotation velocity, close to the maximum possible. This suggests a tight physical link between a galaxy's light output and its rotation.

\looseness=1 Similarly, main sequence stars in the Hertzsprung-Russell diagram show a very strong correlation between temperature and luminosity. This correlation reflects the physics of stellar structure, because stars are highly ordered systems governed by precise physical laws.

\paragraph{The Covariance Matrix} For\index{covariance matrix|(} several random variables at once, we use a random vector $\mathbf{X} = (X_1, \ldots, X_D)^T$. The variance then generalizes to a covariance\index{covariance matrix} matrix, written $\text{Cov}(\mathbf{X})$ or $\boldsymbol{\Sigma}$
\begin{equation}
\text{Cov}(\mathbf{X}) = \mathbb{E}[(\mathbf{X} - \mathbb{E}[\mathbf{X}])(\mathbf{X} - \mathbb{E}[\mathbf{X}])^T].
\end{equation}

The transpose matters for the dimensions. $\mathbf{X} - \mathbb{E}[\mathbf{X}]$ is a $D \times 1$ column vector, and its transpose $(\mathbf{X} - \mathbb{E}[\mathbf{X}])^T$ is a $1 \times D$ row vector. Their product is a $D \times D$ matrix, in which element $(i,j)$ is the expected product of the deviations of the $i$th and $j$th components. This gives
\begin{equation}
\Sigma_{ij} = \mathbb{E}[(X_i - \mathbb{E}[X_i])(X_j - \mathbb{E}[X_j])] = \begin{cases}
\text{Var}(X_i) & \text{if } i = j \\
\text{Cov}(X_i, X_j) & \text{if } i \neq j
\end{cases}.
\end{equation}The covariance matrix $\boldsymbol{\Sigma}$ holds all the pairwise covariances between the variables, so it is the basic summary of their linear relationships in multivariate data.

Just as variance scales as $\text{Var}(aX) = a^2\text{Var}(X)$, the covariance matrix transforms in a fixed way under linear operations. For a linear transformation $\mathbf{Y} = \mathbf{A}\mathbf{X} + \mathbf{b}$, where $\mathbf{A}$ is an $m \times D$ matrix and $\mathbf{b}$ is an $m$-dimensional vector, we have
\begin{equation}
\text{Cov}(\mathbf{Y}) = \mathbf{A}\text{Cov}(\mathbf{X})\mathbf{A}^T.
\end{equation}This ``sandwiching'' of the covariance matrix between $\mathbf{A}$ and $\mathbf{A}^T$ makes the dimensions work. The $D \times D$ covariance matrix of $\mathbf{X}$ becomes an $m \times m$ covariance matrix for $\mathbf{Y}$, matching the dimension of the transformed random vector. We prove it with the same properties of expectation as in the univariate case
\begin{align}
\text{Cov}(\mathbf{Y}) &= \mathbb{E}[(\mathbf{Y} - \mathbb{E}[\mathbf{Y}])(\mathbf{Y} - \mathbb{E}[\mathbf{Y}])^T] \\*
&= \mathbb{E}[(\mathbf{A}\mathbf{X} + \mathbf{b} - (\mathbf{A}\mathbb{E}[\mathbf{X}] + \mathbf{b}))(\mathbf{A}\mathbf{X} + \mathbf{b} - (\mathbf{A}\mathbb{E}[\mathbf{X}] + \mathbf{b}))^T] \\
&= \mathbb{E}[\mathbf{A}(\mathbf{X} - \mathbb{E}[\mathbf{X}])(\mathbf{X} - \mathbb{E}[\mathbf{X}])^T\mathbf{A}^T] \\
&= \mathbf{A}\mathbb{E}[(\mathbf{X} - \mathbb{E}[\mathbf{X}])(\mathbf{X} - \mathbb{E}[\mathbf{X}])^T]\mathbf{A}^T \\*
&= \mathbf{A}\text{Cov}(\mathbf{X})\mathbf{A}^T.
\end{align}

This property has many uses. When we apply principal component analysis (Chapter~10), standardize features, or make another linear transformation, it shows how the covariance structure changes.\index{covariance matrix|)}\index{variance|)}

\section{Sampling Statistics and Uncertainty}\label{sec:ch3_sampling}

Next,\index{standard error|(} we ask how variance behaves when we combine different random variables. As with expectation, we start with sums. They arise whenever we combine several observations or propagate errors through a calculation.

For any two random variables $X$ and $Y$, we have
\begin{equation}
\text{Var}[X + Y] = \text{Var}[X] + 2\text{Cov}(X,Y) + \text{Var}[Y].
\end{equation}

We prove this with the definition of variance
\begin{align}
\text{Var}[X + Y] &= \mathbb{E}[((X + Y) - \mathbb{E}[X + Y])^2] \\*
&= \mathbb{E}[((X - \mathbb{E}[X]) + (Y - \mathbb{E}[Y]))^2] \\
&= \mathbb{E}[(X - \mathbb{E}[X])^2 + 2(X - \mathbb{E}[X])(Y - \mathbb{E}[Y]) + (Y - \mathbb{E}[Y])^2] \\*
&= \text{Var}[X] + 2\text{Cov}(X,Y) + \text{Var}[Y].
\end{align}

Unlike expectation, variance does not simply add, because the variance of a sum includes a covariance term. Expanding the squared deviation of $X + Y$ produces cross terms of the form $(X - \mathbb{E}[X])(Y - \mathbb{E}[Y])$, and their average is the covariance term $2\text{Cov}(X,Y)$. So the uncertainty of a sum depends on the individual uncertainties and on how the variables move together. This is a basic principle of error\index{error propagation} propagation in measurements and predictions.

\begin{figure}[ht!]
    \centering
    \includegraphics[width=0.9\textwidth]{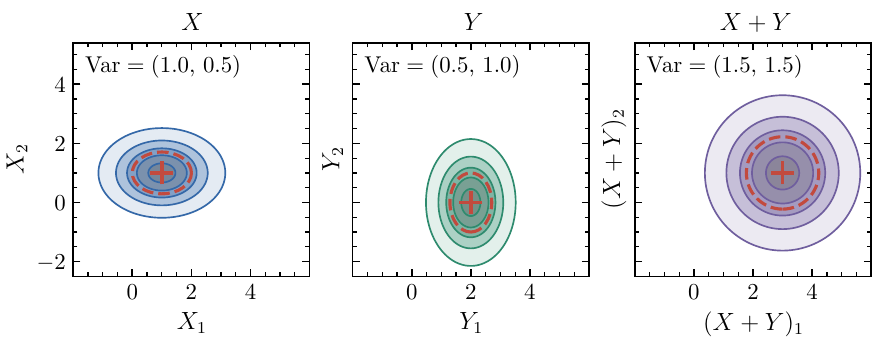}
    \caption{Variances add for independent random variables. The left panel shows the distribution of $X$ (blue contours) and the middle panel shows $Y$ (green contours). The right panel shows the sum $X + Y$ (purple contours). The red dashed ellipses show the variance structure of each distribution. For independent variables the variances add component-wise, so the variances in the right panel are the sums of those in the left and middle panels. Each variable contributes its own randomness to the sum, so the uncertainty grows.}
    \label{fig:variance_additivity}
\end{figure}

For independent variables, $\text{Cov}(X,Y) = 0$, and the formula simplifies to
\begin{equation}
\text{Var}[X + Y] = \text{Var}[X] + \text{Var}[Y] \quad \text{(for independent variables)}.
\end{equation}So the variances of independent random variables add directly (Figure~\ref{fig:variance_additivity}). This underlies much of statistical analysis and shows how variances add up when we combine measurements or build aggregate statistics.

\paragraph{Uncertainty of the Sample Mean}

This additivity gives the uncertainty of a sample mean. In astronomy we often combine several observations of the same quantity to improve precision. Examples are repeated magnitude measurements of a star and repeated spectroscopic observations of a galaxy's redshift.

Consider $N$ independent measurements $\{X_1, ..., X_N\}$ of the same quantity. The sample mean, our best estimate of the true value, is
\begin{equation}
\bar{X} = \frac{1}{N}\sum_{i=1}^N X_i.
\end{equation}

Using the variance properties above, and assuming the measurements are independent with the same variance $\sigma^2$, we find
\begin{align}
\text{Var}[\bar{X}] &= \text{Var}\left[\frac{1}{N}\sum_{i=1}^N X_i\right] \\*
&= \frac{1}{N^2}\text{Var}\left[\sum_{i=1}^N X_i\right] \\
&= \frac{1}{N^2}\sum_{i=1}^N \text{Var}[X_i] \quad \text{(all covariances = 0)} \\
&= \frac{1}{N^2}(N\sigma^2) \\*
&= \frac{\sigma^2}{N}.
\end{align}This gives the well-known result
\begin{equation}
\text{Var}[\bar{X}] = \frac{\sigma^2}{N},
\end{equation}
which says that the variance of the mean decreases inversely with sample size. Taking the square root, the uncertainty decreases as $1/\sqrt{N}$.

This uncertainty, $\sigma/\sqrt{N}$, is called the standard\index{standard error} error of the mean. It is the standard deviation that the sample mean would show over many repeated samples of size $N$. For large $N$, the Central Limit Theorem (Section~2.3) also makes the sample mean nearly Gaussian.

\begin{figure}[ht!]
    \centering
    \includegraphics[width=0.9\textwidth]{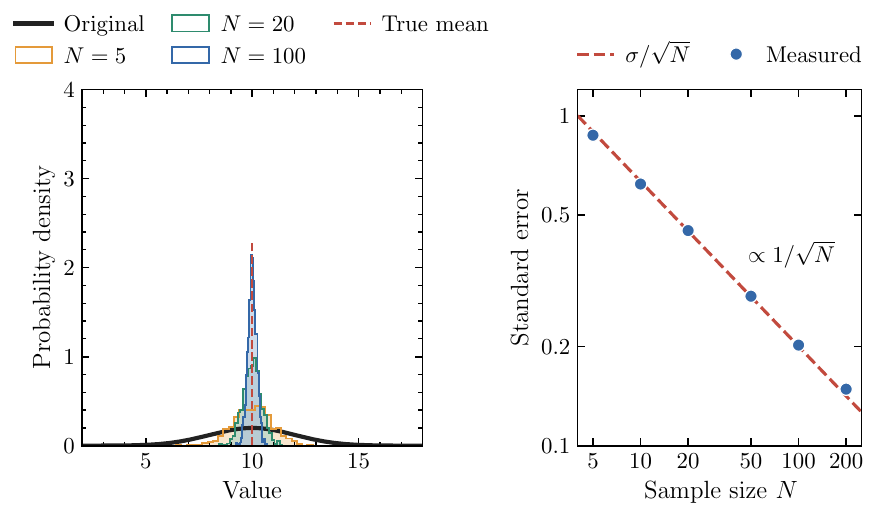}
    \caption{The uncertainty of the sample mean decreases with sample size. The left panel shows the distribution of individual measurements (black curve) with the true mean (red dashed line). Colored histograms show the distributions of sample means for increasing sample size $N$. The sample means concentrate around the true mean as $N$ grows. The right panel shows the standard error, the uncertainty of the sample mean, as a function of $N$. The red dashed line is the theoretical prediction $\sigma/\sqrt{N}$, and the blue points are measured in numerical simulations.}
    \label{fig:sample_mean_scaling}
\end{figure}

If each measurement of a star's brightness has uncertainty $\sigma$, averaging $N$ independent measurements reduces the uncertainty to $\sigma/\sqrt{N}$ (Figure~\ref{fig:sample_mean_scaling}). This is why longer exposures, which act like a larger $N$, give more precise measurements. The same principle explains why larger datasets generally give more reliable parameter estimates. This holds provided the data are independent\index{independent and identically distributed (i.i.d.)} and identically distributed, meaning the values do not influence each other and share one distribution.

\paragraph{Sampling Noise in Moment Estimates}

We return to our original motivation, which is to describe probability distributions with summary statistics. From a finite sample, we estimate moments with sums of the form
\begin{equation}
\hat{\mu}_k = \frac{1}{N}\sum_{i=1}^N x_i^k.
\end{equation}The hat notation shows that this is an estimate of the true moment $\mu_k$. Any number computed from a sample, such as $\hat{\mu}_k$ or $\bar{X}$, is called a statistic. A rule that turns a sample into an estimate is called an estimator.

The sample covariance of $N$ pairs $(x_i,y_i)$ is the sum of the products $(x_i - \bar{X})(y_i - \bar{Y})$ of the deviations from the sample means, divided by $N$. The sample variance is the same expression with $y_i$ replaced by $x_i$. Chapters~6 and~10 use these definitions and write $\mathbb{E}$ for the average over a data set.

The estimate itself varies from sample to sample. This variability is finite sampling noise, the uncertainty from using a finite number of samples to estimate properties of the underlying population. The population is the full distribution from which the sample is drawn. This noise determines how reliable our parameter estimates and inferences are.

Consider estimating the moments of the stellar IMF. Even with perfect mass measurements, $\hat{\mu}_k$ would vary if we repeated the measurement with different random samples of stars. Each star's mass is a random draw from the IMF, so $X^k$ and therefore $\hat{\mu}_k$ are random variables. In general, even with perfect measurements, estimates carry uncertainty because they come from finite samples. We must estimate this uncertainty for reliable inference.

As we found for sample means, moment estimates become more precise with larger samples. For independent measurements, the variance of the $k$th moment estimate scales as $\text{Var}(X^k)/N$, so the sampling noise decreases as $1/\sqrt{N}$. This is one reason modern astronomy carries out deep surveys with wide sky coverage. The large number of objects reduces finite sampling noise, which gives more precise estimates of population statistics and tighter constraints on theoretical models.

The $1/\sqrt{N}$ scaling gives the expected improvement with sample size. The total uncertainty also depends on $\text{Var}(X^k)$, whose size relative to $\mu^{2k}$ grows rapidly with $k$. To see why, expand a fluctuation around the mean. If $X = \mu + \delta$, where $\mu$ is the mean of $X$, a Taylor expansion gives
\begin{align}
X^k &= (\mu + \delta)^k \\*
&= \mu^k + k\mu^{k-1}\delta + \frac{k(k-1)}{2}\mu^{k-2}\delta^2 + \ldots.
\end{align}

If we keep only the linear term $\mu^k + k\mu^{k-1}\delta$, the scaling property of variance gives
\begin{equation}
\text{Var}(X^k) \approx k^2\mu^{2k-2}\text{Var}(X).
\end{equation}Relative to $\mu^{2k}$, this variance is $k^2\,\text{Var}(X)/\mu^2$, so it grows approximately as $k^2$. As in our earlier discussion of higher moments, astronomical distributions often contain a few very massive galaxies or extremely bright quasars. These objects have a large effect on higher-moment estimates, so those estimates are noisier than lower-moment estimates. The same happens in other fields with heavy-tailed distributions.

Yet higher moments often carry the physical information we want. In cosmology, for example, the non-Gaussian features of the galaxy distribution, which appear only in moments beyond the variance, constrain theories of structure formation. This sets up a trade-off. Higher moments carry unique physical information, but they need much larger samples to measure well. We must balance information content against estimation precision when we design studies and analyze data.\index{standard error|)}

\section{Bootstrapping}\label{sec:ch3_bootstrap}

The\index{bootstrap|(} $1/\sqrt{N}$ scaling law shows how uncertainty decreases with sample size, but in practice we often need precise values for the uncertainties. Our analysis of $\text{Var}(X^k)$ kept only the leading term of the Taylor expansion, to build intuition. The neglected higher-order terms can matter, especially for small samples or highly skewed distributions, which are common in astronomy.

For multidimensional or nonlinear statistics, analytical expressions can become intractable. This leads us to empirical methods for estimating uncertainty, which have grown in importance with computational statistics.

\begin{figure}[ht!]
    \centering
    \includegraphics[width=0.9\textwidth]{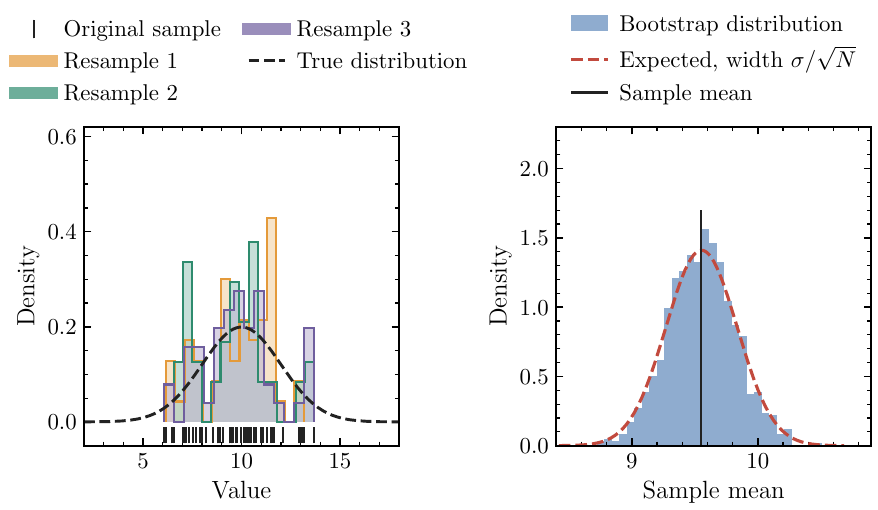}
    \caption{The bootstrap estimates the sampling variance of the first moment. The left panel shows an original sample $\{x_1,...,x_N\}$ (black tick marks) drawn from the true distribution $p(x)$ (black dashed curve). Colored histograms show three bootstrap resamples $\{x_1^*,...,x_N^*\}$ drawn with replacement. The right panel shows the bootstrap estimates $\hat{\mu}_1^* = (1/N)\sum_{i=1}^N x_i^*$ from many resamples (blue histogram) and a Gaussian of the theoretical width $\sigma/\sqrt{N}$ (red dashed curve). The bootstrap variance $\text{Var}(\hat{\mu}_1^*)$ agrees with the theoretical variance $\sigma^2/N$, although the bootstrap does not use $p(x)$. The black vertical line is the original sample estimate $\hat{\mu}_1 = (1/N)\sum_{i=1}^N x_i$.}
    \label{fig:bootstrap_illustration}
\end{figure}

The bootstrap\index{bootstrap} estimates the variance of an estimator when analytical expressions are not available (Figure~\ref{fig:bootstrap_illustration}). Given an original sample $\{x_1, \ldots, x_N\}$, the algorithm is
\begin{enumerate}
\item Draw $N$ values from this set with replacement, so each value can be drawn again. They form a ``bootstrap sample'' $\{x_1^*, \ldots, x_N^*\}$
\item Calculate the statistic of interest, for example the $k$th moment, on this bootstrap sample
   \begin{equation}
   \hat{\mu}_k^* = \frac{1}{N}\sum_{i=1}^N (x_i^*)^k
   \end{equation}
\item Repeat steps 1 and 2 many times (typically thousands)
\item Use the variance of these bootstrap estimates to approximate $\text{Var}(\hat{\mu}_k)$
\end{enumerate}

\looseness=1 For example, if the original dataset is $\{1, 2, 3\}$, one bootstrap sample might be $\{1, 1, 2\}$ and another $\{2, 3, 3\}$. Each bootstrap sample has the same size as the original, but some values appear more than once and others not at all. This resampling mimics the variability we would see if we could collect new datasets of the same size from the same population. It gives a computational solution to what is otherwise a theoretical problem.

\paragraph{Why the Bootstrap Works}

The logic of the bootstrap comes from how we think about sampling. Our sample $\{x_1, ..., x_N\}$ is drawn from an unknown probability distribution $p(x)$. If we knew $p(x)$, a moment would just be an integral, as above. We usually have only a finite sample. The bootstrap uses this sample as a stand-in for the underlying distribution. We can then study the sampling\index{sampling distribution} distribution of our statistics, meaning how they vary from sample to sample.

Resampling with replacement treats the observed sample as a discrete approximation to $p(x)$, called the empirical\index{empirical distribution} distribution, in which each observed value has probability $1/N$. Our original sample was drawn from the true distribution $p(x)$, and the bootstrap samples are drawn from this empirical distribution. If we could draw many new samples from $p(x)$, we could study how $\hat{\mu}_k$ varies between them. That would show its sampling distribution and its uncertainty. The bootstrap approximates this ideal process using only the data we have, so it needs no additional data collection.

As the sample size $N$ grows, the empirical distribution approaches the true distribution $p(x)$. The distribution of bootstrap statistics then approaches the true sampling distribution. This is why the bootstrap gives reliable estimates of sampling uncertainty and why we use it widely for model validation (Section~\ref{sec:cross_validation}) and uncertainty quantification.

The strength of the bootstrap is its generality. Instead of deriving analytical formulas for standard errors, which may be intractable for complex statistics, we get the sampling behavior directly by simulation. It suits astronomy, where we often need the uncertainty of complex measurements such as galaxy morphology classifications or stellar parameters from spectroscopy. It also applies in other fields with complex statistics, from economics to ecology. There it gives uncertainties for quantities from correlation coefficients to diversity indices, which measure the variety of species in a sample.

\paragraph{When the Bootstrap Fails}

The bootstrap can fail in three ways.

First, the sample must be representative. The bootstrap assumes the sample is large enough to approximate the underlying distribution. If it misses features of $p(x)$, such as rare events, the bootstrap misses them too. This applies most to higher moments, which are sensitive to the tails. It also applies to applications that focus on the tails. Examples are extreme value analysis, which studies the largest values of a sample, and risk assessment.

Second, the basic bootstrap resamples individual values and so treats observations as independent. This can mislead when data are correlated, as in time series or spatial data where neighboring measurements are related. In astronomy this can happen for galaxy clustering or stellar variability. Block bootstrapping was designed for this case. It resamples blocks of observations, which preserves the correlation within each block.

Third, for statistics that depend on extreme values, the bootstrap can underestimate uncertainty. It can only resample values that are in the original sample, so it cannot produce new extreme values that future datasets might contain. The uncertainty estimates for statistics that are sensitive to outliers can then be too optimistic. This matters in hydrology and finance, where extreme events carry great importance.

Even so, the bootstrap is among our most useful tools for estimating uncertainty. Its logic is simple, and it suits astronomical measurements that are complex and have unknown underlying distributions. Two related methods exist. Parametric bootstrapping resamples from a fitted parametric distribution instead of the empirical distribution. The jackknife leaves out one observation at a time to measure its influence on the statistic.\index{bootstrap|)}

\section{Law of Total Variance}

Just\index{law of total variance|(} as the law of total expectation breaks down expected values in hierarchical problems, the law\index{law of total variance} of total variance decomposes variability. It shows how much of our uncertainty comes from different sources, such as measurement noise and intrinsic variability, or explainable and unexplainable variation.

For any two random variables $X$ and $Y$, the law of total variance states
\begin{equation}
\text{Var}_X[X] = \mathbb{E}_Y[\text{Var}_X[X|Y]] + \text{Var}_Y[\mathbb{E}_X[X|Y]].
\end{equation}

The total variance of a random variable splits into two parts with clear meanings. The first term, $\mathbb{E}_Y[\text{Var}_X[X|Y]]$, is the expected conditional variance, the average ``unexplained'' variability that remains after we know $Y$. For each value of $Y$, the conditional variance is the spread of possible $X$ values, and we average these spreads over all values of $Y$. For example, even if we know a star's luminosity ($Y$) exactly, its mass ($X$) is still uncertain. Stars of the same luminosity can differ slightly in mass because of factors like age and metallicity. This term gives that remaining uncertainty, averaged over all luminosities, which is irreducible once we use all our information about $Y$.

The second term, $\text{Var}_Y[\mathbb{E}_X[X|Y]]$, is the variability of $X$ that knowing $Y$ explains. It measures how much the average value of $X$ changes as $Y$ changes. In our stellar example, it measures how much the average stellar mass varies with luminosity, which is the strength of the mass-luminosity relation itself. It is the part of the variance that the relationship between $X$ and $Y$ explains.

\paragraph{Proof of the Law}

To prove the law, we write the total variance and split the deviation $(X - \mathbb{E}_X[X])$ into two parts. One is the deviation from the conditional mean, $(X - \mathbb{E}_X[X|Y])$. The other is the deviation of the conditional mean from the overall mean, $(\mathbb{E}_X[X|Y] - \mathbb{E}_X[X])$
\begin{align}
\text{Var}_X[X] &= \mathbb{E}[(X - \mathbb{E}[X])^2] \\*
&= \mathbb{E}[((X - \mathbb{E}[X|Y]) + (\mathbb{E}[X|Y] - \mathbb{E}[X]))^2] \\
&= \mathbb{E}[(X - \mathbb{E}[X|Y])^2] + \mathbb{E}[(\mathbb{E}[X|Y] - \mathbb{E}[X])^2] \\*
&\quad + 2\mathbb{E}[(X - \mathbb{E}[X|Y])(\mathbb{E}[X|Y] - \mathbb{E}[X])].
\end{align}

The cross term vanishes because
\begin{align}
\mathbb{E}[(X - \mathbb{E}[X|Y])(\mathbb{E}[X|Y] - \mathbb{E}[X])] &= \mathbb{E}[\mathbb{E}[(X - \mathbb{E}[X|Y])(\mathbb{E}[X|Y] - \mathbb{E}[X])|Y]] \\*
&= \mathbb{E}[(\mathbb{E}[X|Y] - \mathbb{E}[X])\mathbb{E}[(X - \mathbb{E}[X|Y])|Y]] \\
&= \mathbb{E}[(\mathbb{E}[X|Y] - \mathbb{E}[X]) \cdot 0] \\*
&= 0,
\end{align}
where the first step applies the law of total expectation (Section~3.4) to the product. The second step moves the factor $\mathbb{E}[X|Y] - \mathbb{E}[X]$ outside the inner expectation, because it is a fixed number once $Y$ is given. The third step uses $\mathbb{E}[X - \mathbb{E}[X|Y]\,|\,Y] = 0$, because given $Y$ the deviation of $X$ from its average $\mathbb{E}[X|Y]$ averages to zero.

The remaining terms give
\begin{align}
\text{Var}_X[X] &= \mathbb{E}[(X - \mathbb{E}[X|Y])^2] + \mathbb{E}[(\mathbb{E}[X|Y] - \mathbb{E}[X])^2] \\*
&= \mathbb{E}[\mathbb{E}[(X - \mathbb{E}[X|Y])^2|Y]] + \mathbb{E}[(\mathbb{E}[X|Y] - \mathbb{E}[X])^2] \\*
&= \mathbb{E}[\text{Var}[X|Y]] + \text{Var}[\mathbb{E}[X|Y]],
\end{align}
which is exactly the law of total variance. The second line applies the law of total expectation to the first term. The inner expectation in the second line is the conditional variance $\text{Var}[X|Y]$. The second term is $\text{Var}[\mathbb{E}[X|Y]]$ because $\mathbb{E}[\mathbb{E}[X|Y]] = \mathbb{E}[X]$ by the same law.

\begin{figure}[ht!]
    \centering
    \includegraphics[width=0.8\textwidth]{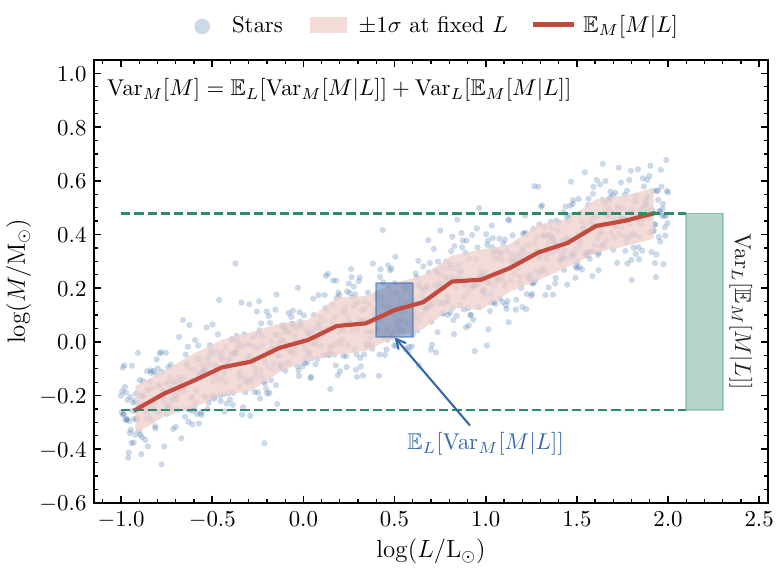}
    \caption{The law of total variance for the stellar mass-luminosity relation. Blue points are individual stars. The red line is the conditional expectation $\mathbb{E}_M[M|L]$, and the red shaded region marks its $\pm 1\sigma$ conditional spread. The blue box at $\log(L/\mathrm{L}_\odot) = 0.5$ marks the scatter in mass at fixed luminosity ($\pm 1\sigma$), the conditional variance that $\mathbb{E}_L[\text{Var}_M[M|L]]$ averages over $L$. The green shaded region and dashed lines on the right show the range of conditional expectations, whose variance is $\text{Var}_L[\mathbb{E}_M[M|L]]$. The total variance in mass is the sum of these two terms.}
    \label{fig:total_variance}
\end{figure}

\paragraph{Interpretation and Extensions}

Figure~\ref{fig:total_variance} illustrates this decomposition with the stellar mass-luminosity relationship. Individual stars are blue points, which show how masses vary at each luminosity. The red line traces the conditional expectation $\mathbb{E}_M[M|L]$, the average mass of stars at each luminosity. The red shaded region shows the conditional spread at each luminosity, which is how much stellar masses typically differ from this average. At $\log(L/\mathrm{L}_\odot) = 0.5$ (the blue box), we see the spread of masses around the conditional mean. Averaging these conditional variances over all luminosities, weighted by how common each luminosity is, gives the first term, $\mathbb{E}_L[\text{Var}_M[M|L]]$.

The green shaded region on the right side of the figure shows the second term. It shows how much the conditional expectations themselves change with luminosity. This variation of $\mathbb{E}_M[M|L]$ (the red line) with $L$ is $\text{Var}_L[\mathbb{E}_M[M|L]]$. It is the part of the total mass variance that the mass-luminosity relationship explains.

Better measurements reduce only the measurement-noise part of the first term. If intrinsic scatter at fixed $Y$ makes up most of that term, or if the second term dominates, the total uncertainty barely changes. In stellar spectroscopy, for example, better resolution or signal-to-noise ratio stops helping beyond a certain point. This happens if the remaining variation comes from intrinsic stellar variability and differences between stellar populations. This matters for experimental design and for how we spend our resources.

The decomposition also shows when an extra variable helps prediction. If the expected conditional variance $\mathbb{E}_Y[\text{Var}_X[X|Y]]$ is much smaller than $\text{Var}_X[X]$, knowing $Y$ greatly reduces our uncertainty about $X$, so $Y$ is a useful predictor. If the two variances are similar, $Y$ gives little predictive information about $X$. It helps us judge how useful different variables are for prediction.

The law of total variance generalizes to the multivariate case
\begin{equation}
\text{Cov}(\mathbf{X}) = \mathbb{E}[\text{Cov}(\mathbf{X}|\mathbf{Y})] + \text{Cov}(\mathbb{E}[\mathbf{X}|\mathbf{Y}]).
\end{equation}The same proof applies, because it used only properties of expectation that hold for vectors. The interpretation is similar. The total covariance is the average of the conditional covariances plus the covariance of the conditional means. This applies to systems of several variables, such as multivariate regression or physical systems with several interacting components.

The law of total variance also applies to predictive uncertainty in Bayesian statistics. Model parameters such as regression coefficients are random variables with uncertainty. To predict a new outcome $y_*$ while including this parameter uncertainty, the law of total variance splits the predictive variance
\begin{equation}
\text{Var}[y_*] = \mathbb{E}_{\theta}[\text{Var}[y_*|\theta]] + \text{Var}_{\theta}[\mathbb{E}[y_*|\theta]].
\end{equation}The first term is the average variance of the predictions at each parameter value $\theta$, weighted by the posterior distribution of $\theta$. The second term is how much our predictions vary because of parameter uncertainty. These two terms are the aleatoric\index{aleatoric uncertainty} uncertainty, the randomness left when the parameters are known, and the epistemic\index{epistemic uncertainty} uncertainty, our limited knowledge of them.

We can now describe a distribution with a few summary statistics, track how their uncertainty shrinks with sample size, and split variability into its sources. The next two chapters use expectations, variances, and covariances to fit a linear regression and to find the uncertainty of its parameters.\index{law of total variance|)}

\section{Tutorial: Finding Baryon Acoustic Oscillations}\label{sec:ch3_tutorial}

This\index{baryon acoustic oscillations|(}\index{summary statistics|(} tutorial applies the covariance and the bootstrap of this chapter to a faint signal in the galaxy distribution, the baryon\index{baryon acoustic oscillations} acoustic oscillation. Cosmologists use it to measure how the universe expanded. The full tutorial can be found at \url{https://statml.org/} as tutorial 3.

In the first 380,000 years after the Big Bang, the universe was a hot plasma of ordinary matter, called baryons, and light. Sound waves moved through it. When atoms formed, the waves froze in place after traveling about 150 million parsecs (in today's units). This left galaxies slightly more likely to be that far apart. This separation is a standard ruler, a length that we know. Its apparent size at each epoch gives the distance to the galaxies, and so how the universe expanded. The signal is faint, because the excess of galaxy pairs at this separation is small.

\begin{figure}[t!]
    \centering
    \includegraphics[width=0.95\textwidth]{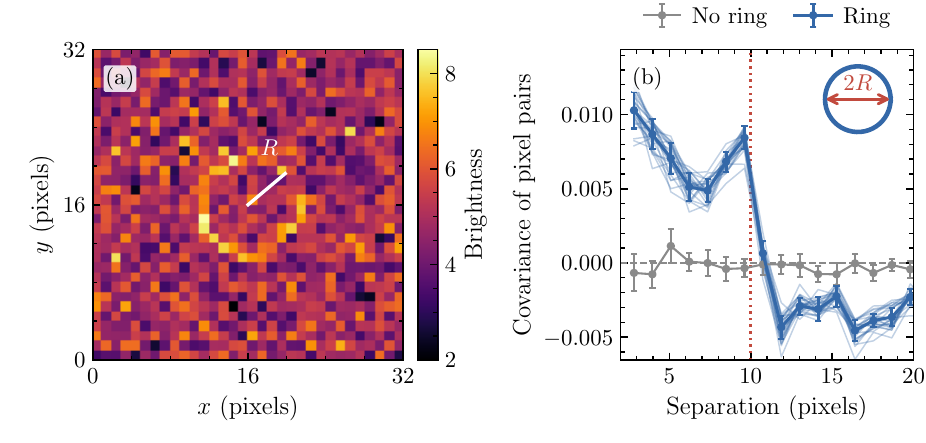}
    \caption{The tutorial searches a simulated map for a ring. Panel~(a) shows a simulated map with a ring of radius $R$ (white line). Panel~(b) shows the covariance of pairs of pixels against their separation, for many maps with rings (blue) and for maps without a ring (grey). Error bars come from the bootstrap, and the thin lines are some of the resamples. The ring maps drop quickly beyond the diameter $2R$ (dotted line), which the inset draws on a ring.}
    \label{fig:tutorial_bao}
\end{figure}

To find such a faint signal, we must first decide what to measure. We observe one universe, which is one random sample from the distribution that theory predicts. The positions of galaxies are therefore random variables, and a map is a finite sample. Theory does not predict where each galaxy sits. It predicts statistical properties, such as how galaxy pairs are correlated at each separation. The data must therefore be compressed into summary statistics, each with a sampling uncertainty.

The first candidates are the mean and the variance, the moments of Section~\ref{sec:ch3_moments}. To test them, the tutorial simulates a toy universe, a map with a ring of radius $R$ in random noise (Figure~\ref{fig:tutorial_bao}a). The ring stands for the preferred separation and takes its brightness from the background, as the real signal moves galaxies and does not create them. The mean and the variance of the map therefore do not change, so these two statistics do not reveal the ring. Shuffling the pixels would leave them unchanged and would destroy the ring.

A statistic that shows the ring must depend on where the pixels lie. The covariance of Section~\ref{sec:ch3_variance_covariance} does, because it compares pairs of pixels. Pixels on the ring are brighter than average, so pairs of them have a positive covariance. Plotted against the separation of the two pixels, this covariance is what astronomers call the correlation\index{correlation function} function (Figure~\ref{fig:tutorial_bao}b). Two points on a ring are at most a diameter $2R$ apart, and most often close to it. The covariance therefore rises toward $2R$ and drops quickly beyond it. Pairs close together on the ring also give a high covariance at small separations.

Whether this feature is real depends on its sampling uncertainty (Section~\ref{sec:ch3_sampling}). We observe the universe once, so the bootstrap of Section~\ref{sec:ch3_bootstrap} uses the sample itself to estimate how the statistic would vary. The drop beyond $2R$ is much larger than the resulting error bars, and maps without a ring show no such feature (Figure~\ref{fig:tutorial_bao}b).

The same reasoning applies to a real survey. A galaxy survey measures the ruler from a similar bump in the correlation function of galaxy positions. The real signal is much weaker than our ring, so surveys need very large volumes to make the sampling noise small.

Theory predicts a summary statistic, the data estimate it from one finite sample, and the bootstrap tells us how far to trust the estimate. The covariance of pairs depends on where things lie and the mean and variance do not, so the statistic must match the signal. The same recipe applies to characteristic scales in noisy spatial data.\index{baryon acoustic oscillations|)}\index{summary statistics|)}

\paragraph{Further Reading} Fisher (1922) developed the ideas of efficiency, sufficiency and consistency that remain central to statistical inference. Cram\'er (1946) gave a mathematical treatment that connects probability theory with statistical methods. Cox and Hinkley (1974) emphasize the general concepts of statistical theory. Casella and Berger (2002) is a graduate text that derives statistical inference from probability. Kendall and Stuart (1977) and Stuart and Ord (1994) are editions of an extensive reference on distribution theory and the properties of moments. Taylor (1982) is a widely used introduction to error analysis and the propagation of uncertainty. Efron (1979) developed the bootstrap, and Davison and Hinkley (1997) cover its theory and applications. Lupton (1993) explains classical statistical theory for scientists. Wall and Jenkins (2003) cover statistical methods for observational astronomy, both frequentist and Bayesian.

\chapter{Linear Regression}

We begin supervised learning with one of the simplest techniques, linear regression, which we use to show the power of Bayesian inference. The method is conceptually straightforward, yet it shows surprising mathematical depth when we view it through Bayesian inference. By treating both the data and the model parameters as random variables, we gain a fuller statistical understanding of the fitting process.

We choose linear regression for two reasons. Its simplicity gives an ideal setting to visualize Bayesian inference in action, because line fitting makes abstract statistical concepts concrete and practical. Its capabilities and limitations also lead to many other concepts in machine learning. What we learn here extends to more sophisticated methods, from logistic regression to Gaussian processes and neural networks in later chapters.

\section{Linear Regression Fundamentals}\label{sec:ch4_fundamentals}

Linear\index{linear regression}\index{linear regression|(} regression maps an input variable $x$ to a continuous output variable $y$ through a linear relationship. In its basic form, the relationship is
\begin{equation}
y = w_0 + w_1x,
\end{equation}
where $w_0$ is the intercept and $w_1$ is the slope. This equation embodies a strong inductive\index{inductive bias} bias, an assumption about the relationship we want to model. Here we assume that $x$ and $y$ are linearly related.

Linear regression may seem too restrictive, because many astronomical phenomena are nonlinear. Yet it remains widely used in astronomical research, for three reasons that show the power of ``keeping things simple'' while keeping the mathematics sound.

\paragraph{Expressiveness Through Transformations} We can make linear regression surprisingly expressive by transforming the input features, the quantities that the model uses to predict $y$. For example, Kepler's Third Law relates the orbital period ($P$) to the semi-major axis ($a$) of an orbit around a mass $M$ through the equation
\begin{equation}
P^2 = \frac{4\pi^2}{GM}a^3.
\end{equation}
This nonlinear relationship becomes linear when we take logarithms of both sides
\begin{equation}
\log P = \frac{1}{2}\log\left(\frac{4\pi^2}{GM}\right) + \frac{3}{2}\log a.
\end{equation}

The transformation technique is especially useful because many basic relationships in astronomy follow power laws. As we discussed in Chapter 2, power laws appear when physical processes operate similarly at different scales, meaning that the system has no preferred scale. This scale invariance appears in systems from stellar masses to galaxy properties, so power laws are common in our field.

Consider the M-$\sigma$ relation,\index{M-sigma relation@$M$--$\sigma$ relation} which links the mass of a supermassive\index{black holes!supermassive} black hole ($M_{\rm BH}$) to the velocity dispersion ($\sigma_\star$) of its host galaxy's bulge
\begin{equation}
M_{\rm BH} = A \sigma_\star^\gamma.
\end{equation}
Similarly, the Kennicutt-Schmidt\index{Kennicutt-Schmidt law} law relates the star formation rate surface density ($\Sigma_{\rm SFR}$) to the gas surface density ($\Sigma_{\rm gas}$)
\begin{equation}
\Sigma_{\rm SFR} = A \Sigma_{\rm gas}^\gamma.
\end{equation}Both relationships are nonlinear in their original form, but they become linear in logarithmic space
\begin{equation}
\log M_{\rm BH} = \log A + \gamma \log \sigma_\star,
\end{equation}
\begin{equation}
\log \Sigma_{\rm SFR} = \log A + \gamma \log \Sigma_{\rm gas}.
\end{equation}

\paragraph{Physical Interpretability} These examples show one advantage of linear regression, which is that its models are often easier to interpret physically than more complex approaches. The coefficients, or weights in modern machine learning terminology, correspond to physically meaningful quantities. The slope $\gamma$ in these logarithmic relationships is the power-law index, which tells us about the underlying physics. For instance, in the Kennicutt-Schmidt law, $\gamma \approx 1.4$ means a superlinear relationship, so doubling the gas density more than doubles the star formation rate. This suggests that star formation becomes more efficient at higher gas densities. Stronger self-gravity and more frequent molecular cloud collisions in denser environments may explain this.

The intercept ($\log A$) also carries physical meaning. In the M-$\sigma$ relation, $\log A$ is the normalization that connects stellar dynamics to black hole mass. In the Kennicutt-Schmidt law, it measures the global efficiency of star formation. This says how effectively a galaxy turns a given density of molecular gas into new stars.

\paragraph{Mathematical Advantages} Linear regression also has mathematical advantages that follow from Bayesian inference. First, for many objective functions derived from Bayesian inference, it has analytical solutions. We can then find the best model parameters by direct calculation, with no computationally expensive iterative optimization. Second, it fits the Bayesian view of the model parameters as random variables. This lets us quantify the full probability distribution of the parameters, which goes beyond a point estimate. It gives the parameter uncertainties and their correlations, which scientific inference needs.

\section{Mathematical Formalism}\label{sec:ch4_formalism}

We assume that the output $y$ is a scalar, such as a single black hole mass. It is a linear combination of the input features plus a noise term $\varepsilon$. With $D$ input features, we can write this explicitly as
\begin{equation}
y = w_0 + w_1x_1 + w_2x_2 + \ldots + w_Dx_D + \varepsilon.
\end{equation}

For mathematical convenience, we can absorb the intercept $w_0$ into the weight vector by adding a constant 1 to the feature vector
\begin{equation}
\begin{pmatrix} y \end{pmatrix} = \begin{pmatrix} w_0 & w_1 & w_2 & \cdots & w_D \end{pmatrix} \cdot \begin{pmatrix} 1 \\ x_1 \\ x_2 \\ \vdots \\ x_D \end{pmatrix} + \varepsilon.
\end{equation}
With this constant as the first feature, $w_0$ becomes part of the weight vector. We write $\mathbf{w} = (w_0, w_1, \ldots, w_D)^T$ and $\mathbf{x} = (1, x_1, \ldots, x_D)^T$. The single dot product $\mathbf{w}^T \mathbf{x}$ then expresses both the linear transformation and the intercept. This is mathematically equivalent to the explicit form, so we can write the model compactly in vector notation
\begin{equation}
y = \mathbf{w}^T \mathbf{x} + \varepsilon.
\end{equation}

The vector $\mathbf{w}$ holds the model parameters (weights), and $\varepsilon$ represents the noise or error in our measurements. The intercept $w_0$ is also called the bias term. It lets the model shift its predictions up or down, independent of the input features. This constant offset differs from the inductive bias of Section~4.1.

For example, take the M-$\sigma$ relation after taking logarithms. In its simplest form, $y$ is $\log M_{\rm BH}$ and $x_1$ is $\log \sigma_\star$. Here $\sigma_\star$ is the velocity dispersion and differs from the noise level $\sigma$ of Section~4.3. We can add other features that might influence black hole mass, such as galaxy morphology indicators or environmental parameters, which makes $\mathbf{x}$ multidimensional. For instance, we might model
\begin{equation}
\log M_{\rm BH} = w_0 + w_1\log \sigma_\star + w_2\log R_e + w_3(B/T) + \varepsilon,
\end{equation}
where $R_e$ is the effective radius of the galaxy and $B/T$ is the bulge-to-total mass ratio. Here $w_0$ is the baseline value of $\log M_{\rm BH}$ when all other terms are zero.

The model generalizes to vector-valued outputs, such as predicting several galaxy properties at once. We then use a separate set of weights $\mathbf{w}_k$ for each output dimension $y_k$, which amounts to running several regressions in parallel.

We also need not use the raw input features directly. As the power-law examples showed, we can apply transformations of our choice and map $\mathbf{x}$ to a new feature\index{feature space} space through functions $\boldsymbol{\phi}(\mathbf{x})$. This gives a more general form of linear regression
\begin{equation}
y(\mathbf{x}, \mathbf{w}) = \sum_{j=0}^{M-1} w_j \cdot \phi_j(\mathbf{x}) = \mathbf{w}^T \boldsymbol{\phi}(\mathbf{x}),
\end{equation}
where $\mathbf{w} = (w_0, \ldots, w_{M-1})^T$ is the vector of parameters and $\boldsymbol{\phi}(\mathbf{x}) = (\phi_0(\mathbf{x}), \ldots, \phi_{M-1}(\mathbf{x}))^T$ is the vector of basis\index{basis functions} functions. By convention, $\phi_0(\mathbf{x}) = 1$, which is the constant first feature from above. Each basis function is a fixed function of the input, and the model takes a weighted sum of them.

The notation $y(\mathbf{x}, \mathbf{w})$ stresses that the model is a function. It maps an input $\mathbf{x}$ to a predicted output $y$ for given parameters $\mathbf{w}$. This functional view matters because our goal is to find parameters $\mathbf{w}$ that fit the existing data and also predict new, unseen data accurately.

In the M-$\sigma$ example, suitable parameters $\mathbf{w}$ let us estimate the black hole mass of a galaxy from its velocity dispersion and other properties. This predictive view, which turns data into insight that generalizes beyond the training set, is central to machine learning. The training set is the data that we use to fit the model (Section~\ref{sec:cross_validation}).

The basis functions $\phi_j(\mathbf{x})$ can take many forms. In our power-law examples, $\phi_1(\mathbf{x})$ might be $\log(x)$. For polynomial\index{polynomial regression} regression, we might use $\phi_j(\mathbf{x}) = x^j$. When we simply take $\boldsymbol{\phi}(\mathbf{x}) = \mathbf{x}$, we recover our original linear regression form.

The choice of basis functions is an important design decision. It lets us encode domain knowledge about the expected relationships in our data. The model stays linear in the parameters $\mathbf{w}$, which keeps the mathematical advantages of linearity. This is why ``linear'' regression can be nonlinear in the input features. For example, $w_0 + w_1x + w_2x^2$ is linear in $w_0$, $w_1$ and $w_2$.

\section{Maximum Likelihood Estimation}\label{sec:ch4_mle}

Having\index{maximum likelihood estimation (MLE)|(} defined the model, we ask what makes a set of parameters $\mathbf{w}$ ``best'' for our purposes. We want parameters that fit the existing data and also generalize to new observations. We can formalize this balance with Bayesian inference.

Recall from Chapter 2 that we use $\boldsymbol{\theta}$ for the model parameters in general Bayesian inference. For linear regression we use $\mathbf{w}$ for the weight parameters, following the machine learning convention. The posterior distribution of the parameters is proportional to the product of the likelihood and the prior
\begin{equation}
p(\mathbf{w}|\mathcal{D}) \propto p(\mathcal{D}|\mathbf{w})p(\mathbf{w}).
\end{equation}
Here $\mathbf{w}$ is itself a random variable, and the posterior gives its probability distribution. The posterior describes our uncertainty about the true parameter values, given the data $\mathcal{D}$. It tells us the most likely parameter values and also how confident we can be in them.

The complete distribution is informative, but we often want a single ``best'' set of parameters. One common choice maximizes the posterior probability
\begin{equation}
\mathbf{w}_{\text{MAP}} = \arg\max_{\mathbf{w}} p(\mathbf{w}|\mathcal{D}).
\end{equation}
This is the Maximum\index{maximum a posteriori estimate (MAP)} A Posteriori (MAP) estimate. Here $\arg\max_{\mathbf{w}}$ is the value of $\mathbf{w}$ at which the function that follows is largest. Similarly, $\arg\min$ is the value at which it is smallest. We start with the simpler case of minimal prior knowledge, a so-called uninformative prior, and we take $p(\mathbf{w})$ to be approximately constant. Maximizing the posterior is then equivalent to maximizing the likelihood $p(\mathcal{D}|\mathbf{w})$ alone
\begin{equation}
\mathbf{w}_{\text{ML}} = \arg\max_{\mathbf{w}} p(\mathcal{D}|\mathbf{w}).
\end{equation}
This approach is maximum\index{maximum likelihood estimation (MLE)} likelihood estimation (MLE),\index{MLE|see{maximum likelihood estimation}} the simplest way to choose parameters, based purely on how well they explain the observed data.

\paragraph{The Gaussian Likelihood} The likelihood\index{likelihood!Gaussian} $p(\mathcal{D}|\mathbf{w})$ asks how likely we are to observe our actual data, given a proposed set of parameters $\mathbf{w}$, which MLE optimizes. Our data are pairs $(\mathbf{x}_i, t_i)$, where $t_i$ is an observed value.

We assume that each observation relates to the model prediction as
\begin{equation}
t = y(\mathbf{x}, \mathbf{w}) + \varepsilon, \text{ where } \varepsilon \sim \mathcal{N}(0, \sigma^2).
\end{equation}
Here we deliberately distinguish between $t$, the observed value, and $y(\mathbf{x}, \mathbf{w})$, the deterministic model prediction for input $\mathbf{x}$ and parameters $\mathbf{w}$. In Section~4.2, $y$ included the noise $\varepsilon$, but from here on only the observed value $t$ includes it. This noise follows a Gaussian distribution centered at zero with variance $\sigma^2$. It states our assumption that measurements scatter symmetrically around the true model values because of random uncertainties.

This treatment of uncertainty is more than a mathematical convenience, because it is necessary for meaningful Bayesian inference. If our data were perfectly certain ($\sigma = 0$), we would expect to know the model parameters with perfect precision. This does not occur in real astronomical observations. As we discussed in Chapter 2, the precision of our measurements limits our understanding of physical systems. This is why we need Bayesian statistics, to quantify and reason about these unavoidable uncertainties.

\begin{figure}[ht!]
    \centering
    \includegraphics[width=0.9\textwidth]{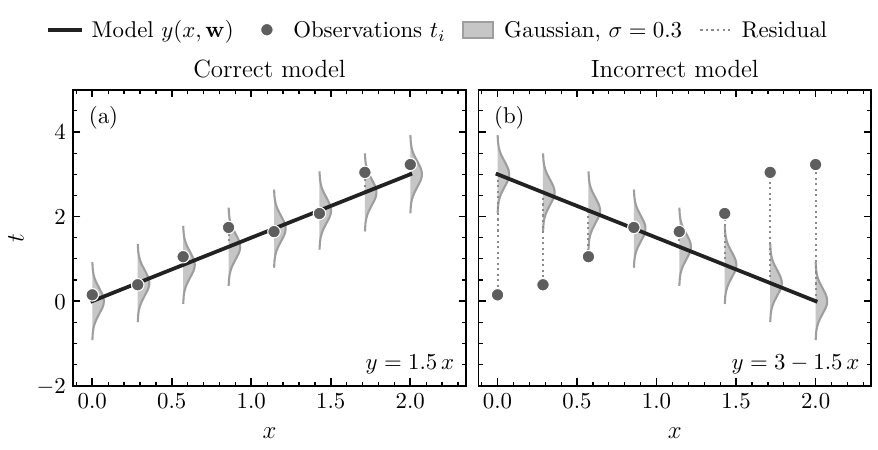}
    \caption{The likelihood of linear regression with Gaussian noise. The solid line is the model prediction $y(x, \mathbf{w})$, and the dots are eight observations $t_i$. At each point, the shaded profile is the Gaussian $\mathcal{N}(t \,|\, y, \sigma^2)$, centered on the prediction, and dotted lines mark the residuals. In panel (a), the model has appropriate coefficients, and the observations scatter within the expected uncertainty. In panel (b), the coefficients are incorrect, so the deviations from the model cannot be explained by the measurement uncertainty alone.}
    \label{fig:likelihood_viz}
\end{figure}

For a single observation, the likelihood of observing value $t$ at input $\mathbf{x}$ follows from our noise model. Because $t = y(\mathbf{x}, \mathbf{w}) + \varepsilon$ with $\varepsilon \sim \mathcal{N}(0, \sigma^2)$, the observation $t$ itself follows a Gaussian distribution centered at the model prediction $y(\mathbf{x}, \mathbf{w})$ with variance $\sigma^2$. We can write this explicitly with the Gaussian probability density function
\begin{align}
p(t | \mathbf{x}, \mathbf{w}, \sigma^2) &= \frac{1}{\sqrt{2\pi\sigma^2}} \exp\left(-\frac{(t - y(\mathbf{x}, \mathbf{w}))^2}{2\sigma^2}\right) \\*
&= \mathcal{N}(t | y(\mathbf{x}, \mathbf{w}), \sigma^2). \label{likelihood-single-data-point}
\end{align}

This equation formalizes our belief about where observations fall relative to the model predictions. The parameter $\sigma^2$ sets how quickly the probability falls off. Larger values of $\sigma^2$ mean greater measurement uncertainty and a more gradual decrease in probability away from the model prediction. With small $\sigma^2$ and the correct parameters $\mathbf{w}$, we expect the observations to cluster tightly around the model predictions, while larger $\sigma^2$ allows more scatter.

With incorrect parameters, we expect systematic deviations, where the predictions consistently miss the data points even after we account for measurement uncertainties. The Gaussians centered on the model predictions then fail to match the pattern of the observations, which tells us to adjust the model parameters.

Figure~\ref{fig:likelihood_viz} shows the likelihood for good and poor model parameters. The dotted lines mark the residuals,\index{residuals} the differences between the observations and the model predictions. In panel (a), the model predictions agree with the data within the expected uncertainties. In panel (b), they systematically miss the observations, which indicates poorly chosen parameters. By maximizing the likelihood, we expect the parameters to converge to values whose predictions best explain the observations, given the known measurement uncertainties.

Next, we consider $N$ data points instead of one. We assume that the observations are independent, so knowing the outcome of one measurement tells us nothing about the others. In the M-$\sigma$ relation, for example, the velocity dispersion we measure for one galaxy does not influence our measurement for another galaxy. This is often a reasonable assumption in astronomy, where we typically observe different objects in separate observations, each with its own random measurement uncertainties.

We also allow each measurement to have a different uncertainty $\sigma_i^2$. The likelihood of the dataset is the product of the individual likelihoods. This follows because independent events occur together with a probability equal to the product of their individual probabilities. Here $\mathbf{t}$ is the vector of all observed target values and $\mathbf{X}$ is the matrix of input features, with one row per data point
\begin{equation}
p(\mathbf{t} | \mathbf{X}, \mathbf{w}, \{\sigma_i^2\}) = \prod_{i=1}^N p(t_i | \mathbf{x}_i, \mathbf{w}, \sigma_i^2) = \prod_{i=1}^N \mathcal{N}(t_i | \mathbf{w}^T \boldsymbol{\phi}(\mathbf{x}_i), \sigma_i^2).
\end{equation}This product form combines the evidence from multiple independent measurements and weights each observation by its own uncertainty $\sigma_i^2$.

To make the product easier to handle, we take the logarithm of both sides. The logarithm is a monotonic function, meaning that it preserves the order of numbers, so maximizing the log-likelihood is equivalent to maximizing the likelihood itself. It also converts the product of probabilities into a sum, which is much easier to work with
\begin{equation}
\ln p(\mathbf{t} | \mathbf{X}, \mathbf{w}, \{\sigma_i^2\}) = \sum_{i=1}^N \ln \mathcal{N}(t_i | \mathbf{w}^T \boldsymbol{\phi}(\mathbf{x}_i), \sigma_i^2).
\end{equation}Expanding this with the explicit form of the Gaussian distribution and heterogeneous noise gives
\begin{equation}
\ln p(\mathbf{t} | \mathbf{X}, \mathbf{w}, \{\sigma_i^2\}) = -\frac{1}{2}\sum_{i=1}^N \ln\left(2\pi\sigma_i^2\right) - \sum_{i=1}^N \frac{\left(t_i - \mathbf{w}^T \boldsymbol{\phi}(\mathbf{x}_i)\right)^2}{2\sigma_i^2}.
\end{equation}

\paragraph{Weighted Least Squares} To\index{least squares!weighted|(} find the parameters $\mathbf{w}$ that maximize this log-likelihood, we need only the terms that depend on $\mathbf{w}$. The first term does not depend on $\mathbf{w}$, so we drop it and minimize the negative of the second term
\begin{equation}
\arg\max_\mathbf{w} \ln p(\mathbf{t} | \mathbf{X}, \mathbf{w}, \{\sigma_i^2\}) = \arg\min_\mathbf{w} \sum_{i=1}^N \frac{(t_i - \mathbf{w}^T \boldsymbol{\phi}(\mathbf{x}_i))^2}{2\sigma_i^2}.
\end{equation}This equation shows that maximizing the likelihood under heterogeneous Gaussian noise is equivalent to minimizing the weighted sum of squared errors. The weights are inversely proportional to the measurement variances, so more precise measurements (smaller $\sigma_i^2$) contribute more to the optimal parameters. These data weights differ from the model weights $\mathbf{w}$, the parameters that we fit. When $\sigma_i^2$ varies, which we call heteroscedastic\index{heteroscedastic noise} errors, standard unweighted least squares would wrongly give equal weight to all measurements, whatever their precision.

For homogeneous noise, where $\sigma_i = \sigma$ for all $i$, which we call homoscedastic,\index{homoscedastic noise} the equation simplifies to
\begin{equation}
\arg\min_\mathbf{w} \sum_{i=1}^N \frac{(t_i - \mathbf{w}^T \boldsymbol{\phi}(\mathbf{x}_i))^2}{2\sigma^2} = \frac{1}{2\sigma^2} \arg\min_\mathbf{w} \sum_{i=1}^N (t_i - \mathbf{w}^T \boldsymbol{\phi}(\mathbf{x}_i))^2.
\end{equation}Since $1/(2\sigma^2)$ is a constant scaling factor, minimizing this expression is equivalent to minimizing the unweighted sum of squared errors. This explains why standard least squares is appropriate when the measurement uncertainties are uniform.

We have seen that weighted\index{least squares!weighted} least squares emerges from maximum likelihood estimation once we account for the measurement uncertainties. Up to a factor of 2, the weighted sum of squared residuals that we derived is $\sum_{i=1}^N (t_i - \mathbf{w}^T \boldsymbol{\phi}(\mathbf{x}_i))^2/\sigma_i^2$, the chi-square\index{chi-square statistic} statistic, written $\chi^2$. It measures the goodness of fit between the model predictions and the data, with each squared residual weighted by its measurement uncertainty. Chi-square is widely used in astronomical model fitting. This is no arbitrary choice, since chi-square is a direct mathematical consequence of the Gaussian noise model.\index{least squares!weighted|)}

\section{Analytical Solution}\label{sec:ch4_analytical}

The quantity that we minimize to fit the model is the loss\index{loss function} function, also called the objective function. To find the optimal parameters $\mathbf{w}$, we start from our weighted least squares loss function
\begin{equation}
E(\mathbf{w}) = \sum_{i=1}^N \frac{(t_i - \mathbf{w}^T \boldsymbol{\phi}(\mathbf{x}_i))^2}{2\sigma_i^2}.
\end{equation}
It can be rewritten in matrix form
\begin{equation}
E(\mathbf{w}) = \frac{1}{2} (\mathbf{t} - \boldsymbol{\Phi}\mathbf{w})^T \mathbf{S}^{-1} (\mathbf{t} - \boldsymbol{\Phi}\mathbf{w}).
\end{equation}
Here $\mathbf{t} = (t_1, \ldots, t_N)^T \in \mathbb{R}^{N \times 1}$ is the vector of observed values and $N$ is the number of data points. The matrix $\boldsymbol{\Phi} \in \mathbb{R}^{N \times M}$ is the design\index{design matrix} matrix of regression analysis. It stacks the features of all data points, so that $\boldsymbol{\Phi}\mathbf{w}$ is the vector of the $N$ model predictions. Each row corresponds to one data point, and each column corresponds to one feature or predictor variable. The entry at row $i$ and column $j$ is the $j$th feature evaluated at the $i$th data point.

More formally, $\boldsymbol{\Phi} = [\boldsymbol{\phi}(\mathbf{x}_1), \ldots, \boldsymbol{\phi}(\mathbf{x}_N)]^T$, where $\boldsymbol{\phi}(\mathbf{x})$ holds the feature transformations (also called basis functions) that map each input $\mathbf{x}$ to an $M$-dimensional feature vector. For a quadratic fit, for example, $\boldsymbol{\phi}(\mathbf{x})$ maps a single input $x$ to the vector $(1, x, x^2)$, so $M=3$. For two input points $x_1$ and $x_2$, $\boldsymbol{\Phi}$ is
\begin{equation}
\boldsymbol{\Phi} = \begin{bmatrix} 1 & x_1 & x_1^2 \\ 1 & x_2 & x_2^2 \end{bmatrix}.
\end{equation}

Finally, $\mathbf{S} \in \mathbb{R}^{N \times N}$ is the diagonal matrix of measurement variances, $\text{diag}(\sigma_1^2, \ldots, \sigma_N^2)$, where $\sigma_i^2$ is the variance of the $i$th measurement. Its inverse is also diagonal, with elements $1/\sigma_i^2$, so $\mathbf{S}^{-1} = \text{diag}(1/\sigma_1^2, \ldots, 1/\sigma_N^2)$.

\paragraph{Deriving the Solution} The quadratic form has two important properties, familiar from simple quadratics such as $y = ax^2 + bx + c$. First, the function is convex (it opens upward when $a > 0$), so it has one global minimum, like a parabola with exactly one lowest point. Second, we can find the minimum analytically by ``completing the square'', a technique from basic algebra that rewrites $ax^2 + bx + c$ as $a(x + b/(2a))^2 + (c - b^2/(4a))$ to reveal the vertex. The same principle applies here, in higher dimensions.

We could complete the square, but taking derivatives is more direct and efficient to compute. It also avoids manipulating large matrices into a perfect square form. Expanding $E(\mathbf{w})$ gives
\begin{equation}
E(\mathbf{w}) = \frac{1}{2} (\mathbf{t}^T\mathbf{S}^{-1}\mathbf{t} - \mathbf{t}^T\mathbf{S}^{-1}\boldsymbol{\Phi}\mathbf{w} - \mathbf{w}^T\boldsymbol{\Phi}^T\mathbf{S}^{-1}\mathbf{t} + \mathbf{w}^T\boldsymbol{\Phi}^T\mathbf{S}^{-1}\boldsymbol{\Phi}\mathbf{w}).
\end{equation}

Since $\mathbf{t}^T\mathbf{S}^{-1}\boldsymbol{\Phi}\mathbf{w}$ is a scalar, it equals its transpose $\mathbf{w}^T\boldsymbol{\Phi}^T\mathbf{S}^{-1}\mathbf{t}$. Two rules of matrix calculus give the gradient. For a constant vector $\mathbf{a}$, the gradient of $\mathbf{a}^T\mathbf{w}$ is $\mathbf{a}$. For a symmetric matrix $\mathbf{A}$, the gradient of $\mathbf{w}^T\mathbf{A}\mathbf{w}$ is $2\mathbf{A}\mathbf{w}$. The matrix $\boldsymbol{\Phi}^T\mathbf{S}^{-1}\boldsymbol{\Phi}$ is symmetric because $\mathbf{S}^{-1}$ is diagonal. Taking the derivative with respect to $\mathbf{w}$ with these rules gives
\begin{equation}
\nabla_\mathbf{w} E(\mathbf{w}) = -\boldsymbol{\Phi}^T\mathbf{S}^{-1}\mathbf{t} + \boldsymbol{\Phi}^T\mathbf{S}^{-1}\boldsymbol{\Phi}\mathbf{w} = -\boldsymbol{\Phi}^T\mathbf{S}^{-1}(\mathbf{t} - \boldsymbol{\Phi}\mathbf{w}).
\end{equation}

Here the nabla symbol $\nabla_\mathbf{w}$ is the gradient operator with respect to the vector $\mathbf{w}$. It gives a vector of the partial derivatives of $E(\mathbf{w})$ with respect to each component of $\mathbf{w}$. For a vector $\mathbf{w} = (w_0, w_1, \ldots, w_{M-1})^T$, the gradient is
\begin{equation}
\nabla_\mathbf{w} E(\mathbf{w}) = \left(\frac{\partial E}{\partial w_0}, \frac{\partial E}{\partial w_1}, \ldots, \frac{\partial E}{\partial w_{M-1}}\right)^T.
\end{equation}

This gradient depends on $\mathbf{w}$. At the minimum of $E(\mathbf{w})$, which is the maximum of the likelihood, it must equal zero. Setting $\nabla_\mathbf{w} E(\mathbf{w}) = 0$ gives $\boldsymbol{\Phi}^T\mathbf{S}^{-1}\boldsymbol{\Phi}\mathbf{w} = \boldsymbol{\Phi}^T\mathbf{S}^{-1}\mathbf{t}$. Solving for $\mathbf{w}$ gives the maximum likelihood estimate
\begin{equation}
\mathbf{w}_{\text{ML}} = (\boldsymbol{\Phi}^T\mathbf{S}^{-1}\boldsymbol{\Phi})^{-1}\boldsymbol{\Phi}^T\mathbf{S}^{-1}\mathbf{t}. \label{eq:wml_weighted}
\end{equation}

For homogeneous noise, where $\sigma_i = \sigma$ for all $i$, the matrix $\mathbf{S}$ simplifies to $\sigma^2\mathrm{Id}$, where $\mathrm{Id}$ is the identity matrix. Substituting this into our solution gives
\begin{align}
\mathbf{w}_{\text{ML}} &= (\boldsymbol{\Phi}^T(\sigma^2\mathrm{Id})^{-1}\boldsymbol{\Phi})^{-1}\boldsymbol{\Phi}^T(\sigma^2\mathrm{Id})^{-1}\mathbf{t} \\*
&= (\boldsymbol{\Phi}^T\frac{1}{\sigma^2}\mathrm{Id}\boldsymbol{\Phi})^{-1}\boldsymbol{\Phi}^T\frac{1}{\sigma^2}\mathrm{Id}\mathbf{t} \\
&= (\frac{1}{\sigma^2}\boldsymbol{\Phi}^T\boldsymbol{\Phi})^{-1}\frac{1}{\sigma^2}\boldsymbol{\Phi}^T\mathbf{t} \\
&= \sigma^2(\boldsymbol{\Phi}^T\boldsymbol{\Phi})^{-1}\frac{1}{\sigma^2}\boldsymbol{\Phi}^T\mathbf{t} \\*
&= (\boldsymbol{\Phi}^T\boldsymbol{\Phi})^{-1}\boldsymbol{\Phi}^T\mathbf{t}.
\end{align}This simplified form is the ordinary\index{least squares!ordinary} least squares solution
\begin{equation}
\mathbf{w}_{\text{ML}} = (\boldsymbol{\Phi}^T\boldsymbol{\Phi})^{-1}\boldsymbol{\Phi}^T\mathbf{t}.
\end{equation}
The general weighted solution thus reduces to the unweighted case when the measurement uncertainties are uniform.

As a concrete example, consider fitting a straight line $y = mx + b$ to data points $(x_i, y_i)$. For this model, $\mathbf{w} = (b,m)^T$ holds the intercept and slope, and $\boldsymbol{\Phi}$ is an $N \times 2$ matrix whose rows are $(1, x_i)$. The vector $\mathbf{t}$ holds the observed $y$ values. The matrix products in the solution then reduce to simple sums over the data
\begin{equation}
\boldsymbol{\Phi}^T\boldsymbol{\Phi} = \begin{pmatrix} N & \sum x_i \\ \sum x_i & \sum x_i^2 \end{pmatrix}, \quad \boldsymbol{\Phi}^T\mathbf{t} = \begin{pmatrix} \sum y_i \\ \sum x_iy_i \end{pmatrix}.
\end{equation}
For uniform uncertainties, inverting this $2 \times 2$ matrix gives the familiar formulas
\begin{equation}
m = \frac{N\sum x_iy_i - (\sum x_i)(\sum y_i)}{N\sum x_i^2 - (\sum x_i)^2}, \quad b = \frac{\sum y_i - m\sum x_i}{N}.
\end{equation}

When the measurement uncertainties $\sigma_i$ vary between points, the weighted least squares solution instead gives
\begin{equation}
m = \frac{\sum \omega_ix_iy_i - (\sum \omega_ix_i)(\sum \omega_iy_i)/\Omega}{\sum \omega_ix_i^2 - (\sum \omega_ix_i)^2/\Omega}, \quad b = \frac{\sum \omega_iy_i - m\sum \omega_ix_i}{\Omega},
\end{equation}
where $\omega_i = 1/\sigma_i^2$ and $\Omega = \sum \omega_i$. This weights more precise measurements (smaller $\sigma_i$) more heavily in the fit.

\paragraph{Properties of the Solution} This analytical solution has several properties that make linear regression useful in astronomy. First, it gives an exact solution without iterative numerical optimization, even with heterogeneous measurement uncertainties. Second, its mathematical form holds whatever the dimensionality of the feature space, as long as the matrix $\boldsymbol{\Phi}^T\mathbf{S}^{-1}\boldsymbol{\Phi}$ can be inverted. The inverse exists only when the columns of $\boldsymbol{\Phi}$ are linearly independent, which requires $N \ge M$. This is especially useful for high-dimensional astronomical data, such as spectroscopic observations where each spectrum contains thousands of flux measurements at different wavelengths.

The cost of the computation also scales well with the size of the dataset. The matrix $\boldsymbol{\Phi}^T$ is $M \times N$, $\mathbf{S}^{-1}$ is $N \times N$, and $\boldsymbol{\Phi}$ is $N \times M$, so their product is an $M \times M$ matrix that we must invert. Since $\mathbf{S}$ is diagonal, forming this product takes a number of operations proportional to $NM^2$, which grows only linearly with the number of data points. The inversion takes a number of operations proportional to $M^3$. This depends on the number of features $M$ and not on the number of data points. With millions of data points and only a few hundred features, for example, both steps remain feasible.

\section{Noise Parameter Estimation}

So\index{noise estimation|(} far we have treated the measurement uncertainties $\sigma_i$ as known. In many astronomical applications, however, we must estimate them from the data. In spectroscopy, for example, the quoted instrumental uncertainties may miss systematic effects from sky subtraction or flux calibration. In photometric time series, the formal uncertainties often underestimate the true scatter because of unmodeled stellar variability or atmospheric effects.

In some cases, the spread in the data also shows physically meaningful variation as well as measurement error. The intrinsic dispersion in a mass-metallicity relation, for example, could indicate real diversity in galaxy evolution histories. Scatter in a period-luminosity relation might reveal real differences in stellar properties. Modeling this intrinsic\index{intrinsic scatter} scatter properly can give useful physical insight beyond accounting for the measurement uncertainties.

From a maximum likelihood perspective, we can include the measurement uncertainties as additional model parameters. If we denote the model parameters by $\mathbf{w}$ and the measurement uncertainties by $\boldsymbol{\sigma}$, we maximize $p(\mathbf{t}|\mathbf{X}, \mathbf{w}, \boldsymbol{\sigma})$ with respect to both $\mathbf{w}$ and $\boldsymbol{\sigma}$ simultaneously. Both the model parameters and the noise\index{noise estimation} characteristics shape the distribution of the observed data.

\begin{figure}[h]
    \centering
    \includegraphics[width=0.9\textwidth]{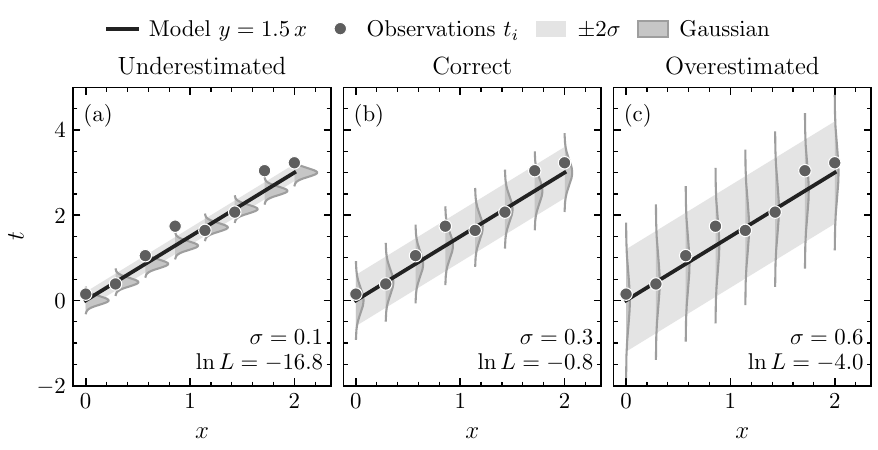}
    \caption{The effect of the assumed uncertainty on the likelihood. In panel (a), $\sigma$ is too small. Points far from the model give very small likelihood terms, so the total likelihood is low. In panel (b), $\sigma$ matches the noise that generated the data and gives the highest likelihood of the three values. In panel (c), $\sigma$ is too large. The Gaussians cover all the points, but their height scales as $1/\sigma$, so each likelihood term is smaller.}
    \label{fig:uncertainty_est}
\end{figure}

Figure~\ref{fig:uncertainty_est} shows that the estimate of the measurement uncertainties matters for the likelihood. When the uncertainties are underestimated (panel a), many data points fall outside the predicted uncertainty bands. Each such point contributes a small likelihood term $\mathcal{N}(t_i|y(\mathbf{x}_i, \mathbf{w}), \sigma^2)$ because the Gaussian falls off rapidly, so the overall likelihood is low. When the uncertainties are overestimated (panel c), the Gaussians are very broad, and more points fall within the expected range. The prefactor $1/\sqrt{2\pi\sigma^2}$ then becomes very small and again lowers the overall likelihood. The maximum likelihood value of $\sigma$ lies close to the choice in panel (b), where these two effects balance.

We must model the uncertainty with care. Letting each data point have its own independent uncertainty $\sigma_i$ would add as many parameters as there are data points and leave the model ill-constrained. To see why, recall that for a Gaussian likelihood each data point contributes a term $(t_i - y_i)^2/\sigma_i^2$ to $\chi^2$. If each $\sigma_i$ could vary independently, setting $\sigma_i = |t_i - y_i|$ would give $\chi^2 = N$ for any model parameters. Each point would then contribute exactly 1, however well or poorly the model fits the data. We could not compare model fits meaningfully, and the model would have no predictive power, because the uncertainties would adjust to fit any predictions.

\paragraph{Estimating a Single Noise Level} To make the problem well defined, we therefore need a noise model with fewer parameters than data points. We might assume, for instance, that the measurement uncertainty varies smoothly with the input features. As an illustration, we consider the simplest scenario, a homogeneous but unknown measurement uncertainty, where all measurements share the same unknown $\sigma$.

We write the log-likelihood with the precision $\beta = 1/\sigma^2$ to simplify the derivation
\begin{equation}
\ln p(\mathbf{t} | \mathbf{X}, \mathbf{w}, \beta) = \frac{N}{2} \ln \beta - \frac{N}{2} \ln(2\pi) - \frac{\beta}{2} \sum_{i=1}^N (t_i - \mathbf{w}^T \boldsymbol{\phi}(\mathbf{x}_i))^2.
\end{equation}

Previously, we treated $\sigma$ as a fixed, known quantity while optimizing for $\mathbf{w}$. Now we treat $\sigma$ as another parameter of the model, just like $\mathbf{w}$.

To find the optimal value of $\beta$, we follow the same approach as for $\mathbf{w}$. We differentiate the log-likelihood with respect to $\beta$ and set the result to zero
\begin{equation}
\frac{\partial}{\partial \beta} \ln p(\mathbf{t} | \mathbf{X}, \mathbf{w}, \beta) = \frac{N}{2\beta} - \frac{1}{2} \sum_{i=1}^N (t_i - \mathbf{w}^T \boldsymbol{\phi}(\mathbf{x}_i))^2 = 0.
\end{equation}Solving for $\beta$ and converting back to $\sigma^2$ gives
\begin{equation}
\frac{1}{\beta_{\text{ML}}} = \sigma^2_{\text{ML}} = \frac{1}{N} \sum_{i=1}^N (t_i - \mathbf{w}^T \boldsymbol{\phi}(\mathbf{x}_i))^2.
\end{equation}

This equation gives an intuitive estimate of the measurement variance, the average of the squared residuals between the observations and the model predictions. If the model fits the data well, the remaining discrepancies between predictions and observations should reflect the underlying measurement noise. Averaging the squared residuals then estimates the typical squared deviation (variance) in the measurements. This is analogous to estimating the uncertainty from repeated measurements of one quantity, where we look at the scatter around the mean. Here we compare with the model predictions instead of a mean, but the principle is the same. Small residuals suggest precise measurements (small $\sigma^2$), and larger residuals indicate more uncertain ones (large $\sigma^2$).

\paragraph{Estimating the Weights and the Noise Together} However, the likelihood $p(\mathbf{t} | \mathbf{X}, \mathbf{w}, \beta)$ depends jointly on $\mathbf{w}$ and $\beta$. To find the true maximum likelihood solution, we need to maximize over both parameters simultaneously. In mathematical terms, we seek
\begin{equation}
(\mathbf{w}_{\text{ML}}, \beta_{\text{ML}}) = \arg\max_{\mathbf{w}, \beta} p(\mathbf{t} | \mathbf{X}, \mathbf{w}, \beta).
\end{equation}

Our problem has a special structure that makes this tractable. We have already shown that for any fixed $\beta$ we can find $\mathbf{w}_{\text{ML}}(\beta)$ analytically. Similarly, for any fixed $\mathbf{w}$ we can find $\beta_{\text{ML}}(\mathbf{w})$ analytically. This suggests the following iterative approach.

\begin{enumerate}
    \item Start with an initial guess for $\beta$
    \item Find $\mathbf{w}_{\text{ML}}(\beta)$ with the least squares solution
    \item Use this $\mathbf{w}_{\text{ML}}(\beta)$ to find $\beta_{\text{ML}}(\mathbf{w})$ with the variance estimation formula
    \item Repeat steps 2-3 until convergence, meaning until $\mathbf{w}$ and $\beta$ stop changing
\end{enumerate}

\begin{figure}[ht!]
    \centering
    \includegraphics[width=0.7\textwidth]{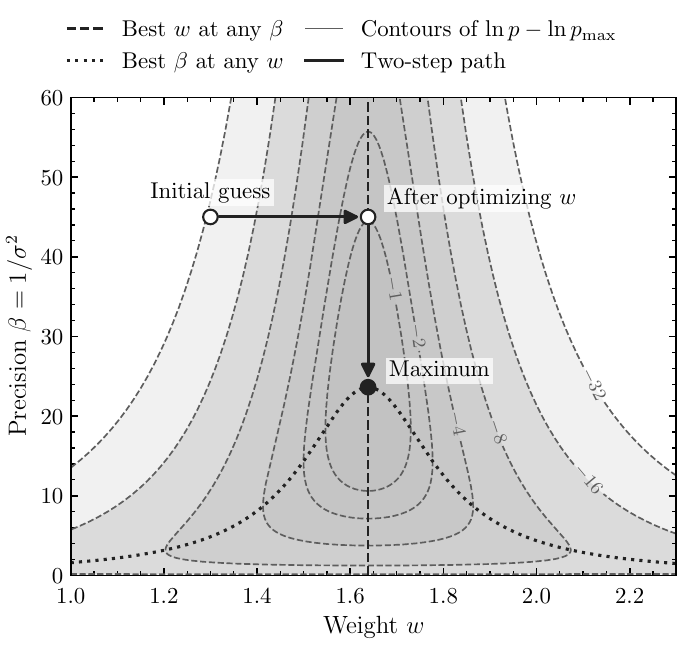}
    \caption{The exact log-likelihood of the model $t = wx$ for the eight observations of the previous figures, as a function of the weight $w$ and the precision $\beta$. The surface has a single maximum. The best $w$ is the same for every $\beta$, and the best $\beta$ for a given $w$ lies on the dotted curve. Optimizing $w$ at any starting $\beta$ and then optimizing $\beta$ reaches the maximum in two steps.}
    \label{fig:likelihood_contour}
\end{figure}

In this specific case, the iteration converges whatever the initial guess for $\beta$. To see why, we examine the structure of the log-likelihood
\begin{equation}
\ln p(\mathbf{t} | \mathbf{X}, \mathbf{w}, \beta) = \frac{N}{2} \ln \beta - \frac{N}{2} \ln(2\pi) - \frac{\beta}{2} \sum_{i=1}^N (t_i - \mathbf{w}^T \boldsymbol{\phi}(\mathbf{x}_i))^2.
\end{equation}

First, each step of the iteration has a single best answer. With $\sigma^2$ fixed, the exponent contains squared distances between the predictions and the observations. The log-likelihood is therefore concave in $\mathbf{w}$ and has a single peak. Away from the peak, the likelihood falls off as a Gaussian.

Similarly, varying $\sigma^2$ while holding $\mathbf{w}$ fixed gives another single peak. A very small $\sigma^2$ makes the Gaussians too narrow, so points that do not match the predictions exactly get exponentially small likelihoods. A very large $\sigma^2$ makes the Gaussians too wide, and the normalization term $1/\sqrt{2\pi\sigma^2}$ drives the likelihood toward zero. The likelihood must therefore peak at an intermediate value where these competing effects balance.

The two parameters are also connected in a simple way. The optimal $\mathbf{w}_{\text{ML}}$ depends only on the relative weighting of the squared residuals, which is independent of the scale of $\beta$. For homogeneous noise, the best $\mathbf{w}$ is therefore the same for every $\beta$, so the global maximum must have $\mathbf{w} = \mathbf{w}_{\text{ML}}$. The optimal $\beta$ then follows directly from this $\mathbf{w}$, so one pass through steps 2 and 3 reaches the single global maximum. This takes two steps, as Figure~\ref{fig:likelihood_contour} shows.

Putting these results together, we obtain the maximum likelihood solution in a simple two-step process. First, we estimate $\mathbf{w}_{\text{ML}}$ with the least squares solution
\begin{equation}
\mathbf{w}_{\text{ML}} = (\boldsymbol{\Phi}^T\boldsymbol{\Phi})^{-1}\boldsymbol{\Phi}^T\mathbf{t}.
\end{equation}Then, using this $\mathbf{w}_{\text{ML}}$, we compute $\sigma^2_{\text{ML}}$ directly as
\begin{equation}
\sigma^2_{\text{ML}} = \frac{1}{N} \sum_{i=1}^N (t_i - \mathbf{w}_{\text{ML}}^T \boldsymbol{\phi}(\mathbf{x}_i))^2.
\end{equation}This estimate of $\sigma^2$ is the average squared deviation between the observations and the model predictions. It gives the typical size of the noise in our data, whether instrumental or intrinsic. This assumes that the residual differences come mainly from noise rather than from model inadequacies, which is reasonable for a model that fits well.

A common case in astronomy has known measurement uncertainties $\sigma_i$ and an intrinsic\index{intrinsic scatter} scatter $\sigma_{\rm int}$ that we must estimate. The total variance of point $i$ is then $\sigma_i^2 + \sigma_{\rm int}^2$. We replace $\sigma_i^2$ by this sum in the heterogeneous log-likelihood and maximize over $\mathbf{w}$ and $\sigma_{\rm int}$. For a fixed $\sigma_{\rm int}$, the weights are $1/(\sigma_i^2 + \sigma_{\rm int}^2)$, and $\mathbf{w}$ follows from weighted least squares. For a fixed $\mathbf{w}$, $\sigma_{\rm int}$ has no closed form, so we alternate the two steps or optimize numerically. If the known uncertainties explain the scatter, the maximum lies at\index{maximum likelihood estimation (MLE)|)}\index{noise estimation|)} $\sigma_{\rm int} = 0$.

\section{Limitations of Maximum Likelihood Approach}\label{sec:ch4_limitations}

Maximum likelihood estimation gives a well-defined way to find the best model parameters and even to estimate the measurement uncertainties. It also has three serious limitations.

The first and most critical limitation is that MLE alone reports a best-fit model, which falls short of a basic principle of science. In science, we want predictive models and also a clear understanding of their uncertainties and limits. We need the uncertainty of the model parameters and of the predictions we make with them.

When we compute $\mathbf{w}_{\text{ML}}$, we obtain a single point estimate, our best guess for the model parameters. This estimate does not tell us how confident we should be in it. Are the parameters well constrained by the data, or might a wide range of values be equally plausible? Without uncertainty quantification, we cannot answer these questions.

Consider our earlier M-$\sigma$ relation example. Suppose we find $\gamma = 4.0$ for the power-law index. MLE alone does not tell us whether the data constrain this value to within $\pm 0.1$ or allow a range from $3.0$ to $5.0$. We need this uncertainty to compare our results with theoretical predictions or with measurements from other studies.

Residuals and goodness-of-fit statistics tell us how well the model fits, but they do not give uncertainties on the parameters. Without them, we cannot propagate errors through later calculations or compare models with different numbers of parameters. Other tools can add them, such as the bootstrap of Chapter 3, which resamples the data. Such tools are extra steps on top of the point estimate. The bootstrap, for example, has practical limits that Chapter 5 discusses.

The second limitation is that MLE assumes no prior knowledge and treats the data alone as sufficient to determine the parameters. In astronomy, we often have prior knowledge that could improve our estimates. We might know physical constraints on the parameter values, or have results from previous studies that suggest likely ranges. In stellar astrophysics, for example, stellar masses must be positive and typically fall within ranges set by stellar evolution theory. In cosmology, nucleosynthesis or structure formation can limit the parameter ranges. MLE has no way to include such knowledge.

The third limitation is that MLE does not guard against overfitting\index{overfitting} on its own. It can produce models that fit the training data well but generalize poorly to new observations. Overfitting is most likely when there are nearly as many parameters as data points, or when the basis functions $\boldsymbol{\phi}(\mathbf{x})$ can follow very complex curves. Our analytical solutions only find the parameters that best explain the observed data. They do not check how well the predictions hold for new data.

The first two limitations need a more complete approach, which we introduce at the end of this chapter. For the third, the next section shows how to detect overfitting with test data.

\section{Model Evaluation and Cross-Validation}
\label{sec:cross_validation}

This section addresses the third limitation of maximum likelihood, overfitting. Overfitting can arise with other approaches too. Each model needs a choice of complexity, such as the number of basis functions, and a test of whether it generalizes. Cross-validation is a standard tool for model evaluation and selection, and we illustrate it with linear regression.

A central question in machine learning is what makes a good model. We explore it with polynomial regression. Recall from our discussion of basis functions that we can write a linear regression model as
\begin{equation}
y(\mathbf{x}, \mathbf{w}) = \sum_{j=0}^{M-1} w_j \cdot \phi_j(\mathbf{x}) = \mathbf{w}^T \boldsymbol{\phi}(\mathbf{x}).
\end{equation}

For polynomial regression, we choose polynomial basis functions $\phi_0(x) = 1$, $\phi_1(x) = x$, $\phi_2(x) = x^2$, and so on. This gives
\begin{equation}
y(x, \mathbf{w}) = w_0 + w_1x + w_2x^2 + \ldots + w_{M-1}x^{M-1},
\end{equation}
where the weights $\mathbf{w}$ come from our maximum likelihood solution $\mathbf{w}_{\text{ML}} = (\boldsymbol{\Phi}^T\boldsymbol{\Phi})^{-1}\boldsymbol{\Phi}^T\mathbf{t}$. This is the linear regression we derived, with polynomial basis functions. This choice of transformations is called feature\index{feature engineering} engineering.

How many polynomial terms should the basis functions $\boldsymbol{\phi}(\mathbf{x})$ include? We choose the dimension $M$ of the feature space, and so the structure of the design matrix $\boldsymbol{\Phi}$. This question is an example of the bias-variance tradeoff, which we explain below. It also connects to the overfitting limitation of maximum likelihood.

We can measure how well the model does with the mean squared error
\begin{equation}
\text{MSE} = \frac{1}{N}\sum_{i=1}^N(t_i - y(\mathbf{x}_i, \mathbf{w}))^2.
\end{equation}
This is the squared-error loss that follows from our homoscedastic Gaussian noise assumption in maximum likelihood estimation.

\begin{figure}[ht!]
    \centering
    \includegraphics[width=0.9\textwidth]{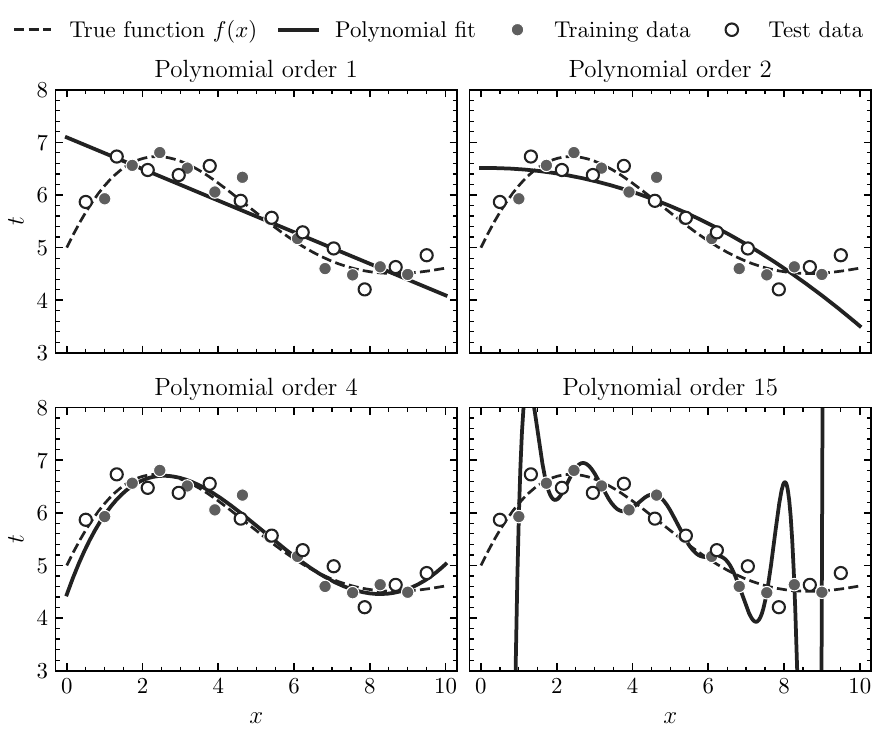}
    \caption{Polynomial fits of increasing order to 12 training points. The fits of orders 1 and 2 are too rigid to follow the pattern in the data. Order 4 gives a good balance. Order 15 follows the noise in the training data and swings far from the data between the points. It has more coefficients than training points, so the fit is not unique, and the curve shown is the one that a least squares routine returns.}
    \label{fig:polynomial_fits}
\end{figure}

\paragraph{Underfitting and Overfitting} Modeling\index{overfitting|(} uses prior knowledge and assumptions, and learning finds patterns in data. Machine learning balances the two. As we add polynomial basis\index{basis functions} functions, we increase the complexity of the feature space, and we see three regimes (Figure~\ref{fig:polynomial_fits}).

First, underfitting\index{underfitting} occurs when the basis functions are too limited to follow the patterns in the data. A purely linear model applied to a nonlinear astronomical relationship is an example. Second, in the optimal balance the model follows the true underlying relationship, and measurement noise has little effect on the fit. Finally, overfitting occurs when the feature space becomes so complex that the model fits the measurement noise instead of the signal.

Good models generalize to new observations, and this is what separates the three regimes. This principle is basic to scientific inference. Our models should make accurate predictions about future data, and low residuals on existing measurements are not enough.\index{overfitting|)}

\paragraph{Training and Test Sets} The solution is to split our dataset into separate portions. We use a training\index{training set} set to fit the model parameters. For each choice of basis functions, we apply our maximum likelihood solution $\mathbf{w}_{\text{ML}} = (\boldsymbol{\Phi}^T\boldsymbol{\Phi})^{-1}\boldsymbol{\Phi}^T\mathbf{t}$ to find the optimal weights. We hold out a test\index{test set} set during training to evaluate how well the model generalizes. We compute the mean\index{mean squared error} squared error on each set, which gives the training\index{training error} error and the test\index{test error} error.

The test data played no role in the parameter estimation. The test error is therefore an unbiased estimate of how the model will do on future data.

This split shows why the errors on training and test data behave differently. On the training data, adding degrees of freedom (more polynomial basis functions) does not worsen the fit and usually improves it. With more terms, the larger feature space gives the maximum likelihood solution more freedom to reduce the training error. At some point, however, we begin to overfit the specific characteristics of the training set, even though the fit to the training data keeps improving.

When overfitting occurs, the model memorizes noise and peculiarities of the training data instead of the true underlying relationship. These noise patterns are artifacts of our particular measurements and say nothing about the physical system we are studying. They do not generalize to new observations, which leads to larger errors on the test set.

\begin{figure}[ht!]
    \centering
    \includegraphics[width=0.9\textwidth]{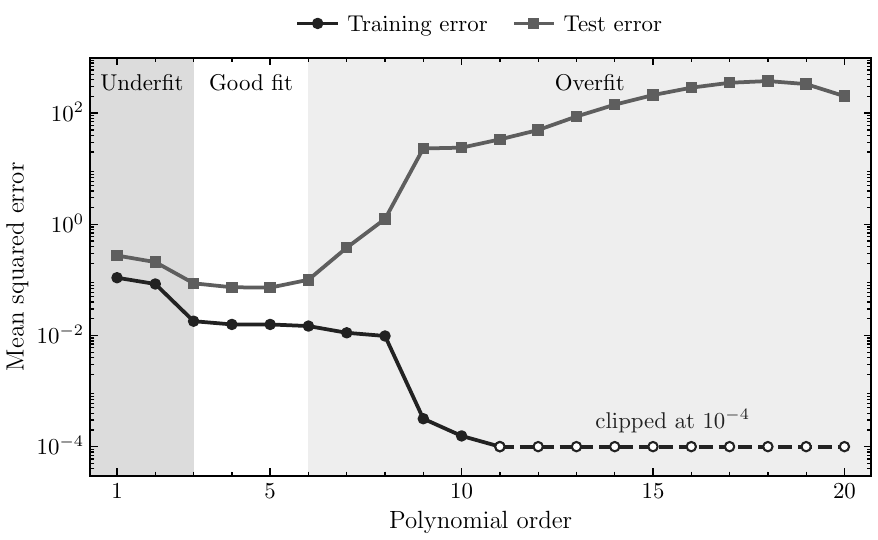}
    \caption{Training and test error as a function of polynomial order, on a logarithmic scale. Shading marks the three regimes, underfit (orders 1 and 2), good fit (3 to 6), and overfit (above 6). The training error decreases with order, and the test error is U-shaped with its minimum at order 5. In the overfit regime, the training error becomes very small while the test error grows. The training curve is clipped from order 11 (open markers), where the polynomial passes through all 12 training points.}
    \label{fig:learning_curves}
\end{figure}

Figure~\ref{fig:learning_curves} shows the result. The training error does not increase as we add basis functions, but the test error follows a U-shaped curve. With too few basis functions (underfitting), both errors are high because the model is too rigid. In the optimal regime, the test error reaches its minimum while the training error is already low. In the overfitting regime, the training error keeps decreasing while the test error increases.

\paragraph{The Bias-Variance Tradeoff} The\index{bias-variance tradeoff|(} bias-variance\index{bias-variance tradeoff} tradeoff explains this U-shape. Bias is the systematic error that remains when the model family cannot represent the true relationship. It is not the intercept $w_0$, which we also called the bias term. A straight line fitted to a curved relation has high bias, and more data do not remove it. Variance measures how much the fitted curve changes from one training set to another. Each training set has its own measurement noise, so a model with many parameters gives a different curve for each set.

Low-order polynomials have high bias and low variance, so they underfit (orders 1 and 2 in Figure~\ref{fig:polynomial_fits}). High-order polynomials have low bias and high variance, so they overfit (order 15). Higher orders lower the bias and raise the variance. The test\index{test error} error depends on both, so it is lowest at an intermediate order (order 5 in Figure~\ref{fig:learning_curves}).

More complex models fit the training data better but generalize worse. This tension is basic to machine learning. Recent work on neural networks has challenged this picture. The ``double\index{double descent} descent'' phenomenon shows that highly overparameterized models can sometimes generalize better than expected, a topic we explore in Chapter 15. Such a model has more parameters than data points. For polynomial regression and many classical methods, however, the U-shaped test error curve remains standard.\index{bias-variance tradeoff|)}

\paragraph{Three-Way Data Splitting} For a more careful procedure, we can use a three-way split. The training set fits the weight vector $\mathbf{w}$ with our maximum likelihood solution for each choice of basis functions. The validation\index{validation set} set determines the dimensionality of the feature space, meaning how many polynomial terms to include in $\boldsymbol{\phi}(\mathbf{x})$. The test set then gives a final evaluation of the chosen model on unseen data. Model selection thus rests on the validation set, and the test set gives an unbiased estimate of how well the model generalizes.

There is a tension between having enough training data for reliable parameter estimation and enough validation or test data for trustworthy evaluation. A common split uses 60\% of the data for training, 20\% for validation, and 20\% for testing. The proportions should depend on the total dataset size and the specific problem.

In astronomy, data can be scarce, so we often use only training and test sets. This is a compromise, because model selection then depends on the test set. It is still a reasonable balance between statistical accuracy and practical constraints.

\paragraph{Cross-Validation for Limited Data} When\index{cross-validation|(} datasets are small, cross-validation\index{cross-validation} is an alternative that uses the available data more efficiently. In Leave-One-Out\index{cross-validation!leave-one-out (LOOCV)} Cross-Validation (LOOCV),\index{LOOCV|see{cross-validation, leave-one-out}} we apply our maximum likelihood solution $N$ times. Each time, we train on $N-1$ observations and hold out the remaining one as the test set. Each observation is held out exactly once, and we average the test errors to estimate how well the model predicts.

This approach makes full use of the data, since nearly all observations train each model. It also gives $N$ estimates of how well the model generalizes. However, LOOCV usually trains $N$ models, which can be slow. Because each test set has only one observation, individual test results can be noisy. The training sets differ by one observation, so the $N$ results are highly correlated. Their average can vary widely between datasets.

K-fold\index{cross-validation!K-fold} cross-validation is a practical compromise. It divides the data into $K$ folds, trains on $K-1$ folds, and tests on the remaining fold. We rotate through the folds and average the test errors. This reduces computation while keeping many advantages of LOOCV.

Cross-validation in astronomy often needs care with the structure of the data. For time-series observations, random splitting can leak information from the future into predictions of the past. Time-aware splitting may be more appropriate. When observations cluster spatially, random splitting may not test predictions for new regions of the sky. Selection effects and systematic uncertainties also affect how we interpret the results.

Cross-validation is a practical tool for any fit. Later chapters give other ways to compare models, such as information criteria for mixture models in Chapter 11. Information criteria balance how well a model fits against its number of parameters. Cross-validation only checks generalization. It gives no parameter uncertainties and ignores priors.

A more complete approach is Bayesian inference. MLE asks which parameter values best explain the observed data. Bayesian inference treats the parameters as random variables and asks for their complete probability distribution, given the data and our prior knowledge. Instead of a point estimate like $\mathbf{w}_{\text{ML}}$, it gives the posterior distribution $p(\mathbf{w}|\mathbf{t})$. The posterior gives the uncertainties in the model parameters in one calculation, with no separate recipe. We then understand the precision and limits of our results.

The posterior also gives predictions with uncertainty estimates, through the posterior predictive distribution (Section~5.7), the distribution of a new observation given our data. It separates aleatoric uncertainty, the randomness of the measurements, from epistemic uncertainty, our limited knowledge of the parameters (Section~3.8). Bayesian inference includes prior knowledge formally, so we can build on previous results and theoretical constraints. The same priors act as regularization, a penalty on large parameter values that guards against overfitting. The width of the prior is a hyperparameter, and we still set it on validation data (Section~\ref{sec:cross_validation}). In astronomy, understanding the uncertainties is often as important as the estimates themselves, and Chapter 5 develops this approach.\index{cross-validation|)}\index{linear regression|)}

\section{Tutorials: Stellar Spectra and Calibration}\label{sec:ch4_tutorials}

Two tutorials apply the least squares fit of this chapter to astronomy. The first predicts the temperature of a star from its spectrum, with a prior and test stars. The second removes instrumental and atmospheric effects from radial velocities, with the bootstrap. The full tutorials can be found at \url{https://statml.org/} as 4a and 4b.

\begin{figure}[ht!]
    \centering
    \includegraphics[width=0.95\textwidth]{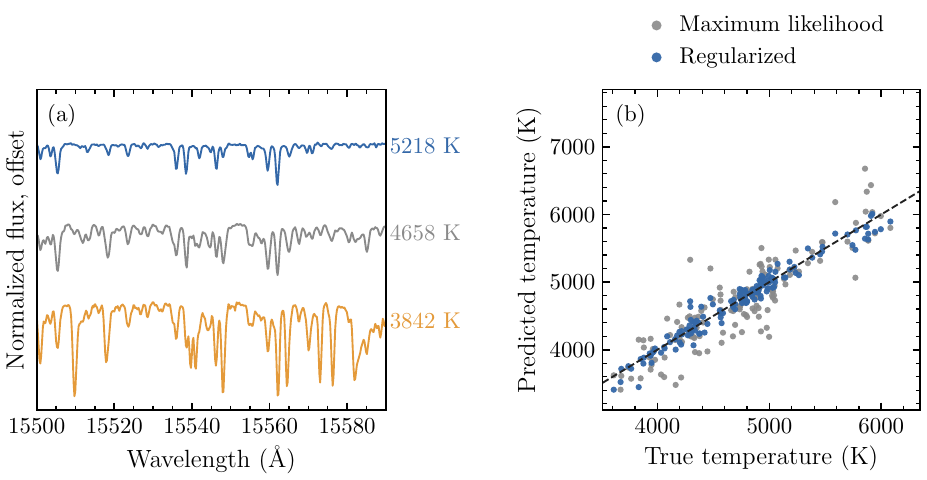}
    \caption{The first tutorial predicts the temperature of a star from its spectrum. Panel~(a) shows the spectra of three stars, labeled with their temperatures, in a short wavelength range, offset vertically. Panel~(b) shows predicted against true (pipeline) temperatures of the test stars. The dashed line marks perfect prediction. The maximum likelihood solution (grey) is poor, and the regularized solution (blue) stays closer to the line.}
    \label{fig:tutorial_spectra}
\end{figure}

\paragraph{Stellar Temperatures from Spectra} A\index{stellar spectra|(}\index{stellar temperatures|(} spectroscopic survey such as APOGEE\index{APOGEE survey} records the infrared spectra\index{stellar spectra} of hundreds of thousands of stars. The strength of the absorption lines in a spectrum depends on the temperature, the gravity and the composition of the stellar atmosphere. Temperature, for example, sets the ionization state of the atoms and so the strength of their lines. Such parameters for many stars map the chemistry of the Milky Way. The lines are deeper in the spectra of cooler stars (Figure~\ref{fig:tutorial_spectra}a).

Linear regression predicts the temperature from the spectrum. The flux at each wavelength is a feature, and the temperature is the quantity to predict. The prediction is a weighted sum of the fluxes, with a weight for each wavelength. The analytical solution of Section~\ref{sec:ch4_analytical} gives these weights without iteration. The training stars have temperatures from the survey pipeline, and the fitted model predicts the temperatures of other stars from their spectra.

A spectrum has so many wavelengths that the model has almost as many weights as there are training stars. With so many weights, the model can follow each training star, including its noise. Section~\ref{sec:ch4_limitations} explains why maximum likelihood does not guard against overfitting here. We hold out some stars as test stars, as in Section~\ref{sec:cross_validation}. The maximum likelihood fit matches the training stars almost exactly but predicts the test stars less well (Figure~\ref{fig:tutorial_spectra}b).

The remedy is a prior on the weights that makes large values unlikely. As the end of Section~\ref{sec:cross_validation} explains, such a prior discourages large weights, which is called regularization. In practice, a third group of stars, the validation stars, sets how much large weights are discouraged. The test stars then remain an unbiased check (Section~\ref{sec:cross_validation}). The regularized fit predicts the test stars better (Figure~\ref{fig:tutorial_spectra}b). Regularization guards against overfitting, especially with many input features (Section\index{stellar spectra|)}\index{stellar temperatures|)}~\ref{sec:cross_validation}).

\paragraph{Calibrating Telescope Measurements} Many\index{telescope calibration|(} planets are found through the tiny wobbles that they cause in their host stars. The radial velocities of the host stars, their velocities along the line of sight, must be very precise. At this precision the measured velocity depends on the observing conditions. The changes from the instrument and the atmosphere are called systematic effects. Uncorrected, they can exceed the wobbles, so they must be removed.

\begin{figure}[ht!]
    \centering
    \includegraphics[width=0.95\textwidth]{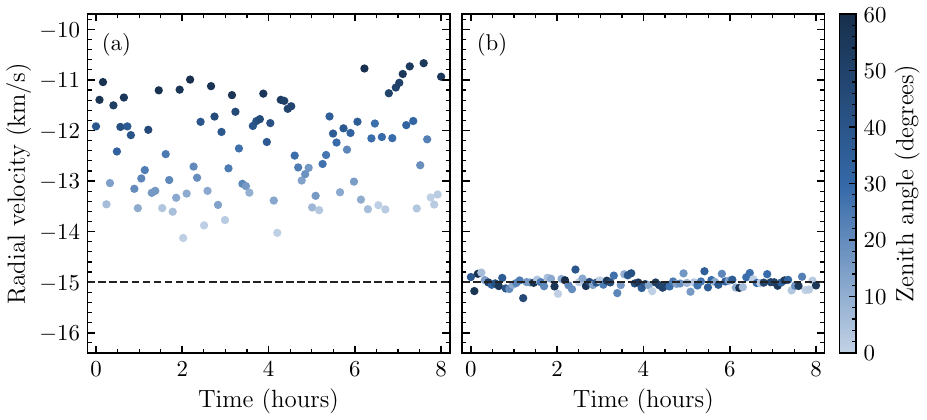}
    \caption{The second tutorial calibrates simulated radial velocities of a standard star. Panel~(a) shows the velocities measured against time, each colored by the zenith angle. The dashed line is the true velocity. Panel~(b) shows the same observations on the same axes after the fitted systematic effects are removed. The velocity rises with the zenith angle in panel~(a) and scatters by the noise around the true value in panel~(b).}
    \label{fig:tutorial_calibration}
\end{figure}

In the simulation we assume that four recorded conditions change the velocity, and a real instrument has many more. The zenith angle is the angle of the star from the vertical, and a star far from the vertical is seen through more air. A change of temperature expands the optics, and humidity changes the refractive index of the air. The instrument itself drifts during the night. A standard star has a constant velocity, so observing it under changing conditions shows how large these effects are (Figure~\ref{fig:tutorial_calibration}a).

We treat calibration as a linear regression. The recorded conditions are the features, and the weights say how much each condition changes the velocity. The velocity of the star does not change from one observation to the next, but the conditions do. The fit therefore treats the part that varies with the conditions as the instrument and the constant part as the star. The constant part is the intercept (Section~\ref{sec:ch4_analytical}). After we subtract the part that the conditions explain, the velocities of the simulated observations scatter around the true value by the noise (Figure~\ref{fig:tutorial_calibration}b).

Scaled by how much each condition varies, the fitted weights also show which condition matters most. This guides the design of the instrument and the planning of observations. Real observations differ in quality, because clouds and poor seeing add noise. Measurements with unequal uncertainties enter through the weighted fit of Section~\ref{sec:ch4_mle}, which counts the noisy measurements less. We also need the uncertainty of the fitted velocity. The bootstrap of Section~\ref{sec:ch3_bootstrap} shows how it shrinks with the number of observations, and so how much telescope time a calibration needs.

The same idea scales to a network of telescopes. Each telescope responds differently to the conditions, but they observe the same standard stars. After calibration, their measurements must agree on the velocity of each star. These common stars calibrate the whole network in one regression with a larger design matrix. The method applies to measurements whose systematic effects depend on recorded quantities, such as the dimming of starlight by the atmosphere at large zenith angles. Surveys that combine telescopes need such cross-calibration to compare their measurements.

Linear models remain useful in astronomy despite their simplicity, especially with well-chosen features. Transformations such as logarithms and polynomials let a model that is linear in its weights describe many nonlinear relations (Section~\ref{sec:ch4_formalism}). The two tutorials use the measurements themselves as features, the flux at each wavelength and the observing conditions. Three tools make such a fit trustworthy. Test stars show how well the model predicts new data, a prior limits overfitting, and the bootstrap gives uncertainties.\index{telescope calibration|)}

\paragraph{Further Reading} Gauss (1823) gave an account of the theory of least squares. Its Gauss-Markov theorem shows that least squares has the lowest variance among unbiased linear estimators. Bevington (1969) is a widely used reference on weighted least squares and error analysis in the physical sciences. Faber and Jackson (1976) found that the luminosity of elliptical galaxies scales roughly with the fourth power of the velocity dispersion. Tully and Fisher (1977) related the luminosity of spiral galaxies to the width of their neutral hydrogen line, a proxy for rotation velocity. Ferrarese and Merritt (2000) and Gebhardt et~al.\ (2000) found tight correlations between supermassive black hole masses and the velocity dispersions of their host galaxies. The intrinsic scatter is small. Kennicutt (1998) characterized the star formation law that relates gas density to star formation rate. Isobe et~al.\ (1990) compared regression methods for astronomical data. Akritas and Bershady (1996) developed an estimator for correlated errors, unequal uncertainties and intrinsic scatter. Stone (1974) used cross-validation to choose among statistical predictors and to test them.

\chapter{Bayesian Linear Regression}

\begin{figure}[ht!]
    \centering
    \includegraphics[width=0.9\textwidth]{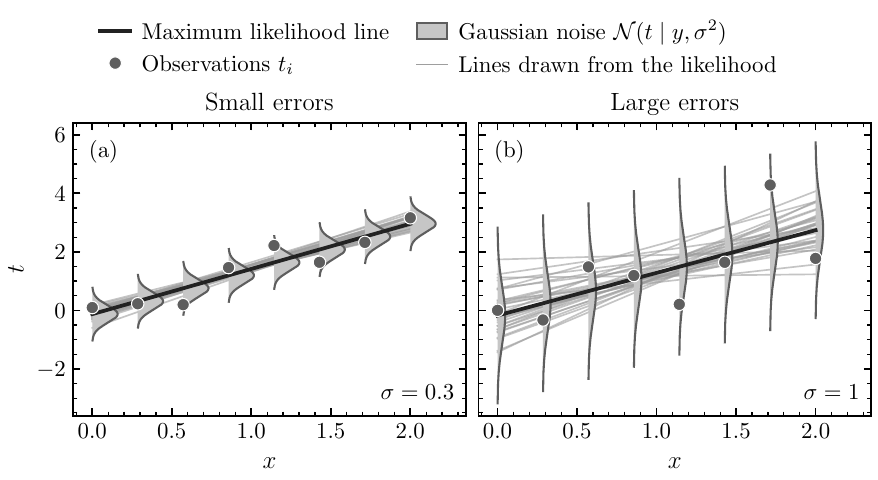}
    \caption{Two data sets of observations $t_i$ (grey dots) drawn around the same true line, $y = 1.5\,x$ (not drawn). Errors are small in panel (a) and large in panel (b). The Gaussian noise $\mathcal{N}(t \,|\, y, \sigma^2)$ is centered on the maximum likelihood line at each $x_i$ and drawn sideways at the same density scale in both panels. The thin grey lines are drawn in proportion to the likelihood. The two fits are similar, but the data in (b) allow many more lines than the data in (a).}
    \label{fig:mle_limitation}
\end{figure}

Suppose we fit a straight line to the M-$\sigma$ relation. The fit gives a slope and an intercept, but a theorist who compares the slope with a prediction needs more than these two numbers. An observer who predicts the black hole mass of a new galaxy needs more too. Both need to know how well the data constrain the line. Two data sets can give similar best-fit lines and still constrain them very differently, as Figure~\ref{fig:mle_limitation} shows.

Frequentist methods can supply this uncertainty, for example with the bootstrap of Chapter~3. They can work well for large, well-sampled data sets. The bootstrap is harder to use in astronomy. Samples are often small or expensive to obtain, and different measurements can have very different uncertainties. The bootstrap also has no direct way to include prior knowledge.

The Bayesian approach takes a different view. It treats the model parameters as random variables with probability distributions, instead of fixed quantities whose sampling distribution we must estimate. This probabilistic view has three advantages for astronomy.

First, it gives the uncertainties of the parameters and of the predictions directly, and it needs no resampling. Second, it lets us include prior knowledge from earlier observations, theoretical constraints or physical intuition. Third, an informative prior acts as a regularizer. It penalizes very complex models unless the data require them, so it protects against overfitting without any ad hoc constraint.

We build the method in steps. Conjugate priors give posteriors in closed form for simple problems, and they show how Bayesian updating works. We then extend conjugacy to the multivariate Gaussians that linear regression needs. Next we derive the posterior of the regression weights and show that the prior acts as a regularizer. Finally we predict with full uncertainties through the posterior predictive distribution.

\section{From Maximum Likelihood to Bayesian Inference}

Bayes'\index{Bayes' theorem}\index{Bayesian linear regression|(} theorem gives the complete probability distribution of the model parameters. In symbols, with $\mathbf{w}$ the parameters and $\mathcal{D}$ the data, it reads
\begin{equation}
p(\mathbf{w}|\mathcal{D}) = \frac{p(\mathcal{D}|\mathbf{w})p(\mathbf{w})}{p(\mathcal{D})}.
\end{equation}This short equation tells us how to learn from data. Each part has a clear interpretation.

The likelihood $p(\mathcal{D}|\mathbf{w})$ is the function we used in maximum likelihood estimation. It describes what the data tell us about the parameters. The prior $p(\mathbf{w})$ expresses our beliefs about the parameters before we see the data, such as earlier results or physical constraints. The evidence $p(\mathcal{D})$ is the integral of the likelihood times the prior over all parameter values, and it normalizes the posterior. The evidence is also called the marginal likelihood. The posterior $p(\mathbf{w}|\mathcal{D})$ is our updated belief after we combine the prior knowledge with the data.

Bayes' theorem balances different sources of information. With strong prior knowledge and weak data, the posterior stays close to the prior. With a weak prior and strong data, the likelihood dominates the posterior. The posterior also shows our confidence. A narrow posterior means well-constrained parameters, and a broad posterior signals high uncertainty.

For the M-$\sigma$ relation, the posterior might give a slope of $4.0 \pm 0.3$ instead of the single value 4.0. For a Gaussian posterior, $4.0 \pm 0.3$ is a 68\% credible\index{credible interval} interval, the range that holds 68\% of the posterior probability. Unlike a confidence interval (Section~2.6), it gives our degree of belief in the parameter itself. With such an interval we can plan follow-up observations and propagate the uncertainty through later calculations.

\section{Conjugate Priors}

Bayes'\index{conjugate prior|(} theorem says what to compute. It does not say how. The evidence is an integral over all parameter values, and the posterior mean and variance need integrals too. These integrals can be intractable for many combinations of priors and likelihoods.

Conjugate\index{conjugate prior}\index{prior!conjugate|see{conjugate prior}} priors solve this problem for many useful cases. A prior is conjugate to a likelihood when the posterior belongs to the same family as the prior. Only the parameters change. For example, a Gamma prior and a Poisson likelihood give a Gamma posterior. Conjugacy allows analytical solutions to inference problems that would otherwise be hard.

The evidence does not depend on $\mathbf{w}$, so Bayes' theorem in proportional form reads
\begin{equation}
p(\mathbf{w}|\mathcal{D}) \propto p(\mathcal{D}|\mathbf{w})p(\mathbf{w}).
\end{equation}

With a conjugate prior, the abstract idea of ``updating beliefs'' becomes a concrete update of parameters that we can compute analytically.

We build intuition with two examples that often arise in astronomical data analysis.

\paragraph{Poisson-Gamma Conjugacy}

Suppose we count stars in different regions of the sky to measure the stellar density. Let $\lambda$ be the expected number of stars in a region, which is the average count we would find over many similar regions.

When we observe one region, the probability of counting exactly $n$ stars follows a Poisson\index{conjugate prior!Poisson-gamma@Poisson-Gamma} distribution
\begin{equation}
p(n|\lambda) = \frac{\lambda^n e^{-\lambda}}{n!}.
\end{equation}This distribution arises when we count rare, independent events, as we do for star counts in a field of view. The parameter $\lambda$ depends on the stellar density of the region and on the selection effects that set how many stars we detect.

We do not know the true value of $\lambda$. We want to learn it from the observations. Before we observe, we may hold prior beliefs about $\lambda$ from earlier surveys, theoretical models of star formation, or the known properties of the region.

We can express these prior beliefs with a Gamma\index{gamma distribution@Gamma distribution} distribution of shape $\alpha$ and rate $\beta$
\begin{equation}
p(\lambda) = \frac{\beta^\alpha}{\Gamma(\alpha)}\lambda^{\alpha-1}e^{-\beta\lambda}.
\end{equation}The parameters $\alpha$ and $\beta$ connect directly to our prior knowledge. The ratio $\alpha/\beta$ is our prior expectation for the star count, and the prior variance is $\alpha/\beta^2$. The individual values of $\alpha$ and $\beta$ set our confidence in this expectation. We can think of $\alpha$ as the total number of stars ``effectively observed'' in earlier surveys, and of $\beta$ as the number of regions surveyed. If earlier surveys of 2 regions suggest about 5 stars per region, we might use $\alpha = 10$ and $\beta = 2$.

Conjugacy pays off when we update our beliefs with a new observation. Suppose we count $n$ stars in a new region. Bayes' theorem gives
\begin{equation}
p(\lambda|n) \propto p(n|\lambda)p(\lambda) \propto \lambda^n e^{-\lambda} \cdot \lambda^{\alpha-1}e^{-\beta\lambda}.
\end{equation}

Combining the terms gives
\begin{equation}
p(\lambda|n) \propto \lambda^{(n+\alpha-1)}e^{-(\beta+1)\lambda}.
\end{equation}This is another Gamma distribution. The posterior is $\text{Gamma}(\alpha + n, \beta + 1)$, so the parameters update by two simple rules. The parameter $\alpha$ increases by the number of stars observed, $n$. The parameter $\beta$ increases by 1, because we have surveyed one more region.

\begin{figure}[ht!]
    \centering
    \includegraphics[width=0.9\textwidth]{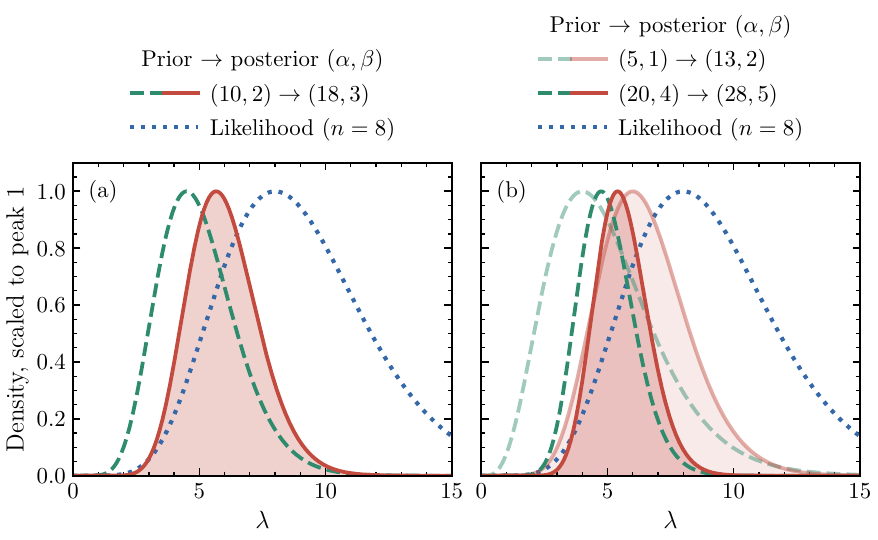}
    \caption{Poisson-Gamma conjugacy in star count analysis. Panel (a) shows a Gamma prior, the Poisson likelihood of one observed count, and the Gamma posterior, which is their product. Panel (b) keeps the same likelihood and compares a broad prior (light curves) with a narrow prior (dark curves). The narrow prior keeps the posterior close to the prior mean. The broad prior lets the likelihood pull the posterior further. The curves peak at the modes, the values where each density is highest, which lie slightly below the means quoted in the text.}
    \label{fig:poisson_gamma}
\end{figure}

Our prior $\text{Gamma}(10,2)$ suggests about 5 stars per region from earlier surveys of 2 regions. If we observe 8 stars in a new region, the posterior is $\text{Gamma}(18,3)$. This suggests a slightly higher average of about 6 stars per region ($18/3 = 6$). The estimate uses both the prior knowledge and the new evidence. Figure~\ref{fig:poisson_gamma}(a) shows this update.

Panel (b) repeats the update for a broader and a narrower prior, both with a mean of 5 stars per region. The broad prior, $\text{Gamma}(5,1)$, stands for one earlier region, and the narrow prior, $\text{Gamma}(20,4)$, stands for four. More earlier surveys mean a stronger belief. The narrow prior therefore holds the posterior mean at 5.6, close to the prior mean. The broad prior lets the new count pull it up to 6.5.

The update extends directly to multiple observations. If we count $\{n_1, n_2, \ldots, n_N\}$ stars in $N$ independent regions, the posterior is $\text{Gamma}(\alpha + \sum_{i=1}^N n_i, \beta + N)$. This matches our intuition, because the estimate uses the total star count and the number of regions.

\paragraph{Bernoulli-Beta Conjugacy}

Consider a second task in astronomy, the classification of galaxies as spiral or elliptical. This binary classification shows conjugacy in a different setting.

For a single galaxy, the classification has two outcomes, spiral or elliptical. Such an event follows a Bernoulli\index{Bernoulli distribution} distribution, in which each outcome has a fixed probability. Let $\theta$ be the probability that a randomly selected galaxy in our survey region is spiral. Then
\begin{equation}
p(x|\theta) = \theta^x(1-\theta)^{1-x},
\end{equation}
where $x = 1$ for a spiral galaxy, which has probability $\theta$, and $x = 0$ for an elliptical galaxy, which has probability $1-\theta$.

We can express our prior beliefs about $\theta$ with a Beta\index{beta distribution@Beta distribution}\index{conjugate prior!Bernoulli-beta@Bernoulli-Beta} distribution
\begin{equation}
p(\theta) = \frac{\Gamma(\alpha+\beta)}{\Gamma(\alpha)\Gamma(\beta)}\theta^{\alpha-1}(1-\theta)^{\beta-1}.
\end{equation}Like the Gamma parameters, $\alpha$ and $\beta$ have a simple meaning. They act like counts of earlier classifications. If we had classified $\alpha + \beta$ galaxies before, with $\alpha$ spiral and $\beta$ elliptical, the Beta prior would encode this history.

Now we observe $N$ new galaxies, with $\sum x_i$ spiral and $N - \sum x_i$ elliptical. The likelihood is
\begin{equation}
p(\{x_i\}|\theta) = \theta^{\sum x_i}(1-\theta)^{N-\sum x_i}.
\end{equation}

The posterior is then
\begin{equation}
p(\theta|\{x_i\}) \propto \theta^{\sum x_i}(1-\theta)^{N-\sum x_i} \cdot \theta^{\alpha-1}(1-\theta)^{\beta-1} = \theta^{(\alpha+\sum x_i)-1}(1-\theta)^{(\beta+N-\sum x_i)-1}.
\end{equation}Once again we obtain a distribution from the same family, $\text{Beta}(\alpha + \sum x_i, \beta + N - \sum x_i)$. The update rules are equally simple. The parameter $\alpha$ increases by the number of spiral galaxies observed. The parameter $\beta$ increases by the number of elliptical galaxies.

These examples show the pattern that makes conjugate priors useful. We can update our beliefs analytically, and each parameter update has a clear meaning as accumulated evidence. This simplicity matters when we extend the idea to linear regression.

\section{Gaussian-Gaussian Conjugacy in One Dimension}

Poisson-Gamma\index{conjugate prior!Gaussian-Gaussian|(} and Bernoulli-Beta conjugacy handle discrete data. Linear regression has continuous data with Gaussian\index{conjugate prior!Gaussian-Gaussian} noise, so we need Gaussian-Gaussian conjugacy.

We start with the simplest case, estimating the mean of a Gaussian distribution when the variance is known. This may seem artificial, but it contains the elements we need for linear regression. Its simplicity lets us focus on the mathematics.

Consider one observation $x$ from a Gaussian with unknown mean $\mu$ and known variance $\sigma^2$
\begin{equation}
p(x|\mu) = \mathcal{N}(x|\mu, \sigma^2) = \frac{1}{\sqrt{2\pi\sigma^2}} \exp\left(-\frac{(x-\mu)^2}{2\sigma^2}\right).
\end{equation}This likelihood tells us how probable the value $x$ is for any proposed mean $\mu$. The quadratic term in the exponent makes the likelihood fall quickly as $\mu$ moves away from $x$. The measurement variance $\sigma^2$ sets the rate of this decrease.

To complete the Bayesian model, we need a prior for $\mu$. A common choice is another Gaussian distribution
\begin{equation}
p(\mu) = \mathcal{N}(\mu|\mu_0, \tau_0^2) = \frac{1}{\sqrt{2\pi\tau_0^2}} \exp\left(-\frac{(\mu-\mu_0)^2}{2\tau_0^2}\right),
\end{equation}
where $\mu_0$ is our prior belief about the mean and $\tau_0^2$ is our prior uncertainty.

Why choose a Gaussian\index{prior!Gaussian} prior? It is mathematically convenient, and it is also a sensible default when we are uncertain about a continuous parameter. A Gaussian gives the highest probability to values near our best guess $\mu_0$ and lower probability further away. The variance $\tau_0^2$ controls how confident we are in the prior.

Now we combine the prior and the likelihood. Both are Gaussian, so their product is also Gaussian. This is the property that conjugacy needs. We show it by multiplying the distributions and dropping normalization constants
\begin{align}
p(\mu|x) &\propto p(x|\mu)p(\mu) \\*
&\propto \exp\left(-\frac{(x-\mu)^2}{2\sigma^2}\right) \exp\left(-\frac{(\mu-\mu_0)^2}{2\tau_0^2}\right) \\*
&\propto \exp\left(-\frac{1}{2}\left[\frac{(x-\mu)^2}{\sigma^2} + \frac{(\mu-\mu_0)^2}{\tau_0^2}\right]\right).
\end{align}

To show that the product is Gaussian in $\mu$, we expand the exponent and collect powers of $\mu$
\begin{align}
-\frac{1}{2}\left[\frac{(x-\mu)^2}{\sigma^2} + \frac{(\mu-\mu_0)^2}{\tau_0^2}\right] &= -\frac{1}{2}\left[\frac{\mu^2-2x\mu+x^2}{\sigma^2} + \frac{\mu^2-2\mu_0\mu+\mu_0^2}{\tau_0^2}\right] \\*
&= -\frac{1}{2}\left[\left(\frac{1}{\sigma^2}+\frac{1}{\tau_0^2}\right)\mu^2 - 2\left(\frac{x}{\sigma^2}+\frac{\mu_0}{\tau_0^2}\right)\mu + \text{const}\right].
\end{align}This quadratic form in $\mu$ confirms that the posterior is Gaussian. To find its parameters, we compare it with the standard form $\mathcal{N}(\mu|\mu_1,\tau_1^2)$, which has the exponent
\begin{equation}
-\frac{1}{2\tau_1^2}(\mu-\mu_1)^2 = -\frac{1}{2}\left[\frac{1}{\tau_1^2}\mu^2 - \frac{2\mu_1}{\tau_1^2}\mu + \frac{\mu_1^2}{\tau_1^2}\right].
\end{equation}Comparing coefficients gives
\begin{align}
\frac{1}{\tau_1^2} &= \frac{1}{\sigma^2} + \frac{1}{\tau_0^2}, \\*
\frac{\mu_1}{\tau_1^2} &= \frac{x}{\sigma^2} + \frac{\mu_0}{\tau_0^2}.
\end{align}

Solving for the posterior parameters gives
\begin{align}
\tau_1^2 &= \left(\frac{1}{\sigma^2} + \frac{1}{\tau_0^2}\right)^{-1}, \\*
\mu_1 &= \tau_1^2\left(\frac{x}{\sigma^2} + \frac{\mu_0}{\tau_0^2}\right).
\end{align}These update equations show the structure of Gaussian conjugacy. The posterior precision (inverse variance) $1/\tau_1^2$ is the sum of the likelihood precision $1/\sigma^2$ and the prior precision $1/\tau_0^2$. Our total confidence comes from combining our prior knowledge with what we learn from the data.

\begin{figure}[ht!]
    \centering
    \includegraphics[width=0.9\textwidth]{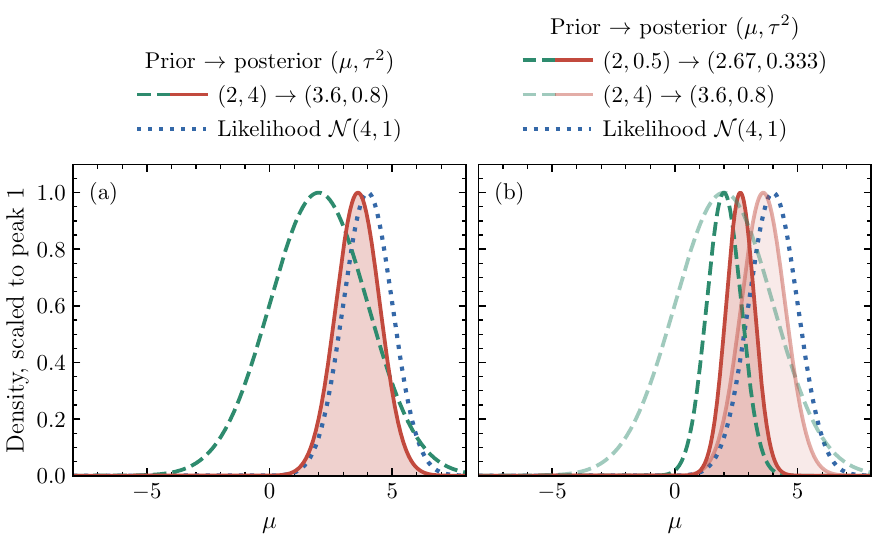}
    \caption{Gaussian conjugacy for the mean $\mu$ of a Gaussian with known variance. The likelihood is $\mathcal{N}(x \,|\, \mu, \sigma^2)$ for one observation, and the prior is a Gaussian for $\mu$. Panel (a) shows the posterior, which is their product. Panel (b) compares a precise prior (dark curves) with a broad prior (light curves). The precise prior keeps the posterior close to the prior mean. The broad prior leaves the posterior close to the likelihood.}
    \label{fig:gaussian_conjugacy}
\end{figure}

The posterior\index{posterior mean} mean $\mu_1$ is a weighted average of the observation $x$ and the prior mean $\mu_0$. The weights are proportional to their precisions. More precise information has more influence on the final estimate, as we would expect. Figure~\ref{fig:gaussian_conjugacy} shows the update for two priors. To build intuition, we consider two extreme cases.

\paragraph{Uninformative Prior ($\tau_0^2 \to \infty$)} When we are very uncertain about our prior\index{uninformative prior} belief, the posterior becomes
\begin{align}
\tau_1^2 &\to \sigma^2, \\*
\mu_1 &\to x.
\end{align}
The data dominate the posterior. Our final estimate equals the observation, and its uncertainty equals the measurement error.

\paragraph{Very Uncertain Measurement ($\sigma^2 \to \infty$)} When our measurement is very noisy, the posterior becomes
\begin{align}
\tau_1^2 &\to \tau_0^2, \\*
\mu_1 &\to \mu_0.
\end{align}
We learn almost nothing from such a measurement, so the posterior stays close to our prior beliefs.

\paragraph{Multiple Observations}

For multiple observations $x_1, x_2, \ldots, x_N$, we assume that each is drawn independently from $\mathcal{N}(\mu, \sigma^2)$. The likelihood is then a product of single-observation likelihoods. We write $\bar{x} = (1/N)\sum_{i=1}^N x_i$ for the sample mean. The sum of squares splits as
\begin{equation}
\sum_{i=1}^N (x_i - \mu)^2 = N(\mu - \bar{x})^2 + \sum_{i=1}^N (x_i - \bar{x})^2,
\end{equation}
and only the first term depends on $\mu$. Hence
\begin{equation}
p(x_1, \ldots, x_N | \mu) \propto \exp\left(-\frac{N(\mu - \bar{x})^2}{2\sigma^2}\right),
\end{equation}
which is the likelihood of a single observation $\bar{x}$ with variance $\sigma^2/N$.

We can therefore reuse the single-observation results, with $x$ replaced by $\bar{x}$ and $\sigma^2$ by $\sigma^2/N$. The posterior parameters become
\begin{align}
\frac{1}{\tau_N^2} &= \frac{1}{\tau_0^2} + \frac{N}{\sigma^2}, \\*
\mu_N &= \frac{\frac{\mu_0}{\tau_0^2} + \frac{N\bar{x}}{\sigma^2}}{\frac{1}{\tau_0^2} + \frac{N}{\sigma^2}}.
\end{align}Each observation adds information, and the posterior precision grows with the amount of data. For $N = 1$ these equations reduce to the single-observation results.

This one-dimensional example shows the logic of Bayesian updating. We combine sources of information according to their precision. The same principle applies to the multivariate case that linear regression needs.\index{conjugate prior!Gaussian-Gaussian|)}\index{conjugate prior|)}

\section{Multivariate Gaussian Conjugacy}

We\index{conjugate prior!multivariate Gaussian|(} now extend this pattern to many dimensions, as linear regression requires. The step from one dimension to many may seem hard, but the principles stay the same.

A one-dimensional Gaussian is set by its mean $\mu$ and variance $\sigma^2$, and a multivariate\index{conjugate prior!multivariate Gaussian} Gaussian by its mean vector $\boldsymbol{\mu}$ and covariance matrix $\boldsymbol{\Sigma}$. The diagonal of $\boldsymbol{\Sigma}$ holds the uncertainty of each parameter. The off-diagonal terms show how the parameters vary together. This matters in linear regression, because slope and intercept estimates are often correlated.

The multivariate Gaussian density generalizes the one-dimensional form
\begin{equation}
\mathcal{N}(\mathbf{x}|\boldsymbol{\mu},\boldsymbol{\Sigma}) = \frac{1}{(2\pi)^{D/2}\det(\boldsymbol{\Sigma})^{1/2}} \exp\left(-\frac{1}{2}(\mathbf{x}-\boldsymbol{\mu})^T\boldsymbol{\Sigma}^{-1}(\mathbf{x}-\boldsymbol{\mu})\right),
\end{equation}
where $D$ is the dimension of $\mathbf{x}$ and $\det(\boldsymbol{\Sigma})$ is the determinant of $\boldsymbol{\Sigma}$. The quadratic form $(\mathbf{x}-\boldsymbol{\mu})^T\boldsymbol{\Sigma}^{-1}(\mathbf{x}-\boldsymbol{\mu})$ generalizes the squared distance $(x-\mu)^2/\sigma^2$ of one dimension, and it includes the correlations between the dimensions.

The main result of one dimension carries over to many dimensions. The product of two Gaussians is, up to a constant, again a Gaussian. This is what makes Bayesian linear regression analytically tractable.

Consider two multivariate Gaussian distributions
\begin{align}
p_1(\mathbf{x}) &= \mathcal{N}(\mathbf{x}|\boldsymbol{\mu}_1,\boldsymbol{\Sigma}_1), \\*
p_2(\mathbf{x}) &= \mathcal{N}(\mathbf{x}|\boldsymbol{\mu}_2,\boldsymbol{\Sigma}_2).
\end{align}

As in one dimension, we multiply the distributions and keep only the exponential terms
\begin{equation}
p_1(\mathbf{x})p_2(\mathbf{x}) \propto \exp\left(-\frac{1}{2}\left[(\mathbf{x}-\boldsymbol{\mu}_1)^T\boldsymbol{\Sigma}_1^{-1}(\mathbf{x}-\boldsymbol{\mu}_1) + (\mathbf{x}-\boldsymbol{\mu}_2)^T\boldsymbol{\Sigma}_2^{-1}(\mathbf{x}-\boldsymbol{\mu}_2)\right]\right).
\end{equation}

To show that the product is Gaussian, we must bring the combined exponent into the standard quadratic form. We expand both quadratic terms
\begin{align}
&-\frac{1}{2}\left[\mathbf{x}^T\boldsymbol{\Sigma}_1^{-1}\mathbf{x} - \boldsymbol{\mu}_1^T\boldsymbol{\Sigma}_1^{-1}\mathbf{x} - \mathbf{x}^T\boldsymbol{\Sigma}_1^{-1}\boldsymbol{\mu}_1 + \boldsymbol{\mu}_1^T\boldsymbol{\Sigma}_1^{-1}\boldsymbol{\mu}_1 \right. \nonumber \\*
&\quad\quad \left. + \mathbf{x}^T\boldsymbol{\Sigma}_2^{-1}\mathbf{x} - \boldsymbol{\mu}_2^T\boldsymbol{\Sigma}_2^{-1}\mathbf{x} - \mathbf{x}^T\boldsymbol{\Sigma}_2^{-1}\boldsymbol{\mu}_2 + \boldsymbol{\mu}_2^T\boldsymbol{\Sigma}_2^{-1}\boldsymbol{\mu}_2\right].
\end{align}

The covariance matrices are symmetric, so the cross terms $\boldsymbol{\mu}_i^T\boldsymbol{\Sigma}_i^{-1}\mathbf{x}$ and $\mathbf{x}^T\boldsymbol{\Sigma}_i^{-1}\boldsymbol{\mu}_i$ are equal scalars. Collecting terms by powers of $\mathbf{x}$ gives
\begin{align}
&-\frac{1}{2}\left[\mathbf{x}^T(\boldsymbol{\Sigma}_1^{-1} + \boldsymbol{\Sigma}_2^{-1})\mathbf{x} - 2(\boldsymbol{\mu}_1^T\boldsymbol{\Sigma}_1^{-1} + \boldsymbol{\mu}_2^T\boldsymbol{\Sigma}_2^{-1})\mathbf{x} + \text{const}\right].
\end{align}This has the quadratic form of a multivariate Gaussian. Comparing it with the standard form lets us identify the parameters of the result. If the result is $\mathcal{N}(\mathbf{x}|\boldsymbol{\mu}_{\rm prod},\boldsymbol{\Sigma}_{\rm prod})$, then the quadratic and linear terms give
\begin{align}
\boldsymbol{\Sigma}_{\rm prod}^{-1} &= \boldsymbol{\Sigma}_1^{-1} + \boldsymbol{\Sigma}_2^{-1}, \\*
\boldsymbol{\Sigma}_{\rm prod}^{-1}\boldsymbol{\mu}_{\rm prod} &= \boldsymbol{\Sigma}_1^{-1}\boldsymbol{\mu}_1 + \boldsymbol{\Sigma}_2^{-1}\boldsymbol{\mu}_2.
\end{align}

\begin{figure}[ht!]
    \centering
    \includegraphics[width=0.6\textwidth]{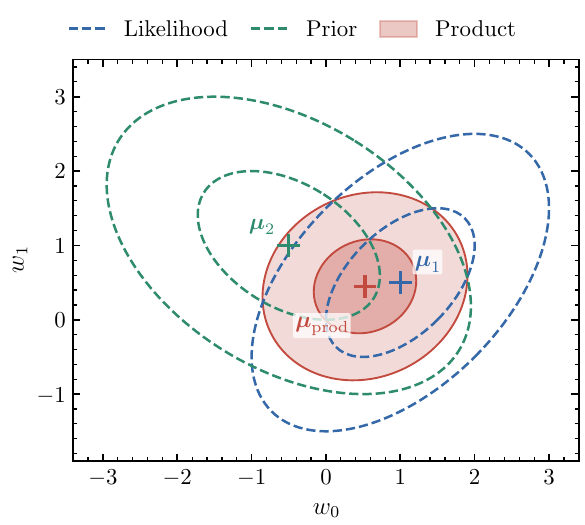}
    \caption{Product of two Gaussian distributions in two parameters, $w_0$ and $w_1$. The blue dashed contours show a Gaussian likelihood with mean $\boldsymbol{\mu}_1 = (1, 0.5)$, and the green dashed contours show a Gaussian prior with mean $\boldsymbol{\mu}_2 = (-0.5, 1)$. All contours are at 1 and 2 standard deviations from the mean, and the plus markers show the means. The filled red contours show the product, a Gaussian with mean $\boldsymbol{\mu}_{\rm prod} = (0.525, 0.45)$. The mean of the product is the precision-weighted average of the two input means, and the product is narrower than either input.}
    \label{fig:gaussian_product}
\end{figure}

Solving for the parameters of the product gives
\begin{align}
\boldsymbol{\Sigma}_{\rm prod} &= (\boldsymbol{\Sigma}_1^{-1} + \boldsymbol{\Sigma}_2^{-1})^{-1}, \\*
\boldsymbol{\mu}_{\rm prod} &= \boldsymbol{\Sigma}_{\rm prod}(\boldsymbol{\Sigma}_1^{-1}\boldsymbol{\mu}_1 + \boldsymbol{\Sigma}_2^{-1}\boldsymbol{\mu}_2).
\end{align}These expressions generalize the one-dimensional results. The precision\index{precision matrix} matrix (inverse covariance) of the product is the sum of the input precision matrices. The mean of the product is a precision-weighted average of the input means. The matrix algebra looks more complex, but the pattern is the same, because we combine information weighted by its precision. Figure~\ref{fig:gaussian_product} shows two Gaussians and their product.

The Gaussian product rule matters for linear regression. A Gaussian prior on the parameters and a Gaussian likelihood give a Gaussian posterior. We then have analytical expressions for the updated estimates and their uncertainties.

The same approach covers one parameter or dozens of regression coefficients. In each case we combine prior knowledge and observational evidence, weighted by precision. We use this again in the more complex models of later chapters.\index{conjugate prior!multivariate Gaussian|)}

\section{Mathematics of Bayesian Linear Regression}\label{sec:ch5_blr_math}

\begin{figure}[ht!]
    \centering
    \includegraphics[width=0.9\textwidth]{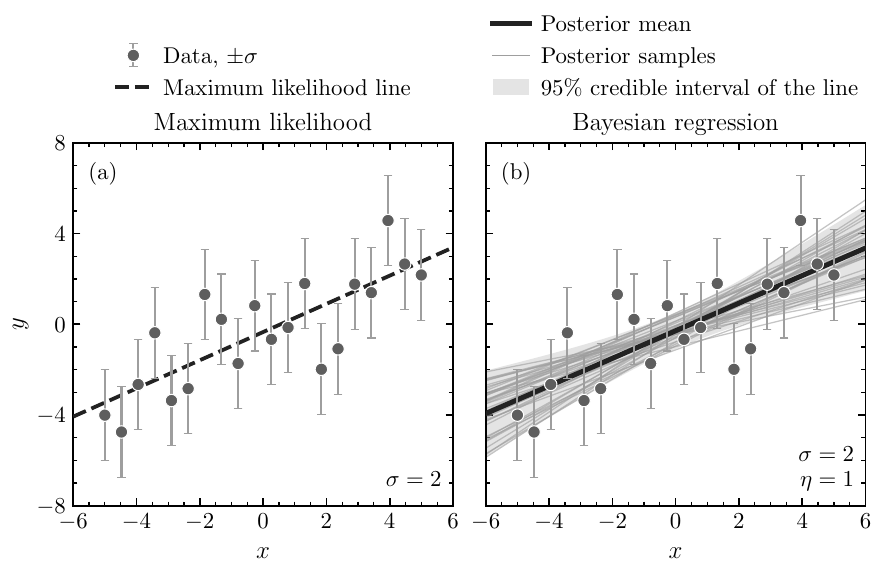}
    \caption{Maximum likelihood and Bayesian fits to the same observations, each with an error bar of $\pm\sigma$. Panel (a) shows the maximum likelihood line. Panel (b) shows the Bayesian solution for a zero-mean Gaussian prior with standard deviation $\eta$ on each weight. The shaded band is the 95\% credible interval of the regression line at each $x$. Both approaches find the same trend. The Bayesian band also shows how well the data constrain the line, and it widens away from the center of the data.}
    \label{fig:regression_comparison}
\end{figure}

We can now derive the full Bayesian solution for linear regression. Our goal is the posterior distribution $p(\mathbf{w}|\mathbf{t},\mathbf{X},\mathbf{S})$ of the parameters. We condition on the observed data $\mathbf{t}$, the input features $\mathbf{X}$ and the measurement variances, which we collect in the diagonal matrix $\mathbf{S}$. Together they form the data $\mathcal{D}$. The posterior gives our best estimates of the regression coefficients, our confidence in them, and how they correlate. Figure~\ref{fig:regression_comparison} compares the Bayesian result with the maximum likelihood fit.

We start again from Bayes' theorem
\begin{equation}
p(\mathbf{w}|\mathbf{t},\mathbf{X},\mathbf{S}) \propto p(\mathbf{t}|\mathbf{X},\mathbf{w},\mathbf{S})p(\mathbf{w}).
\end{equation}

Chapter~4 showed that, for Gaussian measurement uncertainties, the likelihood is
\begin{equation}
p(\mathbf{t}|\mathbf{X},\mathbf{w},\mathbf{S}) = \mathcal{N}(\mathbf{t}|\boldsymbol{\Phi}\mathbf{w},\mathbf{S}),
\end{equation}
where $\boldsymbol{\Phi}$ is the design matrix. It holds the basis functions $\boldsymbol{\phi}(\mathbf{x})$ evaluated at each input point, with one row for each point and one column for each feature. For a polynomial of degree 2 and two input points $x_1$ and $x_2$, it is
\begin{equation}
\boldsymbol{\Phi} = \begin{bmatrix} 
1 & x_1 & x_1^2 \\
1 & x_2 & x_2^2
\end{bmatrix}.
\end{equation}For a straight-line fit of the M-$\sigma$ relation, row $i$ is $(1, \log\sigma_{\star,i})$.

For the prior, we choose a multivariate Gaussian distribution, the natural conjugate prior for our Gaussian likelihood
\begin{equation}
p(\mathbf{w}) = \mathcal{N}(\mathbf{w}|\mathbf{m}_0,\mathbf{S}_0).
\end{equation}
Here $\mathbf{m}_0$ is our prior belief about the most likely parameter values, and $\mathbf{S}_0$ holds our prior uncertainties and correlations between the parameters. A common choice is $\mathbf{S}_0 = \eta^2\mathrm{Id}$, where $\eta$ is the prior standard deviation of each weight and $\mathrm{Id}$ is the identity matrix. We call such a prior isotropic. It gives the weights the same variance and no correlation with one another.

The choice of prior needs care in astronomy. For the M-$\sigma$ relation, we could set $\mathbf{m}_0$ from earlier studies, for example a slope near 4 and an intercept that agrees with theoretical predictions. Then $\mathbf{S}_0$ would hold our uncertainty in them. The off-diagonal term of $\mathbf{S}_0$ would hold the expected correlation between the slope and the intercept.

\paragraph{Rewriting the Likelihood}

To use the multivariate conjugacy result, we need the likelihood and the prior as Gaussians in the parameter vector $\mathbf{w}$. The prior already has this form. The likelihood is a Gaussian in the observations $\mathbf{t}$, so we must rewrite it as a Gaussian in $\mathbf{w}$.

We start with the likelihood and write out its quadratic form
\begin{equation}
p(\mathbf{t}|\mathbf{w}) = \mathcal{N}(\mathbf{t}|\boldsymbol{\Phi}\mathbf{w},\mathbf{S}) \propto \exp\left(-\frac{1}{2}(\mathbf{t}-\boldsymbol{\Phi}\mathbf{w})^T\mathbf{S}^{-1}(\mathbf{t}-\boldsymbol{\Phi}\mathbf{w})\right).
\end{equation}Expanding the quadratic form gives
\begin{equation}
(\mathbf{t}-\boldsymbol{\Phi}\mathbf{w})^T\mathbf{S}^{-1}(\mathbf{t}-\boldsymbol{\Phi}\mathbf{w}) = \mathbf{t}^T\mathbf{S}^{-1}\mathbf{t} - 2\mathbf{t}^T\mathbf{S}^{-1}\boldsymbol{\Phi}\mathbf{w} + \mathbf{w}^T\boldsymbol{\Phi}^T\mathbf{S}^{-1}\boldsymbol{\Phi}\mathbf{w}.
\end{equation}The first term does not depend on $\mathbf{w}$, so we absorb it into the proportionality. The terms that depend on $\mathbf{w}$ give
\begin{equation}
p(\mathbf{t}|\mathbf{w}) \propto \exp\left(-\frac{1}{2}\mathbf{w}^T(\boldsymbol{\Phi}^T\mathbf{S}^{-1}\boldsymbol{\Phi})\mathbf{w} + \mathbf{w}^T(\boldsymbol{\Phi}^T\mathbf{S}^{-1}\mathbf{t})\right).
\end{equation}

Comparing this with the standard Gaussian form in $\mathbf{w}$, we can write it as
\begin{equation}
p(\mathbf{t}|\mathbf{w}) \propto \mathcal{N}(\mathbf{w}|\boldsymbol{\mu}_1,\boldsymbol{\Sigma}_1),
\end{equation}
where
\begin{align}
\boldsymbol{\Sigma}_1^{-1} &= \boldsymbol{\Phi}^T\mathbf{S}^{-1}\boldsymbol{\Phi}, \\*
\boldsymbol{\Sigma}_1^{-1}\boldsymbol{\mu}_1 &= \boldsymbol{\Phi}^T\mathbf{S}^{-1}\mathbf{t}.
\end{align}Solving for the likelihood parameters gives
\begin{align}
\boldsymbol{\Sigma}_1 &= (\boldsymbol{\Phi}^T\mathbf{S}^{-1}\boldsymbol{\Phi})^{-1}, \\*
\boldsymbol{\mu}_1 &= (\boldsymbol{\Phi}^T\mathbf{S}^{-1}\boldsymbol{\Phi})^{-1}\boldsymbol{\Phi}^T\mathbf{S}^{-1}\mathbf{t}.
\end{align}Notice that $\boldsymbol{\mu}_1$ is the maximum likelihood estimate that we derived in Chapter~4. The likelihood, viewed as a function of the parameters, has the shape of a Gaussian centered on that estimate.

\paragraph{The Posterior of the Weights}

Now we apply the multivariate conjugacy result. The posterior after $N$ data points is the product of two Gaussians
\begin{equation}
p(\mathbf{w}|\mathbf{t}) \propto \mathcal{N}(\mathbf{w}|\boldsymbol{\mu}_1,\boldsymbol{\Sigma}_1)\mathcal{N}(\mathbf{w}|\mathbf{m}_0,\mathbf{S}_0) = \mathcal{N}(\mathbf{w}|\mathbf{m}_N,\mathbf{S}_N),
\end{equation}
where, by the conjugacy formulas,
\begin{align}
\mathbf{S}_N^{-1} &= \boldsymbol{\Sigma}_1^{-1} + \mathbf{S}_0^{-1} = \boldsymbol{\Phi}^T\mathbf{S}^{-1}\boldsymbol{\Phi} + \mathbf{S}_0^{-1}, \\*
\mathbf{m}_N &= \mathbf{S}_N(\boldsymbol{\Sigma}_1^{-1}\boldsymbol{\mu}_1 + \mathbf{S}_0^{-1}\mathbf{m}_0) = \mathbf{S}_N(\boldsymbol{\Phi}^T\mathbf{S}^{-1}\mathbf{t} + \mathbf{S}_0^{-1}\mathbf{m}_0).
\end{align}The complete Bayesian\index{Bayesian linear regression} linear regression solution is therefore
\begin{equation}
p(\mathbf{w}|\mathbf{t},\mathbf{X},\mathbf{S}) = \mathcal{N}(\mathbf{w}|\mathbf{m}_N,\mathbf{S}_N),
\end{equation}
where
\begin{align}
\mathbf{S}_N &= (\mathbf{S}_0^{-1} + \boldsymbol{\Phi}^T\mathbf{S}^{-1}\boldsymbol{\Phi})^{-1}, \\*
\mathbf{m}_N &= \mathbf{S}_N(\mathbf{S}_0^{-1}\mathbf{m}_0 + \boldsymbol{\Phi}^T\mathbf{S}^{-1}\mathbf{t}).
\end{align}These expressions generalize our one-dimensional results. The posterior precision $\mathbf{S}_N^{-1}$ is the sum of the prior precision $\mathbf{S}_0^{-1}$ and the data precision $\boldsymbol{\Phi}^T\mathbf{S}^{-1}\boldsymbol{\Phi}$. The posterior mean $\mathbf{m}_N$ is a precision-weighted combination of our prior belief $\mathbf{m}_0$ and the data term $\boldsymbol{\Phi}^T\mathbf{S}^{-1}\mathbf{t}$, which contains the maximum likelihood estimate.

A draw from $\mathcal{N}(\mathbf{w}|\mathbf{m}_N,\mathbf{S}_N)$ is one plausible parameter vector, and so one plausible line. The thin lines in Figure~\ref{fig:regression_comparison}(b) are such draws. The shaded band is the 95\% credible interval of the line. Section~\ref{sec:posterior_predictive} explains how we compute it.

\paragraph{Limiting Cases}

Limiting cases build intuition about this general solution. The first has minimal prior knowledge. We take the isotropic prior $\mathbf{S}_0 = \eta^2\mathrm{Id}$ with $\eta^2 \to \infty$, and $\mathbf{m}_0 = \mathbf{0}$. In this uninformative limit we get
\begin{align}
\mathbf{S}_N &\to (\boldsymbol{\Phi}^T\mathbf{S}^{-1}\boldsymbol{\Phi})^{-1}, \\*
\mathbf{m}_N &\to (\boldsymbol{\Phi}^T\mathbf{S}^{-1}\boldsymbol{\Phi})^{-1}\boldsymbol{\Phi}^T\mathbf{S}^{-1}\mathbf{t}.
\end{align}The posterior mean reduces to the maximum likelihood estimate, so the Bayesian and frequentist point estimates agree when prior knowledge is minimal. On top of the point estimate, we obtain the covariance matrix $\mathbf{S}_N$, which describes our uncertainty about the parameters given the data.

In a simple special case, with homogeneous measurement uncertainties, $\mathbf{S} = \sigma^2\mathrm{Id}$, and an isotropic prior, $\mathbf{S}_0 = \eta^2\mathrm{Id}$ with $\mathbf{m}_0 = \mathbf{0}$, the expressions simplify to
\begin{align}
\mathbf{S}_N &= \left(\frac{1}{\eta^2}\mathrm{Id} + \frac{1}{\sigma^2}\boldsymbol{\Phi}^T\boldsymbol{\Phi}\right)^{-1}, \\*
\mathbf{m}_N &= \left(\frac{\sigma^2}{\eta^2}\mathrm{Id} + \boldsymbol{\Phi}^T\boldsymbol{\Phi}\right)^{-1}\boldsymbol{\Phi}^T\mathbf{t}.
\end{align}Compare this with the ordinary least squares solution $\mathbf{w}_{\text{ML}} = (\boldsymbol{\Phi}^T\boldsymbol{\Phi})^{-1}\boldsymbol{\Phi}^T\mathbf{t}$ of Chapter~4. The Bayesian solution has an extra regularization term $(\sigma^2/\eta^2)\mathrm{Id}$, which discourages large weights (Section~5.6). This ratio of measurement variance to prior variance sets the tradeoff between fitting the data and our prior beliefs.

A second limiting case is a large number of observations. As $N$ grows, the data term $\boldsymbol{\Phi}^T\mathbf{S}^{-1}\boldsymbol{\Phi}$ typically grows in proportion to $N$, so $\mathbf{S}_N$ shrinks roughly as $1/N$. Parameter uncertainty therefore decreases as we collect more data. The standard deviation of each parameter falls roughly as $1/\sqrt{N}$, as for the sample mean in Chapter~3. For $N \to \infty$, the posterior concentrates around the true parameter values if the linear model describes the data.

The Bayesian solution is analytical. Unlike many advanced methods, it needs no iterative numerical optimization. It gives the parameter uncertainties with little computation. Astronomical applications often need such uncertainties for scientific inference. The off-diagonal terms of $\mathbf{S}_N$ matter in practice.

\paragraph{Correlation Between Slope and Intercept}

\looseness=1 Consider a mass-luminosity relation for stars, $L = A M^\gamma$. In logarithmic form, \mbox{$\log L = \gamma \log M + \log A$}. The power-law index $\gamma$ tells us how luminosity scales with mass, and the normalization $A$ sets the overall luminosity scale. The posterior covariance matrix $\mathbf{S}_N$ typically shows that these parameters are correlated. For mostly positive inputs, a lower normalization can compensate for a slightly steeper slope and fit the same data. We must understand these degeneracies when we use such relations to make predictions or to compare with theoretical models.

The size of this correlation depends on where we place the intercept. It is strong when the intercept, the value of the line at $x = 0$, lies far outside the data, as in the M-$\sigma$ relation. We can remove it by writing the model as $y = w_0 + w_1(x - x_{\rm p})$, so that $w_0$ is the value of the line at a pivot $x_{\rm p}$. The data precision matrix then reads
\begin{equation}
\boldsymbol{\Phi}^T\mathbf{S}^{-1}\boldsymbol{\Phi} = \begin{bmatrix} \Omega & \sum_i \omega_i (x_i - x_{\rm p}) \\ \sum_i \omega_i (x_i - x_{\rm p}) & \sum_i \omega_i (x_i - x_{\rm p})^2 \end{bmatrix},
\end{equation}
where $\omega_i = 1/\sigma_i^2$ and $\Omega = \sum_i \omega_i$.

The off-diagonal entry vanishes when $x_{\rm p}$ is the inverse-variance weighted mean of the inputs, $x_{\rm p} = \sum_i \omega_i x_i / \Omega$. The matrix is then diagonal, and so is its inverse $\mathbf{S}_N$ in the uninformative limit. The parameters $w_0$ and $w_1$ are therefore uncorrelated, and $w_0$ has variance $1/\Omega$. This is the smallest variance of $w_0$ for any choice of $x_{\rm p}$, because the data constrain the line best at their weighted mean. Astronomers use this idea when they write the M-$\sigma$ relation with $\log(\sigma_\star/200\,\mathrm{km\,s^{-1}})$.

\section{Prior as Regularization}\label{sec:ch5_regularization}

The\index{prior|(}\index{regularization|(} examples above have only a few parameters. Many problems in astronomy and machine learning have far more, so they need regularization,\index{regularization} a penalty on large parameter values that guards against overfitting.

For a stellar spectrum, we might use hundreds or thousands of spectral features as basis functions to estimate the temperature, surface gravity and chemical abundances. Examples are absorption line strengths and continuum measurements. The number of weights can approach the number of data points, so such a model overfits easily. A high-degree polynomial does the same, because it can fit a few data points exactly and oscillate between them. Regularization helps in such cases. It finds solutions that generalize better, because it stops the model from over-weighting any single spectral feature.

\paragraph{MAP Estimation and Ridge Regression}

We now link maximizing the posterior probability to minimizing a regularized loss function. Instead of the full posterior distribution, we look for the single most probable parameter values, the maximum\index{maximum a posteriori estimate (MAP)} a posteriori (MAP)\index{MAP|see{maximum a posteriori estimate}} estimate
\begin{equation}
\mathbf{w}_{\text{MAP}} = \arg\max_{\mathbf{w}} p(\mathbf{w}|\mathbf{t},\mathbf{X},\sigma^2).
\end{equation}

Here $\arg\max_{\mathbf{w}}$ means the value of $\mathbf{w}$ at which the expression after it is largest. The logarithm is a monotonic function, so maximizing the log-posterior is equivalent to maximizing the posterior itself. Bayes' theorem gives
\begin{equation}
\ln p(\mathbf{w}|\mathbf{t},\mathbf{X},\sigma^2) = \ln p(\mathbf{t}|\mathbf{w},\mathbf{X},\sigma^2) + \ln p(\mathbf{w}) + \text{const}.
\end{equation}

We take homogeneous noise, $\mathbf{S} = \sigma^2\mathrm{Id}$, and a zero-mean isotropic prior, $p(\mathbf{w}) = \mathcal{N}(\mathbf{w}|\mathbf{0},\eta^2\mathrm{Id})$. We then evaluate each term explicitly.

The likelihood term measures the fit of the model predictions to the data
\begin{align}
\ln p(\mathbf{t}|\mathbf{w},\mathbf{X},\sigma^2) &= \ln \mathcal{N}(\mathbf{t}|\boldsymbol{\Phi}\mathbf{w},\sigma^2\mathrm{Id}) \\*
&= -\frac{1}{2\sigma^2}(\mathbf{t} - \boldsymbol{\Phi}\mathbf{w})^T(\mathbf{t} - \boldsymbol{\Phi}\mathbf{w}) + \text{const} \\*
&= -\frac{1}{2\sigma^2}\|\mathbf{t} - \boldsymbol{\Phi}\mathbf{w}\|^2 + \text{const}.
\end{align}This is the negative of the sum of squared residuals that we minimized in maximum likelihood estimation, scaled by $1/(2\sigma^2)$. Here $\|\mathbf{v}\|$, the L2 norm, is the Euclidean length of a vector $\mathbf{v}$, so $\|\mathbf{v}\|^2 = \mathbf{v}^T\mathbf{v}$.

The prior term describes our belief about the parameter values
\begin{align}
\ln p(\mathbf{w}) &= \ln \mathcal{N}(\mathbf{w}|\mathbf{0},\eta^2\mathrm{Id}) \\*
&= -\frac{1}{2\eta^2}\mathbf{w}^T\mathbf{w} + \text{const} \\*
&= -\frac{1}{2\eta^2}\|\mathbf{w}\|^2 + \text{const}.
\end{align}

A zero-mean prior expresses a specific belief. Before we see the data, we expect each coefficient to be near zero. This follows Occam's\index{Occam's razor} razor, the principle that among competing explanations we prefer the one with the fewest assumptions.

A large coefficient means that the model leans heavily on one feature, and a coefficient near zero means a weak dependence. When several models fit the data equally well, the prior favors the one without unnecessary dependence on single features, unless the evidence demands it. This assumption suits problems with many features that each have a small effect, such as spectra with thousands of pixels. Shrinking the coefficients adds some bias, a systematic offset of the estimates from the true values. It also lowers their variance, the scatter of the estimates from one data set to another. This exchange is the bias-variance tradeoff of Section~\ref{sec:cross_validation}.

\looseness=1 If we expect large coefficients for physical reasons, we use a wider prior or a nonzero mean. If we expect only a few important features, we use a prior that favors exact zeros.

We combine the two terms and define $\lambda = \sigma^2/\eta^2$
\begin{align}
\ln p(\mathbf{w}|\mathbf{t},\mathbf{X},\sigma^2) &= -\frac{1}{2\sigma^2}\|\mathbf{t} - \boldsymbol{\Phi}\mathbf{w}\|^2 - \frac{1}{2\eta^2}\|\mathbf{w}\|^2 + \text{const} \\*
&= -\frac{1}{2\sigma^2}\left[\|\mathbf{t} - \boldsymbol{\Phi}\mathbf{w}\|^2 + \lambda\|\mathbf{w}\|^2\right] + \text{const}.
\end{align}Maximizing this log-posterior is equivalent to minimizing the regularized loss function
\begin{equation}
E(\mathbf{w}) = \|\mathbf{t} - \boldsymbol{\Phi}\mathbf{w}\|^2 + \lambda\|\mathbf{w}\|^2.
\end{equation}This result links Bayesian inference and regularization. The first term, $\|\mathbf{t} - \boldsymbol{\Phi}\mathbf{w}\|^2$, measures how well the model predictions match the observations. The second term, $\lambda\|\mathbf{w}\|^2$, is L2\index{regularization!L2} regularization, also called ridge\index{ridge regression} regression or Tikhonov\index{Tikhonov regularization|see{ridge regression}} regularization. It implements the Occam's razor preference for solutions that depend weakly on the features. It penalizes large coefficients unless they give a much better fit to the data.

The posterior is Gaussian, so its peak, the MAP estimate, coincides with its mean. The minimizer of $E(\mathbf{w})$ is the posterior mean $\mathbf{m}_N$ of the last section, so the term $(\sigma^2/\eta^2)\mathrm{Id}$ found there is the ridge penalty. Machine learning uses the same penalty widely, called weight\index{weight decay} decay in neural networks, and the prior gives it a statistical basis (Chapter~15).

\begin{figure}[ht!]
    \centering
    \includegraphics[width=0.9\textwidth]{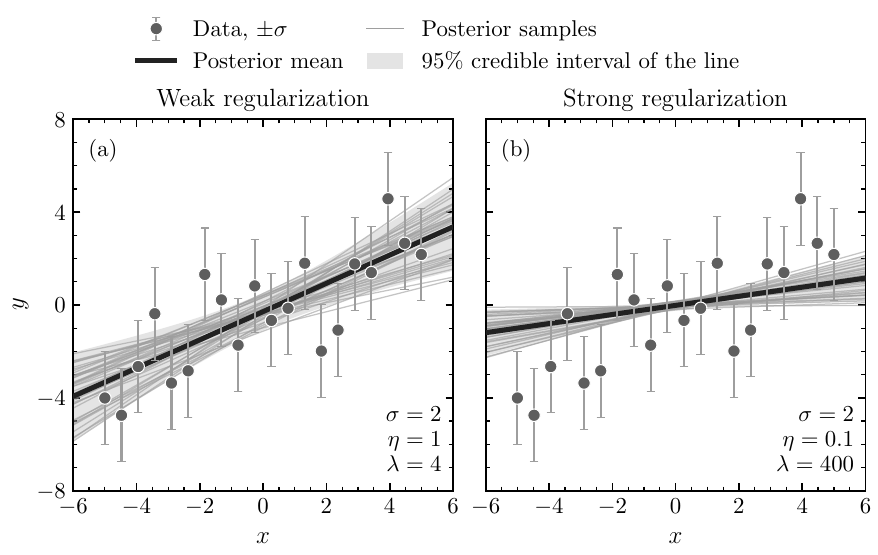}
    \caption{Effect of the regularization strength $\lambda = \sigma^2/\eta^2$ on the same data. Panel (a) shows weak regularization, and panel (b) shows strong regularization. The shaded band is the 95\% credible interval of the regression line. Strong regularization shrinks the posterior mean toward zero and narrows the band. The parameter $\lambda$ sets the tradeoff between following the data and keeping the parameters small.}
    \label{fig:regularization_comparison}
\end{figure}

The regularization parameter $\lambda = \sigma^2/\eta^2$ is the ratio of the measurement variance to the prior variance. When $\lambda$ is large (noisy measurements or a confident prior), the regularization term dominates. It forces the solution toward simpler models with smaller parameter values. When $\lambda$ is small (precise measurements or an uncertain prior), the data term dominates. The model can then follow the observations more closely with more complex fits. Figure~\ref{fig:regularization_comparison} shows the effect of $\lambda$.

We can choose $\lambda$ by cross-validation (Section~\ref{sec:cross_validation}), which scores each trial value on data held out of the fit. The prior width $\eta$ is a hyperparameter of the Bayesian model too, and we still set it on validation data. The Bayesian view ties $\lambda$ to the ratio $\sigma^2/\eta^2$, so we can also set it from the measurement uncertainties and the plausible parameter ranges. This helps in astronomy, where physical constraints often give strong guidance on plausible parameters. We can also choose $\lambda$ by maximizing the evidence $p(\mathcal{D})$, as Chapter~14 does for Gaussian processes.

\paragraph{Laplace Prior}

Different priors lead to different regularization schemes. A Laplace\index{Laplace prior} prior centered at zero, with scale $\kappa$, has the probability density $p(w_j) = \exp(-|w_j|/\kappa)/(2\kappa)$ for each weight. For independent weights, \mbox{$p(\mathbf{w}) \propto \exp(-\|\mathbf{w}\|_1/\kappa)$}, where $\|\mathbf{w}\|_1 = \sum_j |w_j|$ is the L1 norm. Minimizing the negative log-posterior then gives L1\index{regularization!L1} regularization, also called the lasso,\index{lasso regression} with the loss function
\begin{equation}
E(\mathbf{w}) = \|\mathbf{t} - \boldsymbol{\Phi}\mathbf{w}\|^2 + \lambda\|\mathbf{w}\|_1,
\end{equation}
where $\lambda = 2\sigma^2/\kappa$.

The Laplace prior peaks sharply at zero, whereas a Gaussian prior is smooth there. The L1 penalty has a kink at zero, so the MAP estimate can set some coefficients exactly to zero. A solution with many zero coefficients is called sparse.\index{sparsity} This fits problems in which only a few features matter. Only the MAP estimate is sparse. The full posterior of a Laplace prior is not sparse and has no simple closed form.

Several common regularizers have a Bayesian reading, with a Gaussian prior for ridge regression and a Laplace prior for the lasso. The Bayesian view guides the choice of regularization from domain knowledge and the structure we expect in good solutions. Sparsity constraints in image reconstruction and smoothness penalties in function approximation also correspond to suitable priors.

The MAP estimate is a regularized point solution. The full posterior holds more information. 

This difference matters in astronomy. For the M-$\sigma$ relation, the MAP estimate gives a best-fit slope and intercept. The full posterior also tells us how uncertain these parameters are and how they correlate.\index{prior|)}\index{regularization|)}

\section{Posterior Predictive Distribution}
\label{sec:posterior_predictive}

So\index{posterior predictive distribution|(} far we have found the distribution of the model parameters. In many applications the goal is to predict new observations. An astronomer who measures the velocity dispersion of a galaxy and wants its black hole mass from the M-$\sigma$ relation needs more than parameter estimates. They need the full uncertainty of the prediction.

The posterior\index{posterior predictive distribution} predictive distribution provides this. It puts measurement noise and parameter uncertainty into one prediction. Rather than a point prediction, it gives a complete probability distribution.

For a new input $\mathbf{x}_*$ with output $y_*$, we define the posterior predictive distribution as
\begin{equation}
p(y_* | \mathbf{x}_*, \mathcal{D}) = \int p(y_* | \mathbf{x}_*, \mathbf{w}) p(\mathbf{w} | \mathcal{D})\,\mathrm{d}\mathbf{w}.
\end{equation}This integral has a simple meaning. For each parameter set $\mathbf{w}$ consistent with our data, we predict the distribution of $y_*$. We then average over all sets, weighted by their probabilities given the data. The prediction thus includes our parameter uncertainty.

A Gaussian likelihood, a Gaussian posterior and a linear model allow us to evaluate this integral analytically. Both factors under the integral are Gaussian. The predictive likelihood is $p(y_* | \mathbf{x}_*, \mathbf{w}) = \mathcal{N}(y_* | \boldsymbol{\phi}(\mathbf{x}_*)^T\mathbf{w}, \sigma^2)$. Here $\sigma^2$ is the measurement variance of the new observation. The posterior is $p(\mathbf{w} | \mathcal{D}) = \mathcal{N}(\mathbf{w} | \mathbf{m}_N, \mathbf{S}_N)$.

By the product rule of Section~2.5, the product of the two factors is the joint distribution of $y_*$ and $\mathbf{w}$. It is Gaussian, because $y_*$ is a linear function of $\mathbf{w}$ plus Gaussian noise. When we marginalize a joint Gaussian distribution, that is, integrate out some of its variables, the result is again Gaussian. The posterior predictive distribution is therefore Gaussian, so we only need its mean and variance. The first appendix of this chapter proves this property.

We find the mean and variance with two tools from Chapter~3. The law of total expectation is $\mathbb{E}[y_*] = \mathbb{E}_{\mathbf{w}}[\mathbb{E}[y_*|\mathbf{w}]]$. For a fixed parameter vector $\mathbf{w}$, our model predicts $y_* = \boldsymbol{\phi}(\mathbf{x}_*)^T\mathbf{w}$ plus zero-mean noise. Taking the expectation over the posterior gives
\begin{align}
\mathbb{E}[y_*] &= \mathbb{E}_{\mathbf{w}}[\mathbb{E}[y_*|\mathbf{w}]] \\*
&= \mathbb{E}_{\mathbf{w}}[\boldsymbol{\phi}(\mathbf{x}_*)^T\mathbf{w}] \\
&= \boldsymbol{\phi}(\mathbf{x}_*)^T\mathbb{E}_{\mathbf{w}}[\mathbf{w}] \\*
&= \boldsymbol{\phi}(\mathbf{x}_*)^T\mathbf{m}_N.
\end{align}This result agrees with intuition. Our best prediction at $\mathbf{x}_*$ uses our best parameter estimate $\mathbf{m}_N$ with the basis functions $\boldsymbol{\phi}(\mathbf{x}_*)$.

The law\index{law of total variance} of total variance, $\text{Var}[y_*] = \mathbb{E}_{\mathbf{w}}[\text{Var}[y_*|\mathbf{w}]] + \text{Var}_{\mathbf{w}}[\mathbb{E}[y_*|\mathbf{w}]]$, shows two sources of uncertainty in our predictions. The first term is the aleatoric\index{aleatoric uncertainty} uncertainty, the randomness of the measurements. Even if we knew the model parameters exactly, repeated observations would scatter because of instrumental noise, photon statistics and other random effects. For our model this term is
\begin{equation}
\mathbb{E}_{\mathbf{w}}[\text{Var}[y_*|\mathbf{w}]] = \mathbb{E}_{\mathbf{w}}[\sigma^2] = \sigma^2.
\end{equation}More data cannot reduce this part, because each new observation still scatters by $\sigma$. Averaging many observations only shrinks the uncertainty of their mean.

The second term is the epistemic\index{epistemic uncertainty} uncertainty, our imperfect knowledge of the true model parameters. It arises because the data are finite and noisy. We evaluate it as follows
\begin{align}
\text{Var}_{\mathbf{w}}[\mathbb{E}[y_*|\mathbf{w}]] &= \text{Var}_{\mathbf{w}}[\boldsymbol{\phi}(\mathbf{x}_*)^T\mathbf{w}] \\*
&= \boldsymbol{\phi}(\mathbf{x}_*)^T\text{Var}_{\mathbf{w}}[\mathbf{w}]\boldsymbol{\phi}(\mathbf{x}_*) \\*
&= \boldsymbol{\phi}(\mathbf{x}_*)^T\mathbf{S}_N\boldsymbol{\phi}(\mathbf{x}_*).
\end{align}This is the variance of the line at $\mathbf{x}_*$. Unlike the aleatoric part, this component shrinks as we collect more training data, because more observations constrain the parameters.

Combining the two contributions gives the posterior predictive distribution
\begin{equation}
p(y_* | \mathbf{x}_*, \mathcal{D}) = \mathcal{N}(y_* | \boldsymbol{\phi}(\mathbf{x}_*)^T \mathbf{m}_N, \sigma^2 + \boldsymbol{\phi}(\mathbf{x}_*)^T \mathbf{S}_N \boldsymbol{\phi}(\mathbf{x}_*)).
\end{equation}This expression includes both sources of predictive uncertainty. The mean $\boldsymbol{\phi}(\mathbf{x}_*)^T \mathbf{m}_N$ is our best estimate. The variance $\sigma^2 + \boldsymbol{\phi}(\mathbf{x}_*)^T \mathbf{S}_N \boldsymbol{\phi}(\mathbf{x}_*)$ is the sum of the measurement noise and the parameter uncertainty.

\begin{figure}[ht!]
    \centering
    \includegraphics[width=0.8\textwidth]{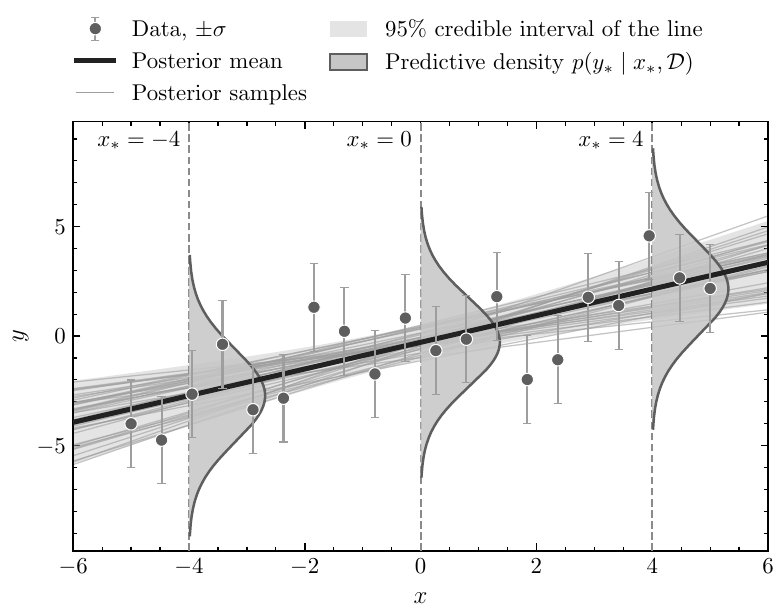}
    \caption{Posterior predictive distribution for the data and posterior of Figure~\ref{fig:regression_comparison}(b). The shaded band is the 95\% credible interval of the regression line, which depends on $\boldsymbol{\phi}(\mathbf{x}_*)^T\mathbf{S}_N\boldsymbol{\phi}(\mathbf{x}_*)$ alone. At $x_* = -4$, $0$ and $4$ (dashed lines), the curves drawn sideways show the predictive distribution $p(y_* | x_*, \mathcal{D})$. Its variance, $\sigma^2 + \boldsymbol{\phi}(\mathbf{x}_*)^T\mathbf{S}_N\boldsymbol{\phi}(\mathbf{x}_*)$, also includes the measurement noise, so each curve is much wider than the band at that input. The curves at $x_* = \pm 4$ are slightly wider than the curve at $x_* = 0$, because the parameter uncertainty grows away from the center of the data.}
    \label{fig:predictive_distribution}
\end{figure}

The two terms give two different intervals, and Figure~\ref{fig:predictive_distribution} shows both. The line value $\boldsymbol{\phi}(\mathbf{x}_*)^T\mathbf{w}$ is a linear function of the Gaussian $\mathbf{w}$, so it is Gaussian with mean $\boldsymbol{\phi}(\mathbf{x}_*)^T\mathbf{m}_N$ and variance $\boldsymbol{\phi}(\mathbf{x}_*)^T\mathbf{S}_N\boldsymbol{\phi}(\mathbf{x}_*)$. A new observation $y_*$ adds the noise $\sigma^2$. A Gaussian has 95\% of its probability within 1.96 standard deviations of its mean, so the 95\% intervals are
\begin{align}
\text{line:} &\quad \boldsymbol{\phi}(\mathbf{x}_*)^T\mathbf{m}_N \pm 1.96\sqrt{\boldsymbol{\phi}(\mathbf{x}_*)^T\mathbf{S}_N\boldsymbol{\phi}(\mathbf{x}_*)}, \\*
\text{new observation:} &\quad \boldsymbol{\phi}(\mathbf{x}_*)^T\mathbf{m}_N \pm 1.96\sqrt{\sigma^2 + \boldsymbol{\phi}(\mathbf{x}_*)^T\mathbf{S}_N\boldsymbol{\phi}(\mathbf{x}_*)}.
\end{align}

The first interval is the shaded band in Figures~\ref{fig:regression_comparison} to~\ref{fig:predictive_distribution}. It describes the line alone. The second interval, the predictive interval for a new observation, is wider because it also includes the noise $\sigma^2$. The sideways curves in Figure~\ref{fig:predictive_distribution} show the full predictive distribution behind it.

The decomposition shows how the uncertainty varies with the input. The parameter uncertainty term $\boldsymbol{\phi}(\mathbf{x}_*)^T \mathbf{S}_N \boldsymbol{\phi}(\mathbf{x}_*)$ typically grows as we move away from the training data. Extrapolation beyond the observed range therefore carries a larger uncertainty than interpolation within it. For the M-$\sigma$ relation, we should be more cautious about black hole mass predictions for galaxies with velocity dispersions far from our training sample.

The quadratic form $\boldsymbol{\phi}(\mathbf{x}_*)^T \mathbf{S}_N \boldsymbol{\phi}(\mathbf{x}_*)$ depends on the features at the prediction point and on the parameter covariance. Regions of feature space that the training data constrain poorly have a higher predictive uncertainty. This shows where new observations help most.

To carry this uncertainty into a larger analysis, we can draw samples from the predictive distribution instead of using a point prediction. This matters in astronomical pipelines, where uncertainties must be tracked through several processing stages.

So far we have assumed that the inputs $\mathbf{x}$ are exact and that only the outputs carry measurement errors. In astronomy, both often carry errors. Ignoring the errors in the inputs flattens the fitted slope. Chapter~6 corrects this.\index{Bayesian linear regression|)}\index{posterior predictive distribution|)}

\section{Appendix: Marginal Distribution of a Multivariate Gaussian}

We\index{Gaussian distribution!marginal|(} asserted that the posterior predictive distribution is Gaussian. The integrand is a joint Gaussian in $y_*$ and $\mathbf{w}$, and integrating out $\mathbf{w}$ leaves a marginal\index{Gaussian distribution!marginal} Gaussian in $y_*$. This follows from a general theorem. Integrating out some variables of a joint multivariate Gaussian leaves a Gaussian.

This result underlies many calculations in Bayesian inference, so we prove it. We show that for a joint multivariate Gaussian over partitioned variables, the marginal distribution of any subset is Gaussian.

\paragraph{Statement of the Result}

Consider a joint multivariate Gaussian distribution over partitioned variables
\begin{equation}
\begin{bmatrix}
\mathbf{x}_1 \\
\mathbf{x}_2
\end{bmatrix}
\;\sim\;
\mathcal{N}\!\left(
\begin{bmatrix}
\boldsymbol{\mu}_1\\
\boldsymbol{\mu}_2
\end{bmatrix},\;
\begin{bmatrix}
\boldsymbol{\Sigma}_{11} & \boldsymbol{\Sigma}_{12}\\
\boldsymbol{\Sigma}_{21} & \boldsymbol{\Sigma}_{22}
\end{bmatrix}
\right),
\end{equation}
where $\mathbf{x}_1 \in \mathbb{R}^{D_1}$ and $\mathbf{x}_2 \in \mathbb{R}^{D_2}$ are vectors of $D_1$ and $D_2$ real numbers. We partition the covariance matrix accordingly.

We prove that the marginal distribution of $\mathbf{x}_1$ is
\begin{equation}
\mathbf{x}_1 \;\sim\; \mathcal{N}\left(\boldsymbol{\mu}_1,\;\boldsymbol{\Sigma}_{11}\right).
\end{equation}

\paragraph{Proof}

The joint probability density function is
\begin{equation}
p(\mathbf{x}_1,\mathbf{x}_2) = \frac{1}{(2\pi)^{(D_1+D_2)/2}\sqrt{\det(\boldsymbol{\Sigma})}} \exp\!\left(-\frac{1}{2}\begin{bmatrix}\mathbf{x}_1 - \boldsymbol{\mu}_1\\ \mathbf{x}_2 - \boldsymbol{\mu}_2\end{bmatrix}^T\boldsymbol{\Sigma}^{-1}\begin{bmatrix}\mathbf{x}_1 - \boldsymbol{\mu}_1\\ \mathbf{x}_2 - \boldsymbol{\mu}_2\end{bmatrix}\right).
\end{equation}

To find the marginal distribution $p(\mathbf{x}_1)$, we integrate over $\mathbf{x}_2$
\begin{equation}
p(\mathbf{x}_1) = \int_{\mathbb{R}^{D_2}} p(\mathbf{x}_1,\mathbf{x}_2)\,\mathrm{d}\mathbf{x}_2.
\end{equation}

We define $\mathbf{u} = \mathbf{x}_1 - \boldsymbol{\mu}_1$ and $\mathbf{v} = \mathbf{x}_2 - \boldsymbol{\mu}_2$ for simplicity. We partition the precision matrix $\boldsymbol{\Sigma}^{-1}$ as
\begin{equation}
\boldsymbol{\Sigma}^{-1} = \begin{bmatrix}
\boldsymbol{\Lambda}_{11} & \boldsymbol{\Lambda}_{12}\\
\boldsymbol{\Lambda}_{21} & \boldsymbol{\Lambda}_{22}
\end{bmatrix},
\end{equation}
where the dimensions match the partitioning of $\boldsymbol{\Sigma}$.

The quadratic form in the exponent becomes
\begin{equation}
\begin{bmatrix}\mathbf{u}\\ \mathbf{v}\end{bmatrix}^T\begin{bmatrix}\boldsymbol{\Lambda}_{11} & \boldsymbol{\Lambda}_{12}\\ \boldsymbol{\Lambda}_{21} & \boldsymbol{\Lambda}_{22}\end{bmatrix}\begin{bmatrix}\mathbf{u}\\ \mathbf{v}\end{bmatrix} = \mathbf{u}^T\boldsymbol{\Lambda}_{11}\mathbf{u} + \mathbf{u}^T\boldsymbol{\Lambda}_{12}\mathbf{v} + \mathbf{v}^T\boldsymbol{\Lambda}_{21}\mathbf{u} + \mathbf{v}^T\boldsymbol{\Lambda}_{22}\mathbf{v}.
\end{equation}

The matrix $\boldsymbol{\Sigma}^{-1}$ is symmetric, as the inverse of a symmetric matrix. Hence $\boldsymbol{\Lambda}_{12} = \boldsymbol{\Lambda}_{21}^T$, and
\begin{equation}
\mathbf{u}^T\boldsymbol{\Lambda}_{12}\mathbf{v} + \mathbf{v}^T\boldsymbol{\Lambda}_{21}\mathbf{u} = 2\mathbf{v}^T\boldsymbol{\Lambda}_{21}\mathbf{u}.
\end{equation}

The integral becomes
\begin{equation}
p(\mathbf{x}_1) = \frac{1}{(2\pi)^{(D_1+D_2)/2}\sqrt{\det(\boldsymbol{\Sigma})}} \exp\left(-\frac{1}{2}\mathbf{u}^T\boldsymbol{\Lambda}_{11}\mathbf{u}\right) \int_{\mathbb{R}^{D_2}} \exp\left(-\frac{1}{2}[\mathbf{v}^T\boldsymbol{\Lambda}_{22}\mathbf{v} + 2\mathbf{v}^T\boldsymbol{\Lambda}_{21}\mathbf{u}]\right)\,\mathrm{d}\mathbf{v}.
\end{equation}

To evaluate the integral, we complete the square in $\mathbf{v}$. We define $\mathbf{b} = \boldsymbol{\Lambda}_{21}\mathbf{u}$, so the quadratic form in $\mathbf{v}$ is
\begin{equation}
\frac{1}{2}\mathbf{v}^T\boldsymbol{\Lambda}_{22}\mathbf{v} + \mathbf{b}^T\mathbf{v} = \frac{1}{2}(\mathbf{v} + \boldsymbol{\Lambda}_{22}^{-1}\mathbf{b})^T\boldsymbol{\Lambda}_{22}(\mathbf{v} + \boldsymbol{\Lambda}_{22}^{-1}\mathbf{b}) - \frac{1}{2}\mathbf{b}^T\boldsymbol{\Lambda}_{22}^{-1}\mathbf{b}.
\end{equation}The integral over the completed square is a standard Gaussian integral
\begin{equation}
\int_{\mathbb{R}^{D_2}} \exp\left(-\frac{1}{2}(\mathbf{v} + \boldsymbol{\Lambda}_{22}^{-1}\mathbf{b})^T\boldsymbol{\Lambda}_{22}(\mathbf{v} + \boldsymbol{\Lambda}_{22}^{-1}\mathbf{b})\right)\,\mathrm{d}\mathbf{v} = (2\pi)^{D_2/2}[\det(\boldsymbol{\Lambda}_{22})]^{-1/2}.
\end{equation}

Combining the results gives
\begin{align}
p(\mathbf{x}_1) &= \frac{(2\pi)^{D_2/2}}{(2\pi)^{(D_1+D_2)/2}\sqrt{\det(\boldsymbol{\Sigma})}} \frac{1}{\sqrt{\det(\boldsymbol{\Lambda}_{22})}} \exp\left(-\frac{1}{2}\mathbf{u}^T\boldsymbol{\Lambda}_{11}\mathbf{u} + \frac{1}{2}\mathbf{b}^T\boldsymbol{\Lambda}_{22}^{-1}\mathbf{b}\right) \\*
&= \frac{1}{(2\pi)^{D_1/2}} \frac{1}{\sqrt{\det(\boldsymbol{\Sigma})}\sqrt{\det(\boldsymbol{\Lambda}_{22})}} \exp\left(-\frac{1}{2}\mathbf{u}^T[\boldsymbol{\Lambda}_{11} - \boldsymbol{\Lambda}_{12}\boldsymbol{\Lambda}_{22}^{-1}\boldsymbol{\Lambda}_{21}]\mathbf{u}\right).
\end{align}The exponent contains the matrix $\boldsymbol{\Lambda}_{11} - \boldsymbol{\Lambda}_{12}\boldsymbol{\Lambda}_{22}^{-1}\boldsymbol{\Lambda}_{21}$, the Schur\index{Schur complement} complement of $\boldsymbol{\Lambda}_{22}$. For a block matrix, the upper left block of the inverse is the inverse of the Schur complement of the lower right block. Here the block matrix is $\boldsymbol{\Lambda}$ and its inverse is $\boldsymbol{\Sigma}$, so
\begin{equation}
\boldsymbol{\Sigma}_{11} = (\boldsymbol{\Lambda}_{11} - \boldsymbol{\Lambda}_{12}\boldsymbol{\Lambda}_{22}^{-1}\boldsymbol{\Lambda}_{21})^{-1}.
\end{equation}

The Schur complement also gives a determinant relation. The determinant of $\boldsymbol{\Sigma}^{-1}$ is $\det\boldsymbol{\Lambda}_{22}$ times the determinant of the Schur complement. Both block matrix results follow by factoring the block matrix into block triangular pieces. By the previous equation, the Schur complement has determinant $1/\det\boldsymbol{\Sigma}_{11}$, and hence
\begin{equation}
\sqrt{\det(\boldsymbol{\Sigma})}\sqrt{\det(\boldsymbol{\Lambda}_{22})} = \sqrt{\det(\boldsymbol{\Sigma}_{11})}.
\end{equation}Substituting these relationships gives
\begin{equation}
p(\mathbf{x}_1) = \frac{1}{(2\pi)^{D_1/2}\sqrt{\det(\boldsymbol{\Sigma}_{11})}} \exp\left(-\frac{1}{2}(\mathbf{x}_1-\boldsymbol{\mu}_1)^T\boldsymbol{\Sigma}_{11}^{-1}(\mathbf{x}_1-\boldsymbol{\mu}_1)\right).
\end{equation}This is the probability density function of $\mathcal{N}(\mathbf{x}_1|\boldsymbol{\mu}_1,\boldsymbol{\Sigma}_{11})$, which completes the proof.

This result matters for Bayesian calculations. When we integrate over the parameter uncertainty to obtain predictions, the result keeps the Gaussian form with analytically computable parameters. We can also compute the marginal likelihood (evidence) analytically by integrating over the parameter priors, which allows model comparison (Chapter~14).

Together with conjugacy, the Gaussian form under marginalization makes tractable many Bayesian calculations that would otherwise need numerical integration.

\section[Appendix: Conditional Distribution of a Multivariate Gaussian]{Appendix: Conditional Distribution of a\\ Multivariate Gaussian}

Chapters~13 and 14 use the conditional distribution of one part of a joint Gaussian given the other part. We derive it here for the joint Gaussian of Section\index{Gaussian distribution!marginal|)}~5.8.

\paragraph{Statement of the Result}

Consider\index{Gaussian distribution!conditional|(} again the joint multivariate Gaussian distribution over partitioned variables
\begin{equation}
\begin{bmatrix}
\mathbf{x}_1 \\
\mathbf{x}_2
\end{bmatrix}
\;\sim\;
\mathcal{N}\!\left(
\begin{bmatrix}
\boldsymbol{\mu}_1\\
\boldsymbol{\mu}_2
\end{bmatrix},\;
\begin{bmatrix}
\boldsymbol{\Sigma}_{11} & \boldsymbol{\Sigma}_{12}\\
\boldsymbol{\Sigma}_{21} & \boldsymbol{\Sigma}_{22}
\end{bmatrix}
\right).
\end{equation}
We prove that the conditional distribution of $\mathbf{x}_1$ given $\mathbf{x}_2$ is
\begin{equation}
p(\mathbf{x}_1|\mathbf{x}_2) = \mathcal{N}\left(\mathbf{x}_1\,\Big|\,\boldsymbol{\mu}_1 + \boldsymbol{\Sigma}_{12}\boldsymbol{\Sigma}_{22}^{-1}(\mathbf{x}_2 - \boldsymbol{\mu}_2),\; \boldsymbol{\Sigma}_{11} - \boldsymbol{\Sigma}_{12}\boldsymbol{\Sigma}_{22}^{-1}\boldsymbol{\Sigma}_{21}\right).
\end{equation}
The mean is a linear function of $\mathbf{x}_2$, and the covariance does not depend on $\mathbf{x}_2$.

\paragraph{Proof}

As in Section~5.8, we define $\mathbf{u} = \mathbf{x}_1 - \boldsymbol{\mu}_1$ and $\mathbf{v} = \mathbf{x}_2 - \boldsymbol{\mu}_2$. We write $\boldsymbol{\Lambda}_{ij}$ for the blocks of the precision matrix $\boldsymbol{\Lambda} = \boldsymbol{\Sigma}^{-1}$, which is symmetric, so $\boldsymbol{\Lambda}_{12} = \boldsymbol{\Lambda}_{21}^T$. The joint density is proportional to the exponential of $-\tfrac{1}{2}$ times the quadratic form
\begin{equation}
\begin{bmatrix}\mathbf{u}\\ \mathbf{v}\end{bmatrix}^T\begin{bmatrix}\boldsymbol{\Lambda}_{11} & \boldsymbol{\Lambda}_{12}\\ \boldsymbol{\Lambda}_{21} & \boldsymbol{\Lambda}_{22}\end{bmatrix}\begin{bmatrix}\mathbf{u}\\ \mathbf{v}\end{bmatrix} = \mathbf{u}^T\boldsymbol{\Lambda}_{11}\mathbf{u} + 2\mathbf{u}^T\boldsymbol{\Lambda}_{12}\mathbf{v} + \mathbf{v}^T\boldsymbol{\Lambda}_{22}\mathbf{v}.
\end{equation}

We keep $\mathbf{x}_2$ fixed and look at the density as a function of $\mathbf{x}_1$. The term with $\boldsymbol{\Lambda}_{22}$ does not contain $\mathbf{u}$, so only the first two terms matter. We complete the square in $\mathbf{u}$
\begin{equation}
\mathbf{u}^T\boldsymbol{\Lambda}_{11}\mathbf{u} + 2\mathbf{u}^T\boldsymbol{\Lambda}_{12}\mathbf{v} = (\mathbf{u} + \boldsymbol{\Lambda}_{11}^{-1}\boldsymbol{\Lambda}_{12}\mathbf{v})^T\boldsymbol{\Lambda}_{11}(\mathbf{u} + \boldsymbol{\Lambda}_{11}^{-1}\boldsymbol{\Lambda}_{12}\mathbf{v}) - \mathbf{v}^T\boldsymbol{\Lambda}_{21}\boldsymbol{\Lambda}_{11}^{-1}\boldsymbol{\Lambda}_{12}\mathbf{v}.
\end{equation}
The last term does not depend on $\mathbf{x}_1$. The density of $\mathbf{x}_1$ is therefore a Gaussian with covariance $\boldsymbol{\Lambda}_{11}^{-1}$ and mean $\boldsymbol{\mu}_1 - \boldsymbol{\Lambda}_{11}^{-1}\boldsymbol{\Lambda}_{12}(\mathbf{x}_2 - \boldsymbol{\mu}_2)$.

We now write both in terms of $\boldsymbol{\Sigma}$. The relation $\boldsymbol{\Lambda}\boldsymbol{\Sigma} = \mathrm{Id}$ gives
\begin{align}
\boldsymbol{\Lambda}_{11}\boldsymbol{\Sigma}_{12} + \boldsymbol{\Lambda}_{12}\boldsymbol{\Sigma}_{22} &= \mathbf{0}, \\*
\boldsymbol{\Lambda}_{11}\boldsymbol{\Sigma}_{11} + \boldsymbol{\Lambda}_{12}\boldsymbol{\Sigma}_{21} &= \mathrm{Id}.
\end{align}
The first relation gives $-\boldsymbol{\Lambda}_{11}^{-1}\boldsymbol{\Lambda}_{12} = \boldsymbol{\Sigma}_{12}\boldsymbol{\Sigma}_{22}^{-1}$, and the second then gives $\boldsymbol{\Lambda}_{11}^{-1} = \boldsymbol{\Sigma}_{11} - \boldsymbol{\Sigma}_{12}\boldsymbol{\Sigma}_{22}^{-1}\boldsymbol{\Sigma}_{21}$. Substituting both into the mean and the covariance gives the stated result, which completes the proof.\index{Gaussian distribution!conditional|)}

\section{Tutorial: Stellar Temperature Uncertainties}\label{sec:ch5_tutorial}

This\index{stellar temperatures|(} tutorial applies the posterior and the predictive distribution of this chapter to stellar temperatures predicted from spectra. Errors in these temperatures propagate into stellar ages and abundances. Each prediction therefore needs an uncertainty. The full tutorial can be found at \url{https://statml.org/} as tutorial 5.

\begin{figure}[t!]
    \centering
    \includegraphics[width=0.95\textwidth]{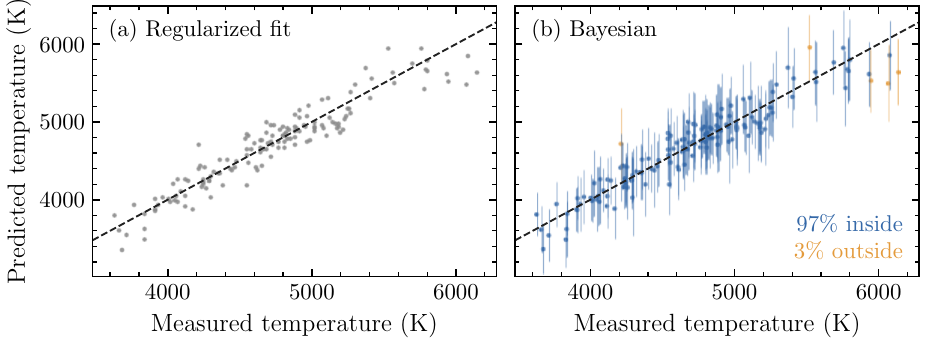}
    \caption{The tutorial predicts temperatures from spectra. Panel~(a) shows the predictions for test stars, and panel~(b) adds a 95 percent interval to each. Blue bars touch the dashed line of perfect prediction and so contain the measured temperature, and orange bars do not. The labels give the percentages of blue and orange bars.}
    \label{fig:ch5_tutorial_intervals}
\end{figure}

The tutorial returns to the APOGEE\index{APOGEE survey} spectra of Section~\ref{sec:ch4_tutorials}. A regularized fit predicts each temperature (Figure~\ref{fig:ch5_tutorial_intervals}a), but it does not say how far to trust it. The tutorial adds simulated noise of known size to the pipeline temperatures and treats the result as the measured temperatures.

The regularized fit gives the peak of the posterior of the weights for a Gaussian prior (Section~\ref{sec:ch5_regularization}). The Bayesian treatment keeps the whole posterior, which has a closed form (Section~\ref{sec:ch5_blr_math}). Different weights that the data allow give different predictions for a new star, and their spread is the uncertainty. The posterior predictive distribution of Section~\ref{sec:posterior_predictive} adds the noise of the new star. The tutorial adds a scatter for the misfit of the linear model. Each star then has a 95 percent interval of its own (Figure~\ref{fig:ch5_tutorial_intervals}b).

We can check these intervals on the test stars, after the validation stars set the prior width. An interval is honest when it contains the measured temperature as often as it claims, as the intervals of Figure~\ref{fig:ch5_tutorial_intervals}b do. Honest intervals let us carry the temperature errors into stellar ages and abundances. The prior keeps the fit from overfitting, and the posterior predictive distribution turns each prediction into such an interval.\index{stellar temperatures|)}

\paragraph{Further Reading} Laplace (1820) treated inverse probability, an early form of Bayesian inference. Zellner (1971) applies Bayesian methods to regression problems in econometrics. Lindley (1965) covers probability and statistics from a Bayesian viewpoint. Raiffa and Schlaifer (1961) developed conjugate priors, which make Bayesian regression analytically tractable, and DeGroot (1970) treats their properties. Anderson (1958) develops the multivariate Gaussian theory that underlies Bayesian linear regression. Tikhonov (1963) developed regularization methods, and Hoerl and Kennard (1970) developed ridge regression. Tibshirani (1996) introduced the lasso, the sparse MAP estimate under a Laplace prior. MacKay (1992) showed how to set regularization parameters with the Bayesian evidence, which links machine learning and Bayesian inference. Kass and Raftery (1995) review Bayes factors and ways to compute the marginal likelihood for model comparison.

\chapter{Linear Regression with Input Uncertainties}

In many real-life machine learning applications, the uncertainty of the inputs is negligible. Image classifiers, for example, often see photographs with a signal-to-noise ratio (SNR), the measured signal divided by its uncertainty, above 100 or even 1000. The input uncertainty then plays almost no role in the inference.

In astronomy, the situation is often different. Both the inputs $x$ and the outputs $y$ can carry large uncertainties. Astronomical images and spectra often have an SNR near 10, and sometimes near 1. The input uncertainty can then reach 10\% or even 100\% of the signal. Derived quantities, such as the black hole masses and velocity dispersions of the M-$\sigma$ relation,\index{M-sigma relation@$M$--$\sigma$ relation} can have large uncertainties as well. These uncertainties often reach about 10\% of the full range of the observed values.

Chapters~4 and~5 ignored the input uncertainties. When we ignore them, the estimated slope is biased toward zero, which weakens (attenuates) the relation. This effect is called attenuation\index{attenuation bias} bias, or regression\index{regression dilution|see{attenuation bias}} dilution. Statisticians have studied it at length under the names measurement\index{measurement error models|see{errors-in-variables regression}} error models and errors-in-variables\index{errors-in-variables regression} regression, that is, regression with uncertain inputs. It also arises in other machine learning models with uncertain inputs. Linear regression is a good place to study it, because the algebra is exact and the geometry is clear.

\looseness=1 We first build intuition for why input uncertainty weakens the relation, then derive the bias and see how large it is in astronomy. We start with equal uncertainties for all points and let them differ at the end. Next we treat the true $x$-values as hidden, or latent, variables and infer them together with the model parameters. This leads to Deming regression, which accounts for uncertainties in both variables and corrects the underestimate of the slope that ordinary\index{least squares!ordinary} least squares (OLS)\index{OLS|see{least squares, ordinary}} gives. The method is linear, but what we learn about handling input uncertainties applies broadly across statistical inference in astronomy.

\section{Attenuation Bias}\label{sec:ch6_attenuation}

\begin{figure}[ht!]
    \centering
    \includegraphics[width=0.9\textwidth]{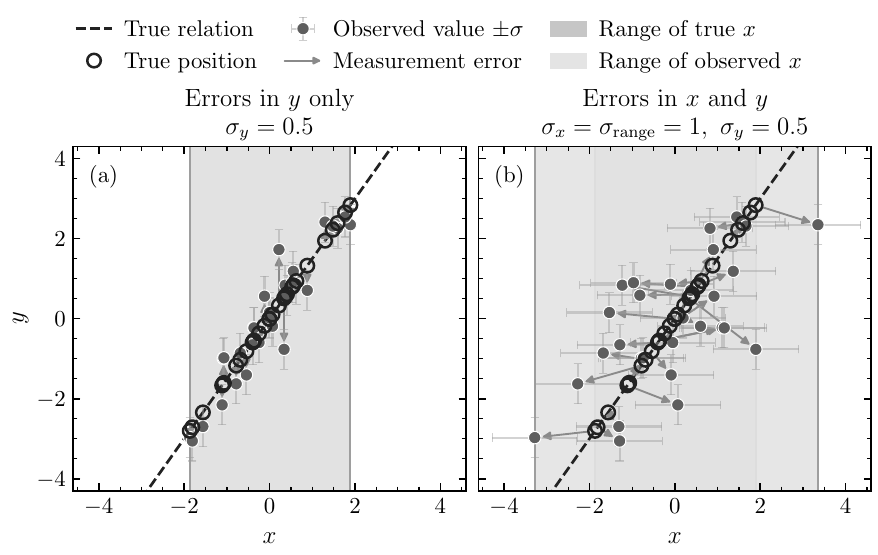}
    \caption{The same true points on the relation $y = 1.5x$, observed with errors in $y$ only (a) and in both $x$ and $y$ (b). The $y$-errors have standard deviation $\sigma_y$ in both panels. In panel~(b) the $x$-errors have standard deviation $\sigma_x$ equal to $\sigma_{\text{range}}$, the standard deviation of the true $x$-values. The dark grey band spans the true $x$-values and the light grey band spans the observed $x$-values. In panel~(a) the two bands coincide. In panel~(b) the observed range is wider than the true range.}
    \label{fig:attenuation_bias}
\end{figure}

Consider\index{attenuation bias|(} a linear relation between $y$ and $x$, such as the M-$\sigma$ relation. If we measured $x$ perfectly but had uncertainties in $y$, the data would scatter vertically around the true relation, as in Figure~\ref{fig:attenuation_bias}(a). As in the previous chapters, maximum likelihood or Bayesian methods can recover the true relation when they account for these $y$-uncertainties.

Now suppose that $x$ is uncertain too. Each point then scatters both horizontally and vertically from its true position, as in panel~(b). The horizontal scatter inflates the range of $x$, the interval between its smallest and largest values, as the bands in Figure~\ref{fig:attenuation_bias} show. The range along $y$ has no matching increase.

Let $\sigma_x$ be the standard deviation of the $x$-errors and $\sigma_{\text{range}}$ the standard deviation of the true $x$-values. For a small error, such as $\sigma_x/\sigma_{\text{range}} = 0.1$, the observed range grows only modestly. For $\sigma_x/\sigma_{\text{range}} = 1$, as in panel~(b), the inflation is large.

In the extreme case, the $x$-uncertainty grows without limit and the $x$-measurements become pure noise with no correlation to their true values. Any relation is then hidden, and the fitted line becomes horizontal.

Why does the vertical scatter not cause the same problem? In regression, $x$ and $y$ play different roles. A fit asks what $y$-value to expect for a given $x$-value. Uncertainty in $y$ adds noise to this prediction, which averages out with enough data. Uncertainty in $x$ changes the question. We now predict $y$ from a noisy version of $x$, which flattens the relation that we find.

\section{Mathematical Model}

To make this intuition precise, we need a model of the true relation and of the measurement errors. We take the simplest case first, univariate (one input variable) linear regression with homogeneous, or homoscedastic, uncertainties. This case is restrictive, but its analytical results show how attenuation bias arises. Section~6.8 treats the heteroscedastic case, in which the uncertainties differ from point to point.

The true relation that we want to measure is
\begin{equation}
    y_{\text{true}} = \gamma x_{\text{true}},
\end{equation}
where $\gamma$ is the true slope. In the M-$\sigma$ relation, $x_{\text{true}}$ is the logarithm of the true velocity dispersion and $y_{\text{true}}$ is the logarithm of the true black hole mass.

We do not measure these true values directly. The measurements contain errors
\begin{align}
    x_{\text{obs}} &= x_{\text{true}} + \delta_x \\*
    y_{\text{obs}} &= y_{\text{true}} + \delta_y = \gamma x_{\text{true}} + \delta_y.
\end{align}

To analyze this model, we make three assumptions about the errors.

First, the errors are unbiased
        \begin{align}
            \mathbb{E}[\delta_x] &= 0 \\*
            \mathbb{E}[\delta_y] &= 0.
        \end{align}
        This means that our measurements do not systematically over- or underestimate the true values. In astronomy, we calibrate our instruments and correct for known systematic effects to ensure this property.

    Second, the errors have homogeneous variances
        \begin{align}
            \text{Var}(\delta_x) &= \sigma_x^2 \\*
            \text{Var}(\delta_y) &= \sigma_y^2.
        \end{align}
        These variances do not change from point to point.

    Third, the errors are independent of the true values and of each other. This is often approximately true in astronomy. Photon noise in different measurements, for example, is typically independent. Instrumental effects usually do not depend on the signal that we measure.

It is also useful to work with centered variables, which are shifted to have zero mean. For the inputs, centering puts the pivot of Section~5.5 at the mean of the $x$-values, which makes the intercept and the slope uncorrelated. We redefine the observed variables by subtracting their means
\begin{align}
    x_{\text{obs}} &\leftarrow x_{\text{obs}} - \mathbb{E}[x_{\text{obs}}] \\*
    y_{\text{obs}} &\leftarrow y_{\text{obs}} - \mathbb{E}[y_{\text{obs}}].
\end{align}Centered variables simplify the algebra and focus attention on how changes in $x$ relate to changes in $y$.

\section{Maximum Likelihood with Input Uncertainties}\label{sec:ch6_mle}

We\index{maximum likelihood estimation (MLE)|(} now connect this model to maximum likelihood estimation. In Chapter~4 we showed that, for homogeneous noise, the maximum likelihood estimator has a matrix form. Here both variables have observed and true values. We therefore write the observed outputs, called $\mathbf{t}$ there, as $\mathbf{y}_{\text{obs}}$, like $x_{\text{obs}}$. The estimator reads
\begin{equation}
    \mathbf{w}_{\text{ML}} = (\boldsymbol{\Phi}^T\boldsymbol{\Phi})^{-1}\boldsymbol{\Phi}^T\mathbf{y}_{\text{obs}}.
\end{equation}

Simple linear regression with one input variable is a special case of this formula. Because the variables are centered, the intercept is zero and we estimate only the slope. The design matrix $\boldsymbol{\Phi}$ is then a vector of the $N$ observed $x$-values
\begin{equation}
    \boldsymbol{\Phi} = \begin{bmatrix} x_{\text{obs},1} & x_{\text{obs},2} & \cdots & x_{\text{obs},N} \end{bmatrix}^T.
\end{equation}The two matrix products reduce to sums
\begin{align}
    \boldsymbol{\Phi}^T\boldsymbol{\Phi} &= \begin{bmatrix} x_{\text{obs},1} & x_{\text{obs},2} & \cdots & x_{\text{obs},N} \end{bmatrix} \begin{bmatrix} x_{\text{obs},1} \\ x_{\text{obs},2} \\ \vdots \\ x_{\text{obs},N} \end{bmatrix} = \sum_{i=1}^N x_{\text{obs},i}^2 \\*
    \boldsymbol{\Phi}^T\mathbf{y}_{\text{obs}} &= \begin{bmatrix} x_{\text{obs},1} & x_{\text{obs},2} & \cdots & x_{\text{obs},N} \end{bmatrix} \begin{bmatrix} y_{\text{obs},1} \\ y_{\text{obs},2} \\ \vdots \\ y_{\text{obs},N} \end{bmatrix} = \sum_{i=1}^N x_{\text{obs},i}y_{\text{obs},i}.
\end{align}The matrix equation then reduces to the scalar OLS slope estimate $\hat{\gamma}$, where the hat distinguishes the estimate from the true slope $\gamma$
\begin{equation}
    \hat{\gamma} = \frac{\sum_{i=1}^N x_{\text{obs},i}y_{\text{obs},i}}{\sum_{i=1}^N x_{\text{obs},i}^2}.
\end{equation}

We can also write this solution as a ratio of a covariance to a variance
\begin{equation}
    \hat{\gamma} = \frac{\text{Cov}(x_{\text{obs}}, y_{\text{obs}})}{\text{Var}(x_{\text{obs}})}.
\end{equation}In this form, the numerator measures how much $y$ changes as $x$ varies. The denominator measures the spread of the $x$-values. Their ratio is the change in $y$ per unit change in $x$, the slope. This interpretation holds only if the denominator is the variance of the true $x$-values.

To see how the uncertainties affect the estimate, we analyze the numerator and the denominator separately. For a large sample, the sample covariance and variance approach their expectations, which we compute. We start with the numerator, the covariance of the observed quantities. We use the expressions for the centered variables and the linearity of the covariance. Linearity means that $\text{Cov}(X + Y, Z) = \text{Cov}(X, Z) + \text{Cov}(Y, Z)$ for random variables $X$, $Y$ and $Z$, and that constant factors come out of either argument
\begin{align}
    \text{Cov}(x_{\text{obs}}, y_{\text{obs}}) &= \text{Cov}(x_{\text{true}} + \delta_x, \gamma x_{\text{true}} + \delta_y) \\*
    &= \gamma \text{Cov}(x_{\text{true}} + \delta_x, x_{\text{true}}) + \text{Cov}(x_{\text{true}} + \delta_x, \delta_y) \\
    &= \gamma [\text{Cov}(x_{\text{true}}, x_{\text{true}}) + \text{Cov}(\delta_x, x_{\text{true}})] + [\text{Cov}(x_{\text{true}}, \delta_y) + \text{Cov}(\delta_x, \delta_y)] \\
    &= \gamma [\text{Var}(x_{\text{true}}) + 0] + [0 + 0] \\*
    &= \gamma \text{Var}(x_{\text{true}}).
\end{align}

The expansion in the third line has four contributions. The first, $\text{Cov}(x_{\text{true}}, x_{\text{true}})$, is the spread of the true values, which is $\text{Var}(x_{\text{true}})$. The second, $\text{Cov}(\delta_x, x_{\text{true}})$, links the measurement errors to the true values. The third, $\text{Cov}(x_{\text{true}}, \delta_y)$, links the $y$-errors to the true $x$-values. The fourth, $\text{Cov}(\delta_x, \delta_y)$, links the $x$-errors to the $y$-errors.

Our independence assumptions make the last three terms zero, because measurement errors do not depend on the true values or on each other. Independent variables have zero covariance (Section~3.6). Therefore
\begin{equation}
    \text{Cov}(x_{\text{obs}}, y_{\text{obs}}) = \gamma \text{Var}(x_{\text{true}}) \equiv \gamma \sigma_{\text{range}}^2.
\end{equation}
Here $\sigma_{\text{range}}^2$ is the variance of the true $x$-values, so $\sigma_{\text{range}}$ is their standard deviation. In the M-$\sigma$ relation, it is the spread of the true velocity dispersions of the galaxies in the sample.

For the denominator, we need the variance of the observed $x$-values
\begin{align}
    \text{Var}(x_{\text{obs}}) &= \text{Var}(x_{\text{true}} + \delta_x) \\*
    &= \text{Var}(x_{\text{true}}) + \text{Var}(\delta_x) \\*
    &= \sigma_{\text{range}}^2 + \sigma_x^2.
\end{align}
The second line follows because $x_{\text{true}}$ and $\delta_x$ are independent, and the variances of independent random variables add.

\paragraph{The Attenuation Factor}

An estimator is biased if its average value over many data sets differs from the true value. We therefore compute the average of $\hat{\gamma}$. For a large sample, the average estimate is the covariance $\gamma\sigma_{\text{range}}^2$ divided by the variance $\sigma_{\text{range}}^2 + \sigma_x^2$, so
\begin{equation}
    \mathbb{E}[\hat{\gamma}] = \gamma \frac{\sigma_{\text{range}}^2}{\sigma_{\text{range}}^2 + \sigma_x^2} = \gamma \frac{1}{1 + (\sigma_x/\sigma_{\text{range}})^2}.
\end{equation}This is the main result. The estimate $\hat{\gamma}$ is related to the true slope $\gamma$ through an attenuation factor
\begin{equation}
    \alpha \equiv \frac{1}{1 + (\sigma_x/\sigma_{\text{range}})^2}.
\end{equation}

\begin{figure}[ht!]
    \centering
    \includegraphics[width=0.9\textwidth]{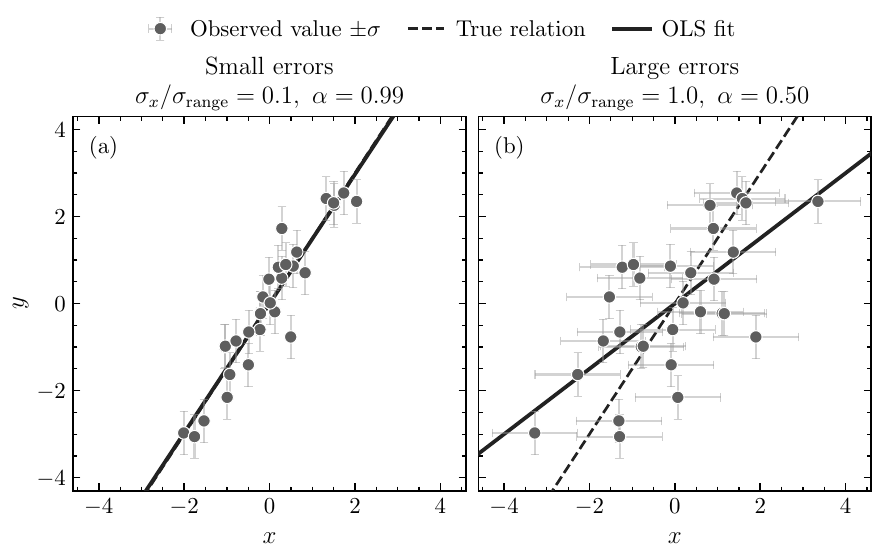}
    \caption{Observed values with error bars of $\pm\sigma_x$ and $\pm\sigma_y$ are shown for small $x$-errors (a) and large $x$-errors (b). The dashed line is the true relation $y = 1.5x$, the solid line is the OLS fit, and $\sigma_y = 0.5$. Panel~(b) shows the data of panel~(b) in Figure~\ref{fig:attenuation_bias}, and panel~(a) uses the same true points with smaller $x$-errors. The panel titles give $\sigma_x/\sigma_{\text{range}}$ and the attenuation factor $\alpha$. The fitted slopes are 1.49 in panel~(a), where the two lines nearly coincide, and 0.75 in panel~(b), half the true slope of 1.5. Both match $\alpha\gamma$.}
    \label{fig:attenuation_slope}
\end{figure}

The average slope estimate is therefore $\alpha\gamma$. The factor $\alpha$ is below one unless $\sigma_x = 0$. In Chapters~4 and~5 the inputs were exact, so $\alpha = 1$ and the slope was unbiased. As $\sigma_x$ grows, $\alpha$ falls and the estimated relation weakens. For $\sigma_x \to \infty$, $\alpha \to 0$, which is the extreme case above with a horizontal fit. Figure~\ref{fig:attenuation_slope} shows the fitted slope for two values of $\sigma_x/\sigma_{\text{range}}$.

\begin{figure}[ht!]
    \centering
    \includegraphics[width=0.8\textwidth]{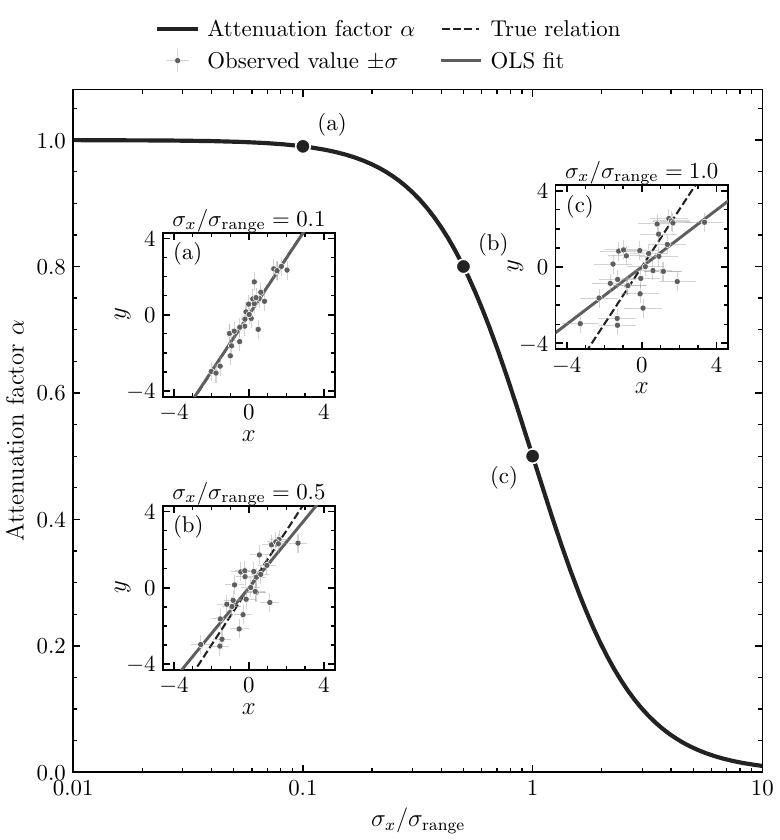}
    \caption{The attenuation factor $\alpha = 1/(1 + (\sigma_x/\sigma_{\text{range}})^2)$ against $\sigma_x/\sigma_{\text{range}}$, on a logarithmic horizontal axis. It is close to 1 for small ratios, equals 1/2 at a ratio of 1, and falls toward 0 for large ratios. The three insets show the same points at three ratios, marked (a), (b) and (c) on the curve. Each inset has the error bars, true relation and OLS fit of Figure~\ref{fig:attenuation_slope}. The fitted slopes are 1.49, 1.20 and 0.75, which equal $\alpha\gamma$ for $\gamma = 1.5$.}
    \label{fig:attenuation_factor}
\end{figure}

The bias depends on the ratio $\sigma_x/\sigma_{\text{range}}$, which compares the errors with the true spread, as Figure~\ref{fig:attenuation_factor} shows. When $\sigma_x$ equals $\sigma_{\text{range}}$, the observed $x$-values have twice the variance of the true values, so their spread is $\sqrt{2}$ times larger. The covariance stays the same while the variance doubles, so the slope falls to half of its true value, $\alpha = 1/2$.

Because $\alpha$ depends only on this ratio and not on the sample size, more data cannot remove the bias of the OLS slope. The $x$-errors are random, but they add the variance $\sigma_x^2$ to the denominator, and averaging over more points does not remove a variance. The $y$-errors add only terms to the covariance that average to zero, so $\alpha$ does not depend on $\sigma_y$. The bias follows directly from the input uncertainties and affects regressions with uncertain independent variables.\index{maximum likelihood estimation (MLE)|)}

\section{Practical Implications of Attenuation Bias}

To gauge the size of the bias in practice, suppose that the uncertainty in $x$ is 10\% of the spread of the true values. Then $\sigma_x/\sigma_{\text{range}} = 0.1$ and $\alpha = 0.99$, a 1\% bias in the slope. This may look small, but even a 1\% systematic bias can matter. Large samples shrink the random errors to the percent level, so a 1\% systematic bias then becomes comparable to them.

In cosmology, for example, photometric redshifts of individual galaxies may have 10\% uncertainties. Yet we aim to constrain cosmological parameters, such as the Hubble constant or the dark energy equation of state, to percent-level precision. The same problem appears in many astronomical measurements.

Several examples below give the uncertainty as a fraction of the full range of a sample. The full range is typically several times $\sigma_{\text{range}}$, so these examples have larger ratios $\sigma_x/\sigma_{\text{range}}$ and larger biases than the case above.

Stellar age uncertainties are often 1 Gyr or more. Only the few stars with precise ages from asteroseismology, the study of stellar oscillations, do better. An uncertainty of 1 Gyr is at least 7\% of the 14 Gyr age of the Universe, the full range of possible ages. Yet we hope to determine the star formation history of galaxies far more precisely from large samples. Relations that depend on age can therefore suffer from attenuation bias.

Careful spectroscopic analyses of chemical abundances typically reach 0.05 to 0.1 dex, where 1 dex is a factor of 10. This is about 10\% of the full metallicity range that we observe in our Galaxy. Such an uncertainty can dilute relations that depend on metallicity.

Black hole mass measurements typically carry uncertainties of 0.3 to 0.5 dex. This is about 10\% of the 3 to 5 orders of magnitude that black hole masses span. When the mass is the independent variable, the resulting attenuation can bias black hole scaling relations.

Stellar and gas masses typically carry uncertainties of 0.1 to 0.2 dex. When they are the independent variables of scaling relations, such as star formation efficiency against gas mass, attenuation bias can weaken the slopes.

Attenuation bias is a systematic effect that we must correct as we push toward more precise inference. Even a very large sample leaves the 1\% bias found above. We therefore need methods that account for uncertainties in both variables.\index{attenuation bias|)}

\section{Latent Variable Formalism}\label{sec:ch6_latent}

To\index{latent variable|(} account for uncertainties in both variables, we change the likelihood. The new likelihood extends the likelihood of Chapters~4 and~5. There we treated the $x$-measurements as fixed, known quantities and modeled only the probability of the $y$-values. Now we model the joint probability of $x$ and $y$, given their true\index{latent variable} values and the model parameters.

Recall the model. A true linear relation links the two variables
\begin{equation}
    y_{\text{true}} = \gamma x_{\text{true}}.
\end{equation}The measurements contain errors
\begin{align}
    x_{\text{obs}} &= x_{\text{true}} + \delta_x \\*
    y_{\text{obs}} &= y_{\text{true}} + \delta_y \\*
    &= \gamma x_{\text{true}} + \delta_y.
\end{align}

For the likelihood we also assume that the errors are normal
\begin{align}
    \delta_x &\sim \mathcal{N}(0, \sigma_x^2) \\*
    \delta_y &\sim \mathcal{N}(0, \sigma_y^2).
\end{align}

\looseness=1 Suppose that we know the true value $x_{\text{true}}$. Then any deviation of $x_{\text{obs}}$ from $x_{\text{true}}$ is due to $\delta_x$ alone, and any deviation of $y_{\text{obs}}$ from $\gamma x_{\text{true}}$ is due to $\delta_y$ alone. Therefore, when we condition on $x_{\text{true}}$, $x_{\text{obs}}$ follows a normal distribution centered at $x_{\text{true}}$ with variance $\sigma_x^2$. Likewise, $y_{\text{obs}}$ follows a normal distribution centered at $\gamma x_{\text{true}}$ with variance $\sigma_y^2$.

Because the two errors are independent, the joint probability is a product
\begin{equation}
    p(x_{\text{obs}}, y_{\text{obs}} \mid x_{\text{true}}, \gamma) = \mathcal{N}(x_{\text{obs}} \mid x_{\text{true}}, \sigma_x^2) \, \mathcal{N}(y_{\text{obs}} \mid \gamma x_{\text{true}}, \sigma_y^2).
\end{equation}In the case of the M-$\sigma$ relation, the first factor describes the scatter from the uncertainty in the velocity dispersion. The second describes the scatter from the uncertainty in the black hole mass.

However, we conditioned everything on $x_{\text{true}}$, and we do not know the true values. If we did, we would be back in the simpler setting of Chapters~4 and~5, with uncertainties only in $y$. Instead, we treat the unknown true values as extra parameters of the model.

The full likelihood therefore depends on the slope $\gamma$ and on the true value $x_{\text{true},i}$ of each observation. We estimate the slope and the true values together. In the M-$\sigma$ relation these are the slope and the true velocity dispersion of each galaxy. Written out, the likelihood is
\begin{equation}
    \mathcal{L}(\gamma, \{x_{\text{true},i}\}) = \prod_{i=1}^N \mathcal{N}(x_{\text{obs},i} \mid x_{\text{true},i}, \sigma_x^2) \, \mathcal{N}(y_{\text{obs},i} \mid \gamma x_{\text{true},i}, \sigma_y^2).
\end{equation}

At first glance this task looks hard. We added $N$ new parameters, one $x_{\text{true},i}$ for each observation. In many problems such additional parameters lead to overfitting (Section~\ref{sec:cross_validation}) or to computational difficulties. Here the inference remains possible, because we have $2N$ measurements for $N+1$ parameters and we know the sizes of the errors.

Treating the unknown true values as parameters to infer gives a hierarchical structure. At the top is the model parameter $\gamma$. Below it are the true values $x_{\text{true},i}$, and at the bottom are the measurements $(x_{\text{obs},i}, y_{\text{obs},i})$.

This hierarchical view is common in astronomy and machine learning. Stellar population studies infer individual stellar parameters together with population-level relations. In cosmology, individual galaxy properties are estimated together with the parameters of large-scale structure.\index{latent variable|)}

\section{Deming Regression}

We\index{Deming regression|(} now solve for both $x_{\text{true}}$ and $\gamma$. Recall that the likelihood of all the observations is
\begin{equation}
    \mathcal{L}(\gamma, \{x_{\text{true},i}\}) = \prod_{i=1}^N \mathcal{N}(x_{\text{obs},i} \mid x_{\text{true},i}, \sigma_x^2) \, \mathcal{N}(y_{\text{obs},i} \mid \gamma x_{\text{true},i}, \sigma_y^2).
\end{equation}Expanding the normal distributions gives
\begin{equation}
    \mathcal{L}(\gamma, \{x_{\text{true},i}\}) = \prod_{i=1}^N \frac{1}{\sqrt{2\pi\sigma_x^2}} \exp\left(-\frac{(x_{\text{obs},i} - x_{\text{true},i})^2}{2\sigma_x^2}\right) \, \frac{1}{\sqrt{2\pi\sigma_y^2}} \exp\left(-\frac{(y_{\text{obs},i} - \gamma x_{\text{true},i})^2}{2\sigma_y^2}\right).
\end{equation}Taking the negative logarithm and dropping constant terms gives the negative log-likelihood
\begin{equation}
    E(\{x_{\text{true},i}\}, \gamma) = \sum_{i=1}^N \left[\frac{(x_{\text{obs},i} - x_{\text{true},i})^2}{2\sigma_x^2} + \frac{(y_{\text{obs},i} - \gamma x_{\text{true},i})^2}{2\sigma_y^2}\right].
\end{equation}The slope $\gamma$ appears in all terms, but each $x_{\text{true},i}$ appears only in the $i$th term. We can therefore optimize each $x_{\text{true},i}$ independently for a fixed $\gamma$, and then use the optimized values to solve for $\gamma$.

For a single observation $i$, the negative log-likelihood is
\begin{equation}
    E(x_{\text{true},i}, \gamma) = \frac{(x_{\text{obs},i} - x_{\text{true},i})^2}{2\sigma_x^2} + \frac{(y_{\text{obs},i} - \gamma x_{\text{true},i})^2}{2\sigma_y^2}.
\end{equation}

To find the optimal $x_{\text{true},i}$, we set the derivative with respect to $x_{\text{true},i}$ to zero
\begin{align}
    \frac{\partial E}{\partial x_{\text{true},i}} &= -\frac{(x_{\text{obs},i} - x_{\text{true},i})}{\sigma_x^2} - \frac{\gamma(y_{\text{obs},i} - \gamma x_{\text{true},i})}{\sigma_y^2} = 0.
\end{align}Rearranging gives
\begin{align}
    \frac{x_{\text{true},i}}{\sigma_x^2} + \frac{\gamma^2 x_{\text{true},i}}{\sigma_y^2} &= \frac{x_{\text{obs},i}}{\sigma_x^2} + \frac{\gamma y_{\text{obs},i}}{\sigma_y^2} \\*
    x_{\text{true},i}\left(\frac{1}{\sigma_x^2} + \frac{\gamma^2}{\sigma_y^2}\right) &= \frac{x_{\text{obs},i}}{\sigma_x^2} + \frac{\gamma y_{\text{obs},i}}{\sigma_y^2}.
\end{align}Multiplying both sides by $\sigma_x^2\sigma_y^2$ gives
\begin{align}
    x_{\text{true},i}(\sigma_y^2 + \gamma^2\sigma_x^2) &= \sigma_y^2x_{\text{obs},i} + \gamma\sigma_x^2y_{\text{obs},i}.
\end{align}

\begin{figure}[ht!]
    \centering
    \includegraphics[width=0.9\textwidth]{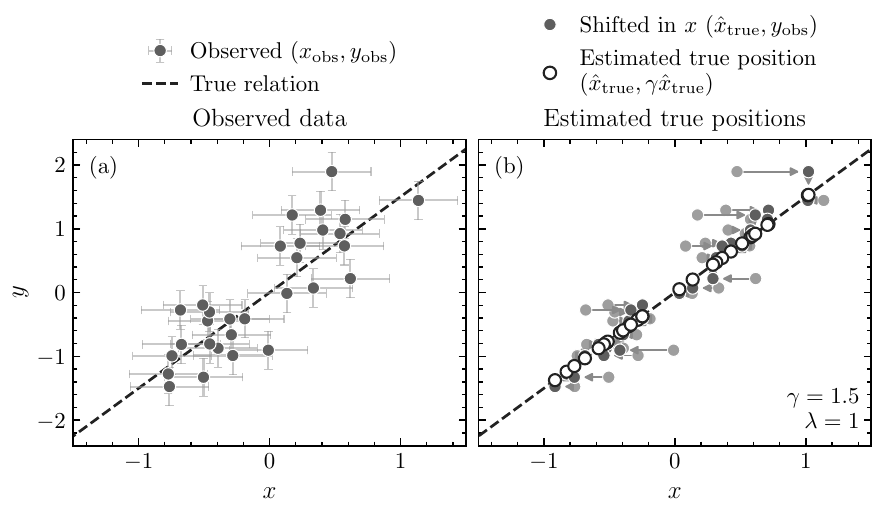}
    \caption{Estimated true positions for data with $\sigma_x = \sigma_y = 0.3$, so $\lambda = \sigma_y^2/\sigma_x^2 = 1$. Panel~(a) shows the observed data with error bars and the true relation (dashed line). In panel~(b), each observed point (light grey) first shifts horizontally to $(\hat{x}_{\text{true}}, y_{\text{obs}})$ (dark grey) and then vertically to the model line at $(\hat{x}_{\text{true}}, \gamma\hat{x}_{\text{true}})$ (open circle), the estimated true position. Here $\hat{x}_{\text{true}}$ is the estimate of the true $x$-value, with $\gamma$ held at its true value. The size of each shift depends on $\sigma_x$, $\sigma_y$ and $\gamma$.}
    \label{fig:xtrue_estimation}
\end{figure}

We define the variance ratio $\lambda = \sigma_y^2/\sigma_x^2$ and divide both sides by $\sigma_x^2(\lambda + \gamma^2)$, which gives
\begin{equation}
    \hat{x}_{\text{true},i} = \frac{\lambda x_{\text{obs},i} + \gamma y_{\text{obs},i}}{\lambda + \gamma^2}.
\end{equation}With this expression we estimate the true $x$-values for a given slope $\gamma$. We write the estimate with a hat, as for the slope. It combines the observed $x$-value, $x_{\text{obs},i}$, with the value that the $y$-measurement predicts, $y_{\text{obs},i}/\gamma$. We thus use the linear relation between $x$ and $y$ to constrain the true values in both dimensions at once. Figure~\ref{fig:xtrue_estimation} shows, for each observation, the shift from $x_{\text{obs},i}$ to the estimate $\hat{x}_{\text{true},i}$. For $\lambda = 1$ the point of the model line at $\hat{x}_{\text{true},i}$ is the one closest to the observed point, and the method is called orthogonal\index{orthogonal regression} regression.

With explicit weights, the formula reads
\begin{equation}
    \hat{x}_{\text{true},i} = \omega_x x_{\text{obs},i} + \omega_y \frac{y_{\text{obs},i}}{\gamma},
\end{equation}
where $\omega_x = \lambda/(\lambda + \gamma^2)$ and $\omega_y = \gamma^2/(\lambda + \gamma^2)$ are weights that sum to 1. Hence $\hat{x}_{\text{true},i}$ is a weighted average of $x_{\text{obs},i}$ and $y_{\text{obs},i}/\gamma$.

The weights depend on the relative uncertainties. The $y$-measurement predicts $x$ as $y_{\text{obs},i}/\gamma$, with uncertainty $\sigma_y/|\gamma|$. If $\sigma_x$ is small compared with this uncertainty, so that $\lambda \gg \gamma^2$, the weight $\omega_x$ approaches 1 and $\hat{x}_{\text{true},i} \approx x_{\text{obs},i}$, so we trust the $x$-measurement. In the opposite limit $\lambda \ll \gamma^2$, $\omega_y$ approaches 1 and $\hat{x}_{\text{true},i} \approx y_{\text{obs},i}/\gamma$, so we infer $x$ from $y$. This weighting is what allows Deming\index{Deming regression} regression to correct the attenuation bias. It balances the direct $x$-measurements against the constraint of the linear model, so we recover estimates of both the slope and the true values.

\section{Analytical Solution}
\enlargethispage{10pt}

To find $\gamma$, we return to the negative log-likelihood. The estimates $\hat{x}_{\text{true},i}$ depend on $\gamma$, so we substitute them below.
\begin{equation}
    E(\{x_{\text{true},i}\}, \gamma) = \sum_{i=1}^N \left[\frac{(x_{\text{obs},i} - x_{\text{true},i})^2}{2\sigma_x^2} + \frac{(y_{\text{obs},i} - \gamma x_{\text{true},i})^2}{2\sigma_y^2}\right].
\end{equation}Setting the derivative with respect to $\gamma$ to zero at the estimates $\hat{x}_{\text{true},i}$ gives
\begin{equation}
    \frac{\partial E}{\partial \gamma} = -\sum_{i=1}^N \frac{x_{\text{true},i}(y_{\text{obs},i} - \gamma x_{\text{true},i})}{\sigma_y^2} = 0.
\end{equation}

Because the derivative with respect to each $x_{\text{true},i}$ vanishes at the estimates, the chain rule makes this the derivative of the optimized negative log-likelihood. The expression for $\hat{x}_{\text{true},i}$ also gives the second factor of each term
\begin{align}
    y_{\text{obs},i} - \gamma\hat{x}_{\text{true},i} &= y_{\text{obs},i} - \frac{\gamma(\lambda x_{\text{obs},i} + \gamma y_{\text{obs},i})}{\lambda + \gamma^2} \\*
    &= \frac{(\lambda + \gamma^2)y_{\text{obs},i} - \gamma(\lambda x_{\text{obs},i} + \gamma y_{\text{obs},i})}{\lambda + \gamma^2} \\*
    &= \frac{\lambda(y_{\text{obs},i} - \gamma x_{\text{obs},i})}{\lambda + \gamma^2}.
\end{align}
Substituting both expressions, the condition becomes
\begin{equation}
    -\sum_{i=1}^N \frac{1}{\sigma_y^2} \left(\frac{\lambda x_{\text{obs},i} + \gamma y_{\text{obs},i}}{\lambda + \gamma^2}\right)\left(\frac{\lambda(y_{\text{obs},i} - \gamma x_{\text{obs},i})}{\lambda + \gamma^2}\right) = 0.
\end{equation}Dividing by the factor $-\lambda/[\sigma_y^2(\lambda + \gamma^2)^2]$, which does not depend on $i$, gives
\begin{equation}
\sum_{i=1}^N (\lambda x_{\text{obs},i} + \gamma y_{\text{obs},i})(y_{\text{obs},i} - \gamma x_{\text{obs},i}) = 0.
\end{equation}Expanding the products gives
\begin{equation}
\sum_{i=1}^N [\lambda x_{\text{obs},i}y_{\text{obs},i} - \lambda\gamma x_{\text{obs},i}^2 + \gamma y_{\text{obs},i}^2 - \gamma^2 x_{\text{obs},i}y_{\text{obs},i}] = 0.
\end{equation}

As in Chapter~3, we describe the data by summary statistics, here the sums of squares and cross products. Here $\mathbb{E}$ is the sample mean, and the sums are defined by
\begin{align}
    S_{xx} &= \sum_{i=1}^N (x_{\text{obs},i} - \mathbb{E}[x_{\text{obs}}])^2 \\*
    S_{yy} &= \sum_{i=1}^N (y_{\text{obs},i} - \mathbb{E}[y_{\text{obs}}])^2 \\*
    S_{xy} &= \sum_{i=1}^N (x_{\text{obs},i} - \mathbb{E}[x_{\text{obs}}])(y_{\text{obs},i} - \mathbb{E}[y_{\text{obs}}]).
\end{align}These sums differ from the sample variance and covariance only by a factor of $N$
\begin{align}
    \text{Var}(x_{\text{obs}}) &= \frac{1}{N}S_{xx} \\*
    \text{Var}(y_{\text{obs}}) &= \frac{1}{N}S_{yy} \\*
    \text{Cov}(x_{\text{obs}},y_{\text{obs}}) &= \frac{1}{N}S_{xy}.
\end{align}For centered variables, where $\mathbb{E}[x_{\text{obs}}] = \mathbb{E}[y_{\text{obs}}] = 0$, they simplify to
\begin{align}
    S_{xx} &= \sum_{i=1}^N x_{\text{obs},i}^2 \\*
    S_{yy} &= \sum_{i=1}^N y_{\text{obs},i}^2 \\*
    S_{xy} &= \sum_{i=1}^N x_{\text{obs},i}y_{\text{obs},i}.
\end{align}

\paragraph{The Deming Estimator}

With these statistics, the equation takes the compact form
\begin{equation}
\lambda S_{xy} - \lambda\gamma S_{xx} + \gamma S_{yy} - \gamma^2 S_{xy} = 0.
\end{equation}Rearranging into standard quadratic form gives
\begin{equation}
\gamma^2 S_{xy} - \gamma(S_{yy} - \lambda S_{xx}) - \lambda S_{xy} = 0.
\end{equation}This is a quadratic equation of the form $a\gamma^2 + b\gamma + c = 0$, with solutions $\gamma = (-b \pm \sqrt{b^2 - 4ac})/(2a)$. Here $a = S_{xy}$, $b = -(S_{yy} - \lambda S_{xx})$ and $c = -\lambda S_{xy}$. The solution is the Deming regression estimator, again written $\hat{\gamma}$, with one root for each sign
\begin{equation}
    \hat{\gamma} = \frac{S_{yy} - \lambda S_{xx} \pm \sqrt{(S_{yy} - \lambda S_{xx})^2 + 4\lambda S_{xy}^2}}{2S_{xy}}.
\end{equation}

The two roots have opposite signs, because their product is $-\lambda$. The maximum likelihood estimate is the root that minimizes the negative log-likelihood. Once the $x_{\text{true},i}$ are optimized, the negative log-likelihood tends to the same constant as $\gamma \to \pm\infty$, because $\hat{x}_{\text{true},i}$ goes to zero while $\gamma\hat{x}_{\text{true},i}$ goes to $y_{\text{obs},i}$. Its derivative, the condition above, vanishes only at the two roots, so one is a minimum and the other a maximum.

\looseness=1 At $\gamma = 0$, between the roots, the derivative is $-S_{xy}/\sigma_y^2$, so the function falls toward the root with the sign of $S_{xy}$, which is the minimum. The root with the plus sign has this sign, because the square root exceeds $|S_{yy} - \lambda S_{xx}|$. We take this root. The plus root also gives the correct limit for $\sigma_x \to 0$, where we recover the OLS slope, as we show below.

We obtain a direct solution without iteration, although both the true values $x_{\text{true},i}$ and the slope $\gamma$ are unknown. The reason is that the problem splits into two steps.

Recall from Section~4.4 that a convex function opens upward and has a single lowest point. The negative log-likelihood is not convex in all the variables jointly. For any fixed $\gamma$, however, it is quadratic, and hence convex, in each $x_{\text{true},i}$, so its optimal values $\hat{x}_{\text{true},i}$ have a closed form. It is also convex in $\gamma$ for fixed $x_{\text{true},i}$. Substituting the optimal $\hat{x}_{\text{true},i}$ back reduces the problem to one dimension, the quadratic in $\gamma$.

\begin{figure}[ht!]
    \centering
    \includegraphics[width=0.9\textwidth]{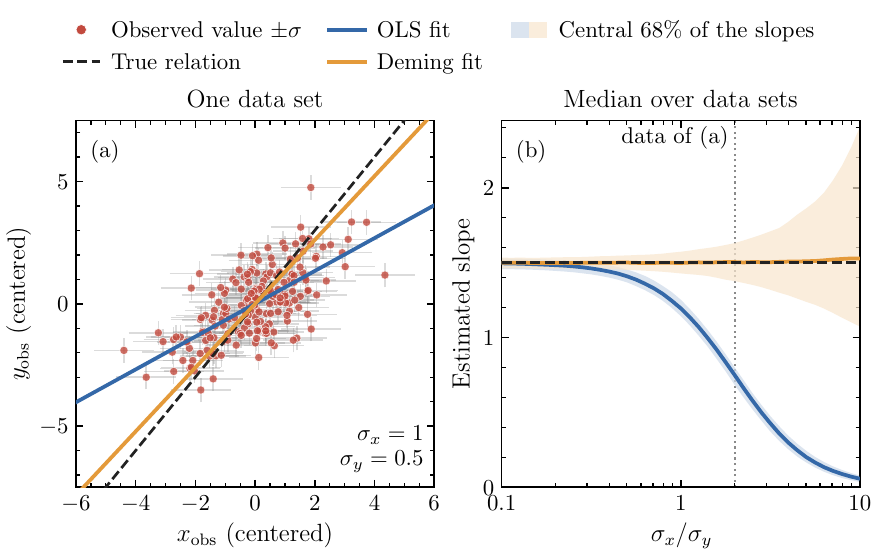}
    \caption{OLS and Deming regression for data sets drawn from $y = 1.5x$ with $\sigma_y = 0.5$ and $\sigma_{\text{range}} = 1$. Panel~(a) shows one data set with $\alpha = 1/2$, and the dotted line in panel~(b) marks it. The OLS slope is 0.67, close to the attenuated value $\alpha\gamma = 0.75$, while the Deming slope is 1.31. Panel~(b) shows the median slope over repeated data sets at each ratio $\sigma_x/\sigma_y$, and the shading marks the central 68\% of the slopes. For $\sigma_x$ much smaller than $\sigma_{\text{range}}$, as at the left end of panel~(b), both methods agree. As $\sigma_x/\sigma_y$ grows, the OLS slope falls increasingly below the true slope, while the median Deming slope stays close to 1.5.}
    \label{fig:deming_correction}
\end{figure}

\paragraph{Recovering the OLS Slope}

Figure~\ref{fig:deming_correction} compares OLS and Deming regression on simulated data. We now check analytically that the Deming estimate $\hat{\gamma}$ reduces to the OLS slope, also written $\hat{\gamma}$, when $\sigma_x \to 0$, or equivalently when $\lambda = \sigma_y^2/\sigma_x^2 \to \infty$.

First we note that
\begin{equation}
    S_{yy} - \lambda S_{xx} = \lambda\left(\frac{S_{yy}}{\lambda} - S_{xx}\right).
\end{equation}The square root can be rewritten as
\begin{equation}
    \sqrt{(S_{yy} - \lambda S_{xx})^2 + 4\lambda S_{xy}^2} = \sqrt{\lambda^2\left(\frac{S_{yy}}{\lambda} - S_{xx}\right)^2 + 4\lambda S_{xy}^2}.
\end{equation}Factoring $\lambda$ out of the square root gives
\begin{equation}
    \sqrt{(S_{yy} - \lambda S_{xx})^2 + 4\lambda S_{xy}^2} = \lambda \sqrt{\left(\frac{S_{yy}}{\lambda} - S_{xx}\right)^2 + \frac{4S_{xy}^2}{\lambda}}.
\end{equation}This allows us to write the estimator as
\begin{equation}
    \hat{\gamma} = \frac{S_{yy} - \lambda S_{xx} + \sqrt{(S_{yy} - \lambda S_{xx})^2 + 4\lambda S_{xy}^2}}{2S_{xy}}
\end{equation}
\begin{equation}
    \hat{\gamma} = \frac{\lambda\left(\frac{S_{yy}}{\lambda} - S_{xx} + \sqrt{\left(\frac{S_{yy}}{\lambda} - S_{xx}\right)^2 + \frac{4S_{xy}^2}{\lambda}}\right)}{2S_{xy}}.
\end{equation}

For large $\lambda$, we expand the square root with the approximation
\begin{equation}
    \sqrt{1+\epsilon} \approx 1+\frac{\epsilon}{2},
\end{equation}
where $\epsilon$ is small. We write the term under the square root as
\begin{equation}
    \left(S_{xx} - \frac{S_{yy}}{\lambda}\right)^2 + \frac{4S_{xy}^2}{\lambda} = S_{xx}^2\left(1 + \left(-\frac{2S_{yy}}{\lambda S_{xx}} + \frac{S_{yy}^2}{\lambda^2 S_{xx}^2} + \frac{4S_{xy}^2}{\lambda S_{xx}^2}\right)\right).
\end{equation}

As $\lambda \to \infty$, the term $S_{yy}^2/(\lambda^2 S_{xx}^2)$ is of second order in $1/\lambda$, and we drop it. The other two terms are of first order. We keep them, because the leading terms cancel below and the factor $\lambda$ lifts them to order one. Then
\begin{equation}
    \epsilon = -\frac{2S_{yy}}{\lambda S_{xx}} + \frac{S_{yy}^2}{\lambda^2 S_{xx}^2} + \frac{4S_{xy}^2}{\lambda S_{xx}^2} \approx -\frac{2S_{yy}}{\lambda S_{xx}} + \frac{4S_{xy}^2}{\lambda S_{xx}^2},
\end{equation}
which becomes small as $\lambda$ grows. This gives
\begin{equation}
    \sqrt{\left(S_{xx} - \frac{S_{yy}}{\lambda}\right)^2 + \frac{4S_{xy}^2}{\lambda}} \approx S_{xx} - \frac{S_{yy}}{\lambda} + \frac{2S_{xy}^2}{\lambda S_{xx}}.
\end{equation}

Substituting back gives
\begin{equation}
    \hat{\gamma} = \frac{\lambda\left[-S_{xx} + \frac{S_{yy}}{\lambda} + \left(S_{xx} - \frac{S_{yy}}{\lambda} + \frac{2S_{xy}^2}{\lambda S_{xx}}\right)\right]}{2S_{xy}}.
\end{equation}The $S_{xx}$ terms and the $S_{yy}/\lambda$ terms cancel, which leaves
\begin{equation}
    \hat{\gamma} = \frac{\lambda\left(\frac{2S_{xy}^2}{\lambda S_{xx}}\right)}{2S_{xy}} = \frac{S_{xy}}{S_{xx}} = \frac{\text{Cov}(x_{\text{obs}},y_{\text{obs}})}{\text{Var}(x_{\text{obs}})}.
\end{equation}In the limit $\lambda \to \infty$, which is $\sigma_x \to 0$, the Deming slope therefore converges to the OLS slope $S_{xy}/S_{xx}$. This is the maximum likelihood estimator that we derived for uncertainty in $y$ only.

\section{Extension to Heteroscedastic Uncertainties}\label{sec:ch6_heteroscedastic}

So far we assumed homogeneous uncertainties, so that $\sigma_x$ and $\sigma_y$ are constant across the observations. This simplification gave us intuition and analytical results. Real astronomical measurements often have uncertainties that differ by large factors from point to point within one data set. In the M-$\sigma$ relation, for example, the velocity dispersions may be more precise for brighter galaxies. Nearby galaxies also typically have more precise measurements than distant ones. The black hole mass uncertainties may depend on the measurement technique.

The same pattern appears in many other astronomical contexts. Brighter objects generally have more precise measurements than fainter ones, and different measurement techniques often have different uncertainties.

For heteroscedastic uncertainties, where $\sigma_{x,i}$ and $\sigma_{y,i}$ vary with $i$, the derivation follows the same steps. We start from the likelihood, but now each data point has its own uncertainties
\begin{equation}
\mathcal{L}(\gamma, \{x_{\text{true},i}\}) = \prod_{i=1}^N \mathcal{N}(x_{\text{obs},i} \mid x_{\text{true},i}, \sigma_{x,i}^2) \, \mathcal{N}(y_{\text{obs},i} \mid \gamma x_{\text{true},i}, \sigma_{y,i}^2).
\end{equation}Taking the negative logarithm and dropping constant terms gives the negative log-likelihood
\begin{equation}
    E(\{x_{\text{true},i}\}, \gamma) = \sum_{i=1}^N \left[\frac{(x_{\text{obs},i} - x_{\text{true},i})^2}{2\sigma_{x,i}^2} + \frac{(y_{\text{obs},i} - \gamma x_{\text{true},i})^2}{2\sigma_{y,i}^2}\right].
\end{equation}

As before, we first solve for each $x_{\text{true},i}$ by setting the derivative with respect to $x_{\text{true},i}$ to zero
\begin{align}
    \frac{\partial E}{\partial x_{\text{true},i}} &= -\frac{(x_{\text{obs},i} - x_{\text{true},i})}{\sigma_{x,i}^2} - \frac{\gamma(y_{\text{obs},i} - \gamma x_{\text{true},i})}{\sigma_{y,i}^2} = 0.
\end{align}Rearranging gives
\begin{align}
    \frac{x_{\text{true},i}}{\sigma_{x,i}^2} + \frac{\gamma^2 x_{\text{true},i}}{\sigma_{y,i}^2} &= \frac{x_{\text{obs},i}}{\sigma_{x,i}^2} + \frac{\gamma y_{\text{obs},i}}{\sigma_{y,i}^2} \\*
    x_{\text{true},i}\left(\frac{1}{\sigma_{x,i}^2} + \frac{\gamma^2}{\sigma_{y,i}^2}\right) &= \frac{x_{\text{obs},i}}{\sigma_{x,i}^2} + \frac{\gamma y_{\text{obs},i}}{\sigma_{y,i}^2}.
\end{align}Solving for $x_{\text{true},i}$ gives
\begin{equation}
\hat{x}_{\text{true},i} = \frac{\sigma_{y,i}^2 x_{\text{obs},i} + \gamma \sigma_{x,i}^2 y_{\text{obs},i}}{\sigma_{y,i}^2 + \gamma^2 \sigma_{x,i}^2}.
\end{equation}This expression is analogous to the homoscedastic result, but the weights now depend on the individual uncertainties $\sigma_{x,i}$ and $\sigma_{y,i}$. It is a weighted average of the direct measurement $x_{\text{obs},i}$, with weight $\sigma_{y,i}^2$, and the model-inferred value $y_{\text{obs},i}/\gamma$, with weight $\gamma^2\sigma_{x,i}^2$. When a point has a large uncertainty in one dimension, the estimate relies more on the other dimension.

Setting the derivative with respect to $\gamma$ to zero gives
\begin{equation}
\frac{\partial E}{\partial \gamma} = -\sum_{i=1}^N \frac{x_{\text{true},i}(y_{\text{obs},i} - \gamma x_{\text{true},i})}{\sigma_{y,i}^2} = 0.
\end{equation}Here $y_{\text{obs},i} - \gamma\hat{x}_{\text{true},i} = \sigma_{y,i}^2(y_{\text{obs},i} - \gamma x_{\text{obs},i})/(\sigma_{y,i}^2 + \gamma^2\sigma_{x,i}^2)$, which follows as in the homoscedastic case. Substituting both expressions, the condition becomes
\begin{equation}
    -\sum_{i=1}^N \frac{1}{\sigma_{y,i}^2} \left(\frac{\sigma_{y,i}^2 x_{\text{obs},i} + \gamma \sigma_{x,i}^2 y_{\text{obs},i}}{\sigma_{y,i}^2 + \gamma^2 \sigma_{x,i}^2}\right)\left(\frac{\sigma_{y,i}^2(y_{\text{obs},i} - \gamma x_{\text{obs},i})}{\sigma_{y,i}^2 + \gamma^2 \sigma_{x,i}^2}\right) = 0.
\end{equation}

The factor $\sigma_{y,i}^2$ cancels the weight $1/\sigma_{y,i}^2$, which gives the function $F(\gamma)$ whose zero we need
\begin{equation}
F(\gamma) = \sum_{i=1}^N \frac{(\sigma_{y,i}^2 x_{\text{obs},i} + \gamma \sigma_{x,i}^2 y_{\text{obs},i})(y_{\text{obs},i} - \gamma x_{\text{obs},i})}{(\sigma_{y,i}^2 + \gamma^2 \sigma_{x,i}^2)^2} = 0.
\end{equation}Unlike the quadratic equation of the homoscedastic case, the equation $F(\gamma) = 0$ has no closed-form solution in general. Each term has a weight that depends on $\gamma$ and on its own uncertainties. We therefore solve it numerically.

As in the homoscedastic case, $F$ is minus the derivative of the negative log-likelihood after the $x_{\text{true},i}$ are optimized. The estimate $\hat{\gamma}$ is therefore the zero where $F$ changes from positive to negative, and there the negative log-likelihood is a minimum.

\begin{figure}[ht!]
    \centering
    \includegraphics[width=0.9\textwidth]{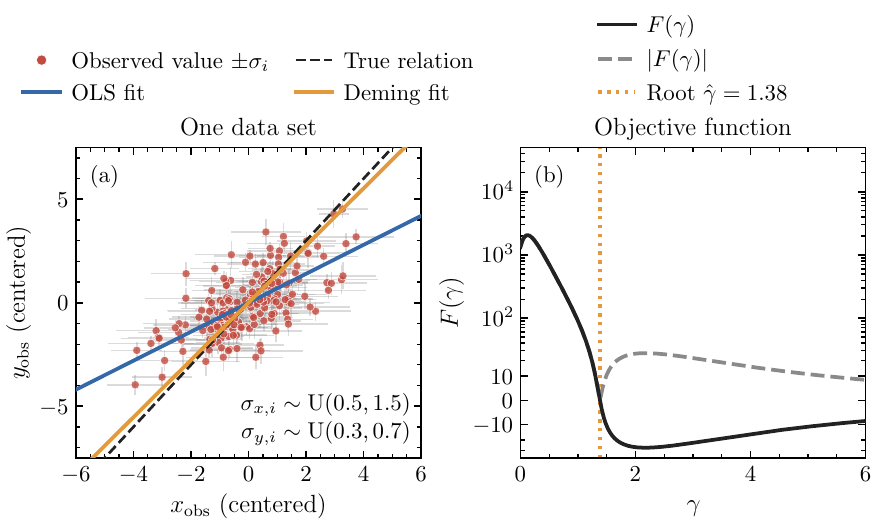}
    \caption{Heteroscedastic Deming regression for data drawn from $y = 1.5x$, with a different error $\sigma_{x,i}$ and $\sigma_{y,i}$ for each point. Panel~(a) shows the data with their individual error bars and the true line. It also shows the OLS fit (slope 0.70) and the Deming fit (slope 1.38). Panel~(b) shows $F(\gamma)$ and $|F(\gamma)|$ on a grid of $\gamma$-values, on a scale that is linear for $|F| < 10$ and logarithmic above. The dotted line marks the zero of $F$, found with Brent's method, where $|F|$ reaches its minimum.}
    \label{fig:heteroscedastic_deming}
\end{figure}

\paragraph{Grid Search} The simplest approach evaluates $F(\gamma)$ on a grid\index{grid search} of $\gamma$-values and finds where it crosses zero. This is computationally inefficient, but reliable and easy to implement. Figure~\ref{fig:heteroscedastic_deming}(b) shows $F(\gamma)$ evaluated on such a grid.

\paragraph{Newton's Method} Newton's\index{Newton's method} method replaces $F(\gamma)$ by its tangent line at the current guess $\gamma_n$ and moves to where that line crosses zero. Repeating this step approaches the root through the iteration
\begin{equation}
\gamma_{n+1} = \gamma_n - \frac{F(\gamma_n)}{F'(\gamma_n)}.
\end{equation}
Near the root, one step roughly doubles the number of correct digits, which is called quadratic convergence. However, the method needs the derivative $F'(\gamma)$.

\paragraph{Library Root Finders} Standard numerical libraries offer reliable root\index{root finding} finders, such as Brent's\index{Brent's method} method. It keeps an interval around the root and shrinks it by bisection, secant steps and inverse quadratic interpolation. The starting interval must have opposite signs of $F$ at its two ends, so a root lies inside. Such root finders converge well without derivatives.

Like the homoscedastic solution, the heteroscedastic solution corrects attenuation bias, but the correction is now specific to each data point and its uncertainties. Points with smaller uncertainties carry more weight in determining $\gamma$, so we use all the data while the most precise measurements dominate the inference. When all $\sigma_{x,i} \to 0$, we recover the weighted least squares solution of Chapter~4, as in the homoscedastic limit above.

This ends our study of linear regression. Many astronomical problems have a discrete label as the output instead, such as star or galaxy. Chapter~7 treats these classification problems.\index{Deming regression|)}

\section{Tutorial: Fitting the Tully-Fisher Relation}\label{sec:ch6_tutorial}

This\index{Tully-Fisher relation|(} tutorial applies the attenuation factor and Deming regression of this chapter to the Tully-Fisher\index{Tully-Fisher relation} relation. It links the rotation velocity of a galaxy to its baryonic mass, the mass of stars and gas. Its slope matters for galaxy formation and for distances. The full tutorial can be found at \url{https://statml.org/} as tutorial 6.

\begin{figure}[t!]
    \centering
    \includegraphics[width=0.95\textwidth]{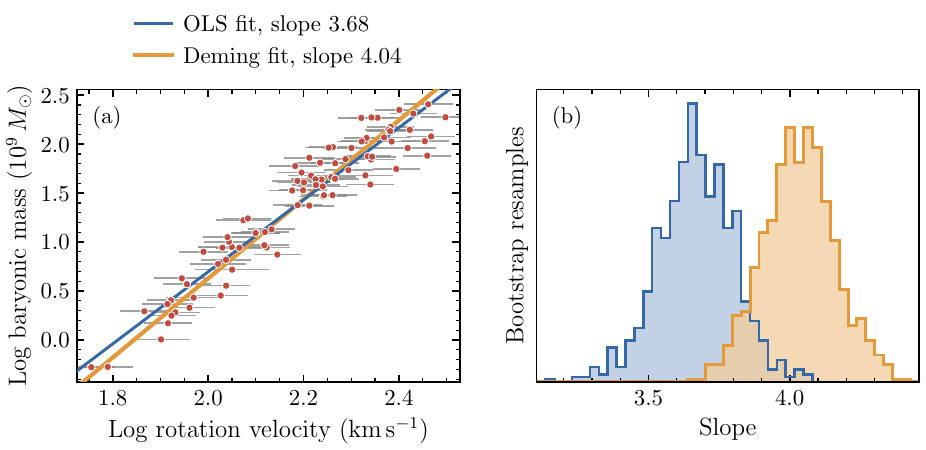}
    \caption{The tutorial fits the Tully-Fisher relation to real nearby galaxies. Panel~(a) shows the galaxies (red) with their velocity errors (grey, including a systematic uncertainty) and two fits, with the slopes in the legend. Ordinary least squares (OLS, blue) treats the rotation velocity as exact, and Deming regression (orange) allows errors in both variables. Panel~(b) shows the slope of each fit in bootstrap resamples, in the same colors. The Deming slope is steeper by more than the spread of either distribution.}
    \label{fig:ch6_tutorial_tully_fisher}
\end{figure}

A galaxy with more baryons rotates faster, because it sits in a more massive halo of dark matter, whose gravity sets the rotation speed. In nearby disk galaxies, we measure the rotation velocity from spectra and the baryonic mass from the radiation of stars and gas. On logarithmic axes the two lie close to a straight line (Figure~\ref{fig:ch6_tutorial_tully_fisher}a). Its slope links the stars and gas to the halo, and so constrains how galaxies assemble their baryons. The relation also gives distances, because the rotation velocity predicts how much light a galaxy emits.

Both uses depend on the slope, so the fit matters. In Chapters~4 and~5 the inputs were exact and the errors were in the output. Here the rotation velocity, our input, carries errors, for example from the tilt of the disk. Ordinary least squares (OLS) treats the input as exact. Errors in the velocity, which spread the points sideways, then flatten the OLS slope by the attenuation factor (Section~\ref{sec:ch6_mle}).

To account for this, Deming regression allows errors in both variables, so it uses the errors in the velocity that OLS ignores. As in Section~\ref{sec:ch6_latent}, it treats the true velocity of each galaxy as a latent variable and infers it together with the slope. The errors differ from galaxy to galaxy, so the fit uses the heteroscedastic form of Section~\ref{sec:ch6_heteroscedastic}.

The tutorial compares the two fits for real galaxies. It adds a systematic uncertainty to the velocity errors, so that they match the scatter of the galaxies. In Figure~\ref{fig:ch6_tutorial_tully_fisher}a the Deming line is steeper than the OLS line, the direction that the flattening of OLS leads us to expect. To judge whether the gap is real, the bootstrap of Section~\ref{sec:ch3_bootstrap} gives the sampling noise of each slope (Figure~\ref{fig:ch6_tutorial_tully_fisher}b). The gap between the two slopes is larger than the noise of either slope.

Noise in the input flattens an OLS fit, and Deming regression allows for that noise. The bootstrap shows whether the difference is larger than the sampling noise. Many scaling relations in astronomy have uncertain inputs. When the input error is not negligible, ignoring it biases the slope, and a biased slope changes the physical conclusions drawn from the relation.\index{Tully-Fisher relation|)}

\paragraph{Further Reading} Adcock (1878) developed what is now called orthogonal regression. Pearson (1901) found the lines of closest fit when both variables contain errors. Deming (1943) popularized the weighted least squares fit for a known ratio of error variances that now carries his name. Golub and Van Loan (1980) analyzed total least squares, a fit with errors in both variables, with the singular value decomposition. Van Huffel and Vandewalle (1991) treat its computation and sensitivity. Fuller (1987) covers the theory of measurement error models. Carroll et~al.\ (2006) extend these methods to nonlinear models. Isobe et~al.\ (1990) compared regression methods for astronomical data. Akritas and Bershady (1996) developed an estimator for errors in both variables, with correlated and unequal uncertainties and intrinsic scatter. Kelly (2007) developed Bayesian methods for heteroscedastic errors, intrinsic scatter and selection effects. Bentler and Weeks (1980) developed a general model of linear relations among latent variables. Carroll and Ruppert (1996) examined the use and misuse of orthogonal regression, which typically overestimates the slope when intrinsic scatter is ignored. Ting (2025, 2026) discussed attenuation bias in the context of astronomy and how to correct it in neural networks. Lelli et~al.\ (2016) published SPARC, the sample of nearby disk galaxies with Spitzer infrared photometry and accurate rotation curves. This tutorial takes its galaxies from it.

\chapter{Classification and Logistic Regression}

\looseness=1 Is an object a star or a galaxy, is a galaxy spiral or elliptical, is a variable star a Cepheid or an RR Lyrae? These are classification problems, the other major category of supervised learning next to regression. Their importance grows with the size of the surveys. Surveys such as the Sloan Digital Sky Survey, Gaia and the Vera C. Rubin Observatory catalog hundreds of millions to billions of objects. Astronomers cannot examine each object, and telescope time for follow-up observations is limited and expensive.

Classification algorithms help prioritize targets. We can select candidates for rare metal-poor stars from photometry before investing in high-resolution spectroscopy. Colors and variability help identify quasar candidates among millions of point sources. Light curves let us classify variable stars and transient events in time for follow-up of interesting phenomena.

A regression model can predict any real value, but a probabilistic classifier must output class probabilities, which lie between 0 and 1. A linear model might predict probabilities above 1 or below 0, which have no meaning, so we cannot simply apply linear regression. Logistic regression solves this problem by mapping the unconstrained linear prediction to a valid probability. Despite its name, it is a classification method.

In this chapter we develop logistic regression from a probabilistic view. We compare discriminative models, which learn to separate the classes directly, with generative models, which learn how each class generates its data. Maximum likelihood then defines the loss of logistic regression. We train the model with gradient descent and treat regularization and other practical issues of training. Logistic regression is the entry point to classification and leads to the more complex methods of later chapters.

\section{Binary Classification Fundamentals}\label{sec:ch7_binary}

How\index{classification!binary|(} do we represent categorical class labels mathematically? Regression labels are numbers, such as stellar mass in solar masses, luminosity in ergs per second or temperature in kelvin. Classes have no numerical value of their own. Galaxy morphologies (spiral, elliptical, irregular), stellar spectral types (O, B, A, F, G, K, M) and object types (star, galaxy, quasar) are examples.

\looseness=1 For binary classification,\index{classification!binary} where we sort objects into one of two classes, the standard approach encodes the classes as 0 and 1. Examples are star (0) versus galaxy (1), non-variable (0) versus variable star (1), and non-transient\index{transients} (0) versus transient event (1). By convention, class 0 is the ``background'' or ``default'' class, and class 1 is the ``target'' or ``positive'' class of interest. In a transient detection system, for instance, we assign 1 to the rare transient events, which we want to find among many non-transient sources.

The binary encoding has a useful symmetry. If we swap the labels, we recover equivalent predictions by taking 1 minus the original probability. A model that gives probability 0.8 to class 1 (transient) gives $1 - 0.8 = 0.2$ to non-transient events, which are class 1 after the swap. Asking for the probability that an object is a transient or that it is not a transient gives mathematically equivalent problems.

With the labels encoded as 0 and 1, a classification algorithm predicts the probability $P(C_1|\mathbf{x})$ that an object belongs to class 1. Here $\mathbf{x}$ represents the features of the object, such as colors, morphology (point-source minus model magnitudes) or Gaia astrometry (proper motions and parallaxes). Since the probabilities of all classes sum to 1, $P(C_0|\mathbf{x}) = 1 - P(C_1|\mathbf{x})$ in binary classification. We assign an object to the class with the higher probability. The decision boundary, the surface in feature space that separates the two classes, lies where $P(C_1|\mathbf{x}) = P(C_0|\mathbf{x}) = 0.5$.

\begin{figure}[ht!]
    \centering
    \includegraphics[width=0.7\textwidth]{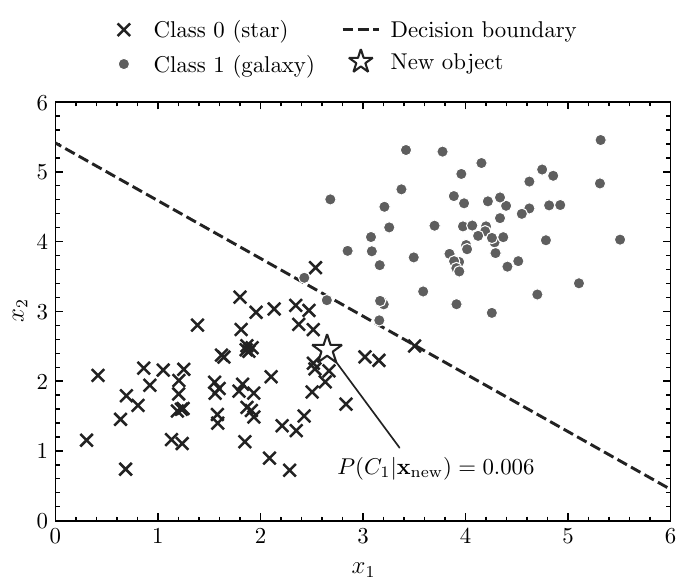}
    \caption{Logistic regression fitted to stars and galaxies in a two-dimensional feature space. The decision boundary is where $P(C_1|\mathbf{x}) = 0.5$. The new object lies on the stellar side.}
    \label{fig:classification_example}
\end{figure}

For a linear classifier, the decision boundary in a two-dimensional feature space is a line. A new object is classified by its side of this line, as Figure~\ref{fig:classification_example} shows.\index{classification!binary|)}

\section{Discriminative vs. Generative Models}\label{sec:ch7_disc_gen}

How\index{discriminative model|(}\index{generative model|(} do we build a model of the probability $P(C_k|\mathbf{x})$ that an object belongs to class $k$? There are two distinct paths, both connected through Bayes' theorem
\begin{equation}
P(C_k|\mathbf{x}) = \frac{p(\mathbf{x}|C_k) \cdot P(C_k)}{p(\mathbf{x})}.
\end{equation}

\paragraph{Discriminative Approach} The discriminative\index{discriminative model} approach models $P(C_k|\mathbf{x})$ directly, the posterior probability of class $k$ given the features. This is the quantity that classification needs. It learns the boundary between the classes. For stars versus galaxies, the model takes the input features $\mathbf{x}$, which may include dozens of measurements. It outputs the probability of each class, for example ``this object has an 80\% chance of being a galaxy.'' It maps complex inputs to class probabilities without modeling how the features are distributed within each class.

\paragraph{Generative Approach} The generative\index{generative model} approach takes the other path through Bayes'\index{Bayes' theorem} theorem. Instead of modeling $P(C_k|\mathbf{x})$ directly, it models the two components needed to compute it indirectly. The first is the distribution $p(\mathbf{x}|C_k)$ of the features within each class, the class-conditional density, also called the likelihood. The second is the prior probability $P(C_k)$ of each class. Bayes' theorem then multiplies the two and normalizes by $p(\mathbf{x})$ to give $P(C_k|\mathbf{x})$.

\begin{figure}[ht!]
    \centering
    \includegraphics[width=0.9\textwidth]{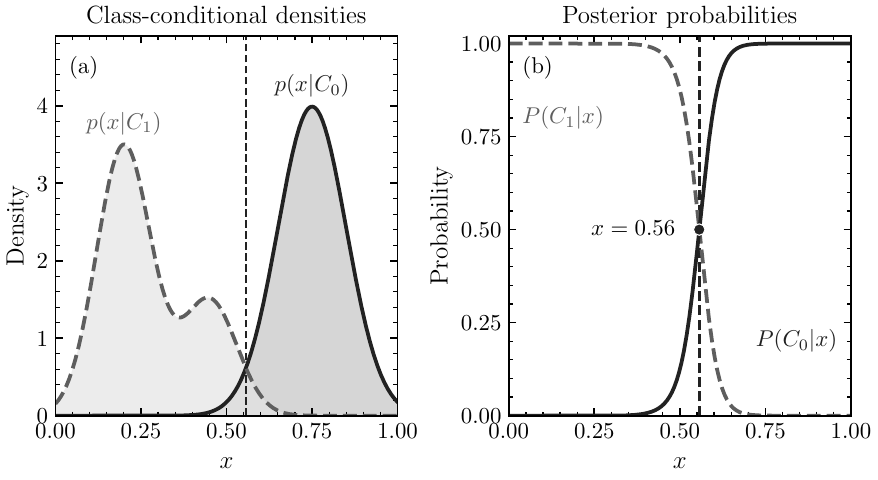}
    \caption{Generative and discriminative views of one feature $x$. Panel~(a) shows the class-conditional densities, and panel~(b) shows the posteriors from Bayes' theorem with equal priors. The posteriors are equal at the dashed line, which is the decision boundary. A discriminative model is simpler to learn, but it learns only the posteriors and misses the two peaks of class 1 in panel~(a).}
    \label{fig:gen_vs_disc}
\end{figure}

The approach is called generative because it models enough about each class to generate new examples. For stars versus galaxies, it learns separately what stars look like, with their colors, point-like appearance and proper motions. It also learns what galaxies look like, with their extended shapes, color distributions and lack of proper motion. Figure~\ref{fig:gen_vs_disc} compares the two approaches for a single feature.

An analogy with language identification shows the difference. A discriminative approach to telling German from Portuguese would focus on features that differ. German might have more hard consonants and Portuguese more nasal sounds. It learns just enough to tell the two languages apart.

A generative approach would instead learn to speak both languages, with their grammar, vocabulary and pronunciation rules. This is much harder but teaches more. Someone who speaks both languages can distinguish them, generate new sentences and recognize dialects. Such a speaker also understands why certain sound combinations are valid in one language but not in the other.

A generative model can also detect outliers or new classes. A discriminative model has no density over $\mathbf{x}$. Someone who only knows how to tell German from Portuguese (the discriminative model) might label Finnish as one of them with a probability close to 1. A fluent speaker of both languages (the generative model) would recognize that Finnish differs from both. In mathematical terms, both $p(\mathbf{x}|\text{German})$ and $p(\mathbf{x}|\text{Portuguese})$ are small.

\looseness=1 A discriminative model trained only on stars and galaxies might label a quasar as one or the other with a probability close to 1. A generative model can flag an object whose features have a low density under both classes as something possibly new and interesting.\index{discriminative model|)}\index{generative model|)}

\section{Logistic Regression and the Sigmoid Function}\label{sec:ch7_logistic}

Logistic\index{logistic regression}\index{logistic regression|(} regression is a discriminative method that models $P(C_1|\mathbf{x})$ directly. Given our experience with linear regression, a simple first attempt is to model this probability as a linear function of the features
\begin{equation}
P(C_1|\mathbf{x}) = \mathbf{w}^T\mathbf{x}.
\end{equation}
Here $\mathbf{w}$ is the vector of model parameters and $\mathbf{x}$ is the vector of input features. A constant 1 appended to $\mathbf{x}$ provides the bias term, the intercept $w_0$ of Chapter~4.

\begin{figure}[ht!]
    \centering
    \includegraphics[width=0.9\textwidth]{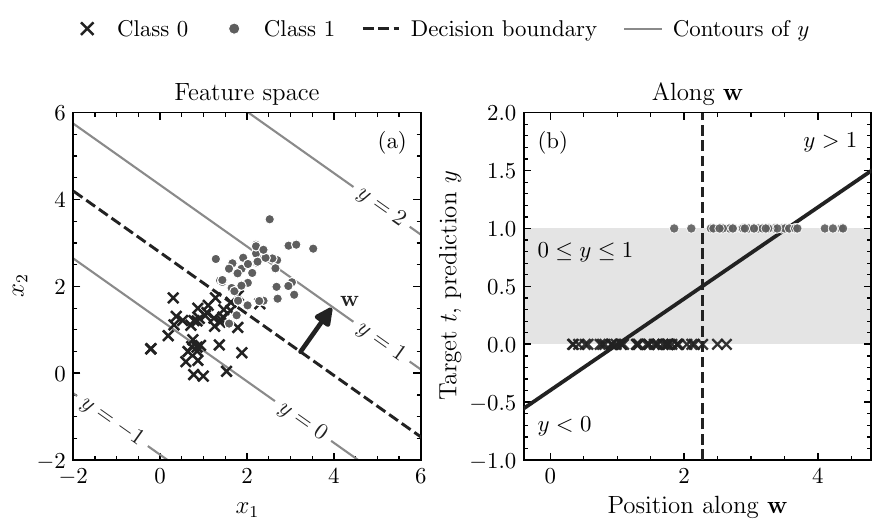}
    \caption{Classification treated as regression. Panel~(a) shows the least-squares predictor $y = \mathbf{w}^T\mathbf{x}$ fitted to the labels 0 and 1, with the decision boundary $y = 0.5$. The weight vector $\mathbf{w}$ is perpendicular to the boundary. Panel~(b) shows the labels and the predictor along the direction of $\mathbf{w}$. The predictor leaves the valid range of 0 to 1.}
    \label{fig:linear_boundary}
\end{figure}

However, a linear model produces unbounded outputs. The predicted value can be any real number, such as $-2.7$ or $5.3$, as Figure~\ref{fig:linear_boundary} shows. A probability must lie in $[0,1]$, so a prediction such as 1.2 or 1.5 has no probabilistic interpretation. In regression, a prediction of 10.2 for a true value of 10.0 is a small error. A probability of 1.2 for a positive example is meaningless, even though it is only 0.2 above 1.

\begin{figure}[ht!]
    \centering
    \includegraphics[width=0.9\textwidth]{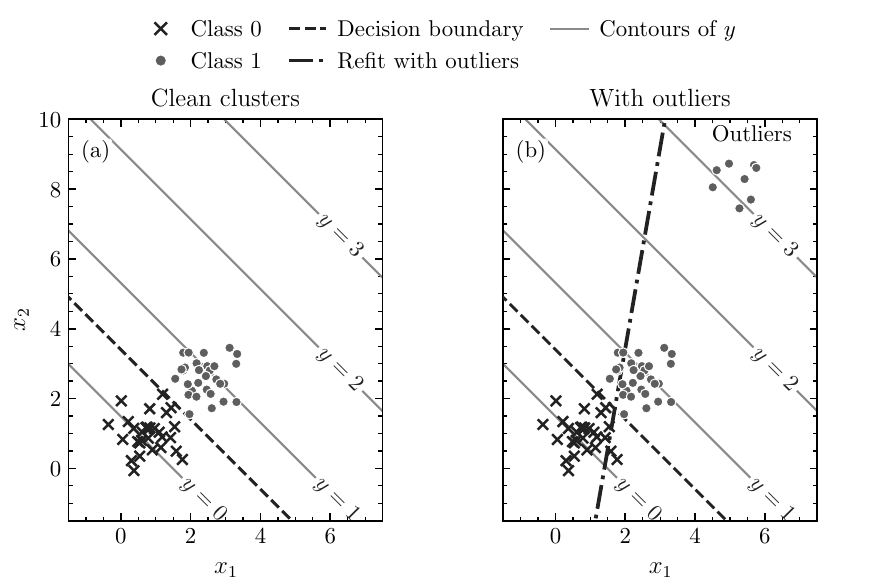}
    \caption{Outliers pull the least-squares boundary. Panel~(a) shows two clusters and the boundary $y = 0.5$ of the least-squares predictor, which separates them. In panel~(b), a few class 1 outliers far from the boundary have large values of $y$ and dominate the squared error. The boundary refitted to all points rotates toward the vertical and misclassifies points of the main clusters. The dashed line and the contours in panel~(b) are those of the fit in panel~(a).}
    \label{fig:boundary_skew}
\end{figure}

The unbounded behavior also distorts the fit. Consider class 1 points far from the boundary, for which a linear model might predict values such as $\mathbf{w}^T\mathbf{x} = 2$. Least squares counts this correct prediction as an error of 1 against the target value of 1, as large as the error of predicting 0. Least squares therefore pulls the boundary toward such points, which can misplace it, as Figure~\ref{fig:boundary_skew} shows.

We need a function that maps the linear predictor $z = \mathbf{w}^T\mathbf{x}$ into the range of probabilities while keeping the decision\index{decision boundary} boundary linear. The sigmoid\index{sigmoid function} function does this
\begin{equation}
\sigma(z) = \frac{1}{1 + e^{-z}}.
\end{equation}

\begin{figure}[ht!]
    \centering
    \includegraphics[width=0.9\textwidth]{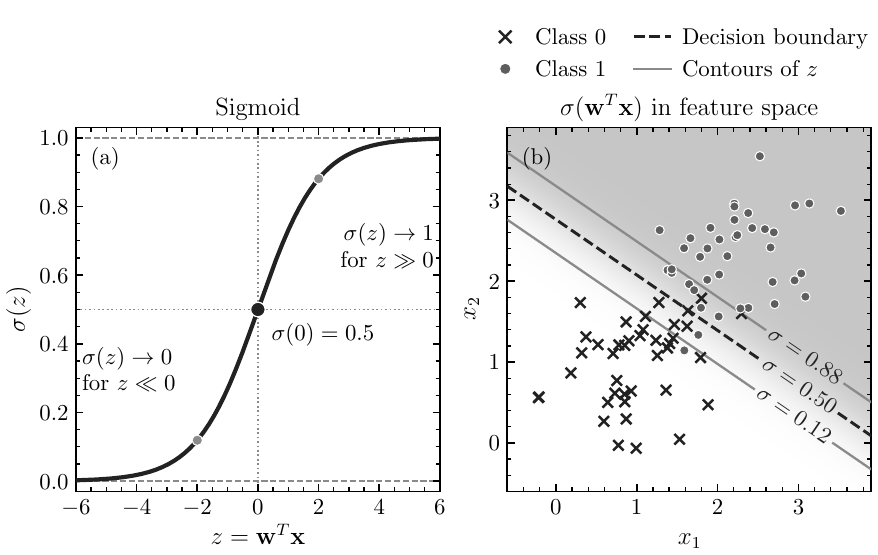}
    \caption{The sigmoid function and its role in logistic regression. Panel~(a) shows the sigmoid function. Panel~(b) shows the probability $\sigma(\mathbf{w}^T\mathbf{x})$ for a fitted logistic regression, with the decision boundary at $z = 0$ and contours of constant $z$.}
    \label{fig:sigmoid}
\end{figure}

\looseness=1 This S-shaped curve maps any real number to the range $(0,1)$, so it can represent a probability. For very negative $z$, $\sigma(z)$ approaches 0, and for very positive $z$, it approaches 1. At $z = 0$, $\sigma(z) = 0.5$, which marks the decision boundary. Points far from the boundary therefore receive probabilities near 0 or 1, and points close to it receive probabilities near 0.5. Figure~\ref{fig:sigmoid} shows the curve and the probability contours that it creates in feature space.

The sigmoid function follows from probabilistic reasoning, through the odds\index{odds} of the two classes. For binary classification, the odds measure the relative belief in the classes
\begin{equation}
\text{odds} = \frac{P(C_1|\mathbf{x})}{P(C_0|\mathbf{x})} = \frac{P(C_1|\mathbf{x})}{1-P(C_1|\mathbf{x})}.
\end{equation}

Probabilities lie between 0 and 1, but the odds range from 0 to infinity. The logarithm of the odds, called the log-odds\index{log-odds} or the ``logit''\index{logit|see{log-odds}} transformation, extends the range to all real numbers
\begin{equation}
\text{logit}(P(C_1|\mathbf{x})) = \ln\left(\frac{P(C_1|\mathbf{x})}{1-P(C_1|\mathbf{x})}\right).
\end{equation}A linear predictor can take any real value, and so can the logit. We therefore model the logit with our linear function
\begin{equation}
\text{logit}(P(C_1|\mathbf{x})) = \mathbf{w}^T\mathbf{x}.
\end{equation}

\begin{figure}[ht!]
    \centering
    \includegraphics[width=0.9\textwidth]{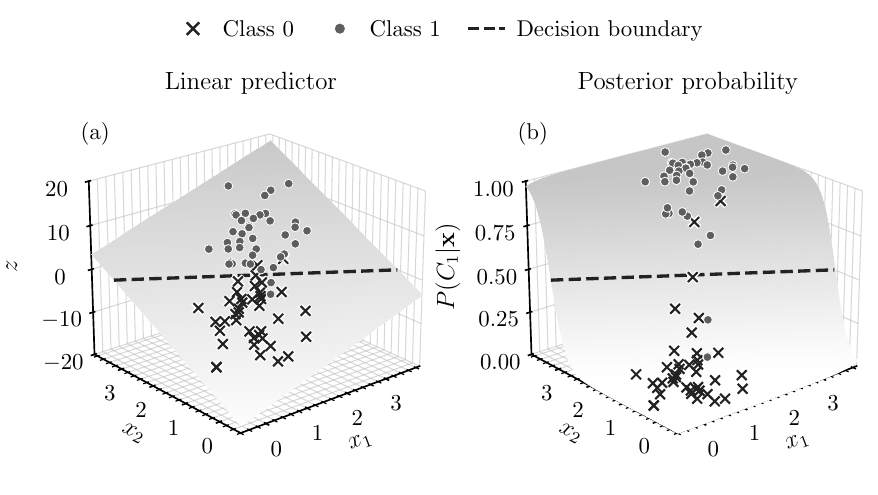}
    \caption{The logistic regression of Figure~\ref{fig:sigmoid} in three dimensions. Panel~(a) shows the log-odds $z = \mathbf{w}^T\mathbf{x}$ over the feature space, a plane that is unbounded in the vertical direction. Panel~(b) shows $P(C_1|\mathbf{x}) = \sigma(z)$, which is bounded between 0 and 1. The decision boundary is where $z = 0$ and $P(C_1|\mathbf{x}) = 0.5$.}
    \label{fig:logistic_transform}
\end{figure}

To find the probability, we reverse the logit transformation. Exponentiating both sides and solving for $P(C_1|\mathbf{x})$ gives
\begin{align}
\ln\left(\frac{P(C_1|\mathbf{x})}{1-P(C_1|\mathbf{x})}\right) &= \mathbf{w}^T\mathbf{x} \\*
\frac{P(C_1|\mathbf{x})}{1-P(C_1|\mathbf{x})} &= e^{\mathbf{w}^T\mathbf{x}} \\*
P(C_1|\mathbf{x}) &= \frac{e^{\mathbf{w}^T\mathbf{x}}}{1 + e^{\mathbf{w}^T\mathbf{x}}} = \frac{1}{1 + e^{-\mathbf{w}^T\mathbf{x}}} = \sigma(\mathbf{w}^T\mathbf{x}).
\end{align}The sigmoid function therefore emerges when we model the log-odds with a linear function. It is also called the logistic\index{logistic function|see{sigmoid function}} function, which gives logistic regression its name. The model is linear in the log-odds, and the sigmoid function maps the log-odds to probabilities. Figure~\ref{fig:logistic_transform} shows this transformation for a linear predictor and its sigmoid.

The linear decision boundary is preserved. The boundary lies where $P(C_1|\mathbf{x}) = 0.5$, which means $\sigma(\mathbf{w}^T\mathbf{x}) = 0.5$, and this happens exactly when $\mathbf{w}^T\mathbf{x} = 0$. The boundary is therefore linear in feature space, as we wanted.

In astronomical terms, the linear predictor $\mathbf{w}^T\mathbf{x}$ is now the log-odds that an object belongs to one class rather than the other. For star-galaxy\index{star-galaxy classification} classification, a log-odds of 2 means that the odds of a galaxy against a star are $e^2 \approx 7.4$ to 1.

\section{From Discriminative to Generative}\label{sec:ch7_gen_link}

Bayes' theorem connects the discriminative view of logistic regression to a generative view. In logistic regression we assumed that the log-odds are a linear function $\mathbf{w}^T \mathbf{x}$ of the features. This creates a hyperplane decision boundary in feature space. A hyperplane is a flat surface with one dimension fewer than the space it lies in, such as a line in 2D or a plane in 3D. Points with $\mathbf{w}^T \mathbf{x} > 0$ belong to class 1, and points with $\mathbf{w}^T \mathbf{x} < 0$ belong to class 0. The vector $\mathbf{w}$ is perpendicular to the boundary and sets its orientation.

Which assumptions about the feature distributions $p(\mathbf{x}|C_k)$ give rise to such a linear boundary?

One answer is that the features of each class follow a multivariate Gaussian distribution, with the same covariance matrix for all classes. This condition is sufficient, and other distributions can also give a linear boundary, so logistic regression does not require Gaussian features.

The Gaussian density of class $k$ is
\begin{equation}
p(\mathbf{x}|C_k) = \frac{1}{(2\pi)^{D/2} \det(\boldsymbol{\Sigma})^{1/2}} \exp\left(-\frac{1}{2} (\mathbf{x} - \boldsymbol{\mu}_k)^T \boldsymbol{\Sigma}^{-1} (\mathbf{x} - \boldsymbol{\mu}_k)\right).
\end{equation}
The dimension $D$ is the number of features. Each class has its own mean vector $\boldsymbol{\mu}_k$, and all classes share the covariance matrix $\boldsymbol{\Sigma}$. Geometrically, each class is a ``cloud'' of points with the same shape and orientation, centered at different locations.

We now show that Gaussian distributions with equal covariance matrices make the log-odds linear in the features. Here we write the bias $b$ explicitly instead of appending a constant 1 to $\mathbf{x}$, because $b$ is where the class priors enter. Bayes' theorem gives
\begin{equation}
\ln\left(\frac{P(C_1|\mathbf{x})}{P(C_0|\mathbf{x})}\right) = \ln\left(\frac{p(\mathbf{x}|C_1)P(C_1)}{p(\mathbf{x}|C_0)P(C_0)}\right).
\end{equation}

The denominator $p(\mathbf{x})$ of Bayes' theorem cancels in the ratio. We now substitute the Gaussian assumption for $p(\mathbf{x}|C_k)$. The normalization constants cancel as well, because the covariance matrix is shared. This gives
\begin{equation}
\ln\left(\frac{P(C_1|\mathbf{x})}{P(C_0|\mathbf{x})}\right) = \ln\left(\frac{P(C_1)}{P(C_0)}\right) + \ln\left(\frac{\exp(-\frac{1}{2}(\mathbf{x}-\boldsymbol{\mu}_1)^T\boldsymbol{\Sigma}^{-1}(\mathbf{x}-\boldsymbol{\mu}_1))}{\exp(-\frac{1}{2}(\mathbf{x}-\boldsymbol{\mu}_0)^T\boldsymbol{\Sigma}^{-1}(\mathbf{x}-\boldsymbol{\mu}_0))}\right).
\end{equation}
The first term, $\ln(P(C_1)/P(C_0))$, represents our prior belief about the relative frequencies of the classes. The second term compares how well the features $\mathbf{x}$ match the Gaussian distribution of each class.

We simplify the second term. By the properties of logarithms, the logarithm of the ratio of Gaussians becomes a difference
\begin{equation}
-\frac{1}{2}(\mathbf{x}-\boldsymbol{\mu}_1)^T\boldsymbol{\Sigma}^{-1}(\mathbf{x}-\boldsymbol{\mu}_1) + \frac{1}{2}(\mathbf{x}-\boldsymbol{\mu}_0)^T\boldsymbol{\Sigma}^{-1}(\mathbf{x}-\boldsymbol{\mu}_0).
\end{equation}
We expand each quadratic form, starting with the first
\begin{equation}
(\mathbf{x}-\boldsymbol{\mu}_1)^T\boldsymbol{\Sigma}^{-1}(\mathbf{x}-\boldsymbol{\mu}_1) = \mathbf{x}^T\boldsymbol{\Sigma}^{-1}\mathbf{x} - \mathbf{x}^T\boldsymbol{\Sigma}^{-1}\boldsymbol{\mu}_1 - \boldsymbol{\mu}_1^T\boldsymbol{\Sigma}^{-1}\mathbf{x} + \boldsymbol{\mu}_1^T\boldsymbol{\Sigma}^{-1}\boldsymbol{\mu}_1.
\end{equation}
The product $\mathbf{x}^T\boldsymbol{\Sigma}^{-1}\boldsymbol{\mu}_1$ is a scalar, so it equals its transpose $\boldsymbol{\mu}_1^T\boldsymbol{\Sigma}^{-1}\mathbf{x}$ because $\boldsymbol{\Sigma}$ is symmetric, and the same holds for the second form. Substituting both expansions into the difference gives
\begin{equation}
\begin{split}
-\frac{1}{2}(&\mathbf{x}^T\boldsymbol{\Sigma}^{-1}\mathbf{x} - 2\boldsymbol{\mu}_1^T\boldsymbol{\Sigma}^{-1}\mathbf{x} + \boldsymbol{\mu}_1^T\boldsymbol{\Sigma}^{-1}\boldsymbol{\mu}_1) \\
&+ \frac{1}{2}(\mathbf{x}^T\boldsymbol{\Sigma}^{-1}\mathbf{x} - 2\boldsymbol{\mu}_0^T\boldsymbol{\Sigma}^{-1}\mathbf{x} + \boldsymbol{\mu}_0^T\boldsymbol{\Sigma}^{-1}\boldsymbol{\mu}_0)
.\end{split}
\end{equation}

The terms with $\mathbf{x}^T\boldsymbol{\Sigma}^{-1}\mathbf{x}$ cancel because they have opposite signs and the same covariance matrix. This cancellation eliminates the quadratic terms in $\mathbf{x}$, which would otherwise prevent a linear boundary. What remains is
\begin{equation}
\boldsymbol{\mu}_1^T\boldsymbol{\Sigma}^{-1}\mathbf{x} - \boldsymbol{\mu}_0^T\boldsymbol{\Sigma}^{-1}\mathbf{x} - \frac{1}{2}\boldsymbol{\mu}_1^T\boldsymbol{\Sigma}^{-1}\boldsymbol{\mu}_1 + \frac{1}{2}\boldsymbol{\mu}_0^T\boldsymbol{\Sigma}^{-1}\boldsymbol{\mu}_0.
\end{equation}
Factoring out $\boldsymbol{\Sigma}^{-1}\mathbf{x}$ gives
\begin{equation}
(\boldsymbol{\mu}_1 - \boldsymbol{\mu}_0)^T\boldsymbol{\Sigma}^{-1}\mathbf{x} - \frac{1}{2}(\boldsymbol{\mu}_1^T\boldsymbol{\Sigma}^{-1}\boldsymbol{\mu}_1 - \boldsymbol{\mu}_0^T\boldsymbol{\Sigma}^{-1}\boldsymbol{\mu}_0).
\end{equation}Adding back the prior term gives the final form
\begin{equation}
\ln\left(\frac{P(C_1|\mathbf{x})}{P(C_0|\mathbf{x})}\right) = \mathbf{w}^T \mathbf{x} + b,
\end{equation}
where the weight vector is
\begin{equation}
\mathbf{w} = \boldsymbol{\Sigma}^{-1}(\boldsymbol{\mu}_1 - \boldsymbol{\mu}_0),
\end{equation}
and the bias is
\begin{equation}
b = \ln\left(\frac{P(C_1)}{P(C_0)}\right) - \frac{1}{2}(\boldsymbol{\mu}_1^T\boldsymbol{\Sigma}^{-1}\boldsymbol{\mu}_1 - \boldsymbol{\mu}_0^T\boldsymbol{\Sigma}^{-1}\boldsymbol{\mu}_0).
\end{equation}

\paragraph{Interpreting Weights and Bias}

The weight vector $\mathbf{w} = \boldsymbol{\Sigma}^{-1}(\boldsymbol{\mu}_1 - \boldsymbol{\mu}_0)$ depends on the class means and the shared covariance. We expect the boundary to lie between the two class centers, so $\mathbf{w}$ needs a component along $(\boldsymbol{\mu}_1 - \boldsymbol{\mu}_0)$. This raw direction is not enough, because we must also account for how the classes are spread in feature space.

\begin{figure}[ht!]
    \centering
    \includegraphics[width=0.9\textwidth]{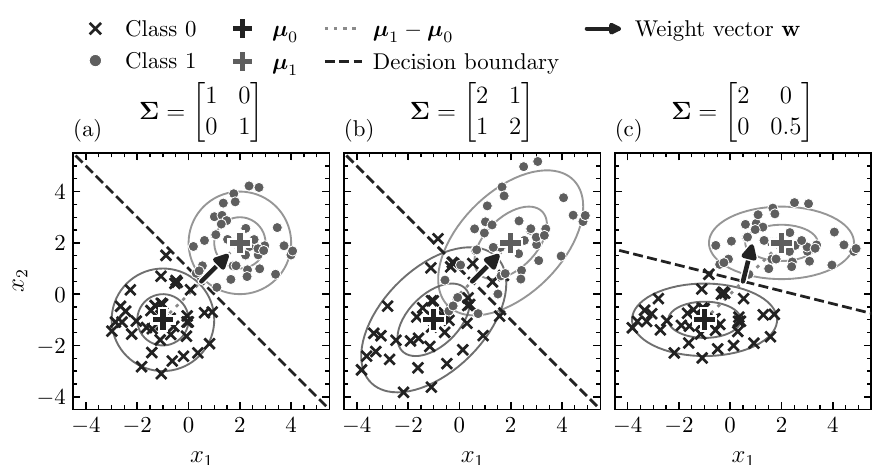}
    \caption{Gaussian classes with equal priors, the same means and a different shared covariance matrix $\boldsymbol{\Sigma}$ in each panel. Off-diagonal terms tilt the contours, and unequal variances stretch them along the axes. The contours show the class densities at one and two standard deviations. The weight vector is $\mathbf{w} \propto \boldsymbol{\Sigma}^{-1}(\boldsymbol{\mu}_1 - \boldsymbol{\mu}_0)$. In panels~(a) and~(b), $\mathbf{w}$ points along $\boldsymbol{\mu}_1 - \boldsymbol{\mu}_0$. In panel~(c), $\boldsymbol{\Sigma}^{-1}$ rotates $\mathbf{w}$ away from $\boldsymbol{\mu}_1 - \boldsymbol{\mu}_0$.}
    \label{fig:covariance_comparison}
\end{figure}

This is the role of $\boldsymbol{\Sigma}^{-1}$. It gives less weight to directions along which the classes spread widely. Depending on the covariance and the mean difference, it can preserve the original direction or rotate it to account for the spread of the classes. Figure~\ref{fig:covariance_comparison} shows this for three covariance matrices.

\begin{figure}[ht!]
    \centering
    \includegraphics[width=0.9\textwidth]{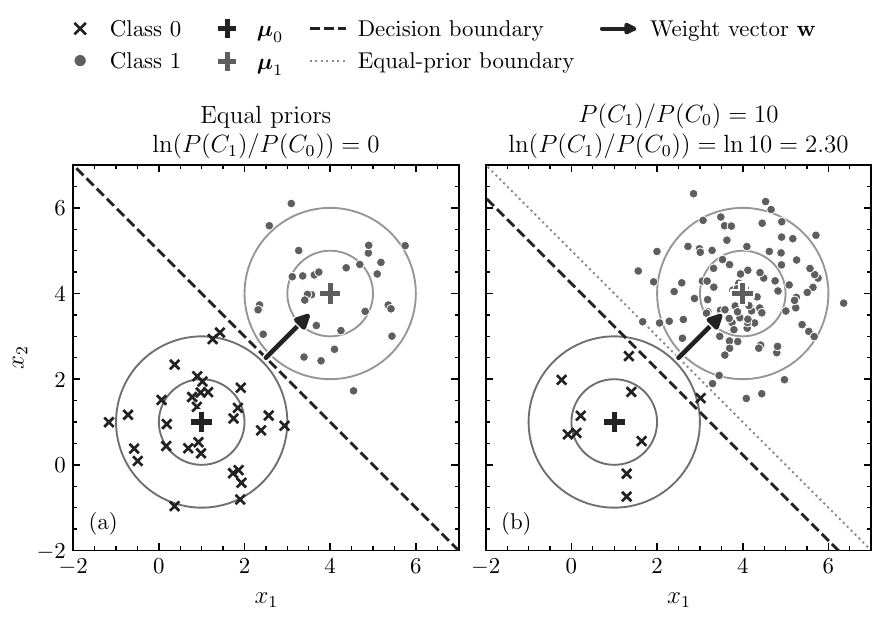}
    \caption{Effect of the class prior on the decision boundary. Both panels show Gaussian classes with the same identity covariance matrix and the same weight vector $\mathbf{w}$. In panel~(a) the priors are equal, and the boundary lies midway between the means. In panel~(b) class 1 is ten times more likely, so the sample has ten times as many class 1 as class 0 points. The boundary moves toward class 0 relative to the equal-prior boundary. The prior changes the bias $b$ and leaves the orientation of the boundary unchanged.}
    \label{fig:prior_effect}
\end{figure}

\looseness=-1 The bias term $b$ has two components. The first, $\ln(P(C_1)/P(C_0))$, is the prior term for class imbalance. In a field where galaxies outnumber stars ten to one, for instance, it equals $\ln(10)$. This moves the decision boundary toward the stars, as Figure~\ref{fig:prior_effect} shows. The second term, $-(1/2)(\boldsymbol{\mu}_1^T\boldsymbol{\Sigma}^{-1}\boldsymbol{\mu}_1 - \boldsymbol{\mu}_0^T\boldsymbol{\Sigma}^{-1}\boldsymbol{\mu}_0)$, sets the position of the boundary, which passes through the midpoint between the class means when the priors are equal. When $\boldsymbol{\Sigma}$ changes, $\mathbf{w}$ and $b$ change together, so the boundary still depends on both the covariance and the class means.

As in the derivation of the sigmoid above, inverting the log-odds $z = \mathbf{w}^T \mathbf{x} + b$ gives
\begin{equation}
P(C_1|\mathbf{x}) = \sigma(z) = \frac{1}{1 + e^{-z}} = \frac{1}{1 + e^{-(\mathbf{w}^T \mathbf{x} + b)}}.
\end{equation}The Gaussian model with equal covariances therefore gives logistic regression with a linear decision boundary. We reached the same model from the discriminative assumption of linear log-odds, so the two views agree. Figures~\ref{fig:covariance_comparison} and~\ref{fig:prior_effect} show how the covariance and the class priors shape the boundary.

\section{Limitations of Logistic Regression}\label{sec:ch7_limitations}

The link between the generative and discriminative views clarifies the inductive bias of logistic regression. This bias is the set of assumptions that shapes how the model learns and generalizes. The model assumes that the classes are best separated by a linear boundary in feature space. When the features of the classes approximately follow Gaussian distributions with similar covariance matrices, this assumption holds and logistic regression is a suitable classifier. In that case the linear boundary is the optimal one, because it assigns each object to the class with the higher posterior probability.

The success of a machine learning method depends on how well its inductive bias matches the structure of the data. The parameters $\mathbf{w}$ and $b$ are tied to the class means, covariances and priors, so they encode the Gaussian inductive bias. Understanding the bias helps us predict when logistic regression succeeds and when it fails.

How realistic are the Gaussian assumptions? In astronomy, different classes often have different covariance structures. Spiral galaxies might show a larger variance in some color indices than elliptical galaxies because their star formation histories vary more. In star-galaxy separation, the spread of the features of galaxies, with their varied morphologies and redshifts, typically differs from the tighter distribution of stellar features. Unequal covariances leave quadratic terms in the log-odds, and a linear boundary can then perform poorly. Poor performance of a linear boundary can hint that the classes are not Gaussian with a shared covariance matrix.

Another limitation appears when a line cannot separate the classes in feature space. A linear boundary cannot follow a curved boundary between the classes, however much data we provide.

\paragraph{Feature Transformations}

Logistic regression remains a useful tool, largely because we can transform the input features. As in linear regression, we are not limited to the raw features $\mathbf{x}$. We can use transformed features $\boldsymbol{\phi}(\mathbf{x})$ that better satisfy our assumptions.

\begin{figure}[ht!]
    \centering
    \includegraphics[width=0.9\textwidth]{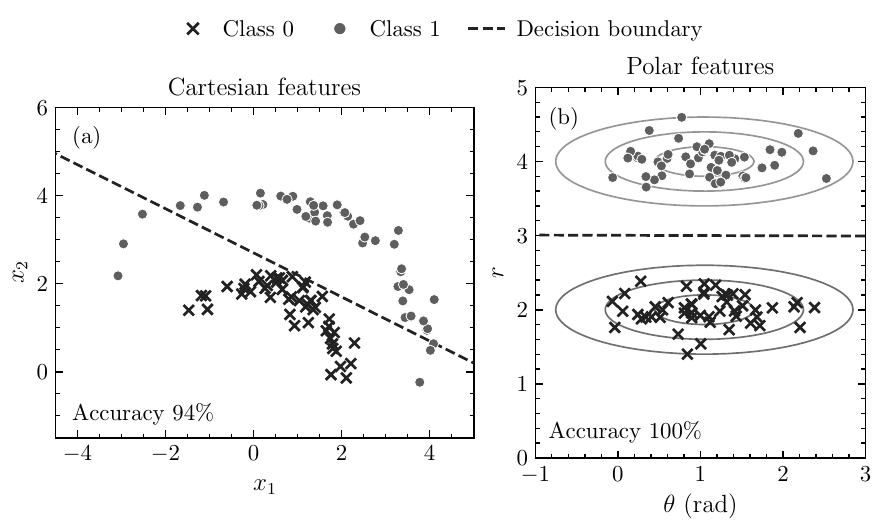}
    \caption{Feature transformation. Panel~(a) shows two classes that form curved arcs in the features $x_1$ and $x_2$. Panel~(b) shows the same points in polar coordinates $(\theta, r)$, where a line separates the classes. In panel~(b), the contours show the Gaussian fitted to each class at 1, 2 and 3 standard deviations.}
    \label{fig:feature_transform}
\end{figure}

Consider separating stars from galaxies with the flux and the half-light radius of an object. Both are positive and span orders of magnitude, so their logarithms often fit the Gaussian picture better than the raw values do. Similarly, classes that form arcs or rings in feature space can become linearly\index{linear separability} separable in polar coordinates, as Figure~\ref{fig:feature_transform} shows. The figure reports the accuracy, which is the fraction of objects that the line classifies correctly.

\looseness=1 Creating new features that separate the classes better is known as feature engineering, and it has been important in astronomical classification. Common transformations include logarithms of fluxes (magnitudes), color indices computed from fluxes, concentration indices from different aperture measurements, and morphological parameters. Each transformation encodes domain knowledge about which features separate the classes better. When we write $\mathbf{w}^T\boldsymbol{\phi}(\mathbf{x})$ instead of $\mathbf{w}^T\mathbf{x}$, we encode this astronomical understanding in the model.

In photometric classification of quasars, for instance, a cut in the $(u-g)$ versus $(g-r)$ plane can help select quasars at certain redshifts. Their ultraviolet excess makes them bluer in $u-g$ than stars of the same $g-r$. The transformation from fluxes to colors creates a feature space where the classes are more separable.

\paragraph{Interpretability}

Logistic regression is also easy to interpret. The linear form $\mathbf{w}^T\boldsymbol{\phi}(\mathbf{x})$ shows how each feature contributes to the decision. When the features are standardized\index{standardization} to zero mean and unit variance and weakly correlated, a larger absolute weight means that the feature matters more. In star-galaxy separation, for example, a larger absolute weight for the concentration index than for $g-r$ suggests that the concentration matters more.

Because the decision boundary is linear, we can also visualize it. In a two-feature space we can draw the line that separates the classes. In higher dimensions we can project onto the features with the largest absolute weights. This lets us check whether the model uses physically sensible criteria. Suppose that we use the concentration index and the color to separate stars from galaxies. We can then check that the boundary lies where physics suggests, since stars are more point-like and follow the stellar locus in color.

When a linear boundary in some feature space separates the classes reasonably well, logistic regression gives good accuracy, interpretability and speed. When it does not, we must find better feature\index{feature transformation} transformations or turn to more complex models. Knowing what a model assumes about the data guides these choices.

\section{Maximum Likelihood and Cross-Entropy Loss}\label{sec:ch7_mle}

How\index{cross-entropy loss|(} do we find the best parameters $\mathbf{w}$ and $b$? For Gaussian classes with equal covariance, the generative derivation gave the closed forms $\mathbf{w} = \boldsymbol{\Sigma}^{-1}(\boldsymbol{\mu}_1 - \boldsymbol{\mu}_0)$ and $b = \ln(P(C_1)/P(C_0)) - (1/2)(\boldsymbol{\mu}_1^T\boldsymbol{\Sigma}^{-1}\boldsymbol{\mu}_1 - \boldsymbol{\mu}_0^T\boldsymbol{\Sigma}^{-1}\boldsymbol{\mu}_0)$.

This solution is limited in practice, because real astronomical data often do not follow Gaussian distributions with equal covariances. We can instead fit the linear boundary directly with a discriminative model. As in linear regression, we use maximum\index{maximum likelihood estimation (MLE)} likelihood estimation to find the parameters that maximize the probability of the observed training data. This principle underlies many loss\index{loss function} functions in machine learning.

Each of the $N$ observations $(\mathbf{x}_i, t_i)$ consists of the features $\mathbf{x}_i$ and a binary label $t_i \in \{0,1\}$. As in linear regression, $t$ denotes the observed target. The model predicts the probability of class 1 through the sigmoid function
\begin{equation}
P(C_1|\mathbf{x}_i, \mathbf{w}) = y_i = \sigma(\mathbf{w}^T \mathbf{x}_i),
\end{equation}
where $\mathbf{w}$ again includes the bias, as in linear regression. In our star-galaxy example, $y_i$ is the probability that the $i$th object is a galaxy.

The likelihood is the probability of the labels of the data set under the model. We assume that the observations are independent, which is standard but can fail in astronomy, where spatial and temporal correlations are common. The likelihood is then the product
\begin{equation}
L(\mathbf{w}) = \prod_{i=1}^N P(t_i|\mathbf{x}_i, \mathbf{w}).
\end{equation}
For each observation, the factor in the product depends on its true class. If $t_i = 1$ (the object is a galaxy), the factor is $y_i$. If $t_i = 0$ (the object is a star), the factor is $1 - y_i$.

We combine the two cases into a single expression
\begin{equation}
L(\mathbf{w}) = \prod_{i=1}^N y_i^{t_i}(1-y_i)^{1-t_i}.
\end{equation}
The exponents $t_i$ and $(1-t_i)$ select the right factor. For a galaxy ($t_i = 1$), the second factor becomes $(1-y_i)^0 = 1$, which leaves $y_i$. For a star ($t_i = 0$), the first factor becomes $y_i^0 = 1$, which leaves $(1-y_i)$.

This is the Bernoulli likelihood from the conjugate priors of Chapter~5, but now a linear model sets the probability through the sigmoid transformation.

As in linear regression, we work with the negative log-likelihood, which turns the product into a sum. We take the negative because most optimization algorithms minimize. The sum is easier to compute. It also avoids numerical underflow, where a product of many small probabilities is rounded to zero because it is too small to store. The problem becomes the minimization of a loss, which measures how poorly the weights fit the training data
\begin{equation}
E(\mathbf{w}) = -\ln L(\mathbf{w}) = -\sum_{i=1}^N [t_i \ln y_i + (1-t_i)\ln(1-y_i)].
\end{equation}

We usually divide by the number of samples $N$, so that the loss does not grow with the size of the data set. From here on, $E(\mathbf{w})$ denotes this average loss
\begin{equation}
E(\mathbf{w}) = -\frac{1}{N}\sum_{i=1}^N [t_i \ln y_i + (1-t_i)\ln(1-y_i)].
\end{equation}
Without this normalization, more data would inflate the loss and require us to retune the learning rate of the optimization algorithm (Section~7.7).

\looseness=1 This expression is the cross-entropy\index{cross-entropy loss} loss, a name that comes from information theory. Entropy measures the average information, or ``surprise'', of a probability distribution. Cross-entropy is the average surprise when we use the model probabilities $y_i$ to describe the true labels $t_i$. Each label $t_i$ defines a distribution that puts probability 1 on the true class, and the model gives class 1 the probability $y_i$. Section~8.6 develops this link to information theory.

To see why this is a suitable measure for classification, consider a single data point. For a galaxy ($t_i = 1$), a correct prediction $y_i \approx 1$ gives $\ln y_i \approx 0$ and thus little penalty. An incorrect prediction $y_i \approx 0$ gives $\ln y_i \to -\infty$ and thus a severe penalty. For a star ($t_i = 0$), a prediction $y_i \approx 0$ gives a small loss, while $y_i \approx 1$ is heavily penalized.

Unlike least squares on the linear predictor, cross-entropy gives a correct prediction with a probability close to 1 almost no loss. Distant points on the correct side therefore no longer dominate the fit. The penalty for a severe mistake grows without bound.

The full optimization problem is
\begin{equation}
\mathbf{w}^* = \text{argmin}_{\mathbf{w}} \left\{-\frac{1}{N}\sum_{i=1}^N \left[t_i \ln \sigma(\mathbf{w}^T \mathbf{x}_i) + (1-t_i)\ln(1-\sigma(\mathbf{w}^T \mathbf{x}_i))\right]\right\},
\end{equation}
where $\text{argmin}_{\mathbf{w}}$ denotes the value of $\mathbf{w}$ that minimizes the expression.

Unlike linear regression, this problem has no closed-form solution, which means no explicit formula for $\mathbf{w}$ like the least-squares solution of Chapter~4. The nonlinearity of the sigmoid function prevents us from solving for $\mathbf{w}$ analytically by setting the derivatives to zero. We therefore need numerical optimization, in particular gradient descent, to find the parameters iteratively. Many models of later chapters need numerical optimization as well. The cross-entropy loss is convex\index{convexity} in $\mathbf{w}$ (Section~4.4), which means that it has a single basin. A local\index{local minima} minimum is a point lower than the points near it, and a convex loss has no local minimum besides the lowest point.\index{cross-entropy loss|)}

\section{Gradient Descent for Logistic Regression}\label{sec:ch7_gd}

\looseness=1 To\index{gradient descent|(} find the weights $\mathbf{w}$ that minimize the cross-entropy loss, we explore the loss surface iteratively. The loss $E(\mathbf{w})$ defines a surface over the space of weight vectors. Each point is a choice of weights, and its height is the loss, which measures how poorly those weights describe the data. Our goal is the lowest point, where the weights describe the data best.

Imagine standing on this surface and trying to reach the valley. A simple strategy is to feel the steepness and direction of the slope and to step downhill by an amount proportional to the steepness. We repeat this until the ground is flat, which corresponds to the minimum for a convex loss.

This intuition translates into mathematics. The slope at a point $\mathbf{w}$ is the gradient $\nabla E(\mathbf{w})$, a vector that points in the direction of steepest ascent in weight space. We therefore move in the negative gradient direction. A parameter $\eta$, the learning rate, controls the size of each step
\begin{equation}
\mathbf{w}^{(\tau+1)} = \mathbf{w}^{(\tau)} - \eta \nabla E(\mathbf{w}),
\end{equation}where $\tau$ indexes the iteration.

\begin{figure}[ht!]
    \centering
    \includegraphics[width=0.7\textwidth]{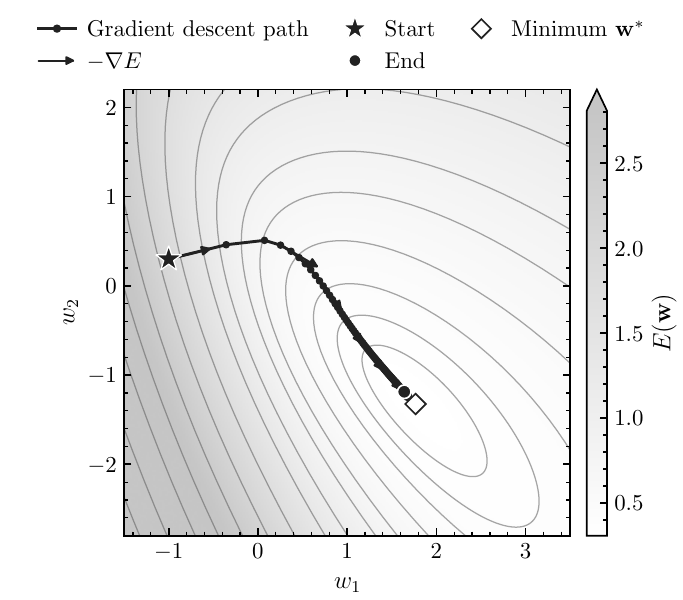}
    \caption{Gradient descent on the loss $E(\mathbf{w})$ of a logistic regression with two features. The path follows $-\nabla E$ and ends close to the minimum. It crosses the contours of $E(\mathbf{w})$ nearly at right angles, as the arrows at selected steps show.}
    \label{fig:gradient_descent}
\end{figure}

Figure~\ref{fig:gradient_descent} shows the path of gradient\index{gradient descent} descent on the loss surface. Starting from a high-loss region, each step follows the negative gradient toward the minimum.

\paragraph{Deriving the Gradient}

To use gradient descent, we need the gradient of the cross-entropy loss with respect to the weights, which we compute step by step. We start with the nonlinearity in our model, the sigmoid function, whose derivative has a simple form
\begin{equation}
\frac{\mathrm{d}\sigma(z)}{\mathrm{d}z} = \sigma(z)(1-\sigma(z)).
\end{equation}We can derive this directly from $\sigma(z) = 1/(1 + e^{-z})$. The chain\index{chain rule} rule gives
\begin{align}
\frac{\mathrm{d}\sigma(z)}{\mathrm{d}z} &= \frac{\mathrm{d}}{\mathrm{d}z}\left(\frac{1}{1 + e^{-z}}\right) \\*
&= \frac{e^{-z}}{(1 + e^{-z})^2} \\
&= \frac{1}{1 + e^{-z}} \cdot \left(1 - \frac{1}{1 + e^{-z}}\right) \\*
&= \sigma(z)(1-\sigma(z)).
\end{align}

When we compute $z = \mathbf{w}^T \mathbf{x}$ and evaluate $\sigma(z)$, the same value gives the derivative $\sigma(z)(1-\sigma(z))$ at no extra cost. With this derivative, we can find the gradient of the loss. To find $\partial E/\partial w_j$, we apply the chain rule step by step.

We first consider how $y_i$ depends on $w_j$
\begin{equation}
\frac{\partial y_i}{\partial w_j} = \frac{\partial}{\partial w_j}\sigma(\mathbf{w}^T \mathbf{x}_i) = \sigma(\mathbf{w}^T \mathbf{x}_i)(1-\sigma(\mathbf{w}^T \mathbf{x}_i))x_{ij} = y_i(1-y_i)x_{ij},
\end{equation}where $x_{ij}$ is the $j$th feature of object $i$.

The derivative of the loss with respect to $w_j$ is
\begin{align}
\frac{\partial E}{\partial w_j} &= -\frac{1}{N}\sum_{i=1}^N \left[t_i \frac{\partial}{\partial w_j}\ln y_i + (1-t_i)\frac{\partial}{\partial w_j}\ln(1-y_i)\right] \\*
&= -\frac{1}{N}\sum_{i=1}^N \left[t_i \frac{1}{y_i}\frac{\partial y_i}{\partial w_j} + (1-t_i)\frac{1}{1-y_i}\frac{\partial (1-y_i)}{\partial w_j}\right] \\*
&= -\frac{1}{N}\sum_{i=1}^N \left[t_i \frac{1}{y_i}\frac{\partial y_i}{\partial w_j} - (1-t_i)\frac{1}{1-y_i}\frac{\partial y_i}{\partial w_j}\right].
\end{align}Substituting the expression for $\partial y_i/\partial w_j$ gives
\begin{align}
\frac{\partial E}{\partial w_j} &= -\frac{1}{N}\sum_{i=1}^N \left[t_i \frac{1}{y_i} - (1-t_i)\frac{1}{1-y_i}\right]y_i(1-y_i)x_{ij} \\*
&= -\frac{1}{N}\sum_{i=1}^N \left[t_i(1-y_i) - (1-t_i)y_i\right]x_{ij} \\*
&= -\frac{1}{N}\sum_{i=1}^N (t_i - y_i)x_{ij}.
\end{align}The gradient vector is therefore
\begin{equation}
\nabla E(\mathbf{w}) = -\frac{1}{N}\sum_{i=1}^N (t_i - y_i) \mathbf{x}_i = \frac{1}{N}\sum_{i=1}^N (y_i - t_i) \mathbf{x}_i.
\end{equation}

\begin{figure}[ht!]
    \centering
    \includegraphics[width=0.7\textwidth]{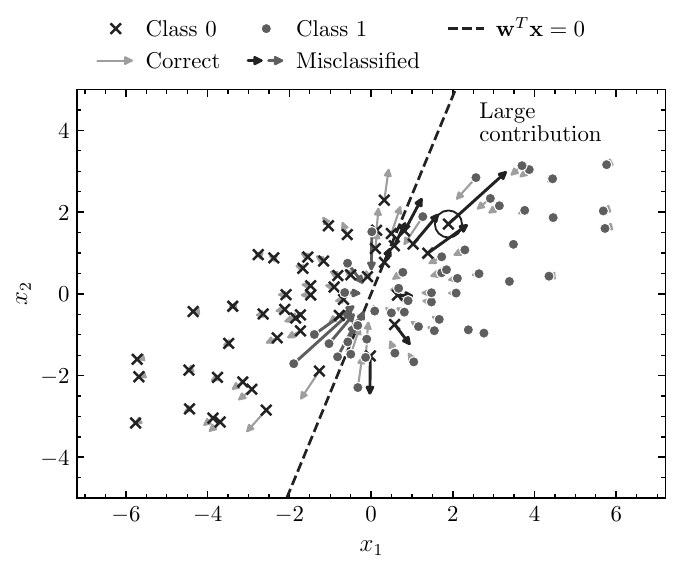}
    \caption{Gradient contributions $(y_i - t_i)\mathbf{x}_i$ of the data points at one step of the gradient descent in Figure~\ref{fig:gradient_descent}. Each arrow starts at its point, and the update subtracts these contributions. The circled class 0 point and its reflection in class 1 are the misclassified points farthest from the boundary and have the largest contributions.}
    \label{fig:logistic_gradients}
\end{figure}

\looseness=1 The gradient is driven by the prediction error $(y_i - t_i)$, scaled by the features $\mathbf{x}_i$ that led to the prediction. When a prediction is close to its label ($y_i \approx t_i$), the sample contributes almost nothing, so weights that already give correct predictions need no adjustment. When the prediction is wrong, the contribution depends on the size of the error and on the strength of the features behind it. Figure~\ref{fig:logistic_gradients} shows these contributions.

A wrong prediction with a high probability for the wrong class, such as $y_i = 0.9$ for a star, gives a larger gradient than one with a lower probability, such as $y_i = 0.6$. A large feature vector $\mathbf{x}_i$ enlarges the gradient further. The model therefore prioritizes fixing its worst errors.

Computing the gradient needs only three operations. We compute the current predictions $y_i = \sigma(\mathbf{w}^T \mathbf{x}_i)$, calculate the prediction errors $(y_i - t_i)$ and scale the features by these errors.

Inserting this gradient into the update rule, we update the weights at iteration $\tau$ according to
\begin{equation}
\mathbf{w}^{(\tau+1)} = \mathbf{w}^{(\tau)} - \eta \frac{1}{N}\sum_{i=1}^N (y_i^{(\tau)} - t_i) \mathbf{x}_i.
\end{equation}

\subsection{Stochastic Gradient Descent}\label{sec:ch7_sgd}

The\index{stochastic gradient descent (SGD)|(} gradient is simple, but each update requires a sum over all $N$ labeled training examples. The cost per iteration therefore grows in proportion to $N$, written $\mathcal{O}(N)$. This becomes expensive for large astronomical data sets, which range from millions of objects to billions of measurements.

Stochastic\index{stochastic gradient descent (SGD)} gradient descent (SGD)\index{SGD|see{stochastic gradient descent}} uses the fact that we do not need the exact gradient. A small random subset of the data, called a mini-batch,\index{mini-batch} approximates the gradient
\begin{equation}
\nabla E(\mathbf{w}) \approx \frac{1}{M}\sum_{i=1}^M (y_i - t_i)\mathbf{x}_i.
\end{equation}
Here $M$ is the batch\index{batch size} size, the number of samples in the mini-batch. It is typically 32 to 128, much smaller than $N$. If the batch is a uniformly random subset of the training set, its gradient is an unbiased estimate of the full gradient. This means that it is correct on average.

\looseness=1 This is like surveying galaxy properties. Measuring all galaxies gives the most accurate statistics, but a random sample gives reasonably good estimates. We can take many such samples in the time needed to measure that population. Similarly, SGD takes many steps in roughly the right direction in the time that gradient descent needs for a few exact steps.

\begin{figure}[ht!]
    \centering
    \includegraphics[width=0.7\textwidth]{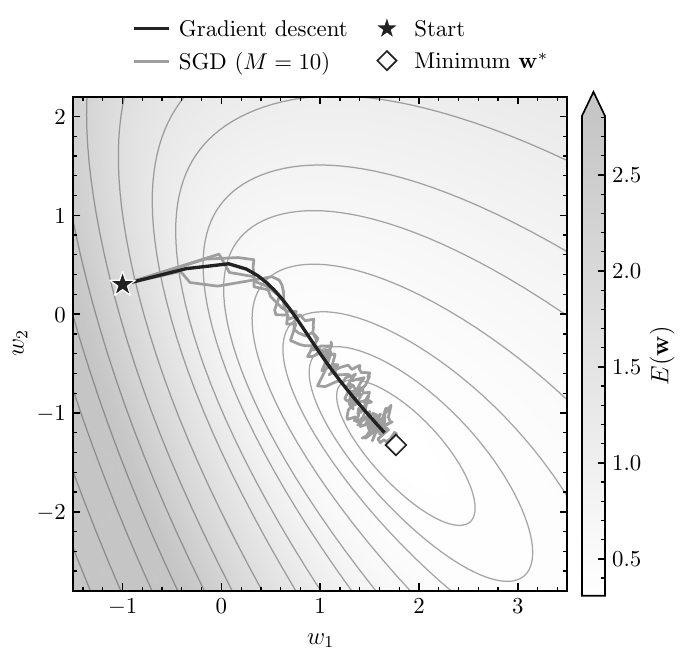}
    \caption{Gradient descent and stochastic gradient descent on the loss of Figure~\ref{fig:gradient_descent}. All paths start at the same point and use $\eta = 0.4$. All paths approach the minimum, and the SGD paths jitter around it.}
    \label{fig:gd_vs_sgd}
\end{figure}

As the $1/\sqrt{N}$ scaling of Chapter~3 shows, larger samples give estimates of lower variance. Mini-batch gradients are noisier than the full gradient, but they allow many more parameter updates for the same computing time. The frequent updates give rapid progress early in training, when the weights are far from optimal. Figure~\ref{fig:gd_vs_sgd} compares a gradient descent path with several SGD paths from the same starting point.

Unlike deterministic gradient descent, SGD can also use its noise to escape poor local minima in the non-convex problems common in machine learning (Chapter~15). The loss of logistic regression has no such minima, so here the main gain of SGD is speed. The noise of mini-batch sampling also acts as a mild form of regularization. The noise in the gradient estimates controls the trade-off between exploitation (following the gradient) and exploration (random perturbations). Smaller batches give noisier estimates.

Using mini-batches, the update rule becomes\index{gradient descent|)}\index{stochastic gradient descent (SGD)|)}
\begin{equation}
\mathbf{w}^{(\tau+1)} = \mathbf{w}^{(\tau)} - \eta \cdot \frac{1}{M}\sum_{i=1}^M (y_i - t_i)\mathbf{x}_i.
\end{equation}

\section{Hyperparameters and Regularization}\label{sec:ch7_hyper}

\paragraph{Learning Rate Considerations} Besides\index{hyperparameter|(}\index{learning rate|(} the mini-batch sampling, the size of each step matters, and the learning\index{learning rate} rate $\eta$ controls it. Think of the learning rate as the size of a wheel rolling down the loss surface. On a bumpy surface, a large wheel (large $\eta$) rolls over small bumps and valleys, which can help it escape local minima. It may also bounce erratically down steep slopes and miss deeper valleys. A small wheel (small $\eta$) follows the terrain precisely and can dig deep for better minima, but it may get stuck in minor depressions. On the single basin of logistic regression, only the speed and the overshoot matter.

\begin{figure}[ht!]
    \centering
    \includegraphics[width=0.9\textwidth]{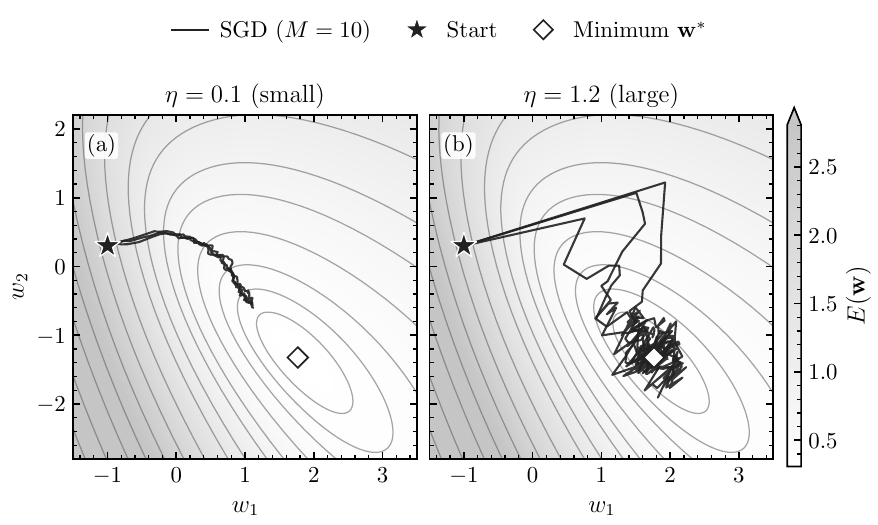}
    \caption{SGD paths for a small and a large learning rate on the loss of Figure~\ref{fig:gradient_descent}. All paths start at the same point. In panel~(a) the paths are smooth but have not yet reached the minimum. In panel~(b) the paths overshoot across the valley at the first step, reach the minimum quickly and then jitter around it.}
    \label{fig:sgd_learning_rates}
\end{figure}

In astronomical terms, the choice resembles slewing a telescope with coarse or fine adjustments. A large $\eta$ is like coarse slewing, which covers large distances quickly but may overshoot the target. A small $\eta$ is like the fine adjustment knobs, which are precise but slow for large movements. Figure~\ref{fig:sgd_learning_rates} shows SGD paths for a small and a large learning rate.

\paragraph{Choosing Hyperparameters}

The learning rate $\eta$ and the batch size $M$ are hyperparameters,\index{hyperparameter} parameters that control the learning process and are not learned from the data. Unlike the weights $\mathbf{w}$, which gradient descent optimizes, hyperparameters must be chosen before training begins. This choice is difficult because we usually do not know the shape of the loss surface in advance. It resembles choosing the wheel size for terrain that we cannot fully see.

For simple problems, a grid\index{grid search} search tries each value of a fixed list and keeps the one with the lowest validation loss (Chapter~4). We typically search learning rates on a logarithmic scale, for example $\{10^{-4}, 10^{-3}, 10^{-2}, 10^{-1}\}$, to cover a wide range of values with few trials. We might try batch sizes that are powers of 2, for example $\{32, 64, 128, 256\}$. In more complex models trained on graphics processing units, powers of 2 are preferred because the memory and compute architecture is optimized for such sizes. For logistic regression the choice matters little, and a simple grid search often suffices, even though more advanced tuning methods exist.

\begin{figure}[ht!]
    \centering
    \includegraphics[width=0.9\textwidth]{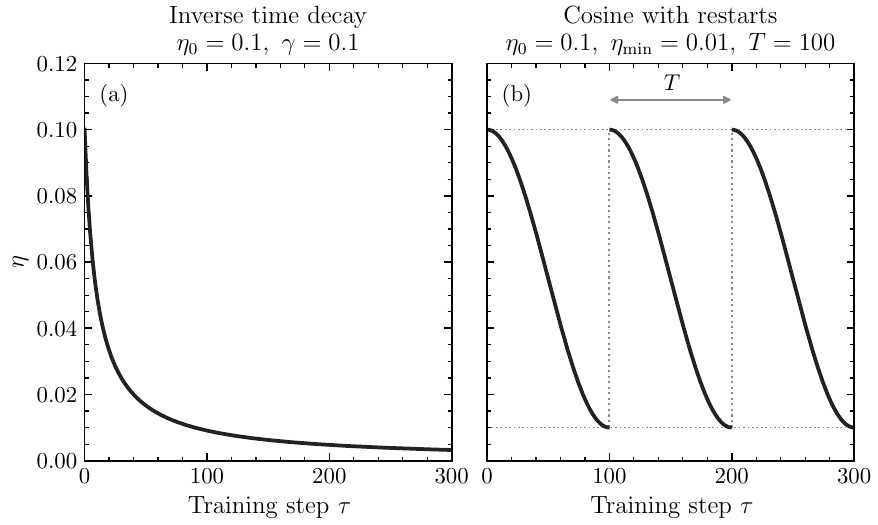}
    \caption{Two learning rate schedules, inverse time decay in panel~(a) and cosine annealing with restarts in panel~(b). The cosine schedule returns to $\eta_0$ at the end of each cycle.}
    \label{fig:lr_schedules}
\end{figure}

\paragraph{Learning Rate Schedules} A fixed learning rate is a compromise between fast progress and a precise result. For more complex problems, we often start with larger steps to reach the right region of parameter space quickly. We then take smaller steps to improve the solution. A simple schedule that does this is the inverse time decay
\begin{equation}
\eta^{(\tau)} = \frac{\eta_0}{1 + \gamma \tau}.
\end{equation}
Here $\eta_0$ is the initial learning rate, $\gamma$ is the decay rate, and $\tau$ is the number of the training step, a whole number. Figure~\ref{fig:lr_schedules}(a) shows this schedule.

\looseness=1 This schedule fits the wheel analogy. At first, $\eta \approx \eta_0$, and we use a big wheel that rolls quickly across the terrain. As the iteration $\tau$ increases, the denominator grows and $\eta$ shrinks, so we switch to a smaller wheel that follows the terrain more precisely. A larger decay rate $\gamma$ gives a quicker transition to refinement, and a smaller $\gamma$ keeps the exploration going longer.

Cosine annealing is a second schedule. It is inspired by annealing in metallurgy, where slowly cooling a heated metal lets its atoms settle into an ordered structure with fewer defects. We lower the learning rate along a cosine curve from $\eta_0$ toward a minimum value $\eta_{\text{min}}$ over a cycle of $T$ steps
\begin{equation}
\eta^{(\tau)} = \eta_{\text{min}} + \frac{1}{2}(\eta_0 - \eta_{\text{min}})\left(1 + \cos\left(\frac{\tau\pi}{T}\right)\right).
\end{equation}
Here $\tau$ counts the steps since the start of the cycle. This gives a smooth transition from high to low learning rates. Many implementations first warm up the learning rate from a small value to $\eta_0$, which we omit here.

In practice, the schedule restarts at $\eta_0$ at the end of each cycle, and $\tau$ restarts at 0. This cyclic pattern alternates exploration, with a high learning rate, and exploitation, with a low learning rate. Repeated cycles of heating and cooling similarly improve the structure of a metal. In non-convex problems it can help the model escape local minima while still allowing precise optimization. Figure~\ref{fig:lr_schedules}(b) shows the cosine schedule with restarts, with the horizontal axis counting all training steps.\index{learning rate|)}

\paragraph{Regularization} As in Chapters~4 and~5, a model with many features can overfit. In the high-dimensional feature spaces that are common in astronomy, the model can easily fit noise in the training data instead of real structure. Regularization\index{regularization} counters this problem by adding a penalty on large weights to the loss. When a line separates the classes perfectly, the loss keeps falling as the weights grow, and the penalty keeps the weights finite.

The standard approach, called L2 regularization or weight decay, adds a penalty proportional to the squared norm of the weight vector
\begin{equation}
E_{\text{reg}}(\mathbf{w}) = E(\mathbf{w}) + \lambda \|\mathbf{w}\|^2 = -\frac{1}{N}\sum_{i=1}^N [t_i \ln y_i + (1-t_i)\ln(1-y_i)] + \lambda \sum_{j=1}^D w_j^2.
\end{equation}
Here $\lambda$ is the regularization strength, which we choose like the other hyperparameters, and $D$ is the number of features. This modification changes the gradient that we use in the update, which becomes
\begin{equation}
\nabla E_{\text{reg}}(\mathbf{w}) = \nabla E(\mathbf{w}) + 2\lambda\mathbf{w} = \frac{1}{N}\sum_{i=1}^N (y_i - t_i) \mathbf{x}_i + 2\lambda\mathbf{w}.
\end{equation}The new term $2\lambda\mathbf{w}$ continuously pulls the weights toward zero during optimization, with a strength proportional to $\lambda$. This keeps single weights from becoming very large. A very large weight would indicate that the model relies too much on one feature, a common sign of overfitting.

From a Bayesian perspective, this regularization corresponds to a Gaussian prior on the weights, as in linear regression (Chapter~5). The posterior distribution of the weights given the training data $\mathcal{D}$ is proportional to the product of the likelihood and the prior
\begin{equation}
p(\mathbf{w}|\mathcal{D}) \propto p(\mathcal{D}|\mathbf{w})p(\mathbf{w}).
\end{equation}Because $E(\mathbf{w})$ is the loss per sample, the negative log-likelihood of the data is $N E(\mathbf{w})$. We take a Gaussian prior on the weights with zero mean and variance $1/(2N\lambda)$, so that the penalty scales with $N$ like the likelihood
\begin{equation}
p(\mathbf{w}) \propto \exp\left(-N\lambda\|\mathbf{w}\|^2\right).
\end{equation}
Maximizing the posterior is then equivalent to minimizing the negative log-posterior
\begin{equation}
-\ln p(\mathbf{w}|\mathcal{D}) = -\ln p(\mathcal{D}|\mathbf{w}) - \ln p(\mathbf{w}) = N\left[E(\mathbf{w}) + \lambda\|\mathbf{w}\|^2\right] + \text{const}.
\end{equation}

Regularization also balances bias and variance (Chapter~4). Bias is the systematic error that remains when the model is too rigid. Variance is how much the fitted model changes from one training set to another. Without regularization, the model might fit the training data closely but generalize poorly to new observations, a case of high variance. With very strong regularization, the model might not even fit the training data well, a case of high bias. The optimal $\lambda$ balances the two, and in the Bayesian view it corresponds to the strength of our prior belief.

Regularization is particularly important for high-dimensional data such as spectra, where the number of features can exceed the number of training examples. When we classify stellar types from spectral features, regularization prevents the model from fitting noise at specific wavelengths. It favors broader patterns that hold across the spectrum.\index{hyperparameter|)}

\section{Implementation Considerations}\label{sec:ch7_implementation}

Even with well-chosen learning rates and batch sizes, an implementation needs safeguards against numerical problems that can corrupt the results.

\paragraph{Numerical Stability} The first problem arises in computing the cross-entropy loss, here averaged over a mini-batch of $M$ samples
\begin{equation}
E(\mathbf{w}) = -\frac{1}{M} \sum_{i=1}^M [t_i \ln y_i + (1-t_i)\ln(1-y_i)].
\end{equation}The logarithm is problematic when a prediction $y_i$ approaches 0 or 1. For $y_i \approx 0$, $\ln(y_i)$ tends to negative infinity, which causes numerical errors. A simple remedy is to add a small constant $\epsilon$, typically $10^{-7}$, so that the logarithm of zero cannot occur
\begin{equation}
E(\mathbf{w}) = -\frac{1}{M}\sum_{i=1}^M [t_i \ln(y_i + \epsilon) + (1-t_i)\ln(1-y_i + \epsilon)].
\end{equation}

We also need a numerically stable implementation of the sigmoid function itself
\begin{equation}
\sigma(z) = \begin{cases}
\frac{1}{1 + e^{-z}} & \text{if } z \geq 0 \\
\frac{e^z}{1 + e^z} & \text{if } z < 0
.\end{cases}
\end{equation}
This piecewise form avoids large exponentials that exceed the largest number the computer can store, an error called overflow.\index{numerical stability} For large positive $z$, computing $e^{-z}$ is safer than computing $e^z$, and the reverse holds for negative $z$.

\paragraph{Optimization Pathologies} Training can encounter two pathologies. The first is vanishing\index{vanishing gradients} gradients, where the length $\|\nabla E(\mathbf{w})\|$ of the gradient becomes so small that the optimization stalls. The derivative of the sigmoid, $\sigma(z)(1-\sigma(z))$, approaches zero for large $|z|$, because $\sigma(z)$ approaches 0 or 1.

For logistic regression with the cross-entropy loss, however, this factor cancels in the gradient derived above. A wrong prediction with a probability close to 1 for the wrong class still gives the large gradient $(y_i - t_i)\mathbf{x}_i$. Vanishing gradients do arise with other losses, such as the squared error. Logistic regression keeps the sigmoid because its linear log-odds model requires it. Neural networks can use alternative activation functions, the nonlinear functions applied to the linear prediction, as we see in Chapter~15.

The second pathology is exploding\index{exploding gradients} gradients, where a ``bad'' batch containing outliers produces a huge step and can cause numerical overflow. For logistic regression, $|y_i - t_i| \le 1$, so the gradient norm cannot exceed the norm of the largest feature vector, $\|\nabla E\| \le \max_i \|\mathbf{x}_i\|$. Each step then moves the weights by at most $\eta \max_i \|\mathbf{x}_i\|$. Explosions therefore arise mainly from features with very large values, which is one reason to standardize the features. Outliers are a common concern in astronomical applications, where they can arise from instrumental artifacts, cosmic rays or rare astrophysical phenomena.

We can detect a potential explosion by monitoring $\max_j |w_j|$ during training. If the weights jump beyond reasonable bounds, say $10^2$, we can apply gradient\index{gradient clipping} clipping, which scales down large gradients and keeps their direction
\begin{equation}
\nabla E_{\text{clipped}} = \min\left(1, \frac{\text{threshold}}{\|\nabla E\|}\right) \nabla E.
\end{equation}
Here $\|\nabla E\|$ is the Euclidean norm of the full gradient vector and the threshold is a chosen constant. Clipping acts as a safety valve. It lets the optimization continue when outliers or other numerical instabilities, which are more common in complex models, temporarily make the gradients very large. Gradient clipping is common in the training of neural networks.

\paragraph{Monitoring and Early Stopping} Unlike linear regression, logistic regression has no closed-form solution, so we must decide when to stop the iterations. Rather than training for a fixed number of iterations, we evaluate the model on a validation set after each epoch.\index{epoch} An epoch is a complete pass through the training data. When the validation error stops decreasing, we have likely reached the limit of what the data can teach. This ``early\index{early stopping} stopping'' strategy often works better than fixed iteration counts or gradient magnitude thresholds, and it provides a defense against overfitting.

\looseness=1 Randomly shuffling the data between epochs keeps each batch representative of the overall distribution. This matters in astronomy, where objects may be ordered by position, brightness or observation time. Without shuffling, consecutive batches would each cover only one part of the sky or one range of brightness, so the gradient estimates would be biased.

\paragraph{Class Imbalance in Astronomy} Severe\index{class imbalance|(} class imbalance\index{class imbalance} is common in astronomical classification, especially in surveys where we search for rare objects among millions of background sources. For rare transient events such as supernovae, or for high-redshift quasars, the target class can be less than 0.1\% of all objects. A model that predicts ``not a transient'' for each object would then reach at least 99.9\% accuracy but fail at its primary task.

To address class imbalance, we can weight the minority class more heavily in the cross-entropy loss
\begin{equation}
E_{\text{weighted}}(\mathbf{w}) = -\frac{1}{N}\sum_{i=1}^N [\omega_1 t_i \ln y_i + \omega_0 (1-t_i)\ln(1-y_i)],
\end{equation}
where $\omega_1$ and $\omega_0$ are class weights. A common approach sets the weights inversely proportional to the class frequencies. If transients make up only 0.1\% of the data set, for example, we might set $\omega_1 = 0.999$ and $\omega_0 = 0.001$. Each transient then counts 999 times more than each non-transient. This ensures that the rare class contributes meaningfully to the gradient despite its rarity.

This ends our study of binary classification. Chapter~8 extends logistic regression to more than two classes, such as stellar spectral types. It also revisits the cross-entropy loss through information theory and mutual information.\index{class imbalance|)}\index{logistic regression|)}

\section{Tutorial: Classifying Red Clump Stars}\label{sec:ch7_tutorial}

This\index{red clump stars|(}\index{red giant branch stars|(} tutorial applies the logistic regression and gradient descent of this chapter to tell red\index{red clump stars} clump stars from red\index{red giant branch stars} giant branch stars. Red clump stars give precise distances across much of the Galaxy, but red giant branch stars look similar and spoil these distances. The full tutorial can be found at \url{https://statml.org/} as tutorial 7.

The two kinds of star are giants at different stages of stellar evolution. After a star exhausts the hydrogen in its core, it ascends the red giant branch, growing larger and more luminous. A larger star has a lower surface gravity, the strength of gravity at its surface. When the core ignites helium, the star becomes a red clump star. Red clump stars have nearly the same luminosity, so they are standard candles, sources of known luminosity whose apparent brightness gives their distance. Red giant branch stars span a wide range of luminosities, so one mistaken for a red clump star can have a less accurate distance.

\begin{figure}[t!]
    \centering
    \includegraphics[width=0.95\textwidth]{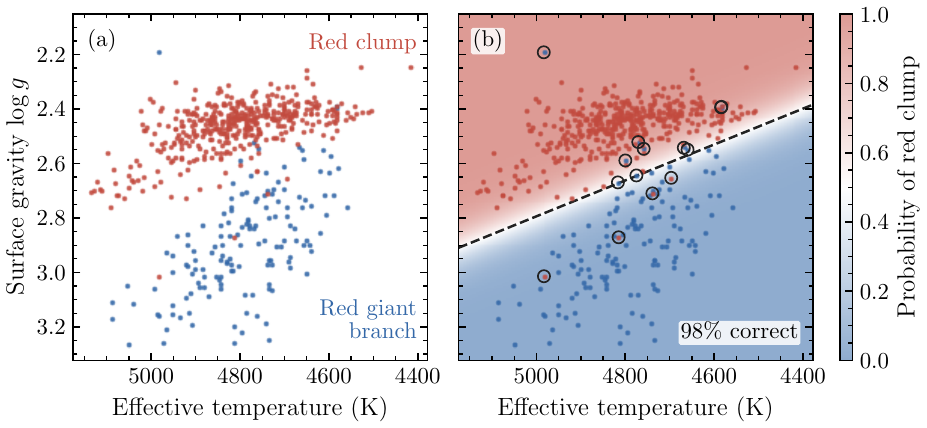}
    \caption{The tutorial classifies real giant stars. Panel~(a) shows test stars colored by their asteroseismic class, red clump in red and red giant branch in blue. Panel~(b) shows the same stars on a plane shaded by the red clump probability of a logistic regression on temperature and gravity. The dots keep their class colors, a dashed line marks probability one half, and rings mark misclassified stars.}
    \label{fig:ch7_tutorial_classes}
\end{figure}

To tell the two classes apart, astronomers use asteroseismology, the study of stellar oscillations, but it needs long monitoring and so covers small fields. Core helium burning makes the two classes differ in temperature and surface gravity, and spectroscopic surveys such as APOGEE measure both for far more stars. The tutorial trains a classifier on stars with known asteroseismic\index{asteroseismology} classes, so that it can classify stars outside those fields. In Section~\ref{sec:ch4_tutorials} the temperature was the output predicted from a spectrum. Here the temperature and gravity are two of eight features, and the output is a class (Section~\ref{sec:ch7_binary}). The classes partly overlap in the temperature and gravity plane (Figure~\ref{fig:ch7_tutorial_classes}a), so a class probability tells more than a label alone.

Logistic regression gives these probabilities. It models the log-odds as linear in the features, and the sigmoid keeps the probability between 0 and 1 (Section~\ref{sec:ch7_logistic}). The weights are the unknowns, and stars with known classes set them.

We fit the weights to these stars by minimizing the cross-entropy loss that maximum likelihood defines. Unlike least squares, this loss barely penalizes stars far on the correct side of the boundary, so they do not dominate the fit (Section~\ref{sec:ch7_mle}). Because the sigmoid is nonlinear, the loss has no closed-form minimum, so we descend it iteratively with gradient descent (Section~\ref{sec:ch7_gd}). Small random batches of stars, as in stochastic gradient descent (Section~\ref{sec:ch7_sgd}), make each step computationally cheap for large samples.

The classifier is meant for stars outside those fields, so we check it on test stars that the fit does not use (Section~\ref{sec:cross_validation}). The check tests the linear decision boundary, a hyperplane that is a line in this plane, where the probability is one half. A line can fail when the classes have unequal covariances (Section~\ref{sec:ch7_limitations}), as their spread in gravity differs (Figure~\ref{fig:ch7_tutorial_classes}a). Here nearly all the test stars lie on the correct side of the line (Figure~\ref{fig:ch7_tutorial_classes}b), so a linear discriminant is enough.

A probability, unlike a label alone, lets a survey choose how many red giant branch stars it accepts. The survey keeps the stars above a chosen probability, and a higher threshold removes more red giant branch stars and some red clump stars too. The kept sample then has few red giant branch stars, so its distances can map the Galaxy. The same method applies to other classes in astronomy when a hyperplane in the feature space separates them well.

In the tutorial, a higher threshold raises the precision,\index{precision (classification)} the share of kept stars that are red clump stars. It lowers the recall,\index{recall} the share of red clump stars that we keep. The false\index{false positive rate} positive rate is the share of red giant branch stars that we keep. The receiver\index{receiver operating characteristic (ROC) curve} operating characteristic (ROC)\index{ROC curve|see{receiver operating characteristic (ROC) curve}} curve plots the recall against the false positive rate for each threshold. It lies near the upper left corner for a good classifier. The area\index{area under the curve (AUC)} under it (AUC)\index{AUC|see{area under the curve}} does not depend on the threshold, so it can compare classifiers and the settings of one classifier.

A classifier turns a small labeled sample into labels for a whole survey. Asteroseismology classifies few stars, and spectroscopy measures far more, so a model trained on the first labels the second. Logistic regression suits this task, because its weights are easy to read and its probabilities say how far to trust each label. The red clump stars that it selects then map distances across the Galaxy, far beyond the reach of parallaxes.\index{red clump stars|)}\index{red giant branch stars|)}

\paragraph{Further Reading} Fisher (1936) developed linear discriminant analysis. Bliss (1934) introduced the probit model for binary responses, and Finney (1947) expanded it. Berkson (1944) applied the logistic function to the response of organisms to a dose, and Cox (1958) analyzed binary responses with several features. Shannon (1948) developed information theory, and Kullback and Leibler (1951) connected it to statistical inference through the KL divergence. Efron (1975) compared the efficiency of logistic regression and normal discriminant analysis. Rubinstein and Hastie (1997) and Ng and Jordan (2001) compared discriminative and generative learning. Cauchy (1847) described gradient descent, and Curry (1944) studied its convergence for nonlinear problems. Robbins and Monro (1951) developed the stochastic approximation behind stochastic gradient descent. Loshchilov and Hutter (2017) proposed a learning rate for stochastic gradient descent that falls along a cosine curve and restarts periodically. The tutorial follows the study of Ting et~al.\ (2018), who selected red clump stars from spectra. Its asteroseismic classes come from the period spacings that Vrard et~al.\ (2016) measured for Kepler red giants.

\chapter{Multi-Class Classification}

\makeatletter\@clubpenalty=10000\makeatother \clubpenalty=10000 \widowpenalty=10000

An astronomical object can be a star, a galaxy or a quasar. A galaxy can be elliptical, spiral or irregular. A star can have the spectral type O, B, A, F, G, K or M, each with distinct physical properties. A single probability cannot describe three or more classes, as it does for two. We need one set of probabilities that adds up to 1 for each object.

A simple idea is to apply binary classification several times and compare each class with all the others. This idea has a flaw, because the separate classifiers can contradict each other. The remedy is to model all classes at once, with a softmax function that generalizes the sigmoid function. Maximum likelihood then gives a loss for $K$ classes, which generalizes the binary cross-entropy.

We first show why binary decisions fail and how one-hot vectors represent the class labels. We then derive the softmax function and the cross-entropy loss. Information theory explains what the cross-entropy loss measures, through entropy and mutual information. Astronomers also use these measures to choose observations that best distinguish between competing models. We then derive the gradient and train the model with stochastic gradient descent (SGD). We end with practical points such as numerical stability and initialization.

\section{Multi-Class Classification vs. Binary Classification}\label{sec:ch8_ovr}

We\index{classification!multi-class|(} want a model for many classes whose predictions do not contradict each other.

\paragraph{One-vs-Rest Limitations}

In the \mbox{One-vs-Rest} strategy,\index{one-vs-rest classification} we train one binary classifier for each class $C_k$. Each classifier separates class $C_k$ from ``not $C_k$'', which combines all other classes.

\begin{figure}[ht!]
    \centering
    \includegraphics[width=0.8\textwidth]{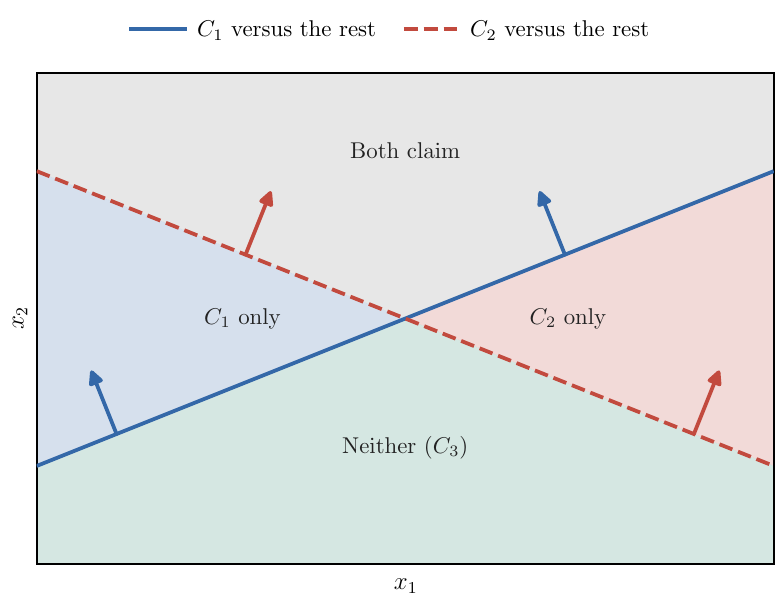}
    \caption{Decision boundaries of two binary classifiers in the plane of two features, for a problem with three classes. Each arrow points to the side on which its classifier predicts its own class. Both classifiers claim the wedge above the crossing, so a point there gets two different classes. In the opposite wedge neither classifier claims the point, so it goes to $C_3$.}
    \label{fig:ovr_classification_pathology}
\end{figure}

This approach can lead to logical inconsistencies. Take three classes $C_1$, $C_2$ and $C_3$. We train one classifier that separates $C_1$ from the rest and another that separates $C_2$ from the rest. These two already show the inconsistency. A classifier claims a point when it predicts its own class, and a point goes to $C_3$ when neither classifier claims it. Each classifier has its own linear decision boundary, and the two boundaries, which cross, divide the feature space into four regions, as Figure~\ref{fig:ovr_classification_pathology} shows.

In the region that both classifiers claim, the $C_1$ classifier predicts $C_1$ and the $C_2$ classifier predicts $C_2$. Both predictions can have high confidence. The point cannot belong to two classes at once. Each classifier is trained on its own question, so their answers can disagree.

\paragraph{One-vs-One Limitations}

The \mbox{One-vs-One} strategy\index{one-vs-one classification} avoids the combined group ``not $C_k$''. It trains a separate binary classifier for each pair of classes, so $K$ classes need $K(K-1)/2$ classifiers. Three classes need three classifiers, which compare $C_1$ with $C_2$, $C_2$ with $C_3$ and $C_1$ with $C_3$. It still produces contradictions.

\begin{figure}[ht!]
    \centering
    \includegraphics[width=0.8\textwidth]{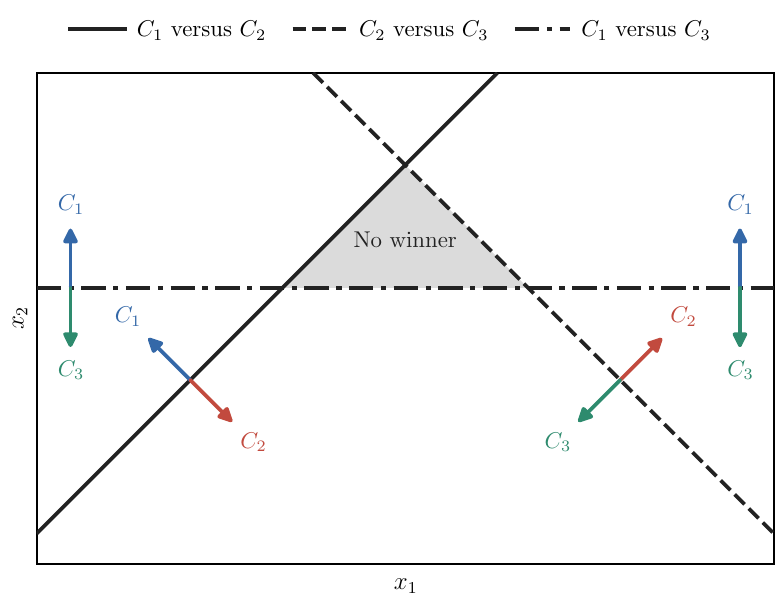}
    \caption{Decision boundaries of three binary classifiers, one for each pair of three classes, in the plane of two features. Each arrow points to the class that wins its pair on that side of the line. In the triangle where the three lines meet, each class beats one rival and loses to the other, so no class wins the vote.}
    \label{fig:ovo_classification_pathology}
\end{figure}

The classifiers decide independently, so their decisions can form a circular contradiction. Imagine three experts who each compare only two classes. For one point, the first expert ranks $C_2$ above $C_1$, the second ranks $C_3$ above $C_2$, and the third ranks $C_1$ above $C_3$. The rankings form a loop without a winner, as Figure~\ref{fig:ovo_classification_pathology} shows. Pairwise votes in an election can form the same loop, which is known as a voting paradox.

In the contradictory regions, the classifiers give no single answer. In practice, the label in such a region reflects the implementation and says little about patterns in the data. A multi-class classifier must therefore treat all classes simultaneously. Galaxy morphology, stellar type identification and exoplanet categorization need predictions without contradictions.

\section{One-Hot Encoding for Multi-Class Problems}\label{sec:ch8_onehot}

The\index{one-hot encoding|(} first step is to rethink how we represent the class labels. A binary problem needs only one scalar label, usually 0 or 1, but this representation breaks down for more than two classes.

A first idea from binary classification\index{classification!multi-class} is to use consecutive integers. Galaxies could be 0 for elliptical, 1 for spiral and 2 for irregular, and stars 0 for O and 1 for B.

Integer labels impose an artificial order on the classes. The sequence 0, 1, 2 suggests that spiral galaxies lie between elliptical and irregular galaxies, although the classes have no such order. In the binary case, we take 1 minus the prediction probability to swap the classes. For several classes, a new numbering changes the predictions of a model that treats the integers as numbers.

We need an encoding that suggests no order. One-hot\index{one-hot encoding} encoding does this with one vector for each class.

A one-hot vector has $K$ entries, one for each class. We put a 1 in the position of the class and 0 in all other positions. Class 1 (for example elliptical galaxies) becomes $[1, 0, 0, \ldots, 0]$, class 2 (spirals) becomes $[0, 1, 0, \ldots, 0]$, and class 3 (irregulars) becomes $[0, 0, 1, \ldots, 0]$.

One-hot encoding treats the classes symmetrically. Each class is a unit vector along a different coordinate axis. Geometrically, the classes are corners of a $K$-dimensional hypercube, which generalizes the square and the cube. Each pair has the same Euclidean distance $\sqrt{2}$.

\looseness=1 One-hot encoding also appears outside astronomy. A language model that predicts the next word of a text solves a multi-class problem. Its classes are the words of a vocabulary of tens or hundreds of thousands of words, and each word is a one-hot vector. At each position of the text, the model predicts a probability distribution over the vocabulary. The softmax function and the cross-entropy loss of this chapter also train such a model.\index{one-hot encoding|)}

\section{Multi-Class Logistic Regression}

The training labels are one-hot vectors, so the model must output a vector as well. It predicts one probability for each class at the same time. This model is also called multinomial\index{multinomial logistic regression|see{logistic regression, multi-class}} logistic\index{logistic regression!multi-class} regression or softmax\index{softmax regression|see{logistic regression, multi-class}} regression.

The model outputs a probability vector \mbox{$\mathbf{p} = [p_1, p_2, \ldots, p_K]$}, where $p_k$ is the probability of belonging to class $k$. Each $p_k$ must lie between 0 and 1, and the probabilities must sum to 1, $\sum_{k=1}^K p_k = 1$.

In the galaxy example, the training label of an elliptical galaxy is the one-hot vector $[1,0,0]$, while the model might output \mbox{$\mathbf{p} = [0.70, 0.25, 0.05]$}. This vector means 70\% elliptical, 25\% spiral and 5\% irregular. The model must map features to such probability vectors and respect their constraints.

A probability vector shows how confident a classification is, and in astronomy the confidence varies with data quality. A bright, well-resolved galaxy might receive $[0.95, 0.04, 0.01]$, a confident elliptical. A faint or poorly resolved galaxy might receive $[0.45, 0.40, 0.15]$, which means that the model is uncertain between elliptical and spiral. A galaxy with both spiral and irregular features might receive $[0.10, 0.45, 0.45]$, which means that the model cannot tell spiral from irregular. Such equal probabilities can also describe a galaxy that is truly intermediate between the two types.

In binary classification, a linear function cannot model a probability directly, because probabilities are bounded and linear functions are not. The logit transformation maps the bounded probability to the unbounded log-odds, which is the logarithm of the ratio of the two class probabilities. We therefore model the log-odds by the linear function $\mathbf{w}^T\mathbf{x}$, with a constant 1 appended to $\mathbf{x}$ for the bias.

With $K$ classes, we must transform the probability vector $\mathbf{p}(\mathbf{x})$ so that a linear function can model it. The transformation must treat all classes together.

The idea is to compare all classes with one reference class, as the binary case compares class 1 with class 0. Probabilities sum to 1, so $K$ classes have only $K-1$ degrees of freedom, and each probability can be expressed relative to one reference class. Any class can be the reference, like the choice of an origin in a coordinate system.

We construct $K-1$ probability ratios by comparing each class $C_k$ with the reference class, which we label class $K$. It plays the role of class $C_0$ in Chapter~7. Bayes' theorem gives
\begin{equation}
\frac{P(C_k|\mathbf{x})}{P(C_K|\mathbf{x})} = \frac{p(\mathbf{x}|C_k)P(C_k)}{p(\mathbf{x}|C_K)P(C_K)} \text{ for } k = 1, ..., K-1.
\end{equation}

As in the binary case, the logarithm of each ratio can take any real value, so a linear function can model it. We model each log-ratio as
\begin{equation}
\ln\left(\frac{P(C_k|\mathbf{x})}{P(C_K|\mathbf{x})}\right) = \mathbf{w}_k^T \mathbf{x}.
\end{equation}This gives $K-1$ weight vectors $\mathbf{w}_k$, one for each non-reference class. We optimize them together, because the probability of each class depends on all of them, as the softmax function below shows. In the \mbox{One-vs-Rest} strategy, each classifier works alone. Adjusting $\mathbf{w}_1$ changes the relation between class 1 and the reference class and, through it, the relation between class 1 and each other class.

In the galaxy example, the reference class is irregular galaxies. Adjusting the weights that separate elliptical galaxies from it also changes how the model separates elliptical from spiral galaxies. No binary classifier for that pair is needed. The decision boundaries therefore depend on one shared set of weight vectors. The shared weights make the boundaries fit together, which independent binary classifiers cannot guarantee.

We now convert the $K-1$ unbounded log-ratios back into $K$ probabilities.

\section{The Softmax Function}\label{sec:ch8_softmax}

In\index{softmax function|(} the binary case, the sigmoid function is the inverse of the logit. It gives the probability of class 1 from the log-odds $\mathbf{w}^T\mathbf{x}$
\begin{equation}
P(C_1|\mathbf{x}) = \frac{1}{1 + \exp(-\mathbf{w}^T \mathbf{x})}.
\end{equation}
The reference class receives the remaining probability, $P(C_0|\mathbf{x}) = 1 - P(C_1|\mathbf{x})$.

For $K$ classes, we need an analogous function that turns the $K-1$ linear predictions into $K$ valid probabilities. This generalization of the sigmoid function is the softmax\index{softmax function} function, and we derive it as in the binary case.

We solve for the individual probabilities. Exponentiating the log-ratio model gives
\begin{equation}
\frac{P(C_k|\mathbf{x})}{P(C_K|\mathbf{x})} = \exp(\mathbf{w}_k^T \mathbf{x}) \text{ for } k = 1, \ldots, K-1,
\end{equation}
and therefore
\begin{equation}
P(C_k|\mathbf{x}) = P(C_K|\mathbf{x})\exp(\mathbf{w}_k^T \mathbf{x}) \text{ for } k = 1, \ldots, K-1.
\end{equation}

The probabilities satisfy
\begin{equation}
\sum_{k=1}^K P(C_k|\mathbf{x}) = 1.
\end{equation}
We substitute the expression for $P(C_k|\mathbf{x})$ in the terms with $k < K$. Factoring out $P(C_K|\mathbf{x})$ then gives\begin{equation}
P(C_K|\mathbf{x})\left(1 + \sum_{k=1}^{K-1} \exp(\mathbf{w}_k^T \mathbf{x})\right) = 1.
\end{equation}
Solving for the reference class gives
\begin{equation}
P(C_K|\mathbf{x}) = \frac{1}{1 + \sum_{k=1}^{K-1} \exp(\mathbf{w}_k^T \mathbf{x})},
\end{equation}
and each other class $k \neq K$ then has the probability
\begin{equation}
P(C_k|\mathbf{x}) = \frac{\exp(\mathbf{w}_k^T \mathbf{x})}{1 + \sum_{j=1}^{K-1} \exp(\mathbf{w}_j^T \mathbf{x})}.
\end{equation}

A more compact form uses a weight vector $\mathbf{w}_K = \mathbf{0}$ for the reference class. This vector is fixed, so we do not optimize it. It gives $\exp(\mathbf{w}_K^T\mathbf{x}) = 1$, and the denominator $1 + \sum_{j=1}^{K-1} \exp(\mathbf{w}_j^T\mathbf{x})$ equals $\sum_{j=1}^{K} \exp(\mathbf{w}_j^T\mathbf{x})$. The reference class then has the same form as all other classes.

The probability of each class $k = 1, 2, \ldots, K$ is therefore
\begin{equation}
P(C_k|\mathbf{x}) = \frac{\exp(\mathbf{w}_k^T \mathbf{x})}{\sum_{j=1}^{K} \exp(\mathbf{w}_j^T \mathbf{x})}.
\end{equation}This function turns $K$ linear predictors into a probability distribution over $K$ classes, and it is called the softmax function. Only the $K-1$ vectors $\mathbf{w}_1, \ldots, \mathbf{w}_{K-1}$ are free parameters.

The softmax function does not change when we subtract the same vector $\mathbf{v}$ from all $K$ weight vectors. The factor $\exp(-\mathbf{v}^T\mathbf{x})$ appears in the numerator and the denominator and cancels. Choosing class $m$ as the reference means $\mathbf{v} = \mathbf{w}_m$. The weight vectors become $\mathbf{w}_k - \mathbf{w}_m$ and describe the same probabilities. For maximum likelihood without a weight penalty, the predictions therefore do not depend on the reference class.

The binary case is the special case $K = 2$. With $\mathbf{w}_2 = \mathbf{0}$ for the reference class, the softmax function gives
\begin{align}
P(C_1|\mathbf{x}) &= \frac{\exp(\mathbf{w}_1^T \mathbf{x})}{\exp(\mathbf{w}_1^T \mathbf{x}) + 1} \\*
&= \frac{1}{1 + \exp(-\mathbf{w}_1^T \mathbf{x})}.
\end{align}This is the sigmoid function of binary classification. The reference class receives the remaining probability $1 - P(C_1|\mathbf{x}) = 1/(1 + \exp(\mathbf{w}_1^T\mathbf{x}))$.

\paragraph{Decision Boundaries}

The model divides the feature space into regions for the different classes. A boundary between two regions is a place where the model is equally uncertain between two classes. Classes $j$ and $k$ are equally likely where
\begin{equation}
P(C_j|\mathbf{x}) = P(C_k|\mathbf{x}).
\end{equation}

The probabilities share the same denominator, so equal probabilities require equal numerators, $\exp(\mathbf{w}_j^T\mathbf{x}) = \exp(\mathbf{w}_k^T\mathbf{x})$. Taking logarithms gives
\begin{equation}
\mathbf{w}_j^T \mathbf{x} = \mathbf{w}_k^T \mathbf{x}.
\end{equation}This equation defines a hyperplane in feature space where classes $j$ and $k$ are equally likely. A hyperplane generalizes a line in two dimensions and a plane in three dimensions (Section~7.4). For $K$ classes there are $\binom{K}{2}$ such hyperplanes, one for each pair of classes. Here $\binom{K}{2}$ means $K$ choose 2, the number of pairs that we can pick from $K$ classes, and it equals $K(K-1)/2$.

\looseness=1 Each hyperplane divides the space into a region where class $j$ is more likely, $\mathbf{w}_j^T\mathbf{x} > \mathbf{w}_k^T\mathbf{x}$, and a region where class $k$ is more likely, $\mathbf{w}_j^T\mathbf{x} < \mathbf{w}_k^T\mathbf{x}$. The feature entries of $\mathbf{w}_j - \mathbf{w}_k$ are normal to the hyperplane, so they set its orientation. The bias entry sets its position. The difference of the two linear predictors changes fastest along this normal.

This resembles the \mbox{One-vs-One} strategy, but here all pairs come from one probability distribution. The decision\index{decision boundary} boundaries are the parts of the hyperplanes where the two equally likely classes are more probable than all other classes.

To find the predicted class at a point, we take
\begin{equation}
    \text{argmax}_k P(C_k|\mathbf{x}).
\end{equation}
This operation returns the class with the highest probability. It gives a definite answer, but the probabilities also show the uncertainty. If two classes have similar probabilities, the model is uncertain between them, even though the argmax still picks one. The probabilities $P(C_k|\mathbf{x})$ therefore tell us more than the boundaries do.

\begin{figure}[ht!]
    \centering
    \includegraphics[width=1.0\textwidth]{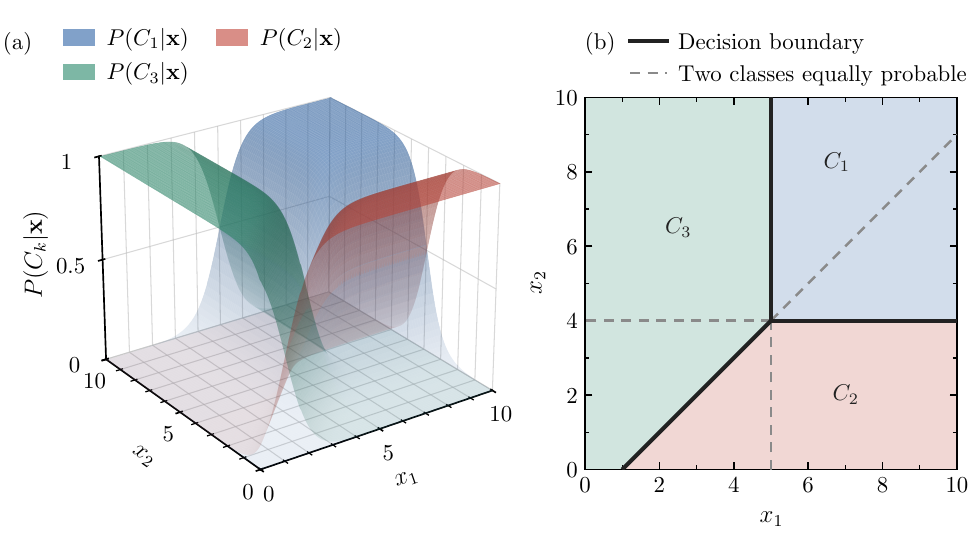}
    \caption{Softmax model with three classes and two features. Panel~(a) shows the probability surfaces of the classes, which add up to one at each point. Panel~(b) shows the model from above, with each point colored by its most probable class. Dashed lines run through the region of a third class, where neither of the two equally probable classes is the most probable.}
    \label{fig:softmax_visualization}
\end{figure}

\begin{figure}[ht!]
    \centering
    \includegraphics[width=1.0\textwidth]{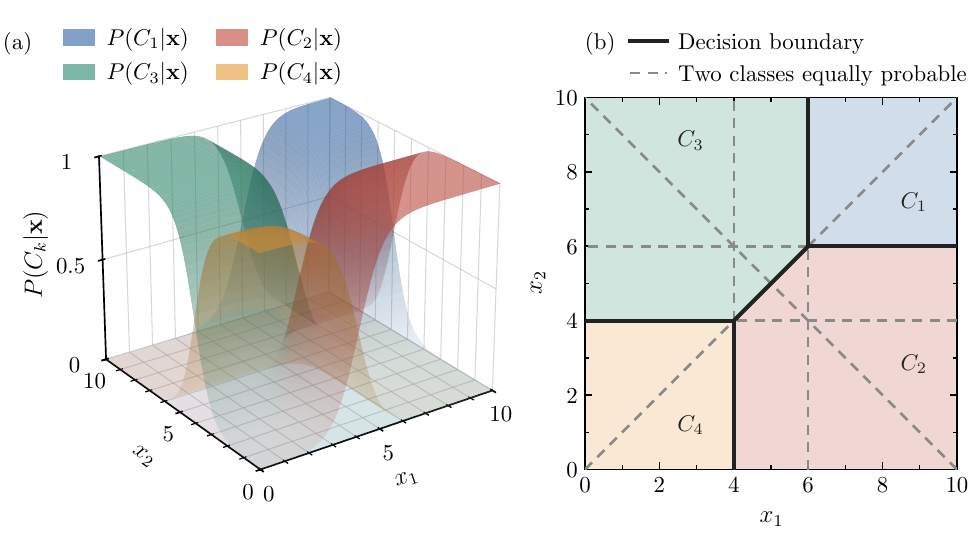}
    \caption{Softmax model with four classes and two features. Panel~(a) shows the four probability surfaces. In panel~(b) the pairwise lines form a window pane.}
    \label{fig:softmax_four_classes}
\end{figure}

Suppose that one linear prediction $\mathbf{w}_m^T\mathbf{x}$ is much larger than the others. Then $\exp(\mathbf{w}_m^T\mathbf{x})$ dominates the denominator, so $P(C_m|\mathbf{x})$ approaches 1 and all other probabilities approach 0. Choosing the class with the largest of the $K$ linear predictions gives probability 1 to that class and 0 to the others. That choice jumps from one class to another where two predictions cross. The softmax function is a smooth and differentiable version of this choice of the maximum, which explains its name. When several linear predictions are similar, the softmax function spreads the probability over the corresponding classes.

In two dimensions, each hyperplane becomes a line, and three classes $C_1$, $C_2$ and $C_3$ have three pairwise lines. Parts of these lines bound the regions in which one class has the highest probability. All probabilities come from one softmax function, so the regions cannot contradict each other as in \mbox{One-vs-One}. The probabilities add up to 1, so no region belongs to two classes or to no class, as in \mbox{One-vs-Rest}. The decision boundaries can form a pattern like a peace sign, as in Figure~\ref{fig:softmax_visualization}, or like the Mercedes-Benz logo. Three regions radiate from a central point where all three lines meet.

Four classes give six pairwise lines, as $\binom{4}{2} = 6$. They can form a pattern like a window pane, as in Figure~\ref{fig:softmax_four_classes}, but some pairs of classes share no border. In this example, $C_1$ and $C_4$ are such a pair, because $C_2$ or $C_3$ is more probable wherever $C_1$ and $C_4$ have equal probability. Each point still has one probability distribution over all classes from the softmax function.\index{softmax function|)}

\section{Multi-Class Cross-Entropy Loss}\label{sec:ch8_loss}

To\index{cross-entropy loss!multi-class|(} find the set of all $K-1$ weight vectors $\{\mathbf{w}_k\}_{k=1}^{K-1}$, we use maximum likelihood as in the binary case. We choose the weights that make the observed data most probable under the model.

The Bernoulli distribution of the binary case has two outcomes, and now each example has $K$ possible outcomes. The distribution over $K$ classes is called the categorical distribution. Let $\mathbf{t}$ be the one-hot label of an example, with $t_k = 1$ if the example belongs to class $k$ and 0 otherwise. For one example, the likelihood is the probability of its observed label given the model
\begin{equation}
L = \prod_{k=1}^K P(C_k|\mathbf{x}, \{\mathbf{w}_k\}_{k=1}^{K-1})^{t_k}.
\end{equation}The factor of a class with $t_k = 0$ equals 1, and only the factor of the true class differs from 1. For an object of class 2, $t_2 = 1$ and all other $t_k = 0$, so the likelihood is $P(C_2|\mathbf{x}, \{\mathbf{w}_k\}_{k=1}^{K-1})$.

For $N$ independent examples, the joint likelihood is the product of the individual likelihoods
\begin{equation}
L = \prod_{i=1}^N \prod_{k=1}^K P(C_k|\mathbf{x}_i, \{\mathbf{w}_k\}_{k=1}^{K-1})^{t_{ik}}.
\end{equation}
Here $i$ indexes the $N$ examples and $k$ the $K$ classes, and $t_{ik}$ is the $(i,k)$ element of the one-hot label matrix. We take the logarithm, which avoids numerical underflow, and change its sign so that we minimize. We also divide by $N$, so that the scale does not depend on the data set size. The result is the loss $E$
\begin{equation}
E(\{\mathbf{w}_k\}_{k=1}^{K-1}) = -\frac{1}{N} \ln L = -\frac{1}{N}\sum_{i=1}^N \sum_{k=1}^K t_{ik} \ln P(C_k|\mathbf{x}_i, \{\mathbf{w}_k\}_{k=1}^{K-1}).
\end{equation}

For $K = 2$, the multi-class cross-entropy\index{cross-entropy loss!multi-class} reads
\begin{align}
E &= -\frac{1}{N}\sum_{i=1}^N \sum_{k=1}^2 t_{ik} \ln P(C_k|\mathbf{x}_i, \{\mathbf{w}_k\}_{k=1}^{1}) \\*
&= -\frac{1}{N} \sum_{i=1}^N \left[t_{i1} \ln \frac{\exp(\mathbf{w}_1^T \mathbf{x}_i)}{1 + \exp(\mathbf{w}_1^T \mathbf{x}_i)} + t_{i2} \ln \frac{1}{1 + \exp(\mathbf{w}_1^T \mathbf{x}_i)}\right].
\end{align}
Each example belongs to one class, so $t_{i1} + t_{i2} = 1$. We write $t_i = t_{i1}$ and $t_{i2} = 1 - t_i$, and the loss becomes
\begin{equation}
E = -\frac{1}{N} \sum_{i=1}^N [t_i \ln y_i + (1-t_i)\ln(1-y_i)],
\end{equation}
where $\mathbf{w} = \mathbf{w}_1$ and $y_i = \exp(\mathbf{w}^T\mathbf{x}_i)/(1 + \exp(\mathbf{w}^T\mathbf{x}_i))$ is the predicted probability of class 1. The probability of class 2 is $1 - y_i = 1/(1 + \exp(\mathbf{w}^T\mathbf{x}_i))$. This is the binary cross-entropy loss of Chapter~7.

For $K$ classes, we write the predicted probability from the softmax function as $y_{ik}$
\begin{equation}
y_{ik} = P(C_k|\mathbf{x}_i) = \frac{\exp(\mathbf{w}_k^T \mathbf{x}_i)}{\sum_{j=1}^{K} \exp(\mathbf{w}_j^T \mathbf{x}_i)},
\end{equation}
where $\mathbf{w}_K = \mathbf{0}$ for the reference class. The loss is then\index{cross-entropy loss!multi-class|)}
\begin{equation}
E = -\frac{1}{N}\sum_{i=1}^N \sum_{k=1}^K t_{ik} \ln y_{ik}.
\end{equation}

\section{Information Theory and Cross-Entropy}

In\index{information theory|(} the previous section we derived the loss $E$ from maximum likelihood. It is called the cross-entropy loss, and information\index{information theory} theory explains this name. A good classifier works as a compression of the data. With good weights the model predicts the class from the features, so the true class of an object can be sent in a shorter message.

We first define the mutual information. We then link it to the loss through the entropy.

\paragraph{Mutual Information}

Let\index{mutual information|(} $X$ and $Y$ be random variables with values in the sets $\mathcal{X}$ and $\mathcal{Y}$. Their mutual\index{mutual information} information $I(X;Y)$ measures how much knowing one of them tells us about the other
\begin{equation}
I(X;Y) = \sum_{x \in \mathcal{X}} \sum_{y \in \mathcal{Y}} P(X=x,Y=y) \log_2\left[\frac{P(X=x,Y=y)}{P(X=x)P(Y=y)}\right].
\end{equation}

The ratio $P(X,Y)/(P(X)P(Y))$ compares the joint probability with the product of the marginal probabilities, which is the joint probability for independent variables. For independent variables the ratio is 1, and its logarithm is 0. For dependent variables, the ratio is larger than 1 where the two variables occur together more often than independence predicts. The logarithm turns the multiplicative comparison into an additive one. The logarithm has base 2, and the example below explains its unit.

\paragraph{Mutual Information Versus Correlation}

From Chapter~3, the Pearson correlation coefficient is
\begin{equation}
\rho_{XY} = \frac{\text{Cov}(X,Y)}{\sigma_X \sigma_Y} = \frac{\mathbb{E}[(X-\mu_X)(Y-\mu_Y)]}{\sigma_X \sigma_Y}.
\end{equation}
Here $\mathbb{E}$ denotes the expectation, $\mu_X$ and $\mu_Y$ are the means, and $\sigma_X$ and $\sigma_Y$ are the standard deviations. The correlation coefficient detects only linear trends.

\begin{figure}[ht!]
    \centering
    \includegraphics[width=1.0\textwidth]{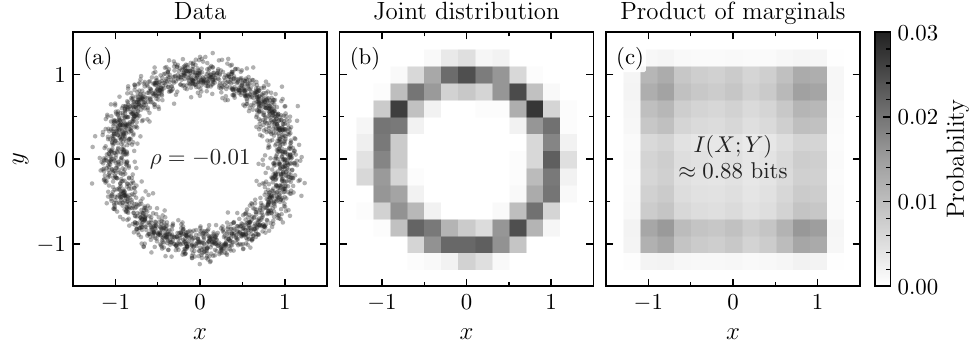}
    \caption{Correlation and mutual information for points scattered around a circle. Panel~(a) shows the sample, whose correlation is near zero. Panels~(b) and~(c) show the joint histogram $p(x,y)$ and the product of its marginals $p(x)p(y)$. Independent variables would give panel~(c), while the probability in panel~(b) lies on the ring. This difference gives the positive mutual information printed in panel~(c), which the correlation misses.}
    \label{fig:mi_vs_corr}
\end{figure}

Consider points that lie close to a circle centered at the origin. Despite the clear pattern, the correlation coefficient is near zero, because moving right sometimes increases $Y$ and sometimes decreases it, which cancels in the average. Knowing $X$ still limits $Y$ to two values near $\pm\sqrt{1-X^2}$ on the unit circle. Mutual information is positive here, because the joint distribution $p(x,y) = P(X=x,Y=y)$ differs from the product of the marginals $p(x)p(y)$. It detects any relationship in which knowing one variable tells us about the other, whatever its shape. Figure~\ref{fig:mi_vs_corr} compares the two measures.

\paragraph{An Example and the Bit}

As an example, let $Y$ be a number drawn uniformly from 1 to 16. Let $X$ tell us whether $Y$ lies in the first half (1 to 8) or the second half (9 to 16). Then $P(Y=y) = 1/16$ for each $y$ and \mbox{$P(X=\text{first}) = P(X=\text{second}) = 1/2$}. Given $X$, we have $P(Y|X=\text{first}) = 1/8$ for $y$ from 1 to 8 and 0 otherwise, and $P(Y|X=\text{second}) = 1/8$ for $y$ from 9 to 16 and 0 otherwise.

Rewriting the ratio inside the logarithm gives
\begin{align}
\frac{P(X,Y)}{P(X)P(Y)} &= \frac{P(Y|X)}{P(Y)}.
\end{align}This ratio compares the probability of $Y$ after we know $X$ with the probability before. Where $P(Y|X) > P(Y)$, the logarithm is positive and we gain information. The mutual information is
\begin{align}
I(X;Y) &= \sum_{x,y} P(X,Y) \log_2\left[\frac{P(Y|X)}{P(Y)}\right] \\*
&= 16 \cdot \frac{1}{16} \log_2\left[\frac{1/8}{1/16}\right] \\*
&= \log_2(2) = 1 \text{ bit}.
\end{align}

The ratio $(1/8)/(1/16) = 2$ means that knowing $X$ halves the number of possible values of $Y$. One bit\index{bit} is the information of a yes or no answer that halves the number of equally likely values. It is also one binary digit, 0 or 1, of a code. The base 2 of the logarithm gives the result in bits, so $X$ provides one bit of information. We still need three more bits to find which of the remaining eight values is correct, and $X$ provides only the first bit. The mutual information therefore counts, on average, the yes or no questions that knowing one variable saves when we specify the other.

The factor $P(X,Y)$ in the sum averages this information gain over all combinations of $X$ and $Y$ according to their joint distribution. Only 16 of the 32 combinations occur, and each has probability $1/16$, so the average is the same one bit.

\paragraph{Mutual Information in Classification}

In classification, $X$ stands for the input features of a galaxy, such as its concentration index or its asymmetry. The class $\mathit{C}$, for example elliptical, spiral or irregular, takes the place of $Y$. The mutual information $I(\mathbf{x};\mathit{C})$ then measures how much the features tell us about the class.

We rewrite the mutual information with conditional probabilities
\begin{equation}
I(\mathbf{x};\mathit{C}) = \sum_{\mathbf{x},\mathit{C}} p(\mathbf{x},\mathit{C}) \ln\left[\frac{p(\mathit{C}|\mathbf{x})}{p(\mathit{C})}\right].
\end{equation}

Here $p(\mathit{C}|\mathbf{x})$ is the probability of class $\mathit{C}$ given the features $\mathbf{x}$. In our classifier, the softmax function computes it as $P(C_k|\mathbf{x})$. From here on we replace $p(\mathit{C}|\mathbf{x})$ by the model probability, and we take $p(\mathbf{x},\mathit{C})$ and $p(\mathit{C})$ from the training data. Then $I(\mathbf{x};\mathit{C})$ is the information that the model has about the class. The base of the logarithm sets the unit, which is bits for base 2 and nats for base $e$. The loss uses base $e$, so we write $\ln$ below and $\log_2$ when we count bits.\index{mutual information|)}

\paragraph{Entropy}

Entropy\index{entropy}\index{entropy|(} measures uncertainty. A code writes each value as a string of bits, and a message is a sequence of such strings. The entropy is the smallest average number of bits per value that a code can reach, when the logarithm has base 2. For a variable $\mathit{C}$ with distribution $p(\mathit{C})$, the entropy in nats is
\begin{equation}
H(\mathit{C}) = -\sum_\mathit{C} p(\mathit{C}) \ln p(\mathit{C}).
\end{equation}
Dividing by $\ln 2$ gives the entropy in bits.

Imagine that we encode the galaxy types of a catalog. If elliptical, spiral and irregular galaxies appear with equal frequency, then $p(\mathit{C}) = 1/3$ for each type, and with the base-2 logarithm
\begin{equation}
H(\mathit{C}) = -3 \cdot \frac{1}{3} \log_2 \frac{1}{3} = \log_2 3 \approx 1.58 \text{ bits}.
\end{equation}We therefore need at least 1.58 bits per galaxy on average, which we approach when we code many galaxies together. A code with equal length for all types needs 2 bits, and the codes 0, 10 and 11 already average 1.67 bits. In these codes no string is the beginning of another string, so a reader can split a message into types without separators.

Entropy is smallest when one class dominates, because most objects then belong to the same class and few bits suffice on average. It is largest when the classes are balanced, because no class is more likely than the others. If the types have unequal frequencies, short codes for common types and long codes for rare types lower the average length. In English text, the letter e is much more common than z, so e gets a shorter code than z. If spiral galaxies are more common than irregular galaxies, for example, the entropy falls below 1.58 bits.\index{entropy|)}

\paragraph{Cross-Entropy as Code Length}

The cross-entropy loss averaged over $N$ training examples is
\begin{equation}
E(\{\mathbf{w}_k\}) = -\sum_{i=1}^N \sum_{k=1}^K \frac{t_{ik}}{N} \ln P(C_k|\mathbf{x}_i).
\end{equation}
The term $t_{ik}/N$ is the weight that example $i$ gives to class $k$ in the empirical\index{empirical distribution} distribution of the training data. This distribution gives each training example the probability $1/N$ (Section~3.7). The sum over examples is therefore an expectation over the empirical joint distribution $p(\mathbf{x},\mathit{C})$ of the training data. Written as a sum over pairs of features and classes, the same loss reads
\begin{equation}
E(\{\mathbf{w}_k\}) = -\sum_{\mathbf{x},\mathit{C}} p(\mathbf{x},\mathit{C}) \ln p(\mathit{C}|\mathbf{x}).
\end{equation}

This loss is called a cross-entropy. To see why, imagine that we send the true class of each object to a receiver who already knows the features. The receiver can use our model probabilities to design an ideal code, and a class with probability $q$ then needs $-\log_2 q$ bits. The average length of these messages is the cross-entropy loss, in bits when the logarithm has base 2. A better model gives shorter messages, which saves bits compared with a code that ignores the features. The code lengths $-\ln p(\mathit{C}|\mathbf{x})$ come from the model and the average comes from the data, which gives the name cross-entropy.

\paragraph{Loss, Entropy and Mutual Information}

Expanding the mutual information gives
\begin{align}
I(\mathbf{x};\mathit{C}) &= \sum_{\mathbf{x},\mathit{C}} p(\mathbf{x},\mathit{C}) \ln p(\mathit{C}|\mathbf{x}) - \sum_{\mathbf{x},\mathit{C}} p(\mathbf{x},\mathit{C}) \ln p(\mathit{C}) \\*
&= -E(\{\mathbf{w}_k\}) - \sum_{\mathit{C}} p(\mathit{C}) \ln p(\mathit{C}).
\end{align}
The first sum is $-E$, as the sum over pairs of features and classes above shows. In the second sum, $\sum_{\mathbf{x}} p(\mathbf{x},\mathit{C}) = p(\mathit{C})$, so the second term is the entropy\index{entropy}\index{mutual information} $H(\mathit{C})$ defined above. It depends only on the class frequencies of the data set. The loss is therefore the entropy minus the information that the model has about the class
\begin{equation}
E(\{\mathbf{w}_k\}) = H(\mathit{C}) - I(\mathbf{x};\mathit{C}).
\end{equation}

In coding terms, $H(\mathit{C})$ is the average length of the best code that ignores the features. The code built from the model probabilities has average length $E$, so $I(\mathbf{x};\mathit{C})$ is the saving. Because the entropy does not depend on the model parameters, minimizing the cross-entropy loss maximizes the information that the model has about the class.

Divide the training examples into groups with the same features. The best code for a group uses the fractions of its examples in each class, and no model code is shorter on average. The loss of any model is therefore at least the average uncertainty about the class that remains given the features. This quantity is the conditional\index{conditional entropy} entropy $H(\mathit{C}|\mathbf{x})$, the entropy of the class within a group, averaged over the examples. Subtracting it from $H(\mathit{C})$ gives the mutual information of the training data, which we write $I_{\text{data}}(\mathbf{x};\mathit{C})$.

The information $I(\mathbf{x};\mathit{C})$ that the model has about the class is therefore at most $I_{\text{data}}(\mathbf{x};\mathit{C})$. The two are equal when the model probabilities equal the fractions of the training examples in each class for the same features. In training, we find the combination of input features that gives the model the most information about the class.

Feature combinations that work well for classification have high mutual information with the class labels. The concentration index measures how concentrated the light of a galaxy is toward its center. It is such a feature for galaxy classification, because it tells us a lot about the morphological class. A feature such as the absolute sky position typically has low mutual information with morphology.

\looseness=1 The entropy term bounds the information about the class that the model can learn. For a perfect prediction, $P(C_k|\mathbf{x}_i) = 1$ whenever $t_{ik} = 1$, so the model probabilities equal the one-hot labels and the loss is $E = 0$. Then $H(\mathit{C}) = I(\mathbf{x};\mathit{C})$, which means the model has extracted all the information that the features hold about the class. Optimizing the cross-entropy therefore brings $I(\mathbf{x};\mathit{C})$ as close as possible to $H(\mathit{C})$. The remaining gap equals the loss, which is the uncertainty about the class that the model leaves.

In our classifier, the entropy term comes from the empirical class distribution of the training data and equals
\begin{equation}
H(\mathit{C}) = -\sum_{k=1}^{K} \frac{N_k}{N} \ln \frac{N_k}{N}.
\end{equation}
Here $N_k$ is the number of examples of class $k$ among the $N$ examples. Neither better features nor better weights can change this term.

\looseness=1 When the loss approaches zero, a receiver who knows the features needs almost no extra bits to recover the class. A large loss suggests room for improvement, either in the choice of features or in how we combine them. If the classes overlap in feature space, however, this model keeps a loss above zero, because the features do not determine the class.\index{information theory|)}

\section{Gradient Descent for Multi-Class Classification}\label{sec:ch8_gd}

To\index{gradient descent|(} minimize the loss with gradient descent, we need its gradient with respect to each weight vector $\mathbf{w}_j$ for $j = 1, 2, \ldots, K-1$. This is harder than in the binary case. A change of one weight vector changes the probabilities of all classes through the normalization term of the softmax function.

The loss is convex in the weights, as in the binary case of Chapter~7. For an example whose true class is $c$, the loss term is $-\ln y_{ic} = -\mathbf{w}_c^T\mathbf{x}_i + \ln\sum_{l=1}^K \exp(\mathbf{w}_l^T\mathbf{x}_i)$. This is a linear function of the weights plus a log-sum-exp function of linear functions of the weights. The log-sum-exp function is the logarithm of a sum of exponentials, and it is a standard convex function. Sums of convex functions are convex, and the loss averages such terms over the examples. A minimum that gradient descent finds is therefore a global minimum.

\paragraph{Deriving the Gradient}

By the chain rule, the gradient with respect to $\mathbf{w}_j$ is
\begin{equation}
\frac{\partial E}{\partial \mathbf{w}_j} = -\frac{1}{N} \sum_{i=1}^N \sum_{k=1}^K t_{ik} \frac{\partial \ln y_{ik}}{\partial \mathbf{w}_j}.
\end{equation}
We must find how a change of $\mathbf{w}_j$ affects each $y_{ik}$. There are two cases, the probability $y_{ij}$ of class $j$ itself and the probabilities $y_{ik}$ of the other classes $k \neq j$.

We start with the derivatives of $y_{ik}$ with respect to the logit $z_{ij} = \mathbf{w}_j^T\mathbf{x}_i$. The logit is the linear prediction of class $j$ before the softmax function, and for two classes it is the log-odds of Chapter~7. The reference class has $z_{iK} = 0$. For $k = j$, both the numerator and the denominator of $y_{ij} = e^{z_{ij}}/\sum_{l=1}^{K} e^{z_{il}}$ depend on $z_{ij}$. The derivative of the denominator is $e^{z_{ij}}$, because only its term $l = j$ depends on $z_{ij}$. The quotient rule $(f/g)' = (f'g - fg')/g^2$ with $f = e^{z_{ij}}$ and $g = \sum_{l=1}^{K} e^{z_{il}}$ gives
\begin{align}
\frac{\partial y_{ij}}{\partial z_{ij}} &= \frac{e^{z_{ij}} \cdot \sum_{l=1}^{K} e^{z_{il}} - e^{z_{ij}} \cdot e^{z_{ij}}}{(\sum_{l=1}^{K} e^{z_{il}})^2} \\*
&= \frac{e^{z_{ij}}}{\sum_{l=1}^{K} e^{z_{il}}} \cdot \frac{\sum_{l=1}^{K} e^{z_{il}} - e^{z_{ij}}}{\sum_{l=1}^{K} e^{z_{il}}} \\*
&= y_{ij} \cdot (1 - y_{ij}).
\end{align}

For $k \neq j$, only the denominator of $y_{ik}$ depends on $z_{ij}$, and we get
\begin{align}
\frac{\partial y_{ik}}{\partial z_{ij}} &= \frac{\partial}{\partial z_{ij}} \frac{e^{z_{ik}}}{\sum_{l=1}^{K} e^{z_{il}}} \\*
&= e^{z_{ik}} \cdot \frac{-e^{z_{ij}}}{(\sum_{l=1}^{K} e^{z_{il}})^2} \\*
&= -y_{ik} \cdot y_{ij}.
\end{align}
Now we need $\partial \ln y_{ik}/\partial \mathbf{w}_j$. The probability $y_{ik}$ depends on the logits $z_{il}$ of the $K-1$ free classes, so the chain rule gives
\begin{align}
\frac{\partial \ln y_{ik}}{\partial \mathbf{w}_j} &= \sum_{l=1}^{K-1} \frac{\partial \ln y_{ik}}{\partial z_{il}} \cdot \frac{\partial z_{il}}{\partial \mathbf{w}_j}.
\end{align}Only one term of this sum is nonzero, because $\partial z_{il}/\partial \mathbf{w}_j = \mathbf{x}_i$ for $l = j$ and 0 for $l \neq j$. The sum therefore reduces to
\begin{align}
\frac{\partial \ln y_{ik}}{\partial \mathbf{w}_j} &= \frac{\partial \ln y_{ik}}{\partial z_{ij}} \cdot \mathbf{x}_i.
\end{align}

For $k = j$, we insert $\partial y_{ij}/\partial z_{ij} = y_{ij}(1-y_{ij})$ and get
\begin{align}
\frac{\partial \ln y_{ij}}{\partial \mathbf{w}_j} &= \frac{1}{y_{ij}} \cdot \frac{\partial y_{ij}}{\partial z_{ij}} \cdot \mathbf{x}_i \\*
&= (1 - y_{ij}) \cdot \mathbf{x}_i.
\end{align}

For $k \neq j$, we insert $\partial y_{ik}/\partial z_{ij} = -y_{ik}\,y_{ij}$ and get
\begin{align}
\frac{\partial \ln y_{ik}}{\partial \mathbf{w}_j} &= \frac{1}{y_{ik}} \cdot \frac{\partial y_{ik}}{\partial z_{ij}} \cdot \mathbf{x}_i \\*
&= -y_{ij} \cdot \mathbf{x}_i.
\end{align}

Substituting these results into the gradient gives
\begin{align}
\frac{\partial E}{\partial \mathbf{w}_j} &= -\frac{1}{N} \sum_{i=1}^N \left[ t_{ij} \cdot (1 - y_{ij}) \cdot \mathbf{x}_i - \sum_{k \neq j} t_{ik} \cdot y_{ij} \cdot \mathbf{x}_i \right].
\end{align}The factors $y_{ij}$ and $\mathbf{x}_i$ do not depend on $k$, so they come out of the sum over $k \neq j$. The label vector is one-hot, so $\sum_{k=1}^K t_{ik} = 1$ and $\sum_{k \neq j} t_{ik} = 1 - t_{ij}$. Therefore
\begin{align}
\frac{\partial E}{\partial \mathbf{w}_j} &= -\frac{1}{N} \sum_{i=1}^N \left[ t_{ij} - t_{ij} \cdot y_{ij} - y_{ij} \cdot (1 - t_{ij}) \right] \mathbf{x}_i \\*
&= -\frac{1}{N} \sum_{i=1}^N \left[ t_{ij} - t_{ij} \cdot y_{ij} - y_{ij} + t_{ij} \cdot y_{ij} \right] \mathbf{x}_i \\*
&= -\frac{1}{N} \sum_{i=1}^N (t_{ij} - y_{ij}) \mathbf{x}_i.
\end{align}

The gradient for each weight vector is therefore
\begin{equation}
\frac{\partial E}{\partial \mathbf{w}_j} = \frac{1}{N} \sum_{i=1}^N (y_{ij} - t_{ij}) \mathbf{x}_i.
\end{equation}Each example contributes the difference between the predicted probability $y_{ij}$ and the true label $t_{ij}$ of class $j$, scaled by $\mathbf{x}_i$. This has the same form as in the binary case, but now we update $K-1$ weight vectors, one for each non-reference class.

\looseness=1 In binary classification, a misclassified example moves one decision boundary in one direction. In multi-class classification, a single misclassified example moves several weight vectors at once. Some weight vectors move toward the features of the example and some away from them, in a coordinated way.

\looseness=1 Consider a galaxy of true class $j \neq K$ that receives a high probability for another class $k \neq K$. For the true class, $y_{ij} - t_{ij} = y_{ij} - 1$ is negative, because $y_{ij} < 1$. The update multiplies the gradient by $-\eta$, where $\eta$ is the learning rate, so it adds a positive multiple of $\mathbf{x}_i$ to $\mathbf{w}_j$. This moves $\mathbf{w}_j$ toward the features of the galaxy. For the wrong class $k$, $y_{ik} - t_{ik} = y_{ik} > 0$, so the update subtracts a multiple of $\mathbf{x}_i$ from $\mathbf{w}_k$.

The size of the corrections depends on how the probability is distributed over the classes. As in binary classification, high-confidence mistakes cause the largest corrections. If $y_{ij} \approx 0$ for the true class, the update $|y_{ij} - 1| \approx 1$ is at its maximum. If the missing probability belongs to one wrong class $k \neq K$ with $y_{ik} \approx 1$, then $|y_{ik} - 0| \approx 1$ as well, and both weight vectors receive large updates. If the probability is spread over many wrong classes with small $y_{ik}$, each update is smaller, although their sum remains large. The reference class receives no update, because $\mathbf{w}_K$ is fixed.

\paragraph{Extending SGD to Multi-Class Problems}

SGD\index{stochastic gradient descent (SGD)} works as in the binary case. We approximate the full gradient with a mini-batch of $M$ examples and update the parameters iteratively.

For $K$ classes, the update rule is
\begin{equation}
\mathbf{w}_j^{(\tau+1)} = \mathbf{w}_j^{(\tau)} - \eta \cdot \frac{1}{M}\sum_{i=1}^M (y_{ij}^{(\tau)} - t_{ij})\mathbf{x}_i \quad \text{for } j = 1, 2, \ldots, K-1,
\end{equation}where $\tau$ counts the iterations. We update the $K-1$ weight vectors together.

Most techniques of the binary case carry over with little change, such as mini-batches, learning rate schedules, adaptive learning rates and regularization. Adaptive learning rates set a separate step size for each parameter (Section~15.3). Weight decay with strength $\lambda$ adds the penalty $\lambda\sum_{j=1}^{K-1}\|\mathbf{w}_j\|^2$ to the loss. It adds the term $2\lambda\mathbf{w}_j$ to the gradient of each weight vector, as in Chapter\index{gradient descent|)}~7.

\section{Implementation Considerations}

\paragraph{Numerical Stability} The softmax function uses exponentials, which can overflow\index{numerical stability} or underflow. The multi-class case computes $K-1$ exponentials of free logits where the binary case computes one, and a single large logit $\mathbf{w}_j^T\mathbf{x}_i$ can overflow the sum.

A standard remedy subtracts the largest logit before the exponentiation
\begin{equation}
y_{ik} = \frac{\exp(\mathbf{w}_k^T \mathbf{x}_i - \max_j \mathbf{w}_j^T \mathbf{x}_i)}{\sum_{j=1}^K \exp(\mathbf{w}_j^T \mathbf{x}_i - \max_j \mathbf{w}_j^T \mathbf{x}_i)}.
\end{equation}This leaves the probabilities unchanged, because the factor $\exp(-\max_j \mathbf{w}_j^T\mathbf{x}_i)$ appears in the numerator and the denominator and cancels. The largest exponent is now 0, so the exponentials are at most 1 and the denominator is at least 1.

\paragraph{Parameter Initialization} We initialize each free weight vector with small values, for example samples from a normal or a uniform distribution. Both distributions are easy to sample in any dimension, also for the many features of astronomical data. Small values keep the logits near zero, so all classes start with probabilities near $1/K$.

Random values are common, because weight vectors that start equal and receive equal gradients would stay equal. This model does not need them. In general, the different labels give each weight vector a different gradient, so even zero initial\index{parameter initialization} values work. The loss is convex, so training reaches the same minimum loss from any initial values, for a suitable learning rate. This holds when the loss has a minimum at finite weights. It has no minimum when a hyperplane separates one class from the others, because the loss then keeps falling as the weights grow (Chapter~7).

The initialization changes only the speed and the path of the optimization. Random restarts, which train several models from different starting points and keep the best one on a validation set, are therefore unnecessary for this model.

\paragraph{Matrix Formulation for Efficiency} For an efficient implementation, we organize the weight vectors as the rows of a matrix $\mathbf{W} \in \mathbb{R}^{K \times D}$. Here $D$ is the feature dimension including the constant 1 for the bias, and the last row $\mathbf{w}_K$ is fixed to zero. This allows vectorized computation, which replaces loops by matrix operations, for all classes and examples at once.

The data matrix $\mathbf{X} \in \mathbb{R}^{N \times D}$ holds $N$ examples with $D$ features each. The product $\mathbf{Z} = \mathbf{X}\mathbf{W}^T \in \mathbb{R}^{N \times K}$ gives all logits $z_{ij}$ of all examples in one operation, and the last column is zero, $z_{iK} = 0$, because $\mathbf{w}_K = \mathbf{0}$. We apply the softmax function to each row of $\mathbf{Z}$, after subtracting the largest logit of the row for numerical stability. This gives all class probabilities $y_{ik}$.

The matrix form suits mini-batches. For a mini-batch $\mathbf{X}_{\text{batch}} \in \mathbb{R}^{M \times D}$, the logits are $\mathbf{Z}_{\text{batch}} = \mathbf{X}_{\text{batch}}\mathbf{W}^T \in \mathbb{R}^{M \times K}$. Linear algebra libraries are optimized for such operations and run them in parallel, which can be orders of magnitude faster than loops.

The gradient also takes a matrix form. Let $\mathbf{Y} \in \mathbb{R}^{N \times K}$ contain the predicted probabilities and $\mathbf{T} \in \mathbb{R}^{N \times K}$ the one-hot labels. Then
\begin{equation}
    \nabla_{\mathbf{W}_{1:K-1}} E = \frac{1}{N}(\mathbf{Y}_{1:K-1} - \mathbf{T}_{1:K-1})^T \mathbf{X},
\end{equation}
where $\mathbf{W}_{1:K-1}$ denotes the first $K-1$ rows of $\mathbf{W}$, and $\mathbf{Y}_{1:K-1}$ and $\mathbf{T}_{1:K-1}$ denote the first $K-1$ columns of $\mathbf{Y}$ and $\mathbf{T}$. The rows of this gradient matrix are the gradients $\partial E/\partial \mathbf{w}_j$ for $j = 1, 2, \ldots, K-1$.

We fitted the weights by maximum likelihood, which gives one set of weights and one probability for each class. The probabilities describe how confident the model is for these weights, but they do not say how uncertain the weights are. Chapter~9 treats this second uncertainty with Bayesian methods for binary logistic regression. The non-Gaussian likelihood of the sigmoid function needs techniques beyond those of regression.\index{classification!multi-class|)}

\section{Tutorial: Classifying Strong Lensing Images}\label{sec:ch8_tutorial}

This\index{classification!multi-class|(}\index{strong gravitational lensing|(} tutorial applies the softmax function and the cross-entropy loss of this chapter to objects in deep images from the James\index{James Webb Space Telescope} Webb Space Telescope. It sorts them into classes, so that strong lensing arcs can be found among them. The full tutorial can be found at \url{https://statml.org/} as tutorial 8.

A massive galaxy or galaxy cluster is a lens\index{strong gravitational lensing} that deflects the light of galaxies behind it, as general relativity predicts. When a background galaxy lies almost behind the lens, it appears as a stretched arc or as several images. Such a strong lens can magnify early-universe galaxies, and the arcs of a cluster measure its mass, mostly dark matter. A survey has too many objects to check by eye, so a classifier can sort them.

\begin{figure}[t!]
    \centering
    \includegraphics[width=0.95\textwidth]{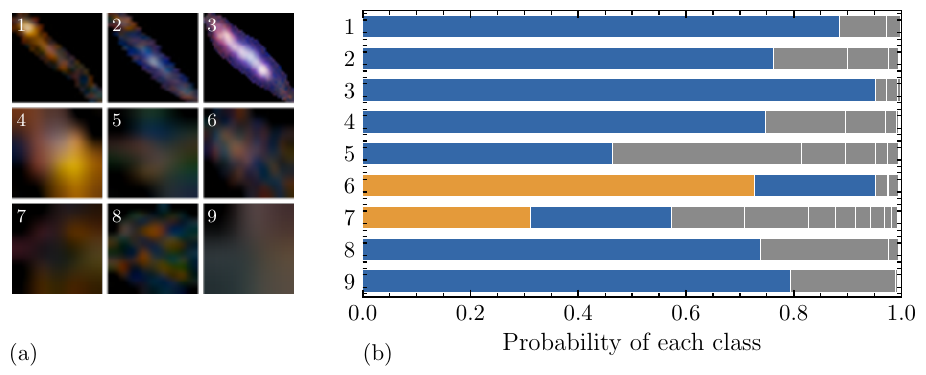}
    \caption{The tutorial classifies objects in James Webb Space Telescope images. Panel~(a) shows nine selected test objects, three arc-like ones on top. Panel~(b) stacks the class probabilities of each object, largest first. Blue is the true class, orange a wrong class with the largest probability, grey the rest. In wrong objects 6 and 7 the true class is second.}
    \label{fig:ch8_tutorial_classes}
\end{figure}

In Section~\ref{sec:ch7_tutorial} one probability separated two classes of stars. The objects in deep images form many classes of look-alikes, such as arcs and blurry diffuse ones (\mbox{Figure~\ref{fig:ch8_tutorial_classes}a}). One probability is not enough for three or more classes, and separate \mbox{One-vs-Rest} classifiers can claim an object twice or leave it unclaimed (Section~\ref{sec:ch8_ovr}). One model must therefore treat the classes together. The softmax function turns one linear prediction for each class into probabilities that add up to one (Section~\ref{sec:ch8_softmax}).

The probabilities need features and targets. A large pre-trained model supplies the features of each object (Section~\ref{sec:ch15_pretrained}), and one-hot labels name the true class without an order (Section~\ref{sec:ch8_onehot}). Maximum likelihood then gives the cross-entropy loss, which is small when the true class has a high probability (Section~\ref{sec:ch8_loss}). Gradient descent minimizes it and so raises that probability, and confident mistakes change the weights most (Section~\ref{sec:ch8_gd}).

We check the classifier on test objects, since it is meant for unlabeled ones. The largest probability usually lies on the true class, as for most of the nine objects in Figure~\ref{fig:ch8_tutorial_classes}b. When the model errs, the true class usually has the second largest probability. A survey can add the probabilities of the arc classes and check the objects with a high sum.\index{classification!multi-class|)}\index{strong gravitational lensing|)}

\paragraph{Further Reading} Dietterich and Bakiri (1991, 1995) developed error-correcting output codes, which give each class a code word of binary classifier outputs. Allwein et~al.\ (2000) unified the \mbox{One-vs-Rest} and \mbox{One-vs-One} reductions to binary classifiers. F\"urnkranz (2002) analyzed round robin classification, with one classifier for each pair of classes. Luce (1959) proposed a model of choice probabilities that leads to the softmax function. McFadden (1974) connected it to choice models in economics and to multinomial logistic regression, and Theil (1969) extended the logit model. Bridle (1990) named the softmax function. Dennis and Schnabel (1983) developed numerical methods for unconstrained optimization. Polyak and Juditsky (1992) showed that averaging the parameter values of successive steps speeds up stochastic gradient methods.

\makeatletter\@clubpenalty=150\makeatother \clubpenalty=150 \widowpenalty=150

\chapter{Bayesian Logistic Regression}

Maximum likelihood finds one set of weights, which gives one decision boundary. For two classes, such as stars\index{star-galaxy classification} and galaxies, this boundary is a straight line or hyperplane in the space of the photometric and morphological features. When two groups in the training data overlap little, many boundaries classify the training data almost equally well. A fit returns only one of them, and its predictions become more confident with distance from that boundary. An unusual object far from the training data can lie far from the boundary and receive a confident label that the data do not support. In astronomy, a confident error on such an object can hide a discovery.

The fitted model also gives a probability for each class. This probability tells us how uncertain the class of an object is for the fitted weights, for example near the boundary. It does not show how uncertain the weights are, because it is computed at one set of weights. The uncertainty of the weights is the topic of this chapter.

Two properties of astronomical data make the problem worse. Labeled data are scarce, because spectroscopic classifications and detailed morphologies need expensive telescope time and expert analysis. The labeled objects are also a biased sample, because astronomers tend to label objects that are easy to classify. Transitional and unusual objects are therefore poorly covered.

\begin{figure}[ht!]
    \centering
    \includegraphics[width=0.95\textwidth]{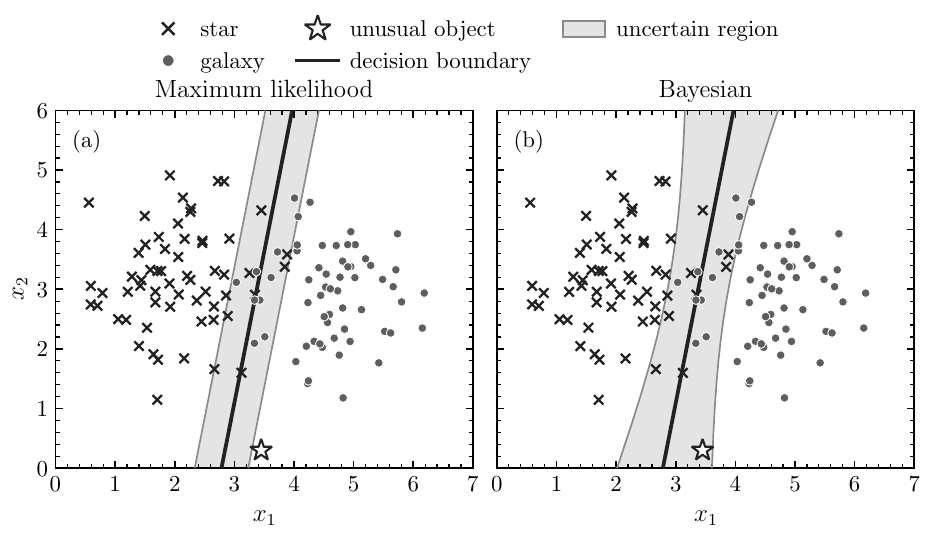}
    \caption{Maximum likelihood and Bayesian logistic regression for two classes, for example stars and galaxies, that overlap slightly in the features $x_1$ and $x_2$. In the shaded region the probability of a galaxy lies between 0.1 and 0.9, so the class is uncertain. Outside it the class is confident. In panel~(b) this region widens away from the data and contains the unusual object, which lies outside the region in panel~(a).}
    \label{fig:bayesian_logistic_regression}
\end{figure}

We want a classifier that reports lower confidence when the data cannot support a confident label. For an unusual object, its probability should then lie closer to 0.5 than the maximum likelihood probability. Bayesian\index{Bayesian logistic regression} logistic regression does this, and Figure~\ref{fig:bayesian_logistic_regression} compares it with maximum likelihood for an unusual object.

Instead of finding a single decision boundary, we characterize our uncertainty about where the boundary should be placed. For this we use the posterior distribution of the weights given the data, as in Bayesian linear regression. In Chapter~5, the posterior $p(\mathbf{w}|\mathcal{D})$ given the data $\mathcal{D}$ made predictions more uncertain far from the training data, and the same happens in classification. Boundaries sampled from the posterior, or from its Gaussian approximation below, show how much predictions vary in different regions of the feature space.

The likelihood contains the sigmoid of a linear combination of the features, so the posterior is not Gaussian. The sigmoid breaks the conjugacy of Bayesian linear regression (Chapter~5). There a Gaussian prior and a Gaussian likelihood gave a Gaussian posterior in closed form. Two approximations give a closed-form prediction. The Laplace approximation replaces the posterior by a Gaussian centered at its mode, the maximum a posteriori (MAP) estimate. The probit approximation then replaces the sigmoid by a function whose integral against a Gaussian has a closed form.

The result is a classifier that reports its confidence along with its prediction. In astronomical surveys, knowing when a model is uncertain can be as important as the classification itself.

\section{The Bayesian Logistic Model}\label{sec:blr_model}

\looseness=1 In\index{Bayesian logistic regression|(} this chapter we treat only binary classification, which simplifies the mathematics. The multi-class model of Chapter~8 can use the sampling methods of Chapters~12 and~13.

Binary logistic regression models the probability of class $C_1$ for a feature vector $\mathbf{x}$ with the sigmoid function. A constant 1 appended to $\mathbf{x}$ makes the bias a component of the weight vector $\mathbf{w}$. The probability is
\begin{equation}
P(C_1|\mathbf{x}, \mathbf{w}) = y = \sigma(\mathbf{w}^T \mathbf{x}) = \frac{1}{1 + e^{-\mathbf{w}^T \mathbf{x}}}.
\end{equation}
The probability of class $C_0$ is $1 - y$.

The likelihood of the labels is a product of Bernoulli probabilities
\begin{equation}
p(\mathbf{t} | \mathbf{w}, \mathbf{X}) = \prod^N_{i=1} y_i^{t_i} (1 - y_i)^{(1-t_i)}.
\end{equation}
Here $N$ is the number of data points and $\mathbf{x}_i$ is the feature vector of data point $i$. Its label $t_i$ is 0 or 1 for the two classes, for example 0 for a star and 1 for a galaxy. The vector $\mathbf{t}$ collects the labels and the matrix $\mathbf{X}$ the feature vectors, so we write $\mathcal{D} = (\mathbf{t}, \mathbf{X})$ for the data. The prediction $y_i = \sigma(\mathbf{w}^T \mathbf{x}_i)$ is the probability of class $C_1$. Maximum likelihood finds the weights that maximize this likelihood, usually by minimizing the cross-entropy loss with gradient descent (Chapter~7).

\paragraph{The Bayesian Perspective}

A finite data set is consistent with several decision boundaries. Some fit the training data slightly better than others, but many are plausible given the data. The Bayesian approach keeps these boundaries and gives each a probability given the data and the prior.

For example, suppose the training data contain two well-separated clusters. Many lines separate them without error. For such a line, larger weights make the sigmoid steeper and push each predicted probability closer to the correct label. The likelihood therefore keeps increasing, and no finite weights maximize it. A penalty on large weights or early stopping, both from Chapter~7, selects one boundary. The Bayesian approach instead averages over the plausible boundaries, and its prior keeps the weights finite.

Uncertainty about the weights becomes uncertainty about the predictions. To classify a new object, we compare the plausible boundaries. If they agree, we can be confident. If they disagree, the prediction is uncertain. Bayesian inference has three steps. We state a prior for the weights, update it with the observed data and use the resulting posterior distribution for predictions.

\paragraph{Prior and Posterior}

We start with a prior that states our beliefs about the weights before we see the data. A common choice is a Gaussian prior
\begin{equation}
p(\mathbf{w}) = \mathcal{N}(\mathbf{w} | \mathbf{m}_0, \mathbf{S}_0).
\end{equation}
Here $\mathbf{m}_0$ is the prior mean, our best guess of the weights. The matrix $\mathbf{S}_0$ is the prior covariance, which describes our uncertainty.

This choice is mathematically convenient and often reasonable in practice. With no strong prior knowledge about which features matter, we can set $\mathbf{m}_0 = \mathbf{0}$, which expresses no preference for any particular decision boundary orientation. The matrix $\mathbf{S}_0$ then controls our confidence in this neutrality. Larger diagonal elements of $\mathbf{S}_0$ mean greater prior uncertainty about the corresponding weights.

By Bayes' theorem, the posterior is proportional to the prior times the likelihood
\begin{equation}
p(\mathbf{w} | \mathbf{t}, \mathbf{X}) \propto p(\mathbf{w}) \cdot p(\mathbf{t} | \mathbf{w}, \mathbf{X}).
\end{equation}
Inserting the prior and the likelihood gives
\begin{equation}
p(\mathbf{w} | \mathbf{t}, \mathbf{X}) \propto \mathcal{N}(\mathbf{w} | \mathbf{m}_0, \mathbf{S}_0) \cdot \prod^N_{i=1} y_i^{t_i} (1 - y_i)^{(1-t_i)}.
\end{equation}

Here we meet the first major challenge. The sigmoid makes the posterior non-Gaussian. This is the price of moving from regression, where outputs can be any real number, to classification, where outputs must lie between 0 and 1. The Gaussian prior is not conjugate to the sigmoid likelihood. We cannot write down a simple formula for it. We also cannot draw independent samples from it directly or compute predictions in closed form.

\paragraph{Why the Posterior Is Non-Gaussian}

To see why, we compare the mathematical forms. The closed form of Chapter~5 depends on the exponent being quadratic in $\mathbf{w}$. In Bayesian linear regression, the posterior was proportional to
\begin{equation}
p(\mathbf{w} | \mathbf{t}, \mathbf{X}) \propto \exp\left(-\frac{1}{2}(\mathbf{w}-\mathbf{m}_0)^T\mathbf{S}_0^{-1}(\mathbf{w}-\mathbf{m}_0)\right) \cdot \exp\left(-\frac{1}{2\sigma^2}(\mathbf{t}-\mathbf{X}\mathbf{w})^T(\mathbf{t}-\mathbf{X}\mathbf{w})\right).
\end{equation}
Here $\sigma^2$ is the noise variance of Chapter~5, and $\mathbf{X}$ is the design matrix, called $\boldsymbol{\Phi}$ in Chapter~5. Both factors are exponentials of quadratic forms in $\mathbf{w}$, so their product is also the exponential of a quadratic form, which is another Gaussian.

In logistic regression, the posterior is proportional to
\begin{equation}
p(\mathbf{w} | \mathbf{t}, \mathbf{X}) \propto \exp\left(-\frac{1}{2}(\mathbf{w}-\mathbf{m}_0)^T\mathbf{S}_0^{-1}(\mathbf{w}-\mathbf{m}_0)\right) \cdot \prod_{i=1}^N \sigma(\mathbf{w}^T\mathbf{x}_i)^{t_i}(1-\sigma(\mathbf{w}^T\mathbf{x}_i))^{1-t_i}.
\end{equation}
The first factor is still the exponential of a quadratic form. The second is a product of sigmoid functions. Its logarithm is a sum of terms with $\ln\sigma(\mathbf{w}^T\mathbf{x}_i)$ and $\ln(1-\sigma(\mathbf{w}^T\mathbf{x}_i))$, which are not quadratic in $\mathbf{w}$, so the posterior is not Gaussian.

\section{Intuition for the Laplace Approximation}\label{sec:ch9_laplace_intuition}

\looseness=1 Without\index{Laplace approximation|(} a closed form, we approximate the posterior by a Gaussian, which is easier to handle. Near its peak, a smooth probability distribution looks like a Gaussian, and the Laplace\index{Laplace approximation} approximation uses this fact. It becomes more accurate as the data set grows.

Consider the posterior $p(\mathbf{w}|\mathbf{t}, \mathbf{X})$ with a peak at $\mathbf{w}_0$. If we zoom in on the peak, the logarithm of the distribution looks like a parabola.

A Gaussian is fixed by its mean and covariance. In the Laplace approximation, the mean is the mode of the posterior, the weights at which the posterior density is largest. The covariance follows from the curvature at the peak, which measures how quickly the density falls away from the mode.

\begin{figure}[ht!]
    \centering
    \includegraphics[width=0.95\textwidth]{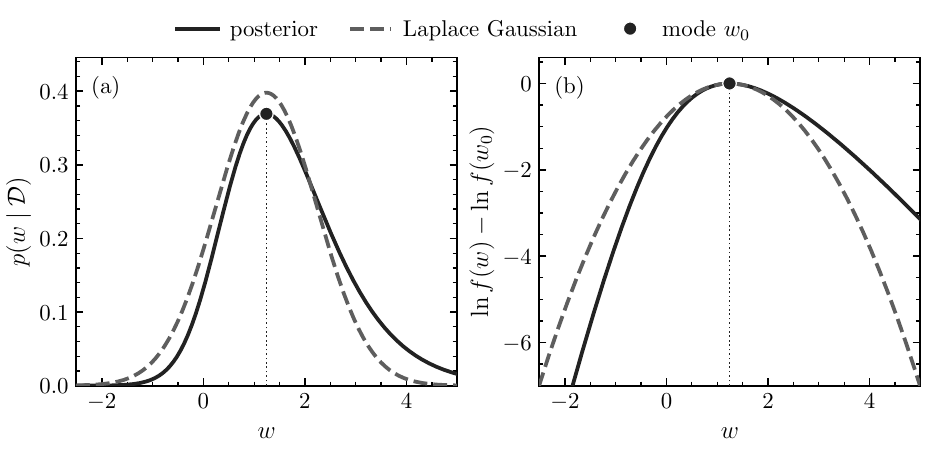}
    \caption{The Laplace approximation of the posterior of one weight $w$. Panel~(a) shows the posterior density and the Gaussian with mean at the mode $w_0$. Panel~(b) shows the logarithms of both, relative to their values at $w_0$. The logarithm of the Gaussian is a parabola that agrees with the logarithm of the posterior in value, slope and curvature at $w_0$.}
    \label{fig:laplace_approximation}
\end{figure}

\looseness=-1 The curvature is the negative second derivative of the log posterior with respect to the weights. A sharper peak has a larger curvature and gives a narrower Gaussian. Figure~\ref{fig:laplace_approximation} shows the posterior of a model with one weight and the Gaussian from its mode and curvature.

The approximation works well under two conditions. The distribution has a single dominant mode, and most of its probability mass lies near that mode.

For Bayesian logistic regression with a Gaussian prior, both conditions are typically satisfied, especially as the data set grows. Well-separated clusters are the exception, as in the example of Section~\ref{sec:blr_model}, because more data do not narrow the posterior along a separating direction. Along such a direction, larger weights give a better fit, so only the prior limits them.

\paragraph{Bernstein-von Mises Theorem} The\index{Bernstein-von Mises theorem|(} Gaussian shape of the posterior for large data sets reflects the Bernstein-von\index{Bernstein-von Mises theorem} Mises theorem. It states that posteriors tend toward Gaussian distributions as the sample size increases. This holds under mild conditions, for example a prior with non-zero density at the true parameter value. It also applies to posteriors of several parameters, as in logistic regression.

The Central Limit Theorem says that sums of many independent random variables approach a Gaussian distribution. The posterior behaves in a similar way, because the log likelihood is a sum with one term per data point. As the data set grows, this sum dominates the prior and narrows the posterior, so a second-order expansion around the mode becomes accurate. The Gaussian is centered at the maximum likelihood estimate, which approaches the true parameter value when the model is correct.

\begin{figure}[ht!]
    \centering
    \includegraphics[width=0.95\textwidth]{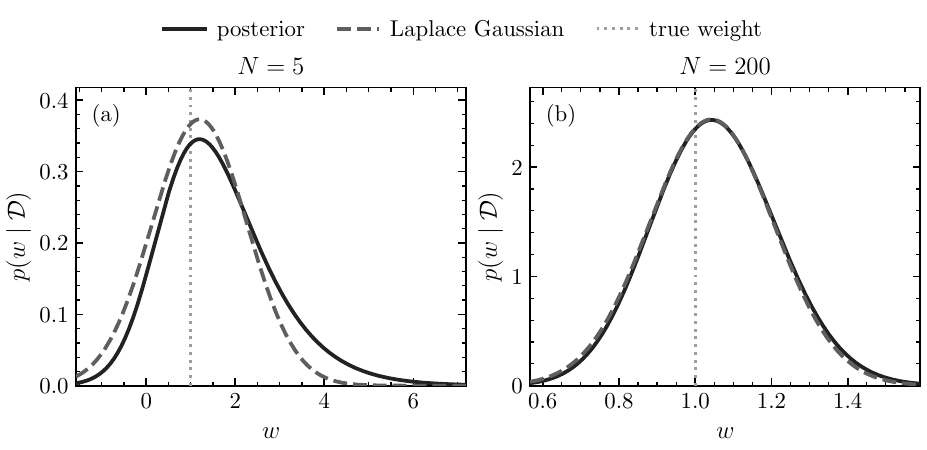}
    \caption{The posterior and its Laplace approximation for the model of Figure~\ref{fig:laplace_approximation}, with $N=5$ data points in panel~(a) and $N=200$ in panel~(b). The posterior is skewed in panel~(a) and almost Gaussian in panel~(b). The axis ranges differ, because the posterior in panel~(b) is much narrower. The dotted vertical line marks the true weight $w=1$ that generated the labels.}
    \label{fig:bernstein_von_mises}
\end{figure}

The Bernstein-von Mises theorem justifies the Laplace approximation for large data sets. Figure~\ref{fig:bernstein_von_mises} shows the posterior of the one-weight model approaching a Gaussian as the number of data points grows.

In astronomical classification, a few observed objects allow many boundaries, so the posterior is spread out. With hundreds or thousands of labeled objects the plausible boundaries narrow. The Gaussian approximation is then good for classes that overlap slightly, and in surveys with thousands or millions of classified objects it is reliable. For small samples the posterior is skewed, as in panel~(a) of Figure~\ref{fig:bernstein_von_mises}, and the Gaussian describes the uncertainty only roughly.\index{Bernstein-von Mises theorem|)}

\section{The Mathematics of the Laplace Approximation}

We now derive the Laplace approximation, first for a single weight $w$. Let $p(w|\mathcal{D})$ be a non-Gaussian posterior distribution with one clear peak. In Bayesian inference we often know it only up to a normalization constant
\begin{equation}
p(w|\mathcal{D}) = \frac{f(w)}{Z}.
\end{equation}
Here $f(w)$ is the unnormalized\index{normalization constant} density and $Z = \int f(w) \mathrm{d}w$ is the normalization constant. In Bayesian statistics this arises from Bayes' theorem
\begin{equation}
p(w|\mathcal{D}) = \frac{p(\mathcal{D}|w)p(w)}{p(\mathcal{D})}.
\end{equation}
The denominator $p(\mathcal{D})$, the evidence, equals $Z$ and is often hard or impossible to compute analytically. The Laplace approximation works directly with the unnormalized density $f(w) = p(\mathcal{D}|w)p(w)$, the likelihood times the prior, and avoids this computation.

We approximate $\ln f(w)$ near its peak by a quadratic function of $w$. The logarithm of a Gaussian has this quadratic form, so exponentiating gives a Gaussian.

\paragraph{Derivation for One Weight} We first locate the mode $w_0$, the point where the density reaches its maximum.

At the mode, the derivative of $f(w)$ vanishes, and so does the derivative of $\ln f(w)$
\begin{equation}
\left.\frac{\mathrm{d}}{\mathrm{d}w}\ln f(w)\right|_{w=w_0} = 0.
\end{equation}
Working with the logarithm is often more convenient, as it transforms products into sums and simplifies our calculations.

We expand the log-density to second order in a Taylor series around the mode
\begin{equation}
\ln f(w) \approx \ln f(w_0) + \left.\frac{\mathrm{d}}{\mathrm{d}w}\ln f(w)\right|_{w=w_0}(w-w_0) + \frac{1}{2}\left.\frac{\mathrm{d}^2}{\mathrm{d}w^2}\ln f(w)\right|_{w=w_0}(w-w_0)^2.
\end{equation}

The first derivative term vanishes because $w_0$ is a maximum. The second derivative at a maximum is negative, which means downward curvature. We define the positive number
\begin{equation}
A = -\left.\frac{\mathrm{d}^2}{\mathrm{d}w^2}\ln f(w)\right|_{w=w_0}.
\end{equation}The number $A$ measures the sharpness of the peak. The quadratic approximation becomes
\begin{equation}
\ln f(w) \approx \ln f(w_0) - \frac{1}{2}A(w-w_0)^2.
\end{equation}The quadratic approximation is the logarithm of a Gaussian with mean $w_0$ and precision $A$, the inverse variance.

Exponentiating the log-density approximation gives
\begin{equation}
f(w) \approx f(w_0)\exp\left(-\frac{1}{2}A(w-w_0)^2\right).
\end{equation}This is an unnormalized Gaussian. The constant factor $f(w_0)$ sets the height and drops out when we normalize.

After normalization, we obtain the Gaussian approximation $q(w)$ of the posterior
\begin{equation}
q(w) = \mathcal{N}(w|w_0, A^{-1}) = \sqrt{\frac{A}{2\pi}}\exp\left(-\frac{1}{2}A(w-w_0)^2\right).
\end{equation}We find the mode $w_0$, compute $A$ from the second derivative of $\ln f(w)$ there, and use the Gaussian with mean $w_0$ and variance $A^{-1}$.

\paragraph{Multivariate Extension} The weights of logistic regression form a vector, so we need several dimensions. The same steps apply, with vectors and matrices in place of scalars.

We find the mode $\mathbf{w}_0$, where the gradient of the log density vanishes
\begin{equation}
\nabla \ln f(\mathbf{w})\Big|_{\mathbf{w}=\mathbf{w}_0} = \mathbf{0}.
\end{equation}

We then expand the log density to second order around the mode
\begin{equation}
\ln f(\mathbf{w}) \approx \ln f(\mathbf{w}_0) + (\mathbf{w}-\mathbf{w}_0)^T \nabla \ln f(\mathbf{w})\Big|_{\mathbf{w}=\mathbf{w}_0} + \frac{1}{2}(\mathbf{w}-\mathbf{w}_0)^T \nabla\nabla \ln f(\mathbf{w})\Big|_{\mathbf{w}=\mathbf{w}_0} (\mathbf{w}-\mathbf{w}_0).
\end{equation}The linear term vanishes at the mode, as in one dimension. The Hessian\index{Hessian matrix} $\mathbf{H}$ is the matrix of second derivatives of $\ln f$ at the mode
\begin{equation}
\mathbf{H} = \nabla\nabla \ln f(\mathbf{w})\Big|_{\mathbf{w}=\mathbf{w}_0}.
\end{equation}
Its elements are $H_{jk} = \partial^2 \ln f/\partial w_j \partial w_k$ at $\mathbf{w} = \mathbf{w}_0$. At a maximum, $\mathbf{H}$ is negative definite, so the log density curves downward along each direction. Negative definite means that $\mathbf{v}^T\mathbf{H}\mathbf{v} < 0$ for any nonzero vector $\mathbf{v}$.

We define $\mathbf{A} = -\mathbf{H}$, the negative Hessian, which is positive definite at a maximum, so its inverse exists. The approximation becomes
\begin{equation}
\ln f(\mathbf{w}) \approx \ln f(\mathbf{w}_0) - \frac{1}{2}(\mathbf{w}-\mathbf{w}_0)^T \mathbf{A} (\mathbf{w}-\mathbf{w}_0).
\end{equation}Exponentiating and normalizing gives the multivariate Gaussian approximation
\begin{equation}
q(\mathbf{w}) = \frac{\det(\mathbf{A})^{1/2}}{(2\pi)^{D/2}}\exp\left(-\frac{1}{2}(\mathbf{w}-\mathbf{w}_0)^T \mathbf{A} (\mathbf{w}-\mathbf{w}_0)\right) = \mathcal{N}(\mathbf{w}|\mathbf{w}_0, \mathbf{A}^{-1}),
\end{equation}
where $\det(\mathbf{A})$ is the determinant of $\mathbf{A}$ and $D$ is the dimension of $\mathbf{w}$.

\looseness=1 The factor $\det(\mathbf{A})^{1/2}/(2\pi)^{D/2}$ makes the density integrate to one. A larger determinant $\det(\mathbf{A})$ means a narrower Gaussian, so the factor grows to keep the integral at one.

\section[The Laplace Approximation for Logistic Regression]{The Laplace Approximation\\for Logistic Regression}\label{sec:laplace_lr}

We now apply the general result to logistic regression. This lets us quantify the uncertainty of the decision boundary, which matters for astronomical data that are sparse or unevenly sampled. Here $f(\mathbf{w})$ is the unnormalized posterior $p(\mathbf{w})\,p(\mathbf{t}|\mathbf{w},\mathbf{X})$ of logistic regression, and the mode $\mathbf{w}_0$ is the MAP estimate $\mathbf{w}_{\text{MAP}}$.

Maximum likelihood maximizes the likelihood $p(\mathcal{D}|\mathbf{w})$. The MAP estimate also uses the prior, because it maximizes the posterior
\begin{equation}
\mathbf{w}_{\text{MAP}} = \arg\max_{\mathbf{w}} p(\mathbf{w}|\mathcal{D}) = \arg\max_{\mathbf{w}} \{p(\mathcal{D}|\mathbf{w})p(\mathbf{w})\}.
\end{equation}For a uniform prior, which treats all weight values alike, the MAP estimate equals the maximum likelihood estimate. For small data sets, an informative prior acts as a regularizer (Chapters~5 and~7).

The Laplace approximation needs the mode and the curvature of the log posterior, so we write the logarithm out first. Taking the logarithm of the product of the Gaussian prior and the Bernoulli likelihood of Section~\ref{sec:blr_model} gives
\begin{align}
\ln p(\mathbf{w}|\mathbf{t},\mathbf{X}) &= \ln p(\mathbf{w}) + \ln p(\mathbf{t}|\mathbf{w},\mathbf{X}) + \text{const} \\*
&= -\frac{1}{2}(\mathbf{w}-\mathbf{m}_0)^T\mathbf{S}_0^{-1}(\mathbf{w}-\mathbf{m}_0) \nonumber \\*
&\quad + \sum_{i=1}^N [t_i\ln\sigma(\mathbf{w}^T\mathbf{x}_i) + (1-t_i)\ln(1-\sigma(\mathbf{w}^T\mathbf{x}_i))] + \text{const}.
\end{align}In the second line, the first term comes from the prior. The second term is the familiar cross-entropy term of maximum likelihood estimation. The constants do not depend on $\mathbf{w}$, so they do not move the mode.

\paragraph{Finding the MAP Estimate}

The\index{maximum a posteriori estimate (MAP)|(} weights $\mathbf{w}_{\text{MAP}}$ that maximize the log posterior make its gradient vanish. The gradient is the sum of the gradient of the log prior and the gradient of the log likelihood.

For the Gaussian prior, the gradient is
\begin{align}
\nabla \ln p(\mathbf{w}) &= -\mathbf{S}_0^{-1}(\mathbf{w}-\mathbf{m}_0).
\end{align}

For the log likelihood, we use the gradient of the loss $E$ from Chapter~7. The loss $E$ is the negative log likelihood divided by $N$. The gradient of the log likelihood is therefore minus $N$ times the gradient of $E$, so
\begin{align}
\nabla \ln p(\mathbf{t}|\mathbf{w},\mathbf{X}) &= \sum_{i=1}^N (t_i - \sigma(\mathbf{w}^T\mathbf{x}_i))\mathbf{x}_i.
\end{align}
Each data point contributes its prediction error $t_i - \sigma(\mathbf{w}^T\mathbf{x}_i)$ times its feature vector $\mathbf{x}_i$.

Combining the two terms and setting the gradient to zero gives
\begin{equation}
\nabla \ln p(\mathbf{w}|\mathbf{t},\mathbf{X}) = -\mathbf{S}_0^{-1}(\mathbf{w}-\mathbf{m}_0) + \sum_{i=1}^N (t_i - \sigma(\mathbf{w}^T\mathbf{x}_i))\mathbf{x}_i = \mathbf{0}.
\end{equation}The first term of this equation is the pull of the prior toward $\mathbf{m}_0$. The second term is the pull of the data. At $\mathbf{w}_{\text{MAP}}$ the two pulls cancel.

Unlike in linear regression, this equation has no closed-form solution. We find $\mathbf{w}_{\text{MAP}}$ by gradient ascent
\begin{equation}
\mathbf{w}^{(\tau+1)} = \mathbf{w}^{(\tau)} + \eta \nabla \ln p(\mathbf{w}^{(\tau)}|\mathbf{t},\mathbf{X}),
\end{equation}
where $\eta$ is the learning rate and $\tau$ counts the iterations, as in Chapter~7. The gradient here is a sum over $N$ data points, while Chapter~7 uses their mean. For the same steps, $\eta$ is therefore $N$ times smaller than there. The sign is positive because we climb the log posterior to its maximum.\index{maximum a posteriori estimate (MAP)|)}

\paragraph{Computing the Hessian}

\looseness=1 Once\index{Hessian matrix|(} gradient ascent has found $\mathbf{w}_{\text{MAP}}$, we characterize the curvature of the log posterior there by computing its Hessian matrix. The Hessian of the log posterior is the sum of the Hessian of the log prior and the Hessian of the log likelihood.

For the Gaussian prior, the Hessian is
\begin{align}
\nabla\nabla \ln p(\mathbf{w}) &= -\mathbf{S}_0^{-1}.
\end{align}

For the log likelihood, we differentiate its gradient again. The derivative of $\sigma(\mathbf{w}^T\mathbf{x}_i)$ with respect to $\mathbf{w}$ is $\sigma(1-\sigma)\mathbf{x}_i$, as in Chapter~7, so
\begin{align}
\nabla\nabla \ln p(\mathbf{t}|\mathbf{w},\mathbf{X}) &= -\sum_{i=1}^N \sigma(\mathbf{w}^T\mathbf{x}_i)(1-\sigma(\mathbf{w}^T\mathbf{x}_i))\mathbf{x}_i\mathbf{x}_i^T.
\end{align}

The Laplace approximation needs the negative Hessian at $\mathbf{w}_{\text{MAP}}$. It is the sum of the two contributions
\begin{equation}
\mathbf{A} = -\mathbf{H}|_{\mathbf{w}=\mathbf{w}_{\text{MAP}}} = \mathbf{S}_0^{-1} + \sum_{i=1}^N \sigma(\mathbf{w}_{\text{MAP}}^T\mathbf{x}_i)(1-\sigma(\mathbf{w}_{\text{MAP}}^T\mathbf{x}_i))\mathbf{x}_i\mathbf{x}_i^T.
\end{equation}The matrix $\mathbf{A}$ is the precision, the inverse covariance, of the Gaussian approximation to the posterior. The first term, the prior precision $\mathbf{S}_0^{-1}$, describes our uncertainty before we see the data, and the second term adds the information from the observations.\index{Hessian matrix|)}

\paragraph{The Final Approximation}

With $\mathbf{w}_{\text{MAP}}$ and $\mathbf{A}$, the Gaussian approximation of the posterior is
\begin{equation}
q(\mathbf{w}) = \mathcal{N}(\mathbf{w} | \mathbf{w}_{\text{MAP}}, \mathbf{A}^{-1}).
\end{equation}This Gaussian describes the plausible decision boundaries, with the MAP\index{maximum a posteriori estimate (MAP)} boundary at the center. The covariance $\mathbf{A}^{-1}$ describes the uncertainty of the weights. Larger diagonal elements of $\mathbf{A}^{-1}$ mean greater uncertainty in the corresponding weights, and off-diagonal elements describe covariances between weights. A larger $\mathbf{A}$ therefore means less uncertainty.

\begin{figure}[ht!]
    \centering
    \includegraphics[width=0.9\textwidth]{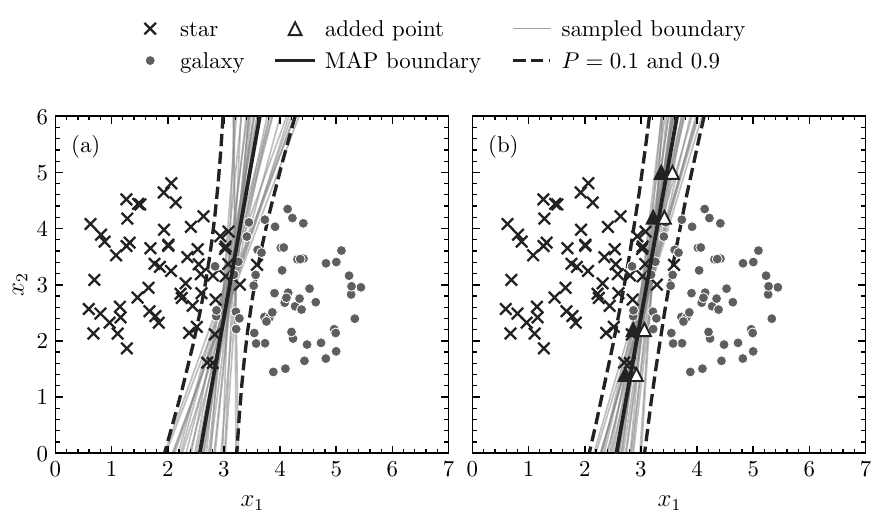}
    \caption{Effect of data points near the decision boundary. Panel~(a) shows boundaries sampled from the Gaussian approximation $q(\mathbf{w})$ of the posterior and contours of the probability of a galaxy. Panel~(b) shows the same quantities after adding points next to the boundary. The sampled boundaries narrow and the contours move closer together.}
    \label{fig:boundary_uncertainty}
\end{figure}

\paragraph{Points Near the Boundary}

Write $y_i = \sigma(\mathbf{w}_{\text{MAP}}^T\mathbf{x}_i)$ for the prediction at the MAP weights. Each data point adds the term $y_i(1-y_i)\mathbf{x}_i\mathbf{x}_i^T$ to $\mathbf{A}$. The factor $y_i(1-y_i)$ has its maximum 0.25 on the decision boundary, where $y_i = 0.5$ and the class probability is most uncertain. It falls toward 0 as a point moves away from the boundary. Points near the boundary therefore add the most to $\mathbf{A}$, and they pin down the boundary best, because a small shift changes their classification. Points far from the boundary add little, because a small shift does not change their classification.

\looseness=1 In Figure~\ref{fig:boundary_uncertainty} the boundaries drawn from $q(\mathbf{w})$ lie closer together after points near the boundary are added. Many points far from the boundary can still leave a large uncertainty about its position. A training set should therefore include points near the possible boundaries between classes, in addition to objects that cover the range of the features.\index{Laplace approximation|)}

\section{The Predictive Distribution}\label{sec:ch9_predictive}

The\index{posterior predictive distribution|(} Gaussian $q(\mathbf{w})$ of the previous section is a distribution over decision boundaries. Each weight vector drawn from it gives a different boundary, and each boundary gives a different probability for a new object.

For a new object, one boundary might give a 70\% probability of class 1 and another only 40\%. This spread comes from our uncertainty about the boundary, and it is largest where the training data are sparse. We could examine the spread of these probabilities, but we usually want a single prediction that includes this uncertainty.

This leads us to the predictive\index{posterior predictive distribution} distribution, the probability of a class label for a new input, as in Section~5.7 for regression. We consider all plausible boundaries simultaneously, and each contributes according to its posterior probability.

\begin{figure}[ht!]
    \centering
    \includegraphics[width=0.9\textwidth]{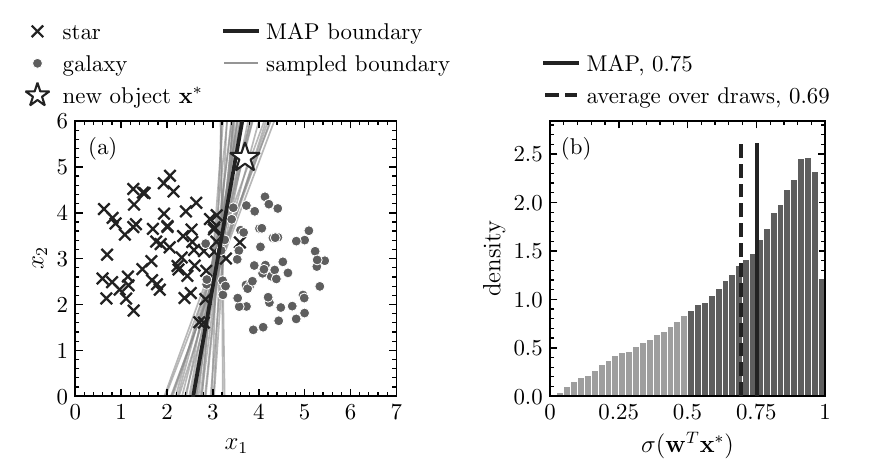}
    \caption{Averaging over decision boundaries. Panel~(a) shows boundaries sampled from the Gaussian approximation $q(\mathbf{w})$ of the posterior and a new object $\mathbf{x}^*$ that different boundaries place on different sides. Panel~(b) shows the distribution of $\sigma(\mathbf{w}^T\mathbf{x}^*)$ over many draws of $\mathbf{w}$ from $q(\mathbf{w})$. Pale bars lie on the star side of 0.5 and dark bars on the galaxy side. The average over the draws estimates the predictive probability and lies below the MAP value.}
    \label{fig:bayesian_averaging}
\end{figure}

For a new input $\mathbf{x}^*$ with unknown label $t^*$, the predictive probability of class 1 is
\begin{equation}
P(t^* = 1 | \mathbf{x}^*, \mathcal{D}) = \int P(t^* = 1 | \mathbf{x}^*, \mathbf{w}) p(\mathbf{w} | \mathcal{D}) \mathrm{d}\mathbf{w}.
\end{equation}The first factor $P(t^* = 1 | \mathbf{x}^*, \mathbf{w})$ is the sigmoid $\sigma(\mathbf{w}^T \mathbf{x}^*)$, and the second factor $p(\mathbf{w} | \mathcal{D})$ is the posterior of the weights. The integral is the average of the sigmoid over the weight vectors, each counted in proportion to its posterior density. Figure~\ref{fig:bayesian_averaging} shows boundaries sampled from the Gaussian approximation and the probabilities that they give for one object. The average of these probabilities estimates the predictive probability.

With the Gaussian approximation of the posterior, this becomes
\begin{equation}
P(t^* = 1 | \mathbf{x}^*, \mathcal{D}) \approx \int \sigma(\mathbf{w}^T \mathbf{x}^*) \mathcal{N}(\mathbf{w} | \mathbf{w}_{\text{MAP}}, \mathbf{A}^{-1}) \mathrm{d}\mathbf{w}.
\end{equation}

\paragraph{Dimensional Reduction}

The integral runs over the high-dimensional weight space, but the prediction needs only one number, the logit $a = \mathbf{w}^T \mathbf{x}^*$ (Chapter~7). The logit determines the class probability through the sigmoid. For a given weight vector, it is proportional to the signed distance of the point from the decision boundary. The distance is positive on the side of class 1 and negative on the other side.

Many weight vectors give the same logit $a$ for a given input $\mathbf{x}^*$. If we add a vector $\mathbf{w}_{\perp}$ perpendicular to $\mathbf{x}^*$ to the weight vector, the logit does not change
\begin{equation}
(\mathbf{w} + \mathbf{w}_{\perp})^T \mathbf{x}^* = \mathbf{w}^T \mathbf{x}^* + \mathbf{w}_{\perp}^T \mathbf{x}^* = \mathbf{w}^T \mathbf{x}^* = a,
\end{equation}
where $\mathbf{w}_{\perp}^T \mathbf{x}^* = 0$ by definition.

\looseness=1 The weight vectors with the same logit $a$ therefore form a hyperplane perpendicular to $\mathbf{x}^*$. The prediction is the same on this hyperplane, so only the distribution of the logit $a$ matters. Under the Gaussian approximation, the weights are Gaussian. The logit $a = \mathbf{w}^T \mathbf{x}^*$ is a linear combination of them, so it is also Gaussian, as for the regression line of Chapter~5.

\begin{figure}[ht!]
    \centering
    \includegraphics[width=0.9\textwidth]{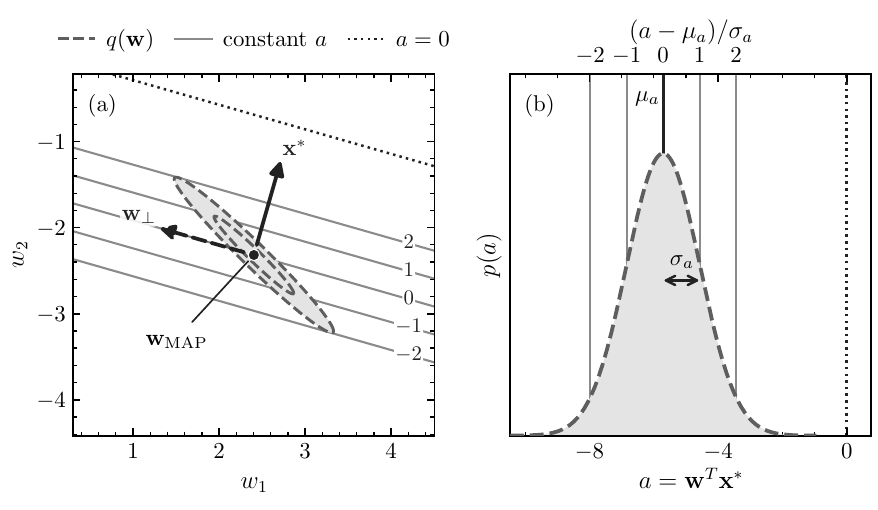}
    \caption{From weight space to logit space for two weights. Panel~(a) shows the Gaussian approximation $q(\mathbf{w})$ and lines of constant logit $a = \mathbf{w}^T\mathbf{x}^*$. Panel~(b) shows the Gaussian distribution of $a$ that follows from projecting $q(\mathbf{w})$ onto $\mathbf{x}^*$.}
    \label{fig:dimensionality_reduction}
\end{figure}

The mean and variance of $a$ follow from the rules for a linear transformation of a random vector (Chapter~3)
\begin{align}
\mu_a &= \mathbf{w}_{\text{MAP}}^T\mathbf{x}^*, \\*
\sigma_a^2 &= \mathbf{x}^{*T}\mathbf{A}^{-1}\mathbf{x}^*.
\end{align}
Therefore $a \sim \mathcal{N}(\mu_a, \sigma_a^2)$. The variance is large when $\mathbf{x}^*$ points along directions in which the weights are poorly determined. Figure~\ref{fig:dimensionality_reduction} shows this for two weights. The Gaussian over the weights in panel~(a) gives the Gaussian distribution of $a$ in panel~(b).

The integral over $\mathbf{w}$ therefore equals an integral over the distribution of $a$. The predictive integral becomes
\begin{equation}
P(t^* = 1 | \mathbf{x}^*, \mathcal{D}) \approx \int \sigma(a) \mathcal{N}(a | \mu_a, \sigma_a^2) \mathrm{d}a.
\end{equation}The mean $\mu_a$ is our best estimate of the logit, and the variance $\sigma_a^2$ is its uncertainty. This projection simplifies the problem. Weight vectors can have hundreds or thousands of components, so a direct integral over the weights is too expensive to compute. The integral over $a$ has only one variable. The one-dimensional integral has no closed form, because the sigmoid does not combine with a Gaussian.\index{posterior predictive distribution|)}

\section{The Probit Approximation}\label{sec:probit}

To\index{probit approximation|(} obtain a closed form, we replace the sigmoid by the probit\index{probit approximation}\index{probit function} function $\Phi(x)$, the cumulative distribution function of the standard Gaussian distribution. The integral of this function against a Gaussian has a closed form.

\begin{figure}[ht!]
    \centering
    \includegraphics[width=0.9\textwidth]{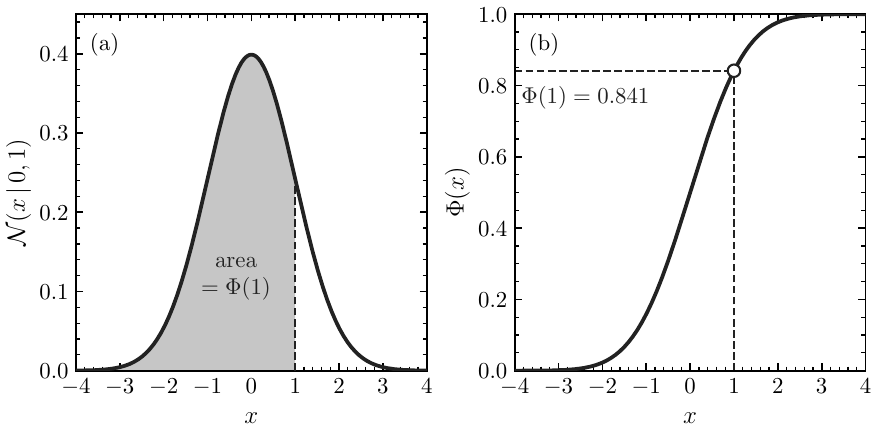}
    \caption{Panel~(a) shows the standard Gaussian density with the area up to a value $x$ shaded. Panel~(b) shows the probit function $\Phi(x)$, which gives this area for each $x$.}
    \label{fig:probit_function}
\end{figure}

The probit function is the integral of the standard Gaussian density
\begin{equation}
\Phi(x) = \int_{-\infty}^{x} \mathcal{N}(u | 0, 1) \mathrm{d}u.
\end{equation}The value $\Phi(x)$ is the probability that a standard Gaussian variable is at most $x$. Figure~\ref{fig:probit_function} shows the density and the cumulative distribution function.

We can scale this function with a parameter $\lambda > 0$, which gives $\Phi(\lambda x)$, the integral of the standard Gaussian density up to $\lambda x$. This scaling is equivalent to working with the random variable $V = U/\lambda$ with $U \sim \mathcal{N}(0,1)$, which has the distribution $V \sim \mathcal{N}(0, 1/\lambda^2)$, because
\begin{equation}
P(V < x) = P(U < \lambda x) = \Phi(\lambda x).
\end{equation}

The probit function and the sigmoid both map the real line to the interval $(0,1)$ and increase along an S-shaped curve from 0 to 1. The two curves look similar, which suggests that one can approximate the other.

\paragraph{The Probit Identity}

For\index{probit function|(} $\lambda > 0$, the integral of the probit function times a Gaussian is again a probit function
\begin{equation}
\int \Phi(\lambda a) \mathcal{N}(a | \mu, \sigma^2) \mathrm{d}a = \Phi\left(\frac{\lambda\mu}{\sqrt{1 + \lambda^2\sigma^2}}\right).
\end{equation}
Here $\mu$ and $\sigma^2$ are the mean and variance of the Gaussian. For the predictive integral, $\mu = \mu_a$ and $\sigma^2 = \sigma_a^2$. Averaging a probit function over a Gaussian gives a probit function with a reduced argument, and the reduction grows with the variance.

We derive the identity with two independent random variables, $X \sim \mathcal{N}(0, 1/\lambda^2)$ and $Y \sim \mathcal{N}(\mu, \sigma^2)$. Here $X$ is distributed like the scaled variable $V$ above, and $Y$ has the density $\mathcal{N}(a|\mu,\sigma^2)$ of the integral. We calculate the probability $P(X < Y)$ in two ways, which must give the same result.

First, we use the difference $X - Y$. The two variables are independent and Gaussian, so the difference is Gaussian with mean $\mathbb{E}[X-Y] = 0 - \mu = -\mu$ and variance $\text{Var}[X-Y] = 1/\lambda^2 + \sigma^2$. Therefore
\begin{equation}
X - Y \sim \mathcal{N}(-\mu, 1/\lambda^2 + \sigma^2).
\end{equation}
The probability $P(X < Y) = P(X - Y < 0)$ is
\begin{align}
P(X - Y < 0) &= \Phi\left(\frac{\mu}{\sqrt{1/\lambda^2 + \sigma^2}}\right) \\*
&= \Phi\left(\frac{\lambda\mu}{\sqrt{1 + \lambda^2\sigma^2}}\right).
\end{align}

Second, we condition on the value of $Y$. For a value $y$, the probability that $X < y$ is $\Phi(\lambda y)$. We weight it by the probability density of $Y = y$ and integrate over $y$. This gives
\begin{equation}
P(X < Y) = \int_{-\infty}^{\infty} \Phi(\lambda y) \mathcal{N}(y | \mu, \sigma^2) \mathrm{d}y.
\end{equation}Both calculations give $P(X < Y)$, so the two results are equal. This proves the identity, with the integration variable $y$ in place of\index{probit function|)} $a$.

\paragraph{Sigmoid Approximation}

To use the identity, we approximate the sigmoid by a scaled probit function
\begin{equation}
\sigma(a) \approx \Phi(\lambda a).
\end{equation}

We fix $\lambda$ by matching the slopes of the sigmoid and the probit function at $a = 0$, the decision boundary, where the probability changes fastest. The derivative of the sigmoid is
\begin{equation}
\frac{\mathrm{d}}{\mathrm{d}a}\sigma(a) = \sigma(a)(1-\sigma(a)) = \frac{e^{-a}}{(1+e^{-a})^2}.
\end{equation}
At $a = 0$ it equals $1/4$. The derivative of the probit function is
\begin{equation}
\frac{\mathrm{d}}{\mathrm{d}a}\Phi(\lambda a) = \lambda\phi(\lambda a) = \frac{\lambda}{\sqrt{2\pi}}e^{-\lambda^2a^2/2},
\end{equation}
where $\phi(x)$ is the density of the standard Gaussian. At $a = 0$ it equals $\lambda/\sqrt{2\pi}$. Equating the two slopes gives
\begin{equation}
\frac{1}{4} = \frac{\lambda}{\sqrt{2\pi}} \implies \lambda = \sqrt{\frac{\pi}{8}}.
\end{equation}

\begin{figure}[ht!]
    \centering
    \includegraphics[width=0.7\textwidth]{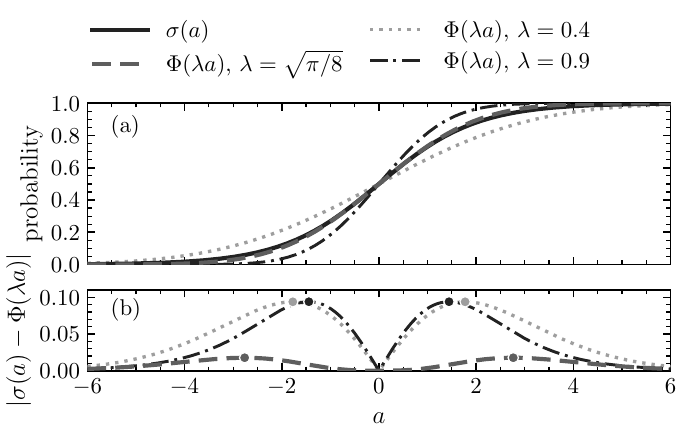}
    \caption{The sigmoid function $\sigma(a)$ and the scaled probit function $\Phi(\lambda a)$. Panel~(a) compares the curves for $\lambda = \sqrt{\pi/8} \approx 0.63$, which matches the slope of the sigmoid at $a = 0$, with two other values of $\lambda$. Panel~(b) shows the absolute error of each curve.}
    \label{fig:sigmoid_probit_approximation}
\end{figure}

With this value, both functions equal 0.5 at $a = 0$ and have the same slope there. This gives a good local approximation at the decision boundary, where it matters most for classification. Figure~\ref{fig:sigmoid_probit_approximation} compares the two functions and shows the error for two other values of $\lambda$.

The approximation is therefore
\begin{equation}
\sigma(a) \approx \Phi\left(\sqrt{\frac{\pi}{8}} a\right).
\end{equation}Substituting $\Phi(\sqrt{\pi/8}\,a)$ for $\sigma(a)$ in the predictive integral and applying the identity with $\lambda = \sqrt{\pi/8}$ gives
\begin{equation}
\int \Phi\left(\sqrt{\frac{\pi}{8}} a\right) \mathcal{N}(a | \mu_a, \sigma_a^2) \mathrm{d}a = \Phi\left(\frac{\sqrt{\pi/8} \cdot \mu_a}{\sqrt{1 + (\pi/8) \sigma_a^2}}\right).
\end{equation}

To write the prediction in the form of standard logistic regression (Chapter~7), we use $\Phi(\sqrt{\pi/8}\,x) \approx \sigma(x)$ with $x = \mu_a/\sqrt{1 + \pi\sigma_a^2/8}$
\begin{equation}
\Phi\left(\frac{\sqrt{\pi/8} \cdot \mu_a}{\sqrt{1 + (\pi/8) \sigma_a^2}}\right) \approx \sigma\left(\frac{\mu_a}{\sqrt{1 + \pi\sigma_a^2/8}}\right).
\end{equation}The predictive probability is therefore
\begin{equation}
P(t^* = 1 | \mathbf{x}^*, \mathcal{D}) \approx \sigma\left(\frac{\mu_a}{\sqrt{1 + \pi\sigma_a^2/8}}\right).
\end{equation}

\paragraph{The Uncertainty Factor}

We\index{uncertainty factor|(} define the uncertainty\index{uncertainty factor} factor $\kappa(\sigma_a^2) = (1 + \pi\sigma_a^2/8)^{-1/2}$ and write the prediction as
\begin{equation}
P(t^* = 1 | \mathbf{x}^*, \mathcal{D}) \approx \sigma(\kappa(\sigma_a^2) \cdot \mu_a).
\end{equation}With $\mu_a$ and $\sigma_a^2$ written out, this is
\begin{equation}
P(t^* = 1 | \mathbf{x}^*, \mathcal{D}) \approx \sigma\left(\kappa(\mathbf{x}^{*T} \mathbf{A}^{-1} \mathbf{x}^*) \cdot \mathbf{w}_{\text{MAP}}^T \mathbf{x}^*\right).
\end{equation}The Bayesian prediction multiplies the logit $\mathbf{w}_{\text{MAP}}^T \mathbf{x}^*$ of standard logistic regression by $\kappa(\sigma_a^2)$, which decreases as the variance $\sigma_a^2$ of the logit grows. This is the behavior we asked for in the introduction, because the prediction is less confident where the training data constrain the weights poorly.\index{probit approximation|)}\index{uncertainty factor|)}

\section{Properties of Bayesian Logistic Regression}

\looseness=1 We now ask where the variance $\sigma_a^2$ is large and what it does to the prediction. The variance $\sigma_a^2 = \mathbf{x}^{*T} \mathbf{A}^{-1} \mathbf{x}^*$ depends on the precision matrix $\mathbf{A}$ of Section~\ref{sec:laplace_lr}, which sums the prior precision and one term $y_i(1-y_i)\mathbf{x}_i\mathbf{x}_i^T$ for each training point. Since $\mathbf{A}^{-1}$ is the covariance of the weights, $\sigma_a^2$ is the uncertainty of the weights along the direction of $\mathbf{x}^*$.

When $\mathbf{x}^*$ is far from the training data, it has low similarity, as measured by inner products, with the training points $\mathbf{x}_i$. The contributions to $\mathbf{A}$ in the direction of $\mathbf{x}^*$ are then small, so the corresponding elements of $\mathbf{A}^{-1}$ are large. This gives a larger $\sigma_a^2$ in regions far from the training data.

The minimum of $\sigma_a^2$ lies near the center of the training points, weighted toward those close to the boundary, which add the most to $\mathbf{A}$. The growth is fastest in the direction in which the training points near the boundary are least spread out. Weights along such a direction are poorly determined, so the variance rises quickly there.

\paragraph{Properties of the Uncertainty Factor}

The\index{uncertainty factor|(} prediction of standard logistic regression, $\sigma(\mathbf{w}_{\text{MAP}}^T\mathbf{x}^*)$, equals the maximum likelihood prediction for a flat prior. Three properties of $\kappa$ show how the Bayesian prediction differs from it.

First, $\kappa$ is positive and at most 1, because $\sigma_a^2 \geq 0$ makes the factor under the square root at least 1. For $\sigma_a^2 = 0$, which means no uncertainty about the logit, $\kappa(0) = 1$ and the Bayesian prediction equals the standard one. As $\sigma_a^2$ grows, $\kappa$ decreases toward 0.

Second, a smaller $\kappa(\sigma_a^2)$ moves the prediction closer to 0.5. Take $\mathbf{w}_{\text{MAP}}^T \mathbf{x}^* = 2$, a strong positive prediction with $\sigma(2) \approx 0.88$. For $\kappa(\sigma_a^2) = 0.5$ we get $\sigma(1) \approx 0.73$, and for $\kappa(\sigma_a^2) = 0.2$ we get $\sigma(0.4) \approx 0.60$. As $\kappa$ approaches 0, the prediction approaches 0.5, the case of maximum uncertainty.

Third, for negative logits, which indicate class 0, the factor raises the prediction toward 0.5.

In a two-dimensional feature space with two overlapping clusters, standard logistic regression has a linear boundary. Its confidence increases with the distance from the boundary, even far from the training data.

\begin{figure}[ht!]
    \centering
    \includegraphics[width=1.0\textwidth]{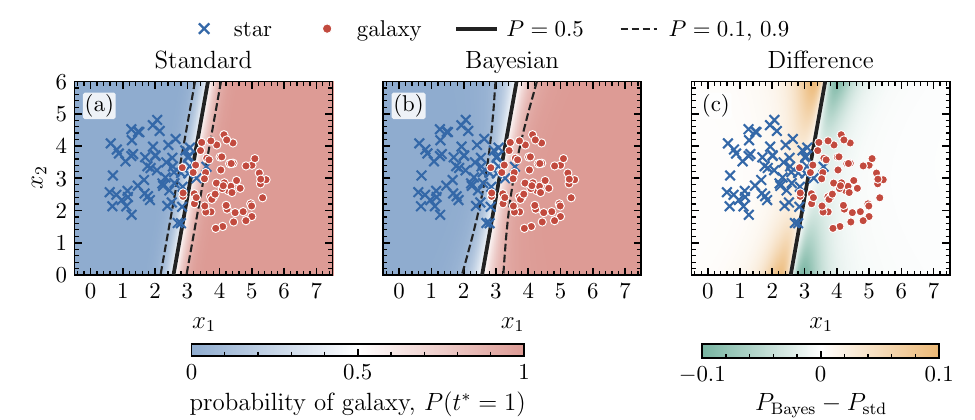}
    \caption{Standard and Bayesian logistic regression fitted to the same data. Panels~(a) and~(b) show the probability of a galaxy for a new object $\mathbf{x}^*$. Away from the data, the contours in panel~(b) spread out. Panel~(c) shows the Bayesian minus the standard probability.}
    \label{fig:uncertainty_comparison}
\end{figure}

In the Bayesian prediction, $\sigma_a^2$ is small near the training points, where the predictions stay close to the standard ones. Away from them $\sigma_a^2$ grows and the predictions move closer to 0.5. The 0.5 contour stays at the boundary of standard logistic regression. The other probability contours spread out, so the boundary becomes soft where evidence is scarce. Figure~\ref{fig:uncertainty_comparison} compares the predictions of the two approaches.\index{uncertainty factor|)}

\paragraph{Astronomical Applications}

In astronomical classification, some objects lie outside the feature ranges of the training data. As an example, take the classification of galaxies as spiral or elliptical from morphological features. Suppose the training set contains mostly typical spiral and elliptical galaxies. Standard logistic regression can then give a peculiar galaxy a 99\% probability, although nothing similar appears in the training data.

Bayesian logistic regression reduces this problem. For a galaxy unlike the training galaxies, the prediction lies closer to 0.5 than the standard one, and $\sigma_a^2$ is large. Such a galaxy may be a new or transitional type.

First, objects with a large $\sigma_a^2$ can be flagged for manual follow-up, while the other classifications can stay automatic. In large surveys, millions of objects can be classified automatically, and overconfident classifications can bias derived statistics or hide unusual objects.

Second, the variance helps to identify boundary cases that may be transitional or new object types. Many discoveries lie between established categories, for example galaxies in transformation, stars in unusual evolutionary states or entirely new classes. A large $\sigma_a^2$ points to these cases.

A probability near 0.5 can also come from overlapping classes, and standard logistic regression already shows this. The variance $\sigma_a^2$ adds the second cause, a poorly constrained boundary. These two causes are the aleatoric and the epistemic uncertainty of Chapter~5. The variance also shows where more labeled objects near the boundary would help most.

\paragraph{Other Sources of Uncertainty}

Two more sources of uncertainty are not part of the model above. The first is the measurement error of the features of the object that we classify. We include it by averaging the probability over the true features $\mathbf{x}$, given the measured features $\mathbf{x}_{\text{obs}}$. The average is $\int P(t^*=1|\mathbf{x})\,p(\mathbf{x}|\mathbf{x}_{\text{obs}})\,\mathrm{d}\mathbf{x}$, where the probability can be the maximum likelihood or the Bayesian one (Chapter~6 treats input errors for regression).

The second is error in the training data, in the labels and in the features. Some of these sources do not shrink as the training set grows. They include the overlap of the classes, the measurement errors of the new object and systematic errors in the training data. The uncertainty of the weights does shrink, as the Bernstein-von Mises theorem says. It therefore matters most for small or unrepresentative training sets and for objects far from the training data.

\medskip Bayesian logistic regression completes our treatment of linear models for regression and classification. In these models labeled data define a map from inputs to outputs, which is called supervised learning. Many astronomical data sets have no target variable to predict, such as galaxy measurements without known groups or survey spectra without known classes. In Chapter~10, we turn to such unsupervised problems, which look for structure in the data without labels. We start with principal component analysis, which finds the main directions of variation in high-dimensional data.\index{Bayesian logistic regression|)}

\section{Tutorial: Uncertainty in Red Clump Classification}\label{sec:ch9_tutorial}

This\index{red clump stars|(}\index{red giant branch stars|(} tutorial adds an uncertainty to the red clump probabilities of Section~\ref{sec:ch7_tutorial} with the posterior and predictive distribution of this chapter. A survey must know how far to trust its probabilities, because the distance of a red giant branch star kept by mistake is less accurate. The full tutorial can be found at \url{https://statml.org/} as tutorial 9.

The probabilities of Section~\ref{sec:ch7_tutorial} come from one set of weights, so they do not say how far to trust them. Other plausible boundaries fit the training stars almost as well and give a star next to the boundary a different probability. The Bayesian treatment thus keeps these boundaries and weighs each by its posterior probability (Section~\ref{sec:blr_model}).

\begin{figure}[t!]
    \centering
    \includegraphics[width=0.95\textwidth]{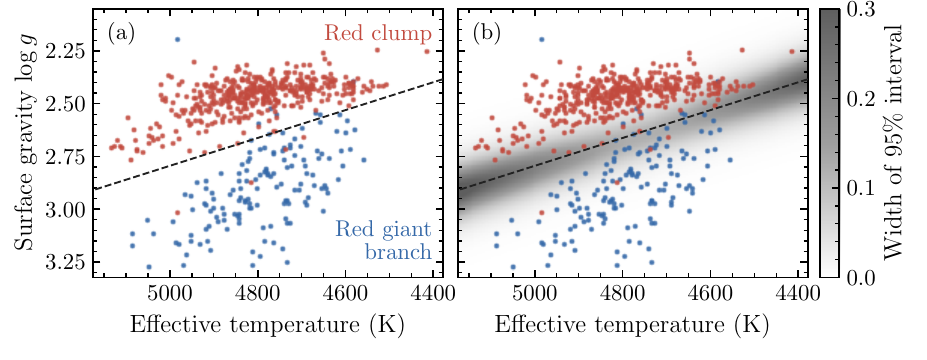}
    \caption{The tutorial adds an uncertainty to the red clump probability of each test star of Figure~\ref{fig:ch7_tutorial_classes}, colored as there. Panel~(a) shows the maximum likelihood boundary (dashed). Panel~(b) repeats the boundary and shades the 95 percent interval width of the probability over the plausible boundaries. The width is largest at the dashed boundary.}
    \label{fig:ch9_tutorial_uncertainty}
\end{figure}

The sigmoid makes the posterior non-Gaussian, so the Laplace approximation replaces it with a Gaussian around the maximum a posteriori weights (Section~\ref{sec:ch9_laplace_intuition}). Weights drawn from this Gaussian give the plausible boundaries, and the predictive distribution averages the probabilities they give each star (Section~\ref{sec:ch9_predictive}).

Maximum likelihood gives one boundary (Figure~\ref{fig:ch9_tutorial_uncertainty}a) and so one probability for each star. The Bayesian treatment gives the average of the plausible probabilities with their 95 percent interval, whose width shows how far to trust it. The width is largest at the dashed boundary, where the sigmoid is steepest, and almost zero far from it (Figure~\ref{fig:ch9_tutorial_uncertainty}b).

The probability from one boundary and its interval answer different questions. The probability shows the overlap of the classes, an aleatoric uncertainty that more training stars do not remove. The interval measures our limited knowledge of the boundary, an epistemic uncertainty that shrinks as training stars are added. A survey that keeps stars above a cut can flag those whose interval reaches below it, because their classification is uncertain.\index{red clump stars|)}\index{red giant branch stars|)}

\paragraph{Further Reading} Zellner and Rossi (1984) gave a Bayesian analysis of binary response models. Albert and Chib (1993) developed latent-variable methods for binary and multi-class response models. Tierney and Kadane (1986) developed Laplace approximations for posterior moments and marginal densities. Walker (1969) examined posterior distributions for large samples, and Le Cam (1986) gave results, including the Bernstein-von Mises theorem. MacKay (2003) shows how to select hyperparameters in Bayesian logistic regression with the evidence. Bishop (2006) treats the Laplace approximation and the predictive distribution (Sections 4.4 and 4.5). McCullagh and Nelder (1989) treat generalized linear models. Kass and Raftery (1995) review Bayes factors and marginal likelihoods.

\chapter{Principal Component Analysis}

A\index{principal component analysis (PCA)|(} single stellar spectrum has measurements at thousands of wavelength bins, and a galaxy image has tens of thousands of pixels. Even a simple stellar catalog can list dozens of elemental abundances. Yet we often suspect that only a few physical parameters determine these measurements. We want to find a few combinations of the measurements that explain most of the variation in the data.

Stellar spectra with thousands of wavelength points are mainly determined by the effective temperature, the surface gravity and the metallicity. Galaxy morphologies look complex but often come from a few formation and evolution processes. The chemical compositions of stars in a cluster, measured in many elements, usually come from a few nucleosynthesis processes.

Principal\index{principal component analysis (PCA)} component analysis (PCA)\index{PCA|see{principal component analysis}} is our first unsupervised method, because it uses no target variable. It finds a lower-dimensional representation of high-dimensional data that keeps as much information as possible. It uses the covariance matrix of Chapter~3 and linear algebra to find the directions of maximum variation.

Suppose that a few underlying factors determine the data. Then most of the variation should lie along a few directions in the data space. PCA finds these directions, which are the principal components, so we can describe each object with far fewer numbers.

Reducing the dimension, also called dimensionality\index{dimensionality reduction} reduction, helps with three problems of astronomical data. First, data with more than two or three dimensions are hard to visualize, so we can inspect them only in projection. Second, the computing cost grows quickly with the number of dimensions. Third, inference becomes harder in many dimensions, because the available data become sparse compared with the volume of the space. This is the curse\index{curse of dimensionality} of dimensionality. For example, 100 objects fill the 10 bins of one feature, but 10 features with 10 bins each make $10^{10}$ bins and most stay empty.

In this chapter we derive PCA as a constrained\index{constrained optimization|see{Lagrange multipliers}} optimization problem (Sections~10.2 and~10.3). We show that the principal components are the eigenvectors of the covariance matrix, ordered by their eigenvalues. The singular value decomposition (SVD) of Section~10.8 makes the computation efficient when there are fewer objects than features.

\section{PCA Fundamentals}\label{sec:ch10_fundamentals}

Consider a data set of $N$ astronomical objects with $D$ measurements each. The data matrix $\mathbf{X} \in \mathbb{R}^{N \times D}$ has one object per row and one feature per column. For stellar spectroscopy, $N$ is the number of observed stars and $D$ is the number of wavelength bins. For galaxy photometry, the features can be magnitudes in different bands.

\looseness=1 Not all directions in this $D$-dimensional feature space are equally informative. Along some directions the objects differ by large amounts, and these directions separate one object from another. Along other directions the data vary little. These directions can hold measurement noise, redundant information or properties that do not distinguish between objects.

\paragraph{Two Perspectives on Dimensionality Reduction}

PCA can be described in two ways that give the same solution. The first looks for the directions in the high-dimensional space along which the data vary the most, and projects the data onto them. Directions of larger variance carry more information about the structure of the data. Here we use information in the loose sense of variance, the spread of the data along a direction. In stellar spectra, directions of maximum variance can match physical differences in temperature, metallicity or surface gravity. In galaxy photometry they can follow differences in mass, star formation rate or morphology.

\looseness=1 The second way looks for a lower-dimensional representation of the data. We map the lower-dimensional representation of each object back to the original space, and the result is its reconstruction. The average squared distance between the data points and their reconstructions should be as small as possible. This view preserves the overall structure of the data cloud. Because the two ways agree, PCA finds the lower-dimensional representation that keeps the most information among linear projections to a given number of dimensions.

\paragraph{Notation and Centering}

We begin with the notation and one simplifying assumption. The observations are the vectors $\mathbf{x}_1, \ldots, \mathbf{x}_N \in \mathbb{R}^D$, and $\mathbf{x}_i^T$ is the $i$-th row of $\mathbf{X}$. We write $\mathbf{x}$ for one object drawn from the data set, for example one star.

To simplify the derivation, we assume that the data are centered, so that their average is zero
\begin{equation}
\mathbb{E}[\mathbf{x}] = \frac{1}{N}\sum_{i=1}^N \mathbf{x}_i = \mathbf{0}.
\end{equation}Here $\mathbb{E}$ denotes the average over the data set. Astronomical data rarely have a clear zero point, so we subtract the sample mean $\overline{\mathbf{x}}$ of the original observations from each observation
\begin{equation}
\mathbf{x}_i' = \mathbf{x}_i - \overline{\mathbf{x}}.
\end{equation}

We took the same step in Chapter~6, where it removed the intercept. From now on, $\mathbf{x}_i$ and $\mathbf{X}$ hold the centered data, and $\overline{\mathbf{x}}$ stays the mean of the original observations. Centered values are deviations from the mean, so we can read the variance directly as a measure of information.

\paragraph{Vector Projections and Basis Decomposition}

PCA rests on vector projection.\index{projection} Projecting a vector $\mathbf{x}$ onto a unit vector $\mathbf{b}$ in $D$ dimensions gives
\begin{equation}
\widehat{\mathbf{x}} = (\mathbf{x}^T\mathbf{b})\mathbf{b}.
\end{equation}

\begin{figure}[ht!]
    \centering
    \includegraphics[width=0.55\textwidth]{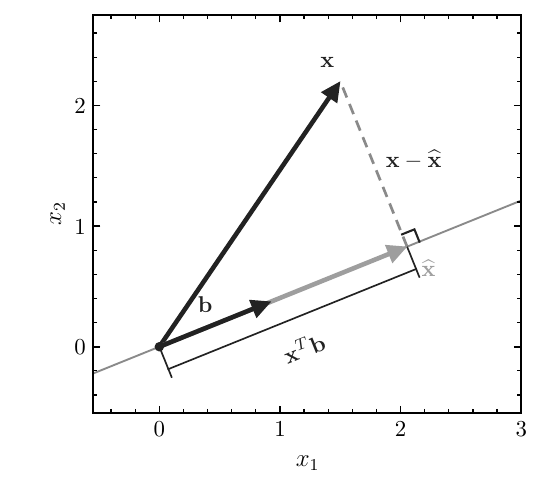}
    \caption{Projection $\widehat{\mathbf{x}}$ of the vector $\mathbf{x}$ onto the unit vector $\mathbf{b}$.}
    \label{fig:vector_projection}
\end{figure}

The scalar $\mathbf{x}^T\mathbf{b}$ measures how far $\mathbf{x}$ extends along $\mathbf{b}$, and multiplying it by $\mathbf{b}$ gives a vector along $\mathbf{b}$ of that length. In Figure~\ref{fig:vector_projection} the bracket along the line has length $\mathbf{x}^T\mathbf{b}$. The dashed segment from $\mathbf{x}$ to $\widehat{\mathbf{x}}$ is perpendicular to $\mathbf{b}$, so $\widehat{\mathbf{x}}$ is the point on the line closest to $\mathbf{x}$. The distance $\|\mathbf{x}-\widehat{\mathbf{x}}\|$ is the reconstruction\index{reconstruction error} error of this projection. This closest-point property matters for PCA, because we seek basis vectors that minimize the sum of squared reconstruction errors across the data points.

This concept extends to multiple dimensions through basis decomposition. With a complete set of orthonormal basis vectors $\{\mathbf{b}_1, \mathbf{b}_2, \ldots, \mathbf{b}_D\}$, which are mutually perpendicular unit vectors, we can write any vector $\mathbf{x}$ as
\begin{equation}
\mathbf{x} = \sum_{m=1}^D (\mathbf{x} \cdot \mathbf{b}_m)\mathbf{b}_m.
\end{equation}This representation uses $D$ coefficients, one for the projection onto each basis vector. In PCA we look for a basis in which most of the variance of the data lies along the first few basis vectors.

In matrix form, let $\mathbf{B}_M$ hold the first $M$ basis vectors as columns
\begin{equation}
\mathbf{B}_M =
\begin{bmatrix}
\mathbf{b}_1 & \mathbf{b}_2 & \cdots & \mathbf{b}_M
\end{bmatrix},
\end{equation}
and let $\mathbf{B}_R$ hold the remaining basis vectors
\begin{equation}
\mathbf{B}_R =
\begin{bmatrix}
\mathbf{b}_{M+1} & \mathbf{b}_{M+2} & \cdots & \mathbf{b}_D
\end{bmatrix}.
\end{equation}

The projection coefficients form the vectors
\begin{equation}
\mathbf{z}_M =
\begin{bmatrix}
\mathbf{x}^T\mathbf{b}_1 \\
\mathbf{x}^T\mathbf{b}_2 \\
\vdots \\
\mathbf{x}^T\mathbf{b}_M
\end{bmatrix}, \quad
\mathbf{z}_R =
\begin{bmatrix}
\mathbf{x}^T\mathbf{b}_{M+1} \\
\mathbf{x}^T\mathbf{b}_{M+2} \\
\vdots \\
\mathbf{x}^T\mathbf{b}_D
\end{bmatrix}.
\end{equation}

We treat $\mathbf{x}$ as a random vector, so the coefficients $\mathbf{z}_M$ and $\mathbf{z}_R$ are random vectors too. The coefficient $z_j = \mathbf{x}^T\mathbf{b}_j$ is the coordinate of $\mathbf{x}$ along $\mathbf{b}_j$. The data are centered, so the coefficients have zero expectation
\begin{equation}
\mathbb{E}[z_j] = \mathbb{E}[\mathbf{x}^T\mathbf{b}_j] = \mathbb{E}[\mathbf{x}]^T \mathbf{b}_j = \mathbf{0}^T \mathbf{b}_j = 0.
\end{equation}The variance of a coefficient is therefore its mean square value, $\mathbb{E}[z_j^2]$, which measures the spread of the data along $\mathbf{b}_j$.

With this notation the complete decomposition is
\begin{equation}
\mathbf{x} = \mathbf{B}_M\mathbf{z}_M + \mathbf{B}_R\mathbf{z}_R.
\end{equation}

PCA keeps only the first term, which gives the lower-dimensional approximation
\begin{equation}
\widehat{\mathbf{x}} = \mathbf{B}_M\mathbf{z}_M.
\end{equation}

\begin{figure}[ht!]
    \centering
    \includegraphics[width=0.85\textwidth]{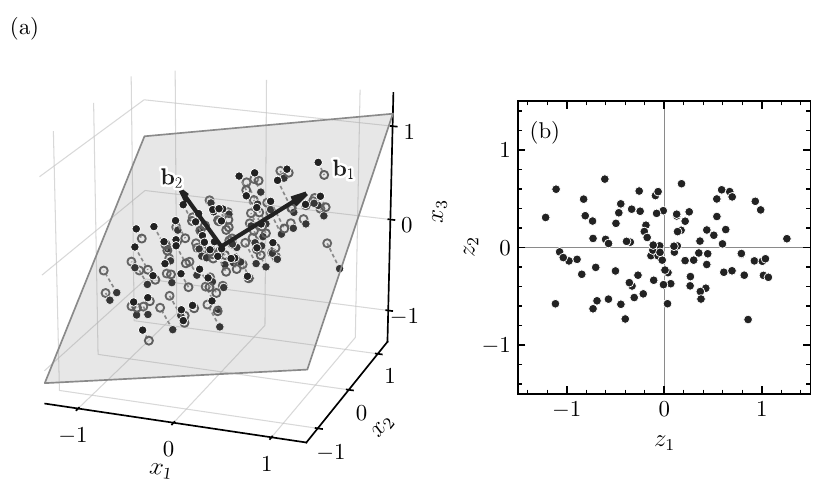}
    \caption{Reduction of three-dimensional data to two dimensions. Panel~(a) shows the data points and their projections onto the plane of the basis vectors $\mathbf{b}_1$ and $\mathbf{b}_2$. Panel~(b) shows the coefficients $z_1$ and $z_2$.}
    \label{fig:pca_dimension_reduction}
\end{figure}

The difference between the data and the approximation is the reconstruction error
\begin{equation}
\mathbf{x} - \widehat{\mathbf{x}} = \mathbf{B}_R\mathbf{z}_R.
\end{equation}Figure~\ref{fig:pca_dimension_reduction} shows an example with $D = 3$ and $M = 2$. The data in panel~(a) lie around a two-dimensional plane in three-dimensional space. The dashed segments are the reconstruction errors. PCA finds the lower-dimensional plane that best preserves the information in the data.

\paragraph{Equivalence of the Two Perspectives}

The equivalence follows from the total variance of the centered data
\begin{equation}
\text{Total Variance} = \mathbb{E}[\|\mathbf{x}\|^2] = \mathbb{E}[\mathbf{x}^T\mathbf{x}].
\end{equation}

In the orthonormal basis $\{\mathbf{b}_1, \ldots, \mathbf{b}_D\}$ we have $\|\mathbf{x}\|^2 = \sum_{m=1}^D z_m^2$, so the total variance becomes
\begin{equation}
\text{Total Variance} = \mathbb{E}\left[\sum_{m=1}^D z_m^2\right] = \sum_{m=1}^M \mathbb{E}[z_m^2] + \sum_{j=M+1}^D \mathbb{E}[z_j^2].
\end{equation}The first term is the variance carried by the first $M$ basis vectors. The second term is the expected squared reconstruction error. The total variance does not depend on the basis, so the two terms add up to a fixed total. The $M$ orthonormal directions that carry the most variance therefore leave the smallest reconstruction error.

\section{PCA as an Optimization Problem}

We now write this goal as an optimization problem. We need basis vectors $\mathbf{b}_1, \ldots, \mathbf{b}_M$ that carry the maximum variance and stay orthonormal. We start with the first principal component, the direction that holds the most variance. Once we solve this case, we can extend the approach to the later components.

Consider the projection of the data onto a vector $\mathbf{b}_1$. For a data point $\mathbf{x}_i$ the projection coefficient is $z_{1i} = \mathbf{x}_i^T\mathbf{b}_1 = \mathbf{b}_1^T\mathbf{x}_i$. We want the direction $\mathbf{b}_1$ that maximizes the variance of these coefficients
\begin{equation}
\text{Var}[z_1] = \frac{1}{N} \sum_{i=1}^N z_{1i}^2.
\end{equation}

For centered data the covariance matrix of Chapter~3 reduces to the second moment
\begin{equation}
\mathbf{S} = \text{Cov}(\mathbf{x}) = \mathbb{E}[\mathbf{x}\mathbf{x}^T] = \frac{1}{N}\sum_{i=1}^N \mathbf{x}_i\mathbf{x}_i^T = \frac{1}{N}\mathbf{X}^T\mathbf{X}.
\end{equation}We write $\mathbf{S}$ instead of the $\boldsymbol{\Sigma}$ of Chapter~3, because $\boldsymbol{\Sigma}$ later holds the singular values.

To connect this variance to the covariance matrix, we use the formula for the covariance of a linear transformation (Chapter~3). If $\mathbf{y} = \mathbf{A}\mathbf{x}$ for a random vector $\mathbf{x}$ and a matrix $\mathbf{A}$, then
\begin{equation}
\text{Cov}(\mathbf{y}) = \mathbf{A} \cdot \text{Cov}(\mathbf{x}) \cdot \mathbf{A}^T.
\end{equation}Here $z_1 = \mathbf{b}_1^T\mathbf{x}$ is such a transformation with $\mathbf{A} = \mathbf{b}_1^T$. The variable $z_1$ is a scalar, so its variance is
\begin{equation}
\text{Var}[z_1] = \mathbf{b}_1^T \cdot \text{Cov}(\mathbf{x}) \cdot \mathbf{b}_1 = \mathbf{b}_1^T \mathbf{S} \mathbf{b}_1.
\end{equation}This expression connects the variance of the projected data to $\mathbf{S}$. Finding the direction of maximum variance reduces to finding the vector $\mathbf{b}_1$ that maximizes the quadratic form $\mathbf{b}_1^T \mathbf{S} \mathbf{b}_1$.

\paragraph{The Need for Constraints}

Without a constraint the problem is ill-defined. Multiplying any vector $\mathbf{b}_1$ by a constant $c > 1$ gives the variance
\begin{equation}
\text{Var}[z_1'] = (c\mathbf{b}_1)^T \mathbf{S} (c\mathbf{b}_1) = c^2 \mathbf{b}_1^T \mathbf{S} \mathbf{b}_1 = c^2 \text{Var}[z_1].
\end{equation}The variance grows without limit as the length of $\mathbf{b}_1$ grows, while the direction stays the same. Only the direction matters for the axis of maximum variance, so we constrain the length.

We require $\mathbf{b}_1$ to be a unit vector
\begin{equation}
\|\mathbf{b}_1\|^2 = \mathbf{b}_1^T\mathbf{b}_1 = 1.
\end{equation}With unit length, the variance depends only on the direction of $\mathbf{b}_1$. The optimization problem becomes
\begin{equation}
\max_{\mathbf{b}_1} \mathbf{b}_1^T \mathbf{S} \mathbf{b}_1 \quad \text{subject to} \quad \mathbf{b}_1^T\mathbf{b}_1 = 1.
\end{equation}Finding the first principal component is a constrained optimization problem. Such problems appear in physics, engineering and economics and are typically solved with the method of Lagrange multipliers.

\section{Lagrange Multipliers}

The\index{Lagrange multipliers|(} method of Lagrange\index{Lagrange multipliers} multipliers finds candidates for the extrema of a function under constraints.

Without constraints we find extrema where the gradient vanishes, $\nabla f(\mathbf{x}) = 0$, because the function stops changing in all directions. Here $\mathbf{x}$ is the vector of all variables of $f$, and it is not the data vector of the previous sections. The function $f$ returns one number. We did this for maximum likelihood estimation in earlier chapters.

\looseness=1 With a constraint we may use only points that satisfy $g(\mathbf{x}) = 0$. A constraint $h(\mathbf{x}) = c$ has this form with $g(\mathbf{x}) = h(\mathbf{x}) - c$. Inequality constraints also exist, but PCA does not need them. The point where $\nabla f(\mathbf{x}) = 0$ may lie off the set $g(\mathbf{x}) = 0$, so we need a new condition.

The constraint $g(\mathbf{x}) = 0$ defines a constraint surface, and we must stay on it. For PCA the constraint is $g(\mathbf{b}_1) = \mathbf{b}_1^T\mathbf{b}_1 - 1 = 0$, which restricts the search to unit vectors. This constraint surface is the unit sphere, and it removes one degree of freedom.

\paragraph{Geometric Intuition}

We find the new condition geometrically. Consider a function $f(x,y)$ of two variables, whose maximum or minimum we want under the constraint $g(x,y) = 0$. We can picture $f$ as a surface with hills and valleys above the $(x,y)$ plane and the constraint as a path in that plane.

The path is one-dimensional, because $g(x,y) = 0$ removes one degree of freedom from the plane. For example, $g(x,y) = x + y - 1 = 0$ is a straight line. We look for the point on the path where $f$ reaches its extremum.

The gradient $\nabla f$ points in the direction of steepest ascent of $f$. If $\nabla f$ had a component along the path, we could stay on the constraint and increase $f$ by moving one way along the path. Moving the other way would decrease $f$. At a constrained extremum no such moves exist, so $\nabla f$ is perpendicular to the path.

\begin{figure}[ht!]
    \centering
    \includegraphics[width=0.6\textwidth]{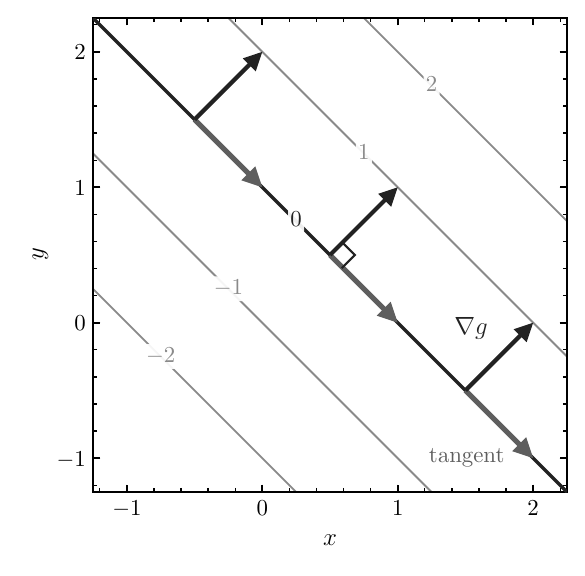}
    \caption{Contours of the constraint function $g(x,y) = x + y - 1$. The contour $g = 0$ is the constraint path. The gradient $\nabla g$ is perpendicular to the contours, including the path. The arrow labelled tangent points along the path.}
    \label{fig:constraint_gradient}
\end{figure}

But what direction is perpendicular to the constraint path? The function $g$ is constant, equal to zero, along the path, so a small step along the path leaves $g$ unchanged to first order. Such a step is perpendicular to $\nabla g$, the direction of steepest ascent of $g$. So $\nabla g$ is perpendicular to the path. Figure~\ref{fig:constraint_gradient} shows this for $g(x,y) = x + y - 1$, where the arrows of $\nabla g$ cross the contours at right angles.

At a constrained extremum, both gradients are perpendicular to the path. The path has one dimension fewer than the space, so the vectors perpendicular to it lie along one line. The two gradients are therefore parallel
\begin{equation}
\nabla f(\mathbf{x}) = \lambda \nabla g(\mathbf{x}),
\end{equation}
where $\lambda$ is the Lagrange multiplier, the scalar that relates them. This parallel condition is the core of the method.

\begin{figure}[ht!]
    \centering
    \includegraphics[width=0.62\textwidth]{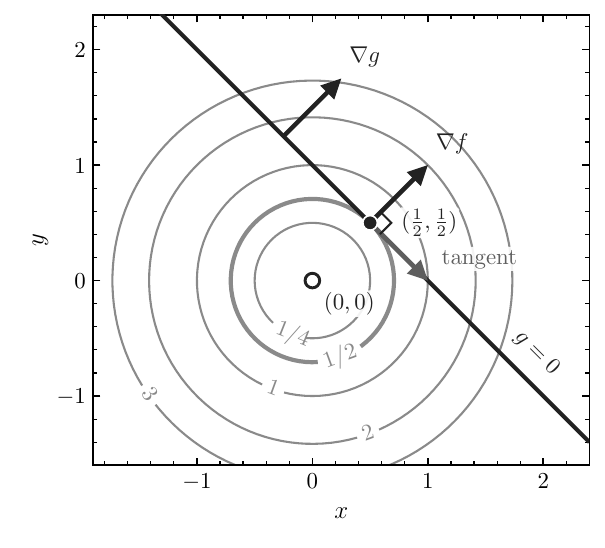}
    \caption{Contours of $f(x,y) = x^2 + y^2$ and the constraint line $g(x,y) = x + y - 1 = 0$. At the minimum of $f$ on the line, the line touches a contour, and $\nabla f$ and $\nabla g$ are parallel. The arrow labelled tangent points along the line.}
    \label{fig:lagrange_multipliers}
\end{figure}

Figure~\ref{fig:lagrange_multipliers} shows both gradients for $f(x,y) = x^2 + y^2$ and the line $g(x,y) = x + y - 1 = 0$. The figure also marks the unconstrained minimum at the origin, which the line does not contain.

\paragraph{The Lagrangian Formulation}

We turn the constrained problem into a search for critical points of a function with one extra variable, the multiplier $\lambda$. This function is the Lagrangian
\begin{equation}
\mathcal{L}(\mathbf{x}, \lambda) = f(\mathbf{x}) - \lambda g(\mathbf{x}).
\end{equation}Its critical points, where all partial derivatives vanish, are the candidate solutions of the constrained problem. We therefore solve an unconstrained problem in a slightly larger space.

The gradient with respect to $\mathbf{x}$ gives
\begin{equation}
\nabla_{\mathbf{x}} \mathcal{L} = \nabla f(\mathbf{x}) - \lambda \nabla g(\mathbf{x}) = 0.
\end{equation}
Rearranging gives $\nabla f(\mathbf{x}) = \lambda \nabla g(\mathbf{x})$ again, which agrees with the geometric argument.

The derivative with respect to $\lambda$ gives
\begin{equation}
\frac{\partial \mathcal{L}}{\partial \lambda} = -g(\mathbf{x}) = 0 \Rightarrow g(\mathbf{x}) = 0.
\end{equation}
This is the constraint itself, so the critical points lie on the constraint surface. Solving the two conditions together gives all candidates for constrained extrema.

\paragraph{A Simple Example}

We now solve the example of Figure~\ref{fig:lagrange_multipliers}. We want the minimum of $f(x,y) = x^2 + y^2$ under the constraint $g(x,y) = x + y - 1 = 0$. The Lagrangian is
\begin{equation}
\mathcal{L}(x,y,\lambda) = x^2 + y^2 - \lambda(x + y - 1).
\end{equation}

Setting the partial derivatives to zero gives
\begin{align}
\frac{\partial \mathcal{L}}{\partial x} &= 2x - \lambda = 0 \Rightarrow x = \frac{\lambda}{2}, \\*
\frac{\partial \mathcal{L}}{\partial y} &= 2y - \lambda = 0 \Rightarrow y = \frac{\lambda}{2}, \\*
\frac{\partial \mathcal{L}}{\partial \lambda} &= -(x + y - 1) = 0 \Rightarrow x + y = 1.
\end{align}The first two equations give $x = y$. The constraint then gives
\begin{equation}
x + x = 1 \Rightarrow x = y = \frac{1}{2}.
\end{equation}

The value at this point is $f(1/2, 1/2) = (1/2)^2 + (1/2)^2 = 1/2$. The Lagrange equations are necessary conditions. All constrained extrema satisfy them, but a point that satisfies them need not be an extremum. Here only one point satisfies them, namely $(1/2, 1/2)$. A minimum exists on the line, because $f$ grows without bound along it in both directions, so this point is the minimum. With several critical points we would need further analysis to find the extremum.

\paragraph{Multiple Constraints}

Lagrange multipliers extend to several constraints $g_1(\mathbf{x}) = 0$, $g_2(\mathbf{x}) = 0$, \ldots, $g_K(\mathbf{x}) = 0$. The condition becomes
\begin{equation}
\nabla f(\mathbf{x}) = \lambda_1 \nabla g_1(\mathbf{x}) + \lambda_2 \nabla g_2(\mathbf{x}) + \ldots + \lambda_K \nabla g_K(\mathbf{x}).
\end{equation}

With several constraints we are restricted to the intersection of all constraint surfaces, a surface of lower dimension called a manifold. At a constrained extremum, $\nabla f(\mathbf{x})$ must be perpendicular to this manifold. Otherwise we could move along the manifold, satisfy all constraints and change $f$.

\looseness=1 A direction tangent to the manifold is tangent to all constraint surfaces, so it is perpendicular to all constraint gradients $\nabla g_k(\mathbf{x})$. The vector $\nabla f(\mathbf{x})$ must therefore lie in the span of these gradients. It is a linear combination of them, with the multipliers $\lambda_k$ as coefficients.

The Lagrangian for this case is
\begin{equation}
\mathcal{L}(\mathbf{x}, \lambda_1, \ldots, \lambda_K) = f(\mathbf{x}) - \lambda_1 g_1(\mathbf{x}) - \ldots - \lambda_K g_K(\mathbf{x}).
\end{equation}
Setting its partial derivatives to zero gives the constraint equations and the gradient condition above.\index{Lagrange multipliers|)}

\section{Solving PCA with Lagrange Multipliers}\label{sec:solving_pca}

We\index{eigenvalue|(}\index{eigenvector|(} now apply the method of Lagrange multipliers to the first principal component. Recall that our objective is to maximize the variance of the projected data. In symbols, we maximize $f(\mathbf{b}_1) = \mathbf{b}_1^T\mathbf{S}\mathbf{b}_1$ under the constraint $g(\mathbf{b}_1) = \mathbf{b}_1^T\mathbf{b}_1 - 1 = 0$.

The Lagrangian of the problem is
\begin{equation}
\mathcal{L}(\mathbf{b}_1, \lambda_1) = \mathbf{b}_1^T\mathbf{S}\mathbf{b}_1 - \lambda_1(\mathbf{b}_1^T\mathbf{b}_1 - 1).
\end{equation}Setting the derivative with respect to $\lambda_1$ to zero gives
\begin{align}
\frac{\partial\mathcal{L}}{\partial\lambda_1} = -(\mathbf{b}_1^T\mathbf{b}_1 - 1) = 0.
\end{align}
This is the constraint $\mathbf{b}_1^T\mathbf{b}_1 = 1$, so $\mathbf{b}_1$ is a unit vector.

For the derivative with respect to $\mathbf{b}_1$ we need two results of matrix calculus. For a symmetric matrix $\mathbf{S}$, the gradient of $\mathbf{b}_1^T\mathbf{S}\mathbf{b}_1$ is $2\mathbf{S}\mathbf{b}_1$. The gradient of $\mathbf{b}_1^T\mathbf{b}_1$ is $2\mathbf{b}_1$. To see the first result, write the quadratic form as $\sum_{j,l} S_{jl} b_{1j} b_{1l}$. Its derivative with respect to $b_{1k}$ is $2\sum_j S_{kj} b_{1j}$, because $S_{jl} = S_{lj}$, and this is the $k$-th component of $2\mathbf{S}\mathbf{b}_1$. Therefore
\begin{align}
\frac{\partial\mathcal{L}}{\partial \mathbf{b}_1} = 2\mathbf{S}\mathbf{b}_1 - 2\lambda_1\mathbf{b}_1 = 0.
\end{align}
Dividing by 2 and rearranging gives
\begin{align}
\mathbf{S}\mathbf{b}_1 = \lambda_1\mathbf{b}_1.
\end{align}

\paragraph{The Eigenvalue Equation}

When $\mathbf{S}$ acts on $\mathbf{b}_1$, the result is $\mathbf{b}_1$ again, multiplied by $\lambda_1$. A vector with this property is called an eigenvector\index{eigenvector} of $\mathbf{S}$, and the multiplier $\lambda_1$ is its eigenvalue.\index{eigenvalue} Any solution $\mathbf{b}_1$ of these equations is therefore an eigenvector of $\mathbf{S}$. A search over all unit vectors becomes a search among the eigenvectors of $\mathbf{S}$, which we can compute with standard linear algebra.

\begin{figure}[ht!]
    \centering
    \includegraphics[width=0.58\textwidth]{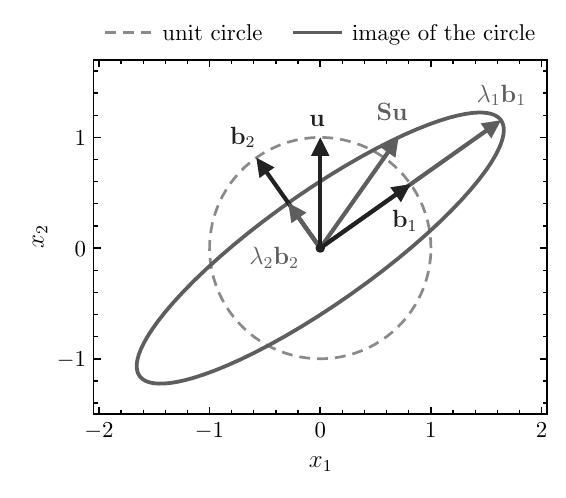}
    \caption{Image of the unit circle under a covariance matrix $\mathbf{S}$. The eigenvectors $\mathbf{b}_1$ and $\mathbf{b}_2$ keep their direction, while a vector $\mathbf{u}$ that is not an eigenvector turns.}
    \label{fig:eigenvector_direction_preservation}
\end{figure}

Eigenvectors of a covariance matrix have a geometric meaning. When we apply the matrix to the vectors of a unit circle, most vectors change both direction and length. An eigenvector keeps its direction, and the matrix only scales it by the eigenvalue. In Figure~\ref{fig:eigenvector_direction_preservation} the eigenvectors $\mathbf{b}_1$ and $\mathbf{b}_2$ of a two-dimensional covariance matrix keep their direction. The vector $\mathbf{u}$ is not an eigenvector, and it turns.

This is what the equation $\mathbf{S}\mathbf{b}_1 = \lambda_1\mathbf{b}_1$ says. This property makes eigenvectors good candidates for our principal components.

A covariance matrix $\mathbf{S}$ is real and symmetric, so its eigenvalues are real and eigenvectors of different eigenvalues are orthogonal. We can also choose $D$ orthonormal eigenvectors that span the $D$-dimensional space. Collecting them as the columns of $\mathbf{B}$ gives the spectral decomposition $\mathbf{S} = \mathbf{B}\boldsymbol{\Lambda}\mathbf{B}^T$. Here $\mathbf{B}$ is the $D \times D$ orthogonal matrix and $\boldsymbol{\Lambda}$ is the diagonal matrix of the eigenvalues. The principal components will come from these eigenvectors, so they form an orthonormal basis.

\paragraph{Connecting Eigenvalues to Variance}

Which of the $D$ orthonormal eigenvectors of $\mathbf{S}$ maximizes the variance?

To decide, we substitute an eigenvector $\mathbf{b}_1$ with eigenvalue $\lambda_1$ into the variance of the projected data
\begin{align}
\text{Var}[z_1] &= \mathbf{b}_1^T\mathbf{S}\mathbf{b}_1 \\*
&= \mathbf{b}_1^T(\lambda_1\mathbf{b}_1) \\
&= \lambda_1\mathbf{b}_1^T\mathbf{b}_1 \\*
&= \lambda_1.
\end{align}
In the last step we used $\mathbf{b}_1^T\mathbf{b}_1 = 1$, because $\mathbf{b}_1$ is a unit vector.

The same steps apply to any unit eigenvector of $\mathbf{S}$, so the variance along an eigenvector equals its eigenvalue. This connection between statistical variance and algebraic eigenvalues justifies our approach to PCA.

\begin{figure}[ht!]
    \centering
    \includegraphics[width=0.58\textwidth]{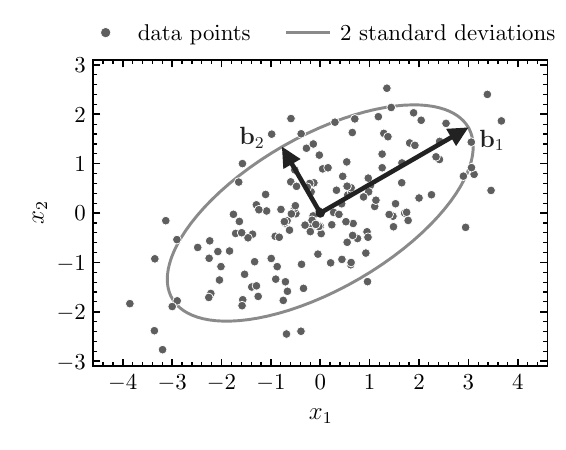}
    \caption{Data points, covariance ellipse and eigenvectors of the sample covariance matrix $\mathbf{S}$. The ellipse has semi-axes $2\sqrt{\lambda_m}$ along $\mathbf{b}_m$, and the arrows have the same lengths.}
    \label{fig:pca_covariance_eigenvectors}
\end{figure}

A maximizer of the variance is an eigenvector, and its variance equals its eigenvalue. The first principal component is therefore an eigenvector with the largest eigenvalue, and the data vary most along it. Figure~\ref{fig:pca_covariance_eigenvectors} shows the data and the covariance ellipse, which extends two standard deviations along each eigenvector, for a two-dimensional example. The eigenvector $\mathbf{b}_1$ lies along the long axis of the ellipse, where the data spread most. The second principal component is orthogonal to the first.\index{eigenvalue|)}\index{eigenvector|)}

\section{Higher-Order Principal Components}

The\index{eigenvalue|(}\index{eigenvector|(} first component rarely represents high-dimensional data well. We therefore look for further components that carry the remaining variance. The second principal component is a direction $\mathbf{b}_2$ with two properties. It carries the maximum variance that the first component has not explained. It is orthogonal to $\mathbf{b}_1$, so that $\mathbf{b}_1^T\mathbf{b}_2 = 0$.

Orthogonality is an assumption of PCA. With only the unit length constraint, the direction of maximum variance is $\mathbf{b}_1$ again, so the second direction needs another condition. Orthogonality keeps the calculation tractable and makes each principal component carry a unique aspect of the variation in the data.

\paragraph{Solving for the Second Component}

The second component solves another constrained optimization problem
\begin{equation}
\max_{\mathbf{b}_2} \mathbf{b}_2^T \mathbf{S} \mathbf{b}_2 \quad \text{subject to} \quad \mathbf{b}_2^T\mathbf{b}_2 = 1 \quad \text{and} \quad \mathbf{b}_1^T\mathbf{b}_2 = 0.
\end{equation}

We use Lagrange multipliers again, now with two constraints. The multiplier $\lambda_2$ belongs to the unit length constraint and $\mu$ to the orthogonality constraint
\begin{equation}
\mathcal{L}(\mathbf{b}_2, \lambda_2, \mu) = \mathbf{b}_2^T\mathbf{S}\mathbf{b}_2 - \lambda_2(\mathbf{b}_2^T\mathbf{b}_2 - 1) - \mu(\mathbf{b}_1^T\mathbf{b}_2).
\end{equation}Setting the derivative with respect to $\mathbf{b}_2$ to zero gives
\begin{equation}
\frac{\partial\mathcal{L}}{\partial \mathbf{b}_2} = 2\mathbf{S}\mathbf{b}_2 - 2\lambda_2\mathbf{b}_2 - \mu\mathbf{b}_1 = 0.
\end{equation}

To solve this equation, we multiply from the left by $\mathbf{b}_1^T$
\begin{equation}
\mathbf{b}_1^T(2\mathbf{S}\mathbf{b}_2 - 2\lambda_2\mathbf{b}_2 - \mu\mathbf{b}_1) = 0.
\end{equation}
Here $\mathbf{b}_1^T\mathbf{b}_2 = 0$ by the orthogonality constraint, and $\mathbf{b}_1^T\mathbf{b}_1 = 1$ because $\mathbf{b}_1$ is a unit vector. The equation becomes
\begin{equation}
2\mathbf{b}_1^T\mathbf{S}\mathbf{b}_2 - \mu = 0.
\end{equation}

Since $\mathbf{b}_1$ is an eigenvector of $\mathbf{S}$, we have $\mathbf{S}\mathbf{b}_1 = \lambda_1\mathbf{b}_1$. The matrix $\mathbf{S}$ is symmetric, so the transpose gives $\mathbf{b}_1^T\mathbf{S} = \lambda_1\mathbf{b}_1^T$. Substituting this gives
\begin{equation}
2\lambda_1\mathbf{b}_1^T\mathbf{b}_2 - \mu = 0.
\end{equation}
Here $\mathbf{b}_1^T\mathbf{b}_2 = 0$, so $\mu = 0$. The derivative equation then reduces to $2\mathbf{S}\mathbf{b}_2 - 2\lambda_2\mathbf{b}_2 = 0$, which gives
\begin{equation}
\mathbf{S}\mathbf{b}_2 = \lambda_2\mathbf{b}_2.
\end{equation}

So $\mathbf{b}_2$ is an eigenvector of the covariance matrix $\mathbf{S}$, and the orthogonality constraint makes it differ from $\mathbf{b}_1$. As before, the variance along $\mathbf{b}_2$ is the eigenvalue $\lambda_2$. To maximize it, we choose $\mathbf{b}_2$ as an eigenvector with the second largest eigenvalue of $\mathbf{S}$.

The pattern we observed extends to the later principal components. For the $m$-th component we would need to solve
\begin{equation}
\max_{\mathbf{b}_m} \mathbf{b}_m^T \mathbf{S} \mathbf{b}_m \quad \text{subject to} \quad \mathbf{b}_m^T\mathbf{b}_m = 1 \quad \text{and} \quad \mathbf{b}_k^T\mathbf{b}_m = 0 \quad \text{for all } k < m.
\end{equation}

\looseness=1 We expect that the $m$-th principal component is an eigenvector of $\mathbf{S}$ with the $m$-th largest eigenvalue. Checking this for each $m$ is long, so in the next section we prove it for all $m$ at once by induction.\index{eigenvalue|)}\index{eigenvector|)}

\section{A Proof by Induction}

Mathematical\index{eigenvector|(} induction proves a statement for all components $1, \ldots, D$ in one argument.

Induction has three steps.
\begin{enumerate}
    \item The base case proves the statement for the first value, usually $m=1$.
    \item The inductive hypothesis assumes the statement for all values below a value $M$.
    \item The inductive step proves that the statement then holds for $M$.
\end{enumerate}

If we complete all three steps, the statement holds for all values. Think of a row of dominoes. The first falls (base case), and each falls when the ones before it have fallen (inductive step), so all fall.

The $m$-th principal component is a unit vector that maximizes $\mathbf{b}^T\mathbf{S}\mathbf{b}$ among the vectors orthogonal to the first $m-1$ components. The statement to prove for all $m$ is that this vector is an eigenvector of $\mathbf{S}$ with the $m$-th largest eigenvalue. We list the eigenvalues in descending order, $\lambda_1 \geq \lambda_2 \geq \ldots \geq \lambda_D$.

\paragraph{Base Case}

We showed above that the first principal component $\mathbf{b}_1$ is an eigenvector of $\mathbf{S}$ with the largest eigenvalue $\lambda_1$.

\paragraph{Inductive Hypothesis}

Assume that for all $m < M$ the $m$-th principal component $\mathbf{b}_m$ is an eigenvector of $\mathbf{S}$ with the $m$-th largest eigenvalue $\lambda_m$.

\paragraph{Inductive Step}

We must show that the $M$-th principal component $\mathbf{b}_M$ is an eigenvector of $\mathbf{S}$ with the $M$-th largest eigenvalue $\lambda_M$.

Finding the $M$-th principal component is equivalent to finding the direction of maximum variance in the residual data. The residual data remain after we remove the projections onto the first $M-1$ components.

To find the projection of the data onto the first $M-1$ components, we recall that $\mathbf{x}$ projects onto a unit vector $\mathbf{b}_m$ as $(\mathbf{x}^T \mathbf{b}_m)\mathbf{b}_m$. Applying it to each row of $\mathbf{X}$ gives $\mathbf{X}\mathbf{b}_m\mathbf{b}_m^T$, the projection of all data points onto the $m$-th principal component.

The projection onto the subspace spanned by the first $M-1$ principal components is the sum
\begin{equation}
\sum_{m=1}^{M-1} \mathbf{X}\mathbf{b}_m\mathbf{b}_m^T.
\end{equation}Subtracting this projection from $\mathbf{X}$ leaves the residual data, the part not explained by these components
\begin{equation}
\mathbf{R} = \mathbf{X} - \sum_{m=1}^{M-1} \mathbf{X}\mathbf{b}_m\mathbf{b}_m^T.
\end{equation}

\begin{figure}[ht!]
    \centering
    \includegraphics[width=0.85\textwidth]{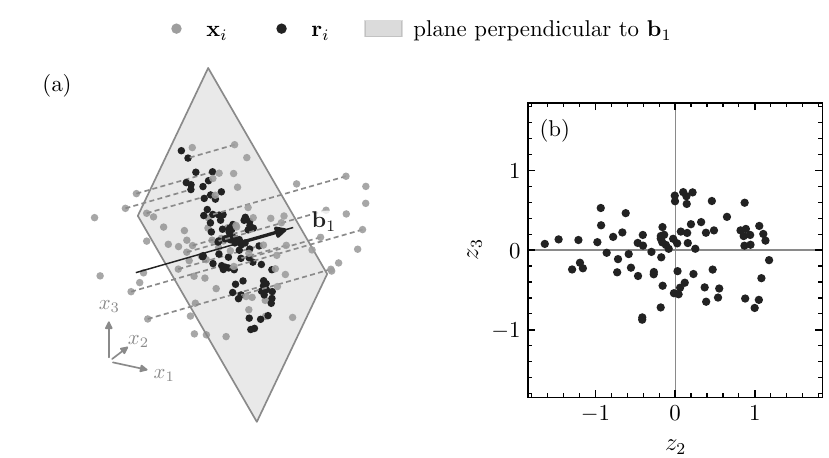}
    \caption{Residual data $\mathbf{R}$ of three-dimensional data after removing the projection onto the first principal component $\mathbf{b}_1$. Panel~(a) shows the data points $\mathbf{x}_i$, their residuals $\mathbf{r}_i$ and the plane through the origin perpendicular to $\mathbf{b}_1$. Panel~(b) shows the residuals in this plane, with the coordinates $z_2$ and $z_3$ along $\mathbf{b}_2$ and $\mathbf{b}_3$.}
    \label{fig:residual_data_subspace}
\end{figure}

Let $\mathbf{B}_{M-1} = [\mathbf{b}_1, \mathbf{b}_2, \ldots, \mathbf{b}_{M-1}]$ be the matrix whose columns are the first $M-1$ principal components. In matrix form the residual is
\begin{equation}
\mathbf{R} = \mathbf{X} - \mathbf{X}\mathbf{B}_{M-1}\mathbf{B}_{M-1}^T.
\end{equation}The matrix $\mathbf{B}_{M-1}\mathbf{B}_{M-1}^T$ projects onto the subspace spanned by the first $M-1$ components. Subtracting this projection leaves the part of each data point that is orthogonal to the subspace. The residual $\mathbf{r}_i$ of the data point $\mathbf{x}_i$ is the $i$-th row of $\mathbf{R}$, written as a column vector. Figure~\ref{fig:residual_data_subspace} shows a three-dimensional example with $M = 2$, where the first \mbox{$M-1 = 1$} component is removed.

To find the direction of maximum variance in the residual data, we need to compute its covariance matrix
\begin{equation}
\mathbf{S}_R = \frac{1}{N}\mathbf{R}^T\mathbf{R}.
\end{equation}Substituting $\mathbf{R}$ and expanding gives
\begin{align}
\mathbf{S}_R &= \frac{1}{N}(\mathbf{X} - \mathbf{X}\mathbf{B}_{M-1}\mathbf{B}_{M-1}^T)^T(\mathbf{X} - \mathbf{X}\mathbf{B}_{M-1}\mathbf{B}_{M-1}^T) \\*
&= \frac{1}{N}(\mathbf{X}^T - \mathbf{B}_{M-1}\mathbf{B}_{M-1}^T\mathbf{X}^T)(\mathbf{X} - \mathbf{X}\mathbf{B}_{M-1}\mathbf{B}_{M-1}^T) \\
&= \frac{1}{N}(\mathbf{X}^T\mathbf{X} - \mathbf{X}^T\mathbf{X}\mathbf{B}_{M-1}\mathbf{B}_{M-1}^T - \mathbf{B}_{M-1}\mathbf{B}_{M-1}^T\mathbf{X}^T\mathbf{X} \nonumber \\*
&\quad + \mathbf{B}_{M-1}\mathbf{B}_{M-1}^T\mathbf{X}^T\mathbf{X}\mathbf{B}_{M-1}\mathbf{B}_{M-1}^T).
\end{align}Since $(1/N)\mathbf{X}^T\mathbf{X} = \mathbf{S}$, the original covariance matrix, this becomes
\begin{align}
\mathbf{S}_R &= \mathbf{S} - \mathbf{S}\mathbf{B}_{M-1}\mathbf{B}_{M-1}^T - \mathbf{B}_{M-1}\mathbf{B}_{M-1}^T \mathbf{S} + \mathbf{B}_{M-1}\mathbf{B}_{M-1}^T \mathbf{S} \mathbf{B}_{M-1}\mathbf{B}_{M-1}^T.
\end{align}

By the inductive hypothesis, $\mathbf{B}_{M-1}$ holds the first $M-1$ eigenvectors of $\mathbf{S}$, so $\mathbf{S}\mathbf{B}_{M-1} = \mathbf{B}_{M-1}\boldsymbol{\Lambda}_{M-1}$, where $\boldsymbol{\Lambda}_{M-1}$ is the diagonal matrix of the first $M-1$ eigenvalues. Using this relation, the symmetry of $\mathbf{S}$ and $\mathbf{B}_{M-1}^T\mathbf{B}_{M-1} = \mathrm{Id}$, the identity matrix, we find that each of the last three terms equals $\mathbf{B}_{M-1}\boldsymbol{\Lambda}_{M-1}\mathbf{B}_{M-1}^T$. Their signs are minus, minus and plus, so they sum to $-\mathbf{B}_{M-1}\boldsymbol{\Lambda}_{M-1}\mathbf{B}_{M-1}^T$, which is $-\sum_{m=1}^{M-1}\lambda_m\mathbf{b}_m\mathbf{b}_m^T$. Therefore
\begin{equation}
\mathbf{S}_R = \mathbf{S} - \sum_{m=1}^{M-1} \lambda_m \mathbf{b}_m\mathbf{b}_m^T.
\end{equation}The residual covariance matrix is the original matrix $\mathbf{S}$ with the variance explained by the first $M-1$ components subtracted.

To understand this relation, we examine how $\mathbf{S}_R$ acts on the eigenvectors of $\mathbf{S}$. The first $M-1$ eigenvectors are $\mathbf{b}_1, \ldots, \mathbf{b}_{M-1}$, and the others are $\mathbf{b}_M, \ldots, \mathbf{b}_D$ with the eigenvalues $\lambda_M \geq \ldots \geq \lambda_D$. For one of the first $M-1$ eigenvectors, $\mathbf{b}_k$ with $k < M$, we find
\begin{align}
\mathbf{S}_R\mathbf{b}_k &= \mathbf{S}\mathbf{b}_k - \sum_{m=1}^{M-1} \lambda_m \mathbf{b}_m\mathbf{b}_m^T\mathbf{b}_k \\*
&= \lambda_k\mathbf{b}_k - \lambda_k \mathbf{b}_k = 0,
\end{align}
where we used the orthonormality of the eigenvectors.

For any other eigenvector, $\mathbf{b}_j$ with $j \geq M$, we find
\begin{align}
\mathbf{S}_R\mathbf{b}_j &= \mathbf{S}\mathbf{b}_j - \sum_{m=1}^{M-1} \lambda_m \mathbf{b}_m\mathbf{b}_m^T\mathbf{b}_j \\*
&= \lambda_j\mathbf{b}_j - 0 = \lambda_j\mathbf{b}_j.
\end{align}Therefore $\mathbf{b}_1, \ldots, \mathbf{b}_{M-1}$ have eigenvalue 0 in $\mathbf{S}_R$, and $\mathbf{b}_M, \ldots, \mathbf{b}_D$ have the same eigenvalues in $\mathbf{S}_R$ as in $\mathbf{S}$. In Figure~\ref{fig:residual_data_subspace} the residuals have no extent along $\mathbf{b}_1$, which matches the eigenvalue 0. In this figure, the largest eigenvalue of $\mathbf{S}_R$ is therefore the second largest eigenvalue of $\mathbf{S}$.

The eigenvalues of $\mathbf{S}_R$ are
\begin{equation}
\{0, 0, \ldots, 0, \lambda_M, \lambda_{M+1}, \ldots, \lambda_D\},
\end{equation}
where the first $M-1$ eigenvalues are zero and the others keep their values from $\mathbf{S}$. The $M$-th principal component is the direction of maximum variance of the residual data. By the result for the first component, it is therefore the eigenvector of $\mathbf{S}_R$ with the largest eigenvalue, which is $\lambda_M$. This eigenvector is $\mathbf{b}_M$, the eigenvector of $\mathbf{S}$ with the $M$-th largest eigenvalue.

This completes the induction. The principal components are the eigenvectors of $\mathbf{S}$ in the order of their eigenvalues, from the largest to the smallest.\index{eigenvector|)}

\section{Implementing PCA}\label{sec:ch10_implementing}

The derivation of PCA is long, but the algorithm is short. PCA takes six steps, and a single computation of the eigenvectors of $\mathbf{S}$ gives all principal components at once.

\begin{enumerate}
    \item Center the data matrix $\mathbf{X}$ of size $N \times D$.
    \item Compute the covariance matrix $\mathbf{S} = (1/N)\mathbf{X}^T\mathbf{X}$.
    \item Compute the eigenvectors and eigenvalues\index{eigenvalue} of $\mathbf{S}$.
    \item Sort the eigenvectors by decreasing eigenvalue.
    \item Take the first $M$ eigenvectors as the columns of $\mathbf{B}_M = [\mathbf{b}_1, \mathbf{b}_2, \ldots, \mathbf{b}_M]$.
    \item Project the data onto this basis with $\mathbf{Z} = \mathbf{X}\mathbf{B}_M$.
\end{enumerate}

\looseness=1 The matrix $\mathbf{Z}$ has size $N \times M$, and each row holds the coefficients of one data point in the principal component coordinates. These coefficients form the reduced representation. We can reconstruct approximations of the centered data with $\widehat{\mathbf{X}} = \mathbf{Z}\mathbf{B}_M^T = \mathbf{X}\mathbf{B}_M\mathbf{B}_M^T$, the projection of $\mathbf{X}$ onto the $M$-dimensional subspace of the principal components. The difference between $\widehat{\mathbf{X}}$ and $\mathbf{X}$ shows how much information the reduction loses.

\paragraph{Selecting the Number of Components}

We must decide how many principal components to keep. For $M = D$ the dimension does not decrease, and for a small $M$ we lose information about the data. Our theory gives a criterion for selecting $M$. By the induction proof, the principal components are eigenvectors of $\mathbf{S}$, and the variance along a component equals its eigenvalue (Section~\ref{sec:solving_pca}). The eigenvalues therefore rank the components by the variance they carry.

The total variance is the sum of the eigenvalues
\begin{equation}
\text{Total Variance} = \sum_{m=1}^D \lambda_m.
\end{equation}
The variance carried by the first $M$ components is
\begin{equation}
\text{Captured Variance} = \sum_{m=1}^M \lambda_m.
\end{equation}
The explained\index{explained variance} variance ratio of the first $M$ components is the fraction of the total variance that they carry
\begin{equation}
\text{Explained Variance Ratio} = \frac{\sum_{m=1}^M \lambda_m}{\sum_{m=1}^D \lambda_m}.
\end{equation}This ratio is a quantitative measure of how much information we retain after the dimension reduction. We can keep the smallest $M$ for which this ratio reaches a threshold, typically 0.90, 0.95 or 0.99. The threshold depends on the requirements of the application.

A scree\index{scree plot} plot shows the eigenvalues, or their ratios to the total variance, against their index. The curve usually falls steeply at first and then flattens. The elbow, where the slope changes, can suggest a cutoff for $M$.

\begin{figure}[ht!]
    \centering
    \includegraphics[width=0.85\textwidth]{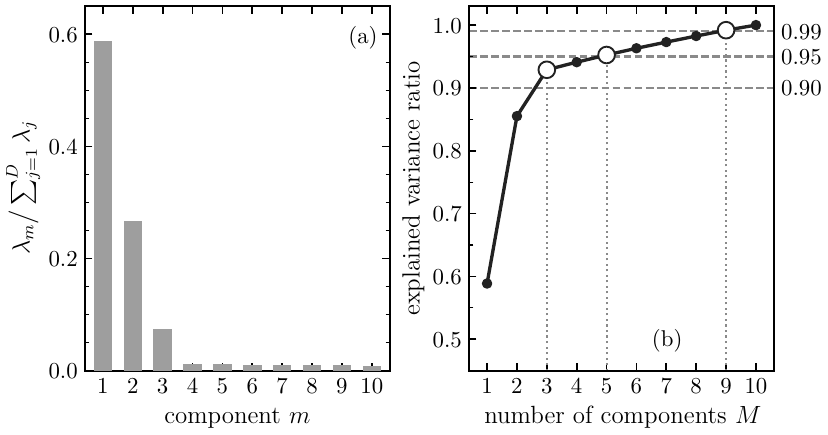}
    \caption{Scree plot for simulated spectra. Panel~(a) shows the ratio $\lambda_m / \sum_j \lambda_j$ of each eigenvalue to the total variance. Panel~(b) shows the explained variance ratio of the first $M$ components, with the smallest $M$ for each threshold marked.}
    \label{fig:scree_plot}
\end{figure}

Figure~\ref{fig:scree_plot} shows a scree plot in panel~(a) and the explained variance ratio of the first $M$ components in panel~(b). In panel~(a) the ratios fall steeply up to the third component and are nearly flat after it. In panel~(b) the thresholds 0.90, 0.95 and 0.99 need 3, 5 and 9 components.

\paragraph{Physical Interpretation of Components}

The output of PCA gives information beyond dimension reduction. The eigenvalues show how much variance a component carries and how important a direction of variation is. The eigenvectors can often be interpreted in terms of the original features, since they show which combinations of measurements contribute most to the variation.

A spectrum has one entry per wavelength bin, so an eigenvector can be plotted like a spectrum. For stellar spectra, the first component might follow the overall flux level or the temperature. Later components might follow the metallicity or particular absorption lines. This ability to interpret components makes PCA useful for exploratory data analysis in astronomy.

PCA can help to identify the physical processes behind astronomical data. If a few components explain most of the variance, as in Figure~\ref{fig:scree_plot}, a few underlying parameters may determine the variation.

\section{SVD for Efficient PCA}\label{sec:ch10_svd}

The\index{principal component analysis (PCA)!singular value decomposition|(}\index{singular value decomposition (SVD)|(} algorithm above needs the $D \times D$ covariance matrix and its eigenvectors. For high-dimensional data the cost of this step is large. A spectrum can have 10,000 wavelength bins, and an image can have tens of thousands of pixels. A time series can have thousands of time points. Diagonalizing a $D \times D$ matrix needs of the order of $D^3$ operations, written $\mathcal{O}(D^3)$, so the cost grows quickly with $D$. For $D = 10^4$ the covariance matrix has $10^8$ entries, which need about 800 megabytes of memory in double precision, where each number takes 8 bytes.

The SVD\index{SVD|see{singular value decomposition}}\index{singular value decomposition (SVD)} gives the same principal components at a lower cost when there are fewer objects than features. Many astronomical data sets have fewer objects than features.

\paragraph{The SVD Decomposition}

The SVD theorem states that any matrix $\mathbf{X} \in \mathbb{R}^{N \times D}$ is the product of three matrices
\begin{equation}
\mathbf{X} = \mathbf{U\Sigma V}^T.
\end{equation}The three matrices have these properties.
\begin{itemize}
    \item $\mathbf{U} \in \mathbb{R}^{N \times N}$ has orthonormal columns, the left singular vectors.
    \item $\boldsymbol{\Sigma} \in \mathbb{R}^{N \times D}$ is a rectangular matrix whose entries are zero except on the main diagonal. The diagonal entries $\Sigma_{mm} = \sigma_m$ are the singular values $\sigma_1 \geq \sigma_2 \geq \ldots \geq \sigma_{\min(N,D)} \geq 0$, and we set $\sigma_m = 0$ for $m > \min(N,D)$.
    \item $\mathbf{V} \in \mathbb{R}^{D \times D}$ has orthonormal columns, the right singular vectors.
\end{itemize}

Each column of $\mathbf{V}$ is a pattern of weights on the features. The matching column of $\mathbf{U}$ gives the relative strength of that pattern in each object, and $\sigma_m$ gives its overall strength.

Geometrically, the SVD splits any linear transformation into three simpler steps. The first is a rotation or reflection in the input space ($\mathbf{V}^T$). The second is a scaling along the coordinate axes ($\boldsymbol{\Sigma}$). The third is a rotation or reflection in the output space ($\mathbf{U}$).

\paragraph{Cost of the SVD}

The SVD of an $N \times D$ matrix takes about $\mathcal{O}(\min(ND^2, N^2D))$ operations when we compute only the first $\min(N,D)$ columns of $\mathbf{U}$ and $\mathbf{V}$, which PCA needs. This reduced form is called the economy-size SVD, and software routines provide it as an option. For a tall matrix with $N > D$ the cost is $\mathcal{O}(ND^2)$, and for a wide matrix with $N < D$ it is $\mathcal{O}(N^2D)$. In both cases we square the smaller dimension and multiply by the larger one. In this form, modern SVD routines adapt to the shape of the matrix automatically.

Take a spectroscopic survey with $N = 1{,}000$ spectra and $D = 10{,}000$ wavelength bins, so there are fewer objects than features. Diagonalizing the covariance matrix costs $\mathcal{O}(D^3) = \mathcal{O}(10^{12})$ operations, about 17 minutes at $10^9$ operations per second. The SVD costs $\mathcal{O}(N^2D) = \mathcal{O}(10^{10})$ operations, about 10 seconds.

The advantage grows as the number of objects falls. With $N = 100$ spectra and the same $D$, the SVD costs $\mathcal{O}(N^2D) = \mathcal{O}(10^8)$ operations, about 0.1 seconds. This is four orders of magnitude fewer than the $10^{12}$ operations of the diagonalization.

\paragraph{Connection to PCA}

With the centered data matrix $\mathbf{X} \in \mathbb{R}^{N \times D}$, PCA needs the covariance matrix $\mathbf{S} = (1/N)\mathbf{X}^T\mathbf{X}$.

Substituting the SVD of $\mathbf{X}$ into this expression gives
\begin{align}
\mathbf{S} &= \frac{1}{N}(\mathbf{U\Sigma V}^T)^T(\mathbf{U\Sigma V}^T) \\*
&= \frac{1}{N}\mathbf{V\Sigma}^T\mathbf{U}^T\mathbf{U\Sigma V}^T.
\end{align}Since $\mathbf{U}$ has orthonormal columns, $\mathbf{U}^T\mathbf{U} = \mathrm{Id}$, and the expression simplifies to
\begin{align}
\mathbf{S} &= \frac{1}{N}\mathbf{V\Sigma}^T\mathbf{\Sigma V}^T.
\end{align}The matrix $\boldsymbol{\Sigma}^T\boldsymbol{\Sigma}$ is diagonal with the entries $\sigma_m^2$, where $\sigma_m$ are the singular values, because $\boldsymbol{\Sigma}$ is a rectangular diagonal matrix. Therefore
\begin{align}
\mathbf{S} &= \mathbf{V}\left(\frac{1}{N}\boldsymbol{\Sigma}^T\boldsymbol{\Sigma}\right)\mathbf{V}^T \\*
&= \mathbf{V}\left(\frac{1}{N}\text{diag}(\sigma_1^2, \sigma_2^2, \ldots, \sigma_D^2)\right)\mathbf{V}^T.
\end{align}

This is the spectral decomposition $\mathbf{S} = \mathbf{B}\boldsymbol{\Lambda}\mathbf{B}^T$ of Section~\ref{sec:solving_pca} with $\mathbf{B} = \mathbf{V}$ and $\boldsymbol{\Lambda} = \boldsymbol{\Sigma}^T\boldsymbol{\Sigma}/N$. The columns of $\mathbf{V}$ are therefore the eigenvectors of $\mathbf{S}$, which are the principal components, and the eigenvalues are $\lambda_m = \sigma_m^2/N$. Larger singular values belong to directions of larger variance. The singular values are in descending order, so the eigenvalues are in descending order too. This matches the order of the principal components by explained variance.

\paragraph{Practical Implementation}

This relation lets us compute PCA from the SVD without forming the covariance matrix, in four steps.
\begin{enumerate}
    \item Center the data matrix $\mathbf{X}$.
    \item Compute its SVD $\mathbf{X} = \mathbf{U\Sigma V}^T$.
    \item Use the columns of $\mathbf{V}$ as the principal components.
    \item Project the data onto these components to get the lower-dimensional representation.
\end{enumerate}

To reduce the dimension to $M < D$, let $\mathbf{V}_M$ hold the first $M$ columns of $\mathbf{V}$. The projection of the data is
\begin{equation}
\mathbf{Z} = \mathbf{X} \mathbf{V}_M.
\end{equation}

The SVD already contains the coefficients, so we can obtain them without the matrix product $\mathbf{X}\mathbf{V}_M$. Substituting $\mathbf{X} = \mathbf{U\Sigma V}^T$ gives
\begin{equation}
\mathbf{Z} = \mathbf{X} \mathbf{V}_M = (\mathbf{U\Sigma V}^T) \mathbf{V}_M.
\end{equation}Since $\mathbf{V}$ is orthogonal, $\mathbf{V}^T\mathbf{V}_M$ is the $D \times M$ matrix with the identity $\mathrm{Id}_M$ on top and zeros below. It selects the first $M$ columns of $\mathbf{U\Sigma}$, so
\begin{equation}
\mathbf{Z} = \mathbf{U}_M \boldsymbol{\Sigma}_M.
\end{equation}
Here $\mathbf{U}_M$ holds the first $M$ columns of $\mathbf{U}$, and $\boldsymbol{\Sigma}_M$ is the diagonal matrix of the first $M$ singular values.

Once the SVD is known, the coefficients $\mathbf{Z} = \mathbf{U}_M \boldsymbol{\Sigma}_M$ need only a scaling of the columns of $\mathbf{U}_M$, which costs $\mathcal{O}(NM)$ operations. Computing $\mathbf{X}\mathbf{V}_M$ directly multiplies an $N \times D$ matrix by a $D \times M$ matrix and costs $\mathcal{O}(NDM)$ operations.

\paragraph{Reconstruction and Explained Variance}

PCA also lets us reconstruct an approximation of the data from the reduced representation. In terms of the SVD of the centered data matrix $\mathbf{X}$, the coefficients are $\mathbf{Z} = \mathbf{U}_M\boldsymbol{\Sigma}_M$ and the reconstruction of the centered data is $\widehat{\mathbf{X}} = \mathbf{Z}\mathbf{V}_M^T = (\mathbf{U}_M \boldsymbol{\Sigma}_M)\mathbf{V}_M^T$. Adding the mean vector $\overline{\mathbf{x}}$ to each row of $\widehat{\mathbf{X}}$ gives the reconstruction in the original space.

To choose the number of components to retain, we can also use the singular values to calculate the proportion of variance explained by a component. For the $m$-th component it is
\begin{equation}
\text{Explained Variance Ratio}_m = \frac{\sigma_m^2}{\sum_{j=1}^D \sigma_j^2}.
\end{equation}This equals $\lambda_m / \sum_j \lambda_j$, because $\lambda_m = \sigma_m^2/N$. Summing the first $M$ of these ratios gives the explained variance ratio of the first $M$ components. By examining this ratio as we add components, we can find how many components capture a desired percentage of the total variance.

The SVD is also more accurate, even when $N > D$. Forming $\mathbf{S}$ squares the ratio of the largest to the smallest singular value, so rounding errors affect the small eigenvalues more. The SVD of $\mathbf{X}$ avoids this squaring. For these reasons the SVD is the usual choice for computing PCA of high-dimensional data.\index{principal component analysis (PCA)!singular value decomposition|)}\index{singular value decomposition (SVD)|)}

\section{Limitations of PCA}\label{sec:limitations}

PCA has four limits that matter for astronomical data.

\paragraph{Orthogonality Constraint}

PCA requires orthogonal basis vectors, but the directions in feature space along which different physical processes change the data need not be orthogonal. In stellar spectra, for example, temperature and metallicity change many of the same absorption lines, so orthogonal axes can mix them. The first few components carry the most variance, but each can mix several physical parameters. Because the components are orthogonal, one physical parameter can also spread over several components. We should therefore be careful when we give a single component a physical meaning.

\paragraph{Unimodal Assumption}

PCA describes the data only through its mean and covariance matrix. The mean and the covariance matrix describe a single, roughly elliptical cluster well. For several clusters PCA gives one common set of directions. Data with several clusters have a multimodal distribution, which has several peaks, or modes (Chapter~9). Two clusters make it bimodal.

\begin{figure}[ht!]
    \centering
    \includegraphics[width=0.85\textwidth]{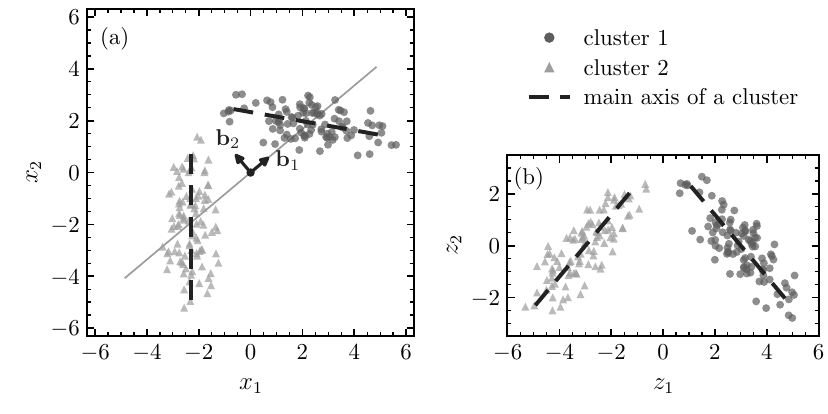}
    \caption{PCA for two clusters of points. Panel~(a) shows the data, the main axis of each cluster, the line through the cluster centers and the principal components $\mathbf{b}_1$ and $\mathbf{b}_2$. Panel~(b) shows the coefficients $z_1$ and $z_2$ of the same points.}
    \label{fig:pca_multimodal}
\end{figure}

Galaxies fall into types such as spiral, elliptical and irregular, and stars form separate groups in parameter space. For such data PCA finds one set of components for all the points together. These components follow the largest total variance, which can differ from the main axis of each cluster. In Figure~\ref{fig:pca_multimodal} the first component follows the line between the two clusters, and the main axis of each cluster points in another direction.

The problem is largest when the variance within the clusters is smaller than the variance between them, as in the figure. As panel~(b) shows, the first principal component mainly carries the variation between the clusters. If we keep only this component, we discard the spread along $z_2$ within each cluster. Such a loss can hide relations inside a single population.

\paragraph{Sensitivity to Outliers}

Because PCA maximizes variance, it is sensitive to outliers.\index{outliers} A few extreme points can change the directions of maximum variance, so that the components describe the outliers instead of the typical pattern.

Astronomical data often contain outliers, and many of them are of scientific interest. They can be rare object types, measurement errors or unusual phenomena, and any of them can affect PCA. We can inspect the outliers first. We can remove those that are errors, or use a variant of PCA that gives outliers less weight.

\paragraph{Scale Dependence}

PCA depends on the relative scales\index{standardization} of the features. A feature with a large numerical range contributes more to the covariance matrix. It can dominate the components, even if it is not the most important feature physically. Changing the unit of one feature changes its variance, and therefore changes the components.

Astronomical measurements often have very different scales, for example magnitudes from 10 to 30 and parallaxes from 0.001 to 0.1. Without rescaling, features with a large range can hide relations in features with a small range. Dividing each centered feature by its standard deviation is a common rescaling, called standardizing.

\paragraph{Alternatives to PCA}

Other methods address some of these limits. Independent\index{independent component analysis} component analysis (ICA) looks for components that are statistically independent, in place of components of maximum variance. Independent variables have zero mutual information (Chapter~8). The components of ICA need not be orthogonal. When the underlying processes are independent, the components can be easier to interpret physically. In astronomy the method can separate the sources in mixed signals.

Autoencoders are neural networks that describe each object with a few numbers and map it back, and they generalize PCA to nonlinear maps (Chapter~15). They need not keep the orthogonality constraint and can represent more complex relations in the data. Their representations may be easier to interpret physically. They also need more computing time and can fit the noise in the data.

For data with clear clusters, we can first find the clusters with a clustering method (Chapter~11) and then apply PCA to each cluster. Each cluster then has its own mean and its own components. This respects the cluster structure of the data and still reduces the dimension within each population.

\medskip PCA describes the data with linear combinations of the features, but astronomical data often form groups. In Chapter~11 we turn to clustering.\index{principal component analysis (PCA)|)}

\section{Tutorial: Principal Components of Galaxies}\label{sec:ch10_tutorial}

This\index{galaxy images|(} tutorial applies principal component analysis (PCA) and the singular value decomposition (SVD) of this chapter to images of galaxies.\index{galaxy images}\index{galaxy morphology|see{galaxy images}} Galaxy shapes depend on how galaxies formed and evolved. The full tutorial can be found at \url{https://statml.org/} as tutorial 10.

\begin{figure}[t!]
    \centering
    \includegraphics[width=0.95\textwidth]{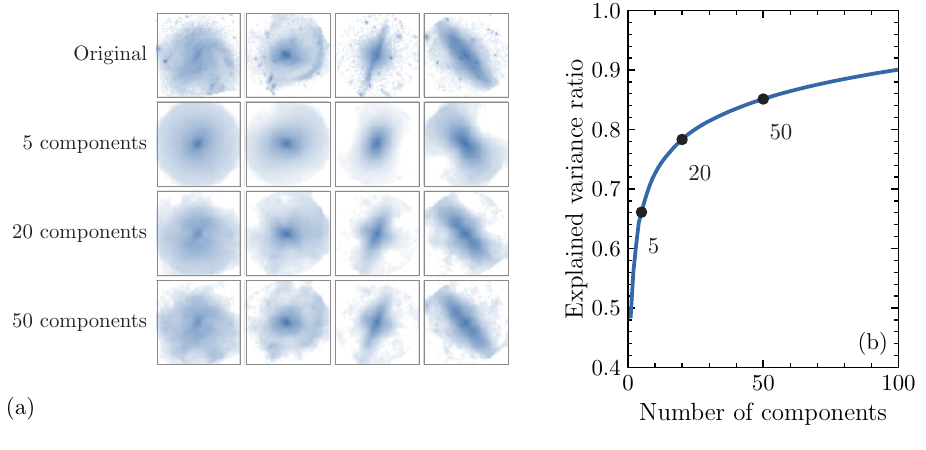}
    \caption{The tutorial describes a sample of real galaxy images with principal components. Panel~(a) shows four selected galaxies and their reconstructions from 5, 20 and 50 components, with darker blue for more light. Panel~(b) shows the share of the sample's variance that a given number of components explains, with 5, 20 and 50 marked. A few components keep the elongation and orientation of a galaxy.}
    \label{fig:ch10_tutorial_pca}
\end{figure}

Mergers tend to make smooth ellipticals, while gas that settles into a rotating disk forms disk galaxies. A survey takes images of more galaxies than we can examine, and each image has thousands of pixels (top row of Figure~\ref{fig:ch10_tutorial_pca}a). We therefore want a few numbers per galaxy, and the images can show which shapes occur.

Learning from images alone is unsupervised learning. The target was a regression value in Section~\ref{sec:ch4_tutorials} and a class in Section~\ref{sec:ch7_tutorial}. Here the images have no target to say which pixels matter, so their variation guides us. We keep the directions in the space of pixels along which the images vary most (Section~\ref{sec:ch10_fundamentals}), because these keep as much information as possible. If a few processes shape the galaxies, most of the variation may lie along a few of these directions.

PCA finds these directions, the principal components, as the eigenvectors of the covariance matrix (Section~\ref{sec:solving_pca}). The tutorial's sample has fewer galaxies than pixels, so the SVD of Section~\ref{sec:ch10_svd} finds them without the large covariance matrix of the pixels.

The SVD gives the full set of components, so we must still decide how many to keep. Without a target there is no prediction error to guide this choice, so we ask how well the first $M$ components reconstruct each galaxy. The variance that they miss is the average squared reconstruction error (Section~\ref{sec:ch10_fundamentals}). The eigenvalues rank the components by variance, so the first $M$ components miss the least. The share of the total variance that they keep is the explained variance ratio (Section~\ref{sec:ch10_implementing}).

A few components reproduce the overall elongation and orientation of a galaxy, and more add detail (Figure~\ref{fig:ch10_tutorial_pca}a). The explained variance ratio rises steeply at first and slowly after, so later components add less (Figure~\ref{fig:ch10_tutorial_pca}b). The number to keep therefore depends on how much detail the application needs.

A physical reading of the components needs caution, because PCA chooses orthogonal directions by variance. A few directions with much variance do not show that a few processes shape the galaxies. Because galaxies have different orientations, elongation changes different pixels in each image, so it can spread over several orthogonal components (Section~\ref{sec:limitations}). Plotted as an image, a component shows a way in which the galaxies vary, and it need not match a physical property.

Once we choose how many components to keep, a galaxy has a few coefficients, its coordinates along them. These summarize it approximately, which is what we wanted. A regression or classification on them has fewer weights than one on the pixels, which can lower the overfitting of Section~\ref{sec:ch4_tutorials}. The same approach applies to spectra and light curves, where a compact description lets us study large populations together.\index{galaxy images|)}

\paragraph{Further Reading} Pearson (1901) found lines and planes of closest fit to points in space. Hotelling (1933) developed principal components as the linear combinations of the variables that carry the most variance. Eckart and Young (1936) showed that the SVD truncated to its largest singular values gives the best least squares approximation with fewer components. Golub and Kahan (1965) and Golub and Reinsch (1970) developed stable algorithms for the SVD, which remain influential in PCA software. Anderson (1958) treats the theory of PCA within multivariate statistics, and Murtagh and Heck (1987) present PCA for astronomers. Jolliffe (2002) covers the theory, variants and uses of PCA. Comon (1994) treats independent component analysis, an alternative for non-Gaussian data. Hinton and Salakhutdinov (2006) explored nonlinear dimension reduction with autoencoders, which go beyond linear PCA. The galaxy images of the tutorial are SDSS images from Galaxy Zoo 2, as in Ma et~al.\ (2026).

\chapter{K-means and Gaussian Mixture Models}

\makeatletter\@clubpenalty=10000\makeatother \clubpenalty=10000 \widowpenalty=10000

Astronomical objects often fall into groups that we do not know in advance. Examples are the types of galaxies and the stellar populations of a globular cluster. Clustering\index{clustering} divides a set of objects into such groups using only their features.

\begin{figure}[ht!]
    \centering
    \includegraphics[width=0.9\textwidth]{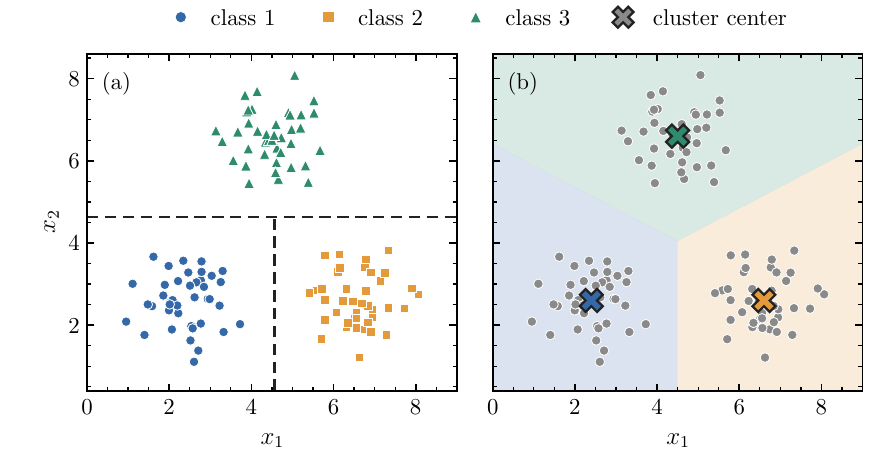}
    \caption{Classification (supervised learning) and clustering (unsupervised learning) of data with two features $x_1$ and $x_2$. Panel~(a) shows three labeled classes and the boundaries that a classifier learns from the labels. Panel~(b) shows the same points without labels. A clustering algorithm finds the groups and their centers from the positions alone, and each shaded region contains the points nearest to one center.}
    \label{fig:clustering_vs_classification}
\end{figure}

Clustering differs from the classification of Chapter~7. A classifier learns the boundaries between classes from objects whose classes are known, which is supervised learning. Clustering is unsupervised learning, so it uses no labels and looks at the data itself to find groups. This suits discovery science, where we may not know which categories to expect. Figure~\ref{fig:clustering_vs_classification} compares the two tasks.

Clustering has three advantages in astronomy. It needs no preconceived categories, so it can reveal unexpected structure. It handles catalogs that are too large for manual classification. It can also find finer differences than the traditional schemes that astronomers developed by visual inspection.

Clustering also has difficulties, because evaluating results without ground truth labels is more subjective than in supervised learning. The number of clusters often depends on the scientific question as well as on the data. Different algorithms can give different clusters for the same data, so we must know their assumptions.

Clustering rests on a simple premise, that similar objects probably belong to the same group and dissimilar objects to different groups. Different definitions of similarity lead to different algorithms.

K-means defines similarity as the distance to a cluster center, which is a geometric view. Gaussian mixture models (GMMs) describe the data with a probability distribution and give each object a probability of belonging to each cluster.

\section{Mathematical Formulation of K-means}

K-means\index{K-means clustering}\index{K-means clustering|(} divides the data into $K$ groups by assigning each data point to the group with the nearest center.

Let the data set consist of $N$ observations $\{\mathbf{x}_1, \mathbf{x}_2, \ldots, \mathbf{x}_N\}$, where each $\mathbf{x}_i$ is a $D$-dimensional feature vector. In a study of stellar populations, the features can be the metallicity and the carbon and nitrogen abundances measured from spectra.

K-means finds $K$ cluster centers $\{\boldsymbol{\mu}_1, \boldsymbol{\mu}_2, \ldots, \boldsymbol{\mu}_K\}$ and assigns each data point to one cluster. The binary indicator $r_{ik} \in \{0, 1\}$ records the assignment, with $r_{ik} = 1$ if $\mathbf{x}_i$ belongs to cluster $k$ and $r_{ik} = 0$ otherwise.

\looseness=1 The values $r_{i1}$ to $r_{iK}$ of a point form the one-hot encoding of Chapter~8, with a single entry equal to 1. For $K=3$ clusters and a data point $\mathbf{x}_5$ in cluster 2, we have $r_{51} = 0$, $r_{52} = 1$ and $r_{53} = 0$. Each point belongs to one cluster, so $\sum_{k=1}^K r_{ik} = 1$ for all $i$.

\paragraph{The K-means Objective Function}

The goal of K-means is to find centers and assignments that make the clusters as compact as possible. We measure compactness by the sum of squared distances between each data point and its assigned center
\begin{equation}
J = \sum_{i=1}^N \sum_{k=1}^K r_{ik} \|\mathbf{x}_i - \boldsymbol{\mu}_k\|^2.
\end{equation}

The quantity $\|\mathbf{x}_i - \boldsymbol{\mu}_k\|^2$ is the squared Euclidean distance between $\mathbf{x}_i$ and $\boldsymbol{\mu}_k$. The factor $r_{ik}$ selects only the distance to the assigned center, because $r_{ik} = 1$ for one $k$ only. Each data point therefore adds one squared distance to $J$. A smaller $J$ means that the points lie closer to their centers, so the clusters are more compact. The Euclidean distance depends on the units of the features. For features with different units, such as colors and morphological indices, we first rescale each feature, for example by its standard deviation (Chapter~10).

\paragraph{The Joint Optimization Problem}

The optimization problem is
\begin{equation}
\min_{\{r_{ik}\}, \{\boldsymbol{\mu}_k\}} J = \sum_{i=1}^N \sum_{k=1}^K r_{ik} \|\mathbf{x}_i - \boldsymbol{\mu}_k\|^2
\end{equation}
subject to $r_{ik} \in \{0, 1\}$ and $\sum_{k=1}^K r_{ik} = 1$ for all $i$.

In regression and classification we optimized one set of parameters. Here we must optimize two sets of variables at once, the cluster centers $\{\boldsymbol{\mu}_k\}$ and the assignments $\{r_{ik}\}$.

\begin{figure}[ht!]
    \centering
    \includegraphics[width=0.65\textwidth]{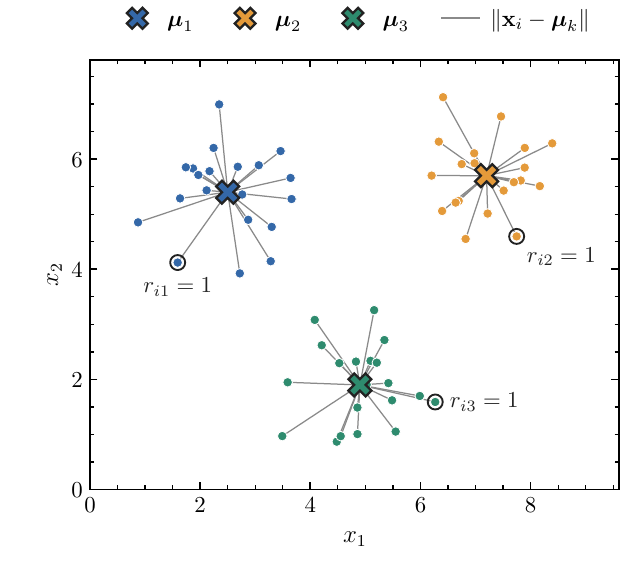}
    \caption{The two sets of variables of K-means for three clusters. The three labeled points belong to clusters 1, 2 and 3, so each has $r_{ik} = 1$ for its own cluster $k$. Each line joins a data point to its assigned center, and $J$ is the sum of the squared line lengths over all data points.}
    \label{fig:kmeans_optimization}
\end{figure}

The two sets of variables have a circular dependency. If we knew the optimal centers, the best assignments would be easy, because each point goes to its nearest center. If we knew the optimal assignments, finding the best centers would be easy, because each center would be the mean of its assigned points. Figure~\ref{fig:kmeans_optimization} shows the two sets of variables and the distances that enter $J$.

This circular dependency makes direct optimization difficult. The binary constraint on $r_{ik}$ adds a combinatorial problem. For $N$ points and $K$ clusters there are $K^N$ possible assignments, a number that grows exponentially with $N$.

The problem has $N \times K$ binary assignment variables and $K \times D$ continuous center coordinates. An exhaustive search over all assignments is too expensive for data sets of realistic size. Gradient descent does not work either, because the binary nature of $r_{ik}$ prevents the use of continuous optimization techniques.

\section{Expectation-Maximization for K-means}\label{sec:ch11_kmeans_em}

We\index{expectation-maximization (EM) algorithm|(} can break the circular dependency by alternating between optimizing one set of variables with the other fixed. This is the Expectation-Maximization\index{expectation-maximization (EM) algorithm} (EM)\index{EM algorithm|see{expectation-maximization}} algorithm. It turns the problem into a sequence of simpler problems that have closed-form solutions.

The procedure is a coordinate descent. It moves alternately along the assignment variables and along the center coordinates. Each step minimizes $J$ over one set of variables, so $J$ decreases or stays the same.

The EM algorithm for K-means has four steps.
\begin{itemize}
\item Choose $K$ initial cluster centers $\boldsymbol{\mu}_k$. Random data points are a simple choice, though more sophisticated strategies often work better.

\item The E-step fixes the centers $\boldsymbol{\mu}_k$ and finds the optimal assignments $r_{ik}$.

\item The M-step fixes the assignments $r_{ik}$ and finds the optimal centers $\boldsymbol{\mu}_k$.

\item Repeat the E-step and the M-step until convergence.
\end{itemize}

The algorithm has converged when the E-step changes no assignment. The names of the steps come from the mixture view of Section~11.6, where the E-step computes expected assignments and the M-step maximizes a log-likelihood. For K-means, maximizing the log-likelihood means minimizing $J$ (Section~11.4).

\paragraph{The Expectation Step}

In the E-step the centers $\boldsymbol{\mu}_k$ are fixed, and we minimize $J$ over the assignments
\begin{equation}
\min_{\{r_{ik}\}} J = \sum_{i=1}^N \sum_{k=1}^K r_{ik} \|\mathbf{x}_i - \boldsymbol{\mu}_k\|^2
\end{equation}
subject to $r_{ik} \in \{0, 1\}$ and $\sum_{k=1}^K r_{ik} = 1$ for all $i$.

With fixed centers the problem is simple, because for each data point we choose which cluster it belongs to. Point $\mathbf{x}_i$ belongs to one cluster, so one $r_{ik}$ equals 1 and the others are 0. Its contribution to the objective is
\begin{equation}
\sum_{k=1}^K r_{ik} \|\mathbf{x}_i - \boldsymbol{\mu}_k\|^2 = \|\mathbf{x}_i - \boldsymbol{\mu}_{k^*}\|^2,
\end{equation}
where $k^*$ is the cluster of point $i$.

This contribution is smallest when $k^*$ is the index of the center closest to $\mathbf{x}_i$. The optimal assignment is therefore
\begin{equation}
r_{ik} = \begin{cases}
1 & \text{if } k = \arg\min_{j} \|\mathbf{x}_i - \boldsymbol{\mu}_j\|^2 \\
0 & \text{otherwise}
.\end{cases}
\end{equation}The E-step therefore assigns each data point to its nearest center.

\paragraph{The Maximization Step}

In the M-step the assignments $r_{ik}$ are fixed, and we minimize $J$ over the centers
\begin{equation}
\min_{\{\boldsymbol{\mu}_k\}} J = \sum_{i=1}^N \sum_{k=1}^K r_{ik} \|\mathbf{x}_i - \boldsymbol{\mu}_k\|^2.
\end{equation}With fixed assignments we group the terms of the sum by cluster
\begin{equation}
J = \sum_{k=1}^K \sum_{i=1}^N r_{ik} \|\mathbf{x}_i - \boldsymbol{\mu}_k\|^2 = \sum_{k=1}^K J_k,
\end{equation}
where $J_k = \sum_{i=1}^N r_{ik} \|\mathbf{x}_i - \boldsymbol{\mu}_k\|^2$ is the contribution of cluster $k$.

Each $J_k$ depends only on $\boldsymbol{\mu}_k$, so we optimize each center independently. For cluster $k$ we minimize
\begin{equation}
J_k = \sum_{i=1}^N r_{ik} \|\mathbf{x}_i - \boldsymbol{\mu}_k\|^2.
\end{equation}

\begin{figure}[ht!]
    \centering
    \includegraphics[width=0.8\textwidth]{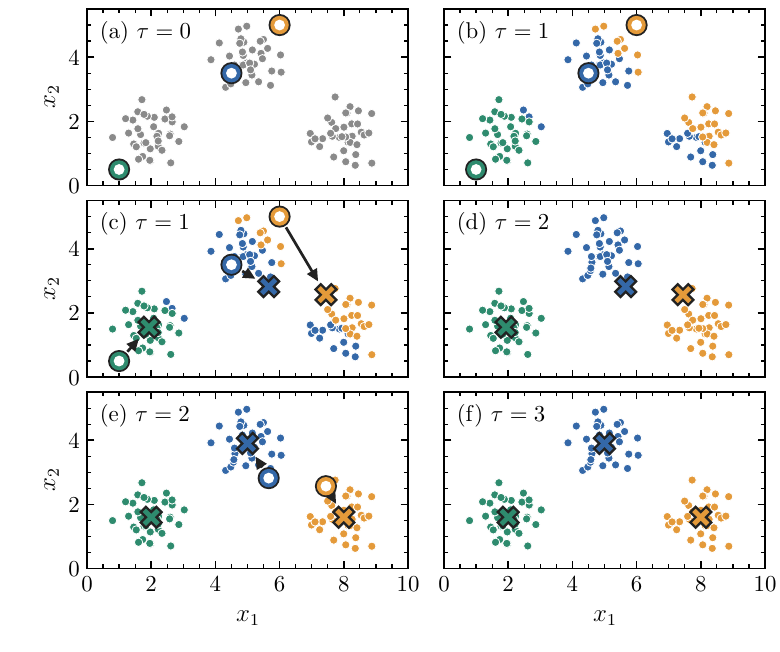}
    \caption{Three iterations of K-means, where $\tau$ counts the iterations. Panel~(a) shows the initial centers. Panels~(b) and (c) show the E-step and the M-step of iteration 1, and panels~(d) and (e) show those of iteration 2. Panel~(f) shows the E-step of iteration 3, which changes no assignment, so the algorithm stops. From panel~(b) on, each point has the color of its nearest center in the latest E-step. In the M-steps, arrows lead from the old centers (open circles) to the new centers (crosses), except for a center that hardly moves.}
    \label{fig:kmeans_em_algorithm}
\end{figure}

Setting the gradient with respect to $\boldsymbol{\mu}_k$ to zero gives
\begin{align}
\frac{\partial J_k}{\partial \boldsymbol{\mu}_k} &= \sum_{i=1}^N r_{ik} \frac{\partial}{\partial \boldsymbol{\mu}_k} \|\mathbf{x}_i - \boldsymbol{\mu}_k\|^2\\*
&= \sum_{i=1}^N r_{ik} (-2)(\mathbf{x}_i - \boldsymbol{\mu}_k)\\*
&= -2 \sum_{i=1}^N r_{ik} \mathbf{x}_i + 2 \boldsymbol{\mu}_k \sum_{i=1}^N r_{ik} = 0.
\end{align}Solving for $\boldsymbol{\mu}_k$ gives
\begin{equation}
\boldsymbol{\mu}_k = \frac{\sum_{i=1}^N r_{ik} \mathbf{x}_i}{\sum_{i=1}^N r_{ik}}.
\end{equation}The optimal center is the mean, or centroid, of the points assigned to the cluster. The denominator $\sum_{i=1}^N r_{ik}$ counts these points, because $r_{ik}$ is 1 for assigned points and 0 otherwise. Figure~\ref{fig:kmeans_em_algorithm} shows how the E-step and the M-step alternate.

\paragraph{The Local Minima Problem}

The EM algorithm converges to a solution, but it typically reaches a local\index{local minima} minimum of $J$ rather than the global minimum. Each step improves $J$ only given the current state, so it gives no guarantee of global optimality. The final solution therefore depends on the initial centers. Poor starting positions can leave the algorithm in a configuration where no step lowers $J$, although much better solutions exist. Since the cluster structure is often unknown, we need initialization methods that raise the chance of a good solution and speed up convergence.

\paragraph{Random Initialization}

\begin{figure}[ht!]
    \centering
    \includegraphics[width=0.9\textwidth]{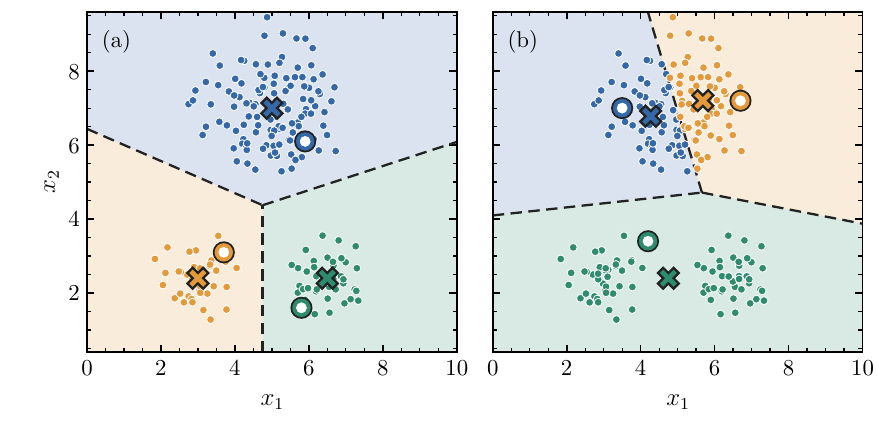}
    \caption{K-means from two different initializations. Open circles are the initial centers, crosses are the final centers, and colors show the final clusters. In panel~(a), each initial center lies near one cluster, and K-means finds the three clusters. In panel~(b), two initial centers lie in the large cluster and one lies between the two small clusters. K-means then splits the large cluster, merges the two small clusters and ends with a larger value of $J$.}
    \label{fig:kmeans_initialization}
\end{figure}

The simplest strategy picks $K$ data points uniformly at random as the initial centers. It is easy to implement, but it can give poor solutions when centers start too close together or in regions of low data density. Figure~\ref{fig:kmeans_initialization} compares a good and a poor initialization. Running K-means several times from different random starts and keeping the result with the smallest $J$ reduces the chance of a poor solution.

\paragraph{K-means++ Initialization}

K-means++\index{K-means++ initialization} has five steps and reduces the chance of poor starting centers by weighted random sampling that tends to spread them over the data.

\begin{itemize}
\item Choose the first center $\boldsymbol{\mu}_1$ uniformly at random from the data points
\begin{equation}
\boldsymbol{\mu}_1 \in \{\mathbf{x}_1, \mathbf{x}_2, \ldots, \mathbf{x}_N\} \text{ with probability } \frac{1}{N}.
\end{equation}

\item For each data point $\mathbf{x}_i$, compute the distance $\delta(\mathbf{x}_i)$ to the nearest chosen center
\begin{equation}
\delta(\mathbf{x}_i) = \min_{j \in \{1,2,\ldots,m\}} \|\mathbf{x}_i - \boldsymbol{\mu}_j\|,
\end{equation}
where $m$ counts the chosen centers.

\item Choose the next center $\boldsymbol{\mu}_{m+1}$ from the data points with probability proportional to $\delta(\mathbf{x}_i)^2$
\begin{equation}
P(\boldsymbol{\mu}_{m+1} = \mathbf{x}_i) = \frac{\delta(\mathbf{x}_i)^2}{\sum_{j=1}^N \delta(\mathbf{x}_j)^2}.
\end{equation}

\item Repeat the previous two steps until $K$ centers are chosen.

\item Run the standard K-means EM algorithm from these initial centers.
\end{itemize}

\begin{figure}[ht!]
    \centering
    \includegraphics[width=0.9\textwidth]{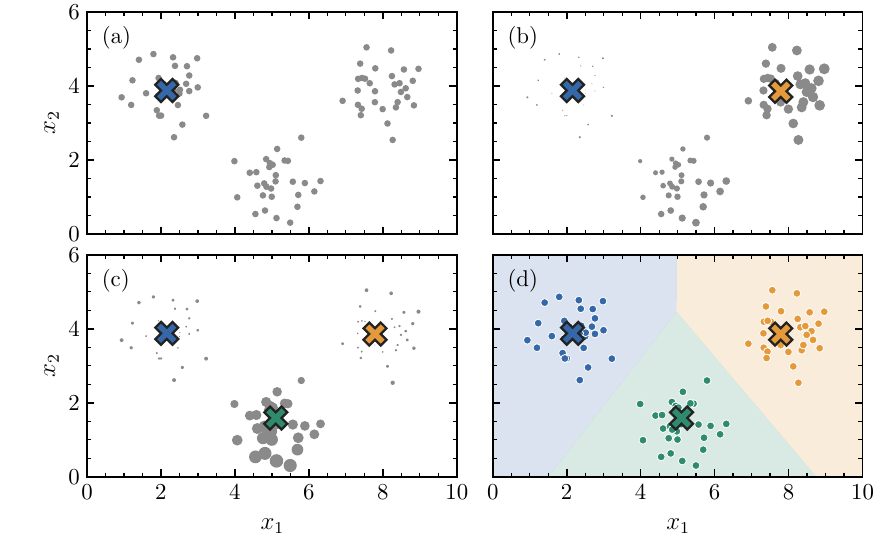}
    \caption{K-means++ initialization for three clusters. Panel~(a) shows the first center, which has the same probability for all points. In panels~(b) and (c), the area of each point is proportional to its probability of becoming the second and the third center. Points far from the existing centers have a high probability. Panel~(d) shows the three chosen centers and the points nearest to each of them. In this draw each cluster contains one center.}
    \label{fig:kmeans_plusplus}
\end{figure}

The squared distance gives a high probability to points far from the existing centers, so the new centers tend to spread over the data. Centers that are far from each other are less likely to lie in the same cluster. Sampling only from the data points places the centers in regions of high data density. Figure~\ref{fig:kmeans_plusplus} shows the probabilities and the chosen centers for three clusters.\index{expectation-maximization (EM) algorithm|)}

\section{Evaluating K-means Clustering}\label{sec:ch11_evaluating}

\paragraph{Cost per Iteration}

The E-step computes the distance between each of the $N$ points and each of the $K$ centers. Each distance needs of the order of $D$ operations, written $\mathcal{O}(D)$ (Section~10.8), so the E-step needs $\mathcal{O}(NKD)$ operations.

The M-step averages the points of each cluster. Assuming roughly equal-sized clusters, each center uses about $N/K$ points. Each mean needs $\mathcal{O}(ND/K)$ operations, so all centers need $\mathcal{O}(ND)$.

The E-step dominates, so one iteration costs $\mathcal{O}(NKD)$. The cost grows linearly with the number of points, clusters and dimensions. K-means is therefore practical for astronomical data sets with millions of objects and hundreds of features. The results are also easy to interpret.

Although K-means is fast, it needs the number of clusters $K$ in advance. The true cluster structure is usually unknown. How do we choose an appropriate value? In astronomy this asks how many types of galaxies exist or how many stellar populations a globular cluster contains.

\paragraph{Inertia and the Elbow Method}

The\index{elbow method|(} most direct approach is to watch the K-means objective as $K$ increases. The within-cluster sum of squares, or inertia,\index{inertia} is the value of $J$ after convergence. It measures how tightly clustered the points are
\begin{equation}
\text{Inertia} = \sum_{k=1}^K \sum_{i \in C_k} \|\mathbf{x}_i - \boldsymbol{\mu}_k\|^2,
\end{equation}
where $C_k$ is the set of points in cluster $k$.

As $K$ increases, the inertia decreases, because the clusters become smaller and more compact. Suppose the data contain $K^*$ clusters. For $K < K^*$ each added cluster separates a real group and lowers the inertia by a large amount. For $K > K^*$ the added clusters split groups that are already compact, so the inertia falls only slowly.

\begin{figure}[ht!]
    \centering
    \includegraphics[width=0.85\textwidth]{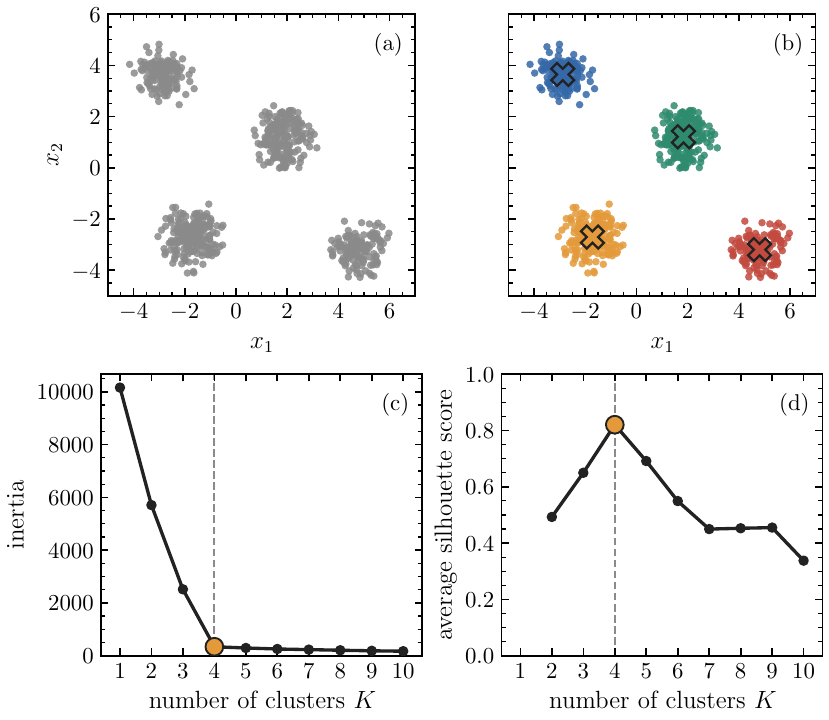}
    \caption{Choice of the number of clusters $K$ for data with four groups of points. Panel~(a) shows the data and panel~(b) shows the result of K-means for $K = 4$. Panel~(c) shows the inertia, which has its elbow at $K = 4$. Panel~(d) shows the average silhouette score, which has its maximum at $K = 4$. The dot and the dashed line mark $K = 4$ in panels~(c) and (d).}
    \label{fig:kmeans_elbow_method}
\end{figure}

We plot the inertia against $K$ and look for the elbow,\index{elbow method} the point where the slope changes sharply. The elbow gives a candidate value of $K$. The choice is subjective when no clear elbow exists. The elbow plot is used like the scree plot of Chapter~10, whose elbow suggests the number of principal components. Panels~(a) to (c) of Figure~\ref{fig:kmeans_elbow_method} show an example with four clusters.\index{elbow method|)}

\paragraph{Silhouette Analysis}

The\index{silhouette score|(} silhouette\index{silhouette score} score is a more quantitative approach. It measures how similar a point is to its own cluster compared with the other clusters. For point $i$ the silhouette score is
\begin{equation}
s(i) = \frac{b(i) - a(i)}{\max\{a(i), b(i)\}},
\end{equation}
where $a(i)$ is the mean distance from point $i$ to the other points of its cluster. The value $b(i)$ is the minimum average distance between point $i$ and the points of any other cluster.

The score lies between $-1$ and 1. A value $s(i) \approx 1$ means that the point is well clustered and far from the neighboring clusters. A value near 0 means that the point lies near a boundary between clusters. A value near $-1$ means that the point may be in the wrong cluster.

We choose the $K$ with the highest average silhouette score over all data points, as panel~(d) of Figure~\ref{fig:kmeans_elbow_method} shows.\index{silhouette score|)}

\section{K-means as a Generative Model}\label{sec:ch11_kmeans_generative}

So far we have described K-means as a geometric optimization problem. It also has a statistical interpretation, which connects K-means to probabilistic models. This connection helps explain the behavior of K-means and points toward clustering methods with fewer restrictions.

In Chapter~7 we distinguished discriminative models, which learn decision boundaries directly, from generative models, which describe how the data of each class are produced. K-means takes the generative view and applies it to the unsupervised setting, where the class labels are unknown.

In this view, the observed data come from a mixture of different underlying processes, and each cluster represents one such process. Our goal is to identify both the processes (the cluster parameters) and which process generated each data point (its cluster assignment).

\paragraph{A Probabilistic Data-Generating Story}

We can describe K-means by a generative model with two steps. First, one of the $K$ clusters, say cluster $k$, is selected with equal probability $1/K$. Then a data point is drawn from a Gaussian distribution with spherical covariance $\sigma^2\mathrm{Id}$ around the cluster center $\boldsymbol{\mu}_k$. Here $\mathrm{Id}$ is the identity matrix, so the variance is the same in all directions.

The likelihood of data point $\mathbf{x}_i$ is
\begin{equation}
p(\mathbf{x}_i) = \sum_{k=1}^K \frac{1}{K} \mathcal{N}(\mathbf{x}_i|\boldsymbol{\mu}_k, \sigma^2\mathrm{Id}),
\end{equation}
where $\mathcal{N}(\mathbf{x}|\boldsymbol{\mu}, \boldsymbol{\Sigma})$ denotes the Gaussian probability density.

\paragraph{From Soft to Hard Assignments}

The\index{hard assignment|(}\index{soft assignment|(} likelihood above describes soft\index{soft assignment} assignment, because each data point has some probability of belonging to each cluster. K-means uses hard\index{hard assignment} assignments, in which each point belongs to one cluster.

We connect the two views by putting the binary assignment variables $r_{ik}$ into the likelihood. For hard assignments the joint probability of $\mathbf{x}_i$ and its assignment becomes
\begin{equation}
p(\mathbf{x}_i, \{r_{ik}\}) = \sum_{k=1}^K r_{ik} \frac{1}{K} \mathcal{N}(\mathbf{x}_i|\boldsymbol{\mu}_k, \sigma^2\mathrm{Id}).
\end{equation}For each data point one $r_{ik}$ equals 1 and the others are 0. The sum therefore reduces to the term of the assigned cluster.

For all data points, the log-likelihood is\index{hard assignment|)}\index{soft assignment|)}
\begin{equation}
\ln L = \sum_{i=1}^N \sum_{k=1}^K r_{ik} \ln \left(\frac{1}{K} \mathcal{N}(\mathbf{x}_i|\boldsymbol{\mu}_k, \sigma^2\mathrm{Id})\right).
\end{equation}

\paragraph{Deriving the K-means Objective}

Maximizing this log-likelihood over the assignments $r_{ik}$ and the centers $\boldsymbol{\mu}_k$ gives the K-means objective. The Gaussian density with spherical covariance is
\begin{equation}
\mathcal{N}(\mathbf{x}|\boldsymbol{\mu}, \sigma^2\mathrm{Id}) = \frac{1}{(2\pi\sigma^2)^{D/2}} \exp\left(-\frac{\|\mathbf{x} - \boldsymbol{\mu}\|^2}{2\sigma^2}\right).
\end{equation}Its logarithm is
\begin{equation}
\ln \mathcal{N}(\mathbf{x}|\boldsymbol{\mu}, \sigma^2\mathrm{Id}) = -\frac{D}{2}\ln(2\pi\sigma^2) - \frac{\|\mathbf{x} - \boldsymbol{\mu}\|^2}{2\sigma^2}.
\end{equation}

Substituting into the log-likelihood gives
\begin{align}
\ln L &= \sum_{i=1}^N \sum_{k=1}^K r_{ik} \left[\ln\frac{1}{K} -\frac{D}{2}\ln(2\pi\sigma^2) - \frac{\|\mathbf{x}_i - \boldsymbol{\mu}_k\|^2}{2\sigma^2}\right].
\end{align}The first two terms are constants with respect to the assignments $r_{ik}$ and the centers $\boldsymbol{\mu}_k$, because $\sum_{k=1}^K r_{ik} = 1$. We ignore them when we maximize the log-likelihood and keep
\begin{equation}
\sum_{i=1}^N \sum_{k=1}^K r_{ik} \left[- \frac{\|\mathbf{x}_i - \boldsymbol{\mu}_k\|^2}{2\sigma^2}\right].
\end{equation}Maximizing this expression is equivalent to minimizing
\begin{equation}
\sum_{i=1}^N \sum_{k=1}^K r_{ik} \|\mathbf{x}_i - \boldsymbol{\mu}_k\|^2.
\end{equation}This is the K-means objective, because the constant factor $1/(2\sigma^2)$ multiplies all terms equally and does not change the optimization.

\paragraph{Understanding K-means Assumptions}

From this derivation we see that K-means corresponds to a specific generative model with three restrictive assumptions. The clusters are spherical, because all have the same covariance matrix $\sigma^2\mathrm{Id}$. The clusters have equal weights, because all have the prior probability $1/K$. The assignments are hard, because each data point belongs to one cluster.

\begin{figure}[ht!]
    \centering
    \includegraphics[width=0.65\textwidth]{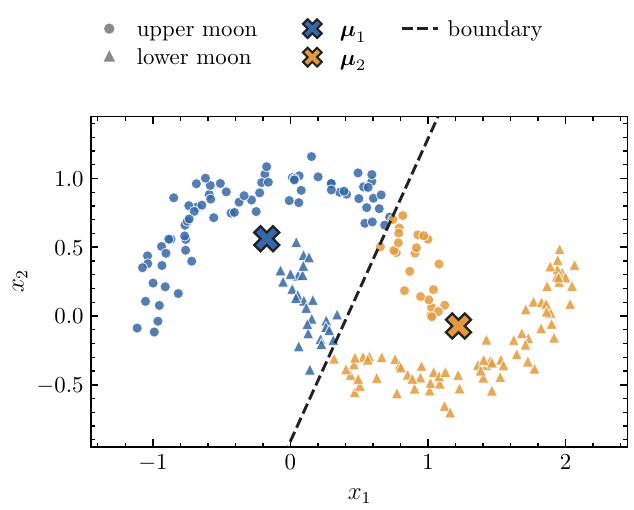}
    \caption{K-means with $K = 2$ for two half-moon clusters. Colors show the clusters found by K-means, and marker shapes show the true moons. The dashed line is the boundary between the clusters. It is the perpendicular bisector of the segment between the centers $\boldsymbol{\mu}_1$ and $\boldsymbol{\mu}_2$, so it is straight, and it cuts through both moons.}
    \label{fig:kmeans_limitations}
\end{figure}

These assumptions explain both the efficiency of K-means and its limits. The spherical assumption makes the algorithm fast but restricts the clusters to roughly round shapes. It also makes the boundary between two clusters a hyperplane, a straight line in two dimensions, as in logistic regression (Chapter~7). Such a boundary cannot separate the two curved clusters of Figure~\ref{fig:kmeans_limitations}, because each K-means cluster is a convex region. A region is convex when the straight segment between any two of its points stays inside it, and a half-moon is not convex.

The equal weight assumption works well when the clusters have similar populations but fails when the populations differ a lot. The hard assignment assumption provides mathematical simplicity but loses information about uncertainty in borderline cases. K-means is also sensitive to outliers, and it works best for well-separated clusters.

\looseness=1 We can relax these assumptions by allowing different covariance matrices, different weights and soft assignments. This gives the Gaussian mixture models of the next section.\index{K-means clustering|)}

\section{Mathematical Formulation of Gaussian Mixture Models}\label{sec:ch11_gmm}

K-means\index{Gaussian mixture model (GMM)|(} assigns each data point to one cluster. GMMs use soft assignments, in which each data point has a probability of belonging to each cluster.

\begin{figure}[ht!]
    \centering
    \includegraphics[width=0.7\textwidth]{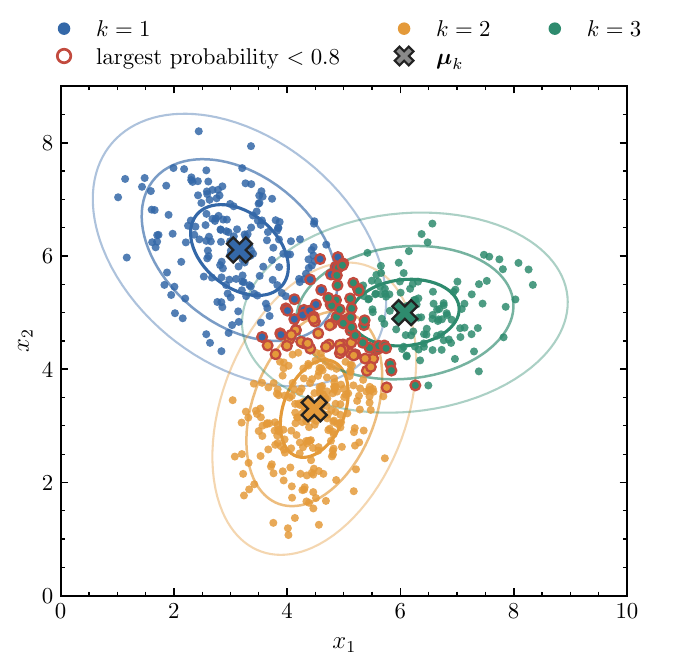}
    \caption{Gaussian mixture with three overlapping clusters. Colors show the most probable cluster of each point. The ellipses show each cluster at one, two and three standard deviations along its axes. Points whose largest cluster probability is below 0.8 lie where the clusters overlap.}
    \label{fig:gmm_soft_assignment}
\end{figure}

In astronomy the boundaries between populations are often fuzzy. Stars near the boundary between the disk and the halo, for example, can share the properties of both populations. Figure~\ref{fig:gmm_soft_assignment} shows soft assignment for three overlapping clusters.

\paragraph{The Mixture Model}

A GMM describes the distribution of the data as a weighted sum of simpler component distributions, and each component describes one cluster. We assume that the data come from a mixture\index{Gaussian mixture model (GMM)}\index{mixture model} of $K$ Gaussian distributions
\begin{equation}
p(\mathbf{x}) = \sum_{k=1}^K \pi_k \mathcal{N}(\mathbf{x}|\boldsymbol{\mu}_k, \boldsymbol{\Sigma}_k).
\end{equation}The function $\mathcal{N}(\mathbf{x}|\boldsymbol{\mu}_k, \boldsymbol{\Sigma}_k)$ is the multivariate Gaussian density of component $k$, with mean vector $\boldsymbol{\mu}_k$ and covariance matrix $\boldsymbol{\Sigma}_k$. It gives the density of observing $\mathbf{x}$ if the data point came from component $k$. The mixture\index{mixture weights} weight $\pi_k$ is the prior probability that a randomly selected data point belongs to component $k$. The weights satisfy $\pi_k \geq 0$ and $\sum_{k=1}^K \pi_k = 1$.

\begin{figure}[ht!]
    \centering
    \includegraphics[width=0.65\textwidth]{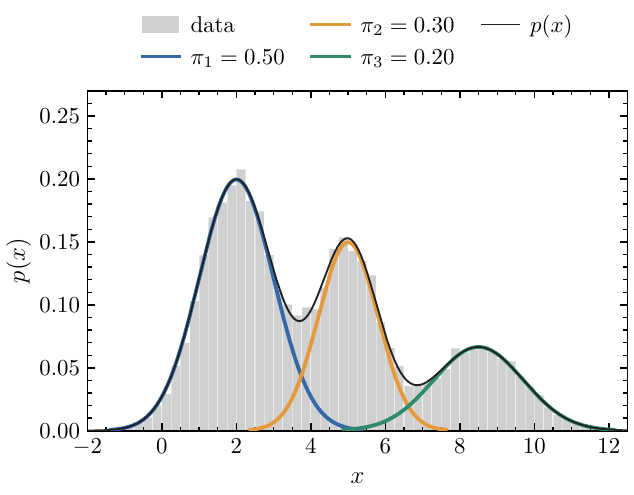}
    \caption{Mixture of three one-dimensional Gaussian components with different weights. The colored curves are the components multiplied by their weights. Their sum is the density $p(x)$, and the histogram shows data drawn from this density.}
    \label{fig:gmm_mixture_weights}
\end{figure}

The weights are the relative abundances of the components in the population. In a galaxy survey, for example, $\pi_1$ could be the fraction of spiral galaxies and $\pi_2$ the fraction of elliptical galaxies. Figure~\ref{fig:gmm_mixture_weights} shows three components with different weights.

\paragraph{Increased Model Flexibility}

\looseness=1 K-means assumes the same spherical covariance for all clusters. A GMM gives each component its own covariance matrix $\boldsymbol{\Sigma}_k$, called a full covariance matrix. It can therefore describe clusters with different shapes, sizes and orientations.

This flexibility needs more parameters. A GMM with $K$ components in $D$ dimensions has $K-1$ free mixture weights, because the weights sum to 1. It has $KD$ mean parameters and $K \cdot D(D+1)/2$ covariance parameters, because each symmetric covariance matrix has $D(D+1)/2$ free parameters. The total is $(K-1) + KD + K \cdot D(D+1)/2$ parameters, which grows quadratically with $D$. This rapid growth can make GMMs computationally challenging in high-dimensional spaces.

\paragraph{Constraining Covariance Complexity}

We can constrain the covariance matrices to reduce the number of parameters, and two choices are common.

A diagonal covariance $\boldsymbol{\Sigma}_k = \text{diag}(\sigma_{k1}^2, \sigma_{k2}^2, \ldots, \sigma_{kD}^2)$ means that the features are uncorrelated within each component. It reduces the covariance parameters from $K \cdot D(D+1)/2$ to $K \cdot D$. The clusters can still be elliptical, but their axes are aligned with the coordinate axes.

\begin{figure}[ht!]
    \centering
    \includegraphics[width=1.0\textwidth]{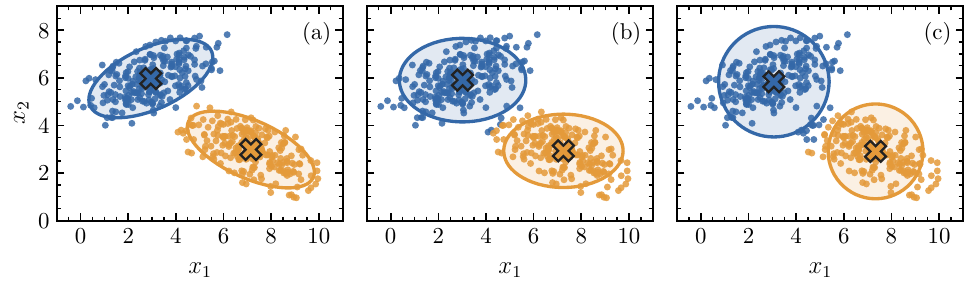}
    \caption{Gaussian mixture with two components fitted to the same data with three covariance types. Panel~(a) shows full covariance matrices, panel~(b) diagonal covariance matrices and panel~(c) spherical covariance matrices. Colors show the most probable component of each point, and the ellipses show two standard deviations of each fitted component. The full covariance follows the tilt of the clusters. The diagonal covariance keeps the axes of the ellipse along $x_1$ and $x_2$, and the spherical covariance gives circles.}
    \label{fig:gmm_covariance_types}
\end{figure}

A spherical covariance $\boldsymbol{\Sigma}_k = \sigma_k^2\mathrm{Id}$ makes each component spherical, with a size $\sigma_k$ that can differ between components. It reduces the covariance parameters to $K$. This keeps the probabilistic form of GMMs while approaching the geometric simplicity of K-means. Figure~\ref{fig:gmm_covariance_types} compares the three covariance types.

\paragraph{The GMM Likelihood Function}

We estimate the parameters by maximum likelihood. For a data set $\{\mathbf{x}_1, \mathbf{x}_2, \ldots, \mathbf{x}_N\}$ of independent observations the likelihood is
\begin{equation}
L = \prod_{i=1}^N p(\mathbf{x}_i) = \prod_{i=1}^N \left[\sum_{k=1}^K \pi_k \mathcal{N}(\mathbf{x}_i|\boldsymbol{\mu}_k, \boldsymbol{\Sigma}_k)\right].
\end{equation}Taking the logarithm turns the product into a sum
\begin{equation}
\ln L = \sum_{i=1}^N \ln\left[\sum_{k=1}^K \pi_k \mathcal{N}(\mathbf{x}_i|\boldsymbol{\mu}_k, \boldsymbol{\Sigma}_k)\right].
\end{equation}

\looseness=1 Direct maximization of this log-likelihood is difficult. The sum inside the logarithm creates a complex likelihood surface with many local maxima. The mixture weights, means and covariances also depend on each other, as in K-means, but in a more complex way.

As an example, consider the gradient with respect to one mean vector $\boldsymbol{\mu}_k$
\begin{align}
\frac{\partial \ln L}{\partial \boldsymbol{\mu}_k} &= \sum_{i=1}^N \frac{1}{\sum_{j=1}^K \pi_j \mathcal{N}(\mathbf{x}_i|\boldsymbol{\mu}_j, \boldsymbol{\Sigma}_j)} \cdot \frac{\partial}{\partial \boldsymbol{\mu}_k}\left[\pi_k \mathcal{N}(\mathbf{x}_i|\boldsymbol{\mu}_k, \boldsymbol{\Sigma}_k)\right].
\end{align}The vector $\boldsymbol{\mu}_k$ appears in both the numerator and the denominator of a complex fraction, so direct optimization is hard. The covariance matrices and the mixture weights are equally hard to find.

\section{Expectation-Maximization for GMMs}\label{sec:ch11_gmm_em}

Direct\index{expectation-maximization (EM) algorithm|(} maximization is hard, so we extend the EM\index{expectation-maximization (EM) algorithm} algorithm of K-means to GMMs.\index{GMM|see{Gaussian mixture model}} The idea is the same, since we alternate between estimating the assignments and updating the model parameters.

\paragraph{Introducing Responsibilities}

The\index{responsibilities|(} central quantity is the responsibility\index{responsibilities} of component $k$ for data point $\mathbf{x}_i$
\begin{equation}
\gamma_{ik} = \frac{\pi_k \mathcal{N}(\mathbf{x}_i|\boldsymbol{\mu}_k, \boldsymbol{\Sigma}_k)}{\sum_{j=1}^K \pi_j \mathcal{N}(\mathbf{x}_i|\boldsymbol{\mu}_j, \boldsymbol{\Sigma}_j)}.
\end{equation}The responsibility $\gamma_{ik}$ is the posterior probability, given the current parameters, that component $k$ generated data point $\mathbf{x}_i$. The responsibility corresponds to the binary assignment variable $r_{ik}$ of K-means. It is also the expected value of a hidden indicator that is 1 if component $k$ generated $\mathbf{x}_i$ and 0 otherwise. This expected value gives the E-step its name.

K-means assignments are 0 or 1, but responsibilities can take any value between 0 and 1. A data point can therefore belong partly to several clusters, which expresses uncertainty in borderline cases. For each data point the responsibilities sum to 1, $\sum_{k=1}^K \gamma_{ik} = 1$, so the total probability of the point is split among the components.\index{responsibilities|)}

\paragraph{Outline and Initialization}

The EM algorithm for GMMs alternates between two steps, as for K-means, but it works with probabilities instead of discrete assignments.

\looseness=1 The algorithm starts from initial values of the parameters $\{\pi_k\}$, $\{\boldsymbol{\mu}_k\}$ and $\{\boldsymbol{\Sigma}_k\}$. A common choice uses the K-means result for the means, uniform mixture weights and identity covariance matrices. Another choice is to partition the data at random and to compute initial statistics from each partition. We can also choose the initial means with K-means++.

In the E-step we compute the responsibilities $\gamma_{ik}$ defined above from the current parameter values. They are our current best estimates of which component generated each data point.

In the M-step we treat the responsibilities $\gamma_{ik}$ as fixed and maximize the responsibility-weighted log-likelihood of the components. Its gradient matches that of $\ln L$ at the current parameters. With fixed responsibilities, we can optimize each set of parameters independently. The E-step was straightforward, because the responsibilities follow from Bayes' theorem (Chapter~2). The M-step needs a separate derivation for the means, the covariance matrices and the mixture weights.

\paragraph{Updating the Mean Vectors}

We start with the means, which have the most direct geometric meaning. For component $k$ we look for the mean $\boldsymbol{\mu}_k$ that maximizes the log-likelihood when all other parameters are held fixed.

Taking the derivative of $\ln L$ with respect to $\boldsymbol{\mu}_k$ gives
\begin{align}
\frac{\partial \ln L}{\partial \boldsymbol{\mu}_k} &= \sum_{i=1}^N \frac{1}{\sum_{j=1}^K \pi_j \mathcal{N}(\mathbf{x}_i|\boldsymbol{\mu}_j, \boldsymbol{\Sigma}_j)} \cdot \frac{\partial}{\partial \boldsymbol{\mu}_k}\left[\pi_k \mathcal{N}(\mathbf{x}_i|\boldsymbol{\mu}_k, \boldsymbol{\Sigma}_k)\right].
\end{align}The chain rule and the fact that $(\partial/\partial \boldsymbol{\mu}_k) \mathcal{N}(\mathbf{x}_i|\boldsymbol{\mu}_k, \boldsymbol{\Sigma}_k) = \mathcal{N}(\mathbf{x}_i|\boldsymbol{\mu}_k, \boldsymbol{\Sigma}_k) \boldsymbol{\Sigma}_k^{-1}(\mathbf{x}_i - \boldsymbol{\mu}_k)$ give
\begin{align}
\frac{\partial \ln L}{\partial \boldsymbol{\mu}_k} &= \sum_{i=1}^N \frac{\pi_k \mathcal{N}(\mathbf{x}_i|\boldsymbol{\mu}_k, \boldsymbol{\Sigma}_k)}{\sum_{j=1}^K \pi_j \mathcal{N}(\mathbf{x}_i|\boldsymbol{\mu}_j, \boldsymbol{\Sigma}_j)} \boldsymbol{\Sigma}_k^{-1}(\mathbf{x}_i - \boldsymbol{\mu}_k).
\end{align}The fraction is the responsibility $\gamma_{ik}$, so
\begin{align}
\frac{\partial \ln L}{\partial \boldsymbol{\mu}_k} &= \sum_{i=1}^N \gamma_{ik} \boldsymbol{\Sigma}_k^{-1}(\mathbf{x}_i - \boldsymbol{\mu}_k).
\end{align}

Setting this to zero and solving for $\boldsymbol{\mu}_k$ gives
\begin{align}
\sum_{i=1}^N \gamma_{ik} \boldsymbol{\Sigma}_k^{-1}(\mathbf{x}_i - \boldsymbol{\mu}_k) &= 0 \\*
\boldsymbol{\Sigma}_k^{-1}\sum_{i=1}^N \gamma_{ik} (\mathbf{x}_i - \boldsymbol{\mu}_k) &= 0 \\*
\sum_{i=1}^N \gamma_{ik} \mathbf{x}_i - \boldsymbol{\mu}_k \sum_{i=1}^N \gamma_{ik} &= 0.
\end{align}
\begin{equation}
\boldsymbol{\mu}_k = \frac{\sum_{i=1}^N \gamma_{ik} \mathbf{x}_i}{\sum_{i=1}^N \gamma_{ik}}.
\end{equation}The new mean is the weighted average of all data points, with the responsibilities as weights. This generalizes the K-means result, where the hard assignments $r_{ik} \in \{0,1\}$ are replaced by soft responsibilities $\gamma_{ik} \in [0,1]$.

The effective number of points assigned to component $k$ is
\begin{equation}
N_k = \sum_{i=1}^N \gamma_{ik}.
\end{equation}With this notation the mean update is
\begin{equation}
\boldsymbol{\mu}_k^{\text{new}} = \frac{1}{N_k}\sum_{i=1}^N \gamma_{ik}\mathbf{x}_i.
\end{equation}

\paragraph{Updating the Covariance Matrices}

The covariance update follows the same principle. We want each component's covariance to reflect the spread of the data points that belong to it, weighted by their responsibilities.

The derivation needs matrix calculus. We differentiate with respect to the precision matrix $\boldsymbol{\Sigma}_k^{-1}$, the inverse of the covariance matrix, and then invert the result.

We start from the multivariate Gaussian density
\begin{equation}
\mathcal{N}(\mathbf{x}|\boldsymbol{\mu}, \boldsymbol{\Sigma}) = \frac{1}{(2\pi)^{D/2}\det(\boldsymbol{\Sigma})^{1/2}} \exp\left(-\frac{1}{2}(\mathbf{x} - \boldsymbol{\mu})^T\boldsymbol{\Sigma}^{-1}(\mathbf{x} - \boldsymbol{\mu})\right).
\end{equation}Its logarithm is
\begin{equation}
\ln \mathcal{N}(\mathbf{x}|\boldsymbol{\mu}, \boldsymbol{\Sigma}) = -\frac{D}{2}\ln(2\pi) - \frac{1}{2}\ln\det(\boldsymbol{\Sigma}) - \frac{1}{2}(\mathbf{x} - \boldsymbol{\mu})^T\boldsymbol{\Sigma}^{-1}(\mathbf{x} - \boldsymbol{\mu}).
\end{equation}Since we differentiate with respect to $\boldsymbol{\Sigma}_k^{-1}$, we write the determinant term with $\det(\boldsymbol{\Sigma}) = 1/\det(\boldsymbol{\Sigma}^{-1})$, which gives $\ln\det(\boldsymbol{\Sigma}) = -\ln\det(\boldsymbol{\Sigma}^{-1})$.

We need two matrix calculus identities
\begin{equation}
\frac{\partial}{\partial \mathbf{A}} \ln\det(\mathbf{A}) = \mathbf{A}^{-1} \quad \text{and} \quad \frac{\partial}{\partial \mathbf{A}} \mathbf{x}^T\mathbf{A}\mathbf{x} = \mathbf{x}\mathbf{x}^T.
\end{equation}The derivative with respect to a matrix is the matrix of derivatives with respect to its elements. The first identity holds for a symmetric matrix such as the precision matrix. In one dimension $\mathbf{A}$ and $\mathbf{x}$ are numbers $a$ and $x$, and the identities read $\mathrm{d}\ln a/\mathrm{d}a = 1/a$ and $\mathrm{d}(a x^2)/\mathrm{d}a = x^2$.

Following the pattern of the means we compute
\begin{align}
\frac{\partial}{\partial \boldsymbol{\Sigma}_k^{-1}} \ln L &= \sum_{i=1}^N \frac{1}{\sum_{j=1}^K \pi_j \mathcal{N}(\mathbf{x}_i|\boldsymbol{\mu}_j, \boldsymbol{\Sigma}_j)} \cdot \frac{\partial}{\partial \boldsymbol{\Sigma}_k^{-1}}\left[\pi_k \mathcal{N}(\mathbf{x}_i|\boldsymbol{\mu}_k, \boldsymbol{\Sigma}_k)\right].
\end{align}The chain rule gives
\begin{align}
\frac{\partial}{\partial \boldsymbol{\Sigma}_k^{-1}}\left[\pi_k \mathcal{N}(\mathbf{x}_i|\boldsymbol{\mu}_k, \boldsymbol{\Sigma}_k)\right] = \pi_k \mathcal{N}(\mathbf{x}_i|\boldsymbol{\mu}_k, \boldsymbol{\Sigma}_k) \cdot \frac{\partial}{\partial \boldsymbol{\Sigma}_k^{-1}} \ln \mathcal{N}(\mathbf{x}_i|\boldsymbol{\mu}_k, \boldsymbol{\Sigma}_k).
\end{align}The derivative of the logarithm of the Gaussian is
\begin{align}
\frac{\partial}{\partial \boldsymbol{\Sigma}_k^{-1}} \ln \mathcal{N}(\mathbf{x}_i|\boldsymbol{\mu}_k, \boldsymbol{\Sigma}_k) &= \frac{\partial}{\partial \boldsymbol{\Sigma}_k^{-1}} \left[-\frac{D}{2}\ln(2\pi) + \frac{1}{2}\ln\det(\boldsymbol{\Sigma}_k^{-1}) - \frac{1}{2}(\mathbf{x}_i - \boldsymbol{\mu}_k)^T\boldsymbol{\Sigma}_k^{-1}(\mathbf{x}_i - \boldsymbol{\mu}_k)\right] \\*
&= \frac{1}{2}\boldsymbol{\Sigma}_k - \frac{1}{2}(\mathbf{x}_i - \boldsymbol{\mu}_k)(\mathbf{x}_i - \boldsymbol{\mu}_k)^T.
\end{align}

Substituting back and using the responsibilities gives
\begin{align}
\frac{\partial}{\partial \boldsymbol{\Sigma}_k^{-1}} \ln L &= \sum_{i=1}^N \gamma_{ik} \left[\frac{1}{2}\boldsymbol{\Sigma}_k - \frac{1}{2}(\mathbf{x}_i - \boldsymbol{\mu}_k)(\mathbf{x}_i - \boldsymbol{\mu}_k)^T\right].
\end{align}Setting this to zero and solving gives
\begin{align}
\sum_{i=1}^N \gamma_{ik} \left[\frac{1}{2}\boldsymbol{\Sigma}_k - \frac{1}{2}(\mathbf{x}_i - \boldsymbol{\mu}_k)(\mathbf{x}_i - \boldsymbol{\mu}_k)^T\right] &= 0 \\*
\frac{1}{2}\boldsymbol{\Sigma}_k \sum_{i=1}^N \gamma_{ik} - \frac{1}{2}\sum_{i=1}^N \gamma_{ik}(\mathbf{x}_i - \boldsymbol{\mu}_k)(\mathbf{x}_i - \boldsymbol{\mu}_k)^T &= 0.
\end{align}
\begin{equation}
\boldsymbol{\Sigma}_k^{\text{new}} = \frac{1}{N_k} \sum_{i=1}^N \gamma_{ik}(\mathbf{x}_i - \boldsymbol{\mu}_k^{\text{new}})(\mathbf{x}_i - \boldsymbol{\mu}_k^{\text{new}})^T.
\end{equation}The new covariance is the responsibility-weighted sample covariance. It generalizes the usual covariance formula by weighting each data point by how much it belongs to the component.

\paragraph{Updating the Mixture Weights}

Finally we derive the update of the mixture\index{mixture weights} weights, which satisfy the constraint $\sum_{k=1}^K \pi_k = 1$. This needs Lagrange multipliers (Chapter~10), which handle optimization under a constraint.

The Lagrangian is
\begin{equation}
\mathcal{L} = \ln L + \lambda\left(1 - \sum_{k=1}^K \pi_k\right),
\end{equation}where $\lambda$ is the Lagrange multiplier.

Setting the derivative with respect to $\pi_k$ to zero gives
\begin{align}
\frac{\partial \mathcal{L}}{\partial \pi_k} &= \sum_{i=1}^N \frac{\mathcal{N}(\mathbf{x}_i|\boldsymbol{\mu}_k, \boldsymbol{\Sigma}_k)}{\sum_{j=1}^K \pi_j \mathcal{N}(\mathbf{x}_i|\boldsymbol{\mu}_j, \boldsymbol{\Sigma}_j)} - \lambda = 0.
\end{align}Multiplying both sides by $\pi_k$ and summing over all components gives
\begin{align}
\sum_{k=1}^K \pi_k \sum_{i=1}^N \frac{\mathcal{N}(\mathbf{x}_i|\boldsymbol{\mu}_k, \boldsymbol{\Sigma}_k)}{\sum_{j=1}^K \pi_j \mathcal{N}(\mathbf{x}_i|\boldsymbol{\mu}_j, \boldsymbol{\Sigma}_j)} &= \lambda \sum_{k=1}^K \pi_k = \lambda.
\end{align}The left side simplifies, because the inner sum equals 1 for each data point
\begin{align}
\sum_{i=1}^N \sum_{k=1}^K \frac{\pi_k\mathcal{N}(\mathbf{x}_i|\boldsymbol{\mu}_k, \boldsymbol{\Sigma}_k)}{\sum_{j=1}^K \pi_j \mathcal{N}(\mathbf{x}_i|\boldsymbol{\mu}_j, \boldsymbol{\Sigma}_j)} = \sum_{i=1}^N 1 = N = \lambda.
\end{align}

Substituting $\lambda = N$ into the equation for the derivative and multiplying by $\pi_k$ turns the fraction into $\gamma_{ik}$, so
\begin{equation}
\sum_{i=1}^N \gamma_{ik} = N \pi_k.
\end{equation}
\begin{equation}
\pi_k^{\text{new}} = \frac{N_k}{N}.
\end{equation}The new mixture weight is the average responsibility of component $k$ over all data points. Components that explain more data points receive larger weights.

\paragraph{The Complete Algorithm}

The complete EM algorithm for GMMs has four steps.

\begin{itemize}
\item Initialize the parameters $\{\pi_k\}$, $\{\boldsymbol{\mu}_k\}$ and $\{\boldsymbol{\Sigma}_k\}$.

\item In the E-step, compute the responsibilities
\begin{equation}
\gamma_{ik} = \frac{\pi_k \mathcal{N}(\mathbf{x}_i|\boldsymbol{\mu}_k, \boldsymbol{\Sigma}_k)}{\sum_{j=1}^K \pi_j \mathcal{N}(\mathbf{x}_i|\boldsymbol{\mu}_j, \boldsymbol{\Sigma}_j)}.
\end{equation}

\item In the M-step, update the parameters
\begin{align}
N_k &= \sum_{i=1}^N \gamma_{ik} \\*
\boldsymbol{\mu}_k^{\text{new}} &= \frac{1}{N_k}\sum_{i=1}^N \gamma_{ik}\mathbf{x}_i \\
\boldsymbol{\Sigma}_k^{\text{new}} &= \frac{1}{N_k}\sum_{i=1}^N \gamma_{ik}(\mathbf{x}_i - \boldsymbol{\mu}_k^{\text{new}})(\mathbf{x}_i - \boldsymbol{\mu}_k^{\text{new}})^T \\*
\pi_k^{\text{new}} &= \frac{N_k}{N}.
\end{align}

\item Repeat the E-step and the M-step until convergence.
\end{itemize}

\begin{figure}[ht!]
    \centering
    \includegraphics[width=0.9\textwidth]{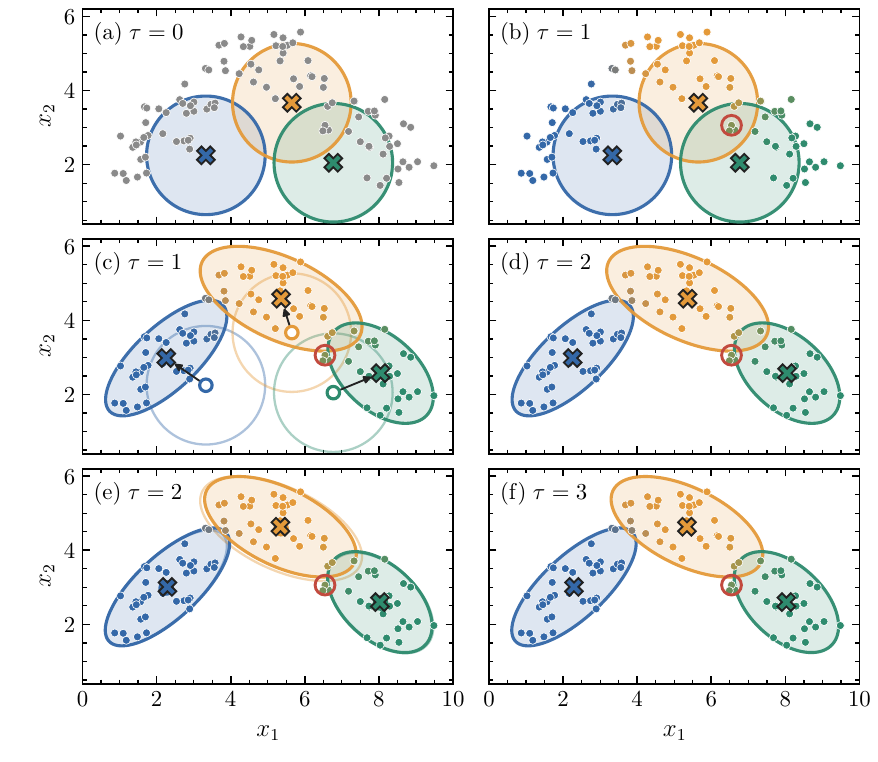}
    \caption{Two iterations and a third E-step of the EM algorithm for a mixture of three Gaussian components, where $\tau$ counts the iterations. Panel~(a) shows the initial parameters, and panels~(b) and (c) show the E-step and the M-step of iteration 1. Panels~(d) and (e) show those of iteration 2, and panel~(f) shows the E-step of iteration 3. The ellipses show each component at two standard deviations, and from panel~(b) on mixed colors show the responsibilities in the latest E-step. In the M-steps, open circles and light ellipses show the previous parameters, which differ little from the new ones in panel~(e). The ringed point is split between two components in the three E-steps.}
    \label{fig:gmm_em_algorithm}
\end{figure}

Figure~\ref{fig:gmm_em_algorithm} shows two iterations and a third E-step of this algorithm for three components.

\paragraph{Convergence and Local Optima}

\begin{figure}[ht!]
    \centering
    \includegraphics[width=0.95\textwidth]{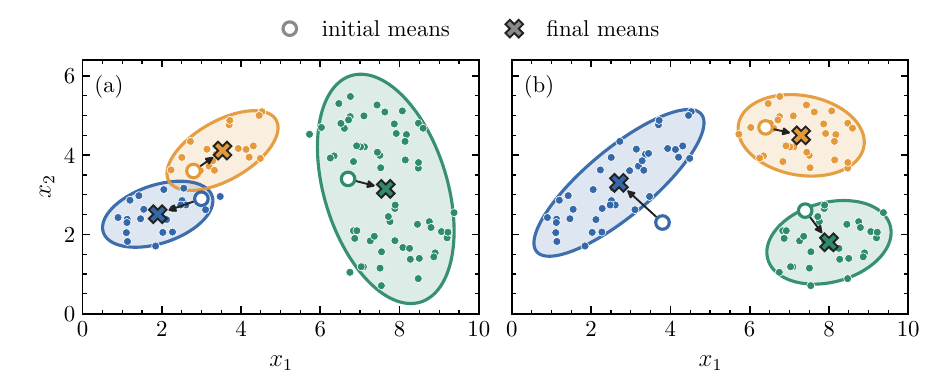}
    \caption{Two runs of the EM algorithm for a mixture of three components, started from different initial means. Arrows join each initial mean to its final mean, the ellipses show two standard deviations, and colors show the most probable component of each point. Panel~(a) ends in a local maximum that splits the correlated cluster and merges the two round clusters. Panel~(b) finds the three clusters and has the higher log-likelihood.}
    \label{fig:gmm_initialization_comparison}
\end{figure}

As for K-means, each iteration of the EM algorithm for GMMs increases the likelihood or leaves it unchanged. We say that the algorithm has converged when the log-likelihood changes by less than a small chosen threshold between two iterations. The algorithm typically converges to a local maximum, which need not be the global maximum, and the final solution depends on the initialization. We therefore run the algorithm several times from different starting points and keep the solution with the highest likelihood. Figure~\ref{fig:gmm_initialization_comparison} shows two runs that end in different solutions.

Initialization matters more for GMMs than for K-means, because the parameter space is larger and the likelihood surface is more complex. A good initial guess, such as the K-means result, helps the algorithm to reach a good solution and speeds up convergence.\index{expectation-maximization (EM) algorithm|)}

\section{Advantages of Gaussian Mixture Models}\label{sec:ch11_gmm_advantages}

GMMs relax the restrictions of K-means and add new uses.

\paragraph{Comparison with K-means}

K-means is a special case of a GMM under three restrictions. First, the mixture weights are equal, $\pi_k = 1/K$. Second, all components have the same spherical covariance, $\boldsymbol{\Sigma}_k = \sigma^2\mathrm{Id}$. Third, we let $\sigma^2 \to 0$, which gives the closest center all the responsibility, so $\gamma_{ik}$ approaches the binary indicator $r_{ik}$ of K-means. K-means is thus the most constrained version of the GMM.

An important advantage of GMMs is soft assignment. The responsibility $\gamma_{ik}$ is the probability that data point $i$ belongs to cluster $k$. It expresses the uncertainty of the cluster membership.

\looseness=1 K-means finds round clusters, because it uses the Euclidean distance. The covariance matrices $\boldsymbol{\Sigma}_k$ let a GMM fit elongated or tilted clusters. The eigenvalues and eigenvectors of a covariance matrix give the size, shape and orientation of its ellipse (Chapter~10).

\paragraph{A Generative Model}

A GMM is a generative model. It describes the probability distribution of the data, so it can do more than assign clusters.

\textbf{Synthetic data.} Once we have fitted a GMM, we can generate new data points that follow the same distribution. We first select a component $k$ with probability $\pi_k$ and then draw a sample from $\mathcal{N}(\boldsymbol{\mu}_k, \boldsymbol{\Sigma}_k)$. A GMM fitted to observed stellar parameters can produce mock catalogs of stellar populations that keep the correlations of the data. Such catalogs help to test survey selection functions and to study observational biases.

\begin{figure}[ht!]
    \centering
    \includegraphics[width=0.95\textwidth]{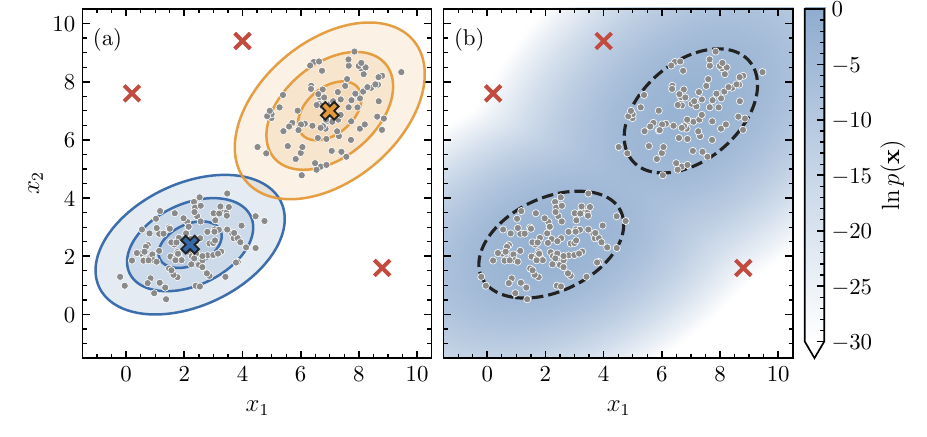}
    \caption{Outlier detection with a mixture of two Gaussian components. The mixture is fitted to the data without the three outliers (red crosses). Panel~(a) shows the data and the fitted components at one, two and three standard deviations. Panel~(b) shows the logarithm of the density $p(\mathbf{x})$ of the mixture. The dashed contour is the level of $\ln p(\mathbf{x})$ below which 5 percent of the data points in the fit lie. The three outliers lie far below this level.}
    \label{fig:gmm_outlier_detection}
\end{figure}

\textbf{Outlier detection.}\index{outliers} A fitted GMM gives the probability density of any data point $\mathbf{x}_*$
\begin{equation}
p(\mathbf{x}_*) = \sum_{k=1}^K \pi_k \mathcal{N}(\mathbf{x}_*|\boldsymbol{\mu}_k, \boldsymbol{\Sigma}_k).
\end{equation}Data points that lie far from all cluster centers or in low-density regions are potential outliers. Stars with unusual abundance patterns can indicate unique nucleosynthetic processes, and galaxies with atypical properties can challenge our understanding of galaxy formation. Figure~\ref{fig:gmm_outlier_detection} shows a density threshold, the 5th percentile of the density of the data, and three outliers below it.

The density tells us how unusual an object is, which goes beyond simple distance-based measures. In astronomy, outliers often represent rare but important objects rather than measurement errors.

\textbf{Density estimation.}\index{density estimation} A GMM estimates the probability density of the data, even when the density has several peaks, called modes. The estimated density shows where the clusters lie and how the probability is spread through the feature space. This can guide the search for the physical processes that produced the populations. It also shows regions of parameter space that hold few objects.\index{Gaussian mixture model (GMM)|)}

\section{Model Selection with Information Criteria}\label{sec:ch11_model_selection}

So\index{model selection|(} far we have assumed that the number of clusters $K$ is known. In practice, choosing $K$ is difficult. It is an example of model\index{model selection} selection, the choice of a model that explains the data well without overfitting.

The elbow method and the silhouette analysis do not use a probability model. A GMM has a likelihood, so we can use the Bayesian\index{information criteria} information criterion (BIC) and the Akaike information criterion (AIC).

\paragraph{The Model Selection Challenge}

As $K$ increases, the model can fit the data more closely, and the log-likelihood increases. The extra parameters may fit noise instead of real structure.

With $K=N$, one cluster per data point, the fit is perfect in terms of likelihood, but we learn nothing about the structure of the data. With $K=1$ the model is as simple as possible, but it fits poorly when the data contain several groups.

Information criteria balance the fit of a model against its complexity. They penalize the likelihood by the number of parameters of the model. They follow the principle of parsimony,\index{parsimony|see{Occam's razor}} often called Occam's\index{Occam's razor} razor. If several models explain the data equally well, we prefer the simplest one. A penalty helps us avoid overfitting.

\subsection{The Bayesian Information Criterion (BIC)}

BIC\index{BIC|see{Bayesian information criterion}}\index{Bayesian information criterion (BIC)}\index{Bayesian information criterion (BIC)|(} comes from a Bayesian approach to model selection. We treat the model itself as a variable. Let $\mathcal{M}_K$ denote a model with $K$ components. The posterior of the model given the data $\mathbf{X}$ is
\begin{equation}
p(\mathcal{M}_K|\mathbf{X}) = \frac{p(\mathbf{X}|\mathcal{M}_K)p(\mathcal{M}_K)}{p(\mathbf{X})}.
\end{equation}

If we have no reason to prefer a model, $p(\mathcal{M}_K)$ is uniform. The posterior probability is then proportional to the marginal\index{marginal likelihood} likelihood, or evidence, $p(\mathbf{X}|\mathcal{M}_K)$ (Chapter~5). The marginal likelihood is the integral over all values of the parameter vector $\boldsymbol{\theta}_K$, which holds the $P$ free parameters of $\mathcal{M}_K$
\begin{equation}
p(\mathbf{X}|\mathcal{M}_K) = \int p(\mathbf{X}|\boldsymbol{\theta}_K, \mathcal{M}_K)p(\boldsymbol{\theta}_K|\mathcal{M}_K)\mathrm{d}\boldsymbol{\theta}_K.
\end{equation}

The marginal likelihood averages the likelihood over the prior. Compare a complex model with many parameters to a simpler one. The complex model spreads its prior probability $p(\boldsymbol{\theta}_K|\mathcal{M}_K)$ thinly over a large parameter space, because the prior density must integrate to 1. The simpler model concentrates its prior in a smaller region and has a higher density there.

If the complex model does not fit the data much better, its likelihood cannot compensate for the thin prior, so its marginal likelihood is lower. Only if it fits much better over a large part of its parameter space does it overcome this disadvantage. Simple models therefore have a higher marginal likelihood unless the data support more complexity. This is Occam's razor, built into the integral.

\paragraph{The Laplace Approximation}

The marginal likelihood integral generally has no closed form. The Laplace\index{Laplace approximation} approximation (Chapter~9) approximates the logarithm of the integrand by a quadratic around its maximum
\begin{equation}
\ln[p(\mathbf{X}|\boldsymbol{\theta}_K, \mathcal{M}_K)p(\boldsymbol{\theta}_K|\mathcal{M}_K)] \approx \ln[p(\mathbf{X}|\hat{\boldsymbol{\theta}}_K, \mathcal{M}_K)p(\hat{\boldsymbol{\theta}}_K|\mathcal{M}_K)] - \frac{1}{2}(\boldsymbol{\theta}_K - \hat{\boldsymbol{\theta}}_K)^T \mathbf{A} (\boldsymbol{\theta}_K - \hat{\boldsymbol{\theta}}_K).
\end{equation}Here $\hat{\boldsymbol{\theta}}_K$ is the maximum of the integrand. The matrix $\mathbf{A}$ is the negative Hessian\index{Hessian matrix} of the logarithm of the integrand at this point, as in Chapter~9. For large samples the likelihood dominates the prior. Then $\hat{\boldsymbol{\theta}}_K$ is close to the maximum likelihood estimate, and $\mathbf{A}$ is approximately the observed information matrix $\mathcal{I}(\hat{\boldsymbol{\theta}}_K) = -\nabla\nabla \ln p(\mathbf{X}|\hat{\boldsymbol{\theta}}_K, \mathcal{M}_K)$, which measures how sharply the log-likelihood peaks.

With this Gaussian approximation the integral becomes
\begin{align}
p(\mathbf{X}|\mathcal{M}_K) &\approx p(\mathbf{X}|\hat{\boldsymbol{\theta}}_K, \mathcal{M}_K)p(\hat{\boldsymbol{\theta}}_K|\mathcal{M}_K) \int \exp\left(-\frac{1}{2}(\boldsymbol{\theta}_K - \hat{\boldsymbol{\theta}}_K)^T \mathbf{A} (\boldsymbol{\theta}_K - \hat{\boldsymbol{\theta}}_K)\right) \mathrm{d}\boldsymbol{\theta}_K \\*
&= p(\mathbf{X}|\hat{\boldsymbol{\theta}}_K, \mathcal{M}_K)p(\hat{\boldsymbol{\theta}}_K|\mathcal{M}_K) \cdot (2\pi)^{P/2}\det(\mathbf{A})^{-1/2}.
\end{align}

\paragraph{Deriving the BIC Penalty}

To see how $\det(\mathbf{A})$ depends on $N$, we start from the log-likelihood, which is a sum over $N$ independent observations
\begin{equation}
\ln p(\mathbf{X}|\boldsymbol{\theta}_K, \mathcal{M}_K) = \sum_{i=1}^N \ln p(\mathbf{x}_i|\boldsymbol{\theta}_K, \mathcal{M}_K).
\end{equation}With the prior neglected, $\mathbf{A}$ is also a sum over $N$ terms
\begin{equation}
\mathbf{A} = -\nabla\nabla \ln p(\mathbf{X}|\boldsymbol{\theta}_K, \mathcal{M}_K) = -\sum_{i=1}^N \nabla\nabla \ln p(\mathbf{x}_i|\boldsymbol{\theta}_K, \mathcal{M}_K).
\end{equation}This means that $\mathbf{A}$ grows linearly with $N$. We write $\mathbf{A} = N \cdot \mathcal{J}(\hat{\boldsymbol{\theta}}_K)$, where $\mathcal{J}(\hat{\boldsymbol{\theta}}_K)$ is the average information matrix per observation.

Scaling a $P \times P$ matrix by a factor $N$ scales its determinant by $N^P$. For the determinant of $\mathbf{A}$ this gives
\begin{align}
\det(\mathbf{A}) &= \det(N \cdot \mathcal{J}(\hat{\boldsymbol{\theta}}_K)) = N^P \cdot \det(\mathcal{J}(\hat{\boldsymbol{\theta}}_K)).
\end{align}

Taking the negative logarithm of the marginal likelihood and multiplying by 2 gives
\begin{align}
-2\ln p(\mathbf{X}|\mathcal{M}_K) &\approx -2\ln p(\mathbf{X}|\hat{\boldsymbol{\theta}}_K, \mathcal{M}_K) - 2\ln p(\hat{\boldsymbol{\theta}}_K|\mathcal{M}_K) - P\ln(2\pi) + \ln\det(\mathbf{A}) \\*
&= -2\ln p(\mathbf{X}|\hat{\boldsymbol{\theta}}_K, \mathcal{M}_K) - 2\ln p(\hat{\boldsymbol{\theta}}_K|\mathcal{M}_K) \\*
&\quad - P\ln(2\pi) + P\ln(N) + \ln\det(\mathcal{J}(\hat{\boldsymbol{\theta}}_K)).
\end{align}The log-likelihood term $-2\ln p(\mathbf{X}|\hat{\boldsymbol{\theta}}_K,\allowbreak \mathcal{M}_K)$ is a sum over the $N$ data points, so it scales as $\mathcal{O}(N)$. The penalty term $P\ln(N)$ scales as $\mathcal{O}(\ln N)$, which grows much more slowly. The remaining terms are the log-prior $-2\ln p(\hat{\boldsymbol{\theta}}_K|\allowbreak\mathcal{M}_K)$, the constant $-P\ln(2\pi)$ and the determinant term $\ln\det(\mathcal{J}(\hat{\boldsymbol{\theta}}_K))$. They approach constants as $N$ grows, so they are $\mathcal{O}(1)$.

For large $N$ the log-likelihood term dominates, followed by the penalty term $P\ln(N)$. The $\mathcal{O}(1)$ terms are small in comparison. Therefore
\begin{align}
-2\ln p(\mathbf{X}|\mathcal{M}_K) &\approx -2\ln p(\mathbf{X}|\hat{\boldsymbol{\theta}}_K, \mathcal{M}_K) + P\ln(N) + \mathcal{O}(1).
\end{align}

We keep the log-likelihood term and the penalty $P\ln(N)$, and we drop the $\mathcal{O}(1)$ terms. This gives BIC
\begin{equation}
\text{BIC} = -2\ln(\hat{L}) + P\ln(N).
\end{equation}
Here $\hat{L}$ is the maximum likelihood of the model, $P$ is the number of free parameters and $N$ is the number of data points. The number of free parameters of a GMM depends on the covariance structure. For full covariance matrices in $D$ dimensions it is
\begin{equation}
P = (K-1) + KD + K\frac{D(D+1)}{2} = K\left(1 + D + \frac{D(D+1)}{2}\right) - 1.
\end{equation}
Diagonal covariance matrices give $P = (K-1) + KD + KD$, and spherical ones give $P = (K-1) + KD + K$.

\paragraph{BIC Properties and Interpretation}

BIC is consistent. Under regularity conditions, it selects the true model as $N \to \infty$, if the true model is among the candidates. The penalty $P\ln(N)$ grows with the sample size, so BIC becomes more demanding as the data set grows. With more data we require stronger evidence for a more complex model.

We select the model with the lowest BIC. For two candidate models $\mathcal{M}_a$ and $\mathcal{M}_b$, the difference $\Delta\text{BIC} = \text{BIC}_a - \text{BIC}_b$ is approximately twice the logarithm of the Bayes\index{Bayes factor} factor of $\mathcal{M}_b$ over $\mathcal{M}_a$. The Bayes factor is the ratio of the marginal likelihoods of the two models, so $\Delta\text{BIC}$ measures the evidence for $\mathcal{M}_b$.

BIC assumes large samples compared with the number of parameters, a well-behaved likelihood and appropriately diffuse priors. In astronomical applications these conditions are not always met, for example with small samples, complex parameter interdependencies or multimodal distributions. BIC is therefore only one piece of evidence, and we must also use scientific knowledge.\index{Bayesian information criterion (BIC)|)}

\subsection{The Akaike Information Criterion (AIC)}

AIC\index{Akaike information criterion (AIC)}\index{Akaike information criterion (AIC)|(} comes from information theory and estimates how well a model predicts new data. It connects to the entropy and the cross-entropy of Chapter~8. It is based on the Kullback-Leibler\index{Kullback-Leibler divergence} divergence, which measures how far the fitted distribution is from the true one. In the coding picture of Chapter~8, this divergence is the extra average code length from using the fitted distribution instead of the true one. Akaike's idea was to choose the model with the smallest distance to the true data-generating process.

If we use the same data to fit a model and to evaluate it, we overestimate how well the model predicts new data. This optimistic bias grows with the number of parameters, because complex models can fit noise.

Akaike showed that for large $N$ this bias is about $P$. We subtract $P$ from the observed log-likelihood to estimate how well the model predicts new data. Multiplying the result by $-2$ gives the AIC
\begin{equation}
\text{AIC} = -2\ln(\hat{L}) + 2P.
\end{equation}

As for BIC, we select the model with the lowest AIC. The AIC penalty $2P$ is smaller than the BIC penalty $P\ln(N)$ when $\ln(N) > 2$. AIC therefore tends to select more complex models than BIC. For Gaussian errors with known variances, $-2\ln\hat{L}$ equals $\chi^2$ up to a constant, so AIC is $\chi^2 + 2P$ up to that constant. A new parameter must therefore lower $\chi^2$ by more than 2.

For small samples we use the corrected AIC, written $\text{AIC}_c$. It is defined as
\begin{equation}
\text{AIC}_c = \text{AIC} + \frac{2P(P+1)}{N-P-1}.
\end{equation}The standard AIC works well for large $N$ but can be biased when the sample is small compared with the number of parameters. The correction term grows as $P$ increases or $N$ decreases, so it penalizes complex models more when data are scarce.

\paragraph{BIC Versus AIC}

BIC\index{Bayesian information criterion (BIC)|(} and AIC follow different views of model selection. BIC assumes that one true model exists among the candidates and searches for it. AIC assumes that all models are approximations and chooses the model that predicts new data best. AIC is efficient. For large $N$ its choice approaches the smallest expected prediction error among the candidates, even when the true model is not among them.

\begin{figure}[ht!]
    \centering
    \includegraphics[width=0.65\textwidth]{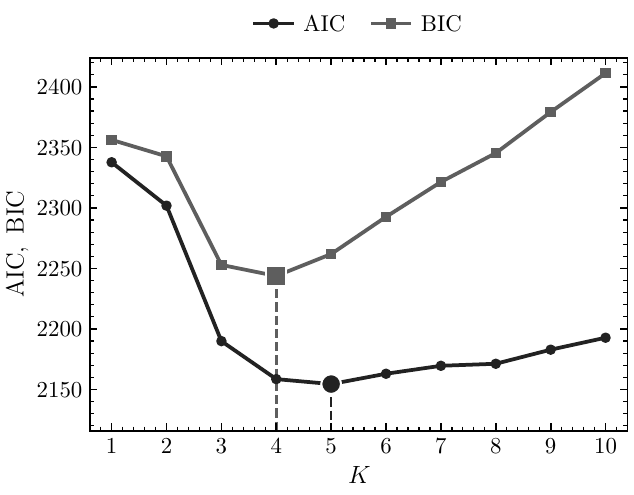}
    \caption{AIC and BIC for Gaussian mixtures with $K$ components, fitted to one data set drawn from four components. Filled dots mark the minimum of each criterion. In this example BIC has its minimum at the true number $K = 4$, and AIC has its minimum at $K = 5$.}
    \label{fig:information_criteria}
\end{figure}

Figure~\ref{fig:information_criteria} shows both criteria for a mixture with a varying number of components. Both fall while the fit improves and rise when the penalty outweighs the gain in fit.

\paragraph{Practical Application to GMMs}

For each $K$ we fit the GMM with the EM algorithm. The likelihood at convergence is $\hat{L}$, and we choose the $K$ with the lowest BIC or AIC. When BIC and AIC agree on the number of components, the choice is well supported. When they disagree, it helps to examine both solutions. The BIC solution typically contains the most prominent, well-separated clusters. The AIC solution can also contain subtler structures that may matter for the scientific question.

BIC and AIC have different aims, so the better criterion depends on the goal of the study. For a first look at the broad population structure, BIC is a good choice, because it selects fewer components. For a detailed study of subpopulations, AIC can help, because it allows more components. Choosing $K$ resembles choosing the number of principal components in Chapter~10, because both set the complexity of the model.

\medskip The M-steps of K-means and of GMMs have closed-form solutions. Complex probabilistic models often have no analytical solution. In Chapter~12 we turn to sampling and Monte Carlo methods for such models.\index{Akaike information criterion (AIC)|)}\index{Bayesian information criterion (BIC)|)}\index{model selection|)}

\section{Tutorials: Stellar Populations and Quasars}\label{sec:ch11_tutorials}

Principal component analysis (PCA) of Chapter~10 summarizes each object by a few coefficients. Clustering adds another summary, the group of each object, and a group can match a physical population. A mixture model also describes the whole distribution of the objects. Two tutorials apply K-means and Gaussian mixture models. The first looks for the stellar\index{stellar populations} populations of a dwarf galaxy, and the second models quasar spectra after dimension reduction. The full tutorials can be found at \url{https://statml.org/} as 11a and 11b.

\paragraph{Populations of a Dwarf Galaxy} Stars keep the composition of the gas that formed them, and supernovae change that gas. Massive stars explode soon after forming and release magnesium and some iron. Type\index{Type Ia supernovae} Ia supernovae, explosions of white dwarfs that gain mass from a binary companion, occur much later and release a large amount of iron. If stars form in bursts, these late explosions add iron between bursts, so the magnesium to iron ratio of the gas, [Mg/Fe], falls. The next burst forms stars from this gas, with a higher iron abundance,\index{chemical abundances} [Fe/H], and a lower [Mg/Fe] than the last burst. Each burst can therefore leave a group of stars in the plane of [Fe/H] and [Mg/Fe], and the gap between groups marks a quiet period.

\begin{figure}[ht!]
    \centering
    \includegraphics[width=0.95\textwidth]{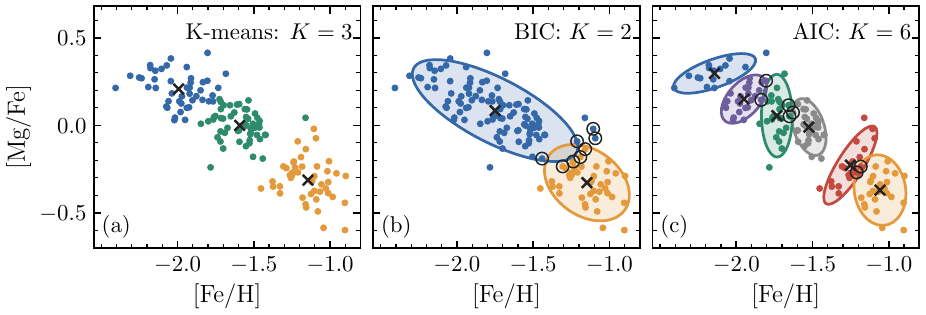}
    \caption{The first tutorial clusters stars of a dwarf galaxy by two measured abundance ratios relative to the Sun. Panel~(a) shows K-means with three clusters, and panels~(b) and~(c) fit Gaussian mixtures of one to eight components. The Bayesian information criterion (BIC) selects the mixture of panel~(b), and the Akaike information criterion (AIC), which penalizes less, selects that of panel~(c). Colors show the cluster or the most probable component of each star. Crosses mark the cluster centers and the component means, and ellipses show two standard deviations. Rings mark stars whose highest probability of belonging is below 0.8.}
    \label{fig:ch11_tutorial_populations}
\end{figure}

We look for these groups in an APOGEE\index{APOGEE survey} sample of the Sculptor dwarf galaxy. K-means puts each star in the cluster of the nearest of $K$ centers, as in Figure~\ref{fig:ch11_tutorial_populations}a. The elbow of its inertia, the total squared distance to the centers, points to three clusters (Section~\ref{sec:ch11_evaluating}). K-means also assumes round clusters of equal weight, each star in one cluster (Section~\ref{sec:ch11_kmeans_generative}), so its three clusters in Figure~\ref{fig:ch11_tutorial_populations}a have similar sizes. Burst populations can differ in weight and shape and can overlap, so we use a Gaussian mixture model (GMM). Each component describes one population, with a weight and covariance matrix, and each star has a probability of belonging to each (Section~\ref{sec:ch11_gmm}).

A GMM still needs its number of components $K$, but the likelihood rises as components are added and so does not choose it. We therefore penalize the number of parameters. The Bayesian information criterion (BIC) and the Akaike information criterion (AIC) are two such penalties. BIC penalizes more than AIC does, and more so as the sample grows, so it tends to select fewer components (Section~\ref{sec:ch11_model_selection}).

On these stars, BIC selects two broad populations (Figure~\ref{fig:ch11_tutorial_populations}b), and AIC, which penalizes less, selects more and smaller ones (Figure~\ref{fig:ch11_tutorial_populations}c). BIC gives the coarse picture, and AIC splits it into smaller groups that may be separate bursts or noise. The number of groups to keep therefore depends on the detail that the science needs.

K-means and the mixture model with AIC are consistent with star formation in repeated bursts. Each group would be one burst, and the gaps between the groups would be the quiet periods, so the number of groups counts the bursts. The same recipe applies to any sample whose populations overlap in a plane of properties. A criterion sets the number of groups, and we read it against the science.

\paragraph{A Mixture Model of Quasar Spectra} A\index{quasar spectra|(} mixture can also model a whole distribution of objects, so the second tutorial turns to quasars.\index{quasar spectra} A quasar is a supermassive black hole that accretes gas and shines brightly. Its spectrum has a smooth part from the accretion disk and broad emission lines from fast gas near the black hole. The strengths of the lines and the overall slope of the spectrum differ from quasar to quasar. We therefore want a model of the distribution of spectra, that is, which spectra occur and how often, so that we can find unusual ones. The tutorial uses simulated spectra with thousands of wavelengths each.

\begin{figure}[ht!]
    \centering
    \includegraphics[width=0.95\textwidth]{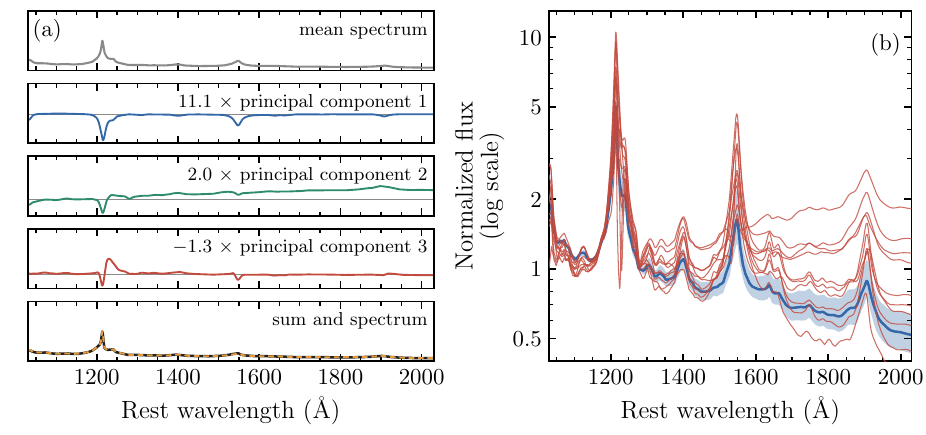}
    \caption{The second tutorial describes simulated quasar spectra with principal components and a Gaussian mixture. Panel~(a) shows one spectrum as the mean spectrum plus three principal components, each multiplied by its coefficient. The sum (orange) follows the spectrum (black), and each curve has its own vertical scale. In panel~(b), the blue band holds the middle two thirds of the spectra at each wavelength, and the blue line is their median. The red lines are the ten spectra of lowest density under the mixture, and each leaves the band at many wavelengths.}
    \label{fig:ch11_tutorial_quasars}
\end{figure}

A GMM can model this distribution, but the parameters of its covariance matrices grow quadratically with the dimension (Section~\ref{sec:ch11_gmm}). One Gaussian component would then need more parameters than there are spectra. Principal components solve this, because each spectrum is the mean spectrum plus a few numbers times the components, as in Figure~\ref{fig:ch11_tutorial_quasars}a. The mixture then models these numbers, the PCA coefficients, instead of the wavelengths. The explained variance ratio (Section~\ref{sec:ch10_implementing}) guides how many coefficients to keep, and BIC the number of Gaussian components.

Because the mixture describes the whole distribution, it can flag unusual spectra and generate synthetic ones (Section~\ref{sec:ch11_gmm_advantages}). The probability density of the coefficients of a spectrum measures how typical it is. The ten spectra of lowest density each leave the band of typical spectra at many wavelengths (Figure~\ref{fig:ch11_tutorial_quasars}b). Coefficients drawn from the mixture, multiplied by the principal components and added to the mean spectrum, give synthetic spectra (Figure~\ref{fig:ch11_tutorial_quasars}a). A survey can then follow up unusual spectra and test methods on mock catalogs.

In both tutorials, a criterion and the goal of the study set how many groups or components to keep. Once this number is chosen, a mixture gives a probability model of the population. It answers how typical an object is, which group it most likely joins and what a new sample would look like. These answers let us test a theory of how a population formed against what we observe. The same questions arise for stars, galaxies and transients.\index{quasar spectra|)}

\paragraph{Further Reading} Steinhaus (1956) formulated the division of a set into homogeneous groups, and Pearson (1894) studied mixtures of distributions. Lloyd (1982) gave the iterative algorithm that K-means implementations use. MacQueen (1967) introduced the name K-means. Hartigan and Wong (1979) gave a widely adopted K-means algorithm. Arthur and Vassilvitskii (2007) introduced K-means++, a careful choice of the starting centers. Day (1969) estimated the components of a mixture of normal distributions, and Wolfe (1970) recast clustering as fitting a mixture of multivariate distributions. Dempster et~al.\ (1977) gave the general form of the EM algorithm, a method for maximum likelihood from incomplete data. Redner and Walker (1984) reviewed mixture densities and their maximum likelihood estimation with EM. Titterington et~al.\ (1985) treat the statistical analysis of finite mixtures. Akaike (1974) and Schwarz (1978) developed AIC and BIC, and Bozdogan (1987) extended the theory of AIC. Kaufman and Rousseeuw (1990) give practical guidance on finding groups in data. Dunn (1973) and Bezdek (1981) developed fuzzy c-means, a K-means variant with soft assignments. Selim and Ismail (1984) proved a convergence theorem for K-means type algorithms. Hartigan (1975) is a classic reference on clustering algorithms, and Duda et~al.\ (2001) place them in the field of pattern classification. Ting and Ji (2025) used mixture models to measure bursty star formation in dwarf galaxies, the context and the sample of the first tutorial. Sun et~al.\ (2023) learned a latent factor model of quasar spectra, which generates the spectra of the second tutorial.

\makeatletter\@clubpenalty=150\makeatother \clubpenalty=150 \widowpenalty=150

\chapter{Sampling and Monte Carlo Methods}

\makeatletter\@clubpenalty=10000\makeatother \clubpenalty=10000 \widowpenalty=10000

Bayesian inference needs integrals over the posterior distribution of the model parameters. Earlier chapters computed them in closed form for simple models, for example Bayesian linear regression with a Gaussian prior and a Gaussian likelihood (Chapter~5). The posterior is then Gaussian, so we can compute with it directly or draw samples with standard packages. Models with non-linearities, hierarchical structures (Section~6.5) or complex priors quickly lose their closed forms. Even modest departures from the simple cases make the computation harder.

Bayes' theorem gives the posterior up to a normalization constant
\begin{equation}
p(\boldsymbol{\theta}|\mathcal{D}) \propto p(\mathcal{D}|\boldsymbol{\theta}) \cdot p(\boldsymbol{\theta}),
\end{equation}
where $\boldsymbol{\theta}$ are the parameters and $\mathcal{D}$ is the data.

The unnormalized posterior is often not enough for practical use. Most questions in Bayesian inference ask for an expectation with respect to the posterior
\begin{equation}
\mathbb{E}[f(\boldsymbol{\theta})] = \int f(\boldsymbol{\theta}) p(\boldsymbol{\theta}|\mathcal{D}) \mathrm{d}\boldsymbol{\theta},
\end{equation}where $f(\boldsymbol{\theta})$ is a function of the parameters. These integrals arise when we predict new observations, summarize parameter values or integrate out the nuisance parameters that we do not need to report.

\looseness=-1 We often chose Gaussian distributions and linear transformations because they are easy to handle mathematically. Logistic regression has a non-linearity, which made its posterior much harder to obtain. We could write down the unnormalized posterior, but we could not easily sample from it without approximations. We used the Laplace approximation for the posterior and the probit function in place of the logistic function for predictions (Chapter~9). These approximations become inadequate as models grow more complex.

Monte Carlo methods approximate these expectations with samples. If we can generate samples $\boldsymbol{\theta}^{(1)}, \boldsymbol{\theta}^{(2)}, \ldots, \boldsymbol{\theta}^{(N)}$ from the target distribution $p(\boldsymbol{\theta}|\mathcal{D})$, the posterior that we want to describe, then
\begin{equation}
\mathbb{E}[f(\boldsymbol{\theta})] \approx \frac{1}{N} \sum_{i=1}^{N} f(\boldsymbol{\theta}^{(i)}).
\end{equation}We used this idea earlier when we estimated means and variances from finite samples. Those estimates are simple Monte Carlo approximations, with the observed data as the samples.

The approximation becomes more accurate as the number of samples $N$ increases. By the law of large numbers, a sample average approaches the expectation as $N \to \infty$. By the central limit theorem, the error decreases in proportion to $1/\sqrt{N}$, written $\mathcal{O}(1/\sqrt{N})$, as for the sample mean (Section~3.6). This rate does not depend on the dimension, which helps for the high-dimensional parameter spaces of many astronomical models.

The name ``Monte Carlo'' comes from the casino in Monaco and refers to the role of chance. Scientists of the Manhattan Project coined it in the 1940s, when they used these methods for nuclear physics calculations.

Monte Carlo approximation works for any distribution from which we can generate samples. These methods do not depend on a particular model and serve regression, classification and clustering alike. We therefore need effective methods to generate samples from complex distributions. We study four ways to obtain samples or expectations. They are the grid approach, the inverse cumulative distribution function (CDF) method, rejection sampling and importance sampling.

\section{Historical Context}

People\index{Buffon's needle|(}\index{Monte Carlo methods|(} used random experiments to answer mathematical questions long before computers existed. One of the earliest examples is the needle experiment of Georges-Louis Leclerc, Comte de Buffon,\index{Buffon's needle} from the 18th century.

\looseness=1 The experiment estimates the constant $\pi$ from random physical events. We drop needles of length $L$ on a floor with parallel lines spaced a distance $T$ apart, where $L \leq T$. By counting the needles that cross a line, we estimate $\pi$ without any complex calculation.

Two random variables describe where a needle lands. The first is the perpendicular distance $x$ from the center of the needle to the nearest line, with $0 \leq x \leq T/2$. The second is the angle $\theta$ between the needle and the lines, with $0 \leq \theta < \pi$.

The two variables are independent and uniformly distributed. The distance $x$ is uniform over $[0, T/2]$ with probability density $2/T$, and the angle $\theta$ is uniform over $[0, \pi]$ with probability density $1/\pi$. A needle crosses a line when $x$ is at most half the projection of the needle onto the perpendicular direction, that is when
\begin{equation}
x \leq \frac{L}{2}\sin\theta.
\end{equation}The probability of crossing is then
\begin{align}
p(\text{crossing}) &= \int_{0}^{\pi} \int_{0}^{\frac{L}{2}\sin\theta} \frac{2}{T} \cdot \frac{1}{\pi} \, \mathrm{d}x \, \mathrm{d}\theta \\*
&= \int_{0}^{\pi} \frac{L\sin\theta}{T\pi} \, \mathrm{d}\theta \\*
&= 2 \frac{L}{\pi T}.
\end{align}

This result connects the probability of a crossing to the value of $\pi$. We drop $N$ needles in total and observe $n_{\rm c}$ crossings. The empirical probability is $n_{\rm c}/N$, and setting it equal to the theoretical probability gives
\begin{align}
\frac{n_{\rm c}}{N} &\approx 2 \frac{L}{\pi T} \\*
\pi &\approx 2 \frac{L}{T} \frac{N}{n_{\rm c}}.
\end{align}

\begin{figure}[ht!]
    \centering
    \includegraphics[width=0.95\textwidth]{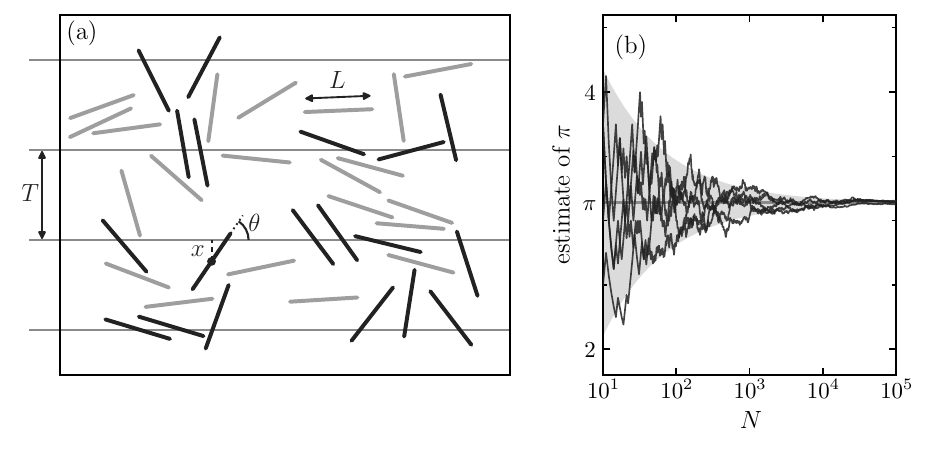}
    \caption{Buffon's needle experiment. Panel~(a) shows needles of length $L = 0.75T$ dropped on a floor with parallel lines spaced $T$ apart. Needles that cross a line are black and the others are grey. For the marked needle, $x$ is the distance from its center to the nearest line and $\theta$ is its angle with the lines. Panel~(b) shows the estimate of $\pi$ for five independent runs against the number of needles $N$. The gray band is one standard deviation of the estimate for large $N$, and its width decreases as $1/\sqrt{N}$.}
    \label{fig:buffon_needle}
\end{figure}

Figure~\ref{fig:buffon_needle} shows a simulated experiment that illustrates two principles that remain central to Monte\index{Monte Carlo methods} Carlo methods. First, random sampling estimates a quantity that would otherwise need complex calculations, here $\pi$. Second, the estimate becomes more accurate as the number of needles $N$ increases, but the error decreases only as $\mathcal{O}(1/\sqrt{N})$.

Buffon's experiment and modern sampling rely on the same principle, random samples that approximate an integral. Buffon used a physical experiment out of necessity. A computer generates random numbers much faster.\index{Buffon's needle|)}

\section{The Grid Approach}\label{sec:ch12_grid}

The\index{grid approach|(} grid\index{grid approach} approach turns the continuous problem of sampling from a distribution into a discrete problem that we solve by counting, as in Buffon's experiment. It replaces the distribution by a histogram, which introduces an approximation. The approximation vanishes as the grid becomes infinitely fine.

\looseness=1 A pinball machine illustrates the idea. A ball drops from the top, bounces through a series of pegs and lands in one of several bins at the bottom. Each bin collects balls with a frequency proportional to the number of paths that lead to it. In sampling, the bins are grid points, and the frequency in a bin corresponds to the probability of the grid point.

Consider a random variable $X$ with probability density function (PDF) $p(x)$ on a bounded domain $[a, b]$. We divide the domain into $n$ intervals of equal width, which gives the $n+1$ grid points
\begin{equation}
x_i = a + i\Delta x, \quad i = 0, 1, \ldots, n,
\end{equation}
where $\Delta x = (b-a)/n$ is the grid spacing.

At each grid point $x_i$ we evaluate the PDF, $p_i = p(x_i)$. These values do not sum to 1, so we normalize them. The normalization constant is $Z = \sum_{i=0}^{n} p_i$, and the normalized probabilities are $\tilde{p}_i = p_i/Z$. They are the probability masses of the grid points. Together they form a histogram approximation of the PDF.

Sampling from this discrete distribution is now straightforward. We partition the unit interval $[0,1]$ into segments with lengths proportional to the probabilities $\tilde{p}_i$.

For this we compute the cumulative probabilities
\begin{equation}
s_i = \sum_{j=0}^{i} \tilde{p}_j, \quad i = 0, 1, \ldots, n.
\end{equation}The cumulative probabilities are thresholds $\{s_0, s_1, \ldots, s_n\}$ with $s_0 = \tilde{p}_0$ and $s_n = 1$. The sampling procedure has three steps.
\begin{enumerate}
\item Generate a uniform random number $u \sim \mathcal{U}(0, 1)$, which means that intervals of equal length in $[0,1]$ have equal probability.
\item Find the index $i$ with $s_{i-1} < u \leq s_i$.
\item Return $x_i$ as the sample.
\end{enumerate}

This procedure ensures that the probability of sampling any grid point $x_i$ is $\tilde{p}_i$, as required.

\paragraph{Bimodal Gaussian Mixture Example}

\begin{figure}[ht!]
    \centering
    \includegraphics[width=0.85\textwidth]{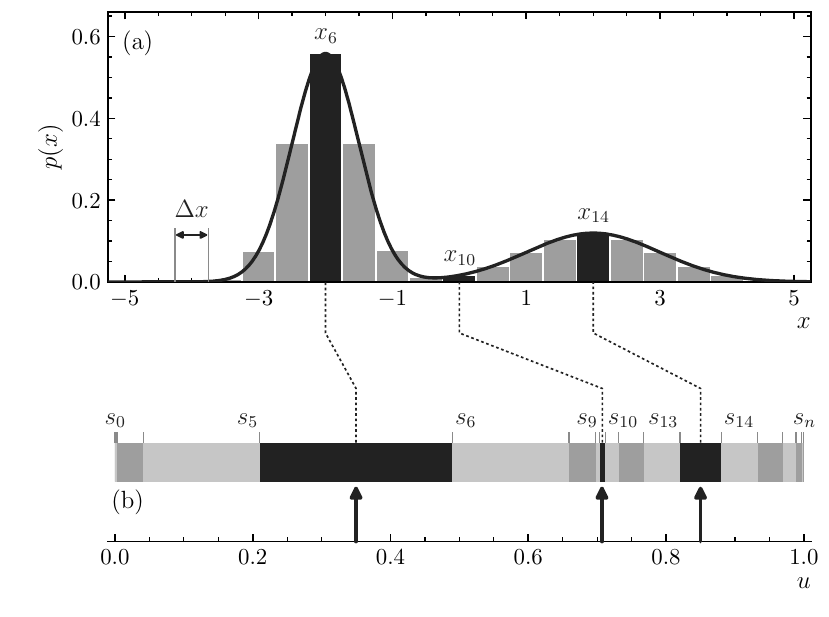}
    \caption{Grid sampling of the bimodal mixture on a coarse grid. Panel~(a) shows the exact density and the values $p_i$ at the grid points $x_i$ as bars of width $\Delta x$. Panel~(b) shows the unit interval divided into segments of length $\tilde{p}_i$, with the thresholds $s_i$ at their ends. Each arrow marks a uniform random number $u$. The black segment that contains $u$ returns the black grid point in panel~(a).}
    \label{fig:grid_sampling}
\end{figure}

Consider a mixture of two Gaussians
\begin{equation}
p(x) = 0.7 \cdot \mathcal{N}(x | -2, 0.5^2) + 0.3 \cdot \mathcal{N}(x | 2, 1^2).
\end{equation}This distribution has two modes. The dominant mode at $x = -2$ is narrower and taller, and the secondary mode at $x = 2$ is broader and shorter. We could sample from the mixture by selecting a component with probabilities equal to the mixture weights (Section~11.5) and then sampling from that Gaussian. The grid approach samples directly from the mixture. Figure~\ref{fig:grid_sampling} shows the grid, the thresholds and three samples for a coarse grid of $n = 20$ intervals.

The grid approach handles the two modes, provided that the grid resolution is fine enough to resolve the features of the distribution.

\paragraph{Extension to Multiple Dimensions}

The grid approach extends to several dimensions, but the number of grid points grows quickly. For a $D$-dimensional random variable $\mathbf{X} = (X_1, X_2, \ldots, X_D)$ with joint PDF $p(\mathbf{x})$, the discretization produces a $D$-dimensional grid
\begin{equation}
\mathbf{x}_{i_1, i_2, \ldots, i_D} = (a_1 + i_1\Delta x_1, a_2 + i_2\Delta x_2, \ldots, a_D + i_D\Delta x_D),
\end{equation}
where $i_j \in \{0, 1, \ldots, n_j\}$ for each dimension $j = 1, 2, \ldots, D$.

We evaluate the joint PDF at each grid point and normalize over all grid points. The sampling procedure is then the same as in the one-dimensional case.

\paragraph{The Curse of Dimensionality}

The main limit of the grid approach is the curse\index{curse of dimensionality} of dimensionality. With $n$ grid points per dimension, the number of grid points scales as $n^D$, which grows exponentially with the dimension $D$. The grid approach is therefore not feasible in high dimensions.

With 100 grid points per dimension, a one-dimensional problem needs 100 evaluations, a two-dimensional problem needs 10,000 and a five-dimensional problem needs 10 billion. At ten dimensions, the grid has $10^{20}$ points, far more than we can evaluate in practice. The grid approach remains practical in one or two dimensions.

The grid approach also has a discretization error, because the samples are restricted to the grid points. This error may be too large for applications that need precise samples from the continuous distribution. A finer grid reduces the error, but in high dimensions the number of grid points grows exponentially. This trade-off between accuracy and computing time motivates more sophisticated methods.\index{grid approach|)}

\section{Inverse CDF Method}\label{sec:ch12_inverse_cdf}

The\index{cumulative distribution function (CDF)|(}\index{inverse CDF method|(} grid approach partitions the unit interval by probability masses and uses a uniform random number to pick a grid point. Let the grid spacing go to zero. The cumulative probabilities then become the CDF\index{CDF|see{cumulative distribution function}}\index{cumulative distribution function (CDF)} of the continuous distribution. The result is the inverse\index{inverse CDF method} CDF method, also called inverse transform sampling, which samples from continuous distributions without discretization error.

For a random variable $X$ with PDF $p(x)$, the CDF $F(x)$ is defined as
\begin{equation}
F(x) = \int_{-\infty}^{x} p(t) \, \mathrm{d}t.
\end{equation}The function $F(x)$ is the probability that $X$ takes a value less than or equal to $x$. It has three properties that make it useful for sampling.
\begin{itemize}
\item $F(x)$ is non-decreasing, so $F(x_1) \leq F(x_2)$ if $x_1 < x_2$.
\item $\lim_{x \to -\infty} F(x) = 0$ and $\lim_{x \to \infty} F(x) = 1$.
\item For continuous distributions, $p(x) = \mathrm{d}F(x)/\mathrm{d}x$.
\end{itemize}

The first two properties mean that the CDF maps the support of the distribution, the set of $x$ with $p(x) > 0$, to the interval $[0,1]$. For example, the CDF of the standard Gaussian distribution is the function $\Phi(x)$ of Section~9.6.

The inverse CDF method uses this map in reverse. If we can compute the CDF and find its inverse function, we generate samples by transforming uniform random numbers.
\begin{enumerate}
\item Generate a uniform random number $u \sim \mathcal{U}(0,1)$.
\item Compute $x = F^{-1}(u)$, where $F^{-1}$ is the inverse of the CDF.
\item Return $x$ as a sample from the distribution with CDF $F$.
\end{enumerate}

The CDF rises quickly where the PDF is high, because outcomes are likely there. It rises slowly where the PDF is low. In this way, equal probability mass maps to segments of equal length in $[0,1]$.

More formally, suppose that $X$ has the CDF $F_X(x)$, and define $Y = F_X(X)$. We show that $Y$ is uniform on $[0,1]$ by computing its CDF. For any $y \in [0,1]$
\begin{align}
P(Y \leq y) &= P(F_X(X) \leq y) \\*
&= P(X \leq F_X^{-1}(y)) \\
&= F_X(F_X^{-1}(y)) \\*
&= y.
\end{align}This is the CDF of a uniform distribution on $[0,1]$. The inverse CDF method uses this result in reverse. If $U \sim \mathcal{U}(0,1)$, then $X = F_X^{-1}(U)$ follows the distribution with the CDF $F_X(x)$.

\begin{figure}[ht!]
    \centering
    \includegraphics[width=0.8\textwidth]{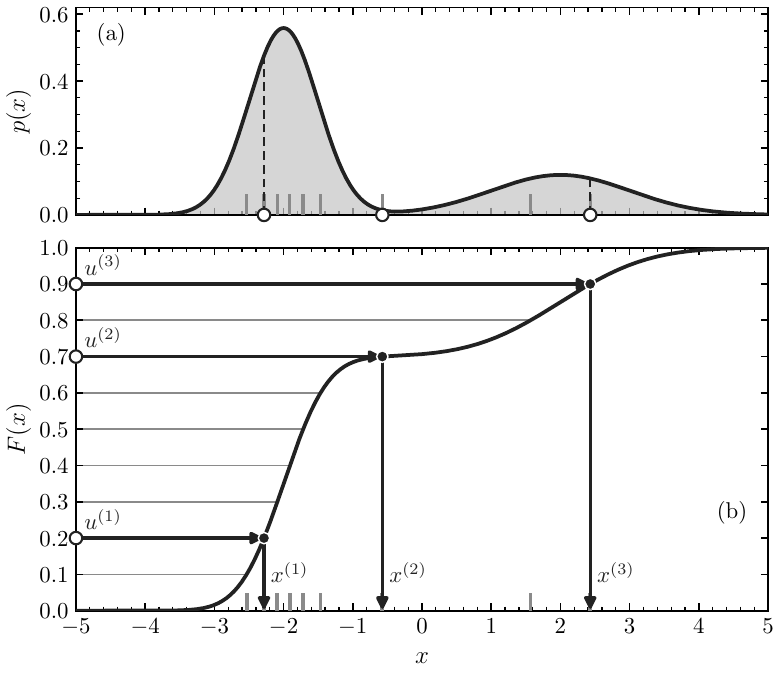}
    \caption{Inverse CDF method for the bimodal mixture. Panel~(a) shows the PDF $p(x)$ and panel~(b) shows the CDF $F(x)$. Each black arrow starts at a uniform random number $u^{(i)}$ and ends at the sample $x^{(i)} = F^{-1}(u^{(i)})$, which an open circle marks in panel~(a). The black arrows are three of the equally spaced heights $u = 0.1, 0.2, \ldots, 0.9$. Gray lines mark the other six, and the gray ticks on the horizontal axes are the projections of all nine heights. Most ticks lie close together near the peak of the left mode, where the PDF is high, because the CDF is steep there.}
    \label{fig:inverse_cdf_method}
\end{figure}

We can also read the procedure on the plot of the CDF. A uniform random number $u$ is a height on the vertical axis of the CDF plot. We move horizontally from height $u$ to the CDF curve and project down to the horizontal axis, which gives the sample $F^{-1}(u)$. Where the CDF is steep, the uniform heights are compressed on the horizontal axis and many samples fall there. Figure~\ref{fig:inverse_cdf_method} shows this for the bimodal mixture.

\paragraph{Exponential Distribution}

For\index{exponential distribution|(} the exponential\index{exponential distribution} distribution with PDF $p(x) = \lambda e^{-\lambda x}$ for $x \geq 0$, the CDF is
\begin{equation}
F(x) = 1 - e^{-\lambda x}.
\end{equation}
Solving for $x$ in terms of $u = F(x)$ gives
\begin{align}
u &= 1 - e^{-\lambda x} \\*
e^{-\lambda x} &= 1 - u \\*
x &= -\frac{1}{\lambda} \ln(1-u).
\end{align}Since $1-u$ is also uniform on $[0,1]$ when $u$ is uniform, we can simplify to\index{exponential distribution|)}
\begin{equation}
x = -\frac{1}{\lambda} \ln(u).
\end{equation}

\paragraph{Power Law Distribution}

Power\index{power law distribution} law distributions appear frequently in astronomy. Examples are the mass function of stars, the distribution of galaxy luminosities and the energy distribution of cosmic rays. For a power law on $[x_{\min}, \infty)$ with PDF $p(x) = (\alpha - 1)x_{\min}^{\alpha-1} x^{-\alpha}$ and $\alpha > 1$, the CDF is
\begin{equation}
F(x) = 1 - \left(\frac{x_{\min}}{x}\right)^{\alpha-1}.
\end{equation}Inverting this for $x$ in terms of $u = F(x)$ gives
\begin{align}
u &= 1 - \left(\frac{x_{\min}}{x}\right)^{\alpha-1} \\*
\left(\frac{x_{\min}}{x}\right)^{\alpha-1} &= 1 - u \\*
x &= x_{\min} (1-u)^{-\frac{1}{\alpha-1}}.
\end{align}Again $1-u \sim \mathcal{U}(0,1)$ when $u \sim \mathcal{U}(0,1)$, so
\begin{equation}
x = x_{\min} u^{-\frac{1}{\alpha-1}}.
\end{equation}

\paragraph{Advantages and Limitations}

The inverse CDF method has three advantages over the grid approach. It generates samples from the target distribution with no discretization error. Each sample needs one uniform random number and one function evaluation, which is efficient when the inverse CDF is analytic. The inverse CDF method is also easy to implement for distributions with a known inverse CDF.

The inverse CDF method also has limits. It needs a closed form for the inverse CDF $F^{-1}(u)$, which many distributions lack. Even when the CDF has a closed form, we may be unable to invert it analytically. A numerical inverse needs a root-finding algorithm for each sample, which can be expensive.

For multivariate distributions, the inverse CDF method meets additional difficulties. The CDF of $D$ variables gives one number for a point with $D$ coordinates. Many points give the same number, so one uniform random number $u$ does not determine a point. The Rosenblatt transformation samples the first variable from its marginal distribution. It then samples each later variable from its conditional distribution. This needs all the conditional distributions and their inverses in closed form, which is rarely possible.

The inverse CDF method remains the method of choice when the inverse CDF is analytic. Libraries implement it for standard distributions, often with optimizations for numerical stability and speed. For many posterior distributions, in particular those from complex likelihoods or hierarchical models, it does not apply, so we need other approaches.\index{cumulative distribution function (CDF)|)}\index{inverse CDF method|)}

\section{Rejection Sampling}\label{sec:ch12_rejection}

The\index{rejection sampling|(} inverse CDF method gives exact samples, but it needs an analytic CDF and an analytic inverse. Many distributions in Bayesian inference and astronomical modeling have neither. Rejection\index{rejection sampling} sampling handles any distribution that we can evaluate, even when we cannot integrate or invert it analytically.

We accept or reject candidate points according to whether they fall under the curve of the target density.

Think of throwing darts at a board. We throw darts randomly at a rectangular region that contains the curve, keep the darts that land below the curve and discard the others. The kept darts follow the shape of the curve, because more darts land below it where it is high.

Let $p(x)$ be the PDF of the target distribution. Rejection sampling uses a proposal\index{proposal distribution} distribution with PDF $q(x)$ that has two properties.

\begin{enumerate}
\item We can generate samples from $q(x)$ efficiently.
\item The scaled proposal covers the target. There is a constant $M \geq 1$ such that $M q(x) \geq p(x)$ for all $x$ in the domain.
\end{enumerate}

The second property makes $p(x)/(M q(x))$ a valid acceptance probability, because this ratio lies between 0 and 1 for all $x$.

The rejection sampling algorithm has four steps.
\begin{enumerate}
\item Generate a sample $x_* \sim q(x)$ from the proposal distribution.
\item Compute the acceptance probability $\alpha = p(x_*)/(M q(x_*))$.
\item Generate a uniform random number $u \sim \mathcal{U}(0, 1)$.
\item If $u \leq \alpha$, accept $x_*$ as a sample from $p(x)$. Otherwise reject it and return to step 1.
\end{enumerate}

Rejection sampling gives samples from $p(x)$ because we accept each candidate $x_*$ with the probability $p(x_*)/(M q(x_*))$. Candidates where $p(x_*)$ is large compared with $q(x_*)$ are accepted more often, and candidates where it is small are rejected more often.

\paragraph{Geometric Interpretation}

\begin{figure}[ht!]
    \centering
    \includegraphics[width=0.8\textwidth]{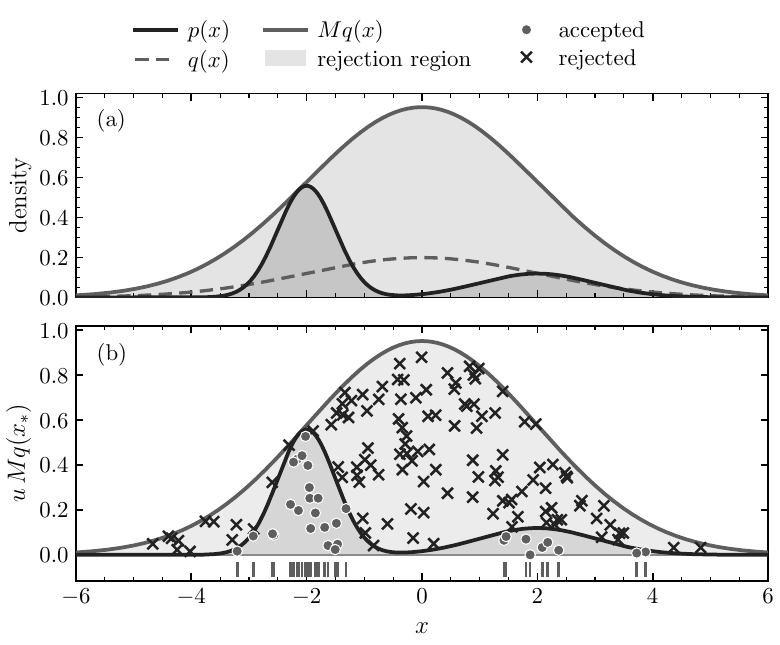}
    \caption{Rejection sampling for the bimodal mixture. Panel~(a) shows the target $p(x)$, the envelope $M q(x)$ and the proposal $q(x)$, which is a broad Gaussian. The constant $M \approx 4.8$ is the smallest value with $M q(x) \geq p(x)$. Panel~(b) shows the candidates $(x_*, u M q(x_*))$. Candidates below the curve $p(x)$ are accepted and the others are rejected. The marks on the horizontal axis are the accepted samples.}
    \label{fig:rejection_sampling}
\end{figure}

Rejection sampling works in a two-dimensional space, the domain of $x$ and a vertical axis for the probability density. The scaled proposal $M q(x)$ is an envelope above the target density. We generate the point $(x_*, u M q(x_*))$, where $u$ is uniform between 0 and 1, and accept it if it falls below the target density curve. Points between the target density and the envelope are rejected. Figure~\ref{fig:rejection_sampling} shows the accepted and rejected points for the bimodal mixture.

\looseness=-1 This is why rejection sampling is often described as ``sculpting'' the proposal distribution into the target. We start with samples from $q(x)$ and keep those that follow the shape of $p(x)$. The acceptance probability sets the fraction of candidates that we keep at each point.

\paragraph{Acceptance Rate and Efficiency}

The efficiency of rejection sampling depends on the proposal distribution and on the constant $M$. The expected number of iterations for one accepted sample equals $M$. Equivalently, the acceptance\index{acceptance rate} rate, which is the fraction of candidates that we accept, equals $1/M$. We therefore want $M$ as small as possible while $M q(x)$ stays above $p(x)$.

The best proposal has the shape of the target, with $M = 1$ and all samples accepted. This proposal needs samples from the target itself, which is the problem that we want to solve. In practice we balance ease of sampling, similarity to the target and coverage of the target by the scaled proposal.

\paragraph{Extension to Multiple Dimensions}

Unlike the inverse CDF method, rejection sampling extends to many dimensions. For a $D$-dimensional target PDF $p(\mathbf{x})$ we use a $D$-dimensional proposal PDF $q(\mathbf{x})$ and follow the same procedure. Rejection sampling needs only values of the density, which may be known only up to a normalization constant. This suits Bayesian inference, where we can write the unnormalized posterior but cannot sample from it directly.

\paragraph{The Curse of Dimensionality}

Rejection sampling becomes less efficient as the dimension increases, which is the curse\index{curse of dimensionality} of dimensionality. Consider a target and a proposal whose coordinates are independent and identically distributed, for example spherical Gaussians with different widths. If the acceptance rate in one dimension is $c < 1$, then in $D$ dimensions it is approximately $c^D$. It decreases exponentially with the dimension.

Heavy tails cause a second problem. If the tails of $p(x)$ decay more slowly than those of $q(x)$, the constant $M$ may be very large or infinite. For example, if $p(x)$ has polynomial tails and $q(x)$ has exponential tails, no finite $M$ satisfies $M q(x) \geq p(x)$ for all $x$.

Even so, rejection sampling remains useful in low dimensions and when the target and the proposal match well. It also applies in many cases where the inverse CDF method fails.\index{rejection sampling|)}

\section{Importance Sampling}\label{sec:ch12_importance}

The\index{importance sampling|(} inverse CDF method and rejection sampling generate samples that follow the target distribution. Often we want only an expectation with respect to the target, such as a posterior mean or variance, and importance\index{importance sampling} sampling estimates it directly.

Rejection sampling discards candidates, while importance sampling keeps all samples and weights each one to account for the difference between the proposal and the target. This can be more efficient for the same number of candidates.

Recall that many problems in Bayesian inference involve computing expectations of the form
\begin{equation}
\mathbb{E}_{p}[f] = \int f(x) p(x) \mathrm{d}x.
\end{equation}
Here $f(x)$ is a function of interest and $p(x)$ is the target PDF. If we could draw samples $\{x^{(1)}, x^{(2)}, \ldots, x^{(N)}\}$ directly from $p(x)$, we would approximate the expectation with the Monte Carlo estimator
\begin{equation}
\mathbb{E}_{p}[f] \approx \frac{1}{N} \sum_{i=1}^{N} f(x^{(i)}).
\end{equation}

Importance sampling needs no samples from $p(x)$. We sample from a different distribution $q(x)$ that is easier to work with, and we correct the calculation.

We rewrite the expectation with the proposal distribution $q(x)$
\begin{align}
\mathbb{E}_{p}[f] &= \int f(x) p(x) \mathrm{d}x \\*
&= \int f(x) \frac{p(x)}{q(x)} q(x) \mathrm{d}x \\*
&= \mathbb{E}_{q}\left[f(x) \frac{p(x)}{q(x)}\right].
\end{align}The subscripts $p$ and $q$ show which distribution each expectation is taken with respect to. We have changed an expectation under $p(x)$, which is hard to sample, into an expectation under $q(x)$, which is easy to sample.

This gives the importance sampling estimator
\begin{equation}
\mathbb{E}_{p}[f] \approx \frac{1}{N} \sum_{i=1}^{N} w(x^{(i)}) f(x^{(i)}),
\end{equation}
where $x^{(i)} \sim q(x)$ and $w(x) = p(x)/q(x)$ is the importance weight.

Points that are more likely under $p(x)$ than under $q(x)$ receive higher weights, and points that are less likely receive lower weights.

\paragraph{Handling Unnormalized Distributions}

Like rejection sampling, importance sampling handles unnormalized\index{normalization constant} distributions. Suppose that we know $p(x)$ only up to a constant, $p(x) = \tilde{p}(x)/Z_p$, where $\tilde{p}(x)$ is the unnormalized density and $Z_p = \int \tilde{p}(x) \mathrm{d}x$ is the normalization constant. Similarly, $q(x) = \tilde{q}(x)/Z_q$. The tilde here marks an unnormalized density, unlike the normalized masses $\tilde{p}_i$ of the grid approach.

The expectation becomes
\begin{align}
\mathbb{E}_{p}[f] &= \frac{1}{Z_p} \int f(x) \tilde{p}(x) \mathrm{d}x \\*
&= \frac{Z_q}{Z_p} \mathbb{E}_{q}\left[f(x) \frac{\tilde{p}(x)}{\tilde{q}(x)}\right].
\end{align}The second line uses $\tilde{q}(x) = Z_q q(x)$. We do not need $Z_p$ and $Z_q$ separately. We need only their ratio, which the samples provide.

We can also write $Z_p$ as
\begin{align}
Z_p &= \int \tilde{p}(x) \mathrm{d}x \\*
&= Z_q \mathbb{E}_{q}\left[\frac{\tilde{p}(x)}{\tilde{q}(x)}\right].
\end{align}Therefore
\begin{align}
\frac{Z_p}{Z_q} = \mathbb{E}_{q}\left[\frac{\tilde{p}(x)}{\tilde{q}(x)}\right],
\end{align}
and we estimate this ratio with the samples from $q(x)$ and the unnormalized weights $\tilde{w}(x) = \tilde{p}(x)/\tilde{q}(x)$
\begin{align}
\frac{Z_p}{Z_q} \approx \frac{1}{N} \sum_{i=1}^{N} \frac{\tilde{p}(x^{(i)})}{\tilde{q}(x^{(i)})} = \frac{1}{N} \sum_{i=1}^{N} \tilde{w}(x^{(i)}).
\end{align}

The factor $Z_q/Z_p$ is the inverse of the ratio that we just estimated. Substituting its estimate and the sample estimate of the expectation under $q(x)$ gives
\begin{align}
\mathbb{E}_{p}[f] &= \frac{Z_q}{Z_p} \mathbb{E}_{q}\left[f(x) \frac{\tilde{p}(x)}{\tilde{q}(x)}\right] \\*
&\approx \frac{\sum_{i=1}^{N} \tilde{w}(x^{(i)}) f(x^{(i)})}{\sum_{i=1}^{N} \tilde{w}(x^{(i)})}.
\end{align}The factors $1/N$ cancel. This is the self-normalized\index{importance sampling!self-normalized} importance sampling estimator.

For the posterior, the normalization constant is the evidence $p(\mathcal{D}) = \int p(\mathcal{D}|\boldsymbol{\theta})p(\boldsymbol{\theta})\mathrm{d}\boldsymbol{\theta}$ in the denominator of Bayes' theorem. It is typically intractable, but the self-normalized estimator does not need it.

\paragraph{Understanding Efficiency Through Weights}

The importance sampling estimator is valid for any proposal $q(x)$ that covers the support of $p(x)$, which means that $q(x) > 0$ wherever $p(x) \neq 0$. Its practical quality depends on how well $q(x)$ approximates $p(x)$.

If $q(x)$ matches $p(x)$, all importance weights are equal and all samples contribute equally. If the two distributions match only in part, some weights are larger than others, and the estimate relies on fewer samples.

The main problem is a proposal that undershoots the target, which means that it has too little probability mass where the target is high. There the weights $w(x) = p(x)/q(x)$ can become very large. A few samples with very large weights can dominate the estimate, even when $N$ is large.\index{importance sampling|)}

\section{Effective Sample Size}\label{sec:ch12_ess}

The\index{effective sample size (ESS)|(} effective\index{effective sample size (ESS)} sample size (ESS) measures how much information the weighted samples contain. Although we draw $N$ samples from the proposal, the variance of the estimator can equal that of far fewer direct samples from $p(x)$. We use the ESS to design good proposals and to find bad ones.

\paragraph{Intuitive Understanding}

In a poll of public opinion, each person's vote should carry about the same weight. If a few votes count as thousands and most votes count as fractions, a few people dominate the estimate. Even when we poll many people, the estimate is unreliable. Importance sampling has the same problem when a few samples have very large weights.

The ESS is the number of direct samples from $p(x)$ that would give the same precision as our $N$ weighted samples from $q(x)$.

\paragraph{The Effective Sample Size Formula}

The appendix derives the effective sample size from the variance of the self-normalized estimator, using approximations. The result, which does not depend on $f$, is
\begin{equation}
\text{ESS} \approx \frac{(\sum_{i=1}^{N} \tilde{w}(x^{(i)}))^2}{\sum_{i=1}^{N} \tilde{w}(x^{(i)})^2},
\end{equation}
where $\tilde{w}(x^{(i)})$ are the unnormalized importance weights.

\looseness=-1 The ESS is $N$ times the ratio of the squared mean of the weights to the mean of their squares. It equals $N$, the maximum, when all weights are equal. It decreases as the weights differ more, and it drops toward 1 when one weight dominates.

\paragraph{A Concrete Example}

We estimate the mean of a standard normal distribution $\mathcal{N}(0,1)$ that is truncated to the region $[5,\infty)$. Truncation keeps only the part with $x \geq 5$ and renormalizes it. The true mean is about 5.19. This could be the average brightness of the brightest stars in a population whose brightness follows a normal distribution.

The target distribution is
\begin{equation}
p(x) = \frac{\mathcal{N}(x|0,1) \cdot \mathbf{1}(x \geq 5)}{\int_{5}^{\infty} \mathcal{N}(t|0,1) \mathrm{d}t},
\end{equation}
where $\mathbf{1}(x \geq 5)$ is the indicator function, which equals 1 for $x \geq 5$ and 0 otherwise. The numerator is $\tilde{p}(x)$ and the integral in the denominator is $Z_p$.

A naive proposal is the standard normal distribution, $q(x) = \mathcal{N}(x|0,1)$. The unnormalized importance weights are then
\begin{equation}
\tilde{w}(x) = \frac{\mathcal{N}(x|0,1) \cdot \mathbf{1}(x \geq 5)}{\mathcal{N}(x|0,1)} = \mathbf{1}(x \geq 5).
\end{equation}Only the samples that fall in $[5,\infty)$ have a non-zero weight. A sample from a standard normal distribution falls there with a probability of about $2.87 \times 10^{-7}$, so we need millions of samples to obtain a few. With $N = 10^7$ samples we expect about 3 samples with non-zero weights.

Suppose that 3 samples have non-zero weights. The unnormalized weights are
\begin{equation}
\tilde{w}(x^{(i)}) =
\begin{cases}
1 & \text{for } i \in \{i_1, i_2, i_3\} \text{ (the 3 samples in } [5,\infty) \text{)} \\
0 & \text{for all other } i
.\end{cases}
\end{equation}The ESS formula gives
\begin{align}
\text{ESS} &\approx \frac{(\sum_{i=1}^{N} \tilde{w}(x^{(i)}))^2}{\sum_{i=1}^{N} \tilde{w}(x^{(i)})^2} \\*
&= \frac{(1 + 1 + 1)^2}{1^2 + 1^2 + 1^2} \\*
&= \frac{9}{3} = 3.
\end{align}Here the ESS equals the number of samples with a non-zero weight. The ESS is limited by the rarity of the region of interest, so the estimator is very inefficient.

A better proposal puts more probability mass in the region of interest. For example, we can use an exponential distribution that is shifted to start at 5
\begin{equation}
q(x) = \lambda e^{-\lambda(x-5)} \cdot \mathbf{1}(x \geq 5),
\end{equation}
where $\lambda > 0$ is a parameter. With this proposal, all samples fall in the region of interest $[5,\infty)$ and have non-zero weights. For $\lambda = 3$ the proposal has a similar shape to the target in the region of interest, so the importance weights remain relatively balanced. For large $N$ the sums in the ESS formula approach $N$ times expectations under $q(x)$, so
\begin{align}
\frac{\text{ESS}}{N} &\approx \frac{(\mathbb{E}_q[\tilde{w}(x)])^2}{\mathbb{E}_q[\tilde{w}(x)^2]} \\*
&= \frac{\left(\int_5^\infty \mathcal{N}(x|0,1)\,\mathrm{d}x\right)^2}{\int_5^\infty \mathcal{N}(x|0,1)^2/q(x)\,\mathrm{d}x} \approx \frac{(2.867 \times 10^{-7})^2}{1.014 \times 10^{-13}} \approx 0.81.
\end{align}Here $\tilde{w}(x) = \mathcal{N}(x|0,1)/q(x)$ on $[5,\infty)$, and numerical integration shows that the ESS is about 80\% of $N$.

\begin{figure}[ht!]
    \centering
    \includegraphics[width=0.95\textwidth]{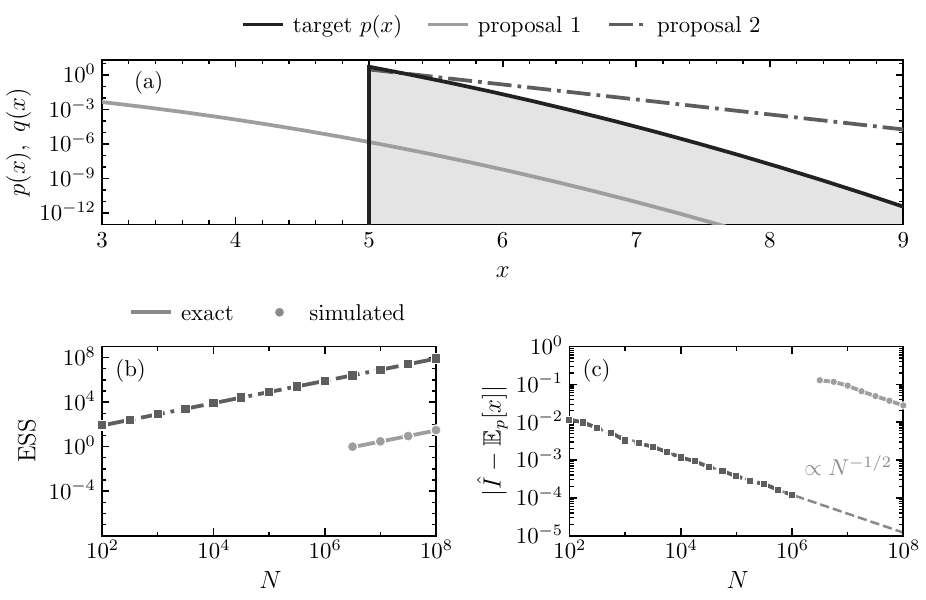}
    \caption{Importance sampling for the mean of a standard normal distribution truncated to $x \geq 5$. Panel~(a) shows the target, the standard normal proposal 1 and the shifted exponential proposal 2 on a logarithmic axis. Panel~(b) shows the effective sample size, which is a large fraction of $N$ for proposal 2 and tiny for proposal 1. Panel~(c) shows the error $|\hat{I} - \mathbb{E}_p[x]|$ of the self-normalized estimate $\hat{I}$ of the mean, averaged over repeated runs. In panels~(b) and (c), proposal 1 has no curve for small $N$, where many runs have no sample in the tail. The dashed line is a fit to proposal 2 that decreases as $1/\sqrt{N}$.}
    \label{fig:importance_sampling_proposals}
\end{figure}

Figure~\ref{fig:importance_sampling_proposals} compares the two proposals. A proposal that matches the target in the region of interest gives a smaller error in the estimated mean for the same $N$. The shifted exponential proposal converges quickly with few samples, while the standard normal proposal needs billions of samples for a comparable error.

\paragraph{Practical Guidelines}

\looseness=-1 When the ESS is small, the estimates have a high variance. The standard error of such an estimate, which is its standard deviation over repeated runs, scales as $\mathcal{O}(1/\sqrt{\text{ESS}})$ instead of $\mathcal{O}(1/\sqrt{N})$. We therefore want a proposal with a high ESS.

When we choose a proposal, we follow four guidelines.
\begin{itemize}
    \item The support of $q(x)$ must include the support of $p(x)$.

    \item Ideally $q(x)$ is proportional to $|f(x)|p(x)$, which minimizes the variance of the estimator $(1/N)\sum_{i=1}^{N} w(x^{(i)}) f(x^{(i)})$. The ESS does not depend on $f$ and is largest for $q(x) = p(x)$, whereas the choice above gives the lowest variance for one particular $f$.

    \item $q(x)$ should have heavier tails than $p(x)$ (Section~12.4), so that extreme weights in the tails do not reduce the effective sample size.

    \item $q(x)$ should be easy to sample from and to evaluate. Otherwise we gain little compared with sampling from $p(x)$ directly.
\end{itemize}

\medskip Astronomical models often have dozens or hundreds of parameters. The grid approach and rejection sampling become impractical in many dimensions, because their computing time grows exponentially with the dimension. Importance sampling is also difficult in many dimensions, because a proposal that covers the target is hard to find. We describe Markov Chain Monte Carlo methods in Chapter~13. They generate correlated samples that explore the parameter space through local moves guided by the target distribution.\index{effective sample size (ESS)|)}

\section{Appendix: Derivation of Effective Sample Size}

This\index{effective sample size (ESS)|(} appendix derives the effective sample size formula for importance sampling. The derivation uses the delta\index{delta method} method to analyze the variance of the self-normalized importance sampling estimator.

We start with our self-normalized importance sampling estimator
\begin{equation}
\hat{I} = \frac{\sum_{i=1}^{N} \tilde{w}(x^{(i)}) f(x^{(i)})}{\sum_{i=1}^{N} \tilde{w}(x^{(i)})}.
\end{equation}

For clarity, we define the normalized weights
\begin{equation}
w_i = \frac{\tilde{w}(x^{(i)})}{\sum_{j=1}^{N} \tilde{w}(x^{(j)})},
\end{equation}
where the weights sum to 1, $\sum_{i=1}^{N} w_i = 1$. Our estimator becomes
\begin{equation}
\hat{I} = \sum_{i=1}^{N} w_i f(x^{(i)}).
\end{equation}

For comparison, suppose that we could sample directly from $p(x)$. The variance of such a direct Monte Carlo estimator is
\begin{equation}
\text{Var}_{\text{direct}}(\hat{I}) = \frac{\text{Var}_p(f(x))}{N},
\end{equation}where $\text{Var}_p$ is the variance under $p(x)$. In the same way, $\text{Var}_q$ and $\text{Cov}_q$ below are the variance and covariance under $q(x)$.

\paragraph{Variance by the Delta Method}

Our self-normalized importance sampling estimator is a ratio of two random quantities
\begin{equation}
\hat{I} = \frac{A}{B},
\end{equation}
where
\begin{align}
A &= \frac{1}{N}\sum_{i=1}^{N} \tilde{w}(x^{(i)})f(x^{(i)}) \\*
B &= \frac{1}{N}\sum_{i=1}^{N} \tilde{w}(x^{(i)}).
\end{align}

For large $N$, both $A$ and $B$ converge to their expected values
\begin{align}
\mathbb{E}[A] &= \mathbb{E}_q[\tilde{w}(x)f(x)] \\*
\mathbb{E}[B] &= \mathbb{E}_q[\tilde{w}(x)].
\end{align}

To derive the variance of our estimator $\hat{I} = A/B$, we apply the delta method. The delta method approximates the variance of a function of random variables with a first-order Taylor expansion. We expand $\hat{I} = A/B$ around the expected values $\mathbb{E}[A]$ and $\mathbb{E}[B]$
\begin{align}
\hat{I} &\approx \frac{\mathbb{E}[A]}{\mathbb{E}[B]} + \frac{\partial}{\partial A}\left(\frac{A}{B}\right)\bigg|_{A=\mathbb{E}[A],B=\mathbb{E}[B]} (A - \mathbb{E}[A]) + \frac{\partial}{\partial B}\left(\frac{A}{B}\right)\bigg|_{A=\mathbb{E}[A],B=\mathbb{E}[B]} (B - \mathbb{E}[B]) \\*
&= \frac{\mathbb{E}[A]}{\mathbb{E}[B]} + \frac{1}{\mathbb{E}[B]}(A - \mathbb{E}[A]) - \frac{\mathbb{E}[A]}{\mathbb{E}[B]^2}(B - \mathbb{E}[B]).
\end{align}

We take the variance of this approximation and note that $\mathbb{E}[A]/\mathbb{E}[B]$ is a constant
\begin{align}
\text{Var}(\hat{I}) &\approx \text{Var}\left(\frac{1}{\mathbb{E}[B]}(A - \mathbb{E}[A]) - \frac{\mathbb{E}[A]}{\mathbb{E}[B]^2}(B - \mathbb{E}[B])\right) \\*
&= \frac{1}{\mathbb{E}[B]^2}\text{Var}(A) + \frac{\mathbb{E}[A]^2}{\mathbb{E}[B]^4}\text{Var}(B) - 2\frac{\mathbb{E}[A]}{\mathbb{E}[B]^3}\text{Cov}(A,B).
\end{align}

Recalling that $A = (1/N)\sum_{i=1}^{N} \tilde{w}(x^{(i)})f(x^{(i)})$ and $B = (1/N)\sum_{i=1}^{N} \tilde{w}(x^{(i)})$, and that $\mathbb{E}[A]/\mathbb{E}[B]$ equals $\mathbb{E}_p[f]$, which we write as $I$, we can express this in the original variables
\begin{align}
\text{Var}(\hat{I}) &\approx \frac{1}{N}\frac{\text{Var}_q(\tilde{w}(x)f(x)) + I^2\text{Var}_q(\tilde{w}(x)) - 2I\text{Cov}_q(\tilde{w}(x)f(x),\tilde{w}(x))}{(\mathbb{E}_q[\tilde{w}(x)])^2} \\*
&= \frac{1}{N}\frac{\text{Var}_q(\tilde{w}(x)f(x) - I \cdot \tilde{w}(x))}{(\mathbb{E}_q[\tilde{w}(x)])^2}.
\end{align}The last step uses $\text{Var}(X-Y) = \text{Var}(X) + \text{Var}(Y) - 2\text{Cov}(X,Y)$, where $X = \tilde{w}(x)f(x)$ and $Y = I \cdot \tilde{w}(x)$.

This formula shows how the uncertainty in both the numerator and the denominator of our estimator contributes to the overall variance. The term $\tilde{w}(x)f(x) - I\,\tilde{w}(x)$ equals $\tilde{w}(x)(f(x) - I)$. Its mean under $q(x)$ is $\mathbb{E}[A] - I\,\mathbb{E}[B]$, which is zero. Its variance is therefore the mean of its square.

The variance of the estimator becomes
\begin{equation}
\text{Var}(\hat{I}) \approx \frac{1}{N} \frac{\mathbb{E}_q[\tilde{w}(x)^2 (f(x) - I)^2]}{(\mathbb{E}_q[\tilde{w}(x)])^2}.
\end{equation}

\paragraph{Equating the Variances}

To find the effective sample size, we set this variance equal to the variance of direct sampling
\begin{equation}
\frac{\text{Var}_p(f(x))}{\text{ESS}} = \frac{1}{N} \frac{\mathbb{E}_q[\tilde{w}(x)^2 (f(x) - I)^2]}{(\mathbb{E}_q[\tilde{w}(x)])^2}.
\end{equation}

Solving for the ESS gives
\begin{equation}
\text{ESS} = N \frac{\text{Var}_p(f(x)) \cdot (\mathbb{E}_q[\tilde{w}(x)])^2}{\mathbb{E}_q[\tilde{w}(x)^2 (f(x) - I)^2]}.
\end{equation}

Since we typically do not know $\text{Var}_p(f(x))$ and $f(x)$ differs between applications, we derive a function-independent approximation. We replace $(f(x) - I)^2$ in the denominator by its mean under $p(x)$, which is $\text{Var}_p(f(x))$. This treats $(f(x) - I)^2$ as unrelated to the weights, and $\text{Var}_p(f(x))$ cancels. The result is
\begin{equation}
\text{ESS} \approx N \frac{(\mathbb{E}_q[\tilde{w}(x)])^2}{\mathbb{E}_q[\tilde{w}(x)^2]}.
\end{equation}

Using sample estimates, this becomes\index{Monte Carlo methods|)}\index{effective sample size (ESS)|)}
\begin{equation}
\text{ESS} \approx N \frac{(\frac{1}{N}\sum_{i=1}^{N} \tilde{w}(x^{(i)}))^2}{\frac{1}{N}\sum_{i=1}^{N} \tilde{w}(x^{(i)})^2} = \frac{(\sum_{i=1}^{N} \tilde{w}(x^{(i)}))^2}{\sum_{i=1}^{N} \tilde{w}(x^{(i)})^2}.
\end{equation}

\section{Tutorial: Sampling Stellar Masses}\label{sec:ch12_tutorial}

Astronomers\index{initial mass function (IMF)|(} often need random samples from a distribution that standard generators do not cover, to simulate a population or to explore a posterior. This tutorial draws stellar masses from the initial\index{initial mass function (IMF)} mass function (IMF).\index{IMF|see{initial mass function}}\index{initial mass function (IMF)} It uses the three methods of this chapter, inversion of the cumulative distribution function (CDF), rejection sampling and importance sampling. The full tutorial can be found at \url{https://statml.org/} as tutorial 12.

The IMF is the distribution of stellar masses at birth. A star's mass sets its luminosity, its lifetime and the elements it returns to the gas. The IMF therefore shapes how a stellar population shines, ages and enriches its surroundings. We cannot observe it directly, partly because massive stars die young. Astronomers instead simulate populations from a model IMF and compare them with observations, which needs many random masses.

\begin{figure}[b!]
    \centering
    \includegraphics[width=0.95\textwidth]{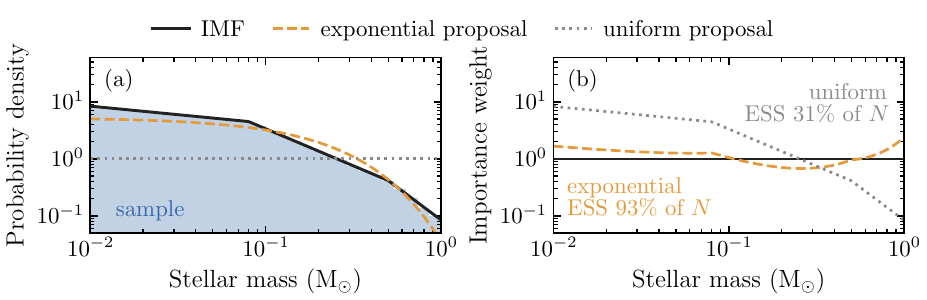}
    \caption{The tutorial samples the initial mass function (IMF). Panel~(a) shows the IMF, a histogram of simulated masses drawn by inverse transform sampling, and two proposals for importance sampling. Panel~(b) shows the weights of the proposals, with a thin line for equal weights. The labels give the effective sample size (ESS) as a share of the $N$ samples. The exponential proposal follows the IMF more closely and has the larger ESS.}
    \label{fig:ch12_tutorial_sampling}
\end{figure}

The most direct route inverts the CDF, the probability of a mass below a given value (Section~\ref{sec:ch12_inverse_cdf}). Each uniform random number from the computer then maps to one mass. Below one solar mass the IMF is a power law in each of a few mass ranges. The CDF of a power law inverts in closed form. Every random number gives an exact mass, and the histogram in Figure~\ref{fig:ch12_tutorial_sampling}a follows the IMF.

Many distributions have no CDF that we can invert, and rejection sampling needs only the density (Section~\ref{sec:ch12_rejection}). It draws candidates from a simpler proposal, here a decaying exponential scaled to lie above the IMF, and keeps those that fall under the IMF. The closer the proposal follows the IMF, the fewer candidates are discarded for each sample that we keep. Its samples follow the same IMF as those of the inversion.

Some questions need an average over the IMF rather than samples, such as the mean mass, which turns a star count into a total mass. Importance sampling estimates such an average from a convenient proposal and weights each sample by the ratio of the IMF to the proposal (Section~\ref{sec:ch12_importance}). As in rejection sampling, the proposal matters. Where it lies far below the IMF, a few large weights dominate the estimate. The effective sample size (ESS) counts how many direct draws from the IMF the weighted samples are worth (Section~\ref{sec:ch12_ess}).

The tutorial compares an exponential and a uniform proposal. The exponential proposal roughly follows the IMF, while the uniform one lies far below it at low masses, where most stars are (Figure~\ref{fig:ch12_tutorial_sampling}a). The uniform weights therefore vary widely, and the exponential weights stay closer to equal (Figure~\ref{fig:ch12_tutorial_sampling}b). The exponential proposal keeps the larger ESS, as the labels show.

All three methods turn uniform random numbers into samples of a distribution that we care about. Inversion is exact when the CDF inverts, and otherwise the proposal decides how efficient the sampling is. The same choice returns for posteriors that we can write down but not sample directly (Chapter~13). There the ESS tells us how many independent samples a weighted estimate is worth.\index{initial mass function (IMF)|)}

\paragraph{Further Reading} Von Neumann (1951) described rejection sampling and other ways to generate random numbers. Box and Muller (1958) turned uniform random numbers into normally distributed ones without inverting the normal CDF. Forsythe (1972) generalized von Neumann's method of comparing random numbers. Kinderman and Monahan (1977) introduced a rejection method based on the ratio of uniform random numbers. Marsaglia (1977) described the squeeze method, which adds simple bounds that save evaluations of the density. Goertzel (1949), Kahn (1950) and Kahn and Harris (1951) applied early importance sampling to particle transport. Kahn and Marshall (1953) presented importance sampling as a way to reduce variance, and Kahn (1955) discussed such techniques. Rubinstein (1981), Kalos and Whitlock (1986) and Fishman (1996) wrote books on Monte Carlo methods. Devroye (1986) covers random number generation. Kroupa (2001) gave the IMF we sample.

\makeatletter\@clubpenalty=150\makeatother \clubpenalty=150 \widowpenalty=150

\chapter{Markov Chain Monte Carlo}

\makeatletter\@clubpenalty=10000\makeatother \clubpenalty=10000 \widowpenalty=10000

Consider\index{Markov chain Monte Carlo (MCMC)|(} a typical astronomical inference problem, the estimation of the physical properties of an exoplanet\index{exoplanets} system from radial velocity and transit data. Such models often have 10 to 20 parameters, including planetary masses, radii, orbital periods, eccentricities and stellar properties. The posterior distribution of these parameters has complex correlations. It can also have several modes, which represent different valid solutions.

Chapter~12 introduced the grid method, the inverse CDF method, rejection sampling and importance sampling. Each of them suffers from some form of the curse of dimensionality, where the computing requirements grow exponentially with the number of parameters. They reach practical limits for the complex, high-dimensional problems of modern astronomy.

With 10 parameters and 100 grid points per dimension, a grid needs $10^{20}$ evaluations. Inverse transform sampling faces a different challenge, because computing and inverting the cumulative distribution function is analytically intractable for complex posteriors. Rejection sampling suffers from exponentially decreasing acceptance rates as the dimension increases. Even importance sampling has difficulty finding proposal distributions that adequately cover high-dimensional parameter spaces.

This chapter introduces Markov Chain Monte Carlo (MCMC), a family of algorithms that overcome these dimensionality limits. They have transformed Bayesian inference in astronomy over the past three decades. Unlike the global approaches of Chapter~12, it moves through the parameter space in sequential, local steps. The sequence of positions that it visits is called a chain. Each step depends only on the current position and not on the history of the chain. This local view is useful for the complex models of later chapters, where MCMC can support advanced techniques like Gaussian processes and neural networks.

The main idea is to construct a random\index{random walk} walk through the parameter space instead of sampling directly from the posterior. A random walk is a path made of random steps. We design the rules of this random walk so that the chain eventually settles into a pattern that matches our target posterior distribution.

Imagine an explorer who wants to map the population distribution of Columbus, Ohio. A full census of the city would be analogous to a grid method. Instead, the explorer follows a simple strategy. Each day, the explorer randomly chooses a nearby neighborhood to visit, based only on the current location. The explorer chooses neighborhoods of higher population density more often.

At first, the explorer may wander somewhat randomly while learning about the city. After many days, a pattern emerges. The explorer spends more time in densely populated areas like the Short North and less time in sparse suburbs. The frequency of visits to each neighborhood eventually matches the true population distribution. Recording the daily locations over many weeks gives a representative sample of the population density of Columbus. We do not need a street-by-street survey.

\begin{figure}[ht!]
    \centering
    \includegraphics[width=0.6\textwidth]{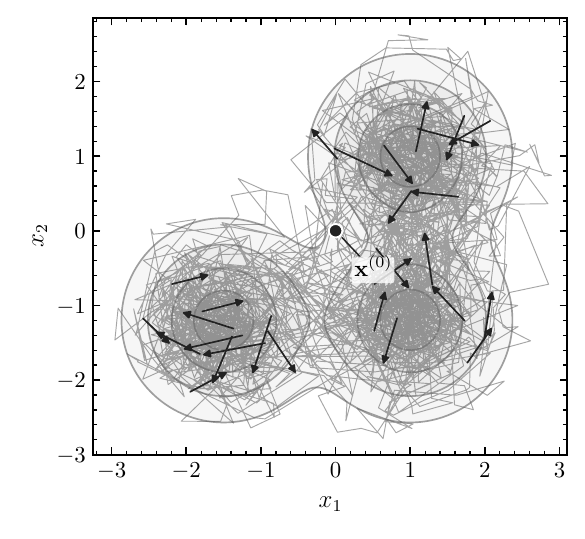}
    \caption{Path of a Markov chain on a posterior with three modes, starting at $\mathbf{x}^{(0)}$. The grey contours show the posterior. The thin grey line is the path of the chain, and the black arrows mark single steps. The chain moves between the modes and spends more of its steps where the density is high.}
    \label{fig:mcmc_sampling}
\end{figure}

In this strategy, the neighborhoods are points in our parameter space and the population density is the posterior probability. The explorer's path is our Markov chain. Each day's destination depends only on today's location, and this sequential nature gives the method its name, Markov Chain Monte Carlo. Figure~\ref{fig:mcmc_sampling} shows a chain on a posterior with three modes.

MCMC has several advantages for astronomical applications. It needs to evaluate the posterior only up to a proportionality constant, so we avoid difficult normalization calculations. It handles correlations between parameters, because it moves through the parameter space step by step. It can discover and sample from multiple modes of the distribution when we design the algorithm appropriately. Most important, it scales to high dimensions, because it spends its computing effort on regions of high probability and does not attempt an exhaustive search.

MCMC also has difficulties that direct sampling does not have. The exploration is sequential, so successive samples are correlated with each other, which reduces their effective information content. We cannot observe directly whether the chain has converged to the target distribution (Section~13.4), so we need careful diagnostics (Section~13.13). The efficiency of the exploration depends on design choices of the algorithm, in particular on how we propose new moves through the parameter space.

\section{Markov Chains and the Core Idea of MCMC}\label{sec:ch13_markov}

A\index{Markov chain|(} process is Markovian when its future depends only on its present state, and not on how it reached that state. A state is one possible position of the process, such as a neighborhood in our analogy or a point in the parameter space. The set of possible states is the state space. The term comes from the Russian mathematician Andrey Markov, who studied such sequences of events. This ``memoryless'' property distinguishes Markov processes from processes in which the history matters.

\looseness=1 In our Columbus analogy, the explorer visits one location each day, such as a café, a museum, a park or a neighborhood. A guidebook suggests where to go next, based only on the current location. The guidebook does not care whether the explorer came from downtown yesterday or from the suburbs last week. This selective memory defines the Markov\index{Markov property} property.

We denote the sequence of locations by $\{X_0, X_1, X_2, \ldots\}$. The Markov property then says that
\begin{equation}
P(X_{t+1} = x | X_0 = x_0, X_1 = x_1, ..., X_t = x_t) = P(X_{t+1} = x | X_t = x_t).
\end{equation}

The Markov property is computationally useful, because the transition probabilities between states are all we need to describe the system. The transition\index{transition matrix} matrix $T(\mathbf{x}, \mathbf{x}')$ determines the behavior of a Markov\index{Markov chain} chain. It gives the probability density of moving to state $\mathbf{x}'$ when the current state is $\mathbf{x}$. In our analogy, $T$ is the guidebook that gives the probability of visiting each location from the current one.

Suppose the explorer is in the Short North, an arts district. The guidebook might suggest a 40\% chance of staying in the Short North, because it has many local attractions. It might suggest a 30\% chance of moving to nearby German Village and a 20\% chance of visiting downtown. The chance of traveling to more distant neighborhoods is 10\%. These percentages sum to 100\%, because the explorer ends each day in some neighborhood, possibly the same one. The transition probabilities from any state must therefore sum to 1, which is a requirement for any valid transition matrix.

\looseness=1 We now follow all the people of Columbus instead of a single explorer. This shift from individual trajectories to population dynamics gives insight into how MCMC\index{MCMC|see{Markov chain Monte Carlo}}\index{Markov chain Monte Carlo (MCMC)} works. They move around the city according to the rules of our guidebook. Our goal is a guidebook for which the population distribution over the neighborhoods settles at a target distribution. The target could be the actual population density of Columbus. We then view the transition matrix $T$ as an operator that transforms one population distribution into another.

The transition operator maps a distribution $p(\mathbf{x})$ of people across the city to the distribution after everyone follows the guidebook for one step. It is defined by
\begin{equation}
(T p)(\mathbf{x'}) = \int p(\mathbf{x}) T(\mathbf{x}, \mathbf{x'}) \mathrm{d}\mathbf{x}.
\end{equation}

The quantity $(Tp)(\mathbf{x}')$ is the probability density of being at state $\mathbf{x}'$ after applying the transition operator to $p(\mathbf{x})$. The integral adds up all the ways of reaching $\mathbf{x}'$ from any starting state $\mathbf{x}$, each weighted by the probability of starting there. For a finite number of states, $T$ is a matrix whose entry $T(i,j)$ is the probability of moving from state $i$ to state $j$. We write $p$ as a row of probabilities, so that $Tp$ is the row $p$ multiplied by the matrix $T$ and $(Tp)(j) = \sum_i p(i)\,T(i,j)$.

How do we create a guidebook that leads to our desired distribution? We need a transition matrix for which the population distribution converges to the target and then stays there. We call a distribution $p(\mathbf{x})$ stationary with respect to the transition operator $T$ if applying $T$ to $p$ returns the same distribution
\begin{equation}
(T p)(\mathbf{x}) = p(\mathbf{x}).
\end{equation}

\looseness=1 In our analogy, suppose the population is already distributed according to the target distribution. When everyone follows the guidebook, the overall population distribution remains unchanged. The stationary\index{stationary distribution} distribution is the equilibrium state of our system.

\begin{figure}[ht!]
    \centering
    \includegraphics[width=0.9\textwidth]{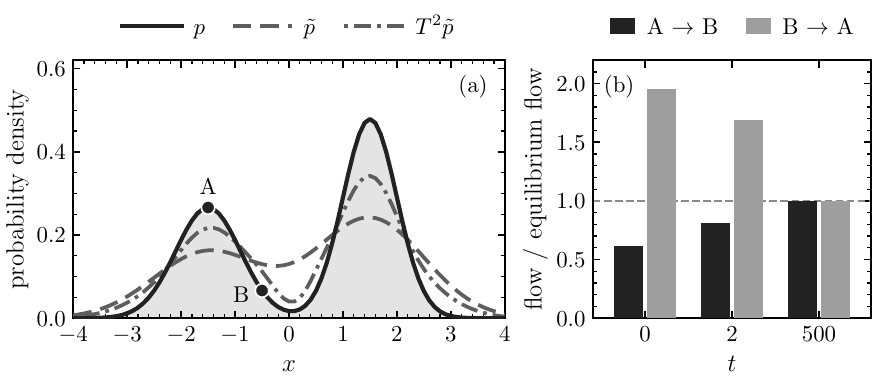}
    \caption{Convergence of a chain on a grid of states to its stationary distribution. Panel (a) shows the target $p$, an initial distribution $\tilde{p}$ and the distribution $T^2\tilde{p}$ after two steps. Panel (b) shows the flows $p^{(t)}(i)T(i,j)$ with $p^{(t)} = T^t\tilde{p}$ between states A and B at three steps $t$, divided by their values at equilibrium. The transition probabilities do not change with $t$, and $T(\mathrm{B},\mathrm{A})$ is about four times $T(\mathrm{A},\mathrm{B})$. Early on, state B holds more of the population than at equilibrium, so the flow from B to A is larger than at equilibrium. At equilibrium the two flows are equal.}
    \label{fig:stationary_distribution}
\end{figure}

Suppose now that we start with any initial distribution of people across Columbus. They follow the guidebook repeatedly. We want the population to redistribute itself to match the target pattern and then to stay that way. Once the population reaches this equilibrium, the flows of people balance. For each neighborhood, as many people arrive as leave. Figure~\ref{fig:stationary_distribution}(a) shows an initial distribution and the distribution after two steps, which is closer to the target distribution.

The central idea of MCMC connects the two viewpoints, the individual explorer and the population distribution. Suppose we design our guidebook so that the target posterior $p(\mathbf{x}|\mathcal{D})$ is the stationary distribution of the resulting Markov chain. A single explorer who follows this guidebook for a long time then visits locations with frequencies proportional to their posterior probabilities. This link between the population and a single explorer relies on a principle called ergodicity, which we study in Section~13.5.

By recording where the explorer goes over many steps, we sample from the posterior distribution. We use these samples for the Bayesian calculations we need, such as computing posterior means, estimating uncertainties, making predictions and comparing models. This approach turns an intractable sampling problem into the design of a Markov chain with the right properties.\index{Markov chain|)}

\section{Detailed Balance}\label{sec:ch13_detailed_balance}

How\index{detailed balance|(} do we construct a transition matrix whose stationary distribution is our target posterior? Detailed\index{detailed balance} balance gives a practical condition for this. For clarity, we write the posterior $p(\mathbf{x}|\mathcal{D})$ simply as $p(\mathbf{x})$. It is the probability distribution of the model parameters $\mathbf{x}$ given the observed data $\mathcal{D}$.

Imagine that we want the Short North to have a bustling population and Upper Arlington to remain a quieter area. To achieve this distribution, our guidebook needs to encourage more people to move from Upper Arlington to the Short North than the other way around. Over time, the Short North population grows. As more people gather in the Short North, the population difference creates a counterbalance. There are more people in the Short North who might move to Upper Arlington, even if each person is less likely to move.

These opposing forces eventually reach equilibrium. People keep moving, and the flows between neighborhoods balance, so the population distribution stays constant. This balance occurs when the number of people moving from the Short North to Upper Arlington equals the number moving the other way. This condition is called detailed balance, and we write it as
\begin{equation}
p(\mathbf{x}) T(\mathbf{x}, \mathbf{x'}) = p(\mathbf{x'}) T(\mathbf{x'}, \mathbf{x}).
\end{equation}

To read this equation, we interpret each term in our Columbus analogy. The quantity $p(\mathbf{x})$ is the population of neighborhood $\mathbf{x}$, here the Short North. The transition matrix $T(\mathbf{x}, \mathbf{x}')$ is the probability that someone in the Short North moves to Upper Arlington, written $\mathbf{x}'$. The left side $p(\mathbf{x}) T(\mathbf{x}, \mathbf{x}')$ is therefore the ``flow'' of people from the Short North to Upper Arlington.

\begin{figure}[ht!]
    \centering
    \includegraphics[width=0.7\textwidth]{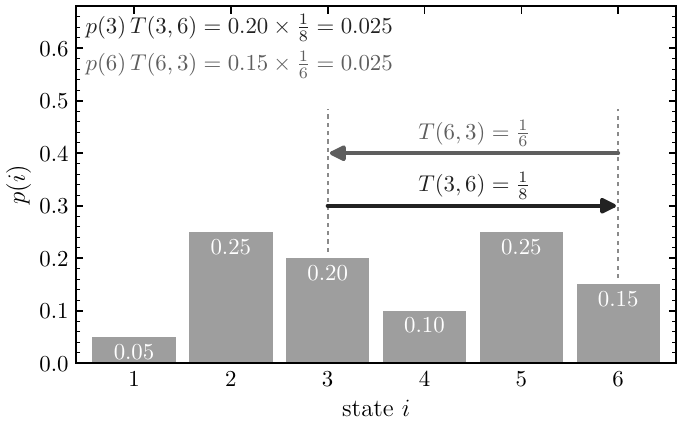}
    \caption{Detailed balance for a chain with six states. The bars show the stationary probabilities $p(i)$. The arrows show the transition probabilities between states 3 and 6. The two transition probabilities differ, but the two flows are equal.}
    \label{fig:detailed_balance_histogram}
\end{figure}

The right side is the flow in the opposite direction. Detailed balance says that these two flows are equal for each pair of neighborhoods in our city. This is a strong condition. A stationary distribution only needs the total flow into each neighborhood to equal the total flow out of it. Figure~\ref{fig:detailed_balance_histogram} shows the two flows between a pair of states for a distribution on discrete states. Figure~\ref{fig:stationary_distribution}(b) shows these flows for two states during the approach to equilibrium.

Detailed balance gives us a concrete way to design the transition matrix. We can evaluate the target distribution $p(\mathbf{x})$ up to a constant, because the posterior is proportional to the likelihood times the prior. We can therefore use $p(\mathbf{x})$ to construct a guidebook that leads to this distribution.

We now prove that detailed balance makes $p(\mathbf{x})$ a stationary distribution. We must show that, if detailed balance holds, then $p(\mathbf{x})$ satisfies the stationary distribution equation
\begin{equation}
p(\mathbf{x'}) = \int p(\mathbf{x}) T(\mathbf{x}, \mathbf{x'}) \mathrm{d}\mathbf{x}.
\end{equation}

We integrate both sides of the detailed balance equation over all states $\mathbf{x}$
\begin{align}
\int p(\mathbf{x}) T(\mathbf{x}, \mathbf{x'}) \mathrm{d}\mathbf{x} &= \int p(\mathbf{x'}) T(\mathbf{x'}, \mathbf{x}) \mathrm{d}\mathbf{x}.
\end{align}

On the right side, $p(\mathbf{x}')$ does not depend on $\mathbf{x}$, so we can move it outside the integral
\begin{align}
\int p(\mathbf{x}) T(\mathbf{x}, \mathbf{x'}) \mathrm{d}\mathbf{x} &= p(\mathbf{x'}) \int T(\mathbf{x'}, \mathbf{x}) \mathrm{d}\mathbf{x}.
\end{align}

The integral $\int T(\mathbf{x}', \mathbf{x}) \mathrm{d}\mathbf{x}$ equals 1, because the probabilities of moving from state $\mathbf{x}'$ to all possible states sum to 1. In our analogy, the probabilities of moving from Upper Arlington to all neighborhoods, including staying in Upper Arlington, add up to 100\%.

We therefore obtain
\begin{equation}
\int p(\mathbf{x}) T(\mathbf{x}, \mathbf{x'}) \mathrm{d}\mathbf{x} = p(\mathbf{x'}) \cdot 1 = p(\mathbf{x'}).
\end{equation}

This is the definition of a stationary distribution. If detailed balance holds, the target distribution $p(\mathbf{x})$ is therefore a stationary distribution of the Markov chain with transition matrix $T$. In our analogy, balanced flows between any two neighborhoods keep the population of each neighborhood constant.

\looseness=1 This gives us a practical strategy for constructing MCMC algorithms. We design transition matrices that satisfy detailed balance with respect to our target posterior distribution. The task is then to construct transition rules that balance the flows between neighborhoods.\index{detailed balance|)}

\section{The Metropolis Algorithm}\label{sec:ch13_metropolis}

Many\index{Metropolis algorithm|(} approaches in the MCMC literature construct such transition rules. The Metropolis\index{Metropolis algorithm} algorithm is the simplest and most intuitive solution. It shows that a suitable transition matrix exists and gives a simple way to construct it.

In the Metropolis algorithm, we split the transition into two steps. First we propose a new state, and then we decide whether to accept or reject the proposal. In our analogy, the explorer first consults a guidebook that suggests a neighborhood to visit, which is the proposal. The explorer then decides personally whether to follow the suggestion, which is the acceptance or rejection step. This two-step process lets us design a transition mechanism that satisfies detailed balance.

We write the transition matrix as
\begin{equation}
T(\mathbf{x}, \mathbf{x}') = q(\mathbf{x}' | \mathbf{x}) A(\mathbf{x}, \mathbf{x}'),
\end{equation}
where $q(\mathbf{x}' | \mathbf{x})$ is the proposal distribution and $A(\mathbf{x}, \mathbf{x}')$ is the acceptance\index{acceptance probability} probability of the proposed move. The proposal distribution is the probability density of suggesting a move to state $\mathbf{x}'$ when we are at state $\mathbf{x}$. The formula for $T(\mathbf{x}, \mathbf{x}')$ holds for $\mathbf{x}' \neq \mathbf{x}$, because a rejected proposal keeps the chain at $\mathbf{x}$ and adds the probability of rejection to $T(\mathbf{x}, \mathbf{x})$.

The Metropolis algorithm requires a symmetric proposal distribution, which means that
\begin{equation}
q(\mathbf{x}' | \mathbf{x}) = q(\mathbf{x} | \mathbf{x}').
\end{equation}

In our analogy, the guidebook is as likely to suggest a trip from the Short North to Upper Arlington as the reverse trip. A common example in astronomical applications is a multivariate Gaussian centered at the current position
\begin{equation}
q(\mathbf{x}' | \mathbf{x}) = \mathcal{N}(\mathbf{x}' | \mathbf{x}, \boldsymbol{\Sigma}),
\end{equation}
where $\boldsymbol{\Sigma}$ is the covariance matrix. The matrix $\boldsymbol{\Sigma}$ controls the step size and direction of our proposals. This proposal is symmetric, because the distance from $\mathbf{x}$ to $\mathbf{x}'$ is the same as the distance from $\mathbf{x}'$ to $\mathbf{x}$.

For this symmetric proposal, the Metropolis algorithm defines the acceptance probability as
\begin{equation}
A(\mathbf{x}, \mathbf{x}') = \min\left(1, \frac{p(\mathbf{x}')}{p(\mathbf{x})}\right).
\end{equation}

\begin{figure}[t]
    \centering
    \includegraphics[width=0.7\textwidth]{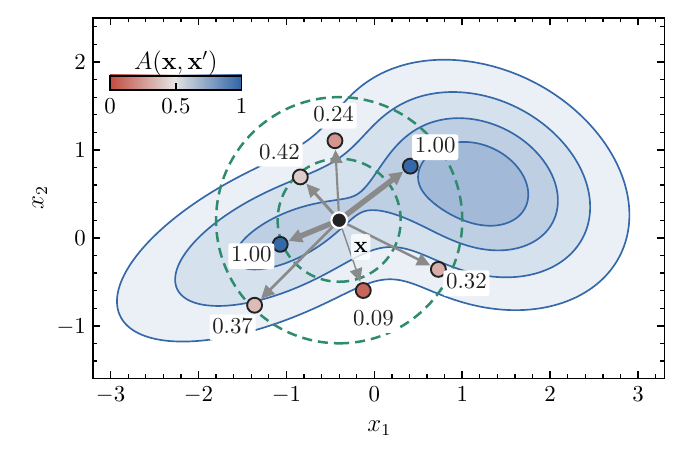}
    \caption{Proposals of the Metropolis algorithm from one current state on an elongated posterior. The blue contours show the posterior. The dashed green circles show the Gaussian proposal around the current state $\mathbf{x}$ at one and two standard deviations. Each dot marks a proposed state $\mathbf{x}'$ with its acceptance probability $A$. The arrow width grows with $A$.}
    \label{fig:transition_dynamics}
\end{figure}

\looseness=1 Suppose the proposed state $\mathbf{x}'$ has a higher posterior probability than the current state $\mathbf{x}$, so that $p(\mathbf{x}') \geq p(\mathbf{x})$. The ratio $p(\mathbf{x}')/p(\mathbf{x})$ is then at least 1, and we accept the proposal with probability 1. If the proposed state has a lower posterior probability, so that $p(\mathbf{x}') < p(\mathbf{x})$, the ratio is below 1. We then accept the proposal with a probability equal to this ratio. For example, if the proposed state has 70\% of the probability density of the current state, we accept the move with 70\% probability. Figure~\ref{fig:transition_dynamics} shows a Gaussian proposal around a current state and the acceptance probabilities of several proposals.

In our analogy, the explorer goes more often to densely populated neighborhoods. Suppose the explorer is in the Short North and the guidebook suggests visiting Upper Arlington. The explorer checks the population densities of both neighborhoods. If Upper Arlington is denser, the explorer goes there. If Upper Arlington has only 70\% of the density of the Short North, the explorer goes there with a 70\% chance. If the explorer decides not to go, they stay in the Short North for another day.

\paragraph{Checking Detailed Balance}

We now verify that this acceptance probability satisfies detailed\index{detailed balance} balance with respect to our target posterior distribution $p(\mathbf{x})$. Detailed balance requires that
\begin{equation}
p(\mathbf{x}) T(\mathbf{x}, \mathbf{x}') = p(\mathbf{x}') T(\mathbf{x}', \mathbf{x}).
\end{equation}

Substituting the transition matrix gives
\begin{equation}
p(\mathbf{x}) q(\mathbf{x}' | \mathbf{x}) A(\mathbf{x}, \mathbf{x}') = p(\mathbf{x}') q(\mathbf{x} | \mathbf{x}') A(\mathbf{x}', \mathbf{x}).
\end{equation}

The proposal distribution is symmetric, with $q(\mathbf{x}' | \mathbf{x}) = q(\mathbf{x} | \mathbf{x}')$. The condition therefore simplifies to
\begin{equation}
p(\mathbf{x}) A(\mathbf{x}, \mathbf{x}') = p(\mathbf{x}') A(\mathbf{x}', \mathbf{x}).
\end{equation}

We consider two cases. If $p(\mathbf{x}') \geq p(\mathbf{x})$, then
\begin{align}
A(\mathbf{x}, \mathbf{x}') &= \min\left(1, \frac{p(\mathbf{x}')}{p(\mathbf{x})}\right) = 1 \\*
A(\mathbf{x}', \mathbf{x}) &= \min\left(1, \frac{p(\mathbf{x})}{p(\mathbf{x}')}\right) = \frac{p(\mathbf{x})}{p(\mathbf{x}')}.
\end{align}

Substituting these into the detailed balance equation gives
\begin{equation}
p(\mathbf{x}) \cdot 1 = p(\mathbf{x}') \cdot \frac{p(\mathbf{x})}{p(\mathbf{x}')},
\end{equation}
which simplifies to $p(\mathbf{x}) = p(\mathbf{x})$ and is true.

If $p(\mathbf{x}') < p(\mathbf{x})$, then
\begin{align}
A(\mathbf{x}, \mathbf{x}') &= \min\left(1, \frac{p(\mathbf{x}')}{p(\mathbf{x})}\right) = \frac{p(\mathbf{x}')}{p(\mathbf{x})} \\*
A(\mathbf{x}', \mathbf{x}) &= \min\left(1, \frac{p(\mathbf{x})}{p(\mathbf{x}')}\right) = 1.
\end{align}

Substituting these into the detailed balance equation gives
\begin{equation}
p(\mathbf{x}) \cdot \frac{p(\mathbf{x}')}{p(\mathbf{x})} = p(\mathbf{x}') \cdot 1,
\end{equation}
which simplifies to $p(\mathbf{x}') = p(\mathbf{x}')$ and is also true.

\looseness=1 The transition matrix of the Metropolis algorithm therefore satisfies detailed balance with respect to our target posterior distribution. The posterior is thus a stationary distribution of the chain. Once the chain has converged to the target distribution, detailed balance keeps the flows between any two neighborhoods equal, so the equilibrium stays in place. If the chain has not yet converged, the net flows of people move the population distribution toward equilibrium. Neighborhoods with higher posterior probability gradually gain more residents, and neighborhoods with lower probability become less populated.

The Metropolis algorithm also has a practical advantage for Bayesian inference in astronomy. It requires only the ratio of posterior probabilities. We therefore avoid calculating the normalization constant, the Bayesian evidence, which is often intractable for complex models with many parameters. Written with the likelihood and the prior, the posterior ratio is
\begin{equation}
\frac{p(\mathbf{x}'|\mathcal{D})}{p(\mathbf{x}|\mathcal{D})} = \frac{p(\mathcal{D}|\mathbf{x}')p(\mathbf{x}')}{p(\mathcal{D})}\frac{p(\mathcal{D})}{p(\mathcal{D}|\mathbf{x})p(\mathbf{x})} = \frac{p(\mathcal{D}|\mathbf{x}')p(\mathbf{x}')}{p(\mathcal{D}|\mathbf{x})p(\mathbf{x})}.
\end{equation}

The evidence $p(\mathcal{D})$ cancels in the ratio. The remaining terms are the likelihood times the prior, which we can typically evaluate directly.\index{Metropolis algorithm|)}

\section{Convergence and Uniqueness in MCMC}

\looseness=-1 Detailed\index{Markov chain Monte Carlo (MCMC)!convergence|(}\index{stationary distribution|(} balance shows only that the target posterior is one possible stationary distribution of the chain. It does not show that the chain converges. It also does not show that a converging chain ends at the correct distribution. In our Columbus analogy, a population that matches the true neighborhood densities is then a possible equilibrium. When we run MCMC in research, we need confidence that we will get the correct posterior. We therefore need two more properties, convergence and uniqueness.

Convergence means that the chain eventually settles into a pattern that matches our target posterior, whatever initial distribution we start with. We write this as
\begin{equation}
\lim_{n\to\infty} T^n \tilde{p}(\mathbf{x}) = p(\mathbf{x}),
\end{equation}
where $\tilde{p}(\mathbf{x})$ is any starting distribution, $T^n$ applies our transition matrix $n$ times, and $p(\mathbf{x})$ is our target distribution.

In the Columbus picture, we give the Metropolis guidebook to thousands of people who are spread over the city in various ways. Convergence means that, given enough time, they redistribute themselves over the neighborhoods to match the true population densities, wherever they started.

Uniqueness means that our chain has only one stationary distribution. Two stationary distributions $p_1(\mathbf{x})$ and $p_2(\mathbf{x})$ satisfy
\begin{equation}
(T p_1)(\mathbf{x}) = p_1(\mathbf{x}) \quad \text{and} \quad (T p_2)(\mathbf{x}) = p_2(\mathbf{x}),
\end{equation}
and uniqueness requires that $p_1(\mathbf{x}) = p_2(\mathbf{x})$. Detailed balance has shown that our target distribution is a stationary distribution. Uniqueness makes sure that it is the only one.

In the Columbus picture, the guidebook allows only one stable population spread over the neighborhoods. Without uniqueness, the population could settle into a distribution that does not match the true neighborhood densities, even after a very long run.

Two transition matrices fail to produce the sampling behavior we need for Bayesian inference, even though detailed balance holds. First, imagine a guidebook that simply says, ``stay where you are.'' Its transition matrix is $T = \mathrm{Id}$, the identity matrix. This matrix satisfies detailed balance with respect to any distribution, because no one moves. If we start with everyone in the Short North, they all stay there, and a uniform population spread over Columbus stays uniform. The system ``converges'' at once, but the distribution it converges to depends on our starting point. This violates uniqueness, since we need the sampling process to converge to the correct posterior wherever we start.

For a second example, take a Columbus with only two neighborhoods, the Short North and German Village. The guidebook says, ``If you are in the Short North today, go to German Village tomorrow. If you are in German Village today, go to the Short North tomorrow.'' This guidebook has the transition matrix
\begin{equation}
T = \begin{bmatrix} 0 & 1 \\ 1 & 0 \end{bmatrix}.
\end{equation}

This transition matrix has the stationary distribution $p^*(\mathbf{x}) = [0.5, 0.5]$. If half of the population starts in each neighborhood, the split stays at half and half as people swap back and forth. However, for any other starting distribution, the system does not converge. Instead, we get
\begin{equation}
p^{(t)}(\mathbf{x}) = \begin{cases}
\tilde{p}(\mathbf{x}), & \text{if } t \text{ is even} \\
1 - \tilde{p}(\mathbf{x}), & \text{if } t \text{ is odd}
.\end{cases}
\end{equation}

Suppose we start with 70\% of the people in the Short North and 30\% in German Village. The next day the split flips to 30\% in the Short North and 70\% in German Village, and it keeps oscillating in this way.

\looseness=1 These examples show two different failures. The identity matrix has many stationary distributions, so uniqueness fails. The swap matrix has one stationary distribution, but the chain does not converge to it from other starting distributions, so convergence fails. Most practical MCMC methods, including the Metropolis algorithm, have transition matrices built to avoid these pathological behaviors. These matrices satisfy an additional property called ergodicity, which guarantees both convergence and uniqueness.

\section{Ergodicity Overview}

The\index{ergodicity|(}\index{irreducibility|(} term ergodicity may be familiar from statistical mechanics, but the concept applies beyond physics. In essence, ergodicity means that a system can explore all its possible states over time and reach a stable equilibrium distribution. An ergodic\index{ergodicity} Markov chain converges to a unique stationary distribution, whatever its starting point. A Markov chain is ergodic if it satisfies three properties.

\begin{figure}[ht!]
    \centering
    \includegraphics[width=0.85\textwidth]{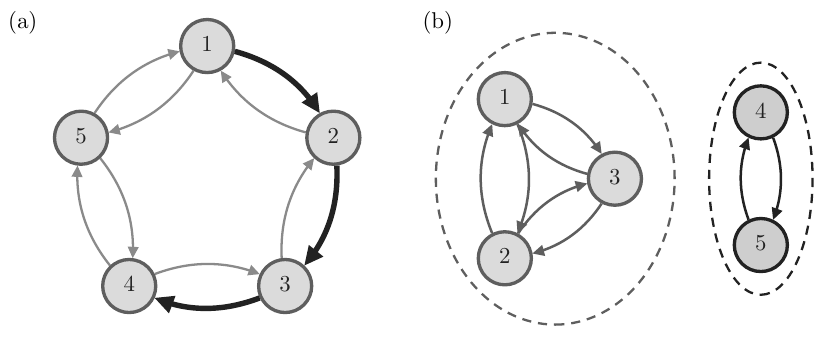}
    \caption{Irreducibility. Panel (a) shows a chain on five states in which any state can reach any other state. The thick arrows mark one route from state 1 to state 4. Panel (b) shows a chain with two groups of states, 1 to 3 and 4 to 5, with no transition between the groups. This chain is reducible.}
    \label{fig:irreducibility}
\end{figure}

The first property is irreducibility. The chain can go from any state to any other state in a finite number of steps. In our Columbus analogy, you can travel from any neighborhood to any other neighborhood given enough time. No area of the parameter space is isolated, so the chain cannot get trapped in one region and miss the rest of the posterior distribution. Figure~\ref{fig:irreducibility} compares an irreducible chain with a reducible chain whose states form two groups that cannot reach each other.

The second property is aperiodicity. The chain does not visit states in a predictable cyclic pattern. This prevents the oscillation that we saw in our Short North and German Village example, where people swapped back and forth.

The third property is positive recurrence. A chain is recurrent when, from any state, it returns to that state with probability 1. It is positive recurrent when the expected time to return is also finite, which prevents the chain from wandering off to infinity.

\begin{figure}[ht!]
    \centering
    \includegraphics[width=0.85\textwidth]{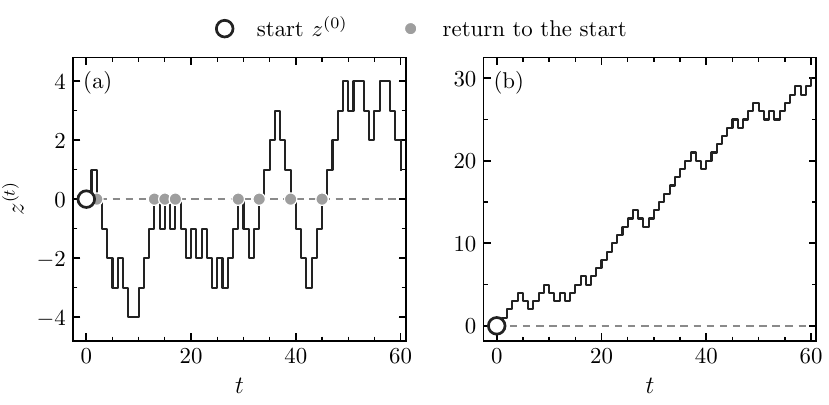}
    \caption{Positive recurrence. Panel (a) shows a random walk on nine states that returns to its starting state again and again. Panel (b) shows a random walk on the integers that moves up with probability 3/4 and down with probability 1/4. It drifts away from its starting state and does not return in the steps shown. A chain on an unbounded set of states can still return in a finite expected time. This depends on its transition probabilities.}
    \label{fig:positive_recurrence}
\end{figure}

In our Columbus analogy, travelers come back to a neighborhood they have visited within a finite expected time. They do not wander to suburbs farther and farther from the city center and take ever longer to return. Without positive recurrence, our population distribution would not settle into the desired equilibrium. In Figure~\ref{fig:positive_recurrence}, a walk on a bounded set of states keeps returning to its start, while a walk with drift moves away from it.

The three conditions together rule out the pathological cases that we met earlier. Irreducibility prevents the chain from getting stuck in isolated regions, like the ``always stay where you are'' example. Aperiodicity eliminates the cyclic oscillation of the Short North and German Village swapping. Positive recurrence stops the chain from drifting away, so that returns to previously visited areas take a finite expected time.

For a chain that satisfies all three conditions, we can prove two results. First, a unique stationary distribution exists. The chain has only one equilibrium distribution, so there is no ambiguity about the distribution we sample from. Second, the chain converges to this stationary distribution from any starting point. Wherever we initialize the chain, it reaches the same equilibrium distribution and stays there. A proof of convergence needs advanced probability theory, so the next sections give basic ideas and intuition before we return to applications and algorithms.

\looseness=1 Convergence to a unique stationary distribution is what we need for MCMC to work correctly in our astronomical applications. It is analogous to gas particles in a closed room, which explore the available space until they reach equilibrium in a Maxwell-Boltzmann distribution. Most MCMC methods used in astronomy, including the Metropolis algorithm, are built to satisfy the three conditions.

However, irreducibility can be violated if the posterior has separated modes with zero probability between them. This is like having two rooms with a vacuum between them, where particles that start in one room cannot reach the other. In astronomy, this can happen when a parameter space has several disconnected high-probability regions, separated by regions with nearly zero probability. Two equally valid solutions with no connecting path are an example. Widely separated modes can also leave the chain reducible in practice, so that it cannot explore the full posterior.

\looseness=1 Aperiodicity is usually satisfied by default in standard MCMC implementations, unless we explicitly design a periodic transition scheme. The random acceptance or rejection step of the Metropolis algorithm typically gives aperiodicity. It does so through self-loops, which occur when a proposed move is rejected and the chain stays at the current state.

Positive recurrence is typically ensured when the priors limit the parameters to a finite range. This is like keeping particles in a finite room instead of an infinite universe, where they could wander without end. In astronomical models, this usually means choosing priors based on physically reasonable parameter ranges.

Astronomers often do not check for ergodicity explicitly in their MCMC runs. Knowing the three conditions still helps us diagnose problems when a method does not behave as expected. Suppose that multiple chains with different starting points fail to converge to the same distribution, although each chain looks fine on its own. This often indicates a violation of one of the conditions, most commonly irreducibility because of disconnected modes in the posterior.\index{ergodicity|)}\index{irreducibility|)}

\section{The Uniqueness of the Stationary Distribution}

The condition that guarantees uniqueness is irreducibility. We prove that an irreducible Markov chain has at most one stationary distribution.

An irreducible Markov chain can reach any state from any other state through a finite sequence of transitions. We define irreducibility as
\begin{equation}
\text{for all states } i, j \text{ there is a } t_{ij} \geq 0 \text{ such that } p(\mathbf{x}^{t_{ij}} = j | \mathbf{x}^0 = i) > 0.
\end{equation}
Here $i$ and $j$ are any two states, and $\mathbf{x}^t$ is the state of the chain at step $t$. The number $t_{ij}$ is the number of steps needed to go from state $i$ to state $j$. The expression $p(\mathbf{x}^{t_{ij}} = j | \mathbf{x}^0 = i)$ is the probability of being in state $j$ after $t_{ij}$ steps, given that we started in state $i$.

The term irreducible comes from graph theory, where all nodes in the graph are connected. This connectivity allows us to prove uniqueness. We treat a finite state space. The result extends to general state spaces under appropriate technical conditions.

Since the chain is irreducible, there is an integer $N$ such that the composite transition matrix
\begin{equation}
\tilde{T} \equiv \sum_{i=1}^{N} T^i
\end{equation}
has all strictly positive entries. Transitions over up to $N$ steps therefore reach any state from any other state with positive probability.

\paragraph{Proof of Uniqueness}

We argue by contradiction. Suppose that $T$ has two distinct stationary distributions $p$ and $q$ (here $q$ is not the proposal), with $p \neq q$, so that
\begin{equation}
Tp = p, \quad Tq = q.
\end{equation}

To see how $\tilde{T}$ acts on $p$, note that $Tp = p$ gives $T^2p = T(Tp) = Tp = p$ and, by induction, $T^ip = p$ for any positive integer $i$. Therefore
\begin{equation}
\tilde{T}p = \sum_{i=1}^{N} T^ip = \sum_{i=1}^{N} p = Np.
\end{equation}

Since we work with probability distributions, we divide $\tilde{T}$ by $N$ and keep the name $\tilde{T}$ for the normalized matrix. The same steps apply to $q$. Both $p$ and $q$ are then stationary distributions of $\tilde{T}$, so that
\begin{equation}
\tilde{T}p = p, \quad \tilde{T}q = q.
\end{equation}

We write the first of these equations element by element as
\begin{equation}
p(\mathbf{x}) = \sum_{\mathbf{y}} \tilde{T}(\mathbf{y}, \mathbf{x}) p(\mathbf{y}).
\end{equation}

Let $\mathbf{x}$ be the state that minimizes the ratio $p(\mathbf{y})/q(\mathbf{y})$ over all states. This minimum exists because the state space is finite. We can then rewrite the equation for $p(\mathbf{x})$ as
\begin{equation}
p(\mathbf{x}) = \sum_{\mathbf{y}} \tilde{T}(\mathbf{y}, \mathbf{x}) q(\mathbf{y}) \frac{p(\mathbf{y})}{q(\mathbf{y})}.
\end{equation}

Because $\mathbf{x}$ minimizes the ratio $p(\mathbf{y})/q(\mathbf{y})$, we know that
\begin{equation}
\frac{p(\mathbf{y})}{q(\mathbf{y})} \geq \frac{p(\mathbf{x})}{q(\mathbf{x})} \text{ for all } \mathbf{y}.
\end{equation}

This gives us
\begin{equation}
p(\mathbf{x}) = \sum_{\mathbf{y}} \tilde{T}(\mathbf{y}, \mathbf{x}) q(\mathbf{y}) \frac{p(\mathbf{y})}{q(\mathbf{y})} \geq \frac{p(\mathbf{x})}{q(\mathbf{x})} \sum_{\mathbf{y}} \tilde{T}(\mathbf{y}, \mathbf{x}) q(\mathbf{y}) = \frac{p(\mathbf{x})}{q(\mathbf{x})} q(\mathbf{x}) = p(\mathbf{x}).
\end{equation}

The expression starts and ends with $p(\mathbf{x})$, so the inequality is an equality, which means that
\begin{equation}
\frac{p(\mathbf{y})}{q(\mathbf{y})} = \frac{p(\mathbf{x})}{q(\mathbf{x})} \text{ for all } \mathbf{y} \text{ where } \tilde{T}(\mathbf{y}, \mathbf{x}) > 0.
\end{equation}

Because $\tilde{T}$ has all positive entries, a consequence of irreducibility, this implies that
\begin{equation}
\frac{p(\mathbf{y})}{q(\mathbf{y})} = \frac{p(\mathbf{x})}{q(\mathbf{x})} \text{ for all } \mathbf{y}.
\end{equation}

The ratio of the two distributions is therefore the same for all states. Both $p$ and $q$ are probability distributions, which means that
\begin{equation}
\sum_{\mathbf{y}} p(\mathbf{y}) = \sum_{\mathbf{y}} q(\mathbf{y}) = 1.
\end{equation}

A constant ratio and equal sums leave only one possibility, $p(\mathbf{y}) = q(\mathbf{y})$ for all $\mathbf{y}$. This contradicts our assumption that $p$ and $q$ are distinct. Hence an irreducible Markov chain with a finite number of states has only one stationary distribution.

\looseness=1 We can also understand this result through linear algebra. The Perron-Frobenius\index{Perron-Frobenius theorem} theorem says that an irreducible non-negative matrix, such as our transition matrix, has a largest real eigenvalue that occurs only once. For a transition matrix this eigenvalue equals 1. Its right eigenvector $\mathbf{v}$ is a column with $T\mathbf{v} = \mathbf{v}$, its left eigenvector $\mathbf{u}$ is a row with $\mathbf{u}T = \mathbf{u}$, and both are positive. Our definition of stationarity sums $p(\mathbf{x}) T(\mathbf{x}, \mathbf{x}')$ over $\mathbf{x}$, which multiplies $T$ from the left, so it involves the left eigenvector. We normalize the left eigenvector so that it sums to 1, and it is then the stationary distribution of the Markov chain.

\section{Convergence to the Stationary Distribution}

Irreducibility gives at most one stationary distribution. We now ask whether the chain converges to it from any starting point. Uniqueness alone does not guarantee this, and we need additional conditions. To see why, we revisit the oscillating example.

The transition matrix of this example is
\begin{equation}
T = \begin{bmatrix} 0 & 1 \\ 1 & 0 \end{bmatrix}.
\end{equation}

\paragraph{Period and Aperiodicity}

This\index{aperiodicity|(} chain has the stationary distribution $p^*(\mathbf{x}) = [0.5, 0.5]$, but it converges to this distribution only if we start with it. If everyone starts in state 1, the population moves to state 2 on the next step, then back to state 1, and so on. The chain is periodic, since a state can be revisited only at regular intervals. Aperiodicity is our second condition for convergence.

We define the period $k_i$ of a state $i$ as the greatest common divisor (gcd) of all possible return times
\begin{equation}
k_i = \text{gcd}\{t : p(z^{(t)} = i | z^{(0)} = i) > 0\}.
\end{equation}
Here $z^{(t)}$ is the state of the chain at step $t$. The expression $p(z^{(t)} = i | z^{(0)} = i)$ is the probability of returning to state $i$ after $t$ steps, given that we started at state $i$. The gcd is the largest integer that divides all of these return times.

In the oscillating example, a traveler who starts in state 1 can return only after an even number of steps. This follows from
\begin{align}
p(z^{(1)} = 1 | z^{(0)} = 1) &= 0 \\*
p(z^{(2)} = 1 | z^{(0)} = 1) &> 0 \\
p(z^{(3)} = 1 | z^{(0)} = 1) &= 0 \\*
p(z^{(4)} = 1 | z^{(0)} = 1) &> 0.
\end{align}
The return times are 2, 4, 6, ..., so the gcd is 2 and the period is $k = 2$.

A Markov chain is aperiodic\index{aperiodicity} if all states have period 1, that is, if the possible return times have no common divisor larger than 1. In a chain with period $k > 1$, the distribution after a large number of steps $n$ depends on the remainder of $n$ divided by $k$. The chain then cycles through $k$ different distributions instead of settling into a single stationary distribution. In an aperiodic chain, this cycling disappears. The distribution depends less and less on the exact number of steps and can settle into the stationary distribution.

\begin{figure}[ht!]
    \centering
    \includegraphics[width=0.85\textwidth]{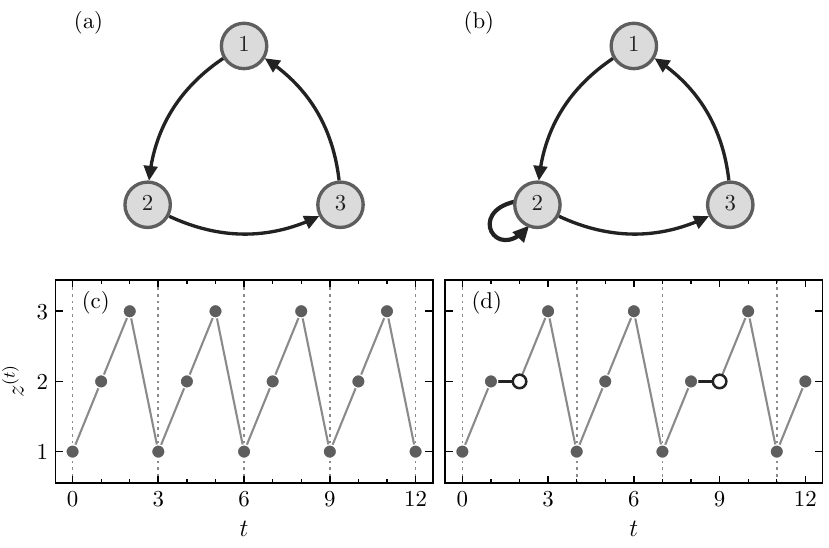}
    \caption{Periodicity. Panel (a) shows a chain with period 3, in which the states follow the cycle 1, 2, 3. Panel (b) shows the same chain after we add a self-loop at state 2. Panels (c) and (d) show the states that each chain visits over time. The dotted lines mark the visits to state 1, and the open marks in (b) and (d) are the self-loop. In (c) the chain returns to state 1 at intervals of three steps, and in (d) the self-loop breaks this pattern.}
    \label{fig:aperiodicity_simple}
\end{figure}

One way to make a chain aperiodic is to allow self-loops, which are transitions from a state back to itself. In our Columbus analogy, travelers would have the option to stay in their current neighborhood with some non-zero probability. Even a single self-loop in a chain can break periodicity, which helps convergence. A state with a self-loop can be revisited after 1 step, 2 steps, 3 steps, and so on. The gcd of the set $\{1,2,3,...\}$ is 1, so the state is aperiodic. Figure~\ref{fig:aperiodicity_simple} shows a chain with period 3 and the same chain after we add a self-loop to state 2.

The Metropolis algorithm produces self-loops because we may reject a proposed move. We reject a proposal with probability $1-A$, and the chain then remains at the current state $\mathbf{x}$. This self-loop means that the Metropolis algorithm typically satisfies the aperiodicity condition.\index{aperiodicity|)}

\paragraph{Positive Recurrence}

Irreducibility\index{positive recurrence|(} and aperiodicity together do not guarantee that a stationary distribution exists, or that the chain converges to it. To see this, consider a simple random\index{random walk} walk on the integers with the transition probabilities
\begin{equation}
T_{ij} = \begin{cases}
1/3, & j \in \{i-1, i, i+1\} \\
0, & \text{otherwise}
.\end{cases}
\end{equation}

This chain is irreducible, since we can reach any integer from any other integer by moving one position at a time. It is also aperiodic because of its self-loops. From any state $i$ we can return to $i$ in 1 step (with probability $1/3$), in 2 steps, in 3 steps, and so on. The gcd of these return times is 1.

Does this chain have a stationary distribution? Suppose that $\pi = (\pi_0, \pi_1, \pi_2, ...)$ is the stationary distribution, where $\pi_i$ is the probability of being in state $i$ after a long run. By definition
\begin{equation}
\pi_i = \frac{1}{3}\pi_{i-1} + \frac{1}{3}\pi_i + \frac{1}{3}\pi_{i+1}.
\end{equation}

Rearranging gives
\begin{equation}
\pi_{i-1} - 2\pi_i + \pi_{i+1} = 0.
\end{equation}

This is a second-order linear recurrence relation, which gives $\pi_{i+1} = 2\pi_i - \pi_{i-1}$. Working out the first terms, we find $\pi_2 = 2\pi_1 - \pi_0$, $\pi_3 = 3\pi_1 - 2\pi_0$ and $\pi_4 = 4\pi_1 - 3\pi_0$. The pattern gives the general solution
\begin{equation}
\pi_i = i\pi_1 - (i-1)\pi_0.
\end{equation}

We now check whether this solution can be a probability distribution. If $\pi_1 = \pi_0 > 0$, then $\pi_i = \pi_0$ for all $i$, a uniform distribution over infinitely many states with $\sum_i \pi_i = \infty$ instead of 1. If $\pi_1 = \pi_0 = 0$, then $\pi_i = 0$ for all $i$ and $\sum_i \pi_i = 0$, which also violates the normalization. If $\pi_1 > \pi_0$, then $\pi_i$ grows linearly with $i$ and $\sum_i \pi_i = \infty$. If $\pi_1 < \pi_0$, then $\pi_i$ eventually becomes negative for large $i$, which violates the requirement that probabilities are non-negative.

This contradiction shows that the random walk has no valid stationary distribution. The issue is that the chain can move away to infinity.

Unlike Columbus, which has a finite number of neighborhoods, this random walk moves through an unbounded space. It has no ``gravitational force'' that pulls the population toward a central region. At each step, a traveler moves left, stays put or moves right, each with probability $1/3$. As time progresses, the population disperses more and more widely over the infinite state space, with no stable configuration. The probability mass spreads thinner and thinner as an expanding frontier of states fills up.

The random walk shows why we need a third condition, positive\index{positive recurrence} recurrence, to guarantee convergence. With positive recurrence, the chain has a well-defined stationary distribution and does not escape to infinity. A state $i$ is positive recurrent if its expected return time, starting from $i$, is finite, which means that
\begin{equation}
\mathbb{E}[t_i] < \infty, \text{ where } t_i = \inf\{t \geq 1 : z^{(t)} = i | z^{(0)} = i\}.
\end{equation}Here $t_i$ is the return time, the number of steps until the chain first comes back to state $i$. The notation $\inf$ means the smallest value, here the smallest $t \geq 1$ with $z^{(t)} = i$ when the chain starts in state $i$.

\looseness=1 In the Columbus picture, a traveler who starts in any neighborhood returns to that neighborhood in a finite expected time. The random walk on the integers is recurrent, since a traveler returns to each state with probability 1. Yet any state has an infinite expected return time, because a traveler can wander arbitrarily far away before returning. This is why the chain fails to converge, although it is irreducible and aperiodic.\index{positive recurrence|)}

\paragraph{Ergodic Chains}

The combination of irreducibility, aperiodicity and positive recurrence defines an ergodic Markov chain. For ergodic chains we can prove several results. First, the long-run fraction of time spent in each state converges to a unique stationary distribution, whatever the starting state. Second, the stationary probability of a state is the inverse of its expected return time, $p^*(i) = 1/\mathbb{E}[t_i]$. Third, for any initial distribution $\tilde{p}(\mathbf{x})$, repeated application of the transition operator converges to the unique stationary distribution, $\lim_{n \to \infty} T^n \tilde{p}(\mathbf{x}) = p^*(\mathbf{x})$.

A traveler walks through the state space for a long time. The footprints that mark each visited location trace out a density map proportional to the stationary distribution, with denser clusters at more frequently visited states. With the target posterior as the stationary distribution, we can generate samples from complex distributions by running the chain and recording its states.\index{Markov chain Monte Carlo (MCMC)!convergence|)}\index{stationary distribution|)}

\section{Implementing Metropolis Algorithm}

The\index{Metropolis algorithm|(} acceptance probability of the Metropolis algorithm ensures detailed balance for our target distribution. Ergodicity, which we typically have in practice, then makes the chain converge to this target. The algorithm has the following steps.

\begin{enumerate}
    \item \textbf{Initialize.} Choose a starting point $\mathbf{x}^{(0)}$ in the parameter space.

    \item \textbf{For each iteration} $t = 1, 2, \ldots, N$, where $N$ is the number of iterations.
    \begin{enumerate}
        \item \textbf{Propose.} Generate a proposed state $\mathbf{x}'$ from a symmetric proposal distribution centered at the current state, $\mathbf{x}' \sim q(\mathbf{x}'|\mathbf{x}^{(t-1)})$, where $q(\mathbf{x}'|\mathbf{x}) = q(\mathbf{x}|\mathbf{x}')$.

        \item \textbf{Compute.} {\sloppy Calculate the acceptance probability $A(\mathbf{x}^{(t-1)}, \mathbf{x}') = \min(1, p(\mathbf{x}')/p(\mathbf{x}^{(t-1)}))$.\par}

        \item \textbf{Accept or reject.} Draw a uniform random number $u \sim \mathcal{U}(0,1)$. If $u \leq A(\mathbf{x}^{(t-1)}, \mathbf{x}')$, accept the proposal and set $\mathbf{x}^{(t)} = \mathbf{x}'$. Otherwise, reject it and stay at the current position, $\mathbf{x}^{(t)} = \mathbf{x}^{(t-1)}$.
    \end{enumerate}

    \item \textbf{Return.} The sequence $\{\mathbf{x}^{(0)}, \mathbf{x}^{(1)}, \ldots, \mathbf{x}^{(N)}\}$ forms the samples from the target distribution. Every iteration adds an entry to this sequence. If we reject the proposal, the entry repeats the current state. We typically discard an initial portion as burn-in, which we explain below.
\end{enumerate}

\begin{figure}[ht!]
    \centering
    \includegraphics[width=0.8\textwidth]{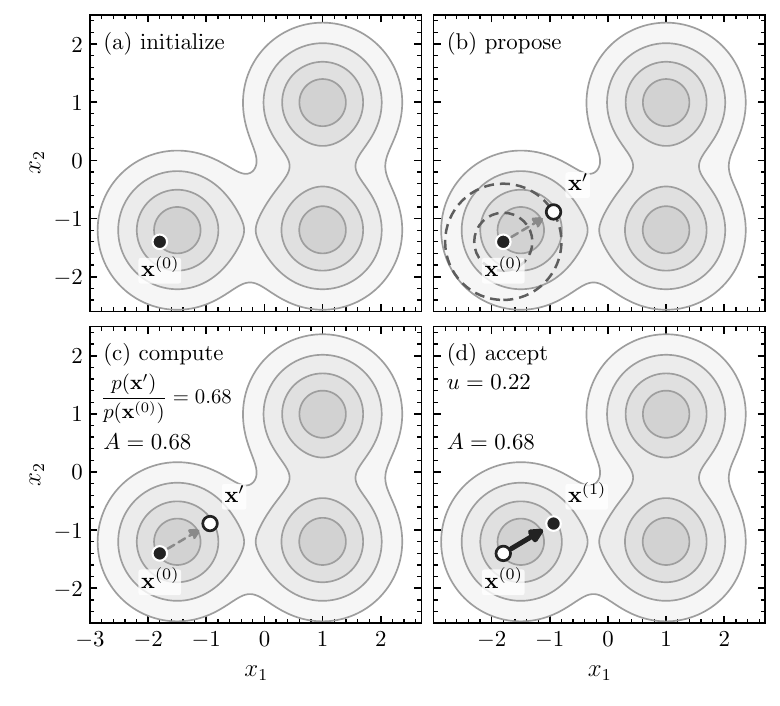}
    \caption{The four steps of one Metropolis update on a posterior with three modes. Panel (a) shows the starting state $\mathbf{x}^{(0)}$ on the posterior, drawn as grey contours. Panel (b) shows a proposed state $\mathbf{x}'$ from a Gaussian proposal around the current state, with dashed circles at one and two standard deviations. Panel (c) shows the ratio $p(\mathbf{x}')/p(\mathbf{x}^{(0)})$ and the acceptance probability $A$. In panel (d) the uniform random number $u$ is smaller than $A$, so we accept the proposal and set $\mathbf{x}^{(1)} = \mathbf{x}'$. The proposal has a lower density than the current state, and we still accept it.}
    \label{fig:metropolis_steps}
\end{figure}

Figure~\ref{fig:metropolis_steps} shows the starting state and the steps Propose, Compute and Accept or reject on a multimodal distribution.

Each state of a Markov chain depends only on the preceding state and not on the earlier history. This property leads to two practical questions. First, how do we handle the early samples, taken before the chain has reached its stationary distribution (the target posterior)? Second, once the chain has converged, how do we account for the correlation between consecutive samples, which reduces their statistical efficiency?

The first question concerns burn-in, the warm-up period of the chain. The chain starts from an arbitrary position $\mathbf{x}^{(0)}$ that may be far from the high-probability regions of the distribution. The initial samples then depend on this starting position and do not represent the target distribution. The chain converges to the target distribution in the end, but this takes time.

\begin{figure}[ht!]
    \centering
    \includegraphics[width=0.7\textwidth]{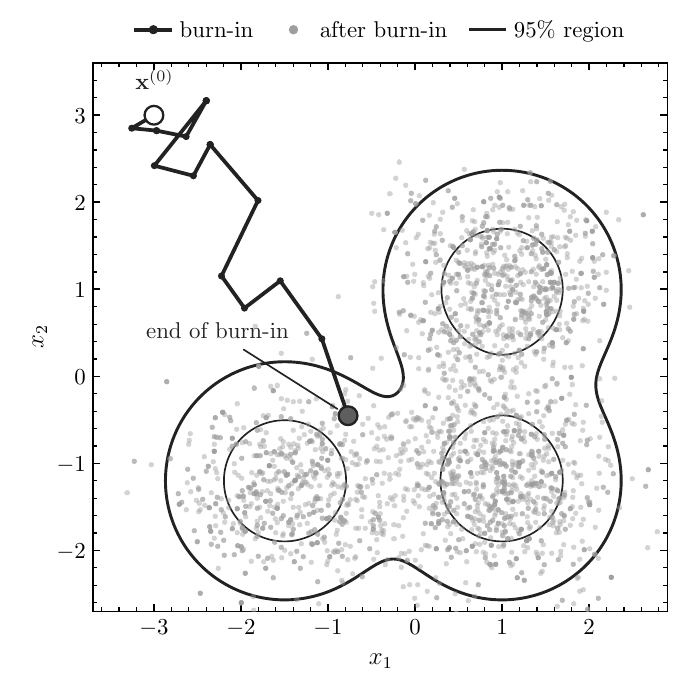}
    \caption{Burn-in of a Metropolis chain on a posterior with three modes. The chain starts at $\mathbf{x}^{(0)}$, far from all modes. The contours show the posterior. The burn-in ends where the chain first enters the 95 percent region, which lies inside the thick contour. We can use this criterion here only because we know the posterior.}
    \label{fig:mcmc_burnin}
\end{figure}

\looseness=1 Imagine that our explorer starts from the outskirts of an unfamiliar city. The first movements go into finding the city center, and only then can the explorer begin to explore the important areas. The chain likewise needs time to find the high-probability regions. Figure~\ref{fig:mcmc_burnin} shows a chain that starts in a low-probability region and settles into the high-probability regions after the burn-in. We cannot compare the chain with the target in practice, so the burn-in section below shows how to find its end and shorten it.

The second question concerns the efficiency of the sampler. Given enough time, MCMC can in theory sample any distribution, but even when ergodicity holds the time to convergence can be prohibitively long. Astronomical research has limited computing resources and project timelines. The question therefore shifts from ``Can we sample this distribution?'' to ``How efficiently can we sample it?''\index{Metropolis algorithm|)}

\section{Proposal Mechanisms and Sampling Efficiency}\label{sec:ch13_proposal}

The\index{Markov chain Monte Carlo (MCMC)!efficiency|(}\index{proposal distribution|(} efficiency of an MCMC method depends on how well the chain covers the parameter space. We control this exploration through the mechanism that proposes new positions for the chain to visit. This mechanism differs between MCMC algorithms but follows similar principles. Sequential steps turn difficult sampling tasks into manageable random walks, so MCMC can handle complex, high-dimensional problems. However, this same property requires us to choose with care how the random walk moves.

In most MCMC algorithms, a transition consists of a proposal\index{proposal distribution} followed by an acceptance or rejection step. The details differ between the Metropolis algorithm above and the Metropolis-Hastings algorithm and Gibbs sampling, which we introduce later. All of them must balance local exploration against global coverage of the parameter space.

For a symmetric proposal distribution, the Metropolis transition probability is
\begin{equation}
T(\mathbf{x}, \mathbf{x}') = q(\mathbf{x}'|\mathbf{x})A(\mathbf{x}, \mathbf{x}'),
\end{equation}
where $q(\mathbf{x}'|\mathbf{x})$ is the proposal distribution and $A(\mathbf{x}, \mathbf{x}')$ is the acceptance probability. The choice of proposal mechanism affects how efficiently we sample from the posterior, in any MCMC variant.

In astronomy, a common proposal is a multivariate Gaussian,
\begin{equation}
q(\mathbf{x}' | \mathbf{x}) = \mathcal{N}(\mathbf{x}' | \mathbf{x}, \boldsymbol{\Sigma}),
\end{equation}
where $\boldsymbol{\Sigma}$ is the covariance matrix that sets the scale and orientation of the proposals. The choice of $\boldsymbol{\Sigma}$ directly affects the sampling efficiency.

In practice, any MCMC method must balance small moves against large moves. Small moves are likely to be accepted but explore slowly. Large moves explore faster but are more likely to be rejected. This balance affects how quickly the chain converges to the target distribution and how efficiently it samples once converged.

\begin{figure}[ht!]
    \centering
    \includegraphics[width=0.95\textwidth]{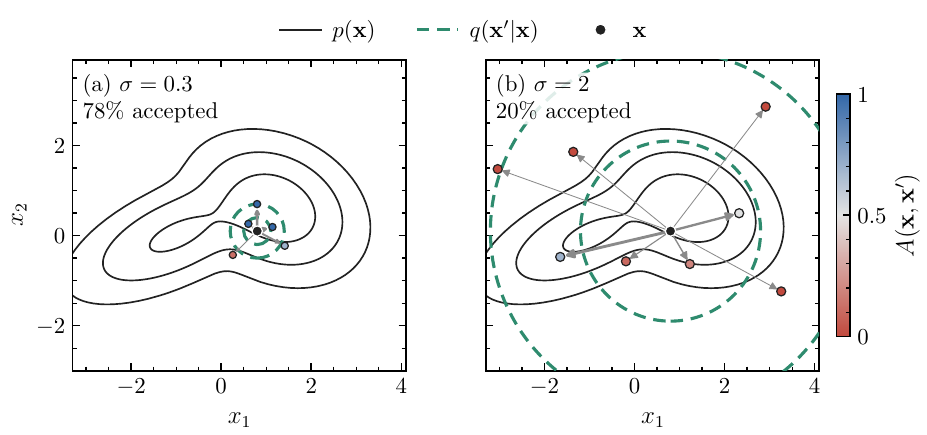}
    \caption{Two proposal widths for a Metropolis chain on an elongated posterior. The current state $\mathbf{x}$ is the same in both panels, and the dashed circles mark the proposal at one and two standard deviations $\sigma$. The arrow width grows with the acceptance probability $A$ of the proposed state. The narrow proposal in panel (a) is accepted most of the time but moves only a short distance. The wide proposal in panel (b) jumps farther, but most proposals land in regions of low density and are rejected. The percentages in the panels are the acceptance rates of a long chain with that $\sigma$.}
    \label{fig:mcmc_proposal_width}
\end{figure}

\looseness=1 If the proposal mechanism is too conservative, the chain has a high acceptance rate but moves through the parameter space very slowly. In our Columbus analogy, we take tiny steps. We accept nearly all of them because we stay close to where we are, but exploring the city takes an impractically long time. If the proposal mechanism is too aggressive, most proposals land in low-probability regions and are rejected. We then attempt giant leaps across the city, but most of them land in unlikely areas, so we often remain in the same location. Figure~\ref{fig:mcmc_proposal_width} compares a narrow and a wide proposal distribution.

\paragraph{Acceptance Rate and Tuning}

The acceptance\index{acceptance rate} rate is the proportion of proposed moves that the chain accepts during sampling. It is a simple diagnostic of the efficiency of many MCMC algorithms. It gives a first idea of how well the chain moves through the parameter space.

A proposal mechanism that is too conservative gives a very high acceptance rate. This may look desirable, but it leads to inefficient sampling. The chain moves through the parameter space in tiny steps, so the samples are highly correlated. We accept nearly all steps, but we cover little ground and gather largely redundant information. The chain needs many iterations to explore the full posterior distribution, so we say that it mixes\index{mixing of a Markov chain} poorly. A chain mixes well when it moves quickly through the region of high posterior probability, so that its samples soon look independent.

A proposal mechanism that is too aggressive gives a very low acceptance rate. Most proposals jump to regions with much lower posterior probability and are rejected. The chain stays at the same position for many iterations, which again gives highly correlated samples. This resembles exploring a city by teleporting across vast distances. Most attempts land in unlikely places, so we stand still most of the time.

The best case typically occurs when the proposal mechanism is tuned to the geometry of the posterior distribution. For random walk proposals like the Gaussian proposal above, acceptance rates around 20 to 40\% tend to give the most efficient sampling. At these rates the moves are large enough to explore the parameter space effectively, but not so large that nearly all proposals are rejected.

In astronomy, models are complex and computing resources are limited. Proper tuning of the proposal mechanism can mean that reliable posterior inferences take hours instead of days. Finding the best tuning often takes some experimentation. We can start with an educated guess based on prior knowledge of the problem. We then adjust it based on the acceptance rate in a short pilot run.

The shape of the proposal matters as much as its size. A round Gaussian proposal wastes steps when the parameters have different scales or are correlated. A better choice sets $\boldsymbol{\Sigma}$ proportional to the covariance of the posterior, so that the steps match its scales and correlations. We estimate this covariance from the samples of the pilot run. We then adjust its overall scale with the acceptance rate, as above.

Acceptance rates are a useful rule of thumb for this initial tuning, but they do not fully describe the sampling efficiency. To see how well the chain covers the parameter space and how much independent information we obtain, we need the correlation between successive samples.\index{proposal distribution|)}

\section{Autocorrelation in MCMC Sampling}\label{sec:ch13_autocorrelation}

\looseness=1 Successive\index{autocorrelation|(} samples of an MCMC chain are not independent, because each new state depends on the previous one. This correlation is called autocorrelation.\index{autocorrelation} It reduces the effective information content of the chain compared to independent samples, and we need to measure it for reliable statistical inference. For $N$ independent and identically distributed samples, the standard error of our estimate decreases as $1/\sqrt{N}$. Correlated samples give a larger error for the same $N$. Each new correlated sample carries less new information about the target distribution than an independent sample, so the estimates are less precise.

The autocorrelation at lag $k$ of $\{X_0, X_1, X_2, ...\}$ is defined as
\begin{equation}
R(k) = \frac{\mathbb{E}[(X_t - \mu)(X_{t+k} - \mu)]}{\sigma^2},
\end{equation}
where $\mu = \mathbb{E}[X_t]$ and $\sigma^2 = \mathbb{E}[(X_t - \mu)^2]$ are the mean and variance of the stationary distribution. This definition follows the correlation coefficient of Chapter~3. Standard correlation measures the linear relationship between two different variables. Autocorrelation measures the relationship between observations of the same variable at different points in the sequence. The lag $k$ is the number of steps between these points.

In a Markov chain, we can view $X_t$ and $X_{t+k}$ as two realizations of the same random variable. Once the chain has reached its stationary distribution, each state it visits is a draw from that distribution. Both $X_t$ and $X_{t+k}$ approximately follow the target distribution, and the evolution of the chain over $k$ steps connects them.

This view rests on the ergodic properties of a well-designed MCMC algorithm. For an ergodic chain, the average over time along the chain, the time average, becomes equivalent to the ensemble average. The ensemble average is the average over many independent samples from the target distribution. This equivalence lets us use a single chain to characterize a probability distribution.

\begin{figure}[ht!]
    \centering
    \includegraphics[width=0.9\textwidth]{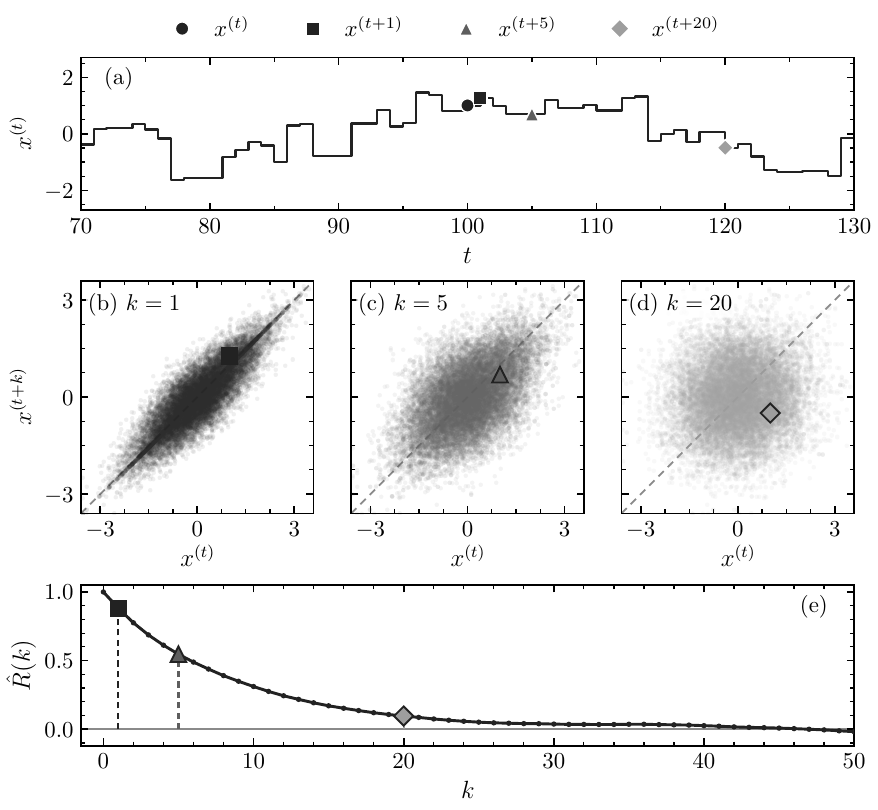}
    \caption{Autocorrelation of a Metropolis chain. Panel (a) shows a section of the chain, the reference point $x^{(100)}$ and the values at lags 1, 5 and 20. Panels (b) to (d) show $x^{(t+k)}$ against $x^{(t)}$ for all pairs of the chain at these lags. The marked point is the pair that starts at the reference point. The points scatter more widely around the diagonal as the lag grows. Panel (e) shows the estimated autocorrelation function with the three lags marked.}
    \label{fig:mcmc_autocorrelation}
\end{figure}

\looseness=1 The dependency between samples is the trade-off we accept for sampling complex distributions through random walks. The results are still correct in the long run, but the sampling noise is larger than for independent samples. We need to understand this correlation to judge the uncertainty of our estimates and to choose the chain length. Figure~\ref{fig:mcmc_autocorrelation} shows a chain and how the correlation between its values at steps $t$ and $t+k$ decreases as the lag $k$ grows. Once the lag is large enough, points of the chain are approximately independent samples of the target distribution. This holds for a chain that mixes well.

In practice, we estimate the autocorrelation from the finite chain. For a chain of length $N$, the autocorrelation at lag $k$ is estimated by
\begin{equation}
r(k) = \frac{\sum_{t=1}^{N-k} (X_t - \bar{X})(X_{t+k} - \bar{X})}{\sum_{t=1}^{N} (X_t - \bar{X})^2},
\end{equation}
where $\bar{X}$ is the sample mean of the chain. This estimator replaces the mean and variance of the definition by sample values. The numerator is the covariance of samples separated by lag $k$, and the denominator normalizes it by the sample variance. Only $N-k$ pairs of points are available at lag $k$, and their number falls as $k$ increases. The estimate therefore becomes less reliable at high lags.

\looseness=1 To build intuition for autocorrelation, we look at different proposal mechanisms. When the proposal distribution is too conservative (too narrow), each new sample is very close to the previous one. The chain moves like a cautious explorer taking baby steps and stays near the same location for many iterations. The value at position $t$ then predicts the values at positions $t+1$, $t+2$ and so on, which gives high autocorrelation even at large lags. We have many samples, but they hold the same information repeated with slight variations.

When the proposal distribution is too aggressive (too wide), most proposals are rejected because they jump to low-probability regions. The chain stays at the same value for many iterations, which creates high autocorrelation through a different mechanism, the literal repetition of the same value. When jumps do occur they can be substantial, but they are too infrequent for good exploration. A well-tuned proposal moves enough to explore effectively and keeps a reasonable acceptance rate. The autocorrelation then typically decays more rapidly with the lag, which means that each new sample brings more unique information about the target distribution. This is the balance we aim for when we tune an MCMC algorithm for efficient sampling.

\looseness=1 The autocorrelation function shows how efficient the sampler is. Independent samples have no relationship with each other, so knowing one value tells us nothing about another and $R(k) = 0$ for $k > 0$. In MCMC, we typically observe an autocorrelation that decays as the lag increases. The rate of this decay sets how efficiently the chain covers the parameter space. A rapidly decaying autocorrelation means that the chain quickly ``forgets'' its starting position and samples the distribution more independently. A faster decay is desirable, because it leads to more efficient parameter estimation and uncertainty quantification.\index{autocorrelation|)}

\section{Effective Sample Size}\label{sec:ch13_ess}

To\index{effective sample size (ESS)|(} measure the loss of information caused by autocorrelation, we use the effective\index{effective sample size (ESS)} sample size (ESS)\index{ESS|see{effective sample size}} that we met for importance sampling in Chapter~12. The ESS is the number of independent samples that would carry the same amount of information as our autocorrelated chain. More samples do not necessarily mean more useful information if the samples are highly correlated.

Imagine an explorer mapping the population distribution of Columbus. A path with high autocorrelation revisits the same neighborhoods for many days or follows predictable patterns. This gives a poor picture of the population distribution of the city, even after a long exploration. A path with low autocorrelation moves through the city in varied, seemingly random ways and gives a more representative view in less time. In MCMC terms, lower autocorrelation means more efficient sampling of the target distribution and a larger effective sample size.

\paragraph{Deriving the Effective Sample Size}

We derive the ESS from the variance of the sample mean of a correlated chain.

The sample mean of our MCMC chain is
\begin{equation}
\bar{X} = \frac{1}{N}\sum_{i=1}^{N}X_i.
\end{equation}

The variance of this estimator is
\begin{equation}
\text{Var}(\bar{X}) = \text{Var}\left(\frac{1}{N}\sum_{i=1}^{N}X_i\right) = \frac{1}{N^2}\text{Var}\left(\sum_{i=1}^{N}X_i\right).
\end{equation}

For our MCMC chain, we must account for all pairwise relationships between samples. Expanding the variance of the sum gives
\begin{equation}
\text{Var}\left(\sum_{i=1}^{N}X_i\right) = \sum_{i=1}^{N}\sum_{j=1}^{N}\text{Cov}(X_i, X_j).
\end{equation}

The covariance between two samples of a stationary Markov chain depends only on the lag between them. We call a chain stationary once it has reached its stationary distribution, because its statistical properties then no longer change from step to step. Once the chain has converged, we have
\begin{equation}
\text{Cov}(X_i, X_j) = \text{Cov}(X_1, X_{1+|i-j|}) = \sigma^2 R(|i-j|),
\end{equation}
where $|i-j|$ is the lag between the samples and $\sigma^2$ is the variance of each sample. The function $R(k)$ is the autocorrelation at lag $k$. Note that $R(0) = 1$, because a sample correlates with itself.

We reorganize the double sum by grouping all pairs that have the same lag. For example, the pairs $(1,3)$, $(2,4)$ and $(3,5)$ all have a lag of 2. For a chain of length $N$, there are $N-k$ pairs with lag $k$. With this count, we can rewrite the variance as
\begin{equation}
\text{Var}(\bar{X}) = \frac{\sigma^2}{N^2}\left[N + 2\sum_{k=1}^{N-1}(N-k)R(k)\right].
\end{equation}

\looseness=1 The first term $N$ counts the pairs with lag 0 (that is, $i=j$), where $R(0)=1$. The factor of 2 in the sum appears because we count both the pairs $(i,i+k)$ and $(i+k,i)$ for each positive lag $k$. These pairs have the same autocorrelation due to stationarity.

We divide the numerator and the denominator by $N$ and simplify further to
\begin{equation}
\text{Var}(\bar{X}) = \frac{\sigma^2}{N}\left[1 + 2\sum_{k=1}^{N-1}\left(1-\frac{k}{N}\right)R(k)\right].
\end{equation}

For large $N$, the term $k/N$ becomes negligible for any fixed lag $k$, so we can approximate
\begin{equation}
\text{Var}(\bar{X}) \approx \frac{\sigma^2}{N}\left(1 + 2\sum_{k=1}^{\infty}R(k)\right).
\end{equation}

This result shows that the variance of our sample mean is inflated by the factor $\left(1 + 2\sum_{k=1}^{\infty}R(k)\right)$ compared to independent samples. This inflation factor measures the information loss caused by autocorrelation in our chain.

The mean of ESS independent samples has the variance $\sigma^2/\text{ESS}$, which we equate with our MCMC variance
\begin{equation}
\frac{\sigma^2}{\text{ESS}} = \frac{\sigma^2}{N}\left(1 + 2\sum_{k=1}^{\infty}R(k)\right).
\end{equation}

Solving for the ESS gives
\begin{equation}
\text{ESS} = \frac{N}{1 + 2\sum_{k=1}^{\infty}R(k)}.
\end{equation}

\looseness=1 The denominator $1 + 2\sum_{k=1}^{\infty}R(k)$ is the autocorrelation\index{autocorrelation time} time $\tau$, the average number of correlated MCMC samples needed to obtain the equivalent of one independent sample. If our samples were independent, as in direct sampling methods, all autocorrelations $R(k)$ would be zero for $k > 0$. The denominator would then equal 1 and $\text{ESS} = N$, which means that all our samples carry full information. When samples are highly correlated, as is typical in MCMC, the autocorrelations $R(k)$ remain positive for many lags. The sum in the denominator is then large, so the denominator exceeds 1 and the ESS falls below $N$. The stronger and more persistent these autocorrelations are, the smaller the ESS becomes relative to $N$.

\looseness=1 For example, if the autocorrelation time equals 20, we need about 20 MCMC samples to get the information of one independent sample. A chain of 10,000 samples then has an ESS of only about 500, so our precision equals that of 500 independent samples. This is why tuning the proposal distribution is so important. A well-tuned proposal reduces autocorrelation and increases the ESS for a given computational budget.

As a rule of thumb, we aim for ESS $> 100$ for reliable posterior estimates. More demanding applications may need higher values. An ESS of this size gives enough independent information to estimate posterior means and standard deviations.

Calculating the autocorrelation function $R(k)$ directly takes $O(N^2)$ operations, which is slow for long chains. The fast Fourier transform gives a much faster method. The power spectrum of the chain is the squared magnitude of its Fourier transform. It shows how much of the variation of the chain sits at each frequency. The Wiener-Khinchin\index{Wiener-Khinchin theorem} theorem says that the autocorrelation function is the inverse Fourier transform of the power spectrum. The cost then falls to $O(N \log N)$, which is practical for the long chains typical of astronomical applications.\index{effective sample size (ESS)|)}

\section{Thinning MCMC Chains}

\looseness=1 Given\index{thinning|(} the autocorrelation in MCMC samples, a common practice is ``thinning'',\index{thinning} which keeps only every $m$-th sample of the chain to reduce correlation. For example, from a chain of 10,000 samples we might keep every 10th sample, which gives a thinned chain of 1,000 samples. If successive samples are highly correlated, keeping every sample may give redundant information. Thinning also reduces the storage and computation needed for the later analysis.

Thinning can be beneficial in several contexts. When storing or processing samples is costly, thinning reduces this cost while keeping a representative sample of the posterior. When the autocorrelation is very strong, thinning makes the data more manageable for analysis and visualization. Certain Monte Carlo estimators also perform better with less correlated samples, which makes thinning advantageous in those applications. However, thinning discards some information, because even correlated samples contain some independent information about the target distribution. For posterior\index{posterior mean} means and credible intervals, the full chain after burn-in generally gives more precise estimates than a thinned chain of the same computing cost.

We can compare the variance of estimators computed from thinned and from full chains. Consider a chain of length $N$ with autocorrelation function $R(k)$. If we thin by keeping every $m$-th sample, we obtain a chain of length $N/m$ with a modified autocorrelation structure. Samples separated by lag $k$ in the thinned chain are separated by lag $mk$ in the original chain. The thinned chain therefore has the autocorrelation function $R_{\text{thin}}(k) = R(mk)$.

The effective sample size of the thinned chain is
\begin{equation}
\text{ESS}_{\text{thin}} = \frac{N/m}{1 + 2\sum_{k=1}^{\infty}R(mk)}.
\end{equation}

Comparing this to the ESS of the full chain shows the trade-off of thinning. Thinning reduces the autocorrelation between consecutive samples in the thinned chain, but it also reduces the total number of samples. The net effect on the statistical efficiency depends on the autocorrelation structure of the original chain.

\looseness=1 To understand when thinning might be beneficial, consider two extreme cases. If the original chain has very weak autocorrelation, the samples are nearly independent. Then $R(k) \approx 0$ for $k > 0$ and the ESS of the full chain is about $N$. Thinning this chain by a factor $m$ gives an ESS of about $N/m$, which is worse than using the full chain. If instead the original chain has very strong autocorrelation that persists for many lags, thinning might remove redundant information without reducing the ESS much. The thinned chain might then keep most of the information of the full chain and be much more manageable computationally.

The decision to thin should therefore balance computational constraints against statistical precision. In many astronomical applications, we are mainly interested in posterior summaries and uncertainties. We should then apply thinning with care and not as a default practice.

The balance between storage costs and statistical efficiency matters most in astronomical applications where chains may contain millions of samples across hundreds of parameters. In such cases, thinning saves computing time and storage. These savings may outweigh the loss of statistical precision, especially if the original chain has very high autocorrelation. However, modern computing resources and efficient storage formats often make it feasible to work with full chains. Thinning is therefore less necessary than in earlier decades of MCMC applications. We resort to thinning only when computational constraints make it necessary or when the autocorrelation structure suggests that little information would be lost.

\begin{figure}[ht!]
    \centering
    \includegraphics[width=0.9\textwidth]{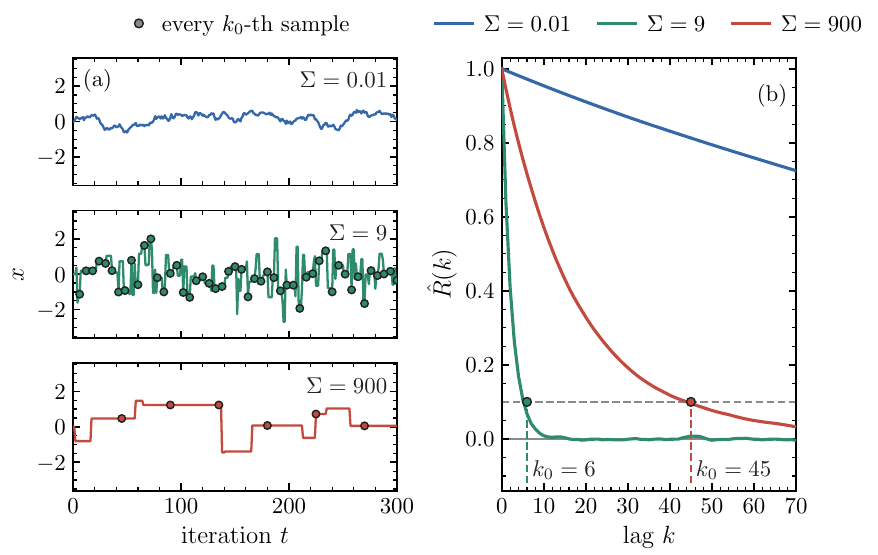}
    \caption{Thinning for Metropolis chains with a narrow, a medium and a wide proposal on a Gaussian target. The proposal variance is $\Sigma$. Panel (a) shows the three chains, and the narrow chain has no dots because its $k_0$ is larger than the plotted range. Panel (b) shows the estimated autocorrelation function of each chain. The dashed lines mark 0.1 and, for the medium and wide chains, the lag $k_0$ at which the function first falls below it. Thinning at $k_0$ keeps about two thirds of the effective sample size of the full chain in all three cases.}
    \label{fig:mcmc_thinning}
\end{figure}

A practical approach bases the thinning factor on the lag $k_0$ at which the autocorrelation function drops to negligible values, say below 0.1. Thinning by a factor of $k_0$ then keeps samples that are approximately independent. Figure~\ref{fig:mcmc_thinning} shows this choice of the thinning factor for chains with three proposal widths.\index{Markov chain Monte Carlo (MCMC)!efficiency|)}\index{thinning|)}

\section{Burn-in}\label{sec:ch13_burnin}

The\index{Markov chain Monte Carlo (MCMC)!convergence diagnostics|(}\index{burn-in|(} discussion of autocorrelation and effective sample size assumed that the chain had already converged to its stationary distribution, the target posterior. Samples taken before convergence would bias the posterior estimates and lead to incorrect inferences. We therefore need to know when the burn-in\index{burn-in} is over, that is, when the chain has converged enough to start collecting useful samples.

In our Columbus analogy, observations that the explorer makes during the burn-in would skew the picture of the demographics of Columbus. We cannot compare the chain with samples of the target distribution, which is why we use MCMC in the first place. We therefore rely on diagnostics that examine the behavior of the chain itself to tell when burn-in has completed.

The simplest and most common approach is visual inspection of trace\index{trace plot} plots, which show the parameter values at each iteration of the chain. A well-converged chain should settle around a central value, fluctuating around a stable mean without obvious drift or trends. The width of the fluctuations stays roughly constant. The chain does not get stuck at particular values for extended periods, because such ``stickiness'' indicates poor mixing. It also covers the posterior properly, visiting all relevant regions of the parameter space with frequencies proportional to their posterior probabilities.

Beyond visual inspection, the ESS that we discussed earlier is another convergence diagnostic. The ESS depends on the autocorrelation structure of the chain, so monitoring how it changes as we include more samples gives insight into convergence. For a well-converged chain, the ESS should grow approximately linearly with the number of samples after burn-in. Plotted against the number of iterations, it should show an initial nonlinear phase during burn-in and a roughly linear increase once the chain has converged. A low ESS relative to the chain length may indicate poor mixing and convergence issues. A high ESS does not guarantee convergence, because a chain can mix well in one region without fully exploring the target.

\looseness=1 Visual inspection and ESS calculations provide useful initial diagnostics. They become impractical for complex models with many parameters and may miss subtle convergence problems. We then need quantitative diagnostics that detect when chains have reached their target distributions. The Gelman-Rubin and Geweke diagnostics are two standard examples.

\subsection{Gelman-Rubin Diagnostic}

\looseness=1 Convergence\index{Gelman-Rubin diagnostic|(} can be checked by running multiple chains from different starting points and comparing their behavior. The starting points should be dispersed over the parameter space. If all chains converge to the same distribution, they should be statistically indistinguishable from each other after some burn-in period. The Gelman-Rubin\index{Gelman-Rubin diagnostic} diagnostic formalizes this intuition. For a parameter $x$, we run $M$ chains, typically at least 3 or 4, each of length $2N$. We consider discarding the first $N$ iterations of each chain as burn-in and compare the within-chain variance with the between-chain variance.

Let $x_{ij}$ be the $i$-th sample of parameter $x$ in the $j$-th chain after the burn-in. For each chain $j$, the mean $\bar{x}_j$ is
\begin{equation}
\bar{x}_j = \frac{1}{N}\sum_{i=1}^{N}x_{ij},
\end{equation}
and the variance $s_j^2$ is
\begin{equation}
s_j^2 = \frac{1}{N}\sum_{i=1}^{N}(x_{ij} - \bar{x}_j)^2.
\end{equation}

The within-chain variance $W$ is the average of these individual variances
\begin{equation}
W = \frac{1}{M}\sum_{j=1}^{M}s_j^2.
\end{equation}
If the chains have converged, $W$ should directly estimate the variance $\sigma^2$ of the target distribution.

For the between-chain variance, we first take the variance of the chain means
\begin{equation}
\text{Var}(\bar{x}_j) = \frac{1}{M}\sum_{j=1}^{M}(\bar{x}_j - \bar{x})^2,
\end{equation}
where $\bar{x} = (1/M)\sum_{j=1}^{M}\bar{x}_j$ is the overall mean over all chains.

\looseness=1 The variance of the chain means is not directly comparable to $W$. Each mean $\bar{x}_j$ is based on $N$ samples. Statistical theory gives $\text{Var}(\bar{x}_j) = \sigma^2/N$ when the chain has converged to the target distribution. This is $N$ times smaller than the variance of individual samples.

To make the between-chain variance comparable to the within-chain variance, we define $B$ by multiplying by $N$
\begin{equation}
B = N \cdot \text{Var}(\bar{x}_j) = \frac{N}{M}\sum_{j=1}^{M}(\bar{x}_j - \bar{x})^2.
\end{equation}

For converged chains, $\text{Var}(\bar{x}_j) = \sigma^2/N$ gives $B = N \cdot \sigma^2/N = \sigma^2$. The within-chain variance $W$ also estimates $\sigma^2$, so we expect
\begin{equation}
B \approx W.
\end{equation}

\looseness=1 Chains that have not converged explore different regions of the parameter space. Their means then differ by much more than for converged chains, so $B$ is much larger than $W$. This extra spread indicates additional uncertainty due to incomplete convergence.

The Gelman-Rubin statistic $\hat{R}$ compares $B$ and $W$ through
\begin{equation}
\hat{R} = \sqrt{1 + \frac{B-W}{NW}}.
\end{equation}

The expression under the square root is the ratio of $W + (B-W)/N$ to $W$. The numerator is our estimate of the marginal posterior variance, which includes both the within-chain and the between-chain variability. Its second term $(B-W)/N$ is the additional variance due to the variability between the chains. Dividing by $W$ compares this total variance with the within-chain variance alone. The ratio has no units, so $\hat{R}$ can be interpreted in the same way for parameters of different scales.

\begin{figure}[ht!]
    \centering
    \includegraphics[width=0.7\textwidth]{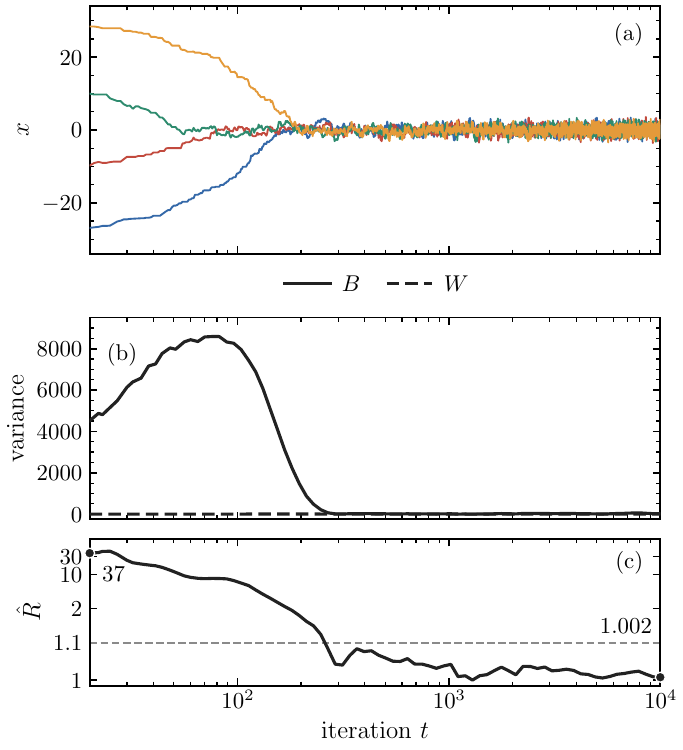}
    \caption{Gelman-Rubin diagnostic for four Metropolis chains that start at widely separated points of a Gaussian posterior. Panel (a) shows the traces, and all panels share a logarithmic iteration axis. Panel (b) shows the between-chain variance $B$ and the within-chain variance $W$, computed from the second half of the first $t$ iterations, so $N = t/2$. As the chains converge, $B$ falls from thousands to the scale of $W$, and $W$ stays near the variance $\sigma^2$ of the target. Panel (c) shows the statistic $\hat{R}$, which approaches 1, with a dashed line at 1.1 and a vertical scale that is stretched near 1.}
    \label{fig:gelman_rubin_components}
\end{figure}

\looseness=1 As the chains converge, $\hat{R}$ approaches 1 from above, because the excess $B-W$ is divided by $N$. Values substantially greater than 1 indicate that the chains have not yet converged. The criterion typically used is $\hat{R} < 1.1$ for acceptable convergence, or $\hat{R} < 1.05$ for a more stringent requirement. These thresholds are practical compromises between computing cost and accuracy. The value $\hat{R} = 1$ is rarely reached with finite samples. Figure~\ref{fig:gelman_rubin_components} shows $B$, $W$ and $\hat{R}$ for four chains that start from widely separated points.

To see the diagnostic intuitively, think of $M$ explorers who start from different parts of Columbus. If they have all learned the true population distribution of the city, they should report similar statistics about the neighborhood populations. This means a low between-explorer variance. If their reports differ widely, this suggests that they have not fully explored the city or are still affected by their starting locations.

The diagnostic does not use the target distribution. It only compares the behavior of different chains. It does require several chains, which increases the computing cost.

The statistic $\hat{R}$ measures the factor by which credible intervals might shrink with more sampling. For example, $\hat{R} = 1.2$ suggests that credible intervals from the current chains might be about 20\% wider than after full convergence.

In multivariate problems, we apply the diagnostic to each parameter separately. This shows which parameters may have convergence issues. However, the diagnostic may miss problems in parameters that we do not check. It also cannot detect complex interactions between parameters that indicate convergence problems even when each parameter looks well-behaved.\index{Gelman-Rubin diagnostic|)}

\subsection{Geweke Diagnostic}

The\index{Geweke diagnostic|(} Gelman-Rubin diagnostic needs multiple chains. Limited computational resources or other practical constraints can rule this out. In such cases, we can use the Geweke\index{Geweke diagnostic} diagnostic, which tests convergence with a single MCMC chain. If a chain has converged to its stationary distribution, different segments of the chain should have similar statistical properties. In particular, the means of different portions should be approximately equal, because all portions sample from the same target distribution.

To implement the Geweke diagnostic, we first run a single chain and discard the initial burn-in period. We then compare two segments of the remaining $N$ samples, segment $A$ and segment $B$. Segment $A$ holds the first $N_A$ samples, typically the first 10\% of the post-burn-in chain, and has mean $\bar{x}_A$. Segment $B$ holds the last $N_B$ samples, typically the last 50\% of the post-burn-in chain, and has mean $\bar{x}_B$. These proportions give each segment enough samples for a reliable estimate. They also keep the two segments far enough apart in the chain to detect remaining non-convergence.

The two means are
\begin{equation}
\bar{x}_A = \frac{1}{N_A}\sum_{i=1}^{N_A}x_i,
\end{equation}
and
\begin{equation}
\bar{x}_B = \frac{1}{N_B}\sum_{i=N-N_B+1}^{N}x_i.
\end{equation}

\paragraph{Accounting for Autocorrelation}

If the chain has converged, the two means should be equal except for random sampling variation. To judge whether an observed difference exceeds this variation, we need the uncertainty of our estimates of the means. However, the samples of the chain are not independent. As the sections on autocorrelation and effective sample size show, consecutive samples are correlated, and this affects the standard error of the sample mean. The standard formula for independent samples would underestimate the true uncertainty and make our test too sensitive.

We therefore need the variance of the sample mean of an autocorrelated chain. We saw earlier that autocorrelation inflates it by a factor that depends on the autocorrelation function, so that
\begin{equation}
\text{Var}(\bar{x}) = \frac{\sigma^2}{N}\left(1 + 2\sum_{k=1}^{\infty}R(k)\right) \equiv S.
\end{equation}The symbol $S$ denotes this variance of the sample mean, which includes the effect of the autocorrelation at all lags.

A chain has finite length, so we cannot sum an infinite number of autocorrelation terms. Estimates of the autocorrelation at high lags are also increasingly noisy, because fewer pairs of observations are available. We therefore use a windowed estimator
\begin{equation}
\hat{S} = \frac{\hat{\sigma}^2}{N}\left(1 + 2\sum_{k=1}^{l} w(k, l) r(k)\right),
\end{equation}
where $\hat{\sigma}^2$ is the sample variance of the segment and $r(k)$ the estimated autocorrelation. Here $N$ is the length of the segment, so $N = N_A$ for $\hat{S}_A$ and $N = N_B$ for $\hat{S}_B$. The window function $w(k, l)$ does not increase with $k$, for example $1 - k/l$, and $l$ is the maximum lag. The term $\hat{\sigma}^2$ estimates the marginal variance of the target distribution from the available samples, while the remaining terms adjust for the autocorrelation structure.

The windowed estimator can be viewed as integrating over a triangular region in lag space. For the window $1 - k/l$, the weight decreases linearly with $k$ and is zero beyond lag $l$. Small lags receive the largest weight, because they matter more and are more reliable in describing the autocorrelation structure of the chain.

With the estimates $\hat{S}_A$ and $\hat{S}_B$ of this variance for the two segments, the Geweke statistic is the Z-score
\begin{equation}
Z = \frac{\bar{x}_A - \bar{x}_B}{\sqrt{\hat{S}_A + \hat{S}_B}}.
\end{equation}

\begin{figure}[ht!]
    \centering
    \includegraphics[width=0.7\textwidth]{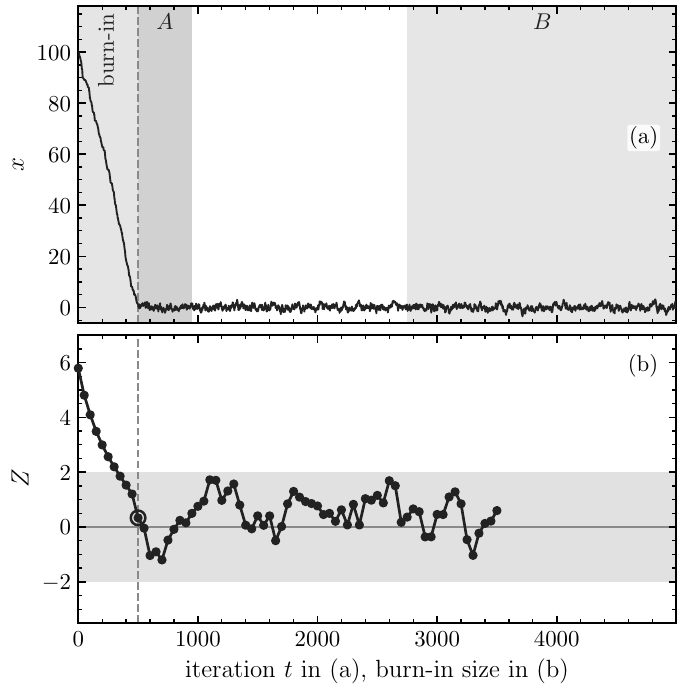}
    \caption{Geweke diagnostic for a chain that starts far from its target. Panel (a) shows the chain and its burn-in. Segment $A$ holds the first 10 percent of the samples after the burn-in, and segment $B$ holds the last 50 percent. Panel (b) shows the Z-score for burn-in sizes up to the point where too few samples remain for the two segments. The circle and the dashed line mark the burn-in of panel (a), and the band marks $|Z| < 2$.}
    \label{fig:geweke_diagnostic}
\end{figure}

\looseness=1 The Z-score is the difference between the means, divided by the standard error of that difference, which is adjusted for autocorrelation. Under the null hypothesis that the chain has converged, meaning that both segments sample the same distribution, $Z$ should approximately follow a standard normal distribution. This follows from the Central Limit Theorem for Markov chains, which says that the mean of a long chain is approximately normal. In practice, we typically consider the chain converged if $|Z| < 2$, which contains about 95\% of the standard normal distribution. Larger Z-scores suggest that the chain may not have reached its stationary distribution yet and may need more burn-in. Figure~\ref{fig:geweke_diagnostic} shows a single chain with the two segments and the Z-scores for different burn-in sizes.

In our Columbus analogy, the Geweke diagnostic compares the observations that the explorer makes in different phases of the journey. If the explorer has learned the population distribution of the city, early and late observations should give statistically similar estimates of the neighborhood populations. The early observations come after the initial familiarization, and the late ones from the end of the journey.

The Geweke diagnostic applies to individual parameters separately. This helps identify which aspects of the model have convergence issues. In complex astronomical models, some parameters often converge more slowly than others, and the diagnostic can highlight them.

The diagnostic only compares two specific segments of the chain and can therefore miss convergence issues in the middle portions. It is also sensitive to the choice of the two segments. It relies on asymptotic properties of the autocorrelation, which can be less reliable for shorter chains. The posterior may have multiple modes, and the chain may not have explored all of them. The diagnostic can then wrongly suggest convergence from a single mode.

In astronomical applications with computationally expensive models, a practical approach is to run a single long chain first. We use the Geweke diagnostic to check convergence and to choose the burn-in period. If resources permit, we then run several shorter chains from dispersed starting points for the final analysis and verify convergence with the Gelman-Rubin diagnostic.

The diagnostics tell us when the burn-in ends, and a good starting point makes it shorter. We can start the chain at the maximum of the posterior, which an optimizer such as gradient descent finds by minimizing the negative log posterior. This works well when the posterior has one dominant mode. The typical posterior samples can still lie some distance from the maximum, so a short burn-in remains. For the Gelman-Rubin diagnostic we then scatter the starting points of the chains around this maximum.\index{Geweke diagnostic|)}\index{Markov chain Monte Carlo (MCMC)!convergence diagnostics|)}\index{burn-in|)}

\section{The Metropolis-Hastings Algorithm}

The\index{Metropolis-Hastings algorithm|(} Metropolis algorithm is widely applied in astronomical research for posterior sampling. Its symmetric proposal can, however, limit the sampling efficiency in practice. This happens when the geometry of the posterior distribution is complex, for instance when parameters have different scales or are highly correlated. In such cases, we can benefit from a more general method, the Metropolis-Hastings\index{Metropolis-Hastings algorithm} algorithm.

\looseness=1 The Metropolis-Hastings algorithm generalizes the Metropolis algorithm by dropping the symmetry requirement for the proposal distribution. It keeps detailed balance, so the target remains a stationary distribution of the chain. The Metropolis algorithm is restricted to symmetric proposals with $q(\mathbf{x}'|\mathbf{x}) = q(\mathbf{x}|\mathbf{x}')$, typically a Gaussian centered at the current position. Asymmetric proposals with $q(\mathbf{x}'|\mathbf{x}) \neq q(\mathbf{x}|\mathbf{x}')$ give more flexibility in exploring the parameter space. This becomes useful when we have prior knowledge about the geometry or the preferred directions of movement within the posterior distribution.

Consider sampling from a highly skewed distribution, such as a gamma or log-normal distribution. A symmetric Gaussian proposal needs very small steps to keep a reasonable acceptance rate in the narrow tail region. This makes the exploration of the wider region very slow. An asymmetric proposal can adapt to the local geometry, with smaller steps in the tail and larger steps in the wider region.

What we can directly control in MCMC is the transition matrix $T(\mathbf{x}, \mathbf{x}')$. In both algorithms, it is the product of the proposal distribution $q(\mathbf{x}'|\mathbf{x})$ and the acceptance probability $A(\mathbf{x}, \mathbf{x}')$, so that
\begin{equation}
T(\mathbf{x}, \mathbf{x}') = q(\mathbf{x}'|\mathbf{x})A(\mathbf{x}, \mathbf{x}').
\end{equation}

Asymmetric proposals require an adjusted acceptance probability to maintain detailed balance. The acceptance probability of the Metropolis-Hastings algorithm is defined as
\begin{equation}
A(\mathbf{x}, \mathbf{x}') = \min\left(1, \frac{p(\mathbf{x}')q(\mathbf{x}|\mathbf{x}')}{p(\mathbf{x})q(\mathbf{x}'|\mathbf{x})}\right).
\end{equation}

\looseness=1 Compared to the Metropolis algorithm, the acceptance probability has the additional factor $q(\mathbf{x}|\mathbf{x}')/q(\mathbf{x}'|\mathbf{x})$. This ratio corrects for the asymmetry of the proposal distribution.

\begin{figure}[ht!]
    \centering
    \includegraphics[width=0.9\textwidth]{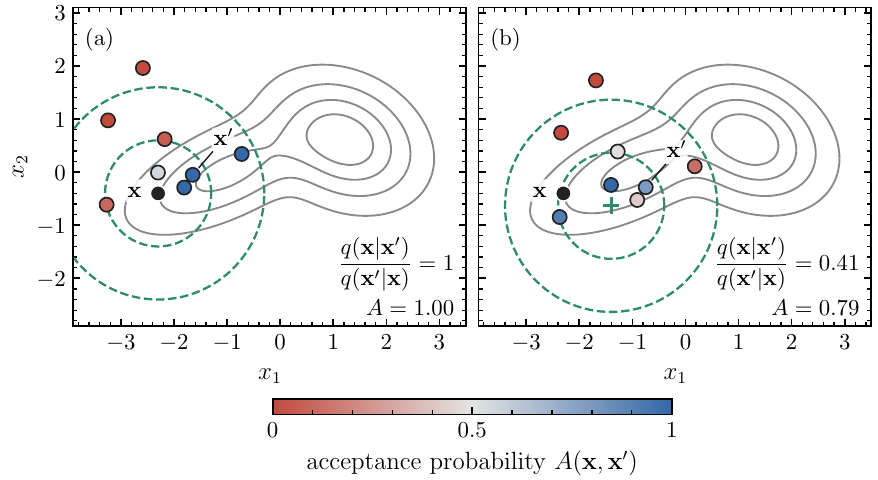}
    \caption{A symmetric and an asymmetric proposal for the same current state $\mathbf{x}$ on a posterior with two lobes. The gray contours show the posterior, and the dashed circles mark the proposal at one and two standard deviations. In panel (a) the proposal is a Gaussian centered at $\mathbf{x}$, and the factor $q(\mathbf{x}|\mathbf{x}')/q(\mathbf{x}'|\mathbf{x})$ equals 1. In panel (b) the Gaussian proposal is centered at the plus sign, a point of higher density than $\mathbf{x}$, and the factor differs from 1. Each dot is a proposed state, and its color shows $A$ including the factor. The factor and $A$ printed in each panel are those of the marked state $\mathbf{x}'$.}
    \label{fig:metropolis_vs_mh}
\end{figure}

To see this intuitively, consider the Columbus explorer again. In the Metropolis algorithm, the explorer was as likely to propose traveling from the Short North to German Village as the reverse. With Metropolis-Hastings, the explorer might be more likely to propose traveling downhill than uphill, or toward the city center than away from it. The factor $q(\mathbf{x}|\mathbf{x}')/q(\mathbf{x}'|\mathbf{x})$ in the acceptance probability compensates for this directional bias. Figure~\ref{fig:metropolis_vs_mh} shows this factor and the acceptance probability for a symmetric and an asymmetric proposal at the same current state.

\paragraph{Checking the Acceptance Ratio}

We check detailed balance, as we did for the Metropolis algorithm. Detailed balance requires that
\begin{equation}
p(\mathbf{x})T(\mathbf{x}, \mathbf{x}') = p(\mathbf{x}')T(\mathbf{x}', \mathbf{x}).
\end{equation}

With the transition probability of the Metropolis-Hastings algorithm, this becomes
\begin{equation}
p(\mathbf{x})q(\mathbf{x}'|\mathbf{x})A(\mathbf{x}, \mathbf{x}') = p(\mathbf{x}')q(\mathbf{x}|\mathbf{x}')A(\mathbf{x}', \mathbf{x}).
\end{equation}

The main difference from the Metropolis algorithm is the proposal distribution. The Metropolis algorithm required $q(\mathbf{x}'|\mathbf{x}) = q(\mathbf{x}|\mathbf{x}')$, so the proposal terms canceled in the acceptance\index{detailed balance} ratio. Here we allow $q(\mathbf{x}'|\mathbf{x}) \neq q(\mathbf{x}|\mathbf{x}')$.

We check detailed balance in two cases. In case 1, the ratio $p(\mathbf{x}')q(\mathbf{x}|\mathbf{x}')/(p(\mathbf{x})q(\mathbf{x}'|\mathbf{x}))$ is at least 1. The two acceptance probabilities are then
\begin{align}
A(\mathbf{x}, \mathbf{x}') &= \min\left(1, \frac{p(\mathbf{x}')q(\mathbf{x}|\mathbf{x}')}{p(\mathbf{x})q(\mathbf{x}'|\mathbf{x})}\right) = 1 \\*
A(\mathbf{x}', \mathbf{x}) &= \min\left(1, \frac{p(\mathbf{x})q(\mathbf{x}'|\mathbf{x})}{p(\mathbf{x}')q(\mathbf{x}|\mathbf{x}')}\right) = \frac{p(\mathbf{x})q(\mathbf{x}'|\mathbf{x})}{p(\mathbf{x}')q(\mathbf{x}|\mathbf{x}')}.
\end{align}

Inserting these into the detailed balance equation gives
\begin{align}
p(\mathbf{x})q(\mathbf{x}'|\mathbf{x}) \cdot 1 &= p(\mathbf{x}')q(\mathbf{x}|\mathbf{x}') \cdot \frac{p(\mathbf{x})q(\mathbf{x}'|\mathbf{x})}{p(\mathbf{x}')q(\mathbf{x}|\mathbf{x}')} \\*
p(\mathbf{x})q(\mathbf{x}'|\mathbf{x}) &= p(\mathbf{x})q(\mathbf{x}'|\mathbf{x}),
\end{align}
which is true.

In case 2, the ratio $p(\mathbf{x}')q(\mathbf{x}|\mathbf{x}')/(p(\mathbf{x})q(\mathbf{x}'|\mathbf{x}))$ is less than 1. The two acceptance probabilities are then
\begin{align}
A(\mathbf{x}, \mathbf{x}') &= \min\left(1, \frac{p(\mathbf{x}')q(\mathbf{x}|\mathbf{x}')}{p(\mathbf{x})q(\mathbf{x}'|\mathbf{x})}\right) = \frac{p(\mathbf{x}')q(\mathbf{x}|\mathbf{x}')}{p(\mathbf{x})q(\mathbf{x}'|\mathbf{x})} \\*
A(\mathbf{x}', \mathbf{x}) &= \min\left(1, \frac{p(\mathbf{x})q(\mathbf{x}'|\mathbf{x})}{p(\mathbf{x}')q(\mathbf{x}|\mathbf{x}')}\right) = 1.
\end{align}

Inserting these into the detailed balance equation gives
\begin{align}
p(\mathbf{x})q(\mathbf{x}'|\mathbf{x}) \cdot \frac{p(\mathbf{x}')q(\mathbf{x}|\mathbf{x}')}{p(\mathbf{x})q(\mathbf{x}'|\mathbf{x})} &= p(\mathbf{x}')q(\mathbf{x}|\mathbf{x}') \cdot 1 \\*
p(\mathbf{x}')q(\mathbf{x}|\mathbf{x}') &= p(\mathbf{x}')q(\mathbf{x}|\mathbf{x}'),
\end{align}
which is also true.

This verifies that the Metropolis-Hastings algorithm satisfies detailed balance with respect to the target posterior distribution. Like the Metropolis algorithm, it also requires the chain to be irreducible, aperiodic and positive recurrent.

In practice, astronomical applications usually satisfy these conditions, with the same caveats as for the Metropolis algorithm. It is generally advisable to use bounded priors. They keep the chain out of regions of the parameter space that are physically meaningless or computationally problematic. We should also avoid regions of zero probability in the parameter space, because they can create barriers that prevent the chain from being irreducible. When these conditions are met, the chain eventually converges to sampling from the target distribution. The long-term visitation pattern of the explorer then matches the population distribution, even though the proposals are directionally biased.

\paragraph{Common Special Cases}

The Metropolis-Hastings algorithm has several special cases that are common in practice. The random walk Metropolis-Hastings algorithm uses $q(\mathbf{x}'|\mathbf{x}) = g(\mathbf{x}' - \mathbf{x})$, where $g$ is a distribution of steps, so it proposes the current position plus a random step. A common choice is a Gaussian $g$ centered at zero, which gives the symmetric proposal of the Metropolis algorithm.

The independence\index{independence sampler} sampler uses a proposal $q(\mathbf{x}'|\mathbf{x}) = g(\mathbf{x}')$ that does not depend on the current state. The proposed state comes from the same distribution $g$ at each step, for example a Gaussian fitted to the posterior. It can be effective when we have good prior knowledge about the target distribution, so that $g$ resembles it.

\looseness=1 A Langevin\index{Langevin proposal} proposal uses gradient information from the target distribution to propose moves toward regions of higher probability. This makes it particularly effective for smooth, differentiable target distributions. The proposal is a Gaussian with step size $\epsilon$, centered at $\mathbf{x} + (\epsilon^2/2)\nabla \ln p(\mathbf{x})$. The gradient of the log target points toward higher probability, so the center lies uphill from $\mathbf{x}$. Panel~(b) of Figure~\ref{fig:metropolis_vs_mh} shows a proposal of this kind.\index{Metropolis-Hastings algorithm|)}

\section{Gibbs Sampling}\label{sec:ch13_gibbs}

Gibbs\index{Gibbs sampling}\index{Gibbs sampling|(} sampling is an MCMC method that updates one parameter, or one block of parameters, at a time. The Metropolis and Metropolis-Hastings algorithms instead treat the parameter vector as a single unit. In many astronomical problems, sampling from the joint posterior $p(\mathbf{x})$ is hard, but the conditional distributions of subsets of parameters are often tractable. We can split a parameter vector $\mathbf{x} = (x_1, x_2, \ldots, x_D)$ into blocks, and the conditional distributions of these blocks often have standard forms that we can sample directly.

Gibbs sampling uses this structure. It breaks the high-dimensional sampling problem into a sequence of lower-dimensional sampling tasks. This is a ``divide and conquer'' approach, and it can improve the efficiency a great deal when the conditional distributions are well-behaved.

When the blocks are single parameters, the procedure is simple. Starting from a state $\mathbf{x}^{(t)} = (x_1^{(t)}, x_2^{(t)}, \ldots, x_D^{(t)})$, one iteration proceeds as follows.
\begin{itemize}
\item Sample $x_1^{(t+1)} \sim p(x_1 | x_2^{(t)}, \ldots, x_D^{(t)})$
\item Sample $x_2^{(t+1)} \sim p(x_2 | x_1^{(t+1)}, x_3^{(t)}, \ldots, x_D^{(t)})$
\item $\vdots$
\item Sample $x_i^{(t+1)} \sim p(x_i | x_1^{(t+1)}, \ldots, x_{i-1}^{(t+1)}, x_{i+1}^{(t)}, \ldots, x_D^{(t)})$
\item $\vdots$
\item Sample $x_D^{(t+1)} \sim p(x_D | x_1^{(t+1)}, \ldots, x_{D-1}^{(t+1)})$
\end{itemize}

\begin{figure}[ht!]
    \centering
    \includegraphics[width=0.7\textwidth]{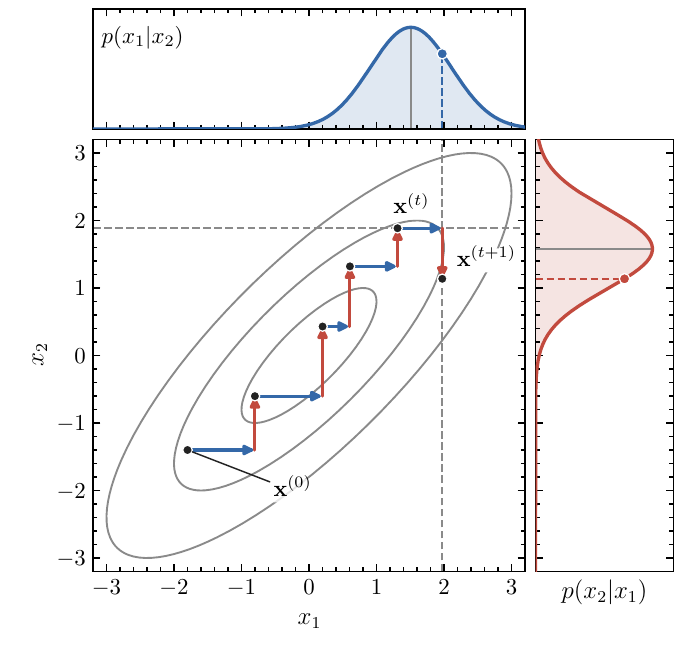}
    \caption{Gibbs sampling for a bivariate Gaussian with unit variances and correlation $\rho = 0.8$. The path starts at $\mathbf{x}^{(0)}$. Blue horizontal steps update $x_1$ given $x_2$, and red vertical steps update $x_2$ given the new $x_1$. The side panels show the two conditional densities of the iteration from $\mathbf{x}^{(t)}$ to $\mathbf{x}^{(t+1)}$. The gray dashed lines in the main panel are the slices on which the conditionals are taken. The solid gray lines mark the means of the conditionals, and the colored dashed lines mark the sampled values.}
    \label{fig:gibbs_sampling}
\end{figure}

Each step samples one parameter $x_i$ directly from its conditional distribution. We use the new value at once when we condition the later parameters. Figure~\ref{fig:gibbs_sampling} shows these steps for two parameters. Sampling $x_1$ given $x_2$ moves the chain horizontally, and sampling $x_2$ given the new $x_1$ moves it vertically, so the chain follows a stair-step path.

Gibbs sampling has an advantage when the conditional distributions have standard forms, such as Gaussian, gamma or beta. Then we can use exact sampling methods, for example the inverse CDF method of Chapter~12 or special random number generators. This avoids proposal distributions and acceptance or rejection steps. If some conditional distributions have no standard form, we can apply an MCMC method to those parameters only and sample the others directly.

\paragraph{Gibbs Sampling as Metropolis-Hastings}

It may seem puzzling that Gibbs sampling does not sample the joint distribution directly and uses only conditional distributions of blocks of parameters. We show that it works because it is a special case of the Metropolis-Hastings algorithm. Unlike the Metropolis algorithm, Gibbs sampling needs no arbitrary proposal distribution such as a Gaussian. It uses the conditional distributions of the target itself as the proposal. When we update the $i$-th component, the proposal is
\begin{equation}
q_i(\mathbf{x}' | \mathbf{x}) = p(x_i' | x_1, \ldots, x_{i-1}, x_{i+1}, \ldots, x_D) \cdot \prod_{j \neq i} \delta(x_j' - x_j),
\end{equation}
where $\delta$ is the Dirac delta function, which keeps all components except the $i$-th unchanged. So we propose new values only for the $i$-th component and draw them directly from its conditional distribution.

This proposal is valid within the Metropolis-Hastings algorithm if the acceptance probability gives detailed balance. We find the acceptance probability of this proposal from the standard Metropolis-Hastings acceptance probability
\begin{equation}
A(\mathbf{x}, \mathbf{x}') = \min\left(1, \frac{p(\mathbf{x}')q(\mathbf{x}|\mathbf{x}')}{p(\mathbf{x})q(\mathbf{x}'|\mathbf{x})}\right).
\end{equation}

We evaluate this acceptance probability for the Gibbs sampler. Since we update only the $i$-th component, we have $x_j' = x_j$ for all $j \neq i$. The product rule of probability gives the joint distribution of $\mathbf{x}$ as
\begin{equation}
p(\mathbf{x}) = p(x_i | x_1, \ldots, x_{i-1}, x_{i+1}, \ldots, x_D) \cdot p(x_1, \ldots, x_{i-1}, x_{i+1}, \ldots, x_D).
\end{equation}
Similarly, for $\mathbf{x}'$ we have
\begin{equation}
p(\mathbf{x}') = p(x_i' | x_1, \ldots, x_{i-1}, x_{i+1}, \ldots, x_D) \cdot p(x_1, \ldots, x_{i-1}, x_{i+1}, \ldots, x_D).
\end{equation}

The second factor is the same in both expressions, because all components except the $i$-th are the same in $\mathbf{x}$ and $\mathbf{x}'$. The proposal for the move from $\mathbf{x}$ to $\mathbf{x}'$ is
\begin{equation}
q(\mathbf{x}'|\mathbf{x}) = p(x_i' | x_1, \ldots, x_{i-1}, x_{i+1}, \ldots, x_D).
\end{equation}
The proposal for the reverse move is
\begin{equation}
q(\mathbf{x}|\mathbf{x}') = p(x_i | x_1, \ldots, x_{i-1}, x_{i+1}, \ldots, x_D).
\end{equation}

We write $\mathbf{x}_{-i}$ for all components of $\mathbf{x}$ except the $i$-th, so $\mathbf{x}_{-i} = (x_1, \ldots,\allowbreak x_{i-1}, x_{i+1}, \ldots,\allowbreak x_D)$. The conditional distributions then take the compact form $p(x_i | \mathbf{x}_{-i})$.

Substituting these expressions into the acceptance probability gives
\begin{align}
A(\mathbf{x}, \mathbf{x}')
&= \min\left(1, \frac{p(\mathbf{x}')q(\mathbf{x}|\mathbf{x}')}{p(\mathbf{x})q(\mathbf{x}'|\mathbf{x})}\right) \\*
&= \min\left(1, \frac{p(x_i' | \mathbf{x}_{-i}) \cdot p(\mathbf{x}_{-i}) \cdot p(x_i | \mathbf{x}_{-i})}{p(x_i | \mathbf{x}_{-i}) \cdot p(\mathbf{x}_{-i}) \cdot p(x_i' | \mathbf{x}_{-i})}\right) \\*
&= \min(1, 1) = 1.
\end{align}

The acceptance probability of Gibbs sampling is 1. Standard Metropolis-Hastings uses an arbitrary proposal distribution, whereas Gibbs sampling draws each parameter from its exact conditional distribution given all the others. No rejection is needed, so Gibbs sampling is often called a ``rejection-free'' MCMC method.

Gibbs sampling is a special case of Metropolis-Hastings. Metropolis-Hastings satisfies detailed balance and, if the chain is ergodic, converges to the target distribution. So Gibbs sampling also converges to the posterior distribution.

\paragraph{Bivariate Gaussian Example}

To see when Gibbs sampling works well and when it works poorly, we look at how the structure of the target distribution affects the efficiency. A bivariate Gaussian distribution is a useful example, because we can derive its conditional distributions analytically. We can also observe directly how the efficiency of Gibbs sampling varies with the correlation structure of the target. We sample from a bivariate Gaussian distribution with zero mean and correlation coefficient $\rho$
\begin{equation}
\begin{pmatrix} x_1 \\ x_2 \end{pmatrix} \sim \mathcal{N}\left(\mathbf{0}, \begin{bmatrix} 1 & \rho \\ \rho & 1 \end{bmatrix}\right).
\end{equation}

Gibbs sampling needs the conditional distributions $p(x_1 | x_2)$ and $p(x_2 | x_1)$. For a bivariate Gaussian distribution, we can apply the standard formula for conditional distributions, which Section~5.9 derives. Suppose we have a joint Gaussian distribution
\begin{equation}
\begin{pmatrix} x_1 \\ x_2 \end{pmatrix} \sim \mathcal{N}\left(\begin{pmatrix} \mu_1 \\ \mu_2 \end{pmatrix}, \begin{bmatrix} \Sigma_{11} & \Sigma_{12} \\ \Sigma_{21} & \Sigma_{22} \end{bmatrix}\right).
\end{equation}

The conditional distribution of $x_1$ given $x_2$ is then
\begin{equation}
p(x_1 | x_2) = \mathcal{N}(\mu_1 + \Sigma_{12}\Sigma_{22}^{-1}(x_2 - \mu_2), \Sigma_{11} - \Sigma_{12}\Sigma_{22}^{-1}\Sigma_{21}).
\end{equation}

In our case, $\mu_1 = \mu_2 = 0$, $\Sigma_{11} = \Sigma_{22} = 1$ and $\Sigma_{12} = \Sigma_{21} = \rho$. Substituting these values gives
\begin{align}
p(x_1 | x_2) &= \mathcal{N}(0 + \rho \cdot 1^{-1} \cdot (x_2 - 0), 1 - \rho \cdot 1^{-1} \cdot \rho) \\*
&= \mathcal{N}(\rho x_2, 1 - \rho^2).
\end{align}

The last row of the display above gives the conditional distribution of $x_1$. By symmetry, the conditional distribution of $x_2$ is
\begin{equation}
p(x_2 | x_1) = \mathcal{N}(\rho x_1, 1-\rho^2).
\end{equation}

These closed-form expressions make the distribution well suited to Gibbs sampling. At each step we sample directly from the conditional distributions and need no rejection step. The sampler alternates between the following two steps.
\begin{enumerate}
\item Sample $x_1^{(t+1)} \sim \mathcal{N}(\rho x_2^{(t)}, 1-\rho^2)$
\item Sample $x_2^{(t+1)} \sim \mathcal{N}(\rho x_1^{(t+1)}, 1-\rho^2)$
\end{enumerate}

We vary the correlation coefficient $\rho$ to see its effect on the efficiency of the sampler. For $\rho = 0$, which means no correlation, the conditional distributions simplify to
\begin{equation}
p(x_1 | x_2) = \mathcal{N}(0, 1) \quad \text{and} \quad p(x_2 | x_1) = \mathcal{N}(0, 1).
\end{equation}

The Gibbs sampler then generates independent samples in each dimension. Each conditional sampling step gives a new value unrelated to the previous samples. The chain moves through the joint distribution with high efficiency. The bivariate Gaussian has circular contours, and the sampler moves freely in both dimensions.

As $|\rho|$ increases, the conditional relationship between $x_1$ and $x_2$ becomes stronger. For high correlation, for example $\rho = 0.95$, the conditional distributions become
\begin{equation}
p(x_1 | x_2) = \mathcal{N}(0.95 x_2, 0.0975) \quad \text{and} \quad p(x_2 | x_1) = \mathcal{N}(0.95 x_1, 0.0975).
\end{equation}

\begin{figure}[ht!]
    \centering
    \includegraphics[width=0.9\textwidth]{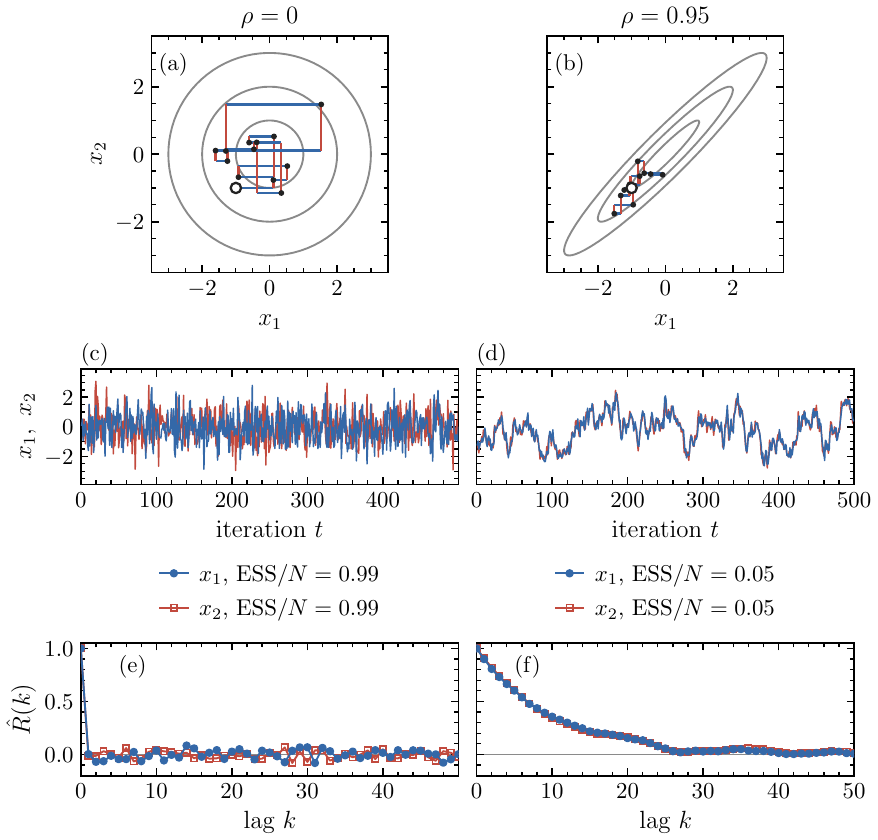}
    \caption{Gibbs sampling for a bivariate Gaussian with correlation $\rho = 0$ (left) and $\rho = 0.95$ (right). Panels (a) and (b) show the first iterations from the open circle. Panels (c) and (d) show the traces of $x_1$ (blue) and $x_2$ (red), and panels (e) and (f) show the estimated autocorrelation functions. $\mathrm{ESS}/N$ is the effective sample size per iteration, and it is much lower for $\rho = 0.95$. The acceptance probability is 1 in both cases.}
    \label{fig:gibbs_correlation}
\end{figure}

The means now follow the other variable closely, and the variance drops to 0.0975, which equals $1-\rho^2$ for $\rho = 0.95$. The successive samples of the Gibbs sampler now have a high autocorrelation. The bivariate Gaussian has a highly elongated elliptical contour, and the sampler moves in a narrow stair-step pattern along this ellipse. The narrow pattern shows how hard it is to explore the major axis of the distribution. Figure~\ref{fig:gibbs_correlation} compares the sampling paths, the chains with their effective sample sizes and the autocorrelation functions for $\rho = 0$ and $\rho = 0.95$.

The case $\rho = 0.95$ shows a limitation of standard Gibbs sampling. It can perform poorly with highly correlated parameters, which are common in astronomical models. The chain may need many iterations to explore the posterior. It then has a high autocorrelation and a low effective sample size, although the acceptance rate is 1. A rejection-free method does not guarantee efficient exploration of the parameter space.

In the Columbus analogy, a Gibbs sampler with highly correlated parameters is an explorer who can move only north-south or east-west. The important roads in the city run diagonally. Each step is valid, but the overall exploration is inefficient.

Correlated parameters lead to several strategies for improving the efficiency of Gibbs sampling. One approach is block\index{Gibbs sampling!block|(} Gibbs sampling, where we update groups of correlated parameters together rather than one at a time. In our bivariate Gaussian example, we could update $x_1$ and $x_2$ together instead of alternating between them. Updating them together samples directly from the joint distribution, which makes our example trivial. The example still shows a general principle, that grouping correlated parameters often lets us sample more efficiently.

Suppose we partition the parameter vector $\mathbf{x}$ into $B$ blocks $(\mathbf{x}_{(1)}, \mathbf{x}_{(2)}, \ldots, \mathbf{x}_{(B)})$. The block Gibbs sampler then proceeds as follows.
\begin{enumerate}
\item Sample $\mathbf{x}_{(1)}^{(t+1)} \sim p(\mathbf{x}_{(1)} | \mathbf{x}_{(2)}^{(t)}, \ldots, \mathbf{x}_{(B)}^{(t)})$
\item Sample $\mathbf{x}_{(2)}^{(t+1)} \sim p(\mathbf{x}_{(2)} | \mathbf{x}_{(1)}^{(t+1)}, \mathbf{x}_{(3)}^{(t)}, \ldots, \mathbf{x}_{(B)}^{(t)})$
\item $\vdots$
\item Sample $\mathbf{x}_{(B)}^{(t+1)} \sim p(\mathbf{x}_{(B)} | \mathbf{x}_{(1)}^{(t+1)}, \ldots, \mathbf{x}_{(B-1)}^{(t+1)})$
\end{enumerate}

Block Gibbs sampling is also a special case of the Metropolis-Hastings algorithm with an acceptance probability of 1. To see this, consider the proposal distribution for updating block $\mathbf{x}_{(i)}$
\begin{equation}
q(\mathbf{x}_{(i)}^{\prime} | \mathbf{x}_{(i)}, \mathbf{x}_{(-i)}) = p(\mathbf{x}_{(i)}^{\prime} | \mathbf{x}_{(-i)}),
\end{equation}
where $\mathbf{x}_{(-i)}$ represents all blocks except the $i$-th. The Metropolis-Hastings acceptance probability, which we write $\alpha$ here, becomes
\begin{align}
\alpha &= \min\left(1, \frac{p(\mathbf{x}_{(i)}^{\prime}, \mathbf{x}_{(-i)})}{p(\mathbf{x}_{(i)}, \mathbf{x}_{(-i)})} \cdot \frac{q(\mathbf{x}_{(i)} | \mathbf{x}_{(i)}^{\prime}, \mathbf{x}_{(-i)})}{q(\mathbf{x}_{(i)}^{\prime} | \mathbf{x}_{(i)}, \mathbf{x}_{(-i)})}\right) \\*
&= \min\left(1, \frac{p(\mathbf{x}_{(i)}^{\prime} | \mathbf{x}_{(-i)}) \cdot p(\mathbf{x}_{(-i)})}{p(\mathbf{x}_{(i)} | \mathbf{x}_{(-i)}) \cdot p(\mathbf{x}_{(-i)})} \cdot \frac{p(\mathbf{x}_{(i)} | \mathbf{x}_{(-i)})}{p(\mathbf{x}_{(i)}^{\prime} | \mathbf{x}_{(-i)})}\right) \\*
&= \min(1, 1) = 1.
\end{align}

So the convergence result of the Metropolis-Hastings algorithm carries over to block\index{Gibbs sampling!block|)} Gibbs sampling. This generalizes the single-component case to blocks of arbitrary size. The proposal is the exact conditional distribution, so the acceptance ratio simplifies to 1 and the algorithm is rejection-free.

\paragraph{Hierarchical Models}

The\index{exoplanets|(} conditional sampling of Gibbs sampling is well suited to hierarchical Bayesian models. In these models the parameters are organized in layers, and higher-level parameters, the hyperparameters, determine the distributions of lower-level parameters. These hyperparameters are inferred from the data, unlike those of Section~7.8, such as the learning rate, which we set before training. For example, consider a model for a population of exoplanets, where we have observations of several planetary systems. At the lowest level we have parameters for individual planets, such as masses, radii and orbital parameters. At a higher level, hyperparameters describe the population-wide distributions of these properties, such as the mean and variance of planetary masses over all systems.

\looseness=1 The conditional structure of such models is often simple. Given the hyperparameters, the parameters of the individual planets are conditionally independent of each other, so we can update each planet without considering the others. Conditionally independent means that, once we know the hyperparameters, the parameters of one planet tell us nothing about those of another. Given the individual planet parameters, the hyperparameters typically have simple conditional distributions that depend only on the collection of individual parameters.

This structure is ideal for Gibbs sampling, which can explore the joint posterior efficiently by alternating between two steps. First, we update all individual planet parameters, possibly in parallel, because they are conditionally independent given the hyperparameters. Then we update the population hyperparameters from the current values of all individual parameters.

The same structure appears in many astronomical applications. In stellar population modeling, cluster-wide hyperparameters determine the properties of individual stars, creating a hierarchy. In galaxy clustering studies, the properties of individual galaxies depend on their host dark matter halos, forming another hierarchy. Cosmological parameter estimation often combines different probes in a hierarchical\index{hierarchical model} model, where Gibbs sampling can handle the complex parameter space efficiently. In these cases Gibbs sampling uses the conditional\index{conditional independence} independence of the model. It is often much more efficient than generic Metropolis-Hastings, which treats all parameters as a single block.\index{exoplanets|)}

\paragraph{Software and Further Methods}

Astronomers rarely code these samplers from scratch, because well-tested packages implement them. Some packages run an ensemble\index{ensemble samplers} of walkers, each of which traces its own chain. Each walker proposes steps from the positions of the other walkers, so these packages need little tuning.

\looseness=1 Hamiltonian\index{Hamiltonian Monte Carlo} Monte Carlo treats the negative log posterior as a potential energy and gives each parameter a momentum. The gradient of the log posterior acts as a force, as in classical mechanics. We let this system move for a short time and propose its end point, which a Metropolis-Hastings step accepts or rejects. Because the motion nearly conserves energy, the proposal can lie far from the start and still have a high acceptance probability. Nested sampling computes the evidence and handles posteriors with several modes.

Previously intractable problems have become routine applications of MCMC. Examples are characterizing exoplanet populations, inferring cosmological parameters from several kinds of observations and modeling stellar evolution with uncertain physics. MCMC methods supply the computations behind many modeling techniques in statistics, machine learning, physics and other fields where complex probability distributions arise. The journey from the simple grid methods of Chapter~12 to the MCMC algorithms of this chapter shows a recurring theme. As problems become more complex, we must often abandon exact solutions for approximate methods that scale.

Some models still allow analytical calculations, and these are often preferable. Chapter~14 treats Gaussian processes, an advanced technique that often gives analytical solutions for the posterior predictions and the marginal likelihood (evidence). Analytical and numerical approaches each have their place in the toolkit of the modern statistician.\index{Gibbs sampling|)}\index{Markov chain Monte Carlo (MCMC)|)}

\section{Tutorials: Deconvolving Noisy Data}\label{sec:ch13_tutorial}

Most posteriors in astronomy have no closed form, so we sample them with Markov chain Monte Carlo (MCMC). The first of two tutorials builds the tools of this chapter, and the second uses them on noisy\index{deconvolution} stellar masses. The full tutorials can be found at \url{https://statml.org/} as 13a and 13b.

\begin{figure}[t!]
    \centering
    \includegraphics[width=0.95\textwidth]{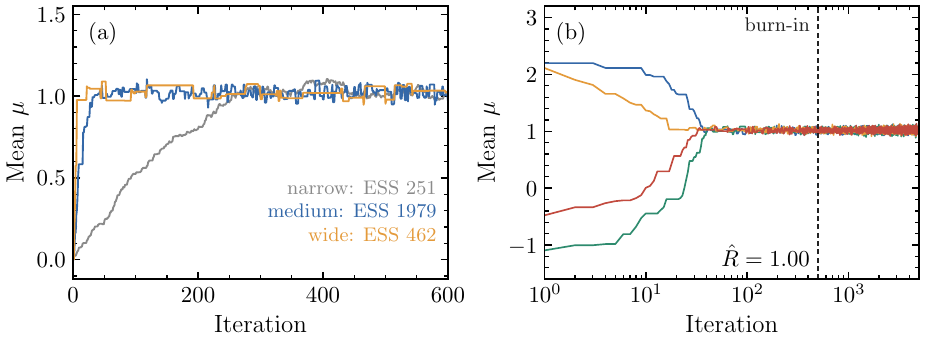}
    \caption{The first tutorial introduces MCMC on a toy posterior, the mean $\mu$ of simulated measurements. Panel~(a) shows the start of three Metropolis chains with a narrow, a medium and a wide proposal, labeled with their effective sample size (ESS). Panel~(b) shows four chains from different starts, with the burn-in (dashed) and the Gelman-Rubin statistic $\hat R$.}
    \label{fig:ch13_tutorial_mcmc}
\end{figure}

The first tutorial implements the Metropolis algorithm together with the checks that decide whether a chain can be trusted (Figure~\ref{fig:ch13_tutorial_mcmc}). The proposal width sets how fast the chain moves, since a narrow proposal takes tiny steps and a wide one is mostly rejected (Section~\ref{sec:ch13_proposal}). Even a well-tuned chain gives correlated samples, so the effective sample size counts how many independent samples it is worth (Section~\ref{sec:ch13_ess}). The chain also starts away from the posterior, so we discard its first steps as burn-in. To check the burn-in, the Geweke test compares parts of one chain and the Gelman-Rubin test compares chains from different starts (Section~\ref{sec:ch13_burnin}). Finally, Gibbs sampling updates one parameter at a time without a proposal to tune, but it slows down when the parameters are correlated (Section~\ref{sec:ch13_gibbs}).

\begin{figure}[t!]
    \centering
    \includegraphics[width=0.95\textwidth]{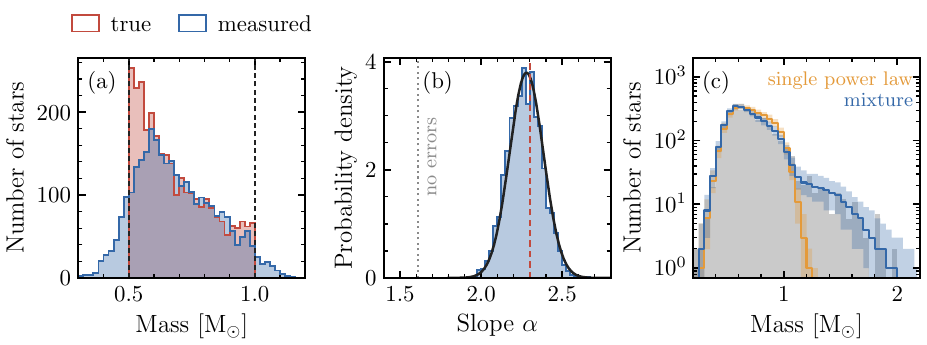}
    \caption{The second tutorial infers the slope of the stellar mass function from noisy masses. In panel~(a) the errors smear the sharp limits of the true masses (dashed lines), and some measured masses fall outside them. In panel~(b) the posterior that models the errors contains the true slope (dashed), and a fit without them finds a much shallower one (dotted). The histogram comes from a Metropolis chain and the curve from a direct numerical integral. In panel~(c) some stars are unresolved binaries. Catalogs simulated from a single power law (orange) miss the high-mass tail of the data, and those from a mixture (blue) reproduce it.}
    \label{fig:ch13_tutorial_deconv}
\end{figure}

The second tutorial asks how measurement errors bias the study of a population. It simulates stellar masses, each with its own error, and fits the slope of their mass\index{stellar mass function} function, a power law with sharp limits. The errors scatter each mass up or down, so the measured masses spread past both limits of the power law. A fit that treats them as true sees a wider, flatter distribution and finds a much shallower slope (Figure~\ref{fig:ch13_tutorial_deconv}a and b). The errors bias the slope much as input errors attenuate a regression slope (Section~\ref{sec:ch6_attenuation}).

To avoid this bias, the likelihood must treat the true mass of each star as unknown. If we knew it, the likelihood would be the power law at that mass. Since we do not, we consider every true mass that could have produced the measurement. Each candidate counts in proportion to how common the power law makes it and how likely its error is to give the measured value. The likelihood of a star is this average over its possible true masses.

The measurement already tells us roughly where the true mass lies, so importance sampling suits this average (Section~\ref{sec:ch12_importance}). It draws possible true masses near the measurement and weights each by how common the power law makes it. MCMC then varies the slope, and at each step the averaged likelihood scores how well that slope explains the measured masses. Together the two account for the errors that the naive fit ignored, and the posterior contains the true slope (Figure~\ref{fig:ch13_tutorial_deconv}b).

This model assumes that every star follows the power law. Real catalogs also contain unresolved binaries, which look like single stars with too high a mass. With them in the data, the chain still converges, but to a narrow posterior far below the true slope. A well-behaved chain does not guarantee a correct answer, because here the model itself is wrong. A posterior predictive check (Section~\ref{sec:posterior_predictive}) catches this, since catalogs simulated from the posterior lack the high-mass tail of the data (Figure~\ref{fig:ch13_tutorial_deconv}c). A mixture model (Section~\ref{sec:ch11_gmm}) of the power law and a Gaussian for the binaries passes the check and recovers the slope.

The same approach applies to any catalog whose errors differ from object to object. Writing the likelihood in terms of the unobserved true values recovers the distribution before the errors blurred it. A posterior predictive check then tests the model itself.

\paragraph{Further Reading} Metropolis et~al.\ (1953) introduced the algorithm in calculations of the equation of state, and Hastings (1970) generalized it to asymmetric proposals. Tierney (1994) treats Markov chains for posterior distributions. Roberts and Smith (1994) give simple conditions for the convergence of the Gibbs sampler and the Metropolis-Hastings algorithm. Geman and Geman (1984) introduced Gibbs sampling in image restoration, and Gelfand and Smith (1990) showed that it can estimate marginal densities. Christensen et~al.\ (2001) proposed MCMC for cosmological parameters from the cosmic microwave background, and Lewis and Bridle (2002) wrote the CosmoMC package. Verde et~al.\ (2003) describe the MCMC analysis of the WMAP microwave background data, and Tegmark et~al.\ (2004) add SDSS galaxy clustering. Ford (2005) and Gregory (2005a) used MCMC for the orbits of extrasolar planets. Gelman and Rubin (1992) and Geweke (1992) proposed convergence diagnostics, and Cowles and Carlin (1996) compared them. Skilling (2006) presented nested sampling, and Feroz and Hobson (2008) extended it to multimodal distributions. Foreman-Mackey et~al.\ (2013) describe emcee, a Python sampler that evolves many chains together and is widely used in astronomy. Abril-Pla et~al.\ (2023) describe PyMC, a Python package for Bayesian inference. Gilks et~al.\ (1996), Robert and Casella (1999) and Liu (2001) are books on Markov chain and Monte Carlo methods. Sharma (2017) reviews MCMC in astronomy, and Hogg and Foreman-Mackey (2018) give practical advice.

\makeatletter\@clubpenalty=150\makeatother \clubpenalty=150 \widowpenalty=150

\chapter{Gaussian Processes}

\makeatletter\@clubpenalty=10000\makeatother \clubpenalty=10000 \widowpenalty=10000

\looseness=1 Consider\index{Gaussian process (GP)|(} the modeling of stellar\index{stellar variability} variability from photometric time series. A linear model can describe it if we design features such as polynomial terms, Fourier components, or physical basis functions. But how do we choose the right features? A star can show pulsations on short timescales, rotational modulation on intermediate timescales, and long-term evolutionary changes. Designing appropriate features for such multi-timescale behavior is very difficult.

We could use Markov chain Monte Carlo (MCMC, Chapter~13) instead, to fit a complex parametric model with several physical processes. Such a model needs many parameters for the different types of variability. MCMC sampling is then expensive and can be unstable. We face the bias-variance tradeoff of Section~\ref{sec:cross_validation}, which balances the systematic error of a rigid model against the variance of a flexible model. Simple models may be too rigid, while complex models may overfit or become intractable.

Gaussian\index{Gaussian process (GP)} processes (GPs)\index{GP|see{Gaussian process}} offer a different approach. Instead of defining a parametric model explicitly, we state our beliefs about the function through a kernel function. Such beliefs concern its smoothness, periodicity and characteristic length scales. The data then determine the functional form within these constraints. A GP does not force a choice between simplicity and flexibility, because the complexity of the model emerges from the structure of the data. This suits astronomical problems in which we understand the underlying processes qualitatively but find them difficult to model explicitly.

For stellar variability, we might expect smooth changes on certain timescales combined with periodic modulation, but we do not know the exact functional form. In exoplanet transit analysis, the intrinsic variability of the star creates complex patterns that resist simple parameterization. Quasar light curves show stochastic variations driven by chaotic accretion processes, which would be very difficult to model explicitly. The multiscale structure of galaxy formation is equally difficult to model.

\looseness=1 GPs handle such cases with kernels, which are functions that measure the similarity between data points. To predict the brightness of a star at a given time, we use observations from nearby times. We weight them by their similarity, as measured by the kernel.

\begin{figure}[t]
    \centering
    \includegraphics[width=0.85\textwidth]{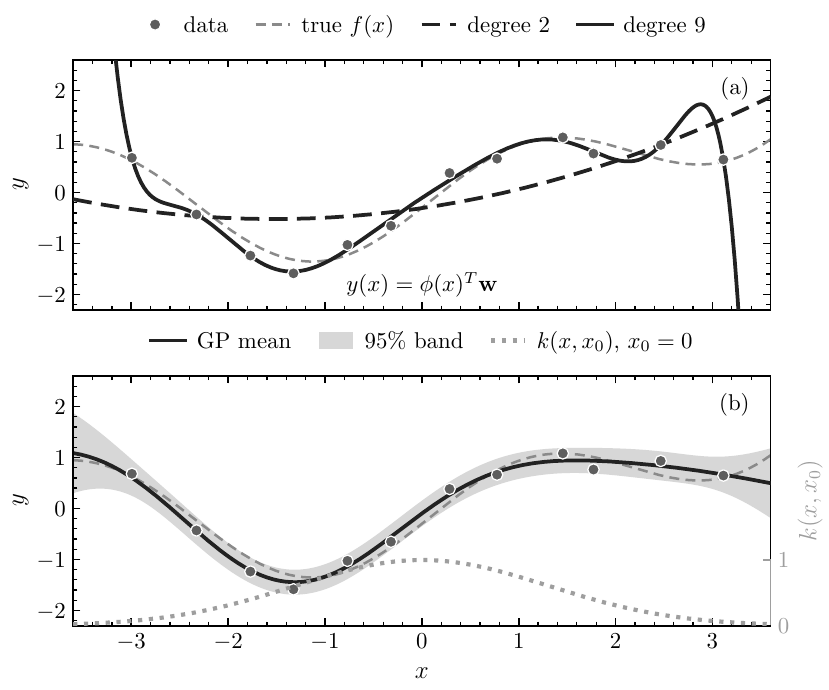}
    \caption{Regression with explicit features and with a Gaussian process for the same noisy training points. Panel~(a) shows least-squares fits with the polynomial features $\boldsymbol{\phi}(x)$ of a low and a high degree. The low-degree fit underfits, and the high-degree fit follows the noise and runs off the plot outside the data. Panel~(b) shows the GP mean and its 95 percent band for a kernel that decreases smoothly with the distance between two inputs. The band excludes the observation noise. The right axis shows this kernel $k(x, x_0)$ for the reference point $x_0 = 0$, scaled to a maximum of 1.}
    \label{fig:gaussian_processes_comparison}
\end{figure}

GPs have several practical advantages for astronomical applications. They follow the observations where these are dense, and they give a wide uncertainty where data are sparse. The predicted uncertainty of a new observation includes both the measurement noise and the uncertainty of the model. They also remove the need for explicit feature engineering, because we specify similarity through kernels instead of designing transformations of our inputs. Figure~\ref{fig:gaussian_processes_comparison} compares a fit with explicit features and a GP fit with its uncertainty band.

We can derive GPs in two ways. The kernel trick extends linear regression to feature spaces of possibly infinite dimension. The function-space view treats a GP as a distribution over functions instead of over parameters. Each view gives its own insight into why GPs work.

The chapter has four parts. Sections~14.1 to 14.5 derive GP regression with the kernel trick and compare it with Bayesian linear regression. Sections~14.6 and 14.7 derive the same equations in function space. Sections~14.8 to 14.10 show how to compute predictions, fit the kernel hyperparameters and choose a kernel. Sections~14.11 to 14.13 extend GPs to classification. Readers who prefer the function-space view can read Sections~14.1, 14.3 and 14.6 to 14.13 first and the rest later.

\section{The Kernel Trick}

The\index{kernel trick|(} kernel\index{kernel trick} trick underlies many machine learning methods, among them Gaussian processes. We first meet kernels in a familiar problem from statistics.

\looseness=1 Consider the estimation of a probability distribution from data. In many astronomical applications, the underlying distribution does not follow a standard parametric form. Measurements of stellar masses, galaxy luminosities, or exoplanet orbital periods can have complex, multimodal, or skewed distributions that resist simple mathematical description.

The traditional approaches are parametric models and histograms. Parametric distributions such as Gaussians work well when the data follow familiar patterns, but they fail when the true distribution has a complex shape. Histograms are flexible, but they are highly sensitive to the choice of bin width. They also give discontinuous estimates, which represent smooth underlying processes poorly.

Kernel\index{kernel density estimation} density estimation provides an alternative that should be familiar to astronomers. It is a non-parametric method that estimates a probability distribution without assuming a specific functional form. We place a kernel function, typically a Gaussian, at each observed data point and sum their contributions. For a one-dimensional $x$, the estimate is
\begin{equation}
p(x) = \frac{1}{N} \sum_{i=1}^N \frac{1}{(2\pi h^2)^{1/2}} \exp\left(-\frac{\|x - x_i\|^2}{2h^2}\right), \label{eq:kde}
\end{equation}where $N$ is the number of data points. The bandwidth $h$ controls the width of each kernel, which sets how much influence each data point has on its surroundings.

\begin{figure}[ht!]
    \centering
    \includegraphics[width=0.85\textwidth]{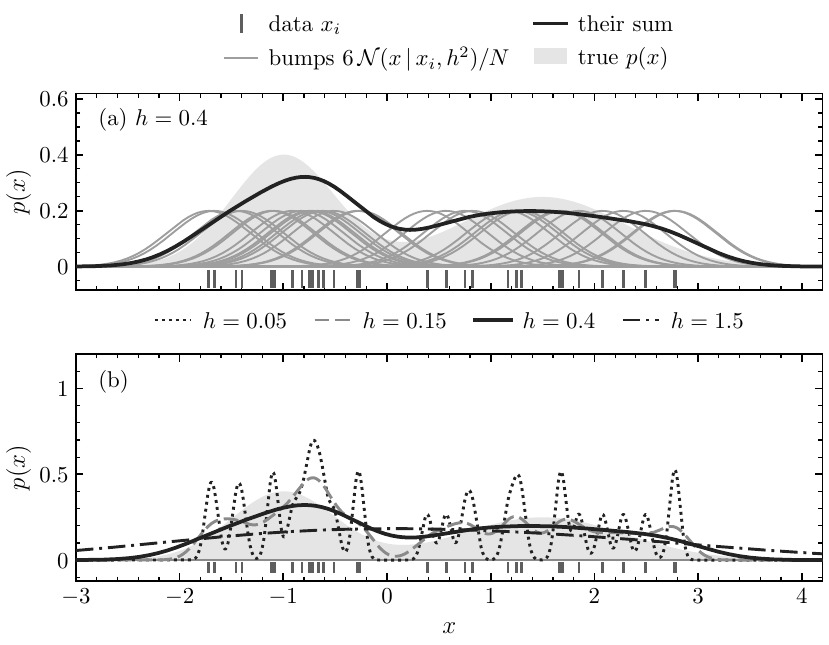}
    \caption{Kernel density estimation for a sample from a bimodal density. Panel~(a) shows the estimate for $h = 0.4$ as the sum of one Gaussian bump per point. Panel~(b) shows the estimates for four bandwidths $h$. The smallest bandwidth gives a spiky estimate with high variance, and the largest one smooths away the two modes. For this sample, $h = 0.4$ has the smallest squared error against the true density.}
    \label{fig:kernel_density_estimation}
\end{figure}

Each data point contributes a ``bump'' to the overall density, and we specify no parametric form for $p(x)$. The only parameter we need to specify is $h$, which controls the smoothness of the estimate. A small $h$ gives a spiky distribution that may overfit to noise. A larger $h$ gives a smoother distribution that may miss important local variations. This choice of $h$ is another example of the bias-variance tradeoff. Figure~\ref{fig:kernel_density_estimation} shows the estimate for four values of $h$.

The same principle appears in N-body simulations and in galaxy density mapping. When we calculate gravitational forces or smooth density fields, kernels distribute the mass or influence of each particle over a finite region. This gives continuous approximations to discrete systems.

The kernel idea extends to regression. In regression we have paired data, with an output value $y_i$ for each input $x_i$. This pairing lets us use similarity in a more directed way, because it connects input similarities to output predictions.

We often assume that similar inputs should produce similar outputs. In astronomy, measurements taken at similar times often show similar behavior. Observations from nearby spatial regions tend to be correlated. Objects with similar physical properties frequently show comparable observational signatures.

To apply this similarity principle, we predict the luminosity of a quasar at time $t_*$ from the luminosities observed at nearby times. Their weighted\index{kernel regression} average is
\begin{equation}
y(t_*) = \frac{\sum_{i=1}^N w_i y_i}{\sum_{i=1}^N w_i},
\end{equation}
where $w_i$ is a weight that decreases with the temporal distance between $t_*$ and $t_i$. We take a Gaussian bump in this distance, as in Equation~\ref{eq:kde}.

\begin{figure}[ht!]
    \centering
    \includegraphics[width=0.85\textwidth]{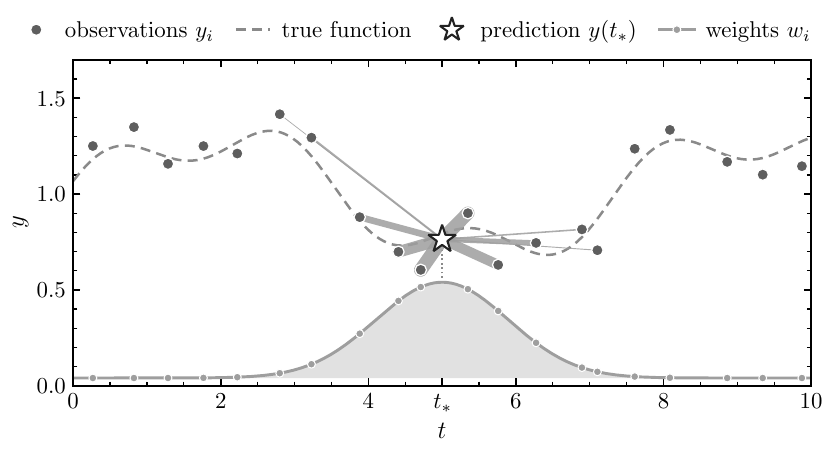}
    \caption{Kernel regression as a weighted average. Lines join the prediction $y(t_*)$ to the observations. The width of each line is proportional to the share $w_i / \sum_j w_j$ of the total weight. Observations with a very small share have no visible line. The weights $w_i$ follow a Gaussian kernel centered at $t_*$, drawn at the bottom on its own vertical scale.}
    \label{fig:kernel_regression}
\end{figure}

In this section, $t$ is a time, $y_i$ is the luminosity at time $t_i$, and $w_i$ is the weight of observation $i$. From Section~14.2 on, the input is called $x$, the target is $t_i$, and $\mathbf{w}$ holds the weights of the features. The light-curve kernels later in this chapter use $t$ for the time again. Figure~\ref{fig:kernel_regression} shows the weights for one prediction.

This approach assumes that the luminosity is continuous in time. We do not model the accretion physics that drives quasar variability, and we do not solve the equations of magnetohydrodynamics around black holes. We use only the empirical pattern that nearby times tend to have similar brightness.

A related kernel-based form is
\begin{equation}
f(x_*) = \sum_{i=1}^N k(x_*, x_i) \alpha_i, \label{eq:kernel_form}
\end{equation}
where $k(x_*, x_i)$ measures the similarity between the prediction point and training point $i$. We can picture each training point as the center of a bump, and the coefficient $\alpha_i$ sets the height of that bump. The data fix these coefficients, and Section~14.2 shows how. The kernel function encodes our beliefs about what makes inputs similar.

\looseness=1 Instead of asking which mathematical function describes the relationship, we ask how to measure the similarity between observations. This question is often easier to answer, because we can define a meaningful similarity measure even when we cannot easily specify an explicit functional form. For stellar light curves, observations close in time are more similar, and for galaxy surveys the similarity can depend on the angular separation. We can choose the similarity measure that suits the domain, so the approach applies to many astronomical problems.

\section{Kernelizing Linear Regression}

\looseness=1 In Section~14.1 we predicted from similarities between inputs. Linear\index{linear regression!kernelized} regression is a method we understand well, and it makes predictions through features. We now rewrite its prediction so that the features appear only through inner products of feature vectors, without designing or computing the features explicitly. An inner product is large when two feature vectors point in similar directions, so it measures the similarity between two inputs. This transformation is called kernelizing. It joins our intuitive similarity-based approach with the statistical tools for parametric models.

An input $x$ is what we know about one observation, for example the time of a measurement. It can be one number, a scalar, or a list of numbers, a vector. We write the input in plain type, even when it is a vector. The feature\index{feature map} map $\boldsymbol{\phi}$ turns $x$ into a vector $\boldsymbol{\phi}(x)$ of $M$ numbers, for example $\boldsymbol{\phi}(x) = (1, x, x^2)^T$ with $M = 3$ for a quadratic fit. In linear regression, the output $y$ is one number, a linear function of these features,
\begin{equation}
y(x) = \boldsymbol{\phi}(x)^T \mathbf{w},
\end{equation}
where $\mathbf{w} \in \mathbb{R}^M$ holds the weights of the $M$ features.

The training data are $N$ inputs $x_1, \ldots, x_N$ with $N$ measured targets $t_1, \ldots, t_N$, which form the vector $\mathbf{t} \in \mathbb{R}^N$. Row $i$ of the design matrix $\mathbf{\Phi} \in \mathbb{R}^{N\times M}$ is $\boldsymbol{\phi}(x_i)^T$. A new input is called $x_*$.

As we derived in Section~4.4 (Equation~\ref{eq:wml_weighted}), the maximum likelihood solution for heteroscedastic noise is
\begin{equation}
\mathbf{w}^* = (\mathbf{\Phi}^T \mathbf{S}^{-1} \mathbf{\Phi})^{-1} \mathbf{\Phi}^T \mathbf{S}^{-1} \mathbf{t},
\end{equation}
where $\mathbf{S}$ is the noise covariance matrix. Heteroscedastic noise means that each data point has its own noise variance $\sigma_i^2$, and $\mathbf{S}$ is diagonal with these variances on its diagonal. This weighted least squares solution gives more weight to reliable measurements and less influence to noisy data points.

To keep the algebra short, we take the homoscedastic case with constant noise variance, $\mathbf{S} = \sigma_n^2\mathrm{Id}$. The heteroscedastic case follows the same principles but needs more complex matrix manipulations. The prediction at a new input $x_*$ is then
\begin{equation}
y(x_*) = \boldsymbol{\phi}(x_*)^T \mathbf{w}^* = \boldsymbol{\phi}(x_*)^T (\mathbf{\Phi}^T \mathbf{\Phi})^{-1} \mathbf{\Phi}^T \mathbf{t}.
\end{equation}

This formula inverts the $M \times M$ matrix $\mathbf{\Phi}^T\mathbf{\Phi}$, whose rank is at most $N$. The inverse does not exist when $M > N$. A Gaussian prior on the weights removes this problem, because it adds a ridge penalty (Section~5.6). For a prior variance of 1 on each weight, the penalty strength is $\sigma_n^2$. The ridge estimate is
\begin{equation}
\mathbf{w}^* = (\mathbf{\Phi}^T \mathbf{\Phi} + \sigma_n^2\mathrm{Id})^{-1} \mathbf{\Phi}^T \mathbf{t}.
\end{equation}

The prediction at $x_*$ is then
\begin{equation}
y(x_*) = \boldsymbol{\phi}(x_*)^T (\mathbf{\Phi}^T \mathbf{\Phi} + \sigma_n^2\mathrm{Id})^{-1} \mathbf{\Phi}^T \mathbf{t}.
\end{equation}

\paragraph{A Matrix Identity}

The ridge prediction contains the $M \times M$ matrix $\mathbf{\Phi}^T\mathbf{\Phi}$, which sums over the $N$ training points and so compares features with each other. The $N \times N$ matrix $\mathbf{\Phi}\mathbf{\Phi}^T$ sums over the $M$ features instead. Its entry $(i,j)$ is the inner product $\boldsymbol{\phi}(x_i)^T\boldsymbol{\phi}(x_j)$, the similarity of inputs $i$ and $j$. An identity trades the $M \times M$ inverse for an $N \times N$ inverse and brings these similarities into the prediction. It says that
\begin{equation}
(\mathbf{\Phi}^T \mathbf{\Phi} + \sigma_n^2\mathrm{Id}_M)^{-1} \mathbf{\Phi}^T = \mathbf{\Phi}^T(\mathbf{\Phi}\mathbf{\Phi}^T + \sigma_n^2\mathrm{Id}_N)^{-1}.
\end{equation}
Here $\mathrm{Id}_M$ and $\mathrm{Id}_N$ are identity matrices of sizes $M$ and $N$, and both inverses exist for $\sigma_n^2 > 0$. In the other equations, $\mathrm{Id}$ has the size that the equation needs.

To prove the identity, we first multiply out
\begin{equation}
\mathbf{\Phi}^T(\mathbf{\Phi}\mathbf{\Phi}^T + \sigma_n^2\mathrm{Id}_N) = \mathbf{\Phi}^T\mathbf{\Phi}\mathbf{\Phi}^T + \sigma_n^2\mathbf{\Phi}^T = (\mathbf{\Phi}^T \mathbf{\Phi} + \sigma_n^2\mathrm{Id}_M)\mathbf{\Phi}^T.
\end{equation}
We then multiply this equation by $(\mathbf{\Phi}^T \mathbf{\Phi} + \sigma_n^2\mathrm{Id}_M)^{-1}$ from the left and by $(\mathbf{\Phi}\mathbf{\Phi}^T + \sigma_n^2\mathrm{Id}_N)^{-1}$ from the right. This gives the identity.

For one training point, $N = 1$, the matrix $\mathbf{\Phi}\mathbf{\Phi}^T$ is the single number $\boldsymbol{\phi}(x_1)^T\boldsymbol{\phi}(x_1)$. The identity then says that $(\mathbf{\Phi}^T\mathbf{\Phi} + \sigma_n^2\mathrm{Id}_M)^{-1}\boldsymbol{\phi}(x_1) = \boldsymbol{\phi}(x_1)/(\boldsymbol{\phi}(x_1)^T\boldsymbol{\phi}(x_1) + \sigma_n^2)$. The $M \times M$ inverse becomes a division by one number.

\paragraph{The Kernel Function}

With the identity, the ridge prediction needs the features only through inner products. Besides the matrix $\mathbf{\Phi}\mathbf{\Phi}^T$, it needs the vector $\boldsymbol{\phi}(x_*)^T\mathbf{\Phi}^T$ of inner products between the test point and the training points. We write each such inner product with the kernel\index{kernel function} function, defined as
\begin{equation}
k(x, x') = \boldsymbol{\phi}(x)^T \boldsymbol{\phi}(x').
\end{equation}

\looseness=1 The kernel measures the similarity between two inputs through their inner product in feature space. We can often compute a kernel without constructing the feature vectors $\boldsymbol{\phi}(x)$ explicitly. Working with kernels in place of explicit features is the kernel trick.

For $N$ training points, we define the $N \times N$ kernel\index{kernel matrix} matrix
\begin{equation}
\mathbf{K} = \mathbf{\Phi}\mathbf{\Phi}^T \in \mathbb{R}^{N \times N},
\end{equation}
where $K_{ij} = k(x_i, x_j)$ is the similarity between points $i$ and $j$.

We define the kernel vector of similarities between $x_*$ and all training points as
\begin{equation}
\mathbf{k}_* = [k(x_1, x_*), k(x_2, x_*), \ldots, k(x_N, x_*)]^T = \mathbf{\Phi}\boldsymbol{\phi}(x_*).
\end{equation}

Substituting the identity and the kernel vector $\mathbf{k}_*$ into the ridge prediction gives
\begin{align}
y(x_*) &= \boldsymbol{\phi}(x_*)^T (\mathbf{\Phi}^T \mathbf{\Phi} + \sigma_n^2\mathrm{Id})^{-1} \mathbf{\Phi}^T \mathbf{t} \\*
&= \boldsymbol{\phi}(x_*)^T \mathbf{\Phi}^T(\mathbf{K} + \sigma_n^2\mathrm{Id})^{-1} \mathbf{t} \\*
&= \mathbf{k}_*^T (\mathbf{K} + \sigma_n^2\mathrm{Id})^{-1} \mathbf{t}.
\end{align}

We have kernelized linear regression. The prediction for a new point $x_*$ is now a weighted sum of the training targets. The weights are determined by the similarities between $x_*$ and the training points, modulated by the inverse of the kernel matrix plus noise, $(\mathbf{K} + \sigma_n^2\mathrm{Id})^{-1}$. With $\boldsymbol{\alpha} = (\mathbf{K} + \sigma_n^2\mathrm{Id})^{-1}\mathbf{t}$, the prediction is $\sum_{i=1}^N k(x_*, x_i)\alpha_i$, which has the form of Equation~\ref{eq:kernel_form}. The data fix the coefficients $\alpha_i$.

\looseness=1 The derivation used $\boldsymbol{\phi}$, because $\mathbf{K}$ and $\mathbf{k}_*$ came from $\mathbf{\Phi}$. The final prediction needs only the numbers $k(x_i, x_j)$ and $k(x_i, x_*)$ and the targets $\mathbf{t}$. If a formula gives these numbers, we do not have to write down $\boldsymbol{\phi}(x)$, and we work directly with similarity measures between data points. The radial basis function (RBF) kernel of the next section is such a case.

The kernel form also has a computational advantage. In the feature formulation, we invert an $M \times M$ matrix, where $M$ is the number of features. With kernels, we invert an $N \times N$ matrix, where $N$ is the number of training points. The kernel approach is preferable for flexible models with many features ($M \gg N$), and it is necessary when $M$ is infinite.

This kernel form of linear regression leads to Gaussian processes. First we need to know which functions are valid kernels.\index{kernel trick|)}

\section{Mercer's Theorem and Valid Kernels}\label{sec:ch14_kernels}

In\index{Gaussian process (GP)!kernels|(}\index{kernel function|(} Section~14.2 we started from a feature mapping $\boldsymbol{\phi}(x)$ and derived the kernel $k(x,x') = \boldsymbol{\phi}(x)^T \boldsymbol{\phi}(x')$, the inner product of two feature vectors. Kernel methods are most useful in the opposite direction, where we define kernels directly without specifying features. This raises the question of when a function $k(x,x')$ is a valid kernel, that is, when it equals $\boldsymbol{\phi}(x)^T \boldsymbol{\phi}(x')$ for some feature mapping $\boldsymbol{\phi}$. Not all similarity functions can be expressed as a dot product in some feature space.

\looseness=1 Mercer's\index{Mercer's theorem} theorem and its extensions answer this question. They state that a function $k(x,x')$ is a valid kernel if and only if it has two properties. The first is symmetry, $k(x,x') = k(x',x)$ for all $x$ and $x'$ in the input space $\mathcal{X}$. The second is that, for any finite set of points $\{x_1, ..., x_N\}$, the kernel matrix $\mathbf{K}$ with entries $K_{ij} = k(x_i, x_j)$ is positive\index{positive semi-definite matrix} semi-definite. In other words, $\mathbf{u}^T \mathbf{K} \mathbf{u} \geq 0$ for any vector $\mathbf{u} \in \mathbb{R}^N$. A matrix is positive definite when this inequality is strict for nonzero vectors $\mathbf{u}$, and such a matrix has an inverse.

If $k$ has these properties, there exists a Hilbert space $\mathcal{H}$. This is a vector space with an inner product $\langle \cdot, \cdot \rangle_{\mathcal{H}}$, and it may have infinite dimension. There also exists a mapping $\boldsymbol{\phi}: \mathcal{X} \rightarrow \mathcal{H}$ such that for all $x, x' \in \mathcal{X}$
\begin{equation}
k(x, x') = \langle \boldsymbol{\phi}(x), \boldsymbol{\phi}(x') \rangle_{\mathcal{H}}.
\end{equation}

The proof of Mercer's theorem lies beyond the scope of this chapter. The idea is that a positive semi-definite kernel can be represented through its eigendecomposition. A symmetric matrix can be diagonalized as $\mathbf{A} = \mathbf{Q}\mathbf{\Lambda}\mathbf{Q}^T$, where $\mathbf{\Lambda}$ contains the eigenvalues and $\mathbf{Q}$ the eigenvectors. In the same way, a kernel function can be decomposed as $k(x,x') = \sum_{i=1}^{\infty} \lambda_i \psi_i(x) \psi_i(x')$, where $\lambda_i$ are the eigenvalues and $\psi_i$ the eigenfunctions. This decomposition allows us to construct a feature mapping $\boldsymbol{\phi}(x) = (\sqrt{\lambda_1}\psi_1(x), \sqrt{\lambda_2}\psi_2(x), \ldots)$ that reproduces the kernel through the inner product.

A symmetric and positive semi-definite kernel function therefore corresponds to some feature space, even if we do not construct that space explicitly. This allows us to design kernels from our domain knowledge about similarity in a particular problem, without having to find the corresponding features.

Now we look at some specific kernels. The value $k(x,x')$ measures how similar two inputs $x$ and $x'$ are. The linear, polynomial, RBF and Matérn kernels below accept an input $x$ that is a scalar or a vector.

\paragraph{Linear and Polynomial Kernels}

The simplest kernel function is the linear\index{kernel function!linear} kernel, defined as
\begin{equation}
k(x_i, x_j) = x_i^T x_j.
\end{equation}

\looseness=1 This kernel corresponds to the identity feature map $\boldsymbol{\phi}(x) = x$. It computes the dot product in the original input space. This looks trivial, but it shows that we can work with features or with kernels, and both give the same result. For a single number $x$, the linear kernel describes a straight line through the origin with an unknown slope.

The polynomial\index{kernel function!polynomial} kernel of degree $d$ extends the linear kernel. It is defined as
\begin{equation}
k(x_i, x_j) = (x_i^T x_j + c)^d,
\end{equation}
where $c \geq 0$ is a constant that trades off the influence of higher-order and lower-order terms.

As an example, we take the polynomial kernel of degree 2, the quadratic kernel, with $c = 0$. It is
\begin{equation}
k(x_i, x_j) = (x_i^T x_j)^2.
\end{equation}
For a two-dimensional input $x_i = (x_{i1}, x_{i2})$, this kernel corresponds to the feature mapping
\begin{equation}
\boldsymbol{\phi}(x_i) = (x_{i1}^2, \sqrt{2}x_{i1}x_{i2}, x_{i2}^2).
\end{equation}
The dot product of these feature vectors reproduces our kernel, since
\begin{equation}
\boldsymbol{\phi}(x_i)^T \boldsymbol{\phi}(x_j) = x_{i1}^2 x_{j1}^2 + 2x_{i1}x_{i2}x_{j1}x_{j2} + x_{i2}^2 x_{j2}^2 = (x_i^T x_j)^2.
\end{equation}

This example shows that a polynomial kernel implicitly maps our data to a higher-dimensional feature space of monomials. For $c > 0$, this space includes all monomials of degree $d$ or less. For $c = 0$, as above, it holds only the monomials of degree $d$, and degree 2 in two dimensions gives three features. As the input dimensionality or the polynomial degree increases, the feature space grows rapidly.

\paragraph{The Gaussian (RBF) Kernel}

Perhaps the most widely used kernel\index{kernel function!Gaussian (RBF)} is the radial basis function (RBF)\index{RBF kernel|see{kernel function, Gaussian (RBF)}} kernel, also known as the Gaussian\index{Gaussian kernel|see{kernel function, Gaussian (RBF)}} kernel, which is
\begin{equation}
k(x_i, x_j) = \exp\left(-\frac{\|x_j - x_i\|^2}{2\ell^2}\right).
\end{equation}

This kernel measures the similarity between the points $x_j$ and $x_i$ from their Euclidean distance, scaled by a length\index{length scale} scale parameter $\ell$. Points that are close together have kernel values near 1, and points far apart have kernel values near 0. The parameter $\ell$ controls the ``width'' of this similarity function. A smaller $\ell$ means that points must be very close to be considered similar. In practice we multiply the kernel by an amplitude $\sigma_f^2$, which sets the size of the variations.

To see which feature space the RBF kernel corresponds to, we expand $\|x_i - x_j\|^2 = \|x_i\|^2 - 2x_i^T x_j + \|x_j\|^2$. This lets us write the kernel as $\exp(-\|x_i\|^2/2\ell^2)\allowbreak\exp(-\|x_j\|^2/2\ell^2)\allowbreak\exp(x_i^T x_j/\ell^2)$. We expand the last factor with the Taylor series of the exponential function
\begin{equation}
\exp(x) = 1 + x + \frac{x^2}{2!} + \frac{x^3}{3!} + \ldots.
\end{equation}

Applying this expansion gives an infinite series of terms, and each term $(x_i^T x_j)^m/\allowbreak(m!\,\ell^{2m})$ is a polynomial kernel with its own features. The two factors in front only rescale the features of each term. The RBF kernel therefore corresponds to an infinite-dimensional feature space. Using it is equivalent to mapping our data to this space and performing linear regression there.

\paragraph{The Matérn Kernel}

The Matérn kernel\index{kernel function!Matern@Matérn} generalizes the RBF kernel and is popular in spatial statistics and Gaussian process modeling. It is
\begin{equation}
k(x_i, x_j) = \frac{2^{1-\nu}}{\Gamma(\nu)}\left(\frac{\sqrt{2\nu}\|x_i - x_j\|}{\ell}\right)^{\nu}K_{\nu}\left(\frac{\sqrt{2\nu}\|x_i - x_j\|}{\ell}\right),
\end{equation}
where $\Gamma$ is the gamma function and $K_{\nu}$ is the modified Bessel function of the second kind. The Bessel function $K_{\nu}$ is unrelated to the kernel matrix $\mathbf{K}$. The parameter $\nu$ controls the smoothness of the resulting functions, and $\ell$ is the length scale parameter.

\begin{figure}[ht!]
    \centering
    \includegraphics[width=0.95\textwidth]{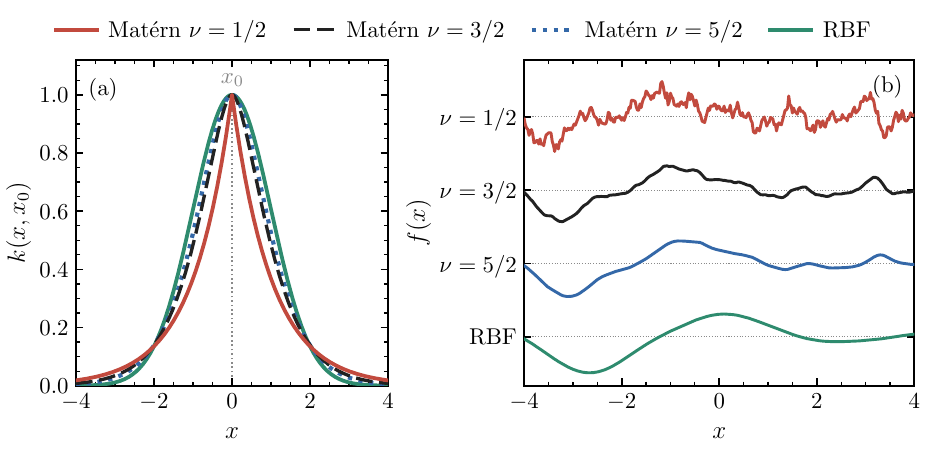}
    \caption{Kernels and sample functions. Panel~(a) shows the RBF kernel and three Mat\'ern kernels, all with the same length scale, for the reference point $x_0 = 0$. Panel~(b) shows one random function from a GP with each kernel, shifted vertically, all drawn from the same random numbers. The function-space perspective later in this chapter explains such draws. The function for $\nu = 1/2$ is continuous but not differentiable. The functions for $\nu = 3/2$ and $5/2$ are once and twice differentiable, and the RBF function is infinitely differentiable.}
    \label{fig:kernel_comparison}
\end{figure}

\looseness=1 Common choices for $\nu$ are 1/2, 3/2 and 5/2, and each gives functions with a different degree of differentiability. Different physical processes in astronomy can have different degrees of smoothness, which makes this control useful. When $\nu \to \infty$, the Matérn kernel converges to the RBF kernel, so the RBF kernel is a limiting case of the Matérn kernel. Figure~\ref{fig:kernel_comparison} compares the RBF kernel with Matérn kernels for these values of $\nu$.

\paragraph{Combining Kernels}

By Mercer's theorem, a kernel\index{kernel function!combining} function is valid if and only if its kernel matrix is positive semi-definite for any set of inputs. Positive semi-definite matrices remain positive semi-definite under addition. The sum of valid kernels is therefore a valid kernel, and we can build complex models by combining simpler components.

For example, we might model stellar variability, with the time $t$ as the input, by a composite kernel
\begin{equation}
k(t, t') = \theta_0 \exp\left(-\frac{\theta_1}{2} |t - t'|^2\right) + \theta_2 + \theta_3 \exp\left(-\frac{2\sin^2(\pi|t-t'|/\theta_4)}{\theta_5^2}\right). \label{eq:composite_kernel}
\end{equation}

This kernel combines three components. The RBF component handles the smooth evolution of active regions. The constant component models the baseline brightness level. The periodic\index{kernel function!periodic} component describes rotational modulation. We can also multiply kernels, because the elementwise product of two positive semi-definite kernel matrices is again positive semi-definite, or compose them in other ways. This lets us tailor the kernel to specific physical processes.

\looseness=1 {\sloppy When we choose kernels for astronomical data, we consider both physical considerations and empirical performance. Physical considerations include the characteristic timescales of the phenomena we study, the expected smoothness of the underlying processes, and any known periodicities or trends. For quasar variability, we might expect an irregular, stochastic process with correlations that decay over characteristic timescales. Section~14.10 shows how to compare candidate kernels with the\index{Gaussian process (GP)!kernels|)}\index{kernel function|)} data.\par}

\section{Kernelizing Bayesian Linear Regression}

\looseness=1 The kernelized linear regression of Section~14.2 estimates one set of weights, so it gives only point predictions. For astronomical applications, we also need uncertainties to make reliable scientific inferences. When we measure stellar brightness, detect exoplanet transits, or analyze galaxy spectra, the uncertainty of a prediction is as important as the prediction itself. We also need to know how this uncertainty changes over the input space.

\looseness=1 By Mercer's theorem (Section~14.3), a valid kernel is an inner product of feature vectors, possibly with infinitely many entries. A model with such a kernel is a linear regression in feature space, so a Gaussian process generalizes the Bayesian linear regression of Chapter~5. We therefore kernelize that treatment first, which gives a formal understanding of the predictions. We obtain the same uncertainties while working implicitly with complex feature spaces. The result is a Gaussian process.

In Section~5.5 we found the posterior distribution over the weights. The prior is $p(\mathbf{w}) = \mathcal{N}(\mathbf{w}|\mathbf{m}_0, \mathbf{S}_0)$ with prior mean $\mathbf{m}_0$ and prior covariance $\mathbf{S}_0$. The likelihood is $p(\mathbf{t}|\mathbf{X}, \mathbf{w}) = \mathcal{N}(\mathbf{t}|\mathbf{\Phi}\mathbf{w}, \sigma_n^2\mathrm{Id})$, where $\mathbf{X}$ holds the $N$ inputs. The posterior is
\begin{align}
p(\mathbf{w}|\mathbf{X}, \mathbf{t}) &= \mathcal{N}(\mathbf{w}|\mathbf{m}_N, \mathbf{S}_N) \\*
\mathbf{S}_N &= \left(\frac{1}{\sigma_n^2}\mathbf{\Phi}^T\mathbf{\Phi} + \mathbf{S}_0^{-1}\right)^{-1} \\*
\mathbf{m}_N &= \mathbf{S}_N\left(\mathbf{S}_0^{-1}\mathbf{m}_0 + \frac{1}{\sigma_n^2}\mathbf{\Phi}^T\mathbf{t}\right).
\end{align}

\looseness=1 Our goal is a predictive distribution in which the features enter only through kernel values. The algebra takes three steps. We apply the Woodbury\index{Woodbury identity} identity, define the kernel quantities and then use a matrix identity that cancels one term.

We use a zero-mean prior $\mathbf{m}_0 = \mathbf{0}$ for the weights, which is a common choice and keeps the algebra short. Before we see any data, we have no reason to believe that the function is biased in any particular direction. A zero prior mean does not mean that the targets average to zero, as Section~14.7 explains.

With this choice, the posterior mean simplifies to
\begin{align}
\mathbf{m}_N &= \mathbf{S}_N\frac{1}{\sigma_n^2}\mathbf{\Phi}^T\mathbf{t}.
\end{align}

Given the training data $\mathcal{D} = \{(x_i, t_i)\}_{i=1}^N$, the predictive distribution of the latent function value $f_* = \boldsymbol{\phi}(x_*)^T\mathbf{w}$ at $x_*$, the function value without noise, is
\begin{equation}
p(f_* | x_*, \mathcal{D}) = \mathcal{N}(f_* | \boldsymbol{\phi}(x_*)^T \mathbf{m}_N, \boldsymbol{\phi}(x_*)^T \mathbf{S}_N \boldsymbol{\phi}(x_*)).
\end{equation}

\paragraph{Kernelizing the Predictive Mean}

We kernelize the mean as in Section~14.2, and we now kernelize the variance as well. We start with the predictive mean
\begin{align}
\boldsymbol{\phi}(x_*)^T \mathbf{m}_N &= \boldsymbol{\phi}(x_*)^T \mathbf{S}_N \frac{1}{\sigma_n^2}\mathbf{\Phi}^T\mathbf{t}.
\end{align}

To kernelize this expression, we need a form of $\mathbf{S}_N$ that involves kernel matrices. The matrix $\mathbf{S}_N$ contains $\mathbf{\Phi}^T\mathbf{\Phi}$, which compares features, but we want products between inputs. We use the Woodbury matrix identity for the inverse of a sum of matrices,
\begin{equation}
(\mathbf{A} + \mathbf{U}\mathbf{C}\mathbf{V})^{-1} = \mathbf{A}^{-1} - \mathbf{A}^{-1}\mathbf{U}(\mathbf{C}^{-1} + \mathbf{V}\mathbf{A}^{-1}\mathbf{U})^{-1}\mathbf{V}\mathbf{A}^{-1}, \label{eq:woodbury}
\end{equation}
with $\mathbf{A} = \mathbf{S}_0^{-1}$, $\mathbf{U} = \mathbf{\Phi}^T$, $\mathbf{C} = \sigma_n^{-2}\mathrm{Id}_N$ and $\mathbf{V} = \mathbf{\Phi}$. This gives
\begin{align}
\mathbf{S}_N &= \left(\frac{1}{\sigma_n^2}\mathbf{\Phi}^T\mathbf{\Phi} + \mathbf{S}_0^{-1}\right)^{-1} \\*
&= \mathbf{S}_0 - \mathbf{S}_0\mathbf{\Phi}^T (\sigma_n^2\mathrm{Id}_N + \mathbf{\Phi}\mathbf{S}_0\mathbf{\Phi}^T)^{-1}\mathbf{\Phi}\mathbf{S}_0. \label{eq:sn_woodbury}
\end{align}

Substituting this form of $\mathbf{S}_N$ into the predictive mean gives
\begin{align}
\boldsymbol{\phi}(x_*)^T \mathbf{m}_N &= \boldsymbol{\phi}(x_*)^T \left[\mathbf{S}_0 - \mathbf{S}_0\mathbf{\Phi}^T(\sigma_n^2\mathrm{Id}_N + \mathbf{\Phi}\mathbf{S}_0\mathbf{\Phi}^T)^{-1}\mathbf{\Phi}\mathbf{S}_0\right] \frac{1}{\sigma_n^2}\mathbf{\Phi}^T\mathbf{t} \\*
&= \frac{1}{\sigma_n^2}\boldsymbol{\phi}(x_*)^T\mathbf{S}_0\mathbf{\Phi}^T\mathbf{t} - \frac{1}{\sigma_n^2}\boldsymbol{\phi}(x_*)^T\mathbf{S}_0\mathbf{\Phi}^T(\sigma_n^2\mathrm{Id}_N + \mathbf{\Phi}\mathbf{S}_0\mathbf{\Phi}^T)^{-1}\mathbf{\Phi}\mathbf{S}_0\mathbf{\Phi}^T\mathbf{t}.
\end{align}

The features now enter only through products of the form $\boldsymbol{\phi}(x)^T\mathbf{S}_0\boldsymbol{\phi}(x')$. Each such product measures the similarity of two inputs. We therefore include the prior covariance in the kernel function, so that it becomes
\begin{align}
k(x,x') &= \boldsymbol{\phi}(x)^T\mathbf{S}_0\boldsymbol{\phi}(x'). \label{eq:k_prior}
\end{align}

With this definition, the kernel matrix of our training data is
\begin{align}
[\mathbf{K}]_{ij} &= k(x_i, x_j) = \boldsymbol{\phi}(x_i)^T\mathbf{S}_0\boldsymbol{\phi}(x_j).
\end{align}

This means that $\mathbf{K} = \mathbf{\Phi}\mathbf{S}_0\mathbf{\Phi}^T$. For $\mathbf{S}_0 = \mathrm{Id}$, this is the matrix $\mathbf{K}$ of Section~14.2. We also define the vector of kernel evaluations between the test point and all training points as
\begin{align}
\mathbf{k}_* &= [k(x_1, x_*), k(x_2, x_*), \ldots, k(x_N, x_*)]^T = \mathbf{\Phi}\mathbf{S}_0\boldsymbol{\phi}(x_*). \label{eq:kstar_prior}
\end{align}

With these definitions, the predictive mean becomes
\begin{align}
\boldsymbol{\phi}(x_*)^T \mathbf{m}_N &= \frac{1}{\sigma_n^2}\boldsymbol{\phi}(x_*)^T\mathbf{S}_0\mathbf{\Phi}^T\mathbf{t} - \frac{1}{\sigma_n^2}\boldsymbol{\phi}(x_*)^T\mathbf{S}_0\mathbf{\Phi}^T(\sigma_n^2\mathrm{Id}_N + \mathbf{K})^{-1}\mathbf{K}\mathbf{t}.
\end{align}

This looks complex, but the second term cancels part of the first. A matrix identity shows how
\begin{equation}
(\sigma_n^2\mathrm{Id}_N + \mathbf{K})^{-1}\mathbf{K} = \mathrm{Id}_N - \sigma_n^2(\sigma_n^2\mathrm{Id}_N + \mathbf{K})^{-1}.
\end{equation}

To prove the identity, we multiply both sides by $(\sigma_n^2\mathrm{Id}_N + \mathbf{K})$ from the left. This gives
\begin{align}
(\sigma_n^2\mathrm{Id}_N + \mathbf{K})(\sigma_n^2\mathrm{Id}_N + \mathbf{K})^{-1}\mathbf{K} &= (\sigma_n^2\mathrm{Id}_N + \mathbf{K})[\mathrm{Id}_N - \sigma_n^2(\sigma_n^2\mathrm{Id}_N + \mathbf{K})^{-1}] \\*
\mathbf{K} &= (\sigma_n^2\mathrm{Id}_N + \mathbf{K}) - \sigma_n^2(\sigma_n^2\mathrm{Id}_N + \mathbf{K})(\sigma_n^2\mathrm{Id}_N + \mathbf{K})^{-1} \\
\mathbf{K} &= \sigma_n^2\mathrm{Id}_N + \mathbf{K} - \sigma_n^2\mathrm{Id}_N \\*
\mathbf{K} &= \mathbf{K}.
\end{align}

Using this identity in the predictive mean gives
\begin{align}
\boldsymbol{\phi}(x_*)^T \mathbf{m}_N &= \frac{1}{\sigma_n^2}\boldsymbol{\phi}(x_*)^T\mathbf{S}_0\mathbf{\Phi}^T\mathbf{t} - \frac{1}{\sigma_n^2}\boldsymbol{\phi}(x_*)^T\mathbf{S}_0\mathbf{\Phi}^T[\mathrm{Id}_N - \sigma_n^2(\sigma_n^2\mathrm{Id}_N + \mathbf{K})^{-1}]\mathbf{t} \\*
&= \frac{1}{\sigma_n^2}\boldsymbol{\phi}(x_*)^T\mathbf{S}_0\mathbf{\Phi}^T\mathbf{t} - \frac{1}{\sigma_n^2}\boldsymbol{\phi}(x_*)^T\mathbf{S}_0\mathbf{\Phi}^T\mathbf{t} + \boldsymbol{\phi}(x_*)^T\mathbf{S}_0\mathbf{\Phi}^T(\sigma_n^2\mathrm{Id}_N + \mathbf{K})^{-1}\mathbf{t} \\
&= \boldsymbol{\phi}(x_*)^T\mathbf{S}_0\mathbf{\Phi}^T(\sigma_n^2\mathrm{Id}_N + \mathbf{K})^{-1}\mathbf{t} \\*
&= \mathbf{k}_*^T(\sigma_n^2\mathrm{Id}_N + \mathbf{K})^{-1}\mathbf{t}.
\end{align}

The mean has the same form as the point-estimate prediction of Section~14.2, but its kernel includes the prior covariance $\mathbf{S}_0$ over the weights. The kernel definition $k(x,x') = \boldsymbol{\phi}(x)^T\mathbf{S}_0\boldsymbol{\phi}(x')$ weights the feature inner products according to our prior beliefs about the importance of the parameters. If $\mathbf{S}_0$ has larger values for certain dimensions of the weight space, the corresponding features may have larger coefficients. They then have more influence on the similarity of two inputs. When $\mathbf{S}_0 = \mathrm{Id}$ (independent weights with unit variance), the kernel reduces to the standard inner product $\boldsymbol{\phi}(x)^T\boldsymbol{\phi}(x')$.

\paragraph{Kernelizing the Predictive Variance}

The Bayesian approach has an important advantage over the point estimate. We can also kernelize the predictive variance, which gives the uncertainty of our predictions. The predictive variance of the latent function is
\begin{equation}
\boldsymbol{\phi}(x_*)^T \mathbf{S}_N \boldsymbol{\phi}(x_*) = \boldsymbol{\phi}(x_*)^T \left(\frac{1}{\sigma_n^2}\mathbf{\Phi}^T\mathbf{\Phi} + \mathbf{S}_0^{-1}\right)^{-1} \boldsymbol{\phi}(x_*).
\end{equation}

Substituting the expression for $\mathbf{S}_N$ from Equation~\ref{eq:sn_woodbury} into the variance gives
\begin{align}
\boldsymbol{\phi}(x_*)^T\mathbf{S}_N\boldsymbol{\phi}(x_*) &= \boldsymbol{\phi}(x_*)^T\left[\mathbf{S}_0 - \mathbf{S}_0\mathbf{\Phi}^T(\sigma_n^2\mathrm{Id}_N + \mathbf{\Phi}\mathbf{S}_0\mathbf{\Phi}^T)^{-1}\mathbf{\Phi}\mathbf{S}_0\right]\boldsymbol{\phi}(x_*) \\*
&= \boldsymbol{\phi}(x_*)^T\mathbf{S}_0\boldsymbol{\phi}(x_*) - \boldsymbol{\phi}(x_*)^T\mathbf{S}_0\mathbf{\Phi}^T(\sigma_n^2\mathrm{Id}_N + \mathbf{\Phi}\mathbf{S}_0\mathbf{\Phi}^T)^{-1}\mathbf{\Phi}\mathbf{S}_0\boldsymbol{\phi}(x_*).
\end{align}

With the kernel $k$, the vector $\mathbf{k}_*$ and the matrix $\mathbf{K}$ defined above, this becomes
\begin{align}
\boldsymbol{\phi}(x_*)^T\mathbf{S}_N\boldsymbol{\phi}(x_*) &= \boldsymbol{\phi}(x_*)^T\mathbf{S}_0\boldsymbol{\phi}(x_*) - \boldsymbol{\phi}(x_*)^T\mathbf{S}_0\mathbf{\Phi}^T(\sigma_n^2\mathrm{Id}_N + \mathbf{K})^{-1}\mathbf{\Phi}\mathbf{S}_0\boldsymbol{\phi}(x_*) \\*
&= k(x_*, x_*) - \mathbf{k}_*^T(\mathbf{K} + \sigma_n^2\mathrm{Id})^{-1}\mathbf{k}_*.
\end{align}

Both the mean and the variance now depend only on kernel evaluations. The complete Bayesian predictive distribution therefore needs no explicit feature representations, and by Mercer's theorem it holds for infinitely many features. For the latent function value, without noise, it is
\begin{equation}
p(f_* | x_*, \mathcal{D}) = \mathcal{N}(f_* | \mu_{f_*}, \sigma_{f_*}^2),
\end{equation}
where
\begin{align}
\mu_{f_*} &= \mathbf{k}_*^T (\mathbf{K} + \sigma_n^2 \mathrm{Id})^{-1} \mathbf{t} \\*
\sigma_{f_*}^2 &= k(x_*, x_*) - \mathbf{k}_*^T(\mathbf{K} + \sigma_n^2 \mathrm{Id})^{-1}\mathbf{k}_*.
\end{align}

For the observed output, which includes the noise, the predictive distribution is
\begin{equation}
p(y_* | x_*, \mathcal{D}) = \mathcal{N}(y_* | \mu_{f_*}, \sigma_{f_*}^2 + \sigma_n^2).
\end{equation}

\paragraph{Interpreting the Predictions}

\begin{figure}[ht!]
    \centering
    \includegraphics[width=0.75\textwidth]{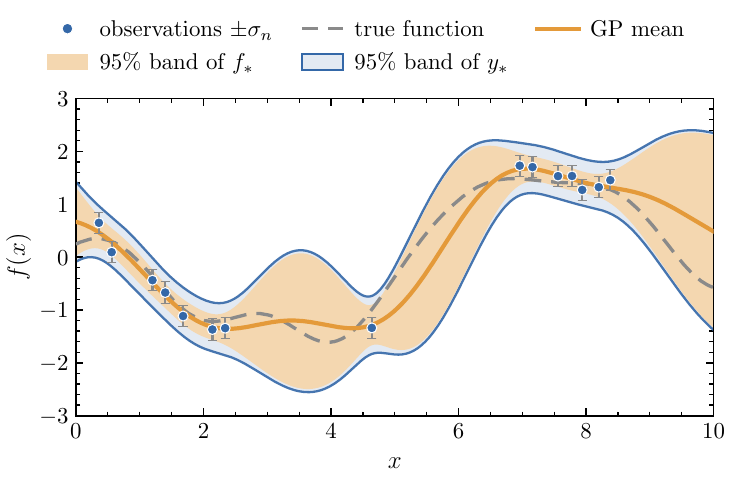}
    \caption{GP regression with an RBF kernel for irregularly spaced observations. The band of a new noisy observation $y_*$ is wider than that of the latent function value $f_*$ because of the observation noise. Their variances satisfy $\mathrm{Var}(y_*) = \mathrm{Var}(f_*) + \sigma_n^2$. The band of $f_*$ is narrow near the observations and wide in the gaps. Beyond the last observation its width approaches the width of the prior band. The two bands nearly merge where the uncertainty of $f_*$ is much larger than $\sigma_n$.}
    \label{fig:gp_f_vs_y}
\end{figure}

We examine $p(f_* | x_*, \mathcal{D})$ when we want to understand the underlying pattern in our data. We use $p(y_* | x_*, \mathcal{D})$ when we want to predict future observations with their noise. In astronomical applications, this distinction allows us to separate the intrinsic variability of celestial objects from measurement uncertainties. Figure~\ref{fig:gp_f_vs_y} shows the uncertainty bands of both distributions.

\looseness=1 In $\sigma_{f_*}^2$, the term $k(x_*,x_*)$ is our prior uncertainty about the function value at $x_*$, the variability we expect before we see any data. For most common kernels, like the RBF, it has the same value for all inputs. This value expresses our prior belief about the overall scale of function variations. The term $\mathbf{k}_*^T(\mathbf{K} + \sigma_n^2 \mathrm{Id})^{-1}\mathbf{k}_*$ measures how much the observations reduce this uncertainty. It is non-negative, so observations can only decrease our uncertainty.

The reduction depends on how close $x_*$ is to the training data. When $x_*$ is close to many training points, the vector $\mathbf{k}_*$ contains large values. The subtraction term then approaches $k(x_*, x_*)$, so $\sigma_{f_*}^2$ is small and the predictions are confident. As we move away from the training data, the elements of $\mathbf{k}_*$ decrease, because kernel functions typically decrease with distance. The subtraction term becomes smaller, so the uncertainty grows smoothly toward the prior variance $k(x_*, x_*)$. In a gap between observations the band therefore widens, as Figure~\ref{fig:gp_f_vs_y} shows.

The mean prediction $\mu_{f_*} = \mathbf{k}_*^T (\mathbf{K} + \sigma_n^2 \mathrm{Id})^{-1} \mathbf{t}$ is a weighted sum of the training outputs $\mathbf{t}$. The weight of each training point depends on its similarity to the test point and also on the correlations among the training points. If two training points are very similar, they provide less independent information than two dissimilar points would. The matrix $(\mathbf{K} + \sigma_n^2 \mathrm{Id})^{-1}$ has two roles. It accounts for the noise in the observations through $\sigma_n^2$ and for the redundancy in the training data through the correlations in $\mathbf{K}$. As the noise increases, the influence of the data on our predictions decreases.

Near dense, reliable data the mean follows the observations closely. In sparse regions $\mathbf{k}_*$ is small, so the mean reverts toward the prior mean, which is zero in our case. Through the matrix $(\mathbf{K} + \sigma_n^2 \mathrm{Id})^{-1}$, the variance follows the varying density of the training points. Parametric models may overfit sparse data regions, but a GP lets the uncertainty grow there. We are therefore less likely to draw overly confident conclusions from limited data. This is valuable in astronomy, where observations are often irregularly sampled and have gaps.

\section{Gaussian Processes vs. Bayesian Linear Regression}

Gaussian processes and the Bayesian linear regression of Section~5.5 are both Bayesian approaches to function approximation. They give the same predictions after some matrix manipulations, as Section~14.4 shows. They differ in how we specify the model and in their cost.

In Bayesian linear regression, we design the basis functions $\phi_1(x), \phi_2(x), \ldots, \phi_M(x)$ by hand and transform the inputs explicitly. The quality of the model depends on this choice. If the features miss the pattern in the data, the fit is poor for any choice of weights. This approach gives direct control and interpretability, because each feature has a weight that measures its importance.

In Gaussian processes, we define a kernel that encodes our beliefs about the function, such as its smoothness or periodicity. We specify no explicit features. This gives more flexibility, because predefined features no longer restrict the model. However, it may give less direct interpretability about which input dimensions matter most.

The two approaches also differ in their cost. In Bayesian linear regression, the posterior over the weights requires inverting an $M \times M$ matrix with $\mathcal{O}(M^3)$ complexity. Once we have the posterior, a prediction needs $\mathcal{O}(M)$ operations for the mean and $\mathcal{O}(M^2)$ for the variance.

In Gaussian processes, the predictive distribution requires inverting an $N \times N$ matrix with $\mathcal{O}(N^3)$ complexity. A single prediction then needs $\mathcal{O}(N)$ operations for the mean and $\mathcal{O}(N^2)$ for the variance. Bayesian linear regression is cheaper than a GP when the number of features $M$ is much smaller than $N$.

The cost of a GP comes with an advantage, because its complexity adapts automatically to the data. The kernel hyperparameters, such as the length scale, control this adaptation, and longer length scales give smoother functions. They sit one level above the function values, and we fit them to the data. This helps when the appropriate complexity is difficult to determine in advance.

\looseness=1 Bayesian linear regression suits cases with strong prior knowledge about the functional form, which we can encode directly in the choice of features. It also suits cases where the interpretability of single feature weights matters for scientific understanding, because the weights show the importance of the model components. It is often better for very large data sets, where the $\mathcal{O}(N^3)$ cost of a GP is prohibitive. It is also often better for problems with a high input dimension that carefully chosen features can model well.

Gaussian processes work well when the functional form is uncertain or complex. They are useful when the uncertainty of the predictions is important. They work best for data sets of moderate size but complex structure, especially for functions whose behavior changes over the input space. Light curves with several timescales of variability are an example. They also work best when the input dimension is low.

Many modern GP packages use approximations with a cost much lower than $\mathcal{O}(N^3)$, usually at the price of some exactness or flexibility. Python packages such as GPyTorch handle much larger data sets than older implementations. GPs therefore become more practical for astronomical applications.

\section{The Function-Space Perspective}

So\index{function-space view|(} far we have kernelized the linear regression of the earlier chapters, a mechanical route that shows how Gaussian processes connect to them. There is a second way to think about Gaussian processes, the function-space\index{function-space view} view. It defines a probability distribution directly over functions. It is more conceptual at first, but it reaches the same results with less algebra.

\paragraph{Connection to Stochastic Processes}

This perspective uses stochastic\index{stochastic process} processes, and the Markov chains of Section~13.1 are one example. A stochastic process is a collection of random variables indexed by time or space. A Gaussian process is another type of stochastic process, with properties that make it useful for regression. MCMC uses the Markov chain as a computational tool to sample from complex distributions. A GP uses its stochastic process as a model that represents an unknown function.

In earlier chapters, we worked with finite-dimensional random variables, vectors $\mathbf{z} \in \mathbb{R}^D$ with a distribution $p(\mathbf{z})$. Each element $z_i$ is a single random variable, and the integers $1, 2, \ldots, D$ index these variables. A stochastic process extends this indexing to sets that can be uncountable. We index random variables by the elements of a set $\mathcal{T}$ instead of by integers. The set might hold time points, spatial locations or inputs to a function. For each $s \in \mathcal{T}$ we have a random variable $X_s$, and the collection $\{X_s : s \in \mathcal{T}\}$ forms a stochastic process.

The index differs between our two contexts. In MCMC, the index is the iteration number, or time step, of the sampling algorithm. The sequence $\{X_1, X_2, X_3, \ldots\}$ holds the state $X_t$ of the Markov chain at iteration $t$. The Markov property says that $P(X_{t+1} | X_1, X_2, \ldots, X_t) = P(X_{t+1} | X_t)$, so the next state depends only on the current state. In a Gaussian process, the index is the input of our function. For a function $f: \mathcal{X} \rightarrow \mathbb{R}$, we have a random variable $f(x)$ for each $x \in \mathcal{X}$, and the collection $\{f(x) : x \in \mathcal{X}\}$ forms our Gaussian process.

MCMC indices are typically discrete, while GP indices can be continuous, for example spatial coordinates or time. This allows GPs to model smooth functions over continuous domains. When $\mathcal{T}$ is uncountable, like the real line $\mathbb{R}$, we effectively have an infinite-dimensional random variable. We do not need all of its dimensions at once. A finite-dimensional vector has marginal distributions for subsets of its components. In the same way, we can examine the joint distribution of $\{X_{s_1}, X_{s_2}, \ldots, X_{s_N}\}$ for any finite subset of indices $\{s_1, s_2, \ldots, s_N\} \subset \mathcal{T}$.

\paragraph{Defining Gaussian Processes}

A Gaussian process is a stochastic process, a collection of random variables $\{f(x) : x \in \mathcal{X}\}$ in which any finite subset follows a multivariate Gaussian distribution.

This definition extends the familiar multivariate Gaussian distribution to infinite dimensions. A random vector $\mathbf{z} \in \mathbb{R}^D$ with a multivariate Gaussian distribution has a mean vector $\boldsymbol{\mu} \in \mathbb{R}^D$ with $\mu_i = \mathbb{E}[z_i]$ and a covariance matrix $\mathbf{\Sigma} \in \mathbb{R}^{D \times D}$ with $\Sigma_{ij} = \mathbb{E}[(z_i - \mu_i)(z_j - \mu_j)]$.

A Gaussian process follows the same principle for functions. Instead of discrete indices $i \in \{1,2,...,D\}$, we have a continuous index set $x \in \mathcal{X}$. To specify a Gaussian process, we need a mean\index{mean function} function and a covariance\index{covariance function} function.

\begin{itemize}
    \item The mean function $m(x)$ gives the expected value at each input and is defined by
    \begin{equation}
    m(x) = \mathbb{E}[f(x)].
    \end{equation}

    \item The covariance function, or kernel, $k(x, x')$ specifies how function values at different inputs relate. It is defined by
    \begin{equation}
    k(x, x') = \mathbb{E}[(f(x) - m(x))(f(x') - m(x'))].
    \end{equation}
\end{itemize}

We denote the Gaussian process by
\begin{equation}
f \sim \mathcal{GP}(m, k).
\end{equation}

A covariance matrix is positive semi-definite, so a covariance function must be a valid kernel of Section~14.3. The covariance function encodes our assumptions about the smoothness of the function. The covariance function resembles the autocorrelation of Section~13.10. There we examined how the samples of a chain at different lags are correlated, which told us about the ``memory'' of the chain.

\begin{figure}[ht!]
    \centering
    \includegraphics[width=0.9\textwidth]{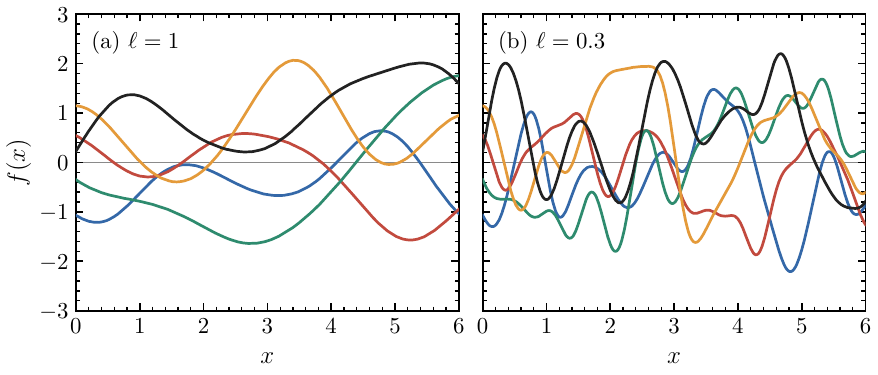}
    \caption{Five functions drawn from a zero-mean GP prior with the kernel $k(x, x') = \sigma_f^2 \exp(-(x - x')^2 / (2\ell^2))$. Panel~(a) has a long length scale $\ell$, and panel~(b) has a short one. Both panels use the same random numbers, so only the length scale differs. The shorter length scale gives faster fluctuations.}
    \label{fig:gp_function_space}
\end{figure}

\looseness=1 Suppose that $k(x, x')$ decreases rapidly as $|x - x'|$ increases. We then model functions that can change quickly, because $f(x)$ tells us little about $f(x')$ when $x$ and $x'$ are far apart. If $k(x, x')$ decreases slowly with distance, we model smoother functions in which distant points remain correlated. A slowly decaying autocorrelation function in MCMC similarly indicates a chain that moves slowly through the parameter space. Figure~\ref{fig:gp_function_space} shows random functions drawn from a GP prior with an RBF kernel for a long and a short length scale.

\paragraph{Finite-Dimensional Projections}

For a Gaussian process, the function values at any finite set of inputs $\mathbf{X} = [x_1,\allowbreak x_2,\allowbreak \ldots,\allowbreak x_N]^T$ follow a multivariate Gaussian
\begin{equation}
\mathbf{f} = [f(x_1), f(x_2), \ldots, f(x_N)]^T \sim \mathcal{N}(\mathbf{m}, \mathbf{K}),
\end{equation}
where $\mathbf{m} = [m(x_1), m(x_2), \ldots, m(x_N)]^T$ and $\mathbf{K}_{ij} = k(x_i, x_j)$. To draw a function from the prior, we choose a grid of inputs and draw $\mathbf{f}$ from this Gaussian.

\looseness=1 A GP defines a distribution over an infinite-dimensional space of functions, but any finite ``slice'' of this distribution is a regular multivariate Gaussian. Any marginal distribution of a Gaussian is also Gaussian, as we showed in Section~5.8. This marginalization property makes GPs computationally tractable despite their infinite-dimensional nature.

For a stellar light curve, we have observations at a finite set of time points, but we believe that there is an underlying continuous process. Through the marginalization property, a GP lets us evaluate this process at our observation times.

Marginalization alone does not let us learn from data. For that we use a second property of Gaussian distributions, conditioning. If a random vector $\mathbf{z} = [\mathbf{z}_A, \mathbf{z}_B]^T$ follows a multivariate Gaussian, then the conditional distribution of $\mathbf{z}_B$ given $\mathbf{z}_A$ is also Gaussian, as Section~5.9 shows. Through conditioning, we can update our prior beliefs about functions after we observe data.\index{function-space view|)}

\section{Conditioning the Gaussian Process}\label{sec:ch14_conditioning}

How\index{Gaussian process (GP)!conditioning|(} do we learn from data with a Gaussian process? We have noisy observations of an unknown function and want to predict the function at new points. A parametric model such as linear regression fits parameters. A Gaussian process conditions its prior distribution over functions on the observed data, and the result is a posterior distribution over functions. For this update we must account for our prior beliefs, the noise in the observations and the correlations that the kernel encodes.

We can treat this update as a standard conditioning problem for a multivariate Gaussian, even though a GP is infinite-dimensional. We need only function values at a finite set of points, our training locations and our test location. On this finite subset, we can apply the conditioning formulas that we already know.

To build intuition, consider a multivariate Gaussian distribution over three variables $(z_1, z_2, z_3)$. We observe the values of $z_1$ and $z_2$, and the conditioning property gives the distribution of $z_3$ given these observations. This conditional distribution is also Gaussian, and its mean depends on the observed values and on the covariances between all variables. Its variance does not depend on the observed values.

\begin{figure}[ht!]
    \centering
    \includegraphics[width=0.95\textwidth]{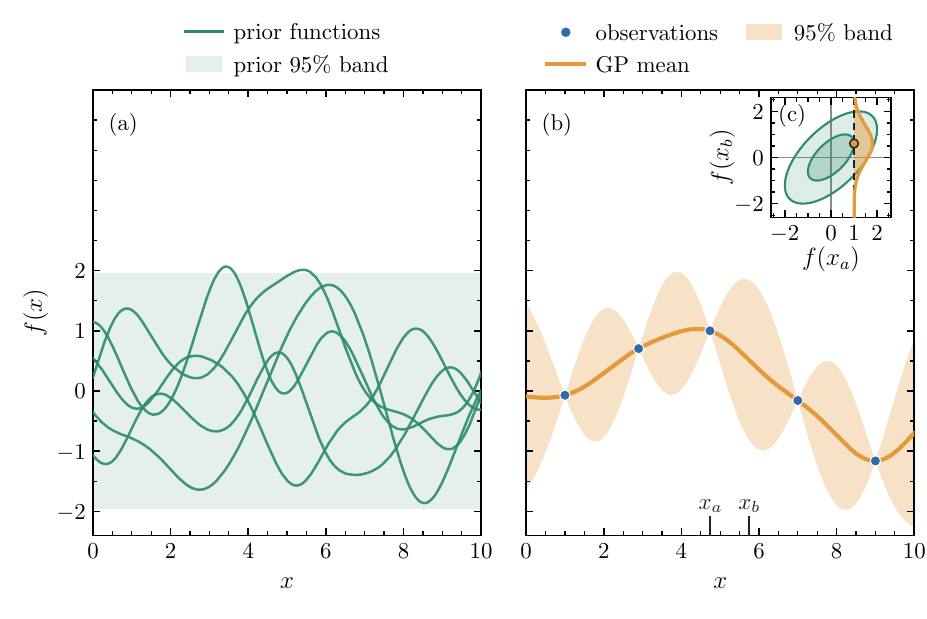}
    \caption{Panel~(a) shows five functions drawn from the GP prior and its 95 percent band. Panel~(b) shows the posterior after conditioning on five noiseless observations, with its mean and 95 percent band. The inset~(c) shows the joint prior density of $f(x_a)$ and $f(x_b)$ with contours at one and two standard deviations. It also shows the line $f(x_a) = 1$ and the conditional density of $f(x_b)$ given $f(x_a) = 1$, with its mean as a dot. Given $f(x_a) = 1$, the value $f(x_b)$ has a higher mean and a smaller standard deviation than in the prior. The inset uses only $f(x_a)$, so it differs from panel~(b), which uses all five observations.}
    \label{fig:gp_prior_posterior}
\end{figure}

A Gaussian process works in the same way, with function values at possibly many input points instead of three variables. Figure~\ref{fig:gp_prior_posterior} shows a GP prior, its posterior after conditioning on observations, and the conditioning of two function values.

From here on we write inputs in bold, as vectors. The training data are $N$ input-output pairs $\mathcal{D} = \{(\mathbf{x}_i, t_i)\}_{i=1}^N$. We assume that each output $t_i$ is a noisy observation of an underlying function $f(\mathbf{x})$, so that
\begin{equation}
t_i = f(\mathbf{x}_i) + \epsilon_i,
\end{equation}
where $\epsilon_i \sim \mathcal{N}(0, \sigma_n^2)$ is Gaussian noise.

Our goal is to infer the function value at an arbitrary test point $\mathbf{x}_*$, that is, to compute $p(f(\mathbf{x}_*) | \mathcal{D})$. To find this posterior, we construct the joint distribution of the relevant function values and then apply Gaussian conditioning.

We begin with the prior
\begin{equation}
f \sim \mathcal{GP}(m, k).
\end{equation}
For simplicity, we use a zero-mean prior, $m(\mathbf{x}) = 0$. A zero prior mean is a statement about the function before we see data. It does not say that the targets $t_i$ have mean zero.

For a kernel that decays with distance, the posterior mean follows the observations and returns to zero only far from them. A constant term in the kernel adds an unknown offset, so the data set the overall level. For a known offset we use a non-zero $m(\mathbf{x})$, which we treat below.

The central step is the joint distribution of the training function values $\mathbf{f} = [f(\mathbf{x}_1),\allowbreak \ldots,\allowbreak f(\mathbf{x}_N)]^T$ and the test function value $f_* = f(\mathbf{x}_*)$. By the definition of a Gaussian process, this joint distribution is multivariate Gaussian, and we can write it as
\begin{equation}
\begin{bmatrix} \mathbf{f} \\ f_* \end{bmatrix} \sim \mathcal{N}\left( \begin{bmatrix} \mathbf{0} \\ 0 \end{bmatrix}, \begin{bmatrix} \mathbf{K} & \mathbf{k}_* \\ \mathbf{k}_*^T & k_{**} \end{bmatrix} \right).
\end{equation}

Here $\mathbf{K}$ is the $N \times N$ covariance matrix of the function values at the training inputs, with entries $K_{ij} = k(\mathbf{x}_i, \mathbf{x}_j)$. The vector $\mathbf{k}_*$ is the $N \times 1$ vector of covariances between the test point and all training points, with entries $[\mathbf{k}_*]_i = k(\mathbf{x}_i, \mathbf{x}_*)$. The scalar $k_{**} = k(\mathbf{x}_*, \mathbf{x}_*)$ is the prior variance at the test point. This joint distribution contains the correlations between function values at different points, as the kernel encodes them.

\paragraph{Observation Noise}

We do not observe the latent function values $\mathbf{f}$ directly. We observe noisy values $\mathbf{t} = [t_1, \ldots, t_N]^T$, and a realistic model must account for this difference before we condition. With independent Gaussian noise $\epsilon_i \sim \mathcal{N}(0, \sigma_n^2)$ on each observation, we write
\begin{equation}
\mathbf{t} = \mathbf{f} + \boldsymbol{\epsilon}, \quad \boldsymbol{\epsilon} \sim \mathcal{N}(\mathbf{0}, \sigma_n^2 \mathrm{Id}).
\end{equation}
The joint distribution of the observations $\mathbf{t}$ and the test function value $f_*$ is then
\begin{equation}
\begin{bmatrix} \mathbf{t} \\ f_* \end{bmatrix} \sim \mathcal{N}\left( \begin{bmatrix} \mathbf{0} \\ 0 \end{bmatrix}, \begin{bmatrix} \mathbf{K} + \sigma_n^2 \mathrm{Id} & \mathbf{k}_* \\ \mathbf{k}_*^T & k_{**} \end{bmatrix} \right).
\end{equation}

The noise adds $\sigma_n^2$ to the diagonal of the covariance matrix. It does not change the covariances between different observations. Two repeated measurements at the same input therefore have a correlation below 1, because each carries its own measurement noise.

We now apply the standard conditioning formula for multivariate Gaussians, which Section~5.9 derives. That section calls the two parts $\mathbf{x}_1$ and $\mathbf{x}_2$, and they are $\mathbf{z}_B$ and $\mathbf{z}_A$ here. Suppose that a random vector with parts $\mathbf{z}_A$ and $\mathbf{z}_B$ has the joint Gaussian distribution
\begin{equation}
\begin{bmatrix} \mathbf{z}_A \\ \mathbf{z}_B \end{bmatrix} \sim \mathcal{N}\left( \begin{bmatrix} \boldsymbol{\mu}_A \\ \boldsymbol{\mu}_B \end{bmatrix}, \begin{bmatrix} \mathbf{\Sigma}_{AA} & \mathbf{\Sigma}_{AB} \\ \mathbf{\Sigma}_{BA} & \mathbf{\Sigma}_{BB} \end{bmatrix} \right).
\end{equation}

Then the conditional distribution of $\mathbf{z}_B$ given $\mathbf{z}_A$ is also Gaussian and is given by
\begin{equation}
p(\mathbf{z}_B|\mathbf{z}_A) = \mathcal{N}(\boldsymbol{\mu}_B + \mathbf{\Sigma}_{BA}\mathbf{\Sigma}_{AA}^{-1}(\mathbf{z}_A - \boldsymbol{\mu}_A), \mathbf{\Sigma}_{BB} - \mathbf{\Sigma}_{BA}\mathbf{\Sigma}_{AA}^{-1}\mathbf{\Sigma}_{AB}).
\end{equation}

For our problem, $\mathbf{z}_A = \mathbf{t}$ and $z_B = f_*$. The blocks are $\boldsymbol{\mu}_A = \mathbf{0}$, $\mu_B = 0$, $\mathbf{\Sigma}_{AA} = \mathbf{K} + \sigma_n^2 \mathrm{Id}$, $\mathbf{\Sigma}_{AB} = \mathbf{k}_*$, $\mathbf{\Sigma}_{BA} = \mathbf{k}_*^T$ and $\Sigma_{BB} = k_{**}$. The formula gives
\begin{align}
p(f_* | \mathbf{t}) &= \mathcal{N}(\mu_{f_*}, \sigma_{f_*}^2) \\*
\mu_{f_*} &= 0 + \mathbf{k}_*^T (\mathbf{K} + \sigma_n^2 \mathrm{Id})^{-1} (\mathbf{t} - \mathbf{0}) \\
&= \mathbf{k}_*^T (\mathbf{K} + \sigma_n^2 \mathrm{Id})^{-1} \mathbf{t} \\*
\sigma_{f_*}^2 &= k_{**} - \mathbf{k}_*^T (\mathbf{K} + \sigma_n^2 \mathrm{Id})^{-1} \mathbf{k}_*.
\end{align}

These equations are the same as those that we derived with the kernel trick in Section~14.4. The prior and the posterior are both Gaussian, so the computation has a closed form.

Given this equivalence, why do we not start from the function-space view and avoid the matrix manipulations of the kernel trick? The function-space view is the more direct route to the final equations. The kernel-trick view shows that GP regression is linear regression in a (potentially infinite-dimensional) feature space, with kernels in place of explicit features. This links parametric models such as linear regression to non-parametric models such as GPs.

\paragraph{Non-Zero Mean and Heteroscedastic Noise}

The function-space view also helps when we extend the model. So far, we have used a zero-mean prior and homoscedastic noise, which means the same noise variance for all observations. We now derive the extensions to non-zero mean functions and heteroscedastic noise.

For a non-zero mean function $m(\mathbf{x})$, we start with the prior
\begin{equation}
f \sim \mathcal{GP}(m, k).
\end{equation}
The joint distribution of the training function values $\mathbf{f}$ and the test function value $f_*$ is then
\begin{equation}
\begin{bmatrix} \mathbf{f} \\ f_* \end{bmatrix} \sim \mathcal{N}\left( \begin{bmatrix} \mathbf{m} \\ m_* \end{bmatrix}, \begin{bmatrix} \mathbf{K} & \mathbf{k}_* \\ \mathbf{k}_*^T & k_{**} \end{bmatrix} \right),
\end{equation}
where $\mathbf{m} = [m(\mathbf{x}_1), \ldots, m(\mathbf{x}_N)]^T$ and $m_* = m(\mathbf{x}_*)$.

With noisy observations $\mathbf{t} = \mathbf{f} + \boldsymbol{\epsilon}$, the joint distribution becomes
\begin{equation}
\begin{bmatrix} \mathbf{t} \\ f_* \end{bmatrix} \sim \mathcal{N}\left( \begin{bmatrix} \mathbf{m} \\ m_* \end{bmatrix}, \begin{bmatrix} \mathbf{K} + \sigma_n^2 \mathrm{Id} & \mathbf{k}_* \\ \mathbf{k}_*^T & k_{**} \end{bmatrix} \right).
\end{equation}

The same blocks, with $\boldsymbol{\mu}_A = \mathbf{m}$ and $\mu_B = m_*$, give
\begin{align}
p(f_* | \mathbf{t}) &= \mathcal{N}(\mu_{f_*}, \sigma_{f_*}^2) \\*
\mu_{f_*} &= m_* + \mathbf{k}_*^T (\mathbf{K} + \sigma_n^2 \mathrm{Id})^{-1} (\mathbf{t} - \mathbf{m}) \label{eq:mean_nonzero} \\*
\sigma_{f_*}^2 &= k_{**} - \mathbf{k}_*^T (\mathbf{K} + \sigma_n^2 \mathrm{Id})^{-1} \mathbf{k}_*. \label{eq:var_nonzero}
\end{align}

We start from the prior mean $m_*$ at the test point and adjust it by the deviation $(\mathbf{t} - \mathbf{m})$ of the observations from their prior means. The weights of this adjustment combine the similarity to the training points with a correction for the correlations among them, through the inverse matrix.

For heteroscedastic noise, each observation has its own noise variance $\sigma_i^2$, and the observation model becomes
\begin{equation}
t_i = f(\mathbf{x}_i) + \epsilon_i, \quad \epsilon_i \sim \mathcal{N}(0, \sigma_i^2).
\end{equation}

The covariance matrix of the noise vector $\boldsymbol{\epsilon}$ is now a general covariance matrix $\mathbf{\Sigma}_n$ in place of $\sigma_n^2 \mathrm{Id}$. It is the matrix $\mathbf{S}$ of Section~14.2. For independent noise, $\mathbf{\Sigma}_n$ is diagonal with entries $\sigma_i^2$, but the formula holds for any positive definite covariance matrix. The joint distribution then becomes
\begin{equation}
\begin{bmatrix} \mathbf{t} \\ f_* \end{bmatrix} \sim \mathcal{N}\left( \begin{bmatrix} \mathbf{m} \\ m_* \end{bmatrix}, \begin{bmatrix} \mathbf{K} + \mathbf{\Sigma}_n & \mathbf{k}_* \\ \mathbf{k}_*^T & k_{**} \end{bmatrix} \right).
\end{equation}

The conditioning formula again gives
\begin{align}
p(f_* | \mathbf{t}) &= \mathcal{N}(\mu_{f_*}, \sigma_{f_*}^2) \\*
\mu_{f_*} &= m_* + \mathbf{k}_*^T (\mathbf{K} + \mathbf{\Sigma}_n)^{-1} (\mathbf{t} - \mathbf{m}) \\*
\sigma_{f_*}^2 &= k_{**} - \mathbf{k}_*^T (\mathbf{K} + \mathbf{\Sigma}_n)^{-1} \mathbf{k}_*.
\end{align}

This handles varying noise levels, because it gives less weight to noisier observations. In astronomy, measurement uncertainties can differ widely between instruments and observing conditions.

\paragraph{Predicting New Observations}

So far, we have derived $p(f_* | \mathbf{t})$, the distribution of the underlying function. To predict a new observation with its noise, such as what a telescope will measure, we need $p(y_* | \mathbf{t})$ as in Section~14.4. Since $y_* = f_* + \epsilon_*$ with $\epsilon_* \sim \mathcal{N}(0, \sigma_{n*}^2)$ independent of $f_*$, we integrate over $f_*$ and obtain
\begin{align}
p(y_* | \mathbf{t}) &= \int p(y_* | f_*) p(f_* | \mathbf{t}) \mathrm{d}f_* \\*
&= \int \mathcal{N}(y_* | f_*, \sigma_{n*}^2) \mathcal{N}(f_* | \mu_{f_*}, \sigma_{f_*}^2) \mathrm{d}f_*.
\end{align}
This integral is the density of the sum $f_* + \epsilon_*$ of two independent Gaussians. As in Section~5.7, the result is another Gaussian, given by
\begin{align}
p(y_* | \mathbf{t}) &= \mathcal{N}(y_* | \mu_{f_*}, \sigma_{f_*}^2 + \sigma_{n*}^2).
\end{align}
The mean does not change, and the variance increases by $\sigma_{n*}^2$ to account for the observation noise.

The extensions to non-zero means and heteroscedastic noise follow from the conditioning formula without new matrix identities. The posterior variance contains our prior uncertainty and the information gained from the observations. The function-space view therefore gives an uncertainty estimate directly, with no additional modeling. Together, the two perspectives show that a GP is both a linear model in a (potentially infinite-dimensional) feature space and a stochastic process.\index{Gaussian process (GP)!conditioning|)}

\section{Practical Implementation of Gaussian Processes}\label{sec:ch14_practical}

How do we implement Gaussian processes for real data? Our predictive formulas depend on terms like $(\mathbf{K} + \mathbf{\Sigma}_n)^{-1}$, so we need a way to compute them efficiently and with numerical stability.

The predictive\index{posterior predictive distribution} distribution of Section~14.7 at a test input $\mathbf{x}_*$ is
\begin{equation}
p(f_* | \mathbf{x}_*, \mathbf{X}, \mathbf{t}) = \mathcal{N}(\mu_{f_*}, \sigma_{f_*}^2),
\end{equation}
with mean and variance given by
\begin{align}
\mu_{f_*} &= m(\mathbf{x}_*) + \mathbf{k}_*^T (\mathbf{K} + \mathbf{\Sigma}_n)^{-1} (\mathbf{t} - \mathbf{m}) \\*
\sigma_{f_*}^2 &= k(\mathbf{x}_*, \mathbf{x}_*) - \mathbf{k}_*^T (\mathbf{K} + \mathbf{\Sigma}_n)^{-1} \mathbf{k}_*.
\end{align}
Here $\mathbf{t}$ holds the $N$ observed targets and $\mathbf{X}$ holds the $N$ training inputs. The vector $\mathbf{m} = [m(\mathbf{x}_1), \ldots, m(\mathbf{x}_N)]^T$ holds the prior mean at the training inputs. The $N \times N$ matrix $\mathbf{\Sigma}_n$ is the noise covariance, which equals $\sigma_n^2\mathrm{Id}$ for equal noise variances. The $N \times N$ matrix $\mathbf{K}$ and the vector $\mathbf{k}_*$ hold the kernel values $k(\mathbf{x}_i, \mathbf{x}_j)$ and $k(\mathbf{x}_i, \mathbf{x}_*)$. The numbers $m(\mathbf{x}_*)$ and $k(\mathbf{x}_*, \mathbf{x}_*)$ are scalars.

Here is the path from data to prediction.
\begin{enumerate}
    \item We start with the training inputs $\mathbf{X}$, the targets $\mathbf{t}$ and the measurement variances, which form the noise covariance $\mathbf{\Sigma}_n$.
    \item We choose a kernel (Sections~14.3 and 14.10) and fix its hyperparameters, the kernel parameters, by maximizing the marginal likelihood of Section~14.9.
    \item We factor $\mathbf{K} + \mathbf{\Sigma}_n$ once into a triangular matrix and its transpose, the Cholesky decomposition, and compute the vector $\boldsymbol{\alpha}$ of coefficients once.
    \item For each new input $\mathbf{x}_*$, we compute $\mathbf{k}_*$.
    \item The mean then needs one dot product with $\boldsymbol{\alpha}$, and the variance needs one forward substitution with the triangular matrix.
\end{enumerate}

Both the mean and the variance require the inverse matrix $(\mathbf{K} + \mathbf{\Sigma}_n)^{-1}$. For a dataset with $N$ observations, this is an $N \times N$ matrix, and direct inversion requires $\mathcal{O}(N^3)$ operations. In astronomy we often work with thousands of data points, and this computation can become expensive. Modern packages such as GPyTorch reduce this cost with approximations, usually at the price of some exactness. Here we treat the exact computation.

Direct matrix inversion is also numerically unstable. A positive semi-definite matrix is poorly conditioned when the ratio of its largest to its smallest eigenvalue is very large. When the kernel matrix is poorly conditioned, small numerical errors can lead to large errors in the final predictions. This can happen when data points are very close together or when the kernel parameters are poorly chosen. We need a more stable approach.

\paragraph{Cholesky Decomposition}

\looseness=1 The\index{Cholesky decomposition|(} Cholesky\index{numerical stability} decomposition suits our problem because $\mathbf{K} + \mathbf{\Sigma}_n$ has a special structure. The matrix is symmetric and positive definite, because valid kernels make $\mathbf{K}$ positive semi-definite and the noise covariance matrix $\mathbf{\Sigma}_n$ is positive definite.

Any symmetric positive definite matrix, and so $\mathbf{K} + \mathbf{\Sigma}_n$, can be decomposed as
\begin{equation}
\mathbf{K} + \mathbf{\Sigma}_n = \mathbf{L}\mathbf{L}^T.
\end{equation}
Here $\mathbf{L}$ is a lower triangular matrix called the Cholesky factor. This decomposition has three advantages. It needs only about $N^3/3$ operations, compared with about $2N^3$ for inverting a general matrix. It is more stable numerically, especially for poorly conditioned matrices. It also gives the predictive mean and variance efficiently.

The cost still grows as $\mathcal{O}(N^3)$. Compared with explicit inversion, this factor of six can turn an hour-long calculation into a ten-minute one, which helps during exploratory data analysis.

\paragraph{Predictive Mean and Variance}

We now use the Cholesky factor $\mathbf{L}$ to compute predictions efficiently without forming the inverse matrix. Instead of computing $(\mathbf{K} + \mathbf{\Sigma}_n)^{-1}$ and multiplying it by vectors, we solve triangular systems of equations, which keeps round-off errors small. For the predictive mean, we need to compute
\begin{equation}
\mu_{f_*} = m(\mathbf{x}_*) + \mathbf{k}_*^T (\mathbf{K} + \mathbf{\Sigma}_n)^{-1} (\mathbf{t} - \mathbf{m}).
\end{equation}

We define $\boldsymbol{\alpha} = (\mathbf{K} + \mathbf{\Sigma}_n)^{-1}(\mathbf{t} - \mathbf{m})$ without computing the inverse. Since $\mathbf{K} + \mathbf{\Sigma}_n = \mathbf{L}\mathbf{L}^T$, we have
\begin{equation}
\mathbf{L}\mathbf{L}^T \boldsymbol{\alpha} = \mathbf{t} - \mathbf{m}.
\end{equation}

We solve this system in two stages
\begin{align}
\mathbf{L}\mathbf{v} &= \mathbf{t} - \mathbf{m} \quad \text{(solve for $\mathbf{v}$ using forward substitution)} \\*
\mathbf{L}^T\boldsymbol{\alpha} &= \mathbf{v} \quad \text{(solve for $\boldsymbol{\alpha}$ using backward substitution)}.
\end{align}

Forward substitution solves a lower triangular system by working from top to bottom. For a system $\mathbf{L}\mathbf{v} = \mathbf{b}$, it computes
\begin{align}
v_1 &= b_1/L_{11} \\*
v_i &= \frac{1}{L_{ii}}\left(b_i - \sum_{j=1}^{i-1} L_{ij}v_j\right) \quad \text{for } i = 2, 3, \ldots, N.
\end{align}

Backward substitution solves an upper triangular system by working from bottom to top. For a system $\mathbf{L}^T\boldsymbol{\alpha} = \mathbf{v}$, it computes
\begin{align}
\alpha_N &= v_N/L_{NN} \\*
\alpha_i &= \frac{1}{L_{ii}}\left(v_i - \sum_{j=i+1}^{N} L_{ji}\alpha_j\right) \quad \text{for } i = N-1, N-2, \ldots, 1.
\end{align}

Both operations cost $\mathcal{O}(N^2)$ and are much faster than matrix inversion. Once we have $\boldsymbol{\alpha}$, the predictive mean is
\begin{equation}
\mu_{f_*} = m(\mathbf{x}_*) + \mathbf{k}_*^T \boldsymbol{\alpha}.
\end{equation}This has the form of Equation~\ref{eq:kernel_form}, with the prior mean added. The vector $\boldsymbol{\alpha}$ holds the coefficients $\alpha_i$ of the training points.

For the predictive variance, we follow a similar approach. We need
\begin{equation}
\sigma_{f_*}^2 = k(\mathbf{x}_*, \mathbf{x}_*) - \mathbf{k}_*^T (\mathbf{K} + \mathbf{\Sigma}_n)^{-1} \mathbf{k}_*.
\end{equation}

We first solve $\mathbf{L}\mathbf{q} = \mathbf{k}_*$ for a new vector $\mathbf{q}$ by forward substitution. We then compute
\begin{equation}
\sigma_{f_*}^2 = k(\mathbf{x}_*, \mathbf{x}_*) - \mathbf{q}^T\mathbf{q}.
\end{equation}

This works because
\begin{align}
\mathbf{q}^T\mathbf{q} &= \mathbf{k}_*^T (\mathbf{L}^T)^{-1} \mathbf{L}^{-1} \mathbf{k}_* \\*
&= \mathbf{k}_*^T (\mathbf{L}\mathbf{L}^T)^{-1} \mathbf{k}_* \\*
&= \mathbf{k}_*^T (\mathbf{K} + \mathbf{\Sigma}_n)^{-1} \mathbf{k}_*.
\end{align}

Thus we get the same answer as direct matrix inversion in exact arithmetic. In floating point, this form is more accurate for the variance, while the explicit inverse can give a negative variance for a poorly conditioned matrix.

We assume that the kernel and its hyperparameters are fixed. The complete algorithm for Gaussian process prediction then has three steps.

\begin{enumerate}
    \item \textbf{Cholesky decomposition.} Compute $\mathbf{L} = \text{cholesky}(\mathbf{K} + \mathbf{\Sigma}_n)$.

    \item \textbf{Predictive mean.}
    \begin{itemize}
        \item Solve $\mathbf{L}\mathbf{v} = \mathbf{t} - \mathbf{m}$ for $\mathbf{v}$ by forward substitution
        \item Solve $\mathbf{L}^T\boldsymbol{\alpha} = \mathbf{v}$ for $\boldsymbol{\alpha}$ by backward substitution
        \item Compute $\mu_{f_*} = m(\mathbf{x}_*) + \mathbf{k}_*^T \boldsymbol{\alpha}$
    \end{itemize}

    \item \textbf{Predictive variance.}
    \begin{itemize}
        \item Solve $\mathbf{L}\mathbf{q} = \mathbf{k}_*$ for $\mathbf{q}$ by forward substitution
        \item Compute $\sigma_{f_*}^2 = k(\mathbf{x}_*, \mathbf{x}_*) - \mathbf{q}^T\mathbf{q}$
    \end{itemize}
\end{enumerate}

Once the Cholesky decomposition and $\boldsymbol{\alpha}$ are computed, each new point needs only $\mathcal{O}(N)$ operations for the mean and $\mathcal{O}(N^2)$ for the variance. This makes it practical to predict at many test points, for example to plot the predicted function or to compute derived quantities in astronomical applications.

\looseness=1 The same formulas can serve as an emulator,\index{emulator} a fast stand-in for a slow simulator. The input $\mathbf{x}$ holds the parameters of the simulator, and $\mathbf{t}$ holds one output number from each of $N$ runs. A deterministic simulator has no measurement noise. Then $\mathbf{K} + \mathbf{\Sigma}_n$ can be numerically singular and the Cholesky decomposition can fail, so we put a small value on the diagonal of $\mathbf{\Sigma}_n$. To emulate a function with many outputs, such as a power spectrum in a cosmological study, we repeat the prediction for each output number.\index{Cholesky decomposition|)}

\section{Optimizing Hyperparameters}\label{sec:ch14_hyperparameters}

How\index{Gaussian process (GP)!hyperparameters|(}\index{marginal likelihood|(} do we select the optimal hyperparameters of our kernel function? Our derivations have assumed that these parameters are known, but in practice we must learn them from data. The choice of hyperparameters changes the functions that our GP can represent. To see why, consider modeling stellar variability with the composite kernel of Section~14.3, in which the time $t$ is the input
\begin{equation}
k(t, t') = \theta_0 \exp\left(-\frac{\theta_1}{2} |t - t'|^2\right) + \theta_2 + \theta_3 \exp\left(-\frac{2\sin^2(\pi|t-t'|/\theta_4)}{\theta_5^2}\right).
\end{equation}

The bold $\mathbf{t}$ below is the vector of targets. The parameters $\theta_0$ and $\theta_1$ belong to the RBF term, $\theta_2$ to the constant term, and $\theta_3$ to $\theta_5$ to the periodic term. Each parameter controls a different aspect of the function.

\begin{figure}[ht!]
    \centering
    \includegraphics[width=0.95\textwidth]{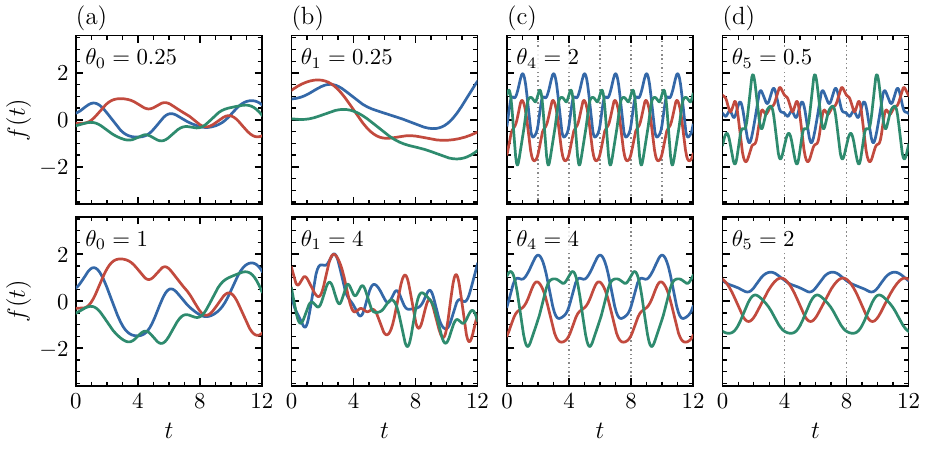}
    \caption{Panels~(a) and (b) use only the RBF term of the composite kernel, and panels~(c) and (d) use only its periodic term. In each panel one parameter takes a small value (top) and a large value (bottom), and the same random numbers generate both rows. The coefficient $\theta_0$ is the variance of the RBF term, so the vertical scale is $\sqrt{\theta_0}$, and $\theta_1$ is the inverse squared length scale. The period $\theta_4$ sets the spacing of the repetitions, marked by dotted lines. A small $\theta_5$ gives sharp features within each period, and a large $\theta_5$ gives nearly a sinusoid.}
    \label{fig:kernel_parameters}
\end{figure}

The coefficient $\theta_0$ is the variance of the RBF term and sets the size of the variations, and $\theta_1 = 1/\ell^2$ is the inverse squared length\index{length scale} scale. Small values of $\theta_1$ give smooth functions, and large values give rapid variations. The parameter $\theta_4$ sets the spacing of the periodic repetitions, and $\theta_5$ sets the shape of each period. Small values of $\theta_5$ give sharp features, and large values give nearly a sinusoid. Figure~\ref{fig:kernel_parameters} shows how these parameters shape the functions.

For stellar data, we want parameter values that make our observations likely under the resulting distribution of functions. If the data show rapid oscillations and we choose parameters that allow only smooth functions, the fit is poor. If we choose parameters that are too flexible for slowly varying data, the GP may overfit the noise.

Gaussian processes use the marginal likelihood to select the parameter values that best capture the patterns in our data. The idea is to treat hyperparameter selection as a model comparison problem. For any choice of hyperparameters $\boldsymbol{\theta}$, we ask how probable our observations are under this choice. This probability is the marginal likelihood
\begin{equation}
p(\mathbf{t} | \mathbf{X}, \boldsymbol{\theta}) = \mathcal{N}(\mathbf{t} | \mathbf{m}_{\boldsymbol{\theta}}, \mathbf{K}_{\boldsymbol{\theta}} + \mathbf{\Sigma}_n).
\end{equation}
Here $\mathbf{K}_{\boldsymbol{\theta}}$ is the kernel matrix, which depends on the hyperparameters, and $\mathbf{m}_{\boldsymbol{\theta}}$ holds the values of the mean function. This is the distribution of $\mathbf{t}$ in the joint Gaussian distribution of Section~14.7, now with the hyperparameters $\boldsymbol{\theta}$ made explicit.

The term ``marginal'' means that we have integrated out the latent function values. This leaves a direct expression for the probability of the data. The marginal likelihood, called the evidence in Section~11.8, balances model fit against complexity.

To see why, we examine its logarithm, which is
\begin{equation}
\ln p(\mathbf{t} | \mathbf{X}, \boldsymbol{\theta}) = -\frac{1}{2} (\mathbf{t} - \mathbf{m}_{\boldsymbol{\theta}})^T (\mathbf{K}_{\boldsymbol{\theta}} + \mathbf{\Sigma}_n)^{-1} (\mathbf{t} - \mathbf{m}_{\boldsymbol{\theta}}) - \frac{1}{2} \ln \det(\mathbf{K}_{\boldsymbol{\theta}} + \mathbf{\Sigma}_n) - \frac{N}{2} \ln(2\pi).
\end{equation}

This expression has two competing terms. The first, the data fit term, rewards models that explain the observations well. The second, the complexity penalty, penalizes overly flexible models. When the model gains flexibility, for example when we decrease the length scales, the determinant increases, and this reduces the likelihood.

The square root of the determinant sets the volume of data space that the model finds probable. A flexible model spreads its probability over a larger volume, so the observed data receive a lower density. This balance of fit and flexibility embodies Occam's\index{Occam's razor} razor. We find the simplest model that explains our data adequately. The BIC and AIC of Section~11.8 also trade a fit term against a complexity penalty, and BIC approximates this marginal likelihood for large samples. In BIC a determinant becomes the penalty $P\ln(N)$, while here the determinant of $\mathbf{K}_{\boldsymbol{\theta}} + \mathbf{\Sigma}_n$ appears directly.

\paragraph{Efficient Computation}

Computing the log marginal likelihood has the same difficulty as prediction. It needs a matrix inverse and a determinant, both expensive for large datasets. The Cholesky decomposition again solves the problem. With $\mathbf{K}_{\boldsymbol{\theta}} + \mathbf{\Sigma}_n = \mathbf{L} \mathbf{L}^T$, we compute the data fit term and the determinant efficiently.

For the data fit term, we reuse $\boldsymbol{\alpha} = (\mathbf{K}_{\boldsymbol{\theta}} + \mathbf{\Sigma}_n)^{-1}(\mathbf{t} - \mathbf{m}_{\boldsymbol{\theta}})$ from prediction, so that
\begin{equation}
(\mathbf{t} - \mathbf{m}_{\boldsymbol{\theta}})^T (\mathbf{K}_{\boldsymbol{\theta}} + \mathbf{\Sigma}_n)^{-1} (\mathbf{t} - \mathbf{m}_{\boldsymbol{\theta}}) = (\mathbf{t} - \mathbf{m}_{\boldsymbol{\theta}})^T \boldsymbol{\alpha}.
\end{equation}

For the determinant, we use the diagonal elements of the Cholesky factor, which give
\begin{equation}
\ln \det(\mathbf{K}_{\boldsymbol{\theta}} + \mathbf{\Sigma}_n) = 2\sum_{i=1}^N \ln L_{ii}.
\end{equation}

Together these results give the final efficient form
\begin{equation}
\ln p(\mathbf{t} | \mathbf{X}, \boldsymbol{\theta}) = -\frac{1}{2} (\mathbf{t} - \mathbf{m}_{\boldsymbol{\theta}})^T \boldsymbol{\alpha} - \sum_{i=1}^N \ln L_{ii} - \frac{N}{2} \ln(2\pi).
\end{equation}

\paragraph{Optimization Strategies}

\looseness=1 With an efficient way to compute the log marginal likelihood, we can maximize it over the hyperparameters $\boldsymbol{\theta}$. The best method depends on the dimensionality of the hyperparameter space and on the available computational resources. For a low-dimensional hyperparameter space, typically with 1 to 3 parameters, grid\index{grid search} search is a straightforward approach. We evaluate the log marginal likelihood on a predefined grid of hyperparameter values and select the combination with the highest value. For example, our kernel may have two hyperparameters, the amplitude $\sigma_f$ and the length scale $\ell$. We might then evaluate the likelihood on a $20 \times 20$ grid, which requires 400 evaluations.

\begin{figure}[ht!]
    \centering
    \includegraphics[width=0.9\textwidth]{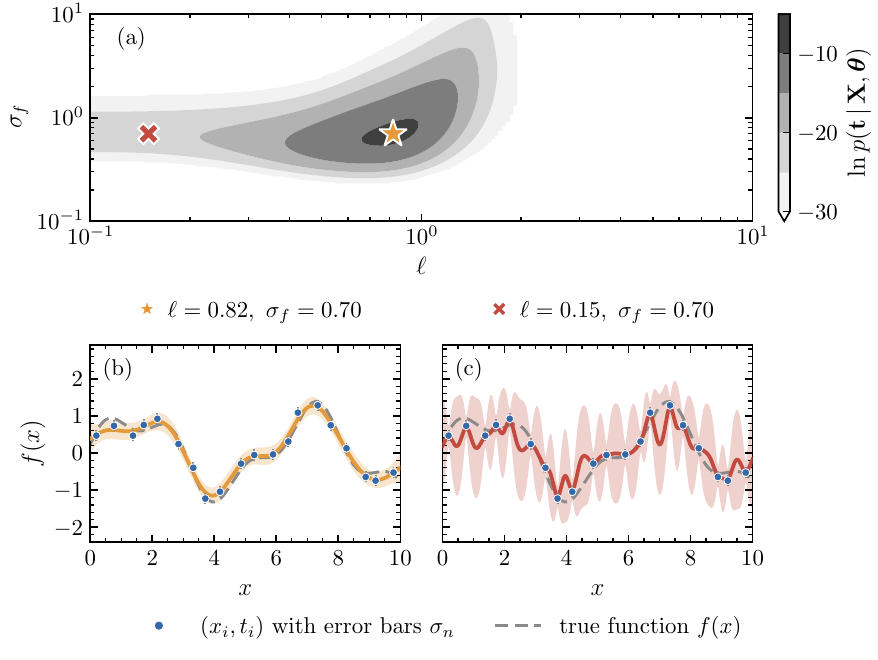}
    \caption{Hyperparameter optimization for the RBF kernel $k(x, x') = \sigma_f^2 \exp(-(x - x')^2 / (2\ell^2))$ with amplitude $\sigma_f$ and length scale $\ell$. Panel~(a) shows the log marginal likelihood on a grid of $\ell$ and $\sigma_f$, and darker regions have a higher likelihood. The star marks the maximum, and the cross marks a poor choice with the same amplitude and a shorter length scale. Panels~(b) and (c) show the GP mean and the 95 percent band of the latent function for the two choices. The mean in panel~(b) follows the true function that generated the noisy observations. In panel~(c) the shorter length scale overfits, with a spiky mean that returns toward the prior mean and a wide band between the observations.}
    \label{fig:gp_hyperparameter_optimization}
\end{figure}

Grid search has three advantages. It is simple to implement and to parallelize on several processors. It visualizes the likelihood surface, which can reveal features such as several peaks. It also finds the best point on the grid. The visualization is especially useful for understanding how different hyperparameters interact and for diagnosing problems with the optimization. Figure~\ref{fig:gp_hyperparameter_optimization} shows this surface for an RBF kernel, together with GP fits for a good and a poor choice of hyperparameters.

However, grid search becomes too expensive as the number of hyperparameters increases. With $P$ hyperparameters and $G$ grid points per dimension, we need $G^P$ evaluations. This exponential growth quickly becomes impractical.

For more than three hyperparameters, we use gradient-based optimization. The basic approach, gradient ascent (Section~9.4), updates the hyperparameters iteratively in the direction of the gradient
\begin{equation}
\boldsymbol{\theta}^{(\tau+1)} = \boldsymbol{\theta}^{(\tau)} + \eta_\tau \nabla_{\boldsymbol{\theta}} \ln p(\mathbf{t} | \mathbf{X}, \boldsymbol{\theta}^{(\tau)}).
\end{equation}
Here $\eta_\tau$ is the learning rate at iteration $\tau$. This approach scales much better than grid search, because the cost of one iteration grows only linearly with the number $P$ of hyperparameters.

For specific kernels, we can compute the gradient of the log marginal likelihood analytically, but this becomes tedious for complex composite kernels. Modern implementations therefore often use automatic\index{automatic differentiation} differentiation to compute the gradients. Automatic differentiation splits the calculation into elementary operations, such as additions and exponentials, and applies the chain rule to each. The derivatives are exact up to rounding.

Python libraries such as TensorFlow, PyTorch,\index{PyTorch} and JAX compute the gradients of arbitrary calculations automatically. Automatic gradients also make it practical to experiment with complex kernel structures. Section~15.4 treats automatic differentiation in much greater detail for neural networks, where it plays a central role in training.

\looseness=1 {\sloppy Hyperparameter optimization shows an advantage of Gaussian processes. The marginal likelihood is an objective criterion for choosing the hyperparameters, and it balances the complexity of the model against the fit to the data. We can therefore often choose the hyperparameters without ad hoc regularization terms or complex cross-validation procedures, which many other machine learning methods\index{Gaussian process (GP)!hyperparameters|)}\index{marginal likelihood|)} need.\par}

\section{Model Selection for Gaussian Process Regression}\label{sec:ch14_model_selection}

Hyperparameter\index{Gaussian process (GP)!model selection|(} optimization works within a fixed kernel structure. How do we choose the kernel structure itself?

Stellar variability might be driven by several physical processes on different timescales, such as pulsations, rotation, magnetic activity, and evolutionary changes. The choice of kernel structure connects directly to the model selection principles of earlier chapters (Sections~4.8 and 11.8). We must balance the complexity of a model against its explanatory power. We must avoid underfitting, which misses important patterns, and overfitting, which models noise as signal.

For modeling a stellar light curve, we might consider these four possibilities.

\begin{enumerate}
    \item A simple RBF kernel, $k(t, t') = \sigma_f^2 \exp\left(-(t-t')^2/(2\ell^2)\right)$
    \item A periodic kernel with period $p$, $k(t, t') = \sigma_f^2 \exp\left(-2\sin^2(\pi|t-t'|/p)/\ell^2\right)$
    \item A composite kernel with RBF, constant, and periodic components
    \item A Matérn kernel with different smoothness parameters
\end{enumerate}

Each choice makes different assumptions about the underlying physical processes. The RBF kernel assumes smooth, aperiodic variations. The periodic kernel assumes regular, repeating patterns. The composite kernel allows several processes to operate at the same time.

A further difficulty is that a more complex kernel can fit the data at least as well as a simpler one. A composite kernel with many components can model the same patterns as a simple kernel, and more. Without a penalty for complexity, we would choose the most flexible option, and this leads to overfitting.

Gaussian processes address this problem with Bayesian model comparison. For each candidate kernel structure $\mathcal{M}_i$, we can compute the marginal\index{marginal likelihood} likelihood
\begin{equation}
p(\mathbf{t} | \mathbf{X}, \mathcal{M}_i) = \int p(\mathbf{t} | \mathbf{X}, \boldsymbol{\theta}, \mathcal{M}_i) p(\boldsymbol{\theta} | \mathcal{M}_i) \mathrm{d}\boldsymbol{\theta}.
\end{equation}
Here $p(\boldsymbol{\theta} | \mathcal{M}_i)$ is the prior over the hyperparameters of model $\mathcal{M}_i$. In Section~14.9 the marginal likelihood is conditioned on fixed hyperparameters, whereas this one averages over them.

In practice, we typically approximate this integral by the marginal likelihood at the optimal hyperparameters
\begin{equation}
p(\mathbf{t} | \mathbf{X}, \mathcal{M}_i) \approx p(\mathbf{t} | \mathbf{X}, \boldsymbol{\theta}^*_i).
\end{equation}
Here $\boldsymbol{\theta}^*_i$ maximizes the marginal likelihood of model $\mathcal{M}_i$. The full integral also penalizes models with more hyperparameters, because it averages the likelihood over a larger space of hyperparameter values. The maximum drops the penalty for the number of hyperparameters. A kernel that contains a simpler kernel as a limit therefore does not score lower.

\looseness=1 The marginal likelihood still penalizes flexible covariance through the determinant term that we examined in Section~14.9. Adding a covariance component raises the determinant when the other parameters stay fixed. For kernels that do not contain one another, the determinant term is the only penalty for complexity. Kernel selection thus adds a second step to hyperparameter optimization. We first optimize the hyperparameters of each structure and then compare the structures with the same objective function.

\paragraph{Information Criteria}

The marginal likelihood is one Bayesian approach to model comparison. But we sometimes want additional safeguards against overfitting, or we must work with approximations to the full Bayesian treatment. Information\index{information criteria} criteria add an explicit penalty for the number of hyperparameters, which the maximized marginal likelihood lacks.

Two simple options are the Bayesian information criterion (BIC) and the Akaike information criterion (AIC) of Section~11.8, which are defined as
\begin{align}
\text{BIC} &= -2\ln(\hat{L}) + P\ln(N) \\*
\text{AIC} &= -2\ln(\hat{L}) + 2P.
\end{align}
Here $\hat{L}$ is the maximized marginal likelihood, $P$ is the number of hyperparameters, and $N$ is the sample size. We choose the model with the lowest criterion value.

The BIC penalty depends on the sample size, so BIC penalizes complexity more for large datasets. For Gaussian processes, these criteria are particularly useful when we compare models with very different numbers of hyperparameters. A simple RBF kernel has 2 to 3 hyperparameters, while a complex composite kernel might have 10 or more.

\paragraph{Cross-Validation}

Cross-validation is another approach. It measures the predictive performance directly and does not rely on theoretical penalties for complexity. For Gaussian processes, we typically use $K$-fold cross-validation (Section~\ref{sec:cross_validation}). It divides the dataset into $K$ subsets, called folds, trains on $K-1$ folds, and tests on the remaining fold. We average a score over the $K$ test folds.

A suitable score for GPs is the negative\index{negative log predictive density (NLPD)} log predictive density (NLPD)\index{NLPD|see{negative log predictive density}} of a test fold, which is
\begin{equation}
\text{NLPD} = -\frac{1}{|\mathcal{D}_{\text{test}}|} \sum_{(\mathbf{x}_i, t_i) \in \mathcal{D}_{\text{test}}} \ln p(t_i | \mathbf{x}_i, \mathcal{D}_{\text{train}}).
\end{equation}Here $|\mathcal{D}_{\text{test}}|$ is the number of test points of the fold, and $p(t_i | \mathbf{x}_i, \mathcal{D}_{\text{train}})$ is the density that the GP trained on $\mathcal{D}_{\text{train}}$ assigns to $t_i$. A lower NLPD means better predictions. NLPD evaluates both the point predictions and the uncertainty estimates, which suits Gaussian processes, where uncertainty quantification is an important advantage. A model with confident but wrong predictions is penalized more heavily than a model that expresses appropriate uncertainty.

Cross-validation for GPs can be computationally expensive, because each fold requires training a model on approximately $(K-1)/K \times N$ data points. However, it gives direct evidence of the predictive performance, and it can reveal overfitting that marginal likelihood comparisons alone may not show. We use the marginal likelihood to find the model that best explains the observed data. We use cross-validation to find the model that is most likely to perform well on future observations.

In practice, effective model selection for Gaussian processes combines several approaches. A typical workflow starts with simple kernels (RBF, periodic, linear) and compares them with the marginal likelihood. We then construct composite models by combining the promising components. Cross-validation provides the final validation of the predictive performance.

\looseness=1 For astronomical applications, we focus on kernel combinations that correspond to the plausible physical processes of Section~14.3, instead of testing many possible combinations. This reduces the computational burden while keeping the models scientifically interpretable.\index{Gaussian process (GP)!model selection|)}

\section{Gaussian Process Classification}\label{sec:ch14_classification}

Gaussian\index{Gaussian process (GP)!classification|(} process regression predicts continuous outputs, such as stellar brightness or radial velocity. Many astronomical problems need classification\index{Gaussian process (GP)!classification} instead, which assigns an object to a discrete category. We might classify stars by spectral type, separate stars from galaxies in photometric surveys, or identify different types of variable objects from their light curves. Can we extend Gaussian processes to classification? The step resembles the move from linear regression to logistic regression (Section~7.3), because we again need a special technique for binary outcomes.

\looseness=1 In regression, the observations are continuous values that we model as Gaussian random variables. We write $t_i = f(\mathbf{x}_i) + \epsilon_i$, where $f$ is the latent\index{latent function} function and $\epsilon_i$ is Gaussian noise. In classification, the observations are discrete class labels, typically 0 and 1 for binary problems. We cannot write $t_i = f(\mathbf{x}_i) + \epsilon_i$ for class labels, because a continuous function plus continuous noise gives continuous values. Such values cannot be the discrete labels 0 and 1. We need a different relationship between the latent function and the observed outcomes.

The solution, familiar from logistic regression, is a link\index{link function} function, such as the sigmoid, that connects continuous function values to class probabilities. We assume an underlying continuous function $a(\mathbf{x})$, often called the ``logit'' or ``score'' function, which determines the class probabilities through the link function. We keep the Gaussian process for the latent function, which can take any real value, and we handle the discrete observations through an appropriate likelihood.

\paragraph{From Linear to Nonlinear Boundaries}

To understand the motivation for Gaussian process classification, we first recall the limitation of logistic regression (Section~7.5). Logistic regression models the class probability as
\begin{equation}
p(t=1|\mathbf{x}, \mathbf{w}) = \sigma(\mathbf{w}^T\boldsymbol{\phi}(\mathbf{x})),
\end{equation}
where $\sigma(a) = 1/(1+e^{-a})$ is the sigmoid function and $\boldsymbol{\phi}(\mathbf{x})$ represents the feature mapping.

The main limitation concerns the decision\index{decision boundary!nonlinear} boundary, the set of points where $p(t=1|\mathbf{x}) = 0.5$. It lies where $\mathbf{w}^T\boldsymbol{\phi}(\mathbf{x}) = 0$, which defines a hyperplane in feature space. This restricts us to boundaries that are linear in the feature space. Even with carefully designed features, many real classification problems need nonlinear boundaries that are hard to describe with explicit feature mappings.

Consider the separation of stars from galaxies by their photometric colors. The true boundary between the two classes in color space is often curved and complex. It depends on redshift effects, stellar populations and observational selection functions. A linear classifier might capture the overall separation but miss important substructure.

Gaussian process classification lifts this restriction. It replaces the linear function $\mathbf{w}^T\boldsymbol{\phi}(\mathbf{x})$ with a nonlinear function $a(\mathbf{x})$ that follows a Gaussian process prior
\begin{equation}
a(\mathbf{x}) \sim \mathcal{GP}(0, k(\mathbf{x}, \mathbf{x}')).
\end{equation}

For simplicity, we use a zero mean function, which can be generalized. The sigmoid turns the logit $a(\mathbf{x})$ into a probability, so that
\begin{equation}
p(t=1|\mathbf{x}, a) = \sigma(a(\mathbf{x})).
\end{equation}

\begin{figure}[ht!]
    \centering
    \includegraphics[width=0.95\textwidth]{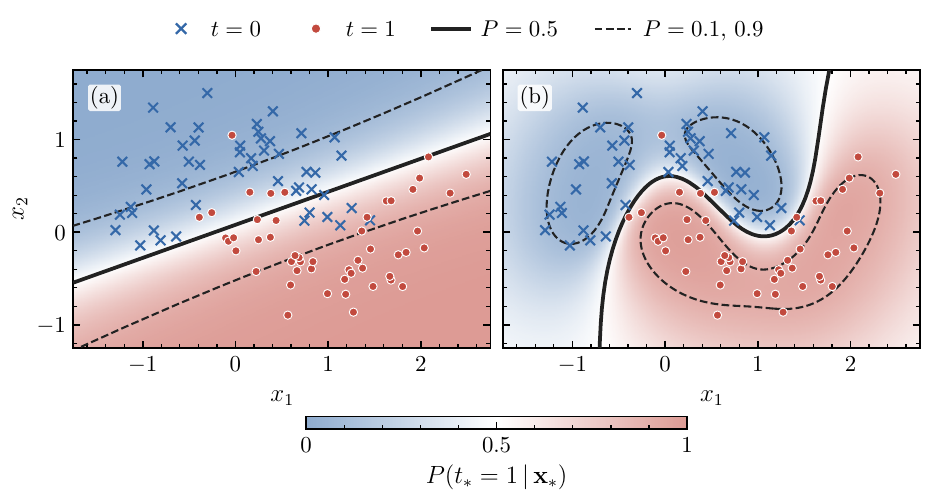}
    \caption{Classification of two noisy moon-shaped classes with (a) a linear classifier, Bayesian logistic regression of Chapter~9, and (b) a Gaussian process classifier. The color shows the predicted probability $P(t_* = 1 | \mathbf{x}_*)$ of class 1. The solid line is the decision boundary. The linear classifier has a straight boundary, cannot follow the moons, and stays confident far from the data on both sides of the boundary. The GP classifier has a curved boundary, and its probability returns toward 0.5 far from the data.}
    \label{fig:gp_classification_comparison}
\end{figure}

The kernel function gives the flexibility to learn complex decision boundaries, and the probabilistic model gives the uncertainty of each classification. Figure~\ref{fig:gp_classification_comparison} compares the decision boundary and the predicted class probability of a linear classifier with those of a Gaussian process classifier.

The model passes information from the input features to the class labels in three steps.

\begin{enumerate}
    \item The Gaussian process maps the input features $\mathbf{x}_i$ to latent function values $a_i = a(\mathbf{x}_i)$. These latent values can take any real number and represent the ``score'' for class membership before the conversion to probabilities.

    \item The sigmoid function converts the latent values to class probabilities $p_i = \sigma(a_i)$, which lie between 0 and 1.

    \item We observe discrete class labels $t_i$ that are sampled from Bernoulli distributions with probabilities $p_i$.
\end{enumerate}

\begin{figure}[ht!]
    \centering
    \includegraphics[width=0.95\textwidth]{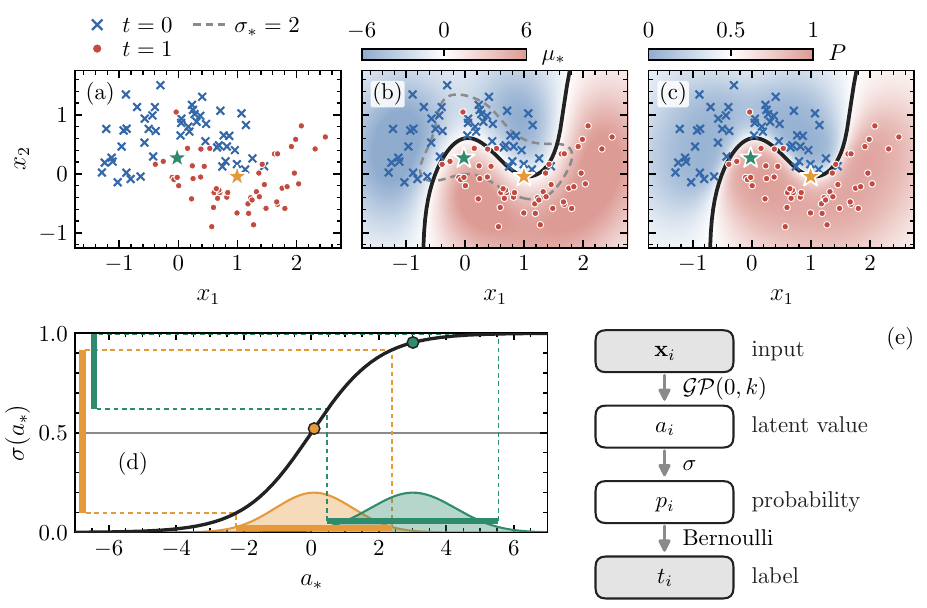}
    \caption{Information flow in GP classification for the data of Figure~\ref{fig:gp_classification_comparison}, from the input to the label in panel~(e). Panels~(a) to (c) mark two test points with stars, an orange one on the decision boundary and a green one inside class 1. Panels~(b) and (c) show the posterior mean $\mu_*$ of the latent value and the probability of class 1, from the Laplace method below. The solid curve is the decision boundary, and the dashed curve in panel~(b) marks where the posterior standard deviation $\sigma_*$ is 2. Panel~(d) shows the posterior of $a_*$ at the two stars, the interval $\mu_* \pm 2\sigma_*$ (horizontal bars) and its image under the sigmoid (vertical bars). Both stars have a similar $\sigma_*$, yet the interval is nearly symmetric at the orange star and narrower and asymmetric at the green one.}
    \label{fig:gp_classification_flow}
\end{figure}

In the first step, the kernel function encodes our assumptions about which inputs should receive similar scores. The uncertainty of the latent function, given by the posterior variance of the Gaussian process, carries over to the class probability through the sigmoid. Near a decision boundary, where $a(\mathbf{x}) \approx 0$, small changes in the latent function produce large changes in the class probability. The model then expresses high uncertainty. Far from the boundary, where $|a(\mathbf{x})|$ is large, the sigmoid saturates, and the uncertainty of the latent function has less effect on the class probability. Figure~\ref{fig:gp_classification_flow} shows the three steps in panel~(e) and this effect in panel~(d).

This propagation of uncertainty makes Gaussian process classification useful for astronomical applications that must identify objects that are difficult to classify. Instead of forcing a discrete classification, the model can express uncertainty about ambiguous cases. Follow-up observations can then target the cases where they help most.

The nonlinear link function makes exact Bayesian inference intractable. In GP regression, a Gaussian likelihood keeps the posterior Gaussian, a property called conjugacy (Section~5.2). The sigmoid creates a non-Gaussian posterior that needs approximation techniques. This complexity is the price of nonlinear decision boundaries.

\section{The Predictive Distribution in GP Classification}\label{sec:ch14_predictive}

The\index{posterior predictive distribution|(} computation of predictions needs more care, because the latent values relate nonlinearly to the class probabilities. Our goal is the probability that a new test point $\mathbf{x}_*$ belongs to class 1, given the training data $\{\mathbf{X}, \mathbf{t}\}$. This is the predictive distribution
\begin{equation}
p(t_*=1|\mathbf{x}_*, \mathbf{X}, \mathbf{t}).
\end{equation}

To see why classification is harder than regression, we follow the four steps of the inference.

\begin{enumerate}
    \item We observe inputs $\mathbf{X}$ and discrete class labels $\mathbf{t} \in \{0,1\}^N$.
    \item We use these data to infer the distribution of the logit values $\mathbf{a}$ at the training points.
    \item The trained Gaussian process predicts the logit value $a_*$ at the test point $\mathbf{x}_*$.
    \item We convert the predicted logit to a class probability with the sigmoid function.
\end{enumerate}

The problem comes from the sigmoid of step 2. We observe discrete labels, which connect to the latent function through the nonlinear sigmoid. The posterior over the latent function values is then non-Gaussian, and we cannot compute it analytically.

To formalize the challenge, we apply the sum and product rules of probability to the predictive distribution. We do not observe the logit values directly, so we marginalize over all possible values of $a_*$ and obtain
\begin{equation}
p(t_*=1|\mathbf{x}_*, \mathbf{X}, \mathbf{t}) = \int p(t_*=1|a_*) p(a_*|\mathbf{x}_*, \mathbf{X}, \mathbf{t}) \mathrm{d}a_*.
\end{equation}

The first term in the integral is straightforward, $p(t_*=1|a_*) = \sigma(a_*)$. The challenge is the second term, $p(a_*|\mathbf{x}_*, \mathbf{X}, \mathbf{t})$, which requires us to infer the latent function distribution from discrete observations.

The sum rule again, now over the latent values $\mathbf{a}$ at the training points, gives
\begin{equation}
p(a_*|\mathbf{x}_*, \mathbf{X}, \mathbf{t}) = \int p(a_*|\mathbf{a}, \mathbf{x}_*, \mathbf{X}) p(\mathbf{a}|\mathbf{X}, \mathbf{t}) \mathrm{d}\mathbf{a}.
\end{equation}

The first factor is a standard GP prediction
\begin{equation}
p(a_*|\mathbf{a}, \mathbf{x}_*, \mathbf{X}) = \mathcal{N}(a_*|\mu_{a_*|\mathbf{a}}, \sigma_{a_*|\mathbf{a}}^2),
\end{equation}
where the mean and variance follow from Gaussian conditioning, as in Section~14.7 without noise.

The problematic factor is $p(\mathbf{a}|\mathbf{X}, \mathbf{t})$, the posterior distribution of the latent function values given the observed class labels. Bayes' rule gives
\begin{equation}
p(\mathbf{a}|\mathbf{X}, \mathbf{t}) = \frac{p(\mathbf{t}|\mathbf{a}) p(\mathbf{a}|\mathbf{X})}{p(\mathbf{t}|\mathbf{X})}.
\end{equation}

The prior $p(\mathbf{a}|\mathbf{X}) = \mathcal{N}(\mathbf{0}, \mathbf{K})$ is Gaussian. The likelihood $p(\mathbf{t}|\mathbf{a}) = \prod_{i=1}^N \sigma(a_i)^{t_i}(1-\sigma(a_i))^{1-t_i}$ is a product of sigmoid functions. A Gaussian prior times this non-Gaussian likelihood is no longer Gaussian.

\paragraph{The Laplace Approximation}

We cannot remove the sigmoid, so we approximate the non-Gaussian posterior $p(\mathbf{a}|\mathbf{X}, \mathbf{t})$ by a Gaussian. We use the Laplace\index{Laplace approximation} approximation, which we introduced in Sections~9.3 and 9.4 for Bayesian logistic regression.

The Laplace approximation has three steps.
\begin{enumerate}
    \item We find the mode $\mathbf{a}_{\text{MAP}}$ of the posterior distribution, the maximum a posteriori (MAP) estimate.
    \item We compute the Hessian\index{Hessian matrix} matrix, the matrix of second derivatives, of the negative log posterior at this mode.
    \item We construct a Gaussian approximation from the mode and the Hessian.
\end{enumerate}

To find the mode, we maximize the log posterior, which is given by
\begin{equation}
\ln p(\mathbf{a}|\mathbf{X}, \mathbf{t}) = \sum_{i=1}^N [t_i \ln \sigma(a_i) + (1-t_i) \ln (1-\sigma(a_i))] - \frac{1}{2}\mathbf{a}^T \mathbf{K}^{-1} \mathbf{a} + \text{const}.
\end{equation}

The first term of the log posterior is the log likelihood of the labels, and the second term comes from the Gaussian prior. We set the gradient to zero. Both the gradient and the Hessian use the derivative of the sigmoid, which Section~7.7 derived. For the likelihood term, the partial derivative with respect to each $a_i$ is
\begin{align}
\frac{\partial}{\partial a_i} [t_i \ln \sigma(a_i) + (1-t_i) \ln (1-\sigma(a_i))] &= t_i \frac{\sigma'(a_i)}{\sigma(a_i)} + (1-t_i) \frac{-\sigma'(a_i)}{1-\sigma(a_i)}.
\end{align}

Using $\sigma'(a_i) = \sigma(a_i)(1-\sigma(a_i))$, this simplifies to
\begin{align}
&= t_i \frac{\sigma(a_i)(1-\sigma(a_i))}{\sigma(a_i)} - (1-t_i) \frac{\sigma(a_i)(1-\sigma(a_i))}{1-\sigma(a_i)} \\*
&= t_i(1-\sigma(a_i)) - (1-t_i)\sigma(a_i) \\
&= t_i - t_i\sigma(a_i) - \sigma(a_i) + t_i\sigma(a_i) \\*
&= t_i - \sigma(a_i).
\end{align}

Setting the full gradient to zero gives
\begin{equation}
\nabla_{\mathbf{a}} \ln p(\mathbf{a}|\mathbf{X}, \mathbf{t}) = \mathbf{t} - \boldsymbol{\sigma}(\mathbf{a}) - \mathbf{K}^{-1}\mathbf{a} = \mathbf{0},
\end{equation}
where $\boldsymbol{\sigma}(\mathbf{a}) = [\sigma(a_1), \sigma(a_2), \ldots, \sigma(a_N)]^T$.

The sigmoid is nonlinear, so this equation has no closed-form solution. We solve it iteratively by gradient ascent with learning rate $\eta$, which gives the update
\begin{equation}
\mathbf{a}^{(\tau+1)} = \mathbf{a}^{(\tau)} + \eta \left(\mathbf{t} - \boldsymbol{\sigma}(\mathbf{a}^{(\tau)}) - \mathbf{K}^{-1}\mathbf{a}^{(\tau)}\right).
\end{equation}

\looseness=-1 This optimization problem is concave. The functions $\ln \sigma(a_i)$ and $\ln (1 - \sigma(a_i))$ are concave, so the log-likelihood term is concave. The term $-\mathbf{a}^T \mathbf{K}^{-1} \mathbf{a}/2$ is strictly concave, because $\mathbf{K}$ is positive definite, which we assume so that $\mathbf{K}^{-1}$ exists. The sum is strictly concave, so the maximum is unique and global. The iteration converges to it for a small enough step $\eta$.

The Hessian matrix gives the local curvature of the log posterior. For the diagonal elements of the likelihood term, the second derivative is
\begin{align}
\frac{\partial^2}{\partial a_i^2} \sum_{j=1}^N [t_j \ln \sigma(a_j) + (1-t_j) \ln (1-\sigma(a_j))] &= \frac{\partial}{\partial a_i} (t_i - \sigma(a_i)) \\*
&= -\sigma'(a_i) \\*
&= -\sigma(a_i)(1-\sigma(a_i)).
\end{align}

For the off-diagonal elements, each term in the sum depends on a single $a_i$, so the mixed partial derivatives of the likelihood term are zero. The prior term $-\mathbf{a}^T \mathbf{K}^{-1} \mathbf{a}/2$ contributes $\mathbf{K}^{-1}$. The Hessian of the negative log posterior at the mode is
\begin{equation}
\mathbf{H} = -\nabla_{\mathbf{a}}\nabla_{\mathbf{a}}^T \ln p(\mathbf{a}|\mathbf{X}, \mathbf{t})|_{\mathbf{a} = \mathbf{a}_{\text{MAP}}} = \mathbf{W} + \mathbf{K}^{-1},
\end{equation}
where $\mathbf{W} = \text{diag}(\sigma(a_1)(1-\sigma(a_1)), \ldots, \sigma(a_N)(1-\sigma(a_N)))$ contains the second derivatives of the negative log-likelihood at $\mathbf{a}_{\text{MAP}}$.

The Laplace approximation of the posterior is then
\begin{equation}
p(\mathbf{a}|\mathbf{X}, \mathbf{t}) \approx \mathcal{N}(\mathbf{a}|\mathbf{a}_{\text{MAP}}, \mathbf{H}^{-1}).
\end{equation}

With a Gaussian approximation for $p(\mathbf{a}|\mathbf{X}, \mathbf{t})$, we can now carry out step 3 and compute the predictive distribution of the logit. It is the integral
\begin{equation}
p(a_*|\mathbf{x}_*, \mathbf{X}, \mathbf{t}) = \int p(a_*|\mathbf{x}_*, \mathbf{X}, \mathbf{a}) p(\mathbf{a}|\mathbf{X}, \mathbf{t}) \mathrm{d}\mathbf{a}.
\end{equation}

Substituting the GP prediction of $a_*$ given $\mathbf{a}$ and the Laplace approximation gives
\begin{equation}
p(a_*|\mathbf{x}_*, \mathbf{X}, \mathbf{t}) \approx \int \mathcal{N}(a_*|\mathbf{k}_*^T \mathbf{K}^{-1} \mathbf{a}, k_{**} - \mathbf{k}_*^T \mathbf{K}^{-1} \mathbf{k}_*) \mathcal{N}(\mathbf{a}|\mathbf{a}_{\text{MAP}}, \mathbf{H}^{-1}) \mathrm{d}\mathbf{a}.
\end{equation}

The integral combines two Gaussians into another Gaussian, so that
\begin{align}
p(a_*|\mathbf{x}_*, \mathbf{X}, \mathbf{t}) &\approx \mathcal{N}(a_*|\mu_*, \sigma_*^2) \\*
\mu_* &= \mathbf{k}_*^T \mathbf{K}^{-1} \mathbf{a}_{\text{MAP}} \\*
\sigma_*^2 &= k_{**} - \mathbf{k}_*^T \mathbf{K}^{-1} \mathbf{k}_* + \mathbf{k}_*^T \mathbf{K}^{-1} \mathbf{H}^{-1} \mathbf{K}^{-1} \mathbf{k}_*.
\end{align}

The gradient condition at $\mathbf{a}_{\text{MAP}}$ simplifies the mean. The Woodbury\index{Woodbury identity} identity of Equation~\ref{eq:woodbury}, with $\mathbf{A} = \mathbf{K}$, $\mathbf{U} = \mathbf{V} = \mathrm{Id}$ and $\mathbf{C} = \mathbf{W}^{-1}$, simplifies the variance. This gives
\begin{align}
\mu_* &= \mathbf{k}_*^T (\mathbf{t} - \boldsymbol{\sigma}(\mathbf{a}_{\text{MAP}})) \\*
\sigma_*^2 &= k_{**} - \mathbf{k}_*^T (\mathbf{W}^{-1} + \mathbf{K})^{-1} \mathbf{k}_*.
\end{align}

We now have the distribution of the logit $a_*$, but step 4 needs the class probability, which is
\begin{equation}
p(t_*=1|\mathbf{x}_*, \mathbf{X}, \mathbf{t}) = \int \sigma(a_*) p(a_*|\mathbf{x}_*, \mathbf{X}, \mathbf{t}) \mathrm{d}a_*.
\end{equation}

This integral has no closed form for the sigmoid. However, we can use the probit\index{probit approximation} approximation of Section~9.6. It approximates the logistic sigmoid closely by the probit function $\Phi(\lambda a)$, where $\Phi$ is the standard normal cumulative distribution function and $\lambda = \sqrt{\pi/8}$. This $\Phi$ is unrelated to the design matrix $\mathbf{\Phi}$.

Inserting this approximation into the integral and using the probit identity, we obtain
\begin{equation}
p(t_*=1|\mathbf{x}_*, \mathbf{X}, \mathbf{t}) \approx \sigma\left(\frac{\mu_*}{\sqrt{1 + \pi \sigma_*^2/8}}\right).
\end{equation}

\paragraph{Interpreting the Class Probabilities}

The factor $1/\sqrt{1 + \pi \sigma_*^2/8}$ is an uncertainty moderation term. It is the same mechanism that we met in Bayesian logistic regression in Sections~9.6 and 9.7. When $\sigma_*^2$ is large, for example far from the training data, the latent function is uncertain, and the factor shrinks $\mu_*$ toward zero. The prediction then moves toward $\sigma(0) = 0.5$, which is maximum uncertainty. Near many training points, $\sigma_*^2$ is smaller, the factor is closer to 1, and the prediction is closer to $\sigma(\mu_*)$. This moderation is an advantage of the full Bayesian treatment, because simpler methods can make overconfident predictions.

The mean $\mu_* = \mathbf{k}_*^T (\mathbf{t} - \boldsymbol{\sigma}(\mathbf{a}_{\text{MAP}}))$ is a weighted sum of the residuals at the training points. A training point whose label is far from the MAP prediction has a large residual and contributes more to the prediction at a new point. The weights are the kernel similarities $\mathbf{k}_*$ between the new point and the training points. The residuals carry the correlations among the training points, because $\mathbf{a}_{\text{MAP}}$ depends on the kernel matrix.

\looseness=1 The variance $\sigma_*^2 = k_{**} - \mathbf{k}_*^T (\mathbf{W}^{-1} + \mathbf{K})^{-1} \mathbf{k}_*$ has the form of the GP regression variance, with $\mathbf{W}^{-1}$ in place of the noise covariance $\mathbf{\Sigma}_n$. The matrix $\mathbf{W}$ measures how sensitive each observation is to changes in the latent function. Points near decision boundaries carry the most information about where the boundaries lie, so they reduce the uncertainty the most.\index{posterior predictive distribution|)}

\section{Extending to Multiclass Classification}

\looseness=1 Many astronomical applications need more than two classes, for example stellar spectral types, galaxy morphologies or classes of variable stars. Can we extend Gaussian process classification beyond the binary case? Naive approaches such as one-vs-rest can give inconsistent decision boundaries and conflicting predictions where the binary classifiers disagree (Section~8.1). Astronomical applications that need uncertainty estimates require a coherent model whose probabilities sum to one and whose decision boundaries agree for all classes.

As with multi-class logistic regression, where Chapter~9 covers only the binary case, we opt out of a full Bayesian treatment of multiple classes. It is conceptually straightforward, but a few points need care, and we sketch them here. A complete development would need another chapter, which this book does not include.

The multinomial probit model extends the binary case to $C$ classes, with a probit link in place of the sigmoid. It uses $C-1$ latent functions, and one function is fixed to zero as the reference class. Each of the other functions is the score of class $c$ relative to the reference class, with the prior
\begin{equation}
f_c(\mathbf{x}) \sim \mathcal{GP}(0, k_c(\mathbf{x}, \mathbf{x}')).
\end{equation}

We can use the same kernel for all classes or give each class its own kernel to reflect class-specific characteristics. The class assignment follows the simple rule
\begin{equation}
t = \arg\max_c f_c(\mathbf{x}),
\end{equation}where the maximum includes the reference class. This gives consistent decision boundaries by construction. The boundary between any two classes $i$ and $j$ is the set of points where $f_i(\mathbf{x}) = f_j(\mathbf{x})$. The input space is therefore partitioned without gaps or overlaps. The class probabilities also sum to one, because one class wins.

The inference is more complex than in the binary case. With $C-1$ functions and $N$ training points, we must infer $(C-1) \times N$ latent variables simultaneously. The likelihood for a training point $(\mathbf{x}, t=c)$ is the probability that class $c$ has the largest score, which is
\begin{equation}
p(t=c|\mathbf{f}(\mathbf{x})) = p(f_c(\mathbf{x}) > f_j(\mathbf{x}) \text{ for all } j \neq c).
\end{equation}

This probability involves integrating over a truncated multivariate Gaussian distribution, which has no convenient closed form. Unlike the binary case, where the probit approximation gave closed-form solutions, the multiclass setting needs numerical approximations.

The Laplace approximation extends to this setting. We find the mode of the posterior over the latent variables and approximate the posterior by a Gaussian. Even then, the integrals that give the predictive class probabilities need additional approximations. The Hessian matrix of the Laplace approximation grows as $\mathcal{O}((C-1)^2 N^2)$. For many classes and large $N$, the memory and the computing time both become too large.

Nevertheless, for moderate problems that need uncertainty estimates, multiclass GP classification keeps the probabilistic treatment of the binary case. It expresses uncertainty about objects that fall between well-defined classes. Where misclassification has serious consequences, this can justify the additional computing.

The $\mathcal{O}(N^3)$ cost of a Gaussian process limits its direct use to moderate problems, typically with fewer than several thousand observations. Approximation methods can scale Gaussian processes to larger data sets, and they remain an active area of research. Neural networks (Chapter~15) are another route to flexible nonlinear modeling. They scale to far larger data sets, but their uncertainty estimates need extra methods. Both approaches have been used in astronomy, and they can complement each other.\index{Gaussian process (GP)!classification|)}\index{Gaussian process (GP)|)}

\section{Tutorials: Stellar Light Curves and Star-Galaxy Classification}\label{sec:ch14_tutorials}

A Gaussian process (GP) infers a function from data without a fixed parametric form, and it states how well the data fix the function. Two tutorials use GPs, one for regression and one for classification. The first learns the timescale of stellar brightness variations from Kepler light curves. The second separates stars from galaxies by their Gaia colors. The full tutorials can be found at \url{https://statml.org/} as 14a and 14b.

\paragraph{Timescales of Stellar Brightness} Convection\index{light curves|(}\index{stellar variability|(} and oscillations make the brightness of a star vary. The variations are slower in an expanded, evolved star of low surface gravity, $\log g$, than in a compact one. A timescale measured from a light\index{light curves} curve could therefore tell us how evolved a star is. The tutorial infers this timescale with a GP from Kepler\index{Kepler mission} light curves and compares it with the asteroseismic\index{asteroseismology} $\log g$ of each star.

A fixed shape, such as a sinusoid, would decide in advance how the brightness varies. A GP with an RBF kernel (Section~\ref{sec:ch14_kernels}) assumes only that the brightness varies smoothly. Its length\index{length scale} scale $\ell$ is the time over which the brightness stays correlated, which is the timescale that we want.

\begin{figure}[t!]
    \centering
    \includegraphics[width=0.95\textwidth]{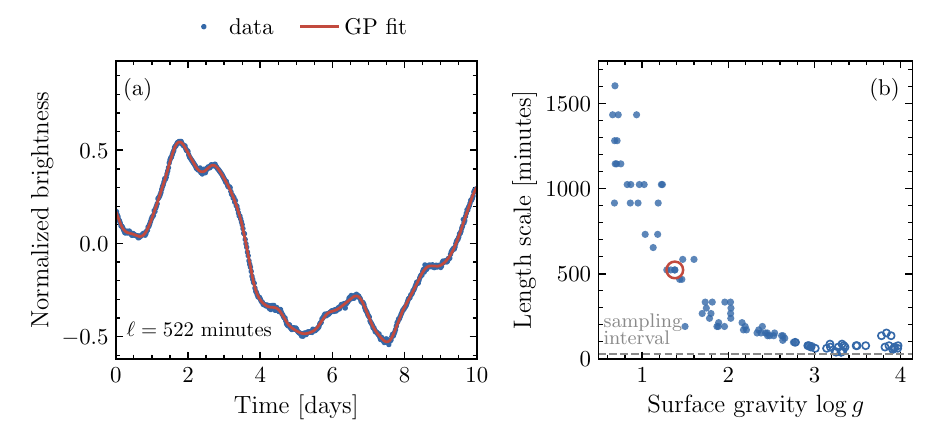}
    \caption{The first tutorial fits Gaussian processes to Kepler light curves. Panel~(a) shows one star (blue points) and its Gaussian process fit (red line). Panel~(b) shows, for each star, the length scale of the kernel that maximizes the marginal likelihood against its surface gravity. Open circles mark the most compact stars, whose brightness varies as fast as the sampling. The fit treats such variations as noise, so their length scales are not reliable. The red ring marks the star of panel~(a), and the dashed line the sampling interval of the light curves.}
    \label{fig:ch14_tutorial_light_curves}
\end{figure}

Each star gets its own length scale, together with a noise level for the measurement errors (Section~\ref{sec:ch14_conditioning}). We choose both by maximizing the marginal likelihood (Section~\ref{sec:ch14_hyperparameters}). A short length scale lets the GP follow every wiggle, including the noise, and the marginal likelihood penalizes this extra flexibility. It thus balances the fit against the complexity of the model. In Figure~\ref{fig:ch14_tutorial_light_curves}a the fit to a star of low gravity traces its slow, smooth variations, and its length scale is long.

Across the stars, the length scale falls as the surface gravity rises (Figure~\ref{fig:ch14_tutorial_light_curves}b), as the slower variations of evolved stars predict. Calibrated on stars of known gravity, the length scale could estimate $\log g$ from a light curve alone.

A GP describes a signal without fixing its shape in advance, and its kernel turns the data into quantities with a physical meaning. Here one number, the length scale, tracks how evolved a star is. The same approach models other signals of unknown shape, such as the variability of quasars or the stellar activity that hides planets.\index{light curves|)}\index{stellar variability|)}

\paragraph{Separating Stars from Galaxies} The\index{star-galaxy classification|(} second tutorial predicts a class instead of a number. Imaging surveys detect far more objects than spectroscopic surveys can observe, so we must separate stars from galaxies\index{star-galaxy classification} by photometry. Faint stars and galaxies have similar colors, the brightness differences between filters, so the classes overlap. An object near their border needs a probability, so that spectroscopy goes to the uncertain ones. The tutorial classifies faint Gaia\index{Gaia} sources of known class by their colors BP$-$RP and G$-$RP.

\begin{figure}[t!]
    \centering
    \includegraphics[width=0.95\textwidth]{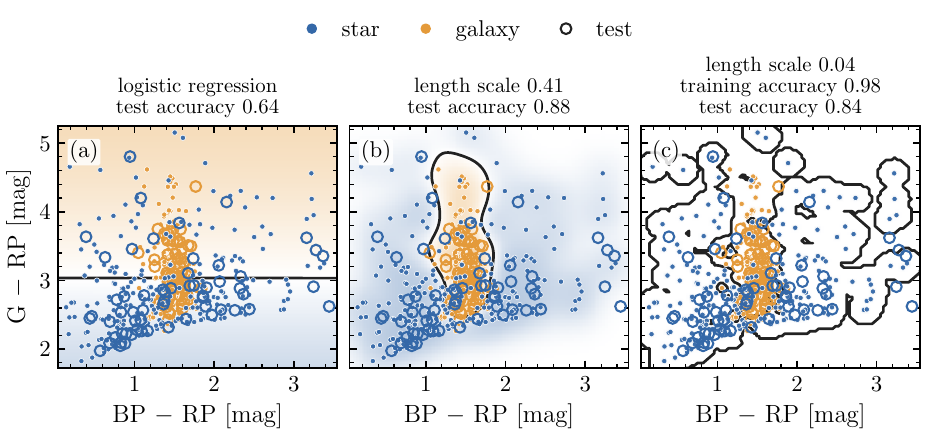}
    \caption{The second tutorial separates stars (blue dots) from galaxies (orange dots) by two Gaia colors. Rings mark the test objects, and the black contour at one half is the decision boundary. The shading is the probability of a star, from orange through white at one half to blue. Panel~(a) uses logistic regression, and panels~(b) and (c) use Gaussian process classifiers with the length scales shown, in standard deviations of the colors. The labels give the accuracy on the test objects and, in panel~(c), on the training objects. The length scale of panel~(b) gives the best test accuracy, and the shorter one of panel~(c) overfits.}
    \label{fig:ch14_tutorial_classification}
\end{figure}

A straight line, as in logistic regression (Section~\ref{sec:ch7_limitations}), separates these classes poorly, because the stars lie on both sides of the galaxies. It misclassifies many of the test objects (Figure~\ref{fig:ch14_tutorial_classification}a). A GP classifier (Section~\ref{sec:ch14_classification}) replaces the line by a latent function, a hidden score that the sigmoid turns into a probability. We give the latent function a GP prior, so that its boundary can bend around the galaxies. The likelihood of a label is not Gaussian, so the posterior of the latent function has no closed form. We approximate it by a Gaussian at its peak, the Laplace approximation, and the probit approximation then gives the probability (Section~\ref{sec:ch14_predictive}).

The length\index{length scale} scale of the kernel sets how quickly the boundary can bend. The marginal likelihood has no closed form here, so we choose the length scale by cross-validation (Section~\ref{sec:ch14_model_selection}). Cross-validation fits the classifier on part of the objects and tests it on the rest, for each length scale, and we keep the best one. The boundary then encloses the galaxies (Figure~\ref{fig:ch14_tutorial_classification}b), and most of the test objects are classified correctly.

With a much shorter length scale (Figure~\ref{fig:ch14_tutorial_classification}c), the boundary follows individual objects. The classifier then gets nearly all the training objects right but fewer test objects than the best length scale does. A much longer length scale makes the boundary too smooth and lowers the test accuracy as well.

The GP classifier gives a probability for each object. The probability is one half on the boundary. It also returns to one half far from every object (Figure~\ref{fig:ch14_tutorial_classification}b). There the kernel carries no information, and the probit approximation pulls uncertain predictions toward one half. An object with a probability near one half is therefore on the border or unlike the training objects. These are the objects to observe spectroscopically.

In both tutorials the kernel leaves the shape of the function open, and the length scale is what we must still learn. For the light curves it is a timescale, and the marginal likelihood finds it for each star. For the classifier it sets how quickly the boundary can bend, and cross-validation chooses it because the marginal likelihood has no closed form there. The cost of a GP grows quickly with the number of objects (Section~\ref{sec:ch14_practical}), so it suits samples of hundreds to a few thousand objects. In return, a GP states what it does not know, as the probability that returns to one half far from the objects.\index{star-galaxy classification|)}

\paragraph{Further Reading} Aronszajn (1950) developed the theory of reproducing kernels, the mathematics behind kernel methods. Matheron (1963) and Cressie (1993) connected Gaussian processes to spatial statistics. O'Hagan (1978) gave a Bayesian treatment of curve fitting with GP priors. Kimeldorf and Wahba (1970) and Wahba (1990) related splines and smoothing to them. Rasmussen and Williams (2006) wrote the standard modern treatment. Neal (1996) showed that GPs are the limit of one-hidden-layer neural networks of infinite width. Csat\'o and Opper (2002) and Snelson and Ghahramani (2005) developed sparse approximations that summarize the data with a few points. Qui\~nonero-Candela and Rasmussen (2005) gave a common view of them. Titsias (2009) chose these points by maximizing a lower bound on the likelihood. Foreman-Mackey et~al.\ (2017) developed celerite, which computes one-dimensional Gaussian processes with suitable kernels in linear time, and applied it to stellar rotation and asteroseismology. Williams and Barber (1998) used the Laplace approximation for classification. Minka (2001) presented expectation propagation, another Gaussian approximation of the posterior, and Kuss and Rasmussen (2005) compared them. MacKay (1992) described the evidence method for hyperparameters. Kennedy and O'Hagan (2001) used GPs to calibrate computer models. Stein (1999) treated interpolation and the Mat\'ern class. Aigrain and Foreman-Mackey (2023) review Gaussian processes for astronomical time series, from kernel choice and scalable computation to applications to exoplanets and active galactic nuclei. Hughes et~al.\ (2022) classified Gaia sources as quasars, galaxies and stars with labels from the Sloan Digital Sky Survey. The classes of the second tutorial follow their work.

\makeatletter\@clubpenalty=150\makeatother \clubpenalty=150 \widowpenalty=150

\chapter{Neural Networks}

\makeatletter\@clubpenalty=10000\makeatother \clubpenalty=10000 \widowpenalty=10000

\looseness=1 Neural\index{neural network|(} networks are the most flexible models in this book. They are not new, since the concept was proposed decades before their recent prominence. At first, theoretical arguments suggested that they would not work well, and limited computing power hindered their practical use. Their empirical success since about 2012 has changed machine learning and led to advances in many fields, even though the theory remains incomplete.

This book has a classical focus, so we limit our coverage of neural networks, for two reasons. First, astronomical data are usually less complex than real-life data. Astronomical phenomena are usually governed by a few main physical principles. In many cases we can then reach excellent results with simpler techniques and can focus on careful statistics. Second, science requires uncertainty quantification. Neural networks have made progress here, but at the time of writing a full Bayesian approach is less well established than for classical techniques.

In astronomy, neural networks have particular value when we model complex, nonlinear systems where simpler techniques are insufficient. Many astronomical phenomena follow complicated nonlinear relationships. Stellar spectra depend nonlinearly on temperature, pressure and chemical composition. Galaxy morphologies vary in complicated ways in image space. Gravitational dynamics create non-Gaussian structure in the cosmic web. Neural networks offer a way forward when linear assumptions break down and the kernel choice for a Gaussian process (Chapter~14) is unclear.

The main innovation of neural networks is feature learning. They discover useful representations directly from the data, without explicit human design. A representation is a new description of the input, given by numbers that the network computes from it. Astronomers need not engineer features by hand, as in linear models (Chapter~4), or specify a kernel function, as in Gaussian processes (Chapter~14). This helps when useful features are hard to specify by hand, as in galaxy classification, where neural networks often find patterns that experts might miss.

\looseness=1 Neural networks and deep\index{deep learning} learning deserve a textbook of their own, but even this classical book should give a basic understanding of them. We cover only the basics, not the full complexity of modern deep learning. We keep the discussion practical, so that you can begin working with neural networks on astronomical problems. We also connect many concepts to the statistical methods of earlier chapters. This follows the philosophy of the book, which starts with simpler methods before it moves to more complex ones. Given the statistical tools we already have, we should often consider neural networks only when simpler methods are insufficient for particularly complex problems.

Neural networks build directly on the statistical principles of the earlier chapters. Their loss functions extend the maximum likelihood principle of regression (Chapter~4) and classification (Chapters~7 and~8). Their optimization uses the stochastic gradient descent of logistic regression (Chapter~7), with enhanced algorithms such as Adam. Their regularization also parallels Bayesian inference. Weight decay corresponds to a Gaussian prior on the parameters, as in ridge regression and Bayesian linear regression (Chapter~5).

The chapter moves from feedforward networks and backpropagation to architectures and Bayesian approaches. It ends with applications such as autoencoders, density estimation and simulation-based inference.

\section{From Linear Regression to Neural Networks}

\looseness=1 We have studied linear models in which the output $y$ depends linearly on some transformation of the input features. For regression this takes the form $y = \mathbf{w}^T \boldsymbol{\phi}(\mathbf{x})$ (Chapter~4). For classification we use $y = \sigma(\mathbf{w}^T \boldsymbol{\phi}(\mathbf{x}))$, where $\sigma$ is the logistic function (Chapter~7).

The main advantage of linear models is their mathematical tractability. For linear regression, we derived closed-form solutions for maximum likelihood estimation, regularized optimization, posterior distributions and predictive distributions. The optimization problem is convex, so a local minimum is also a global minimum.

Linear models are not limited to strictly linear relationships between the inputs and the output. With appropriate feature transformations $\boldsymbol{\phi}(\mathbf{x})$ we can describe nonlinear patterns. For example, we can transform a single input $x$ into the polynomial features $[x, x^2, x^3]$, or create interaction terms such as $x_1x_2$. The output $y$ is a linear combination of the parameters $\mathbf{w}$, so the model stays linear. We write this as
\begin{equation}
y = \sum_i w_i \phi_i(\mathbf{x}),
\end{equation}
where each $\phi_i(\mathbf{x})$ can be any nonlinear function of the inputs.

Linearity refers to how the parameters $w_i$ combine with the features. They appear only as multiplicative coefficients, and not inside nonlinear functions such as $\sin(w_i)$ or $e^{w_i}$. This linear parameter structure keeps the optimization problem convex, even when the relationship between the inputs $\mathbf{x}$ and the outputs $y$ is highly nonlinear.

These methods share a limitation. They depend on our ability to define appropriate feature transformations $\boldsymbol{\phi}(\mathbf{x})$. For complex problems, especially high-dimensional data with nonlinear relationships, designing suitable transformations by hand is challenging. We must somehow guess the right nonlinear features before we apply the linear model. This task needs substantial domain expertise and often considerable trial and error.

Gaussian processes (Chapter~14) offer one solution. They replace explicit feature engineering with a kernel function, which defines features implicitly through a similarity measure between data points. This lets them describe complex nonlinear relationships without constructing features explicitly. However, they scale poorly to large data sets.

Neural\index{neural network} networks take a different approach. Instead of using kernels or manually engineered features, they learn the feature transformations directly from the data through a composition of simple nonlinear functions.

Neural networks also have a theoretical connection to Gaussian processes. Under certain conditions, a randomly initialized neural network of infinite\index{infinite-width limit} width is a Gaussian process whose kernel is set by the architecture of the network. Here the width is the number of neurons in a layer, and we draw the initial weights at random.

Before we see how neural networks compose simple nonlinear functions, we first show why stacking linear functions is not enough. We write two successive linear transformations, with weights $w$ and bias terms $b$, as
\begin{equation}
y^{(1)}_j = \sum_{i} w^{(1)}_{ji} x_i + b^{(1)}_j.
\end{equation}
\begin{equation}
y^{(2)} = \sum_{j} w^{(2)}_j y^{(1)}_j + b^{(2)}.
\end{equation}
Substituting the first equation into the second gives
\begin{equation}
y^{(2)} = \sum_{j} w^{(2)}_j (\sum_{i} w^{(1)}_{ji} x_i + b^{(1)}_j) + b^{(2)} = \sum_{i} (\sum_{j} w^{(2)}_j w^{(1)}_{ji}) x_i + (\sum_{j} w^{(2)}_j b^{(1)}_j + b^{(2)}).
\end{equation}
The two bracketed terms are a new weight $w_i$ for each input and a new bias $b$, so the result is another linear transformation. No matter how many linear transformations we stack, the result stays a linear combination of the form $y = \sum_i w_i x_i + b$.

The main insight behind neural networks is that we must introduce nonlinearity between the linear transformations. We do this by sandwiching nonlinear activation functions between linear layers, so the models can represent complex nonlinear relationships. Alternating linear transformations with nonlinear activations is the core of neural networks. With a proper structure, it lets them approximate complex functions.

\paragraph{The Artificial Neuron}

The basic computational unit of a neural network is the artificial\index{artificial neuron} neuron, which is loosely inspired by biological neurons. In its elementary form, a neuron applies a linear transformation to its inputs, followed by a nonlinear activation\index{activation function} function. For a neuron with the inputs $\mathbf{x} = [x_1, x_2, \ldots, x_D]^T$, the output is
\begin{equation}
y(\mathbf{x}, \mathbf{w}) = f\left(\sum_{j=1}^D w_j x_j + b\right) = f(\mathbf{w}^T\mathbf{x} + b).
\end{equation}
Here $\mathbf{w}$ is a vector of weights, $b$ is a bias term, and $f$ is the activation function.

Frank Rosenblatt proposed an early artificial neuron, the perceptron,\index{perceptron} in 1958. The perceptron used a step function as its activation, given by
\begin{equation}
f(z) =
\begin{cases}
1 & \text{if } z \geq 0 \\
0 & \text{if } z < 0
.\end{cases}
\end{equation}
This binary activation was inspired by the all-or-nothing firing of biological neurons and matched the binary computing of that era. With it, a single perceptron can classify linearly separable patterns, because it divides the input space with a hyperplane (Section~7.4). Two classes are linearly separable when one hyperplane puts them on opposite sides.

Marvin Minsky and Seymour Papert's 1969 book ``Perceptrons'' tempered the initial excitement. It showed the limits of single-layer perceptrons. In particular, they cannot learn the XOR\index{XOR problem} function, which outputs 1 when its two binary inputs differ. The same holds for any pattern that is not linearly separable. This critique and the limited computing resources of the time led to a decline in neural network research, often called the ``artificial intelligence (AI) winter.''

\looseness=1 Interest in neural networks resurged in the 1980s with the development of backpropagation, an efficient algorithm for training multilayer neural networks. We explore backpropagation in detail later in this chapter. Its main innovation was to make it possible to train neural networks with several layers, which overcame the limits of single-layer perceptrons.

\begin{figure}[ht!]
    \centering
    \includegraphics[width=0.9\textwidth]{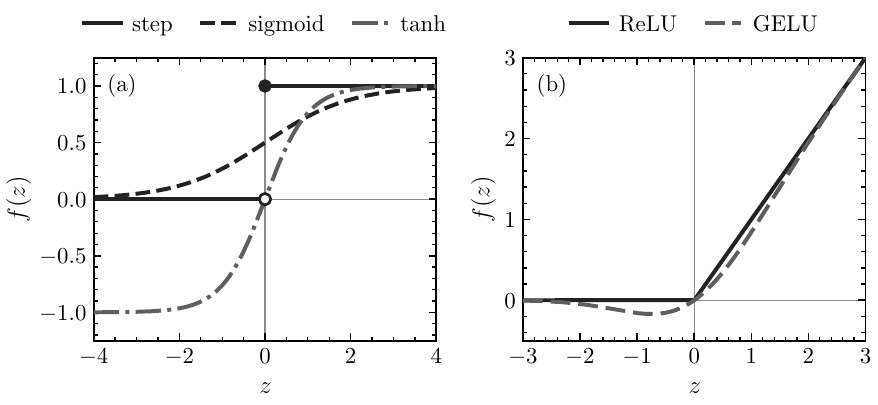}
    \caption{Activation functions. Panel~(a) shows bounded functions and panel~(b) unbounded ones. The gradient of the step function of the perceptron is zero wherever it is defined, so gradient-based learning cannot use it. The hyperbolic tangent is a sigmoid rescaled to the range from $-1$ to $1$. ReLU is piecewise linear, and GELU is a smooth variant of it.}
    \label{fig:activation_functions}
\end{figure}

These neural networks replaced the discontinuous step function with continuous, differentiable activation functions, most commonly the sigmoid function
\begin{equation}
f(z) = \frac{1}{1 + e^{-z}}.
\end{equation}
Gradient-based optimization methods require differentiable functions, so this change was necessary. The sigmoid connects directly to logistic regression (Chapter~7). In effect, each neuron becomes a logistic regression model whose inputs are transformations of the original features. The sigmoid became common because neural networks began in classification tasks, although modern neural networks also handle regression, generation and many other problems. Figure~\ref{fig:activation_functions} compares the step function and the sigmoid with other activation functions, most of which we meet later in this chapter.

\paragraph{Multi-Layer Composition}

Neural networks gain their expressive capacity (the range of functions a model can represent) through composition, which combines neurons in parallel and in several layers. This lets the network learn more complex representations of the input data. A layer between the input and the output is a hidden\index{hidden layer} layer, and its neurons are hidden units. The value that a neuron outputs is its activation.

\begin{figure}[ht!]
    \centering
    \includegraphics[width=0.9\textwidth]{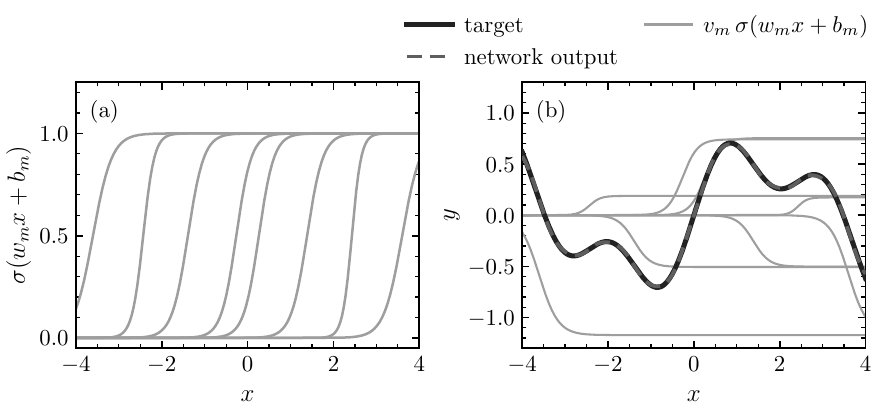}
    \caption{A neural network of $M = 8$ sigmoid neurons fitted to a target made of two sine waves of different frequencies. Panel~(a) shows the neurons $\sigma(w_m x + b_m)$, each with its own weight $w_m$ and bias $b_m$. Panel~(b) shows the weighted neurons $v_m \sigma(w_m x + b_m)$, the target and the network output, which is the sum of the weighted neurons plus a constant.}
    \label{fig:neural_network_composition}
\end{figure}

For a neural network with a single hidden layer of $M$ neurons and a single output, the function becomes
\begin{equation}
y(\mathbf{x}, \mathbf{W}^{(1)}, \mathbf{W}^{(2)}) = f^{(2)}\left(\sum_{j=1}^M w^{(2)}_j f^{(1)}\left(\sum_{i=1}^D w^{(1)}_{ji} x_i + b^{(1)}_j\right) + b^{(2)}\right).
\end{equation}
Here $\mathbf{W}^{(1)}$ is the weight matrix of the first layer, with elements $w^{(1)}_{ji}$. The matrix $\mathbf{W}^{(2)}$ holds the weights of the output layer, and $b^{(1)}_j$ and $b^{(2)}$ are the respective bias terms. Figure~\ref{fig:neural_network_composition} shows a single hidden layer of sigmoid neurons that approximates a target made of two sine waves by combining simple nonlinear curves.

\begin{figure}[ht!]
    \centering
    \includegraphics[width=0.95\textwidth]{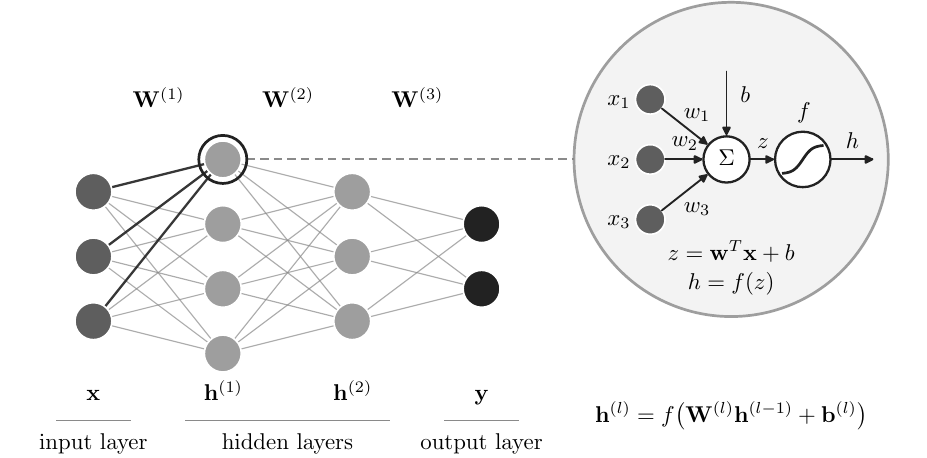}
    \caption{A fully connected feedforward network with two hidden layers, in which information flows from left to right. The matrix $\mathbf{W}^{(l)}$ holds the weights of the connections into layer $l$, and the output is $\mathbf{y} = \mathbf{h}^{(3)}$. The magnified neuron is the top neuron of the first hidden layer, and the dark lines are its three inputs.}
    \label{fig:neural_network_architecture}
\end{figure}

We extend this to neural networks with several outputs and additional hidden layers by further composition. A deep neural network with $L$ layers obeys the recursion
\begin{equation}
\mathbf{h}^{(l)} = f^{(l)}(\mathbf{W}^{(l)}\mathbf{h}^{(l-1)} + \mathbf{b}^{(l)}).
\end{equation}
Here $\mathbf{h}^{(0)} = \mathbf{x}$ is the input, $\mathbf{h}^{(L)}$ is the output, and $\mathbf{h}^{(l)}$ holds the activations of the $l$-th hidden layer. Figure~\ref{fig:neural_network_architecture} shows a neural network with two hidden layers, in which each neuron connects to all neurons of the next layer.

We call this architecture a feedforward\index{feedforward neural network} neural network, because information flows in one direction from the inputs to the outputs. Here it is also fully\index{fully connected network|see{feedforward neural network}} connected, since every neuron receives the outputs of all neurons of the previous layer. Other feedforward architectures connect only some neurons, as the Convolutional Neural Networks of this chapter do. Fully connected feedforward networks are also often called multilayer\index{multilayer perceptron|see{feedforward neural network}} perceptrons. This is technically a misnomer, since they use differentiable activation functions instead of the step function of the original perceptron. The name is historical, because these neural networks developed from the perceptron.

By applying simple nonlinear transformations iteratively, we can approximate complex functions. To see why this works, consider how each layer transforms the input space. The first layer creates a set of basic nonlinear transformations. They carve the input space into regions bounded by the activation boundary of each neuron, the set where its linear transformation equals zero. The second layer combines these regions in new ways and creates more complex patterns. Each additional layer makes more detailed combinations of these patterns, much as we build complex shapes from simple geometric forms.

\section{Theory of Neural Networks}

We\index{neural network!theory|(}\index{universal approximation theorem|(} now turn to a mathematical result about what these neural networks can represent in principle. This result gives part of the reason why neural networks are useful in many scientific applications, although their building blocks are simple.

The main theoretical result is the Universal\index{universal approximation theorem} Approximation Theorem. Let the activation function not be a polynomial. Then, given enough hidden units, a neural network with one hidden layer can approximate any continuous function on a closed, bounded domain to arbitrary accuracy. We call this the universal approximation property. The formal proof needs advanced mathematics beyond the scope of this book.

To see the theorem intuitively, consider how neural networks transform data. Each added neuron has its own activation function, weights and bias, and so adds one more building block to the approximation. Each neuron contributes its own ``bump'' or ``edge'' to the space of functions. By combining these basic shapes carefully, we can construct continuous functions to a chosen accuracy.

Readers who know real analysis may recognize the Stone-Weierstrass theorem. It shows that polynomials can approximate a continuous function uniformly on a closed and bounded interval. In physics, Fourier series let us represent a periodic function as a sum of sines and cosines. The universal approximation property extends these ideas to a different class of basis functions, the outputs of individual neurons, which build complex approximations.

In principle, a neural network could approximate a continuous relationship in nature over a bounded range of inputs. Examples range from the gravitational interactions of celestial bodies to the spectral features of distant stars. This assumes that the phenomenon can be represented as a continuous function.

\paragraph{Practical Limitations}

\looseness=1 The universal approximation theorem has important practical limitations. It guarantees that a solution exists, but it gives no guidance on how to find it efficiently. The appropriate weights lie somewhere in a vast, high-dimensional space, so the search for them is a challenge. The theorem also proves only that a sufficiently large neural network exists. It does not give the minimum size needed for a particular problem.

This gap between theoretical capability and practical use is part of the reason for the historical skepticism toward neural networks. For decades after their proposal, researchers had two main concerns. First, the computational demands seemed prohibitive, since even modest neural networks needed computing resources that were not widely available for many years. Second, the large number of parameters seemed to invite overfitting, as when we fit high-degree polynomials to limited data (Section~\ref{sec:cross_validation}). According to statistical learning theory, models with more parameters than training examples should generalize poorly to new data. Yet neural networks often generalize well in this regime.

At first glance, neural networks resembled epicycles in astronomy, which add complexity without necessarily improving understanding. Astronomers added epicycles to patch the geocentric model of the solar system instead of adopting the simpler heliocentric view. Neural networks seemed likewise to add layers of parameters instead of developing better theories. Their lack of interpretability reinforced this impression. In linear models the coefficients have clear meanings, but neural network parameters often resist a straightforward interpretation.

Several things changed this view. Researchers kept working on neural networks, which succeeded in many domains. Other factors were computational advances (particularly graphics processing units, GPUs, originally designed for video games) and larger data sets. These successes surprised many machine learning practitioners, who were still pursuing kernel-based techniques with firmer statistical roots, such as Gaussian processes.

Despite their empirical success, some criticisms of neural networks persist, particularly in astronomy. Some of them are justified, but many are not. A legitimate limitation, as we will see, concerns the Bayesian treatment. We know how to optimize neural networks, which is analogous to finding the maximum likelihood solution in linear regression. Fully Bayesian neural networks, however, remain challenging.

\looseness=1 Many common criticisms are less well founded. For instance, critics call neural networks ``black boxes,'' yet feature engineering in linear regression is often equally non-physical and based on trial and error. Kernel-based methods can be just as opaque in practice.

\begin{figure}[ht!]
    \centering
    \includegraphics[width=0.9\textwidth]{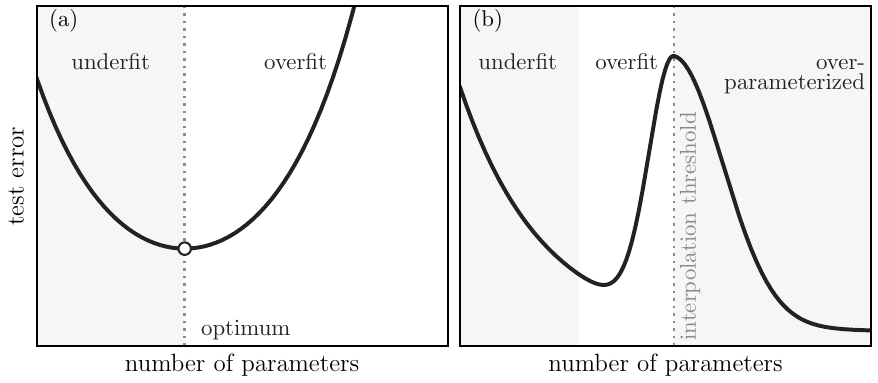}
    \caption{The classical bias-variance picture (a) and the modern double descent picture (b) of the test error against the number of parameters. In the classical picture, the test error falls in the underfitting region, reaches an optimum and then rises in the overfitting region. In the modern picture, the test error rises to a peak at the interpolation threshold, where the number of parameters equals the number of training points. It then falls again in the overparameterized regime. This second descent helps explain why neural networks with far more parameters than training points can generalize well.}
    \label{fig:double_descent}
\end{figure}

The empirical success of neural networks has also led to new areas of mathematics that aim to understand their structure and behavior. This parallels how we built steam engines before we fully understood the laws of thermodynamics. Thermodynamic theory then allowed us to improve engine efficiency substantially. In the same way, research on the mathematics of neural networks, the emerging ``Science of AI,'' now gives lessons that improve their applications. The details of this field are beyond this introductory book, but theoretical work is addressing many of the criticisms.

Two results of this theoretical work are double descent and the Neural Tangent Kernel. In double descent, the test error falls, rises and falls again as the model grows more complex (Figure~\ref{fig:double_descent}). This runs against the classical bias-variance picture of how complexity affects performance (Section~\ref{sec:cross_validation}). The Neural Tangent Kernel connects neural networks to the kernel methods that we studied with Gaussian processes (Chapter~14). We return to both later in this chapter. Both bear on why neural networks work well in practice, although they seem to violate classical statistical wisdom.

For now we leave the details of this theory aside. We focus on the practical implementation of neural networks. We start with training, particularly backpropagation, which makes the efficient optimization of these complex models possible. We also examine applications in astronomy, which connect these advanced methods back to the statistical principles of this book.\index{neural network!theory|)}\index{universal approximation theorem|)}

\section{Training Neural Networks}\label{sec:ch15_training}

A\index{neural network!training|(} neural network has a nested structure that looks complex, but the gradient-based methods of Chapters~7 and~8 train it. Training extends the optimization principles that we already know.

A neural network can map $\mathbb{R}^D$ to $\mathbb{R}^K$, that is, vectors of $D$ real numbers to vectors of $K$ real numbers. This works for any $D$ and $K$, because a neural network consists of matrix multiplications and nonlinear transformations. One layer computes
\begin{equation}
\mathbf{h} = f(\mathbf{W}\mathbf{x} + \mathbf{b}).
\end{equation}
Here $\mathbf{W} \in \mathbb{R}^{M \times D}$ is a weight matrix that maps $D$ input dimensions to $M$ hidden dimensions, and $\mathbf{b} \in \mathbb{R}^M$ is a bias vector. The function $f$ is a nonlinear activation function applied element-wise.

The hidden layers can have any number of neurons $M$, and each layer can have its own dimension. The output layer can also have any size. For example, a spectrum with $D$ flux values can map to $K = 3$ stellar properties, namely temperature, pressure and chemical composition. The free choice of the output size also suits multi-class classification, where we need $C$ outputs (one per class) whatever the input dimension. A galaxy image with $D$ pixel values can map to the probabilities of $C$ morphological classes.

\paragraph{Loss Functions}

As in Chapters~4 and~7, we follow maximum\index{maximum likelihood estimation (MLE)} likelihood estimation and choose the parameters that maximize the probability of the observed data. We combine all weights $\mathbf{W}^{(l)}$ and biases $\mathbf{b}^{(l)}$ into one vector $\mathbf{w}$, as we did for the intercept in Chapter~4. The loss\index{loss function} functions of neural networks follow from this principle.

For regression with inputs $\mathbf{x}_i \in \mathbb{R}^D$ and outputs $y_i \in \mathbb{R}$, we assume homoscedastic Gaussian noise, as in linear regression (Chapter~4). Maximizing the likelihood then means minimizing the mean squared error (MSE)
\begin{equation}
L_{\text{MSE}}(\mathbf{w}) = \frac{1}{N}\sum_{i=1}^N (y_i - f_{\mathbf{w}}(\mathbf{x}_i))^2.
\end{equation}

When our astronomical observations have heteroscedastic measurement uncertainties $\sigma_i$, the likelihood gives the weighted MSE
\begin{equation}
L_{\text{weighted MSE}}(\mathbf{w}) = \frac{1}{N}\sum_{i=1}^N \frac{(y_i - f_{\mathbf{w}}(\mathbf{x}_i))^2}{\sigma_i^2}.
\end{equation}
This is, up to constant factors, the weighted loss that we derived in Chapter~4 for linear regression with heteroscedastic Gaussian noise.

The statistical principles stay the same as in linear regression. Only the form of $f_{\mathbf{w}}(\mathbf{x}_i)$ changes. Instead of a simple linear function, it is now the output of the neural network, a flexible function composed of multiple transformations
\begin{equation}
f_{\mathbf{w}}(\mathbf{x}) = f^{(L)}(\mathbf{W}^{(L)}f^{(L-1)}(\mathbf{W}^{(L-1)}...(f^{(1)}(\mathbf{W}^{(1)}\mathbf{x} + \mathbf{b}^{(1)})...)+ \mathbf{b}^{(L-1)}) + \mathbf{b}^{(L)}).
\end{equation}Here $f_{\mathbf{w}}$ depends on $\mathbf{w}$ through the weight matrices $\mathbf{W}^{(l)}$ and the bias vectors $\mathbf{b}^{(l)}$ of all layers. For the losses below, the last activation $f^{(L)}$ is the identity, so that $f_{\mathbf{w}}$ is the raw network output.

For binary classification we follow the likelihood principles of logistic regression (Chapter~7). We model the probability of class membership with the sigmoid function
\begin{equation}
p(y=1|\mathbf{x}) = \sigma(f_{\mathbf{w}}(\mathbf{x})) = \frac{1}{1 + e^{-f_{\mathbf{w}}(\mathbf{x})}}.
\end{equation}
Maximizing the likelihood of the observed binary outcomes then means minimizing the binary cross-entropy
\begin{equation}
L_{\text{BCE}}(\mathbf{w}) = -\frac{1}{N}\sum_{i=1}^N [y_i \ln(\sigma(f_{\mathbf{w}}(\mathbf{x}_i))) + (1-y_i) \ln(1-\sigma(f_{\mathbf{w}}(\mathbf{x}_i)))].
\end{equation}

For $C$ classes the same principle applies (Chapter~8). We model the class probabilities with the softmax function, a generalization of the sigmoid to several classes. Maximizing the likelihood gives the cross-entropy loss
\begin{equation}
L_{\text{CE}}(\mathbf{w}) = -\frac{1}{N}\sum_{i=1}^N \sum_{c=1}^C y_{ic} \ln(p_{\mathbf{w}}(\mathbf{x}_i)_c).
\end{equation}
Here $y_{ic}$ is 1 if example $i$ belongs to class $c$ and 0 otherwise. The predicted probability $p_{\mathbf{w}}(\mathbf{x}_i)_c$ of class $c$ is
\begin{equation}
p_{\mathbf{w}}(\mathbf{x}_i)_c = \frac{e^{f_{\mathbf{w}}(\mathbf{x}_i)_c}}{\sum_{j=1}^C e^{f_{\mathbf{w}}(\mathbf{x}_i)_j}}.
\end{equation}

\paragraph{Gradient Descent}

We find the parameters $\mathbf{w}$ that minimize these potentially complex, non-convex loss functions with gradient descent (Chapter~7). The basic update rule is unchanged
\begin{equation}
\mathbf{w}^{(t+1)} = \mathbf{w}^{(t)} - \eta \nabla_{\mathbf{w}} L(\mathbf{w}^{(t)}).
\end{equation}
Here $\eta$ is the learning rate and $\nabla_{\mathbf{w}} L(\mathbf{w}^{(t)})$ is the gradient of the loss with respect to all parameters at iteration $t$. The activation functions $f$ of this chapter are fixed in advance and have no parameters to train. Gradient descent adjusts only the weights and the biases in $\mathbf{w}$.

In logistic regression, the chain rule gives a compact gradient (Chapter~7). For a single example of binary classification it is
\begin{equation}
\nabla_{\mathbf{w}} L(\mathbf{w}) = (\sigma(\mathbf{w}^T\mathbf{x}) - y)\mathbf{x}.
\end{equation}

In a neural network the same chain rule applies, but the gradient calculation is considerably more complex. We must propagate derivatives through many layers of nonlinear transformations. This creates a cascade of derivatives that flows backward through the network. The backpropagation algorithm solves this problem, and we describe it in the next section.

Computing the gradient over the full training set for each update would be too expensive for large data sets, as in logistic regression. We therefore use stochastic\index{stochastic gradient descent (SGD)} gradient descent (SGD), which approximates the gradient with a small random subset of the training examples, a ``mini-batch''.\index{mini-batch} The update becomes
\begin{equation}
\mathbf{w}^{(t+1)} = \mathbf{w}^{(t)} - \eta \nabla_{\mathbf{w}} L_B(\mathbf{w}^{(t)}).
\end{equation}
Here $L_B$ is the loss computed on a mini-batch $B$ of examples sampled at random from the full data set.

Mini-batches have two advantages. First, they reduce the cost of each update. We can then update the parameters more often, and the training converges faster in wall-clock time. Second, the noise from sampling different mini-batches acts as a form of regularization. This noise helps the optimization escape shallow local minima and saddle points of the loss surface, which can lead to better solutions.

\looseness=1 A historical criticism of neural networks was that their non-convex loss surfaces contain many local minima that trap the optimization. Recent theoretical work, such as the Neural Tangent Kernel of Section~15.8, suggests that the surface may be easier to optimize than feared. The noise of SGD still helps against such optimization problems.

The mini-batch size sets a trade-off, which we can see with the analogy of a vehicle on rough terrain. A large batch is like a big, heavy vehicle. It follows the gradient more precisely, but it may not fit through the narrow paths that lead to better solutions. A small batch is like a light, agile vehicle. It may not move in the best direction at each step, but it can take more varied paths and sometimes reach a better final position. Common batch sizes range from 32 to 256 examples, and the best choice depends on the problem and on the available computing resources.

The learning rate $\eta$ also matters. A value that is too large can make the optimization diverge, and a value that is too small gives slow convergence. Practitioners therefore use learning rate schedules. Chapter~7 describes two of them and plots them in Figure~\ref{fig:lr_schedules}. Inverse time decay gives quick initial progress followed by more precise updates. Cosine annealing with restarts lowers the rate along a cosine curve and returns it to its starting value at the end of each cycle.

These cycles of exploration (high learning rate) and exploitation (low learning rate) help the optimization escape local minima. The restarts suit the complex loss surfaces of deep neural networks.

\begin{figure}[ht!]
    \centering
    \includegraphics[width=0.95\textwidth]{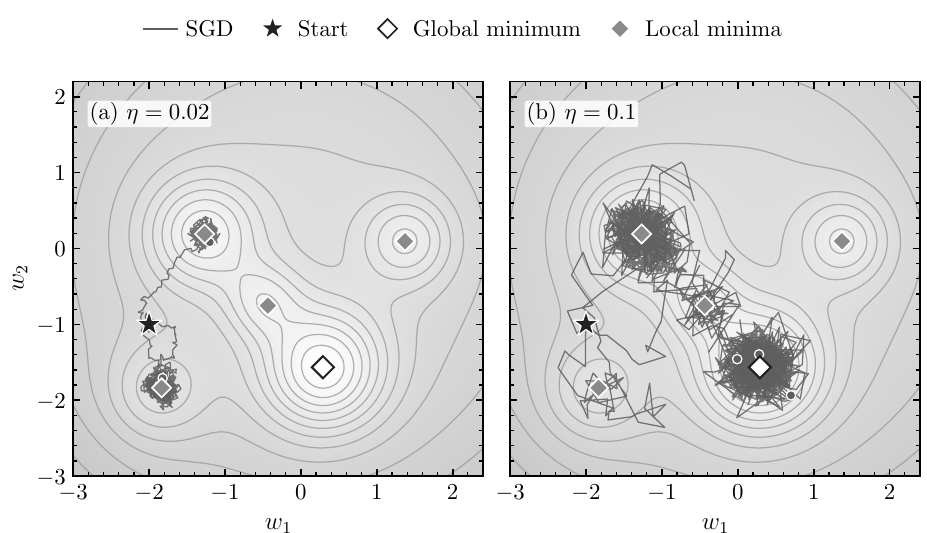}
    \caption{Effect of the learning rate $\eta$ on gradient descent with noisy gradients, on a loss with five minima. Panel~(a) uses a small learning rate and panel~(b) a large one, and darker shading marks larger loss. Each step uses the gradient plus random noise, a stand-in for the noise of mini-batches. Each panel shows three runs from the same start, and the dots mark where they end. In repeated runs, few end in the basin of the global minimum with the small rate and most with the large rate. Without the noise, both rates end in a local minimum.}
    \label{fig:sgd_neural_networks}
\end{figure}

With noisy gradients, a large learning rate helps the optimization escape a shallow local minimum. The noise enters each step multiplied by the learning rate, so a large rate gives larger random moves over the barrier. A small learning rate follows the gradient closely and may stay in the minimum. Figure~\ref{fig:sgd_neural_networks} shows the effect of the learning rate on a non-convex loss with noisy gradients.

\paragraph{Adam}

The\index{Adam optimizer|(} most widely used optimizer beyond basic SGD is Adam\index{Adam optimizer} (Adaptive Moment Estimation). It has two main ideas. Momentum keeps a moving average of past gradients, so that the updates continue in promising directions and oscillations are damped. Adaptive learning rates set the step of each parameter from its gradient history.

Adam estimates the first and second moments of the gradients with exponential moving averages
\begin{align}
m_t &= \beta_1 m_{t-1} + (1 - \beta_1) g_t \\*
v_t &= \beta_2 v_{t-1} + (1 - \beta_2) g_t^2.
\end{align}
Here $g_t$ is the current gradient. The variable $m_t$ is the moving average of the gradient, the first moment. The variable $v_t$ is the moving average of the squared gradient, the second moment.

The parameters $\beta_1$ and $\beta_2$ (typically 0.9 and 0.999) act like window sizes of the moving averages. Larger values give more weight to past gradients. The exponential decay gives recent gradients more influence and keeps some memory of past updates.

We initialize the moving averages at zero, with $m_0 = 0$ and $v_0 = 0$, so they are biased toward zero in the early steps of training. After one step the first moment estimate is $m_1 = (1-\beta_1)g_1$, which is much smaller than the gradient $g_1$ because $\beta_1$ is close to 1. This bias fades as training proceeds, but it can slow down the early steps.

Both estimates therefore need a correction. Adam rescales the estimates
\begin{align}
\hat{m}_t &= \frac{m_t}{1-\beta_1^t} \\*
\hat{v}_t &= \frac{v_t}{1-\beta_2^t}.
\end{align}
These correction terms depend on the number of steps $t$ taken so far. When $t$ is small, the denominators $1-\beta_1^t$ and $1-\beta_2^t$ are also small, so the estimates grow and compensate for the bias. As $t$ increases, the denominators approach 1 and the correction becomes negligible.

The bias-corrected moments give the parameter update
\begin{equation}
\mathbf{w}^{(t+1)} = \mathbf{w}^{(t)} - \eta \frac{\hat{m}_t}{\sqrt{\hat{v}_t} + \epsilon}.
\end{equation}Here $\epsilon$ is a small constant that keeps the denominator away from zero.

\begin{figure}[ht!]
    \centering
    \includegraphics[width=0.8\textwidth]{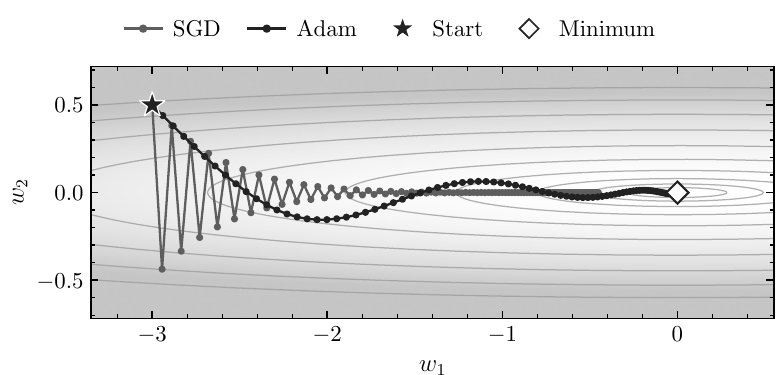}
    \caption{SGD and Adam with fixed learning rates on a quadratic loss that curves much more steeply in one direction than in the other. Both start at the same point and use exact gradients, so SGD is plain gradient descent here, and each dot is one step. SGD oscillates across the valley, whereas Adam crosses it only a few times and gets much closer to the minimum. Darker shading marks larger loss. The valley is aligned with the coordinate axes, which suits the per-coordinate steps of Adam.}
    \label{fig:adam_optimizer}
\end{figure}

Adam adapts the step of each parameter to the size of its recent gradients. Parameters with consistent gradients take larger steps. Parameters with highly variable gradients take smaller, more cautious steps. Figure~\ref{fig:adam_optimizer} compares SGD and Adam on a loss surface that curves much more steeply in one direction than in the other.\index{Adam optimizer|)}

\paragraph{Regularization}

Neural networks can overfit, as the linear models of Chapters~4 and~5 do. Many techniques that prevent overfitting parallel the classical statistical methods of those chapters. Modern deep learning often rediscovers and builds on classical statistical principles.

Weight\index{weight decay} decay, also called L2 regularization, is one of the regularization techniques of neural networks. It is the principle of ridge regression (Chapter~5). As before, we add to the loss a penalty term that grows quadratically with the Euclidean length $\|\mathbf{w}\|_2$ of the weight vector
\begin{equation}
L_{\text{reg}}(\mathbf{w}) = L(\mathbf{w}) + \lambda \|\mathbf{w}\|_2^2.
\end{equation}
The gradient update then contains a decay term for each weight, as in ridge regression
\begin{equation}
\mathbf{w}^{(t+1)} = \mathbf{w}^{(t)} - \eta(\nabla L(\mathbf{w}^{(t)}) + 2\lambda\mathbf{w}^{(t)}) = (1-2\eta\lambda)\mathbf{w}^{(t)} - \eta\nabla L(\mathbf{w}^{(t)}).
\end{equation}
The multiplicative factor $(1-2\eta\lambda)$ gives weight decay its name. The penalty keeps the weights small, in line with our preference for simpler models.

For $0 < \eta\lambda < 1$ the factor has magnitude below 1, so each training step shrinks the weights when no opposing gradients act. This form holds for plain gradient descent. With Adam, the penalty gradient enters $m_t$ and $v_t$ like any other gradient, so the decay is no longer this factor.

As in ridge regression, weight decay also has a Bayesian reading. If $L$ is the negative log-likelihood summed over the training examples, the penalty is equivalent to a zero-mean Gaussian prior on the weights
\begin{equation}
p(\mathbf{w}) \propto \exp(-\lambda \|\mathbf{w}\|_2^2).
\end{equation}
The losses of this section are averages over $N$ examples, so for the cross-entropy losses $L$ is the negative log-likelihood divided by $N$. The prior is then $\exp(-N\lambda\|\mathbf{w}\|_2^2)$, with variance $1/(2N\lambda)$, as Chapter~7 shows. For the mean squared error with noise variance $\sigma^2$, the prior is $\exp(-N\lambda\|\mathbf{w}\|_2^2/(2\sigma^2))$. In a neural network the penalty acts on the weights of all layers.

Early stopping (Chapter~7) is another common technique. It applies the validation principle of Chapter~4. We monitor the training loss and the validation loss, which we compute on validation data that the gradient updates do not use. The validation loss first decreases together with the training loss. When the model starts to overfit, the validation loss increases while the training loss keeps decreasing. We keep the parameters with the lowest validation loss and stop the training when the validation loss no longer improves.

Early stopping saves computing time and gives a simpler model, because the network has less time to fit the noise. The validation set decides when to stop, so it selects the complexity of the model. Ideally the validation set is separate from the test set. With limited astronomical data, the two sets are sometimes combined. This compromise is practical when data are scarce. The estimate of the test error is then optimistic, because we chose the stopping step on the same data.

For a large enough neural network and data set, double descent suggests that we can train as long as the computing budget allows (Section~15.8). Astronomical data sets are often too small for this. Early stopping then remains a useful choice.

Training a neural network can seem more art than science, because we must make many choices, from the optimizer to the regularization. This complexity is not unique to neural networks. Linear models also needed choices of features or kernel functions. The learning rate, the regularization strength, the network architecture, the activation functions and the batch size are hyperparameters of a neural network. For many astronomical applications with modest data sizes, a simple grid or random search over the most important hyperparameters is often sufficient and practical. A grid search tries each value of a fixed list, and a random search tries values picked at random from chosen ranges.

The most important hyperparameters are typically the learning rate and the network architecture (depth and width), which have the largest effect on performance. Other choices, such as the regularization strength and the optimizer, often have a more moderate effect. Optimizer parameters such as the momentum values can usually stay at established defaults.

Some hyperparameter choices perform well on a wide range of tasks. For example, the cosine annealing schedule with restarts has been effective on a range of problems and reduces the need for extensive tuning. Such settings are good starting points for many astronomical applications. Systematic search then tunes them.\index{neural network!training|)}

\section{Backpropagation and Efficient Gradient Calculation}\label{sec:ch15_backprop}

Each\index{backpropagation|(} training step moves the parameters a small distance in the direction that reduces the loss. The update is
\begin{equation}
\mathbf{w}^{(t+1)} = \mathbf{w}^{(t)} - \eta \nabla L(\mathbf{w}^{(t)}).
\end{equation}
This update rule is simple, but it hides a computational challenge. We must compute the gradient $\nabla L(\mathbf{w}^{(t)})$ efficiently for a neural network with potentially millions of parameters.

At first this task seems out of reach, because we need the change of the loss with respect to each parameter in the network. However, the backpropagation\index{backpropagation} algorithm makes the calculation possible and efficient. It underlies the training of neural networks.

Backpropagation rests on the chain\index{chain rule} rule of calculus, which has been known for centuries. A neural network, however complex, is a nested composition of simpler functions. The chain rule breaks the derivative of such a composition into manageable pieces. For a composite function $f(g(x))$ the chain rule says that
\begin{equation}
\frac{\mathrm{d}}{\mathrm{d}x}f(g(x)) = \frac{\mathrm{d}f}{\mathrm{d}g} \cdot \frac{\mathrm{d}g}{\mathrm{d}x}.
\end{equation}

For the multivariate functions of a neural network, the rule extends to partial derivatives and matrix calculus. For a modest neural network, a derivation by hand yields expressions that span many pages, with many dependencies between parameters in different layers. The number of parameters and the nested structure make the gradient practically impossible to derive without computational techniques that exploit this structure. To see why backpropagation matters, we first look at two traditional approaches to differentiation.

\paragraph{Symbolic Differentiation} In calculus we learn the rules for elementary functions. The derivative of $x^n$ is $nx^{n-1}$ and the derivative of $\sin(x)$ is $\cos(x)$. For compositions of functions we apply the chain rule. To differentiate $f(x) = \sin(x^2)$ we write $f(x) = \sin(g(x))$ with $g(x) = x^2$, and the chain rule gives
\begin{equation}
f'(x) = \frac{\mathrm{d}}{\mathrm{d}x}\sin(g(x)) = \cos(g(x)) \cdot g'(x) = \cos(x^2) \cdot 2x.
\end{equation}
This symbolic\index{symbolic differentiation} approach derives explicit formulas with the differentiation rules. We learn it in calculus courses, and computer algebra systems such as Wolfram Alpha implement it.

However, as functions become more complex, with several nested compositions, even the chain rule becomes unwieldy to apply by hand or symbolically. Consider a moderately complex neural network with just a few layers, given by
\begin{equation}
f(x) = W^{(3)}\sigma(W^{(2)}\sigma(W^{(1)}x + b^{(1)}) + b^{(2)}) + b^{(3)}.
\end{equation}
Deriving $\partial f/\partial W^{(1)}$ symbolically requires a careful application of the chain rule through several nested functions. The result is an expression with numerous terms that can grow exponentially as the network gets deeper. For modern neural networks with dozens of layers and millions to billions of parameters, the symbolic expressions would be impossibly long and inefficient to evaluate. We also do not need the explicit symbolic form of the gradient, only its numerical value at specific points for the gradient descent updates.

\paragraph{Finite Differences} A second approach uses the definition of the derivative from calculus, which is a limit
\begin{equation}
\frac{\mathrm{d}f}{\mathrm{d}x} = \lim_{\Delta x \to 0} \frac{f(x + \Delta x) - f(x)}{\Delta x}.
\end{equation}
Keeping a small but non-zero $\Delta$ instead of the limit gives an approximation of the derivative by finite\index{finite differences} differences
\begin{equation}
\frac{\partial f}{\partial w_i} \approx \frac{f(w_i + \Delta) - f(w_i)}{\Delta}.
\end{equation}
Finite differences are a classical numerical method that scientific computing has used widely, from fluid dynamics simulations to astrophysical modeling. The approximation approaches the true derivative as $\Delta$ decreases.

Finite differences have several drawbacks. For a neural network with $P$ parameters, the full gradient needs at least $P+1$ forward evaluations. One is the baseline evaluation of $f(w)$. The others perturb one parameter each while keeping all other parameters fixed, as in $f(w_1 + \Delta, w_2, ..., w_P)$ and $f(w_1, w_2 + \Delta, ..., w_P)$. This quickly becomes prohibitively expensive as the number of parameters grows.

\begin{figure}[ht!]
    \centering
    \includegraphics[width=0.95\textwidth]{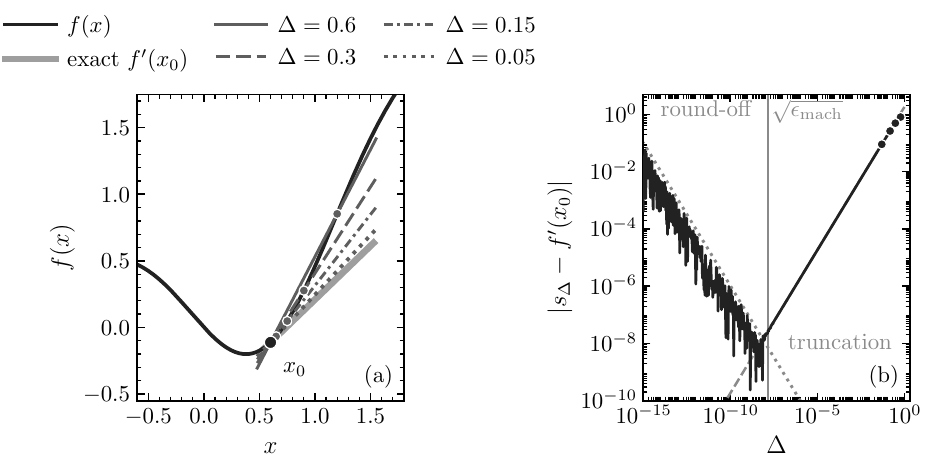}
    \caption{Forward finite differences of $f(x) = x^2/2 - 0.3\sin(3x)$ at $x_0$. Panel~(a) shows the exact slope $f'(x_0)$ and the secants through $x_0$ and $x_0 + \Delta$ for four values of $\Delta$. The secant slope $s_\Delta$ approaches the exact slope as $\Delta$ decreases. Panel~(b) shows the error of $s_\Delta$ in double precision, with the four values of $\Delta$ as dots. The error falls in proportion to $\Delta$ down to about $\sqrt{\epsilon_{\mathrm{mach}}} \approx 10^{-8}$, where $\epsilon_{\mathrm{mach}}$ is the machine precision, and then grows because of round-off. The gray lines are estimates of the truncation and round-off errors.}
    \label{fig:finite_difference_gradients}
\end{figure}

Choosing $\Delta$ is also a delicate balance. If $\Delta$ is too large, the approximation is inaccurate. If $\Delta$ is too small, the finite precision of computer arithmetic causes numerical errors. For a neural network with millions or billions of parameters, a single gradient needs millions or billions of forward passes. Figure~\ref{fig:finite_difference_gradients} shows the secants for several values of $\Delta$ in panel~(a) and the error of the forward difference in panel~(b).

\paragraph{Automatic Differentiation}

Automatic\index{automatic differentiation|(} differentiation combines the numerical efficiency of computation with the accuracy of symbolic approaches. Like the symbolic approach, it applies the chain rule to elementary operations, such as addition, multiplication and exponentials. Unlike it, it does not build an expression for the derivative. It stores the derivative of each elementary operation, and the chain rule is applied implicitly as numbers pass through the computation. For $f(x) = \sin(x^2)$ it tracks how a small change in $x$ propagates through each operation, first $x^2$ and then $\sin$. The derivatives are exact up to rounding, unlike the approximations of finite differences, and the computation is efficient.

Automatic differentiation has two main forms, forward mode and reverse mode. Imagine computing derivatives through a chain of operations. Forward mode starts at the inputs and works forward, computing how a change in each input affects the intermediate values and the final output. Think of it as following a ripple from its source. Reverse mode first computes the function value and then works backward from the output to determine how each input contributed to the final result. It is like tracing back the path that led to an outcome.

To make this concrete, we take a simple neural network layer with a loss function. For a neural network with a single hidden unit we might have
\begin{equation}
L = E(W, b) = \frac{1}{2}(y - \hat{y})^2 = \frac{1}{2}(y - \sigma(Wx + b))^2.
\end{equation}
Here $x$ is the input, $W$ is the weight and $b$ is the bias. The function $\sigma$ is an activation function like the sigmoid, $\hat{y}$ is the network's prediction and $y$ is the target value. For now we take $x$, $W$, $b$, $\hat{y}$ and $y$ as scalars, so $Wx + b$ is a number. Below we generalize $x$, $W$ and $b$ to vectors and matrices.

\begin{figure}[ht!]
    \centering
    \includegraphics[width=0.95\textwidth]{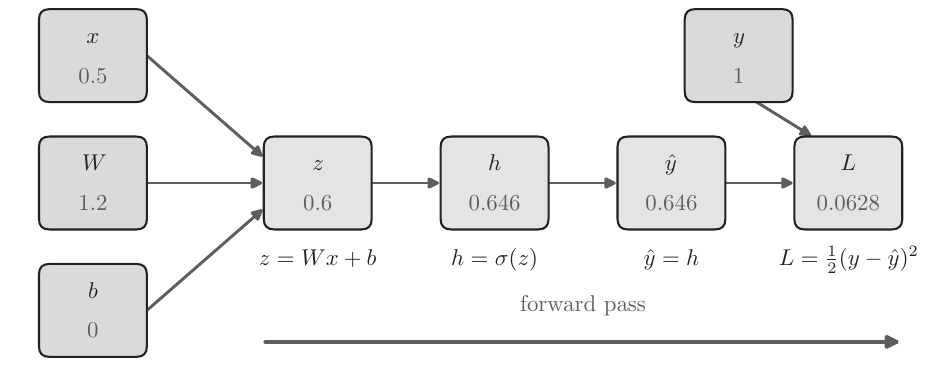}
    \caption{Computational graph of the example $L = (1/2)(y - \sigma(Wx + b))^2$. The leaves are the inputs $x$, $W$ and $b$ and the target $y$. Each node stores the value that the forward pass gives. The backward pass computes the gradients at these stored values.}
    \label{fig:computational_graph}
\end{figure}

We can represent this function as a computational\index{computational graph} graph, a directed acyclic graph. Its nodes represent variables, both inputs and intermediate values, and its edges show which variables each variable depends on. The edges are arrows, and no path along the arrows returns to its starting node. Figure~\ref{fig:computational_graph} shows the graph of our example for the forward pass.

Each node in this graph stores the value of its variable at specific inputs, together with the operation that produces it. The graph makes the structure of the calculation explicit. In the forward pass information flows from the inputs to the outputs, and in backpropagation gradients flow from the output back to the inputs.

The nodes of the graph correspond to a sequence of simpler operations
\begin{align}
z &= Wx + b\\*
h &= \sigma(z)\\
\hat{y} &= h\\*
L &= \frac{1}{2}(y - \hat{y})^2.
\end{align}

To update the parameters $W$ and $b$ by gradient descent, we need $\partial L/\partial W$ and $\partial L/\partial b$. The chain rule gives
\begin{align}
&\frac{\partial L}{\partial W} = \frac{\partial L}{\partial \hat{y}} \cdot \frac{\partial \hat{y}}{\partial h} \cdot \frac{\partial h}{\partial z} \cdot \frac{\partial z}{\partial W}\\*
&\frac{\partial L}{\partial b} = \frac{\partial L}{\partial \hat{y}} \cdot \frac{\partial \hat{y}}{\partial h} \cdot \frac{\partial h}{\partial z} \cdot \frac{\partial z}{\partial b}.
\end{align}

At their specific evaluation points the terms are
\begin{align}
&\frac{\partial L}{\partial \hat{y}}\bigg|_{\hat{y}=\hat{y}^*} = -(y - \hat{y}^*)\\*
&\frac{\partial \hat{y}}{\partial h}\bigg|_{h=h^*} = 1\\
&\frac{\partial h}{\partial z}\bigg|_{z=z^*} = \sigma'(z^*)\\
&\frac{\partial z}{\partial W}\bigg|_{W=W^*, x=x^*} = x^{*T}\\*
&\frac{\partial z}{\partial b}\bigg|_{b=b^*} = 1.
\end{align}

Multiplying the terms of each chain gives
\begin{align}
&\frac{\partial L}{\partial W}\bigg|_{W=W^*, x=x^*, z=z^*, \hat{y}=\hat{y}^*} = -(y - \hat{y}^*) \cdot \sigma'(z^*) \cdot x^{*T}\\*
&\frac{\partial L}{\partial b}\bigg|_{b=b^*, z=z^*, \hat{y}=\hat{y}^*} = -(y - \hat{y}^*) \cdot \sigma'(z^*).
\end{align}

This calculation shows that automatic differentiation needs two ingredients. We must know how to differentiate each primitive operation, and we store these derivatives. We also need the intermediate values $z^*$, $h^*$ and $\hat{y}^*$ to evaluate these derivatives at specific points. The chain rule then becomes a product of numbers, without a symbolic formula for the gradient.

A gradient is the slope of the function at a particular location in parameter space. To compute it, we must first evaluate the function at that point and store all intermediate values. For example, to compute $\partial h/\partial z|_{z=z^*} = \sigma'(z^*)$ we need the value $z^*$ from the forward pass, and to compute $-(y - \hat{y}^*)$ we need $\hat{y}^*$.

\paragraph{The Forward and Backward Passes}

\begin{figure}[ht!]
    \centering
    \includegraphics[width=0.95\textwidth]{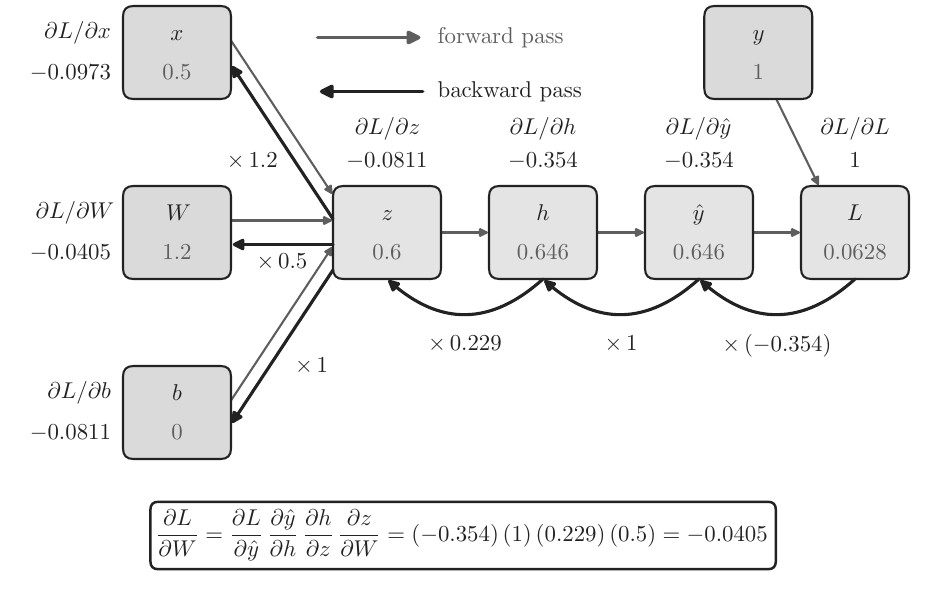}
    \caption{Backward pass for the graph of Figure~\ref{fig:computational_graph}. It starts at the loss and multiplies the local derivatives, written on the thick black arrows, from right to left. The black numbers outside the boxes are the gradients of the loss at the nodes, with the value 1 at the loss itself. The box at the bottom shows the chain of local derivatives that gives $\partial L/\partial W$. All other gradients are negative, because the prediction is below the target and the remaining local derivatives are positive.}
    \label{fig:backpropagation}
\end{figure}

Backpropagation\index{backward pass|(}\index{forward pass|(} implements reverse mode automatic differentiation in two phases. It takes its name from the second phase, in which gradients propagate backward from the output to the input layer.
\begin{enumerate}
    \item \textbf{Forward pass.}\index{forward pass} We follow the computational graph from the inputs to the outputs. We evaluate each operation in sequence and store all intermediate values ($z^*$, $h^*$, $\hat{y}^*$). This pass fixes the specific point at which we evaluate the gradients.

    \item \textbf{Backward pass.}\index{backward pass} We propagate gradients backward from the loss to each parameter. We apply the chain rule step by step and so traverse the computational graph in reverse. At each node we compute how a change in the output of that node affects the final loss. We start with $\partial L/\partial \hat{y}$ and work backward through the graph, calculating quantities such as $\partial L/\partial h = (\partial L/\partial \hat{y}) \cdot (\partial \hat{y}/\partial h)$ and so on.
\end{enumerate}
Figure~\ref{fig:backpropagation} shows the backward pass for our example.

This two-phase approach is a form of dynamic programming, which stores the results of small subproblems and reuses them. The forward pass stores the intermediate results, so the backward pass avoids redundant computation.\index{backward pass|)}\index{forward pass|)}

\paragraph{Why Reverse Mode Is Efficient} To see why backpropagation is efficient, we examine the dimensions of the matrices in our calculation. Let $x \in \mathbb{R}^D$ be the input, $W \in \mathbb{R}^{M \times D}$ the weight matrix of $M$ hidden units, and $z, h \in \mathbb{R}^M$ the hidden layer activations. We let $\hat{y}$ be a scalar that depends on $h$, as $\hat{y} = h$ does for $M = 1$. The dimensions are then
\begin{align}
\frac{\partial L}{\partial \hat{y}} &\in \mathbb{R} \quad \text{(scalar)}\\*
\frac{\partial \hat{y}}{\partial h} &\in \mathbb{R}^{1 \times M} \quad \text{(row vector)}\\
\frac{\partial h}{\partial z} &\in \mathbb{R}^{M \times M} \quad \text{(diagonal matrix)}\\*
\frac{\partial z}{\partial W} &\in \mathbb{R}^{M \times (M \times D)} \quad \text{(tensor)}.
\end{align}

Vectors are one-dimensional arrays and matrices are two-dimensional arrays. Tensors generalize them to any number of dimensions. For $M$ hidden units, $\partial z/\partial W$ is a third-order tensor, because it records how each element of $z$ changes with each element of the matrix $W$. Its element $(i,j,k)$ is the partial derivative of the $i$th element of $z$ with respect to the $(j,k)$th element of $W$.

We use the backward pass instead of forward mode because of the cost of the computation. In forward mode we would compute
\begin{equation}
\frac{\partial L}{\partial W} = \frac{\partial}{\partial W} \left[\frac{1}{2}(y - \sigma(Wx + b))^2\right].
\end{equation}
For $M = 1$, the forward chain rule first gives $\partial z/\partial W = x^T$ and then $\partial h/\partial W = (\partial h/\partial z) \cdot (\partial z/\partial W) = \sigma'(z) \cdot x^T$. For $M$ units, $\partial h/\partial W$ already has dimensions $M \times (M \times D)$. In a deep neural network the Jacobians, the matrices of partial derivatives, are taken with respect to the weights of all earlier layers. They grow with the number of parameters and become intractably large.

In contrast, the dimensions stay small when we work backward, because we start from the gradient of the loss, which is a scalar. This simple change of direction makes the computation faster by orders of magnitude, even though the mathematical operations are the same. Starting from the scalar $\partial L/\partial \hat{y}$, reverse mode computes
\begin{enumerate}
    \item $\partial L/\partial h = \partial L/\partial \hat{y} \cdot \partial \hat{y}/\partial h \in \mathbb{R}^{1 \times M}$ (row vector)
    \item $\partial L/\partial z = \partial L/\partial h \cdot \partial h/\partial z \in \mathbb{R}^{1 \times M}$ (row vector)
    \item $\partial L/\partial W = \partial L/\partial z \cdot \partial z/\partial W \in \mathbb{R}^{M \times D}$ (matrix)
\end{enumerate}

\begin{figure}[ht!]
    \centering
    \includegraphics[width=0.95\textwidth]{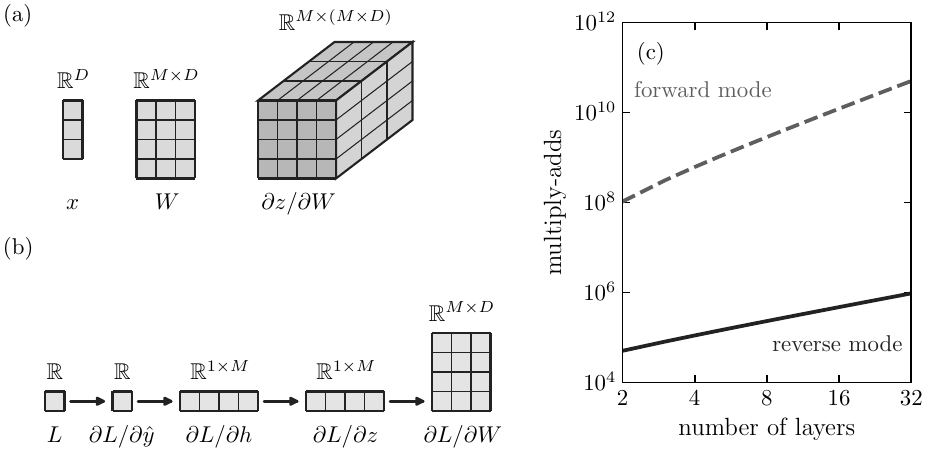}
    \caption{Forward and reverse mode for the example with $D$ inputs and $M$ hidden units. Panel~(a) shows the shapes in forward mode, where $\partial z/\partial W$ has the shape $M \times (M \times D)$ and most entries are zero. Panel~(b) shows the shapes in reverse mode, where $\partial L/\partial \hat{y}$ is a scalar. Panel~(c) shows the number of multiply-adds for the full gradient of a neural network with $M$ units per layer. Each layer adds $M^2$ weights, so forward mode grows with the square of the number of layers and reverse mode grows linearly.}
    \label{fig:backpropagation_efficiency}
\end{figure}

At each step the dimensions are manageable, so the computation stays tractable even for very deep neural networks. Figure~\ref{fig:backpropagation_efficiency} compares the dimensions in the two directions. Panel~(c) shows the cost. Forward mode needs far more multiply-adds than reverse mode, and the gap grows with the number of layers. One forward pass and one backward pass give the gradient with respect to all parameters, whatever their number. Together they cost only a small multiple of one forward pass, whereas finite differences need $P + 1$ forward evaluations.

Backpropagation has another advantage, the reuse of intermediate calculations. The calculations of $\partial L/\partial W$ and $\partial L/\partial b$ share common terms
\begin{align}
\frac{\partial L}{\partial W} &= \frac{\partial L}{\partial z} \cdot \frac{\partial z}{\partial W} = \frac{\partial L}{\partial z} \cdot x^T\\*
\frac{\partial L}{\partial b} &= \frac{\partial L}{\partial z} \cdot \frac{\partial z}{\partial b} = \frac{\partial L}{\partial z} \cdot 1.
\end{align}
For $M$ units, $\partial L/\partial z$ is a row. The product with $x^T$ is the outer product of the column $(\partial L/\partial z)^T$ and the row $x^T$, and $\partial L/\partial b$ equals the row $\partial L/\partial z$. Once we have $\partial L/\partial z$, we can reuse it for several parameter gradients, which increases the efficiency further. The computational graph illustrates this reuse, because the same gradient information flows backward through shared paths in the network.

{\makeatletter\@clubpenalty=10000\makeatother \clubpenalty=10000 \widowpenalty=10000 \displaywidowpenalty=10000 For deeper neural networks, the backward pass applies the same chain rule layer by layer. Take a two-layer neural network with $z^{(l)} = W^{(l)}h^{(l-1)} + b^{(l)}$ and $h^{(l)} = \sigma(z^{(l)})$, where $h^{(0)} = x$. The gradient of the loss with respect to $W^{(1)}$ is
\begin{equation}
\frac{\partial L}{\partial W^{(1)}} = \frac{\partial L}{\partial h^{(2)}} \cdot \frac{\partial h^{(2)}}{\partial z^{(2)}} \cdot \frac{\partial z^{(2)}}{\partial h^{(1)}} \cdot \frac{\partial h^{(1)}}{\partial z^{(1)}} \cdot \frac{\partial z^{(1)}}{\partial W^{(1)}}.
\end{equation}
Each extra layer adds one more factor to this chain, but the mechanics stay the same. In reverse mode every factor multiplies the row vector from the previous step, so the cost grows linearly with the number of layers. Forward mode must carry derivatives with respect to all earlier weights, so the gain of reverse mode grows with\index{automatic differentiation|)} depth.\par}

\paragraph{Hardware and Software} Backpropagation is also well matched to modern hardware accelerators such as GPUs,\index{graphics processing unit (GPU)} which excel at matrix multiplication. GPUs were created for rendering graphics in video games. These games need fast parallel matrix operations for transforming coordinates and computing pixel colors, which are the same operations that neural network training needs. When researchers found that these gaming processors speed up neural network training a great deal, deep learning grew quickly. Neural networks, one might say, won the ``hardware lottery'', because hardware built for graphics happened to suit their training.

What we call ``deep learning packages'', such as PyTorch,\index{PyTorch} JAX and TensorFlow, are at their core automatic differentiation systems. They are often described as tools for neural networks. Their automatic differentiation also applies to a much broader range of scientific computing applications, from physics simulations to optimization problems in various domains. Efficient gradient computation makes them useful for any application that needs derivatives of complex functions.

One such application is the derivative of a neural network with respect to its inputs, in addition to its parameters. Some networks use these derivatives to add a physical law to the loss. The loss then penalizes violations of a differential equation, so the network fits the data and obeys the physics.

Researchers define their models in terms of differentiable operations, and the package computes the gradients. This allows rapid experimentation with complex architectures without deriving gradients by hand, and it has enabled sophisticated architectures that go far beyond simple feedforward networks. The term ``deep learning'' itself stems from the ability to train neural networks with many layers. Multilayer neural networks existed for decades. The combination of automatic differentiation, modern optimization techniques and architectural innovations made the training of such deep networks feasible.\index{backpropagation|)}

\section{Gradient Flow and Activation Functions}

However,\index{activation function|(}\index{gradient flow|(} writing down a model and calculating its gradients does not guarantee successful training or convergence to a good solution. We can construct and differentiate more complex neural networks, yet our theoretical understanding of why they succeed or fail remains incomplete. Understanding gradient flow remains important for training neural networks well. Gradient flow is the way in which the gradient of the loss travels backward from the output through the layers. Its size can change along the way. It matters especially when we diagnose vanishing\index{vanishing gradients} gradients, which become very small, or exploding gradients, which become very large.

The chain rule explains the vanishing gradient problem. If several terms in the chain have small derivatives, as with sigmoid activations in their saturated regions, their product can become vanishingly small. This nearly stops the gradient flow to the earlier layers. The choice of activation function may look like a minor implementation detail. Its derivative enters this product, so it affects how a neural network trains and how well it performs. Without nonlinear activation functions the network reduces to a linear model, and the universal approximation theorem would not hold.

Figure~\ref{fig:activation_functions} compares the common activation functions. The sigmoid was a prominent early choice, and it is
\begin{equation}
\sigma(x) = \frac{1}{1 + e^{-x}}.
\end{equation}
It maps any real input to the range $(0, 1)$, which makes it appealing for representing probabilities or normalized activations. Despite its historical importance, the sigmoid is rarely used for the hidden units of modern neural networks, because of problems that appear in its gradient.

\begin{figure}[ht!]
    \centering
    \includegraphics[width=0.9\textwidth]{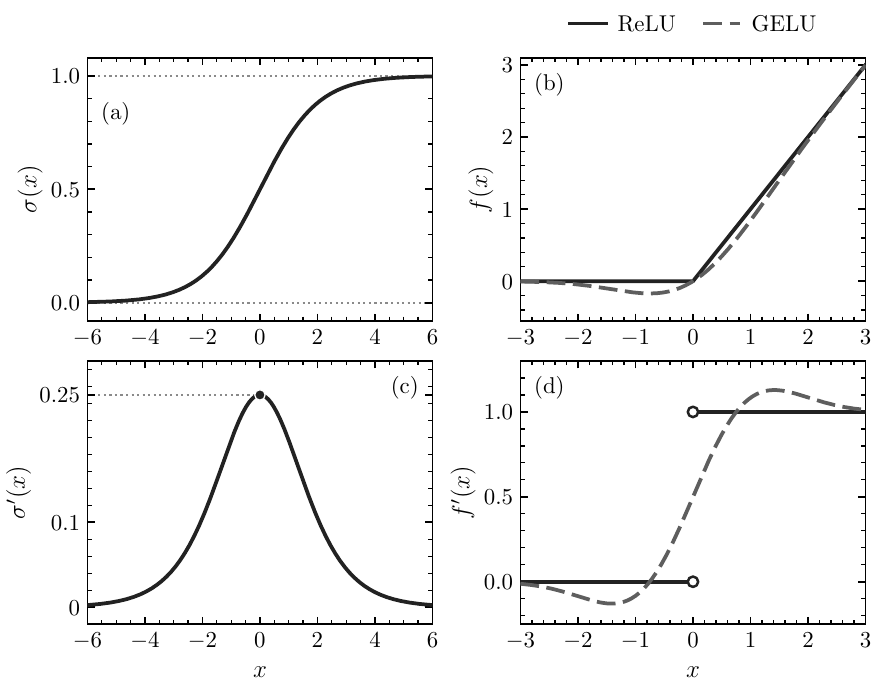}
    \caption{Panel~(a) shows the sigmoid and panel~(b) shows ReLU and GELU, with their derivatives in panels~(c) and (d). The derivative of the sigmoid has its maximum 0.25 at $x = 0$ and approaches zero for large $|x|$. GELU is a smooth variant of ReLU. Its derivative exceeds 1 for some positive inputs and is slightly negative for $x$ below about $-0.75$. The open circles mark that the derivative of ReLU is undefined at $x = 0$.}
    \label{fig:sigmoid_relu_comparison}
\end{figure}

As Chapter~7 shows for logistic regression, the derivative of the sigmoid is
\begin{equation}
\frac{\mathrm{d}\sigma(x)}{\mathrm{d}x} = \sigma(x)(1 - \sigma(x)).
\end{equation}
This derivative has a maximum value of 0.25 at $x = 0$ and approaches zero as $|x|$ increases, as panels~(a) and (c) of Figure~\ref{fig:sigmoid_relu_comparison} show. This creates problems for gradient-based learning. When $|x|$ is large, the gradient becomes very small and learning nearly stops in those regions.

This small gradient resembles excessive politeness in communication, where the message is too subtle to be effective. In optimization we often need more decisive signals. The small gradients of the sigmoid lead to long training times and possible convergence problems.

\paragraph{ReLU and GELU} The\index{Gaussian error linear unit (GELU)|(}\index{rectified linear unit (ReLU)|(} sigmoid has these problems, so researchers looked for alternative activation functions. The most prominent is the Rectified Linear Unit (ReLU),\index{ReLU|see{rectified linear unit}}\index{rectified linear unit (ReLU)} which became dominant in the 2010s and is defined as
\begin{equation}
\text{ReLU}(x) = \max(0, x).
\end{equation}
Its derivative is
\begin{equation}
\frac{\mathrm{d}\text{ReLU}(x)}{\mathrm{d}x} =
\begin{cases}
1 & \text{if } x > 0 \\
0 & \text{if } x < 0
.\end{cases}
\end{equation}

ReLU has several advantages over the sigmoid. For positive inputs the gradient is 1, so it does not shrink with the size of the input as the gradient of the sigmoid does. For negative inputs ReLU outputs zero, so for a given input many neurons are inactive and output exactly zero. We call these sparse activations. They can save computation and may reduce overfitting. Most practically, both the function and its derivative are simple to compute, so ReLU networks are faster to train and evaluate.

However, ReLU has its own limitation, the ``dying ReLU'' problem. When the weight updates of a neuron push the input of its activation function consistently negative, the gradient becomes zero. The neuron is then ``dying'' and stops learning. This led to variants of ReLU that try to remove the problem and keep its advantages.

More recently, the Gaussian Error Linear Unit (GELU),\index{GELU|see{Gaussian error linear unit}}\index{Gaussian error linear unit (GELU)} introduced in 2016, has gained popularity, particularly in Transformer architectures. It is defined as
\begin{equation}
\text{GELU}(x) = x \cdot \Phi(x),
\end{equation}
where $\Phi(x)$ is the cumulative distribution function of the standard normal distribution. GELU smoothly interpolates between passing and blocking signals. It avoids the non-differentiability of ReLU at zero and keeps ReLU's advantages for large positive values.

Since $\Phi(x)$ approaches 1 for large positive $x$, GELU behaves almost like ReLU there, with gradients close to 1. For large negative $x$, $\Phi(x)$ approaches 0 and the output is close to zero, as for ReLU. Figure~\ref{fig:sigmoid_relu_comparison} shows the sigmoid, ReLU and GELU functions and their derivatives.

Most neural networks use ReLU, GELU or a close variant in their hidden layers, whatever the task. The output layer is different, because its activation must give the kind of number that the task needs. For regression the output can be any real number, so we use the identity. For two classes the output is a probability between 0 and 1, so we apply the sigmoid. For several classes the outputs are probabilities that sum to one, so we apply the softmax.\index{Gaussian error linear unit (GELU)|)}\index{rectified linear unit (ReLU)|)}

\paragraph{Avoiding Superlinear Growth} A function $f$ is superlinear if it grows faster than a linear function as $|x|$ increases, which means that
\begin{equation}
\lim_{|x| \to \infty} \left|\frac{f(x)}{x}\right| = \infty.
\end{equation}The activation functions above are not superlinear. The ratio $|f(x)/x|$ is at most 1 for ReLU and GELU and tends to 0 for the sigmoid.

Growing at most linearly does not make a neural network linear. ReLU is linear on each side of zero and has a kink at zero. GELU curves smoothly near zero. Different neurons put the kink or the curve at different inputs, because their weights and biases differ. The sum of many such units is then typically a nonlinear function of the input. The growth condition limits only how fast an activation grows, and the quadratic and cubic functions are nonlinear but grow faster than linearly.

\begin{figure}[ht!]
    \centering
    \includegraphics[width=0.95\textwidth]{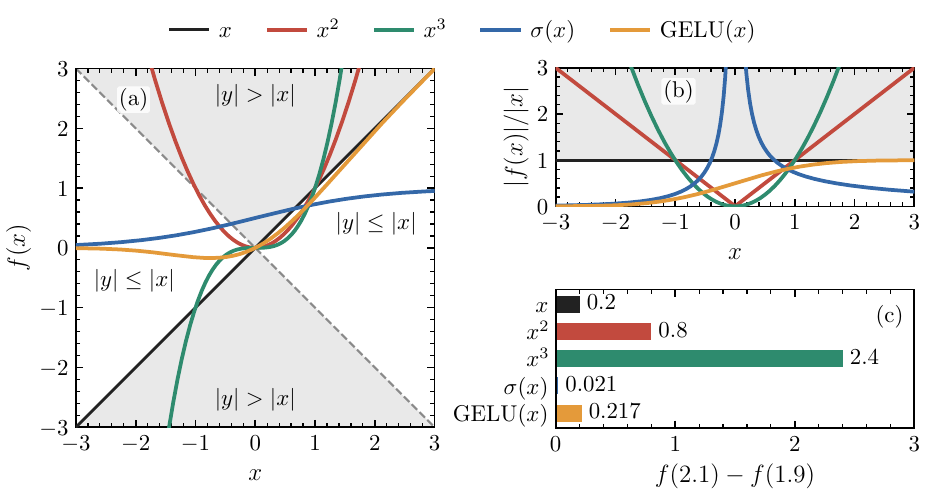}
    \caption{Superlinear and non-superlinear functions. Panel~(a) shows $x$, $x^2$, $x^3$, the sigmoid and GELU, with the region $|y| > |x|$ shaded. Panel~(b) shows $|f(x)|/|x|$, which exceeds 1 in the shaded region. At large $|x|$ the ratio grows without limit for $x^2$ and $x^3$ and tends to 0 for the sigmoid. For GELU it tends to 1 at large positive $x$ and to 0 at large negative $x$. Panel~(c) shows $f(2.1) - f(1.9)$ for each function, the change of the output for an input error of $\pm 0.1$ at $x = 2$.}
    \label{fig:superlinear_functions}
\end{figure}

The absence of superlinearity has a statistical reason. As Chapter~3 shows, the power $X^k$ of a random variable $X$ with mean $\mu$ has about $k^2\mu^{2k-2}$ times the variance of $X$. We compare three polynomial functions of increasing order, namely
\begin{equation}
f_1(x) = x \quad \text{(linear)}.
\end{equation}
\begin{equation}
f_2(x) = x^2 \quad \text{(quadratic)}.
\end{equation}
\begin{equation}
f_3(x) = x^3 \quad \text{(cubic)}.
\end{equation}Figure~\ref{fig:superlinear_functions} shows these polynomials together with the sigmoid and GELU.

With noisy data and values above 1 in size, the error in the function value grows with the polynomial order. Consider a data set with values around $x = 2$ and a small measurement error of $\pm 0.1$, which changes the function values as follows.
\begin{itemize}
\item The linear function differs by $2.1 - 1.9 = 0.2$.
\item The quadratic function differs by $2.1^2 - 1.9^2 = 0.8$.
\item The cubic function differs by $2.1^3 - 1.9^3 \approx 2.4$.
\end{itemize}

This example shows how each additional power amplifies the impact of measurement errors near $x = 2$. In a deep neural network the amplification compounds from layer to layer. Superlinear activation functions would therefore amplify small input variations and lead to unstable gradients and numerical problems during training.

Activation functions with at most linear growth help to keep training stable, so we can stack many layers and build deep representations of complex patterns. Large weights can still make gradients explode, so such activation functions do not guarantee stable training. We choose the activation function to balance expressive power with training stability. We do not yet fully understand many properties of neural networks. Still, this connection between activation functions and the variance of noisy inputs shows that some design choices follow statistical principles.

Architectural innovations also help keep the gradient flow stable. Two examples are residual connections and layer normalization.

\paragraph{Residual Connections} A residual\index{residual connection} connection adds the input of a block of layers to the output of the block. If $F$ is the function that the block computes, the output of block $l$ is
\begin{equation}
\mathbf{h}^{(l)} = \mathbf{h}^{(l-1)} + F(\mathbf{h}^{(l-1)}).
\end{equation}
The addition gives the gradients a shortcut through very deep neural networks. By the chain rule, the derivative of the output with respect to the input is
\begin{equation}
\frac{\partial \mathbf{h}^{(l)}}{\partial \mathbf{h}^{(l-1)}} = \mathrm{Id} + \frac{\partial F}{\partial \mathbf{h}^{(l-1)}},
\end{equation}
where $\mathrm{Id}$ is the identity matrix. When the derivatives of $F$ are small, each factor in the chain stays close to the identity. The product of many such factors then does not shrink toward zero, so the gradients can still reach the earliest layers.

\paragraph{Layer Normalization} Layer normalization\index{layer normalization} rescales the activations of a layer. For the $M$ activations $z_j$ of a layer, it first computes their mean and variance
\begin{equation}
\bar{z} = \frac{1}{M}\sum_{j=1}^M z_j, \qquad s^2 = \frac{1}{M}\sum_{j=1}^M (z_j - \bar{z})^2.
\end{equation}
It then normalizes the activations and applies a learned scale and shift
\begin{equation}
\hat{z}_j = \frac{z_j - \bar{z}}{\sqrt{s^2 + \epsilon}}, \qquad y_j = \gamma_j \hat{z}_j + \delta_j.
\end{equation}
The normalized values $\hat{z}_j$ have zero mean and unit variance, up to the small constant $\epsilon$. The learned $\gamma_j$ and $\delta_j$ let the network undo the normalization where this helps. Activations in a fixed range help keep the gradients from growing or shrinking from layer to layer.

These innovations look simple in hindsight, yet they made it possible to train much deeper neural networks than before. A full treatment of these techniques is beyond this chapter. They show that an understanding of gradient flow still guides research in deep learning.\index{activation function|)}\index{gradient flow|)}

\section{Architectures and Inductive Biases}

So\index{neural network!architectures|(} far we have focused on fully connected feedforward networks, also called multilayer perceptrons, but neural networks form a much broader family of architectures. This book concentrates on the principles of statistical machine learning, so we do not give the engineering details or complete specifications of these architectures. We define only the core operation of each one. Still, the conceptual differences between architectures and their inductive biases matter for applications in astronomy. A fully connected network is one architectural pattern, with neurons arranged in layers and each neuron connected to every neuron of the previous layer. This structure is not always the most efficient or effective choice for a specific type of data.

Machine learning methods rely on inductive bias. It is the set of assumptions that guide learning when the data alone cannot determine the solution. Linear regression assumes that relationships are linear in the feature space (Chapter~4). Logistic regression assumes a linear decision boundary between classes (Chapter~7). Gaussian processes assume smoothness governed by a specific kernel function (Chapter~14).

Neural networks are no exception. The universal approximation theorem makes feedforward networks very flexible, but this flexibility is both a blessing and a curse. Without appropriate inductive biases, a model might need far more data and computation to learn patterns that a more constrained architecture finds more easily. A constrained architecture, on the other hand, might miss patterns that do not match its built-in assumptions. Machine learning balances modeling, where we build in prior knowledge through inductive biases, and learning, where we extract patterns directly from data. Different architectures encode different inductive biases.

You have likely met Convolutional Neural Networks (CNNs), Recurrent Neural Networks (RNNs), Transformers and Graph Neural Networks (GNNs) in the scientific literature or popular discourse. These architectures use the same backpropagation algorithm as feedforward networks. The theory of double descent applies to them as well. They differ in their structure and in the inductive biases they encode. They suit different kinds of data. CNNs suit images on a grid, RNNs suit ordered sequences, Transformers suit long-range dependencies, and GNNs suit objects with relations but no grid or order.

\paragraph{Convolutional Neural Networks}

Convolutional\index{convolutional neural network (CNN)|(} Neural Networks encode two inductive biases that suit image data, translation\index{translation equivariance} equivariance and hierarchical local structure. Translation equivariance means that shifting the image shifts the output of a convolutional layer by the same amount. The layer therefore detects a pattern wherever it appears. Hierarchical local structure means that nearby pixels are more closely related than distant ones. These local patterns combine to form more complex features at larger scales.

Convolutional layers implement these inductive biases. In a convolutional layer, small filters (kernels) slide across the input and apply the same transformation at each position. Here a kernel is a small array of weights and differs from the kernel function of Chapter~14. The convolution\index{convolutional neural network (CNN)} of an image $X$ with a $k \times k$ kernel $K$ at $(i,j)$ is
\begin{equation}
(X * K)_{i,j} = \sum_{m=1}^{k} \sum_{n=1}^{k} X_{i+m-1,j+n-1} \cdot K_{m,n},
\end{equation}
where $X_{i,j}$ is a pixel and $K_{m,n}$ a kernel weight.

The operation computes a weighted sum of the pixels in the $k \times k$ patch whose top left corner is $(i,j)$. Strictly, this sum is a cross-correlation, because it does not flip the kernel as a convolution in Fourier analysis does. Deep learning still calls it convolution, and flipping would only reorder the learned weights. Centering the kernel on the patch would only shift the index of the output.

This parameter sharing reduces the number of learnable parameters compared with a fully connected network. It also encodes the assumption that the same feature detectors apply throughout the image. For an input image of size $n \times n$ and a kernel of size $k \times k$, a convolutional layer needs only $k^2$ parameters (plus bias). A fully connected layer with the same $(n-k+1)^2$ outputs connects each output to all $n^2$ pixels, so it would need $n^2 (n-k+1)^2$ weights. The convolutional layer needs far fewer because it reuses the same kernel weights at all positions of the input.

One convolutional layer sees only a small patch of the image. To exploit the hierarchical local structure of images, we stack several convolutional layers. The first layer responds to simple patterns such as edges, and each later layer combines the patterns of the layer before. The region of the image that a unit depends on, its receptive field, therefore grows with the depth. A unit in a deep layer can then respond to a complex feature.

A CNN\index{CNN|see{convolutional neural network}} often places a pooling\index{pooling layer} layer after a convolutional layer to reduce the size of the feature map. Max pooling splits the map into small blocks and keeps only the largest value in each block. The pooled map is smaller, and it changes little when a feature moves by a small shift.

\begin{figure}[ht!]
    \centering
    \includegraphics[width=1.0\textwidth]{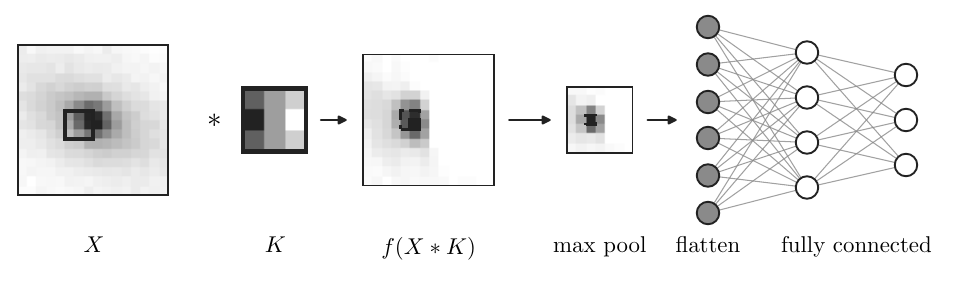}
    \caption{A CNN applied to a toy image $X$ of a galaxy, where $K$ is the kernel and $f$ is the activation function. A thick frame marks the kernel window and its output pixel. Max pooling keeps the largest value in each block, and a dashed square marks a pooled block. Pooling adds invariance to small shifts. The fully connected layers turn the pooled map into the output.}
    \label{fig:cnn_architecture}
\end{figure}

Take a CNN that classifies images of galaxies. The convolutional layers respond to local features such as edges and bright spots, and the pooling layers reduce the size of the maps. The fully connected layers combine the pooled features, and the output layer gives a probability for each galaxy class. Figure~\ref{fig:cnn_architecture} shows the structure of a CNN.

The mathematics of CNNs has led to the theory of scattering transforms, which we do not cover in this book. Roughly, a scattering transform filters an image repeatedly at different scales and orientations, then takes absolute values and averages. This gives features that change little under small shifts and deformations of the image. Scattering transforms give insight into why CNNs work so well for image data. It is an example of the ``Science of AI'' that we mentioned earlier in this chapter.

Scattering transforms have also become a rising summary statistic in cosmology. They analyze the large-scale structure of the universe and extract features from galaxy surveys.\index{convolutional neural network (CNN)|)}

\paragraph{Recurrent Neural Networks}

A\index{recurrent neural network (RNN)|(} CNN expects data on a grid. Many astronomical data are instead sequences in which the order matters, such as the light\index{light curves} curve of a variable star. An RNN\index{RNN|see{recurrent neural network}} reads such a sequence one observation at a time and keeps a memory of what it has seen. This suits time series that come from evolving physical processes. To understand the brightness of a pulsating star, we need its pulsation phase as well as the flux at one instant. The sequence encodes this phase.

An RNN processes a sequence $\mathbf{x}_1, \ldots, \mathbf{x}_n$ with a hidden state $\mathbf{h}_t$ that summarizes the observations up to time $t$. At each step the network computes the new state from the previous state and the new observation,
\begin{equation}
\mathbf{h}_t = f(\mathbf{W}_{hh}\mathbf{h}_{t-1} + \mathbf{W}_{xh}\mathbf{x}_t + \mathbf{b}_h),
\end{equation}
where $f$ is an activation function. The network learns the weights $\mathbf{W}_{hh}$ and $\mathbf{W}_{xh}$ and the bias $\mathbf{b}_h$ from data, and the initial state is $\mathbf{h}_0$. Think of $\mathbf{h}_t$ as the notebook of the network at time $t$, which changes with each new observation.

\begin{figure}[ht!]
    \centering
    \includegraphics[width=1.0\textwidth]{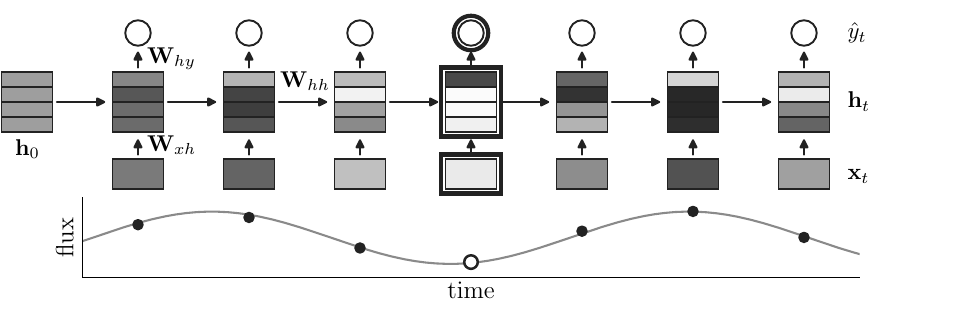}
    \caption{An RNN unrolled over the light curve of a pulsating star. The same weights act at each step, and a thick outline marks one time step. The grey levels show the hidden state of an untrained neural network. The input $\mathbf{x}_t$ is the flux, shown in grey, and the output $\hat{y}_t$ is a prediction at each step. The initial state $\mathbf{h}_0$ is zero.}
    \label{fig:rnn_architecture}
\end{figure}

To make a prediction, the network turns its current state into an output,
\begin{equation}
\hat{y}_t = g(\mathbf{W}_{hy}\mathbf{h}_t + \mathbf{b}_y),
\end{equation}
where $g$ is an output activation function. The output at each step can be the next flux value. A variability class describes the whole light curve of the star. We then apply the output layer to the final state $\mathbf{h}_n$ or to the average of the states. Figure~\ref{fig:rnn_architecture} shows the structure of an RNN unrolled in time.

The same weights apply at each time step, in the way that physical laws stay the same from one moment to the next. This sharing of weights over time is the inductive bias of an RNN. It mirrors the sharing over space in a CNN, where the same kernel applies at each position. An RNN needs no fixed input length, because it runs for as many steps as the sequence has. It also handles streaming data, since the state updates as each new observation arrives. This suits real-time uses, in which the classification changes as data accumulate.

The recurrence also limits the memory of an RNN. Information from early observations passes through many updates, like a message in the game of telephone. During backpropagation each step multiplies the gradient, so it can vanish or explode as in deep feedforward networks. The network then has difficulty learning long-term dependencies. Variants such as the long short-term memory (LSTM) and the gated recurrent\index{recurrent neural network (RNN)} unit (GRU) add gates that decide what to keep and forget. They give information a direct path through time, as residual connections do through layers.

Researchers have used RNNs to classify transients, to predict coronal mass ejections from magnetic fields, and to follow galaxy assembly through merger trees. Transformers have replaced them in many of these tasks.\index{recurrent neural network (RNN)|)}

\paragraph{Transformers}

CNNs\index{attention|(}\index{token|(}\index{transformer|(} and RNNs both have difficulty with long-range dependencies. A CNN sees only a small patch of the input in each layer. An RNN stores the whole history in a state of fixed size, which loses fine detail in long sequences. Spectra and time series can carry important information in long-range dependencies and global patterns as well as in local structures.

In a stellar spectrum, the absorption lines of one element appear at many wavelengths, and measuring its abundance requires combining them. A CNN might focus on local spectral features and miss correlations between widely separated spectral lines. Such correlations can indicate specific physical conditions. The kernel of a CNN also treats a line in the same way at each wavelength, although lines at different wavelengths have different physical meanings.

Transformers\index{transformer} avoid these difficulties. They are widely used for language processing and in a growing number of other domains. Transformers encode a different set of inductive biases from CNNs. They do not assume locality and instead use attention\index{attention} mechanisms that connect any two elements of a sequence directly, regardless of their distance.

The central element of a Transformer is self-attention,\index{attention!self-attention} which relates each element of a sequence to all other elements through learned attention weights. Before the attention step, an embedding\index{embedding} maps each element of the sequence, called a token,\index{token} to a vector of higher dimension. For a spectrum, a token represents the flux in one wavelength bin, and the position of the token gives the wavelength. Self-attention itself treats the tokens as an unordered set, so it cannot tell which bin comes first. The embedding therefore also adds a code for the position of each token, which tells the network the wavelength of the bin.

Let the matrix $\mathbf{X}$ hold one embedded token in each row. Self-attention first makes three versions of each token with learned linear maps,
\begin{align}
& \mathbf{Q} = \mathbf{X}\mathbf{W}_Q, \\*
& \mathbf{K} = \mathbf{X}\mathbf{W}_K, \\*
& \mathbf{V} = \mathbf{X}\mathbf{W}_V.
\end{align}
The rows of $\mathbf{Q}$ are queries, which encode what a token is looking for. The rows of $\mathbf{K}$ are keys, which encode what a token contains. The rows of $\mathbf{V}$ are values, which encode the information that a token passes on. Here $\mathbf{K}$ is unrelated to the kernel of a CNN.

The attention sums the values of all tokens, with weights that depend on the queries and the keys,
\begin{equation}
\mathrm{Attention}(\mathbf{Q}, \mathbf{K}, \mathbf{V}) = \mathrm{softmax}\left(\frac{\mathbf{Q}\mathbf{K}^T}{\sqrt{d}}\right)\mathbf{V},
\end{equation}
where $d$ is the dimension of the query and key vectors. For a sequence of length $n$, the product $\mathbf{Q}\mathbf{K}^T$ is an $n \times n$ matrix with a score for each of the $n^2$ pairs of tokens. Element $(i,j)$ is the dot product of the query of token $i$ with the key of token $j$. It measures how relevant token $j$ is for token $i$. The factor $1/\sqrt{d}$ keeps the scores from growing with $d$. The softmax turns each row of scores into positive weights that sum to 1.

The output for token $i$ is the sum of all value vectors, weighted by row $i$ of the attention weights. The model therefore learns which elements should influence each other. If a token gives large weights to the tokens of bins with magnesium lines, for example, its output then draws mostly on the magnesium features.

Like the stacked layers of a CNN, a Transformer stacks attention layers, and each layer works on the output of the one before. Multihead attention runs $h$ attention operations in parallel, each with its own matrices $\mathbf{W}_Q$, $\mathbf{W}_K$ and $\mathbf{W}_V$. A final projection combines the concatenated outputs of all heads. The heads can specialize. In a spectrum with superimposed lines, one head might follow iron lines and another carbon lines.

\begin{figure}[ht!]
    \centering
    \includegraphics[width=1.0\textwidth]{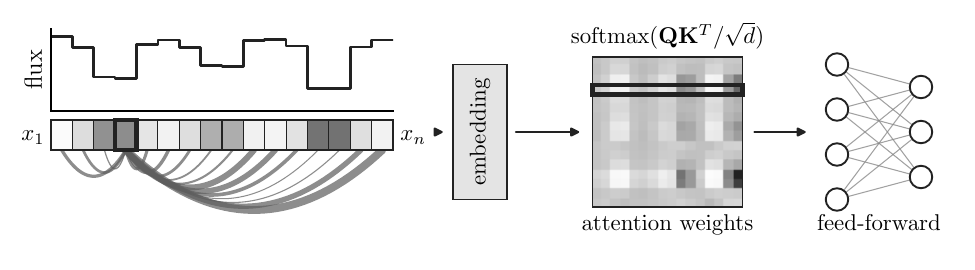}
    \caption{A Transformer applied to a toy spectrum, with one token for the flux in each wavelength bin. The arcs show the attention weights of the marked token (outlined) for all other tokens. The matrix shows the attention weights of one untrained layer, with the outlined row for the marked token. The rows of $\mathbf{Q}$ and $\mathbf{K}$ are query and key vectors of dimension $d$. The feedforward layers act on each vector.}
    \label{fig:transformer_architecture}
\end{figure}

Take a Transformer that reads a stellar spectrum, with one token for the flux in each wavelength bin. After the attention layers, a feedforward network acts on each vector. The output layer then turns the vectors into a stellar parameter such as the temperature, for example after averaging them over the tokens. Figure~\ref{fig:transformer_architecture} shows the structure of a Transformer.

Transformers are promising for astronomy. In a stellar spectrum, a Transformer can directly model relationships between the fluxes at any two wavelengths. This helps to identify complex patterns that indicate specific stellar compositions or evolutionary states. In light curves, Transformers can likewise relate events separated by long time intervals. They may reveal subtle periodicities or evolutionary trends that other architectures might miss.\index{attention|)}\index{token|)}\index{transformer|)}

\paragraph{Graph Neural Networks}

CNNs\index{graph neural network (GNN)|(}\index{message passing|(} need data on a grid, and Transformers need data that we split into tokens. Many astronomical observations are instead discrete objects with relations between them. A galaxy survey maps many galaxies as points in space, each with properties such as mass, star formation rate and morphology. These galaxies cluster hierarchically, interact gravitationally and share evolutionary histories through their environment. Dark matter halos likewise form merger trees that record their assembly history. A GNN\index{GNN|see{graph neural network}} represents such data as a graph,\index{graph neural network (GNN)} with objects as nodes and relations as edges.

A GNN works on a graph $\mathcal{G} = (\mathcal{V}, \mathcal{E})$, where the nodes $\mathcal{V}$ are objects and the edges $\mathcal{E}$ encode relations. Each galaxy node $i$ has features $\mathbf{h}_i$, such as mass, gas fraction and star formation rate. An edge from galaxy $j$ to galaxy $i$ can carry attributes $\mathbf{e}_{ij}$, such as the distance or the relative velocity.

{\makeatletter\@clubpenalty=10000\makeatother \clubpenalty=10000 \widowpenalty=10000 \displaywidowpenalty=10000 The central operation is message\index{message passing} passing. In each layer, a node forms its new representation from its own state and from the messages that it receives from its neighbors,
\begin{equation}
\mathbf{h}_i^{(l+1)} = \phi^{(l)}\left(\mathbf{h}_i^{(l)}, \bigoplus_{j \in \mathcal{N}(i)} \psi^{(l)}(\mathbf{h}_i^{(l)}, \mathbf{h}_j^{(l)}, \mathbf{e}_{ij})\right),
\end{equation}
where $(l)$ is the layer index and $\mathcal{N}(i)$ is the set of neighbors of node $i$.\par}

Think of this as galaxies talking to their neighbors. The message function $\psi$ decides what galaxy $j$ sends to galaxy $i$, such as how its star formation might influence its neighbors. The aggregation $\bigoplus$ collects the messages of all neighbors. The update function $\phi$ then forms the new representation of galaxy $i$ from its current state and the received messages. Both $\psi$ and $\phi$ are neural networks with learnable parameters.

The aggregation is a sum, a mean or a maximum. It does not depend on the order of the neighbors, just as the mean mass of galaxies does not depend on catalog order. This invariance to permutations\index{permutation invariance} is the main inductive bias of a GNN.

\begin{figure}[ht!]
    \centering
    \includegraphics[width=1.0\textwidth]{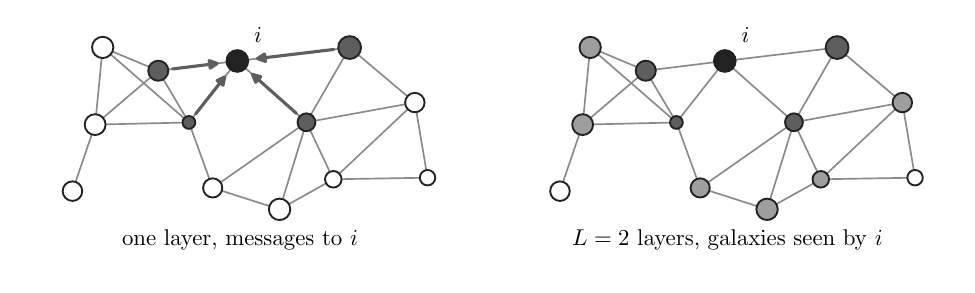}
    \caption{A GNN applied to a toy graph of galaxies that are connected when they are close, with the circle size showing the mass. In the left graph, the neighbors (dark grey) of the black galaxy $i$ send messages to it in one layer. In the right graph, two layers carry information to $i$ from its neighbors (dark grey) and from their neighbors (light grey).}
    \label{fig:gnn_architecture}
\end{figure}

After $L$ layers, each node has received information from objects up to $L$ connections away. The final representations $\mathbf{h}_i^{(L)}$ then give predictions for single nodes, such as galaxy properties. Pooling them over all nodes gives predictions for the whole graph, such as properties of a cluster. Figure~\ref{fig:gnn_architecture} shows the structure of a GNN.

This layered propagation resembles physical processes. Gravity acts through structure, and chemical enrichment spreads between interacting systems. A GNN can discover which relations matter most, perhaps finding that galaxy properties depend more on the dark matter environment than on spatial proximity. The graph can also be fully connected, with every node connected to every other node. This suits point clouds such as galaxy distributions, in which the order of the objects is arbitrary but their collective properties matter.

Researchers have used GNNs to learn galaxy intrinsic alignments, a systematic effect in weak lensing, from the environments of galaxies. They have reconstructed merger histories, in which progenitor systems form trees with a variable number of nodes. They have also inferred cosmological parameters directly from galaxy catalogs by treating galaxies as nodes. Such GNNs can learn which spatial scales and relations constrain the parameters, and they may find features beyond the usual two-point statistics.

The choice of architecture matters for each type of astronomical data, because each type has its own characteristics that suit some architectures better than others. A fully connected network is inefficient for image classification compared with a CNN, which uses the spatial structure of images. Researchers also combine modules, for example convolutional layers for local features with attention layers for long-range dependencies.

Neural network architectures and their inductive biases are becoming more important for astronomers. The field of architectures is evolving rapidly, and new designs combine computational principles with domain-specific knowledge. Astronomers should follow these developments and carefully evaluate their relevance to specific astronomical problems.\index{graph neural network (GNN)|)}\index{message passing|)}\index{neural network!architectures|)}

\section{Bayesian Neural Networks}

Backpropagation\index{Bayesian neural network|(}\index{neural network!uncertainty|(} and SGD train a neural network efficiently. They lead only to a point estimate of the parameters, the vector $\mathbf{w}$ that minimizes the empirical loss (the loss on the training data)
\begin{equation}
\mathbf{w}^* = \arg\min_{\mathbf{w}} L(\mathbf{w}).
\end{equation}

This is analogous to maximum likelihood estimation in linear regression (Chapter~4). It gives the best fit but no information about the uncertainty of the parameters. In Bayesian analysis we characterize the posterior distribution of the parameters, given by
\begin{equation}
p(\mathbf{w}|\mathcal{D}) \propto p(\mathcal{D}|\mathbf{w})p(\mathbf{w}).
\end{equation}

With the posterior we can make predictions that account for the uncertainty of the parameters. We integrate over the parameter values and obtain
\begin{equation}
p(y|\mathbf{x}, \mathcal{D}) = \int p(y|\mathbf{x}, \mathbf{w})p(\mathbf{w}|\mathcal{D})\mathrm{d}\mathbf{w}.
\end{equation}

In astronomy, this uncertainty is often as important as the prediction itself. It helps us tell confident predictions, based on ample evidence, from speculative ones where the model is uncertain.

One main limitation of neural networks, compared with the simpler models of the earlier chapters, is that Bayesian inference is difficult. For Bayesian linear regression we derived the posterior analytically (Chapter~5), so the inference was tractable. For neural networks, the nonlinear structure and the high dimension of the parameter space make the posterior intractable to compute exactly. A modern neural network can have millions or even billions of parameters, so sampling directly from its posterior is too costly.

Researchers have developed several practical approximations to Bayesian inference in neural networks. One of them, dropout,\index{dropout} is simple to apply and also interesting in theory. It was introduced as a regularization technique, and its connections to Bayesian methods make it useful for uncertainty\index{uncertainty quantification} quantification.

\paragraph{Monte Carlo Dropout}

In\index{Monte Carlo dropout|(}\index{dropout|(} machine learning, training fits the weights of a neural network, and inference uses the trained network to predict on new data. During training, dropout sets the output of each neuron to zero with probability $p$. It works in two steps.
\begin{enumerate}
    \item For each forward pass in training, we create a random mask for each neuron $i$ such that
    \begin{equation}
        m_i \sim \text{Bernoulli}(1-p).
    \end{equation}
    Here $m_i = 1$ means that the neuron stays active and $m_i = 0$ means that it is removed for this pass. The Bernoulli distribution takes the value 1 with probability $1-p$ and the value 0 with probability $p$.

    \item We then apply the mask to the outputs of the layer and rescale them to get
    \begin{equation}
        \tilde{\mathbf{h}}^{(l)} = \mathbf{m} \odot \mathbf{h}^{(l)} / (1-p).
    \end{equation}
    The symbol $\odot$ denotes the element-wise product of the mask $\mathbf{m}$ and the layer outputs $\mathbf{h}^{(l)}$. The factor $1/(1-p)$ keeps the expected output of the layer the same in training and in inference. It is similar to the bias correction of the Adam optimizer, where we divide by $1-\beta_1^t$ to correct the initialization bias of the moving averages. In both cases the division restores the expected value.
\end{enumerate}

\begin{figure}[ht!]
    \centering
    \includegraphics[width=0.95\textwidth]{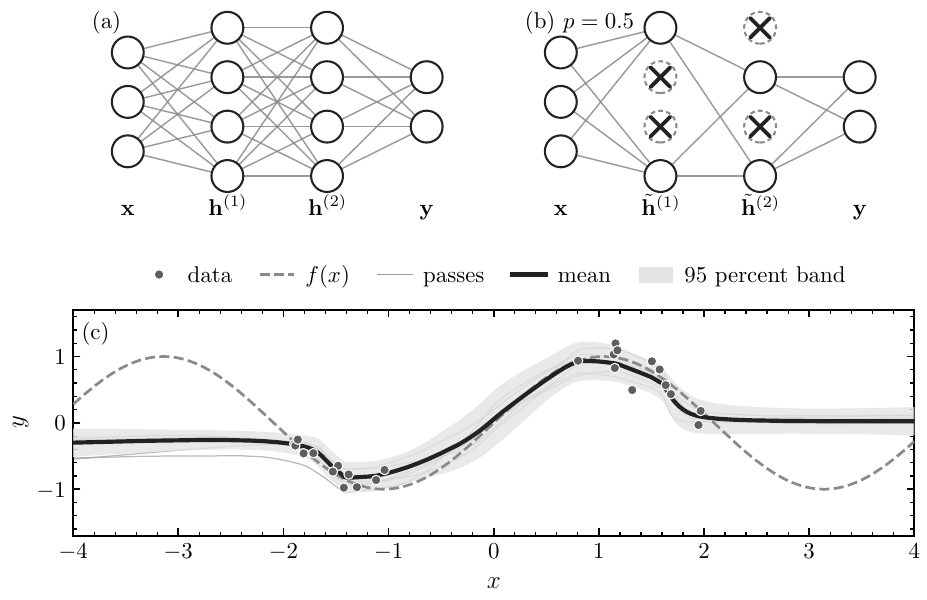}
    \caption{Panel~(a) shows a fully connected network, and panel~(b) shows it with a random mask, which removes the crossed neurons. Panel~(c) shows Monte Carlo dropout for a neural network trained on noisy data with a gap. The thin curves are a few passes with different masks, and the thick curve is the mean of all passes. The band is the 95 percent interval of all passes. It is wider in the gap than in the data regions. Outside the data range the band does not widen, and the mean departs from the function $f(x)$.}
    \label{fig:dropout_visualization}
\end{figure}

Dropout makes the network build redundant paths, so that the loss of a few neurons does little harm. It is like training a team whose members sometimes take sick leave. The team must learn to work well when some members are absent, so that no single person becomes a bottleneck. In the same way, the network does not rely too heavily on any single neuron or path, and this redundancy helps prevent overfitting. Figure~\ref{fig:dropout_visualization} shows a neural network and the same network with some neurons dropped by a mask. Panel~(c) shows Monte\index{Monte Carlo dropout} Carlo dropout, which we describe below.

In standard inference, the trained network predicts on new data with dropout typically turned off, so we use the full network. Training with dropout lets the network tolerate missing neurons, and the full network generally gives the most accurate predictions.

However, we can also keep dropout active in inference and perform multiple forward passes with different random masks. This technique is called Monte Carlo dropout. It approximates the Bayesian predictive integral above and gives uncertainty estimates in addition to predictions.

Monte Carlo dropout lets us sample different network configurations during inference. The predictions vary from pass to pass, and this variation gives insight into how confident the network is. If the predictions agree whichever neurons are dropped, we can trust them more. Large variations between the predictions suggest higher uncertainty.

Dropout has an ensemble learning view and a Bayesian view. In the ensemble view, dropout creates an implicit collection of neural networks. Each mask gives a different ``thinned'' network, a subnetwork of the original architecture. This is similar to bootstrap aggregating (bagging), which trains independent models on different bootstrap resamples of the data (Section~3.7). The spread of their predictions estimates the variance of the prediction, and the average of the predictions reduces it and improves generalization. With Monte Carlo dropout we use the thinned networks in the same way when we predict on new data.

This ensemble view connects directly to the Bayesian view. With dropout kept active in inference, multiple forward passes with different random masks effectively sample from an approximate posterior distribution over network functions. Each dropout mask gives a different neural network configuration, analogous to a sample from the posterior distribution of the weights. Averaging the predictions of these forward passes gives
\begin{equation}
p(y|\mathbf{x}, \mathcal{D}) \approx \frac{1}{T}\sum_{t=1}^T p(y|\mathbf{x}, \mathbf{m}_t).
\end{equation}
Here $T$ is the number of forward passes with different masks $\mathbf{m}_t$.

This is an approximation to the predictive distribution that we would ideally compute by Bayesian model averaging. Bayesian model averaging is the average of the predictions over the possible parameter values, weighted by the posterior. The variance of these predictions is a measure of the model uncertainty, and we estimate it by
\begin{equation}
\text{Var}_{\mathbf{m}}[f(\mathbf{x}, \mathbf{m})] \approx \frac{1}{T}\sum_{t=1}^T (f(\mathbf{x}, \mathbf{m}_t) - \bar{f}(\mathbf{x}))^2.
\end{equation}
The function $f(\mathbf{x}, \mathbf{m}_t)$ is the prediction of the network with the mask $\mathbf{m}_t$, and $\bar{f}(\mathbf{x})$ is the mean of these predictions over the masks.

The uncertainty from Monte Carlo dropout mainly represents model uncertainty, also called epistemic\index{epistemic uncertainty} uncertainty. It stems from the limited knowledge of the model, because of limited training data or model capacity. It differs from aleatoric\index{aleatoric uncertainty} uncertainty, which comes from the noise in the data itself. Even the model uncertainty alone is an improvement over standard deterministic neural networks, which give only point predictions with no indication of confidence.

The spread of the passes, drawn as a band in panel~(c) of Figure~\ref{fig:dropout_visualization}, depends on the dropout probability $p$. A larger $p$ tends to widen the band. The value of $p$ is a choice for the training and does not by itself measure our actual uncertainty. The band shows only the model uncertainty, so we add the noise of the data to it and count how often test observations fall inside. This shows whether a given $p$ gives a band that is too narrow or too wide.\index{Monte Carlo dropout|)}\index{dropout|)}

\paragraph{Beyond Dropout}

Dropout is only an approximation to Bayesian inference. It can fail to reproduce the complete predictive distribution, especially far from the training data or when the uncertainty is highly structured. Panel~(c) of Figure~\ref{fig:dropout_visualization} shows this far from the data, where the band stays narrow but the mean departs from the function. The dropout masks add diversity to the predictions, but they do not explore the parameter space well enough to represent the posterior distribution. This has spurred research into full Bayesian approaches to neural networks.

One direction of this research adapts the MCMC methods of Chapter~13 to sample from the posterior distribution of the network weights. However, the high dimension of the weight space makes these methods computationally costly.

An alternative is variational\index{variational inference} inference, which turns a sampling problem into an optimization problem. We assume a simple parametric form for the posterior distribution of the weights, such as a Gaussian, and optimize its parameters instead of sampling. We can then sample weights from this distribution to predict with uncertainty. Variational inference is a central idea in statistical inference, but its detailed application to neural networks is technical, so we only describe the idea.

Standard and Bayesian\index{Bayesian neural network} neural networks show a trade-off that we met in Chapter~14, between low computing cost and uncertainty quantification. Standard neural networks are efficient, because backpropagation and SGD handle massive data sets and complex functions. The main limitation of neural networks is their uncertainty quantification. It is not well understood, or at least not easy to achieve or to derive analytically. The blanket criticisms of their interpretability matter less.

The lack of uncertainty quantification has led modern research in two directions. One is to make classical Bayesian models like Gaussian processes more scalable. For instance, neural networks can take over some computational components, so that we pay their cost once in training. The core of such a model remains Bayesian linear regression in a feature space (Chapter~14). The other is to develop full Bayesian approaches for neural networks themselves. Both directions have progressed in recent years.\index{Bayesian neural network|)}\index{neural network!uncertainty|)}

\section{Beyond A Fancy Interpolator}

A\index{neural network!generalization|(} common misconception is that neural networks are merely fancy interpolation methods that only fit the training points, without theoretical grounding. This view overlooks the mathematical properties of neural networks and the theoretical advances that explain in part why and how they work. Giving the proofs is beyond the scope of this book, but we highlight three developments. They give insight into learning, generalization and computation in neural networks.

\paragraph{Double Descent}

Double\index{double descent}\index{double descent|(} descent became widely known in deep learning around 2019. It runs against classical statistical learning theory and our intuitions about generalization and the complexity of a model. In classical statistical learning we expect a U-shaped test error curve, as in panel~(a) of Figure~\ref{fig:double_descent} (Section~\ref{sec:cross_validation}). As the complexity increases, the performance first improves, because the model fits the real patterns in the data. It then deteriorates, because the model starts to overfit. This is the bias-variance tradeoff between underfitting and overfitting.

Neural networks can violate this classical expectation. After the initial overfitting phase, the performance can improve again as the complexity increases, which gives a double descent pattern in the test error curve. Models can then have far more parameters than training examples, and we call them overparameterized.\index{overparameterization} Even so, they can avoid the severe overfitting that classical theory predicts. We illustrated this phenomenon earlier in panel~(b) of Figure~\ref{fig:double_descent}.

One explanation of this phenomenon concerns how the solution space changes as we add parameters. In classical optimization, adding parameters may create more local\index{local minima} minima, so the global minimum becomes harder to find. In neural networks, however, the solution space may change at some point. Beyond that point the solutions may be more connected, with paths between what were isolated local minima. Gradient descent can then move between solutions more easily and gradually find better minima that may have been out of reach in lower dimensions. This may let the network find better solutions even with many more parameters than training examples.

\begin{figure}[ht!]
    \centering
    \includegraphics[width=0.8\textwidth]{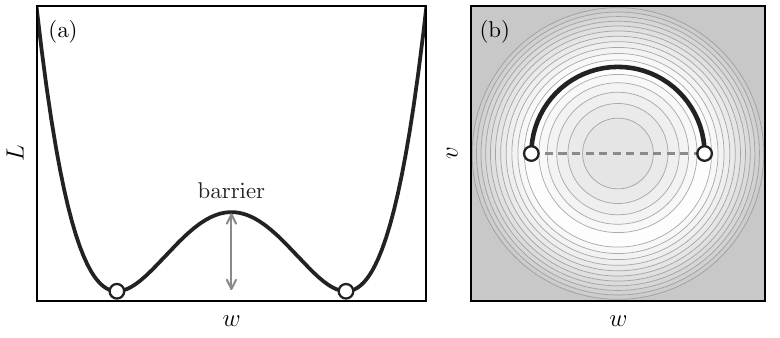}
    \caption{Toy loss surfaces. Panel~(a) shows a loss of one parameter with two isolated minima and a barrier between them. In panel~(b) one more parameter turns the minima into a ring-shaped valley, the brim of a Mexican hat. The solid path along the valley avoids the hill that the dashed straight path must climb.}
    \label{fig:toy_loss_surfaces}
\end{figure}

Figure~\ref{fig:toy_loss_surfaces} gives a toy example. In panel~(a) one parameter gives two isolated minima with a barrier between them. In panel~(b) a second parameter turns them into a ring-shaped valley, like the brim of a Mexican hat. A path along the valley avoids the barrier.\index{double descent|)}

\paragraph{Neural Tangent Kernel}

Another\index{neural tangent kernel|(} theoretical advance, popular since about 2018, is the Neural\index{neural tangent kernel} Tangent Kernel (NTK). It connects neural networks to kernel methods such as the Gaussian processes of Chapter~14. Recall that a kernel is the inner product of feature vectors, $k(\mathbf{x}, \mathbf{x}') = \boldsymbol{\phi}(\mathbf{x})^T \boldsymbol{\phi}(\mathbf{x}')$. For a neural network with one output the features are the gradients of the output with respect to the weights. The NTK is the inner product of these gradients at the initial weights for two inputs, $\Theta(\mathbf{x}, \mathbf{x}') = \nabla_{\mathbf{w}} f_{\mathbf{w}}(\mathbf{x})^T \nabla_{\mathbf{w}} f_{\mathbf{w}}(\mathbf{x}')$.

Consider the limit of infinite\index{infinite-width limit} width, where the number of neurons in each hidden layer goes to infinity. In this limit the behavior of a neural network during training can be described by a kernel function. This kernel depends only on the network architecture and the initialization. It does not change with the parameter values during training.

The NTK gives one reason why gradient descent in overparameterized neural networks can find good solutions although the loss function is non-convex. In the infinite-width limit, each parameter changes only slightly during training. The network output is then linear in the parameters near the initial values, so the squared-error loss is convex. Gradient descent can then reach a global minimum of the training loss.

In the infinite-width limit, a neural network behaves like a kernel method with a fixed kernel. Its kernel is defined by the architecture, whereas for Gaussian processes we choose the kernel a priori. As for Gaussian processes, a kernel can implicitly map the inputs to an infinite-dimensional feature space and still keep the computation tractable. A neural network of finite width can change its features during training and so learn the feature representation from the data.

The NTK also helps explain why some architectures work better than others for specific problems. Different architectures induce different kernel functions, which have different inductive biases. This lets us reason about the design of a neural network in terms of the properties of the induced kernel, as well as empirically. From a statistical perspective, the NTK gives a way to analyze the convergence and generalization properties of neural networks.\index{neural tangent kernel|)}

\paragraph{Grokking}

Grokking\index{grokking}\index{grokking|(} is a phenomenon that bears on a question about learning. The question is whether neural networks only memorize the training examples or extract the underlying patterns. The term ``grokking'' borrows the concept of deep intuitive understanding from the science fiction author Robert A. Heinlein. Grokking suggests the second possibility.

One definition of intelligence is the ability to generalize beyond what was taught. It means to extract underlying principles and apply them to new situations. This ability separates rote memorization from learning the principle behind the examples, and it is characteristic of human cognition. Grokking suggests that neural networks can show a similar ability.

Consider neural networks trained on algorithmic tasks, for example modular addition, $(a + b) \mod c$, the remainder of $a + b$ after division by $c$. At first, these neural networks appear to memorize the training examples. They reach 100 percent training accuracy but generalize poorly to unseen examples. With continued training and weight decay, the networks suddenly ``grok'' the underlying pattern and reach a test accuracy close to 100 percent.

\begin{figure}[ht!]
    \centering
    \includegraphics[width=0.95\textwidth]{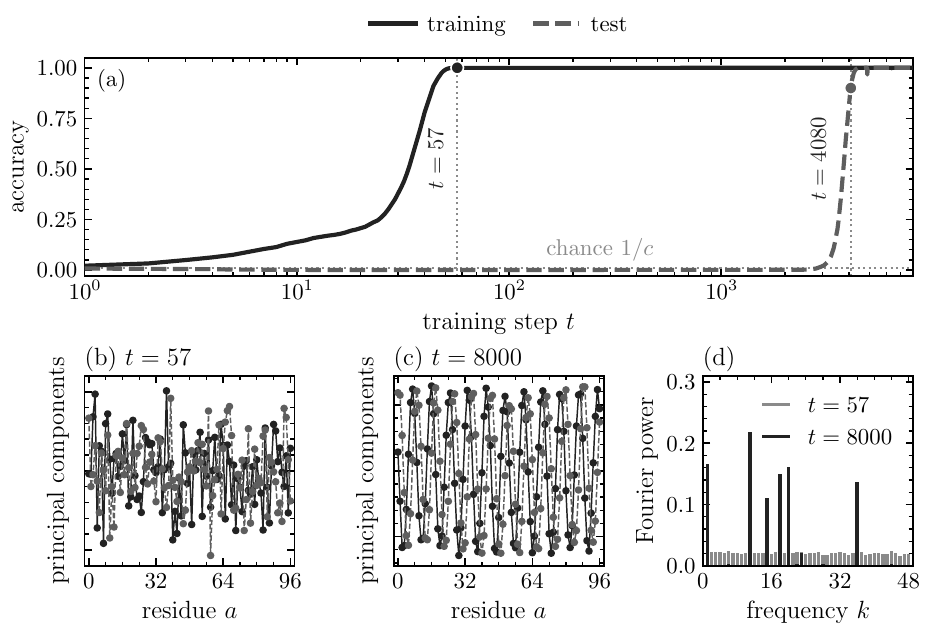}
    \caption{Grokking of a neural network that learns addition modulo a number, trained with weight decay. Panel~(a) shows the training and test accuracy against the training step on a logarithmic axis. The training accuracy reaches its maximum early, and the test accuracy rises only much later. Panels~(b) and (c) show the two leading principal components of the learned embedding of the residue $a$ (black and dashed grey) at two steps. Panel~(d) shows the share of the Fourier power of the embedding table at each frequency $k$. After the jump it concentrates on a few frequencies.}
    \label{fig:grokking_phenomenon}
\end{figure}

The internal representations change at this transition. While a neural network memorizes, the embedding of the inputs, which gives each input value a learned vector, tends to look irregular. After it groks the pattern, the embedding becomes structured and matches the mathematical properties of the task. For modular arithmetic, this embedding becomes periodic in the residue and puts most of its Fourier power on a few frequencies. This indicates that the network has discovered the mathematical structure instead of memorizing individual cases. Figure~\ref{fig:grokking_phenomenon} shows the accuracy and the internal representations before and after grokking.

Grokking occurs even when neural networks have far more parameters than training examples, a regime where classical theory would predict memorization. Yet these neural networks eventually discover generalizable patterns, which suggests that learning in neural networks goes beyond interpolation between training points. The implications of grokking reach past simple arithmetic tasks. We may see related behavior in large language models, which can generalize in unexpected ways. For instance, models trained mostly on English text can generate coherent responses in languages that they saw very little in training, given relatively few examples. This suggests that these models extract linguistic patterns that hold across languages, instead of memorizing the training data.

Neural networks can also perform counterfactual generalization, which means combining concepts that did not occur together in training. Humans can imagine new combinations of concepts, and modern neural networks can generate coherent representations of such combinations. This recombination of learned elements is more than memorization.

Phenomena like grokking suggest that neural networks can go beyond memorization and learn the underlying patterns and principles of the data. Used with an understanding of how they work, they are useful tools for astronomical research and may find patterns and relationships that standard methods miss.\index{grokking|)}\index{neural network!generalization|)}

\section{Foundation Models}\label{sec:ch15_pretrained}

Grokking\index{foundation model|(}\index{neural network!foundation models|(} suggests that neural networks can generalize beyond their training examples. We can use this ability in two steps. First, we pre-train one large model on extensive and varied data, usually without labels because labels are scarce in astronomy. Second, we fine-tune it, which means that we continue its training briefly on the data of the specific application. This replaces training a specialized neural network from scratch for each task, and we pay the high computing cost of pre-training only once. Reusing a pre-trained model for a new task in this way is called transfer\index{transfer learning} learning.

A pre-trained model is a foundation\index{foundation model} model only if its representation transfers well enough to new tasks for few-shot learning, which we describe below.

\begin{figure}[ht!]
    \centering
    \includegraphics[width=0.95\textwidth]{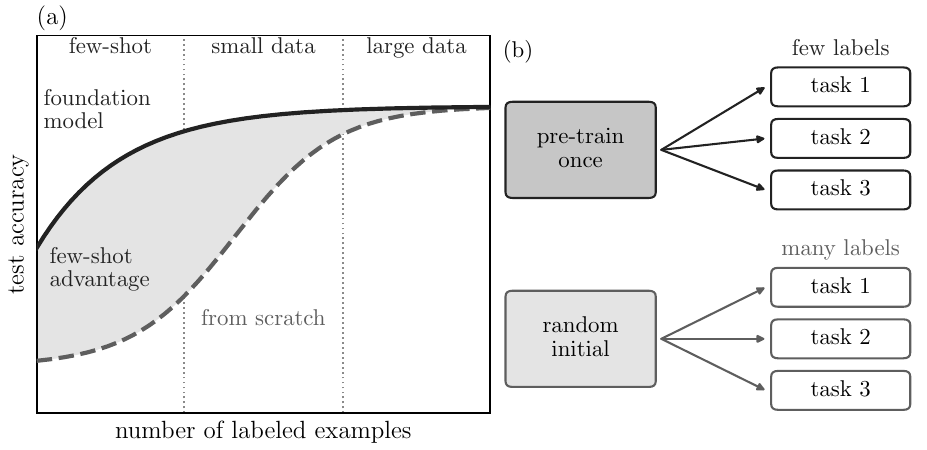}
    \caption{Schematic comparison of a foundation model with few-shot learning and a model trained from scratch. Panel~(a) shows a conceptual performance curve against the number of labeled examples. The foundation model (solid) reaches high accuracy with few examples, and the model trained from scratch (dashed) needs far more. The shaded area is the few-shot advantage, which is largest in the few-shot region and vanishes in the large data region. Panel~(b) sketches the two routes to several tasks.}
    \label{fig:foundation_models}
\end{figure}

\paragraph{Few-Shot Learning} A foundation model can learn from very few examples. This ability is called few-shot\index{few-shot learning} learning, and zero-shot learning when no example is needed. On a new task it can need far fewer examples than a model trained from scratch. A model pre-trained on a vast corpus of text and images can adapt to classify astronomical objects after a handful of examples from each category. This suggests that pre-training gives the model general knowledge about the structure of data in many domains, beyond a memory of its training data. Figure~\ref{fig:foundation_models} compares this route with training a separate model from scratch for each task.

Many pre-trained models do not reach few-shot performance, because pre-training alone does not give a representation that is useful for a new task. Testing transfer needs care. A model trained and tested on objects from the same survey can score well. This shows generalization to similar data and does not show transfer. For transfer we test on data that differ in a way that matters, such as a different instrument, a different population or a different task.

Where labels are plentiful, a model trained directly on the task can already do well. It is then the baseline that a foundation model has to beat. The case for foundation models is strongest where labels are scarce.

Few-shot learning helps in astronomy, where labeled examples of rare phenomena can be scarce. A model pre-trained on broader astronomical data might need only a few examples of a rare event to identify other instances. Rare events include a particular type of stellar explosion and an unusual galaxy merger. This could speed up the discovery and classification of rare phenomena in the massive data sets of modern observatories. Simulations can supply labels where real ones are rare. A model trained on them must still cross the gap to real observations, which is itself a transfer problem.

\paragraph{Scientific Discovery} Pre-trained models also open possibilities for scientific discovery. When we model a complex phenomenon with a neural network, we want more than a fit of the observed data points. We want the underlying physical laws or patterns.

Astronomers have long read their data through heuristics built by hand, such as colors, line indices and fitted templates. Each is a compression chosen in advance. A representation learned from the data could in principle recover the compact description that the physics implies and that we have not written down. Present models fall well short of this. It remains the right aim, because it separates automating the measurements we already make from finding the variables we should have been measuring.

For astronomers, neural networks could then give more than efficient data processing. We would examine what the networks learn, both their outputs and their internal representations. We might then gain new insight into physical laws that are hard to find by traditional analysis.

\paragraph{A Common Error} The label foundation model is easy to misuse. A common error is to train a Transformer or another large architecture on one task and to call the result a foundation model. The architecture and the size are only the means. What counts is a representation that transfers to new instruments, populations or tasks. Without it, the model is only pre-trained and large, and the name foundation model overstates what it does.\index{foundation model|)}\index{neural network!foundation models|)}

\section{Autoencoders: Nonlinear Dimension Reduction}\label{sec:ch15_autoencoders}

A\index{autoencoder|(}\index{neural network!autoencoders|(} foundation model learns a representation of its training data, usually without labels. Learning a representation without labels is unsupervised learning. So far we used neural networks for supervised learning, where we learn a mapping from inputs $\mathbf{x}$ to targets $y$. In unsupervised learning we have the inputs $\mathbf{x}$ alone and no targets $y$.

Chapters~10 and~11 taught classical methods for this case. Principal component analysis (PCA) reduces the dimension, K-means clusters the data and Gaussian mixture models estimate densities. These methods rest on linear algebra and statistical principles, and their assumptions limit them. Neural networks extend them, keeping their ideas and relaxing their restrictive assumptions.

\paragraph{PCA as the Linear Case} PCA is the linear method that the autoencoder generalizes. For centered data $\mathbf{x} \in \mathbb{R}^D$ it finds a linear map $\mathbf{W} \in \mathbb{R}^{M \times D}$ to a representation $\mathbf{z} \in \mathbb{R}^M$ with $M < D$. We obtain this map from the directions of maximum variance, which are the eigenvectors of the data covariance matrix
\begin{equation}
\mathbf{S} = \frac{1}{N}\sum_{i=1}^N \mathbf{x}_i \mathbf{x}_i^T.
\end{equation}
The rows of $\mathbf{W}$ are the eigenvectors with the $M$ largest eigenvalues, so $\mathbf{W}$ is the transpose of the matrix $\mathbf{B}_M$ of Chapter~10. The projection
\begin{equation}
\mathbf{z} = \mathbf{W}\mathbf{x}
\end{equation}
keeps as much variance as possible in the lower-dimensional space.

The reconstruction of PCA uses the transpose of the projection matrix
\begin{equation}
\hat{\mathbf{x}} = \mathbf{W}^T \mathbf{z}.
\end{equation}
This is the best linear approximation of the original data from the compressed representation. Its reconstruction error is
\begin{equation}
E = \frac{1}{N}\sum_{i=1}^N \|\mathbf{x}_i - \hat{\mathbf{x}}_i\|^2 = \frac{1}{N}\sum_{i=1}^N \|\mathbf{x}_i - \mathbf{W}^T \mathbf{W}\mathbf{x}_i\|^2.
\end{equation}
This error is the information that the compression loses. Minimizing it is equivalent to maximizing the variance that the projection preserves, as Chapter~10 shows.

For $M = D$ the rows of $\mathbf{W}$ are $D$ orthonormal eigenvectors, so $\mathbf{W}$ is orthogonal and $\mathbf{W}^T \mathbf{W}$ is the identity. The error is then zero and nothing is compressed. The strategy of PCA is therefore to keep only some of the components, with $M < D$, and still most of the information. The error measures the small loss of information that we accept for a large gain in compression.

PCA has a closed-form solution through the eigendecomposition, but it is limited to linear transformations. This limit matters when the data lie on a nonlinear manifold, a curved surface of lower dimension in the high-dimensional space. For example, the Swiss roll is a two-dimensional sheet curled up in three dimensions. Linear projections have difficulty unfolding it. An autoencoder can use nonlinear transformations to describe the roll with two numbers for each point.

\paragraph{The Autoencoder} Autoencoders\index{autoencoder} replace the linear transformations of PCA with nonlinear neural networks. They keep the goal of PCA, a low-dimensional representation with minimal reconstruction error. The nonlinear transformations can represent complex geometric relationships in the data.

PCA encodes with $\mathbf{z} = \mathbf{W}\mathbf{x}$ and decodes with $\hat{\mathbf{x}} = \mathbf{W}^T\mathbf{z}$. An autoencoder has the same two parts, an encoder and a decoder.

The encoder $e(\mathbf{x})$ maps the input to a low-dimensional representation
\begin{equation}
\mathbf{z} = e(\mathbf{x}).
\end{equation}

The decoder $d(\mathbf{z})$ gives the reconstruction of the input from this representation
\begin{equation}
\hat{\mathbf{x}} = d(\mathbf{z}).
\end{equation}

\begin{figure}[ht!]
    \centering
    \includegraphics[width=0.95\textwidth]{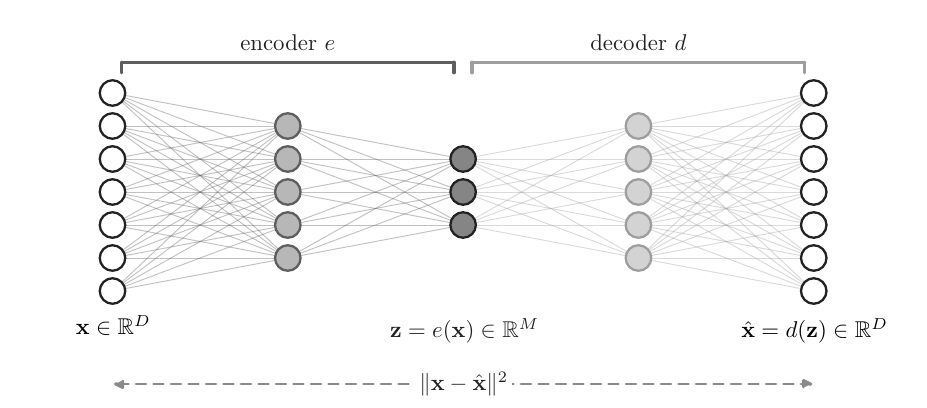}
    \caption{An autoencoder. The encoder $e$ compresses the input to a bottleneck $\mathbf{z}$ of lower dimension $M < D$, and the decoder $d$ maps it back to the input space. Training minimizes the reconstruction error between the input and its reconstruction.}
    \label{fig:autoencoder_architecture}
\end{figure}

Both functions $e(\mathbf{x})$ and $d(\mathbf{z})$ are neural networks, and their architecture can vary with the data type and the task. Figure~\ref{fig:autoencoder_architecture} shows an autoencoder made of fully connected layers.

A simple encoder with two layers is
\begin{equation}
e(\mathbf{x}) = \sigma_2(\mathbf{W}_2\sigma_1(\mathbf{W}_1\mathbf{x} + \mathbf{b}_1) + \mathbf{b}_2),
\end{equation}
where $\sigma_1$ and $\sigma_2$ are nonlinear activation functions such as GELU. The matrices $\mathbf{W}_1$ and $\mathbf{W}_2$ and the vectors $\mathbf{b}_1$ and $\mathbf{b}_2$ are the learnable parameters. The decoder $d(\mathbf{z})$ has a similar structure with its own weight matrices and bias vectors. The encoder and the decoder are learnable functions, and their nonlinear layers can represent complex relationships in the data.

This is only one possible implementation. In practice we often use Convolutional Neural Networks for image data and Transformers for sequential data. The architecture encodes the inductive biases that we want for the data. The autoencoder concept stays the same for any building blocks.

We train an autoencoder with an objective that parallels the reconstruction error of PCA
\begin{equation}
\mathcal{L} = \frac{1}{N}\sum_{i=1}^N \|\mathbf{x}_i - d(e(\mathbf{x}_i))\|^2.
\end{equation}
This is the average squared distance between the input and its reconstruction after passing through the encoder and the decoder. We optimize all the weights of both networks.

Backpropagation works on any differentiable computational graph, so it gives the gradients of this loss with respect to all parameters. We update them by gradient descent as in supervised learning. Only the loss differs, because we minimize a reconstruction error.

Suppose that the encoder and the decoder are single-layer linear networks without activation functions
\begin{align}
e(\mathbf{x}) &= \mathbf{W}_e\mathbf{x} \\*
d(\mathbf{z}) &= \mathbf{W}_d\mathbf{z},
\end{align}
and that we minimize the reconstruction error. The encoder then converges to the PCA projection matrix up to an invertible $M \times M$ matrix. The product $\mathbf{W}_d\mathbf{W}_e$ converges to the PCA reconstruction map $\mathbf{W}^T\mathbf{W}$. In this sense a linear autoencoder is equivalent to PCA. The nonlinear autoencoder extends PCA to more complex data structures.

An autoencoder has the shape of an hourglass, as Figure~\ref{fig:autoencoder_architecture} shows, because the bottleneck dimension $M$ is smaller than $D$. The encoder reduces the dimension layer by layer until the bottleneck layer, which holds the low-dimensional latent representation $\mathbf{z}$ (Section~6.5). The decoder then expands the dimension symmetrically back to the original one. The bottleneck is the constraint that drives the learning. Without it the network could copy the inputs to the outputs and extract no useful features. With it the network must keep the aspects of the data that matter most for an accurate reconstruction.

This compression resembles the way in which PCA finds principal components. PCA keeps the directions of maximum variance, and the bottleneck keeps the structure that matters most for the reconstruction. Autoencoders add nonlinear transformations to this idea.

The compressed representations can reveal features that we can interpret, much as PCA vectors can show physically meaningful directions in data space. In face images the latent dimensions might encode pose, lighting, expression and identity. In galaxy images they could correspond to physical properties such as morphological type, size, orientation and brightness. The training does not ask for these features.

We can examine a latent\index{latent space} space in two simple ways. We can plot two latent coordinates for many objects and color each point by a measured property, such as the stellar mass of a galaxy. We can also change one latent coordinate, decode the result and see which features of the image change. A coordinate that follows a known property may carry a physical meaning.

\paragraph{Applications in Astronomy}

Autoencoders have practical applications in astronomy beyond dimension reduction. One of them is denoising.\index{denoising autoencoder} Consider astronomical images with noise from cosmic rays, readout noise or sky background fluctuations. Noise typically lacks the structured correlations of the signal, so the bottleneck cannot encode it efficiently. The autoencoder then tends to keep the structured signal in its reconstruction and to discard the random noise.

{\makeatletter\@clubpenalty=10000\makeatother \clubpenalty=10000 \widowpenalty=10000 \displaywidowpenalty=10000 A denoising autoencoder takes noisy inputs and targets clean outputs. We need pairs of noisy and clean examples, which can come from simulations or controlled observations. With such pairs we minimize
\begin{equation}
\mathcal{L} = \frac{1}{N}\sum_{i=1}^N \|\mathbf{x}_i^{\text{clean}} - d(e(\mathbf{x}_i^{\text{noisy}}))\|^2.
\end{equation}
This loss trains the network to remove the noise and to keep the underlying signal.\par}

A denoising autoencoder does not create new information or ``generate more photons'' from the observations. In Bayesian terms, a neural network trained with the squared error estimates the posterior mean of the clean signal given the noisy data. The network weights encode a prior on the signal through the training on similar examples. Denoising is useful for visual analysis and pattern recognition. The cleaned output is our best guess from prior knowledge, and it adds no observational data.

In astronomy it is tempting to read denoised results as ``super-resolution'' or ``enhanced signal-to-noise''. Denoising has worked well for cleaning astronomical images, spectra and time series. We must not treat the results as if they held more observational information than the original data.

Autoencoders can also find anomalies or outliers. We train the autoencoder on a data set of ``normal'' examples. Instances with a large reconstruction error are then likely outliers, because they do not follow the regularities that the autoencoder learned. The technique has been used to detect unusual astronomical objects, instrument failures and data quality problems.

In all these applications the advantage of autoencoders over PCA is that they can represent nonlinear relationships. Astronomical data often have nonlinear structure. Galaxy morphologies form complex manifolds in image space, and light curves of variable stars show nonlinear variation in time. Autoencoders can find this structure, and this can lead to more efficient compression, better denoising and more accurate anomaly detection.\index{autoencoder|)}\index{neural network!autoencoders|)}

\section{Domain Transfer and Cross-Modal Learning}

An autoencoder maps an input back to itself through a bottleneck. The same encoder-decoder structure can go beyond this self-reconstruction and relate different data domains to each other. The extension rests on the bottleneck layer, which holds a meaningful representation of the main structure of the input. This latent representation drops superficial details and keeps the patterns that define the data. If it encodes the ``knowledge'' or ``semantics'' of the data, it should help with tasks beyond simple reconstruction.

We generalize the autoencoder by keeping the encoder $e(\mathbf{x}) \rightarrow \mathbf{z}$ and allowing different uses of the latent representation $\mathbf{z}$. The decoder no longer needs to reconstruct the original input. We can train it to produce outputs in different domains or modalities, which are types of data such as images and spectra. The encoder $e(\cdot)$ and the decoder $d(\cdot)$ remain neural networks, but we may design their architectures for the specific input and output. This gives two uses in astronomy, domain transfer and cross-modal learning.

\paragraph{Domain Transfer}

Domain\index{domain transfer}\index{domain transfer|(} transfer maps between different representations of similar data. It differs from transfer learning (Section~\ref{sec:ch15_pretrained}), where we reuse a pre-trained model for a new task. An autoencoder follows the pattern $\mathbf{x} \mapsto \mathbf{z} \mapsto \mathbf{x}$. Domain transfer follows the pattern $\mathbf{x} \mapsto \mathbf{z} \mapsto \mathbf{y}$, where $\mathbf{x}$ and $\mathbf{y}$ belong to related but distinct domains. The two steps are
\begin{align}
\mathbf{z} &= e(\mathbf{x}) \\*
\mathbf{y} &= d(\mathbf{z}).
\end{align}

One application in astronomy reduces the gap between simulated and observed data. In stellar spectroscopy, theoretical atmospheric models predict stellar spectra from physical parameters such as temperature, surface gravity and metallicity. However, these simulations often miss some physical processes. Missing opacity sources, incomplete line broadening mechanisms and simplified atmospheric assumptions create systematic differences between predicted and observed spectra.

An encoder-decoder network can learn these systematic differences from pairs of simulated and observed spectra. The encoder obtains the physical information from the simulated spectrum. The decoder translates this information into a more realistic observed spectrum. The latent space $\mathbf{z}$ still encodes the physical parameters, but the output $d(\mathbf{z})$ matches the observations better than the direct simulation does.

Our simulations are imperfect approximations of physical processes. Instead of discarding these models, domain transfer can correct the systematic differences and keep the physical basis of the models. The resulting hybrid model joins the theory built into the simulations with empirical corrections learned from data.

We can use this correction to measure the parameters of stars from their observed spectra. We choose calibration stars whose physical parameters, such as temperature, surface gravity and metallicity, we trust. For each of them we pair the simulated spectrum, computed from its parameters, with the observed spectrum, and the network learns the correction. For a new star we vary the parameters until the spectrum that we simulate and correct with the network matches the observed spectrum. The parameters that give the best match are our estimates. They should be better than those of a fit with the pure simulations, if the new star resembles the calibration stars.

Domain transfer also gives predictions across wavelengths. We can train a neural network to predict infrared galaxy spectra from optical observations, or to generate radio emission maps from optical images. These predictions are educated guesses based on learned correlations and carry no new physical information. They can still guide observing plans and help identify promising targets for follow-up studies.\index{domain transfer|)}

\paragraph{Cross-Modal Learning}

Cross-modal\index{cross-modal learning}\index{cross-modal learning|(} learning requires several data modalities to share one latent representation. Instead of a mapping between two spaces, different data types encode into and decode from the same latent space. For two modalities, the encoders and decoders are
\begin{align}
\mathbf{z}_1 &= e_1(\mathbf{x}_1) \\*
\mathbf{z}_2 &= e_2(\mathbf{x}_2) \\
\hat{\mathbf{x}}_1 &= d_1(\mathbf{z}) \\*
\hat{\mathbf{x}}_2 &= d_2(\mathbf{z}).
\end{align}

Each modality has its own encoder and decoder, but all modalities share the same latent representation $\mathbf{z}$. The training objective typically asks for the reconstruction of each modality from itself. For one object seen in both modalities we require $\mathbf{z}_1 = \mathbf{z}_2 = \mathbf{z}$, or we add a penalty on $\|\mathbf{z}_1 - \mathbf{z}_2\|^2$ to the loss, so that $\mathbf{z}_1 \approx \mathbf{z}_2 \approx \mathbf{z}$. The network must then find one representation of the information that the modalities share.

\begin{figure}[ht!]
    \centering
    \includegraphics[width=0.95\textwidth]{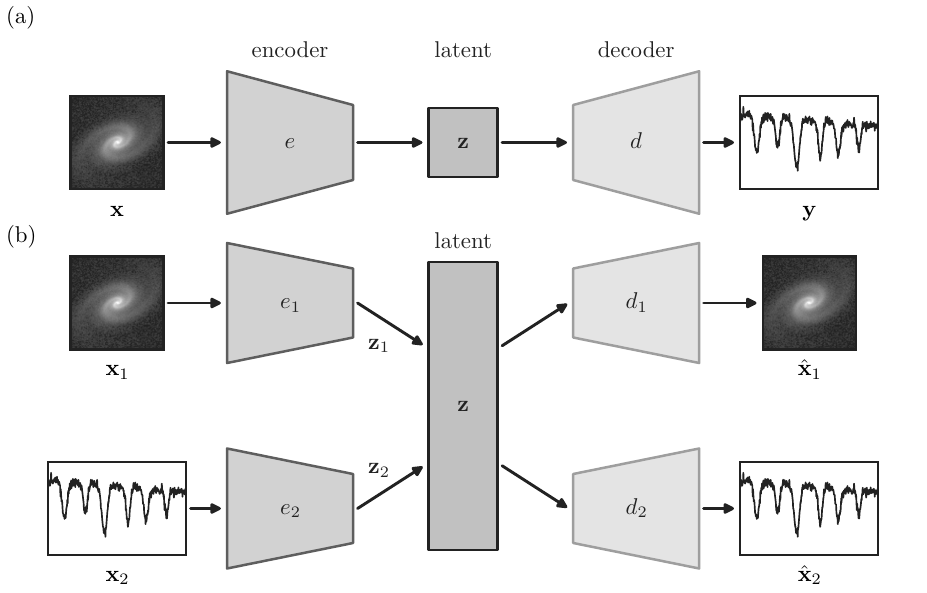}
    \caption{Panel~(a) shows domain transfer, where the output $\mathbf{y}$ lies in a different domain from the input $\mathbf{x}$, here a spectrum predicted from a galaxy image. Panel~(b) shows a shared latent space, where the encoders of two modalities map into the same $\mathbf{z}$.}
    \label{fig:encoder_decoder_architectures}
\end{figure}

Consider galaxy images and galaxy spectra that we encode into the same latent space. The image encoder $e_1$ must extract information from morphological features, colors and spatial structure. The spectrum encoder $e_2$ must extract information from spectral lines, continuum shapes and emission features. Both encoders map to the same latent space. Figure~\ref{fig:encoder_decoder_architectures} compares this shared latent space with domain transfer.

The shared representation can reflect the physical properties that affect both modalities. For instance, the stellar population age influences both the color of a galaxy in images and the strength of specific absorption features in its spectrum. The stellar mass affects both the apparent size of the galaxy and its overall spectral energy distribution. Because the latent space must represent both modalities, the network can find these physical relationships without explicit supervision.

Cross-modal learning can do three things. Each modality can be reconstructed from itself, as in an autoencoder. The network can predict one modality from the other, for example by generating spectra from images or the reverse. It can also discover latent variables that correspond to shared physical properties.

The tests that we described for autoencoders are more informative in a shared latent space. We can change one latent coordinate and decode it into every modality. This shows how different observational signatures of the same phenomenon are related.

Combining observations from many instruments, surveys and wavelength regimes is a long-standing problem in astronomy. Cross-modal learning addresses it, because objects are routinely observed with multiple instruments. Large surveys already provide both imaging and spectroscopic data for millions to billions of galaxies, and future facilities will generate multi-wavelength observations at larger scales. By learning shared representations from these observations, we analyze the data types together instead of in isolation. We can then discover the physical properties that show up in the different modalities.\index{cross-modal learning|)}

\section{Mixture Density Networks}\label{sec:ch15_mdn}

Neural\index{mixture density network (MDN)|(}\index{neural network!density estimation|(} networks can also model a density. Instead of fitting a deterministic curve, a neural network can model the full probabilistic structure of the data. Standard supervised neural networks have a limitation. They assume a deterministic mapping that gives one output $\mathbf{y}$ for each input $\mathbf{x}$. This assumption frequently breaks down in scientific applications, where one-to-many mappings arise.

\paragraph{One-to-Many Relationships}

We\index{one-to-many relationship|(} take a cross-section of the Swiss roll, but now as a supervised learning task. The data points form a folded structure in which a single input value $x$ corresponds to several distinct output values $y$. A standard neural network that predicts $y$ from $x$ with the mean squared error fails to represent this structure.

\begin{figure}[ht!]
    \centering
    \includegraphics[width=1.0\textwidth]{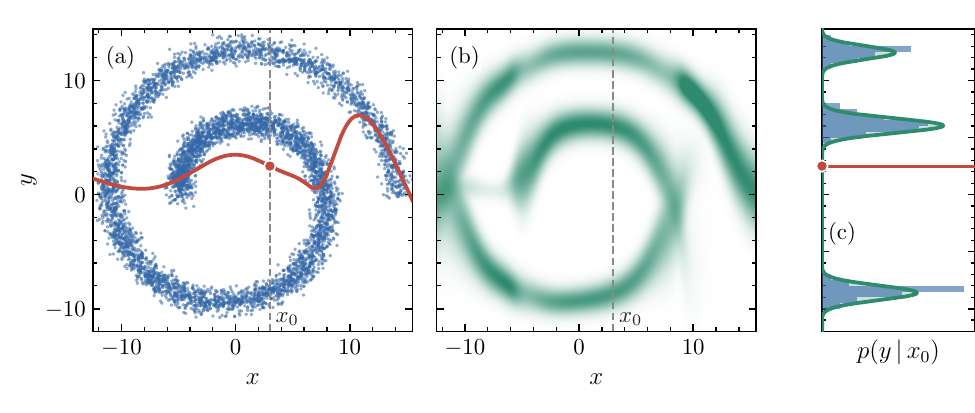}
    \caption{A one-to-many mapping. Panel~(a) shows noisy data on a spiral cross-section of a Swiss roll and the prediction of a neural network trained with the mean squared error. At the input $x_0$ of the dashed line the prediction lies between the branches of the spiral, where there are no data. Panel~(b) shows the conditional density $p(y|x)$ of a Mixture Density Network with three components, which changes with $x$. Darker green is higher density, and since the density at each $x$ adds up to one, a branch is darker where fewer branches share it. Panel~(c) shows $p(y|x_0)$ with its three peaks, the histogram of the data near $x_0$ and the prediction of panel~(a) as a red line.}
    \label{fig:density_estimation_swiss_roll}
\end{figure}

The network learns the conditional mean $\mathbb{E}[y|x]$ at each input, which averages over all possible outputs for that input. This averaging smooths out the multimodal structure of the true distribution. The prediction may then correspond to no realistic outcome. At an input where three distinct output values are possible, the network predicts their average weighted by probability, which might be an impossible configuration. Panel~(a) of Figure~\ref{fig:density_estimation_swiss_roll} shows this prediction for the Swiss roll.

One-to-many\index{one-to-many relationship} relationships are common in astronomy and other sciences. A cosmological simulation with fixed physical parameters gives a different realization for each random seed, with the same statistics but different individual structures. Several galaxy types can share the same stellar mass and differ in morphology, metallicity or star formation history. These examples show that a single output for each input is rare in complex scientific domains. A deterministic mapping is not enough here. Instead of a function $y = f(x)$ with a single prediction, we need a model of the distribution of all possible outcomes given the input.

This need leads to density estimation, which has two forms. Unconditional density estimation learns the probability density $p(\mathbf{x})$ of the data, as the Gaussian mixture models (GMMs) of Chapter~11 do. Conditional density estimation learns $p(\mathbf{y}|\mathbf{x})$ instead, and supervised problems with one-to-many relationships need it. The conditional density describes how the distribution of $\mathbf{y}$ varies with $\mathbf{x}$. It represents the range of possible outputs for each input through a probability distribution. It is of the same kind as the likelihood of Bayesian inference, a conditional distribution of the data.

Modeling $p(\mathbf{y}|\mathbf{x})$ instead of the joint distribution $p(\mathbf{x},\mathbf{y})$ also has a computational advantage. The joint distribution of a complex structure such as the Swiss roll would be hard to describe. However, if we fix $\mathbf{x}$ and look at the distribution of $\mathbf{y}$, we often find a simpler and more tractable structure.

GMMs describe complex multimodal distributions by representing multiple peaks. This is the structure that we see in one-to-many mappings. However, a standard GMM has one main limitation. It models a single static distribution $p(\mathbf{y})$ instead of a distribution $p(\mathbf{y}|\mathbf{x})$ that varies with the input. Consider one GMM for $p(\mathbf{y})$ at all values of $\mathbf{x}$ in the Swiss roll.

This model must represent all $y$ values that occur anywhere in the range of $x$. It averages out the distinct local patterns at specific $x$ values. We need a GMM whose parameters change as functions of $\mathbf{x}$. These parameters are the means, the covariances and the mixture weights.

These parameters should change gradually and smoothly. When two inputs $\mathbf{x}_1$ and $\mathbf{x}_2$ are similar, the conditional distributions $p(\mathbf{y}|\mathbf{x}_1)$ and $p(\mathbf{y}|\mathbf{x}_2)$ should also be similar. This assumption of local smoothness is physically reasonable, and it makes the computation easier.\index{one-to-many relationship|)}

\paragraph{Model and Training}

This idea leads to Mixture\index{mixture density network (MDN)} Density Networks (MDNs),\index{MDN|see{mixture density network}} which combine neural networks with mixture models for conditional\index{density estimation!conditional} density estimation. A neural network computes the parameters of a GMM as functions of the input $\mathbf{x}$. The resulting model can represent complex multimodal distributions that vary smoothly across the input space. Panels~(b) and (c) of Figure~\ref{fig:density_estimation_swiss_roll} show the conditional density of such a neural network.

An MDN models $p(\mathbf{y}|\mathbf{x})$ as a mixture of Gaussians
\begin{equation}
p(\mathbf{y}|\mathbf{x}) = \sum_{k=1}^K \pi_k(\mathbf{x}) \mathcal{N}(\mathbf{y}|\boldsymbol{\mu}_k(\mathbf{x}), \boldsymbol{\Sigma}_k(\mathbf{x})),
\end{equation}
where $\pi_k(\mathbf{x})$ are the mixture weights, $\boldsymbol{\mu}_k(\mathbf{x})$ the means and $\boldsymbol{\Sigma}_k(\mathbf{x})$ the covariance matrices. All of them vary with the input $\mathbf{x}$.

By the universal approximation property, a neural network can represent these functions, and we learn them directly from data. No prior knowledge of the true functional forms is needed.

\begin{figure}[ht!]
    \centering
    \includegraphics[width=0.95\textwidth]{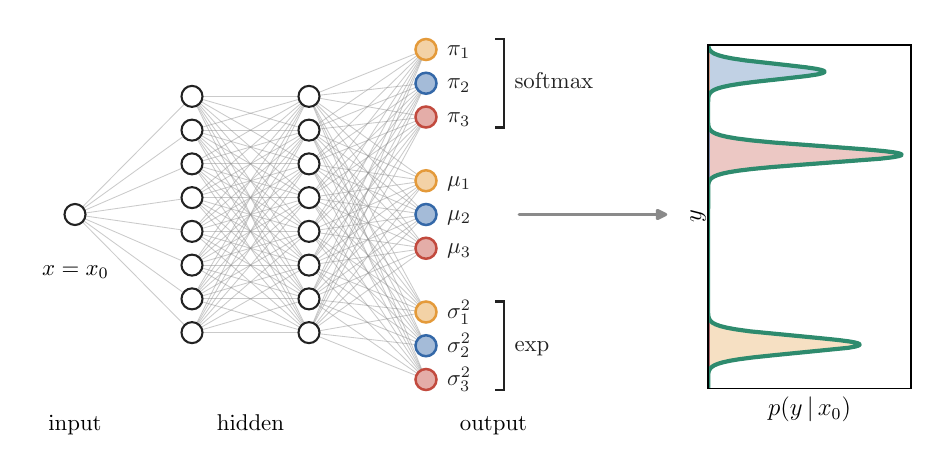}
    \caption{A Mixture Density Network with a scalar input $x$ and $K = 3$ components. For a scalar output $y$ the covariance $\boldsymbol{\Sigma}_k$ is the variance $\sigma_k^2$. A softmax gives the weights, and an exponential keeps the variances positive. The right panel shows the mixture at one input $x_0$ for the trained network of Figure~\ref{fig:density_estimation_swiss_roll}. The output neurons and the filled bump of a component share a color, and the green curve is the mixture.}
    \label{fig:mdn_architecture}
\end{figure}

For a mixture with $K$ components in a $D$-dimensional output space, the network outputs $K$ mixture weights $\pi_k(\mathbf{x})$, which must sum to 1. It outputs $K \times D$ mean values $\boldsymbol{\mu}_k(\mathbf{x})$, one $D$-dimensional mean vector for each component. It also outputs the covariance parameters of each component, and we often simplify the covariances to diagonal matrices. Figure~\ref{fig:mdn_architecture} shows the network and the mixture that it produces.

We train an MDN by maximum likelihood, as in Chapters~4 and 7. Given $N$ input-output pairs $\{(\mathbf{x}_i, \mathbf{y}_i)\}_{i=1}^N$, we maximize the log-likelihood
\begin{equation}
\ell(\mathbf{w}) = \sum_{i=1}^N \ln p(\mathbf{y}_i|\mathbf{x}_i, \mathbf{w}) = \sum_{i=1}^N \ln \left[ \sum_{k=1}^K \pi_k(\mathbf{x}_i) \mathcal{N}(\mathbf{y}_i|\boldsymbol{\mu}_k(\mathbf{x}_i), \boldsymbol{\Sigma}_k(\mathbf{x}_i)) \right],
\end{equation}
where $\mathbf{w}$ are the parameters of the neural network. We optimize them with the gradient-based methods of the earlier sections of this chapter.

The mixture weights $\pi_k(\mathbf{x})$ must be non-negative and sum to one. We apply a softmax function (Chapter~8) to the corresponding network outputs, so that
\begin{equation}
\pi_k(\mathbf{x}) = \frac{\exp(z_k(\mathbf{x}))}{\sum_{j=1}^K \exp(z_j(\mathbf{x}))},
\end{equation}
where $z_k(\mathbf{x})$ are the raw network outputs. The means can take any value, so they need no transformation. The variances must be positive, so we typically apply an exponential function or the softplus function $\ln(1 + e^{z})$ to the relevant network outputs. Numerical stability requires log-probabilities, because the product of many small probabilities can cause underflow (Chapter~7). The objective function is then the negative log-likelihood, which we minimize with gradient descent.

\paragraph{Stellar Applications}

MDNs have many applications in astronomy. In stellar parameter inference, the same photometric observations can arise from different combinations of temperature, gravity and metallicity. Degeneracies and unmodeled effects such as extinction cause this. An MDN can provide posterior distributions over the stellar parameters instead of point estimates. These distributions show the uncertainty and a possible multimodality.

To train such an MDN we need pairs of stellar parameters and photometry. A stellar model can predict the photometry for given parameters, and this forward model generates the training pairs. The MDN then learns the inverse direction, from photometry to a distribution over the parameters. The posterior is only as good as the forward model, because the MDN learns it from the generated pairs.

Color-magnitude diagrams show complex multimodal distributions, because different stellar populations and evolutionary stages overlap. MDNs can model these structures, which allows more accurate stellar population analysis and distance determinations. In cosmological parameter estimation, MDNs can model the conditional distribution of the cosmological parameters given summary statistics of the observations. The conditional distribution represents degeneracies and constraints in parameter space, and it fits the Bayesian analysis that is widely used in modern cosmological inference.

These applications are early examples of simulation-based inference, which we meet in the last section of this chapter. There neural networks learn to approximate complex likelihoods or posteriors from simulated data.\index{mixture density network (MDN)|)}

\section{Modern Neural Density Estimators}\label{sec:ch15_flows}

Mixture Density Networks are one of several neural density estimators. A neural density estimator uses a neural network to estimate a probability density function, either joint or conditional. The other estimators address limitations of MDNs and open new ways to model probabilistic relationships in scientific data. Many scientific applications need the full probabilistic relationship between variables, because it gives us the uncertainty.

MDNs rely on a specific parametric form, the Gaussian mixture model. This form is flexible, but it may need many components to approximate a complex distribution accurately. Each added component increases the number of parameters to estimate, which can lead to overfitting with limited data.

More fundamentally, the Gaussian assumption may not match the structure of scientific data. Cosmic ray hits can give measurement errors with heavy tails. Some physical processes generate distributions with sharp boundaries, which Gaussians represent poorly. Forcing a Gaussian mixture on such data may bias the results.

The limits of MDNs led to neural density estimators that represent complex distributions without a fixed parametric family. They rest on the same idea. Suppose we can transform a complex distribution into a simple one, such as a standard Gaussian. Then we can estimate densities by working in reverse, from the simple distribution back to the complex one. This idea connects to optimal transport theory, which studies efficient ways to transform one distribution into another.

Cosmology suggests an analogy. The universe began as a nearly Gaussian random field, similar to the Gaussian processes of Chapter~14. Gravitational evolution turned it into the complex, non-Gaussian structures we observe today. Neural density estimators attempt to reverse this process. They learn transformations that map complex distributions back to simple ones, or the other way round. Three approaches use this principle, normalizing flows, diffusion models and flow matching.

\paragraph{Normalizing Flows}

A\index{normalizing flow|(} normalizing\index{normalizing flow} flow learns an invertible mapping between a simple base distribution, typically a standard Gaussian, and a complex target distribution. Invertibility is the main requirement, because we must be able to transform in both directions.

The flow rests on the transformation of random variables of Chapter~3. Let $\mathbf{z}$ be a random variable with known density $p_Z(\mathbf{z})$, and let $f$ be an invertible function that gives $\mathbf{y} = f(\mathbf{z})$. The density of $\mathbf{y}$ is then
\begin{equation}
p_Y(\mathbf{y}) = p_Z(f^{-1}(\mathbf{y})) \left| \det\left( \frac{\partial f^{-1}}{\partial \mathbf{y}} \right) \right|.
\end{equation}
The Jacobian\index{Jacobian} determinant term accounts for how the transformation stretches or compresses space, so that probability mass is conserved.

\begin{figure}[ht!]
    \centering
    \includegraphics[width=0.95\textwidth]{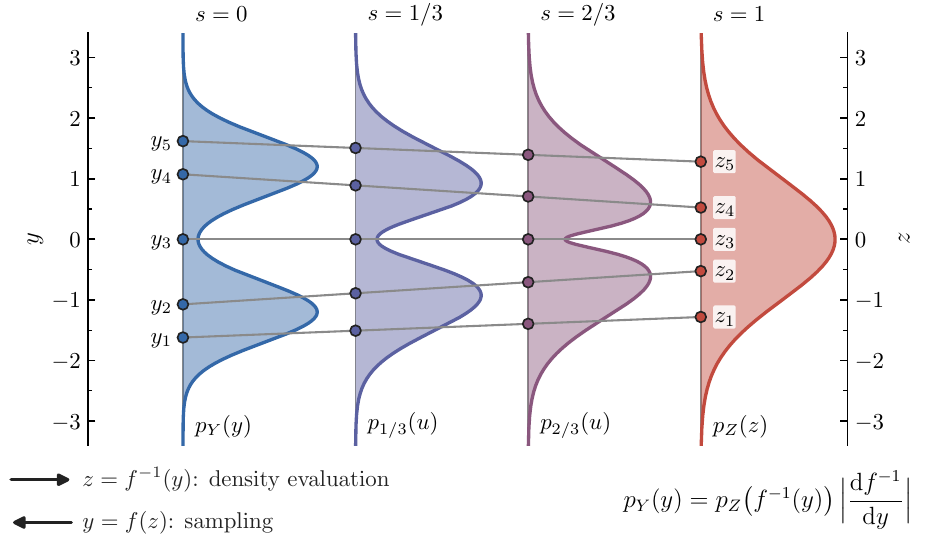}
    \caption{A normalizing flow between a bimodal target density $p_Y$ on the left and a standard Gaussian base density $p_Z$ on the right, with two stages in between. Gray lines follow five points through the stages.}
    \label{fig:normalizing_flows}
\end{figure}

We parameterize $f$ as a neural network that remains invertible, and so we can learn complex distributions directly from data. The base distribution is simple, so we easily evaluate probabilities and generate samples. The network maps it to the target distribution, where evaluation and sampling would otherwise be difficult. Figure~\ref{fig:normalizing_flows} illustrates this mapping for a one-dimensional density.

In one dimension we know such a map already, the inverse CDF method of Chapter~12. For a target with CDF $F_Y$ and a standard Gaussian with CDF $\Phi$, the map $f(z) = F_Y^{-1}(\Phi(z))$ is invertible and carries the Gaussian into the target. Figure~\ref{fig:normalizing_flows} shows this map. In many dimensions the inverse CDF method has difficulties, and a flow learns such a map instead.

A flow gives both exact density evaluation and efficient sampling. To evaluate a density, we map an observed data point back to the base distribution and apply the formula above. To sample, we draw from the base distribution and map the samples to the target distribution.

However, the requirement of invertibility limits the neural networks we can use. Not all transformations are invertible. Among those that are, not all have a tractable inverse or a tractable Jacobian determinant. Flows therefore use specialized architectures that give up some expressivity to stay invertible.

In practice a flow is a stack of many simple invertible layers. Each layer makes a simple change to the density, and together the layers build the full transformation step by step. The intermediate stages of Figure~\ref{fig:normalizing_flows} show such a gradual change.\index{normalizing flow|)}

\paragraph{Diffusion Models}

The\index{diffusion model|(} invertibility requirement leads to an alternative. Diffusion\index{diffusion model} models transform between simple and complex distributions without a single invertible transformation.

A diffusion model gradually turns a complex distribution into a simple Gaussian. It does so in many small steps that each add noise. This forward process is a Markov chain (Chapter~13). A predefined schedule sets the amount of Gaussian noise that each step adds. One step reads
\begin{equation}
x_t = \sqrt{1-\beta_t}x_{t-1} + \sqrt{\beta_t}\varepsilon,
\end{equation}
where $x_t$ is the data at time step $t$ and $\varepsilon \sim \mathcal{N}(0, \mathrm{Id})$ is random Gaussian noise. The noise schedule $\beta_t$ is a small number that typically increases with $t$.

\begin{figure}[ht!]
    \centering
    \includegraphics[width=0.95\textwidth]{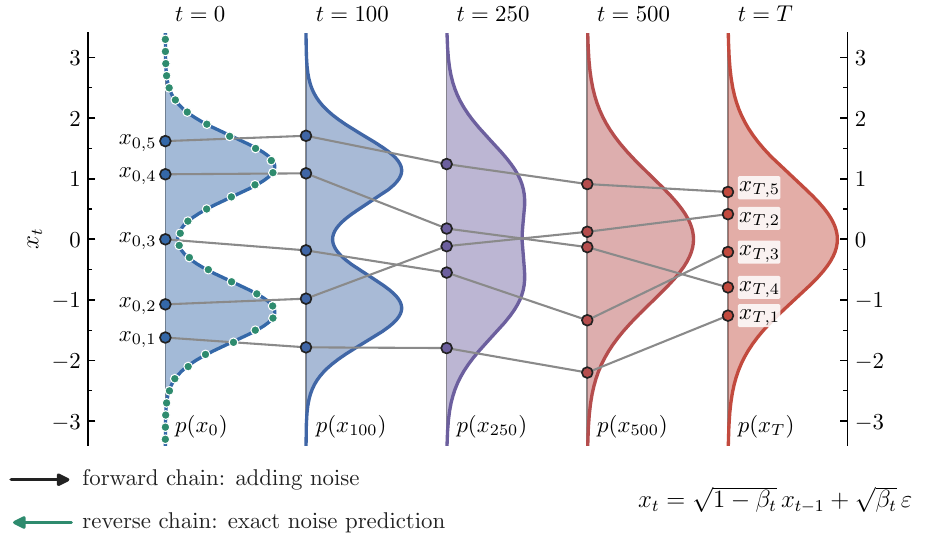}
    \caption{Forward and reverse diffusion for a bimodal density. The forward chain uses a linear noise schedule over $T$ steps. It turns the density at $t = 0$ into almost a standard Gaussian at $t = T$. Gray lines follow five chains through the times shown. The green dots in the first column are the histogram of samples from a reverse chain that starts from a standard Gaussian, and they recover the bimodal density.}
    \label{fig:diffusion_models}
\end{figure}

As $t$ approaches the final time step $T$, the distribution of $x_t$ approaches a standard Gaussian, whatever the initial data distribution is. The noise schedule is designed so that after enough steps the accumulated noise leaves almost no trace of the structure of the original data. Figure~\ref{fig:diffusion_models} shows how repeated noising turns a complex distribution into a Gaussian.

A neural network is trained to reverse this process. The noises of the steps combine into one Gaussian noise, so we compute $x_t$ from $x_0$ in one step with a single draw $\varepsilon \sim \mathcal{N}(0, \mathrm{Id})$. The network learns to predict this $\varepsilon$ from $x_t$ and $t$. The training goal is
\begin{equation}
\hat{\varepsilon}_{\mathbf{w}}(x_t, t) \approx \varepsilon,
\end{equation}
where $\hat{\varepsilon}_{\mathbf{w}}$ is the neural network with parameters $\mathbf{w}$.

This is the denoising task of Section~15.10, where an autoencoder learns to remove noise, here repeated at many noise levels. By predicting and removing this noise step by step, the model can turn a pure Gaussian sample back into a sample from the target distribution. Figure~\ref{fig:diffusion_models} shows this reverse chain for the bimodal density.

The main idea is to break the transformation into many small steps. This lets us turn a complex distribution into a Gaussian without requiring invertibility of the full transformation. Given the original data, the noised data follow a known Gaussian distribution, so each step only needs to model the noise component. Standard neural network architectures then work without invertibility constraints, and diffusion models allow a wider choice of architecture than normalizing flows.

The trade-off has two parts. Sampling from a diffusion model typically takes many steps, often hundreds or thousands of denoising iterations for one sample. This is expensive compared with the single pass of a normalizing flow. The random steps also give no direct reading of the density, which a normalizing flow provides exactly. However, the quality of the samples often justifies these costs for many applications.

A diffusion model can be trained on simulated pairs of parameters and observations and then generate parameter values for a given observation. Bayesian inference, however, often needs the density of the posterior or the likelihood, which a diffusion model does not give directly. The denoising network also estimates the score, the gradient of the log density. The score gives a route to the density, but only with extra computing and approximations. This limits the use of diffusion models in simulation-based inference (Section\index{diffusion model|)}~\ref{sec:ch15_sbi}).

\paragraph{Flow Matching}

Normalizing\index{flow matching|(} flows and diffusion models trade expressivity against a direct reading of the likelihood. A normalizing flow gives the exact density but restricts the network, and a diffusion model allows any network but gives no direct density. Flow\index{flow matching} matching takes an idea from each. It moves points deterministically from a Gaussian to the target distribution, as a normalizing flow does. It trains a free choice of network by a simple regression, as a diffusion model does. This has made flow matching a popular alternative to diffusion models.

Flow matching moves points along a velocity field $v(x, t)$, which gives the velocity at position $x$ and time $t$. Moving points along this field solves an ordinary differential equation, and the resulting flow is invertible by construction. The network that models $v$ therefore needs no invertibility constraint.

Training uses pairs of a data point $x_0$ and a noise point $\varepsilon \sim \mathcal{N}(0, \mathrm{Id})$. For a time $t$ between 0 and 1, we join them by a straight line
\begin{equation}
x_t = (1 - t)\,x_0 + t\,\varepsilon.
\end{equation}
Along this line the point moves from the data point at $t = 0$ to the noise point at $t = 1$, with the constant velocity $\varepsilon - x_0$. A network $v_{\mathbf{w}}(x_t, t)$ learns to predict this velocity by minimizing the squared error
\begin{equation}
\mathcal{L}(\mathbf{w}) = \mathbb{E}\left[ \left\| v_{\mathbf{w}}(x_t, t) - (\varepsilon - x_0) \right\|^2 \right],
\end{equation}
where the average runs over data points, noise points and times.

\begin{figure}[ht!]
    \centering
    \includegraphics[width=0.95\textwidth]{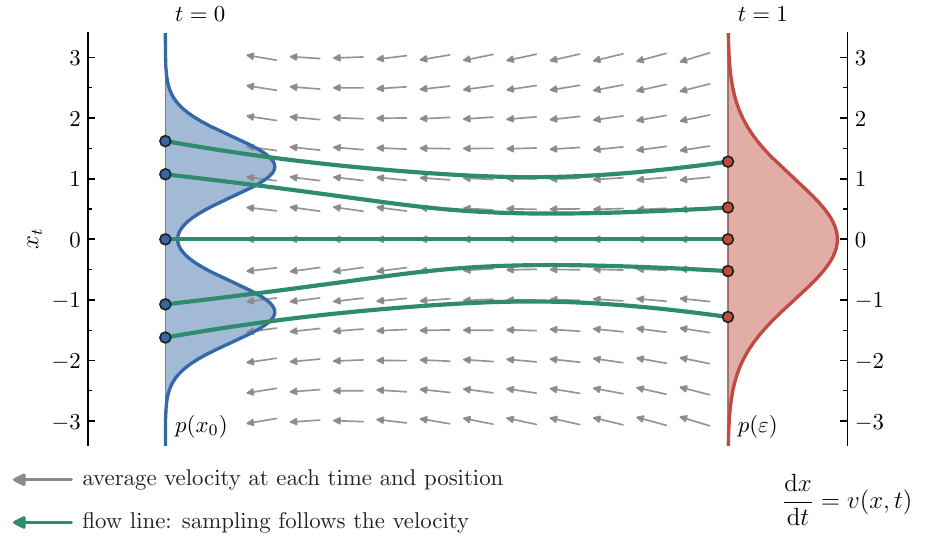}
    \caption{Flow matching for a bimodal density. The arrows show the average velocity at each time and position, in the direction of sampling. The green lines follow five points from the Gaussian at $t = 1$ to the bimodal density at $t = 0$. They do not cross. The curves at the sides show the data density at $t = 0$ and the Gaussian at $t = 1$.}
    \label{fig:flow_matching}
\end{figure}

At a given position $x$ and time $t$, the straight lines of many different pairs cross, each with its own velocity. The network sees only $x$ and $t$, not which line the point is on. Its best prediction is therefore the average velocity of all lines through that point. This average is the field the network learns, and following it carries the Gaussian into the data distribution. The paths that follow this field are the flow lines, and unlike the straight lines they cannot cross. Figure~\ref{fig:flow_matching} shows this field and its flow lines for a bimodal density.

To sample, we draw a point from the Gaussian at $t = 1$ and follow the differential equation $\mathrm{d}x/\mathrm{d}t = v_{\mathbf{w}}(x, t)$ back to $t = 0$ with a numerical solver. Only sampling needs this solver. Training is the regression above. The flow lines are gentle curves, so sampling often needs fewer steps than a diffusion model. Because the flow is invertible, we can also compute the density of a sample, as a normalizing flow does.

Flow matching and diffusion models are closely related. A diffusion model follows a stochastic process, in which each step adds random noise. Flow matching follows an ordinary differential equation with a velocity field, so its paths are deterministic. Both connect data to noise, but flow matching also allows other choices of path, such as straight lines. Flow matching is still little explored in astronomy.

MDNs, normalizing flows, diffusion models and flow matching all use neural networks to model probability distributions. Their implementations differ, but they have the same goal, which is density estimation and sampling for complex distributions.

In astronomy, these methods can model complex observational processes, account for selection effects, and generate synthetic data sets with the statistical properties of real observations.

Normalizing flows have been used to model the distribution of galaxy properties in large surveys. They describe complex correlations between mass, color and morphology that simple parametric models miss. Diffusion models have given encouraging results for generating synthetic astronomical images. These images can keep realistic noise properties and subtle correlations across wavelengths.

Neural density estimators are becoming standard in modern astrostatistics. This holds especially for simulation-based inference, where complex physical models must be connected with observational data. These methods can work with high-dimensional data directly or with summary statistics. With the full data they can extract more information and still keep a sound statistical basis.

The field evolves rapidly, and new architectures and training methods appear regularly. Neural networks that learn complex probability distributions give scientific inference more flexibility than parametric methods do. They keep the probabilistic treatment of uncertainty.\index{flow matching|)}\index{neural network!density estimation|)}

\section{Simulation-Based Inference}\label{sec:ch15_sbi}

Simulation-based\index{neural network!simulation-based inference|(}\index{simulation-based inference (SBI)}\index{simulation-based inference (SBI)|(} inference (SBI)\index{SBI|see{simulation-based inference}} is an application of neural density estimators. It connects complex scientific theories with observations. SBI addresses a challenge of modern astronomy, which is Bayesian inference when the theoretical models are so complex that analytical likelihood functions become intractable.

\paragraph{The Likelihood Problem}

Bayesian inference (Chapter~2) needs a likelihood\index{likelihood!intractable} function $p(\mathbf{x}|\boldsymbol{\theta})$, which gives the probability of the data $\mathbf{x}$ given the parameters $\boldsymbol{\theta}$. We combine it with a prior $p(\boldsymbol{\theta})$ to compute the posterior distribution
\begin{equation}
p(\boldsymbol{\theta}|\mathbf{x}) \propto p(\mathbf{x}|\boldsymbol{\theta})p(\boldsymbol{\theta}).
\end{equation}

This works well when we can write down the likelihood function explicitly. As scientific models become more sophisticated, an analytical likelihood becomes hard or impossible to specify. Modern cosmological simulations are an example. Given cosmological\index{cosmological parameters} parameters, we can run numerical simulations that generate synthetic observations such as galaxy distributions or weak lensing maps. Typically, however, we cannot write a closed-form expression for the probability of a specific observed outcome.

The core issue is dimensionality and distributional complexity. A galaxy density field represented as an image is high dimensional. A modest $512 \times 512$ pixel image lives in a space with 262,144 dimensions. In such a space, characterizing the distribution $p(\mathbf{x}|\boldsymbol{\theta})$ explicitly becomes intractable. The distribution typically has complex multimodality, strong correlations and heavy tails, which resist a simple parametric representation.

Over the past decades, astronomy has made much progress in simulating complex physical systems, from cosmological structure formation to stellar evolution. Advances in high-performance computing and numerical methods made this possible. The simulations include more detailed physics and reach higher resolution. A gap has remained in how to connect the simulations to observational data in a statistically sound way.

Traditionally, astronomers reduce the dimensionality with summary\index{summary statistics} statistics. Low-dimensional summaries such as power spectra or correlation functions condense the high-dimensional observations into a more manageable statistical problem. By the central limit theorem, these summary statistics often have distributions closer to Gaussian. Techniques such as the Laplace approximation (Chapter~9) then apply more readily.

This convenience has a cost, which is information loss. The power\index{power spectrum} spectrum describes only certain aspects of the data, namely the second moments of the distribution. For a Gaussian random field, the power spectrum is a complete statistical description. Cosmic structures and other astronomical phenomena, however, are non-Gaussian. Their higher-order moments and phase correlations carry information that the power spectrum discards.

This is a dilemma between tractability and completeness. Full high-dimensional observations preserve all available information, but the likelihood becomes practically incalculable. Summary statistics give tractable inference problems, but they may sacrifice information. For cosmic structure no single summary statistic keeps all the information, but whether a few together keep nearly all of it is less clear. Astronomers have therefore made pragmatic compromises, typically favoring tractability over statistical power.

SBI resolves this dilemma by using an asymmetry common to many scientific domains. The forward problem, which generates observations from parameters by simulation, is tractable. The inverse problem, which infers parameters from observations, is hard. SBI uses neural networks to learn surrogates, which are fast approximations, of the main components of the Bayesian inference pipeline directly from simulated data.

SBI does not force a choice between the two sides of this dilemma. We can apply it to the full observations, and cosmologists call such an analysis field-level inference. We can also apply it to summary statistics whose likelihood we cannot write down. SBI also differs from emulation, which is common in cosmology. An emulator typically predicts the mean observable for given parameters, for example a power spectrum, and replaces a slow simulation. SBI instead learns the distribution of the observable or of the parameters.

The SBI workflow begins with simulation. Consider a cosmological example. We want to infer parameters from observations of the cosmic web. The parameters include the matter density $\Omega_m$, the dark energy equation of state $w$ and the amplitude of fluctuations $\sigma_8$. We generate parameter-observation pairs $\{(\boldsymbol{\theta}_i, \mathbf{x}_i)\}_{i=1}^N$ in two steps.

\begin{enumerate}
    \item Sample cosmological parameters $\boldsymbol{\theta}_i = (\Omega_m, w, \sigma_8, ...)_i$ from a prior distribution $p(\boldsymbol{\theta})$.
    \item Run cosmological simulations with these parameters to generate synthetic galaxy density fields $\mathbf{x}_i$.
\end{enumerate}

\begin{figure}[ht!]
    \centering
    \includegraphics[width=0.85\textwidth]{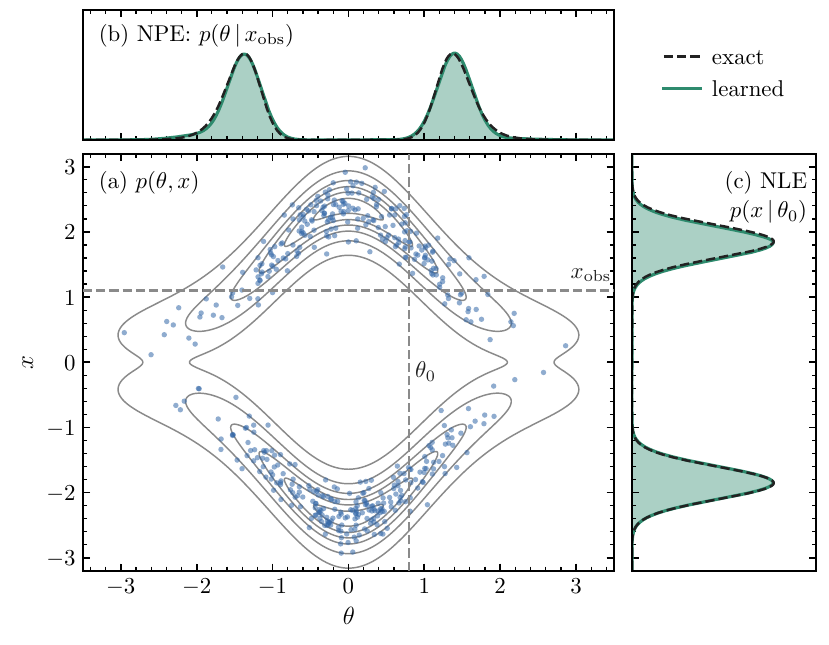}
    \caption{Simulation-based inference for a toy simulator with one parameter $\theta$ and one observation $x$. Panel~(a) shows the joint density $p(\theta, x)$ by its contours and a random subset of the simulations by dots. Panel~(b) shows the posterior $p(\theta | x_{\text{obs}})$ and panel~(c) the likelihood $p(x | \theta_0)$, obtained by conditioning at the dashed lines. The exact curves are normalized slices of the known joint density. The learned curves come from two Mixture Density Networks trained on the simulations, one for the posterior (NPE) and one for the likelihood (NLE). They agree closely with the exact curves.}
    \label{fig:sbi_visualization}
\end{figure}

With simulations first, we need no analytical likelihood function. We use the simulated data to train neural networks that approximate either the likelihood $p(\mathbf{x}|\boldsymbol{\theta})$ or the posterior $p(\boldsymbol{\theta}|\mathbf{x})$ directly. Figure~\ref{fig:sbi_visualization} shows the simulated pairs and the two distributions that we can learn from them.

\paragraph{Neural Likelihood Estimation}

Neural\index{neural likelihood estimation (NLE)}\index{neural likelihood estimation (NLE)|(} likelihood estimation (NLE)\index{NLE|see{neural likelihood estimation}} is the most straightforward approach to SBI in principle. It learns the likelihood function $p(\mathbf{x}|\boldsymbol{\theta})$ directly. This keeps the structure of standard Bayesian inference and extends it to problems that were intractable. The learned likelihood replaces the analytical one and fits into the rest of the Bayesian analysis.

We can implement NLE with the neural density estimators of the previous section. With an MDN, for example, we model the likelihood as
\begin{equation}
p(\mathbf{x}|\boldsymbol{\theta}) = \sum_{k=1}^K \pi_k(\boldsymbol{\theta}) \mathcal{N}(\mathbf{x}|\boldsymbol{\mu}_k(\boldsymbol{\theta}), \boldsymbol{\Sigma}_k(\boldsymbol{\theta})).
\end{equation}

Training follows maximum likelihood. Given the simulated parameter-observation pairs, we maximize
\begin{equation}
\ell(\mathbf{w}) = \sum_{i=1}^N \ln p(\mathbf{x}_i|\boldsymbol{\theta}_i, \mathbf{w}),
\end{equation}
where $\mathbf{w}$ denotes the neural network parameters.

For high-dimensional observations such as galaxy density fields, we might use normalizing flows, diffusion models or flow matching instead of MDNs. The principle stays the same. We learn a mapping from the parameters to a distribution of observations, which encodes complex physics in a differentiable probabilistic model.

NLE preserves the modular nature of Bayesian inference. We can change the prior, combine the likelihood with other likelihood terms, or use a different sampling method without retraining the neural likelihood. We still run MCMC for each new observation, but each likelihood evaluation is now a fast forward pass through the network. However, high-dimensional observations are costly to model. Predicting distributions in such a high-dimensional observation space needs enormous neural networks.\index{neural likelihood estimation (NLE)|)}

\paragraph{Neural Posterior Estimation}

This\index{neural posterior estimation (NPE)|(} dimensionality challenge leads to neural\index{neural posterior estimation (NPE)} posterior estimation (NPE).\index{NPE|see{neural posterior estimation}} NPE takes a more direct route and approximates the posterior $p(\boldsymbol{\theta}|\mathbf{x})$ with a conditional density estimator. By modeling the posterior instead of the likelihood, we only predict distributions in the much lower-dimensional parameter space.

The observation $\mathbf{x}$ might be a $512 \times 512$ pixel density field, while $\boldsymbol{\theta}$ might hold just 5 to 10 cosmological parameters. The network takes the observations $\mathbf{x}$ as input and outputs a distribution over the parameters $\boldsymbol{\theta}$. This makes NPE computationally more efficient and statistically more feasible.

The two methods differ in what they model. NLE asks, ``Given parameters $\boldsymbol{\theta}$, what observations $\mathbf{x}$ might we see?'' NPE asks, ``Given observations $\mathbf{x}$, what parameters $\boldsymbol{\theta}$ might have generated them?''

\looseness=-1 NPE gives posteriors directly, without evaluating a likelihood or running MCMC. Inference is therefore very fast once the network is trained. However, the network implicitly contains the prior used in training. To explore other priors later, we must retrain the network or correct the result with importance sampling (Chapter~12). This trade-off suits applications with well-established priors that are unlikely to change. It also suits applications where inference speed matters most, such as analyzing millions of galaxy spectra from large surveys.\index{neural posterior estimation (NPE)|)}

\paragraph{Amortized Inference}

\looseness=-1 Both NLE and NPE enable amortized\index{amortized inference} inference. Traditionally we run MCMC again for each new observation, which is expensive when each likelihood evaluation needs extensive calculation. In amortized inference, we train a neural network once to approximate either the likelihood or the posterior. We then reuse the approximation for many observations without retraining. The word amortized means that we pay the training cost once and spread it over later observations, so the advantage grows with their number.

Inference without explicit likelihood functions is an old idea. Approximate\index{approximate Bayesian computation} Bayesian computation had similar goals. It compares summary statistics of simulated and observed data and accepts the parameter values whose simulations are ``close enough'' to the observations. However, these methods scaled poorly and often needed careful manual design of the summary statistics and distance metrics. Modern SBI methods differ because neural density estimation has advanced. The deep learning innovations behind generative AI make it possible to approximate high-dimensional, multimodal distributions reliably and at scale.

Early neural approaches often suffered from mode\index{mode collapse} collapse. In mode collapse, the network generates samples from only one or a few modes of a multimodal distribution. It misses other equally valid regions of high probability. For a distribution with three distinct peaks, for example, a neural network with mode collapse might represent one peak and ignore the others. Normalizing flows, diffusion models and flow matching reduce this problem. Their architectures and training procedures better preserve the multimodal structure.

SBI applies well beyond cosmology, to other areas with complex forward models but intractable likelihoods. In stellar population synthesis, we can infer stellar population properties from integrated galaxy spectra. In exoplanet characterization, we can determine atmospheric compositions from transmission spectra. In gravitational wave analysis, we can infer properties of merging compact objects from the signals.

\paragraph{Caveats}

SBI has caveats. The neural surrogate is valid only within the parameter space used for training. Its accuracy depends on how well the training simulations represent reality. When we apply SBI to the full observations, we use information at all scales and also include more systematic effects. For example, cosmological simulations may not model baryon physics accurately at all scales. Applying SBI blindly, especially in field-level inference, might then give overconfident inferences that depend on these systematic effects.

All simulation-based methods face the same problem. We address it with testing and validation, namely benchmarking, blind tests and systematic error analysis. We generate additional simulations with varied physical assumptions to assess the sensitivity to modeling choices. We can also train on one set of simulations and test on others, which measures the impact of systematic effects.

\looseness=-1 Using the full observations does not always increase the systematic errors compared with summary statistics. In some cases, biases might average out. This remains an active research area. These problems call for careful scientific methodology, and we can still use SBI.

A last caveat concerns the network itself. Cross-validation estimates the prediction error of a neural network, but it does not fully account for the uncertainty of the network parameters themselves. Even so, the approximation errors of modern neural density estimators are often smaller than the observational noise and the systematic effects of the simulations. We can then focus on the physical questions and on the familiar tests above, now with more information from the observations.\index{neural network!simulation-based inference|)}\index{neural network|)}\index{simulation-based inference (SBI)|)}

\section{Tutorials: Backpropagation, Autoencoders and Neural Density Estimators}\label{sec:ch15_tutorials}

\looseness=-1 The four tutorials build up from the mechanics of training to networks that predict and model whole densities. The first writes backpropagation by hand and then uses PyTorch, a Python library that computes gradients automatically. The second compresses galaxy images with an autoencoder. The third predicts the lithium abundance of stars with a mixture density network. The fourth models the elemental abundances of stars with a normalizing flow. The full tutorials can be found at \url{https://statml.org/} as 15a to 15d.

\paragraph{Backpropagation and PyTorch} \looseness=-1 Gradient descent needs the gradient of the loss with respect to every weight, and a network can have millions of weights. Backpropagation instead computes all the gradients in one backward pass (Section~\ref{sec:ch15_backprop}). The pass applies the chain rule from the loss back to each weight and reuses the numbers that the forward pass stored. The first tutorial writes this backward pass by hand for a small network that fits a noisy cubic.

\begin{figure}[t!]
    \centering
    \includegraphics[width=0.95\textwidth]{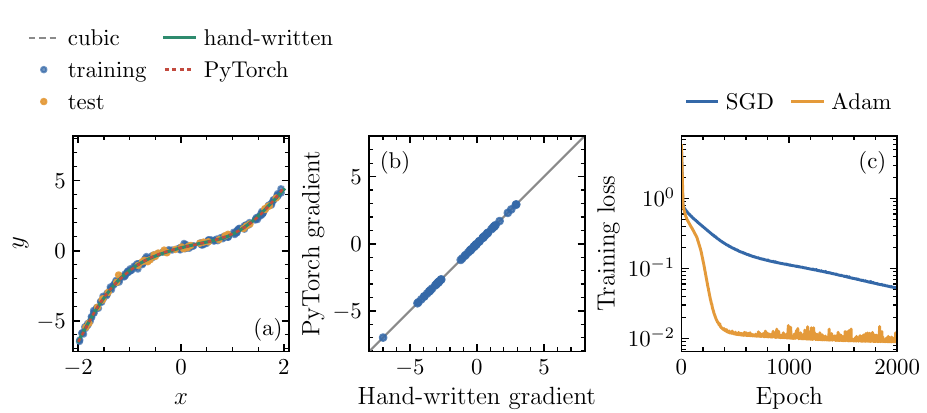}
    \caption{The first tutorial fits a noisy cubic polynomial with a network whose backward pass is written by hand. In panel~(a) the hand-written and the PyTorch networks both follow the cubic through the training and test points. In panel~(b) the hand-written gradients equal those of PyTorch, so the backward pass is correct. In panel~(c) Adam lowers the training loss much faster than stochastic gradient descent.}
    \label{fig:ch15_tutorial_backprop}
\end{figure}

\looseness=-1 PyTorch\index{PyTorch} then computes the same gradients by automatic differentiation, from the forward pass alone. The hand-written gradients agree with those of PyTorch for the same weights (Figure~\ref{fig:ch15_tutorial_backprop}b), and both networks follow the cubic (Figure~\ref{fig:ch15_tutorial_backprop}a). PyTorch also provides common optimizers, so we can compare them on the same network. With the same mini-batches and number of epochs, Adam\index{Adam optimizer} (Section~\ref{sec:ch15_training}) reaches a much lower loss than stochastic gradient descent (Figure~\ref{fig:ch15_tutorial_backprop}c). Adam rescales the step of each parameter by the size of its recent gradients, so parameters with small gradients still move.

\looseness=-1 The chain rule, the two passes and a gradient-based optimizer are most of what even very large networks need. A library supplies the backward pass for any model we write, so the other three tutorials can focus on the models.

\paragraph{Autoencoders for Galaxy Images} \looseness=-1 Principal component analysis (PCA) of Chapter~10 compresses an image with a linear map. An elongated or tilted galaxy, however, spreads over several components (Section~\ref{sec:ch10_tutorial}). An autoencoder\index{galaxy images} (Section~\ref{sec:ch15_autoencoders}) replaces the linear map with neural networks. An encoder compresses each image into a few numbers, and a decoder rebuilds the image from them. Because the networks are nonlinear, these numbers can follow curved structure that a linear map misses.

\begin{figure}[t!]
    \centering
    \includegraphics[width=0.95\textwidth]{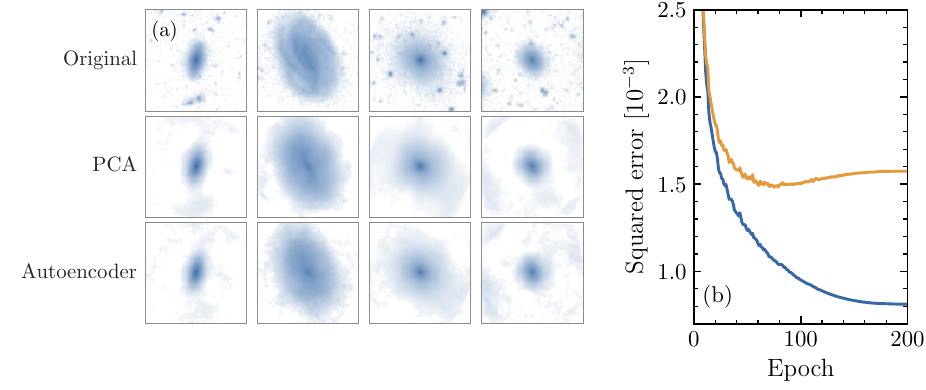}
    \caption{The second tutorial compresses galaxy images to a short code. In panel~(a) test galaxies rebuilt by PCA and by the autoencoder, with codes of the same length, keep their size, shape and orientation. The autoencoder is slightly sharper, and darker blue is more light. In panel~(b) the training error keeps falling, while the validation error turns up once the network starts to memorize the training images.}
    \label{fig:ch15_tutorial_autoencoder}
\end{figure}

\looseness=-1 A network this flexible can also memorize its training images. We therefore stop the training where the error on separate validation images is lowest (Section~\ref{sec:ch15_training}). On test images the reconstructions keep the size, shape and orientation of each galaxy, as with PCA, and are slightly sharper (Figure~\ref{fig:ch15_tutorial_autoencoder}a). The training error ends below that of any linear map of the same size, so the network represents structure that PCA misses (Figure~\ref{fig:ch15_tutorial_autoencoder}b).

\looseness=-1 An autoencoder thus learns a compact, nonlinear description of large images from the images alone, without labels. Galaxies that it rebuilds poorly differ from its training sample, so the reconstruction error can also flag unusual objects.

\paragraph{Lithium Abundance in Open Clusters} \looseness=-1 Lithium\index{lithium abundance}\index{lithium abundance|(}\index{open clusters|(} burns in the hot layers below the surface, so a star destroys the lithium that convection carries there. The lithium abundance A(Li) of a star therefore depends on its temperature, which sets the depth of convection, and on its age. It also depends on unmeasured properties such as rotation, so stars of the same temperature and age differ in A(Li). The third tutorial uses stars of open\index{open clusters} clusters, whose ages are known.

\begin{figure}[t!]
    \centering
    \includegraphics[width=0.95\textwidth]{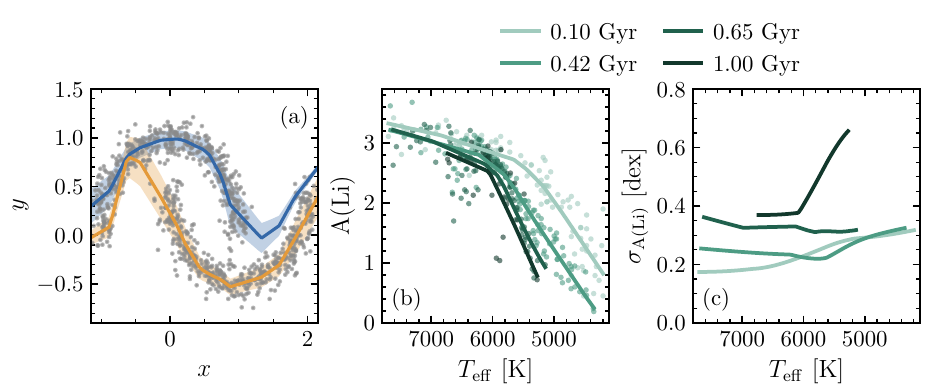}
    \caption{The third tutorial predicts densities with mixture density networks. In panel~(a) each of the two components (blue and orange, mean and standard deviation) follows one of the two moons (grey points). In panel~(b) the predicted mean of A(Li) (lines) follows the cluster stars (points) of the same age, shown in the same color. Lithium falls toward cooler stars and in older clusters. In panel~(c) the predicted width grows with age, most for the cool stars.}
    \label{fig:ch15_tutorial_mdn}
\end{figure}

\looseness=-1 A standard network predicts one value of A(Li) for each temperature and age. With the squared error as its loss, that value is the mean of the stars with these inputs. The mean loses the scatter and any second branch of the data (Section~\ref{sec:ch15_mdn}). A mixture density network (MDN) instead predicts the mixture weights, means and widths of a Gaussian mixture as functions of the inputs. We train it by maximizing the likelihood of the observed abundances.

\looseness=-1 The tutorial first tests the network on the two moons, two interleaved half circles. Some values of $x$ have two values of $y$, one on each moon. A standard network would predict their average, on neither moon, while the MDN gives each moon its own component (Figure~\ref{fig:ch15_tutorial_mdn}a).

\looseness=-1 For the stars, the network predicts a density of A(Li) at each temperature and age. Its mean falls toward cooler stars, whose convection reaches deeper, and with age, because the burning continues (Figure~\ref{fig:ch15_tutorial_mdn}b). Its width, which holds what temperature and age leave unexplained, grows with age, most for the cool stars (Figure~\ref{fig:ch15_tutorial_mdn}c).

\looseness=-1 A network can therefore predict a whole probability density rather than one value. Many quantities in astronomy have a scatter that changes with the inputs, or more than one possible value. For them the predicted width is itself a result.\index{lithium abundance|)}\index{open clusters|)}

\begin{figure}[t!]
    \centering
    \includegraphics[width=0.95\textwidth]{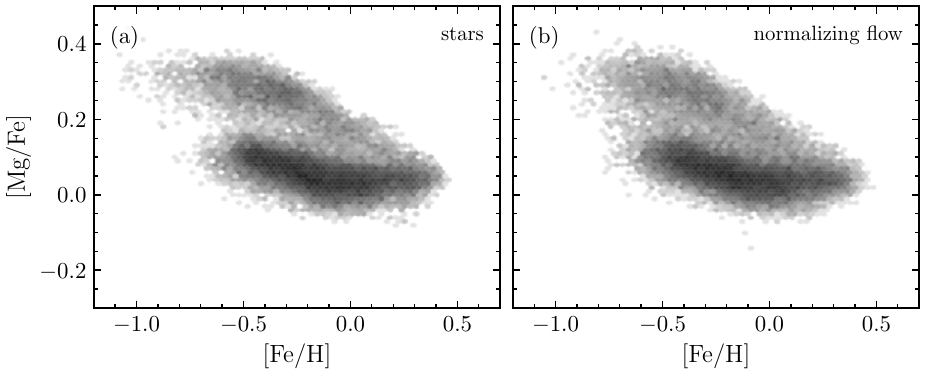}
    \caption{The fourth tutorial models the elemental abundances of stars with a normalizing flow. Panel~(a) shows the two sequences of the stars in the plane of [Fe/H] and [Mg/Fe], with darker grey for more stars. In panel~(b) an equal number of samples from the flow follow both sequences and keep a lower density between them.}
    \label{fig:ch15_tutorial_flow}
\end{figure}

\paragraph{Chemical Abundances with a Normalizing Flow} \looseness=-1 The\index{chemical abundances|(} elemental abundances\index{chemical abundances} of stars in the Milky Way form sequences, such as the two sequences of [Mg/Fe] against [Fe/H] (Figure~\ref{fig:ch15_tutorial_flow}a). Massive-star supernovae make magnesium and mainly enrich the upper sequence, while Type Ia supernovae also add iron to the gas of the lower one. The fourth tutorial models 13 abundances of APOGEE\index{APOGEE survey} stars together, as a density in 13 dimensions. A Gaussian has one peak, however, and a mixture needs many components for curved sequences in many dimensions.

\looseness=-1 A normalizing flow instead transforms a standard Gaussian into the density of the stars with an invertible network (Section~\ref{sec:ch15_flows}). The change of variables gives its exact likelihood, which we maximize on the stars. We build its most basic form, RealNVP, whose layers freeze some abundances and scale and shift the others with networks of the frozen ones. The inverse needs no inverted network, and the Jacobian is triangular and cheap.

\looseness=-1 Samples of the flow follow both sequences and keep a lower density between them (Figure~\ref{fig:ch15_tutorial_flow}b). This plane is one view of the 13 dimensions, and the samples follow the stars in the other views too.

\looseness=-1 A normalizing flow represents a density without a fixed parametric family, unlike the Gaussian mixture of an MDN. It gives both the likelihood of a star and mock stars, and it scales to many dimensions. This makes it a basis for density estimation in astronomy and for simulation-based inference (Section\index{chemical abundances|)}~\ref{sec:ch15_sbi}).

\paragraph{Further Reading} \looseness=-1 McCulloch and Pitts (1943) described neural computation with a mathematical model. Rosenblatt (1958) introduced the perceptron, and Minsky and Papert (1969) analyzed the limits of single-layer networks. Rumelhart et~al.\ (1986) presented backpropagation. Kingma and Ba (2015) introduced the Adam optimizer, which uses moving averages of the gradient and its square. Cybenko (1989) and Hornik et~al.\ (1989) studied universal approximation, and Vapnik (1995) placed neural networks in statistical learning theory. MacKay (1992) applied the Bayesian evidence to regularization and model comparison. Neal (1996) gave a full Bayesian treatment of networks, and Williams (1996) related them to Gaussian processes. LeCun et~al.\ (1998) described convolutional networks, Elman (1990) recurrent networks and Hochreiter and Schmidhuber (1997) long short-term memory. He et~al.\ (2016) introduced residual networks, whose shortcut connections make very deep networks easier to train. Vaswani et~al.\ (2017) introduced the Transformer, a network built on attention without recurrence or convolution. Bishop (1994) introduced mixture density networks, and Hinton et~al.\ (2012) and Srivastava et~al.\ (2014) developed dropout. Astronomers applied neural networks early. Storrie-Lombardi et~al.\ (1992) classified galaxy morphology, and von Hippel et~al.\ (1994) classified stellar spectra. Lahav et~al.\ (1995) compared networks with human classifiers, and Bailer-Jones et~al.\ (1998) classified spectral type and luminosity class. Belkin et~al.\ (2019) described double descent, Nakkiran et~al.\ (2019) showed it in deep learning, and Adlam and Pennington (2020) analyzed it. Jacot et~al.\ (2018) introduced the neural tangent kernel, and Lee et~al.\ (2017) and Matthews et~al.\ (2018) related wide networks to Gaussian processes. Cranmer et~al.\ (2020) review simulation-based inference, and Papamakarios et~al.\ (2018) examined neural likelihood estimation. Greenberg et~al.\ (2019) and Gon\c{c}alves et~al.\ (2020) developed neural posterior estimation. Bommasani et~al.\ (2021) review large pre-trained models. Brown et~al.\ (2020) showed that large language models learn from few examples. Wei et~al.\ (2022) analyzed abilities that appear only in large models. Devlin et~al.\ (2018) and Raffel et~al.\ (2020) examined pre-training and transfer learning. Kingma and Welling (2013) introduced variational autoencoders. Goodfellow et~al.\ (2014) introduced generative adversarial networks, in which two networks compete, and Rezende and Mohamed (2015) normalizing flows. Dinh et~al.\ (2016) and Durkan et~al.\ (2019) developed them further. Ho et~al.\ (2020) introduced denoising diffusion models, and Song et~al.\ (2020) unified them with score-based generative models. Ting and Weinberg (2022) used a normalizing flow on the APOGEE stars of the fourth tutorial to study the intrinsic dimensionality of the abundances. Sun et~al.\ (2025) trained a mixture density network on the open clusters of the third tutorial and studied lithium in comoving pairs of stars. The second tutorial uses the Galaxy Zoo~2 images of Ma et~al.\ (2026), who generated new ones with a conditional diffusion model.

\makeatletter\@clubpenalty=150\makeatother \clubpenalty=150 \widowpenalty=150

\chapter{Afterword: A Personal Note}

Writing this textbook has been an interesting personal journey. Two experiences motivated it, and many researchers at the interface of statistics, AI and astronomy will recognize both. First, I repeatedly discussed techniques with colleagues and often ended up scribbling LaTeX notes to share my thoughts. Over time, these scattered notes helped my collaborators, and they also helped me to organize and clarify my own understanding. Writing forces a clarity of thought that conversation does not give.

The second experience matters more to me. I have repeatedly been drawn into arguments about the usefulness of machine learning in astrophysics. In these arguments, ``machine learning'' often means different things to different people. The responses I encounter range from too pessimistic to unrealistically optimistic, and I find both positions difficult to defend.

These discussions touch on questions that go far beyond technical implementation. What does it mean for a method to be trustworthy? What makes something interpretable? Is deep learning always a black box? Are simpler models morally better? What is our philosophy of science, and what should be the ultimate goals of science?

I find these conversations fascinating because they show how much our choices of method rest on deeper beliefs about the nature of scientific knowledge. Writing this textbook has helped me organize my thoughts on these questions. It has also given me a more coherent view of the role of different statistical approaches in astronomical research. The process has been as much about understanding my own views as about communicating them to others.

An informal poll that I ran on social media encouraged me to start this project. The overwhelming support of colleagues convinced me to take it on. The step from scattered notes and lecture materials to a coherent textbook proved more demanding than I expected, especially for a non-native English speaker.

When I began this project, I ambitiously planned a single volume. It would cover both classical machine learning (Chapters~4 to 14) and deep learning (Chapter~15) in equal detail. This turned out to be wildly optimistic. The classical methods alone expanded to fill the current volume. I hope that I have given a sufficient introduction to neural networks and deep learning. A full treatment of these topics will need a separate volume, a project I look forward to once I have recovered from this one.

This has been one of my largest academic undertakings. From the first idea to the first draft, each page took about one hour of focused work. Many cups of coffee and the occasional dessert kept me motivated, and rounds of editing and revision added another hour per page. For the original 650-page version, the total time approached 1,300 hours. It consumed most of my free time and weekends during the 2024-2025 academic year.

Trimming the book to about 500 pages added another few hundred hours. The companion tutorials come from many years of my teaching, and curating them took another few hundred hours. Many of their examples are drawn from or inspired by my own research. This time commitment made me more sure that the deep learning volume should wait.

I should acknowledge the role that large language models (LLMs) played in making this project feasible. The preliminary writing, the organization, the mathematical content and the overall perspective represent years of my own thinking and research. The first draft of much of the lecture notes began in 2020, before LLMs, and I taught the course until recently.

I used LLM assistance mainly for language editing. It made the text as readable as possible while keeping my professional voice and, I hope, my occasional attempts at humor. Without this support, the language polishing alone would have made the project too time-consuming for a non-native speaker. The irony is not lost on me that a textbook partly about machine learning was itself made possible by machine learning.

I also thank David Weinberg for his enthusiastic support, his very careful reading of the book and his many insightful comments. I cannot believe that anyone would have the patience to read through such a gigantic volume, but David did.

My own research is on deep learning applications in astronomy. Even so, I remain convinced that a solid grounding in classical statistical methods covers a large part of the analytical needs of our field. Some astronomical problems do benefit from deep learning, and this is why I find the research area so interesting. Still, domain knowledge combined with solid statistical reasoning can solve many problems that first seem to need more complex approaches.

We should choose between classical and modern methods according to the nature of the problem, rather than fashion or technological enthusiasm. My goal is to provide the grounding needed to make such choices thoughtfully, knowing the trade-offs.

As I said in the preface (Chapter~1), I hope that this textbook adds depth to our understanding of how machine learning techniques have evolved. I also hope that it shows how they connect to the statistical ideas that astronomers have long used. If it contributes to more meaningful discussions about the role of these methods in astronomy, the considerable time investment will have been worthwhile.

Accessibility mattered to me from the start. I grew up in Malaysia, where educational resources in fields such as statistical astronomy can be hard to find. I know well how barriers to access can hold people back. I therefore chose to release this book on arXiv, freely available to anyone with an internet connection.

Later, Cambridge University Press came to me and, despite the arXiv version, was kind enough to want to publish the book. I am surprised by this interest and grateful for it, and it has led to the final published version.

This textbook is dedicated to my family, who supported my somewhat unusual intellectual path with great patience. My parents put up with a child who preferred solving linear algebra problems to playing with friends. My father (Baba) even paid me to go outside and play, hoping to encourage more conventional childhood activities. Their tolerance let me explore the mathematical topics that have long fascinated me and later helped me develop expertise in physics, statistics and artificial intelligence.

My father passed away while I was polishing the second version for Cambridge University Press. I have since taken a long break at home in Malaysia to finish this book. When I tidied up what he left behind, I found that statistics and calculus were his worst subjects at university. Perhaps this is why he paid me to go outside and play.

{\makeatletter\@clubpenalty=10000\makeatother \clubpenalty=10000 \widowpenalty=10000 \displaywidowpenalty=10000 The dedication also extends to my grandmother (Popo), who had no formal education herself and was amused by my mathematics scribble books. When we visited, she sewed blankets while I sat in her back room solving equations. She would occasionally peek over my shoulder with curiosity and pride. Her quiet support and the warmth of those childhood memories have stayed with me throughout this journey.\par}

Looking back, I realize that writing a textbook is itself an exercise in the kind of thinking it teaches. Like fitting a statistical model, it requires a balance of completeness and clarity, of rigor and accessibility, and of personal perspective and objective presentation. The process has taught me as much as I hope the book will teach its readers.

To those who work through this material, I hope that you learn something from it. I also hope that it changes how you think about the relationship between data and knowledge. I expect the future of astronomy to need ever more sophisticated statistical thinking. I hope that this grounding helps you, whichever specific techniques prove most useful for your research problems.

The conversation about machine learning in astronomy is far from over. It should continue to evolve as our computing power and our theoretical understanding grow. What matters most is that it rests on careful thinking. We need to ask what we are trying to achieve and why some approaches reach those goals better than others.

Thank you for joining me on this journey through statistical machine learning for astronomy. The real work begins when you close this book and start applying these ideas to your own research questions.

\cleardoublepage
\phantomsection
\addcontentsline{toc}{chapter}{Bibliography}
{\small
\setlength{\bibsep}{4.9pt plus 2pt}

}

\index{confidence interval|seealso{credible interval}}\index{feature engineering|seealso{basis functions; feature transformation}}\index{feature transformation|seealso{basis functions}}\index{ridge regression|seealso{weight decay}}\index{errors-in-variables regression|seealso{attenuation bias; Deming regression}}\index{dimensionality reduction|seealso{autoencoder; principal component analysis}}\index{density estimation|seealso{kernel density estimation; mixture density network; normalizing flow}}\index{covariance function|seealso{kernel function}}\index{Jacobian|seealso{transformation of random variables}}\index{clustering|seealso{Gaussian mixture model; K-means clustering}}\index{classification|seealso{Bayesian logistic regression; Gaussian process, classification; logistic regression}}\index{uncertainty quantification|seealso{aleatoric uncertainty; credible interval; epistemic uncertainty; posterior predictive distribution}}\index{linear regression|seealso{Bayesian linear regression; errors-in-variables regression}}\index{logistic regression|seealso{Bayesian logistic regression}}\index{black holes|seealso{$M$--$\sigma$ relation}}
\cleardoublepage
\phantomsection
\addcontentsline{toc}{chapter}{Index}
{\raggedright\printindex}
\cleardoublepage

\phantomsection
\addcontentsline{toc}{part}{Companion Tutorials}
\part*{Companion Tutorials}
\sloppy
\tutorialfloats

\tutorialchapter{2a}{Poisson Statistics for Cluster Detection}
\noindent\textit{Companion material for Chapter 2}

\section*{Introduction}

In Chapter 2 we built the theory of probabilistic reasoning, and we saw how prior knowledge and observed data combine into a posterior belief. This tutorial applies these ideas to a practical problem, the detection of stellar clusters from the spatial pattern of stars.

Stars in our Galaxy follow a smooth distribution, set by the gravitational potential and the formation history of the Milky Way. Against this background, a stellar cluster is a local overdensity, a region where the density of stars is far above the expectation. This is a good setting to learn how probability theory measures deviations from the expectation for a random field and finds physically meaningful structures.

Such overdensities connect to basic questions in astrophysics. Consider the search for intermediate-mass black holes (IMBHs) in the Milky Way halo, which lie between stellar-mass and supermassive black holes. Theory predicts that IMBHs should keep bound clusters of stars, which create detectable overdensities. \href{https://ui.adsabs.harvard.edu/abs/2021ApJ...917...17G/abstract}{Greene et al.~(2021)} searched for stellar overdensities that could indicate such black holes.

We compare the observed star counts with the expectation from the background, and so we find candidate clusters. When we find no overdensity, the same comparison limits the number of IMBHs. Each one would leave a star count that is very unlikely under the background.

The same ideas apply beyond stellar astronomy. Examples are galaxy clusters in large-scale structure surveys and excess high-energy particle events that could signal dark matter annihilation. In each case, we want to tell real physical structures from random fluctuations of the background.

A quantitative approach has an advantage over visual inspection, which becomes difficult in high-dimensional spaces. We replace the judgment of the eye with a number, the likelihood of the star count at each position under the background model. This lets us make objective statements about the presence or absence of clusters.

\section*{Prerequisites and Learning Objectives}

\subsection*{Prerequisites}

This tutorial assumes the following.

\begin{itemize}
\tightlist
\item
  Python programming with NumPy arrays and matplotlib plots
\item
  Probability theory, in particular probability distributions and discrete distributions
\item
  Basic knowledge of random variables, expectation values and independence
\end{itemize}

\subsection*{Learning Objectives}

After this tutorial you can do the following.

\begin{itemize}
\tightlist
\item
  Model the star counts of a field with the Poisson distribution
\item
  Check with joint and conditional distributions that the counts of two regions are independent
\item
  Find overdensities as counts that are very unlikely under the background model
\item
  Estimate the expected count with maximum likelihood estimation
\item
  Handle edge effects and spatial correlations in a two-dimensional point pattern
\end{itemize}

\section*{Setup}

We begin by importing the libraries, setting the plot style and fixing the random seed.

\tutorialbarrier
\begin{codecell}

\end{codecell}

\section*{Modeling the Stellar Background Field}

We begin with the simplest case, a uniform random distribution of stars in a two-dimensional field. This idealization is our null hypothesis, the expected distribution in the absence of clusters. We model it as a spatial Poisson process. The number of stars in any region then follows the Poisson distribution of Chapter 2 (Section 2.3).

This model has three assumptions.

\begin{enumerate}
\def\labelenumi{\arabic{enumi}.}
\tightlist
\item
  \textbf{Spatial homogeneity.} The background density \(\rho\) is the same everywhere in the field.
\item
  \textbf{Spatial stationarity.} The statistical properties do not change under a translation.
\item
  \textbf{Complete spatial randomness.} A star at one location does not change the probability of finding stars at other locations.
\end{enumerate}

With these assumptions, the probability that the number of stars \(X\) in a region of area \(A\) equals \(k\) is

\[P(X = k|\lambda) = \frac{\lambda^k e^{-\lambda}}{k!},\]

where \(\lambda = \rho A\) is the expected number of stars in the region.

The uniform model is an idealization. In real astronomical applications, the stellar density \(\rho(l,b)\) changes with the Galactic coordinates. The causes are the structure of the Galactic disk, the spiral arms, differential extinction and the selection effects of the observations. The uniform model is still our starting point, and we can extend the method to realistic densities with coordinate transformations or local density estimates.

We first simulate a field of 10,000 stars and plot it.

\tutorialbarrier
\begin{codecell}

\end{codecell}
\tutorialplot{\includegraphics[width=4.249in,height=0.75\textheight,keepaspectratio]{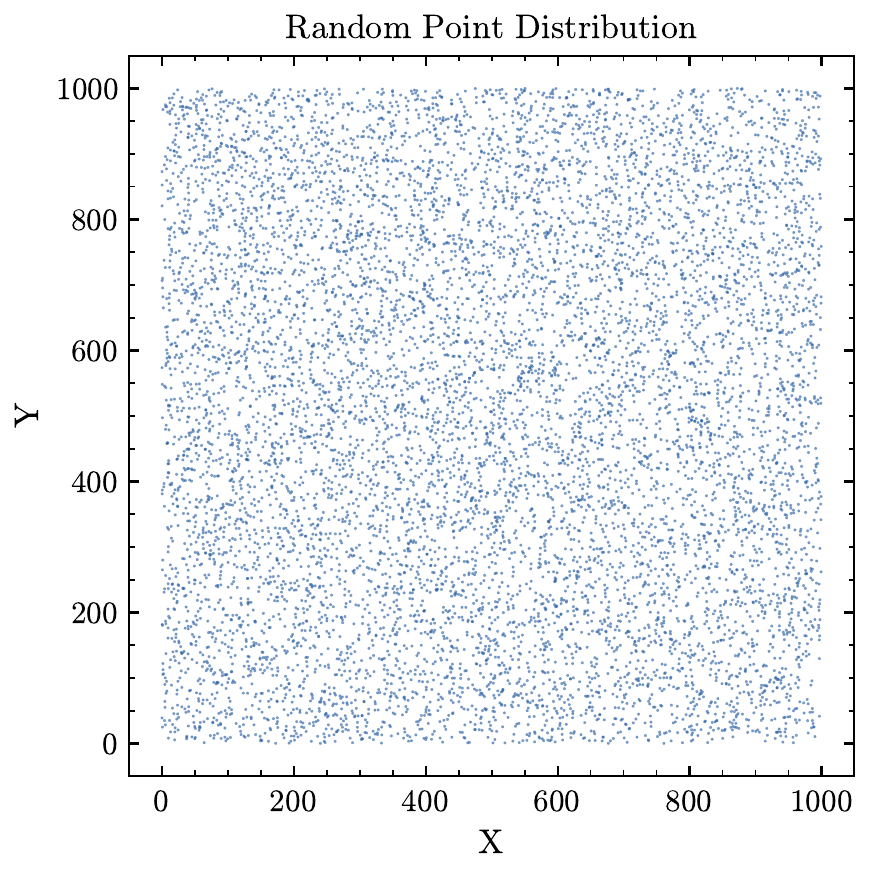}}%
  {\includegraphics[height=\tutorialheight]{tutorials/2a/fig01.pdf}}

\section*{Random Variables and Counting Statistics}

The difference between deterministic and random variables is clear in a stellar field. Consider the average stellar density in a region of our simulated field. It is set by the underlying properties of the stellar distribution, so it is a deterministic variable. The actual number of stars \(X\) that we observe in a region is a random variable (Section 2.1). It fluctuates around the expected value, because the star positions are random.

This is like many astronomical measurements. The true orbital period of a planet is deterministic, but our measurements of it contain random uncertainties. In the same way, the true stellar density is fixed, but our counts vary randomly. For counting observations, this randomness has a specific mathematical form.

We count discrete objects such as stars. If each detection is independent and the average rate is constant, the counts follow the Poisson distribution above. Here \(\lambda = \rho A\) is the expected number of stars for the background density \(\rho\) and the area \(A\) of the counting region. The Poisson distribution arises when three conditions hold.

\begin{enumerate}
\def\labelenumi{\arabic{enumi}.}
\tightlist
\item
  The events occur independently.
\item
  The average rate is constant over the observation interval.
\item
  Two events cannot occur at exactly the same location.
\end{enumerate}

To see how the expected density relates to the observed counts, we analyze the distribution of stars in circular annuli. An annulus is the region between two radii, and the first one starts at radius 0. By dividing the analysis region into concentric rings, we see how the counts change with the distance from a center. We draw a new field and mark two annuli around a center.

\tutorialbarrier
\begin{codecell}

\end{codecell}
\tutorialplot{\includegraphics[width=6.171in,height=0.75\textheight,keepaspectratio]{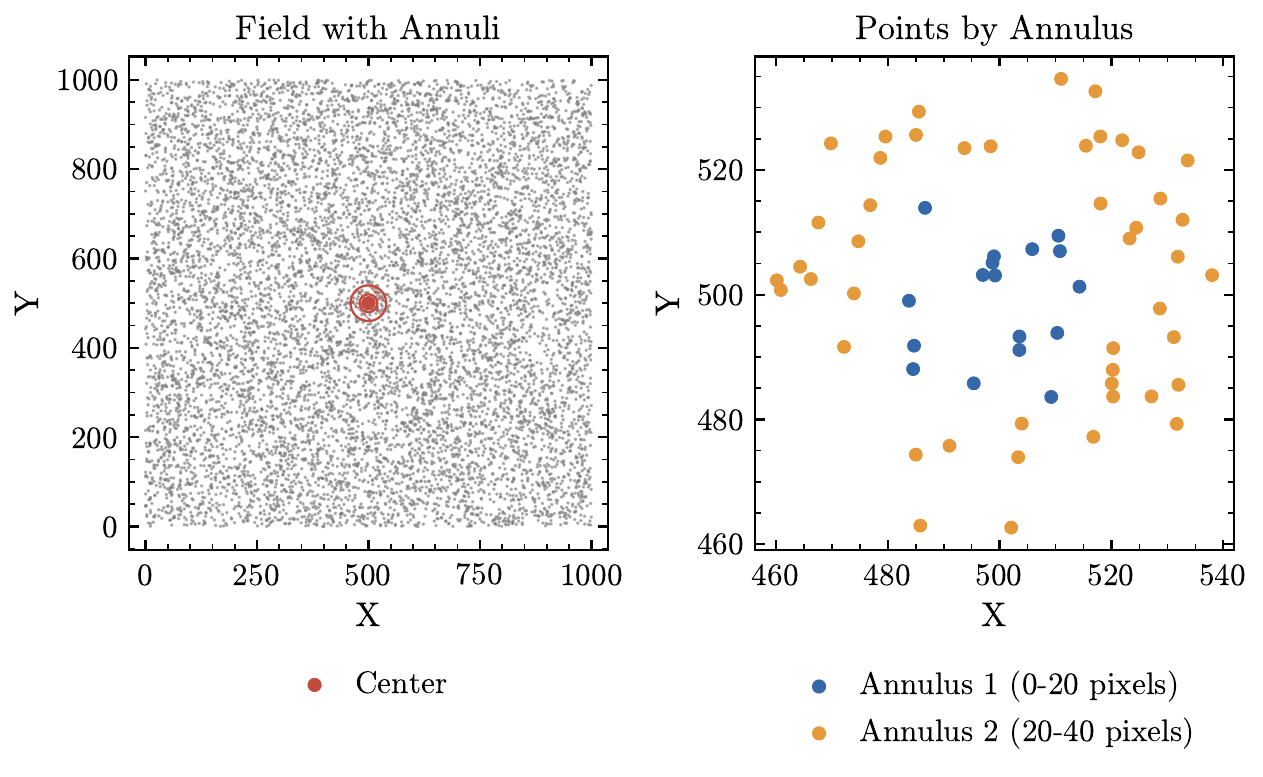}}%
  {\includegraphics[height=\tutorialheight]{tutorials/2a/fig02.pdf}}

This visualization shows how the random variable \(X\), the number of stars, relates to the deterministic background density in different regions. The annular geometry lets us sample different radial distances from a given center. The second annulus covers a larger area, so it holds more stars.

For a uniform random field, the expected number of stars \(\lambda\) in an annulus scales with its area. For an inner radius \(r_1\) and an outer radius \(r_2\), the area is

\[A = \pi(r_2^2 - r_1^2).\]

With a uniform background density \(\rho\) (stars per unit area), the expected count is

\[\lambda = \rho A = \rho\pi(r_2^2 - r_1^2).\]

We generated 10,000 stars in a field of 1000 by 1000 pixels, so the background density is \(\rho = 0.01\) stars per pixel\(^2\). This known density lets us predict the expected counts in each annulus and check that the observations follow the theoretical Poisson distribution.

For this check, we use each star in the field as a center and count its neighbors in both annuli. This gives many samples, and it prepares us for cluster detection. The stars of a cluster will show deviations from the Poisson expectation in their neighbor counts.

\tutorialbarrier
\begin{codecell}

\end{codecell}

We compare the neighbor counts of each annulus with a Poisson distribution in histograms.

\tutorialbarrier
\begin{codecell}

\end{codecell}
\tutorialplot{\includegraphics[width=4.221in,height=0.75\textheight,keepaspectratio]{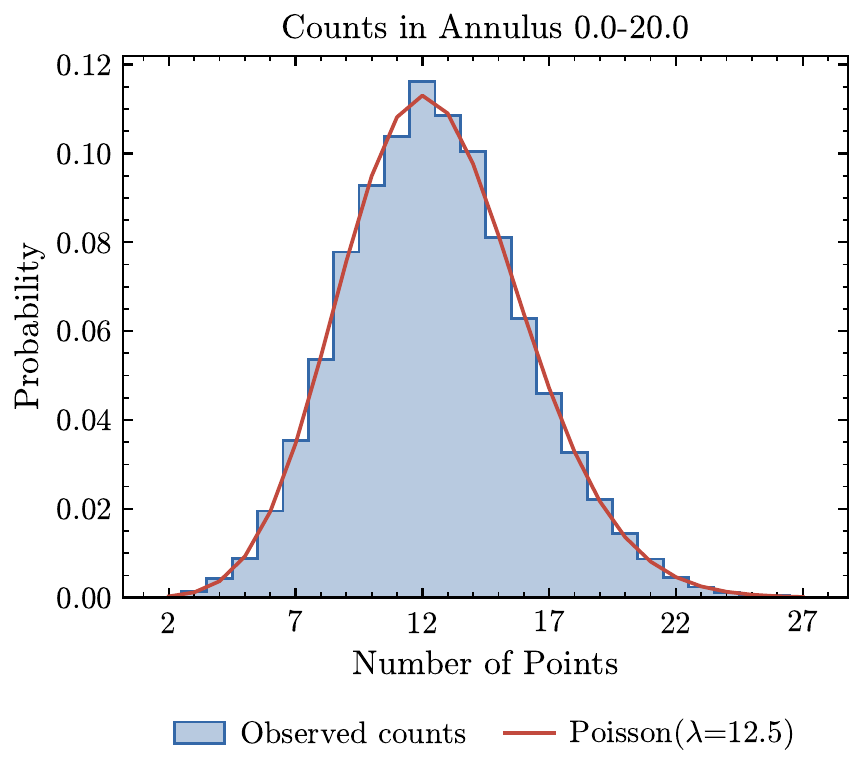}}%
  {\includegraphics[height=\tutorialheight]{tutorials/2a/fig03.pdf}}
\begin{Verbatim}[fontsize=\footnotesize, xleftmargin=5mm]
Expected λ from area = 12.57
Observed mean count = 12.54
Observed variance = 12.56
\end{Verbatim}
\tutorialplot{\includegraphics[width=4.221in,height=0.75\textheight,keepaspectratio]{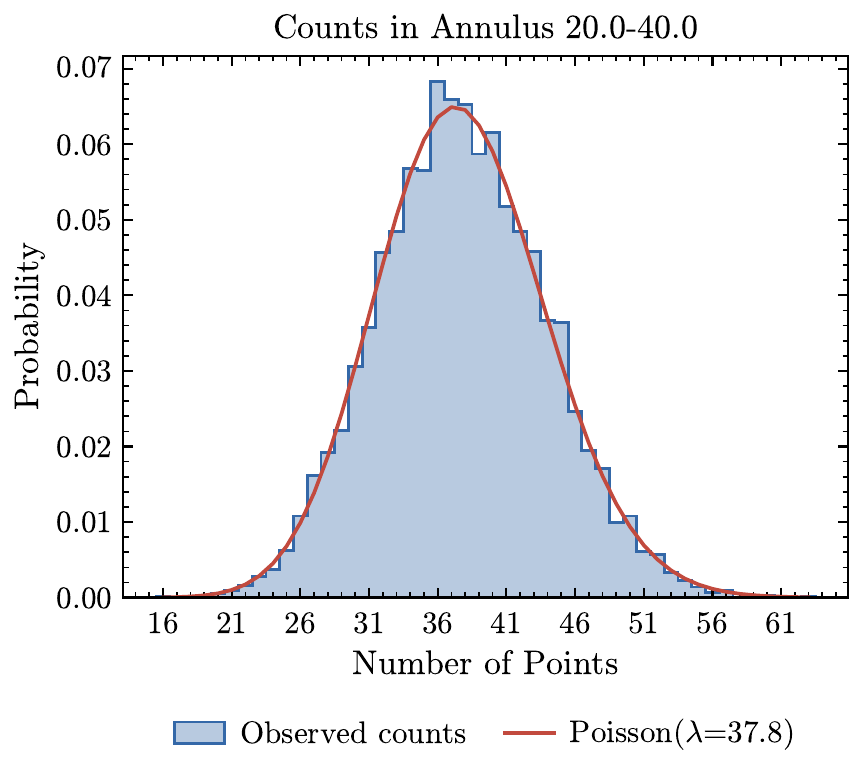}}%
  {\includegraphics[height=\tutorialheight]{tutorials/2a/fig04.pdf}}
\begin{Verbatim}[fontsize=\footnotesize, xleftmargin=5mm]
Expected λ from area = 37.70
Observed mean count = 37.78
Observed variance = 37.03
\end{Verbatim}

Our implementation uses several techniques that deserve a comment. First, we use every star in the field as a center and count its neighbors, without the star itself. This gives 10,000 overlapping measurements of the local density distribution, far more than a single center gives. The Poisson distribution is then well characterized.

Second, we use periodic boundary conditions, so the field is a torus. Without them, stars near the edges would have fewer neighbors than stars in the interior, which gives a wrong density. When the x-distance between two stars is larger than half the field size, the shorter path wraps around the boundary. The same holds for the y-distance. The wrap connects the left edge to the right edge and the top to the bottom, so every star has a complete neighborhood. We use it for simplicity, while real astronomical applications typically exclude the edge regions or apply window functions.

The histograms confirm our expectation. Each annulus shows a count distribution that matches the Poisson distribution with \(\lambda\) equal to the observed mean. The observed mean is also close to the expected \(\lambda\) from the geometry. The variance is close to the mean, as it is for a Poisson distribution. This agreement confirms both our uniform field and our understanding of how a Poisson process appears in a spatial point pattern.

\section*{Joint and Conditional Distributions in Point Counts}

The distributions that we examined for the individual annuli are marginal distributions. They describe the counts in a single region and ignore the relation between regions. We get the marginal distribution of \(X_1\), the count in the first annulus, from the joint distribution. We sum over all values of \(X_2\), the count in the second annulus (Section 2.5)

\[p(X_1) = \sum_{X_2} p(X_1,X_2),\]

where \(p(X_1,X_2)\) is the joint probability (Section 2.4) of \(X_1\) stars in the first annulus and \(X_2\) stars in the second.

Joint distributions matter for cluster detection. Marginal distributions describe the overall density field, and joint distributions show spatial correlations. In a truly random field, we expect no correlation between the counts of non-overlapping regions. A cluster, however, creates a positive correlation, because a high count in one annulus predicts a high count in the nearby annulus.

The joint distribution \(p(X_1,X_2)\) of our two annuli describes this relation. We estimate it with a two-dimensional histogram of the observed pairs of counts. Each star gives one pair \((X_1, X_2)\). Here \(X_1\) counts the neighbors within 20 pixels, and \(X_2\) counts those between 20 and 40 pixels.

\tutorialbarrier
\begin{codecell}

\end{codecell}
\tutorialplot{\includegraphics[width=4.385in,height=0.75\textheight,keepaspectratio]{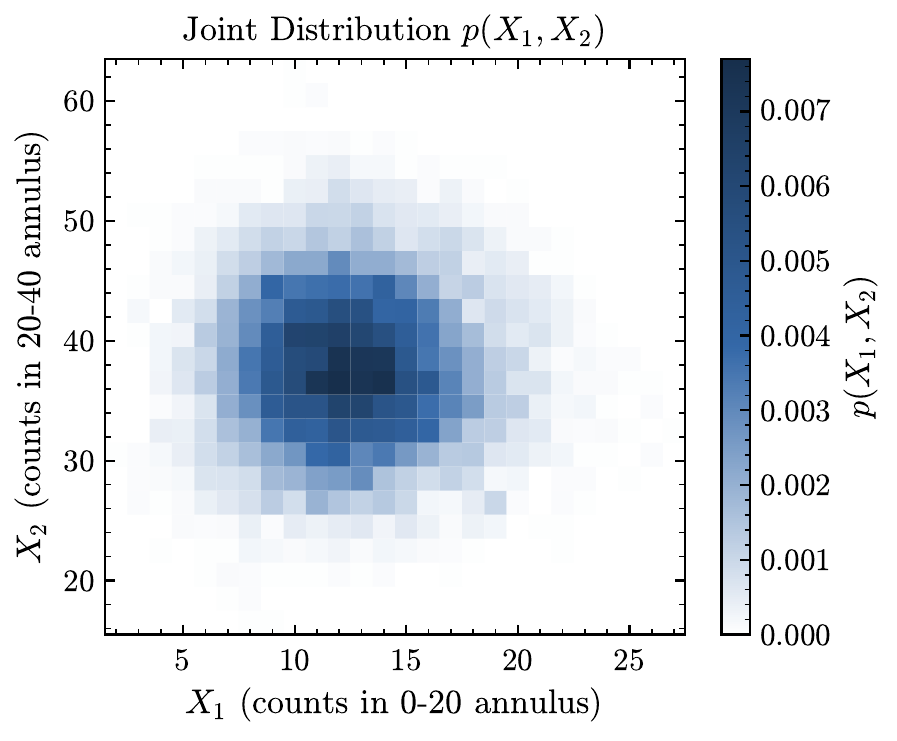}}%
  {\includegraphics[height=\tutorialheight]{tutorials/2a/fig05.pdf}}

The joint distribution shows the relation between the counts of our two annuli. For a truly random point process, we expect no correlation structure. The count in one annulus should not predict the count in the other, and the plot shows a single blob without a tilt.

To examine this independence more carefully, we look at the conditional distribution \(p(X_2|X_1)\). It is the distribution of the counts in the outer annulus for a given count in the inner annulus. If the annuli are statistically independent, then

\[p(X_2|X_1) = p(X_2).\]

This equality says that knowing \(X_1\) gives no information about \(X_2\). We can test it by selecting the stars with a specific \(X_1\) and looking at their \(X_2\) distribution. Because the counts are discrete, we can choose exact integer values of \(X_1\).

By the product rule (Section 2.5),

\[p(X_1,X_2) = p(X_2|X_1)p(X_1).\]

When the counts are independent, this becomes

\[p(X_1,X_2) = p(X_2)p(X_1).\]

This factorization has a practical consequence. We can analyze each annulus separately without losing information about their joint behavior. This reduces the dimension of the problem. It helps when we extend the analysis to many annuli, or when the data are few and a high-dimensional joint distribution is hard to estimate. Below we compare the \(X_2\) distributions for two values of \(X_1\).

\tutorialbarrier
\begin{codecell}

\end{codecell}
\tutorialplot{\includegraphics[width=4.254in,height=0.75\textheight,keepaspectratio]{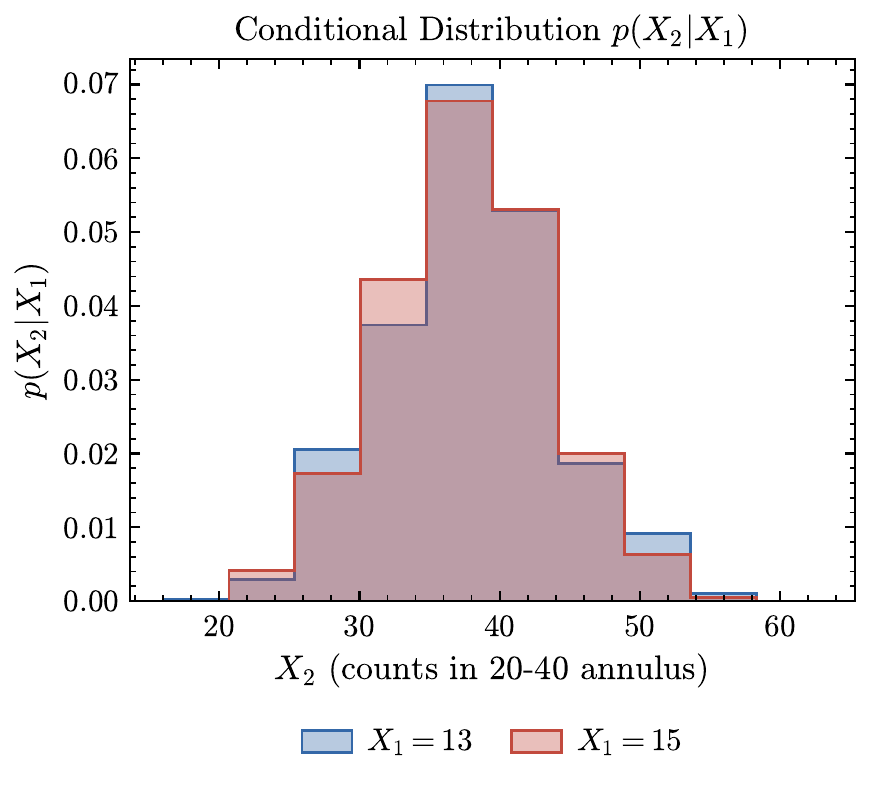}}%
  {\includegraphics[height=\tutorialheight]{tutorials/2a/fig06.pdf}}

The conditional distributions \(p(X_2|X_1=13)\) and \(p(X_2|X_1=15)\) are similar. This is empirical evidence for statistical independence. Despite the different \(X_1\), the distribution of \(X_2\) stays the same. A truly random point process behaves this way, because the number of stars in one annulus does not predict the number in a non-overlapping annulus.

Independence is the basis of our cluster detection strategy. In a random field, a high neighbor count around a star is a statistical fluctuation that does not extend to larger radii. A real cluster instead creates correlated high counts in several annuli around the same star. By measuring the deviations from independence, we can tell random fluctuations from true overdensities.

Independence also makes the analysis simpler. The joint distribution over many annuli is high-dimensional and hard to estimate from finite data. If independence holds, it factors into a product of simple marginal distributions

\[p(X_1, X_2, \dots, X_k) = p(X_1)p(X_2)\cdots p(X_k).\]

This reduces the number of parameters to estimate from exponential in \(k\) to linear in \(k\). The analysis then stays manageable for many annuli. The simplification applies only to our null hypothesis of a random field, and to detect clusters we must find where the independence breaks.

\section*{Detecting Clusters Through Likelihood Analysis}

We have shown that our background field follows Poisson statistics, with independent counts in non-overlapping regions. We can now build a method for cluster detection. A stellar cluster is a region where the local neighbor count \(X\) is far above the Poisson expectation for a random field.

For each star in our field, we count the neighbors within a given radius. We compare the count with our null hypothesis, the expected distribution in a random field. The stars of a cluster have higher neighbor counts, so we can tell them from the background stars.

To test the method, we add simulated clusters to our random field. We model each cluster as a two-dimensional Gaussian distribution of stars. It is simple, and it is a reasonable approximation of many real stellar clusters. The cluster centers are placed uniformly at random in the field, and the position of each member star follows

\[\mathbf{x} \sim \mathcal{N}(\boldsymbol{\mu}_c, \sigma^2\mathrm{Id}_2),\]

where \(\mathbf{x} = (x, y)\) is the position of a member star and \(\boldsymbol{\mu}_c\) is the cluster center. Here \(\mathrm{Id}_2\) is the \(2 \times 2\) identity matrix, and \(\sigma\) sets the spatial extent of the cluster. We choose \(\sigma = 20\) pixels for the first cluster (200 stars) and \(\sigma = 15\) pixels for the second (100 stars). Both are close to the radius of our neighbor count.

\tutorialbarrier
\begin{codecell}

\end{codecell}

We plot the field with a zoom on the first cluster, and we print the cluster parameters.

\tutorialbarrier
\begin{codecell}

\end{codecell}
\tutorialplot{\includegraphics[width=6.173in,height=0.75\textheight,keepaspectratio]{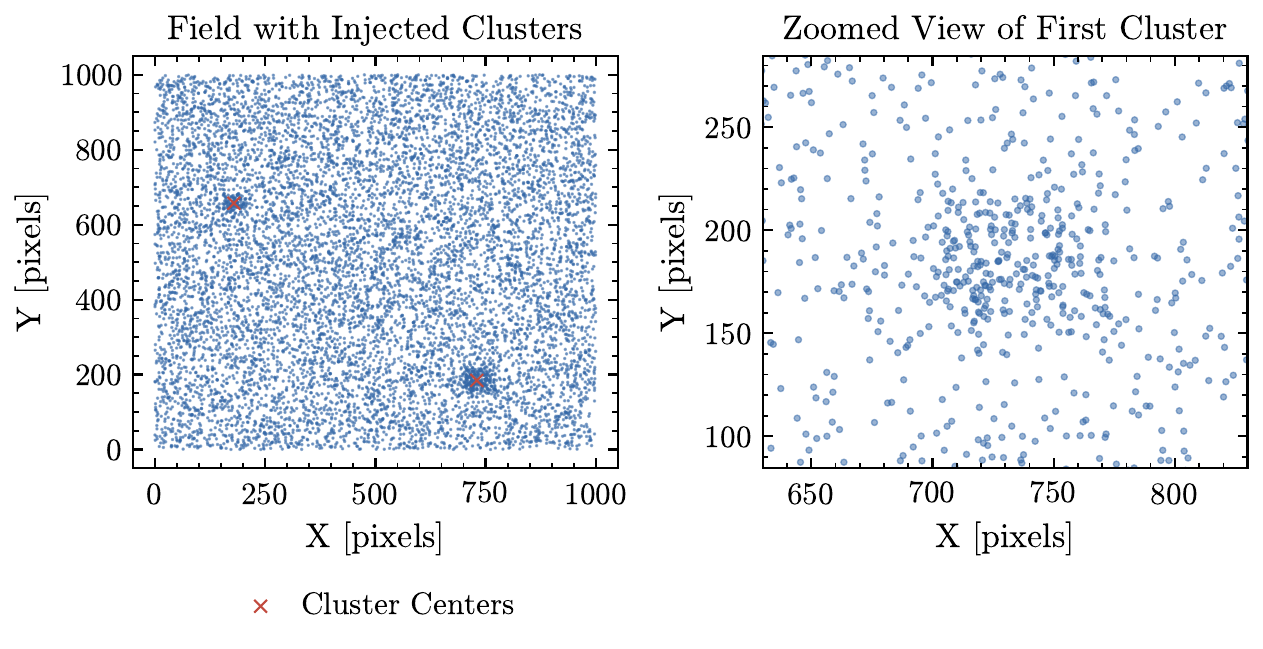}}%
  {\includegraphics[height=\tutorialheight]{tutorials/2a/fig07.pdf}}
\begin{Verbatim}[fontsize=\footnotesize, xleftmargin=5mm]
First cluster = 200 points at (730.0, 184.5)
Second cluster = 100 points at (179.0, 657.2)
\end{Verbatim}

\section*{Likelihood Analysis}

With the clusters in our field, we can develop a statistical method to detect them. For each star, we count the neighbors \(X_1\) within a radius \(r\) of 20 pixels. Under the null hypothesis of a random field, this count follows a Poisson distribution with the expected count \(\lambda = \rho\pi r^2\) of the background stars. The probability of exactly \(k\) neighbors is

\[P(X_1 = k|\lambda) = \frac{\lambda^k e^{-\lambda}}{k!}.\]

This probability is the likelihood \(\mathcal{L}\) (Section 2.7) of observing exactly \(k\) neighbors under our background model. It tells us how probable our observation is under the model. A low likelihood means that the observation is unlikely under the model assumptions.

The stars of a cluster have unusually high neighbor counts, so they have very low likelihoods under the background-only model. This makes sense, because finding many neighbors around a star is very improbable if the stars are distributed at random. Clusters produce neighbor counts that are incompatible with random placement.

The log-likelihood is more convenient

\[\ln \mathcal{L} = k\ln(\lambda) - \lambda - \ln(k!).\]

The logarithm keeps very small probabilities from rounding to zero on the computer, and it turns a product of probabilities into a sum. A more negative log-likelihood means a more improbable observation under the null hypothesis.

By calculating the log-likelihood over the whole field, we create a map that shows regions of unusually high density. Clusters appear as wells in this map, regions where the counts are so high that a random field would very rarely produce them. We now count the neighbors of every star in the new field and compare the background stars with the two clusters.

\tutorialbarrier
\begin{codecell}

\end{codecell}

We plot the distributions of the neighbor count and of its log-likelihood for the background and the two clusters.

\tutorialbarrier
\begin{codecell}

\end{codecell}
\tutorialplot{\includegraphics[width=6.351in,height=0.75\textheight,keepaspectratio]{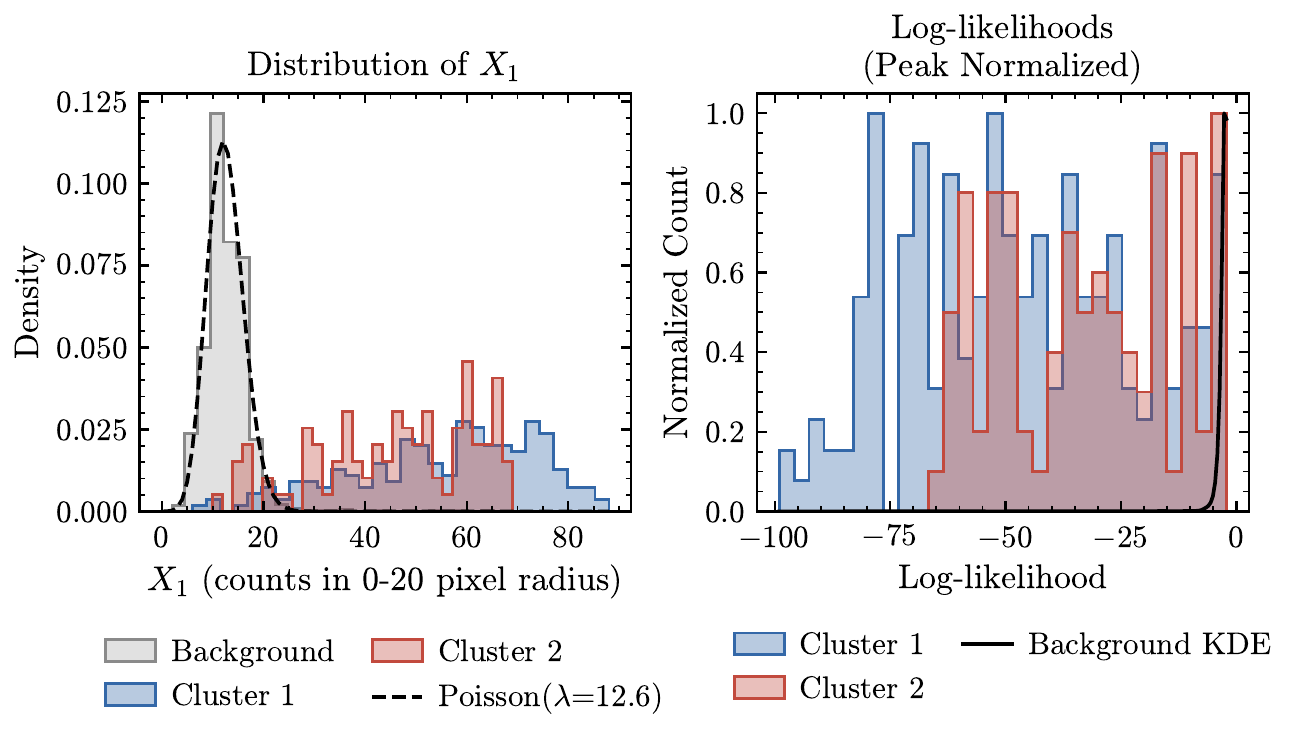}}%
  {\includegraphics[height=\tutorialheight]{tutorials/2a/fig08.pdf}}
\begin{Verbatim}[fontsize=\footnotesize, xleftmargin=5mm]
Mean N1 counts
Background = 12.9
Cluster 1 = 55.6
Cluster 2 = 46.2

Expected from density = 12.6
\end{Verbatim}

The comparison shows large differences between the background and the clusters. In the left panel, the background stars (gray) follow the expected Poisson distribution closely, as our random field model predicts. The stars in the clusters (blue and red) have far higher neighbor counts. Their mean counts are several times the background mean. This follows from the higher local density inside a cluster.

The right panel shows the distributions of the log-likelihood, each scaled to a peak of 1 because the groups differ in size. The clusters are histograms. The black line is a kernel density estimate (KDE) of the background log-likelihood. A histogram of the background would look spiky, because the log-likelihoods come from integer counts. The KDE smooths these values into a continuous density, which shows the overall shape better.

The log-likelihood separates the background from the clusters. Most cluster stars have log-likelihoods below \(-10\), while the background stars lie mostly between \(-2\) and \(-4\). This suggests a detection algorithm that finds the regions where the local density gives an unusually low likelihood under the background model.

To see where the detection works, we create a spatial map of the log-likelihood over the field. We evaluate the likelihood on a grid of positions, which turns the point data into a continuous detection surface. Clusters appear in it as regions of extremely low likelihood.

\tutorialbarrier
\begin{codecell}

\end{codecell}

We plot the map of the log-likelihood.

\tutorialbarrier
\begin{codecell}

\end{codecell}
\tutorialplot{\includegraphics[width=4.406in,height=0.75\textheight,keepaspectratio]{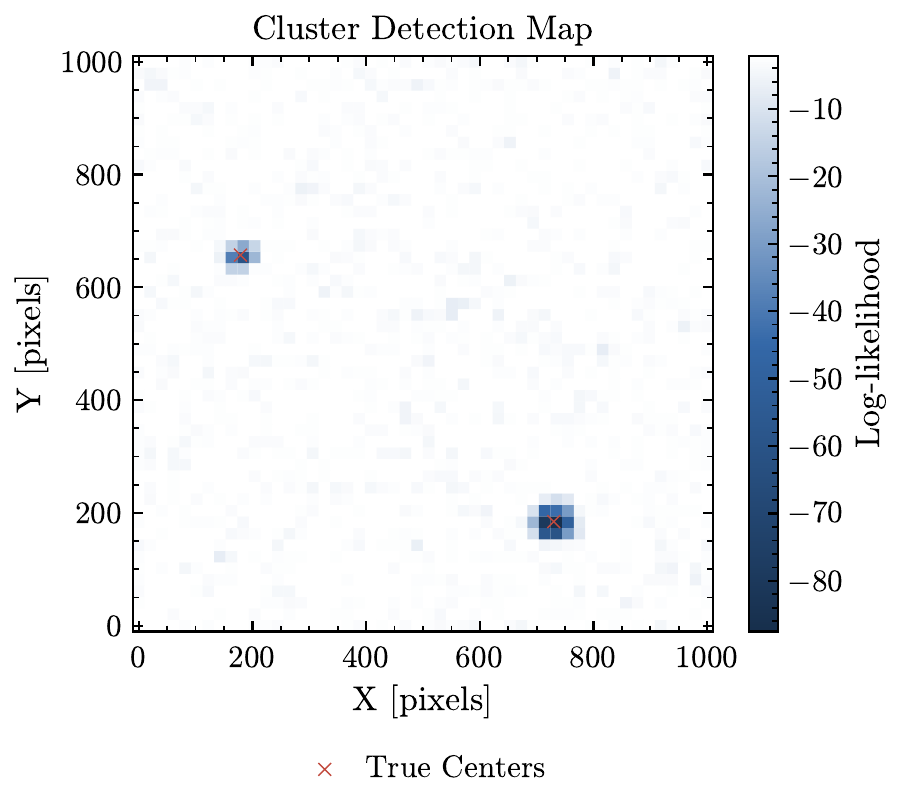}}%
  {\includegraphics[height=\tutorialheight]{tutorials/2a/fig09.pdf}}

The likelihood map shows both injected clusters as regions of extremely low likelihood, the dark regions of the map. The cluster locations stand out as clear minima, with values far below \(-20\) against values between \(-2\) and \(-4\) for the background. This contrast shows that the Poisson likelihood detects overdensities well.

\section*{Considerations on the Method}

Our analysis so far relies on two simplifying choices. Both deserve a closer look.

\textbf{Choice of the Analysis Scale}

We chose a radius of 20 pixels for our neighbor counts, close to the size of our injected clusters. This prior knowledge helps the detection. In practice, we rarely know the cluster sizes in advance, so we need a multi-scale approach. Annuli are more useful than circles here.

\begin{itemize}
\tightlist
\item
  Annuli at different radii are statistically independent for a random field, so we can analyze each scale separately.
\item
  Nested circles are correlated, because a larger circle contains all the stars of a smaller one.
\item
  Real surveys often use a hierarchy of annuli to search for clusters over a range of physical scales.
\end{itemize}

\textbf{Poisson Distribution for the Background}

The Poisson distribution has three advantages for the background counts.

\begin{itemize}
\tightlist
\item
  The single parameter \(\lambda\) has a clear physical meaning as the average count.
\item
  The distribution models discrete counting data.
\item
  For a uniform field, we can estimate \(\lambda\) directly from the observed mean.
\end{itemize}

Real astronomical fields can have more complex statistics. The variance of real star counts can exceed their mean, for example when the background stars are themselves clustered. The negative binomial distribution can model such counts. Whether the simple Poisson model is enough depends on the astronomical context.

\section*{Maximum Likelihood Estimation}

So far we assumed that we know the true background parameter \(\lambda\). In practice, we must estimate it from the data. Maximum likelihood estimation (MLE) is a principled way to find the parameter value that best explains our observations (Section 2.7).

The joint likelihood is the probability of observing all our data points together. For independent observations \(x_1, \dots, x_N\) from a Poisson distribution, it is

\[\mathcal{L}(\lambda|x_1,\dots,x_N) = \prod_{i=1}^N \frac{\lambda^{x_i} e^{-\lambda}}{x_i!}.\]

Each factor is the probability of the count \(x_i\) for the parameter \(\lambda\). The product arises because we require all observations to occur together. The joint probability of independent events is the product of their probabilities.

The logarithm turns the product into a sum, which gives the log-likelihood

\[\ln \mathcal{L}(\lambda|x_1,\dots,x_N) = \sum_{i=1}^N (x_i \ln(\lambda) - \lambda - \ln(x_i!)).\]

The maximum likelihood estimate \(\hat{\lambda}\) is the value that maximizes this function. For the Poisson distribution, we can find it analytically by setting the derivative to zero, which gives \(\hat{\lambda} = \bar{x}\), the sample mean. Here we find it numerically instead, by computing the log-likelihood on a grid of \(\lambda\) values. We apply it to the \(X_1\) counts of the background stars in the field with clusters.

\tutorialbarrier
\begin{codecell}

\end{codecell}
\tutorialplot{\includegraphics[width=4.679in,height=0.75\textheight,keepaspectratio]{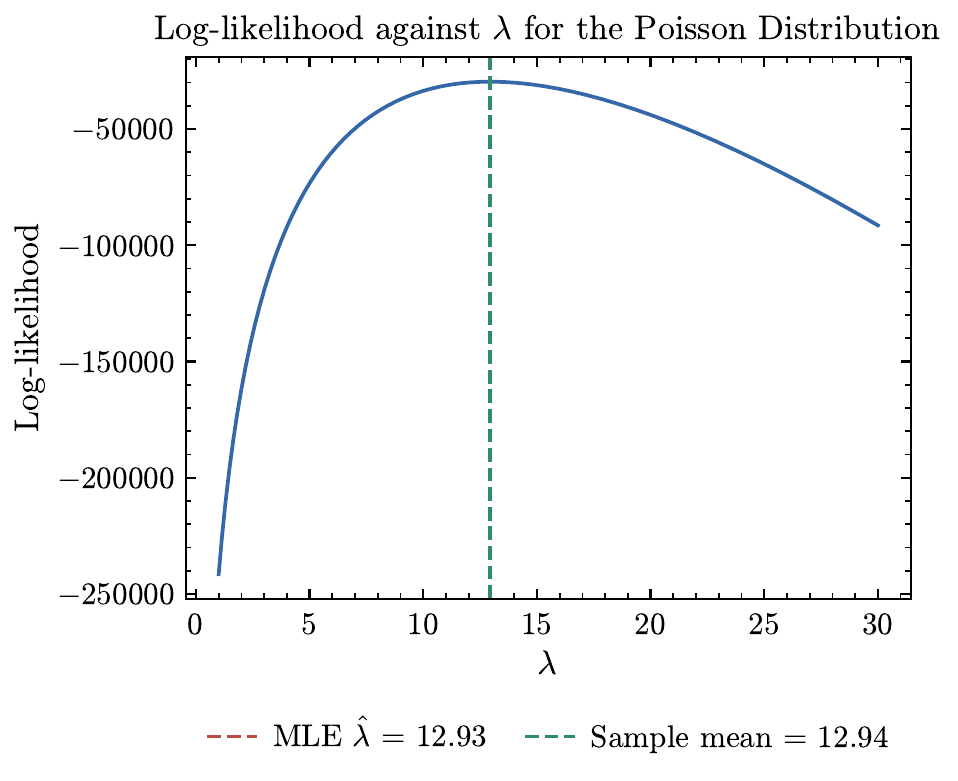}}%
  {\includegraphics[height=\tutorialheight]{tutorials/2a/fig10.pdf}}
\begin{Verbatim}[fontsize=\footnotesize, xleftmargin=5mm]
MLE estimate of lambda = 12.931
Sample mean = 12.938
\end{Verbatim}

The log-likelihood has a well-defined maximum that agrees with the sample mean. This confirms the theoretical result that the MLE of a Poisson distribution is the arithmetic mean of the observations. It also shows that our grid search works.

The grid search is simple, and it illustrates the general principle of maximum likelihood estimation. We evaluate the likelihood on a range of parameter values and select the value that makes our data most probable. For more complex distributions or models with several parameters, this brute-force approach becomes too expensive. This leads to the gradient-based optimization methods of later chapters.

From a Bayesian perspective, our MLE implicitly assumes a uniform prior over \(\lambda\) (Section 2.7). We consider all parameter values equally plausible before we see the data. This uninformative prior is a conservative choice when we have no prior knowledge of the background density. When we have physical constraints or earlier observations, an informative prior can improve the estimate, in particular when the data are few.

\section*{Summary}

In this tutorial we used probability theory to detect structure in a spatial point pattern. Stellar cluster detection illustrates concepts that apply far beyond this one problem.

\begin{enumerate}
\def\labelenumi{\arabic{enumi}.}
\tightlist
\item
  \textbf{Statistical modeling.} We built a null hypothesis from the Poisson distribution, which describes the expected behavior of randomly distributed stars.
\item
  \textbf{Likelihood-based detection.} We compared the observed neighbor counts with the expectation. The log-likelihood of a count is very low for most cluster stars, so it finds the overdensities.
\item
  \textbf{Parameter estimation.} Maximum likelihood estimation gave the expected count from the data. For a Poisson distribution it is the sample mean, and a grid search finds it as well.
\item
  \textbf{Spatial statistics.} The annuli showed how to handle a two-dimensional point pattern. Counts in non-overlapping annuli are independent for a random field, and periodic boundaries remove the edge effects.
\end{enumerate}

The uniform Poisson background is an idealization. Real fields have density variations or clustering in the background, and a grid search becomes too expensive for models with many parameters. The same techniques apply to galaxy clusters in redshift surveys and to excess high-energy particle events. In each case, we measure what we expect under a null hypothesis and then identify the departures that signal real physical phenomena. A detection is therefore a statement about probability, and it depends on the background model.

\tutorialchapter{2b}{Bayesian Inference for Binary Star Eccentricities}
\noindent\textit{Companion material for Chapter 2}

\section*{Introduction}

Binary stars are pairs of stars that are bound by gravity and orbit their common center of mass. They are among the most important objects in astrophysics. Binaries give direct measurements of stellar masses, test theories of stellar evolution and probe the physics of star formation. Yet measuring the orbital properties of a binary is still a hard observational task.

The usual approach needs long monitoring campaigns. Astronomers observe a binary over multiple orbital periods and track the positions and velocities of both stars along their elliptical paths. For orbital periods of years or decades, this takes a large share of telescope time, which is not possible for large populations.

This tutorial studies another approach, inspired by \href{https://ui.adsabs.harvard.edu/abs/2022MNRAS.512.3383H/abstract}{Hwang et al.~(2022)}. We use Bayesian inference to constrain the orbital eccentricity from a single epoch of observation. One observation already carries information about the orbit. At one moment we measure the relative position and velocity of the two stars. The angle \(\gamma\) between these two vectors gives a statistical constraint on the eccentricity \(e\).

We build the Bayesian analysis step by step and learn how the following ideas work.

\begin{itemize}
\tightlist
\item
  The angle \(\gamma\) relates to the eccentricity for a Keplerian orbit.
\item
  A likelihood connects an observable quantity to a physical parameter.
\item
  Marginalization over the unknown orbital phase gives a probabilistic constraint on the eccentricity.
\item
  A measurement error changes the inference.
\end{itemize}

This example shows how Bayesian methods turn limited data into physical constraints, even when a deterministic approach fails.

\section*{Prerequisites and Learning Objectives}

\subsection*{Prerequisites}

This tutorial assumes the following.

\begin{itemize}
\tightlist
\item
  Working knowledge of Python, NumPy, matplotlib
\item
  Probability distributions, Bayes' theorem and the likelihood
\item
  Marginalization and posterior distributions
\item
  Introductory knowledge of stellar properties and orbital mechanics
\end{itemize}

\subsection*{Learning Objectives}

After this tutorial you can do the following.

\begin{itemize}
\tightlist
\item
  Construct a likelihood and a posterior for an astronomical observation
\item
  Build a forward model that connects physical parameters to an observable quantity
\item
  Propagate a measurement error with marginalization
\item
  Compute Bayesian results on a grid for a low-dimensional parameter space
\item
  Connect the mathematics to the orbital mechanics
\end{itemize}

We begin by importing the libraries and setting the plot style.

\tutorialbarrier
\begin{codecell}

\end{codecell}

\section*{Understanding Binary Star Orbits}

Binary stars orbit their common center of mass and follow Kepler's laws, as planets orbit the Sun. The shape and orientation of an orbit carry information about the formation history and the dynamical evolution of the system. To build the Bayesian analysis, we first need to know how orbital parameters give quantities that we can observe.

\subsection*{Orbital Parameters}

Several parameters describe a binary orbit completely, and we use two of them. The orbit is the relative orbit, the path of the secondary star around the primary star. It is an ellipse with the primary at one focus.

\begin{enumerate}
\def\labelenumi{\arabic{enumi}.}
\tightlist
\item
  The semi-major axis \(a\) is the size scale of the orbit. It sets the average separation between the stars.
\item
  The eccentricity \(e\) measures how stretched the orbit is. It runs from 0 (circular) to nearly 1 (highly elongated).
\end{enumerate}

We view the orbit face-on, with the line of sight along the normal of the orbital plane.

The eccentricity is important for understanding how binaries form. Different formation mechanisms predict different eccentricity distributions.

\begin{itemize}
\tightlist
\item
  Capture often produces highly eccentric orbits.
\item
  Disk fragmentation tends to produce nearly circular orbits.
\item
  Dynamical interactions can produce a thermal distribution of eccentricities, with \(p(e) \propto e\), which favors high eccentricities.
\end{itemize}

When we measure the eccentricity distribution of a population of binaries, we can find which mechanisms dominate in different regions. The eccentricity is therefore a useful probe of binary star physics.

We now plot face-on orbits for several eccentricities. The plots mark the periapsis, the point closest to the primary, and the apoapsis, the farthest point.

\subsection*{Effect of Eccentricity on Orbit Shape}

The eccentricity controls the shape of the orbit. A circular orbit (\(e = 0\)) keeps a constant separation between the stars. A highly eccentric orbit (\(e \to 1\)) brings the stars much closer at the periapsis than at the apoapsis. This change in separation affects the velocities and accelerations that we observe along the orbit. We plot orbits with \(e = 0\), 0.5 and 0.8.

\tutorialbarrier
\begin{codecell}

\end{codecell}
\tutorialplot{\includegraphics[width=6.500in,height=0.75\textheight,keepaspectratio]{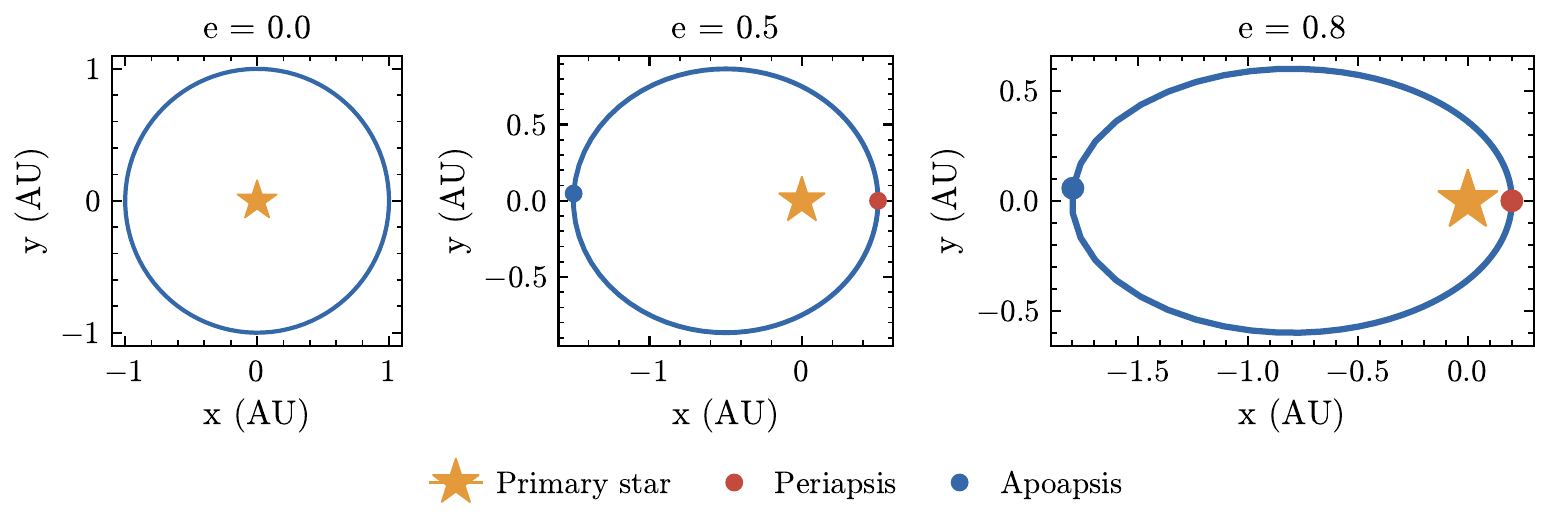}}%
  {\includegraphics[height=\tutorialheight]{tutorials/2b/fig01.pdf}}

\section*{The Physics of Orbital Motion}

To infer the eccentricity, we need to know how the stars move along the orbit. At any point of a binary orbit, the relative position vector \(\mathbf{r}\) and the relative velocity vector \(\mathbf{v}\) form an angle. This angle depends on the eccentricity and on the orbital phase \(\phi\), the angle of \(\mathbf{r}\) measured from the periapsis. The relation lets us infer the eccentricity from a single epoch.

\subsection*{The Velocity-Position Angle}

Consider a point on the elliptical orbit. The position vector points from the primary to the secondary. The velocity vector is the velocity of the secondary relative to the primary, and it points along the direction of motion. The angle \(\gamma\) between the two vectors changes around the orbit.

For a Keplerian orbit viewed face-on, the position and the velocity at the orbital phase \(\phi\) are

\tutorialdisplay{\mathbf{r} = r\,(\cos\phi, \sin\phi), \quad r = \frac{1 - e^2}{1 + e\cos\phi},}{\mathbf{v} = \frac{(-\sin\phi, \; e + \cos\phi)}{\sqrt{1 - e^2}}.}

The velocity is the time derivative of the position along the orbit. The scale of the velocity does not matter for the angle, so we use units with \(a = 1\) and \(G(M_1 + M_2) = 1\).

The change of \(\gamma\) shows the shape of the orbit, so even one observation of \(\gamma\) gives a statistical constraint on the eccentricity. We now compute these vectors and the angle between them. We plot them for a circular orbit and for an orbit with \(e = 0.7\).

For a circular orbit (\(e = 0\)), the velocity is always perpendicular to the position vector, so the angle stays at 90° throughout the orbit. For an eccentric orbit, \(\gamma\) is 90° at the periapsis and the apoapsis. It falls below 90° on one half of the orbit and rises above it on the other half. As the eccentricity increases, the angle varies more, so the geometry at one instant carries information about the shape of the orbit.

\tutorialbarrier
\begin{codecell}

\end{codecell}
\tutorialplot{\includegraphics[width=6.168in,height=0.75\textheight,keepaspectratio]{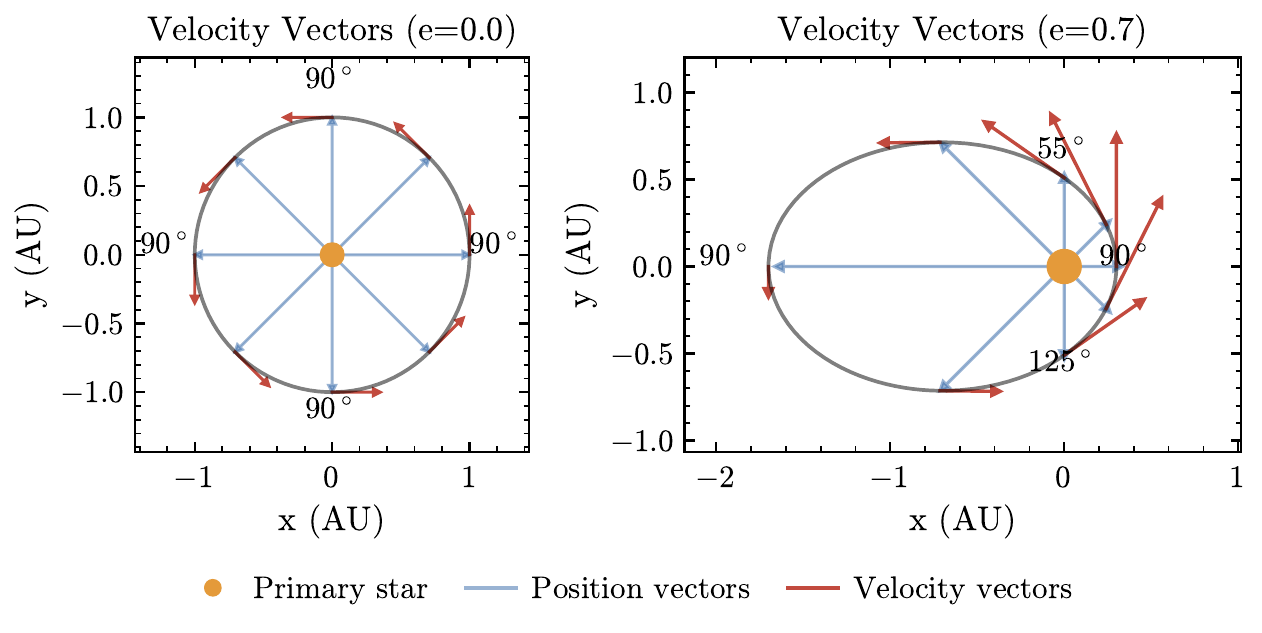}}%
  {\includegraphics[height=\tutorialheight]{tutorials/2b/fig02.pdf}}

\section*{Mathematical Formulation: From Angles to Eccentricity}

The figure shows how \(\gamma\) changes around the orbit. The Bayesian analysis needs \(\gamma\) as a function of the orbital parameters. The dot product of the two vectors above gives it in closed form. Their dot product is \(\mathbf{r}\cdot\mathbf{v} = r e\sin\phi/\sqrt{1 - e^2}\), and their lengths are \(|\mathbf{r}| = r\) and \(|\mathbf{v}| = \sqrt{1 + e^2 + 2e\cos\phi}/\sqrt{1 - e^2}\). The factors of \(r\) and \(\sqrt{1 - e^2}\) cancel, so

\[\cos\gamma = \frac{\mathbf{r}\cdot\mathbf{v}}{|\mathbf{r}||\mathbf{v}|} = \frac{e\sin\phi}{\sqrt{1 + e^2 + 2e\cos\phi}},\]

where

\begin{itemize}
\tightlist
\item
  \(e\) is the eccentricity, with \(0 \le e < 1\),
\item
  \(\phi\) is the orbital phase, with \(0 \le \phi < 2\pi\).
\end{itemize}

This equation gives the full relation between the shape of the orbit and the geometry at one instant. It explains what the figure shows. A circular orbit (\(e = 0\)) has a numerator of zero, so \(\gamma = 90°\) everywhere. At the periapsis (\(\phi = 0\)) and the apoapsis (\(\phi = \pi\)) the sine term vanishes, so \(\gamma = 90°\).

We now implement this relation and explore it.

\tutorialbarrier
\begin{codecell}
\begin{Verbatim}[commandchars=\\\{\},fontsize=\small]
\PYG{k}{def}\PYG{+w}{ }\PYG{n+nf}{cos\PYGZus{}gamma}\PYG{p}{(}\FancyVerbBreakAfterBreak{}\PYG{n}{e}\PYG{p}{,}\FancyVerbBreakAfterBreak{} \PYG{n}{phi}\PYG{p}{)}\PYG{p}{:}
\PYG{+w}{    }\PYG{l+s+sd}{\PYGZdq{}\PYGZdq{}\PYGZdq{}}
\PYG{l+s+sd}{    Cosine of the angle between the position and the velocity vector of a face\PYGZhy{}on orbit.}

\PYG{l+s+sd}{    The inputs are the eccentricity e and the orbital phase phi in radians.}
\PYG{l+s+sd}{    \PYGZdq{}\PYGZdq{}\PYGZdq{}}
    \PYG{n}{numerator} \PYG{o}{=}\FancyVerbBreakAfterBreak{} \PYG{n}{e} \PYG{o}{*} \PYG{n}{np}\PYG{o}{.}\FancyVerbBreakAfterBreak{}\PYG{n}{sin}\PYG{p}{(}\FancyVerbBreakAfterBreak{}\PYG{n}{phi}\PYG{p}{)}
    \PYG{n}{denominator} \PYG{o}{=}\FancyVerbBreakAfterBreak{} \PYG{n}{np}\PYG{o}{.}\FancyVerbBreakAfterBreak{}\PYG{n}{sqrt}\PYG{p}{(}\FancyVerbBreakAfterBreak{}\PYG{l+m+mf}{1.0} \PYG{o}{+}\FancyVerbBreakAfterBreak{} \PYG{n}{e}\PYG{o}{*}\PYG{o}{*}\PYG{l+m+mi}{2} \PYG{o}{+}\FancyVerbBreakAfterBreak{} \PYG{l+m+mf}{2.0}\PYG{o}{*}\PYG{n}{e}\PYG{o}{*}\PYG{n}{np}\PYG{o}{.}\FancyVerbBreakAfterBreak{}\PYG{n}{cos}\PYG{p}{(}\FancyVerbBreakAfterBreak{}\PYG{n}{phi}\PYG{p}{)}\PYG{p}{)}

    \PYG{c+c1}{\PYGZsh{} Avoid a division by zero with a small floor}
    \PYG{n}{cos\PYGZus{}gamma} \PYG{o}{=}\FancyVerbBreakAfterBreak{} \PYG{n}{numerator} \PYG{o}{/} \PYG{n}{np}\PYG{o}{.}\FancyVerbBreakAfterBreak{}\PYG{n}{maximum}\PYG{p}{(}\FancyVerbBreakAfterBreak{}\PYG{n}{denominator}\PYG{p}{,}\FancyVerbBreakAfterBreak{} \PYG{l+m+mf}{1e\PYGZhy{}15}\PYG{p}{)}

    \PYG{c+c1}{\PYGZsh{} Keep the value in the range of a cosine}
    \PYG{k}{return} \PYG{n}{np}\PYG{o}{.}\FancyVerbBreakAfterBreak{}\PYG{n}{clip}\PYG{p}{(}\FancyVerbBreakAfterBreak{}\PYG{n}{cos\PYGZus{}gamma}\PYG{p}{,}\FancyVerbBreakAfterBreak{} \PYG{o}{\PYGZhy{}}\PYG{l+m+mf}{1.0}\PYG{p}{,}\FancyVerbBreakAfterBreak{} \PYG{l+m+mf}{1.0}\PYG{p}{)}
\end{Verbatim}
\end{codecell}

We now plot \(\gamma\) against the orbital phase for five eccentricities.

\tutorialbarrier
\begin{codecell}

\end{codecell}
\tutorialplot{\includegraphics[width=4.399in,height=0.75\textheight,keepaspectratio]{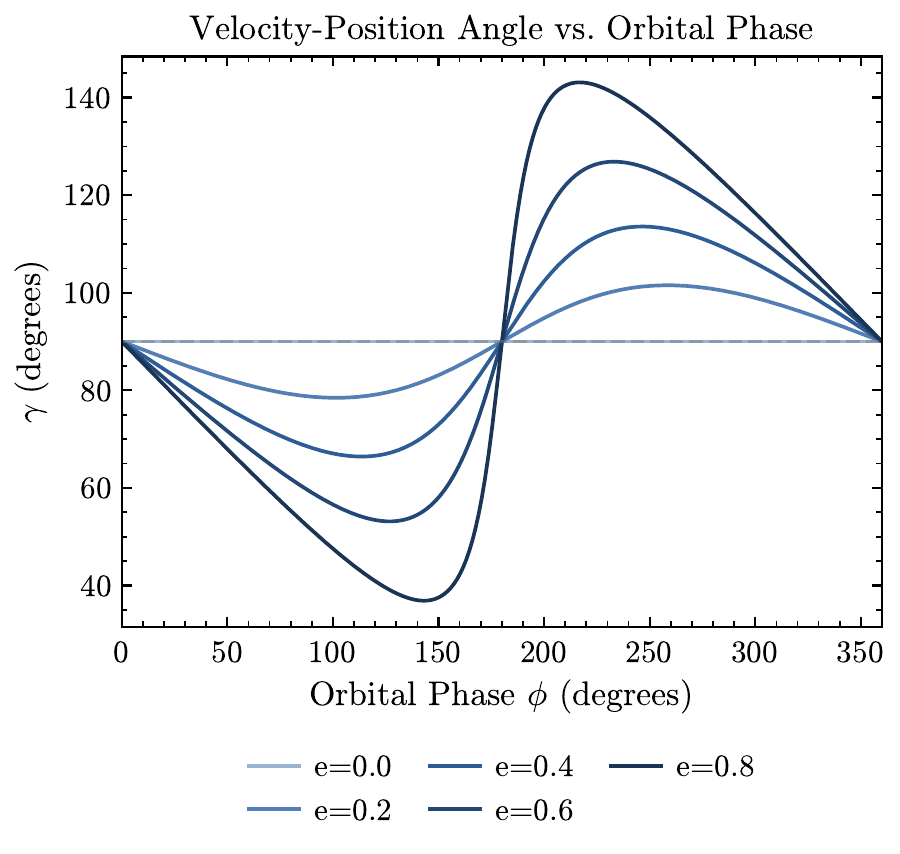}}%
  {\includegraphics[height=\tutorialheight]{tutorials/2b/fig03.pdf}}

The angle varies periodically with the orbital phase, and the eccentricity sets the size of the variation. As the eccentricity increases, we see symmetric oscillations around 90° with a growing amplitude. The extrema occur at intermediate orbital phases and not at the periapsis or the apoapsis. Their location depends on the eccentricity.

For any observed angle \(\gamma \ne 90°\), many combinations of \(e\) and \(\phi\) give the same angle. This degeneracy is part of every single-epoch observation. The curves are also symmetric. Replacing \(\phi\) by \(2\pi - \phi\) turns \(\gamma\) into \(180° - \gamma\), which follows from the symmetry of the orbit.

\section*{Bayesian Inference: From Observations to Eccentricity}

We now have the relation between the angle \(\gamma\) and the eccentricity \(e\). Next we build a Bayesian method that infers \(e\) from observations.

\subsection*{The Forward Problem}

We have already solved the forward problem. Given \(e\) and \(\phi\), the closed form above gives the angle that we expect. In practice, however, we measure \(\gamma\) and want to infer \(e\). This is an inverse problem. Bayesian inference handles the uncertainties that come with it.

\subsection*{Constructing the Likelihood Function}

The first step is to build the likelihood \(p(\gamma|e)\). It gives the probability of observing the angle \(\gamma\) when the true eccentricity is \(e\). The orbital phase \(\phi\) is unknown and random for a single observation, so we marginalize over it (Section 2.5),

\[p(\gamma|e) = \int_0^{2\pi} p(\gamma|e,\phi) p(\phi) \mathrm{d}\phi.\]

We assume a uniform distribution of the phase, \(p(\phi) = 1/(2\pi)\). The integral is then an average over all orbital phases. We evaluate it numerically by drawing random phases. For one eccentricity, the histogram of the angles that these phases give shows \(p(\gamma|e)\) as a function of the angle \(\gamma\).

\tutorialbarrier
\begin{codecell}

\end{codecell}
\tutorialplot{\includegraphics[width=4.393in,height=0.75\textheight,keepaspectratio]{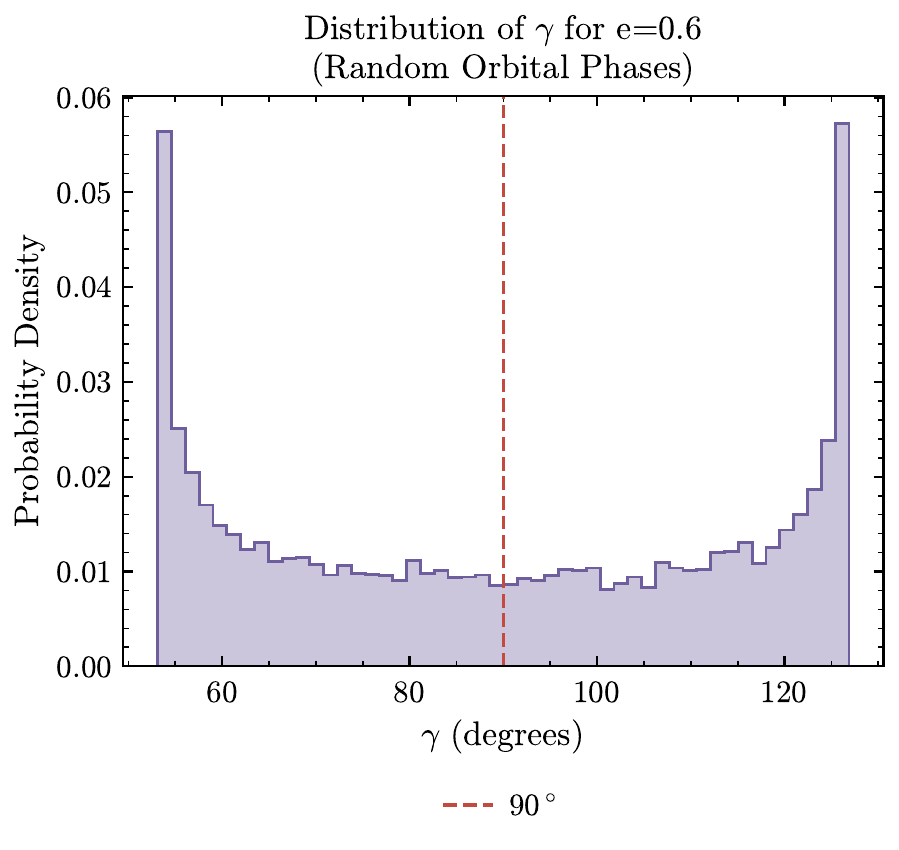}}%
  {\includegraphics[height=\tutorialheight]{tutorials/2b/fig04.pdf}}

The histogram is bimodal, with its peaks away from 90°. The likelihood of an observed angle is the height of the histogram at that angle. We now repeat this for every eccentricity on a grid. For each \(e\) we draw random phases and count the fraction of angles in a bin of 1° around the observed angle.

\tutorialbarrier
\begin{codecell}

\end{codecell}

To see the likelihood for every eccentricity at once, we compute a map of \(p(\gamma|e)\) over the observed angle and the eccentricity.

\tutorialbarrier
\begin{codecell}

\end{codecell}
\tutorialplot{\includegraphics[width=4.416in,height=0.75\textheight,keepaspectratio]{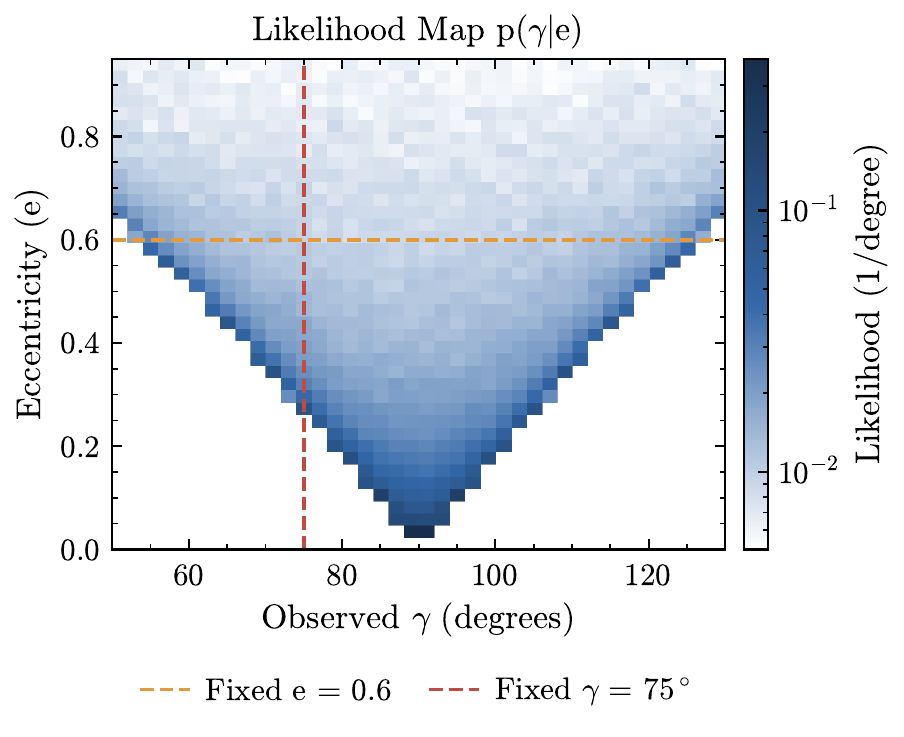}}%
  {\includegraphics[height=\tutorialheight]{tutorials/2b/fig05.pdf}}

The likelihood is nonzero in a V-shaped region. For each eccentricity the angles lie between \(\arccos e\) and \(180° - \arccos e\), so the region widens as \(e\) grows. At \(e = 0\) every angle is 90°. The likelihood is highest along the edges of the V, where the histogram peaks.

The two slices show how we use the map. A horizontal slice at one eccentricity is the histogram above, the probability of each angle for that \(e\). A vertical slice at the observed angle is the likelihood as a function of \(e\). It says how well each eccentricity explains that angle, and it is zero below the lower edge of the V, at \(e = |\cos\gamma|\).

\subsection*{From Likelihood to Posterior: Applying Bayes' Theorem}

With the likelihood in hand, we apply Bayes' theorem to infer the posterior distribution of the eccentricity given the observed angle,

\[p(e|\gamma) = \frac{p(\gamma|e) p(e)}{p(\gamma)},\]

where

\begin{itemize}
\tightlist
\item
  \(p(e|\gamma)\) is the \textbf{posterior}, the probability of the eccentricity given our observation,
\item
  \(p(\gamma|e)\) is the \textbf{likelihood}, which we just defined,
\item
  \(p(e)\) is the \textbf{prior}, our belief about the eccentricity before the observation,
\item
  \(p(\gamma)\) is the \textbf{evidence}, a normalization constant.
\end{itemize}

We use an uninformative uniform prior, \(p(e) = 1\) for \(e \in [0,1]\). It says that all eccentricities are equally likely before we see the data. The evidence \(p(\gamma)\) only makes the posterior integrate to 1. We can therefore ignore it and normalize at the end. With a uniform prior, the posterior is the vertical slice of the likelihood map at the observed angle, normalized to integrate to 1.

\tutorialbarrier
\begin{codecell}

\end{codecell}

We apply the method to four observed angles. For each angle we plot the posterior and mark the maximum a posteriori (MAP) value, which is the eccentricity at the maximum. We also shade the shortest interval that holds 68\% of the probability. It always contains the MAP.

\tutorialbarrier
\begin{codecell}

\end{codecell}
\tutorialplot{\includegraphics[width=6.175in,height=0.75\textheight,keepaspectratio]{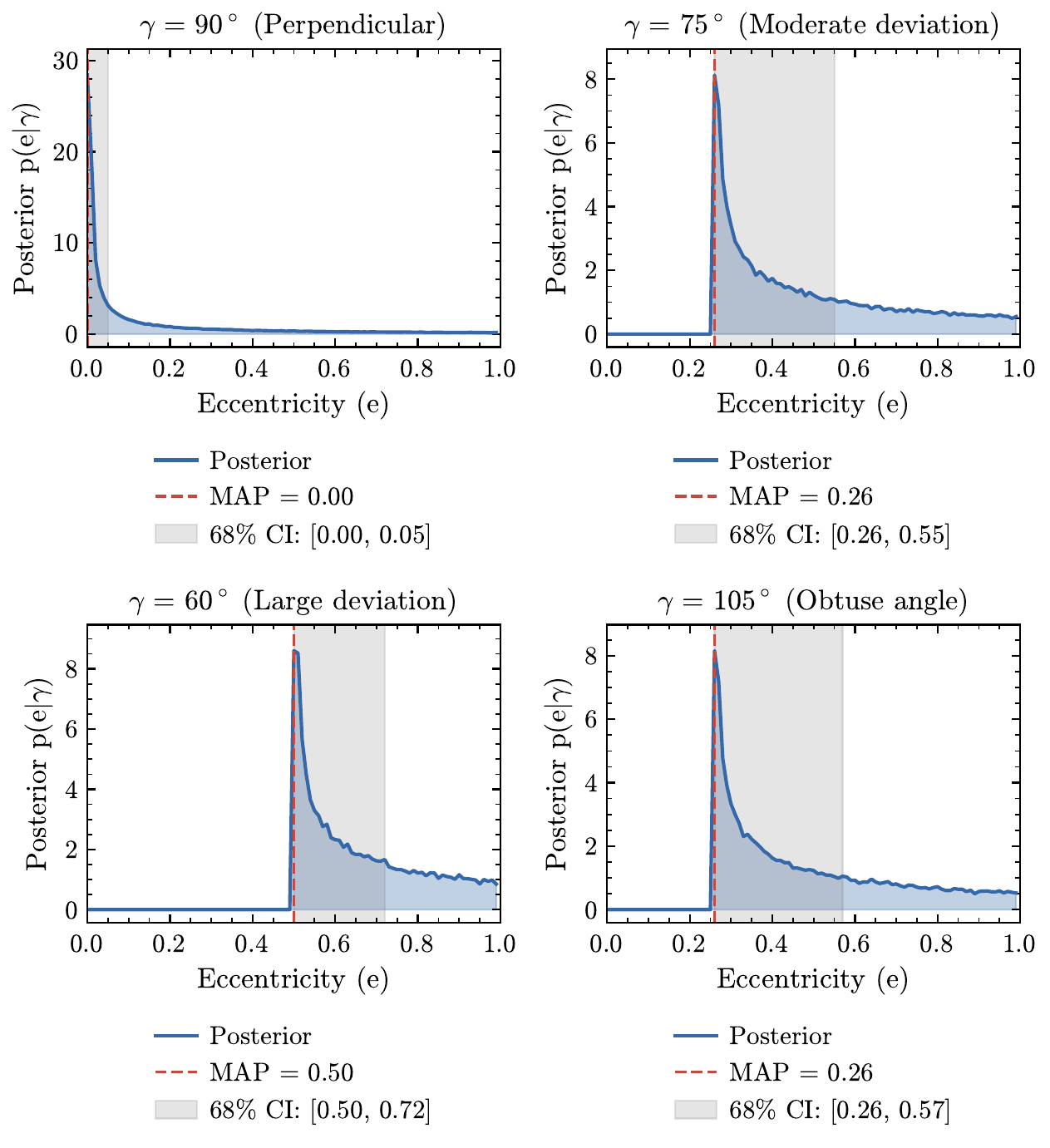}}%
  {\includegraphics[height=\tutorialheight]{tutorials/2b/fig06.pdf}}

The posterior for \(\gamma = 90°\) peaks at \(e = 0\), because a circular orbit always has \(\gamma = 90°\). For \(75°\) and \(60°\) the posterior starts at a minimum eccentricity, because large deviations from 90° need elongated orbits. This minimum grows as the angle moves away from 90°. The posterior also keeps a long tail toward \(e = 1\). One angle therefore sets a lower limit on the eccentricity but does not fix it. The angle \(105°\) gives the same posterior as \(75°\), by symmetry.

Bayesian inference does not always need a complex method. Our inverse problem infers one parameter from one observation, so a grid works well. We evaluate the likelihood on a grid of eccentricities and apply Bayes' theorem directly. This gives the full posterior distribution and the complete uncertainty, and it needs little computing time.

A grid becomes too expensive when the number of parameters grows, because the number of grid points grows quickly. For one or two parameters, a grid remains a good method.

\section*{Including Measurement Uncertainties}

In real observations, we do not measure \(\gamma\) perfectly. The measured angle \(\gamma_{\mathrm{meas}}\) comes with an uncertainty, which we describe with a standard deviation \(\sigma_\gamma\). We must include this uncertainty in the Bayesian analysis.

\subsection*{Marginalizing Over Measurement Uncertainty}

We take a Gaussian measurement error,

\[p(\gamma_{\mathrm{meas}}|\gamma_{\mathrm{true}}) = \frac{1}{\sqrt{2\pi\sigma_\gamma^2}} \exp\left(-\frac{(\gamma_{\mathrm{meas}} - \gamma_{\mathrm{true}})^2}{2\sigma_\gamma^2}\right).\]

The measurement depends on \(e\) only through the true angle, so we marginalize over the unknown true angle (Section 2.5),

\[p(\gamma_{\mathrm{meas}}|e) = \int p(\gamma_{\mathrm{meas}}|\gamma_{\mathrm{true}}) p(\gamma_{\mathrm{true}}|e) \mathrm{d}\gamma_{\mathrm{true}}.\]

This integral is a weighted average of the likelihoods \(p(\gamma_{\mathrm{true}}|e)\) of all possible true angles. The weights come from the measurement uncertainty. We evaluate it as a sum over true angles in steps of 1°, out to four standard deviations on each side.

\tutorialbarrier
\begin{codecell}

\end{codecell}

We first look at how a growing uncertainty widens the posterior for a fixed observation.

\tutorialbarrier
\begin{codecell}

\end{codecell}
\tutorialplot{\includegraphics[width=4.400in,height=0.75\textheight,keepaspectratio]{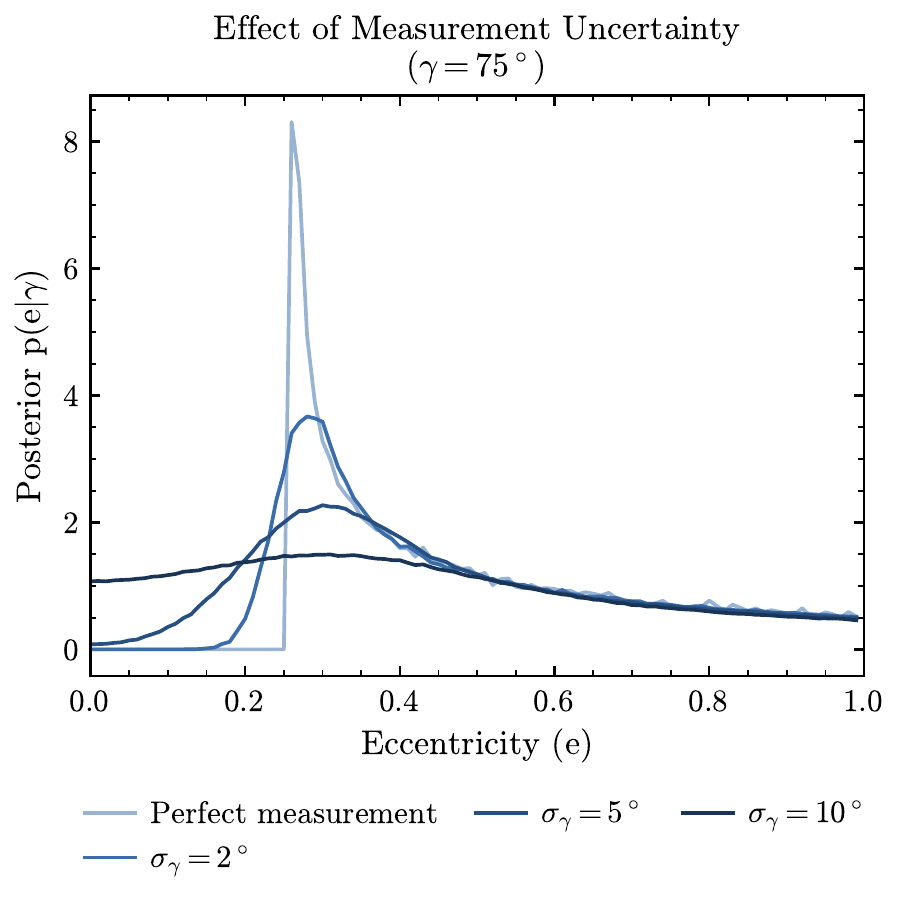}}%
  {\includegraphics[height=\tutorialheight]{tutorials/2b/fig07.pdf}}

A growing uncertainty widens the posterior and makes it harder to find the eccentricity precisely.

Next we compare how the same measurement uncertainty affects different observed angles.

\tutorialbarrier
\begin{codecell}

\end{codecell}
\tutorialplot{\includegraphics[width=4.480in,height=0.75\textheight,keepaspectratio]{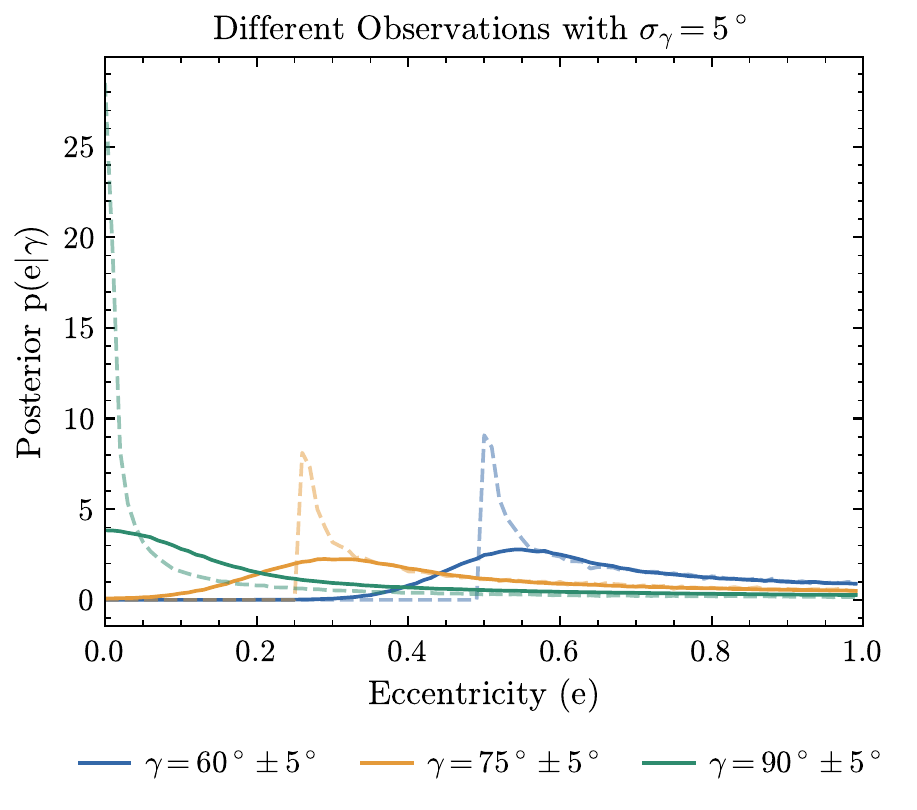}}%
  {\includegraphics[height=\tutorialheight]{tutorials/2b/fig08.pdf}}

To see how the measurement uncertainty enters, we draw the marginalization explicitly. The final posterior is a weighted average over the likelihoods of different true angles \(\gamma_{\mathrm{true}}\). We show five of these likelihoods.

\tutorialbarrier
\begin{codecell}

\end{codecell}
\tutorialplot{\includegraphics[width=4.583in,height=0.75\textheight,keepaspectratio]{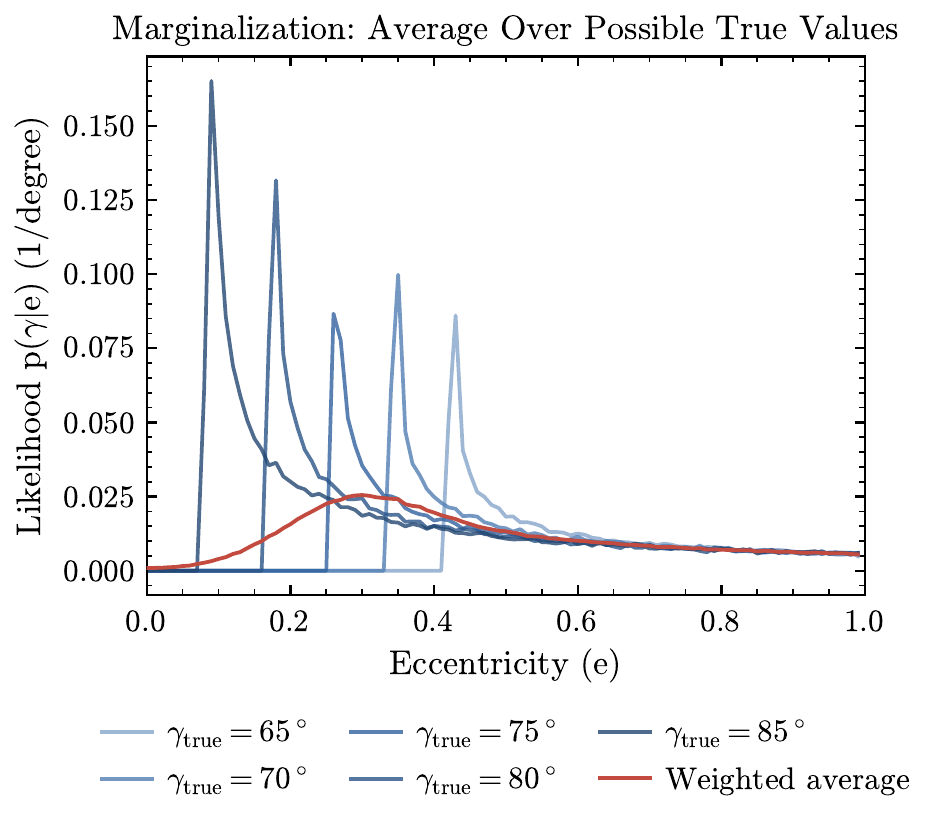}}%
  {\includegraphics[height=\tutorialheight]{tutorials/2b/fig09.pdf}}

A measurement error also widens the posterior and rounds the sharp lower edge, because we average over the possible true angles. Low eccentricities stay unlikely.

\section*{Other Factors in a Real Analysis}

We assumed a face-on orbit. Real orbits have an unknown inclination \(i\), the angle between the normal of the orbital plane and our line of sight. A tilt projects the position and velocity vectors on the sky and changes the observed angle. We can include it with the recipe of this tutorial. We marginalize over a random direction of the line of sight in the frame of the orbit. This is the same step as for the phase and the measurement error.

An unknown inclination widens the posterior further and removes its sharp lower limit on the eccentricity. Projection lets a low-eccentricity orbit show an angle that only a more eccentric face-on orbit shows.

These results show why large surveys with precise measurements matter for population studies. A single observation with a realistic uncertainty gives only a wide range for an individual system. Many such observations together can reveal the eccentricity distribution of the population, which motivated the work of Hwang et al.~(2022).

\section*{Summary}

In this tutorial we used Bayesian inference to infer a physical parameter from limited data. We constrained the eccentricity of a binary orbit from the angle between the position and the velocity at a single epoch. The same ideas apply to many problems of astronomical data analysis.

\begin{enumerate}
\def\labelenumi{\arabic{enumi}.}
\tightlist
\item
  \textbf{Likelihood construction.} We built a forward model that links the eccentricity to the observable angle, and we inverted it with Bayes' theorem.
\item
  \textbf{Grid-based inference.} For this low-dimensional problem, we evaluated the posterior directly on a grid. This gave the complete posterior distribution.
\item
  \textbf{Marginalization.} We included two sources of uncertainty by integrating over unknown quantities. These are the orbital phase and the measurement error.
\item
  \textbf{Information content.} A single observation gives a broad posterior because of degeneracies. This is why population studies need large surveys.
\end{enumerate}

The same recipe applies to other problems, for example measuring a dark matter halo from stellar kinematics or inferring exoplanet properties from transit photometry. We build a likelihood from the physics, apply Bayes' theorem, and marginalize over the uncertainties that we do not need. This gives constraints on the parameters of interest.

\tutorialchapter{3}{Two-Point Statistics for Baryon Acoustic Oscillations}
\noindent\textit{Companion material for Chapter 3}

\section*{Introduction}

Baryon acoustic oscillations (BAO) are one of the main tools of modern cosmology to study the expansion history of the universe. In this tutorial we use the statistical ideas of Chapter 3, which are moments, correlations and the bootstrap, to detect this faint cosmological signal.

In the early universe, about 380,000 years after the Big Bang, matter and radiation were tightly coupled in a hot, dense plasma. Density fluctuations in this plasma made spherical pressure waves, like sound waves in air. Then the universe cooled enough for electrons and protons to combine into neutral hydrogen, in an event that we call recombination. The waves froze in place and left a characteristic scale of about 150 million parsecs in the matter distribution. Cosmologists write this scale as about 100 Mpc/h, where h is the Hubble constant in units of 100 km/s/Mpc, which is about 0.7. This unit keeps distances independent of the exact value of h.

Today this frozen acoustic scale appears as a slight statistical excess of galaxy pairs at this separation. It is a standard ruler in the cosmic web. The pattern is not obvious to detect, because the signal is faint. This is like a faint ring in a noisy image, where the ring is barely brighter than the random fluctuations.

The detection connects directly to the statistical concepts that we studied. We must analyze correlations over very large cosmic volumes to get enough statistical power. We treat galaxy positions as realizations of an underlying random process. Physics predicts an excess of galaxy pairs at the BAO separation, but the pairs themselves are random. Here we need random variables, correlations and statistical uncertainties.

\section*{Prerequisites and Learning Objectives}

\subsection*{Prerequisites}

This tutorial uses the statistical ideas of Chapter 3 and assumes the following.

\begin{itemize}
\tightlist
\item
  Probability theory, with random variables, probability distributions and their properties
\item
  Statistical moments, with expectation, variance and higher moments of a distribution
\item
  Correlation and covariance, as measures of the relation between random variables
\item
  The bootstrap, as a resampling method to estimate uncertainties
\item
  Python, with NumPy arrays, matplotlib plots and SciPy functions
\end{itemize}

\subsection*{Learning Objectives}

After this tutorial you can do the following.

\begin{itemize}
\tightlist
\item
  Treat pixel intensities as random variables and analyze their collective properties
\item
  Implement two-point statistics to find spatial structure in noisy data
\item
  Analyze a discrete random field and its statistical properties
\item
  Use the bootstrap to estimate confidence intervals for a cosmological measurement
\item
  Connect abstract statistical concepts to the task of finding a faint physical effect
\end{itemize}

\section*{A Simplified Model: Discrete Pixel Grid}

To make the BAO detection problem easier, we begin with a simplified discrete model. Each pixel in our grid represents a small patch of sky. Its intensity \(X_{ij}\) is the total brightness of all galaxies in that patch, so brighter pixels generally mean more galaxies in that region. This links the statistics to physical observations.

This is a large simplification of reality. The real cosmic web has complex structure from galaxy clustering on many scales, with clusters, filaments and voids. Real observations cover different patches of sky, each with its own noise and systematic effects, from atmospheric distortions to instrumental artifacts. Even so, the simplified model keeps the main statistical problem of BAO detection.

\section*{Generating a Mock BAO Field}

Our first task is to create a simulated dataset with the main features of BAO detection. We start with a 32 × 32 pixel grid with a ring at the center, where each pixel intensity is a random variable. This simple geometry lets us build intuition before we go to more realistic cases.

Our simulated field has three components.

\begin{enumerate}
\def\labelenumi{\arabic{enumi}.}
\tightlist
\item
  \textbf{Baseline.} A constant background intensity, which stands for the normal galaxy distribution.
\item
  \textbf{BAO ring.} A circular feature at the radius \(R\), which stands for the characteristic BAO scale.
\item
  \textbf{Gaussian noise.} Random fluctuations that simulate observational uncertainties and cosmic variance.
\end{enumerate}

In the real universe, BAO does not add new galaxies. It only changes how the existing galaxies are distributed in space. We therefore conserve the mean and the variance of the field. We lower the mean and the variance of the baseline by the mean and the variance of the ring. The sum then keeps the mean and the variance of a field of pure noise.

We begin by importing the libraries and setting the plot style.

\tutorialbarrier
\begin{codecell}

\end{codecell}

We now write a function that generates the field. It adds the ring to a Gaussian baseline whose mean and variance are lowered by those of the ring.

\tutorialbarrier
\begin{codecell}

\end{codecell}

We now generate one field with the ring at the center and display it.

\tutorialbarrier
\begin{codecell}

\end{codecell}
\tutorialplot{\includegraphics[width=5.665in,height=0.75\textheight,keepaspectratio]{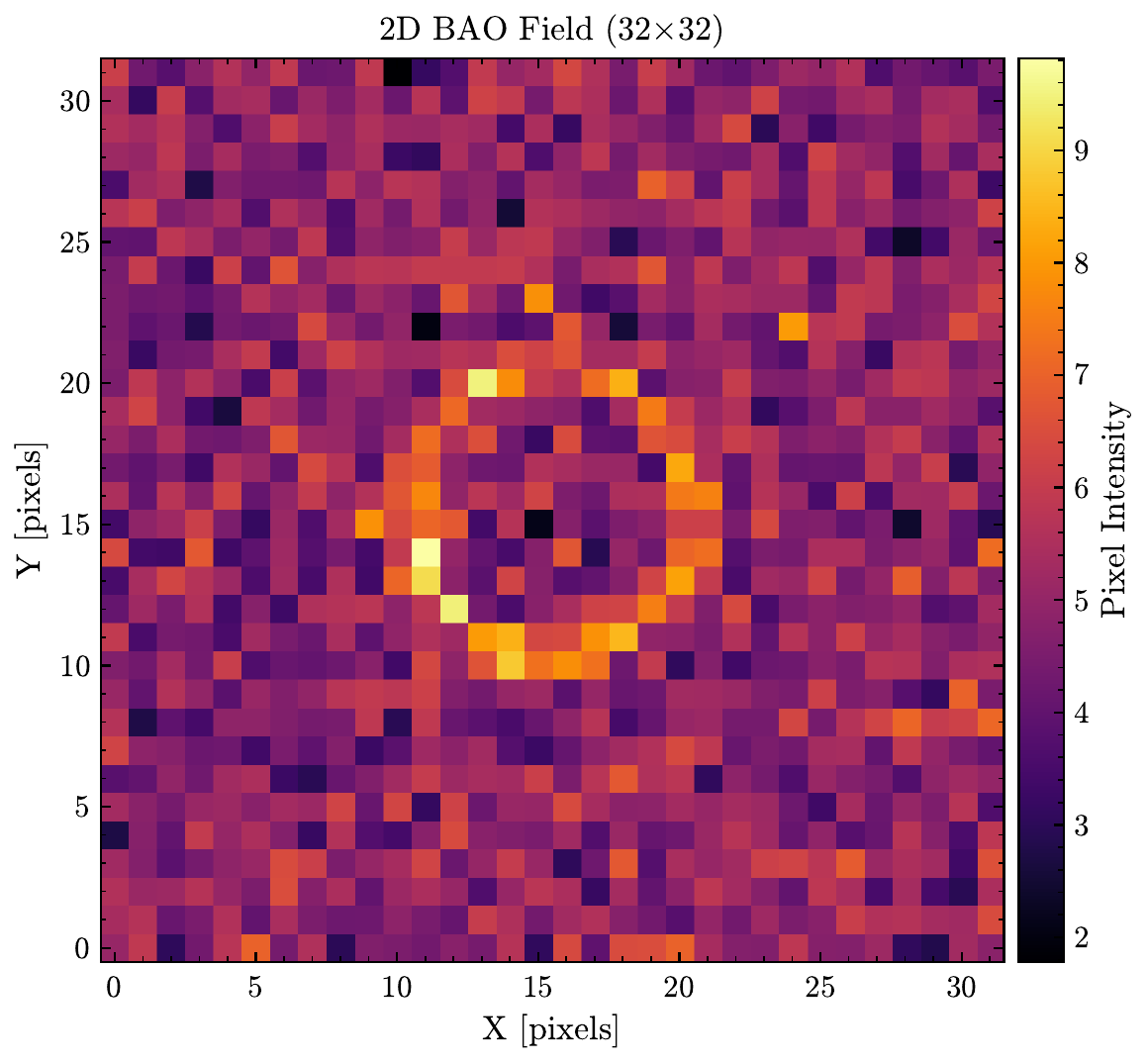}}%
  {\includegraphics[height=\tutorialheight]{tutorials/3/fig01.pdf}}

The image shows a 32 × 32 pixel grid. Each pixel intensity is a random variable \(X_{ij}\) with a contribution from each of the three components.

On top of the background we place the BAO signal, the feature that we want to detect. It is a faint ring-like enhancement of the pixel intensities. The ring is centered in the field with a radius of \(R = 5.0\) pixels. Its strength is the amplitude \(A = 3.0\), and the width of its profile is \(\sigma_{\mathrm{ring}} = 0.5\) pixel. We chose these parameters so that the ring is present but not obvious to detect. In real observations the signal is weaker still.

Each pixel intensity is the sum of a random baseline \(b_{ij}\) and the ring \(s_{ij}\),

\tutorialdisplay{X_{ij} = b_{ij} + s_{ij},}{s_{ij} = A \exp\left[-\frac{(r_{ij} - R)^2}{2\sigma_{\mathrm{ring}}^2}\right],}

where \(r_{ij}\) is the distance of the pixel from the center. The baseline is Gaussian noise. A field of pure noise would have the mean \(\mu_0 = 5.0\) and the standard deviation \(\sigma_0 = 1.0\).

We give the baseline the mean \(\mu_0 - \bar{s}\) and the variance \(\sigma_0^2 - \mathrm{Var}[s]\). Here \(\bar{s}\) and \(\mathrm{Var}[s]\) are the mean and the variance of the ring over the pixels. The whole field then keeps \(\mu_0\) and \(\sigma_0^2\). The ring moves brightness and adds none. In the image, brighter regions are pixels with higher intensity.

\section*{One-Point Statistics}

In our generated field, each pixel intensity \(X_{ij}\) is a random variable. We can analyze the pixels one by one, but we now want their collective properties. We can treat many random variables as one collection when they are independent and identically distributed (i.i.d.). In our field, the noise has both properties.

\begin{itemize}
\tightlist
\item
  \textbf{Independent.} The noise in each pixel is generated independently.
\item
  \textbf{Identically distributed.} The noise in each pixel follows the same Gaussian distribution with a standard deviation of 1.0.
\end{itemize}

As in Chapter 3, we describe the distribution of a random variable \(X\) with its moments. The first moment, the mean, gives the central tendency

\[\mu = \mathbb{E}[X] = \int x p(x) \mathrm{d}x.\]

The second moment about the mean, the variance, measures the spread

\[\sigma^2 = \mathrm{Var}[X] = \mathbb{E}[(X - \mu)^2] = \int (x - \mu)^2 p(x) \mathrm{d}x.\]

The standard deviation \(\sigma = \sqrt{\mathrm{Var}[X]}\) gives the spread in the same units as \(X\).

For i.i.d. observations, we estimate these moments with sample statistics. The grid has \(N \times N\) pixels, so \(N_\mathrm{pix} = N^2\). We treat the \(N_\mathrm{pix}\) pixels as separate measurements of the same underlying process. The sample mean estimates the true mean \(\mu\), and it is

\[\hat{\mu} = \frac{1}{N_\mathrm{pix}}\sum_{i,j} X_{ij}.\]

The sample variance estimates the true variance \(\sigma^2\), and it is

\[\widehat{\mathrm{Var}}[X] = \frac{1}{N_\mathrm{pix}}\sum_{i,j}(X_{ij} - \hat{\mu})^2.\]

These estimators describe the overall properties of our field. We compute them next.

\tutorialbarrier
\begin{codecell}

\end{codecell}
\begin{Verbatim}[fontsize=\footnotesize, xleftmargin=5mm]
Global pixel statistics
Mean = 4.984
Variance = 1.096
Std Dev = 1.047
\end{Verbatim}

The measured mean \(\hat{\mu}\) and variance are close to the input values \(\mu_0 = 5.0\) and \(\sigma_0^2 = 1.0\). This is expected, because the ring conserves them. The values \(\hat{\mu}\) and \(\widehat{\mathrm{Var}}\) are sample estimates. They approximate the true parameters \(\mu\) and \(\sigma^2\) of the random process. The hat reminds us that they come from a finite sample and are not the true population values. We now ask how reliable these summary statistics are.

To estimate the uncertainty of the sample statistics, we use the bootstrap (Section 3.7). We resample our data with replacement many times and compute \(\hat{\mu}\) and \(\widehat{\mathrm{Var}}\) for each resample. The 16th and 84th percentiles of these values give a \(1\sigma\) confidence interval. This estimates the standard errors without a strong assumption about the probability distribution.

This is useful in our case, because the pixel intensities do not come from a simple Gaussian background. They come from a more complex mixture of the background noise and the ring. The bootstrap remains valid for such non-Gaussian distributions. It treats the pixels as independent samples from this mixture, which is only approximate, because the ring pixels lie at fixed places.

\tutorialbarrier
\begin{codecell}

\end{codecell}

We plot the distribution of the means and of the variances over the resamples, and we print the intervals and the fractional errors.

\tutorialbarrier
\begin{codecell}

\end{codecell}
\tutorialplot{\includegraphics[width=6.173in,height=0.75\textheight,keepaspectratio]{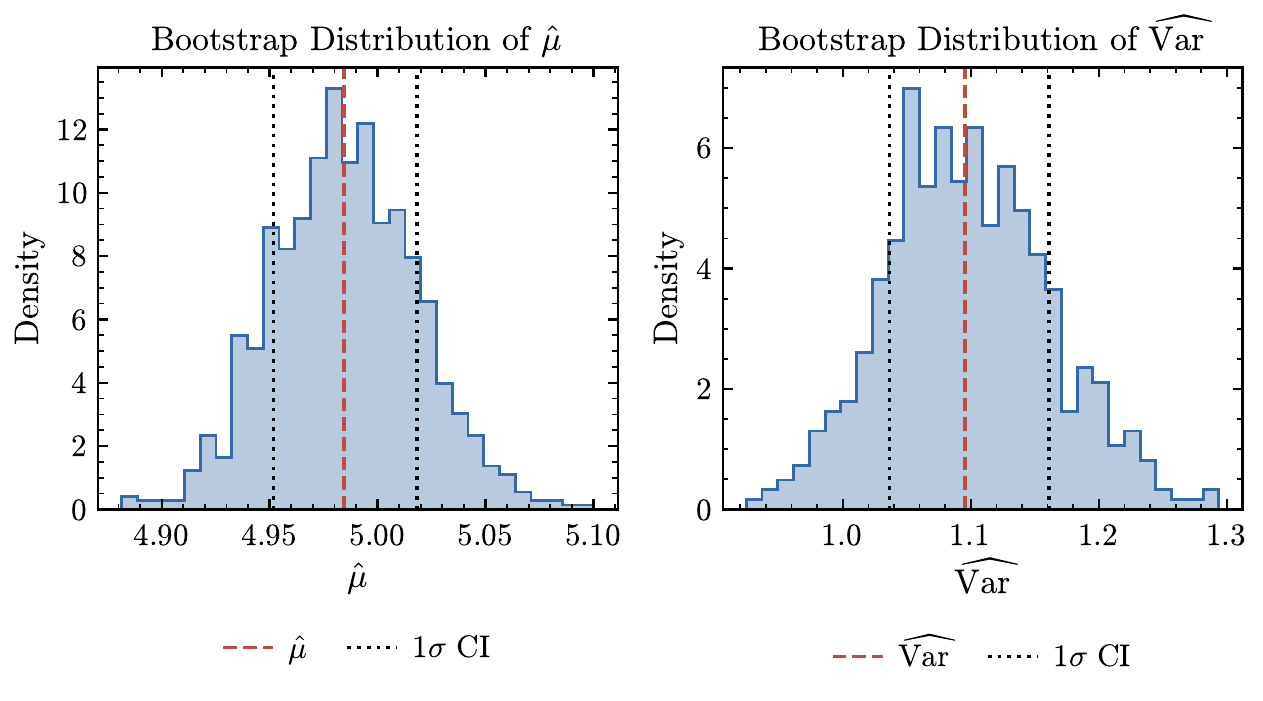}}%
  {\includegraphics[height=\tutorialheight]{tutorials/3/fig02.pdf}}
\begin{Verbatim}[fontsize=\footnotesize, xleftmargin=5mm]
Bootstrap results (with 1 sigma confidence intervals)
Mean = 4.984 [4.952, 5.018]
Variance = 1.096 [1.036, 1.161]

Fractional errors
Mean fractional error = 0.667%
Variance fractional error = 5.684%
\end{Verbatim}

The bootstrap shows the sampling uncertainty of our statistics. The histograms show the distributions of the means and the variances that we can expect if we repeat our measurement many times. The vertical dotted lines mark the confidence intervals, which give a range where we expect the true population parameters to lie. The background mean 5.0 lies inside its interval. The background variance 1.0 lies just below its interval, by less than two standard errors. This single field therefore shows no change of the mean and the variance beyond the sampling uncertainty.

The fractional error of the mean is smaller than the fractional error of the variance. This is a general feature of statistical estimation. Higher-order statistics such as the variance typically have larger uncertainties than lower-order ones such as the mean. This relates to Chapter 3, where higher moments are more sensitive to sampling noise (Section 3.6).

\section*{Sample Size and Finite Sampling Noise}

We used the bootstrap to estimate the uncertainty of our measurements. We now turn to a main question of observational cosmology, which is how the sample size changes the precision of our statistics.

The real universe is more complex than our simplified model. Cosmologists do not observe a single, perfectly centered BAO ring. They must detect a pattern of many rings. To study this problem, we create a more realistic simulated field with several rings at random positions.

\tutorialbarrier
\begin{codecell}

\end{codecell}
\tutorialplot{\includegraphics[width=6.500in,height=0.75\textheight,keepaspectratio]{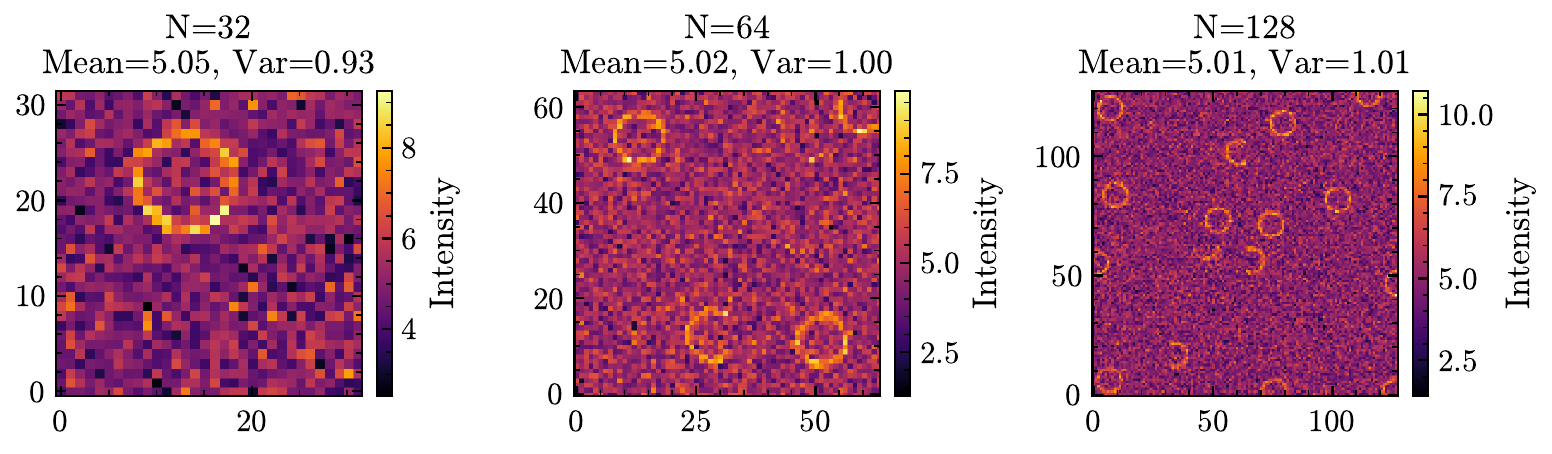}}%
  {\includegraphics[height=\tutorialheight]{tutorials/3/fig03.pdf}}

The fields above are an analogy to real cosmological observations. Each pixel represents a fixed physical area of space, and the BAO ring has a fixed physical size. To simulate a larger survey, we stitch together several 32 × 32 pixel patches, each with its own ring at a random position. When we increase the field from \(N = 32\) to \(N = 128\), we add more patches and so more rings. A real survey does the same when it combines observations of different patches of sky.

For each size we generate 5000 fields and compute the mean and the variance of each. The standard deviation of these 5000 values is the standard error of the statistic. We use many fields, so that the standard errors are precise.

\tutorialbarrier
\begin{codecell}

\end{codecell}
\begin{Verbatim}[fontsize=\footnotesize, xleftmargin=5mm]
Processing 32x32 fields...
Processing 64x64 fields...
Processing 128x128 fields...
\end{Verbatim}

We plot the standard error of the mean and of the variance against the number of pixels, and we print the numbers.

\tutorialbarrier
\begin{codecell}

\end{codecell}
\tutorialplot{\includegraphics[width=6.474in,height=0.75\textheight,keepaspectratio]{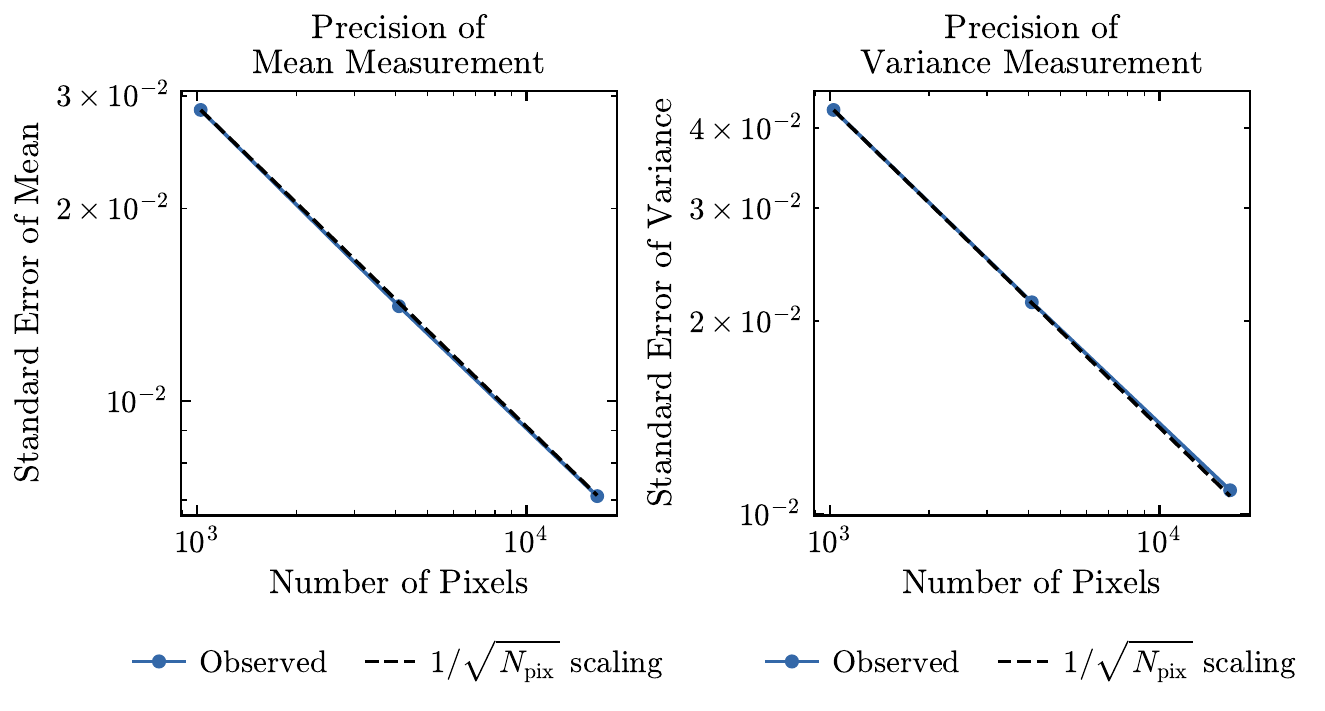}}%
  {\includegraphics[height=\tutorialheight]{tutorials/3/fig04.pdf}}
\begin{Verbatim}[fontsize=\footnotesize, xleftmargin=5mm]

Numerical results

Size Mean±Std.Err. Variance±Std.Err.
------------------------------------
32x32 5.001±0.028 1.000±0.043
64x64 5.000±0.014 0.999±0.021
128x128 5.000±0.007 1.000±0.011
\end{Verbatim}

The plots show how the precision of the mean and the variance increases with the sample size. The standard error of the mean and the standard error of the variance both follow approximately a \(1/\sqrt{N_\mathrm{pix}}\) scaling, as the dashed reference lines show. This agrees with the statistics of i.i.d. random variables. Their uncertainty falls with the square root of the sample size (Section 3.6). The variance of the sample mean scales as \(1/N_\mathrm{pix}\), where \(N_\mathrm{pix}\) is the number of samples, which are pixels here.

The mean and the variance stay at their background values at every survey size, within the standard errors of the table. The ring does not shift them, because the local fluctuations balance out globally. A larger survey gives more precise values, but it does not make the ring visible.

This failure of one-point statistics, which treat each pixel as an independent sample, gives an important lesson about BAO. BAO creates spatially correlated structures, namely rings of a characteristic size. To detect such structures, we need statistical tools that measure how the values at different positions relate to each other.

This leads us to two-point statistics, such as correlation functions and covariance. They measure the relation between pairs of pixels at different separations. This is closer to the techniques of actual BAO analysis in modern cosmology. Before we turn to the correlation function, we build intuition with a simpler case.

\section*{Two-Point Statistics}

We return to a single centered ring, like the field at the beginning.

In a single image, the ring is not always obvious because of the noise, but structure appears when we look at many images together. We generate many images in which the strength of the ring changes from image to image, with a random amplitude between 1 and 7. This variation makes the pixels along the ring brighten and dim together, because a strong ring brightens all its pixels at once. When one part of the ring brightens, other parts follow. This extends to pixels on opposite sides of the ring, at a distance of its diameter, which also brighten and dim together.

Pixels far from the ring show no particular relation to each other and vary almost independently. This systematic relation between pairs of pixels leads us to correlation analysis. Instead of single pixel values, we study how pairs of pixels behave together, to reveal the ring.

\section*{Introducing the Correlation Matrix}

For two pixels in an image, we can ask whether the second pixel tends to be brighter when the first one is brighter than average. A correlation measures exactly this. For the pixels \((i,j)\) and \((k,\ell)\) it is

\[\rho_{ij,k\ell} = \frac{\mathrm{Cov}(X_{ij}, X_{k\ell})}{\sqrt{\mathrm{Var}(X_{ij})\mathrm{Var}(X_{k\ell})}},\]

where

\begin{itemize}
\tightlist
\item
  \(X_{ij}\) is the value at the pixel \((i,j)\)
\item
  \(\mathrm{Cov}(X_{ij}, X_{k\ell})\) measures how much the two pixels vary together
\item
  \(\mathrm{Var}(X_{ij})\) and \(\mathrm{Var}(X_{k\ell})\) measure how much each pixel varies alone
\end{itemize}

A correlation of the following values has these meanings.

\begin{itemize}
\tightlist
\item
  +1 means that the pixels always brighten and dim together
\item
  0 means that the pixels have no linear relation
\item
  -1 means that one pixel dims when the other brightens
\end{itemize}

To estimate these correlations from data, we need many realizations (images) of our field. In cosmology, this is like surveying different patches of sky to study the galaxy distribution and to detect the BAO signal. Each patch is one realization.

\tutorialbarrier
\begin{codecell}

\end{codecell}

We plot the stacked field and the correlation matrix, and we print statistics of the correlations.

\tutorialbarrier
\begin{codecell}

\end{codecell}
\tutorialplot{\includegraphics[width=5.588in,height=0.75\textheight,keepaspectratio]{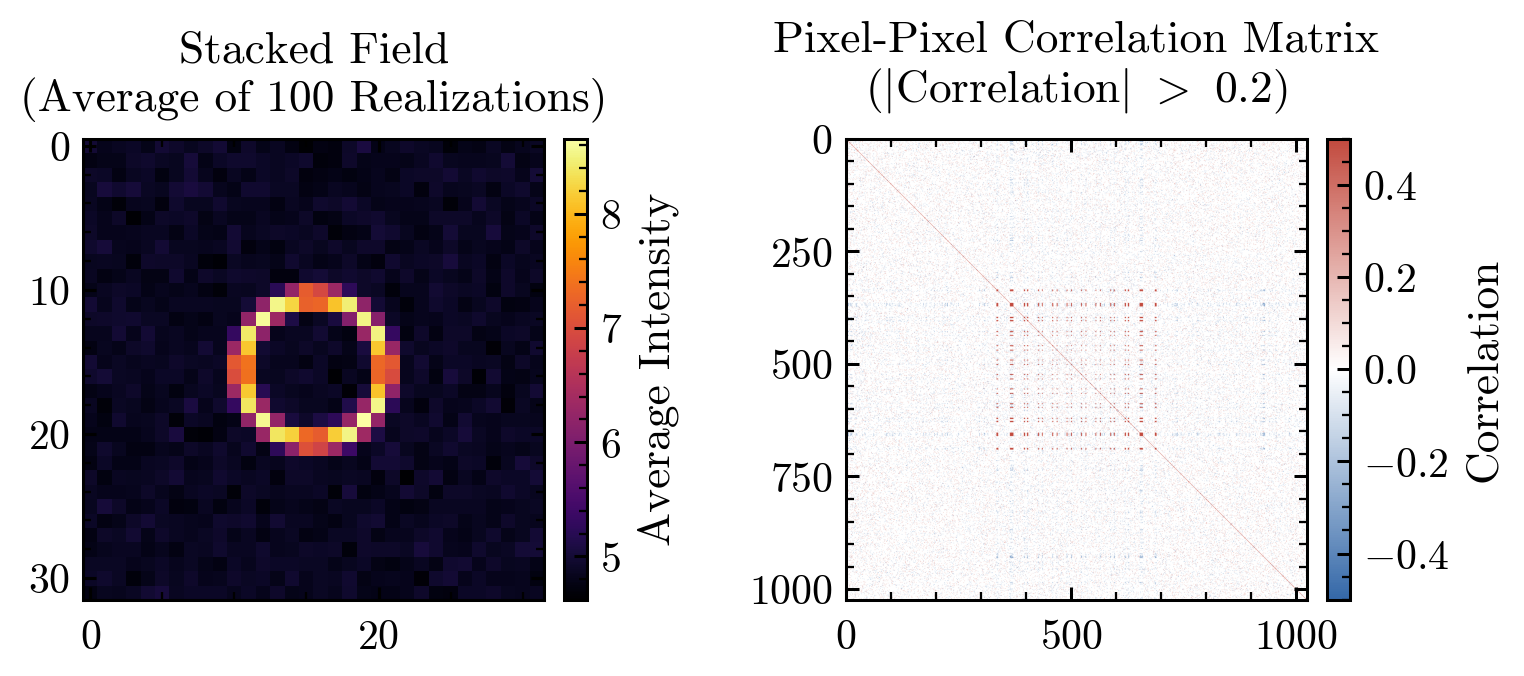}}%
  {\includegraphics[height=\tutorialheight]{tutorials/3/fig05.png}}
\begin{Verbatim}[fontsize=\footnotesize, xleftmargin=5mm]

Correlation statistics
Mean correlation = 0.001
Std deviation = 0.119
Fraction of strong correlations (|corr| > 0.2) = 8.2%
\end{Verbatim}

We generate many versions of our field. Then we analyze how different parts of the image relate to each other over the observations.

The first panel shows the stacked field, the average of all realizations

\[\bar{X}_{ij} = \frac{1}{N_\mathrm{real}} \sum_{n=1}^{N_\mathrm{real}} X^{(n)}_{ij},\]

where \(X^{(n)}_{ij}\) is the value at the pixel \((i,j)\) in the realization \(n\). Stacking helps because the random noise in each pixel follows \(\mathcal{N}(0, \sigma^2)\) and is independent from one realization to the next. When we average \(N_\mathrm{real}\) independent noise realizations, the variance falls to \(\mathrm{Var}[\bar{X}_{ij}] = \sigma^2/N_\mathrm{real}\). The noise level in the stacked image therefore falls by a factor of \(\sqrt{N_\mathrm{real}} = 10\). The ring stays at the same position and keeps its strength, so it stands out in the stacked image.

For the correlation analysis, we first compute the covariance between the pixels \((i,j)\) and \((k,\ell)\)

\[\mathrm{Cov}_{ij,k\ell} = \frac{1}{N_\mathrm{real}} \sum_{n=1}^{N_\mathrm{real}} (X^{(n)}_{ij} - \bar{X}_{ij})(X^{(n)}_{k\ell} - \bar{X}_{k\ell}).\]

Then we compute the variance of each pixel

\[\mathrm{Var}_{ij} = \frac{1}{N_\mathrm{real}} \sum_{n=1}^{N_\mathrm{real}} (X^{(n)}_{ij} - \bar{X}_{ij})^2.\]

The correlation coefficient is then

\[\rho_{ij,k\ell} = \frac{\mathrm{Cov}_{ij,k\ell}}{\sqrt{\mathrm{Var}_{ij}\mathrm{Var}_{k\ell}}}.\]

The second panel shows this correlation matrix, with the pixels in row-major order. The pixel \((i,j)\) then has the index \(i \times N + j\). The diagonal line is the self-correlation of each pixel, which is always 1.0.

We see a grid-like structure in the correlation matrix. It appears because pixels of the ring in the same row of the image are at fixed distances in the flattened index space. This view also hides part of the ring structure. Adjacent pixels of the image in different rows can be far apart in the matrix.

We can organize the analysis differently, with correlations between pixels that depend on their physical separation and not on their array indices. This leads us to radial correlations, which we explore next.

\section*{Radial Correlation Analysis}

Our previous correlation matrix shows the ring, but it does not suit a circular structure. The BAO signal creates a ring, so it makes sense to analyze the data by the radial distance from the center.

Instead of single pixels, we group the pixels by their distance from the center. For a pixel at the position \((i,j)\), the radial distance is

\[r_{ij} = \sqrt{(i - i_c)^2 + (j - j_c)^2},\]

where \((i_c, j_c)\) is the center of our image.

We divide the radial distances into consecutive bins of equal width, starting from the center. Bin 0 holds the pixels closest to the center, bin 1 holds the pixels at the next distances, and so on. For each radial bin \(k\), we compute the average intensity

\[P_k = \frac{1}{N_k} \sum_{r_{ij} \in \mathrm{bin}_k} X_{ij},\]

where \(N_k\) is the number of pixels in the bin \(k\). \(P_k\) is an average of random variables, the pixel intensities, so it is itself a random variable with its own probability distribution.

Radial binning respects the circular symmetry of our signal and reduces the noise through the averaging of pixels. This makes features at the characteristic BAO scale easier to see. It also simplifies the correlation analysis. The \(1024 \times 1024\) matrix of pixel correlations becomes a \(20 \times 20\) matrix of correlations between radial bins.

\tutorialbarrier
\begin{codecell}

\end{codecell}

We plot the stacked field, an example radial profile and the radial correlation matrix, and we print statistics of the radial correlations.

\tutorialbarrier
\begin{codecell}

\end{codecell}
\tutorialplot{\includegraphics[width=6.500in,height=0.75\textheight,keepaspectratio]{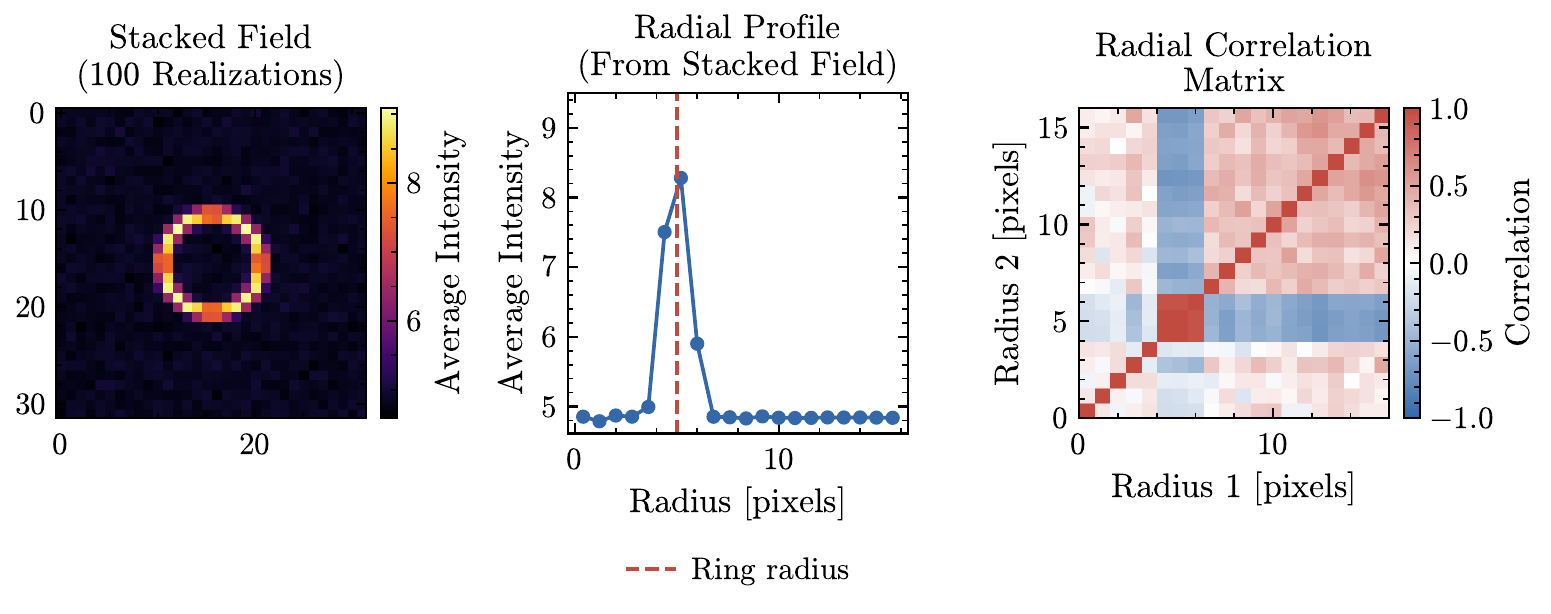}}%
  {\includegraphics[height=\tutorialheight]{tutorials/3/fig06.pdf}}
\begin{Verbatim}[fontsize=\footnotesize, xleftmargin=5mm]

Radial correlation statistics
Mean correlation = 0.062
Std deviation = 0.378
Min correlation = -0.697
Max correlation = 0.986
\end{Verbatim}

The stacked field (left panel) is the average of all realizations.

The radial profile (middle panel) shows the average intensity against the radius. It has a peak at \(r \approx 5\) pixels, which is the BAO ring. A red dashed line marks the ring radius.

The radial correlation matrix (right panel) shows the correlations between different radii. The bins near \(r_1 = r_2 \approx 5\) pixels form a strong block, so the pixels at the ring radius have a high correlation with each other. The printed statistics omit the diagonal elements of the correlation matrix, because they are always 1 by definition.

This radial analysis makes the BAO signal much easier to see. It respects the circular symmetry, reduces the noise through radial averaging and shows the correlations in terms of physical distances and not pixel indices.

\section*{BAO in Cosmological Surveys and Two-Point Correlation Analysis}

Our simplified model has helped us to establish basic statistical ideas, but real cosmological BAO analysis is more complex. In reality, BAO shells can appear at any random location, and many shells overlap, unlike the single centered rings that we used so far. The correlation function that we measure averages over all these possible centers. It gives a statistical view of the overall structure.

To go from our toy model toward reality, we now simulate multiple rings with random centers. The tool for such structure is the two-point correlation function. In cosmology, it is the excess probability, compared with a random distribution, of finding pairs of galaxies at a given separation. For a field \(f(\mathbf{x})\) with the mean \(\bar{f}\), the two-point correlation function is

\[\xi(\mathbf{r}) = \mathbb{E}_{\mathbf{x}}[\delta(\mathbf{x})\delta(\mathbf{x}+\mathbf{r})],\]

where \(\delta(\mathbf{x}) = (f(\mathbf{x}) - \bar{f})/\bar{f}\) is the density contrast, the deviation from the mean in units of the mean. Here \(\mathbf{r}\) is the separation vector and \(\mathbb{E}_{\mathbf{x}}\) is the average over the position \(\mathbf{x}\). The position \(\mathbf{x}\) is any location in the field. By averaging over all positions, we do not assume a special center.

In BAO analysis, we are interested in how \(\xi\) depends on the separation distance \(\Delta r = |\mathbf{r}|\). We first convert our intensity field to the density contrast, \(\delta_{ij} = (X_{ij} - \bar{X})/\bar{X}\). We then find all pairs of pixels in each bin of the separation and compute the average of their products. The function below lists each pair of pixels once with \texttt{np.\allowbreak{}triu\_\allowbreak{}indices}, which gives the entries above the diagonal of the matrix of pairs.

\tutorialbarrier
\begin{codecell}

\end{codecell}

We now apply this to many maps and estimate the error with the bootstrap (Section 3.7).

\tutorialbarrier
\begin{codecell}

\end{codecell}

We now generate maps with rings and, as a control, maps of pure noise with the same mean and variance. We compute the correlation function of both.

\tutorialbarrier
\begin{codecell}

\end{codecell}
\begin{Verbatim}[fontsize=\footnotesize, xleftmargin=5mm]
Computing correlations for all fields...
Bootstrapping...
Computing correlations for all fields...
Bootstrapping...
\end{Verbatim}

We plot the correlation function of both, with a line at the diameter of the ring. We also find the steepest fall of the ring curve near that line.

\tutorialbarrier
\begin{codecell}

\end{codecell}
\tutorialplot{\includegraphics[width=4.918in,height=0.75\textheight,keepaspectratio]{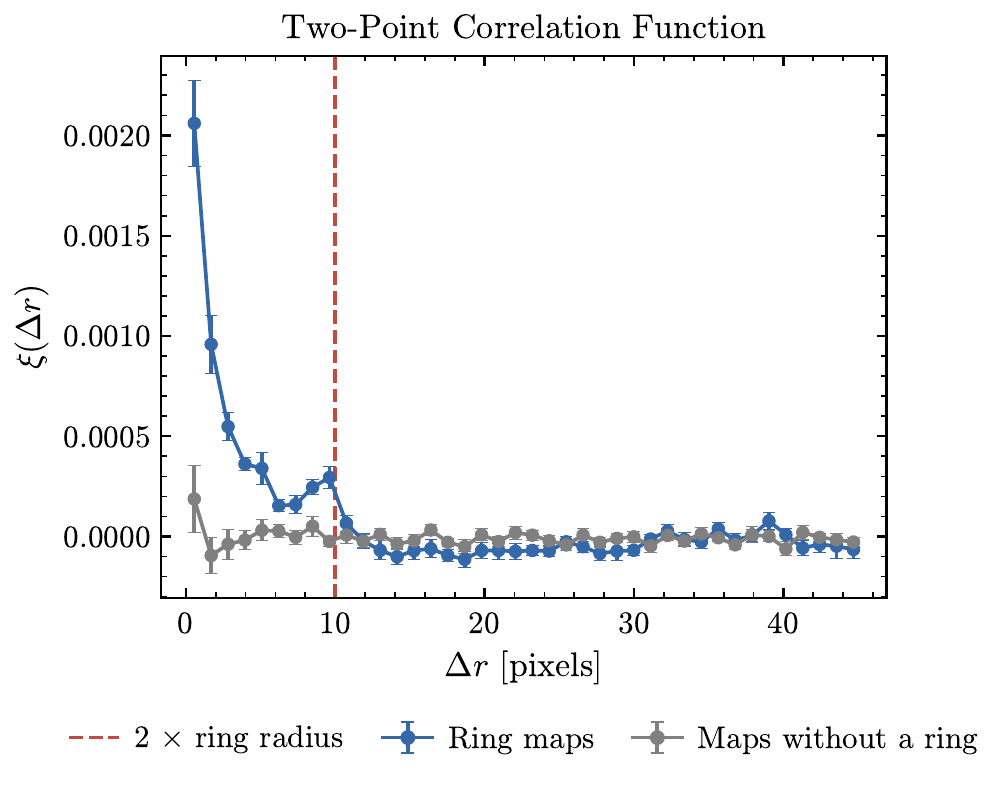}}%
  {\includegraphics[height=\tutorialheight]{tutorials/3/fig07.pdf}}
\begin{Verbatim}[fontsize=\footnotesize, xleftmargin=5mm]

Correlation function statistics
Steepest fall: from r = 9.6 to 10.7 pixels
Expected edge (2×R): 10.0 pixels
\end{Verbatim}

In the plot of the correlation function, we see several features at different scales. At small separations (\(\Delta r < 3\)), the correlation is strong and positive, because nearby pixels of the ring pattern are necessarily similar in value.

The BAO signal appears at intermediate scales (\(3 < \Delta r < 12\)). Two points on one ring are at most a diameter \(2R = 10\) pixels apart, and most pairs of points are close to this distance. We therefore expect a feature at \(\Delta r \approx 2R\). The correlation function stays positive up to \(2R\) and falls steeply just beyond it. The printed result shows that the steepest fall lies at \(2R\). Maps without a ring (gray) show no such feature, and at large separations (\(\Delta r > 15\)) both curves are close to zero.

The error bars come from the bootstrap over the maps. Each of the \(N_\mathrm{bootstrap} = 100\) resamples draws \(N_\mathrm{real}\) maps with replacement and computes the correlation function. The error bars show the standard deviation of the resampled values at each separation \(\Delta r\). This does not assume a probability distribution.

In our example, we chose a ring amplitude comparable to the background fluctuations for clarity. In real cosmological surveys, the BAO signal is much weaker relative to the background noise. A reliable detection needs either very large survey volumes (many pixels) or deeper observations (less noise). This is why modern BAO studies use massive galaxy surveys that cover a large part of the sky.

\section*{Summary}

In this tutorial we explored the main ideas behind BAO detection with the statistics of spatial fields. We started from random variables and statistical moments, and we built up to correlation analysis that detects a faint cosmological signal.

\begin{enumerate}
\def\labelenumi{\arabic{enumi}.}
\tightlist
\item
  \textbf{One-point statistics.} We treated pixel intensities as random variables and described the field with means, variances and their uncertainties. These statistics alone cannot detect a spatially correlated pattern such as BAO, when the ring conserves the mean and the variance.
\item
  \textbf{Bootstrap methods.} We used resampling to estimate the uncertainty of our measurements. It gives error estimates without a specific assumption about the probability distribution.
\item
  \textbf{Two-point statistics.} We developed correlation matrices, radial profiles and the two-point correlation function. They measure spatial relations and show structure that one-point statistics cannot see.
\item
  \textbf{Scale-dependent analysis.} The detectability of a signal depends on the sample size, with uncertainties that scale as \(1/\sqrt{N_\mathrm{pix}}\).
\end{enumerate}

These techniques are the basis of modern cosmological analysis of the large-scale structure of the universe. The real BAO signal is much weaker than our ring, so surveys need very large volumes. The same ideas apply beyond BAO to any problem that needs the detection of characteristic scales in noisy spatial data.

\tutorialchapter{4a}{Linear Regression for Stellar Spectra}
\noindent\textit{Companion material for Chapter 4}

\section*{Introduction}

In Chapter 3 we described distributions with summary statistics and studied the uncertainty of our measurements. We now turn to linear regression, one of the basic techniques that predict a quantity from data. Linear regression is simple, and it helps us to understand more complex machine learning methods. Many sophisticated algorithms build on it.

This tutorial applies linear regression to an astronomical problem, the prediction of the effective temperature of a star from its spectrum. We use data from the Apache Point Observatory Galactic Evolution Experiment (APOGEE), part of the Sloan Digital Sky Survey. APOGEE has obtained high-resolution infrared spectra for hundreds of thousands of stars. The spectra give the chemical compositions, temperatures and kinematics of the stars, and they have improved our understanding of the Milky Way.

Deriving physical parameters from stellar spectra is a central problem of modern astronomy. A spectrum carries a lot of information about the atmosphere of a star. The temperature sets the overall energy distribution and the ionization states. A higher surface gravity gives a higher gas pressure in the atmosphere, which widens the spectral lines. The chemical abundances appear as absorption features at specific wavelengths. Linear regression can turn thousands of flux measurements into physical parameters.

The application also shows concepts of machine learning. We see how maximum likelihood estimation gives an analytical solution for linear regression. We study regularization, which limits overfitting, and we test whether feature engineering helps the predictions. These concepts apply to any problem that predicts a continuous quantity from high-dimensional data.

Linear regression is also easy to interpret. Many modern machine learning methods act as black boxes, but a linear model gives a transparent relation between the inputs and the output. The weights can hint at which spectral regions carry information about the temperature.

\section*{Prerequisites and Learning Objectives}

\subsection*{Prerequisites}

This tutorial assumes the following.

\begin{itemize}
\tightlist
\item
  Linear algebra, with matrix operations, transposes and inverses, which the solution of linear regression uses
\item
  Probability theory, in particular Gaussian distributions and the likelihood
\item
  Maximum likelihood estimation, the principle of finding the parameters that make the observed data most probable
\item
  Python programming, with NumPy arrays and matplotlib plots
\end{itemize}

\subsection*{Learning Objectives}

After this tutorial you can do the following.

\begin{itemize}
\tightlist
\item
  Explain how stellar spectra encode physical information and how a linear regression model can extract it
\item
  Implement the maximum likelihood solution of linear regression for high-dimensional data
\item
  Apply regularization to reduce overfitting and improve generalization
\item
  Use basis functions, here quadratic features, to model nonlinear relations with a model that is linear in its weights
\item
  Evaluate a model with training, validation and test sets
\end{itemize}

We begin by importing the libraries and setting the plot style.

\tutorialbarrier
\begin{codecell}

\end{codecell}

\section*{The APOGEE Dataset}

APOGEE is a large stellar spectroscopic survey. It observes in the near-infrared H band (1.5 to 1.7 \(\mu\)m), where dust absorbs less light than in the optical. APOGEE can therefore see stars in the disk, the bulge and the halo of the Milky Way. Each APOGEE spectrum has thousands of absorption lines that carry information about the atmosphere of the star.

We use a subset of 3,000 spectra from APOGEE Data Release 17 (DR17), with a signal-to-noise ratio (SNR) above 100. The high SNR makes the noise in the fluxes, the inputs of our model, small. We choose the stars at random among those with clean spectra. One third are dwarfs and two thirds are giants.

Full APOGEE spectra have more than 7,000 pixels. We truncate each spectrum to 2,000 pixels. This keeps enough spectral information to predict the temperature and keeps the computation short. The concepts of this tutorial stay the same for 2,000 or 7,000 features. Fewer pixels let us focus on the method.

\subsection*{Physical Parameters}

The APOGEE spectroscopic pipeline gives many parameters for each star. The three main stellar parameters of a spectrum are usually the following.

\begin{enumerate}
\def\labelenumi{\arabic{enumi}.}
\tightlist
\item
  \textbf{Effective temperature (\(T_{\rm eff}\))} is the temperature of a black body that emits the same power per unit surface area as the star. It sets the overall energy distribution and the ionization states of the elements in the atmosphere. We use it as the target of the regression.
\item
  \textbf{Surface gravity (log g)} is the logarithm of the gravitational acceleration at the surface in cgs units. It separates dwarfs (high gravity) from giants (low gravity) and changes the pressure broadening of the spectral lines.
\item
  \textbf{Metallicity ({[}Fe/H{]})} is the base-10 logarithm of the iron to hydrogen number ratio of the star, relative to the Sun. A star with {[}Fe/H{]} = -1 has an iron to hydrogen ratio ten times lower than the Sun. It traces the chemical enrichment history of the gas that formed the star.
\end{enumerate}

The pipeline gives the temperatures that we predict, so the errors below measure the agreement with the pipeline. We treat the noise of these target temperatures as uniform (homoscedastic), a simplification that lets us focus on the core concepts of linear regression. Real spectra also have a measurement uncertainty at each wavelength, which is noise in the inputs. The SNR cut makes it small, so we assume that it is negligible.

We now load the dataset.

\tutorialbarrier
\begin{codecell}

\end{codecell}
\begin{Verbatim}[fontsize=\footnotesize, xleftmargin=5mm]
Dataset overview
Number of stars = 3000
Spectral pixels per star = 2000
Wavelength range = 15168.1 - 15592.9 Å

Ranges of the stellar parameters
Temperature = 3585 - 6433 K
Surface gravity = 0.07 - 4.76 dex
Metallicity = -2.41 - 0.42 dex
\end{Verbatim}

\section*{Visualizing Stellar Spectra}

Before the regression, we look at what stellar spectra are like and how they change with temperature. A spectrum has many absorption features, which are dark lines where atoms and molecules in the stellar atmosphere absorb light at specific wavelengths. The strength of these features depends on the physical parameters of the star.

We compare the spectra of stars with different effective temperatures. This comparison shows why linear regression can extract the temperature, because different temperatures give different spectra. Cooler stars usually have deeper absorption features, both atomic lines and molecular bands. Hotter stars have shallower but still characteristic absorption. These systematic changes with temperature are the signal that the regression model learns to use.

\tutorialbarrier
\begin{codecell}

\end{codecell}
\tutorialplot{\includegraphics[width=4.561in,height=0.75\textheight,keepaspectratio]{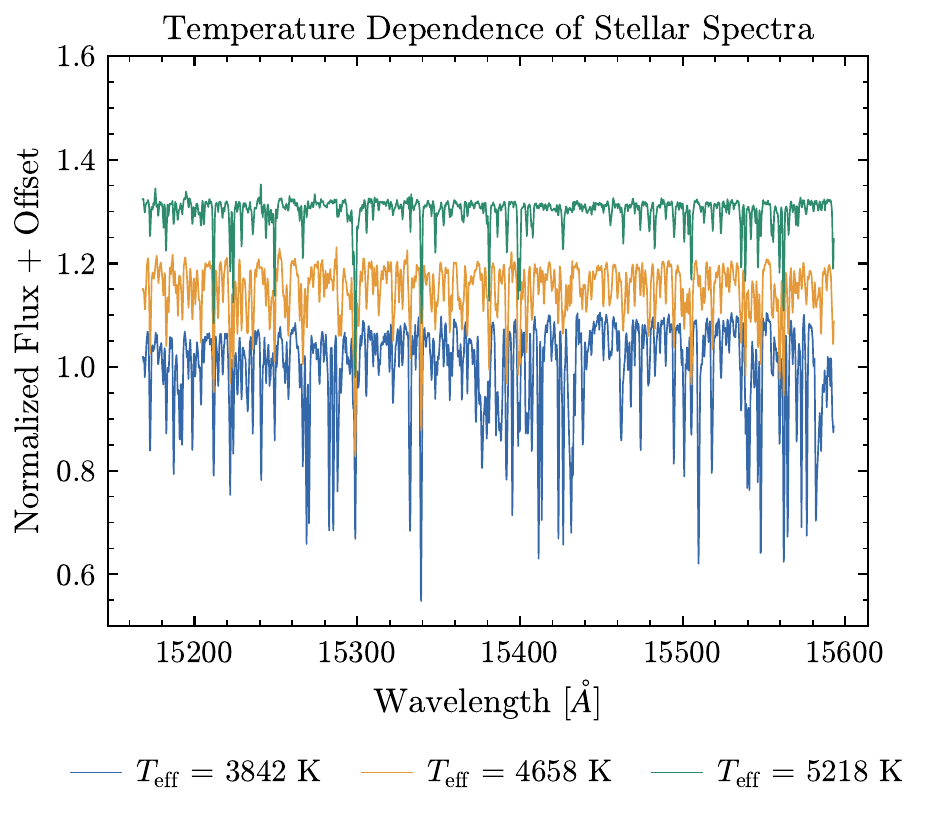}}%
  {\includegraphics[height=\tutorialheight]{tutorials/4a/fig01.pdf}}

\section*{The Kiel Diagram}

We also look at the distribution of the stellar parameters in our sample. The Kiel diagram, a plot of surface gravity against effective temperature, shows the evolutionary states of the stars. Different regions of the diagram belong to different types of stars. Dwarfs (main sequence stars) have high surface gravity, and red giants have lower gravity and cooler temperatures. The color shows the metallicity, which adds a third parameter and shows how the chemical composition varies in the diagram.

\tutorialbarrier
\begin{codecell}

\end{codecell}
\tutorialplot{\includegraphics[width=4.401in,height=0.75\textheight,keepaspectratio]{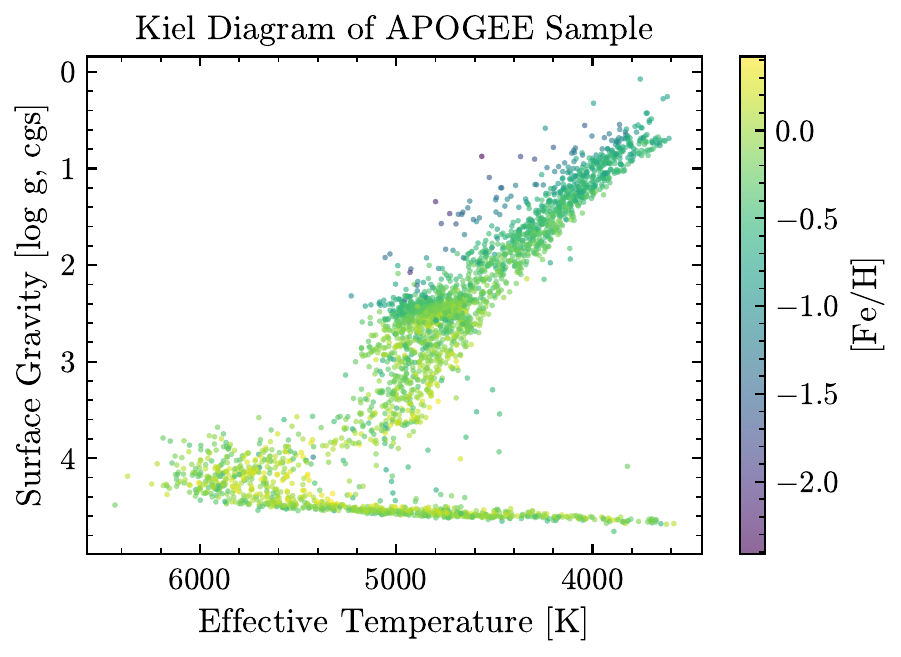}}%
  {\includegraphics[height=\tutorialheight]{tutorials/4a/fig02.pdf}}
\begin{Verbatim}[fontsize=\footnotesize, xleftmargin=5mm]
Parameter space coverage
Temperature = 3585 - 6433 K (Δ = 2848 K)
Surface gravity = 0.07 - 4.76 dex (Δ = 4.68 dex)
Metallicity = -2.41 - 0.42 dex (Δ = 2.83 dex)
\end{Verbatim}

The sample covers both groups of stars. The main sequence of dwarfs lies at high gravity, and the red giant branch lies at low gravity. The dense group near \(\log g = 2.5\) and 4800 K is the red clump, which holds giants that burn helium in their cores.

This distribution matters for the interpretation of the regression results. The model is most reliable where the sample is dense, for temperatures between 3500 and 5500 K. Predictions for hotter stars need care, because the model has few examples of them.

\section*{Linear Regression: Setting Up the Model}

With the data explored, we turn to the core task, the prediction of the effective temperature from the spectrum. Linear regression models the relation between the input features, the flux values of the spectrum, and the target, the effective temperature, as a linear combination

\[T_{\rm eff} = \sum_{j=1}^{D} w_j \phi_j + w_0,\]

where \(D\) is the number of pixels and \(\phi_j\) is the flux at the \(j\)th wavelength. The weights \(w_j\) are what we learn, and \(w_0\) is the bias term.

To simplify the notation and the computation, we absorb the bias term into the weight vector. We append a column of ones to the design matrix \(\boldsymbol{\Phi}\), and the model becomes

\[\mathbf{t} \approx \boldsymbol{\Phi}\mathbf{w},\]

where \(\boldsymbol{\Phi}\) now holds the spectral features and the column of ones, and \(\mathbf{w}\) holds all weights including the bias. The bias is one more parameter, so no step of the computation needs a special rule.

We now build \(\boldsymbol{\Phi}\) from the spectra and keep 20 percent of the stars as test stars. The training stars fit the weights. The test stars show how well the model predicts stars that it has not seen.

\tutorialbarrier
\begin{codecell}

\end{codecell}
\begin{Verbatim}[fontsize=\footnotesize, xleftmargin=5mm]
Design matrix shape = (3000, 2001)
(3000 spectra x 2000 spectral pixels + 1 bias term)

Target variable shape = (3000,)
Temperature range = 3585 - 6433 K

Training set = 2400 spectra
Test set = 600 spectra
\end{Verbatim}

\section*{Maximum Likelihood Estimation for Linear Regression}

The principle of maximum likelihood is a systematic way to find the best model parameters. We assume Gaussian noise with a constant variance \(\sigma^2\). The likelihood of the data for the parameters \(\mathbf{w}\) is

\[p(\mathbf{t}|\boldsymbol{\Phi}, \mathbf{w}, \sigma^2) = \prod_{i=1}^{N} \mathcal{N}(t_i | \mathbf{w}^T\boldsymbol{\phi}_i, \sigma^2),\]

where \(N\) is the number of training stars and \(\boldsymbol{\phi}_i\) is the \(i\)th row of \(\boldsymbol{\Phi}\).

Maximizing the likelihood, or equivalently minimizing the negative log-likelihood, gives the familiar least-squares objective. As Section 4.4 of the textbook shows, this optimization has an analytical solution

\[\mathbf{w}_{\text{ML}} = (\boldsymbol{\Phi}^T \boldsymbol{\Phi})^{-1} \boldsymbol{\Phi}^T \mathbf{t}.\]

This closed-form solution is an advantage of linear regression, because we can find the globally optimal parameters without iterative optimization. It needs only matrix operations. We compute the matrix \(\boldsymbol{\Phi}^T \boldsymbol{\Phi}\), invert it and multiply. We now implement the solution and compute the root mean squared error (RMSE) of the predicted temperatures on the training and test stars.

\tutorialbarrier
\begin{codecell}

\end{codecell}
\begin{Verbatim}[fontsize=\footnotesize, xleftmargin=5mm]
Maximum likelihood fit
Training RMSE = 28.7 K
Test RMSE = 279.5 K
\end{Verbatim}
\tutorialplot{\includegraphics[width=6.173in,height=0.75\textheight,keepaspectratio]{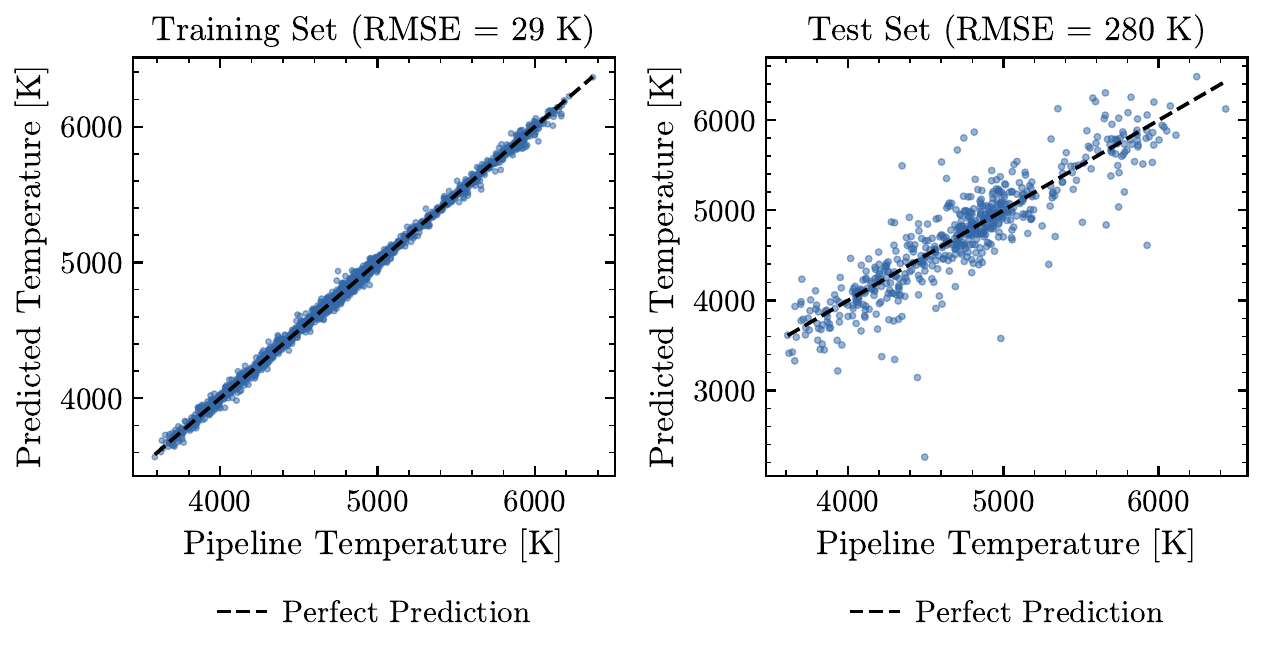}}%
  {\includegraphics[height=\tutorialheight]{tutorials/4a/fig03.pdf}}

The training RMSE is about 30 K, so the model fits the training stars very well. The test RMSE is about 280 K, so the model fails on stars that it has not seen.

This is a typical case of overfitting. The model has memorized the training spectra and has not learned a relation that generalizes. The count of the parameters explains this. We fit 2,001 parameters (2,000 spectral features and the bias) with only 2,400 training stars. When the number of parameters approaches the number of training stars, the model can reproduce the training data almost perfectly. It also learns spurious relations that do not generalize.

This large increase of the error shows that the unregularized model does not learn the true relation between the spectra and the temperature. It fits the noise and the peculiarities of the training stars.

\section*{Regularization: Preventing Overfitting}

Regularization limits overfitting. It constrains the model to prefer simpler solutions that generalize better. We can interpret regularization in a Bayesian way. A zero-mean Gaussian prior with variance \(\eta^2\) on the weights is equivalent to adding a penalty to the objective function. The penalty is \(\lambda\) times the sum of the squared weights.

The regularized solution is

\[\mathbf{w}_{\text{reg}} = (\lambda \mathrm{Id} + \boldsymbol{\Phi}^T\boldsymbol{\Phi})^{-1}\boldsymbol{\Phi}^T\mathbf{t},\]

where \(\mathrm{Id}\) is the identity matrix and \(\lambda = \sigma^2/\eta^2\) is the regularization parameter that sets the strength of the penalty. A larger \(\lambda\) constrains the weights more, so the model fits less of the noise in the training data.

We choose \(\lambda\) on validation stars and not on the test stars. The test stars then stay an unbiased check (Section 4.7 of the textbook). The split is 60 percent fit stars, 20 percent validation stars and 20 percent test stars. For each \(\lambda\) we fit the fit stars and evaluate on the validation stars.

\tutorialbarrier
\begin{codecell}

\end{codecell}

We plot the fit and the validation RMSE against \(\lambda\), and we find the value with the lowest validation RMSE.

\tutorialbarrier
\begin{codecell}

\end{codecell}
\tutorialplot{\includegraphics[width=4.197in,height=0.75\textheight,keepaspectratio]{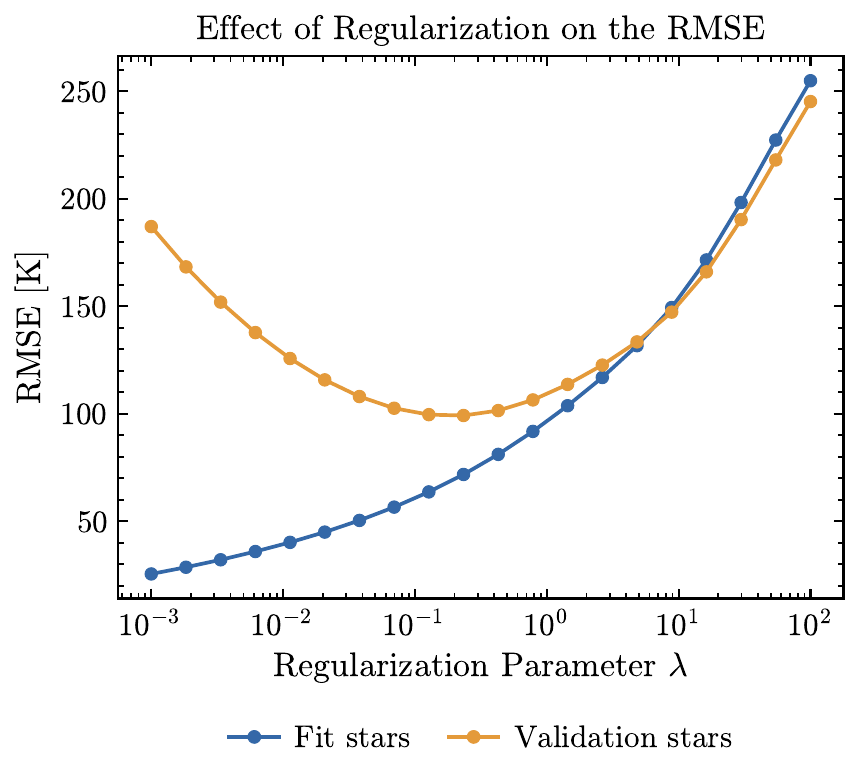}}%
  {\includegraphics[height=\tutorialheight]{tutorials/4a/fig04.pdf}}
\begin{Verbatim}[fontsize=\footnotesize, xleftmargin=5mm]
Optimal regularization parameter λ = 0.234
Validation RMSE with optimal λ = 99.2 K
\end{Verbatim}

The plot of the RMSE against \(\lambda\) shows the bias-variance tradeoff (Section 4.7 of the textbook). It has three regimes.

\begin{enumerate}
\def\labelenumi{\arabic{enumi}.}
\tightlist
\item
  \textbf{Low \(\lambda\).} The model overfits. The RMSE is low on the fit stars and high on the validation stars.
\item
  \textbf{Best \(\lambda\), about 0.2.} The validation RMSE is lowest, so the model generalizes best.
\item
  \textbf{High \(\lambda\).} The constraints are too strong, and the model underfits. The RMSE is high on both groups, because the model cannot learn even the real relations in the data.
\end{enumerate}

The validation RMSE only selects \(\lambda\). We now refit the model on all training stars with the best \(\lambda\) and compute the test RMSE once.

\tutorialbarrier
\begin{codecell}

\end{codecell}
\tutorialplot{\includegraphics[width=6.173in,height=0.75\textheight,keepaspectratio]{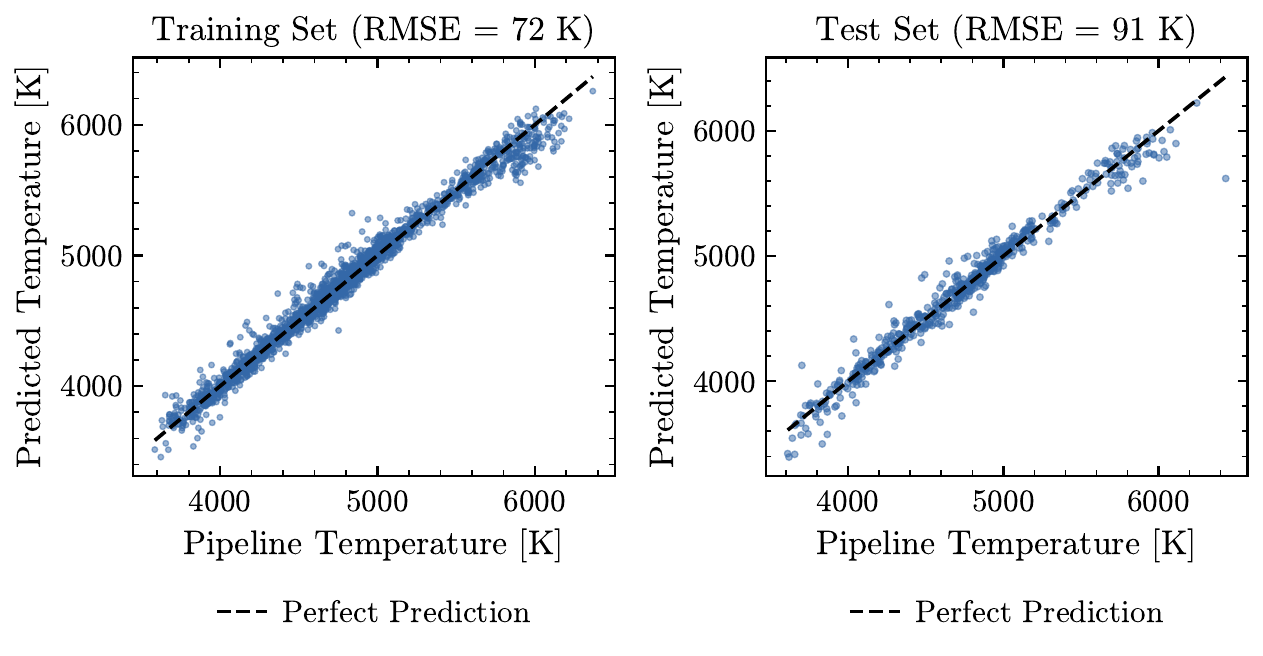}}%
  {\includegraphics[height=\tutorialheight]{tutorials/4a/fig05.pdf}}
\begin{Verbatim}[fontsize=\footnotesize, xleftmargin=5mm]
Model comparison
Unregularized, training = 28.7 K, test = 279.5 K
Regularized, training = 72.0 K, test = 91.0 K
\end{Verbatim}

The regularized model generalizes much better. The test RMSE falls from about 280 K to about 90 K, which turns an unusable model into a useful one. The penalty keeps the weights from growing too large, so the model follows the noise of the training stars less.

The test predictions lie much closer to the diagonal than before, so the model has learned a real relation between the spectra and the temperature. The training and test RMSE are now close, which indicates good generalization. The stars hotter than 6000 K mostly lie below the diagonal, so the model under-predicts them. The penalty pulls the predictions toward the average temperature. Noise in the inputs, which we neglect here, flattens the relation in a similar way. This effect is called attenuation bias, and a later chapter treats it.

\section*{Feature Engineering}

Our regularized linear model works well, but so far it assumes a linear relation between the temperature and the spectral flux. This assumption may be too restrictive, because the physics of stellar atmospheres often gives nonlinear relations between the spectra and the stellar parameters.

We can model nonlinear relations and keep the analytical solution if we transform the input features with basis functions. The relation is then nonlinear in the original features, and the problem stays linear in the transformed features.

A simple extension adds quadratic terms. For each pixel with flux \(\phi_j\), we add the feature \(\phi_j^2\). The model can then learn that the temperature depends on the square of the flux as well as on the flux. This can describe the saturation of strong absorption lines, whose depth grows more slowly than the number of absorbing atoms. It can also describe a nonlinear sensitivity to the temperature.

The extension raises the number of features from 2,001 to 4,001 (2,000 linear features, 2,000 quadratic features and the bias). Overfitting becomes more likely, so regularization matters even more.

\tutorialbarrier
\begin{codecell}

\end{codecell}
\begin{Verbatim}[fontsize=\footnotesize, xleftmargin=5mm]
Original feature dimensions = 2001
Quadratic feature dimensions = 4001
Feature expansion = 2001 → 4001

Testing quadratic features without regularization...
Quadratic model (no regularization)
Training RMSE = 0.0 K
Test RMSE = 247.7 K
\end{Verbatim}

The quadratic model has more weights than training stars, so it needs regularization. Without it, the fit passes through every training star, so the training RMSE is zero. We choose \(\lambda\) for the expanded features on the validation stars as before, and we compute the test RMSE once.

\tutorialbarrier
\begin{codecell}

\end{codecell}
\begin{Verbatim}[fontsize=\footnotesize, xleftmargin=5mm]
Optimal λ for quadratic features = 0.785
Test RMSE with quadratic features = 85.5 K

Comparison
Linear features = 91.0 K
Quadratic features = 85.5 K
\end{Verbatim}

The small improvement from the quadratic features gives three lessons.

\begin{enumerate}
\def\labelenumi{\arabic{enumi}.}
\tightlist
\item
  \textbf{Small gain from more features.} Here doubling the number of features lowers the test RMSE by only a few percent. This suggests that the linear relations between the spectra and the temperature already carry most of the predictable signal.
\item
  \textbf{Regularization matters more.} The quadratic model needs a stronger regularization, a larger \(\lambda\), than the linear model. This is expected, because more features make overfitting more likely.
\item
  \textbf{Domain knowledge matters.} Instead of adding polynomial features blindly, we can use astrophysical knowledge. Examples are the ratio of two absorption lines that depend differently on the temperature, and spectral regions that correlate with the stellar parameters.
\end{enumerate}

This example shows a basic principle of machine learning. A more complex model is not always better. The regularized linear model is simpler and cheaper to compute, and its test RMSE is close to that of the quadratic model. In practice, the choice between simple and complex features depends on whether the small gain justifies a more complex model for the application.

\section*{Summary}

This tutorial showed the strengths and the limits of linear regression in an astronomical application. With real APOGEE spectra we studied concepts that apply far beyond this problem.

\begin{enumerate}
\def\labelenumi{\arabic{enumi}.}
\tightlist
\item
  \textbf{Maximum likelihood estimation.} We implemented the analytical solution of linear regression, which follows from maximizing the Gaussian likelihood. The closed-form solution \(\mathbf{w}_{\text{ML}} = (\boldsymbol{\Phi}^T \boldsymbol{\Phi})^{-1} \boldsymbol{\Phi}^T \mathbf{t}\) gives the global optimum, which we do not have for more complex models.
\item
  \textbf{Overfitting.} The first model had a training RMSE of about 30 K and a test RMSE of about 280 K. This shows why every machine learning application needs an evaluation on test stars.
\item
  \textbf{Regularization.} A penalty on the weights turned this unusable model into one with a test RMSE of about 90 K. The regularization parameter \(\lambda\) sets the bias-variance tradeoff, and we chose it on validation stars. A Bayesian prior has the same kind of hyperparameter, its width, and it also needs validation stars.
\item
  \textbf{Feature engineering.} The quadratic features gave only a small improvement here. This does not make feature engineering unimportant. For other stellar parameters or a wider parameter range, nonlinear features may be needed. The data and the validation error should decide how complex the model is.
\end{enumerate}

The principles of this tutorial are maximum likelihood estimation, regularization and feature engineering. They are the basis of more sophisticated methods, such as Bayesian approaches, nonlinear models and deep learning. The basic problem stays the same, which is to extract useful relations from data and to avoid overfitting.

\tutorialchapter{4b}{Linear Regression for Telescope Calibration}
\noindent\textit{Companion material for Chapter 4}

\section*{Introduction}

Before any scientific analysis, we must understand and correct the systematic effects in our measurements. The atmospheric conditions change during the night, telescopes flex as they track the sky, and optical components expand with the temperature. Uncorrected, these effects can exceed the signals that we want to measure.

This tutorial studies calibration with radial velocity (RV) measurements. The RV method finds planets through the tiny wobbles that they cause in the velocity of their host stars. We observe RV standard stars, whose velocities stay stable for decades, under different conditions. The observations contain systematic errors that correlate with recorded metadata, such as the temperature and the position of the telescope.

The calibration shows how linear regression estimates a quantity of physical interest together with nuisance effects. The tools are design matrices, maximum likelihood estimation, the bootstrap and weighted least squares. We go from the calibration of one telescope to a network of telescopes.

\section*{Prerequisites and Learning Objectives}

\subsection*{Prerequisites}

This tutorial assumes the following.

\begin{itemize}
\tightlist
\item
  The linear regression model and its assumptions
\item
  Maximum likelihood estimation and its analytical solution
\item
  Design matrices, which hold the features of a regression problem
\item
  Bootstrap methods, which use resampling to estimate parameter uncertainties
\item
  Python programming with NumPy arrays and matplotlib plots
\end{itemize}

\subsection*{Learning Objectives}

After this tutorial you can do the following.

\begin{itemize}
\tightlist
\item
  Formulate an instrumental calibration as a linear regression, where the systematic effects depend on observable metadata
\item
  Estimate astrophysical parameters and calibration coefficients together
\item
  Use weighted least squares when the measurements have different uncertainties, as is common in astronomy
\item
  Design and solve a regression with several types of parameters, here stellar properties and telescope systematic effects
\item
  Combine the observations of several instruments to strengthen the calibration of a telescope network
\end{itemize}

We begin by importing the libraries and setting the plot style.

\tutorialbarrier
\begin{codecell}

\end{codecell}

\section*{From Telescope Systematics to Linear Models}

Consider a single RV measurement of a standard star, a star with a constant true velocity \(y_{\text{true}}\). The observed RV \(t\) differs from this true value because of effects that come from the instrument and from the observing conditions

\[t = y_{\text{true}} + \sum_{j=1}^{4} w_j x_j + \varepsilon.\]

Here \(x_j\) is the \(j\)th item of the metadata that describe the observing conditions. These are quantities that we can measure or record at the time of the observation. The coefficient \(w_j\) says how much the \(j\)th item affects our measurement, and \(\varepsilon\) is random measurement noise.

In this example we assume that four recorded conditions influence the measurements. A real instrument can record many more metadata.

\begin{enumerate}
\def\labelenumi{\arabic{enumi}.}
\tightlist
\item
  \textbf{Zenith angle.} The angle of the star from the vertical. At a larger angle the light crosses more atmosphere, and the atmospheric dispersion can bias the velocity.
\item
  \textbf{Temperature.} Thermal expansion of the optics can shift the spectral lines on the detector.
\item
  \textbf{Humidity.} It changes the refractive index of the air and so the optical path.
\item
  \textbf{Instrumental drift.} Flexure and detector instabilities cause velocity offsets that change during the night.
\end{enumerate}

The linear model is simple and easy to interpret. Once we know the coefficients \(w_j\), we can correct any observation with recorded metadata. When we scale the effects to the same unit, they also show which effects dominate our error budget.

We now generate synthetic data that mimic these systematic effects. For simplicity, the zenith angle of each observation is random and does not follow the track of a star.

\tutorialbarrier
\begin{codecell}
\begin{Verbatim}[commandchars=\\\{\},fontsize=\small]
\PYG{k}{def}\PYG{+w}{ }\PYG{n+nf}{generate\PYGZus{}telescope\PYGZus{}data}\PYG{p}{(}\FancyVerbBreakAfterBreak{}\PYG{n}{N}\PYG{o}{=}\FancyVerbBreakAfterBreak{}\PYG{l+m+mi}{100}\PYG{p}{,}\FancyVerbBreakAfterBreak{} \PYG{n}{seed}\PYG{o}{=}\FancyVerbBreakAfterBreak{}\PYG{l+m+mi}{42}\PYG{p}{)}\PYG{p}{:}
\PYG{+w}{    }\PYG{l+s+sd}{\PYGZdq{}\PYGZdq{}\PYGZdq{}Generate synthetic RV data with systematic effects and no noise.\PYGZdq{}\PYGZdq{}\PYGZdq{}}
    \PYG{n}{np}\PYG{o}{.}\FancyVerbBreakAfterBreak{}\PYG{n}{random}\PYG{o}{.}\FancyVerbBreakAfterBreak{}\PYG{n}{seed}\PYG{p}{(}\FancyVerbBreakAfterBreak{}\PYG{n}{seed}\PYG{p}{)}  \PYG{c+c1}{\PYGZsh{} fix the seed so that the data are reproducible}

    \PYG{c+c1}{\PYGZsh{} Metadata matrix with one row per observation}
    \PYG{n}{X} \PYG{o}{=}\FancyVerbBreakAfterBreak{} \PYG{n}{np}\PYG{o}{.}\FancyVerbBreakAfterBreak{}\PYG{n}{zeros}\PYG{p}{(}\FancyVerbBreakAfterBreak{}\PYG{p}{(}\FancyVerbBreakAfterBreak{}\PYG{n}{N}\PYG{p}{,}\FancyVerbBreakAfterBreak{} \PYG{l+m+mi}{4}\PYG{p}{)}\PYG{p}{)}

    \PYG{c+c1}{\PYGZsh{} Feature 1 is the zenith angle in degrees}
    \PYG{c+c1}{\PYGZsh{} A higher angle means more atmosphere and a stronger dispersion effect}
    \PYG{n}{X}\PYG{p}{[}\FancyVerbBreakAfterBreak{}\PYG{p}{:}\PYG{p}{,}\FancyVerbBreakAfterBreak{} \PYG{l+m+mi}{0}\PYG{p}{]} \PYG{o}{=}\FancyVerbBreakAfterBreak{} \PYG{n}{np}\PYG{o}{.}\FancyVerbBreakAfterBreak{}\PYG{n}{random}\PYG{o}{.}\FancyVerbBreakAfterBreak{}\PYG{n}{uniform}\PYG{p}{(}\FancyVerbBreakAfterBreak{}\PYG{l+m+mi}{0}\PYG{p}{,}\FancyVerbBreakAfterBreak{} \PYG{l+m+mi}{60}\PYG{p}{,}\FancyVerbBreakAfterBreak{} \PYG{n}{N}\PYG{p}{)}  \PYG{c+c1}{\PYGZsh{} random angle for each observation}

    \PYG{c+c1}{\PYGZsh{} Feature 2 is the temperature in degrees Celsius}
    \PYG{c+c1}{\PYGZsh{} It follows a cosine during the night, with random fluctuations}
    \PYG{n}{time} \PYG{o}{=}\FancyVerbBreakAfterBreak{} \PYG{n}{np}\PYG{o}{.}\FancyVerbBreakAfterBreak{}\PYG{n}{linspace}\PYG{p}{(}\FancyVerbBreakAfterBreak{}\PYG{l+m+mi}{0}\PYG{p}{,}\FancyVerbBreakAfterBreak{} \PYG{l+m+mi}{8}\PYG{p}{,}\FancyVerbBreakAfterBreak{} \PYG{n}{N}\PYG{p}{)}  \PYG{c+c1}{\PYGZsh{} 8\PYGZhy{}hour observing night}
    \PYG{n}{X}\PYG{p}{[}\FancyVerbBreakAfterBreak{}\PYG{p}{:}\PYG{p}{,}\FancyVerbBreakAfterBreak{} \PYG{l+m+mi}{1}\PYG{p}{]} \PYG{o}{=}\FancyVerbBreakAfterBreak{} \PYG{l+m+mi}{20} \PYG{o}{+}\FancyVerbBreakAfterBreak{} \PYG{l+m+mi}{5} \PYG{o}{*} \PYG{n}{np}\PYG{o}{.}\FancyVerbBreakAfterBreak{}\PYG{n}{cos}\PYG{p}{(}\FancyVerbBreakAfterBreak{}\PYG{l+m+mi}{2}\PYG{o}{*}\PYG{n}{np}\PYG{o}{.}\FancyVerbBreakAfterBreak{}\PYG{n}{pi}\PYG{o}{*}\PYG{n}{time}\PYG{o}{/}\PYG{l+m+mi}{8}\PYG{p}{)} \PYG{o}{+}\FancyVerbBreakAfterBreak{} \PYG{n}{np}\PYG{o}{.}\FancyVerbBreakAfterBreak{}\PYG{n}{random}\PYG{o}{.}\FancyVerbBreakAfterBreak{}\PYG{n}{normal}\PYG{p}{(}\FancyVerbBreakAfterBreak{}\PYG{l+m+mi}{0}\PYG{p}{,}\FancyVerbBreakAfterBreak{} \PYG{l+m+mf}{0.5}\PYG{p}{,}\FancyVerbBreakAfterBreak{} \PYG{n}{N}\PYG{p}{)}

    \PYG{c+c1}{\PYGZsh{} Feature 3 is the humidity in percent}
    \PYG{c+c1}{\PYGZsh{} It varies around 50 percent with a large scatter}
    \PYG{n}{X}\PYG{p}{[}\FancyVerbBreakAfterBreak{}\PYG{p}{:}\PYG{p}{,}\FancyVerbBreakAfterBreak{} \PYG{l+m+mi}{2}\PYG{p}{]} \PYG{o}{=}\FancyVerbBreakAfterBreak{} \PYG{l+m+mi}{50} \PYG{o}{+}\FancyVerbBreakAfterBreak{} \PYG{l+m+mi}{10}\PYG{o}{*}\PYG{n}{np}\PYG{o}{.}\FancyVerbBreakAfterBreak{}\PYG{n}{random}\PYG{o}{.}\FancyVerbBreakAfterBreak{}\PYG{n}{randn}\PYG{p}{(}\FancyVerbBreakAfterBreak{}\PYG{n}{N}\PYG{p}{)}

    \PYG{c+c1}{\PYGZsh{} Feature 4 is the time\PYGZhy{}dependent instrumental drift}
    \PYG{c+c1}{\PYGZsh{} It is linear in time, with small random variations}
    \PYG{n}{X}\PYG{p}{[}\FancyVerbBreakAfterBreak{}\PYG{p}{:}\PYG{p}{,}\FancyVerbBreakAfterBreak{} \PYG{l+m+mi}{3}\PYG{p}{]} \PYG{o}{=}\FancyVerbBreakAfterBreak{} \PYG{n}{np}\PYG{o}{.}\FancyVerbBreakAfterBreak{}\PYG{n}{linspace}\PYG{p}{(}\FancyVerbBreakAfterBreak{}\PYG{l+m+mi}{0}\PYG{p}{,}\FancyVerbBreakAfterBreak{} \PYG{l+m+mi}{1}\PYG{p}{,}\FancyVerbBreakAfterBreak{} \PYG{n}{N}\PYG{p}{)} \PYG{o}{+}\FancyVerbBreakAfterBreak{} \PYG{l+m+mf}{0.1}\PYG{o}{*}\PYG{n}{np}\PYG{o}{.}\FancyVerbBreakAfterBreak{}\PYG{n}{random}\PYG{o}{.}\FancyVerbBreakAfterBreak{}\PYG{n}{randn}\PYG{p}{(}\FancyVerbBreakAfterBreak{}\PYG{n}{N}\PYG{p}{)}

    \PYG{c+c1}{\PYGZsh{} True RV of our simulated standard star}
    \PYG{n}{y\PYGZus{}true} \PYG{o}{=}\FancyVerbBreakAfterBreak{} \PYG{o}{\PYGZhy{}}\PYG{l+m+mf}{15.0}  \PYG{c+c1}{\PYGZsh{} km/s}

    \PYG{c+c1}{\PYGZsh{} True coefficients, which say how each parameter affects the RV}
    \PYG{c+c1}{\PYGZsh{} We chose these values for this example}
    \PYG{n}{w\PYGZus{}true} \PYG{o}{=}\FancyVerbBreakAfterBreak{} \PYG{n}{np}\PYG{o}{.}\FancyVerbBreakAfterBreak{}\PYG{n}{array}\PYG{p}{(}\FancyVerbBreakAfterBreak{}\PYG{p}{[}\FancyVerbBreakAfterBreak{}
        \PYG{l+m+mf}{0.05}\PYG{p}{,}\FancyVerbBreakAfterBreak{}   \PYG{c+c1}{\PYGZsh{} zenith angle, in km/s per degree}
        \PYG{l+m+mf}{0.02}\PYG{p}{,}\FancyVerbBreakAfterBreak{}   \PYG{c+c1}{\PYGZsh{} temperature, in km/s per °C}
        \PYG{l+m+mf}{0.01}\PYG{p}{,}\FancyVerbBreakAfterBreak{}   \PYG{c+c1}{\PYGZsh{} humidity, in km/s per \PYGZpc{}}
        \PYG{l+m+mf}{0.5}     \PYG{c+c1}{\PYGZsh{} drift, in km/s over the night}
    \PYG{p}{]}\PYG{p}{)}

    \PYG{c+c1}{\PYGZsh{} Generate clean velocities (systematic effects but no noise)}
    \PYG{n}{y\PYGZus{}clean} \PYG{o}{=}\FancyVerbBreakAfterBreak{} \PYG{n}{y\PYGZus{}true} \PYG{o}{+}\FancyVerbBreakAfterBreak{} \PYG{n}{X} \PYG{o}{@} \PYG{n}{w\PYGZus{}true}  \PYG{c+c1}{\PYGZsh{} matrix\PYGZhy{}vector product}

    \PYG{k}{return} \PYG{n}{X}\PYG{p}{,}\FancyVerbBreakAfterBreak{} \PYG{n}{y\PYGZus{}clean}\PYG{p}{,}\FancyVerbBreakAfterBreak{} \PYG{n}{y\PYGZus{}true}\PYG{p}{,}\FancyVerbBreakAfterBreak{} \PYG{n}{w\PYGZus{}true}\PYG{p}{,}\FancyVerbBreakAfterBreak{} \PYG{n}{time}

\PYG{c+c1}{\PYGZsh{} Generate example data}
\PYG{n}{X}\PYG{p}{,}\FancyVerbBreakAfterBreak{} \PYG{n}{y\PYGZus{}clean}\PYG{p}{,}\FancyVerbBreakAfterBreak{} \PYG{n}{y\PYGZus{}true}\PYG{p}{,}\FancyVerbBreakAfterBreak{} \PYG{n}{w\PYGZus{}true}\PYG{p}{,}\FancyVerbBreakAfterBreak{} \PYG{n}{time} \PYG{o}{=}\FancyVerbBreakAfterBreak{} \PYG{n}{generate\PYGZus{}telescope\PYGZus{}data}\PYG{p}{(}\FancyVerbBreakAfterBreak{}\PYG{n}{N}\PYG{o}{=}\FancyVerbBreakAfterBreak{}\PYG{l+m+mi}{100}\PYG{p}{)}
\end{Verbatim}
\end{codecell}

The synthetic velocities contain the systematic effects but no noise. Real observations also contain random errors from photon statistics and detector readout. We add Gaussian noise with a standard deviation of 0.1 km/s.

\tutorialbarrier
\begin{codecell}

\end{codecell}
\begin{Verbatim}[fontsize=\footnotesize, xleftmargin=5mm]
True RV of the standard star = -15.00 km/s
Mean observed RV = -12.40 km/s
RV standard deviation = 0.89 km/s
\end{Verbatim}
\tutorialplot{\includegraphics[width=4.372in,height=0.75\textheight,keepaspectratio]{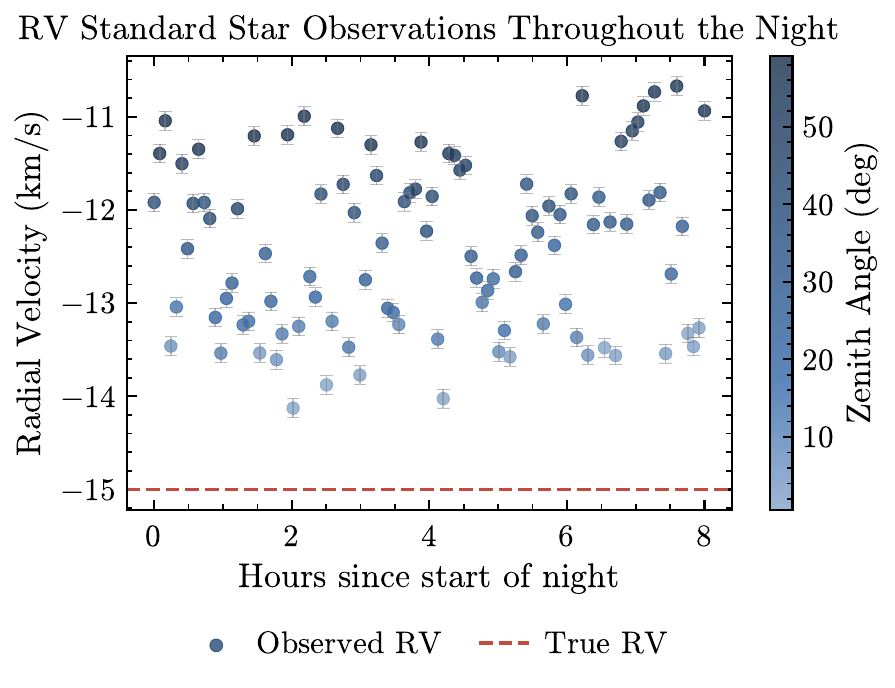}}%
  {\includegraphics[height=\tutorialheight]{tutorials/4b/fig01.pdf}}

This time series shows how important the calibration is. The observed velocities scatter around values that differ a lot from the true RV (red dashed line). The variations far exceed our measurement precision of 0.1 km/s. The systematic offset of about 2.6 km/s and the additional scatter of about 0.9 km/s make uncalibrated observations unfit for precise RV work.

The color shows the zenith angle, and the velocity rises with it. The instrumental drift is part of the data too, but the larger effects hide it in this plot. The task of the calibration is to separate all these overlapping effects with the recorded metadata.

\section*{Calibrating a Single Telescope}

With our dataset in hand, we turn to the calibration problem itself. We need to solve for two types of unknowns at the same time.

\begin{enumerate}
\def\labelenumi{\arabic{enumi}.}
\tightlist
\item
  \textbf{The true RV of our standard star} (\(y_{\text{true}}\))
\item
  \textbf{The calibration coefficients} (\(w_1, \ldots, w_4\)) that relate the metadata to the systematic offsets
\end{enumerate}

In a typical regression we know the true value of what we want to predict. Here the true RV of the star is unknown, and we must infer it from the data. The linear structure of the problem lets us solve for all parameters together with an augmented design matrix.

The next step is to rewrite our model

\[t = y_{\text{true}} + \sum_{j=1}^{4} w_j x_j + \varepsilon\]

as a standard linear regression. As in Section 4.2, we add a constant feature to the metadata and absorb \(y_{\text{true}}\) into the weight vector as the intercept \(w_0\)

\[t = \mathbf{w}^T\boldsymbol{\phi}(\mathbf{x}) + \varepsilon,\]

where \(\boldsymbol{\phi}(\mathbf{x}) = (1, x_1, x_2, x_3, x_4)^T\) and \(\mathbf{w} = (y_{\text{true}}, w_1, w_2, w_3, w_4)^T\). The design matrix \(\boldsymbol{\Phi}\) has one row \(\boldsymbol{\phi}(\mathbf{x}_i)^T\) for each observation \(i\), so it has a column of ones. The maximum likelihood solution then follows from our standard linear regression machinery.

\tutorialbarrier
\begin{codecell}

\end{codecell}
\begin{Verbatim}[fontsize=\footnotesize, xleftmargin=5mm]
Data split = 80 training, 20 test observations
\end{Verbatim}

\subsection*{Maximum Likelihood Solution}

With our augmented design matrix in place, finding the calibration parameters becomes a standard linear regression problem. For Gaussian noise with a constant variance \(\sigma^2\), the maximum likelihood solution is the ordinary least squares solution (Section 4.4)

\[\mathbf{w}_{\text{ML}} = (\boldsymbol{\Phi}^T\boldsymbol{\Phi})^{-1}\boldsymbol{\Phi}^T\mathbf{t},\]

where \(\mathbf{t}\) is the vector of observed RVs. The vector \(\mathbf{w}_{\text{ML}}\) contains our estimate of the true RV and the calibration coefficients.

This formulation uses the linear algebra of Chapter 4. The augmented design matrix treats all unknowns in the same way. The mathematics is the same whether we estimate an astrophysical parameter or an instrumental systematic effect.

We now implement this solution and examine what it tells us about the systematic effects of our telescope.

\tutorialbarrier
\begin{codecell}

\end{codecell}
\begin{Verbatim}[fontsize=\footnotesize, xleftmargin=5mm]
Estimated true RV = -15.006 km/s
True RV (simulated) = -15.000 km/s
Difference = -0.006 km/s

Zenith angle         = 0.050 (true value 0.050) km/s per degree
Temperature          = 0.022 (true value 0.020) km/s per °C
Humidity             = 0.009 (true value 0.010) km/s per %
Instrumental drift   = 0.520 (true value 0.500) km/s per unit
\end{Verbatim}

We now check the calibration. We compute the predictions and the residuals before and after the fit, and we measure their scatter with the root mean square (RMS).

\tutorialbarrier
\begin{codecell}

\end{codecell}
\begin{Verbatim}[fontsize=\footnotesize, xleftmargin=5mm]
RMS velocity scatter
Training set before calibration = 0.892 km/s
Training set after calibration = 0.103 km/s
Test set before calibration = 0.868 km/s
Test set after calibration = 0.099 km/s
Photon noise limit = 0.100 km/s
\end{Verbatim}

\subsection*{Parameter Importance}

We now ask which systematic effects contribute most to the error budget. A direct comparison of the coefficients can mislead, because the metadata have different units and scales. A coefficient of 0.05 km/s per degree can be a larger or a smaller effect than 0.5 km/s per unit of drift. It depends on how much each parameter varies.

To compare the effects, we scale each coefficient with the standard deviation of its metadata, \(w_j \operatorname{std}(x_j)\). This is the typical RV change that one standard deviation of parameter \(j\) causes. It has the unit km/s for all parameters, so we can compare the contributions directly. It shows which effects need the most attention in the design of the instrument and the planning of observations. The figure below shows the test residuals before and after the calibration on the left and this normalized importance on the right.

\tutorialbarrier
\begin{codecell}

\end{codecell}
\tutorialplot{\includegraphics[width=6.178in,height=0.75\textheight,keepaspectratio]{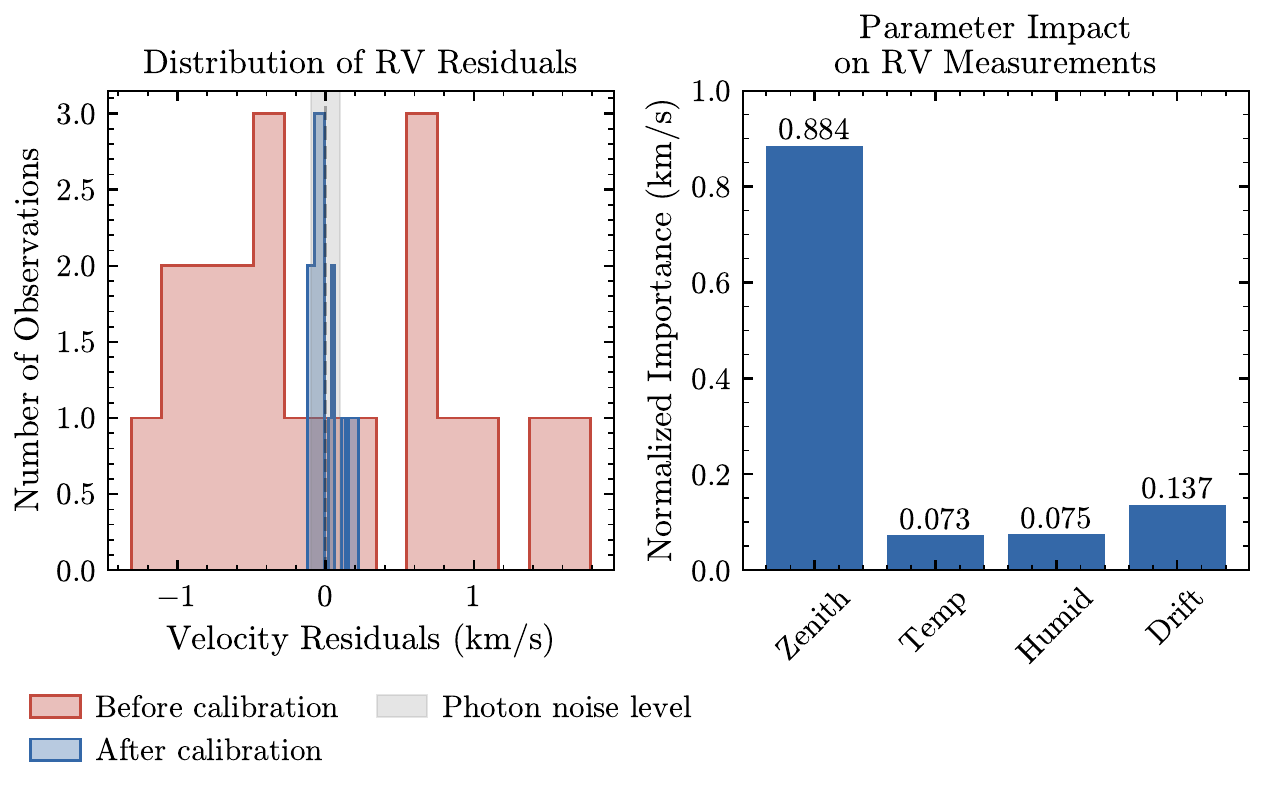}}%
  {\includegraphics[height=\tutorialheight]{tutorials/4b/fig02.pdf}}

\subsection*{Results from Single-Telescope Calibration}

Our calibration analysis gives three main results.

\begin{enumerate}
\def\labelenumi{\arabic{enumi}.}
\tightlist
\item
  \textbf{Successful systematic correction.} We reduced the RMS scatter from about 0.9 km/s to about 0.1 km/s, which is close to the photon noise limit. This improvement of nearly an order of magnitude turns unusable data into data that science can use.
\item
  \textbf{Accurate parameter estimates.} The estimated true RV differs from the true value by only 0.006 km/s. The calibration coefficients match their true values closely, so the joint estimation of all parameters works.
\item
  \textbf{Dominant systematic effect.} The normalized importance shows that the zenith angle dominates our error budget, and the drift comes second. Careful scheduling of observations at small zenith angles, and then mechanical stability, would improve future measurements most.
\end{enumerate}

This single-telescope calibration shows that linear regression can correct systematic errors.

\section*{Statistical Power and Sample Size}

Our analysis used 100 observations for a good calibration. We now ask how many observations we need. The relation between the sample size and the parameter uncertainty matters for an efficient scheduling of the telescope. Too few observations give a poor calibration. Too many observations waste telescope time that we can use for science targets.

The bootstrap lets us explore this trade-off (Section 3.7). We resample our data with replacement, which gives the uncertainty of the recovered RV. We then see how this uncertainty scales with the number of observations. For each number we simulate 20 datasets and average their uncertainties, because the uncertainty also depends on the particular values of the metadata.

\tutorialbarrier
\begin{codecell}

\end{codecell}

We plot the bootstrap distribution of the RV for one dataset with \(N = 100\), and the uncertainty against the number of observations.

\tutorialbarrier
\begin{codecell}

\end{codecell}
\tutorialplot{\includegraphics[width=6.375in,height=0.75\textheight,keepaspectratio]{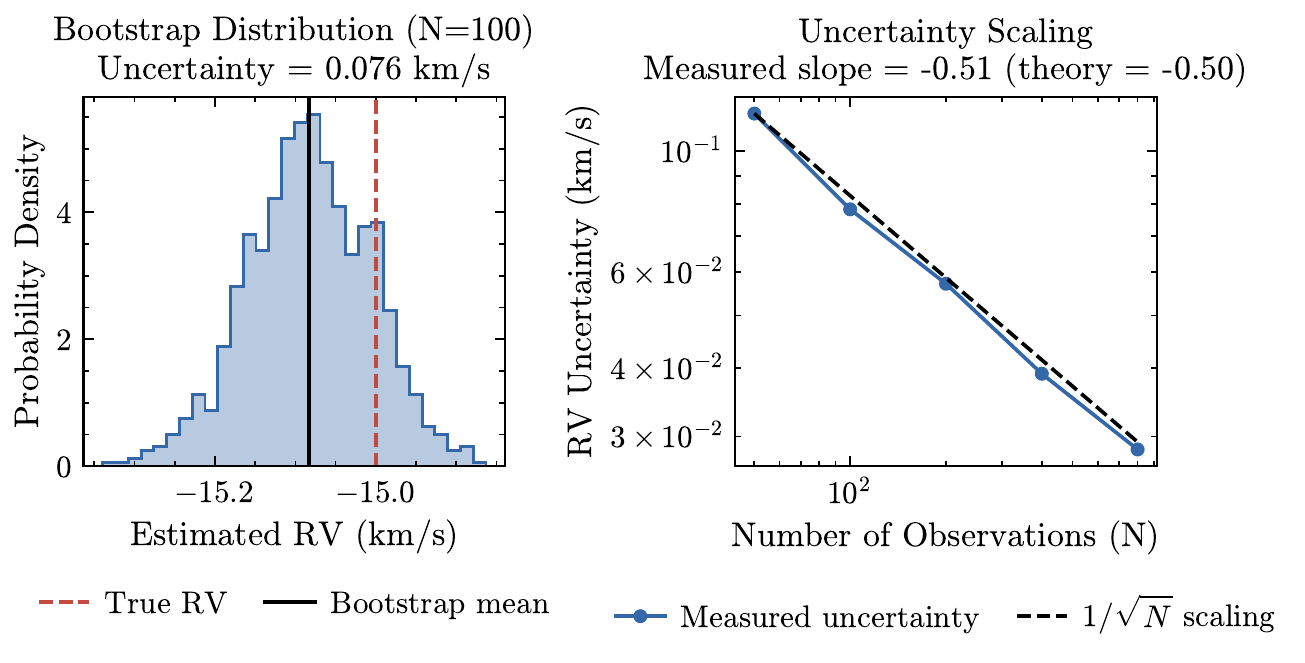}}%
  {\includegraphics[height=\tutorialheight]{tutorials/4b/fig03.pdf}}

\subsection*{Results from Bootstrap Analysis}

The bootstrap analysis shows that the uncertainty of the fitted RV follows the classical scaling law. The left panel shows the distribution of the RV estimates from 1000 bootstrap samples of one dataset with \(N = 100\) observations. The distribution is approximately Gaussian, and its width is the uncertainty of the fitted RV. It is centered on the fit to the sample, so it shows the spread of the estimate and not its offset from the true RV.

The right panel shows the effect of a larger sample, with the uncertainty averaged over the 20 datasets of each size. The measured scaling exponent of about \(-0.5\) is close to the theoretical prediction, which confirms that the uncertainty decreases as \(1/\sqrt{N}\). This has practical consequences.

\begin{itemize}
\tightlist
\item
  \textbf{Smaller gains.} An increase from 50 to 100 observations gives a large improvement, but an increase from 400 to 800 observations gives much smaller gains. This suggests a balance between the telescope time and the precision.
\item
  \textbf{Precision requirements.} If our science needs a precision of 0.05 km/s, the right panel shows that we need 200 to 400 observations. This helps us to plan the telescope time.
\end{itemize}

\section*{Calibration with Variable Measurement Uncertainties}

So far we have assumed that all observations have the same measurement uncertainty (\(\sigma = 0.1\) km/s). This homoscedastic noise assumption rarely holds in practice. Real observations suffer from changing conditions, such as the seeing, clouds and instrumental problems. These conditions cause heteroscedastic uncertainties, which differ from one observation to the next.

When the measurement uncertainties vary, our calibration strategy must adapt. Observations with small uncertainties should carry more weight in the calibration, and noisy measurements should carry less. This leads to weighted least squares, in which the influence of an observation is inversely proportional to its variance (Section 4.3). We now examine how variable uncertainties appear in realistic data.

\tutorialbarrier
\begin{codecell}

\end{codecell}

We plot the measurements with their uncertainties as error bars.

\tutorialbarrier
\begin{codecell}

\end{codecell}
\tutorialplot{\includegraphics[width=4.398in,height=0.75\textheight,keepaspectratio]{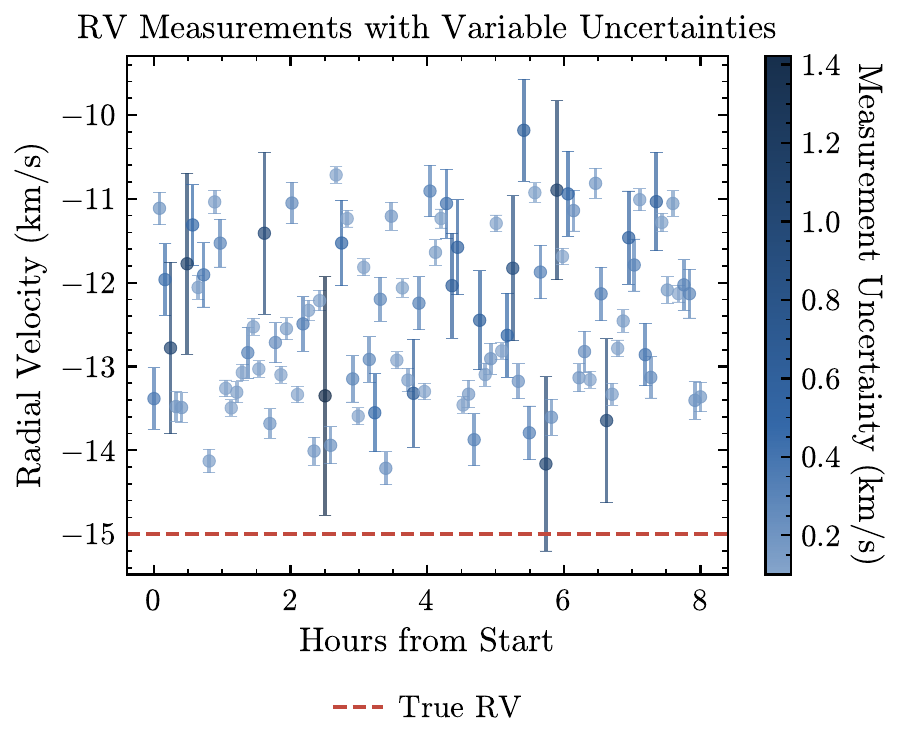}}%
  {\includegraphics[height=\tutorialheight]{tutorials/4b/fig04.pdf}}

\subsection*{Weighted Least Squares: Maximum Likelihood with Heteroscedastic Noise}

The plot above shows the problem. Some observations (the darker points) have uncertainties about ten times larger than our best measurements, or more. These observations should not influence the calibration as much as the precise ones. Weighted least squares gives a principled solution.

When the measurement uncertainties vary, the likelihood of \(N\) independent observations is

\[p(\mathbf{t}|\mathbf{X}, \mathbf{w}, \{\sigma_i^2\}) = \prod_{i=1}^{N} \frac{1}{\sqrt{2\pi\sigma_i^2}} \exp\left(-\frac{(t_i - \mathbf{w}^T\boldsymbol{\phi}(\mathbf{x}_i))^2}{2\sigma_i^2}\right),\]

where each observation \(i\) has its own uncertainty \(\sigma_i\). Minus twice the logarithm of the likelihood, without the constants, gives the weighted least squares objective

\[\chi^2 = \sum_{i=1}^{N} \frac{(t_i - \mathbf{w}^T\boldsymbol{\phi}(\mathbf{x}_i))^2}{\sigma_i^2}.\]

Each observation has the data weight \(\omega_i = 1/\sigma_i^2\), which differs from the model weights \(\mathbf{w}\). Observations with small uncertainties (small \(\sigma_i\)) get large weights, and noisy observations contribute less to the total \(\chi^2\). This is what we want, because precise measurements should dominate our calibration.

In matrix form, we write

\[\chi^2 = (\mathbf{t} - \boldsymbol{\Phi}\mathbf{w})^T \mathbf{S}^{-1} (\mathbf{t} - \boldsymbol{\Phi}\mathbf{w}),\]

where \(\mathbf{S} = \text{diag}(\sigma_1^2, \sigma_2^2, \ldots, \sigma_N^2)\) is the diagonal matrix of the measurement variances. The maximum likelihood solution is

\[\mathbf{w}_{\text{ML}} = (\boldsymbol{\Phi}^T\mathbf{S}^{-1}\boldsymbol{\Phi})^{-1}\boldsymbol{\Phi}^T\mathbf{S}^{-1}\mathbf{t}.\]

This weighted solution generalizes our previous result. When all \(\sigma_i\) are equal, \(\mathbf{S}^{-1}\) is a scalar multiple of the identity matrix \(\mathrm{Id}\), and we recover ordinary least squares. With varying uncertainties, the weight matrix \(\mathbf{S}^{-1}\) makes each observation contribute according to its precision.

We now implement this weighted calibration and compare it with an unweighted approach.

\tutorialbarrier
\begin{codecell}

\end{codecell}
\begin{Verbatim}[fontsize=\footnotesize, xleftmargin=5mm]
Parameter           True   Weighted  Unweighted  Weighted Error
------------------------------------------------------------
RV (km/s)        -15.000    -14.895     -14.853           0.105
Zenith (deg)       0.050      0.050       0.050           0.000
Temp (°C)          0.020      0.017       0.023           0.003
Humidity (%)       0.010      0.009       0.006           0.001
Drift              0.500      0.496       0.455           0.004

Total RMS parameter error
 Weighted = 0.105
 Unweighted = 0.148
 Improvement factor = 1.4x
\end{Verbatim}

The comparison shows that weighted least squares gives estimates closer to the true values for the RV, the humidity and the drift. The zenith angle and the temperature are about equally close. The overall improvement factor of 1.4x is modest in this example.

In real astronomical applications, the benefit of proper weighting becomes larger in the following cases.

\begin{itemize}
\tightlist
\item
  We combine data from different instruments with very different precisions.
\item
  We include observations taken under highly variable conditions.
\item
  We work near the precision limits, where small improvements matter.
\end{itemize}

\section*{Multi-telescope Network Calibration}

A network of telescopes observes the same standard stars. Each telescope has its own systematic effects, for example its own temperature coefficient, but the true velocity of each star is constant. This gives an opportunity for cross-calibration. For a network that observes several standard stars, each measurement depends on the following.

\begin{itemize}
\tightlist
\item
  Which star we observed, which sets the true RV
\item
  Which telescope made the observation, which sets the systematic effects
\item
  The metadata at the time of the observation
\end{itemize}

We model each observation \(i\) as

\[t_i = y_{\text{true},s(i)} + \mathbf{x}_i^T\mathbf{w}_{k(i)} + \varepsilon_i,\]

where \(s(i)\) is the star of observation \(i\) and \(k(i)\) is the telescope that made it. The vector \(\mathbf{w}_{k(i)}\) is the response of that telescope to the metadata.

This formulation describes the structure of the network. Each telescope keeps its own calibration coefficients, and all observations share the stellar RVs. We must construct a design matrix that encodes these relations.

We now explore this multi-telescope scenario with synthetic data for an observing campaign.

\tutorialbarrier
\begin{codecell}

\end{codecell}
\begin{Verbatim}[fontsize=\footnotesize, xleftmargin=5mm]
Number of telescopes = 5
Number of RV standard stars = 20
Observations per star = 50
Total observations = 1000
\end{Verbatim}

\subsection*{Constructing the Design Matrix}

The multi-telescope calibration problem needs a careful construction of the design matrix. In the single-telescope calibration we had one set of calibration coefficients. We now have the following parameters.

\begin{itemize}
\tightlist
\item
  \(N_{\text{tel}} \times 4\) telescope calibration parameters, because each telescope has 4 metadata coefficients
\item
  \(N_{\text{stars}}\) stellar RV parameters
\end{itemize}

Our parameter vector is

\[\mathbf{w} = [\mathbf{w}_1^T, \mathbf{w}_2^T, \ldots, \mathbf{w}_{N_{\text{tel}}}^T, y_{\text{true},1}, y_{\text{true},2}, \ldots, y_{\text{true},N_{\text{stars}}}]^T.\]

For each observation \(i\), we need a row of the design matrix \(\boldsymbol{\Phi}\) that does the following.

\begin{enumerate}
\def\labelenumi{\arabic{enumi}.}
\tightlist
\item
  It places the metadata \(\mathbf{x}_i\) in the columns of telescope \(k(i)\).
\item
  It places a 1 in the column of star \(s(i)\).
\item
  It places 0 everywhere else.
\end{enumerate}

Consider an example with 3 telescopes and 4 stars. Observation \(i\) is star 2 observed by telescope 3, with metadata \(\mathbf{x}_i = [x_1, x_2, x_3, x_4]\). The row of the design matrix is

\[\text{row } i \text{ of } \boldsymbol{\Phi} = [\underbrace{0,0,0,0}_{\text{tel 1}}, \underbrace{0,0,0,0}_{\text{tel 2}}, \underbrace{x_1,x_2,x_3,x_4}_{\text{tel 3}}, \underbrace{0,1,0,0}_{\text{stars}}].\]

This sparse structure shows which parameters affect each observation. Each observation depends on only 5 parameters, which are 4 telescope coefficients and 1 stellar RV. The full parameter vector can have hundreds of parameters.

Once we have constructed the matrix, we apply standard linear regression

\[\mathbf{w}_{\text{ML}} = (\boldsymbol{\Phi}^T\boldsymbol{\Phi})^{-1}\boldsymbol{\Phi}^T\mathbf{t}.\]

The solution gives all stellar RVs and all telescope calibration coefficients together, and it uses all observations of the network.

\tutorialbarrier
\begin{codecell}

\end{codecell}
\begin{Verbatim}[fontsize=\footnotesize, xleftmargin=5mm]
Design matrix shape = (1000, 40)

Matrix sparsity = 87.5%
\end{Verbatim}

We compare the estimated and the true coefficients and RVs in two plots, and we print the errors.

\tutorialbarrier
\begin{codecell}

\end{codecell}
\tutorialplot{\includegraphics[width=6.464in,height=0.75\textheight,keepaspectratio]{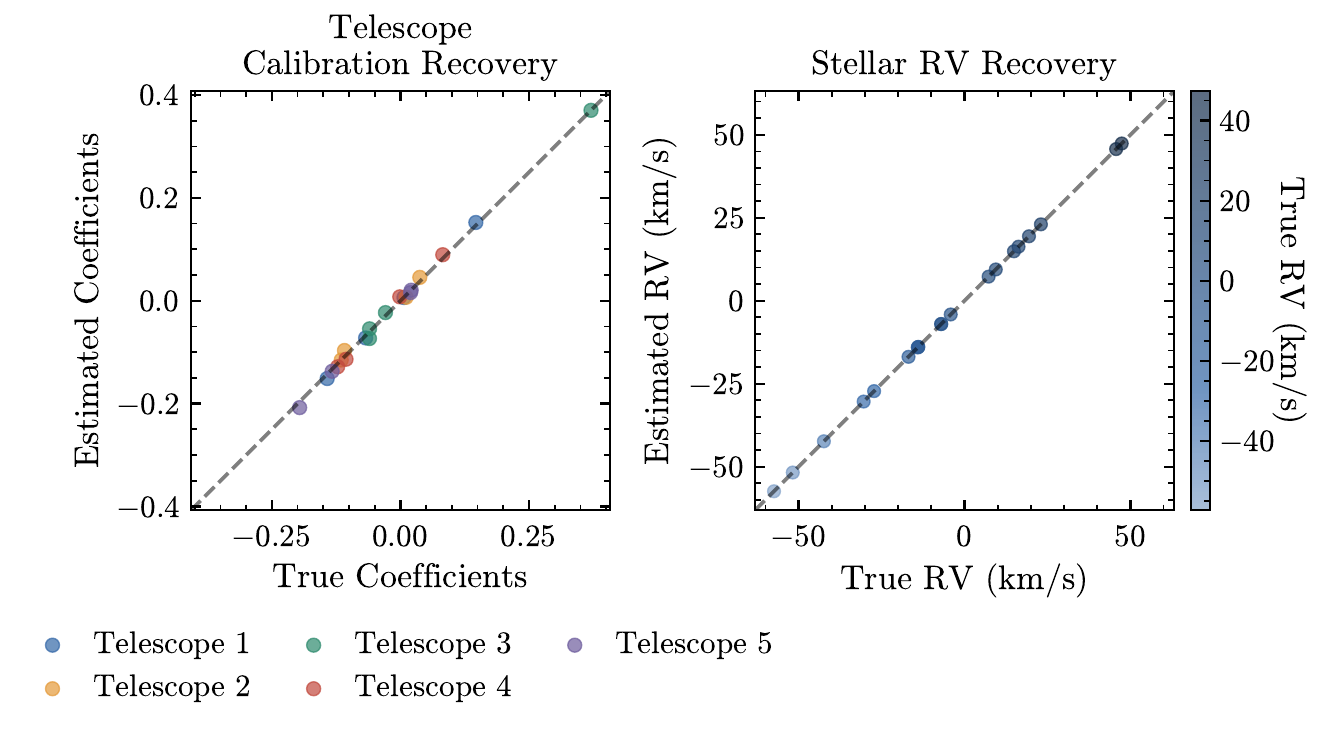}}%
  {\includegraphics[height=\tutorialheight]{tutorials/4b/fig05.pdf}}

We compute the RMS errors of the coefficients and of the stellar RVs, and we print them with the errors of each telescope.

\tutorialbarrier
\begin{codecell}

\end{codecell}
\begin{Verbatim}[fontsize=\footnotesize, xleftmargin=5mm]
Telescope coefficients, RMS error = 0.0071
Stellar RVs, RMS error = 0.014 km/s

Per-telescope coefficient errors (RMS)
 Telescope 1 = 0.0057
 Telescope 2 = 0.0076
 Telescope 3 = 0.0078
 Telescope 4 = 0.0077
 Telescope 5 = 0.0066
\end{Verbatim}

The results show that the multi-telescope calibration works. We recover both the telescope coefficients and the stellar RVs with good precision.

\begin{itemize}
\tightlist
\item
  \textbf{Telescope calibration.} In the left panel all telescope coefficients lie close to the diagonal, so they agree with the true values. We separated the systematic signature of each telescope from the stellar signals.
\item
  \textbf{Stellar RVs.} The right panel confirms that we determined the true velocities of all 20 standard stars well. The typical errors are about 0.01 to 0.02 km/s, much smaller than the measurement noise of 0.1 km/s.
\end{itemize}

This precision comes from the cross-constraints of the network. Suppose that telescope 1 and telescope 3 both observe star 7. After we account for the systematic effects of each telescope, their measurements must agree on the true RV of the star. These overlapping constraints strengthen the whole solution.

The per-telescope errors show another result. All telescopes reach a similar calibration quality, although their systematic effects differ. This suggests that the network approach does not need identical instruments. Different telescopes can calibrate each other if they observe enough common targets.

\section*{Summary}

This tutorial showed how linear regression is a basic tool for the calibration of astronomical measurements, with radial velocity standard stars as the example. We went from observations with one telescope to networks of telescopes, and we saw how to model, understand and correct the systematic instrumental effects.

\begin{enumerate}
\def\labelenumi{\arabic{enumi}.}
\tightlist
\item
  \textbf{Calibration as regression.} We turned the calibration into the estimation of several parameters together. Both the astrophysical quantities (the stellar RVs) and the instrumental systematic effects (the response of the telescopes to the metadata) became regression coefficients. The augmented design matrix handles this joint estimation.
\item
  \textbf{Maximum likelihood for different noise.} With homoscedastic uncertainties we used the standard least squares solution. With heteroscedastic noise, weighted least squares followed from maximum likelihood, so each observation contributes according to its precision.
\item
  \textbf{Statistical scaling and experimental design.} The bootstrap confirmed that the uncertainty of the fitted RV scales as \(1/\sqrt{N}\). This gives quantitative guidance for the planning of observations.
\item
  \textbf{Design matrix construction.} The step from one telescope to many showed how the design matrix encodes the structure of the problem. The sparse block structure of the multi-telescope matrix records which parameters affect each observation.
\end{enumerate}

The techniques apply to other systematic effects that depend on measurable quantities. Examples are the dimming of starlight at large zenith angles and the drift of the wavelength scale of spectra with the temperature. The mathematics stays the same from two small telescopes to a global network.

\tutorialchapter{5}{Bayesian Linear Regression for Stellar Spectra}
\noindent\textit{Companion material for Chapter 5}

\section*{Introduction}

In Chapter 4 and Tutorials 4a and 4b, we explored linear regression through maximum likelihood estimation (MLE). We saw that the fit overfits when the model has many weights and that regularization with validation stars cures it. Regularization has limits, because it gives one best set of weights and one global error. It tells us little about the uncertainty of a prediction. This matters, because the confidence in a prediction is often as important as the prediction itself.

This tutorial shows how Bayesian linear regression, as developed in Chapter 5, gives a principled estimate of this uncertainty. We apply it to APOGEE stellar spectra. We first generalize the fit and the regularization to the known noise of each star. We then move to the full Bayesian treatment, which handles the different sources of uncertainty in our predictions.

The uncertainty of a prediction has two sources. The measurement noise is the aleatoric uncertainty, and more data cannot reduce it. Our imperfect knowledge of the weights is the epistemic uncertainty, and more data reduce it. We see how the Bayesian approach separates the two and gives the size of each.

\section*{Prerequisites and Learning Objectives}

\subsection*{Prerequisites}

This tutorial assumes the following.

\begin{itemize}
\tightlist
\item
  Python programming with NumPy arrays, matplotlib plots and basic numerical computing
\item
  Linear regression and its maximum likelihood solution (Chapter 4 and Tutorials 4a and 4b)
\item
  Gaussian distributions, Bayes' theorem and posterior distributions
\item
  Overfitting and regularization (Chapter 4 and Tutorial 4a)
\end{itemize}

\subsection*{Learning Objectives}

After this tutorial you can do the following.

\begin{itemize}
\tightlist
\item
  Go beyond point estimates and explain why regularization alone does not give the uncertainty of a prediction
\item
  Include measurement uncertainties that differ from star to star
\item
  Compute the posterior distribution of the weights of a linear regression
\item
  Explain how the prior affects predictions and uncertainties, and connect regularization to Bayesian priors
\item
  Implement the predictive distribution with its two parts, the measurement noise and the uncertainty of the weights
\item
  Use the coverage of the intervals to check whether the uncertainty estimates are calibrated
\end{itemize}

We begin by importing the libraries and setting the plot style.

\tutorialbarrier
\begin{codecell}

\end{codecell}

\section*{Loading the APOGEE Stellar Spectra Dataset}

We use stellar spectra from the Apache Point Observatory Galactic Evolution Experiment (APOGEE). It is an infrared spectroscopic survey that has observed hundreds of thousands of stars in the Milky Way. Each spectrum has thousands of flux measurements at different wavelengths. They carry information about the physical properties of the star, such as the temperature, the surface gravity, the chemical composition and the radial velocity.

In this tutorial we focus on determining stellar effective temperatures from spectra. Our linear model is an imperfect approximation of the complex physics that relates a spectrum to a temperature. The temperatures come from the APOGEE pipeline, and we call them the pipeline temperatures.

The file holds the subset of 3,000 spectra from APOGEE Data Release 17 (DR17) of Tutorial 4a, with a signal-to-noise ratio above 100. We truncate each spectrum to 2,000 pixels, which keeps enough spectral information to predict the temperature.

Tutorial 4a left two things open. It treated the noise of the temperatures as uniform, and it gave point predictions without uncertainties. To address both and to show the Bayesian treatment of uncertainty, we add artificial noise of known size to the pipeline temperatures. Known noise lets us check whether the predicted uncertainties are right. The size of the noise differs from star to star, as in real data and in the heteroscedastic case of Tutorial 4b. The Bayesian approach then gives more weight to the more reliable measurements.

The noise has a standard deviation \(\sigma_i\) for star \(i\). It rises linearly with the temperature of the star. We know \(\sigma_i\) exactly, because we simulate it. The formal uncertainties of the pipeline temperatures are much smaller than \(\sigma_i\) at this signal-to-noise ratio, so we ignore them. The noisy value is the measured temperature \(t_i\) that the model predicts.

The high signal-to-noise ratio also makes the noise in the fluxes, the inputs of our model, small, so we treat only the temperatures as noisy. We now load the data and add the noise.

\tutorialbarrier
\begin{codecell}

\end{codecell}
\begin{Verbatim}[fontsize=\footnotesize, xleftmargin=5mm]
Number of stars = 3000
Number of wavelength points = 2000
Wavelength range = 15168.1 - 15592.9 Å
Pipeline temperature range = 3585 - 6433 K
Uncertainty range = 50 - 200 K
\end{Verbatim}

\section*{Preparing the Design Matrix}

As in Tutorial 4a, each star is a row of the design matrix \(\boldsymbol{\Phi}\) with the features \(\boldsymbol{\phi}(\mathbf{x}_i)\) of its spectrum \(\mathbf{x}_i\). The features are the flux at each wavelength and a bias term. The design matrix then has dimensions \(N \times M\), which is \(3000 \times 2001\) for our 3000 stars, 2000 wavelength points and the bias term.

We split the stars at random into 60 percent fit stars, 20 percent validation stars and 20 percent test stars (Section 4.7). This lets us test the model and its uncertainty estimates. We split the uncertainties in the same way, because each temperature measurement has its own uncertainty. The fit stars set the weights. The validation stars choose the regularization strength, so the test stars stay unseen. The fit and validation stars together are the training stars.

\tutorialbarrier
\begin{codecell}

\end{codecell}
\begin{Verbatim}[fontsize=\footnotesize, xleftmargin=5mm]
Fit stars = 1800, validation stars = 600, test stars = 600
Features = 2001
\end{Verbatim}

\section*{Maximum Likelihood and Regularization with Known Noise}

We now generalize the maximum likelihood fit of Tutorial 4a and its regularization to the known noise of each star, the heteroscedastic case. With the known uncertainties \(\sigma_i\), the maximum likelihood solution is the weighted least squares solution of Tutorial 4b (Section 4.3)

\[\mathbf{w}_{\text{ML}} = (\boldsymbol{\Phi}^T\mathbf{S}^{-1}\boldsymbol{\Phi})^{-1}\boldsymbol{\Phi}^T\mathbf{S}^{-1}\mathbf{t},\]

where \(\mathbf{S} = \text{diag}(\sigma_1^2, \dots, \sigma_N^2)\) is the diagonal matrix of the measurement variances. With 2001 weights and 2400 training stars, this solution overfits. Ridge regression adds to the weighted loss a penalty on large parameter values. The loss function is

\[E(\mathbf{w}) = (\mathbf{t} - \boldsymbol{\Phi}\mathbf{w})^T\mathbf{S}^{-1}(\mathbf{t} - \boldsymbol{\Phi}\mathbf{w}) + \lambda\|\mathbf{w}\|^2,\]

and the regularized solution is

\[\mathbf{w}_{\text{MAP}} = (\boldsymbol{\Phi}^T\mathbf{S}^{-1}\boldsymbol{\Phi} + \lambda\mathrm{Id})^{-1}\boldsymbol{\Phi}^T\mathbf{S}^{-1}\mathbf{t}.\]

Here \(\lambda\) controls the strength of the regularization. This modification has a close link to Bayesian inference. It is equivalent to a zero-mean Gaussian prior on the weights with variance \(\eta^2\), where \(\lambda = 1/\eta^2\) because \(\mathbf{S}\) holds the measurement variances (Section 5.6). The minimizer is the maximum a posteriori (MAP) estimate, the peak of the posterior. This equivalence is our first hint that Bayesian methods give a principled approach to uncertainty quantification.

As in Tutorial 4a, we fit each value of \(\lambda\) on the fit stars. We compare the errors on the fit stars and on the validation stars. We measure the error with the normalized error, the RMSE of Tutorial 4a in units of the known uncertainty of each star

\[\text{normalized error} = \sqrt{\frac{1}{N}\sum_{i=1}^N\left(\frac{t_i - \mathbf{w}^T\boldsymbol{\phi}(\mathbf{x}_i)}{\sigma_i}\right)^2},\]

where \(N\) is the number of stars. It is close to 1 when the errors come only from the measurement noise.

\tutorialbarrier
\begin{codecell}

\end{codecell}

We plot the normalized errors against \(\lambda\).

\tutorialbarrier
\begin{codecell}

\end{codecell}
\tutorialplot{\includegraphics[width=4.399in,height=0.75\textheight,keepaspectratio]{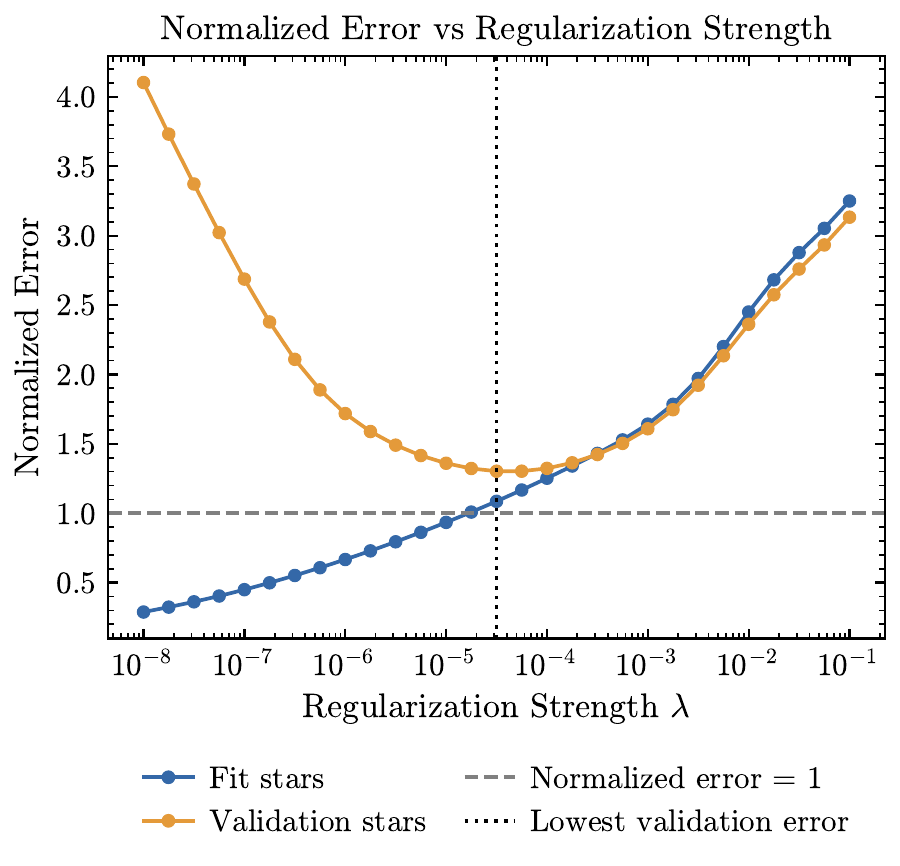}}%
  {\includegraphics[height=\tutorialheight]{tutorials/5/fig01.pdf}}
\begin{Verbatim}[fontsize=\footnotesize, xleftmargin=5mm]
Best λ on the validation stars = 3.2e-05
\end{Verbatim}

The plot shows the tradeoff of Tutorial 4a between overfitting and underfitting, in units of the measurement uncertainty. At small \(\lambda\) the model overfits, and its fit error is far below its validation error. At large \(\lambda\) the two errors agree but are large, so the model underfits. The validation error is lowest near \(\lambda = 3 \times 10^{-5}\), where the fit error is close to 1. Even there the validation error is above 1, so the errors do not come from the measurement noise alone. The values of \(\lambda\) are about \(10^{-4}\) times those of Tutorial 4a, because the data weights \(\omega_i = 1/\sigma_i^2\) are of order \(10^{-4}\) K\(^{-2}\).

The errors show the tradeoff, and the predictions show it too. We plot the predictions for the validation stars at four values of \(\lambda\), which include the lowest validation error.

\tutorialbarrier
\begin{codecell}

\end{codecell}
\tutorialplot{\includegraphics[width=6.173in,height=0.75\textheight,keepaspectratio]{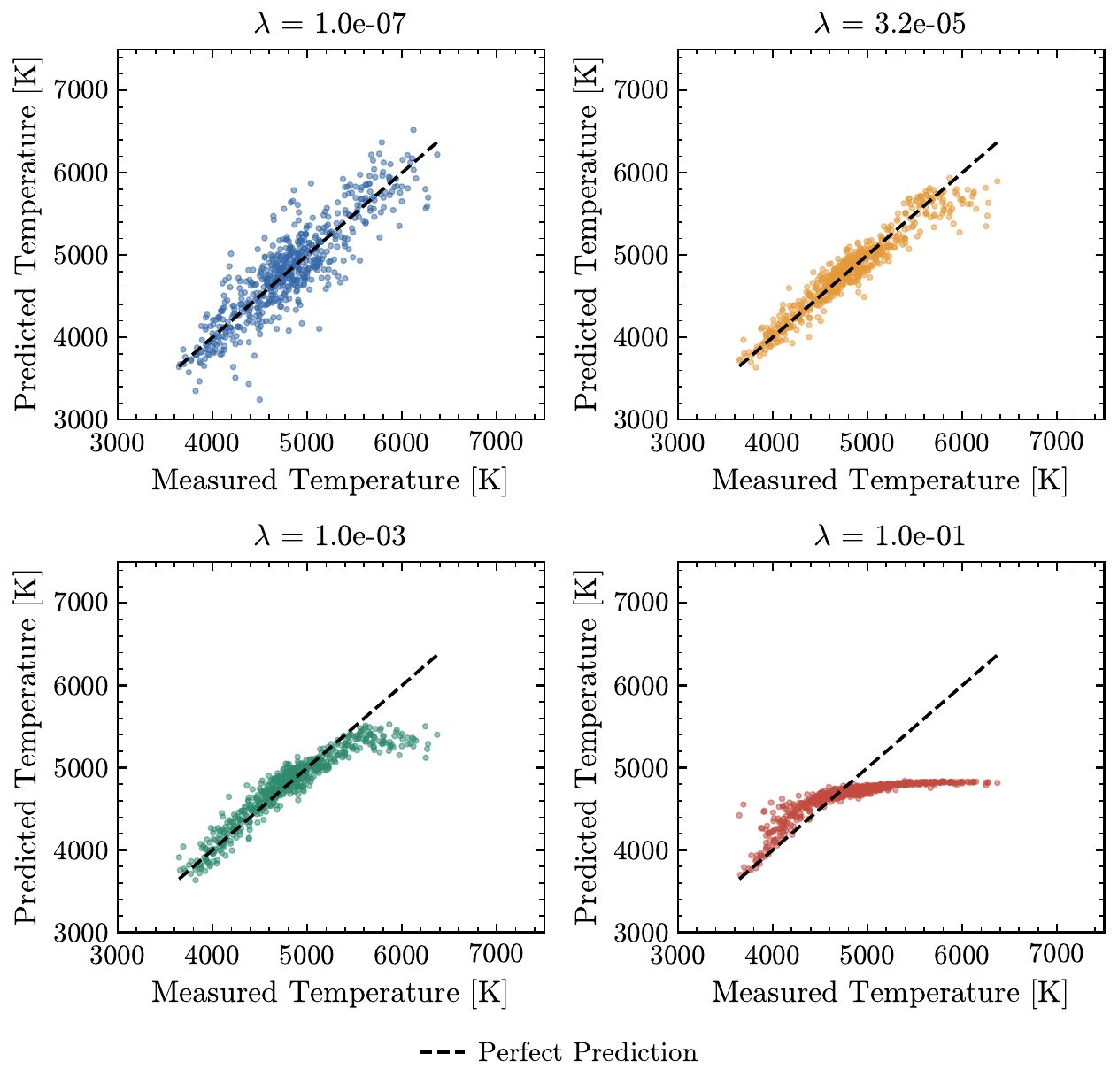}}%
  {\includegraphics[height=\tutorialheight]{tutorials/5/fig02.pdf}}

For the smallest \(\lambda\) the predictions scatter widely around the line, which is overfitting. At the lowest validation error they follow the line closely. For larger \(\lambda\) they flatten toward the mean temperature, which is underfitting. The hottest stars flatten first, because the sample has few of them. The data constrain them least, so the penalty pulls them toward the mean.

Regularization has three limits.

\begin{enumerate}
\def\labelenumi{\arabic{enumi}.}
\tightlist
\item
  \textbf{Choosing \(\lambda\).} We choose \(\lambda\) for the best prediction on the validation stars. This does not optimize the calibration of the uncertainty.
\item
  \textbf{Point estimates only.} We get one best set of weights and no uncertainty of the weights. We cannot tell which weights the data constrain.
\item
  \textbf{A single global error.} The validation error is one number for all predictions. It does not separate the measurement noise from the uncertainty of the weights.
\end{enumerate}

These limits motivate the Bayesian approach of Chapter 5. We treat the weights as random variables with explicit prior distributions.

\section*{Bayesian Linear Regression}

The Bayesian approach changes our perspective on regression. We do not look for single best values of the weights. We compute a full posterior distribution. It describes our uncertainty about every parameter. Bayes' theorem gives

\[p(\mathbf{w}|\mathbf{t}, \mathbf{X}, \mathbf{S}) \propto p(\mathbf{t}|\mathbf{X}, \mathbf{w}, \mathbf{S})p(\mathbf{w}),\]

where \(\mathbf{X}\) holds the spectra. The posterior is proportional to the product of our prior belief \(p(\mathbf{w})\) and the likelihood of the data. The likelihood includes the heteroscedastic measurement uncertainties

\[p(\mathbf{t}|\mathbf{X}, \mathbf{w}, \mathbf{S}) = \mathcal{N}(\mathbf{t}|\boldsymbol{\Phi}\mathbf{w}, \mathbf{S}),\]

where each observation has its own known variance \(\sigma_i^2\) on the diagonal of \(\mathbf{S}\).

The prior \(p(\mathbf{w})\) encodes our belief about the weights before we see any data. A common choice is a zero-mean isotropic Gaussian

\[p(\mathbf{w}) = \mathcal{N}(\mathbf{w}|\mathbf{0}, \mathbf{S}_0), \quad \mathbf{S}_0 = \eta^2\mathrm{Id}.\]

Isotropic means that every weight has the same variance \(\eta^2\) and no correlation with the other weights. The prior says that the weights are generally small, centered at zero. The variance \(\eta^2\) controls how far they can differ from zero. A larger \(\eta^2\) allows larger weights, and a smaller \(\eta^2\) enforces stronger regularization.

For a Gaussian likelihood and a Gaussian prior, the posterior is also Gaussian. This property is called conjugacy. The posterior is

\[p(\mathbf{w}|\mathbf{t}, \mathbf{X}, \mathbf{S}) = \mathcal{N}(\mathbf{w}|\mathbf{m}_N, \mathbf{S}_N),\]

where

\begin{itemize}
\tightlist
\item
  \(\mathbf{S}_N = (\mathbf{S}_0^{-1} + \boldsymbol{\Phi}^T\mathbf{S}^{-1}\boldsymbol{\Phi})^{-1}\) is the posterior covariance,
\item
  \(\mathbf{m}_N = \mathbf{S}_N\boldsymbol{\Phi}^T\mathbf{S}^{-1}\mathbf{t}\) is the posterior mean,
\item
  \(\mathbf{S}_0 = \eta^2\mathrm{Id}\) is the prior covariance.
\end{itemize}

The posterior mean is exactly the weighted regularized solution \(\mathbf{w}_{\text{MAP}}\) of the last section for \(\lambda = 1/\eta^2\). We now implement these expressions and see how the prior strength affects the posterior. We compute the posterior from all training stars for the prior variances \(\eta^2 = 10^3\), \(10^5\) and \(10^7\). We call them the strong, the medium and the weak prior, because a smaller \(\eta^2\) holds the weights closer to zero.

These prior variances correspond to \(\lambda = 10^{-3}\), \(10^{-5}\) and \(10^{-7}\) of the plot above. The prior variance is a hyperparameter, and the posterior does not set it. The validation stars favor \(\lambda \approx 3 \times 10^{-5}\), which is \(\eta^2 \approx 3 \times 10^4\), so the medium prior is the closest of the three.

\tutorialbarrier
\begin{codecell}

\end{codecell}

We plot the posterior mean of the weights with their 95 percent bands for the three priors. For a Gaussian distribution, 95 percent of the probability lies within 1.96 standard deviations of the mean. Each band is therefore the mean \(\pm\) 1.96 standard deviations.

\tutorialbarrier
\begin{codecell}

\end{codecell}
\tutorialplot{\includegraphics[width=6.500in,height=0.75\textheight,keepaspectratio]{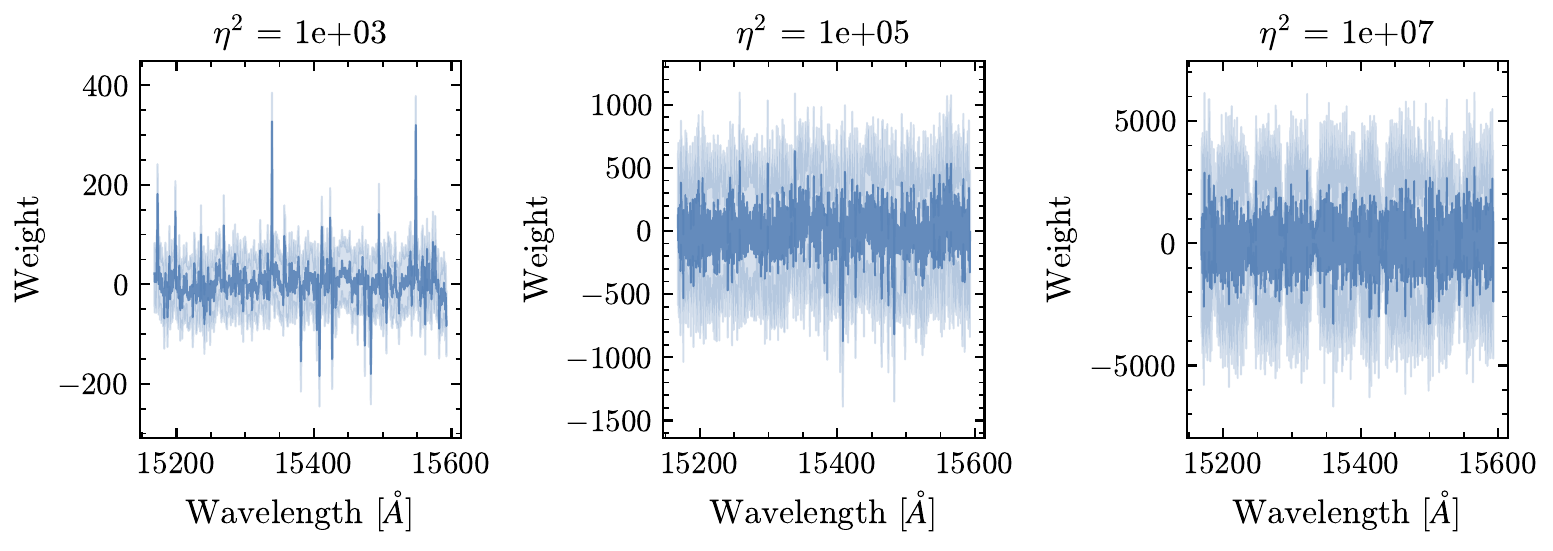}}%
  {\includegraphics[height=\tutorialheight]{tutorials/5/fig03.pdf}}

These plots show how the prior strength \(\eta^2\) affects the spectral weights and their uncertainties.

\textbf{Strong prior (\(\eta^2 = 10^3\)).} The weights are small, up to a few hundred, with narrow uncertainty bands. This prior holds the weights close to zero.

\textbf{Medium prior (\(\eta^2 = 10^5\)).} The weights reach more than a thousand, with larger uncertainty bands. The model can follow the relation between temperature and spectrum, and the prior still regularizes the weights.

\textbf{Weak prior (\(\eta^2 = 10^7\)).} The weights are large, up to several thousand, with wide uncertainty bands. This prior gives almost no regularization and approaches the unregularized solution. The very wide bands suggest that the model may overfit.

From the strong to the weak prior, the weights grow, the bands widen and the model gets more freedom. The middle panel shows a balance, with enough freedom for the fit and not too much uncertainty in the weights.

\section*{Making Predictions with Uncertainty}

The main benefit of Bayesian linear regression appears when we make predictions. We do not get a point estimate with an ad hoc error bar. We get a full predictive distribution. It includes the measurement noise and the uncertainty of the weights.

For a new stellar spectrum \(\mathbf{x}_*\) with output \(y_*\), we marginalize over all possible weights (Section 5.7)

\[p(y_*|\mathbf{x}_*, \mathcal{D}) = \int p(y_*|\mathbf{x}_*, \mathbf{w}, \sigma_*)p(\mathbf{w}|\mathcal{D})\,\mathrm{d}\mathbf{w}.\]

Here \(\mathcal{D}\) is the training data, and \(\sigma_*\) is the measurement uncertainty of the new observation. The integral averages the predictions over all plausible weights, weighted by their posterior probability. For a Gaussian likelihood and a Gaussian posterior, it is a Gaussian with the mean \(\boldsymbol{\phi}(\mathbf{x}_*)^T\mathbf{m}_N\). Its variance is the sum of two terms.

\begin{enumerate}
\def\labelenumi{\arabic{enumi}.}
\tightlist
\item
  \(\sigma_*^2\) is the measurement uncertainty of the new observation (aleatoric).
\item
  \(\boldsymbol{\phi}(\mathbf{x}_*)^T\mathbf{S}_N\boldsymbol{\phi}(\mathbf{x}_*)\) is the uncertainty from our imperfect knowledge of the weights (epistemic).
\end{enumerate}

We predict the test stars for the three priors.

\tutorialbarrier
\begin{codecell}
\begin{Verbatim}[commandchars=\\\{\},fontsize=\small]
\PYG{k}{def}\PYG{+w}{ }\PYG{n+nf}{predict\PYGZus{}with\PYGZus{}uncertainty}\PYG{p}{(}\FancyVerbBreakAfterBreak{}\PYG{n}{Phi\PYGZus{}new}\PYG{p}{,}\FancyVerbBreakAfterBreak{} \PYG{n}{sigma\PYGZus{}new}\PYG{p}{,}\FancyVerbBreakAfterBreak{} \PYG{n}{m\PYGZus{}N}\PYG{p}{,}\FancyVerbBreakAfterBreak{} \PYG{n}{S\PYGZus{}N}\PYG{p}{)}\PYG{p}{:}
\PYG{+w}{    }\PYG{l+s+sd}{\PYGZdq{}\PYGZdq{}\PYGZdq{}Compute the predictive mean and standard deviation for new spectra.\PYGZdq{}\PYGZdq{}\PYGZdq{}}

    \PYG{c+c1}{\PYGZsh{} Compute the predictive mean}
    \PYG{n}{pred\PYGZus{}mean} \PYG{o}{=}\FancyVerbBreakAfterBreak{} \PYG{n}{Phi\PYGZus{}new} \PYG{o}{@} \PYG{n}{m\PYGZus{}N}

    \PYG{c+c1}{\PYGZsh{} Compute the predictive variance, the sum of two terms}
    \PYG{c+c1}{\PYGZsh{} 1. Uncertainty of the weights (epistemic), phi\PYGZca{}T S\PYGZus{}N phi for each star}
    \PYG{n}{weight\PYGZus{}var} \PYG{o}{=}\FancyVerbBreakAfterBreak{} \PYG{n}{np}\PYG{o}{.}\FancyVerbBreakAfterBreak{}\PYG{n}{sum}\PYG{p}{(}\FancyVerbBreakAfterBreak{}\PYG{n}{Phi\PYGZus{}new} \PYG{o}{@} \PYG{n}{S\PYGZus{}N} \PYG{o}{*} \PYG{n}{Phi\PYGZus{}new}\PYG{p}{,}\FancyVerbBreakAfterBreak{} \PYG{n}{axis}\PYG{o}{=}\FancyVerbBreakAfterBreak{}\PYG{l+m+mi}{1}\PYG{p}{)}

    \PYG{c+c1}{\PYGZsh{} 2. Measurement noise of the new star (aleatoric)}
    \PYG{n}{noise\PYGZus{}var} \PYG{o}{=}\FancyVerbBreakAfterBreak{} \PYG{n}{sigma\PYGZus{}new}\PYG{o}{*}\PYG{o}{*}\PYG{l+m+mi}{2}

    \PYG{c+c1}{\PYGZsh{} Total predictive variance}
    \PYG{n}{pred\PYGZus{}std} \PYG{o}{=}\FancyVerbBreakAfterBreak{} \PYG{n}{np}\PYG{o}{.}\FancyVerbBreakAfterBreak{}\PYG{n}{sqrt}\PYG{p}{(}\FancyVerbBreakAfterBreak{}\PYG{n}{weight\PYGZus{}var} \PYG{o}{+}\FancyVerbBreakAfterBreak{} \PYG{n}{noise\PYGZus{}var}\PYG{p}{)}

    \PYG{k}{return} \PYG{n}{pred\PYGZus{}mean}\PYG{p}{,}\FancyVerbBreakAfterBreak{} \PYG{n}{pred\PYGZus{}std}
\end{Verbatim}
\end{codecell}

We now predict the test stars with the three posteriors, and we plot the predictions with their 95 percent intervals.

\tutorialbarrier
\begin{codecell}

\end{codecell}
\tutorialplot{\includegraphics[width=6.500in,height=0.75\textheight,keepaspectratio]{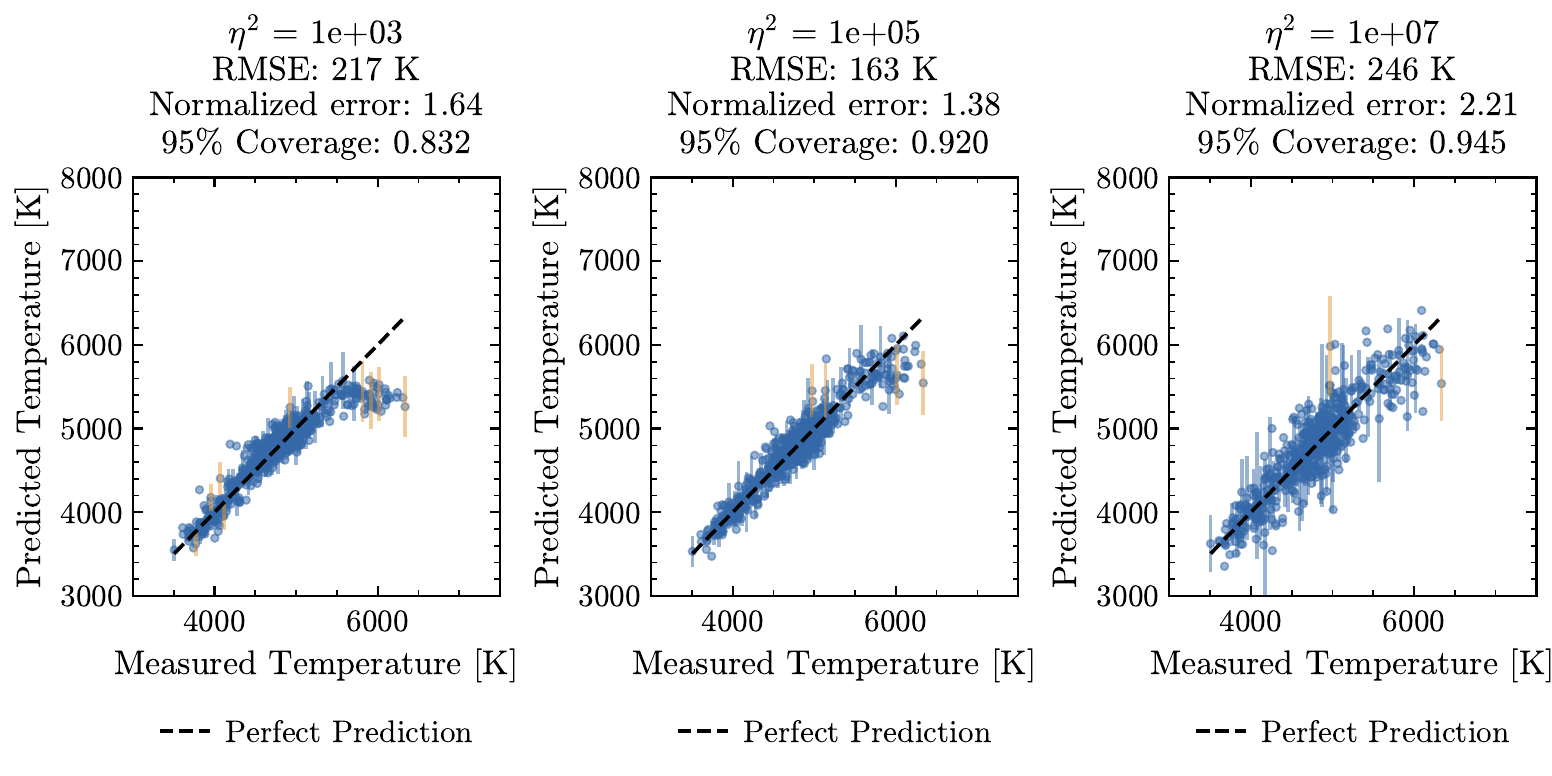}}%
  {\includegraphics[height=\tutorialheight]{tutorials/5/fig04.pdf}}

These results show how the prior strength affects the prediction quality. The title of each panel gives the RMSE, the normalized error and the coverage on all test stars. The bars show the intervals of 50 stars. A blue bar touches the dashed line of perfect prediction, so its interval contains the measured temperature. An orange bar does not.

To evaluate the uncertainty estimates, we use the coverage. This is the fraction of the measured temperatures that fall inside their 95 percent prediction intervals. For calibrated uncertainties, we expect a coverage of 95 percent. A lower coverage means that the uncertainties are underestimated and the model is overconfident. A higher coverage means that the estimates are too conservative.

\textbf{Strong prior (\(\eta^2 = 10^3\)).} The errors are higher than for the medium prior, and the coverage is only about 83 percent. This prior constrains the model too much, so it cannot follow the relation between temperature and spectrum well. The predictions of the hottest stars lie far below the line, and their bars do not reach it. The uncertainties are underestimated, because the strong prior fixes the weights tightly and the uncertainty of the weights is small.

\textbf{Medium prior (\(\eta^2 = 10^5\)).} This prior has the lowest errors, and the coverage is about 92 percent. The model can follow the data, but the prior still shrinks the weights, so the intervals are a little too short.

\textbf{Weak prior (\(\eta^2 = 10^7\)).} The coverage is close to 95 percent, but the errors are higher than for the medium prior. The weak prior leaves the weights free, so their uncertainty is large and the intervals are wide. The model overfits, and the wide intervals show how uncertain the weights are.

The two parts of the uncertainty behave differently. The measurement noise is the same for the three priors, and more data cannot reduce it. The uncertainty of the weights grows with the prior variance, and more data reduce it. The medium prior has the lowest errors, and the weak prior has the best coverage. The prior width is a hyperparameter that the validation stars set, as for regularization. Such a split helps us to decide whether to take more data or to reduce the measurement noise.

\section*{Beyond the Two Parts of the Uncertainty}

The medium prior has the lowest error, but its coverage is 92 percent. The 1.96 standard deviation intervals should contain 95 percent of the measured temperatures if the uncertainty is well calibrated. The two parts explain the uncertainty only if the linear model and the prior are right. A linear model may not describe the relation fully, as the small gain from quadratic features in Tutorial 4a showed. A prior that shrinks the weights also pulls the predictions away from the true relation. This misfit gives an error that neither part contains, so the intervals are too short.

A common remedy in astronomy is a scatter \(\sigma_{\text{model}}\) for the misfit of the model, like the intrinsic scatter of Section 4.5. Its variance adds to the variance of each star

\[\sigma_*^2 + \boldsymbol{\phi}(\mathbf{x}_*)^T\mathbf{S}_N\boldsymbol{\phi}(\mathbf{x}_*) + \sigma_{\text{model}}^2.\]

The scatter is one more hyperparameter, and we estimate it on the validation stars in three steps.

\begin{enumerate}
\def\labelenumi{\arabic{enumi}.}
\tightlist
\item
  Compute the posterior of the fit stars at the \(\lambda\) with the lowest validation error.
\item
  Predict the validation stars with the two parts.
\item
  Take the part of the squared validation error that the two parts do not explain as \(\sigma_{\text{model}}^2\).
\end{enumerate}

We then add \(\sigma_{\text{model}}^2\) to the variance of every test star. We compare the coverage of the three priors without and with it.

\tutorialbarrier
\begin{codecell}

\end{codecell}
\begin{Verbatim}[fontsize=\footnotesize, xleftmargin=5mm]
Model scatter = 92 K

η²  Without scatter With scatter
1e+03   0.832           0.893
1e+05   0.920           0.960
1e+07   0.945           0.968
\end{Verbatim}

The model scatter is 92 K. It widens every interval but does not change the predictions, so the errors stay the same. It raises the coverage of the three priors from 0.83, 0.92 and 0.945 to 0.89, 0.96 and 0.97. The medium prior now has both the lowest error and the coverage closest to 95 percent.

\section*{Summary}

This tutorial showed how Bayesian linear regression gives each prediction an uncertainty. We added simulated heteroscedastic noise to the APOGEE pipeline temperatures and predicted them from the spectra.

\begin{enumerate}
\def\labelenumi{\arabic{enumi}.}
\tightlist
\item
  \textbf{Regularization with known noise.} The weighted maximum likelihood fit overfits, and a penalty on the weights cures it. We chose its strength on validation stars. The regularized fit is the MAP estimate of a Gaussian prior, and it gives point estimates only.
\item
  \textbf{The posterior.} The Bayesian treatment gives a full distribution of the weights. The prior variance sets how much freedom the weights have, and we choose it on validation stars too.
\item
  \textbf{The predictive distribution.} Each prediction has two parts, the measurement uncertainty of the star (aleatoric) and the uncertainty of the weights (epistemic). The prior strength sets the second part, and a prior that is too strong gives intervals that are too short.
\item
  \textbf{The misfit.} Both parts assume that the linear model and the prior are right. A model scatter, set on the validation stars, covers the misfit that they leave.
\end{enumerate}

The coverage on the test stars checks whether the uncertainties are right. Calibrated uncertainties are as important as the predictions themselves.

\tutorialchapter{6}{Deming Regression for the Tully-Fisher Relation}
\noindent\textit{Companion material for Chapter 6}

\section*{Introduction}

In Chapters 4 and 5 we developed techniques to model the relation between two variables when the dependent variable has uncertainties. When the dependent variable \(y\) has measurement errors, we can handle them with weighted least squares or Bayesian inference. But the independent variable \(x\) can carry measurement uncertainties too.

The Tully-Fisher relation is an astronomical laboratory for this problem. This scaling law links the rotation velocity of a galaxy to its total baryonic mass, the mass of its stars and gas. It is a cosmic distance indicator and a probe of the physics of galaxy formation. Both quantities carry uncertainties. The rotation velocities come from spectroscopic observations, and the masses come from photometric measurements.

Ordinary least squares (OLS) is the standard straight-line fit of Chapter 4. It minimizes the sum of the squared vertical distances between the points and the line. This is the maximum likelihood estimate when the noise in \(y\) is Gaussian with constant size. OLS treats the independent variable as exact, so it ignores its uncertainties, and this causes a problem.

The measurement errors in the velocity make the observed distribution of velocities wider than the true distribution. Think of it this way. If the measured velocity of every galaxy includes random noise, the cloud of points spreads out horizontally on our plot. This widening has no matching effect on the vertical spread, so the relation appears shallower than it truly is.

This underestimate of the slope is called \textbf{attenuation bias} or \textbf{regression dilution}. The word ``attenuation'' means weakening, because the measurement uncertainties weaken the observed relation. This is not a random error that averages out with more data. With infinitely many galaxies, we still underestimate the slope by a predictable amount if we ignore the velocity uncertainties.

In this tutorial, we use real galaxy data from the SPARC (Spitzer Photometry and Accurate Rotation Curves) survey (\href{https://ui.adsabs.harvard.edu/abs/2016AJ....152..157L/abstract}{Lelli et al.~2016}). First, we show attenuation bias in simulations and compare it with its mathematical form. The bias depends on the ratio of the measurement uncertainty to the intrinsic spread of the data. Then we implement Deming regression, which treats the true velocities as latent variables and infers them together with the slope. Finally, we compare OLS and Deming regression for the galaxies.

\section*{Prerequisites and Learning Objectives}

\subsection*{Prerequisites}

This tutorial assumes the following.

\begin{itemize}
\tightlist
\item
  Python programming with NumPy arrays and matplotlib plots
\item
  Linear regression, in particular maximum likelihood estimation and residuals
\item
  Summary statistics and transformations of random variables
\item
  The bootstrap, a resampling method to estimate uncertainties
\end{itemize}

\subsection*{Learning Objectives}

After this tutorial you can do the following.

\begin{itemize}
\tightlist
\item
  Show how measurement uncertainties in the independent variable bias an OLS fit
\item
  Fit a line with errors in both variables, with an analytical and a numerical Deming regression
\item
  Compute the attenuation factor and predict the size of the bias from the measurement uncertainties
\item
  Extend Deming regression to uncertainties that differ from point to point
\item
  Compare OLS and Deming regression on real data
\end{itemize}

We begin by importing the libraries and setting the plot style.

\tutorialbarrier
\begin{codecell}

\end{codecell}

\section*{The SPARC Galaxy Dataset}

The SPARC database is a laboratory for the study of the Tully-Fisher relation. It has 175 nearby disk galaxies. They have Spitzer Space Telescope photometry at 3.6 \(\mu\)m. Their rotation curves come from radio observations of neutral hydrogen (HI) and optical observations of ionized hydrogen (H\(\alpha\)). This tutorial uses the 85 spiral galaxies, of types Sa to Sd, with a flat outer rotation curve.

Near-infrared photometry traces the stellar mass well, because dust and recent star formation affect it little. SPARC takes the stellar mass from the 3.6 \(\mu\)m luminosity and the gas mass from the HI observations. The baryonic mass is their sum.

For this tutorial, we extracted four quantities from SPARC. They are the logarithms of the flat rotation velocity \(V\) in km/s and of the baryonic mass \(M_b\) in units of \(10^9\) solar masses. The other two are the measurement uncertainty of each logarithm, in dex. One dex is a factor of 10.

These logarithmic quantities suit the Tully-Fisher relation, which is a power law between velocity and mass.

Velocity measurements face several difficulties. The tilt of the disk, called the inclination, is hard to measure and becomes very uncertain for face-on galaxies. Non-circular motions of the gas may not trace the gravitational potential exactly.

The uncertainties reported by SPARC are much smaller for the mass than for the velocity, as the data below show. The mass uncertainty comes from the 3.6 \(\mu\)m photometry alone, which is precise to about 1\% for these galaxies. It excludes the larger uncertainties of the distance and of the conversion from light to mass. The independent variable, the velocity, thus carries the larger reported uncertainty. This makes the Tully-Fisher relation a good case for Deming regression, which uses the uncertainties of both variables. We now load the data.

\tutorialbarrier
\begin{codecell}

\end{codecell}
\begin{Verbatim}[fontsize=\footnotesize, xleftmargin=5mm]
Median velocity uncertainty = 0.016 dex
Median mass uncertainty = 0.003 dex
\end{Verbatim}

\section*{Visualizing the Tully-Fisher Relation}

The Tully-Fisher relation states that the rotation velocity \(V\) of a galaxy is related to its baryonic mass \(M_b\) through a power law

\[M_b \propto V^{\gamma},\]

where empirical studies typically find \(\gamma \approx 3\) to 4. This power law follows from the virial theorem, which links speed and mass in a bound system, and from the dynamics of rotating disks. For a galaxy in dynamical equilibrium, the rotation velocity at a given radius depends on the enclosed mass. For the characteristic velocity and the total mass, a power law arises. When we take the logarithm of both sides, the power law becomes linear

\[\log M_b = \gamma \log V + c.\]

The input is \(x = \log V\), the output is \(y = \log M_b\), \(\gamma\) is the slope and \(c\) is the intercept. We now plot the galaxies with their error bars, and the errors of both variables relative to the spread of each variable.

\tutorialbarrier
\begin{codecell}

\end{codecell}
\tutorialplot{\includegraphics[width=6.174in,height=0.75\textheight,keepaspectratio]{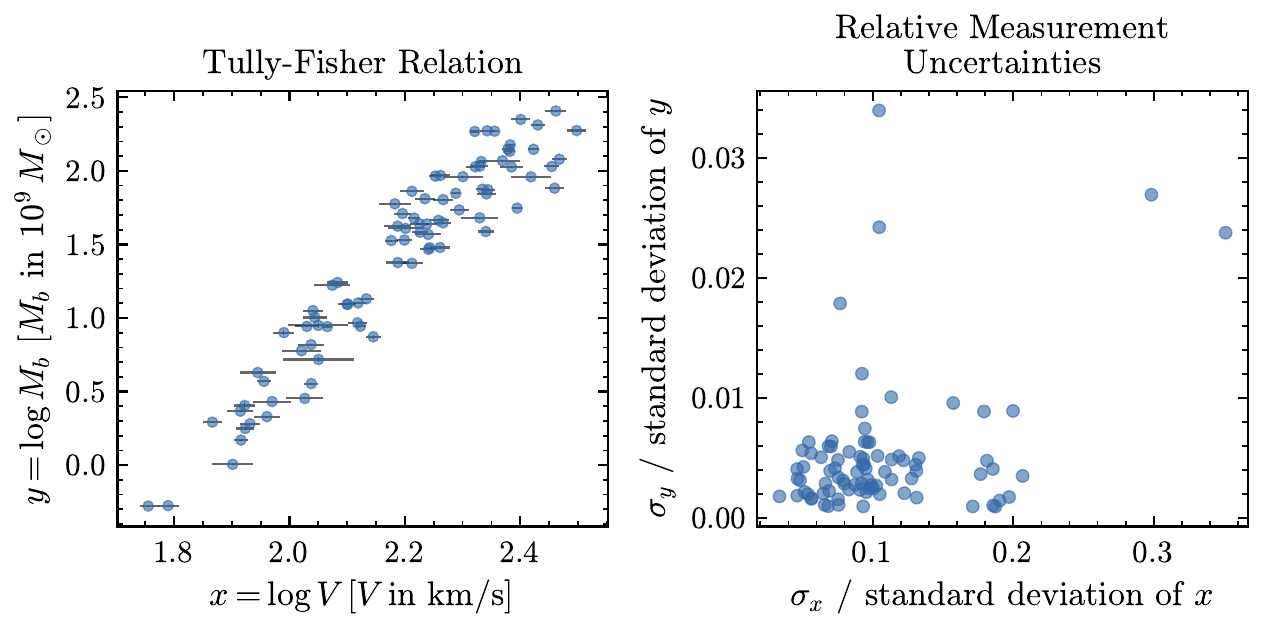}}%
  {\includegraphics[height=\tutorialheight]{tutorials/6/fig01.pdf}}
\begin{Verbatim}[fontsize=\footnotesize, xleftmargin=5mm]

Median relative uncertainties, in units of the standard deviation
X-axis (velocity) = 9.3%
Y-axis (mass) = 0.4%
\end{Verbatim}

The left panel shows the error bars in \(x\) and \(y\), and the error bars in \(y\) are too small to see. The right panel shows that the measurement uncertainties of the two variables are not symmetric. The velocity uncertainties are about 9\% of the standard deviation of \(x\), and the mass uncertainties are about 0.4\% of that of \(y\). The independent variable thus carries much more uncertainty than the dependent variable, and OLS treats it as exact.

\section*{Simulating Attenuation Bias}

To understand how measurement uncertainties affect regression, we create a controlled simulation. We generate data from a known linear relation, add measurement errors and observe how OLS underestimates the true slope.

In our simulation, the true data points follow \(y_{\text{true}} = \gamma x_{\text{true}}\) with \(\gamma = 3\). This is our ground truth. We then add Gaussian noise to both \(x\) and \(y\) to simulate measurement errors, as real astronomical observations contain uncertainties. We fit a line to these noisy observations with OLS and compare the fitted slope with the true slope.

As in Tutorial 4a, the OLS solution is the analytical maximum likelihood solution of Section 4.4, \(\mathbf{w}_{\text{ML}} = (\boldsymbol{\Phi}^T\boldsymbol{\Phi})^{-1}\boldsymbol{\Phi}^T\mathbf{t}\). Consider one input and one output, with the data centered on their means, so that the line passes through the origin. Then \(\boldsymbol{\Phi}\) is the single column of the observed \(x\), and \(\mathbf{t}\) is the column of the observed \(y\). The matrices become sums, and the solution is the slope

\[\hat{\gamma} = \frac{\sum_i x_{\text{obs},i} y_{\text{obs},i}}{\sum_i x_{\text{obs},i}^2}.\]

The code uses this formula for every OLS slope.

The derivation of OLS assumes that the independent variable \(x\) is known exactly and that only \(y\) has measurement errors. When these assumptions hold, OLS recovers the true parameters. When \(x\) has measurement errors, we treat noisy \(x\)-values as the true values, and the OLS slope is too low.

In Section 6.3 we derived the expected OLS slope when \(x\) has measurement errors

\[\mathbb{E}[\hat{\gamma}] = \alpha\gamma,\]

where the attenuation factor is

\[\alpha = \frac{1}{1 + (\sigma_x/\sigma_{\text{range}})^2}.\]

Here \(\sigma_x\) is the measurement uncertainty in \(x\), and \(\sigma_{\text{range}}\) is the intrinsic spread, the standard deviation, of the true \(x\)-values. The measurement uncertainties raise the observed variance of \(x\) from \(\sigma_{\text{range}}^2\) to \(\sigma_{\text{range}}^2 + \sigma_x^2\). They leave the covariance of \(x\) and \(y\) unchanged. The ratio of these two quantities gives the attenuated slope. We now simulate a data set for each of three levels of \(x\)-uncertainty and fit it with OLS.

\tutorialbarrier
\begin{codecell}

\end{codecell}

We plot each data set with the true line and the OLS fit. The title shows the attenuation factor that the formula predicts.

\tutorialbarrier
\begin{codecell}

\end{codecell}
\tutorialplot{\includegraphics[width=6.500in,height=0.75\textheight,keepaspectratio]{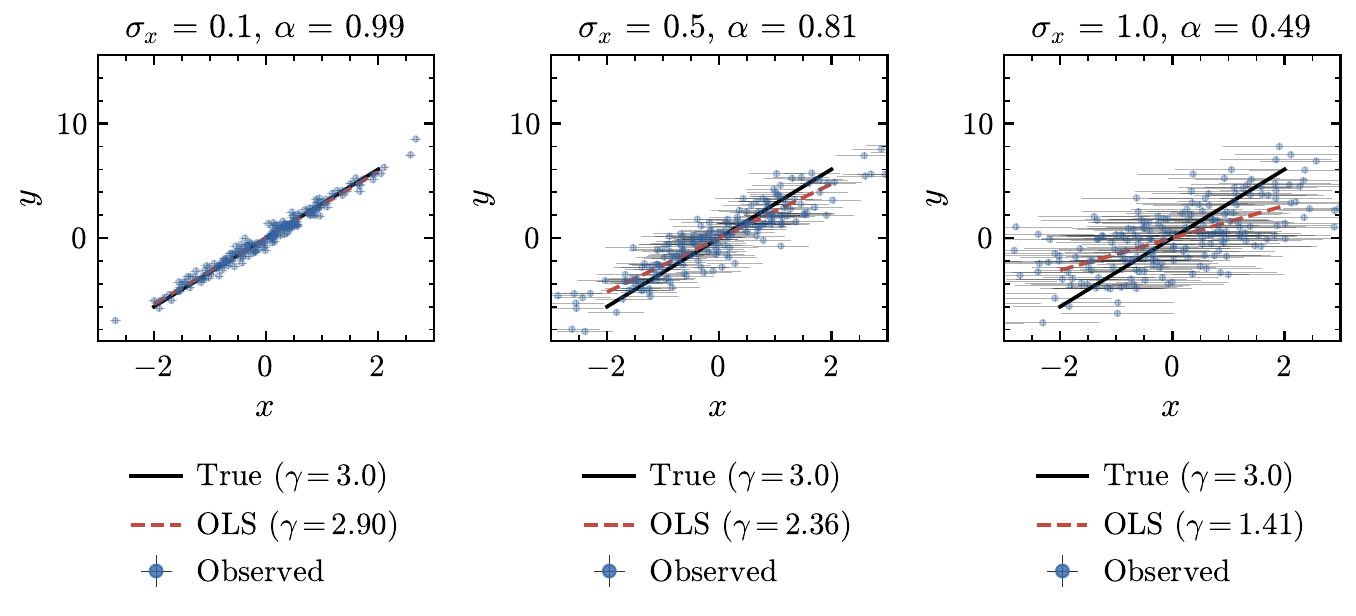}}%
  {\includegraphics[height=\tutorialheight]{tutorials/6/fig02.pdf}}

The simulation shows attenuation bias in action. As the measurement uncertainty in \(x\) increases, the OLS slope underestimates the true relation more and more. The attenuation factor \(\alpha\) measures this bias. The attenuation factor is 0.99 for a ratio \(\sigma_x/\sigma_{\text{range}}\) of 0.1 and 0.92 for a ratio of 0.3. When \(\sigma_x\) equals \(\sigma_{\text{range}}\), we recover only half of the true slope. The bias depends on the ratio \(\sigma_x/\sigma_{\text{range}}\) and not on the sample size, so more data do not remove it.

\section*{The Asymmetry of Measurement Uncertainties}

The attenuation factor depends only on the uncertainty of the independent variable, so only this uncertainty causes attenuation bias. To understand this asymmetry, recall that OLS minimizes the squared residuals in the \(y\)-direction. When \(y\) has measurement errors, they add to the residual variance, which the OLS fit already includes. When \(x\) has errors, we use incorrect \(x\)-values to predict \(y\), and this flattens the observed relation.

We show this asymmetry with a simulation that has errors in \(y\) only. We can compare it with the earlier simulation, where \(x\) had errors.

\tutorialbarrier
\begin{codecell}

\end{codecell}

We plot each data set with the true line and the OLS fit.

\tutorialbarrier
\begin{codecell}

\end{codecell}
\tutorialplot{\includegraphics[width=6.500in,height=0.75\textheight,keepaspectratio]{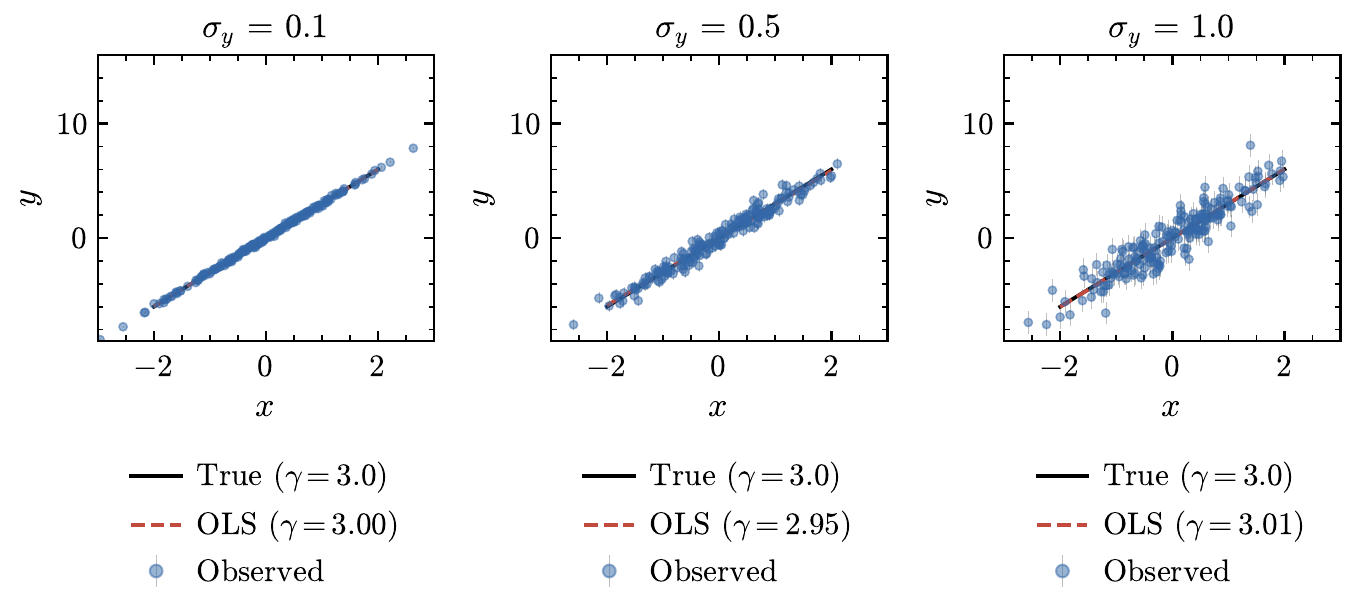}}%
  {\includegraphics[height=\tutorialheight]{tutorials/6/fig03.pdf}}

Even with a large uncertainty in \(y\) (\(\sigma_y = 1.0\)), OLS recovers the true slope almost perfectly. The error bars show the measurement uncertainty in \(y\), yet the fitted slope has no bias. In our earlier simulation, uncertainties in \(x\) of the same size caused a large attenuation.

\section*{Implementing Deming Regression}

Now that we understand attenuation bias, we implement Deming regression to correct for it. Deming regression uses the uncertainties of both variables. It treats the true \(x\)-values as latent variables and infers them together with the slope.

For homoscedastic errors, which are constant uncertainties, Section 6.7 gives the Deming regression solution

\[\hat{\gamma} = \frac{S_{yy} - \lambda S_{xx} + \sqrt{(S_{yy} - \lambda S_{xx})^2 + 4\lambda S_{xy}^2}}{2S_{xy}},\]

where \(S_{xx} = \sum_i x_{\text{obs},i}^2\), \(S_{yy} = \sum_i y_{\text{obs},i}^2\) and \(S_{xy} = \sum_i x_{\text{obs},i} y_{\text{obs},i}\) are the summary statistics of the centered data, and \(\lambda = \sigma_y^2/\sigma_x^2\) is the variance ratio.

The formula reduces to the OLS slope \(S_{xy}/S_{xx}\) when \(\sigma_x \to 0\), which makes \(\lambda \to \infty\). It has a geometric interpretation. OLS minimizes the sum of squared vertical distances from the points to the line. Deming regression instead minimizes perpendicular distances, weighted by the uncertainty ratio. When the uncertainties are equal in both dimensions (\(\sigma_x = \sigma_y\)), this becomes orthogonal regression, which minimizes the true perpendicular distances. When the uncertainties differ, the ``perpendicular'' is stretched according to the uncertainty ratio, and the dimension with the smaller uncertainty gets more weight.

This geometric view explains why Deming regression handles the forward and inverse regression problem. The perpendicular distances are the same whichever axis we call \(x\) or \(y\). We test the method on simulated data where we know the true slope. The data have \(\sigma_x = 0.5\) and \(\sigma_y = 0.3\), so the variance ratio is \(\lambda = \sigma_y^2/\sigma_x^2\).

\tutorialbarrier
\begin{codecell}

\end{codecell}

We plot the observed data with their error bars, the true line and the two fits.

\tutorialbarrier
\begin{codecell}

\end{codecell}
\tutorialplot{\includegraphics[width=4.784in,height=0.75\textheight,keepaspectratio]{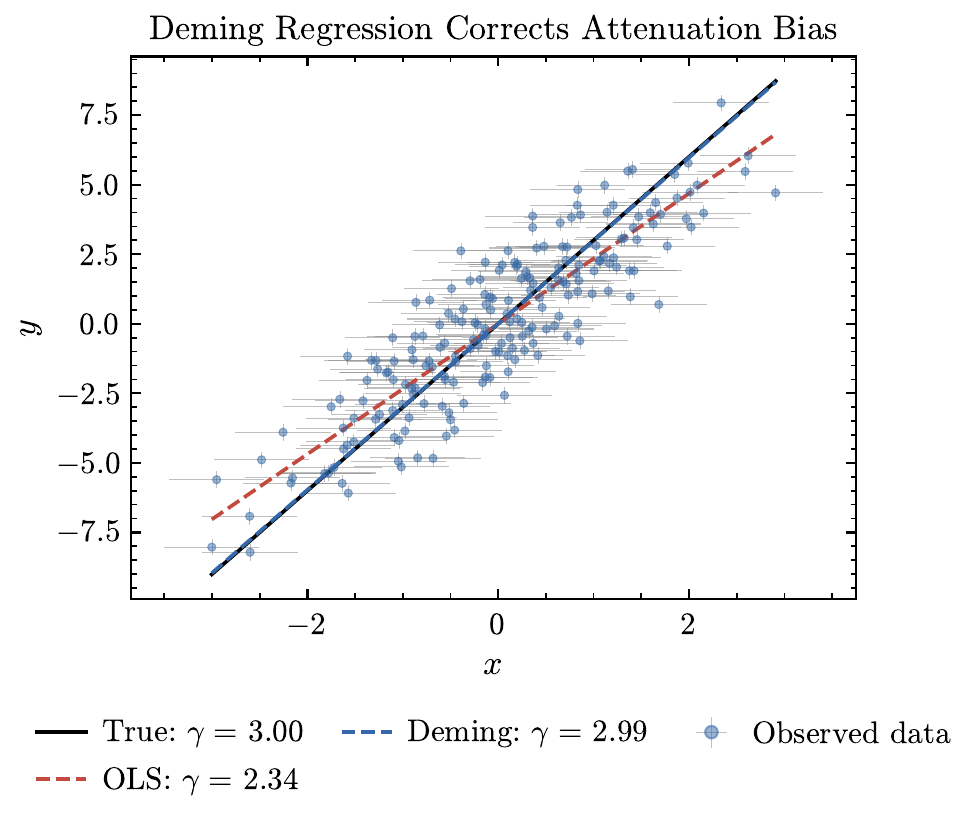}}%
  {\includegraphics[height=\tutorialheight]{tutorials/6/fig04.pdf}}

Deming regression corrects most of the attenuation bias. OLS underestimates the slope by a large amount, and Deming regression recovers a value much closer to the true slope. We want to know whether the remaining difference from the true slope is within the expected statistical variation. For this, we use bootstrap resampling (Section 3.7) to estimate the uncertainty of our estimates. Each resample draws the same number of points with replacement from the simulated data, and we compute both slopes again. We repeat this 1000 times.

\tutorialbarrier
\begin{codecell}

\end{codecell}

We plot the distribution of each slope and print the 95\% intervals, from the 2.5th and 97.5th percentiles.

\tutorialbarrier
\begin{codecell}

\end{codecell}
\tutorialplot{\includegraphics[width=6.176in,height=0.75\textheight,keepaspectratio]{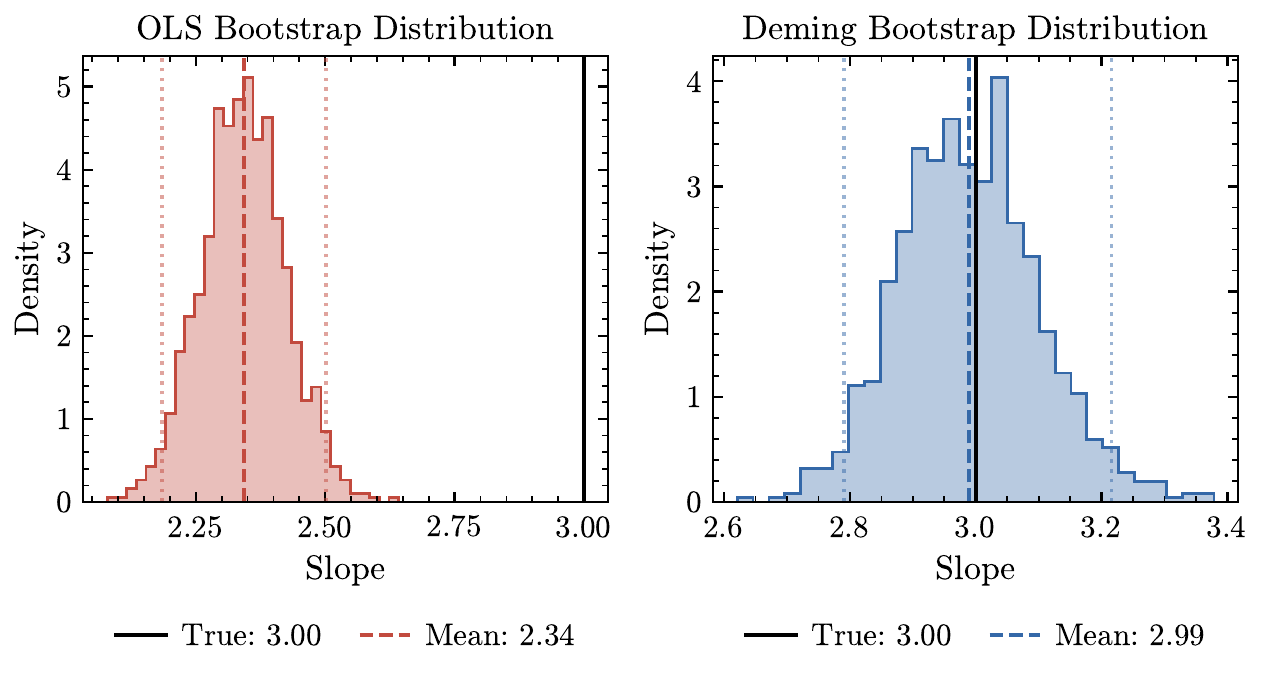}}%
  {\includegraphics[height=\tutorialheight]{tutorials/6/fig05.pdf}}
\begin{Verbatim}[fontsize=\footnotesize, xleftmargin=5mm]
OLS 95% CI = [2.185, 2.502]
Deming 95% CI = [2.790, 3.216]
\end{Verbatim}

The dotted lines mark the 95\% intervals. The OLS distribution is centered far from the true value, and the bias from the attenuation is much larger than the statistical uncertainty. In contrast, the interval of the Deming slopes contains the true slope.

\section*{Heteroscedastic Uncertainties: The Realistic Case}

Our analysis so far assumed constant measurement uncertainties for all data points. Real astronomical data rarely satisfy this assumption. Brighter objects typically have more precise measurements than fainter ones, and different observational techniques give different uncertainties.

For heteroscedastic uncertainties, which vary from point to point, Deming regression extends to the general case (Section 6.8) but loses its simple analytical form. For \(N\) points with the data centered on their means, the slope \(\gamma\) must satisfy

\[F(\gamma) = \sum_{i=1}^N \frac{(\sigma_{y,i}^2 x_{\text{obs},i} + \gamma \sigma_{x,i}^2 y_{\text{obs},i})(y_{\text{obs},i} - \gamma x_{\text{obs},i})}{(\sigma_{y,i}^2 + \gamma^2 \sigma_{x,i}^2)^2} = 0.\]

Section 6.8 derives this condition after it eliminates the true \(x\)-values. Unlike the homoscedastic case, the equation does not reduce to a quadratic, so we must solve it numerically. The slope is one number, so a grid search is enough. We evaluate \(F\) on a grid of slopes and take the slope where \(F\) is closest to zero. The plot of \(F\) below shows this zero.

To test this extension, we simulate data with the same true slope. The uncertainties grow with the distance of a point from the center, as in real data where fainter objects have larger errors. The \(x\)-uncertainty is 0.2 at the center and 0.4 at \(|x_{\text{true}}| = 1\), and the \(y\)-uncertainty is 0.1 and 0.2 at these points.

\tutorialbarrier
\begin{codecell}

\end{codecell}

We now fit the data. OLS ignores the \(x\)-uncertainties. For the Deming fit, we code the function \(F\) and search a grid of slopes for the one where \(F\) is closest to zero. The grid runs from 1 to 10, because we expect a positive slope.

\tutorialbarrier
\begin{codecell}

\end{codecell}

We plot the data with the three lines, and the function \(F(\gamma)\) with the three slopes.

\tutorialbarrier
\begin{codecell}

\end{codecell}
\tutorialplot{\includegraphics[width=6.213in,height=0.75\textheight,keepaspectratio]{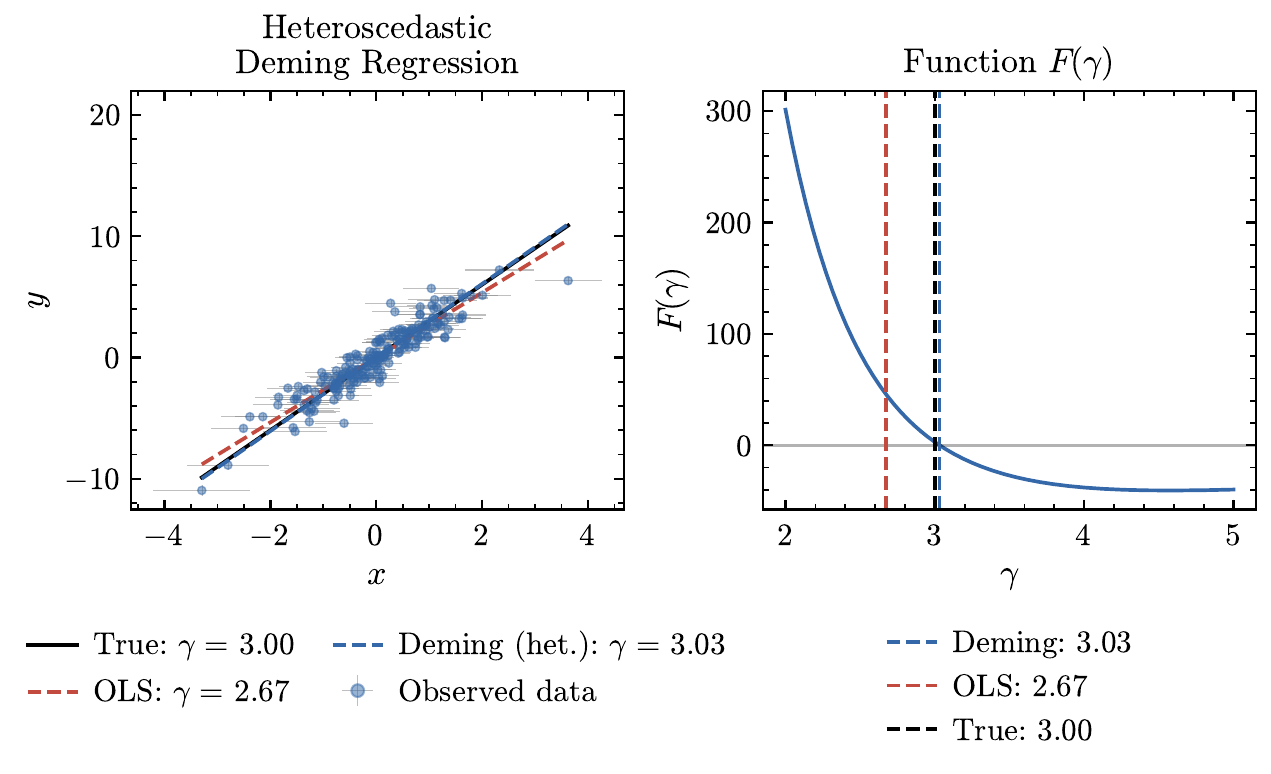}}%
  {\includegraphics[height=\tutorialheight]{tutorials/6/fig06.pdf}}

The Deming fit recovers the true slope well even when the uncertainties differ from point to point. The right panel shows the function \(F(\gamma)\), which crosses zero smoothly at the Deming solution. OLS ignores the \(x\)-uncertainties, and it stays biased whether they are constant or differ from point to point.

\section*{Forward vs Inverse Regression: A Common Confusion}

Before we apply Deming regression to the SPARC data, we address a common source of confusion in astronomical scaling relations. When we fit the Tully-Fisher relation, we can regress the mass on the velocity (\(M\) against \(V\)). We can also regress the velocity on the mass (\(V\) against \(M\)). With OLS, these two choices give different results, and the slopes are not simply reciprocals of each other.

This asymmetry arises because OLS minimizes the residuals in one direction only. Forward regression, with \(V\) as \(x\) and \(M\) as \(y\), minimizes the vertical residuals. Inverse regression, with \(M\) as \(x\) and \(V\) as \(y\), minimizes the horizontal residuals. The choice matters, because it assumes that one variable is measured perfectly.

In the SPARC dataset, the reported mass uncertainty is much smaller than the velocity uncertainty. From a statistical perspective, we should use the mass as the independent variable. However, astronomical convention plots the velocity on the \(x\)-axis of the Tully-Fisher relation. This creates a dilemma between statistical practice and astronomical convention.

Deming regression removes this tension. It treats both variables symmetrically and uses the uncertainties in both dimensions. The result does not depend on which variable we call \(x\) or \(y\), so we get the same relation either way. We show this with the SPARC data.

We center the data and compute four slopes. The forward and inverse fits are OLS with the roles of the variables exchanged, and the Deming fit uses the uncertainties of both variables. We also repeat the Deming fit with \(x\) and \(y\) exchanged.

\tutorialbarrier
\begin{codecell}

\end{codecell}
\begin{Verbatim}[fontsize=\footnotesize, xleftmargin=5mm]
Deming regression, γ = 4.135
Deming regression with x and y swapped, 1/γ = 0.242 → γ = 4.135
\end{Verbatim}

We plot the three fits with the error bars of both variables. In the inverse fit, the velocity is on the vertical axis.

\tutorialbarrier
\begin{codecell}

\end{codecell}
\tutorialplot{\includegraphics[width=6.500in,height=0.75\textheight,keepaspectratio]{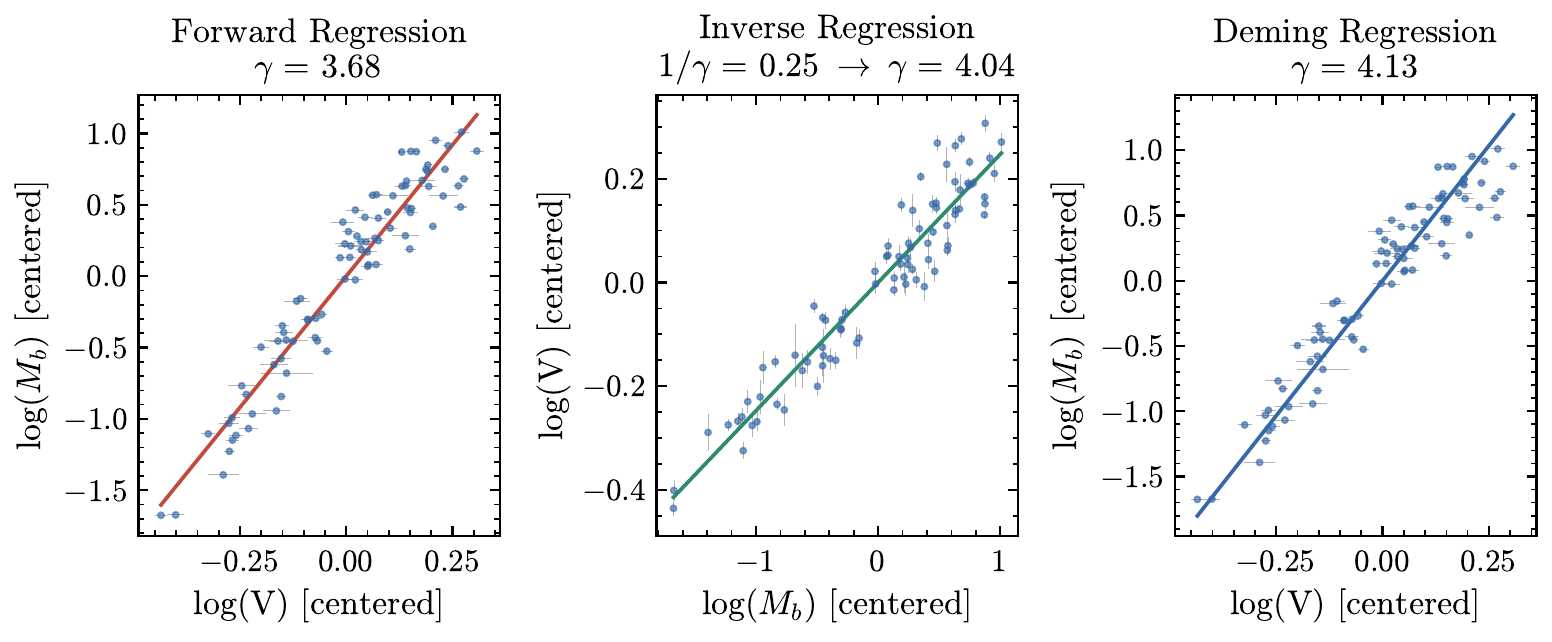}}%
  {\includegraphics[height=\tutorialheight]{tutorials/6/fig07.pdf}}

The forward slope is 3.68 and the inverse slope is 4.04. The Deming slope is about 4.1, close to the inverse slope. The reported mass uncertainty is much smaller than the velocity uncertainty, so Deming regression gives the mass more weight. The Deming fit with \(x\) and \(y\) exchanged gives the same slope after the inversion.

The same reason makes the inverse slope, which has the mass as the independent variable, nearly free of attenuation. Suppose velocity errors alone cause the difference between the slopes. The inverse slope then equals \(\gamma\) and the forward slope is \(\alpha\gamma\), so the data give the attenuation factor,

\[\alpha = \frac{\text{forward slope}}{\text{inverse slope}} = \frac{3.68}{4.04} = 0.91.\]

The reported uncertainties predict a different value. The velocity uncertainties are about 9\% of the standard deviation of \(x\), so

\[\alpha \approx \frac{1}{1 + 0.09^2} \approx 0.99.\]

This is far from 0.91. If the relation is a straight line, the reported uncertainties therefore miss a systematic uncertainty. The galaxies also scatter about the line by more than their reported errors, as the error bars of the figure show. For this study we assume that the missing uncertainty belongs to the velocity, for example systematic errors in the inclination. We add a systematic uncertainty \(\sigma_{\text{sys}}\) in quadrature to the reported velocity uncertainty \(\sigma_{\text{rep},i}\) of every galaxy,

\[\sigma_{x,i}^2 = \sigma_{\text{rep},i}^2 + \sigma_{\text{sys}}^2.\]

\section*{Choosing the Systematic Uncertainty}

We reuse the attenuation factor. In the simulation, we knew the velocity error and predicted how much it flattens the OLS slope. Now we go the other way. The data give the attenuation factor of 0.91, and the same formula gives the error that causes it.

In Section 6.3, \(\alpha\) is the variance of the true velocities over the variance of the observed velocities. The observed variance is \(\sigma_{\text{obs}}^2 = \sigma_{\text{range}}^2 + \sigma_x^2\), so

\[\alpha = \frac{\sigma_{\text{range}}^2}{\sigma_{\text{obs}}^2} = 1 - \frac{\sigma_x^2}{\sigma_{\text{obs}}^2}.\]

When the uncertainties differ between galaxies, \(\sigma_x^2\) is their mean square. We compute \(\sigma_{\text{obs}}^2\) from the observed velocities. The two expressions for \(\alpha\), from the slopes and from the uncertainties, describe the same quantity, so we set them equal and solve for \(\sigma_x^2\). With \(\sigma_{x,i}^2 = \sigma_{\text{rep},i}^2 + \sigma_{\text{sys}}^2\), the result is

\[\sigma_{\text{sys}}^2 = (1 - \alpha)\,\sigma_{\text{obs}}^2 - \sigma_{\text{rep}}^2,\]

where \(\sigma_{\text{rep}}^2\) is the mean square of the reported uncertainties.

We now compute \(\sigma_{\text{sys}}\) from the two slopes.

\tutorialbarrier
\begin{codecell}

\end{codecell}
\begin{Verbatim}[fontsize=\footnotesize, xleftmargin=5mm]
Attenuation factor = 0.91
Systematic velocity uncertainty = 0.048 dex
\end{Verbatim}

\section*{Checking the Uncertainties Against the Scatter}

The two slopes give an attenuation factor of 0.91 and a systematic uncertainty of about 0.05 dex, three times the median reported velocity uncertainty. This step ignores the mass uncertainty \(\sigma_{y,i}\). More properly, the mass uncertainty should enter too.

We check with the residuals of the fit whether the inflated uncertainties describe the scatter. If the centered data follow the line, the residual \(y_i - \gamma x_i\) of galaxy \(i\) is the error in \(y_i\) minus \(\gamma\) times the error in \(x_i\). The two errors are independent. By the rules for sums of random variables in Section 3.5, the variance of the residual is

\[\mathrm{Var}[y_i - \gamma x_i] = \sigma_{y,i}^2 + \gamma^2\sigma_{x,i}^2.\]

The normalized residual is the residual divided by its standard deviation,

\[\frac{y_i - \gamma x_i}{\sqrt{\sigma_{y,i}^2 + \gamma^2\sigma_{x,i}^2}}.\]

If the uncertainties describe the scatter, the horizontal error bars reach the line for about two thirds of the galaxies. The normalized residuals then have a standard deviation of about 1, and the mean of their squares is about 1. This is the square of the normalized error of Tutorial 5, now with errors in both variables.

The slope \(\gamma\) is the Deming slope for the uncertainties. We fit it with the inflated uncertainties and compute the mean squared normalized residual for the reported and for the inflated uncertainties.

\tutorialbarrier
\begin{codecell}

\end{codecell}
\begin{Verbatim}[fontsize=\footnotesize, xleftmargin=5mm]
Deming slope with the inflated uncertainties = 4.036
Mean squared normalized residual, reported uncertainties = 13.83
Mean squared normalized residual, inflated uncertainties = 1.03
\end{Verbatim}

The inflated uncertainties give a mean squared normalized residual of 1.03, against 13.8 for the reported ones, so they describe the scatter.

The check also suggests another way to choose \(\sigma_{\text{sys}}\). Without the estimate from the attenuation, we could search for the value at which the mean squared normalized residual equals 1. This is more principled, because it uses the uncertainty of each galaxy and not their mean, and it includes \(\sigma_{y,i}\) when needed. We do not need the search here, because the estimate already gives a value of about 1.

\section*{Deming Regression with the Inflated Uncertainties}

Given these uncertainties, the Deming fit gives the slope of the relation. The forward OLS fit shows what ignoring the velocity errors gives. We plot the galaxies with the inflated velocity uncertainties and both fits.

\tutorialbarrier
\begin{codecell}

\end{codecell}
\tutorialplot{\includegraphics[width=4.809in,height=0.75\textheight,keepaspectratio]{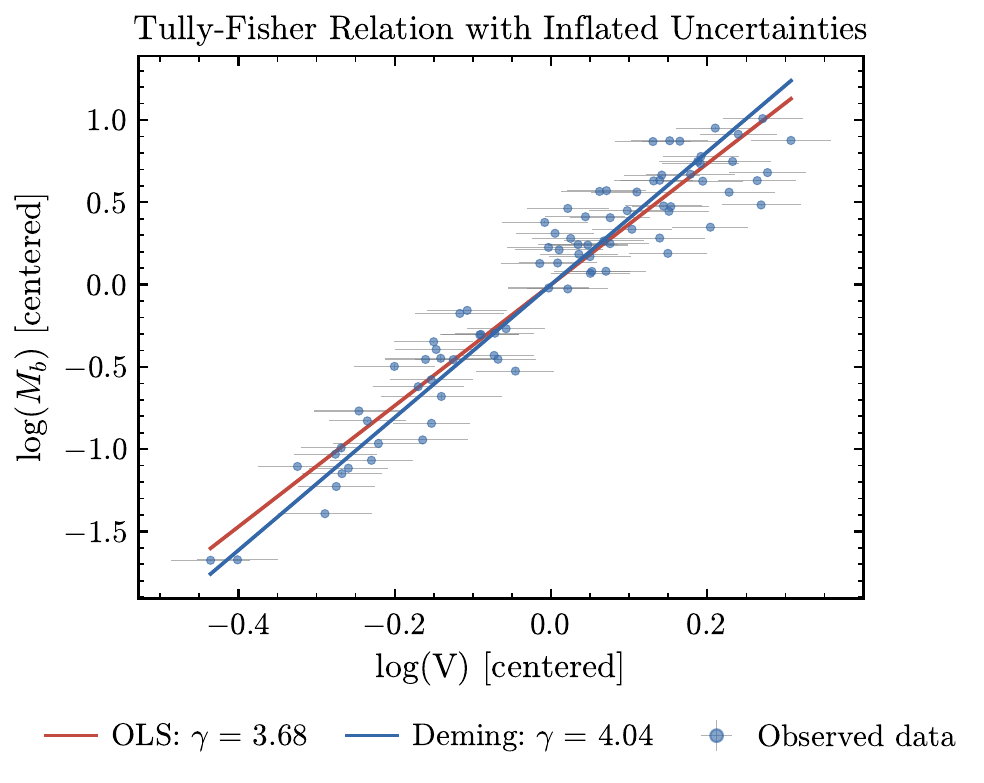}}%
  {\includegraphics[height=\tutorialheight]{tutorials/6/fig08.pdf}}

The inflated error bars reach the lines. Deming regression with the inflated uncertainties gives a slope of 4.04, close to the inverse slope, because the velocity now carries almost all the uncertainty. Attenuation bias therefore explains why the Deming line is steeper than the OLS line, and Deming regression corrects it.

\section*{Bootstrap Uncertainty for the Tully-Fisher Relation}

We now estimate the uncertainty of each slope with bootstrap resampling of the SPARC data. This shows whether the difference between the OLS and Deming slopes is larger than the sampling noise. We run the Deming regression for every resample with the same grid search. It uses the inflated uncertainties, with \(\sigma_{\text{sys}}\) fixed at the value from the full sample. Each resample draws galaxies with replacement together with their uncertainties and computes the two slopes.

\tutorialbarrier
\begin{codecell}

\end{codecell}

We plot the distribution of each slope and print the 95\% intervals and the standard errors.

\tutorialbarrier
\begin{codecell}

\end{codecell}
\tutorialplot{\includegraphics[width=6.178in,height=0.75\textheight,keepaspectratio]{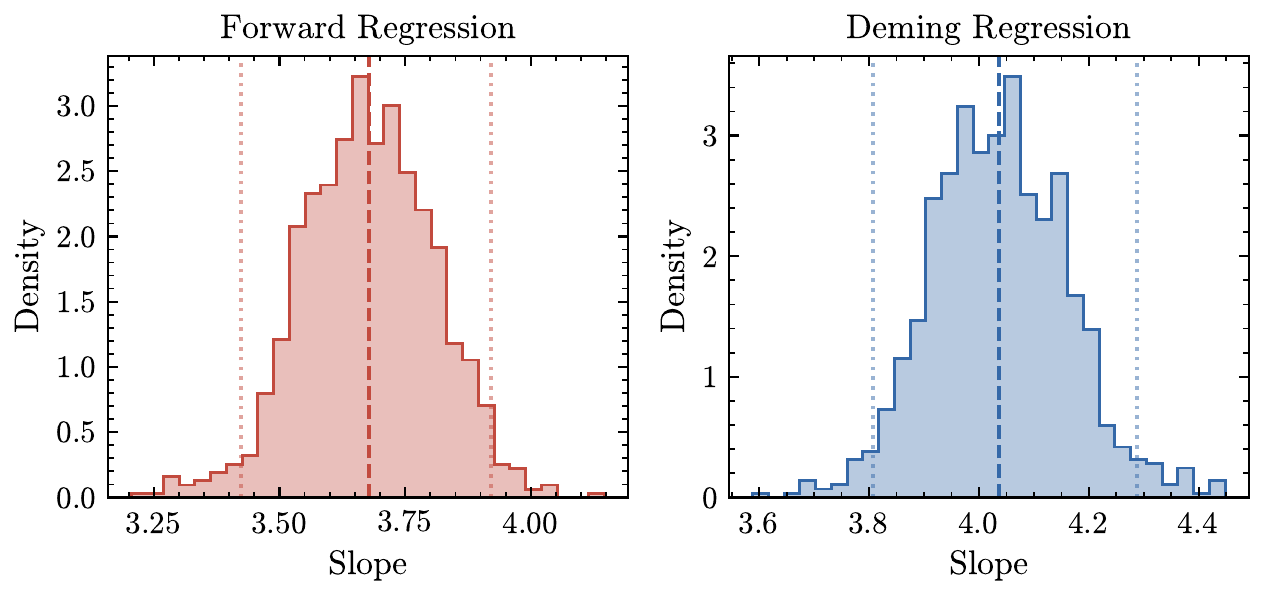}}%
  {\includegraphics[height=\tutorialheight]{tutorials/6/fig09.pdf}}
\begin{Verbatim}[fontsize=\footnotesize, xleftmargin=5mm]
Tully-Fisher slope estimates (95% CI)
Forward = 3.679 [3.424, 3.921]
Deming = 4.036 [3.807, 4.287]

Standard errors
Forward = 0.130
Deming = 0.124
\end{Verbatim}

The dashed lines mark the slopes from the full sample, and the dotted lines mark the 95\% intervals. The forward distribution is centered near 3.7, and the Deming distribution near 4.0. The difference between the forward and the Deming slope is larger than the standard error of either slope. Sampling noise is therefore unlikely to explain it.

For these galaxies, the choice of the fit changes the slope by about 10\%. This matters, because the Tully-Fisher slope constrains galaxy formation and sets distances, as the relation is a rung of the cosmic distance ladder.

The steps of this analysis apply to other relations with errors in both variables. First compute the attenuation factor from the reported uncertainties. If the forward and inverse slopes show more attenuation than the reported uncertainties explain, state where the extra uncertainty lies. Infer its size from the attenuation, and check the result with the normalized residuals. Finally, fit with Deming regression and estimate the uncertainty of the slope with the bootstrap.

\section*{Limits of the Assumption}

These steps rest on one assumption. We assumed that the extra scatter belongs to the velocity. The data alone cannot show this, because a straight-line fit sees only the total scatter and not how it splits between the two variables. The split must come from outside the data.

The mass can contribute to the scatter too. The reported mass uncertainty comes from the photometry alone, and the distance and the conversion from light to mass add more. The relation may also have an intrinsic scatter.

Suppose the extra scatter lay in the mass. Then the forward OLS slope would be closer to the truth, and the Deming slope would be too steep. The slope of about 4 is therefore a result under our assumption.

\section*{Summary}

This tutorial showed how measurement uncertainties in the independent variable lead to attenuation bias, through the Tully-Fisher relation. We explored the following concepts.

\begin{enumerate}
\def\labelenumi{\arabic{enumi}.}
\item
  \textbf{Attenuation bias in regression.} Attenuation bias arises when measurement uncertainties in the independent variable inflate the observed spread and flatten the measured relation. The bias remains whatever the sample size. It depends on the ratio of the measurement uncertainty to the intrinsic spread.
\item
  \textbf{Deming regression solution.} Deming regression corrects the attenuation bias. It treats the true values as latent variables and infers them together with the model parameters. It recovered the true slope in our simulations, where the measurement errors were the only scatter. It reduces to OLS when the uncertainties in \(x\) approach zero.
\item
  \textbf{Practical implementation.} For homoscedastic uncertainties, Deming regression has a closed-form solution. For heteroscedastic uncertainties, which are common in astronomical data, a grid search gives the solution. Bootstrap resampling estimates the uncertainty of the slopes.
\item
  \textbf{Systematic uncertainty.} Reported uncertainties can miss a systematic part. With a proper assumption about where it lies, we can estimate it from the data, and Deming regression can account for it. The fitted slope is then a result under that assumption.
\end{enumerate}

The techniques apply whenever we measure relations between quantities that both carry measurement uncertainties. This is a common situation in observational astronomy, where perfect measurements do not exist.

\tutorialchapter{7}{Logistic Regression for Red Clump Stars}
\noindent\textit{Companion material for Chapter 7}

\section*{Introduction}

In the earlier chapters we predicted continuous quantities, such as stellar parameters. Many questions in astronomy ask for a discrete class instead, because stars occupy distinct stages of evolution and galaxies fall into morphological types. Is this object a star or a galaxy? Is this transient a supernova or an active galactic nucleus? Is this star on the red giant branch, or has it evolved to the red clump phase?

Logistic regression links the linear methods that we know to classification. We apply a simple nonlinear transformation, the sigmoid function, to our linear model. The result is the probability that an object belongs to a class. A probability is useful for noisy astronomical data, where sharp boundaries between classes rarely exist.

In this tutorial we separate red clump (RC) stars from red giant branch (RGB) stars. The reason is the distance. Parallax measurements from missions such as Gaia gave accurate distances for nearby stars, but the fractional error of a parallax grows with distance. For stars beyond a few kiloparsecs, which include most stars of our Galaxy, we need other distance indicators.

RC stars are a good choice. They burn helium in their cores after the helium flash, the sudden ignition of the helium. The core of such a star ignites helium at nearly the same core mass, whatever its earlier evolution. RC stars are therefore standard candles, objects of known intrinsic brightness whose distances we can infer from their apparent magnitudes.

The use of RC stars depends on how well we can tell them from RGB stars. An RGB star climbs the giant branch while it burns hydrogen in a shell around an inert helium core. RGB stars populate a similar region of the temperature and gravity plane, but they have a wide range of luminosities. An RGB star taken for an RC star would get a large distance error. This classification therefore affects how well we can map the three-dimensional structure of our Galaxy.

Asteroseismology, the study of stellar oscillations, is the usual way to identify RC stars. It needs long and frequent photometric monitoring, from missions such as Kepler or TESS, so it covers only small fields. The onset of core helium burning also changes the temperature and the surface gravity of a star. Spectroscopic surveys such as the Apache Point Observatory Galactic Evolution Experiment (APOGEE) measure these properties for far more stars. We therefore train a classifier on stars with asteroseismic classes, using their spectroscopic and photometric measurements as features. It can then classify stars from these measurements alone.

\section*{Prerequisites and Learning Objectives}

\subsection*{Prerequisites}

This tutorial assumes the following.

\begin{itemize}
\tightlist
\item
  Working knowledge of Python, with NumPy arrays and matplotlib plots
\item
  Linear regression, with linear models and model parameters
\item
  Probability and the likelihood
\item
  Splitting data into training, validation and test sets
\end{itemize}

\subsection*{Learning Objectives}

After this tutorial you can do the following.

\begin{itemize}
\tightlist
\item
  Turn a linear model into a classifier with the sigmoid function
\item
  Explain how the sigmoid converts a linear output into a probability
\item
  Write batch and mini-batch gradient descent from scratch
\item
  Evaluate a classifier on classes of unequal size with the recall, the precision and the balanced accuracy
\item
  Set a probability threshold, and read the ROC curve and the AUC
\item
  Tune the learning rate and the batch size on data that the final score does not use
\end{itemize}

We begin by importing the libraries and setting the plot style.

\tutorialbarrier
\begin{codecell}

\end{codecell}

\section*{The APOGEE Data}

We load an APOGEE sample with asteroseismic classes.

\tutorialbarrier
\begin{codecell}
\begin{Verbatim}[commandchars=\\\{\},fontsize=\small]
\PYG{c+c1}{\PYGZsh{} Load the APOGEE dataset with asteroseismic classifications}
\PYG{n}{data} \PYG{o}{=}\FancyVerbBreakAfterBreak{} \PYG{n}{np}\PYG{o}{.}\FancyVerbBreakAfterBreak{}\PYG{n}{load}\PYG{p}{(}\FancyVerbBreakAfterBreak{}\PYG{l+s+s2}{\PYGZdq{}}\PYG{l+s+s2}{../data/dataset\PYGZus{}apogee\PYGZus{}asteroseismic.npz}\PYG{l+s+s2}{\PYGZdq{}}\PYG{p}{)}

\PYG{c+c1}{\PYGZsh{} Feature matrix, with spectroscopic and photometric features, and labels (1 for RC, 0 for RGB)}
\PYG{n}{X} \PYG{o}{=}\FancyVerbBreakAfterBreak{} \PYG{n}{data}\PYG{p}{[}\FancyVerbBreakAfterBreak{}\PYG{l+s+s2}{\PYGZdq{}}\PYG{l+s+s2}{data}\PYG{l+s+s2}{\PYGZdq{}}\PYG{p}{]}
\PYG{n}{t} \PYG{o}{=}\FancyVerbBreakAfterBreak{} \PYG{n}{data}\PYG{p}{[}\FancyVerbBreakAfterBreak{}\PYG{l+s+s2}{\PYGZdq{}}\PYG{l+s+s2}{label}\PYG{l+s+s2}{\PYGZdq{}}\PYG{p}{]}

\PYG{c+c1}{\PYGZsh{} Print the numbers of stars}
\PYG{n+nb}{print}\PYG{p}{(}\FancyVerbBreakAfterBreak{}\PYG{l+s+sa}{f}\PYG{l+s+s2}{\PYGZdq{}}\PYG{l+s+s2}{Stars = }\PYG{l+s+si}{\PYGZob{}}\PYG{n+nb}{len}\PYG{p}{(}\FancyVerbBreakAfterBreak{}\PYG{n}{t}\PYG{p}{)}\PYG{l+s+si}{:}\PYG{l+s+s2}{,}\FancyVerbBreakAfterBreak{}\PYG{l+s+si}{\PYGZcb{}}\FancyVerbBreakAfterBreak{}\PYG{l+s+s2}{, RGB = }\PYG{l+s+si}{\PYGZob{}}\PYG{n}{np}\PYG{o}{.}\FancyVerbBreakAfterBreak{}\PYG{n}{sum}\PYG{p}{(}\FancyVerbBreakAfterBreak{}\PYG{n}{t}\PYG{+w}{ }\PYG{o}{==}\PYG{+w}{ }\PYG{l+m+mi}{0}\PYG{p}{)}\PYG{l+s+si}{:}\PYG{l+s+s2}{,}\FancyVerbBreakAfterBreak{}\PYG{l+s+si}{\PYGZcb{}}\FancyVerbBreakAfterBreak{}\PYG{l+s+s2}{, RC = }\PYG{l+s+si}{\PYGZob{}}\PYG{n}{np}\PYG{o}{.}\FancyVerbBreakAfterBreak{}\PYG{n}{sum}\PYG{p}{(}\FancyVerbBreakAfterBreak{}\PYG{n}{t}\PYG{+w}{ }\PYG{o}{==}\PYG{+w}{ }\PYG{l+m+mi}{1}\PYG{p}{)}\PYG{l+s+si}{:}\PYG{l+s+s2}{,}\FancyVerbBreakAfterBreak{}\PYG{l+s+si}{\PYGZcb{}}\FancyVerbBreakAfterBreak{}\PYG{l+s+s2}{\PYGZdq{}}\PYG{p}{)}
\end{Verbatim}
\end{codecell}
\begin{Verbatim}[fontsize=\footnotesize, xleftmargin=5mm]
Stars = 2,855, RGB = 730, RC = 2,125
\end{Verbatim}

The labels come from Kepler asteroseismology, as in \href{https://ui.adsabs.harvard.edu/abs/2018ApJ...858L...7T/abstract}{Ting, Hawkins \& Rix (2018)}. The period spacing of the mixed oscillation modes, which \href{https://ui.adsabs.harvard.edu/abs/2016A&A...588A..87V/abstract}{Vrard et al.~(2016)} measured, probes the core. It is above 250 s for an RC star, whose core burns helium, and below for an RGB star, whose core is inert. We take these classes as the correct labels. Each star has eight features from three kinds of measurement.

\textbf{Spectroscopic parameters} come from the spectrum of the star, which APOGEE observed.

\begin{itemize}
\tightlist
\item
  \textbf{Effective temperature (Teff)} is the temperature of a black body that emits the same power per unit surface area as the star.
\item
  \textbf{Surface gravity (log g)} is the logarithm of the gravitational acceleration at the photosphere of the star. It relates to the mass and the radius of the star through \(g = GM/R^2\).
\item
  \textbf{Metallicity ({[}Fe/H{]})} is the base-10 logarithm of the iron to hydrogen number ratio of the star, relative to the Sun. A star with {[}Fe/H{]} = -1 has an iron to hydrogen ratio ten times lower than the Sun.
\end{itemize}

\textbf{Photometric measurements} are the near-infrared magnitudes J, H and K, apparent brightnesses in the photometric system of the Two Micron All Sky Survey (2MASS). They depend on both the distance and the intrinsic luminosity of the star.

\textbf{Sky coordinates (RA, Dec)} give the position of the star. We include them to test whether our classifier uses spatial information that it should not use.

We now plot the two populations in the plane of temperature and gravity.

\tutorialbarrier
\begin{codecell}

\end{codecell}
\tutorialplot{\includegraphics[width=4.230in,height=0.75\textheight,keepaspectratio]{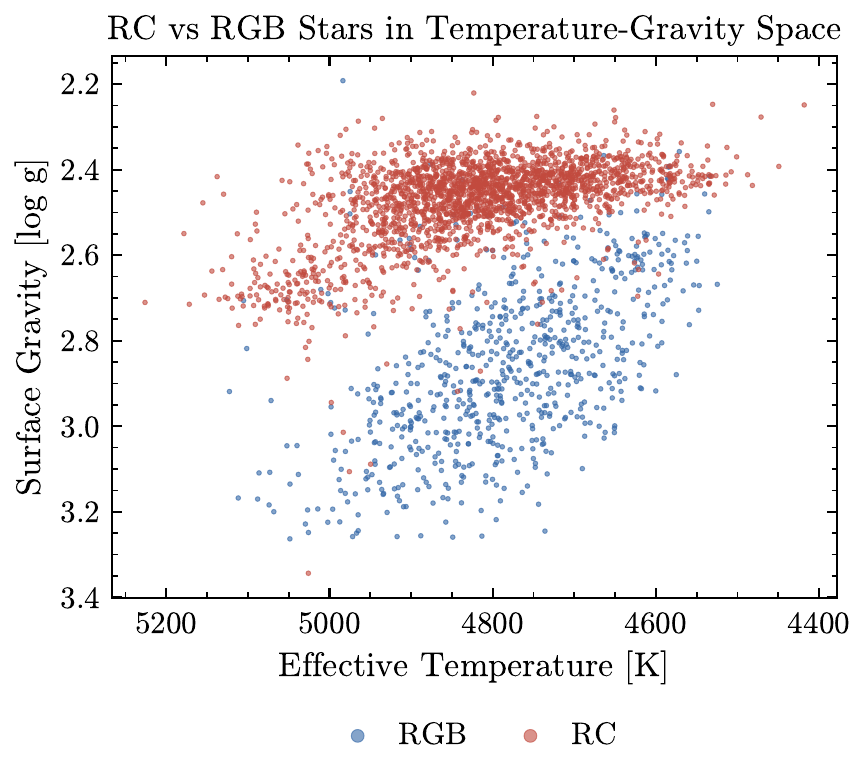}}%
  {\includegraphics[height=\tutorialheight]{tutorials/7/fig01.pdf}}

RC stars (red points) cluster tightly around \(\log g \approx 2.5\), because they all burn helium in their cores. RGB stars (blue points) span a broader range of surface gravities, mostly at a higher gravity than the RC stars. As a star ascends the giant branch, its radius increases and its surface gravity decreases steadily.

The two populations overlap at intermediate gravities, \(\log g \approx 2.5\) to 2.8. There, ascending RGB stars and RC stars occupy nearly the same region. A single threshold in log g or in temperature would misclassify some of these stars, so we use a probabilistic approach. Logistic regression learns the probability that a star with given spectroscopic properties belongs to the RC population.

\section*{From Linear to Logistic Regression}

In earlier chapters we described continuous quantities, such as stellar parameters, as linear combinations of features. Classification needs another approach, because we predict discrete classes and not continuous values.

Consider a direct linear regression of our binary labels, 0 for RGB and 1 for RC. A linear model \(\mathbf{w}^T\mathbf{x}\) can give any real number, but the labels are limited to 0 and 1. Worse, a prediction outside the interval {[}0, 1{]} has no meaning as a probability, because a probability cannot be negative or exceed one.

Logistic regression solves this with the sigmoid function (Section 7.3)

\[\sigma(z) = \frac{1}{1 + e^{-z}}.\]

This transformation maps any real number to the interval (0, 1), so the output is a valid probability. The function has three useful properties.

\begin{itemize}
\tightlist
\item
  \textbf{Monotonicity} means that a larger linear output gives a higher probability.
\item
  \textbf{Symmetry} means that \(\sigma(-z) = 1 - \sigma(z)\).
\item
  \textbf{Smooth gradients} allow an efficient optimization.
\end{itemize}

For stellar classification, our model computes

\[P(\text{RC}|\mathbf{x}, \mathbf{w}) = \sigma(\mathbf{w}^T\mathbf{x}).\]

The linear term \(\mathbf{w}^T\mathbf{x}\) is a weighted sum of the features, and it indicates the type of the star. We now code the sigmoid and plot it.

\tutorialbarrier
\begin{codecell}

\end{codecell}
\tutorialplot{\includegraphics[width=4.183in,height=0.75\textheight,keepaspectratio]{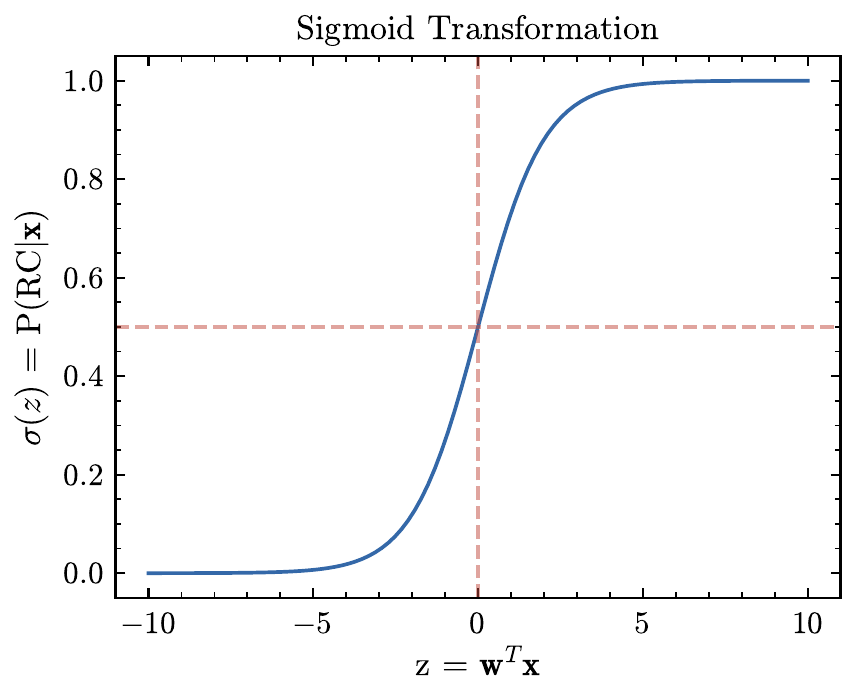}}%
  {\includegraphics[height=\tutorialheight]{tutorials/7/fig02.pdf}}

The sigmoid has properties that suit a classification.

\begin{enumerate}
\def\labelenumi{\arabic{enumi}.}
\tightlist
\item
  \textbf{Decision boundary at \(z = 0\).} When \(\mathbf{w}^T\mathbf{x} = 0\), the probability is 0.5, which is the decision threshold. A star with a positive linear output is more likely RC, and a negative value suggests RGB.
\item
  \textbf{Extreme probabilities for a large \(|z|\).} As \(|z|\) grows, the sigmoid approaches 0 or 1. A star with \(z = 5\) gives \(P(\text{RC}) \approx 0.993\), and a star with \(z = -5\) gives \(P(\text{RC}) \approx 0.007\).
\item
  \textbf{Gradient properties.} The derivative \(\sigma'(z) = \sigma(z)(1-\sigma(z))\) is largest at \(z = 0\) and vanishes for large \(|z|\).
\end{enumerate}

\section*{Maximum Likelihood Formulation}

To learn the best weights \(\mathbf{w}\), we use the maximum likelihood principle (Section 7.6). The training data are \(\{(\mathbf{x}_i, t_i)\}_{i=1}^N\), where \(t_i \in \{0,1\}\) gives the true type of star \(i\). We look for the weights that maximize the probability of the observed labels.

For one star with features \(\mathbf{x}_i\) and true label \(t_i\), the likelihood is

\[p(t_i|\mathbf{x}_i, \mathbf{w}) = y_i^{t_i}(1-y_i)^{1-t_i},\]

where \(y_i = \sigma(\mathbf{w}^T\mathbf{x}_i)\) is the RC probability that our model predicts. This compact expression covers both cases.

\begin{itemize}
\tightlist
\item
  If \(t_i = 1\) (an RC star), the likelihood is \(y_i\).
\item
  If \(t_i = 0\) (an RGB star), the likelihood is \(1-y_i\).
\end{itemize}

For independent observations the total likelihood is the product over all stars. We take the logarithm, which turns the product into a sum and avoids numerical underflow. We also take the negative and divide by \(N\), so that the loss does not grow with the number of stars. This gives the negative log-likelihood loss

\[E(\mathbf{w}) = -\frac{1}{N}\sum_{i=1}^N \left[t_i \ln y_i + (1-t_i)\ln(1-y_i)\right].\]

This cross-entropy loss measures the difference between the predicted probabilities and the true labels. It is the objective of our optimization. We now code it, with a small constant that prevents the logarithm of zero.

\tutorialbarrier
\begin{codecell}
\begin{Verbatim}[commandchars=\\\{\},fontsize=\small]
\PYG{k}{def}\PYG{+w}{ }\PYG{n+nf}{cost}\PYG{p}{(}\FancyVerbBreakAfterBreak{}\PYG{n}{w}\PYG{p}{,}\FancyVerbBreakAfterBreak{} \PYG{n}{X}\PYG{p}{,}\FancyVerbBreakAfterBreak{} \PYG{n}{t}\PYG{p}{,}\FancyVerbBreakAfterBreak{} \PYG{n}{N}\PYG{p}{)}\PYG{p}{:}
\PYG{+w}{    }\PYG{l+s+sd}{\PYGZdq{}\PYGZdq{}\PYGZdq{}Return the average negative log\PYGZhy{}likelihood (the cross\PYGZhy{}entropy loss) of N stars.\PYGZdq{}\PYGZdq{}\PYGZdq{}}
    \PYG{n}{z} \PYG{o}{=}\FancyVerbBreakAfterBreak{} \PYG{n}{X} \PYG{o}{@} \PYG{n}{w}  \PYG{c+c1}{\PYGZsh{} linear output for every star}
    \PYG{n}{y} \PYG{o}{=}\FancyVerbBreakAfterBreak{} \PYG{n}{sigmoid}\PYG{p}{(}\FancyVerbBreakAfterBreak{}\PYG{n}{z}\PYG{p}{)}

    \PYG{c+c1}{\PYGZsh{} Add a small epsilon to prevent log(0)}
    \PYG{n}{epsilon} \PYG{o}{=}\FancyVerbBreakAfterBreak{} \PYG{l+m+mf}{1e\PYGZhy{}15}
    \PYG{k}{return} \PYG{o}{\PYGZhy{}}\PYG{n}{np}\PYG{o}{.}\FancyVerbBreakAfterBreak{}\PYG{n}{sum}\PYG{p}{(}\FancyVerbBreakAfterBreak{}\PYG{n}{t} \PYG{o}{*} \PYG{n}{np}\PYG{o}{.}\FancyVerbBreakAfterBreak{}\PYG{n}{log}\PYG{p}{(}\FancyVerbBreakAfterBreak{}\PYG{n}{y} \PYG{o}{+}\FancyVerbBreakAfterBreak{} \PYG{n}{epsilon}\PYG{p}{)} \PYG{o}{+}\FancyVerbBreakAfterBreak{}
                   \PYG{p}{(}\FancyVerbBreakAfterBreak{}\PYG{l+m+mi}{1} \PYG{o}{\PYGZhy{}} \PYG{n}{t}\PYG{p}{)} \PYG{o}{*} \PYG{n}{np}\PYG{o}{.}\FancyVerbBreakAfterBreak{}\PYG{n}{log}\PYG{p}{(}\FancyVerbBreakAfterBreak{}\PYG{l+m+mi}{1} \PYG{o}{\PYGZhy{}} \PYG{n}{y} \PYG{o}{+}\FancyVerbBreakAfterBreak{} \PYG{n}{epsilon}\PYG{p}{)}\PYG{p}{)} \PYG{o}{/} \PYG{n}{N}
\end{Verbatim}
\end{codecell}

\section*{Gradient-Based Optimization}

To minimize the loss, we use gradient descent. It adjusts the weights iteratively in the direction of the steepest decrease. The gradient of the loss with respect to the weights has a simple form (Section 7.7)

\[\nabla E(\mathbf{w}) = \frac{1}{N}\sum_{i=1}^N (y_i - t_i)\mathbf{x}_i.\]

The gradient is the average prediction error \((y_i - t_i)\), weighted by the feature vector. Suppose that we predict a high RC probability (\(y_i \approx 1\)) for an RGB star (\(t_i = 0\)). The positive error then pushes the weights down for the features that are typical of that star.

The gradient is simple because the derivative of the sigmoid cancels against the derivatives of the logarithms in the cross-entropy loss. The error \((y_i - t_i)\) is close to 1 in magnitude when the model gives the wrong class a probability close to 1. It is close to 0 when the model gives the right class such a probability. We now code the gradient.

\tutorialbarrier
\begin{codecell}
\begin{Verbatim}[commandchars=\\\{\},fontsize=\small]
\PYG{k}{def}\PYG{+w}{ }\PYG{n+nf}{grad}\PYG{p}{(}\FancyVerbBreakAfterBreak{}\PYG{n}{w}\PYG{p}{,}\FancyVerbBreakAfterBreak{} \PYG{n}{X}\PYG{p}{,}\FancyVerbBreakAfterBreak{} \PYG{n}{t}\PYG{p}{,}\FancyVerbBreakAfterBreak{} \PYG{n}{N}\PYG{p}{)}\PYG{p}{:}
\PYG{+w}{    }\PYG{l+s+sd}{\PYGZdq{}\PYGZdq{}\PYGZdq{}Return the gradient of the average negative log\PYGZhy{}likelihood.\PYGZdq{}\PYGZdq{}\PYGZdq{}}
    \PYG{n}{z} \PYG{o}{=}\FancyVerbBreakAfterBreak{} \PYG{n}{X} \PYG{o}{@} \PYG{n}{w}  \PYG{c+c1}{\PYGZsh{} linear output for every star}
    \PYG{n}{y} \PYG{o}{=}\FancyVerbBreakAfterBreak{} \PYG{n}{sigmoid}\PYG{p}{(}\FancyVerbBreakAfterBreak{}\PYG{n}{z}\PYG{p}{)}
    \PYG{k}{return} \PYG{n}{X}\PYG{o}{.}\FancyVerbBreakAfterBreak{}\PYG{n}{T} \PYG{o}{@} \PYG{p}{(}\FancyVerbBreakAfterBreak{}\PYG{n}{y} \PYG{o}{\PYGZhy{}} \PYG{n}{t}\PYG{p}{)} \PYG{o}{/} \PYG{n}{N}  \PYG{c+c1}{\PYGZsh{} average of the error times the features}
\end{Verbatim}
\end{codecell}

\section*{Preparing the Data}

We split the stars into 80 percent training stars and 20 percent test stars. The split function of scikit-learn keeps the class shares equal in both sets. The search for the best settings below splits the training stars once more. The stars then divide into 60 percent fit stars, 20 percent validation stars and 20 percent test stars. The fit stars set the weights, the validation stars choose the settings, and the test stars give the final score.

We standardize the features to zero mean and unit variance, with the mean and the standard deviation of the training stars. A column of ones, the bias term, completes the feature matrix.

Standardizing the features has two purposes.

\begin{enumerate}
\def\labelenumi{\arabic{enumi}.}
\tightlist
\item
  \textbf{Numerical stability.} Stellar parameters span very different scales, from temperatures of thousands of kelvin to metallicities in logarithmic units. Without standardization, a feature with a large scale can dominate the linear combination at the start of the descent.
\item
  \textbf{Efficiency of gradient descent.} Standardized features usually give the loss surface a rounder shape. One learning rate then suits all weights better, with fewer tiny steps along some directions and fewer overshoots along others.
\end{enumerate}

We now split and standardize the stars.

\tutorialbarrier
\begin{codecell}

\end{codecell}
\begin{Verbatim}[fontsize=\footnotesize, xleftmargin=5mm]
Training stars = 2284, test stars = 571
\end{Verbatim}

\section*{Training with Batch Gradient Descent}

We begin with batch gradient descent, which uses the entire training set to compute the gradient at each iteration. This gives the most accurate gradient estimate, but it can be expensive for large data sets. The update rule is the standard formula of gradient descent

\[\mathbf{w}^{(\tau+1)} = \mathbf{w}^{(\tau)} - \eta \nabla E(\mathbf{w}),\]

where \(\tau\) counts the iterations and \(\eta\) is the learning rate, a setting that controls the step size. A learning rate that is too small gives a very slow convergence. A learning rate that is too large may overshoot the minimum or even diverge.

We monitor the training loss and the test loss during the optimization. This gives us three pieces of information.

\begin{itemize}
\tightlist
\item
  The training loss measures how well we fit the labeled data.
\item
  The test loss shows how well the model generalizes to stars that it has not seen.
\item
  Overfitting shows as a divergence between the two losses.
\end{itemize}

We start from random weights and run 2,000 iterations with \(\eta = 0.1\). At every iteration we record both losses.

\tutorialbarrier
\begin{codecell}

\end{codecell}

We now plot the two learning curves.

\tutorialbarrier
\begin{codecell}

\end{codecell}
\tutorialplot{\includegraphics[width=4.150in,height=0.75\textheight,keepaspectratio]{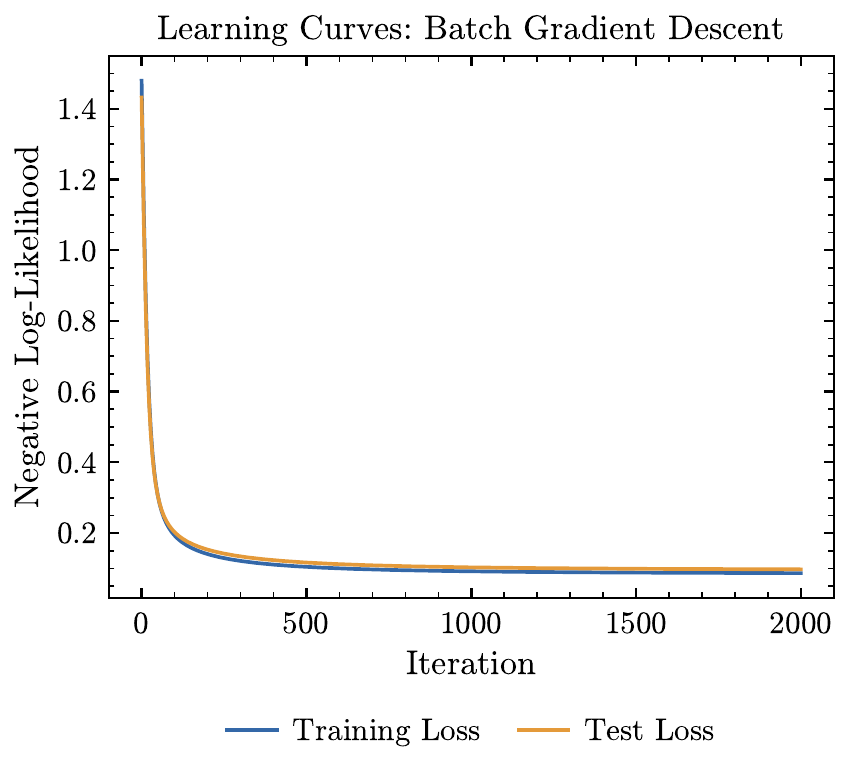}}%
  {\includegraphics[height=\tutorialheight]{tutorials/7/fig03.pdf}}

The learning curves show the typical behavior of gradient descent. Both the training loss and the test loss fall quickly during the first few hundred iterations. The model learns the main separation between the RC and RGB populations in this phase. The curves then become nearly flat. The loss is convex and has a single minimum, so more iterations give little improvement.

The training loss and the test loss stay close together, which suggests that the model generalizes without overfitting. This is expected, because the number of features (8) is small compared with the number of training stars (about 2,300).

\section*{Mini-batch Gradient Descent}

Batch gradient descent converges steadily, but it becomes expensive for large data sets. Modern surveys such as APOGEE observe a very large number of stars. A gradient over the whole data set at each iteration is not practical.

Mini-batch gradient descent is a compromise. Instead of all \(N\) training stars, we sample small random batches of size \(M\) (typically 16 to 256). We compute the gradient over each batch

\[\nabla E_{\text{batch}}(\mathbf{w}) = \frac{1}{M}\sum_{i \in \text{batch}} (y_i - t_i)\mathbf{x}_i.\]

This approach has several advantages.

\begin{itemize}
\tightlist
\item
  \textbf{Speed.} We update the weights \(N/M\) times per epoch, a pass through all training stars, and not once.
\item
  \textbf{Stochastic updates.} The noise of the random batches can help an optimizer escape poor local minima. The loss of logistic regression has no such minima, so here the main gain is speed.
\item
  \textbf{Memory use.} We need to load only a small subset of the data at a time.
\end{itemize}

We write the training loop as a function. Each epoch shuffles the stars, updates the weights on each mini-batch and records the training loss and the loss on a second set of stars. The function also keeps the weights with the lowest loss on the second set.

\tutorialbarrier
\begin{codecell}

\end{codecell}

We train with a learning rate of 0.1 and a batch size of 32 for 200 epochs.

\tutorialbarrier
\begin{codecell}
\begin{Verbatim}[commandchars=\\\{\},fontsize=\small]
\PYG{c+c1}{\PYGZsh{} Train with mini\PYGZhy{}batch gradient descent}
\PYG{n}{w\PYGZus{}minibatch}\PYG{p}{,}\FancyVerbBreakAfterBreak{} \PYG{n}{\PYGZus{}}\PYG{p}{,}\FancyVerbBreakAfterBreak{} \PYG{n}{train\PYGZus{}hist}\PYG{p}{,}\FancyVerbBreakAfterBreak{} \PYG{n}{test\PYGZus{}hist} \PYG{o}{=}\FancyVerbBreakAfterBreak{} \PYG{n}{train\PYGZus{}minibatch}\PYG{p}{(}\FancyVerbBreakAfterBreak{}
    \PYG{n}{X\PYGZus{}train}\PYG{p}{,}\FancyVerbBreakAfterBreak{} \PYG{n}{t\PYGZus{}train}\PYG{p}{,}\FancyVerbBreakAfterBreak{} \PYG{n}{X\PYGZus{}test}\PYG{p}{,}\FancyVerbBreakAfterBreak{} \PYG{n}{t\PYGZus{}test}\PYG{p}{,}\FancyVerbBreakAfterBreak{} \PYG{n}{learning\PYGZus{}rate}\PYG{o}{=}\FancyVerbBreakAfterBreak{}\PYG{l+m+mf}{0.1}\PYG{p}{,}\FancyVerbBreakAfterBreak{} \PYG{n}{n\PYGZus{}epochs}\PYG{o}{=}\FancyVerbBreakAfterBreak{}\PYG{l+m+mi}{200}\PYG{p}{,}\FancyVerbBreakAfterBreak{} \PYG{n}{batch\PYGZus{}size}\PYG{o}{=}\FancyVerbBreakAfterBreak{}\PYG{l+m+mi}{32}
\PYG{p}{)}
\end{Verbatim}
\end{codecell}

We plot the learning curves and print the final losses.

\tutorialbarrier
\begin{codecell}

\end{codecell}
\tutorialplot{\includegraphics[width=4.221in,height=0.75\textheight,keepaspectratio]{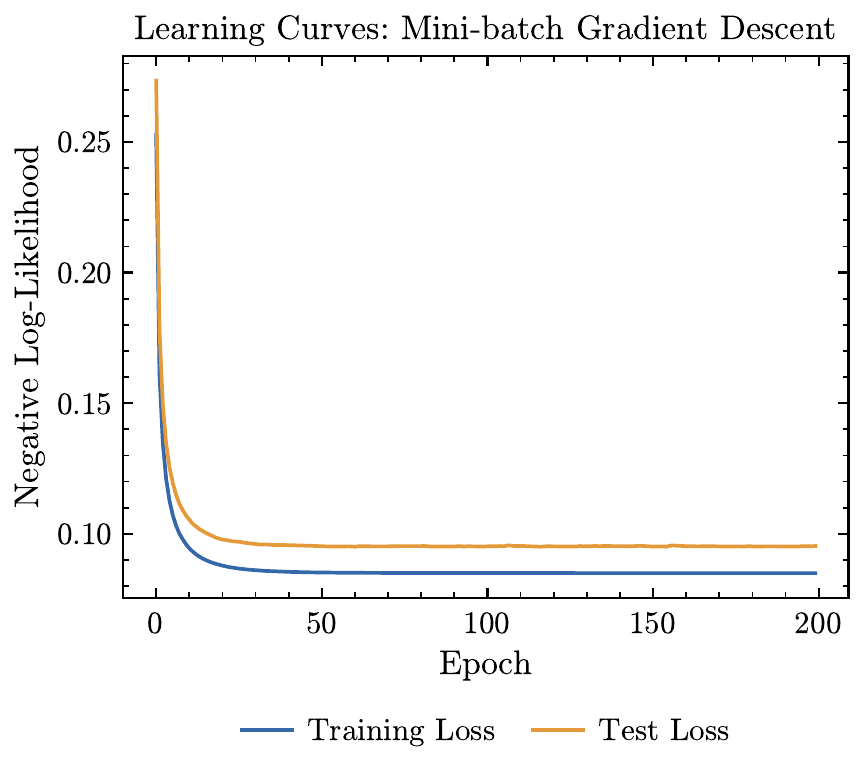}}%
  {\includegraphics[height=\tutorialheight]{tutorials/7/fig04.pdf}}
\begin{Verbatim}[fontsize=\footnotesize, xleftmargin=5mm]
Final training loss = 0.0849, final test loss = 0.0953
\end{Verbatim}

The mini-batch loss falls quickly and is nearly flat after about 30 epochs, which are 30 passes through the training stars. Batch gradient descent needs its 2,000 iterations, each one a pass through the stars, to reach a similar loss. The training loss and the test loss end close together, so the model generalizes.

\section*{Model Evaluation}

We have trained our classifier, so we now test how well it works. We ask two questions about the RC stars. How many of them does the model find? How clean is the sample that the model calls RC? A confusion matrix sorts the predictions into four kinds and answers both. We call the RC class the positive class, because it is the class that we want to find.

\begin{itemize}
\tightlist
\item
  A true positive (TP) is an RC star that the model calls RC.
\item
  A true negative (TN) is an RGB star that the model calls RGB.
\item
  A false positive (FP) is an RGB star that the model calls RC.
\item
  A false negative (FN) is an RC star that the model calls RGB.
\end{itemize}

The first question is the \textbf{recall}. It is the fraction of the RC stars that the model finds, which are the RC stars that it calls RC

\[\text{Recall} = \frac{TP}{TP + FN}.\]

A false negative removes a reliable distance indicator from our sample. To keep the sample large, we want a high recall.

The second question is the \textbf{precision}. It is the fraction of the stars that the model calls RC that truly are RC

\[\text{Precision} = \frac{TP}{TP + FP}.\]

Like the recall, the precision refers to the positive class. A false positive, an RGB star taken for an RC star, gives a systematic error in the distance. A sample of distance indicators needs few RGB stars, so we want to limit the false positives and keep a high precision.

The \textbf{accuracy} gives one number for all predictions. It is the fraction of correct predictions

\[\text{Accuracy} = \frac{TP + TN}{TP + TN + FP + FN}.\]

The accuracy can mislead when one class is much larger than the other. A classifier that always predicts the majority class gets a high accuracy but fails at the task. Here the RC stars outnumber the RGB stars.

The \textbf{balanced accuracy} avoids this problem. It averages the share of RC stars that the model calls RC and the share of RGB stars that it calls RGB. It therefore treats the classes equally whatever their size

\[\text{Balanced accuracy} = \frac{1}{2}\left(\frac{TP}{TP + FN} + \frac{TN}{TN + FP}\right).\]

We now code the prediction and the metrics.

\tutorialbarrier
\begin{codecell}

\end{codecell}

We evaluate the mini-batch model on the test stars. We also score a baseline that calls every star RC, to see what the class sizes alone give.

\tutorialbarrier
\begin{codecell}

\end{codecell}
\begin{Verbatim}[fontsize=\footnotesize, xleftmargin=5mm]
Recall = 0.988, precision = 0.977
Accuracy = 0.974, balanced accuracy = 0.960

               Predicted RGB    Predicted RC
Actual RGB               136              10
Actual RC                  5             420

Every star called RC: accuracy = 0.744, balanced accuracy = 0.500
\end{Verbatim}

The model finds almost all RC stars, with a recall of 0.988. The precision is 0.977, so about 2 percent of the stars that the model calls RC are RGB stars. They would give wrong distances in a sample of RC distance indicators. The accuracy is high, and the balanced accuracy is a little lower, because the model errs more often on the smaller RGB class.

The baseline shows why we need the balanced accuracy. The RC stars are 74 percent of the test stars. Calling every star RC therefore gives an accuracy of 0.744 and a balanced accuracy of only 0.5. The model is far above both values.

We also look at the predicted probabilities. The histogram shows them separately for the true RC stars and the true RGB stars.

\tutorialbarrier
\begin{codecell}

\end{codecell}
\tutorialplot{\includegraphics[width=4.097in,height=0.75\textheight,keepaspectratio]{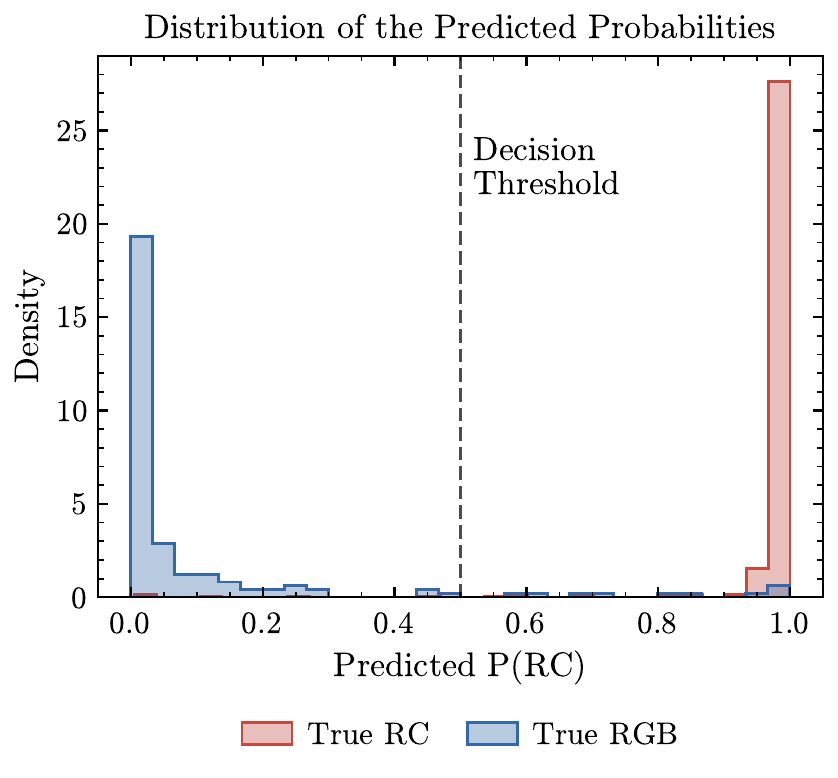}}%
  {\includegraphics[height=\tutorialheight]{tutorials/7/fig05.pdf}}

The distribution of the probabilities shows how well our model separates the two populations. True RC stars (red) cluster near the probability 1, and true RGB stars (blue) concentrate near 0. The few stars between the two peaks are the most difficult cases. Their spectroscopic parameters place them in the region where the populations overlap. They deserve special attention in astronomical applications, because they are the stars most likely to be misclassified and to introduce systematic errors in distances.

\section*{The Probability Threshold and the ROC Curve}

So far we call a star RC when its probability is at least 0.5. The stars that we call RC form our RC sample. This threshold is a choice, and the best value depends on the use of the sample. A sample of distance indicators needs few RGB stars, so it can use a higher threshold. A higher threshold removes RGB stars from the RC sample, but it also loses RC stars.

We measure this trade-off with the recall and the precision. The recall is the fraction of all RC stars that are in the RC sample. The precision is the fraction of the RC sample that are RC stars. A higher threshold raises the precision and lowers the recall.

A third rate counts the RGB stars that we let in. The \textbf{false positive rate} (FPR) is the fraction of the RGB stars that we call RC

\[\text{FPR} = \frac{FP}{FP + TN}.\]

The recall is also called the true positive rate.

We apply five thresholds to the probabilities of the mini-batch model on the test stars, and we list the confusion matrix and the three rates. We only look at the trade-off and do not choose a threshold, so no choice uses the test stars.

\tutorialbarrier
\begin{codecell}

\end{codecell}
\begin{Verbatim}[fontsize=\footnotesize, xleftmargin=5mm]
Threshold    TN   FP   FN    TP   Recall     FPR  Precision
      0.3   133   13    4   421    0.991   0.089      0.970
      0.5   136   10    5   420    0.988   0.068      0.977
      0.7   139    7    9   416    0.979   0.048      0.983
      0.9   142    4   12   413    0.972   0.027      0.990
     0.99   144    2  171   254    0.598   0.014      0.992
\end{Verbatim}

A higher threshold gives fewer false positives, so the false positive rate falls and the precision rises. The precision rises from 0.977 at 0.5 to 0.990 at 0.9. The price is more false negatives, which are RC stars that we lose, so the recall falls. At 0.99 the model loses 171 of the 425 RC stars. A lower threshold does the opposite. It keeps more RC stars and lets in more RGB stars.

The five rows are five points on one curve of this trade-off. The receiver operating characteristic (ROC) curve shows the whole curve. It plots the recall against the false positive rate for every threshold. Both are fractions within one class, so the curve does not change when one class is larger than the other.

The threshold moves the point along the curve. A low threshold calls many stars RC. The recall and the false positive rate are then both high, and the point lies at the upper right. A high threshold calls few stars RC, so both are low and the point lies at the lower left. A classifier that guesses at random gives the diagonal. A good classifier reaches a high recall at a low false positive rate, so its curve lies near the upper left corner.

We now compute the recall and the false positive rate for the test stars.

\tutorialbarrier
\begin{codecell}

\end{codecell}

We now plot the ROC curve, the diagonal of random guesses and the five thresholds of the table.

\tutorialbarrier
\begin{codecell}

\end{codecell}
\tutorialplot{\includegraphics[width=4.620in,height=0.75\textheight,keepaspectratio]{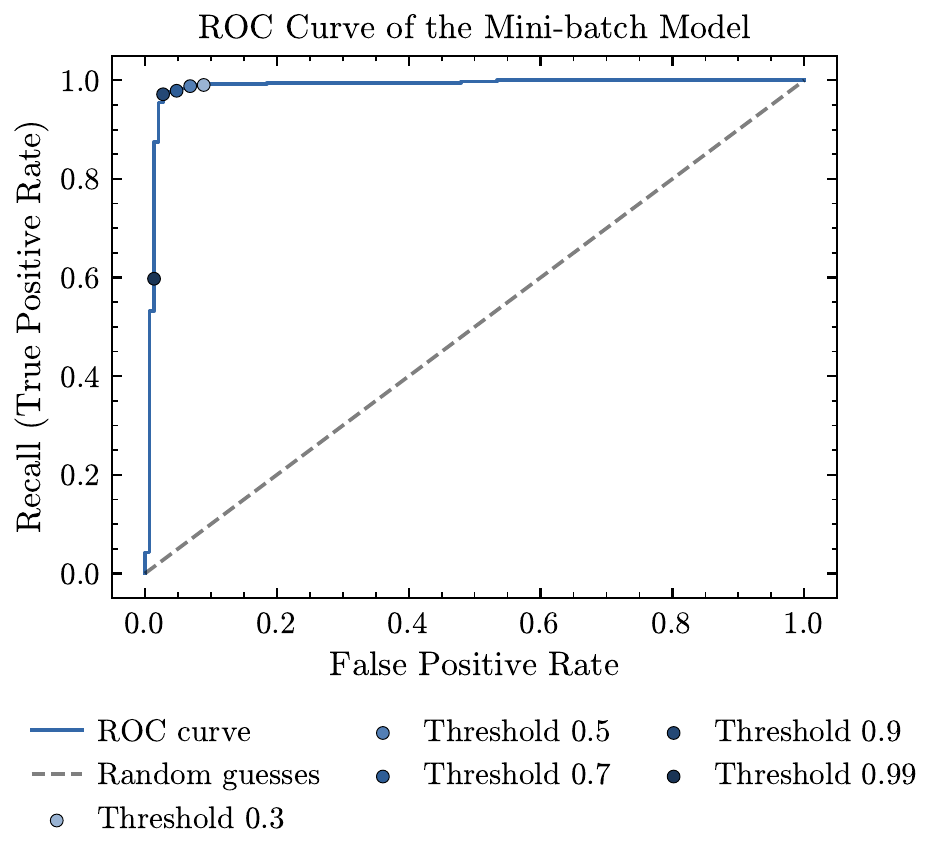}}%
  {\includegraphics[height=\tutorialheight]{tutorials/7/fig06.pdf}}

The curve rises steeply and stays close to the upper left corner, far from the diagonal, so the classifier separates the two classes well. The marks for the thresholds 0.3 to 0.9 lie close together near the corner, with a recall above 0.97. The mark for 0.99 lies far below them, because the model loses many RC stars there.

We can summarize how close the curve lies to the upper left corner by the area under it, the AUC. It is 0.5 for random guesses and 1 for a perfect classifier. The AUC does not depend on the threshold, so it is a good way to compare different classifiers. We use it below to compare the settings of our own classifier.

The AUC has a direct meaning. Take one RC star and one RGB star. A good classifier gives the RC star the higher probability P(RC). The AUC is the fraction of all such pairs in which it does.

We count these pairs. We also sum the thin strips under the curve, to check that the two agree. Each strip lies between two neighboring thresholds. Its width is the step of the false positive rate, and its height is the mean recall at the two ends.

\tutorialbarrier
\begin{codecell}
\begin{Verbatim}[commandchars=\\\{\},fontsize=\small]
\PYG{k}{def}\PYG{+w}{ }\PYG{n+nf}{auc\PYGZus{}score}\PYG{p}{(}\FancyVerbBreakAfterBreak{}\PYG{n}{probabilities}\PYG{p}{,}\FancyVerbBreakAfterBreak{} \PYG{n}{labels}\PYG{p}{)}\PYG{p}{:}
\PYG{+w}{    }\PYG{l+s+sd}{\PYGZdq{}\PYGZdq{}\PYGZdq{}Return the fraction of the pairs of an RC star and an RGB star in which the RC star has the higher probability.\PYGZdq{}\PYGZdq{}\PYGZdq{}}

    \PYG{c+c1}{\PYGZsh{} Split the probabilities by the true class}
    \PYG{n}{rc\PYGZus{}probs} \PYG{o}{=}\FancyVerbBreakAfterBreak{} \PYG{n}{probabilities}\PYG{p}{[}\FancyVerbBreakAfterBreak{}\PYG{n}{labels} \PYG{o}{==} \PYG{l+m+mi}{1}\PYG{p}{]}
    \PYG{n}{rgb\PYGZus{}probs} \PYG{o}{=}\FancyVerbBreakAfterBreak{} \PYG{n}{probabilities}\PYG{p}{[}\FancyVerbBreakAfterBreak{}\PYG{n}{labels} \PYG{o}{==} \PYG{l+m+mi}{0}\PYG{p}{]}
    \PYG{k}{return} \PYG{n}{np}\PYG{o}{.}\FancyVerbBreakAfterBreak{}\PYG{n}{mean}\PYG{p}{(}\FancyVerbBreakAfterBreak{}\PYG{n}{rc\PYGZus{}probs}\PYG{p}{[}\FancyVerbBreakAfterBreak{}\PYG{p}{:}\PYG{p}{,}\FancyVerbBreakAfterBreak{} \PYG{k+kc}{None}\PYG{p}{]} \PYG{o}{\PYGZgt{}} \PYG{n}{rgb\PYGZus{}probs}\PYG{p}{[}\FancyVerbBreakAfterBreak{}\PYG{k+kc}{None}\PYG{p}{,}\FancyVerbBreakAfterBreak{} \PYG{p}{:}\PYG{p}{]}\PYG{p}{)}  \PYG{c+c1}{\PYGZsh{} compare every RC star with every RGB star}

\PYG{c+c1}{\PYGZsh{} The same area from the curve, as thin strips between neighboring thresholds}
\PYG{n}{widths} \PYG{o}{=}\FancyVerbBreakAfterBreak{} \PYG{n}{fpr}\PYG{p}{[}\FancyVerbBreakAfterBreak{}\PYG{p}{:}\PYG{o}{\PYGZhy{}}\PYG{l+m+mi}{1}\PYG{p}{]} \PYG{o}{\PYGZhy{}} \PYG{n}{fpr}\PYG{p}{[}\FancyVerbBreakAfterBreak{}\PYG{l+m+mi}{1}\PYG{p}{:}\PYG{p}{]}  \PYG{c+c1}{\PYGZsh{} step of the false positive rate, which falls as the threshold rises}
\PYG{n}{heights} \PYG{o}{=}\FancyVerbBreakAfterBreak{} \PYG{p}{(}\FancyVerbBreakAfterBreak{}\PYG{n}{tpr}\PYG{p}{[}\FancyVerbBreakAfterBreak{}\PYG{p}{:}\PYG{o}{\PYGZhy{}}\PYG{l+m+mi}{1}\PYG{p}{]} \PYG{o}{+}\FancyVerbBreakAfterBreak{} \PYG{n}{tpr}\PYG{p}{[}\FancyVerbBreakAfterBreak{}\PYG{l+m+mi}{1}\PYG{p}{:}\PYG{p}{]}\PYG{p}{)} \PYG{o}{/} \PYG{l+m+mi}{2}  \PYG{c+c1}{\PYGZsh{} mean recall at the two ends of the step}
\PYG{n+nb}{print}\PYG{p}{(}\FancyVerbBreakAfterBreak{}\PYG{l+s+sa}{f}\PYG{l+s+s2}{\PYGZdq{}}\PYG{l+s+s2}{Fraction of ordered pairs = }\PYG{l+s+si}{\PYGZob{}}\PYG{n}{auc\PYGZus{}score}\PYG{p}{(}\FancyVerbBreakAfterBreak{}\PYG{n}{test\PYGZus{}probs}\PYG{p}{,}\FancyVerbBreakAfterBreak{}\PYG{+w}{ }\PYG{n}{t\PYGZus{}test}\PYG{p}{)}\PYG{l+s+si}{:}\PYG{l+s+s2}{.4f}\PYG{l+s+si}{\PYGZcb{}}\FancyVerbBreakAfterBreak{}\PYG{l+s+s2}{, }\PYG{l+s+s2}{\PYGZdq{}}
      \PYG{l+s+sa}{f}\PYG{l+s+s2}{\PYGZdq{}}\PYG{l+s+s2}{area under the curve = }\PYG{l+s+si}{\PYGZob{}}\PYG{n}{np}\PYG{o}{.}\FancyVerbBreakAfterBreak{}\PYG{n}{sum}\PYG{p}{(}\FancyVerbBreakAfterBreak{}\PYG{n}{widths}\PYG{+w}{ }\PYG{o}{*}\PYG{+w}{ }\PYG{n}{heights}\PYG{p}{)}\PYG{l+s+si}{:}\PYG{l+s+s2}{.4f}\PYG{l+s+si}{\PYGZcb{}}\FancyVerbBreakAfterBreak{}\PYG{l+s+s2}{\PYGZdq{}}\PYG{p}{)}
\end{Verbatim}
\end{codecell}
\begin{Verbatim}[fontsize=\footnotesize, xleftmargin=5mm]
Fraction of ordered pairs = 0.9858, area under the curve = 0.9858
\end{Verbatim}

The two values agree. The AUC is 0.986, close to the value 1 of a perfect classifier, so the model puts almost all pairs in the right order. A survey can read the curve to find the threshold that keeps enough RC stars at the false positive rate that it accepts.

\section*{Hyperparameter Optimization}

Our first model used hyperparameters that we fixed in advance. Hyperparameters are settings that control the training and that the data do not determine. Two of them matter most for mini-batch gradient descent.

\begin{enumerate}
\def\labelenumi{\arabic{enumi}.}
\tightlist
\item
  \textbf{Learning rate (\(\eta\)).} It controls the step size in parameter space. A value that is too small gives a very slow convergence, and a value that is too large causes instability or divergence.
\item
  \textbf{Batch size (\(M\)).} It sets the number of stars that estimate each gradient. Smaller batches increase the stochasticity but reduce the accuracy of the gradient.
\end{enumerate}

These hyperparameters interact. Small batches give noisier gradients, which can need a smaller learning rate for stability. Large batches give accurate gradients but make fewer updates per epoch.

We explore them with a grid search. The threshold is a separate choice, so we score each model by the AUC, which does not depend on it. We train a model for 100 epochs for each combination. We trace the validation loss after every epoch and keep the weights with the lowest validation loss. The combination with the highest AUC of these weights on the validation stars wins (Section 4.7).

We split the training stars for this into 75 percent fit stars and 25 percent validation stars. Of all stars, 60 percent are then fit stars, 20 percent validation stars and 20 percent test stars. The fit stars set the weights. The test stars take no part in the choice, so they stay a fair test of the final model.

\tutorialbarrier
\begin{codecell}

\end{codecell}
\begin{Verbatim}[fontsize=\footnotesize, xleftmargin=5mm]
Best: learning rate 0.01, batch size 32, validation AUC 0.9817
\end{Verbatim}

We show the validation AUC of every combination as a heatmap, and we mark the best combination with a star. The color scale starts at 0.97, so all lower values share the lightest color.

\tutorialbarrier
\begin{codecell}

\end{codecell}
\tutorialplot{\includegraphics[width=6.134in,height=0.75\textheight,keepaspectratio]{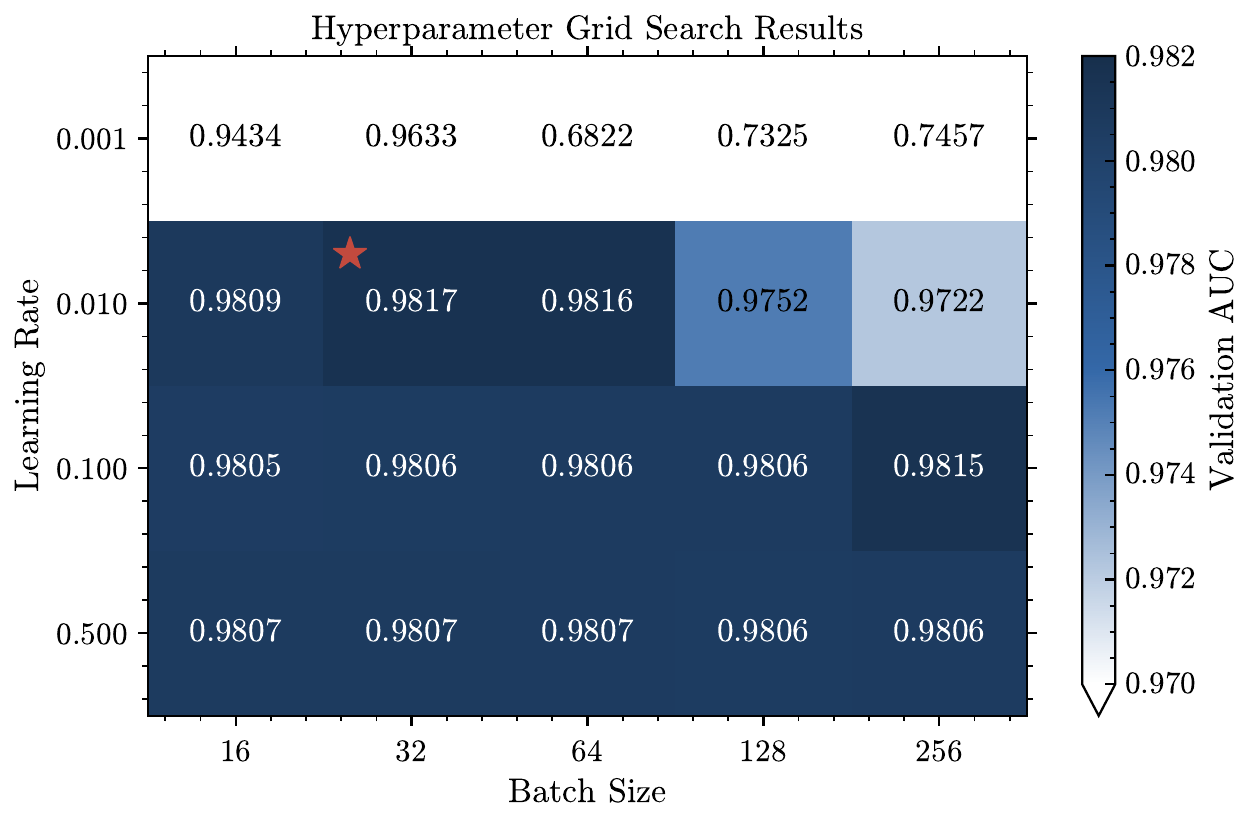}}%
  {\includegraphics[height=\tutorialheight]{tutorials/7/fig07.pdf}}

We read three points from the heatmap.

\begin{enumerate}
\def\labelenumi{\arabic{enumi}.}
\tightlist
\item
  \textbf{Learning rate effects.} The smallest learning rate, 0.001, gives poor results, likely because 100 epochs are too few for the optimizer to make enough progress. The three larger values give an AUC between 0.972 and 0.982.
\item
  \textbf{Batch size effects.} At the small learning rates, 0.001 and 0.01, large batches give a lower AUC than small batches. A larger batch makes fewer updates in a fixed number of epochs. At the learning rates 0.1 and 0.5 the batch size matters little.
\item
  \textbf{Interaction effects.} The best learning rate depends on the batch size. Small batches give noisy gradients and do well with a smaller learning rate, and large batches need a larger one. The interaction is modest for 0.1, which gives an AUC of about 0.98 for all batch sizes.
\end{enumerate}

\section*{Final Model}

The best weights of the best combination are our final model. They come from the fit stars, so the validation stars stay validation stars, and the test stars stay untouched until now. We score the model on the test stars, with the accuracy, the balanced accuracy and the AUC.

\tutorialbarrier
\begin{codecell}
\begin{Verbatim}[commandchars=\\\{\},fontsize=\small]
\PYG{c+c1}{\PYGZsh{} Evaluate the final model on the test stars}
\PYG{n}{test\PYGZus{}pred\PYGZus{}final}\PYG{p}{,}\FancyVerbBreakAfterBreak{} \PYG{n}{test\PYGZus{}probs\PYGZus{}final} \PYG{o}{=}\FancyVerbBreakAfterBreak{} \PYG{n}{predict}\PYG{p}{(}\FancyVerbBreakAfterBreak{}\PYG{n}{w\PYGZus{}final}\PYG{p}{,}\FancyVerbBreakAfterBreak{} \PYG{n}{X\PYGZus{}test}\PYG{p}{)}
\PYG{n}{final\PYGZus{}metrics} \PYG{o}{=}\FancyVerbBreakAfterBreak{} \PYG{n}{compute\PYGZus{}metrics}\PYG{p}{(}\FancyVerbBreakAfterBreak{}\PYG{n}{t\PYGZus{}test}\PYG{p}{,}\FancyVerbBreakAfterBreak{} \PYG{n}{test\PYGZus{}pred\PYGZus{}final}\PYG{p}{)}
\PYG{n+nb}{print}\PYG{p}{(}\FancyVerbBreakAfterBreak{}\PYG{l+s+sa}{f}\PYG{l+s+s2}{\PYGZdq{}}\PYG{l+s+s2}{Accuracy = }\PYG{l+s+si}{\PYGZob{}}\PYG{n}{final\PYGZus{}metrics}\PYG{p}{[}\FancyVerbBreakAfterBreak{}\PYG{l+s+s1}{\PYGZsq{}}\PYG{l+s+s1}{accuracy}\PYG{l+s+s1}{\PYGZsq{}}\PYG{p}{]}\PYG{l+s+si}{:}\PYG{l+s+s2}{.3f}\PYG{l+s+si}{\PYGZcb{}}\FancyVerbBreakAfterBreak{}\PYG{l+s+s2}{, balanced accuracy = }\PYG{l+s+si}{\PYGZob{}}\PYG{n}{final\PYGZus{}metrics}\PYG{p}{[}\FancyVerbBreakAfterBreak{}\PYG{l+s+s1}{\PYGZsq{}}\PYG{l+s+s1}{balanced\PYGZus{}accuracy}\PYG{l+s+s1}{\PYGZsq{}}\PYG{p}{]}\PYG{l+s+si}{:}\PYG{l+s+s2}{.3f}\PYG{l+s+si}{\PYGZcb{}}\FancyVerbBreakAfterBreak{}\PYG{l+s+s2}{, }\PYG{l+s+s2}{\PYGZdq{}}
      \PYG{l+s+sa}{f}\PYG{l+s+s2}{\PYGZdq{}}\PYG{l+s+s2}{AUC = }\PYG{l+s+si}{\PYGZob{}}\PYG{n}{auc\PYGZus{}score}\PYG{p}{(}\FancyVerbBreakAfterBreak{}\PYG{n}{test\PYGZus{}probs\PYGZus{}final}\PYG{p}{,}\FancyVerbBreakAfterBreak{}\PYG{+w}{ }\PYG{n}{t\PYGZus{}test}\PYG{p}{)}\PYG{l+s+si}{:}\PYG{l+s+s2}{.4f}\PYG{l+s+si}{\PYGZcb{}}\FancyVerbBreakAfterBreak{}\PYG{l+s+s2}{\PYGZdq{}}\PYG{p}{)}
\end{Verbatim}
\end{codecell}
\begin{Verbatim}[fontsize=\footnotesize, xleftmargin=5mm]
Accuracy = 0.968, balanced accuracy = 0.945, AUC = 0.9851
\end{Verbatim}

The final model scores comparably to the first model on the test stars, with an accuracy of 0.968 and an AUC of 0.985.

\section*{Visualizing the Decision Boundary}

To see how our model separates RC and RGB stars, we visualize the decision boundary in the plane of temperature and gravity. These are the two main spectroscopic parameters. The model has more features, so we set the other features to their median values. This gives a slice through the high-dimensional feature space.

The decision boundary lies where \(P(\text{RC}) = 0.5\), which corresponds to \(\mathbf{w}^T\mathbf{x} = 0\). We evaluate the predictions of our model on a grid of temperatures and gravities. This maps the regions where the model gives a probability close to 1 for the RC class or the RGB class.

\tutorialbarrier
\begin{codecell}

\end{codecell}

We now draw the probability as filled contours, the decision boundary as a black line and the test stars as points. An orange cross marks each star that the model classifies wrongly.

\tutorialbarrier
\begin{codecell}

\end{codecell}
\tutorialplot{\includegraphics[width=4.408in,height=0.75\textheight,keepaspectratio]{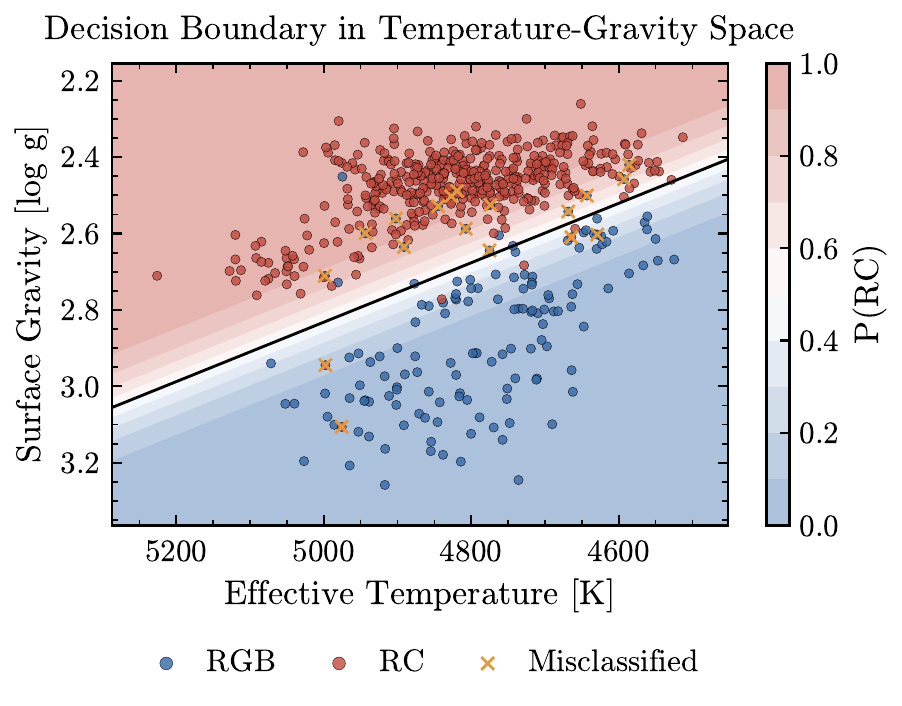}}%
  {\includegraphics[height=\tutorialheight]{tutorials/7/fig08.pdf}}

The black contour marks \(P(\text{RC}) = 0.5\), the decision threshold. In the space of the eight features, the condition \(\mathbf{w}^T\mathbf{x} = 0\) describes a flat surface, a hyperplane. Our plot is a two-dimensional slice of that space, with the other six features fixed. A hyperplane cuts a two-dimensional slice along a straight line, so the contour is a line. The sigmoid does not bend the boundary (Section 7.3).

We read four things from the plot.

\begin{enumerate}
\def\labelenumi{\arabic{enumi}.}
\tightlist
\item
  \textbf{Physical coherence.} RC stars (red) cluster in a compact region around \(\log g \approx 2.5\), and RGB stars (blue) spread over higher gravities. The model learns this separation in gravity without any explicit physical constraint.
\item
  \textbf{Regions of extreme probability.} The darkest red and blue regions mark probabilities close to 1 for one class. The model gives these to the stars far from the boundary, the RGB stars at high gravity and the tightly clustered RC stars.
\item
  \textbf{Ambiguous zone.} The gradient region around the decision boundary holds the most difficult classifications. A small measurement error could move a star to the other side of the boundary here. The probability of such a star lies near 0.5, so its two classes are almost equally probable.
\item
  \textbf{Misclassifications.} Orange crosses mark the stars that the model classifies wrongly. Most of them lie on the side of the line that belongs to the other class, and one lies on the correct side. The model uses all eight features, and the line is only the slice for the median values of the other six. A star with other values has its own slice, a parallel line that lies a little higher or lower in the plot.
\end{enumerate}

\section*{Summary}

This tutorial showed how logistic regression classifies astronomical objects, with two classes of giant stars as the example. We transformed a linear model with the sigmoid function and obtained a probabilistic classifier. It reaches a high accuracy.

Our main lessons follow.

\begin{enumerate}
\def\labelenumi{\arabic{enumi}.}
\tightlist
\item
  \textbf{Probabilistic approach.} Logistic regression gives the probability that an object belongs to a class. When the classes feed a later analysis, these probabilities give the chance of each class for each object, and later steps can use them. A threshold on the probability sets the balance between a clean sample and a complete one. The ROC curve shows this balance for all thresholds.
\item
  \textbf{Optimization methods.} We implemented batch and mini-batch gradient descent from scratch. Mini-batches scale to modern surveys with a very large number of objects, because each update needs only a small part of the data. For this convex loss the main gain is speed.
\item
  \textbf{Evaluation and hyperparameters.} The confusion matrix, the recall and the precision tell more than the accuracy. The balanced accuracy corrects the accuracy for classes of unequal size. The AUC scores a classifier without a choice of threshold. We chose the learning rate and the batch size with a grid search on the AUC of validation stars. The final model has the weights with the lowest validation loss, and we scored it on test stars.
\end{enumerate}

The same approach can apply to other classification problems in astronomy. It needs a linear boundary that separates the classes well. Logistic regression gives a principled and interpretable baseline against which we can evaluate more complex methods.

\tutorialchapter{8}{Multi-Class Logistic Regression for Strong Lensing}
\noindent\textit{Companion material for Chapter 8}

\section*{Introduction}

In Chapter 7, we developed the mathematical description of binary classification with logistic regression. It separates two classes with the sigmoid function and gradient descent. We now extend these concepts to multi-class classification, where we must distinguish among many classes. This situation often arises in astronomy.

This tutorial studies multi-class logistic regression with strong gravitational lensing. The curvature of spacetime near a massive foreground object deflects the light of a distant galaxy, as general relativity predicts. The geometry of the lens and the source decides the pattern. It can be multiple images of the same galaxy, giant arcs, complete Einstein rings or short radial arcs near the center of the lens. Such lenses map the dark matter of galaxy clusters and magnify galaxies of the early universe. The time delays between multiple images also measure cosmological parameters.

The James Webb Space Telescope (JWST) changed the study of strong lensing. With its segmented primary mirror of 6.5 meters and its infrared instruments, JWST reaches an angular resolution of about 0.1 arcseconds in the near-infrared. This is enough to resolve individual star-forming regions in lensed galaxies at cosmic noon, the epoch of peak star formation in the universe. A single deep field can contain dozens of lensing features, such as arcs of different curvature, multiple images and partial Einstein rings. Classification by eye is not practical for large surveys, so we need automatic methods that distinguish between different types of objects.

In this tutorial, we build a complete multi-class classification pipeline, from cutouts of JWST images to the evaluation of the classes. A human or a vision-language model can label an image, but this is slow and costly for every new image. We ask whether a classifier can learn the labels from the features of labeled images and give the same labels to new images. The step from binary to multi-class classification shows how machine learning concepts extend to more complex problems, while the computation stays tractable.

\section*{Prerequisites and Learning Objectives}

\subsection*{Prerequisites}

This tutorial uses the concepts of Chapter 7 and assumes the following.

\begin{itemize}
\tightlist
\item
  Binary logistic regression, with the sigmoid function and gradient descent
\item
  The cross-entropy loss and its probabilistic interpretation
\item
  The evaluation of a binary classifier with the confusion matrix, the recall and the precision
\item
  The gradient of a loss function and its use in optimization
\item
  NumPy operations such as matrix multiplication, broadcasting and array manipulation
\item
  Python programming and vectorized computations
\end{itemize}

\subsection*{Learning Objectives}

After this tutorial you can do the following.

\begin{itemize}
\tightlist
\item
  Extend binary classification to many classes with the softmax function and the multi-class cross-entropy loss
\item
  Convert images into numerical features with a pre-trained neural network
\item
  Implement the softmax function in a numerically stable way
\item
  Implement stochastic gradient descent with mini-batches
\item
  Choose the learning rate and the weight decay on validation objects and score the model on test objects
\item
  Build and read a confusion matrix, compute the precision and the recall of each class and find systematic confusions between classes
\end{itemize}

These skills are the basis of classification pipelines for large surveys.

We begin by importing the libraries and setting the plot style.

\tutorialbarrier
\begin{codecell}

\end{codecell}

\section*{A JWST Observation of Strong Lensing}

We begin with an example of a JWST observation of strong lensing.

\tutorialbarrier
\begin{codecell}
\begin{Verbatim}[commandchars=\\\{\},fontsize=\small]
\PYG{c+c1}{\PYGZsh{} Load and display a JWST strong lensing image}
\PYG{n}{img} \PYG{o}{=}\FancyVerbBreakAfterBreak{} \PYG{n}{Image}\PYG{o}{.}\FancyVerbBreakAfterBreak{}\PYG{n}{open}\PYG{p}{(}\FancyVerbBreakAfterBreak{}\PYG{l+s+s1}{\PYGZsq{}}\PYG{l+s+s1}{../data/dataset\PYGZus{}jwst.png}\PYG{l+s+s1}{\PYGZsq{}}\PYG{p}{)}

\PYG{c+c1}{\PYGZsh{} Show the image without axes}
\PYG{n}{plt}\PYG{o}{.}\FancyVerbBreakAfterBreak{}\PYG{n}{figure}\PYG{p}{(}\FancyVerbBreakAfterBreak{}\PYG{n}{figsize}\PYG{o}{=}\FancyVerbBreakAfterBreak{}\PYG{p}{(}\FancyVerbBreakAfterBreak{}\PYG{l+m+mf}{6.2}\PYG{p}{,}\FancyVerbBreakAfterBreak{} \PYG{l+m+mf}{5.7}\PYG{p}{)}\PYG{p}{)}  \PYG{c+c1}{\PYGZsh{} a photograph, shown at its original size}
\PYG{n}{plt}\PYG{o}{.}\FancyVerbBreakAfterBreak{}\PYG{n}{imshow}\PYG{p}{(}\FancyVerbBreakAfterBreak{}\PYG{n}{img}\PYG{p}{)}
\PYG{n}{plt}\PYG{o}{.}\FancyVerbBreakAfterBreak{}\PYG{n}{axis}\PYG{p}{(}\FancyVerbBreakAfterBreak{}\PYG{l+s+s1}{\PYGZsq{}}\PYG{l+s+s1}{off}\PYG{l+s+s1}{\PYGZsq{}}\PYG{p}{)}
\PYG{n}{plt}\PYG{o}{.}\FancyVerbBreakAfterBreak{}\PYG{n}{title}\PYG{p}{(}\FancyVerbBreakAfterBreak{}\PYG{l+s+s1}{\PYGZsq{}}\PYG{l+s+s1}{Strong Gravitational Lensing Captured by JWST}\PYG{l+s+s1}{\PYGZsq{}}\PYG{p}{,}\FancyVerbBreakAfterBreak{} \PYG{n}{pad}\PYG{o}{=}\FancyVerbBreakAfterBreak{}\PYG{l+m+mi}{20}\PYG{p}{)}
\PYG{n}{plt}\PYG{o}{.}\FancyVerbBreakAfterBreak{}\PYG{n}{tight\PYGZus{}layout}\PYG{p}{(}\FancyVerbBreakAfterBreak{}\PYG{p}{)}
\PYG{n}{plt}\PYG{o}{.}\FancyVerbBreakAfterBreak{}\PYG{n}{show}\PYG{p}{(}\FancyVerbBreakAfterBreak{}\PYG{p}{)}
\end{Verbatim}
\end{codecell}
\tutorialplot{\includegraphics[width=4.363in,height=0.75\textheight,keepaspectratio]{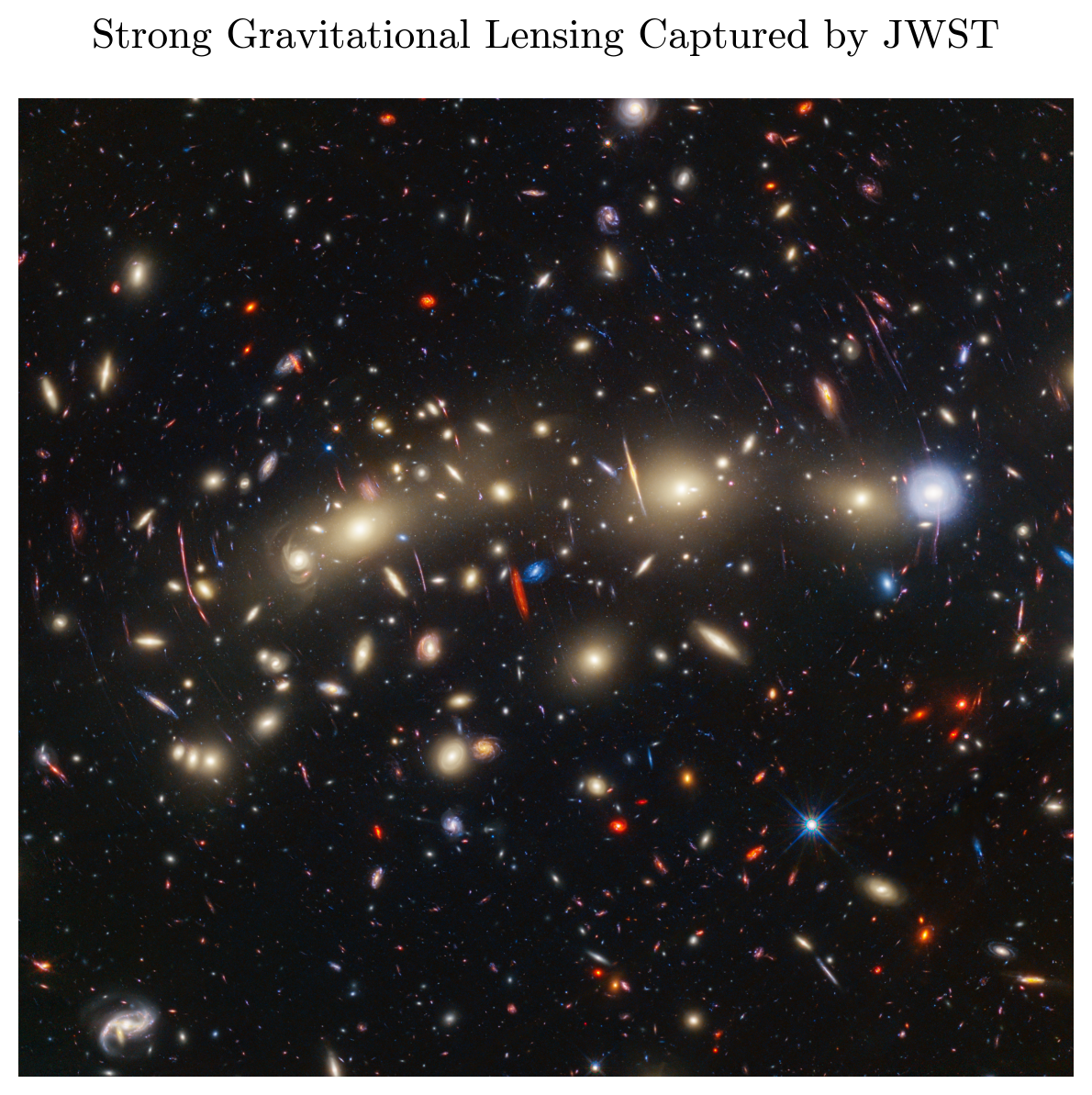}}%
  {\includegraphics[height=\tutorialheight]{tutorials/8/fig01.png}}

\section*{Data Preparation: From Raw Images to Machine Learning Input}

Raw astronomical images are difficult to use in machine learning. A single JWST deep field contains thousands of objects at different distances, with very different angular sizes, surface brightnesses and morphologies. Earlier approaches needed us to identify and extract the regions of interest manually, which does not scale to modern surveys.

For this tutorial, we preprocessed our JWST images with the Segment Anything Model (SAM), which Meta AI Research developed. SAM learned from more than 1 billion masks, which mark the pixels of an object, to find objects in many domains. It separates the objects of an astronomical image and makes precise masks around extended, irregular objects. The segmentation turns a complex field into individual cutouts, and each cutout contains a possible object of interest. A hard image analysis problem thus becomes a classification task in which each cutout belongs to a class.

We assembled a dataset of segmented objects from public JWST observations, and we preprocessed each object to make the data consistent.

\begin{itemize}
\tightlist
\item
  \textbf{Standardized size.} All images have 64×64 pixels, which balances the detail that we keep against the computing time.
\item
  \textbf{Labeling.} Vision-language models, which process images and text, described each class and gave initial ratings of how likely its objects contain gravitational arcs.
\end{itemize}

We now load the preprocessed data.

\tutorialbarrier
\begin{codecell}

\end{codecell}
\begin{Verbatim}[fontsize=\footnotesize, xleftmargin=5mm]
Objects = 2873, classes = 50
\end{Verbatim}

The data have four components.

\begin{itemize}
\tightlist
\item
  \textbf{Image.} A 64×64×3 RGB image array for each object. The size is the same for all objects, which gives a consistent input for our machine learning pipeline.
\item
  \textbf{Class.} An analysis of the image features grouped the objects into classes of similar appearance, numbered from 0. The classes may correspond to different lensing configurations, galaxy morphologies or image artifacts.
\item
  \textbf{Description.} A short text summarizes the appearance of each class.
\item
  \textbf{Gravitational arc confidence.} A rating from 1 to 5 shows how likely it is that the objects of a class contain gravitational arcs. A rating of 1 means definitely no arcs, 3 means uncertain and 5 means definitely arcs.
\end{itemize}

The description and the rating belong to the class and not to the single object. The labels are not perfect. They stand in for a visual inspection by a human, so the accuracy below measures the agreement with these labels.

We use the classes as the classification targets, and we use the arc ratings to understand the astrophysical meaning of the classes. To see what we work with, we now show five random objects for each of five classes. We choose the classes from different parts of the list sorted by arc rating. We label each row with the class, its rating and its description.

\tutorialbarrier
\begin{codecell}

\end{codecell}
\tutorialplot{\includegraphics[width=6.500in,height=0.75\textheight,keepaspectratio]{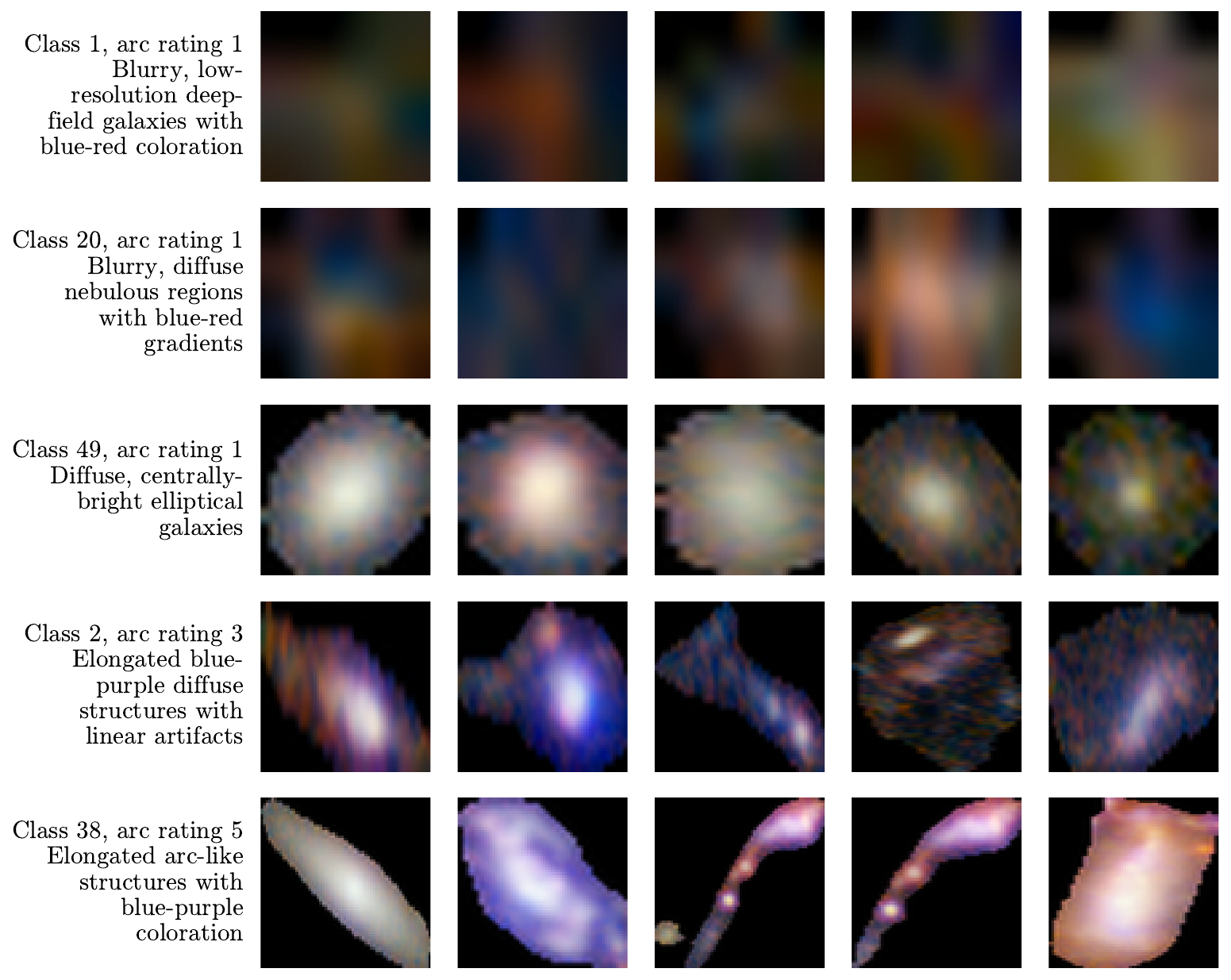}}%
  {\includegraphics[height=\tutorialheight]{tutorials/8/fig02.pdf}}

Objects of one class look alike, and the arc rating changes from the blurry background patches to the elongated galaxies with bright cores. This is what the classifier must learn to tell apart.

\section*{From Pixels to Features: Numerical Representations of Images}

Our 64×64 RGB images contain 12,288 pixel values (64 × 64 × 3). Raw pixels as features for classification have several basic problems. A model with more than 12,000 features needs much memory and computing time. A small translation or rotation changes the pixel values a lot, while the identity of the object stays the same. Pixels also do not represent higher-level concepts such as edges, textures or shapes, so a model needs very large training sets to learn them.

Modern computer vision solves this problem with deep neural networks that train on very large sets of images and learn hierarchical representations. The early layers capture edges, corners and simple textures. The later layers capture more complex shapes and the semantic concepts that matter for object recognition. We can use such a pre-trained network as a feature extractor, which is transfer learning. We pass our images through the network and take the output of one of the later layers. This gives compact numerical representations, called embeddings, typically with 512 to 2048 dimensions.

Pre-trained networks did not train on astronomical images, but the visual features that they learned transfer to our domain. These features detect curves, gradients, symmetries and complex shapes. Strong lensing features such as arcs and rings have geometric properties in common with everyday curved objects, so the networks give meaningful representations.

We use the pre-trained model \texttt{clip-ViT-B-32}, which Hugging Face provides through the sentence-transformers library. It is a Vision Transformer that was trained with the technique CLIP (Contrastive Language-Image Pre-training) on pairs of images and text. An image and its description then get similar embeddings, as in cross-modal learning. The embeddings therefore encode both the visual appearance and the semantic meaning. We now implement the feature extraction.

\tutorialbarrier
\begin{codecell}

\end{codecell}
\begin{Verbatim}[fontsize=\footnotesize, xleftmargin=5mm]
Each image: 12,288 pixel values, 512 embedding values, ratio 24:1
\end{Verbatim}

\section*{Visualizing the Embedding Space}

Our image embeddings live in a 512-dimensional space. We cannot visualize this space directly, but it contains rich structure that organizes the images by their visual and semantic properties. To see how the images relate to each other in this space, we map the data to 2D and try to keep the important relationships.

Principal component analysis (PCA) is a common linear method for such a map. For complex, high-dimensional data such as image embeddings, nonlinear methods often preserve the local structure better. We use UMAP (Uniform Manifold Approximation and Projection), a nonlinear dimensionality reduction method. It preserves both the local neighborhood structure and the global structure. Similar images in the 512-dimensional embedding space should appear near each other in the 2D map.

If the embeddings relate to the presence of arcs, we expect a gradient from low to high ratings. We also expect a separation of the objects with no arcs from the objects with definite arcs. We first compute the map.

\tutorialbarrier
\begin{codecell}

\end{codecell}

Each point of the map is one object. We first color the five classes of the gallery and draw all other objects in grey, because 50 colors are not readable.

\tutorialbarrier
\begin{codecell}

\end{codecell}
\tutorialplot{\includegraphics[width=5.881in,height=0.75\textheight,keepaspectratio]{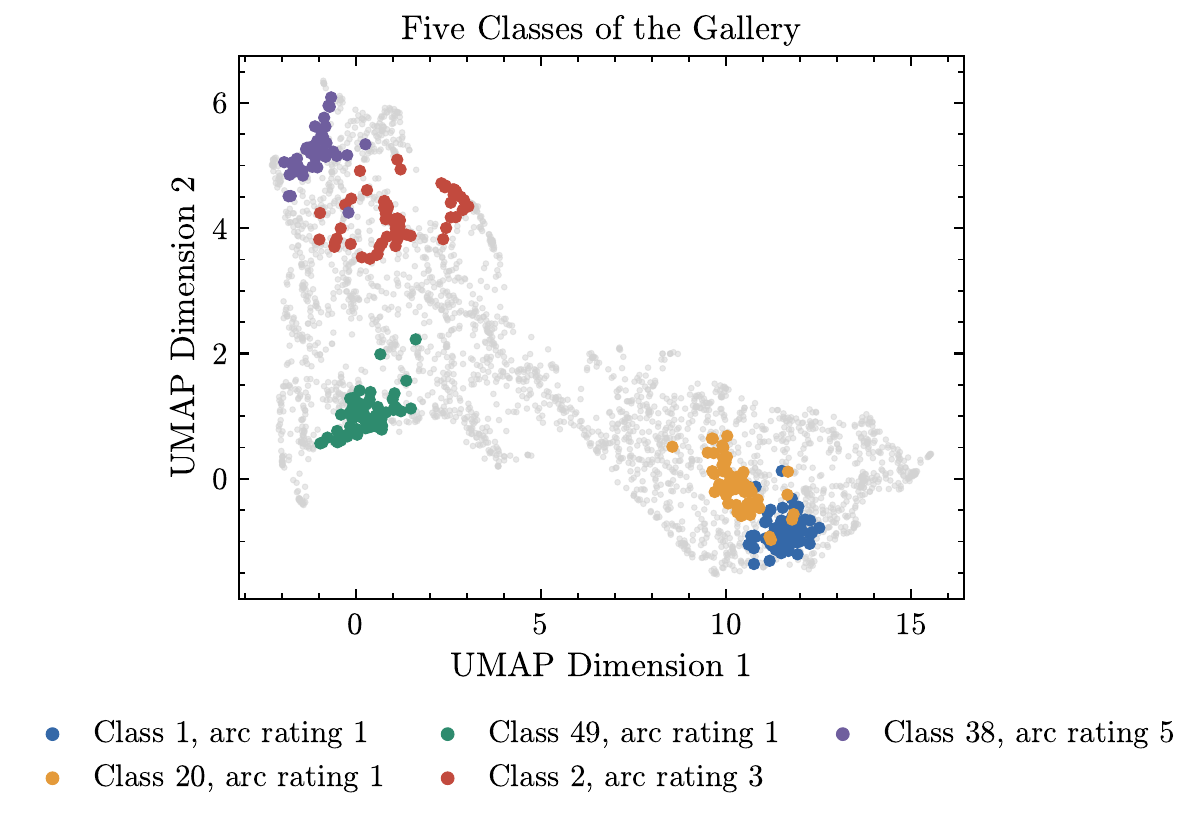}}%
  {\includegraphics[height=\tutorialheight]{tutorials/8/fig03.pdf}}

\textbf{Grouping by visual properties.} Objects of one class lie close together. The five classes of the gallery occupy different regions, except that classes 1 and 20 lie side by side. The grey objects show that other classes overlap, so the map does not separate all of them. A map in two dimensions loses information, and the classification below tests how well the features separate the classes.

The classes alone do not tell us the astrophysical meaning of the groups, so we color the same map by the gravitational arc rating. We also overlay image thumbnails, framed in the color of their rating, which show directly which types of objects occupy different regions.

\tutorialbarrier
\begin{codecell}

\end{codecell}
\tutorialplot{\includegraphics[width=6.188in,height=0.75\textheight,keepaspectratio]{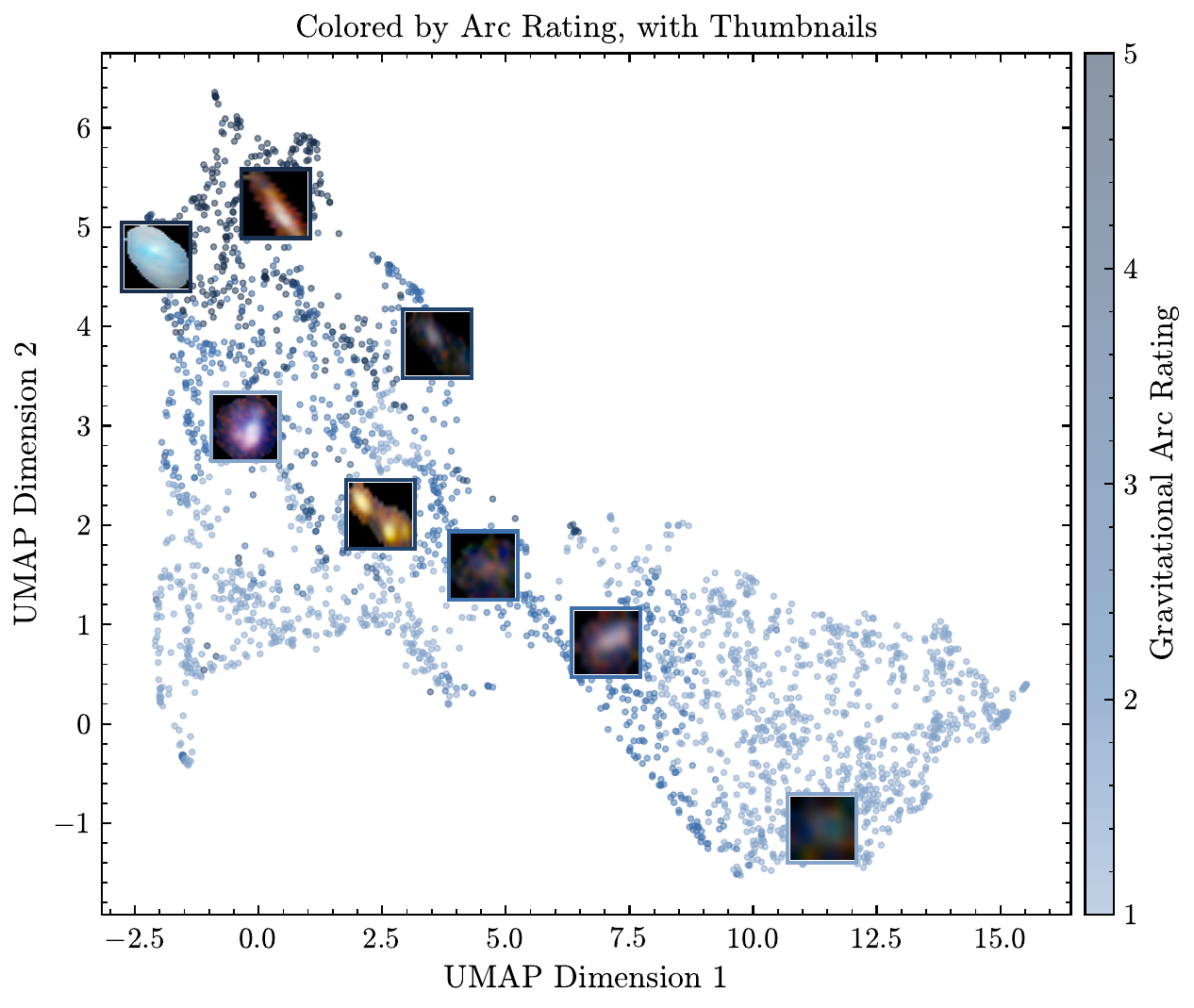}}%
  {\includegraphics[height=\tutorialheight]{tutorials/8/fig04.pdf}}

\textbf{Gradient in the arc ratings.} The map by arc rating shows a gradient in the embedding space. Objects with high ratings (4 and 5) lie mostly in the upper left. Objects with low ratings (1 and 2) lie in the lower part of the map. This suggests that the pre-trained embeddings encode features that are relevant to lensing. The separation is not perfect, and some overlap exists, in particular for the uncertain objects (rating 3).

\textbf{Confirmation with thumbnails.} The thumbnails show that nearby points are objects of similar appearance. Bright elongated structures lie together in one region of the map, and dark blurry patches lie together in another. Objects of intermediate appearance lie between them.

\section*{Multi-class Logistic Regression}

With the images transformed into numerical features, we now turn to the classification task. We predict which of the \(K = 50\) classes each object belongs to. The model for this problem is multi-class logistic regression (Section 8.3). It extends the binary logistic regression of Chapter 7.

\subsection*{From Binary to Multi-class: The Softmax Function}

In binary classification, the sigmoid function maps a real-valued score to a probability between 0 and 1. For \(K\) classes, we need a function that maps \(K\) real-valued scores to a probability distribution over all classes. This function is the softmax function (Section 8.4),

\[\text{softmax}(\mathbf{z})_k = \frac{e^{z_k}}{\sum_{j=1}^{K} e^{z_j}},\]

where \(\mathbf{z} = [z_1, z_2, \ldots, z_K]\) are the raw scores, called logits, of the classes. For an object with features \(\mathbf{x}\), the logit of class \(k\) is the linear prediction \(z_k = \mathbf{w}_k^T\mathbf{x}\). The softmax function has three properties.

\begin{itemize}
\tightlist
\item
  All outputs are positive, \(\text{softmax}(\mathbf{z})_k > 0\).
\item
  The outputs sum to one, \(\sum_{k=1}^{K} \text{softmax}(\mathbf{z})_k = 1\).
\item
  Larger logits give higher probabilities.
\end{itemize}

\subsection*{Numerical Stability in Implementation}

The softmax function needs a careful implementation to avoid numerical overflow. When the logits are large, \(e^{z_k}\) can exceed the limits of floating-point numbers. As in Section 8.8, we use the numerically stable version

\[\text{softmax}(\mathbf{z})_k = \frac{e^{z_k - \max(\mathbf{z})}}{\sum_{j=1}^{K} e^{z_j - \max(\mathbf{z})}}.\]

Subtracting the maximum logit does not change the result, but it prevents overflow. We now code the softmax function and the cross-entropy loss, and we check them on random logits.

\tutorialbarrier
\begin{codecell}

\end{codecell}
\begin{Verbatim}[fontsize=\footnotesize, xleftmargin=5mm]
Row sums of the probabilities = [1. 1. 1.]
Cross-entropy loss = 1.0040
\end{Verbatim}

\subsection*{Computing Gradients for Optimization}

To train our multi-class classifier, we need the gradient of the cross-entropy loss \(E\) (Section 8.5) with respect to the weight matrix. For a single training example with feature vector \(\mathbf{x}_i\), the gradient with respect to the weight vector of class \(k\) is (Section 8.7)

\[\nabla_{\mathbf{w}_k} E = (y_{ik} - t_{ik})\mathbf{x}_i,\]

where \(y_{ik}\) is the predicted probability for class \(k\) and \(t_{ik}\) is the true label, which is 1 for the correct class and 0 otherwise.

For a mini-batch of \(M\) examples, we average these gradients. In matrix notation, we compute this efficiently as

\[\nabla_{\mathbf{W}} E = \frac{1}{M}(\mathbf{Y} - \mathbf{T})^T\mathbf{X},\]

where

\begin{itemize}
\tightlist
\item
  \(\mathbf{X}\) is the data matrix (\(M \times D\)) that contains the \(D\) features of all objects,
\item
  \(\mathbf{Y}\) is the matrix of predicted probabilities (\(M \times K\)),
\item
  \(\mathbf{T}\) is the matrix of true labels in one-hot encoding (\(M \times K\)),
\item
  \(\mathbf{W}\) is the weight matrix (\(K \times D\)).
\end{itemize}

This formulation shows that the gradient is the average outer product of the prediction errors and the features. It is a direct generalization of the binary logistic regression case. We now code the gradient.

\tutorialbarrier
\begin{codecell}
\begin{Verbatim}[commandchars=\\\{\},fontsize=\small]
\PYG{k}{def}\PYG{+w}{ }\PYG{n+nf}{compute\PYGZus{}gradient}\PYG{p}{(}\FancyVerbBreakAfterBreak{}\PYG{n}{X}\PYG{p}{,}\FancyVerbBreakAfterBreak{} \PYG{n}{y\PYGZus{}pred}\PYG{p}{,}\FancyVerbBreakAfterBreak{} \PYG{n}{y\PYGZus{}true}\PYG{p}{)}\PYG{p}{:}
\PYG{+w}{    }\PYG{l+s+sd}{\PYGZdq{}\PYGZdq{}\PYGZdq{}Compute the gradient of the mean cross\PYGZhy{}entropy loss with respect to the weight matrix.\PYGZdq{}\PYGZdq{}\PYGZdq{}}
    \PYG{n}{batch\PYGZus{}size} \PYG{o}{=}\FancyVerbBreakAfterBreak{} \PYG{n}{X}\PYG{o}{.}\FancyVerbBreakAfterBreak{}\PYG{n}{shape}\PYG{p}{[}\FancyVerbBreakAfterBreak{}\PYG{l+m+mi}{0}\PYG{p}{]}

    \PYG{c+c1}{\PYGZsh{} Compute prediction error}
    \PYG{n}{error} \PYG{o}{=}\FancyVerbBreakAfterBreak{} \PYG{n}{y\PYGZus{}pred} \PYG{o}{\PYGZhy{}} \PYG{n}{to\PYGZus{}one\PYGZus{}hot}\PYG{p}{(}\FancyVerbBreakAfterBreak{}\PYG{n}{y\PYGZus{}true}\PYG{p}{,}\FancyVerbBreakAfterBreak{} \PYG{n}{y\PYGZus{}pred}\PYG{o}{.}\FancyVerbBreakAfterBreak{}\PYG{n}{shape}\PYG{p}{[}\FancyVerbBreakAfterBreak{}\PYG{l+m+mi}{1}\PYG{p}{]}\PYG{p}{)}  \PYG{c+c1}{\PYGZsh{} Shape (batch\PYGZus{}size, num\PYGZus{}classes)}

    \PYG{c+c1}{\PYGZsh{} Gradient is the average outer product of error and features}
    \PYG{c+c1}{\PYGZsh{} Matrix product of shapes (num\PYGZus{}classes, batch\PYGZus{}size) and (batch\PYGZus{}size, num\PYGZus{}features)}
    \PYG{n}{gradient} \PYG{o}{=}\FancyVerbBreakAfterBreak{} \PYG{n}{error}\PYG{o}{.}\FancyVerbBreakAfterBreak{}\PYG{n}{T} \PYG{o}{@} \PYG{n}{X} \PYG{o}{/} \PYG{n}{batch\PYGZus{}size}

    \PYG{k}{return} \PYG{n}{gradient}
\end{Verbatim}
\end{codecell}

\subsection*{Stochastic Gradient Descent for Large-Scale Classification}

As in Tutorial 7, we train with mini-batch stochastic gradient descent (SGD), here with \(M = 32\) objects in a batch. We start from small random weights, which keep the logits near zero (Section 8.8). In each epoch we shuffle the training objects and change the weights on each mini-batch with \(\mathbf{W} \leftarrow \mathbf{W} - \eta\nabla_{\mathbf{W}}E\), where \(\eta\) is the learning rate. After each epoch we record the training and validation loss, and we keep the weights with the lowest validation loss.

\subsection*{Regularization for Better Generalization}

To prevent overfitting, we add L2 regularization, also called weight decay (Section 7.8), to our objective

\[E_{\text{reg}} = E + \lambda\|\mathbf{W}\|^2,\]

where \(\|\mathbf{W}\|^2\) is the sum of the squares of all weights. This penalizes large weights and encourages the model to find simpler solutions. The gradient becomes

\[\nabla_{\mathbf{W}}E_{\text{reg}} = \nabla_{\mathbf{W}}E + 2\lambda\mathbf{W}.\]

We record the cross-entropy loss without the penalty, so that runs with different \(\lambda\) have comparable losses.

\subsection*{Implementation Considerations}

We fix one class as a reference class and set its weights to zero (Section 8.4). Here it is the last class. The \(K\) probabilities add up to one, so \(K - 1\) weight vectors are enough. This does not change what the model can express. The features are the 512 embedding values and a constant 1, which gives each class a bias term.

We now implement the complete training pipeline.

\tutorialbarrier
\begin{codecell}

\end{codecell}

\section*{Training the Multi-class Classifier}

With our implementation complete, we train a model that classifies our JWST strong lensing images into their classes. We start with a baseline configuration with \(\eta = 0.1\) and \(\lambda = 0.001\) and then explore how different hyperparameters change the result.

First, we extract the features and the labels. We split the objects at random into 60 percent fit, 20 percent validation and 20 percent test objects (Section 4.7). The fit objects set the weights. The validation objects compare the hyperparameters, so no choice uses the test objects.

We make the split in two steps with the split function of scikit-learn. We first hold out 20 percent of the objects as test objects. We then take 25 percent of the remaining training objects as validation objects. The class shares stay nearly equal in each step.

\tutorialbarrier
\begin{codecell}

\end{codecell}
\begin{Verbatim}[fontsize=\footnotesize, xleftmargin=5mm]
Fit objects = 1723, validation objects = 575, test objects = 575
\end{Verbatim}

We now train the baseline model on the fit objects and record the loss on the validation objects after each epoch.

\tutorialbarrier
\begin{codecell}
\begin{Verbatim}[commandchars=\\\{\},fontsize=\small]
\PYG{c+c1}{\PYGZsh{} Fit the weights on the fit objects and record the loss on the validation objects}
\PYG{n}{W}\PYG{p}{,}\FancyVerbBreakAfterBreak{} \PYG{n}{\PYGZus{}}\PYG{p}{,}\FancyVerbBreakAfterBreak{} \PYG{n}{train\PYGZus{}history}\PYG{p}{,}\FancyVerbBreakAfterBreak{} \PYG{n}{val\PYGZus{}history} \PYG{o}{=}\FancyVerbBreakAfterBreak{} \PYG{n}{train\PYGZus{}multiclass\PYGZus{}logistic\PYGZus{}regression}\PYG{p}{(}\FancyVerbBreakAfterBreak{}
    \PYG{n}{X\PYGZus{}fit}\PYG{p}{,}\FancyVerbBreakAfterBreak{} \PYG{n}{y\PYGZus{}fit}\PYG{p}{,}\FancyVerbBreakAfterBreak{} \PYG{n}{X\PYGZus{}val}\PYG{p}{,}\FancyVerbBreakAfterBreak{} \PYG{n}{y\PYGZus{}val}\PYG{p}{,}\FancyVerbBreakAfterBreak{}
    \PYG{n}{num\PYGZus{}classes}\PYG{o}{=}\FancyVerbBreakAfterBreak{}\PYG{n}{num\PYGZus{}classes}\PYG{p}{,}\FancyVerbBreakAfterBreak{}
    \PYG{n}{num\PYGZus{}epochs}\PYG{o}{=}\FancyVerbBreakAfterBreak{}\PYG{l+m+mi}{150}\PYG{p}{,}\FancyVerbBreakAfterBreak{}
    \PYG{n}{batch\PYGZus{}size}\PYG{o}{=}\FancyVerbBreakAfterBreak{}\PYG{l+m+mi}{32}\PYG{p}{,}\FancyVerbBreakAfterBreak{}
    \PYG{n}{learning\PYGZus{}rate}\PYG{o}{=}\FancyVerbBreakAfterBreak{}\PYG{l+m+mf}{0.1}\PYG{p}{,}\FancyVerbBreakAfterBreak{}
    \PYG{n}{weight\PYGZus{}decay}\PYG{o}{=}\FancyVerbBreakAfterBreak{}\PYG{l+m+mf}{0.001}\PYG{p}{,}\FancyVerbBreakAfterBreak{}
    \PYG{n}{seed}\PYG{o}{=}\FancyVerbBreakAfterBreak{}\PYG{l+m+mi}{42}
\PYG{p}{)}
\end{Verbatim}
\end{codecell}

We plot the learning curves and mark the epoch with the lowest validation loss.

\tutorialbarrier
\begin{codecell}

\end{codecell}
\tutorialplot{\includegraphics[width=4.398in,height=0.75\textheight,keepaspectratio]{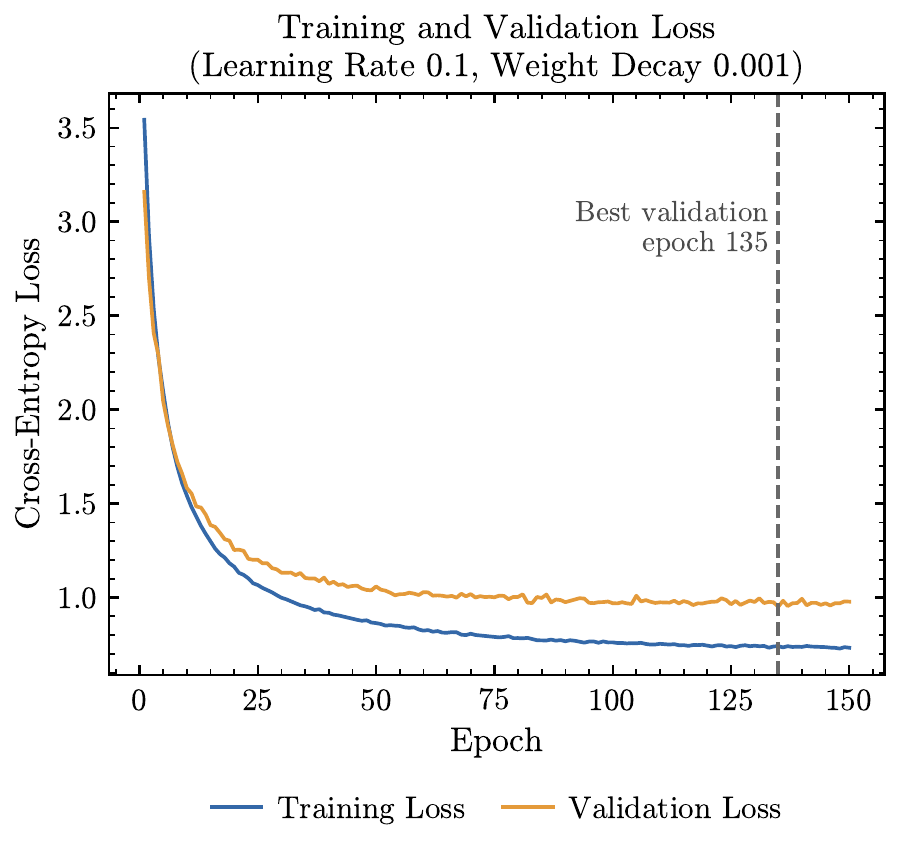}}%
  {\includegraphics[height=\tutorialheight]{tutorials/8/fig05.pdf}}
\begin{Verbatim}[fontsize=\footnotesize, xleftmargin=5mm]
Final training loss = 0.7328, final validation loss = 0.9782
\end{Verbatim}

Both losses fall quickly during the first 20 epochs. The training loss then keeps falling slowly, and the validation loss flattens after about 60 epochs with small fluctuations. The dashed line marks the lowest validation loss, which differs little from the values at the later epochs. The validation loss stays above the training loss, because the weights fit the fit objects better than the validation objects.

\section*{Hyperparameter Optimization}

As in Tutorial 7, we choose the hyperparameters with a grid search on the validation objects. The learning rate \(\eta\) sets the step size. The weight decay \(\lambda\) sets the strength of the L2 regularization. It trades a good fit of the fit objects against small weights that generalize better.

We fix the batch size at 32 and the number of epochs at 150 for all models. We train a model for each combination of the learning rate and the weight decay. We trace the validation loss after every epoch and keep the weights with the lowest validation loss. We select the combination with the lowest validation loss. We now implement the grid search.

\tutorialbarrier
\begin{codecell}

\end{codecell}
\begin{Verbatim}[fontsize=\footnotesize, xleftmargin=5mm]
Best: learning rate 0.1, weight decay 0.0
Lowest validation loss 0.7242 at epoch 141
\end{Verbatim}

We show the lowest validation loss of every combination as a heatmap, and we mark the best combination with a blue frame.

\tutorialbarrier
\begin{codecell}

\end{codecell}
\tutorialplot{\includegraphics[width=5.859in,height=0.75\textheight,keepaspectratio]{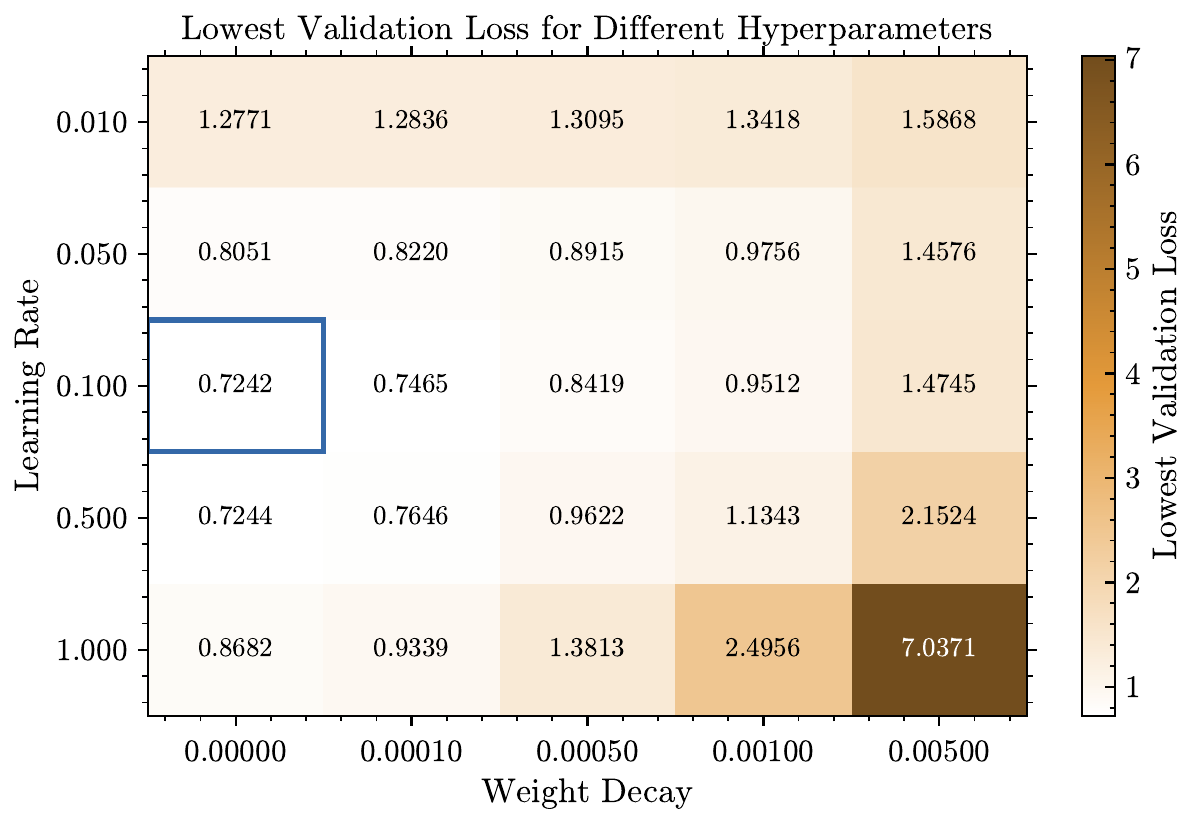}}%
  {\includegraphics[height=\tutorialheight]{tutorials/8/fig06.pdf}}

The heatmap shows that the best learning rate is 0.1 and the best weight decay is 0.0.

\textbf{Learning rate effects.} The result gets much better when the learning rate increases from 0.01 to 0.1, and the best value is 0.1. The learning rate 0.5 gives nearly the same loss, and 1.0 gives a higher loss. The model benefits from large optimization steps up to a point, and then the training becomes noisy.

\textbf{Regularization effects.} Weight decay hurts the result at all learning rates. The degradation is very large at the high learning rates. For this model, the additional L2 penalty does not help.

\section*{Model Evaluation: Confusion Matrix and Accuracy}

The best weights of the best combination are our final model. We test it on the test objects, which took no part in any choice. As in Tutorial 7, a confusion matrix counts the objects of each true class that the model predicts as each class. With 50 classes it shows which classes the model gets right and which classes it confuses. We compute the confusion matrix and the metrics that follow from it.

\section*{Understanding Classification Metrics}

The accuracy is the fraction of correct predictions over all classes. Tutorial 7 defined the recall and the precision for a positive class. With many classes no single class is the positive class, so we compute them for every class \(k\) in turn. Class \(k\) is then the positive class, and all other classes together are the negative class.

\(\text{TP}_k\) is the number of objects of class \(k\) that the model predicts as class \(k\). \(\text{FP}_k\) is the number of objects of other classes that the model predicts as class \(k\). \(\text{FN}_k\) is the number of objects of class \(k\) that the model predicts as another class.

The \textbf{recall} of class \(k\) is the fraction of the objects of class \(k\) that the model finds,

\[\text{Recall}_k = \frac{\text{TP}_k}{\text{TP}_k + \text{FN}_k}.\]

The \textbf{precision} of class \(k\) is the fraction of the objects that the model predicts as class \(k\) and that truly belong to it,

\[\text{Precision}_k = \frac{\text{TP}_k}{\text{TP}_k + \text{FP}_k}.\]

A low recall means that the model misses many objects of the class. A low precision means that its predictions of the class contain many wrong objects. For many classes we average the two over the classes. The per-class values show which types of objects the model gets right and which types cause confusion. We now code the prediction and the confusion matrix.

\tutorialbarrier
\begin{codecell}
\begin{Verbatim}[commandchars=\\\{\},fontsize=\small]
\PYG{k}{def}\PYG{+w}{ }\PYG{n+nf}{predict\PYGZus{}classes}\PYG{p}{(}\FancyVerbBreakAfterBreak{}\PYG{n}{X}\PYG{p}{,}\FancyVerbBreakAfterBreak{} \PYG{n}{W}\PYG{p}{)}\PYG{p}{:}
\PYG{+w}{    }\PYG{l+s+sd}{\PYGZdq{}\PYGZdq{}\PYGZdq{}Predict the class with the largest probability for each row of X.\PYGZdq{}\PYGZdq{}\PYGZdq{}}
    \PYG{n}{logits} \PYG{o}{=}\FancyVerbBreakAfterBreak{} \PYG{n}{X} \PYG{o}{@} \PYG{n}{W}\PYG{o}{.}\FancyVerbBreakAfterBreak{}\PYG{n}{T}
    \PYG{n}{probs} \PYG{o}{=}\FancyVerbBreakAfterBreak{} \PYG{n}{softmax}\PYG{p}{(}\FancyVerbBreakAfterBreak{}\PYG{n}{logits}\PYG{p}{)}

    \PYG{c+c1}{\PYGZsh{} Index of the largest probability in each row}
    \PYG{n}{predictions} \PYG{o}{=}\FancyVerbBreakAfterBreak{} \PYG{n}{np}\PYG{o}{.}\FancyVerbBreakAfterBreak{}\PYG{n}{argmax}\PYG{p}{(}\FancyVerbBreakAfterBreak{}\PYG{n}{probs}\PYG{p}{,}\FancyVerbBreakAfterBreak{} \PYG{n}{axis}\PYG{o}{=}\FancyVerbBreakAfterBreak{}\PYG{l+m+mi}{1}\PYG{p}{)}
    \PYG{k}{return} \PYG{n}{predictions}

\PYG{k}{def}\PYG{+w}{ }\PYG{n+nf}{compute\PYGZus{}confusion\PYGZus{}matrix}\PYG{p}{(}\FancyVerbBreakAfterBreak{}\PYG{n}{y\PYGZus{}true}\PYG{p}{,}\FancyVerbBreakAfterBreak{} \PYG{n}{y\PYGZus{}pred}\PYG{p}{,}\FancyVerbBreakAfterBreak{} \PYG{n}{num\PYGZus{}classes}\PYG{p}{)}\PYG{p}{:}
\PYG{+w}{    }\PYG{l+s+sd}{\PYGZdq{}\PYGZdq{}\PYGZdq{}Compute the confusion matrix. Rows are true classes and columns are predicted classes.\PYGZdq{}\PYGZdq{}\PYGZdq{}}
    \PYG{n}{cm} \PYG{o}{=}\FancyVerbBreakAfterBreak{} \PYG{n}{np}\PYG{o}{.}\FancyVerbBreakAfterBreak{}\PYG{n}{zeros}\PYG{p}{(}\FancyVerbBreakAfterBreak{}\PYG{p}{(}\FancyVerbBreakAfterBreak{}\PYG{n}{num\PYGZus{}classes}\PYG{p}{,}\FancyVerbBreakAfterBreak{} \PYG{n}{num\PYGZus{}classes}\PYG{p}{)}\PYG{p}{,}\FancyVerbBreakAfterBreak{} \PYG{n}{dtype}\PYG{o}{=}\FancyVerbBreakAfterBreak{}\PYG{n+nb}{int}\PYG{p}{)}
    \PYG{n}{np}\PYG{o}{.}\FancyVerbBreakAfterBreak{}\PYG{n}{add}\PYG{o}{.}\FancyVerbBreakAfterBreak{}\PYG{n}{at}\PYG{p}{(}\FancyVerbBreakAfterBreak{}\PYG{n}{cm}\PYG{p}{,}\FancyVerbBreakAfterBreak{} \PYG{p}{(}\FancyVerbBreakAfterBreak{}\PYG{n}{y\PYGZus{}true}\PYG{p}{,}\FancyVerbBreakAfterBreak{} \PYG{n}{y\PYGZus{}pred}\PYG{p}{)}\PYG{p}{,}\FancyVerbBreakAfterBreak{} \PYG{l+m+mi}{1}\PYG{p}{)}  \PYG{c+c1}{\PYGZsh{} count each pair of true class and predicted class}
    \PYG{k}{return} \PYG{n}{cm}
\end{Verbatim}
\end{codecell}

We also code the accuracy and the metrics of each class, which follow from the confusion matrix.

\tutorialbarrier
\begin{codecell}
\begin{Verbatim}[commandchars=\\\{\},fontsize=\small]
\PYG{k}{def}\PYG{+w}{ }\PYG{n+nf}{calculate\PYGZus{}metrics}\PYG{p}{(}\FancyVerbBreakAfterBreak{}\PYG{n}{cm}\PYG{p}{)}\PYG{p}{:}
\PYG{+w}{    }\PYG{l+s+sd}{\PYGZdq{}\PYGZdq{}\PYGZdq{}Calculate the accuracy and the precision and the recall of each class from the confusion matrix.\PYGZdq{}\PYGZdq{}\PYGZdq{}}
    \PYG{n}{tp} \PYG{o}{=}\FancyVerbBreakAfterBreak{} \PYG{n}{np}\PYG{o}{.}\FancyVerbBreakAfterBreak{}\PYG{n}{diag}\PYG{p}{(}\FancyVerbBreakAfterBreak{}\PYG{n}{cm}\PYG{p}{)}  \PYG{c+c1}{\PYGZsh{} objects of each class that are predicted as that class}
    \PYG{n}{fp} \PYG{o}{=}\FancyVerbBreakAfterBreak{} \PYG{n}{cm}\PYG{o}{.}\FancyVerbBreakAfterBreak{}\PYG{n}{sum}\PYG{p}{(}\FancyVerbBreakAfterBreak{}\PYG{n}{axis}\PYG{o}{=}\FancyVerbBreakAfterBreak{}\PYG{l+m+mi}{0}\PYG{p}{)} \PYG{o}{\PYGZhy{}} \PYG{n}{tp}  \PYG{c+c1}{\PYGZsh{} objects of other classes that are predicted as the class}
    \PYG{n}{fn} \PYG{o}{=}\FancyVerbBreakAfterBreak{} \PYG{n}{cm}\PYG{o}{.}\FancyVerbBreakAfterBreak{}\PYG{n}{sum}\PYG{p}{(}\FancyVerbBreakAfterBreak{}\PYG{n}{axis}\PYG{o}{=}\FancyVerbBreakAfterBreak{}\PYG{l+m+mi}{1}\PYG{p}{)} \PYG{o}{\PYGZhy{}} \PYG{n}{tp}  \PYG{c+c1}{\PYGZsh{} objects of the class that are predicted as other classes}

    \PYG{c+c1}{\PYGZsh{} Precision and recall of each class, zero where the denominator is zero}
    \PYG{n}{precision} \PYG{o}{=}\FancyVerbBreakAfterBreak{} \PYG{n}{np}\PYG{o}{.}\FancyVerbBreakAfterBreak{}\PYG{n}{divide}\PYG{p}{(}\FancyVerbBreakAfterBreak{}\PYG{n}{tp}\PYG{p}{,}\FancyVerbBreakAfterBreak{} \PYG{n}{tp} \PYG{o}{+}\FancyVerbBreakAfterBreak{} \PYG{n}{fp}\PYG{p}{,}\FancyVerbBreakAfterBreak{} \PYG{n}{out}\PYG{o}{=}\FancyVerbBreakAfterBreak{}\PYG{n}{np}\PYG{o}{.}\FancyVerbBreakAfterBreak{}\PYG{n}{zeros}\PYG{p}{(}\FancyVerbBreakAfterBreak{}\PYG{n+nb}{len}\PYG{p}{(}\FancyVerbBreakAfterBreak{}\PYG{n}{cm}\PYG{p}{)}\PYG{p}{)}\PYG{p}{,}\FancyVerbBreakAfterBreak{} \PYG{n}{where}\PYG{o}{=}\FancyVerbBreakAfterBreak{}\PYG{p}{(}\FancyVerbBreakAfterBreak{}\PYG{n}{tp} \PYG{o}{+}\FancyVerbBreakAfterBreak{} \PYG{n}{fp}\PYG{p}{)} \PYG{o}{\PYGZgt{}} \PYG{l+m+mi}{0}\PYG{p}{)}
    \PYG{n}{recall} \PYG{o}{=}\FancyVerbBreakAfterBreak{} \PYG{n}{np}\PYG{o}{.}\FancyVerbBreakAfterBreak{}\PYG{n}{divide}\PYG{p}{(}\FancyVerbBreakAfterBreak{}\PYG{n}{tp}\PYG{p}{,}\FancyVerbBreakAfterBreak{} \PYG{n}{tp} \PYG{o}{+}\FancyVerbBreakAfterBreak{} \PYG{n}{fn}\PYG{p}{,}\FancyVerbBreakAfterBreak{} \PYG{n}{out}\PYG{o}{=}\FancyVerbBreakAfterBreak{}\PYG{n}{np}\PYG{o}{.}\FancyVerbBreakAfterBreak{}\PYG{n}{zeros}\PYG{p}{(}\FancyVerbBreakAfterBreak{}\PYG{n+nb}{len}\PYG{p}{(}\FancyVerbBreakAfterBreak{}\PYG{n}{cm}\PYG{p}{)}\PYG{p}{)}\PYG{p}{,}\FancyVerbBreakAfterBreak{} \PYG{n}{where}\PYG{o}{=}\FancyVerbBreakAfterBreak{}\PYG{p}{(}\FancyVerbBreakAfterBreak{}\PYG{n}{tp} \PYG{o}{+}\FancyVerbBreakAfterBreak{} \PYG{n}{fn}\PYG{p}{)} \PYG{o}{\PYGZgt{}} \PYG{l+m+mi}{0}\PYG{p}{)}

    \PYG{c+c1}{\PYGZsh{} Collect the metrics in a dictionary}
    \PYG{k}{return} \PYG{p}{\PYGZob{}}
        \PYG{l+s+s1}{\PYGZsq{}}\PYG{l+s+s1}{accuracy}\PYG{l+s+s1}{\PYGZsq{}}\PYG{p}{:} \PYG{n}{tp}\PYG{o}{.}\FancyVerbBreakAfterBreak{}\PYG{n}{sum}\PYG{p}{(}\FancyVerbBreakAfterBreak{}\PYG{p}{)} \PYG{o}{/} \PYG{n}{cm}\PYG{o}{.}\FancyVerbBreakAfterBreak{}\PYG{n}{sum}\PYG{p}{(}\FancyVerbBreakAfterBreak{}\PYG{p}{)}\PYG{p}{,}\FancyVerbBreakAfterBreak{}
        \PYG{l+s+s1}{\PYGZsq{}}\PYG{l+s+s1}{precision}\PYG{l+s+s1}{\PYGZsq{}}\PYG{p}{:} \PYG{n}{precision}\PYG{p}{,}\FancyVerbBreakAfterBreak{}
        \PYG{l+s+s1}{\PYGZsq{}}\PYG{l+s+s1}{recall}\PYG{l+s+s1}{\PYGZsq{}}\PYG{p}{:} \PYG{n}{recall}\PYG{p}{,}\FancyVerbBreakAfterBreak{}
        \PYG{l+s+s1}{\PYGZsq{}}\PYG{l+s+s1}{avg\PYGZus{}precision}\PYG{l+s+s1}{\PYGZsq{}}\PYG{p}{:} \PYG{n}{np}\PYG{o}{.}\FancyVerbBreakAfterBreak{}\PYG{n}{mean}\PYG{p}{(}\FancyVerbBreakAfterBreak{}\PYG{n}{precision}\PYG{p}{)}\PYG{p}{,}\FancyVerbBreakAfterBreak{}
        \PYG{l+s+s1}{\PYGZsq{}}\PYG{l+s+s1}{avg\PYGZus{}recall}\PYG{l+s+s1}{\PYGZsq{}}\PYG{p}{:} \PYG{n}{np}\PYG{o}{.}\FancyVerbBreakAfterBreak{}\PYG{n}{mean}\PYG{p}{(}\FancyVerbBreakAfterBreak{}\PYG{n}{recall}\PYG{p}{)}
    \PYG{p}{\PYGZcb{}}\FancyVerbBreakAfterBreak{}
\end{Verbatim}
\end{codecell}

We predict the class of every test object with the final model, and we compute the confusion matrix and the metrics.

\tutorialbarrier
\begin{codecell}
\begin{Verbatim}[commandchars=\\\{\},fontsize=\small]
\PYG{c+c1}{\PYGZsh{} Make predictions on the test set, and compute the confusion matrix and the metrics}
\PYG{n}{y\PYGZus{}test\PYGZus{}pred} \PYG{o}{=}\FancyVerbBreakAfterBreak{} \PYG{n}{predict\PYGZus{}classes}\PYG{p}{(}\FancyVerbBreakAfterBreak{}\PYG{n}{X\PYGZus{}test}\PYG{p}{,}\FancyVerbBreakAfterBreak{} \PYG{n}{W\PYGZus{}final}\PYG{p}{)}
\PYG{n}{cm} \PYG{o}{=}\FancyVerbBreakAfterBreak{} \PYG{n}{compute\PYGZus{}confusion\PYGZus{}matrix}\PYG{p}{(}\FancyVerbBreakAfterBreak{}\PYG{n}{y\PYGZus{}test}\PYG{p}{,}\FancyVerbBreakAfterBreak{} \PYG{n}{y\PYGZus{}test\PYGZus{}pred}\PYG{p}{,}\FancyVerbBreakAfterBreak{} \PYG{n}{num\PYGZus{}classes}\PYG{p}{)}
\PYG{n}{metrics} \PYG{o}{=}\FancyVerbBreakAfterBreak{} \PYG{n}{calculate\PYGZus{}metrics}\PYG{p}{(}\FancyVerbBreakAfterBreak{}\PYG{n}{cm}\PYG{p}{)}

\PYG{n+nb}{print}\PYG{p}{(}\FancyVerbBreakAfterBreak{}\PYG{l+s+sa}{f}\PYG{l+s+s2}{\PYGZdq{}}\PYG{l+s+s2}{Accuracy = }\PYG{l+s+si}{\PYGZob{}}\PYG{n}{metrics}\PYG{p}{[}\FancyVerbBreakAfterBreak{}\PYG{l+s+s1}{\PYGZsq{}}\PYG{l+s+s1}{accuracy}\PYG{l+s+s1}{\PYGZsq{}}\PYG{p}{]}\PYG{l+s+si}{:}\PYG{l+s+s2}{.4f}\PYG{l+s+si}{\PYGZcb{}}\FancyVerbBreakAfterBreak{}\PYG{l+s+s2}{\PYGZdq{}}\PYG{p}{)}
\PYG{n+nb}{print}\PYG{p}{(}\FancyVerbBreakAfterBreak{}\PYG{l+s+sa}{f}\PYG{l+s+s2}{\PYGZdq{}}\PYG{l+s+s2}{Average precision = }\PYG{l+s+si}{\PYGZob{}}\PYG{n}{metrics}\PYG{p}{[}\FancyVerbBreakAfterBreak{}\PYG{l+s+s1}{\PYGZsq{}}\PYG{l+s+s1}{avg\PYGZus{}precision}\PYG{l+s+s1}{\PYGZsq{}}\PYG{p}{]}\PYG{l+s+si}{:}\PYG{l+s+s2}{.4f}\PYG{l+s+si}{\PYGZcb{}}\FancyVerbBreakAfterBreak{}\PYG{l+s+s2}{, average recall = }\PYG{l+s+si}{\PYGZob{}}\PYG{n}{metrics}\PYG{p}{[}\FancyVerbBreakAfterBreak{}\PYG{l+s+s1}{\PYGZsq{}}\PYG{l+s+s1}{avg\PYGZus{}recall}\PYG{l+s+s1}{\PYGZsq{}}\PYG{p}{]}\PYG{l+s+si}{:}\PYG{l+s+s2}{.4f}\PYG{l+s+si}{\PYGZcb{}}\FancyVerbBreakAfterBreak{}\PYG{l+s+s2}{\PYGZdq{}}\PYG{p}{)}
\end{Verbatim}
\end{codecell}
\begin{Verbatim}[fontsize=\footnotesize, xleftmargin=5mm]
Accuracy = 0.7617
Average precision = 0.7842, average recall = 0.7663
\end{Verbatim}

The model classifies about three of four test objects correctly. Random guessing among 50 classes gives an accuracy of about 2 percent. The average recall is close to the accuracy, because most classes have the same size of 64 objects. The average precision is a little higher than the average recall.

\tutorialbarrier
\begin{codecell}

\end{codecell}
\tutorialplot{\includegraphics[width=5.962in,height=0.75\textheight,keepaspectratio]{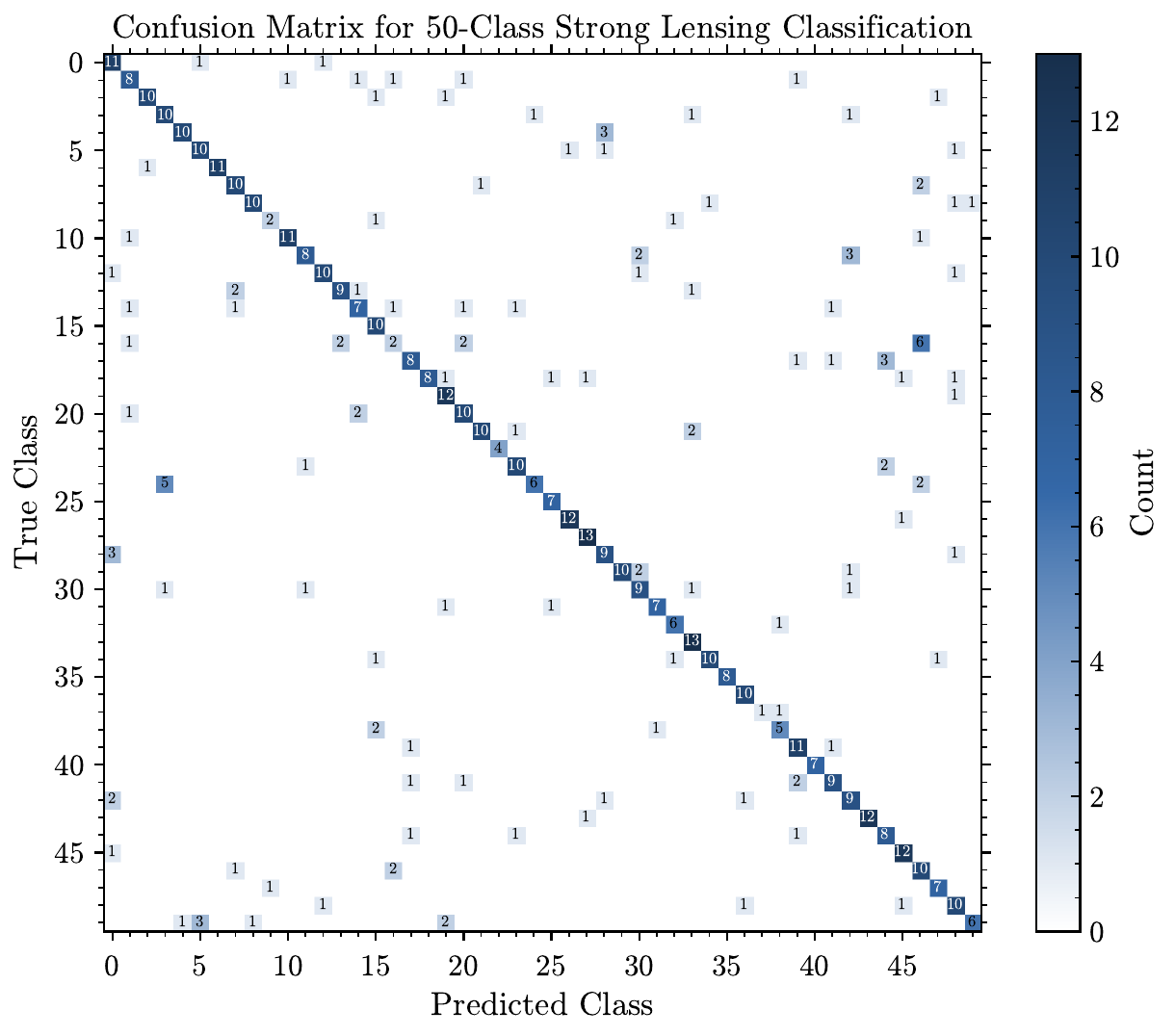}}%
  {\includegraphics[height=\tutorialheight]{tutorials/8/fig07.pdf}}

\section*{Understanding Classification Errors}

The confusion matrix shows which classes our model finds hard to distinguish. When we analyze the most common misclassifications, we can learn about the limits of our feature representation and find possible improvements.

The strong diagonal of the confusion matrix shows a good overall accuracy. The off-diagonal elements show systematic confusions, and some classes are nearly perfect while others have large errors.

We now identify the three worst-performing classes and examine which classes they are confused with.

\tutorialbarrier
\begin{codecell}

\end{codecell}
\begin{Verbatim}[fontsize=\footnotesize, xleftmargin=5mm]
Class 16: recall 0.154, most confused with class 46 (count 6)
- 16: Colorful, blurry nebulous structures with blue-orange contrast
- 46: Extremely blurry, low-resolution red-blue color patches
------------------------------------------------------------
Class 24: recall 0.462, most confused with class 3 (count 5)
- 24: Blurry blue-orange diffuse deep-field objects
- 3: Blurry multicolored cosmic structures with elongated features
------------------------------------------------------------
Class 49: recall 0.462, most confused with class 5 (count 3)
- 49: Diffuse, centrally-bright elliptical galaxies
- 5: Compact elliptical galaxies with bright cores
\end{Verbatim}

We show four random objects of each of these classes, and four of the class that it is confused with. Each row is one pair.

\tutorialbarrier
\begin{codecell}

\end{codecell}
\tutorialplot{\includegraphics[width=6.500in,height=0.75\textheight,keepaspectratio]{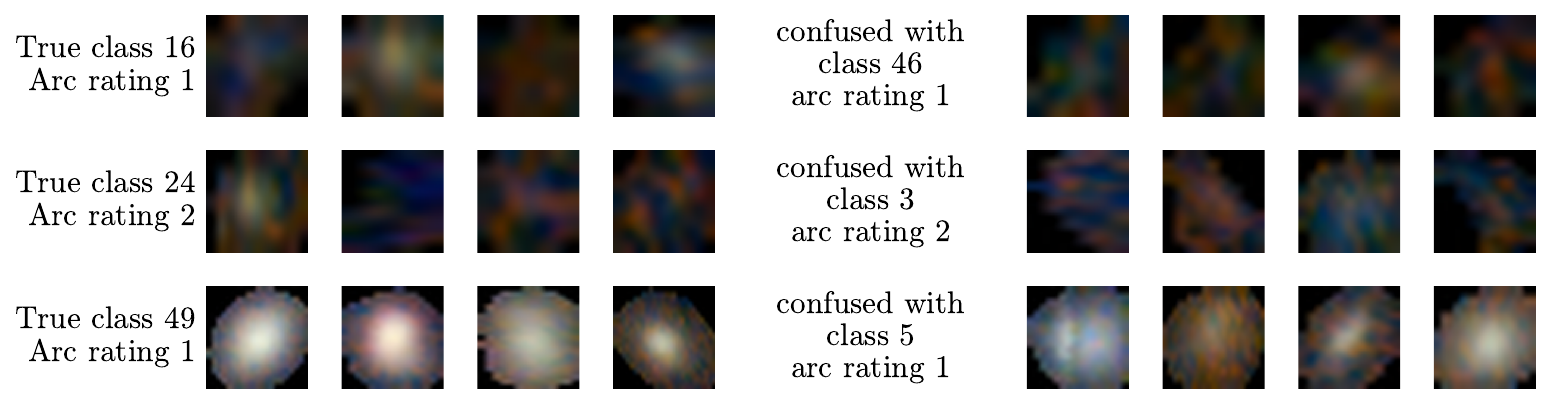}}%
  {\includegraphics[height=\tutorialheight]{tutorials/8/fig08.pdf}}

\subsection*{Misclassifications}

The visual comparison shows that the model confuses classes that look alike.

\textbf{Hard cases.} The first two pairs (classes 16 and 46, and classes 24 and 3) show low-contrast, diffuse structures with subtle color variations. These objects are hard to distinguish. The classes of a pair have the same low arc rating, 1 for the first pair and 2 for the second. The objects therefore have probably no gravitational arcs. The confusion here may come from an ambiguity of the data and not from a failure of the approach.

\textbf{Similar morphologies.} The third pair (classes 49 and 5) shows round elliptical galaxies with bright cores. The two classes differ mainly in how compact the galaxy is and how bright its core is. These are small differences, and the objects of the two classes overlap.

\textbf{Limits of the approach.} The confusion of classes with subtle differences suggests that the embeddings may not encode fine details of the appearance. The model behind the embeddings trained mostly on everyday photographs, so we apply it in a change of domain. Multi-class logistic regression also separates the classes with linear boundaries (Section 8.4), which may miss fine differences.

In all three pairs, the confused classes have the same arc rating. This suggests that the embeddings encode the presence or absence of lensing features at a basic level. For many astronomical applications, finding the presence of strong lensing matters more than separating subtle morphological types.

\section*{Summary}

This tutorial extended the binary logistic regression of Chapter 7 to multi-class classification. We turned images of astronomical objects into features with a pre-trained network. We trained a classifier that gives new images the labels of a labeled sample. The classifier labels most test objects correctly.

Our main lessons follow.

\begin{enumerate}
\def\labelenumi{\arabic{enumi}.}
\tightlist
\item
  \textbf{From two classes to many.} The softmax function generalizes the sigmoid function to one probability for each class. We implemented it in a numerically stable way. The multi-class cross-entropy loss has a gradient of the same form as in the binary case, the prediction error times the features. Matrix operations then make the training efficient.
\item
  \textbf{Features from a pre-trained network.} Embeddings turn images into compact numerical features that carry visual and semantic information. A linear model can then separate the classes, and a two-dimensional map of the embeddings shows how the objects group.
\item
  \textbf{Training and hyperparameters.} As in Tutorial 7, mini-batch stochastic gradient descent trains the model. Weight decay penalizes large weights, and we chose it and the learning rate with a grid search on validation objects. We kept the weights with the lowest validation loss and scored this final model on test objects.
\item
  \textbf{Evaluation with many classes.} The confusion matrix and the recall and precision of each class show more than the accuracy. They reveal which classes the model confuses.
\end{enumerate}

Multi-class classification is common in astronomy, because objects fall into many types and not into two. Multi-class logistic regression on pre-trained features gives a principled baseline that needs little computing time. We can evaluate more complex methods against it.

\tutorialchapter{9}{Bayesian Logistic Regression for Red Clump Stars}
\noindent\textit{Companion material for Chapter 9}

\section*{Introduction}

In Tutorial 7 we separated red clump (RC) stars from red giant branch (RGB) stars with logistic regression. The model gives the probability that a star is an RC star, but it rests on one set of weights. This point estimate, one decision boundary, conceals the uncertainty that finite data leave. RC stars have nearly the same luminosity, so they are distance indicators. An RGB star that we mistake for an RC star leads to a wrong distance estimate.

This matters most for a star whose temperature and gravity place it far from any training star. Standard logistic regression might assign it with confidence to one class or the other, but this confidence is unwarranted. We have too little information in that region for a firm statement. Such overconfidence is a problem in astronomy, where misclassifications can carry over to later work.

Bayesian logistic regression deals with these limits through a different view. Instead of one optimal set of weights, we keep a distribution over all plausible sets of weights that agree with the data. This distribution, the posterior, encodes our best estimates and our uncertainty about them. A prediction averages over the whole posterior, which we call marginalization. Where many training stars constrain the boundary, the plausible weights agree and the prediction stays as it is. Where few do, the weights disagree and the predictive probability moves closer to 0.5.

To compute this posterior, we combine a Gaussian prior over the weights with the likelihood of standard logistic regression. The sigmoid makes the posterior non-Gaussian, so the Laplace approximation replaces it with a Gaussian centered at the maximum a posteriori (MAP) estimate. The probit approximation then gives the predictive distribution in closed form.

We test these steps on the 2,855 APOGEE stars and asteroseismic classes of Tutorial 7, with only the effective temperature and the surface gravity. These are the main spectroscopic indicators of the evolutionary state of a star. In this plane we see the decision boundaries and the uncertainty regions directly, and the two populations overlap in some regions. We therefore need a probabilistic classification. We use the posterior and the predictive distribution to show how far we can trust the probability of each star.

\section*{Prerequisites and Learning Objectives}

\subsection*{Prerequisites}

This tutorial assumes the following.

\begin{itemize}
\tightlist
\item
  Logistic regression, with the sigmoid function and the cross-entropy loss (Chapter 7)
\item
  Bayesian inference, with prior distributions, likelihood functions and posterior inference (Chapters 2 and 5)
\item
  Linear algebra, for matrix operations, eigenvalues and the quadratic forms of the Hessian and covariance matrices
\item
  Gradient descent with mini-batches
\end{itemize}

\subsection*{Learning Objectives}

After this tutorial you can do the following.

\begin{itemize}
\tightlist
\item
  Extend a point estimate to the full posterior distribution of the model parameters
\item
  Implement the Laplace approximation, which expands the log posterior to second order around its mode and so approximates the posterior with a Gaussian
\item
  Compute and visualize the uncertainty of a prediction through marginalization over the posterior
\item
  Explain how the strength of the prior affects the parameter estimates and the confidence of the predictions
\item
  Compute predictive probabilities in closed form with the probit approximation
\item
  Relate the mathematical uncertainty to how far we can trust an astronomical classification
\end{itemize}

We begin by importing the libraries and setting the plot style.

\tutorialbarrier
\begin{codecell}

\end{codecell}

\section*{The APOGEE Data}

We load the stars of Tutorial 7 with their asteroseismic classes as labels, and we keep the effective temperature and the surface gravity.

\tutorialbarrier
\begin{codecell}
\begin{Verbatim}[commandchars=\\\{\},fontsize=\small]
\PYG{c+c1}{\PYGZsh{} Load the APOGEE dataset, with the labels (1 for RC and 0 for RGB)}
\PYG{n}{data} \PYG{o}{=}\FancyVerbBreakAfterBreak{} \PYG{n}{np}\PYG{o}{.}\FancyVerbBreakAfterBreak{}\PYG{n}{load}\PYG{p}{(}\FancyVerbBreakAfterBreak{}\PYG{l+s+s2}{\PYGZdq{}}\PYG{l+s+s2}{../data/dataset\PYGZus{}apogee\PYGZus{}asteroseismic.npz}\PYG{l+s+s2}{\PYGZdq{}}\PYG{p}{)}
\PYG{n}{t} \PYG{o}{=}\FancyVerbBreakAfterBreak{} \PYG{n}{data}\PYG{p}{[}\FancyVerbBreakAfterBreak{}\PYG{l+s+s2}{\PYGZdq{}}\PYG{l+s+s2}{label}\PYG{l+s+s2}{\PYGZdq{}}\PYG{p}{]}

\PYG{c+c1}{\PYGZsh{} Keep the temperature and the surface gravity, the first two features}
\PYG{n}{X} \PYG{o}{=}\FancyVerbBreakAfterBreak{} \PYG{n}{data}\PYG{p}{[}\FancyVerbBreakAfterBreak{}\PYG{l+s+s2}{\PYGZdq{}}\PYG{l+s+s2}{data}\PYG{l+s+s2}{\PYGZdq{}}\PYG{p}{]}\PYG{p}{[}\FancyVerbBreakAfterBreak{}\PYG{p}{:}\PYG{p}{,}\FancyVerbBreakAfterBreak{} \PYG{l+m+mi}{0}\PYG{p}{:}\PYG{l+m+mi}{2}\PYG{p}{]}

\PYG{n+nb}{print}\PYG{p}{(}\FancyVerbBreakAfterBreak{}\PYG{l+s+sa}{f}\PYG{l+s+s2}{\PYGZdq{}}\PYG{l+s+s2}{Stars = }\PYG{l+s+si}{\PYGZob{}}\PYG{n+nb}{len}\PYG{p}{(}\FancyVerbBreakAfterBreak{}\PYG{n}{t}\PYG{p}{)}\PYG{l+s+si}{:}\PYG{l+s+s2}{,}\FancyVerbBreakAfterBreak{}\PYG{l+s+si}{\PYGZcb{}}\FancyVerbBreakAfterBreak{}\PYG{l+s+s2}{, RC = }\PYG{l+s+si}{\PYGZob{}}\PYG{n}{np}\PYG{o}{.}\FancyVerbBreakAfterBreak{}\PYG{n}{sum}\PYG{p}{(}\FancyVerbBreakAfterBreak{}\PYG{n}{t}\PYG{+w}{ }\PYG{o}{==}\PYG{+w}{ }\PYG{l+m+mi}{1}\PYG{p}{)}\PYG{l+s+si}{:}\PYG{l+s+s2}{,}\FancyVerbBreakAfterBreak{}\PYG{l+s+si}{\PYGZcb{}}\FancyVerbBreakAfterBreak{}\PYG{l+s+s2}{, RGB = }\PYG{l+s+si}{\PYGZob{}}\PYG{n}{np}\PYG{o}{.}\FancyVerbBreakAfterBreak{}\PYG{n}{sum}\PYG{p}{(}\FancyVerbBreakAfterBreak{}\PYG{n}{t}\PYG{+w}{ }\PYG{o}{==}\PYG{+w}{ }\PYG{l+m+mi}{0}\PYG{p}{)}\PYG{l+s+si}{:}\PYG{l+s+s2}{,}\FancyVerbBreakAfterBreak{}\PYG{l+s+si}{\PYGZcb{}}\FancyVerbBreakAfterBreak{}\PYG{l+s+s2}{\PYGZdq{}}\PYG{p}{)}
\end{Verbatim}
\end{codecell}
\begin{Verbatim}[fontsize=\footnotesize, xleftmargin=5mm]
Stars = 2,855, RC = 2,125, RGB = 730
\end{Verbatim}

\section*{From Standard to Bayesian Logistic Regression}

In standard logistic regression, as in Tutorial 7, we seek a single weight vector \(\mathbf{w}\) that maximizes the likelihood of our observed data. This point estimate is the maximum likelihood solution. The probability that a star with features \(\mathbf{x}\) is an RC star is given by

\[
P(t=1|\mathbf{x}, \mathbf{w}) = \sigma(\mathbf{w}^T\mathbf{x}),
\]

where \(\sigma(z) = 1/(1 + e^{-z})\) is the sigmoid function. The vector \(\mathbf{x}\) holds the two standardized features and a constant 1, so the last weight is the bias (Section 9.1).

Instead, the Bayesian approach keeps the whole posterior distribution \(p(\mathbf{w}|\mathcal{D})\) of the weights. It represents our beliefs about the weights after we observe the data \(\mathcal{D}\). Bayes' theorem provides the recipe

\[
p(\mathbf{w}|\mathcal{D}) = \frac{p(\mathcal{D}|\mathbf{w})p(\mathbf{w})}{p(\mathcal{D})}.
\]

The likelihood \(p(\mathcal{D}|\mathbf{w})\) is the same as in standard logistic regression, the probability of our stellar classifications for given weights. The addition is the prior \(p(\mathbf{w})\), which encodes our beliefs before we see the data. We choose a Gaussian prior, which regularizes the weights and is simple to compute with

\[
p(\mathbf{w}) = \mathcal{N}(\mathbf{w}|\mathbf{0}, \lambda^{-1}\mathrm{Id}).
\]

This prior is centered at zero, and \(\lambda\) is the precision of each weight. It makes smaller weights more probable, a form of L2 regularization that follows from Bayesian principles. A larger \(\lambda\) encodes a stronger belief that the weights are small, which reduces overfitting.

In the notation of Section 9.1, the mean is \(\mathbf{m}_0 = \mathbf{0}\) and the covariance is \(\mathbf{S}_0 = \lambda^{-1}\mathrm{Id}\), so the prior precision matrix is \(\mathbf{S}_0^{-1} = \lambda\mathrm{Id}\). In Chapter 5 the prior variance of each weight is \(\eta^2\), so \(\lambda = 1/\eta^2\). Chapter 7 writes the penalty per star, so its \(\lambda\) is this one divided by \(2N\).

With this prior, the remaining problem is to compute the posterior. In Bayesian linear regression, a Gaussian prior and a Gaussian likelihood give a Gaussian posterior. The nonlinearity of the sigmoid function prevents this analytical solution here, so the posterior has no simple closed form. We therefore need an approximation, such as the Laplace approximation.

We first code the sigmoid. We then split the stars into 80 percent training stars and 20 percent test stars as in Tutorial 7, with the split function of scikit-learn. We also standardize the features. No setting depends on the test stars, so we do not need validation stars.

\tutorialbarrier
\begin{codecell}

\end{codecell}
\begin{Verbatim}[fontsize=\footnotesize, xleftmargin=5mm]
Training stars = 2284, test stars = 571
\end{Verbatim}

\section*{Finding the Maximum A Posteriori (MAP) Estimate}

The first step in Bayesian logistic regression is finding the maximum a posteriori (MAP) estimate, the mode of the posterior (Section 9.4). The MAP estimate is not fully Bayesian, because it is still a point estimate. It is the basis of the Laplace approximation.

The MAP estimate maximizes the posterior probability

\[
\mathbf{w}_{\text{MAP}} = \arg\max_{\mathbf{w}} p(\mathbf{w}|\mathcal{D}) = \arg\max_{\mathbf{w}} \left[\ln p(\mathcal{D}|\mathbf{w}) + \ln p(\mathbf{w})\right].
\]

For our Gaussian prior, this becomes

\[
\mathbf{w}_{\text{MAP}} = \arg\min_{\mathbf{w}} \left[-\sum_{i=1}^N \left(t_i \ln y_i + (1-t_i)\ln(1-y_i)\right) + \frac{\lambda}{2}\|\mathbf{w}\|^2\right],
\]

where \(y_i = \sigma(\mathbf{w}^T\mathbf{x}_i)\) is the predicted probability of star \(i\) and \(N\) is the number of training stars. The difference from standard logistic regression is the additional term \(\lambda\|\mathbf{w}\|^2/2\) from the Gaussian prior. This term penalizes large weights. For a uniform prior, which treats all weight values alike, the term vanishes (\(\lambda = 0\)). The MAP estimate is then the maximum likelihood estimate, as in Tutorial 7.

The gradient of the negative log posterior is the sum of the gradient from the likelihood and the gradient from the prior

\[
\nabla\left[-\ln p(\mathbf{w}|\mathcal{D})\right] = \sum_{i=1}^N (y_i - t_i)\mathbf{x}_i + \lambda\mathbf{w}.
\]

This form shows how the prior pulls the weights toward zero, while the data pull them toward values that explain the observations. The balance depends on both the prior strength \(\lambda\) and the amount of data. With abundant data the likelihood dominates, and with sparse data the prior has more influence.

As in Tutorial 7, the code works with the objective divided by \(N\). Its likelihood part is the cross-entropy loss \(E(\mathbf{w})\) of Section 7.6, and the gradient of its prior part is \(\lambda\mathbf{w}/N\). We first code the loss.

\tutorialbarrier
\begin{codecell}
\begin{Verbatim}[commandchars=\\\{\},fontsize=\small]
\PYG{k}{def}\PYG{+w}{ }\PYG{n+nf}{cost}\PYG{p}{(}\FancyVerbBreakAfterBreak{}\PYG{n}{w}\PYG{p}{,}\FancyVerbBreakAfterBreak{} \PYG{n}{X}\PYG{p}{,}\FancyVerbBreakAfterBreak{} \PYG{n}{t}\PYG{p}{,}\FancyVerbBreakAfterBreak{} \PYG{n}{N}\PYG{p}{)}\PYG{p}{:}
\PYG{+w}{    }\PYG{l+s+sd}{\PYGZdq{}\PYGZdq{}\PYGZdq{}Average negative log\PYGZhy{}likelihood of N stars.\PYGZdq{}\PYGZdq{}\PYGZdq{}}
    \PYG{n}{z} \PYG{o}{=}\FancyVerbBreakAfterBreak{} \PYG{n}{X} \PYG{o}{@} \PYG{n}{w}
    \PYG{n}{y} \PYG{o}{=}\FancyVerbBreakAfterBreak{} \PYG{n}{sigmoid}\PYG{p}{(}\FancyVerbBreakAfterBreak{}\PYG{n}{z}\PYG{p}{)}

    \PYG{c+c1}{\PYGZsh{} Add a small epsilon to prevent log(0)}
    \PYG{n}{epsilon} \PYG{o}{=}\FancyVerbBreakAfterBreak{} \PYG{l+m+mf}{1e\PYGZhy{}15}
    \PYG{k}{return} \PYG{o}{\PYGZhy{}}\PYG{n}{np}\PYG{o}{.}\FancyVerbBreakAfterBreak{}\PYG{n}{sum}\PYG{p}{(}\FancyVerbBreakAfterBreak{}\PYG{n}{t} \PYG{o}{*} \PYG{n}{np}\PYG{o}{.}\FancyVerbBreakAfterBreak{}\PYG{n}{log}\PYG{p}{(}\FancyVerbBreakAfterBreak{}\PYG{n}{y} \PYG{o}{+}\FancyVerbBreakAfterBreak{} \PYG{n}{epsilon}\PYG{p}{)} \PYG{o}{+}\FancyVerbBreakAfterBreak{}
                   \PYG{p}{(}\FancyVerbBreakAfterBreak{}\PYG{l+m+mi}{1} \PYG{o}{\PYGZhy{}} \PYG{n}{t}\PYG{p}{)} \PYG{o}{*} \PYG{n}{np}\PYG{o}{.}\FancyVerbBreakAfterBreak{}\PYG{n}{log}\PYG{p}{(}\FancyVerbBreakAfterBreak{}\PYG{l+m+mi}{1} \PYG{o}{\PYGZhy{}} \PYG{n}{y} \PYG{o}{+}\FancyVerbBreakAfterBreak{} \PYG{n}{epsilon}\PYG{p}{)}\PYG{p}{)} \PYG{o}{/} \PYG{n}{N}
\end{Verbatim}
\end{codecell}

We now write the optimizer for the MAP estimate. Section 9.4 climbs the log posterior by gradient ascent. Here we descend the negative log posterior, which gives the same estimate, and we use mini-batches. The optimizer shuffles the stars in each epoch. It then updates the weights on each mini-batch with the gradient of the likelihood term and the prior term \(\lambda\mathbf{w}/N\).

\tutorialbarrier
\begin{codecell}

\end{codecell}

We follow the first mini-batch run in Tutorial 7 with a learning rate of 0.1, a batch size of 32 and 200 epochs. We assume the prior precision \(\lambda = 0.1\). We record the training loss and the test loss in each epoch.

\tutorialbarrier
\begin{codecell}
\begin{Verbatim}[commandchars=\\\{\},fontsize=\small]
\PYG{c+c1}{\PYGZsh{} Set the hyperparameters}
\PYG{n}{learning\PYGZus{}rate} \PYG{o}{=}\FancyVerbBreakAfterBreak{} \PYG{l+m+mf}{0.1}
\PYG{n}{n\PYGZus{}epochs} \PYG{o}{=}\FancyVerbBreakAfterBreak{} \PYG{l+m+mi}{200}
\PYG{n}{batch\PYGZus{}size} \PYG{o}{=}\FancyVerbBreakAfterBreak{} \PYG{l+m+mi}{32}
\PYG{n}{lambda\PYGZus{}reg} \PYG{o}{=}\FancyVerbBreakAfterBreak{} \PYG{l+m+mf}{0.1}  \PYG{c+c1}{\PYGZsh{} Prior precision}

\PYG{c+c1}{\PYGZsh{} Find the MAP estimate}
\PYG{n}{w\PYGZus{}map}\PYG{p}{,}\FancyVerbBreakAfterBreak{} \PYG{n}{train\PYGZus{}losses}\PYG{p}{,}\FancyVerbBreakAfterBreak{} \PYG{n}{test\PYGZus{}losses} \PYG{o}{=}\FancyVerbBreakAfterBreak{} \PYG{n}{find\PYGZus{}map\PYGZus{}estimate}\PYG{p}{(}\FancyVerbBreakAfterBreak{}
    \PYG{n}{X\PYGZus{}train}\PYG{p}{,}\FancyVerbBreakAfterBreak{} \PYG{n}{t\PYGZus{}train}\PYG{p}{,}\FancyVerbBreakAfterBreak{} \PYG{n}{X\PYGZus{}test}\PYG{p}{,}\FancyVerbBreakAfterBreak{} \PYG{n}{t\PYGZus{}test}\PYG{p}{,}\FancyVerbBreakAfterBreak{}
    \PYG{n}{learning\PYGZus{}rate}\PYG{p}{,}\FancyVerbBreakAfterBreak{} \PYG{n}{n\PYGZus{}epochs}\PYG{p}{,}\FancyVerbBreakAfterBreak{} \PYG{n}{batch\PYGZus{}size}\PYG{p}{,}\FancyVerbBreakAfterBreak{} \PYG{n}{lambda\PYGZus{}reg}
\PYG{p}{)}
\end{Verbatim}
\end{codecell}

We plot the learning curves.

\tutorialbarrier
\begin{codecell}

\end{codecell}
\tutorialplot{\includegraphics[width=4.302in,height=0.75\textheight,keepaspectratio]{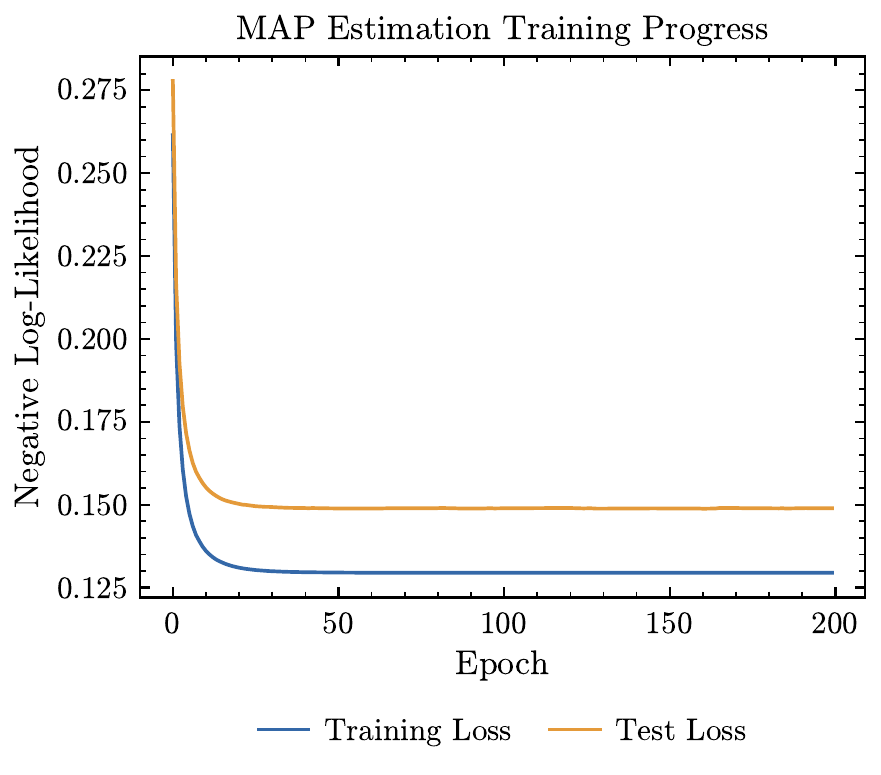}}%
  {\includegraphics[height=\tutorialheight]{tutorials/9/fig01.pdf}}

Both losses fall quickly and flatten after about 25 epochs, so 200 epochs are enough. The test loss stays slightly above the training loss, and the two do not diverge. The objective is convex and has one minimum, so we keep the weights of the last epoch as the MAP estimate.

\section*{Visualizing the MAP Decision Boundary}

The MAP weights define a decision boundary, which lies where the probability is 0.5 and \(\mathbf{w}^T\mathbf{x} = 0\). With two features, this is a line. We evaluate the probability on a grid of temperatures and gravities, and we predict the classes of the test stars.

\tutorialbarrier
\begin{codecell}

\end{codecell}

We now plot the probability contours, the decision boundary and the test stars, and we mark the misclassified stars. We also print the accuracy, the share of test stars with the correct class.

\tutorialbarrier
\begin{codecell}

\end{codecell}
\tutorialplot{\includegraphics[width=5.127in,height=0.75\textheight,keepaspectratio]{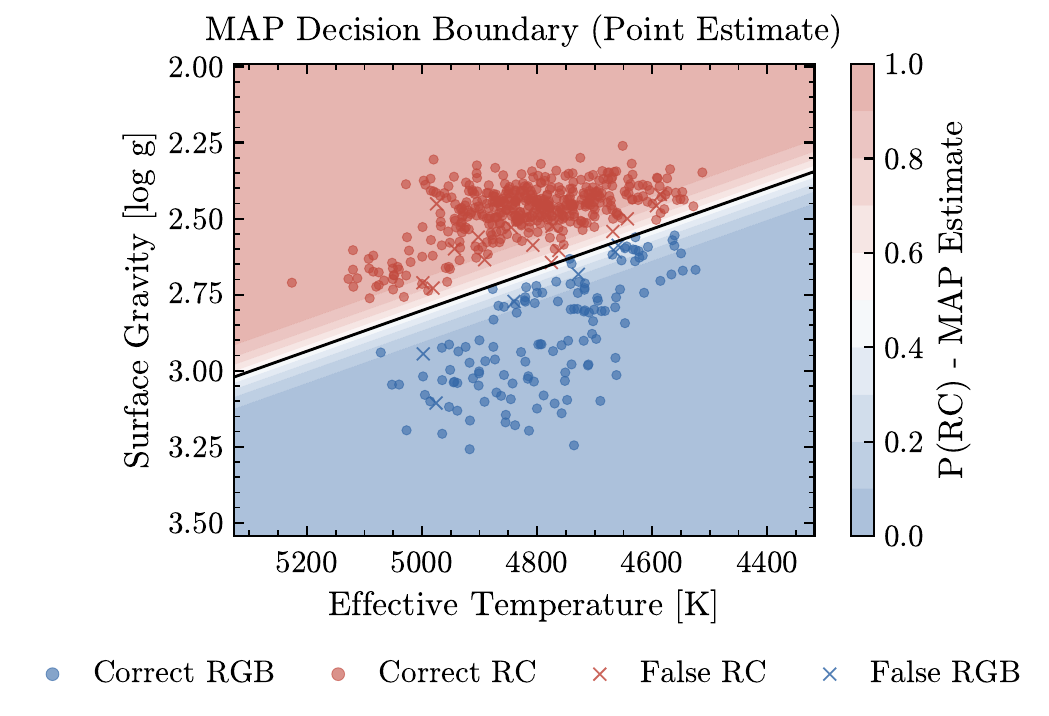}}%
  {\includegraphics[height=\tutorialheight]{tutorials/9/fig02.pdf}}
\begin{Verbatim}[fontsize=\footnotesize, xleftmargin=5mm]
Test set accuracy = 0.960, misclassified stars = 23 out of 571
\end{Verbatim}

The line separates most RC stars from RGB stars, and the misclassified stars lie next to it. The accuracy is comparable to the 0.974 of Tutorial 7 on the same test stars, which used eight features. These two features therefore separate the classes well.

\section*{The Laplace Approximation}

The MAP estimate gives our best single set of weights, but it conveys nothing about the uncertainty of the parameters. In regions with sparse training data we should be less confident about our classifications, yet the MAP probabilities can be as extreme there as elsewhere. To measure this uncertainty, we need the full posterior distribution \(p(\mathbf{w}|\mathcal{D})\).

The Laplace approximation (Section 9.4) approximates the posterior with a Gaussian distribution centered at the MAP estimate. It is based on a second-order Taylor expansion of the log posterior around its mode (Section 9.3). Near the MAP estimate, the log posterior is

\[
\ln p(\mathbf{w}|\mathcal{D}) \approx \ln p(\mathbf{w}_{\text{MAP}}|\mathcal{D}) - \frac{1}{2}(\mathbf{w} - \mathbf{w}_{\text{MAP}})^T \mathbf{A} (\mathbf{w} - \mathbf{w}_{\text{MAP}}),
\]

where \(\mathbf{A}\) is the Hessian matrix of the negative log posterior at \(\mathbf{w}_{\text{MAP}}\). The gradient vanishes at the mode, so the linear term disappears. Exponentiating this quadratic form gives a Gaussian

\[
p(\mathbf{w}|\mathcal{D}) \approx \mathcal{N}(\mathbf{w}|\mathbf{w}_{\text{MAP}}, \mathbf{A}^{-1}).
\]

The Hessian \(\mathbf{A}\) describes the curvature of the log posterior, and it is the precision matrix of this Gaussian. A sharp curvature (large eigenvalues) means that the posterior concentrates tightly around the MAP estimate, and a gentle curvature means a greater uncertainty. For logistic regression with a Gaussian prior, the Hessian has the form (Section 9.4)

\[
\mathbf{A} = \lambda\mathrm{Id} + \sum_{i=1}^N y_i(1-y_i)\mathbf{x}_i\mathbf{x}_i^T,
\]

where \(y_i = \sigma(\mathbf{w}_{\text{MAP}}^T\mathbf{x}_i)\). The first term is the prior precision \(\mathbf{S}_0^{-1} = \lambda\mathrm{Id}\), and it adds the same curvature in all directions. The second term depends on the data, and each star contributes according to the uncertainty of its prediction. Stars near the decision boundary, where \(y_i \approx 0.5\), contribute most to the curvature. We now code the Hessian.

\tutorialbarrier
\begin{codecell}

\end{codecell}

We compute the Hessian at the MAP estimate, and we invert it to get the covariance matrix of the weights. The square roots of its diagonal are the uncertainties of the weights. We print them next to the weights, with the correlation of the temperature and gravity weights.

\tutorialbarrier
\begin{codecell}

\end{codecell}
\begin{Verbatim}[fontsize=\footnotesize, xleftmargin=5mm]
Temperature coefficient = 1.7278 ± 0.1241
Log g coefficient = -4.4663 ± 0.2247
Bias = 2.1635 ± 0.1286
Parameter correlation = -0.6175
\end{Verbatim}

Each weight is much larger than its uncertainty, so the data constrain all three. The correlation of the temperature and gravity weights is negative, and the two weights have opposite signs. A larger temperature weight therefore goes with a more negative gravity weight, so their ratio, which sets the slope of the boundary, varies less.

\section*{Sampling from the Posterior Distribution}

The Gaussian approximation lets us draw samples of the weights and so see the uncertainty as a spread of decision boundaries. Each sample from \(\mathcal{N}(\mathbf{w}_{\text{MAP}}, \mathbf{A}^{-1})\) is a plausible set of weights that is consistent with our data and prior, and it gives its own boundary. We draw 20 samples.

\tutorialbarrier
\begin{codecell}
\begin{Verbatim}[commandchars=\\\{\},fontsize=\small]
\PYG{k}{def}\PYG{+w}{ }\PYG{n+nf}{sample\PYGZus{}from\PYGZus{}posterior}\PYG{p}{(}\FancyVerbBreakAfterBreak{}\PYG{n}{w\PYGZus{}map}\PYG{p}{,}\FancyVerbBreakAfterBreak{} \PYG{n}{cov\PYGZus{}matrix}\PYG{p}{,}\FancyVerbBreakAfterBreak{} \PYG{n}{n\PYGZus{}samples}\PYG{o}{=}\FancyVerbBreakAfterBreak{}\PYG{l+m+mi}{20}\PYG{p}{)}\PYG{p}{:}
\PYG{+w}{    }\PYG{l+s+sd}{\PYGZdq{}\PYGZdq{}\PYGZdq{}Draw weight vectors from the Gaussian approximation of the posterior.\PYGZdq{}\PYGZdq{}\PYGZdq{}}
    \PYG{k}{return} \PYG{n}{np}\PYG{o}{.}\FancyVerbBreakAfterBreak{}\PYG{n}{random}\PYG{o}{.}\FancyVerbBreakAfterBreak{}\PYG{n}{multivariate\PYGZus{}normal}\PYG{p}{(}\FancyVerbBreakAfterBreak{}\PYG{n}{w\PYGZus{}map}\PYG{p}{,}\FancyVerbBreakAfterBreak{} \PYG{n}{cov\PYGZus{}matrix}\PYG{p}{,}\FancyVerbBreakAfterBreak{} \PYG{n}{size}\PYG{o}{=}\FancyVerbBreakAfterBreak{}\PYG{n}{n\PYGZus{}samples}\PYG{p}{)}

\PYG{c+c1}{\PYGZsh{} Draw samples from the posterior}
\PYG{n}{np}\PYG{o}{.}\FancyVerbBreakAfterBreak{}\PYG{n}{random}\PYG{o}{.}\FancyVerbBreakAfterBreak{}\PYG{n}{seed}\PYG{p}{(}\FancyVerbBreakAfterBreak{}\PYG{l+m+mi}{42}\PYG{p}{)}
\PYG{n}{weight\PYGZus{}samples} \PYG{o}{=}\FancyVerbBreakAfterBreak{} \PYG{n}{sample\PYGZus{}from\PYGZus{}posterior}\PYG{p}{(}\FancyVerbBreakAfterBreak{}\PYG{n}{w\PYGZus{}map}\PYG{p}{,}\FancyVerbBreakAfterBreak{} \PYG{n}{cov\PYGZus{}matrix}\PYG{p}{,}\FancyVerbBreakAfterBreak{} \PYG{n}{n\PYGZus{}samples}\PYG{o}{=}\FancyVerbBreakAfterBreak{}\PYG{l+m+mi}{20}\PYG{p}{)}
\end{Verbatim}
\end{codecell}

We plot the decision boundary of each sampled weight vector together with the MAP boundary.

\tutorialbarrier
\begin{codecell}

\end{codecell}
\tutorialplot{\includegraphics[width=5.714in,height=0.75\textheight,keepaspectratio]{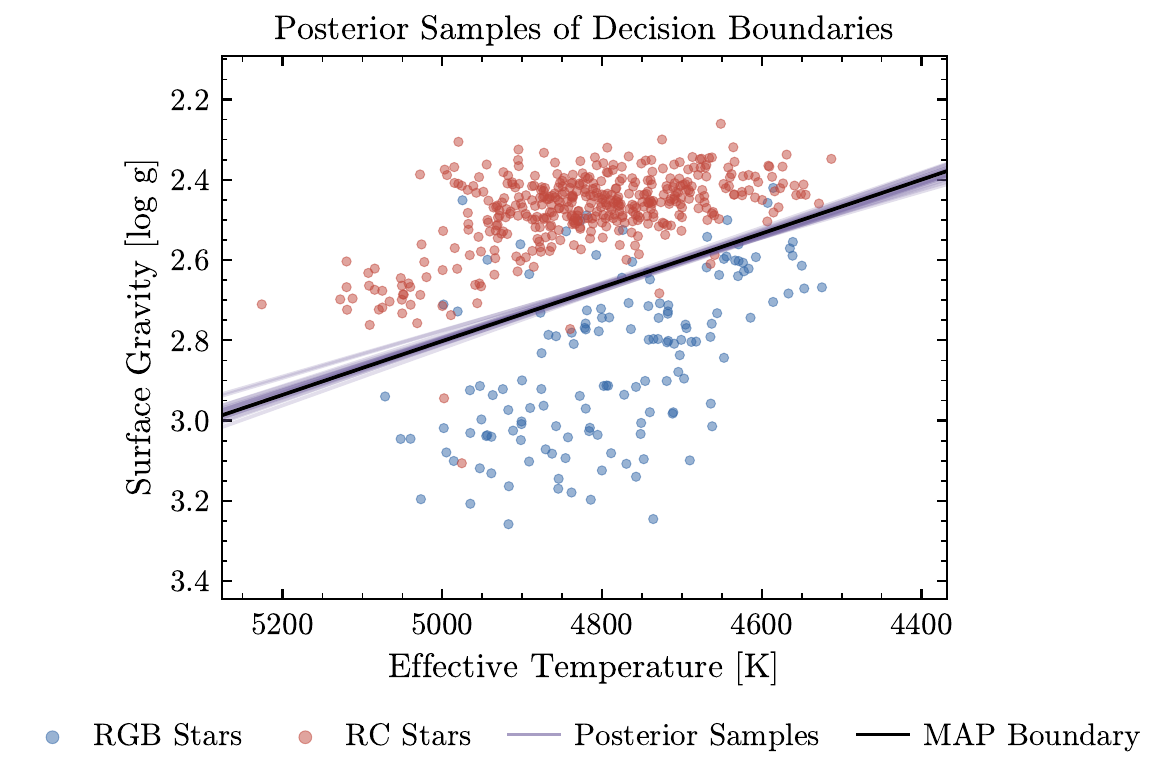}}%
  {\includegraphics[height=\tutorialheight]{tutorials/9/fig03.pdf}}

The sampled decision boundaries show how the uncertainty of the weights becomes uncertainty of the classification. They agree in the middle of the temperature range, where many training stars lie next to the boundary and the data fix it well. They spread at the two ends, where few stars lie and several decision rules remain plausible. Where the boundaries agree, we can be confident in our classifications, and where they spread, we should be more cautious.

To measure this uncertainty in more detail, we compute the predictions for each sampled weight vector and examine their distribution. At any point of the feature space, the variation of the predictions between the samples measures our uncertainty about its classification. We map the mean of the predictions and the width of their 95\% interval, from the 2.5th to the 97.5th percentile. We use 1000 samples.

\tutorialbarrier
\begin{codecell}

\end{codecell}

We map the mean of the predictions on the left and the width of the interval on the right.

\tutorialbarrier
\begin{codecell}

\end{codecell}
\tutorialplot{\includegraphics[width=6.180in,height=0.75\textheight,keepaspectratio]{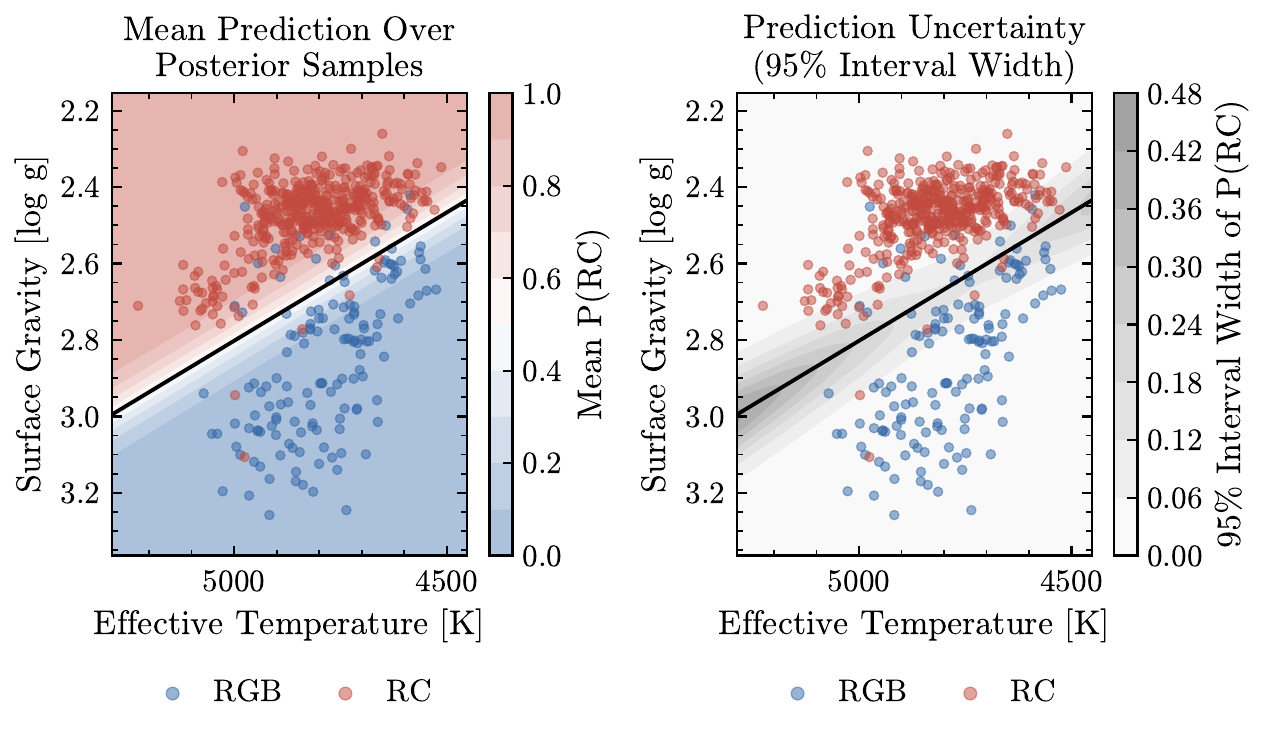}}%
  {\includegraphics[height=\tutorialheight]{tutorials/9/fig04.pdf}}

The mean prediction in the left plot looks like the MAP probability. In the right plot, the interval is narrow far from the boundary. It is widest next to the boundary at the two ends, where the sampled boundaries spread.

\section*{From Sample-Based to Analytical Predictions}

The sampling approach shows the uncertainty, but it needs many samples that we must generate and evaluate. The probit approximation (Section 9.6) is more efficient. It gives a closed-form expression for the predictive distribution that averages over the parameter uncertainty without sampling.

For a new observation \(\mathbf{x}^*\), the predictive probability requires a marginalization over the posterior (Section 9.5)

\[
P(t^*=1|\mathbf{x}^*, \mathcal{D}) = \int \sigma(\mathbf{w}^T\mathbf{x}^*) p(\mathbf{w}|\mathcal{D}) \mathrm{d}\mathbf{w}.
\]

This integral has no closed form. The probit approximation gives a closed-form approximation

\[
P(t^*=1|\mathbf{x}^*, \mathcal{D}) \approx \sigma\left(\frac{\mu_a}{\sqrt{1 + \pi\sigma_a^2/8}}\right),
\]

where

\begin{itemize}
\tightlist
\item
  \(\mu_a = \mathbf{w}_{\text{MAP}}^T \mathbf{x}^*\) is the logit at the MAP weights,
\item
  \(\sigma_a^2 = \mathbf{x}^{*T} \mathbf{A}^{-1} \mathbf{x}^*\) is the variance of the logit.
\end{itemize}

Without the denominator, the formula gives the MAP probability \(\sigma(\mu_a)\), the prediction of the MAP weights alone. We call \(\kappa = 1/\sqrt{1 + \pi\sigma_a^2/8}\) the uncertainty factor. When the parameter uncertainty is high (large \(\sigma_a^2\)), \(\kappa\) becomes small and shrinks the logit toward zero. This pushes the prediction toward 0.5, the maximum uncertainty, in regions where we lack confidence in the parameters. We call a probability more cautious when it lies closer to 0.5, because it then claims less certainty about the class.

For a uniform prior (\(\lambda \to 0\)), \(\mu_a\) is the logit of the logistic regression of Tutorial 7, here with two features. The predictive probability then differs from the probability of that regression only through \(\kappa\). A prior with \(\lambda > 0\) also changes \(\mu_a\) and \(\kappa\).

The model expresses less confidence as a consequence of integrating over the parameter uncertainty and not through an ad hoc rule that we add. We now code the predictive distribution.

\tutorialbarrier
\begin{codecell}

\end{codecell}

We now evaluate this predictive distribution on the grid of temperatures and gravities that we used for the samples.

\tutorialbarrier
\begin{codecell}
\begin{Verbatim}[commandchars=\\\{\},fontsize=\small]
\PYG{c+c1}{\PYGZsh{} Compute the predictive distribution with the probit approximation}
\PYG{n}{pred\PYGZus{}probs}\PYG{p}{,}\FancyVerbBreakAfterBreak{} \PYG{n}{difference}\PYG{p}{,}\FancyVerbBreakAfterBreak{} \PYG{n}{std\PYGZus{}dev}\PYG{p}{,}\FancyVerbBreakAfterBreak{} \PYG{n}{kappa\PYGZus{}values} \PYG{o}{=}\FancyVerbBreakAfterBreak{} \PYG{n}{compute\PYGZus{}predictive\PYGZus{}distribution\PYGZus{}probit}\PYG{p}{(}\FancyVerbBreakAfterBreak{}
    \PYG{n}{grid\PYGZus{}points\PYGZus{}norm}\PYG{p}{,}\FancyVerbBreakAfterBreak{} \PYG{n}{w\PYGZus{}map}\PYG{p}{,}\FancyVerbBreakAfterBreak{} \PYG{n}{cov\PYGZus{}matrix}
\PYG{p}{)}

\PYG{c+c1}{\PYGZsh{} Reshape for plotting, each array to the shape of the grid}
\PYG{n}{pred\PYGZus{}probs} \PYG{o}{=}\FancyVerbBreakAfterBreak{} \PYG{n}{pred\PYGZus{}probs}\PYG{o}{.}\FancyVerbBreakAfterBreak{}\PYG{n}{reshape}\PYG{p}{(}\FancyVerbBreakAfterBreak{}\PYG{n}{xx\PYGZus{}norm}\PYG{o}{.}\FancyVerbBreakAfterBreak{}\PYG{n}{shape}\PYG{p}{)}
\PYG{n}{difference} \PYG{o}{=}\FancyVerbBreakAfterBreak{} \PYG{n}{difference}\PYG{o}{.}\FancyVerbBreakAfterBreak{}\PYG{n}{reshape}\PYG{p}{(}\FancyVerbBreakAfterBreak{}\PYG{n}{xx\PYGZus{}norm}\PYG{o}{.}\FancyVerbBreakAfterBreak{}\PYG{n}{shape}\PYG{p}{)}
\PYG{n}{std\PYGZus{}dev} \PYG{o}{=}\FancyVerbBreakAfterBreak{} \PYG{n}{std\PYGZus{}dev}\PYG{o}{.}\FancyVerbBreakAfterBreak{}\PYG{n}{reshape}\PYG{p}{(}\FancyVerbBreakAfterBreak{}\PYG{n}{xx\PYGZus{}norm}\PYG{o}{.}\FancyVerbBreakAfterBreak{}\PYG{n}{shape}\PYG{p}{)}  \PYG{c+c1}{\PYGZsh{} Standard deviation of the logit}
\PYG{n}{kappa\PYGZus{}values} \PYG{o}{=}\FancyVerbBreakAfterBreak{} \PYG{n}{kappa\PYGZus{}values}\PYG{o}{.}\FancyVerbBreakAfterBreak{}\PYG{n}{reshape}\PYG{p}{(}\FancyVerbBreakAfterBreak{}\PYG{n}{xx\PYGZus{}norm}\PYG{o}{.}\FancyVerbBreakAfterBreak{}\PYG{n}{shape}\PYG{p}{)}  \PYG{c+c1}{\PYGZsh{} Uncertainty factor}
\end{Verbatim}
\end{codecell}

We plot the difference between the predictive and the MAP probability on the left and the uncertainty factor \(\kappa\) on the right.

\tutorialbarrier
\begin{codecell}

\end{codecell}
\tutorialplot{\includegraphics[width=6.406in,height=0.75\textheight,keepaspectratio]{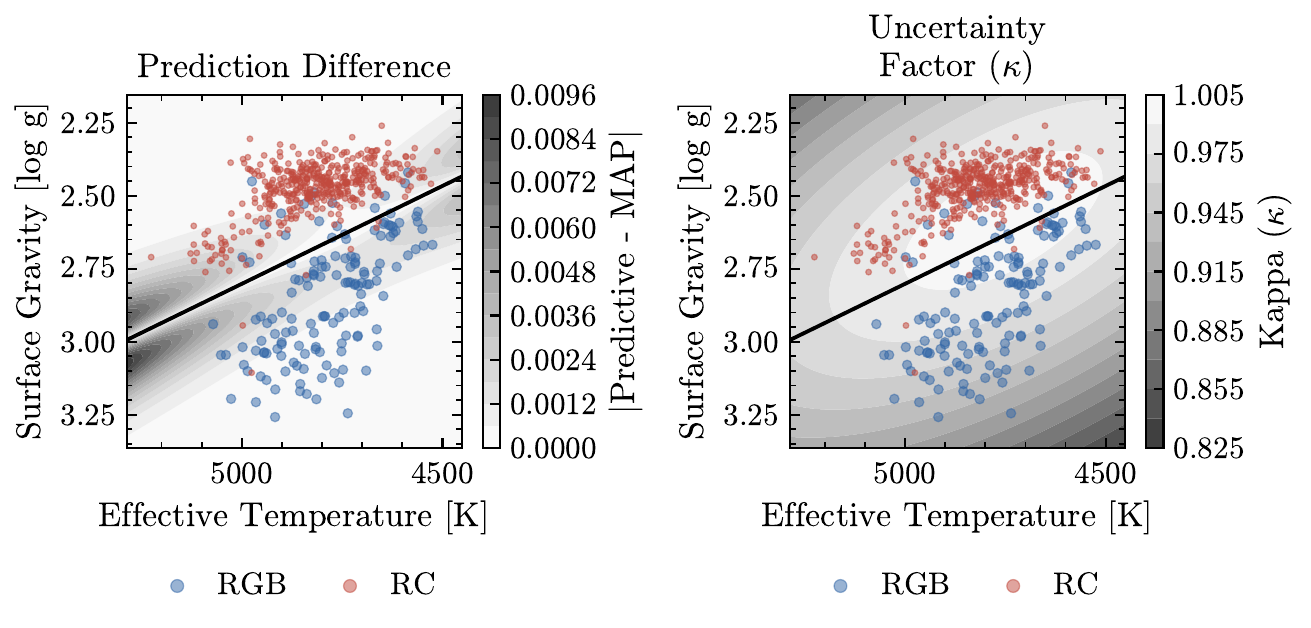}}%
  {\includegraphics[height=\tutorialheight]{tutorials/9/fig05.pdf}}

The two plots show where the uncertainty of the weights changes the predictions.

\begin{enumerate}
\def\labelenumi{\arabic{enumi}.}
\tightlist
\item
  \textbf{Difference of the probabilities.} The left plot shows \(|\text{predictive} - \text{MAP}|\), the difference between the predictive probability and the MAP probability. Larger values mark the areas where the parameter uncertainty changes the prediction most.
\item
  \textbf{Uncertainty factor \(\kappa\).} This factor is between 0 and 1. When \(\kappa\) is close to 1, the predictive probability is similar to the MAP probability. As \(\kappa\) decreases, the prediction moves toward 0.5.
\end{enumerate}

The color bars show that \(\kappa\) stays above 0.83 and that the difference stays below 0.01. The predictive and the MAP probabilities are therefore close for these data.

\section*{Understanding the Relationship Between Kappa and Prediction Changes}

Even far from the training stars, where \(\kappa\) is smaller, the difference is small. The probit formula explains why (Section 9.7). The effect of \(\kappa\) depends on two things.

\begin{enumerate}
\def\labelenumi{\arabic{enumi}.}
\tightlist
\item
  The value of \(\kappa\) itself, which shows how uncertain we are.
\item
  The strength of the original prediction, which is how far it lies from 0.5.
\end{enumerate}

We look at the probit approximation again

\[
P(t^*=1|\mathbf{x}^*, \mathcal{D}) \approx \sigma(\kappa \cdot \mu_a),
\]

where \(\mu_a = \mathbf{w}_{\text{MAP}}^T \mathbf{x}^*\) is the logit from standard logistic regression. For a star far from the decision boundary, \(|\mu_a|\) can be very large, for example 10. Even if \(\kappa\) reduces it by 10 percent (\(\kappa = 0.9\)), the result is still large enough to give a probability very close to 0 or 1.

\begin{itemize}
\tightlist
\item
  The standard prediction with \(\mu_a = 10\) gives \(\sigma(10) \approx 0.99995\).
\item
  The predictive probability with \(\kappa = 0.9\) gives \(\sigma(0.9 \times 10) = \sigma(9) \approx 0.99988\).
\end{itemize}

The impact of \(\kappa\) is most noticeable in two cases.

\begin{enumerate}
\def\labelenumi{\arabic{enumi}.}
\tightlist
\item
  Near the decision boundary, where \(|\mu_a|\) is small.
\item
  When \(\kappa\) is much lower than 1, which means a much higher uncertainty.
\end{enumerate}

This explains why the largest differences between the MAP and the predictive probabilities appear near the decision boundary. We plot the sigmoid for several values of \(\kappa\) to show this.

\tutorialbarrier
\begin{codecell}

\end{codecell}
\tutorialplot{\includegraphics[width=5.569in,height=0.75\textheight,keepaspectratio]{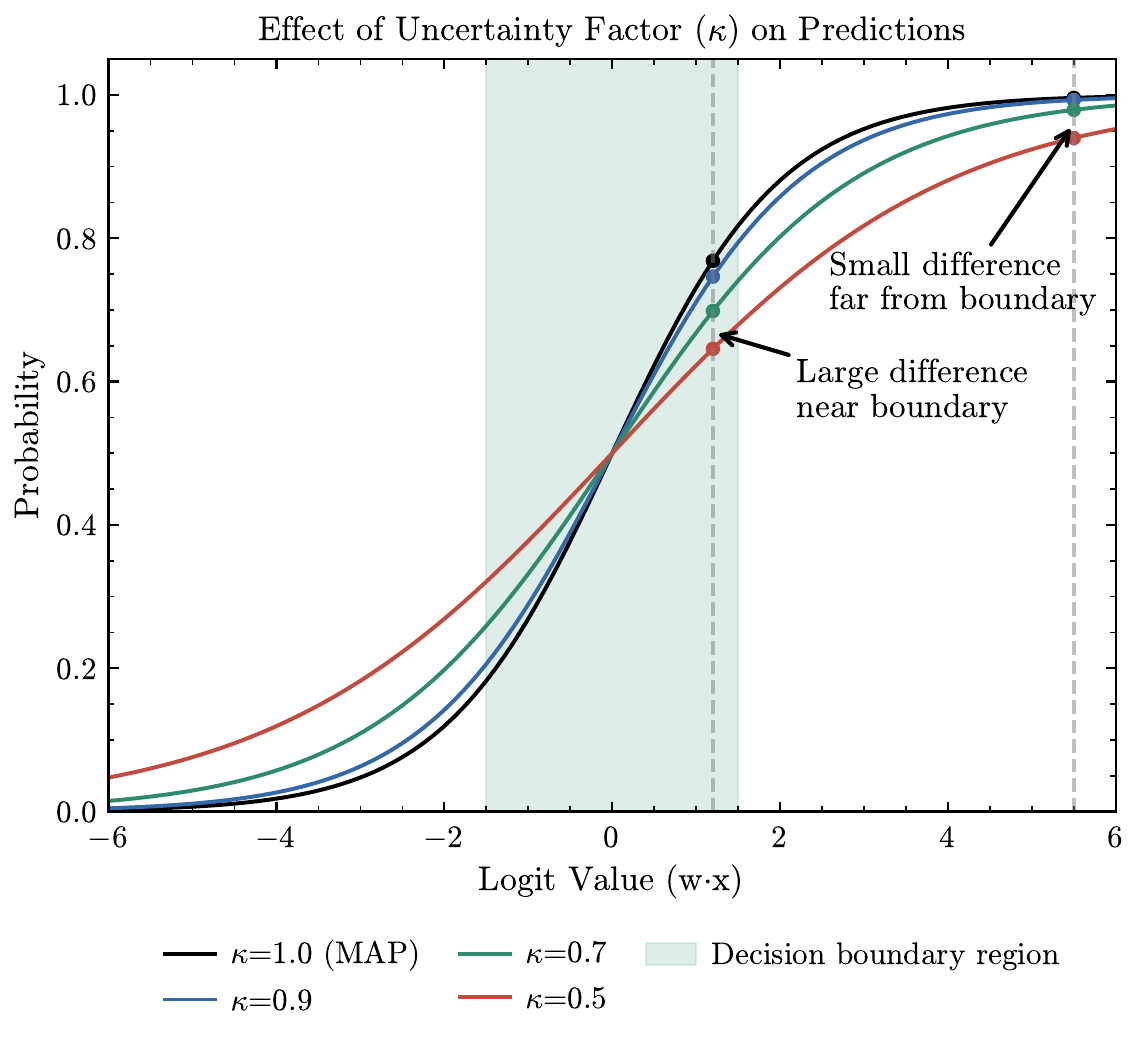}}%
  {\includegraphics[height=\tutorialheight]{tutorials/9/fig06.pdf}}

For the classification of RC and RGB stars, this is the behavior that we want. Stars far from the boundary keep their confident classification, even where the parameter uncertainty is high. The probabilities near the boundary become more cautious.

\section*{The Effect of Prior Strength on Bayesian Logistic Regression}

So far we used the prior precision \(\lambda = 0.1\). We now see what the prior strength does by comparing two extreme values. The prior \(\mathcal{N}(\mathbf{w}|\mathbf{0}, \lambda^{-1}\mathrm{Id})\) gives each weight the standard deviation \(1/\sqrt{\lambda}\).

\begin{itemize}
\tightlist
\item
  A \textbf{weak prior} (\(\lambda = 0.001\), a standard deviation of about 32) is far wider than the weights, which are of order 1 to 4. It leaves them almost free, as in the maximum likelihood fit of Tutorial 7.
\item
  A \textbf{strong prior} (\(\lambda = 100\), a standard deviation of 0.1) is much narrower than the weights. It pulls them toward zero.
\end{itemize}

We want only to see the effect of the prior strength, so we do not tune \(\lambda\) here. We train one model for each prior strength, with the other settings of the first run. We then look at the weights, their uncertainties and the predictions.

\tutorialbarrier
\begin{codecell}

\end{codecell}

We compare the two priors on a grid in the standardized feature space. A function draws one map for each prior, with the training stars and the decision boundary.

\tutorialbarrier
\begin{codecell}

\end{codecell}

The penalty \(\lambda\|\mathbf{w}\|^2/2\) pulls the weights toward zero, so a stronger prior should give smaller weights. We first plot the MAP probability of RC for the two priors, and the titles show the weights.

\tutorialbarrier
\begin{codecell}

\end{codecell}
\tutorialplot{\includegraphics[width=6.184in,height=0.75\textheight,keepaspectratio]{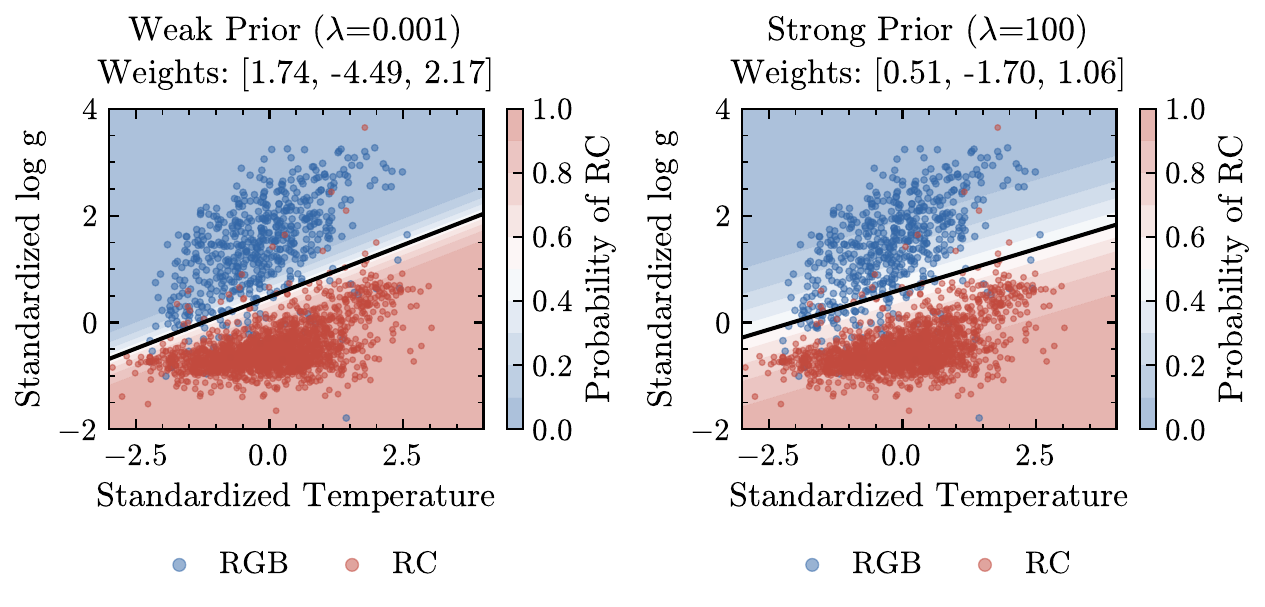}}%
  {\includegraphics[height=\tutorialheight]{tutorials/9/fig07.pdf}}

The plots show how the prior strength affects the MAP decision boundary.

\begin{enumerate}
\def\labelenumi{\arabic{enumi}.}
\tightlist
\item
  \textbf{Smaller weights, wider contours.} A strong prior pulls all three weights toward zero, as the titles show. The logit then grows more slowly across the boundary, so the probability changes more slowly and the contours widen.
\item
  \textbf{Boundary slope.} The boundary lies where the logit is zero, so its slope depends on the ratio of the two weights. The strong prior shrinks the temperature weight by a larger factor, so the boundary is less steep.
\end{enumerate}

\section*{Impact of Prior Strength on Uncertainty Quantification}

The prior also adds \(\lambda\mathrm{Id}\) to the Hessian \(\mathbf{A}\), so a stronger prior should give a smaller covariance of the weights. We compute the covariance matrix of each prior from the Hessian at its MAP estimate. We print each weight with its uncertainty, the standard deviation from the diagonal of the covariance matrix.

\tutorialbarrier
\begin{codecell}

\end{codecell}
\begin{Verbatim}[fontsize=\footnotesize, xleftmargin=5mm]
Weak Prior (λ=0.001): 1.74 ± 0.13, -4.49 ± 0.23, 2.17 ± 0.13
Strong Prior (λ=100): 0.51 ± 0.05, -1.70 ± 0.06, 1.06 ± 0.05
\end{Verbatim}

The strong prior shrinks the uncertainties of the weights as well as the weights, because the added \(\lambda\mathrm{Id}\) narrows the posterior. Smaller uncertainties of the weights give a smaller variance \(\sigma_a^2\) of the logit and so a \(\kappa\) closer to 1. We plot \(\kappa\), the factor that scales the logit, for the two priors.

\tutorialbarrier
\begin{codecell}

\end{codecell}
\tutorialplot{\includegraphics[width=6.185in,height=0.75\textheight,keepaspectratio]{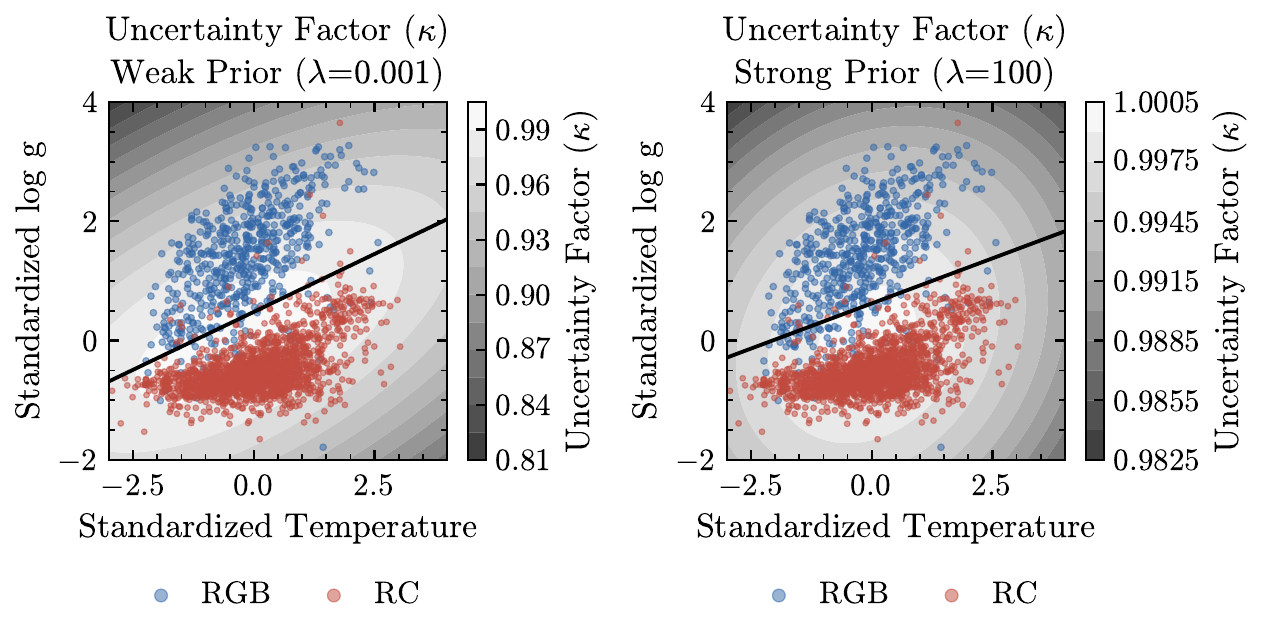}}%
  {\includegraphics[height=\tutorialheight]{tutorials/9/fig08.pdf}}

The color bars show that \(\kappa\) is closer to 1 for the strong prior. The average over the posterior therefore moves its probabilities very little toward 0.5. The smaller weights already do that, as the wider contours in its panel of the probability maps show.

\section*{Comparing Predictive and MAP Probabilities for Different Prior Strengths}

We now plot the predictive probability minus the MAP probability for the two priors. The difference shows where the average over the posterior matters.

\tutorialbarrier
\begin{codecell}

\end{codecell}
\tutorialplot{\includegraphics[width=6.223in,height=0.75\textheight,keepaspectratio]{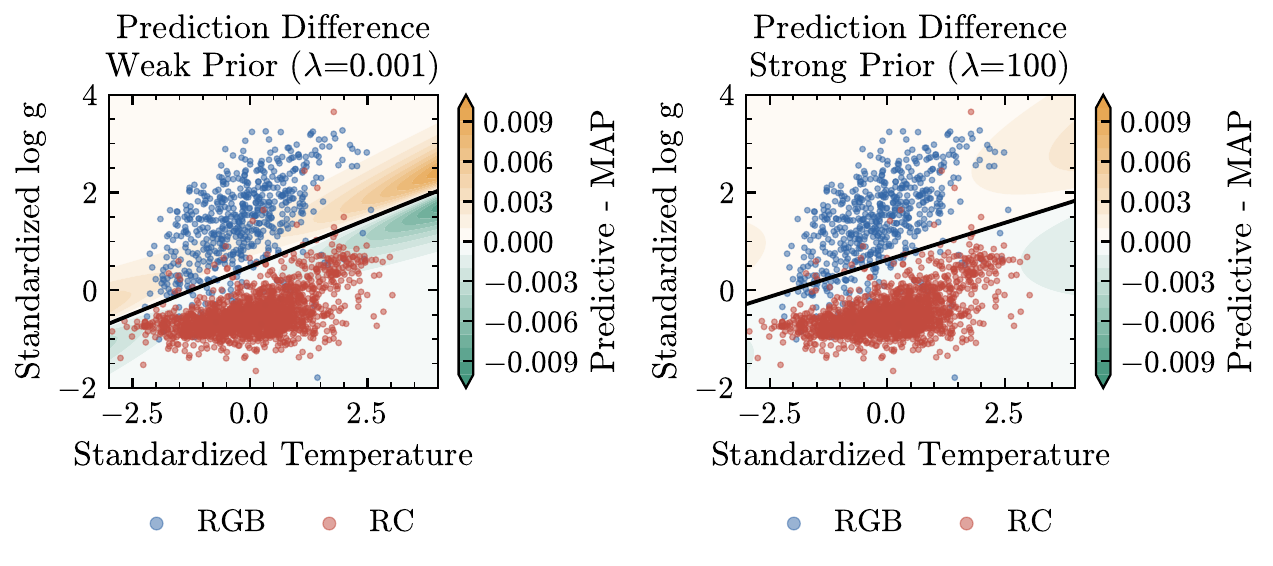}}%
  {\includegraphics[height=\tutorialheight]{tutorials/9/fig09.pdf}}

The colors show the predictive probability minus the MAP probability. Both maps follow the same pattern, and the differences are larger for the weak prior, because its \(\kappa\) is smaller.

\begin{itemize}
\tightlist
\item
  The difference is zero on the boundary, where both probabilities are 0.5.
\item
  Above the boundary, on the RGB side, it is positive (orange). Below it, on the RC side, it is negative (green). In both cases the predictive probability is closer to 0.5 than the MAP probability.
\item
  The difference is largest next to the boundary near the hot end on the right. There the logit is small and few training stars make \(\kappa\) small.
\end{itemize}

The two priors move the probabilities toward 0.5 in different ways. Which one suits depends on how well the data constrain the weights.

\begin{itemize}
\tightlist
\item
  \textbf{Weak prior.} The weights stay as in Tutorial 7, and only the average over the posterior moves the probabilities toward 0.5, through \(\kappa\). It suits cases like ours, where thousands of labeled stars constrain only three weights, so we can let the data fix the weights.
\item
  \textbf{Strong prior.} The weights shrink, which moves the probabilities toward 0.5 from the start. It suits cases where the data constrain the weights poorly. Examples are a few hundred labeled stars in a space of many features, and classes that are nearly linearly separable. In both cases maximum likelihood can give very large weights (Section 9.1).
\end{itemize}

Which of the two cases applies is not always clear, so in practice we set \(\lambda\) by validation (Section 4.7). A prior that is too weak overfits, and a prior that is too strong underfits. Either one generalizes poorly to the test stars. For each candidate \(\lambda\), we find the MAP weights on the training stars and compute a loss on validation stars, for example the cross-entropy loss. We keep the \(\lambda\) with the lowest validation loss, and the test stars give the final score once.

\section*{Summary}

This tutorial developed Bayesian logistic regression for two classes of giant stars. Our main lessons follow.

\begin{enumerate}
\def\labelenumi{\arabic{enumi}.}
\tightlist
\item
  \textbf{From point estimates to distributions.} Standard logistic regression gives a single optimal set of weights. Bayesian logistic regression keeps a distribution over all plausible weights and so describes the parameter uncertainty.
\item
  \textbf{The Laplace approximation.} We approximate the posterior with a Gaussian centered at the MAP estimate. It gives the uncertainty at a small computing cost.
\item
  \textbf{Predictive distributions.} The probit approximation gives closed-form predictions that integrate over the parameter uncertainty.
\item
  \textbf{The role of the uncertainty.} The uncertainty moves the probabilities toward one half where the training data are sparse. The effect is largest near the decision boundary, and objects far from the boundary keep their confident probabilities.
\item
  \textbf{Prior strength.} A strong prior shrinks the weights and so moves the probabilities toward one half from the start. A weak prior leaves the weights as in maximum likelihood, and then only the average over the posterior moves them toward one half.
\end{enumerate}

A prediction that feeds a later analysis or decision should say how far to trust it. Bayesian logistic regression gives each object a probability and an interval. The probability shows how much the classes overlap, and the width of the interval shows how well the data fix the boundary.

\tutorialchapter{10}{Principal Component Analysis for Galaxy Images}
\noindent\textit{Companion material for Chapter 10}

\section*{Introduction}

In the previous chapters we studied supervised learning with regression and classification. We now turn to unsupervised learning, and we begin with one of the most basic and widely used methods, principal component analysis (PCA). Astronomical data are often high-dimensional, and this makes it hard to find meaningful structure. A sky survey produces images with thousands of pixels or more. A spectrum has thousands of wavelength channels, and a light curve from a time-domain survey has hundreds of measurements.

PCA addresses this curse of dimensionality, the difficulty of finding structure in many dimensions. It finds the directions of maximum variance in the data and projects the data onto a lower-dimensional subspace that keeps the most important information. In this way PCA can reveal the hidden structure of a complex data set. This structure can guide the search for the physical processes behind the data, but PCA does not identify them by itself.

Consider galaxy morphology, which carries information about the formation history, the stellar populations and the evolutionary state of a galaxy. Galaxies look very different from each other, but their variation may follow a few basic shapes. PCA looks for such shapes in the data, without a prior choice of the features that matter most. The shapes can show, for example, the elongation of a galaxy and the concentration of its light.

This approach differs from parametric methods, which fit the brightness of a galaxy with a function of the radius, such as a Sérsic profile. These models are very useful, but they limit the analysis to features that we choose in advance. PCA uses only the data, and it can reveal structure that we did not expect. This matters more for fainter objects and subtler effects, where our physical intuition is less reliable.

PCA has many applications in astronomy beyond galaxy morphology. In spectroscopy, it can find the main modes of spectral variation. It can reduce thousands of wavelength channels to a handful of principal components, which are called eigenspectra. These components can relate to the stellar temperature and the metallicity, although they often mix the two (Section 10.9). In time-domain astronomy, PCA helps to separate the intrinsic variability from the systematic effects in light curves.

In this tutorial we apply PCA to images of galaxies. We look at the principal components as images, at the variance that they keep and at the reconstructions from a few components.

\section*{Prerequisites and Learning Objectives}

\subsection*{Prerequisites}

This tutorial assumes the following.

\begin{itemize}
\tightlist
\item
  Linear algebra, with matrices, eigenvalues, eigenvectors and matrix decompositions
\item
  Python programming with NumPy arrays and matplotlib plots
\item
  Statistics, with covariance matrices, variance and correlation
\end{itemize}

\subsection*{Learning Objectives}

After this tutorial you can do the following.

\begin{itemize}
\tightlist
\item
  Implement PCA with the singular value decomposition (SVD) and relate the singular values to the eigenvalues of the covariance matrix
\item
  Display the principal components of images as eigengalaxies and interpret them
\item
  Measure the variance that a number of components keeps, relate it to the reconstruction error and choose the reduced dimension
\item
  Project data into a space of lower dimension and reconstruct them, and judge the trade-off between compression and accuracy
\item
  Apply PCA to real galaxy images and explain why it can separate coherent structure from noise
\end{itemize}

We begin by importing the libraries and setting the plot style.

\tutorialbarrier
\begin{codecell}

\end{codecell}

\section*{Understanding the Data: Galaxy Images from Modern Surveys}

In this tutorial we use a curated data set of galaxy cutouts. They come from the Sloan Digital Sky Survey (SDSS) images of Galaxy Zoo 2, as in \href{https://ui.adsabs.harvard.edu/abs/2026ApJS..282...25M/abstract}{Ma et al.~2026}. The galaxy cutouts have the following properties.

\begin{itemize}
\tightlist
\item
  \textbf{Subsampled.} We subsample the images to 64 × 64 pixels, so the computation is tractable and the morphological features remain.
\item
  \textbf{Normalized.} The pixel values lie between 0 and 1.
\end{itemize}

Each 64 × 64 image has 4,096 pixels, so each galaxy is a point in a space of 4,096 dimensions.

We load the images and look at their shape and pixel values.

\tutorialbarrier
\begin{codecell}
\begin{Verbatim}[commandchars=\\\{\},fontsize=\small]
\PYG{c+c1}{\PYGZsh{} Load the galaxy image dataset}
\PYG{n}{galaxy\PYGZus{}images} \PYG{o}{=}\FancyVerbBreakAfterBreak{} \PYG{n}{np}\PYG{o}{.}\FancyVerbBreakAfterBreak{}\PYG{n}{load}\PYG{p}{(}\FancyVerbBreakAfterBreak{}\PYG{l+s+s1}{\PYGZsq{}}\PYG{l+s+s1}{../data/dataset\PYGZus{}images\PYGZus{}galaxy.npy}\PYG{l+s+s1}{\PYGZsq{}}\PYG{p}{)}

\PYG{c+c1}{\PYGZsh{} Print the shape (galaxies, pixels) and the pixel value statistics}
\PYG{n+nb}{print}\PYG{p}{(}\FancyVerbBreakAfterBreak{}\PYG{l+s+s2}{\PYGZdq{}}\PYG{l+s+s2}{Shape of the galaxy images =}\PYG{l+s+s2}{\PYGZdq{}}\PYG{p}{,}\FancyVerbBreakAfterBreak{} \PYG{n}{galaxy\PYGZus{}images}\PYG{o}{.}\FancyVerbBreakAfterBreak{}\PYG{n}{shape}\PYG{p}{)}
\PYG{n+nb}{print}\PYG{p}{(}\FancyVerbBreakAfterBreak{}\PYG{l+s+sa}{f}\PYG{l+s+s2}{\PYGZdq{}}\PYG{l+s+s2}{Pixel values: minimum = }\PYG{l+s+si}{\PYGZob{}}\PYG{n}{np}\PYG{o}{.}\FancyVerbBreakAfterBreak{}\PYG{n}{min}\PYG{p}{(}\FancyVerbBreakAfterBreak{}\PYG{n}{galaxy\PYGZus{}images}\PYG{p}{)}\PYG{l+s+si}{:}\PYG{l+s+s2}{.3f}\PYG{l+s+si}{\PYGZcb{}}\FancyVerbBreakAfterBreak{}\PYG{l+s+s2}{, maximum = }\PYG{l+s+si}{\PYGZob{}}\PYG{n}{np}\PYG{o}{.}\FancyVerbBreakAfterBreak{}\PYG{n}{max}\PYG{p}{(}\FancyVerbBreakAfterBreak{}\PYG{n}{galaxy\PYGZus{}images}\PYG{p}{)}\PYG{l+s+si}{:}\PYG{l+s+s2}{.3f}\PYG{l+s+si}{\PYGZcb{}}\FancyVerbBreakAfterBreak{}\PYG{l+s+s2}{, }\PYG{l+s+s2}{\PYGZdq{}}
      \PYG{l+s+sa}{f}\PYG{l+s+s2}{\PYGZdq{}}\PYG{l+s+s2}{mean = }\PYG{l+s+si}{\PYGZob{}}\PYG{n}{np}\PYG{o}{.}\FancyVerbBreakAfterBreak{}\PYG{n}{mean}\PYG{p}{(}\FancyVerbBreakAfterBreak{}\PYG{n}{galaxy\PYGZus{}images}\PYG{p}{)}\PYG{l+s+si}{:}\PYG{l+s+s2}{.3f}\PYG{l+s+si}{\PYGZcb{}}\FancyVerbBreakAfterBreak{}\PYG{l+s+s2}{\PYGZdq{}}\PYG{p}{)}
\end{Verbatim}
\end{codecell}
\begin{Verbatim}[fontsize=\footnotesize, xleftmargin=5mm]
Shape of the galaxy images = (1500, 4096)
Pixel values: minimum = 0.000, maximum = 0.899, mean = 0.065
\end{Verbatim}

Before we reduce the dimension, we look at the raw data. The variety of the galaxies helps us to interpret the principal components that we extract. Galaxy morphology covers a wide range, from smooth elliptical galaxies with concentrated light profiles to grand-design spirals with extended disks and prominent arms.

The function below shows many galaxy images in a grid. We use it throughout this tutorial to compare the original images with their PCA reconstructions.

\tutorialbarrier
\begin{codecell}

\end{codecell}
\tutorialplot{\includegraphics[width=6.500in,height=0.75\textheight,keepaspectratio]{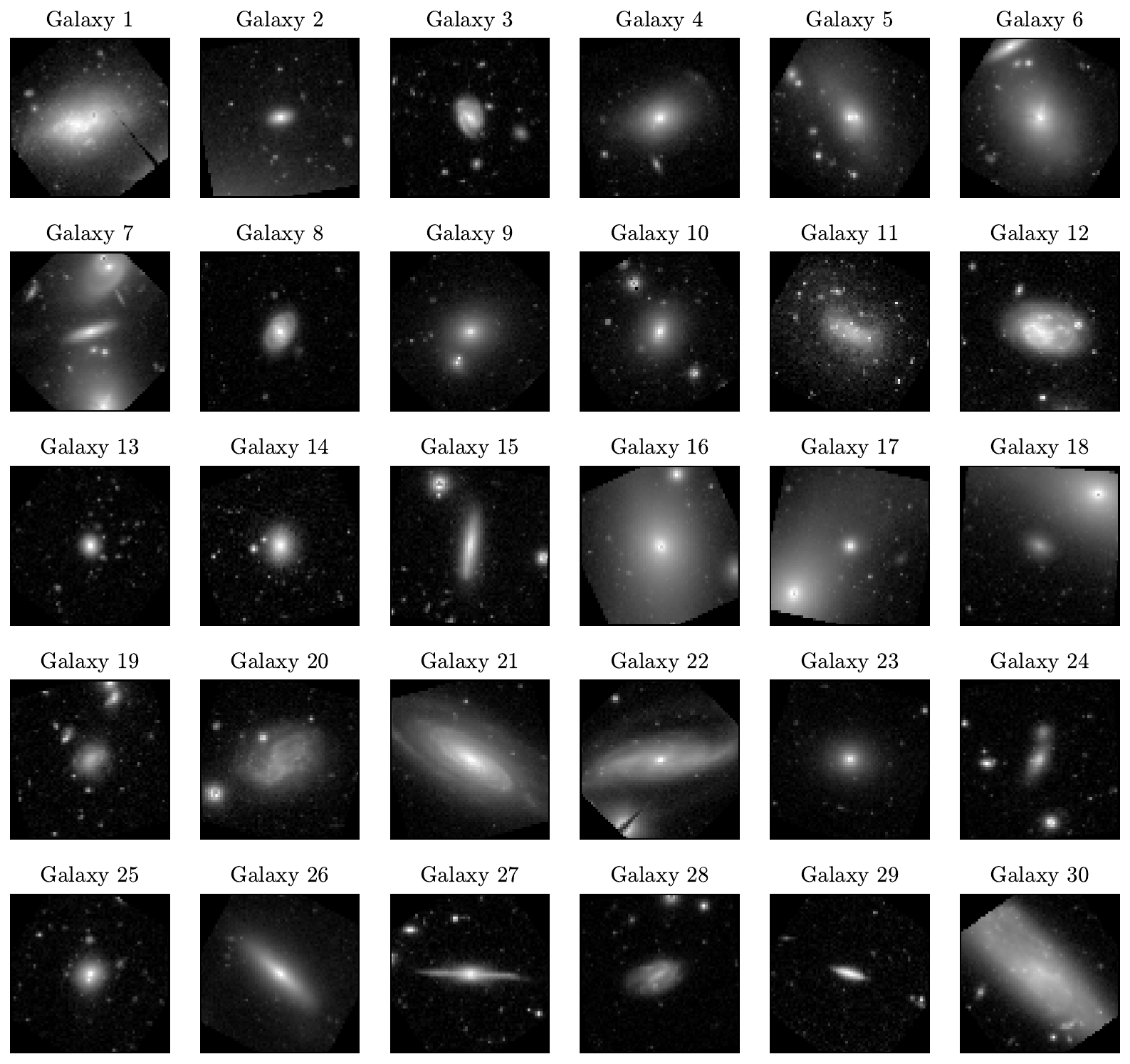}}%
  {\includegraphics[height=\tutorialheight]{tutorials/10/fig01.pdf}}

The galaxies differ in size, shape and brightness. Some cutouts hold bright neighbors, and some have black corners and tilted edges at the border of the image frame.

\section*{The Mathematics of PCA}

The galaxies share some patterns, such as a bright center. PCA finds such patterns. It transforms the galaxy images into a new coordinate system. The axes of this system, the principal components, are ordered by the amount of variance that they carry. This transformation shows the main structure of galaxy morphology and lets us reduce the dimension.

\subsection*{Step 1: Centering the Data}

The first step of PCA is to center the data by subtracting the mean. The data set has \(N\) images \(\{\mathbf{x}_i\}_{i=1}^N\), and each image is a row vector in \(\mathbb{R}^D\) with \(D = 4096\) pixels. We compute the mean galaxy image

\[\overline{\mathbf{x}} = \frac{1}{N}\sum_{i=1}^{N}\mathbf{x}_i.\]

This mean image is the ``average galaxy'' of the sample. Centering makes the principal components describe the variation around this average and not the mean signal itself. We subtract the mean from every image,

\[\mathbf{x}_i \leftarrow \mathbf{x}_i - \overline{\mathbf{x}}.\]

\subsection*{Step 2: Computing the Singular Value Decomposition}

PCA needs the eigenvectors of the covariance matrix \(\mathbf{S} = (1/N)\mathbf{X}^T\mathbf{X}\) of the centered data matrix \(\mathbf{X} \in \mathbb{R}^{N \times D}\). These eigenvectors are the principal components, and the variance along each one is its eigenvalue (Section 10.4). We do not form \(\mathbf{S}\), because it has \(4096 \times 4096\) entries and its eigenvectors need more operations than the SVD of \(\mathbf{X}\) when \(N < D\) (Section 10.8). Our data have \(N = 1500\) galaxies and \(D = 4096\) pixels, so \(N < D\). The SVD gives the same eigenvectors,

\[\mathbf{X} = \mathbf{U}\boldsymbol{\Sigma}\mathbf{V}^T.\]

Here \(\mathbf{U} \in \mathbb{R}^{N \times N}\) holds the left singular vectors, \(\boldsymbol{\Sigma} \in \mathbb{R}^{N \times D}\) is a rectangular diagonal matrix of singular values, and \(\mathbf{V} \in \mathbb{R}^{D \times D}\) holds the right singular vectors. The columns of \(\mathbf{V}\) are the principal components, the directions in pixel space along which the galaxy images vary most.

\subsection*{Step 3: Understanding the Variance Decomposition}

The singular values in \(\boldsymbol{\Sigma}\) give the eigenvalues of \(\mathbf{S}\). For component \(m\), the variance is the eigenvalue (Section 10.8)

\[\lambda_m = \frac{\sigma_m^2}{N},\]

where \(\sigma_m\) is the \(m\)-th singular value. This relation lets us measure how much of the total variation in galaxy morphology each component carries. It guides our choice of how many components to keep.

\subsection*{Step 4: Dimensionality Reduction}

To reduce the galaxy images from 4096 dimensions to \(M\) dimensions, we project them onto the first \(M\) principal components,

\[\mathbf{Z} = \mathbf{X}\mathbf{V}_M = \mathbf{U}_M\boldsymbol{\Sigma}_M,\]

where \(\mathbf{V}_M\) holds the first \(M\) columns of \(\mathbf{V}\). Each row of \(\mathbf{Z}\) is an \(M\)-dimensional representation of a galaxy that keeps the most important morphological information.

We now implement PCA for the galaxy images. We write two functions that use the efficient SVD of NumPy.

\begin{enumerate}
\def\labelenumi{\arabic{enumi}.}
\tightlist
\item
  The function \texttt{pca()} runs the complete PCA. It centers the data, computes the SVD and returns the projected data, the principal components and the SVD for later use. The projected data are the coefficients of each galaxy.
\item
  The function \texttt{svd2pca()} lets us try different numbers of components without a new computation of the expensive SVD.
\end{enumerate}

This design lets us see how the reconstruction quality changes with the number of components that we keep.

NumPy's \texttt{linalg.\allowbreak{}svd()} returns \(\mathbf{U}\), the singular values as a one-dimensional array and \(\mathbf{V}^T\). The principal components are the rows of \(\mathbf{V}^T\), which are the columns of \(\mathbf{V}\). The option \texttt{full\_\allowbreak{}matrices=\allowbreak{}False} returns only the first \(N\) rows of \(\mathbf{V}^T\), which is enough because \(N < D\) (Section 10.8). The other rows of \(\mathbf{V}^T\) belong to zero singular values.

\tutorialbarrier
\begin{codecell}

\end{codecell}

\section*{Eigengalaxies: Visualizing Principal Components}

When we apply PCA to galaxy images, we can display each principal component as an image, which we call an eigengalaxy. An eigengalaxy shows one mode of variation of the data set. Every galaxy of the sample is the mean galaxy plus a weighted sum of these basic images. Astronomers can look at these images directly.

A component has one weight per pixel, so we reshape it to 64 × 64. Each tile has its own gray scale, from the lowest weight (black) to the highest weight (white). The sign of a whole component is arbitrary. The reconstruction stays the same when the weights and the coefficients of a component both change sign.

PCA works well here because the pixel values of galaxies are correlated. Some distributions of light occur often and others are implausible, so the images do not fill the space of all possible pixel values. PCA finds the directions in pixel space along which the galaxies vary most.

We now compute the first 20 principal components of the galaxy data set and display them. The function \texttt{pca()} returns the coefficients of every galaxy and the principal components.

\tutorialbarrier
\begin{codecell}

\end{codecell}
\tutorialplot{\includegraphics[width=6.500in,height=0.75\textheight,keepaspectratio]{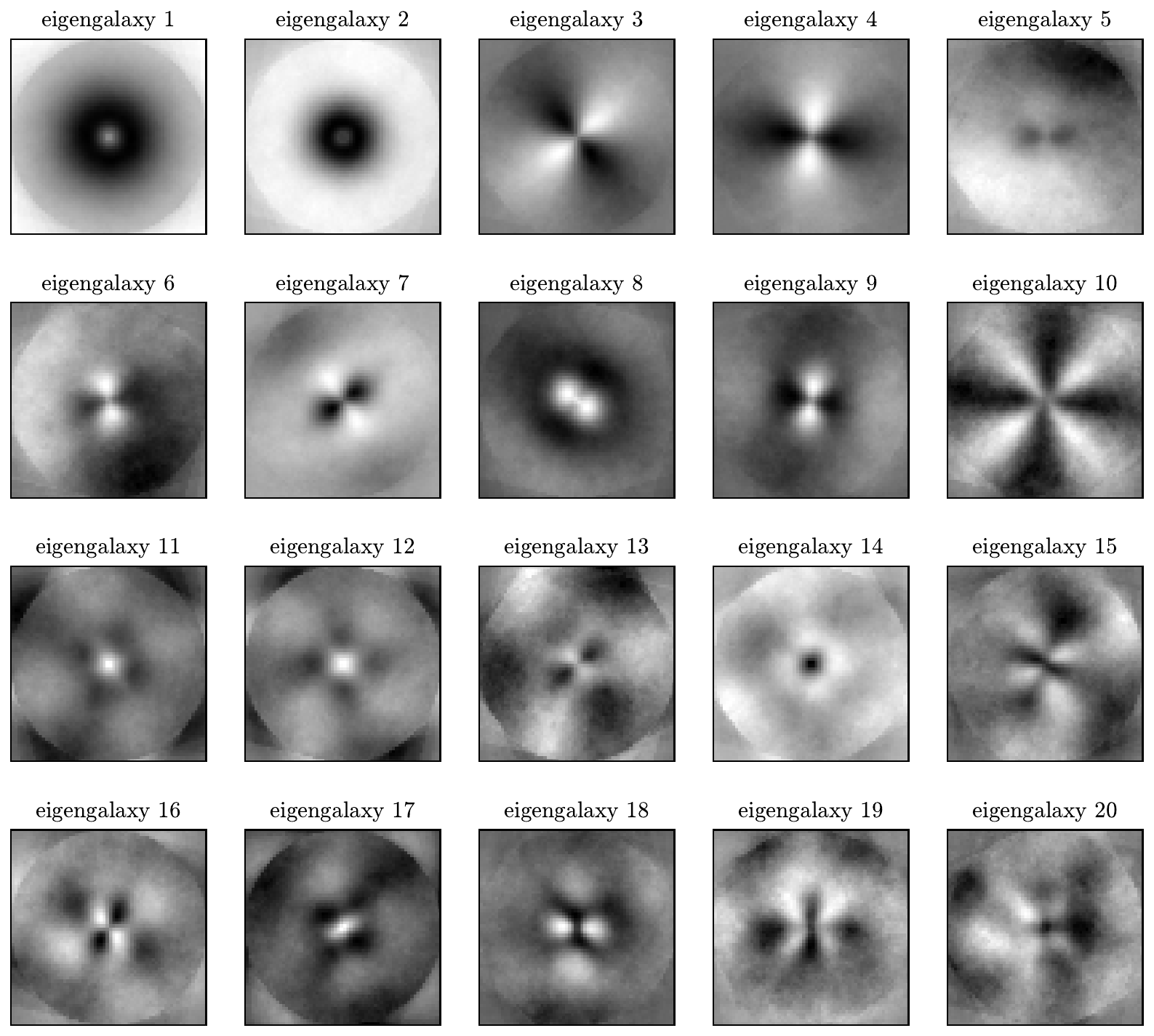}}%
  {\includegraphics[height=\tutorialheight]{tutorials/10/fig02.pdf}}

The eigengalaxies show a hierarchy in the features that they encode.

\textbf{First components (eigengalaxies 1 and 2).} These describe the most dominant variations of the sample. Eigengalaxy 1 is circularly symmetric and has weights of one sign in nearly every pixel, so its coefficient follows the overall brightness of a galaxy. Eigengalaxy 2 is also circularly symmetric, and it encodes the difference between concentrated and extended galaxies.

\textbf{Early components (eigengalaxies 3 to 10).} Eigengalaxies 3 and 4 show two quadrupole shapes, each with two bright and two dark lobes. Eigengalaxy 4 is eigengalaxy 3 turned by about 45 degrees. A mix of the two, with the right coefficients, points along any angle, so together they describe the elongation and the orientation of a galaxy. Eigengalaxies 5 to 10 are more complex. Eigengalaxies 7 to 9 have their structure near the center. Eigengalaxy 5 is a broad gradient, and eigengalaxy 10 has eight lobes over the whole image.

\textbf{Middle components (eigengalaxies 11 to 20).} The shapes become more complex, and they describe more subtle variations.

These shapes are the ways in which the galaxies differ from the mean galaxy. We read them as modes of variation, which need not match a physical property (Section 10.9).

\section*{Measuring Information Content: Variance Analysis}

To decide how many components to keep, we measure how much variance each component carries. A regression has a prediction error that guides the choice of its complexity. PCA has no target and so no prediction error, and the variance of the components guides us instead. In Step 3 the variance along component \(m\) is the eigenvalue \(\lambda_m = \sigma_m^2/N\), where \(N\) is the number of galaxy images.

We first check this relation on the galaxies. The check forms the covariance matrix of the 4,096 pixels once. It compares the five largest eigenvalues of this matrix with \(\sigma_m^2/N\).

\tutorialbarrier
\begin{codecell}

\end{codecell}
\begin{Verbatim}[fontsize=\footnotesize, xleftmargin=5mm]
Covariance matrix: [13.0422  2.0541  1.0255  0.8811  0.4966]
SVD, sigma^2 / N:  [13.0422  2.0541  1.0255  0.8811  0.4966]
\end{Verbatim}

The two lists agree, so the singular values give the eigenvalues without the covariance matrix. The fraction of the total variance that component \(m\) explains is

\[\frac{\lambda_m}{\sum_j \lambda_j}.\]

The running sum of these fractions is the explained variance ratio of the first \(M\) components (Section 10.7). Plotted against \(M\), it is the cumulative variance curve. The curve shows how many components keep a chosen share of the variance. The share that we need depends on the application, because a measurement of galaxy shapes needs more of the information than a fast classification.

We plot the curve with dots at 5, 20 and 50 components. We also print the number of components that reach 80, 90 and 95 percent of the variance.

\tutorialbarrier
\begin{codecell}

\end{codecell}
\tutorialplot{\includegraphics[width=4.576in,height=0.75\textheight,keepaspectratio]{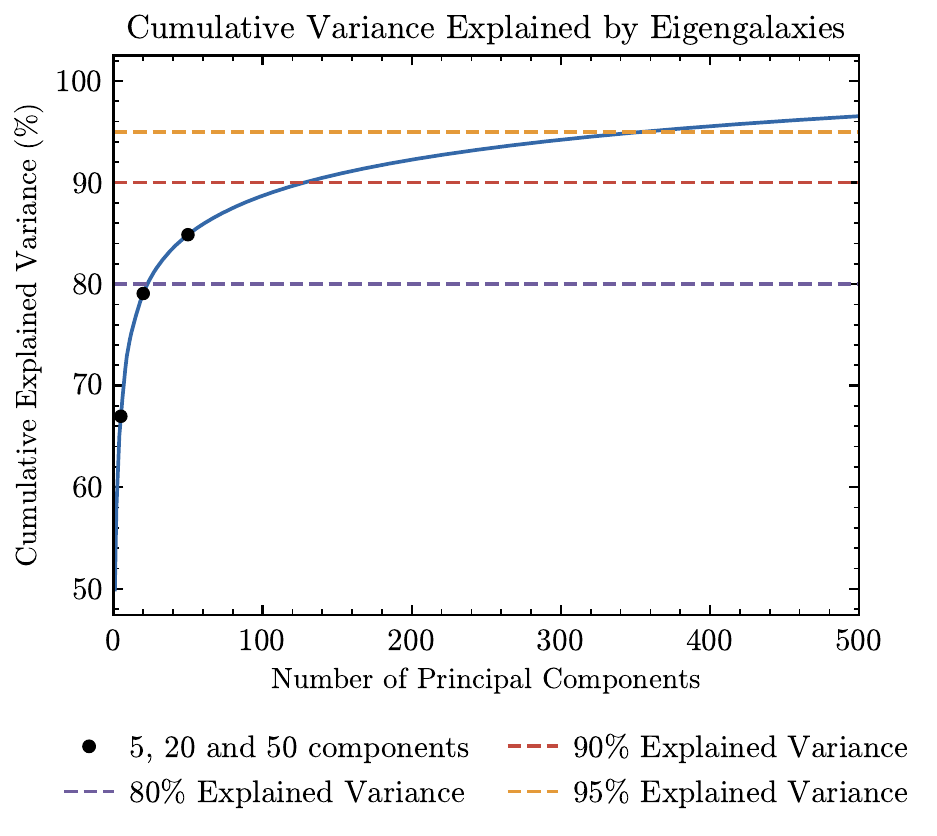}}%
  {\includegraphics[height=\tutorialheight]{tutorials/10/fig03.pdf}}
\begin{Verbatim}[fontsize=\footnotesize, xleftmargin=5mm]
80% of the variance needs 23 components (0.56% of the pixels)
90% of the variance needs 129 components (3.15% of the pixels)
95% of the variance needs 356 components (8.69% of the pixels)
\end{Verbatim}

The cumulative variance curve rises steeply and then slowly. The first component alone carries about half of the variance. With 23 principal components, which are 0.56\% of the 4,096 dimensions, we keep 80\% of the total variance of the galaxy sample. The levels of 90\% and 95\% need 129 and 356 components, many more than the 23 for 80\%. The plot shows the first 500 of the 1,500 components, and the curve reaches 100\% when it includes all of them.

The steep initial rise is typical of data with structure and strong pixel correlations. For random noise, the variance would spread much more evenly over the components. The remaining 20\% of the variance holds smaller details, the peculiarities of individual galaxies and noise.

The curve shows that the galaxy images are highly redundant. It does not show that a few physical processes shape the galaxies, because PCA only finds the directions of large variance (Section 10.9).

\section*{Reconstructing Galaxies from Principal Components}

The explained variance says how much variance we keep. The variance that we drop is the average squared reconstruction error (Section 10.1). The reconstruction is therefore the direct test of a dimensionality reduction. We project the galaxy images onto a reduced set of principal components and then reconstruct them. This shows what morphological information the reduction keeps and what it loses.

The reconstruction follows directly from the PCA decomposition. For a galaxy image \(\mathbf{x}\), we first subtract the mean, \(\mathbf{x}' = \mathbf{x} - \overline{\mathbf{x}}\). The projection onto \(M\) principal components gives the coefficients \(\mathbf{z} = \mathbf{x}'\mathbf{V}_M\), where \(\mathbf{V}_M\) holds the first \(M\) columns of the principal component matrix. The reconstruction is then (Section 10.1)

\[\widehat{\mathbf{x}} = \mathbf{z}\mathbf{V}_M^T + \overline{\mathbf{x}} = \mathbf{x}'\mathbf{V}_M\mathbf{V}_M^T + \overline{\mathbf{x}}.\]

The matrix product \(\mathbf{V}_M\mathbf{V}_M^T\) is a projection operator. It keeps only the part of the image that lies in the \(M\)-dimensional subspace of the kept principal components. When \(M\) reaches the full dimension, the projection becomes the identity and we recover the original image exactly.

We fit the PCA to all galaxies of the sample and reconstruct the same galaxies. We first compute the reconstructions from 50, 20 and 5 principal components (PCs). We then check their error and show them to see the trade-off.

\tutorialbarrier
\begin{codecell}
\begin{Verbatim}[commandchars=\\\{\},fontsize=\small]
\PYG{c+c1}{\PYGZsh{} Calculate the mean of the original galaxy images}
\PYG{n}{mean\PYGZus{}image} \PYG{o}{=}\FancyVerbBreakAfterBreak{} \PYG{n}{np}\PYG{o}{.}\FancyVerbBreakAfterBreak{}\PYG{n}{mean}\PYG{p}{(}\FancyVerbBreakAfterBreak{}\PYG{n}{galaxy\PYGZus{}images}\PYG{p}{,}\FancyVerbBreakAfterBreak{} \PYG{n}{axis}\PYG{o}{=}\FancyVerbBreakAfterBreak{}\PYG{l+m+mi}{0}\PYG{p}{)}

\PYG{c+c1}{\PYGZsh{} Reconstruct with 50, 20 and 5 principal components}
\PYG{n}{reconstructed} \PYG{o}{=}\FancyVerbBreakAfterBreak{} \PYG{p}{\PYGZob{}}\PYG{p}{\PYGZcb{}}\FancyVerbBreakAfterBreak{}
\PYG{k}{for} \PYG{n}{n\PYGZus{}pc} \PYG{o+ow}{in} \PYG{p}{(}\FancyVerbBreakAfterBreak{}\PYG{l+m+mi}{50}\PYG{p}{,}\FancyVerbBreakAfterBreak{} \PYG{l+m+mi}{20}\PYG{p}{,}\FancyVerbBreakAfterBreak{} \PYG{l+m+mi}{5}\PYG{p}{)}\PYG{p}{:}
    \PYG{n}{galaxy\PYGZus{}Z}\PYG{p}{,}\FancyVerbBreakAfterBreak{} \PYG{n}{galaxy\PYGZus{}P} \PYG{o}{=}\FancyVerbBreakAfterBreak{} \PYG{n}{svd2pca}\PYG{p}{(}\FancyVerbBreakAfterBreak{}\PYG{n}{U}\PYG{p}{,}\FancyVerbBreakAfterBreak{} \PYG{n}{Sig}\PYG{p}{,}\FancyVerbBreakAfterBreak{} \PYG{n}{Vt}\PYG{p}{,}\FancyVerbBreakAfterBreak{} \PYG{n}{n\PYGZus{}pc}\PYG{o}{=}\FancyVerbBreakAfterBreak{}\PYG{n}{n\PYGZus{}pc}\PYG{p}{)}
    \PYG{n}{reconstructed}\PYG{p}{[}\FancyVerbBreakAfterBreak{}\PYG{n}{n\PYGZus{}pc}\PYG{p}{]} \PYG{o}{=}\FancyVerbBreakAfterBreak{} \PYG{n}{np}\PYG{o}{.}\FancyVerbBreakAfterBreak{}\PYG{n}{dot}\PYG{p}{(}\FancyVerbBreakAfterBreak{}\PYG{n}{galaxy\PYGZus{}Z}\PYG{p}{,}\FancyVerbBreakAfterBreak{} \PYG{n}{galaxy\PYGZus{}P}\PYG{p}{)} \PYG{o}{+}\FancyVerbBreakAfterBreak{} \PYG{n}{mean\PYGZus{}image}  \PYG{c+c1}{\PYGZsh{} coefficients times components plus the mean}
\end{Verbatim}
\end{codecell}

We now check that the dropped variance equals the average squared reconstruction error. The error of one galaxy is the squared distance between the galaxy and its reconstruction, summed over the 4,096 pixels. We average it over the 1,500 galaxies and compare it with the sum of the eigenvalues that we left out.

\tutorialbarrier
\begin{codecell}
\begin{Verbatim}[commandchars=\\\{\},fontsize=\small]
\PYG{c+c1}{\PYGZsh{} Compare the average squared error with the sum of the dropped eigenvalues}
\PYG{k}{for} \PYG{n}{n\PYGZus{}pc} \PYG{o+ow}{in} \PYG{p}{(}\FancyVerbBreakAfterBreak{}\PYG{l+m+mi}{5}\PYG{p}{,}\FancyVerbBreakAfterBreak{} \PYG{l+m+mi}{20}\PYG{p}{,}\FancyVerbBreakAfterBreak{} \PYG{l+m+mi}{50}\PYG{p}{)}\PYG{p}{:}
    \PYG{n}{error} \PYG{o}{=}\FancyVerbBreakAfterBreak{} \PYG{n}{np}\PYG{o}{.}\FancyVerbBreakAfterBreak{}\PYG{n}{mean}\PYG{p}{(}\FancyVerbBreakAfterBreak{}\PYG{n}{np}\PYG{o}{.}\FancyVerbBreakAfterBreak{}\PYG{n}{sum}\PYG{p}{(}\FancyVerbBreakAfterBreak{}\PYG{p}{(}\FancyVerbBreakAfterBreak{}\PYG{n}{galaxy\PYGZus{}images} \PYG{o}{\PYGZhy{}} \PYG{n}{reconstructed}\PYG{p}{[}\FancyVerbBreakAfterBreak{}\PYG{n}{n\PYGZus{}pc}\PYG{p}{]}\PYG{p}{)}\PYG{o}{*}\PYG{o}{*}\PYG{l+m+mi}{2}\PYG{p}{,}\FancyVerbBreakAfterBreak{} \PYG{n}{axis}\PYG{o}{=}\FancyVerbBreakAfterBreak{}\PYG{l+m+mi}{1}\PYG{p}{)}\PYG{p}{)}
    \PYG{n+nb}{print}\PYG{p}{(}\FancyVerbBreakAfterBreak{}\PYG{l+s+sa}{f}\PYG{l+s+s2}{\PYGZdq{}}\PYG{l+s+si}{\PYGZob{}}\PYG{n}{n\PYGZus{}pc}\PYG{l+s+si}{\PYGZcb{}}\FancyVerbBreakAfterBreak{}\PYG{l+s+s2}{ components: error }\PYG{l+s+si}{\PYGZob{}}\PYG{n}{error}\PYG{l+s+si}{:}\PYG{l+s+s2}{.2f}\PYG{l+s+si}{\PYGZcb{}}\FancyVerbBreakAfterBreak{}\PYG{l+s+s2}{, dropped variance }\PYG{l+s+si}{\PYGZob{}}\PYG{n}{np}\PYG{o}{.}\FancyVerbBreakAfterBreak{}\PYG{n}{sum}\PYG{p}{(}\FancyVerbBreakAfterBreak{}\PYG{n}{eigenvalues}\PYG{p}{[}\FancyVerbBreakAfterBreak{}\PYG{n}{n\PYGZus{}pc}\PYG{p}{:}\PYG{p}{]}\PYG{p}{)}\PYG{l+s+si}{:}\PYG{l+s+s2}{.2f}\PYG{l+s+si}{\PYGZcb{}}\FancyVerbBreakAfterBreak{}\PYG{l+s+s2}{\PYGZdq{}}\PYG{p}{)}
\end{Verbatim}
\end{codecell}
\begin{Verbatim}[fontsize=\footnotesize, xleftmargin=5mm]
5 components: error 8.63, dropped variance 8.63
20 components: error 5.47, dropped variance 5.47
50 components: error 3.96, dropped variance 3.96
\end{Verbatim}

The two numbers agree for each number of components, so the explained variance and the reconstruction error describe the same loss. After the reduction, a galaxy is described by its coefficients, which are its coordinates along the principal components. We take Galaxy 26 and print its first five coefficients. We then add the weighted eigengalaxies to the mean galaxy for 1, 3, 5 and 20 components (Section 10.1).

\tutorialbarrier
\begin{codecell}
\begin{Verbatim}[commandchars=\\\{\},fontsize=\small]
\PYG{c+c1}{\PYGZsh{} Print the first five coefficients of Galaxy 26, the row 25 of the arrays}
\PYG{n+nb}{print}\PYG{p}{(}\FancyVerbBreakAfterBreak{}\PYG{l+s+s2}{\PYGZdq{}}\PYG{l+s+s2}{Galaxy 26, coefficients 1 to 5:}\PYG{l+s+s2}{\PYGZdq{}}\PYG{p}{,}\FancyVerbBreakAfterBreak{} \PYG{n}{np}\PYG{o}{.}\FancyVerbBreakAfterBreak{}\PYG{n}{round}\PYG{p}{(}\FancyVerbBreakAfterBreak{}\PYG{n}{Z}\PYG{p}{[}\FancyVerbBreakAfterBreak{}\PYG{l+m+mi}{25}\PYG{p}{,}\FancyVerbBreakAfterBreak{} \PYG{p}{:}\PYG{l+m+mi}{5}\PYG{p}{]}\PYG{p}{,}\FancyVerbBreakAfterBreak{} \PYG{l+m+mi}{2}\PYG{p}{)}\PYG{p}{)}

\PYG{c+c1}{\PYGZsh{} Add the weighted eigengalaxies to the mean galaxy}
\PYG{n}{build\PYGZus{}up} \PYG{o}{=}\FancyVerbBreakAfterBreak{} \PYG{p}{[}\FancyVerbBreakAfterBreak{}\PYG{n}{mean\PYGZus{}image} \PYG{o}{+}\FancyVerbBreakAfterBreak{} \PYG{n}{Z}\PYG{p}{[}\FancyVerbBreakAfterBreak{}\PYG{l+m+mi}{25}\PYG{p}{,}\FancyVerbBreakAfterBreak{} \PYG{p}{:}\PYG{n}{m}\PYG{p}{]} \PYG{o}{@} \PYG{n}{P}\PYG{p}{[}\FancyVerbBreakAfterBreak{}\PYG{p}{:}\PYG{n}{m}\PYG{p}{]} \PYG{k}{for} \PYG{n}{m} \PYG{o+ow}{in} \PYG{p}{(}\FancyVerbBreakAfterBreak{}\PYG{l+m+mi}{1}\PYG{p}{,}\FancyVerbBreakAfterBreak{} \PYG{l+m+mi}{3}\PYG{p}{,}\FancyVerbBreakAfterBreak{} \PYG{l+m+mi}{5}\PYG{p}{,}\FancyVerbBreakAfterBreak{} \PYG{l+m+mi}{20}\PYG{p}{)}\PYG{p}{]}
\end{Verbatim}
\end{codecell}
\begin{Verbatim}[fontsize=\footnotesize, xleftmargin=5mm]
Galaxy 26, coefficients 1 to 5: [-2.95 -0.96 -2.59 -0.57 -0.01]
\end{Verbatim}

We show Galaxy 26, the mean galaxy and the four partial sums.

\tutorialbarrier
\begin{codecell}
\begin{Verbatim}[commandchars=\\\{\},fontsize=\small]
\PYG{c+c1}{\PYGZsh{} Show the galaxy, the mean galaxy and the partial sums}
\PYG{n}{images} \PYG{o}{=}\FancyVerbBreakAfterBreak{} \PYG{p}{[}\FancyVerbBreakAfterBreak{}\PYG{n}{galaxy\PYGZus{}images}\PYG{p}{[}\FancyVerbBreakAfterBreak{}\PYG{l+m+mi}{25}\PYG{p}{]}\PYG{p}{,}\FancyVerbBreakAfterBreak{} \PYG{n}{mean\PYGZus{}image}\PYG{p}{]} \PYG{o}{+}\FancyVerbBreakAfterBreak{} \PYG{n}{build\PYGZus{}up}
\PYG{n}{titles} \PYG{o}{=}\FancyVerbBreakAfterBreak{} \PYG{p}{[}\FancyVerbBreakAfterBreak{}\PYG{l+s+s2}{\PYGZdq{}}\PYG{l+s+s2}{Galaxy 26}\PYG{l+s+s2}{\PYGZdq{}}\PYG{p}{,}\FancyVerbBreakAfterBreak{} \PYG{l+s+s2}{\PYGZdq{}}\PYG{l+s+s2}{Mean galaxy}\PYG{l+s+s2}{\PYGZdq{}}\PYG{p}{,}\FancyVerbBreakAfterBreak{} \PYG{l+s+s2}{\PYGZdq{}}\PYG{l+s+s2}{Mean + 1 PC}\PYG{l+s+s2}{\PYGZdq{}}\PYG{p}{,}\FancyVerbBreakAfterBreak{} \PYG{l+s+s2}{\PYGZdq{}}\PYG{l+s+s2}{Mean + 3 PCs}\PYG{l+s+s2}{\PYGZdq{}}\PYG{p}{,}\FancyVerbBreakAfterBreak{} \PYG{l+s+s2}{\PYGZdq{}}\PYG{l+s+s2}{Mean + 5 PCs}\PYG{l+s+s2}{\PYGZdq{}}\PYG{p}{,}\FancyVerbBreakAfterBreak{} \PYG{l+s+s2}{\PYGZdq{}}\PYG{l+s+s2}{Mean + 20 PCs}\PYG{l+s+s2}{\PYGZdq{}}\PYG{p}{]}
\PYG{n}{plot\PYGZus{}gallery}\PYG{p}{(}\FancyVerbBreakAfterBreak{}\PYG{n}{images}\PYG{p}{,}\FancyVerbBreakAfterBreak{} \PYG{n}{titles}\PYG{p}{,}\FancyVerbBreakAfterBreak{} \PYG{l+m+mi}{64}\PYG{p}{,}\FancyVerbBreakAfterBreak{} \PYG{l+m+mi}{64}\PYG{p}{,}\FancyVerbBreakAfterBreak{} \PYG{l+m+mi}{1}\PYG{p}{,}\FancyVerbBreakAfterBreak{} \PYG{l+m+mi}{6}\PYG{p}{)}
\end{Verbatim}
\end{codecell}
\tutorialplot{\includegraphics[width=6.500in,height=0.75\textheight,keepaspectratio]{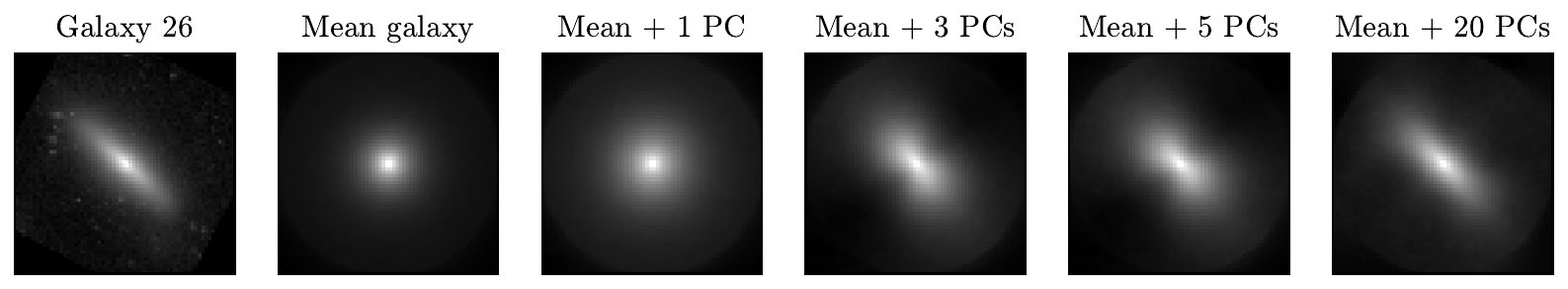}}%
  {\includegraphics[height=\tutorialheight]{tutorials/10/fig04.pdf}}

The mean galaxy is a round blob, and the first component keeps it round. The elongation appears with the third component, whose coefficient of \(-2.59\) is large. More components make the shape more like the galaxy. The first coefficient is negative, because the weights of eigengalaxy 1 are negative and Galaxy 26 has more light than the mean galaxy.

We now show the first 30 galaxies, and then their reconstructions from 50, 20 and 5 principal components. We look at the elongated galaxies 15, 21, 26 and 30, and at galaxies with neighbors, such as 7, 17 and 18.

\tutorialbarrier
\begin{codecell}

\end{codecell}
\tutorialplot{\includegraphics[width=6.500in,height=0.75\textheight,keepaspectratio]{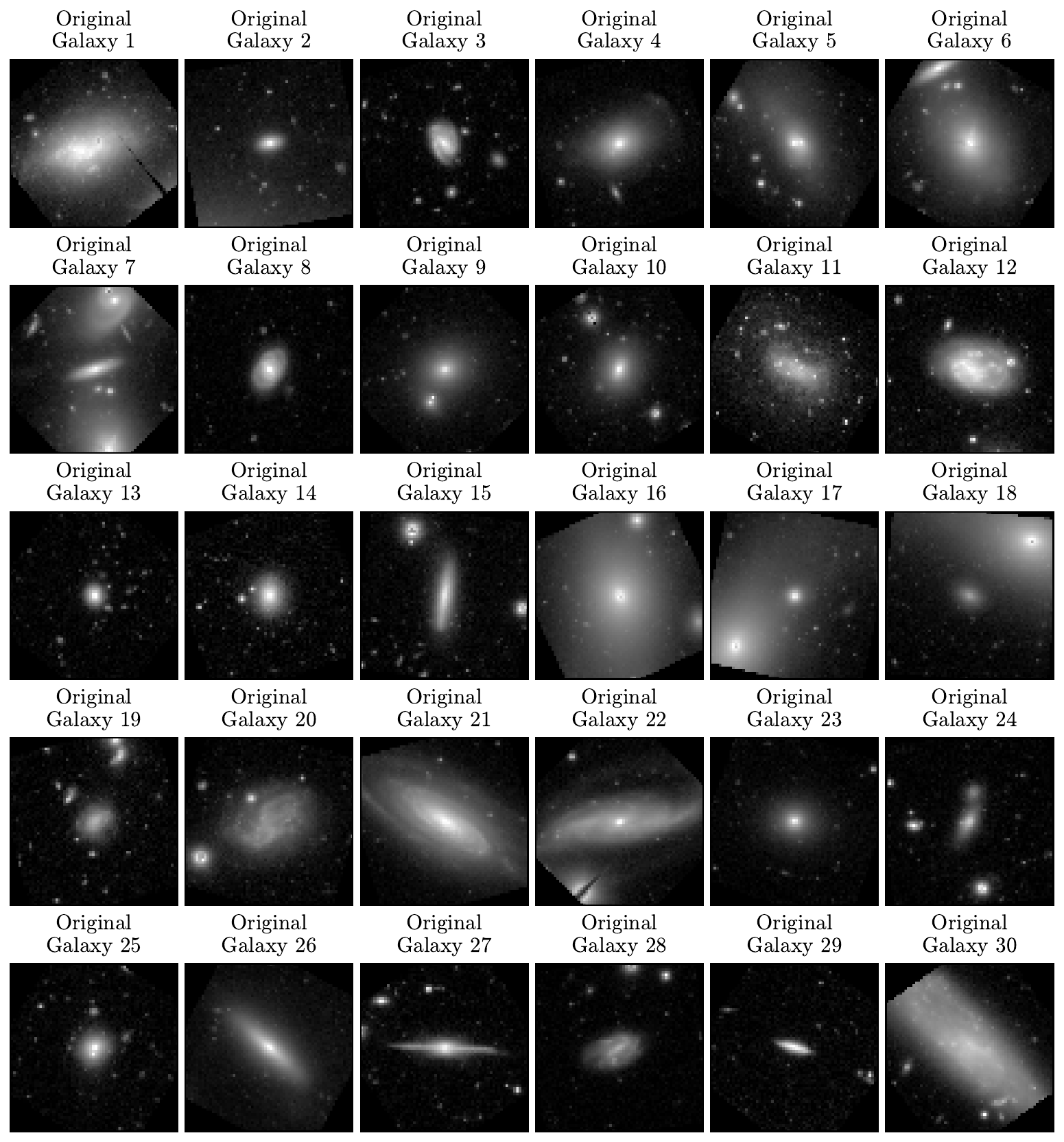}}%
  {\includegraphics[height=\tutorialheight]{tutorials/10/fig05.pdf}}
\tutorialplot{\includegraphics[width=6.500in,height=0.75\textheight,keepaspectratio]{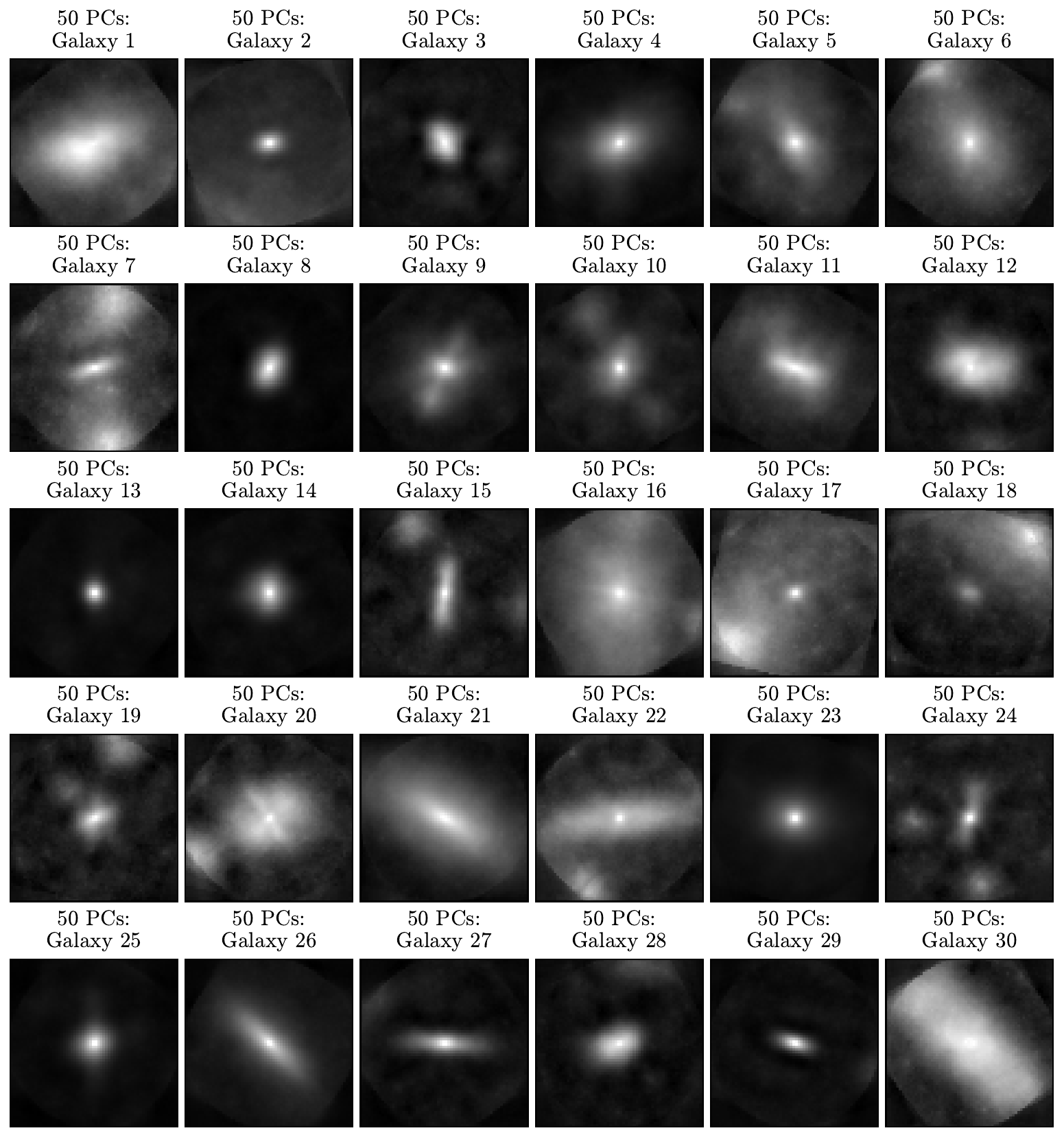}}%
  {\includegraphics[height=\tutorialheight]{tutorials/10/fig06.pdf}}
\tutorialplot{\includegraphics[width=6.500in,height=0.75\textheight,keepaspectratio]{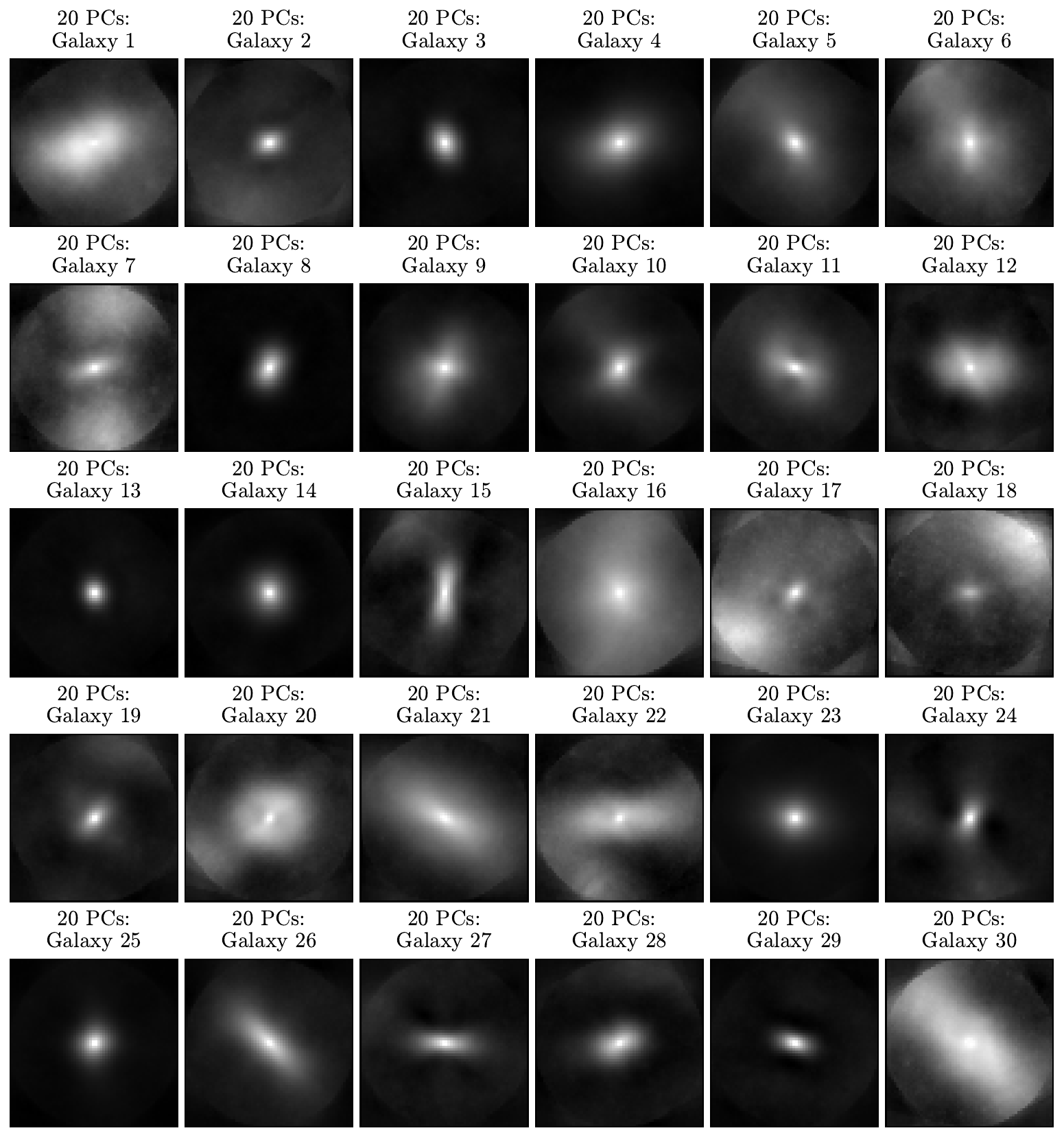}}%
  {\includegraphics[height=\tutorialheight]{tutorials/10/fig07.pdf}}
\tutorialplot{\includegraphics[width=6.500in,height=0.75\textheight,keepaspectratio]{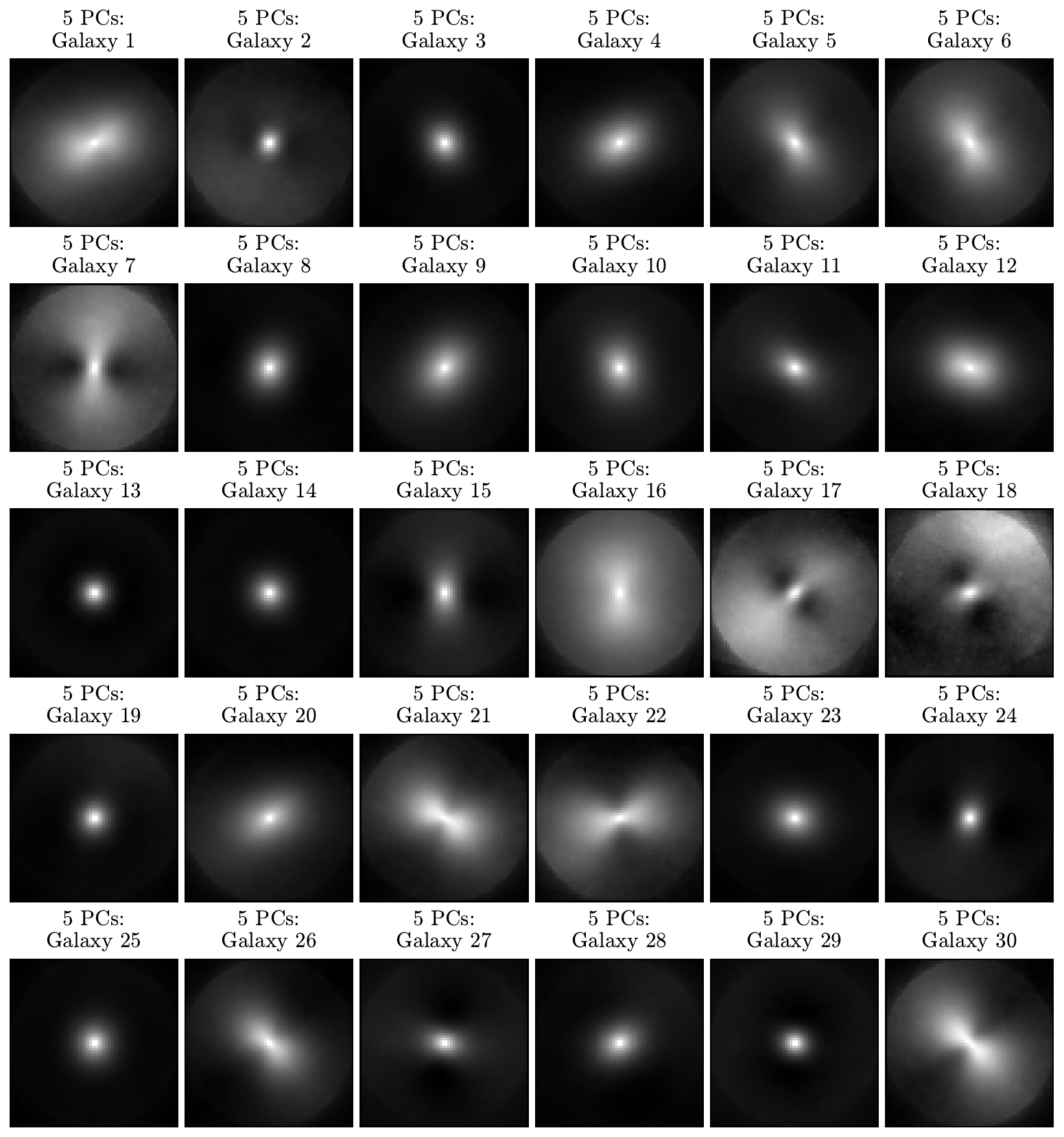}}%
  {\includegraphics[height=\tutorialheight]{tutorials/10/fig08.pdf}}

The comparison shows how the images change when we reduce the dimension a lot.

\textbf{50 principal components.} With 50 components, which carry about 85\% of the variance (the third dot), the reconstructed galaxies keep their main shapes. Elongated galaxies, such as Galaxy 15 and 21, keep their elongated profiles. Round galaxies keep their smooth light distributions.

\textbf{20 principal components.} With 20 components, which carry about 80\% of the variance (the second dot), most galaxies keep their size, elongation and concentration. Round galaxies appear as smooth, centrally concentrated blobs, and elongated galaxies such as Galaxy 26 keep their shape. The fine structure is lost. Cutouts with bright neighbors or frame edges reconstruct poorly, because these features follow no common pattern. Galaxy 7, 17 and 18 show bright patches from their neighbors.

\textbf{5 principal components.} With 5 components, which carry about 67\% of the variance (the first dot), most galaxies reduce to simple blobs. The elongated galaxies 21, 22, 26 and 30 show bow-tie shapes, which come from the quadrupole eigengalaxies 3 and 4. Galaxy 17 and 18 show a pinwheel that comes from the light of their bright neighbors, and the round center of Galaxy 18 looks elongated. This strong compression keeps only the overall brightness, the orientation and the rough shape.

The loss of information goes from the fine details to the basic structure. The early components carry the most variance and keep the large-scale shape. The number of components to keep is a choice between compression and accuracy.

\section*{PCA as a Denoising Filter}

The reconstructions look smoother than the originals. We now ask whether the dropped components hold noise.

PCA tends to separate coherent signal from random noise, because it keeps the directions of large variance. This property helps in astronomical imaging, where several noise sources act, such as photon noise, detector readout noise and atmospheric effects.

Consider how different types of features contribute to the principal components.

\begin{itemize}
\tightlist
\item
  \textbf{Coherent features.} Morphological structures that appear in many galaxies, such as bulges and disks, create correlated variations in the pixel values. These correlations form the early principal components, which describe these repeating structures efficiently.
\item
  \textbf{Random noise.} Pixel-level noise has, by definition, no correlation between images. Each noise realization is unique. Its variance spreads over many later components and does not concentrate in the early ones.
\item
  \textbf{Transient features.} Satellite galaxies, foreground stars and cosmic ray hits appear at random positions in different images. Unlike the central galaxy, which appears in each frame, these features do not create strong correlations in the covariance matrix. Consequently, they tend to appear in higher-order components.
\end{itemize}

The reconstructions agree with the last point for faint neighbors. They are weak at 20 components, and some return at 50 components (Galaxy 10, 19 and 24). Bright neighbors are an exception, because Galaxy 7, 17 and 18 show them already at 20 components.

We can therefore keep only the dominant principal components and so remove part of the noise. The coherent signal of the galaxy remains. The number of components to keep is a trade-off between noise suppression and signal preservation.

How well PCA removes noise depends on the signal-to-noise ratio of the data. Well-exposed images of bright galaxies show a clear separation between signal-dominated and noise-dominated components. For fainter galaxies near the detection limit, this separation is less distinct, and we must choose the truncation point with care.

\section*{Storage Efficiency and Practical Considerations}

Besides the removal of part of the noise, the reduction saves storage. We now count the storage savings of the principal component representation.

For our galaxy data set we have the following.

\begin{itemize}
\tightlist
\item
  Each galaxy needs \(64 \times 64 = 4{,}096\) pixel values.
\item
  The whole data set needs \(N \times 4{,}096\) values for \(N\) galaxies.
\end{itemize}

With \(M\) principal components, we need to store the following.

\begin{enumerate}
\def\labelenumi{\arabic{enumi}.}
\tightlist
\item
  The principal component basis \(\mathbf{V}_M^T \in \mathbb{R}^{M \times 4096}\)
\item
  The projection coefficients \(\mathbf{Z} \in \mathbb{R}^{N \times M}\) for all galaxies
\item
  The mean galaxy \(\overline{\mathbf{x}} \in \mathbb{R}^{4096}\)
\end{enumerate}

The total storage is \(4096 M + N M + 4096 = 4096 (M + 1) + N M\) values. For a survey with 10,000 galaxies, more than our sample, we find the following.

\begin{itemize}
\tightlist
\item
  The original images need \(10{,}000 \times 4{,}096 = 40{,}960{,}000\) values.
\item
  With 50 components, PCA needs \(4{,}096 \times 51 + 10{,}000 \times 50 = 708{,}896\) values, which is a reduction of 98.3\%.
\item
  With 20 components, PCA needs \(4{,}096 \times 21 + 10{,}000 \times 20 = 286{,}016\) values, which is a reduction of 99.3\%.
\end{itemize}

This compression helps even more for larger surveys, because the basis and the mean galaxy take a fixed number of values. For very many galaxies, the original storage divided by the PCA storage approaches \(4096/M\), which is 82 for 50 components. The Vera Rubin Observatory Legacy Survey of Space and Time (LSST) will observe many more galaxies than any survey before.

PCA also makes the computation efficient. Many astronomical studies, such as the clustering of galaxies by morphology, need the similarity between galaxies. In the original 4,096-dimensional space, one similarity needs 4,096 operations. In the 50-dimensional PCA space, it needs only 50 operations once we have projected each galaxy onto the components. This is nearly two orders of magnitude less.

The compressed representation also helps machine learning. A classifier or a regression model trained on the 4,096 pixels often overfits because of the curse of dimensionality. The PCA coefficients form a smaller set of features that keeps the morphological variation. A model on these features has fewer weights, and this can lower the overfitting.

\section*{Limits of PCA}

PCA is a good method to analyze galaxy morphology, but we must know its limits to apply it well.

\textbf{Linearity constraint.} PCA finds only linear relationships in the data. It assumes that we can write the galaxies as linear combinations of the principal components. This works well for many morphological variations. Linear scaling and addition of components can describe elongation, size and brightness. Non-linear transformations such as rotations or complex changes of structure, however, may need many components to approximate what non-linear methods can describe more efficiently.

\textbf{Orthogonality requirement.} The principal components must be orthogonal. This can lead to components that mix several physical features (Section 10.9). It can also spread one feature over several components. Elongation is an example. Each galaxy has its own angle, so elongation uses the two orthogonal eigengalaxies 3 and 4, and no single component describes it. Both effects make a physical interpretation hard.

\textbf{Sensitivity to preprocessing.} The PCA results depend on the choices of the data preprocessing.

\begin{itemize}
\tightlist
\item
  \textbf{Centering.} The way we center the galaxies in their cutouts affects which spatial modes appear in the early components.
\item
  \textbf{Scaling.} The relative scaling of the pixel values affects which features dominate the variance.
\item
  \textbf{Background subtraction.} Residual background gradients can dominate the early components if we do not remove them well.
\end{itemize}

The goals of the analysis must guide these choices. For morphological studies, careful centering of the galaxies and careful background subtraction are necessary. For detection or classification, some preprocessing choices can remove useful information.

\textbf{Sensitivity to outliers.} PCA maximizes variance, so a few extreme images can shape the components (Section 10.9). Bright neighbors and frame edges are not part of the galaxies, yet they carry variance. The bright patches in the reconstructions of Galaxy 7, 17 and 18 are an example.

\textbf{Several populations.} Galaxies fall into types such as ellipticals and disks. PCA gives one set of components for all of them (Section 10.9). These components follow the largest total variance, which can differ from the main axis of each type.

\textbf{Global nature of components.} Each principal component is a global pattern over the whole image. Local features such as individual star-forming regions or tidal streams may be described poorly, unless they occur in the same way in many galaxies.

\section*{Summary}

In this tutorial we applied PCA to galaxy images. We went from the pixel space to a compact representation that keeps most of the variation. Our main lessons follow.

\begin{enumerate}
\def\labelenumi{\arabic{enumi}.}
\tightlist
\item
  \textbf{Few components carry most of the variance.} The galaxy images are highly redundant, so a few components keep most of the variation.
\item
  \textbf{Components are ordered by variance.} The early components encode large-scale features of the images. Later components describe smaller details and noise.
\item
  \textbf{The reduced dimension is a choice.} PCA has no target and no prediction error. The variance of the dropped components equals the average squared reconstruction error, so it shows the trade-off between compression and accuracy.
\item
  \textbf{Denoising and storage.} PCA tends to separate coherent signal from random noise, and the coefficients need far less storage than the pixels.
\item
  \textbf{Components can be viewed.} For images, the principal components are images that we can look at, although they need not match a physical property.
\end{enumerate}

PCA gives a compact description of high-dimensional data and shows which patterns dominate it. It has limits. It is linear, its components are orthogonal and global, and its results depend on the preprocessing. Outliers can shape its components, and it gives one set of components for all groups of objects. PCA is still a good first step in the exploration of a new data set.

\tutorialchapter{11a}{K-means and Gaussian Mixtures for Stellar Populations}
\noindent\textit{Companion material for Chapter 11}

\section*{Introduction}

Stars keep the chemical composition of the gas from which they formed, and supernovae change that gas. Massive stars explode soon after they form and release magnesium and some iron. Type Ia supernovae happen much later and release a large amount of iron. They are explosions of white dwarfs in binary systems, where the white dwarf gains mass from a companion. We describe the composition of a star by two logarithmic abundance ratios.

The ratio {[}X/H{]} is the base-10 logarithm of the number ratio of the element X to hydrogen, relative to the Sun. The ratio {[}Mg/Fe{]} is {[}Mg/H{]} minus {[}Fe/H{]}. A star with {[}Fe/H{]} = -1 has an iron to hydrogen ratio ten times lower than the Sun. A star with a low {[}Fe/H{]} is metal-poor, and a star with a high {[}Fe/H{]} is metal-rich.

If stars form in bursts, the late explosions add iron between the bursts, so {[}Mg/Fe{]} of the gas falls. The next burst forms stars from this gas, with a higher {[}Fe/H{]} and a lower {[}Mg/Fe{]} than the last burst. Each burst can therefore leave a group of stars in the plane of {[}Fe/H{]} and {[}Mg/Fe{]}. The gap between two groups can mark a quiet period. \href{https://ui.adsabs.harvard.edu/abs/2025OJAp....8E...7T/abstract}{Ting \& Ji (2025)} used mixture models to measure bursty star formation in dwarf galaxies.

Finding groups by eye is subjective, in particular when the groups overlap, so we use clustering algorithms (Chapter 11). K-means puts each star in the cluster of the nearest of \(K\) centers. A Gaussian mixture model (GMM) describes the stars as a weighted sum of \(K\) Gaussian distributions. It gives each star a probability of belonging to each group. The expectation-maximization (EM) algorithm fits both models. Both need the number of groups \(K\).

We build K-means and a GMM from scratch and apply them to an APOGEE sample of stars in the Sculptor dwarf galaxy. We then choose the number of groups with the elbow of the K-means inertia. We also use two information criteria, AIC and BIC, which penalize the number of parameters of the mixture.

\section*{Prerequisites and Learning Objectives}

\subsection*{Prerequisites}

This tutorial assumes the following.

\begin{itemize}
\tightlist
\item
  Working knowledge of Python, NumPy and matplotlib
\item
  Covariance matrices (Chapter 3)
\item
  The multivariate Gaussian distribution and Bayes' theorem (Chapter 2)
\item
  Maximum likelihood estimation (Chapter 4)
\end{itemize}

\subsection*{Learning Objectives}

After this tutorial you can do the following.

\begin{itemize}
\tightlist
\item
  Implement K-means and a Gaussian mixture model with the EM algorithm from first principles
\item
  Compare random and K-means++ initialization over many starts
\item
  Follow the convergence of both algorithms over the iterations
\item
  Choose the number of groups with the elbow of the inertia and with the information criteria AIC and BIC
\item
  Fit mixtures from several starts and discard fits with tiny components
\item
  Read groups of objects against a physical hypothesis and judge how far the data support it
\end{itemize}

We begin by importing the libraries and setting the plot style.

\tutorialbarrier
\begin{codecell}

\end{codecell}

\section*{The Sculptor Data}

The Apache Point Observatory Galactic Evolution Experiment (APOGEE) measured high-resolution spectra of stars in the Sculptor dwarf spheroidal galaxy, a satellite of the Milky Way. Our sample is the APOGEE sample that Ting \& Ji (2025) studied. We load the abundances {[}Fe/H{]} and {[}Mg/Fe{]} of its stars and plot them.

\tutorialbarrier
\begin{codecell}

\end{codecell}
\begin{Verbatim}[fontsize=\footnotesize, xleftmargin=5mm]
143 stars, [Fe/H] from -2.41 to -0.90, [Mg/Fe] from -0.60 to 0.41
\end{Verbatim}
\tutorialplot{\includegraphics[width=4.393in,height=0.75\textheight,keepaspectratio]{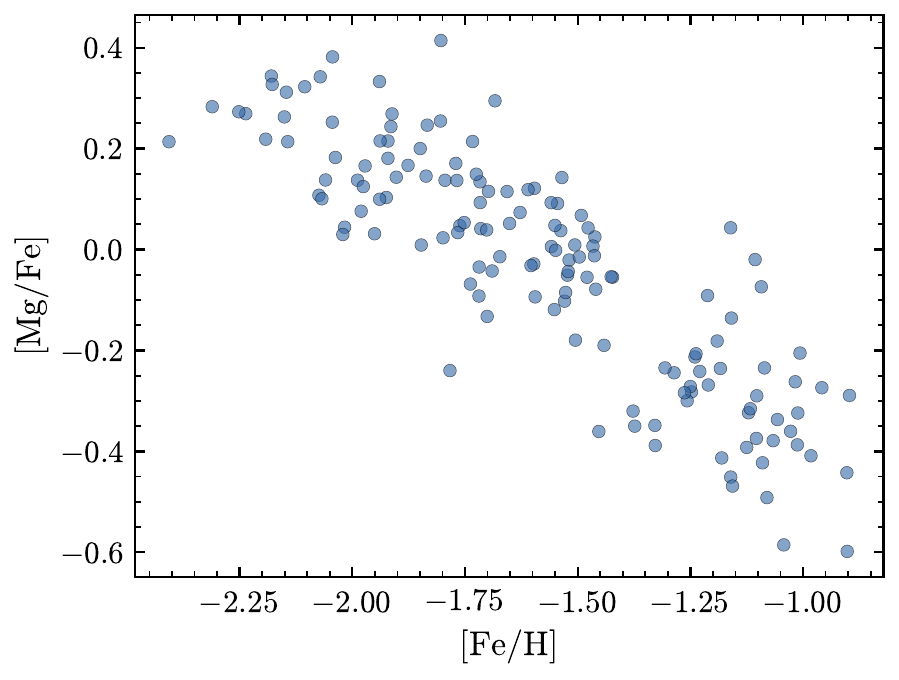}}%
  {\includegraphics[height=\tutorialheight]{tutorials/11a/fig01.pdf}}

{[}Mg/Fe{]} falls as {[}Fe/H{]} rises, so the stars lie along one band. The density along the band is uneven, and the gaps between possible groups are subtle. We therefore need a rule that we can state and test, and we start with the simplest one, K-means.

\section*{K-means Clustering from Scratch}

K-means divides the \(N\) stars, with abundance vectors \(\mathbf{x}_i\), into \(K\) clusters. It minimizes the total squared distance of the stars to the centers of their clusters (Section 11.1)

\[J = \sum_{i=1}^N \sum_{k=1}^K r_{ik} \|\mathbf{x}_i - \boldsymbol{\mu}_k\|^2,\]

where \(r_{ik} \in \{0, 1\}\) is 1 if star \(i\) belongs to cluster \(k\), and \(\boldsymbol{\mu}_k\) is the center of cluster \(k\). Each star belongs to one cluster, so \(\sum_{k=1}^K r_{ik} = 1\). We call the value of \(J\) the inertia (Section 11.3).

A search over all \(K^N\) possible assignments is too expensive. If we know the centers, the best assignments are easy. If we know the assignments, the best centers are easy. K-means therefore alternates between two steps (Section 11.2).

The E-step fixes the centers and gives each star to its nearest center. The M-step fixes the assignments and moves each center to the mean of its stars. This is the expectation-maximization (EM) algorithm for K-means. Each step lowers \(J\) or leaves it unchanged, so the algorithm stops at a local minimum of \(J\), which need not be the lowest value.

K-means uses Euclidean distances, so an abundance with a larger spread weighs more in the distances. The ranges above differ, so {[}Fe/H{]} would weigh more. We therefore standardize both abundances to zero mean and unit variance (Section 11.1), so that they count alike.

\tutorialbarrier
\begin{codecell}
\begin{Verbatim}[commandchars=\\\{\},fontsize=\small]
\PYG{c+c1}{\PYGZsh{} Standardize the abundances to zero mean and unit variance}
\PYG{n}{mean\PYGZus{}X} \PYG{o}{=}\FancyVerbBreakAfterBreak{} \PYG{n}{np}\PYG{o}{.}\FancyVerbBreakAfterBreak{}\PYG{n}{mean}\PYG{p}{(}\FancyVerbBreakAfterBreak{}\PYG{n}{X}\PYG{p}{,}\FancyVerbBreakAfterBreak{} \PYG{n}{axis}\PYG{o}{=}\FancyVerbBreakAfterBreak{}\PYG{l+m+mi}{0}\PYG{p}{)}
\PYG{n}{std\PYGZus{}X} \PYG{o}{=}\FancyVerbBreakAfterBreak{} \PYG{n}{np}\PYG{o}{.}\FancyVerbBreakAfterBreak{}\PYG{n}{std}\PYG{p}{(}\FancyVerbBreakAfterBreak{}\PYG{n}{X}\PYG{p}{,}\FancyVerbBreakAfterBreak{} \PYG{n}{axis}\PYG{o}{=}\FancyVerbBreakAfterBreak{}\PYG{l+m+mi}{0}\PYG{p}{)}
\PYG{n}{X\PYGZus{}std} \PYG{o}{=}\FancyVerbBreakAfterBreak{} \PYG{p}{(}\FancyVerbBreakAfterBreak{}\PYG{n}{X} \PYG{o}{\PYGZhy{}} \PYG{n}{mean\PYGZus{}X}\PYG{p}{)} \PYG{o}{/} \PYG{n}{std\PYGZus{}X}
\end{Verbatim}
\end{codecell}

\section*{Initialization Methods for K-means}

The start decides which local minimum K-means reaches. A poor start can lead to a poor local minimum or to slow convergence. We implement two initializations.

\textbf{Random initialization} picks \(K\) different stars at random as the initial centers. The centers can then lie close together in one region of the data.

\textbf{K-means++} spreads the initial centers over the data (Section 11.2).

\begin{enumerate}
\def\labelenumi{\arabic{enumi}.}
\tightlist
\item
  Choose the first center uniformly at random from the stars.
\item
  For each next center, choose a star with a probability proportional to its squared distance from the nearest center chosen so far.
\item
  Repeat until \(K\) centers exist.
\end{enumerate}

A star that is far from all chosen centers has a high probability of becoming the next center. The initial centers therefore tend to lie far apart. K-means++ also has a guarantee (Arthur \& Vassilvitskii, 2007). Its expected inertia is within a factor of order \(\log K\) of the lowest possible inertia.

We implement both methods.

\tutorialbarrier
\begin{codecell}

\end{codecell}

\section*{The Expectation-Maximization Algorithm for K-means}

We implement the two steps of K-means as functions. The E-step gives star \(i\) to the cluster of its nearest center, so \(r_{ik} = 1\) for that cluster and \(r_{ik} = 0\) for the others. The M-step moves each center to the mean of the set \(C_k\) of stars in its cluster (Section 11.2)

\[\boldsymbol{\mu}_k = \frac{1}{|C_k|} \sum_{\mathbf{x}_i \in C_k} \mathbf{x}_i.\]

The nearest center gives each star its smallest squared distance. The mean is the point with the smallest total squared distance to the stars of a cluster. This is why each step lowers \(J\) or leaves it unchanged. A third function computes the inertia.

\tutorialbarrier
\begin{codecell}
\begin{Verbatim}[commandchars=\\\{\},fontsize=\small]
\PYG{k}{def}\PYG{+w}{ }\PYG{n+nf}{assign\PYGZus{}clusters}\PYG{p}{(}\FancyVerbBreakAfterBreak{}\PYG{n}{X}\PYG{p}{,}\FancyVerbBreakAfterBreak{} \PYG{n}{centroids}\PYG{p}{)}\PYG{p}{:}
\PYG{+w}{    }\PYG{l+s+sd}{\PYGZdq{}\PYGZdq{}\PYGZdq{}E\PYGZhy{}step of K\PYGZhy{}means. Give each star to its nearest center and return the labels.\PYGZdq{}\PYGZdq{}\PYGZdq{}}

    \PYG{c+c1}{\PYGZsh{} Squared Euclidean distance of every star to every center}
    \PYG{n}{distances} \PYG{o}{=}\FancyVerbBreakAfterBreak{} \PYG{n}{np}\PYG{o}{.}\FancyVerbBreakAfterBreak{}\PYG{n}{sum}\PYG{p}{(}\FancyVerbBreakAfterBreak{}\PYG{p}{(}\FancyVerbBreakAfterBreak{}\PYG{n}{X}\PYG{p}{[}\FancyVerbBreakAfterBreak{}\PYG{p}{:}\PYG{p}{,}\FancyVerbBreakAfterBreak{} \PYG{k+kc}{None}\PYG{p}{,}\FancyVerbBreakAfterBreak{} \PYG{p}{:}\PYG{p}{]} \PYG{o}{\PYGZhy{}} \PYG{n}{centroids}\PYG{p}{[}\FancyVerbBreakAfterBreak{}\PYG{k+kc}{None}\PYG{p}{,}\FancyVerbBreakAfterBreak{} \PYG{p}{:}\PYG{p}{,}\FancyVerbBreakAfterBreak{} \PYG{p}{:}\PYG{p}{]}\PYG{p}{)}\PYG{o}{*}\PYG{o}{*}\PYG{l+m+mi}{2}\PYG{p}{,}\FancyVerbBreakAfterBreak{} \PYG{n}{axis}\PYG{o}{=}\FancyVerbBreakAfterBreak{}\PYG{l+m+mi}{2}\PYG{p}{)}

    \PYG{c+c1}{\PYGZsh{} Take the nearest center}
    \PYG{k}{return} \PYG{n}{np}\PYG{o}{.}\FancyVerbBreakAfterBreak{}\PYG{n}{argmin}\PYG{p}{(}\FancyVerbBreakAfterBreak{}\PYG{n}{distances}\PYG{p}{,}\FancyVerbBreakAfterBreak{} \PYG{n}{axis}\PYG{o}{=}\FancyVerbBreakAfterBreak{}\PYG{l+m+mi}{1}\PYG{p}{)}

\PYG{k}{def}\PYG{+w}{ }\PYG{n+nf}{update\PYGZus{}centroids}\PYG{p}{(}\FancyVerbBreakAfterBreak{}\PYG{n}{X}\PYG{p}{,}\FancyVerbBreakAfterBreak{} \PYG{n}{labels}\PYG{p}{,}\FancyVerbBreakAfterBreak{} \PYG{n}{centroids}\PYG{p}{)}\PYG{p}{:}
\PYG{+w}{    }\PYG{l+s+sd}{\PYGZdq{}\PYGZdq{}\PYGZdq{}M\PYGZhy{}step of K\PYGZhy{}means. Move each center to the mean of its stars. An empty cluster keeps its center.\PYGZdq{}\PYGZdq{}\PYGZdq{}}
    \PYG{n}{new\PYGZus{}centroids} \PYG{o}{=}\FancyVerbBreakAfterBreak{} \PYG{n}{centroids}\PYG{o}{.}\FancyVerbBreakAfterBreak{}\PYG{n}{copy}\PYG{p}{(}\FancyVerbBreakAfterBreak{}\PYG{p}{)}

    \PYG{k}{for} \PYG{n}{i} \PYG{o+ow}{in} \PYG{n+nb}{range}\PYG{p}{(}\FancyVerbBreakAfterBreak{}\PYG{n+nb}{len}\PYG{p}{(}\FancyVerbBreakAfterBreak{}\PYG{n}{centroids}\PYG{p}{)}\PYG{p}{)}\PYG{p}{:}
        \PYG{c+c1}{\PYGZsh{} Use the mean of the stars of cluster i as the new center}
        \PYG{k}{if} \PYG{n}{np}\PYG{o}{.}\FancyVerbBreakAfterBreak{}\PYG{n}{any}\PYG{p}{(}\FancyVerbBreakAfterBreak{}\PYG{n}{labels} \PYG{o}{==} \PYG{n}{i}\PYG{p}{)}\PYG{p}{:}
            \PYG{n}{new\PYGZus{}centroids}\PYG{p}{[}\FancyVerbBreakAfterBreak{}\PYG{n}{i}\PYG{p}{]} \PYG{o}{=}\FancyVerbBreakAfterBreak{} \PYG{n}{np}\PYG{o}{.}\FancyVerbBreakAfterBreak{}\PYG{n}{mean}\PYG{p}{(}\FancyVerbBreakAfterBreak{}\PYG{n}{X}\PYG{p}{[}\FancyVerbBreakAfterBreak{}\PYG{n}{labels} \PYG{o}{==} \PYG{n}{i}\PYG{p}{]}\PYG{p}{,}\FancyVerbBreakAfterBreak{} \PYG{n}{axis}\PYG{o}{=}\FancyVerbBreakAfterBreak{}\PYG{l+m+mi}{0}\PYG{p}{)}

    \PYG{k}{return} \PYG{n}{new\PYGZus{}centroids}

\PYG{k}{def}\PYG{+w}{ }\PYG{n+nf}{compute\PYGZus{}inertia}\PYG{p}{(}\FancyVerbBreakAfterBreak{}\PYG{n}{X}\PYG{p}{,}\FancyVerbBreakAfterBreak{} \PYG{n}{labels}\PYG{p}{,}\FancyVerbBreakAfterBreak{} \PYG{n}{centroids}\PYG{p}{)}\PYG{p}{:}
\PYG{+w}{    }\PYG{l+s+sd}{\PYGZdq{}\PYGZdq{}\PYGZdq{}Total squared distance of the stars to the centers of their clusters.\PYGZdq{}\PYGZdq{}\PYGZdq{}}

    \PYG{c+c1}{\PYGZsh{} Each star minus the center of its own cluster}
    \PYG{k}{return} \PYG{n}{np}\PYG{o}{.}\FancyVerbBreakAfterBreak{}\PYG{n}{sum}\PYG{p}{(}\FancyVerbBreakAfterBreak{}\PYG{p}{(}\FancyVerbBreakAfterBreak{}\PYG{n}{X} \PYG{o}{\PYGZhy{}} \PYG{n}{centroids}\PYG{p}{[}\FancyVerbBreakAfterBreak{}\PYG{n}{labels}\PYG{p}{]}\PYG{p}{)}\PYG{o}{*}\PYG{o}{*}\PYG{l+m+mi}{2}\PYG{p}{)}
\end{Verbatim}
\end{codecell}

We now apply one E-step and one M-step to the K-means++ start for \(K = 3\). We take \(K = 3\), the number that the elbow below favors. We plot the stars in the standardized plane, where K-means measures distances. In the first panel, each star has the color of its nearest start center. In the second panel, arrows lead from the start centers (open circles) to the new centers (crosses).

\tutorialbarrier
\begin{codecell}

\end{codecell}
\begin{Verbatim}[fontsize=\footnotesize, xleftmargin=5mm]
Inertia after the E-step = 93.59, after the M-step = 56.32
\end{Verbatim}
\tutorialplot{\includegraphics[width=6.172in,height=0.75\textheight,keepaspectratio]{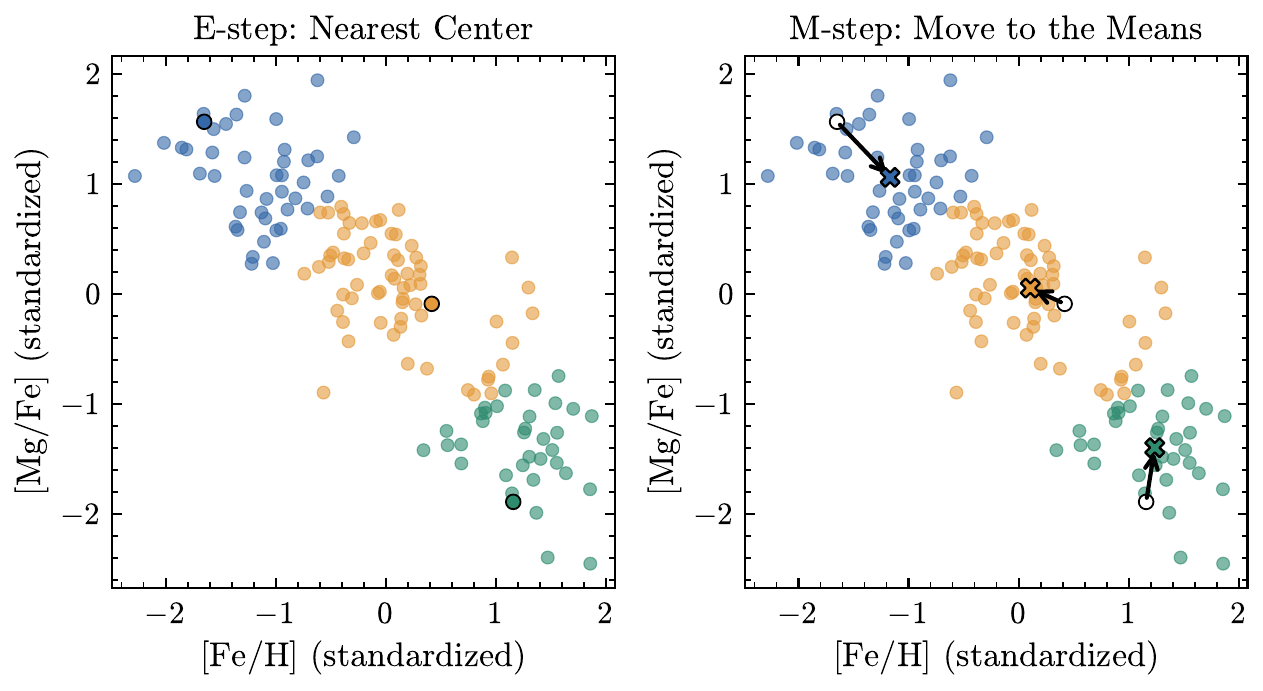}}%
  {\includegraphics[height=\tutorialheight]{tutorials/11a/fig02.pdf}}

The M-step moves each center to the mean of the stars with its color, and the inertia falls. This is only the first step. The centers keep moving until the algorithm converges.

\section*{The Complete K-means Algorithm}

The complete algorithm repeats the E-step and the M-step until the centers stop moving, that is, until their total shift falls below a small tolerance. We store the inertia and the centers of every iteration, to follow the convergence.

We implement the complete algorithm.

\tutorialbarrier
\begin{codecell}

\end{codecell}

We run the complete algorithm for \(K = 3\) with both initializations.

\tutorialbarrier
\begin{codecell}

\end{codecell}
\begin{Verbatim}[fontsize=\footnotesize, xleftmargin=5mm]
Random initialization, 10 iterations, inertia = 47.26
K-means++ initialization, 6 iterations, inertia = 47.26
Cluster sizes: 45 45 53
\end{Verbatim}

We plot the final clusters and the path of each center for both runs. We use the original abundance plane, because the standardized axes have no physical meaning. Circles mark the initial centers and crosses mark the final centers. The right panel shows the inertia of both runs at each iteration. The first segment of each K-means++ path is the M-step above. The first point of its inertia curve is the inertia after the first E-step.

\tutorialbarrier
\begin{codecell}

\end{codecell}
\tutorialplot{\includegraphics[width=6.500in,height=0.75\textheight,keepaspectratio]{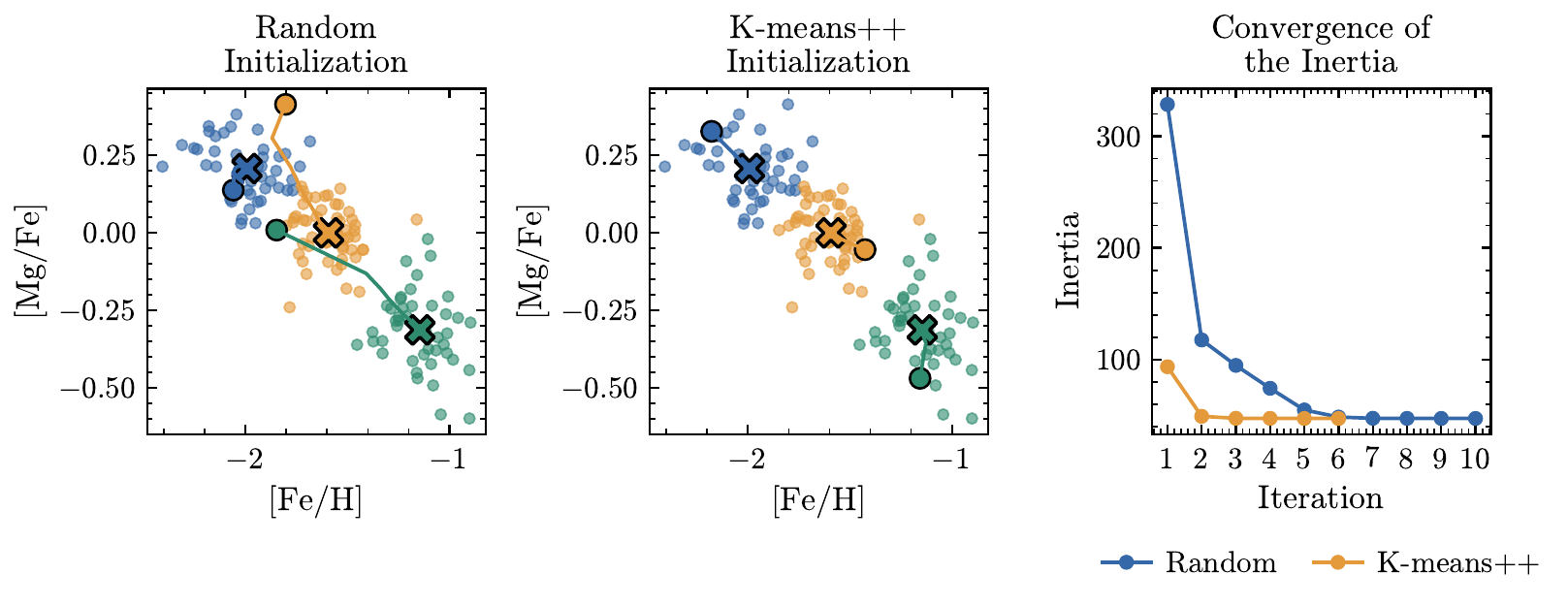}}%
  {\includegraphics[height=\tutorialheight]{tutorials/11a/fig03.pdf}}

Both runs end at the same three clusters and the same inertia, and the inertia never rises. The three clusters have similar sizes. In this run, the K-means++ centers start near their final positions, so they move less, and the inertia starts lower. The K-means++ run stops after fewer iterations than the random start.

\section*{Initialization over Many Starts}

To compare the initializations, we run K-means from 1000 seeds for each initialization, for \(K = 3\) and \(K = 6\). For each case we print the lowest inertia, the mean inertia and the mean number of iterations. We also print the share of runs within 1 percent of the lowest inertia that any run reached for that \(K\).

\tutorialbarrier
\begin{codecell}

\end{codecell}
\begin{Verbatim}[fontsize=\footnotesize, xleftmargin=5mm]
K | initialization | lowest |  mean | within 1% | iterations
3 | random         |  47.26 | 47.30 |      100% |        8.4
3 | kmeans++       |  47.26 | 47.26 |      100% |        7.4
6 | random         |  25.72 | 28.64 |        6% |        9.8
6 | kmeans++       |  25.72 | 27.75 |        9% |        8.7
\end{Verbatim}

For \(K = 3\) every run of both initializations ends within 1 percent of the lowest inertia. For \(K = 6\) the runs differ, and K-means++ gives a lower mean inertia and more runs close to the lowest inertia. A single run can therefore end above the lowest inertia. In practice we run K-means from several starts and keep the run with the lowest inertia, especially for a large \(K\) (Section 11.2).

\section*{Selecting the Optimal Number of Clusters}

K-means needs the number of clusters \(K\) in advance, but we do not know the number of stellar populations in Sculptor. The elbow method looks at how the inertia falls as \(K\) grows (Section 11.3). The inertia always falls as \(K\) grows, and it is zero for \(K = N\). If the data contain groups, we expect a point after which more clusters give little gain. This point is the elbow.

We compute the inertia for \(K = 2\) to 9. For each \(K\) we run K-means++ from 20 starts and keep the lowest inertia.

\tutorialbarrier
\begin{codecell}

\end{codecell}
\tutorialplot{\includegraphics[width=4.390in,height=0.75\textheight,keepaspectratio]{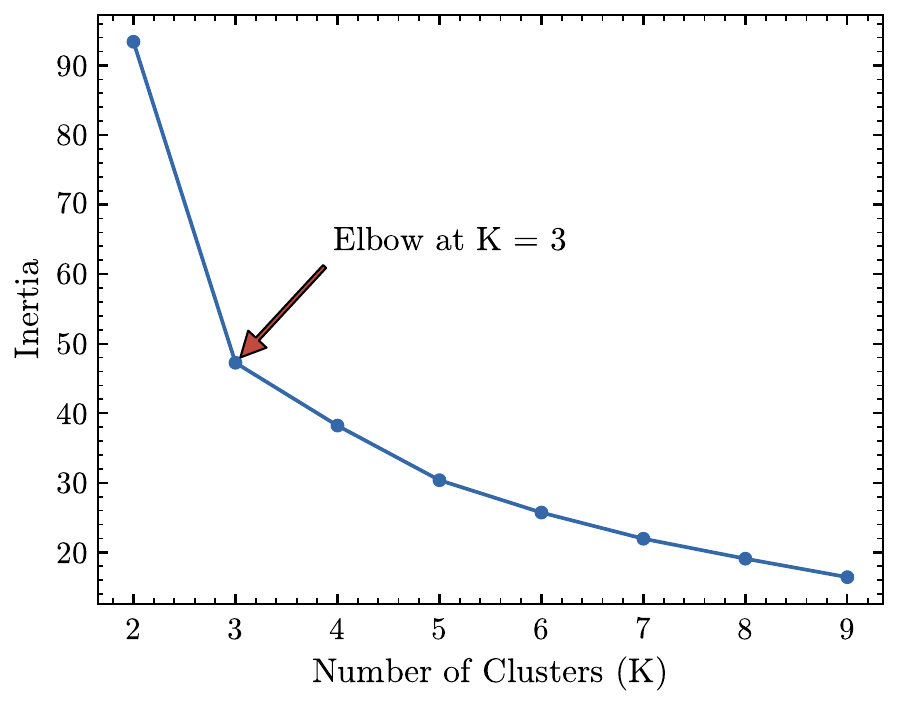}}%
  {\includegraphics[height=\tutorialheight]{tutorials/11a/fig04.pdf}}

The inertia falls steeply from \(K = 2\) to \(K = 3\). After that it falls slowly and smoothly, with no further sharp bend. The elbow is at \(K = 3\), the number of clusters that we used above. The inertia keeps falling as we add clusters, so the choice of \(K\) rests on visual inspection and on the science.

\section*{From K-means to Gaussian Mixture Models}

K-means rests on rigid assumptions. It gives each star to one cluster, it expects clusters of equal weight, and it makes each cluster round. These are the assumptions of its generative model (Section 11.4). Stellar populations can differ in size and shape in elemental abundance space, and they can overlap. A Gaussian mixture model relaxes these assumptions, and its likelihood gives criteria for choosing \(K\) (Section 11.8).

A GMM describes the distribution of the stars as a weighted sum of \(K\) Gaussian components (Section 11.5)

\[p(\mathbf{x}) = \sum_{k=1}^K \pi_k \mathcal{N}(\mathbf{x}|\boldsymbol{\mu}_k, \boldsymbol{\Sigma}_k),\]

where \(\pi_k\) is the weight of component \(k\), with \(\sum_{k=1}^K \pi_k = 1\). The vector \(\boldsymbol{\mu}_k\) and the matrix \(\boldsymbol{\Sigma}_k\) are the mean and the covariance matrix of its Gaussian density \(\mathcal{N}\). Each component can describe one stellar population. The mean gives its location in elemental abundance space. The covariance matrix gives its spread and tilt, and the weight gives its share of the stars. A star near the boundary of two populations has a probability of belonging to each of them.

The log-likelihood of the stars is (Section 11.5)

\[\ln L = \sum_{i=1}^N \ln \left( \sum_{k=1}^K \pi_k \mathcal{N}(\mathbf{x}_i|\boldsymbol{\mu}_k, \boldsymbol{\Sigma}_k) \right).\]

We cannot maximize it in closed form, because it has a sum inside the logarithm. If we knew which component made each star, we could fit each Gaussian alone. We therefore alternate again (Section 11.6). The E-step has the same job as in K-means, and it now gives soft assignments. The M-step now sets a weight and a covariance matrix as well as a mean.

\section*{The E-step and the M-step}

In the E-step we compute the responsibility \(\gamma_{ik}\), the posterior probability that component \(k\) generated star \(i\) (Section 11.6)

\[\gamma_{ik} = \frac{\pi_k \mathcal{N}(\mathbf{x}_i|\boldsymbol{\mu}_k, \boldsymbol{\Sigma}_k)}{\sum_{j=1}^K \pi_j \mathcal{N}(\mathbf{x}_i|\boldsymbol{\mu}_j, \boldsymbol{\Sigma}_j)}.\]

The responsibilities of a star sum to 1 over the components. They replace the 0 and 1 of the K-means assignments \(r_{ik}\), so a star between two populations can have responsibilities of 0.6 and 0.4. The denominator is the density of star \(i\), which is the sum inside \(\ln L\). The E-step therefore also gives \(\ln L\), which we use to stop the iterations.

We implement the E-step with a function for the Gaussian density. We add 1e-6 to the diagonal of every covariance matrix in the initialization, in the M-step and in the density, so that it stays invertible.

\tutorialbarrier
\begin{codecell}

\end{codecell}

In the M-step we update the parameters with the responsibilities as weights (Section 11.6). With the effective number of stars \(N_k = \sum_{i=1}^N \gamma_{ik}\) of component \(k\), the new weight is the share of the stars that the component explains

\[\pi_k^{\text{new}} = \frac{N_k}{N}.\]

The new mean is the average position of the stars, weighted by the responsibilities

\[\boldsymbol{\mu}_k^{\text{new}} = \frac{1}{N_k}\sum_{i=1}^N \gamma_{ik} \mathbf{x}_i.\]

The new covariance matrix is the weighted scatter of the stars around the new mean

\[\boldsymbol{\Sigma}_k^{\text{new}} = \frac{1}{N_k}\sum_{i=1}^N \gamma_{ik} (\mathbf{x}_i - \boldsymbol{\mu}_k^{\text{new}})(\mathbf{x}_i - \boldsymbol{\mu}_k^{\text{new}})^T.\]

A star with a high responsibility for a component has a strong influence on its parameters. For responsibilities of 0 and 1, the mean is the center of K-means. We implement the M-step.

\tutorialbarrier
\begin{codecell}

\end{codecell}

\section*{The Complete EM Algorithm for GMM}

We start the algorithm from a K-means run. K-means gives a sensible first partition, and its clusters give the first weights, means and shapes. The weights are the shares of the stars in the clusters. The means are the centers. The covariance matrices are the scatter of the stars in each cluster. This is richer than the common start of Section 11.6, which takes only the means from K-means and uses equal weights and identity covariance matrices.

The algorithm then repeats the E-step and the M-step until the log-likelihood gains less than a small tolerance. The algorithm reaches a local maximum of the log-likelihood, which need not be the global maximum.

We implement the start.

\tutorialbarrier
\begin{codecell}

\end{codecell}

We store the log-likelihood, the weights and the means of every iteration, to follow the convergence. We implement the complete algorithm and fit a GMM with 6 components. We print the number of iterations, the weights and the first responsibilities.

\tutorialbarrier
\begin{codecell}

\end{codecell}
\begin{Verbatim}[fontsize=\footnotesize, xleftmargin=5mm]
EM converged in 109 iterations
Weights of the components = 0.225, 0.179, 0.146, 0.176, 0.165, 0.109
Responsibilities of the first 5 stars
[[0.006 0.992 0.    0.    0.001 0.   ]
 [0.    1.    0.    0.    0.    0.   ]
 [0.948 0.052 0.    0.    0.    0.   ]
 [0.    0.    0.969 0.031 0.    0.   ]
 [0.    0.    0.    0.    0.    1.   ]]
\end{Verbatim}

Each row of responsibilities sums to 1. The first star belongs almost entirely to component 1. The third star belongs mostly to component 0 and has a small share in component 1.

We draw each component as an ellipse. The eigenvectors of the covariance matrix give the directions of the axes, and the square roots of the eigenvalues give the standard deviations along them. The ellipse shows two standard deviations on each side of the mean. We write one function for the ellipse and one that plots a fitted mixture in the original abundance plane.

The plot function colors the components in the order of their mean {[}Fe/H{]}. It also rings every star whose largest responsibility is below 0.8, because such a star lies where components overlap.

\tutorialbarrier
\begin{codecell}

\end{codecell}

We plot the fitted mixture with the path of each mean from its start to its end. Dashed ellipses show the start of the fit, and rings mark the stars whose largest responsibility is below 0.8. We also plot the log-likelihood and the weights at each iteration.

\tutorialbarrier
\begin{codecell}

\end{codecell}
\tutorialplot{\includegraphics[width=6.500in,height=0.75\textheight,keepaspectratio]{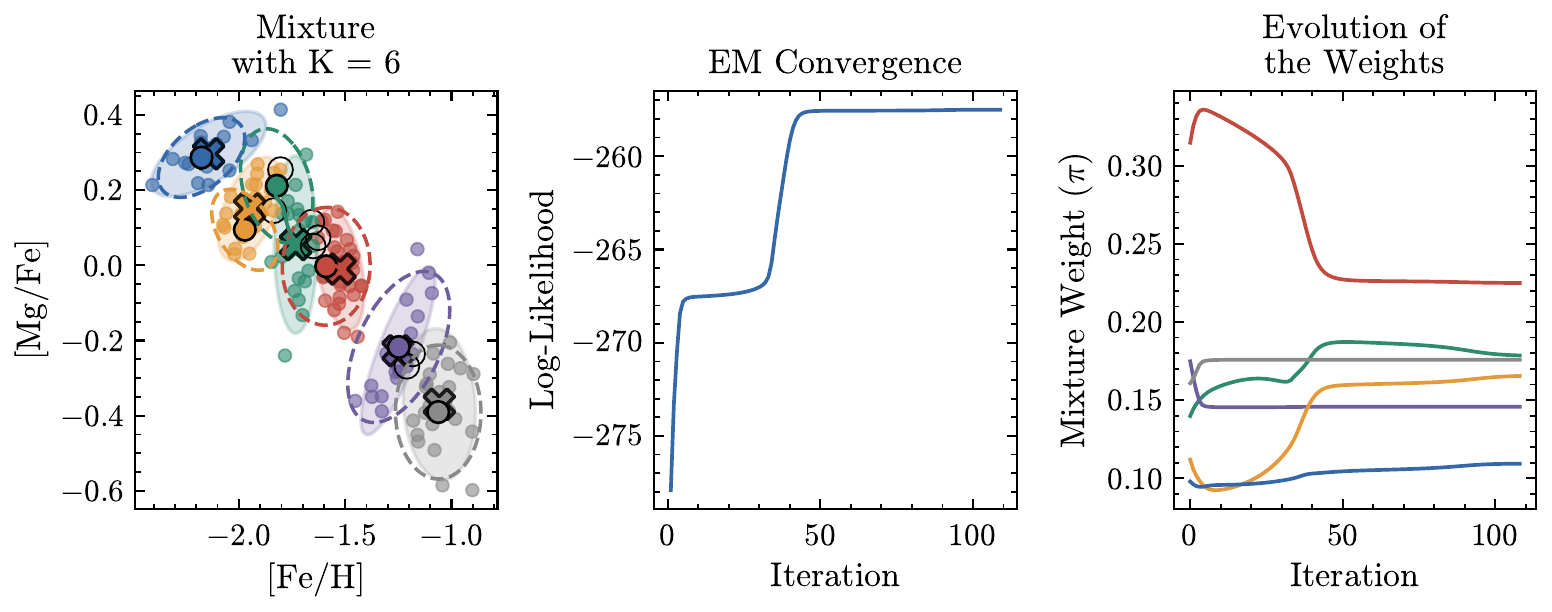}}%
  {\includegraphics[height=\tutorialheight]{tutorials/11a/fig05.pdf}}

The log-likelihood rises quickly at first, stays nearly flat until about iteration 35, and gains again before iteration 50. The weights change at the same iterations. The component with the largest starting share of the stars gives part of it to its neighbors.

The components differ in weight and shape, which K-means does not model. The K-means clusters have similar sizes. The means move little. EM needs many more iterations than K-means.

\section*{Model Selection for GMM: Information Criteria}

The likelihood of the stars rises with every added component, so it cannot choose \(K\). The information criteria correct the likelihood of the fit for the number of parameters \(P\) with a penalty. They therefore replace a test set (Section 11.8). For \(K\) components with full covariance matrices in \(D\) dimensions, the number of parameters is \(P = (K-1) + KD + KD(D+1)/2\) (Section 11.5). The terms count the weights, which sum to 1, the means and the symmetric covariance matrices. For \(D = 2\) this gives \(P = 6K - 1\).

The Bayesian information criterion (BIC) and the Akaike information criterion (AIC) are

\[\text{BIC} = -2\ln(\hat{L}) + P\ln(N),\]

\[\text{AIC} = -2\ln(\hat{L}) + 2P,\]

where \(\hat{L}\) is the maximized likelihood and \(N\) is the number of stars. Lower values are better. For our \(N = 143\) stars, each parameter costs \(\ln N = 4.96\) in BIC and 2 in AIC, so BIC tends to select fewer components. A component is worth adding only if it lowers \(-2\ln(\hat{L})\) by more than its penalty. BIC approximates the evidence of the model, and AIC estimates how well the model predicts new stars (Section 11.8).

\tutorialbarrier
\begin{codecell}

\end{codecell}

EM can end in a local maximum of the log-likelihood (Section 11.6). A component can also shrink onto a few stars, which raises \(\ln L\) without describing a population. We therefore fit each \(K\) from 1 to 8 from 20 starts. We drop the fits with a component of fewer than 6 stars, the number of free parameters of one component in two dimensions. We keep the fit with the highest \(\ln L\) and compute its criteria.

\tutorialbarrier
\begin{codecell}

\end{codecell}

We plot the two criteria and mark the lowest value of each.

\tutorialbarrier
\begin{codecell}

\end{codecell}
\tutorialplot{\includegraphics[width=6.174in,height=0.75\textheight,keepaspectratio]{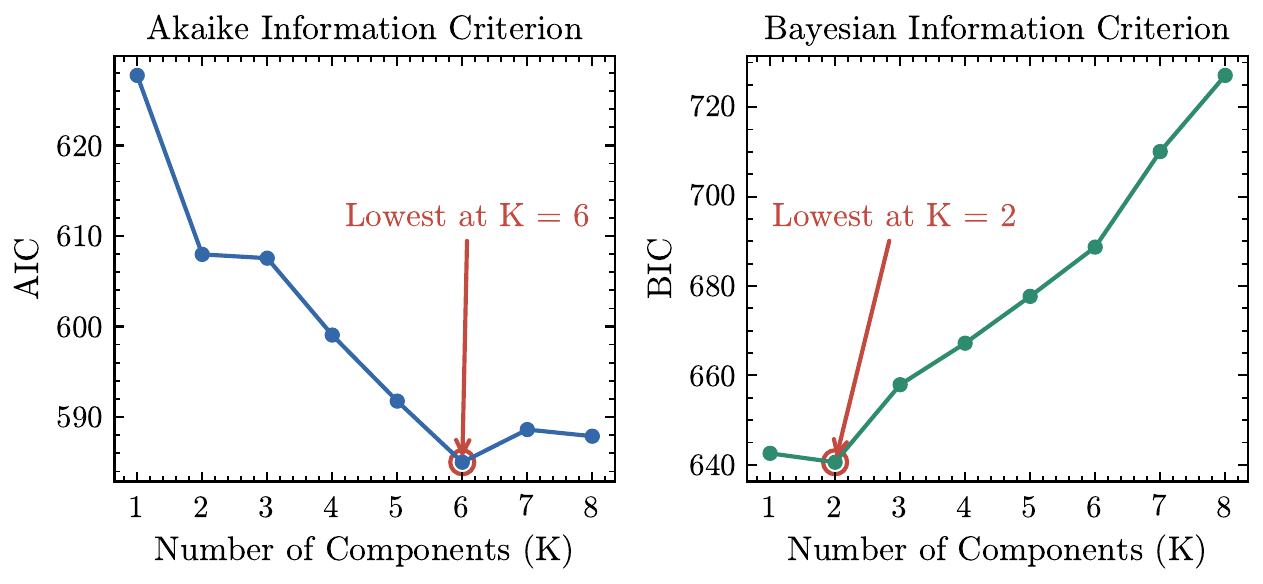}}%
  {\includegraphics[height=\tutorialheight]{tutorials/11a/fig06.pdf}}

\section*{Reading the Model Selection Results}

BIC is lowest at \(K = 2\), but only slightly below BIC at \(K = 1\), so the data weakly prefer two components to one. AIC is lowest at \(K = 6\) and stays nearly flat for larger \(K\).

We plot the BIC model and the AIC model. The BIC model has \(K = 2\), and the AIC model has \(K = 6\). Rings mark the stars whose largest responsibility is below 0.8.

\tutorialbarrier
\begin{codecell}

\end{codecell}
\tutorialplot{\includegraphics[width=6.172in,height=0.75\textheight,keepaspectratio]{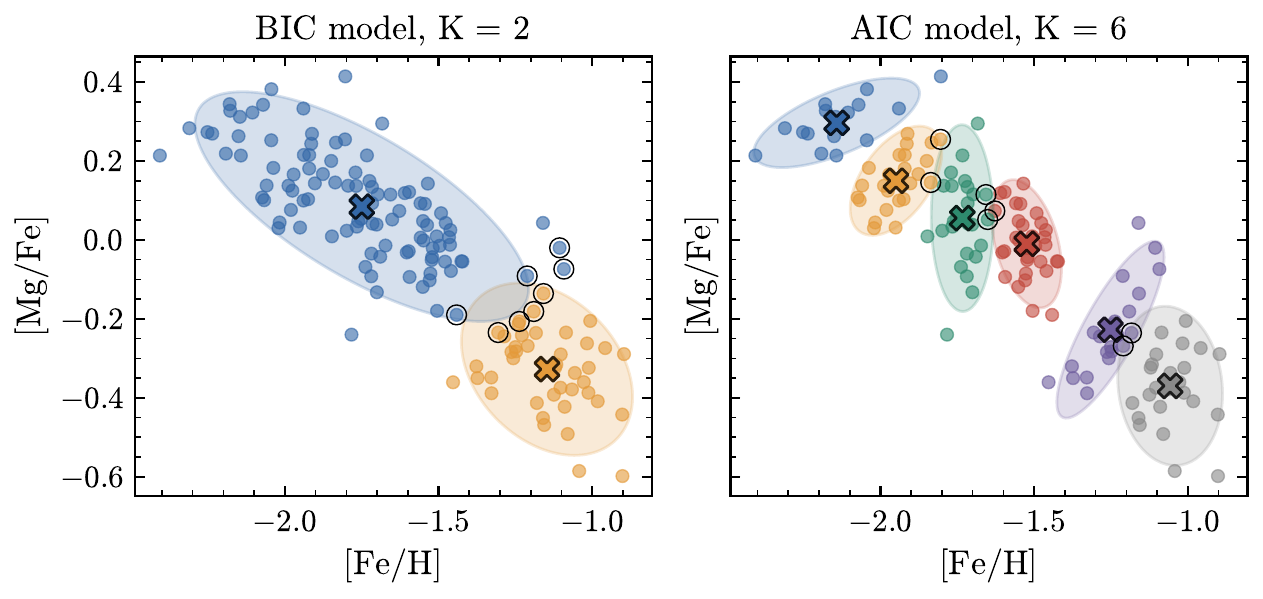}}%
  {\includegraphics[height=\tutorialheight]{tutorials/11a/fig07.pdf}}

The BIC model splits the stars into a metal-poor group with a higher {[}Mg/Fe{]} and a metal-rich group with a lower {[}Mg/Fe{]}. The AIC model has more and smaller components along the same band.

The two criteria follow different views of model selection (Section 11.8). BIC treats the candidates as possible true models and approximates their evidence, so it prefers the simplest model that the sample supports. AIC treats every candidate as an approximation and estimates how well it predicts new stars, so it accepts more components when they improve the fit. Here the stars lie along a continuous band. The two broad components of the BIC model smooth over its shape, while the AIC components follow it more closely. BIC therefore suits a first look at the broad population structure, and AIC suits a detailed study of subpopulations.

The gaps between the groups are subtle, and the AIC mixture resolves them better than the broad BIC components. A mixture with AIC is therefore a good tool for groups of this kind.

\section*{Summary}

In this tutorial we looked for groups in the elemental abundances of stars in a dwarf galaxy. We built K-means and a Gaussian mixture model with the EM algorithm from scratch. Our main lessons follow.

\begin{enumerate}
\def\labelenumi{\arabic{enumi}.}
\tightlist
\item
  \textbf{K-means.} K-means alternates between giving each point to its nearest center and moving each center to the mean of its points. Each step lowers the objective, so the algorithm ends in a local minimum. It assumes round clusters of equal weight and gives hard assignments.
\item
  \textbf{Initialization.} The result of K-means depends on its start, and so does the result of EM. Several starts, with the lowest inertia or the highest log-likelihood, reduce the risk of a poor local solution. K-means++ spreads the initial centers, and it can lower the mean inertia.
\item
  \textbf{Gaussian mixtures.} The EM algorithm fits a mixture with soft assignments. The E-step computes responsibilities, and the M-step sets the weights, means and covariance matrices to responsibility-weighted averages. Each component has its own weight and shape.
\item
  \textbf{The number of groups.} The inertia always falls as \(K\) grows. The elbow is a first guess, and the science guides the choice. AIC and BIC penalize the number of parameters instead of using test data. BIC approximates the evidence and favors the simplest model, and AIC estimates the prediction of new data and allows more components, so they can disagree. A criterion needs a good fit for each \(K\), so we use several starts and drop fits with tiny components.
\item
  \textbf{Reading the groups.} A group of objects can match a physical population, such as one burst of star formation. The number of groups from an algorithm or a criterion does not by itself count the populations, so we read it against the science.
\end{enumerate}

K-means and Gaussian mixtures give a quantitative description of the groups in a sample. The mixture adds the probability that each object belongs to each group. A criterion and the science question together set how many groups to keep.

\tutorialchapter{11b}{Gaussian Mixtures and PCA for Quasar Spectra}
\noindent\textit{Companion material for Chapter 11}

\section*{Introduction}

A quasar is a supermassive black hole that accretes matter at the center of a distant galaxy. The infalling matter forms an accretion disk and radiates over the whole electromagnetic spectrum. The spectrum of a quasar has a smooth part from the disk and broad emission lines from gas clouds near the black hole. The strengths of the lines and the overall slope of the spectrum differ from quasar to quasar.

Spectroscopic surveys such as the Sloan Digital Sky Survey (SDSS) and the Dark Energy Spectroscopic Instrument (DESI) observe hundreds of thousands of quasars. Each spectrum has thousands of wavelength bins, so it is hard to describe and to search a whole population. A few physical properties largely determine a spectrum, for example the black hole mass, the accretion rate and the viewing angle. We can therefore expect that a few numbers describe each spectrum with little loss of information.

We also want a probability model of the whole population of spectra. Such a model lets us draw synthetic spectra and flag the spectra that it finds unlikely, for follow-up study. A Gaussian mixture model (GMM) can describe a distribution. Tutorial 11a fitted a GMM with the expectation-maximization (EM) algorithm to two elemental abundance ratios of stars.

A spectrum has thousands of dimensions. A full covariance matrix at that dimension has more than 4 million parameters, and we have only 1000 spectra (Section 11.5). We therefore compress each spectrum with principal component analysis (PCA, Chapter 10) and fit the mixture to the few coefficients. We use simulated spectra for all of this, and along the way we relate the principal components to the physics of quasar spectra.

\section*{Prerequisites and Learning Objectives}

\subsection*{Prerequisites}

This tutorial assumes the following.

\begin{itemize}
\tightlist
\item
  Working knowledge of Python, NumPy and matplotlib
\item
  Covariance matrices, multivariate Gaussian distributions and maximum likelihood
\item
  Principal component analysis and the singular value decomposition (Chapter 10)
\item
  Gaussian mixture models, the EM algorithm and information criteria (Chapter 11, Tutorial 11a)
\end{itemize}

\subsection*{Learning Objectives}

After this tutorial you can do the following.

\begin{itemize}
\tightlist
\item
  Apply PCA to spectra and relate the principal components to the emission lines and the overall slope
\item
  Choose the number of principal components from the explained variance and the reconstruction error
\item
  Fit a Gaussian mixture model to the PCA coefficients and choose its number of components with the Bayesian information criterion (BIC)
\item
  Draw synthetic spectra from the mixture and compare them with the training spectra
\item
  Use the likelihood of the mixture to flag unusual spectra
\end{itemize}

We begin by importing the libraries and setting the plot style.

\tutorialbarrier
\begin{codecell}

\end{codecell}

\section*{The Quasar Spectral Dataset}

We use 1000 simulated quasar spectra, generated from the quasar factor analysis model of \href{https://ui.adsabs.harvard.edu/abs/2023ApJS..269....4S/abstract}{Sun, Ting \& Cai (2023)}. The authors learned this latent factor model of quasar spectra from data. It describes each spectrum with a few hidden numbers, the latent factors. The spectra are more idealized than real spectra, which also show instrumental effects and absorption by gas along the line of sight. Each spectrum has 2947 bins of the rest-frame wavelength in the ultraviolet. We load the spectra and their wavelengths.

\tutorialbarrier
\begin{codecell}
\begin{Verbatim}[commandchars=\\\{\},fontsize=\small]
\PYG{c+c1}{\PYGZsh{} Load the spectra and their rest\PYGZhy{}frame wavelengths}
\PYG{n}{data} \PYG{o}{=}\FancyVerbBreakAfterBreak{} \PYG{n}{np}\PYG{o}{.}\FancyVerbBreakAfterBreak{}\PYG{n}{load}\PYG{p}{(}\FancyVerbBreakAfterBreak{}\PYG{l+s+s2}{\PYGZdq{}}\PYG{l+s+s2}{../data/dataset\PYGZus{}quasar\PYGZus{}sample.npz}\PYG{l+s+s2}{\PYGZdq{}}\PYG{p}{)}
\PYG{n}{X} \PYG{o}{=}\FancyVerbBreakAfterBreak{} \PYG{n}{data}\PYG{p}{[}\FancyVerbBreakAfterBreak{}\PYG{l+s+s2}{\PYGZdq{}}\PYG{l+s+s2}{continua}\PYG{l+s+s2}{\PYGZdq{}}\PYG{p}{]}
\PYG{n}{wavelength} \PYG{o}{=}\FancyVerbBreakAfterBreak{} \PYG{n}{data}\PYG{p}{[}\FancyVerbBreakAfterBreak{}\PYG{l+s+s2}{\PYGZdq{}}\PYG{l+s+s2}{wavelength}\PYG{l+s+s2}{\PYGZdq{}}\PYG{p}{]}

\PYG{n+nb}{print}\PYG{p}{(}\FancyVerbBreakAfterBreak{}\PYG{l+s+sa}{f}\PYG{l+s+s2}{\PYGZdq{}}\PYG{l+s+si}{\PYGZob{}}\PYG{n}{X}\PYG{o}{.}\FancyVerbBreakAfterBreak{}\PYG{n}{shape}\PYG{p}{[}\FancyVerbBreakAfterBreak{}\PYG{l+m+mi}{0}\PYG{p}{]}\PYG{l+s+si}{\PYGZcb{}}\FancyVerbBreakAfterBreak{}\PYG{l+s+s2}{ spectra, }\PYG{l+s+si}{\PYGZob{}}\PYG{n}{X}\PYG{o}{.}\FancyVerbBreakAfterBreak{}\PYG{n}{shape}\PYG{p}{[}\FancyVerbBreakAfterBreak{}\PYG{l+m+mi}{1}\PYG{p}{]}\PYG{l+s+si}{\PYGZcb{}}\FancyVerbBreakAfterBreak{}\PYG{l+s+s2}{ bins, }\PYG{l+s+s2}{\PYGZdq{}}
      \PYG{l+s+sa}{f}\PYG{l+s+s2}{\PYGZdq{}}\PYG{l+s+si}{\PYGZob{}}\PYG{n}{wavelength}\PYG{o}{.}\FancyVerbBreakAfterBreak{}\PYG{n}{min}\PYG{p}{(}\FancyVerbBreakAfterBreak{}\PYG{p}{)}\PYG{l+s+si}{:}\PYG{l+s+s2}{.0f}\PYG{l+s+si}{\PYGZcb{}}\FancyVerbBreakAfterBreak{}\PYG{l+s+s2}{ to }\PYG{l+s+si}{\PYGZob{}}\PYG{n}{wavelength}\PYG{o}{.}\FancyVerbBreakAfterBreak{}\PYG{n}{max}\PYG{p}{(}\FancyVerbBreakAfterBreak{}\PYG{p}{)}\PYG{l+s+si}{:}\PYG{l+s+s2}{.0f}\PYG{l+s+si}{\PYGZcb{}}\FancyVerbBreakAfterBreak{}\PYG{l+s+s2}{ Å}\PYG{l+s+s2}{\PYGZdq{}}\PYG{p}{)}
\end{Verbatim}
\end{codecell}
\begin{Verbatim}[fontsize=\footnotesize, xleftmargin=5mm]
1000 spectra, 2947 bins, 1030 to 2030 Å
\end{Verbatim}

\subsection*{Visualizing Representative Quasar Spectra}

The disk is hotter near the black hole than far from it. It therefore emits more light at short wavelengths, and the spectrum falls toward long wavelengths. The gas clouds orbit the black hole at thousands of kilometers per second, so the Doppler effect broadens the lines. The two strongest lines are Lyman-\(\alpha\) (Ly\(\alpha\)) at 1215 Å from hydrogen and C IV at 1549 Å from triply ionized carbon. The Roman numeral counts the ionization, so C IV is carbon that has lost three electrons. Three weaker lines are Si IV at about 1400 Å, He II at 1640 Å and C III{]} at 1909 Å.

We plot five spectra at evenly spaced indices. Dashed lines mark these five lines.

\tutorialbarrier
\begin{codecell}

\end{codecell}
\tutorialplot{\includegraphics[width=4.392in,height=0.75\textheight,keepaspectratio]{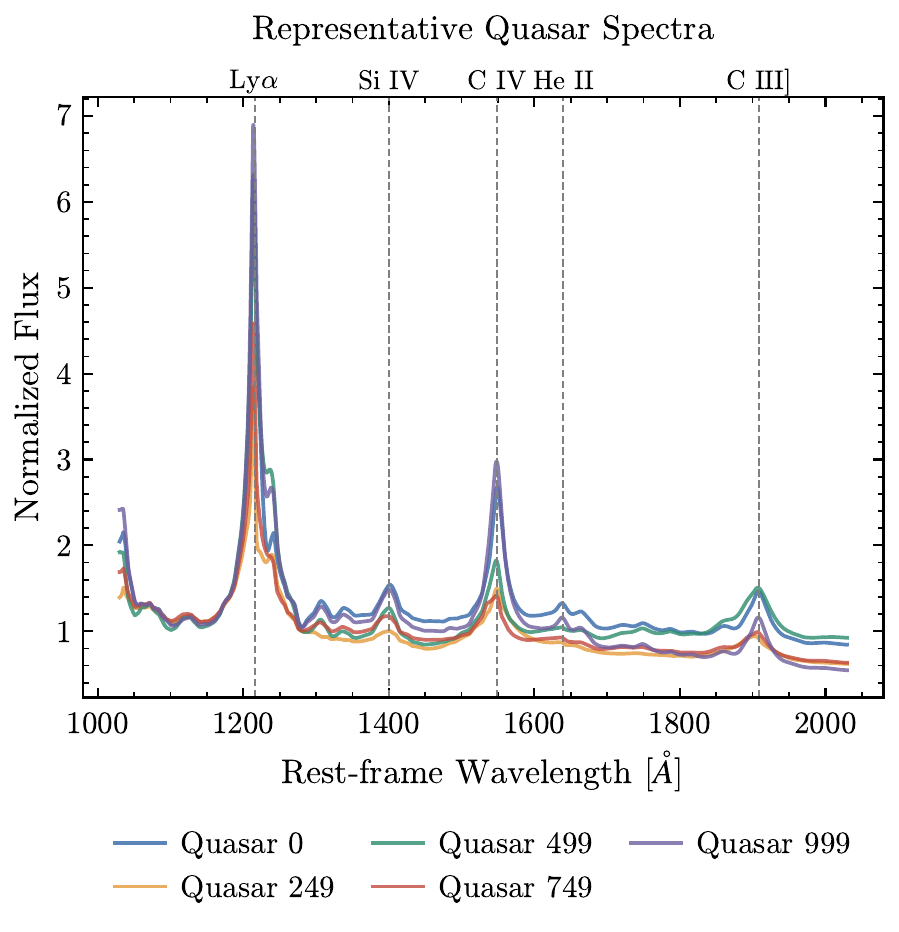}}%
  {\includegraphics[height=\tutorialheight]{tutorials/11b/fig01.pdf}}

The line strengths and the slopes differ between the spectra. A model with few parameters produced these spectra, so we expect a few principal components to carry most of the variance.

\section*{Principal Component Analysis for Spectral Compression}

We now compress the spectra with PCA. Tutorial 10 applied PCA to galaxy images, and the steps are the same for spectra. We center the \(N\) spectra, which have \(D = 2947\) bins each, and compute the singular value decomposition (SVD) of the centered matrix. The right singular vectors are the principal components, and the variance along component \(m\) is \(\sigma_m^2/N\) for the singular value \(\sigma_m\) (Section 10.8).

The first \(M\) components, written as the columns of \(\mathbf{V}_M \in \mathbb{R}^{D \times M}\), map a centered spectrum \(\mathbf{x}' = \mathbf{x} - \overline{\mathbf{x}}\) to its \(M\) coefficients,

\[\mathbf{z} = \mathbf{x}'\mathbf{V}_M,\]

and map the coefficients back to a spectrum,

\[\widehat{\mathbf{x}} = \mathbf{z}\mathbf{V}_M^T + \overline{\mathbf{x}}.\]

The explained variance ratio of a component is its share of the total variance. It guides how many components to keep (Section 10.7). We use the reconstruction formula to test this choice below, and in the generation section to turn coefficients into spectra.

We compute the PCA of the 1000 spectra. We print the number of components that carry 99.99 percent of the variance.

\tutorialbarrier
\begin{codecell}

\end{codecell}
\begin{Verbatim}[fontsize=\footnotesize, xleftmargin=5mm]
Components for 99.99% of the variance = 8
\end{Verbatim}

The spectra come from a model with a few hidden numbers, so they should lie in a space of low dimension. We plot the variance of each component and the cumulative variance.

\tutorialbarrier
\begin{codecell}

\end{codecell}
\tutorialplot{\includegraphics[width=6.500in,height=0.75\textheight,keepaspectratio]{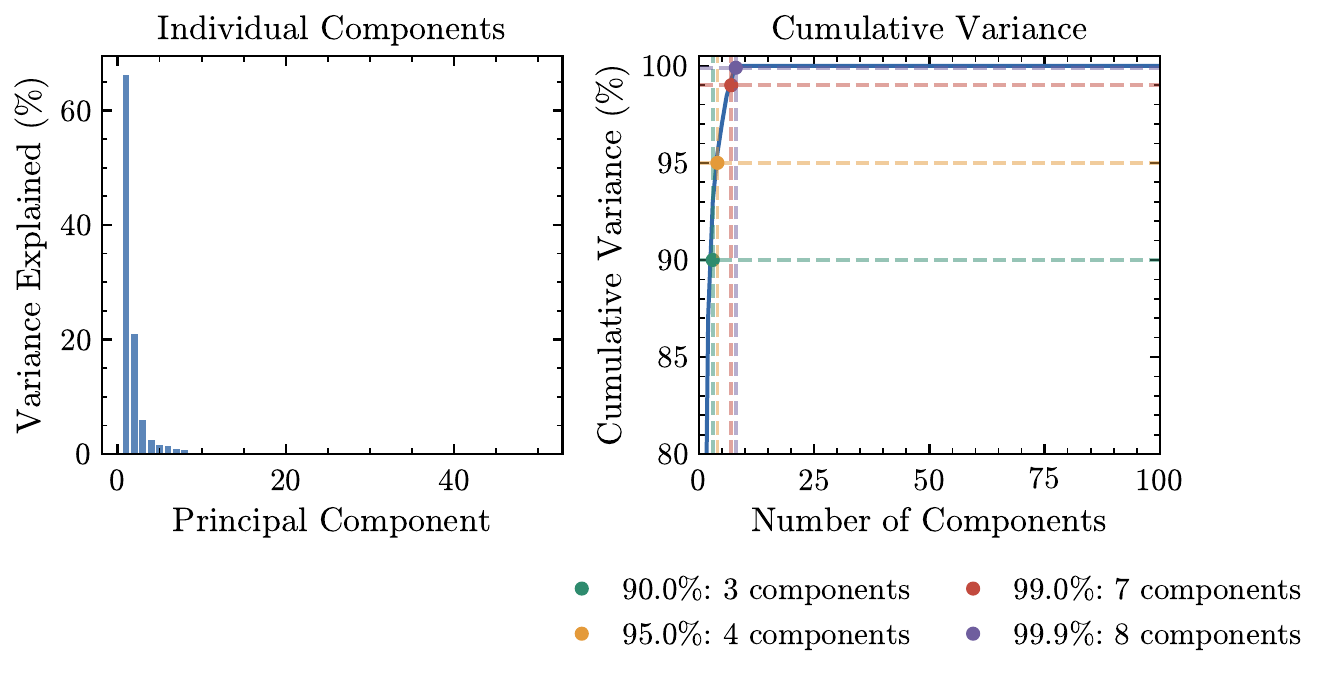}}%
  {\includegraphics[height=\tutorialheight]{tutorials/11b/fig02.pdf}}

The variance is zero after component 8, so these spectra lie in an eight-dimensional space. This holds by construction, because the model has no noise or instrumental effects. In real spectra, noise, instrumental effects and further variation add variance to more components, usually in small shares.

\subsection*{Reconstructing a Spectrum from a Few Components}

How many components do we need? The explained variance is an average over all spectra. The variance of the discarded components is the average squared reconstruction error (Section 10.1), but one spectrum can have a larger error than the average. We therefore rebuild every spectrum from the mean and its first \(M\) coefficients, with the reconstruction formula above. We print the largest absolute error over all spectra and wavelengths for \(M = 1\), 3, 5 and 8.

\tutorialbarrier
\begin{codecell}

\end{codecell}
\begin{Verbatim}[fontsize=\footnotesize, xleftmargin=5mm]
M = 1: largest absolute error = 2.01
M = 3: largest absolute error = 0.784
M = 5: largest absolute error = 0.632
M = 8: largest absolute error = 1.07e-06
\end{Verbatim}

We now look at one spectrum, spectrum 100, and its reconstructions for \(M = 1\), 3, 5 and 8. Dashed lines mark Ly\(\alpha\) and C IV. The blue area is the residual, the spectrum minus its reconstruction. It uses the right axis, which has the same limits in all panels.

\tutorialbarrier
\begin{codecell}

\end{codecell}
\tutorialplot{\includegraphics[width=6.173in,height=0.75\textheight,keepaspectratio]{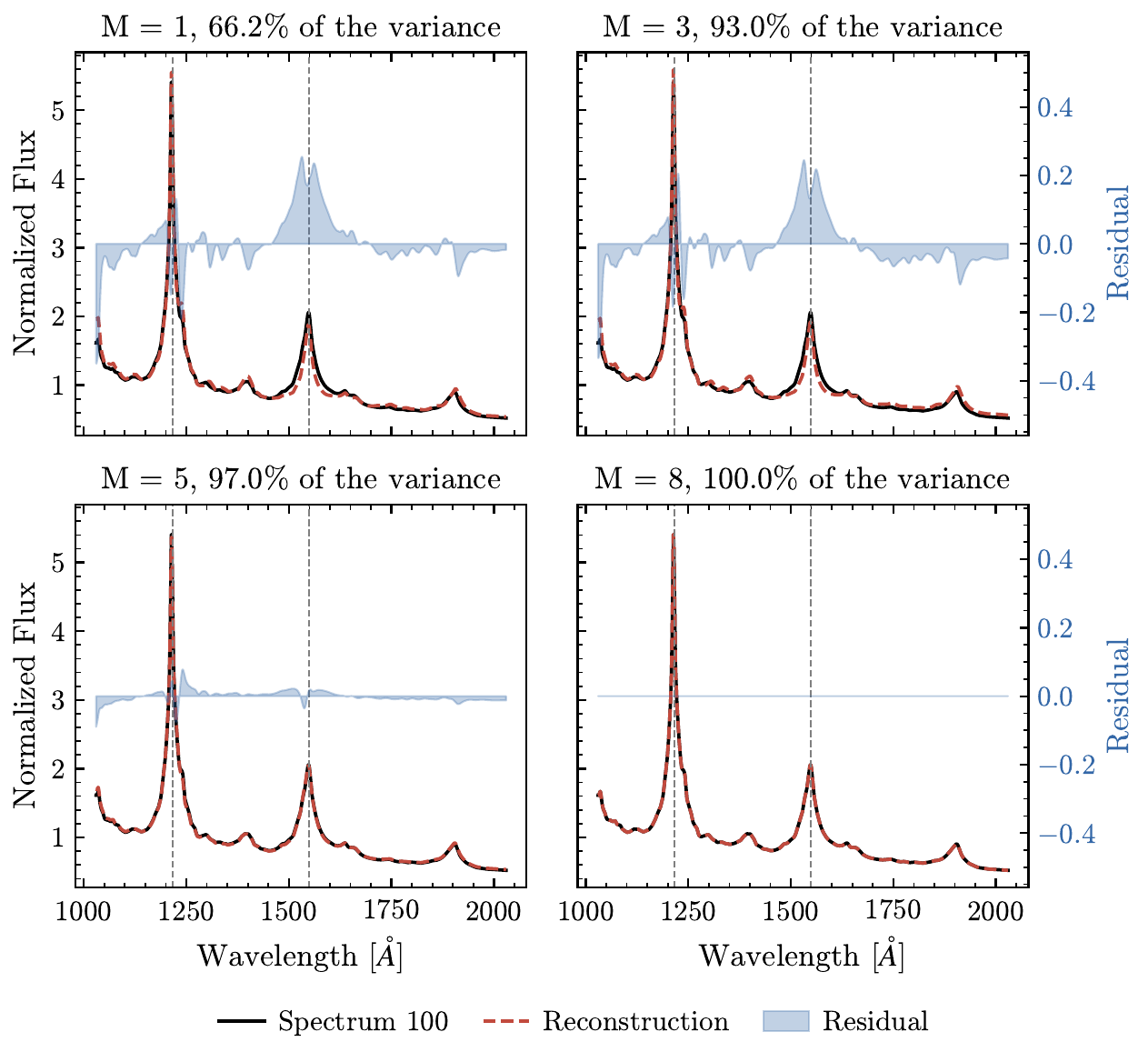}}%
  {\includegraphics[height=\tutorialheight]{tutorials/11b/fig03.pdf}}

With one component, the reconstruction follows the Ly\(\alpha\) line but leaves a residual near C IV. Three components carry 93 percent of the variance, yet the residual near C IV hardly changes. With five components the residual of this spectrum is small. The printed errors show that the worst spectrum still has a large error for five components. Eight components rebuild every flux of every spectrum to numerical precision, so the residual is a flat line. We therefore keep the first eight components in the rest of the tutorial.

\subsection*{Reading the Principal Component Structure}

The principal components are directions in the space of the wavelengths. We plot the mean spectrum and the first three components to see what they describe. Dashed lines mark the five emission lines of the first plot. The shading marks the positive part of a component in its color and the negative part in red.

\tutorialbarrier
\begin{codecell}

\end{codecell}
\tutorialplot{\includegraphics[width=6.179in,height=0.75\textheight,keepaspectratio]{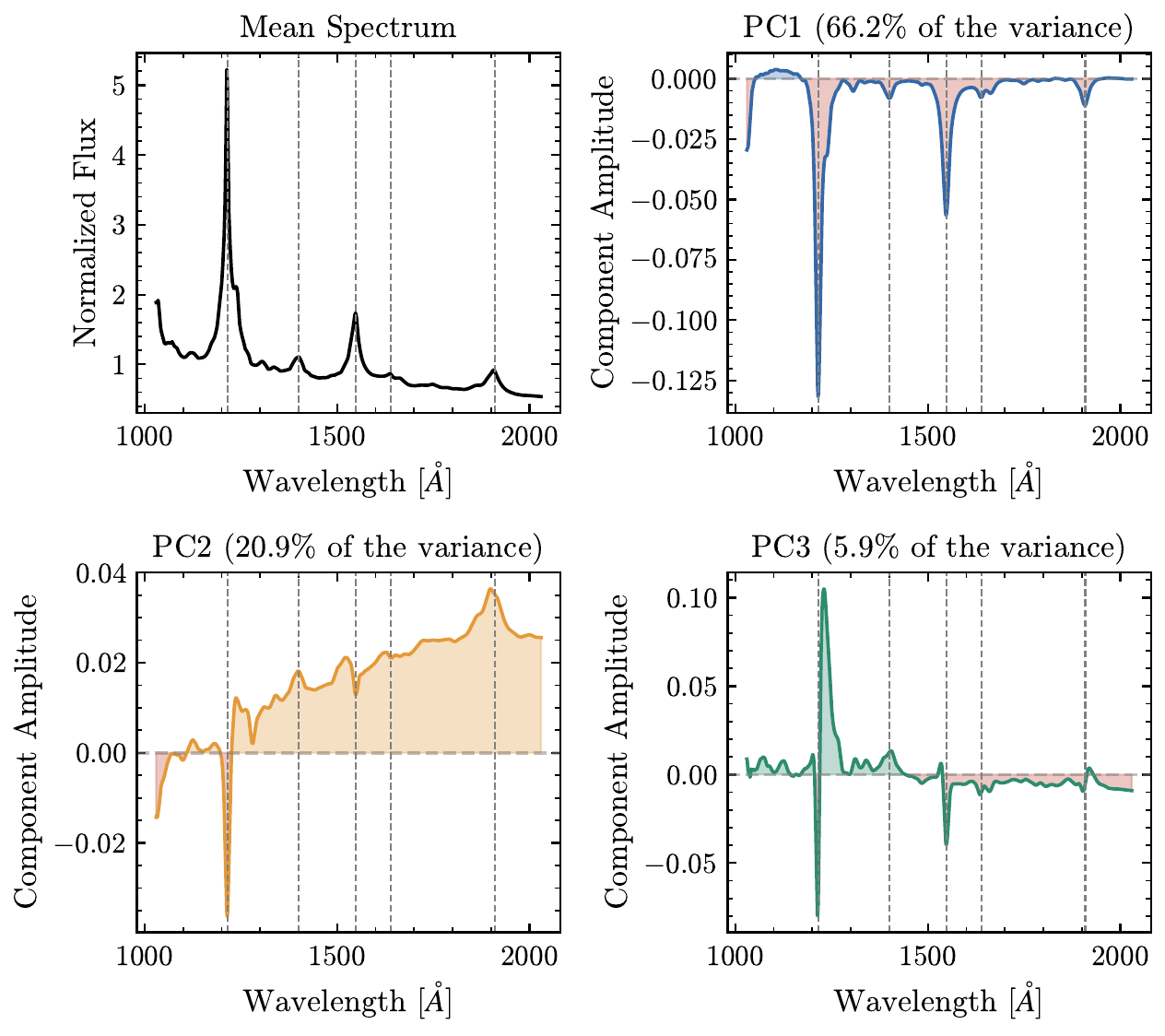}}%
  {\includegraphics[height=\tutorialheight]{tutorials/11b/fig04.pdf}}

A component is a direction in flux space, and it can mix several effects of the model that made the spectra. The sign of a component is arbitrary, so the readings of the signs below hold for the components as plotted.

\textbf{Mean spectrum.} It shows the average quasar spectrum, with the Ly\(\alpha\) and C IV lines.

\textbf{PC1.} It has negative values at all emission lines, with the lowest values at Ly\(\alpha\) and C IV. It resembles a change of the overall line strength relative to the rest of the spectrum. A quasar with a positive PC1 coefficient has weaker lines, and a quasar with a negative coefficient has stronger lines. A change of the ionization state of the gas or of its covering factor could cause this. The covering factor is the fraction of the sky around the black hole that the gas covers.

\textbf{PC2.} It rises toward long wavelengths, so it resembles a change of the overall slope of the spectrum. A quasar with a positive PC2 coefficient is redder, and a quasar with a negative coefficient is bluer. The accretion rate, the black hole mass or the viewing angle could change the slope.

\textbf{PC3.} It has a positive peak on the red side of Ly\(\alpha\) and negative values at Ly\(\alpha\) and at C IV. It changes the shape of the Ly\(\alpha\) line and the strength of the C IV line. The remaining components carry progressively smaller variations.

\section*{Gaussian Mixture Models in PCA Space}

Each spectrum is now described by eight coefficients, and we model their distribution. We project all spectra onto the first eight components and split them at random into 800 training spectra and 200 test spectra. BIC uses the training spectra only (Tutorial 11a), so they fit the mixture and choose its number of components. The 200 test spectra check the final mixture. The PCA above used all spectra.

\tutorialbarrier
\begin{codecell}

\end{codecell}
\begin{Verbatim}[fontsize=\footnotesize, xleftmargin=5mm]
Training spectra = 800, test spectra = 200
\end{Verbatim}

\subsection*{Gaussian Mixtures and Model Selection}

We model the coefficients \(\mathbf{z}\) with the mixture of Tutorial 11a, now in \(M = 8\) dimensions. Each mean has 8 entries and each covariance matrix is \(8 \times 8\) (Section 11.5). We reuse the code of Tutorial 11a, so that this notebook runs on its own. The mixture starts from a K-means run with K-means++ seeds, so we first write the functions of K-means (Section 11.2).

\tutorialbarrier
\begin{codecell}

\end{codecell}

Next come the E-step and the M-step of the mixture (Section 11.6). The start takes the weights, means and covariance matrices from the K-means clusters. EM then repeats the two steps until the log-likelihood gains less than a tolerance.

\tutorialbarrier
\begin{codecell}

\end{codecell}

We choose the number of components \(K\) with the Bayesian information criterion (BIC), as in Tutorial 11a (Section 11.8). We reuse its function for the number of parameters and the criteria.

\tutorialbarrier
\begin{codecell}

\end{codecell}

EM can end in a local maximum, so we fit each \(K\) from 10 K-means++ starts and keep the start with the highest log-likelihood.

\tutorialbarrier
\begin{codecell}

\end{codecell}

We plot BIC and mark its lowest value.

\tutorialbarrier
\begin{codecell}

\end{codecell}
\tutorialplot{\includegraphics[width=4.396in,height=0.75\textheight,keepaspectratio]{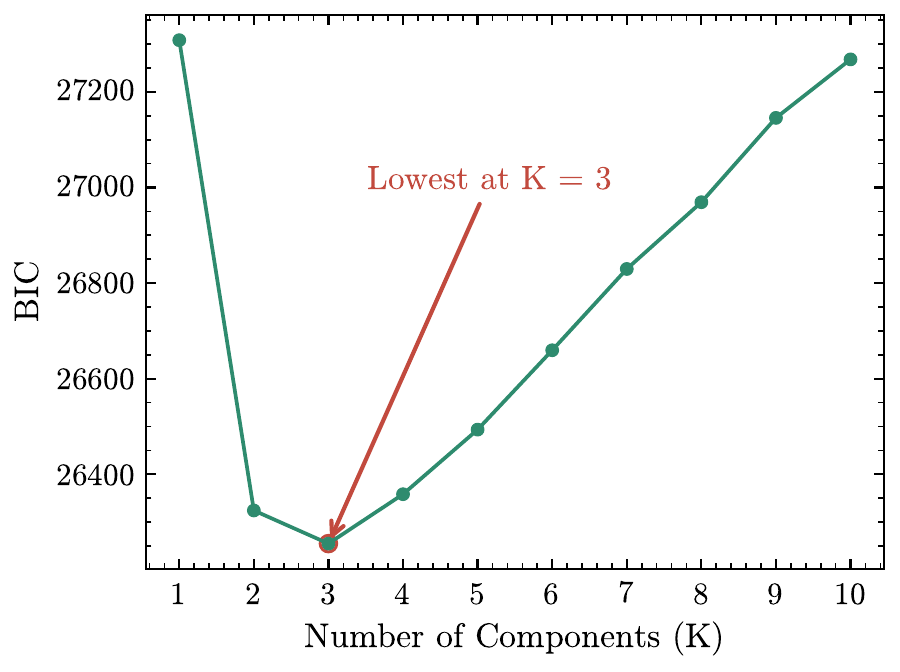}}%
  {\includegraphics[height=\tutorialheight]{tutorials/11b/fig05.pdf}}

BIC is lowest at \(K = 3\), so we use three components. More components raise the log-likelihood, but not enough to pay for their extra parameters.

We check this choice on the test spectra. A mixture that overfits the training spectra gives a lower mean log-likelihood per spectrum for the test spectra than for the training spectra. We plot both values for every \(K\).

\tutorialbarrier
\begin{codecell}

\end{codecell}
\tutorialplot{\includegraphics[width=4.393in,height=0.75\textheight,keepaspectratio]{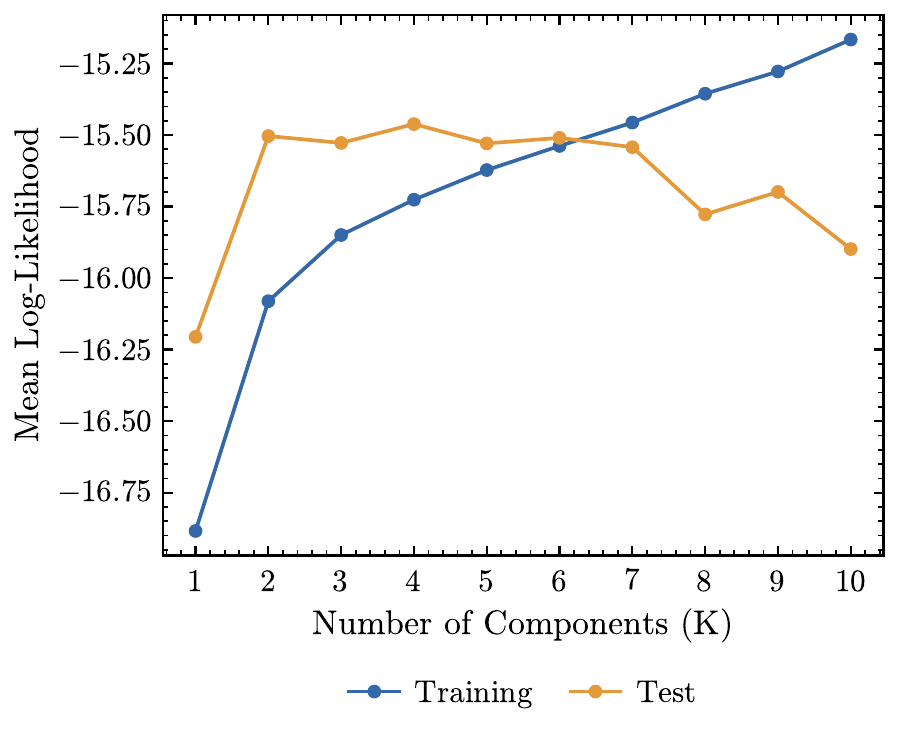}}%
  {\includegraphics[height=\tutorialheight]{tutorials/11b/fig06.pdf}}

The training value keeps rising with \(K\), while the test value stops rising at about \(K = 3\) and falls for \(K\) above 7. The test spectra therefore agree with the three components of BIC.

Eight coefficients cannot be drawn at once, so we plot three planes of the coefficients with the largest variance. The color shows the most probable component of each training spectrum, and the black crosses mark the component means.

\tutorialbarrier
\begin{codecell}

\end{codecell}
\tutorialplot{\includegraphics[width=6.500in,height=0.75\textheight,keepaspectratio]{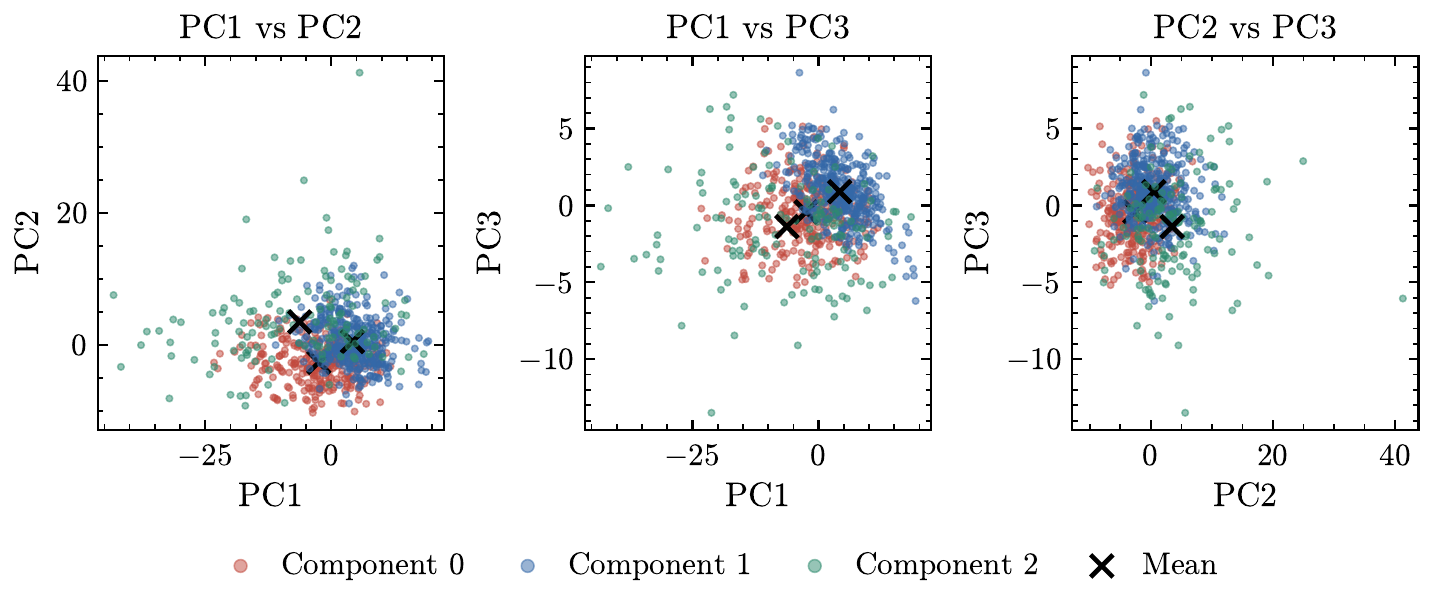}}%
  {\includegraphics[height=\tutorialheight]{tutorials/11b/fig07.pdf}}

The three components overlap a lot in these planes, partly because we project eight dimensions onto two. They do not form distinct clusters, which fits a continuous variation of the spectra. We read each component as a region of the distribution. The distribution has heavier tails than one Gaussian, and the broadest component holds the spectra far from the center. We only need the broad trend of the distribution to draw samples from it, and three components give it. A criterion that rewards fit more, such as AIC, would keep splitting the overlapping cloud into smaller pieces.

\section*{Generating Synthetic Quasar Spectra}

The fitted mixture is a model of the whole distribution, so we can draw synthetic spectra from it (Section 11.7). They can test pipelines for spectroscopic analysis, and they show whether the eight coefficients keep the main features of the spectra. A draw has two stages. First we draw a component \(k\) with the probability \(\pi_k\), and then we draw \(\mathbf{z} \sim \mathcal{N}(\boldsymbol{\mu}_k, \boldsymbol{\Sigma}_k)\).

We map the coefficients back to a spectrum with the reconstruction of the PCA section, \(\widehat{\mathbf{x}} = \mathbf{z}\mathbf{V}_8^T + \overline{\mathbf{x}}\). The synthetic spectra then keep the correlations between wavelengths that the components encode.

We draw 1000 synthetic spectra.

\tutorialbarrier
\begin{codecell}
\begin{Verbatim}[commandchars=\\\{\},fontsize=\small]
\PYG{c+c1}{\PYGZsh{} Seed for the draw, so that every run of the notebook gives the same spectra}
\PYG{n}{np}\PYG{o}{.}\FancyVerbBreakAfterBreak{}\PYG{n}{random}\PYG{o}{.}\FancyVerbBreakAfterBreak{}\PYG{n}{seed}\PYG{p}{(}\FancyVerbBreakAfterBreak{}\PYG{l+m+mi}{0}\PYG{p}{)}
\PYG{n}{n\PYGZus{}synthetic} \PYG{o}{=}\FancyVerbBreakAfterBreak{} \PYG{l+m+mi}{1000}

\PYG{c+c1}{\PYGZsh{} Stage 1, the component of each synthetic spectrum}
\PYG{n}{component\PYGZus{}labels} \PYG{o}{=}\FancyVerbBreakAfterBreak{} \PYG{n}{np}\PYG{o}{.}\FancyVerbBreakAfterBreak{}\PYG{n}{random}\PYG{o}{.}\FancyVerbBreakAfterBreak{}\PYG{n}{choice}\PYG{p}{(}\FancyVerbBreakAfterBreak{}\PYG{n}{k\PYGZus{}bic}\PYG{p}{,}\FancyVerbBreakAfterBreak{} \PYG{n}{size}\PYG{o}{=}\FancyVerbBreakAfterBreak{}\PYG{n}{n\PYGZus{}synthetic}\PYG{p}{,}\FancyVerbBreakAfterBreak{} \PYG{n}{p}\PYG{o}{=}\FancyVerbBreakAfterBreak{}\PYG{n}{weights}\PYG{p}{)}

\PYG{c+c1}{\PYGZsh{} Stage 2, a Gaussian draw in the PCA space from the component of each spectrum}
\PYG{n}{synthetic\PYGZus{}pca} \PYG{o}{=}\FancyVerbBreakAfterBreak{} \PYG{n}{np}\PYG{o}{.}\FancyVerbBreakAfterBreak{}\PYG{n}{array}\PYG{p}{(}\FancyVerbBreakAfterBreak{}\PYG{p}{[}\FancyVerbBreakAfterBreak{}\PYG{n}{np}\PYG{o}{.}\FancyVerbBreakAfterBreak{}\PYG{n}{random}\PYG{o}{.}\FancyVerbBreakAfterBreak{}\PYG{n}{multivariate\PYGZus{}normal}\PYG{p}{(}\FancyVerbBreakAfterBreak{}\PYG{n}{means}\PYG{p}{[}\FancyVerbBreakAfterBreak{}\PYG{n}{k}\PYG{p}{]}\PYG{p}{,}\FancyVerbBreakAfterBreak{} \PYG{n}{covariances}\PYG{p}{[}\FancyVerbBreakAfterBreak{}\PYG{n}{k}\PYG{p}{]}\PYG{p}{)}
                          \PYG{k}{for} \PYG{n}{k} \PYG{o+ow}{in} \PYG{n}{component\PYGZus{}labels}\PYG{p}{]}\PYG{p}{)}

\PYG{c+c1}{\PYGZsh{} Back to the space of the spectra}
\PYG{n}{synthetic\PYGZus{}spectra} \PYG{o}{=}\FancyVerbBreakAfterBreak{} \PYG{n}{synthetic\PYGZus{}pca} \PYG{o}{@} \PYG{n}{components}\PYG{p}{[}\FancyVerbBreakAfterBreak{}\PYG{p}{:}\PYG{n}{n\PYGZus{}components\PYGZus{}pca}\PYG{p}{]} \PYG{o}{+}\FancyVerbBreakAfterBreak{} \PYG{n}{X\PYGZus{}mean}
\end{Verbatim}
\end{codecell}

\subsection*{Validating Generation}

We compare the synthetic spectra with the training spectra, which the mixture was fitted to. The test check showed no overfit, so they are a fair reference. At each wavelength we use the median and the range from the 16th to the 84th percentile. This range holds the middle two thirds of the spectra.

\tutorialbarrier
\begin{codecell}
\begin{Verbatim}[commandchars=\\\{\},fontsize=\small]
\PYG{c+c1}{\PYGZsh{} 16th, 50th and 84th percentile at each wavelength}
\PYG{n}{train\PYGZus{}lower}\PYG{p}{,}\FancyVerbBreakAfterBreak{} \PYG{n}{train\PYGZus{}median}\PYG{p}{,}\FancyVerbBreakAfterBreak{} \PYG{n}{train\PYGZus{}upper} \PYG{o}{=}\FancyVerbBreakAfterBreak{} \PYG{n}{np}\PYG{o}{.}\FancyVerbBreakAfterBreak{}\PYG{n}{percentile}\PYG{p}{(}\FancyVerbBreakAfterBreak{}\PYG{n}{X}\PYG{p}{[}\FancyVerbBreakAfterBreak{}\PYG{n}{train\PYGZus{}indices}\PYG{p}{]}\PYG{p}{,}\FancyVerbBreakAfterBreak{} \PYG{p}{[}\FancyVerbBreakAfterBreak{}\PYG{l+m+mi}{16}\PYG{p}{,}\FancyVerbBreakAfterBreak{} \PYG{l+m+mi}{50}\PYG{p}{,}\FancyVerbBreakAfterBreak{} \PYG{l+m+mi}{84}\PYG{p}{]}\PYG{p}{,}\FancyVerbBreakAfterBreak{} \PYG{n}{axis}\PYG{o}{=}\FancyVerbBreakAfterBreak{}\PYG{l+m+mi}{0}\PYG{p}{)}
\PYG{n}{synthetic\PYGZus{}lower}\PYG{p}{,}\FancyVerbBreakAfterBreak{} \PYG{n}{synthetic\PYGZus{}median}\PYG{p}{,}\FancyVerbBreakAfterBreak{} \PYG{n}{synthetic\PYGZus{}upper} \PYG{o}{=}\FancyVerbBreakAfterBreak{} \PYG{n}{np}\PYG{o}{.}\FancyVerbBreakAfterBreak{}\PYG{n}{percentile}\PYG{p}{(}\FancyVerbBreakAfterBreak{}
    \PYG{n}{synthetic\PYGZus{}spectra}\PYG{p}{,}\FancyVerbBreakAfterBreak{} \PYG{p}{[}\FancyVerbBreakAfterBreak{}\PYG{l+m+mi}{16}\PYG{p}{,}\FancyVerbBreakAfterBreak{} \PYG{l+m+mi}{50}\PYG{p}{,}\FancyVerbBreakAfterBreak{} \PYG{l+m+mi}{84}\PYG{p}{]}\PYG{p}{,}\FancyVerbBreakAfterBreak{} \PYG{n}{axis}\PYG{o}{=}\FancyVerbBreakAfterBreak{}\PYG{l+m+mi}{0}\PYG{p}{)}
\end{Verbatim}
\end{codecell}

We plot the two bands and their medians.

\tutorialbarrier
\begin{codecell}

\end{codecell}
\tutorialplot{\includegraphics[width=5.280in,height=0.75\textheight,keepaspectratio]{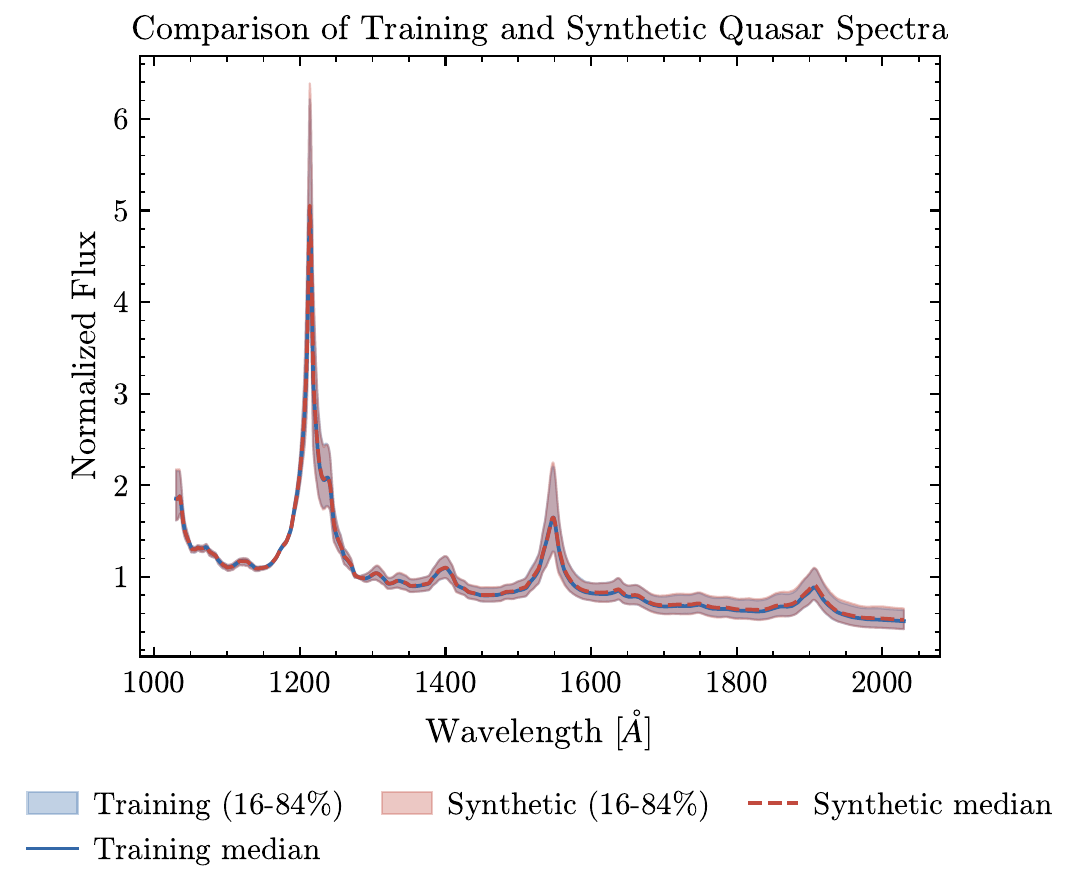}}%
  {\includegraphics[height=\tutorialheight]{tutorials/11b/fig08.pdf}}

The medians overlap and the two ranges match, so the mixture reproduces the center and the spread of the training spectra. The synthetic spectra keep the Ly\(\alpha\) and C IV lines and the overall slope.

We now look at five synthetic spectra that we choose at random.

\tutorialbarrier
\begin{codecell}

\end{codecell}
\tutorialplot{\includegraphics[width=4.970in,height=0.75\textheight,keepaspectratio]{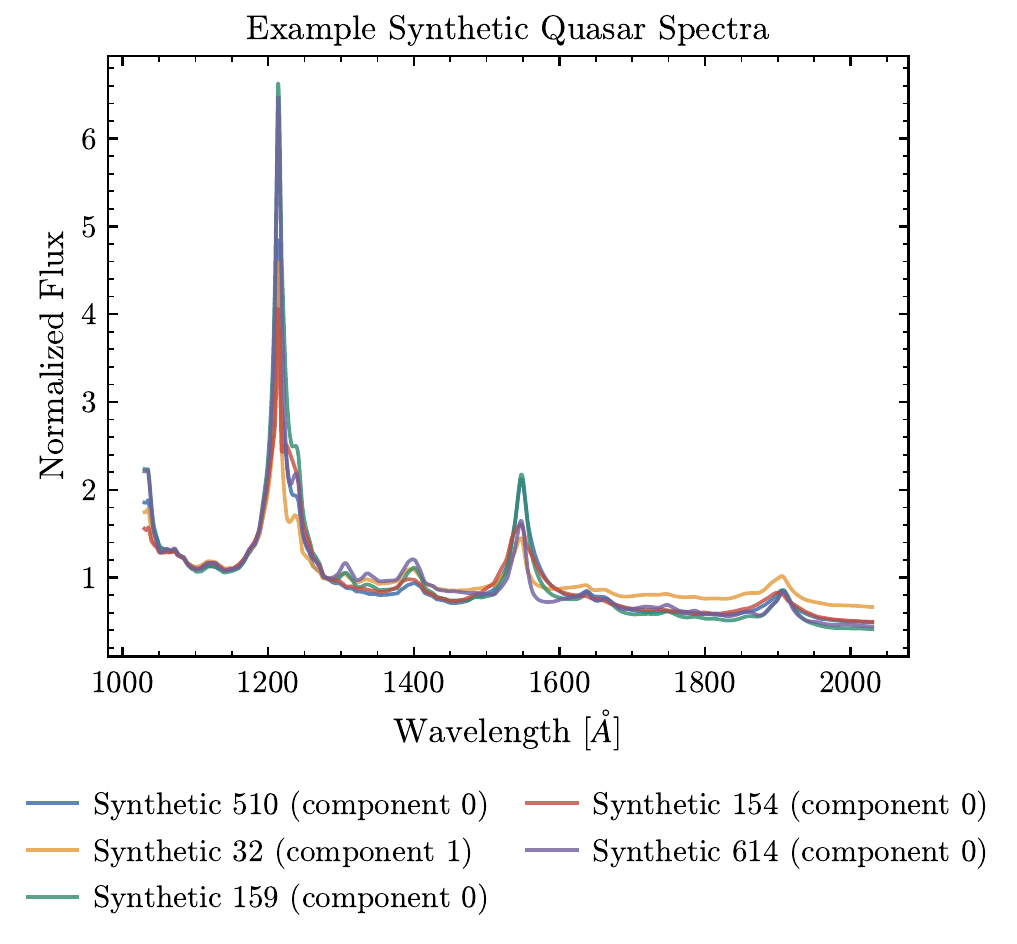}}%
  {\includegraphics[height=\tutorialheight]{tutorials/11b/fig09.pdf}}

The five spectra differ in their line strengths and slopes, as the training spectra do. The legend gives the component of each spectrum.

\section*{Outlier Detection with Gaussian Mixture Models}

The mixture also gives the density \(p(\mathbf{z})\) of any spectrum (Section 11.7). A spectrum with a low density lies in a region that the model finds unlikely, so it is a candidate outlier. We use the log-likelihood \(\ln p(\mathbf{z})\), the logarithm of the density. This differs from the distance to the nearest component mean, because the density takes the covariance into account. A spectrum can be far from every mean and still have a high density when it lies along a direction of large variance.

\subsection*{Implementing GMM-Based Outlier Detection}

We compute the log-likelihood of all 1000 spectra. The threshold is our choice. We flag the 1 percent of the spectra with the lowest log-likelihood, so the method always flags the same share.

\tutorialbarrier
\begin{codecell}

\end{codecell}
\begin{Verbatim}[fontsize=\footnotesize, xleftmargin=5mm]
Log-likelihood threshold = -25.32
\end{Verbatim}

We plot where the outliers lie in the space of the coefficients, relative to the components of the mixture. Each plane projects eight dimensions onto two. An outlier can therefore lie inside the cloud in one plane and still be unusual in other coefficients.

\tutorialbarrier
\begin{codecell}

\end{codecell}
\tutorialplot{\includegraphics[width=6.500in,height=0.75\textheight,keepaspectratio]{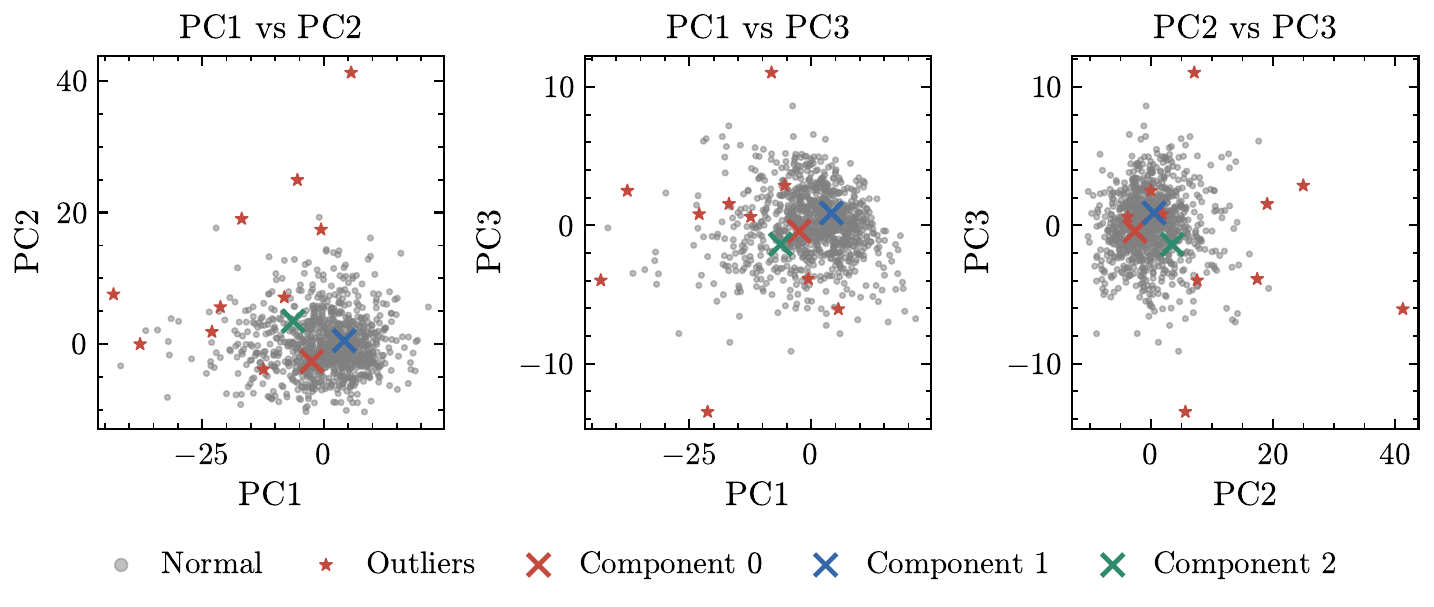}}%
  {\includegraphics[height=\tutorialheight]{tutorials/11b/fig10.pdf}}

Most outliers (red stars) lie in the tails of the distribution, away from the cores of the components, where the model gives a low density. They appear at different places in the different planes, so they can be unusual along several dimensions at the same time.

\subsection*{Inspecting Outlier Spectra}

To see what makes these spectra unusual, we plot four outliers against a band of the normal spectra, which are all spectra except the outliers. The band holds 90 percent of them at each wavelength, from the 5th to the 95th percentile.

\tutorialbarrier
\begin{codecell}
\begin{Verbatim}[commandchars=\\\{\},fontsize=\small]
\PYG{c+c1}{\PYGZsh{} Band of the normal spectra, 5th, 50th and 95th percentile at each wavelength}
\PYG{n}{normal\PYGZus{}lower}\PYG{p}{,}\FancyVerbBreakAfterBreak{} \PYG{n}{normal\PYGZus{}median}\PYG{p}{,}\FancyVerbBreakAfterBreak{} \PYG{n}{normal\PYGZus{}upper} \PYG{o}{=}\FancyVerbBreakAfterBreak{} \PYG{n}{np}\PYG{o}{.}\FancyVerbBreakAfterBreak{}\PYG{n}{percentile}\PYG{p}{(}\FancyVerbBreakAfterBreak{}\PYG{n}{X}\PYG{p}{[}\FancyVerbBreakAfterBreak{}\PYG{o}{\PYGZti{}}\PYG{n}{is\PYGZus{}outlier}\PYG{p}{]}\PYG{p}{,}\FancyVerbBreakAfterBreak{} \PYG{p}{[}\FancyVerbBreakAfterBreak{}\PYG{l+m+mi}{5}\PYG{p}{,}\FancyVerbBreakAfterBreak{} \PYG{l+m+mi}{50}\PYG{p}{,}\FancyVerbBreakAfterBreak{} \PYG{l+m+mi}{95}\PYG{p}{]}\PYG{p}{,}\FancyVerbBreakAfterBreak{} \PYG{n}{axis}\PYG{o}{=}\FancyVerbBreakAfterBreak{}\PYG{l+m+mi}{0}\PYG{p}{)}
\end{Verbatim}
\end{codecell}

We plot the first four outliers by index. Dashed lines mark Ly\(\alpha\) and C IV. The panel titles give their coefficients on PC1 and PC2. Above, PC1 resembled a change of the line strength and PC2 a change of the overall slope.

\tutorialbarrier
\begin{codecell}

\end{codecell}
\tutorialplot{\includegraphics[width=6.175in,height=0.75\textheight,keepaspectratio]{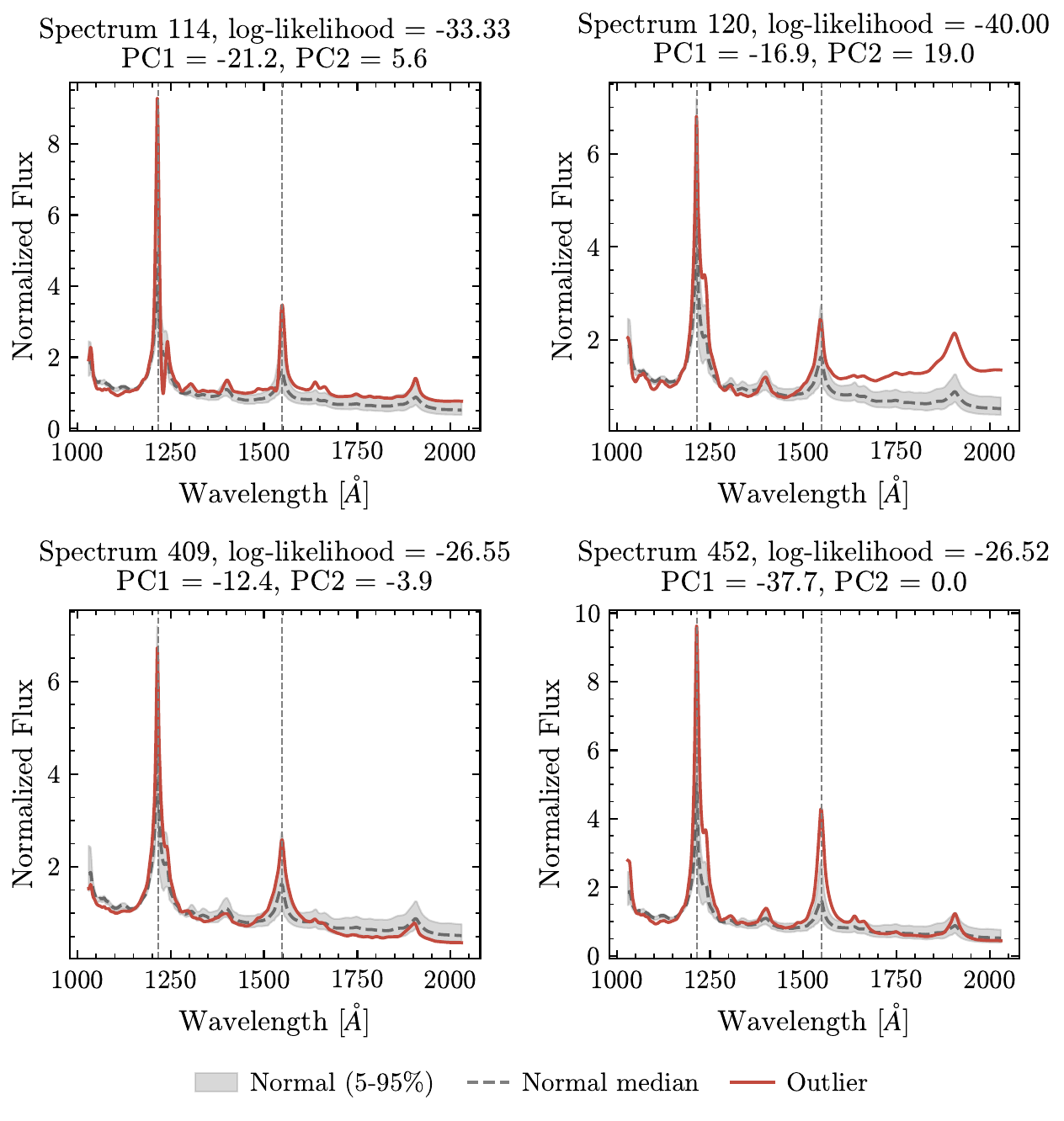}}%
  {\includegraphics[height=\tutorialheight]{tutorials/11b/fig11.pdf}}

The outlier spectra differ from the normal population, in particular in their emission lines. In all four, the Ly\(\alpha\) line is stronger than the median of the normal spectra. Spectra 114 and 452 have Ly\(\alpha\) peaks above the band and a strong C IV line at 1549 Å. Spectrum 120 lies above the band at long wavelengths, and spectrum 409 lies below it. The outliers therefore differ in the line strengths and in some cases in the overall slope.

The coefficients in the titles agree with the readings of the components. All four spectra have a negative PC1 coefficient, which the reading of PC1 links to stronger lines. Spectrum 120 has a large positive PC2 coefficient, which the reading of PC2 links to a redder spectrum.

The planes above show the first three coefficients only. Spectrum 409 lies inside the cloud in them. To see what makes it unusual, we print the coefficients of the four outliers in units of the standard deviation of each coefficient.

\tutorialbarrier
\begin{codecell}

\end{codecell}
\begin{Verbatim}[fontsize=\footnotesize, xleftmargin=5mm]
Spectrum   PC1   PC2   PC3   PC4   PC5   PC6   PC7   PC8
     114  -2.5   1.2  -5.3  -2.7  -1.7  -0.8  -3.4   6.7
     120  -2.0   4.0   0.6  -2.5   1.1   9.6  -0.2  -3.7
     409  -1.5  -0.8   0.2   3.2   2.9   0.5  -3.6   2.5
     452  -4.5   0.0   1.0   2.8  -0.8   0.0  -2.0   2.5
\end{Verbatim}

Spectrum 409 is typical on PC1 to PC3 and deviates by about three standard deviations on PC4, PC5 and PC7. A spectrum can therefore look typical in the first three coefficients and still be unusual in others. Spectrum 120 deviates most on PC6, far more than any other entry, and we do not interpret that component.

In real data, such spectra could come from unusual gas around the black hole, for example with an extreme covering factor or ionization state. Other causes are unusual viewing angles, problems of the data processing and rare objects that deserve a follow-up study. The likelihood threshold only flags the spectra. It does not tell us why they are unusual.

\section*{Summary}

This tutorial combined PCA with a Gaussian mixture to model the distribution of spectra, which have a high dimension. Our main lessons follow.

\begin{enumerate}
\def\labelenumi{\arabic{enumi}.}
\tightlist
\item
  \textbf{Compression before modeling.} A full covariance matrix in thousands of dimensions has far more parameters than we have objects. PCA first describes each object with a few coefficients that carry most of the variance. The explained variance ratio and the error of the reconstruction guide how many to keep.
\item
  \textbf{Principal components as modes of variation.} The components are directions of variation, ordered by the variance that they carry. A component can mix several effects and its sign is arbitrary, so a physical reading is a hypothesis to test.
\item
  \textbf{Generative model.} A Gaussian mixture fitted with EM is a generative model of the coefficients, and BIC guides the number of components.
\item
  \textbf{Generation.} We draw a component, then a Gaussian sample, and map it back with the inverse of the PCA projection. This gives synthetic objects that follow the fitted distribution, and comparing their statistics with the data tests the model.
\item
  \textbf{Outlier detection.} The likelihood under the mixture measures how typical an object is, and a threshold on it flags objects for follow-up. The threshold is a choice, and the likelihood does not say why an object is unusual.
\end{enumerate}

A mixture in the space of PCA coefficients is a probability model of a whole population of objects with a high dimension. It tells us how typical an object is and what a new sample would look like.

\tutorialchapter{12}{Monte Carlo Sampling for Stellar Masses}
\noindent\textit{Companion material for Chapter 12}

\section*{Introduction}

In earlier chapters we treated posterior distributions in closed form for simple models (Chapter 5) and with approximations for harder ones (Chapter 9). Many problems give a distribution that we cannot treat analytically, whether we estimate a posterior, compute predictions or simulate an astronomical phenomenon. This tutorial uses Monte Carlo methods, which replace analytical integration by random sampling. They turn such problems into calculations that a computer can do. The main example of this tutorial is the generation of a stellar population.

The initial mass function (IMF) is the distribution of stellar masses at birth. A star's mass sets its luminosity, its lifetime and the elements that it returns to the gas. The IMF therefore sets how a stellar population shines and ages. It also affects the numbers of white dwarfs, neutron stars and black holes, and the enrichment of the gas. Planet formation and the dynamics of star clusters depend on the IMF as well.

We cannot observe the IMF at birth, so we infer it from observed populations. To test an IMF model, we simulate a population from it and compare that population with observations. This tutorial uses the Kroupa IMF, which Pavel Kroupa introduced in 2001. It summarizes observations of star-forming regions, open clusters and the solar neighborhood as a power law in several mass ranges. Some stellar population synthesis models, which compute the light of a population of stars, adopt it, and the form of the IMF is still debated.

The need to sample from complex distributions reaches beyond stellar masses. One example is the propagation of measurement uncertainties through the non-linear transformations that convert observed fluxes into physical luminosities. Another is the simulation of the photon arrival times of a variable star. The photons of even one star arrive at random times. Their rate changes with time, so a simulated light curve needs random times from that time-dependent rate.

Monte Carlo methods get their name from the casino in Monaco, because they rely on random sampling. By the law of large numbers, sample averages converge to the true expected values. We can therefore learn about a distribution by drawing samples from it. This idea underlies many modern statistical methods.

The tutorial follows the example through three sampling methods of Chapter 12.

\begin{itemize}
\tightlist
\item
  Inverse transform sampling turns uniform random numbers into samples with the cumulative distribution function (CDF). It needs a CDF that we can invert.
\item
  Rejection sampling needs only the density of the distribution and no inversion.
\item
  Importance sampling estimates a property of the population, such as its mean mass, from weighted samples of a simpler distribution.
\end{itemize}

\section*{Prerequisites and Learning Objectives}

\subsection*{Prerequisites}

This tutorial assumes the following.

\begin{itemize}
\tightlist
\item
  Probability distributions and expectations, with probability density functions and the sample mean (Chapters 2 and 3)
\item
  Calculus, with integration and the inversion of functions
\item
  Python programming, with NumPy arrays, matplotlib plots and SciPy functions
\end{itemize}

\subsection*{Learning Objectives}

After this tutorial you can do the following.

\begin{itemize}
\tightlist
\item
  Transform uniform random numbers into samples from a distribution with inverse transform sampling
\item
  Implement rejection sampling with a simple proposal distribution
\item
  Estimate an expectation with importance sampling and weighted samples
\item
  Judge a proposal distribution with the acceptance rate and the effective sample size (ESS)
\item
  Check samples and estimates against the exact distribution and mean
\end{itemize}

We begin by importing the libraries and setting the plot style.

\tutorialbarrier
\begin{codecell}

\end{codecell}

\section*{Inverse Transform Sampling: From Uniform Numbers to Samples}

Monte Carlo simulations need random numbers that follow specific probability distributions. Modern programming languages provide generators for common distributions. Inverse transform sampling is one way to build such a generator. It connects the uniform random numbers that computers generate to a continuous distribution whose CDF we can invert. It is also called the inverse CDF method.

\subsection*{The Probability Integral Transform}

The method rests on the probability integral transform theorem (Section 12.3). Let \(X\) be a continuous random variable with probability density function (PDF) \(p(x)\) and CDF \(F(x)\), the probability that \(X\) is at most \(x\). Then \(U = F(X)\) follows a uniform distribution on \([0,1]\). The converse gives our sampling method. If \(U \sim \mathcal{U}(0,1)\), then \(X = F^{-1}(U)\) follows the distribution with the CDF \(F\).

The CDF maps probability densities to cumulative probabilities, so it ``flattens'' any distribution into the uniform distribution on \([0,1]\). The inverse CDF reverses this process and turns uniform numbers back into samples from the target distribution.

For a continuous distribution with PDF \(p(x)\) the algorithm has three steps.

\begin{enumerate}
\def\labelenumi{\arabic{enumi}.}
\tightlist
\item
  \textbf{Compute the CDF.} \(F(x) = \int_{-\infty}^{x} p(t)\,\mathrm{d}t\).
\item
  \textbf{Derive the inverse CDF.} Solve \(u = F(x)\) for \(x\) to obtain \(x = F^{-1}(u)\).
\item
  \textbf{Generate samples.} For each uniform random number \(u \sim \mathcal{U}(0,1)\), compute \(x = F^{-1}(u)\).
\end{enumerate}

Each uniform random number gives exactly one sample from the target distribution. The method does not reject any sample and does not iterate.

\subsection*{Example: The Exponential Distribution}

We begin with the exponential distribution. It describes waiting times, for example the intervals between photon arrivals from a source of constant brightness. It is a simple first example of inverse transform sampling.

The exponential distribution has the PDF

\[p(x) = \begin{cases} \lambda e^{-\lambda x}, & x \geq 0 \\ 0, & x < 0, \end{cases}\]

where \(\lambda > 0\) sets the scale of the distribution. Larger values of \(\lambda\) concentrate the distribution near zero, and smaller values give more extended distributions.

To apply inverse transform sampling, we first compute the CDF

\[F(x) = \int_{0}^{x} \lambda e^{-\lambda t}\,\mathrm{d}t = 1 - e^{-\lambda x}.\]

Setting \(u = F(x)\) and solving for \(x\) gives the inverse CDF

\[x = F^{-1}(u) = -\frac{1}{\lambda} \ln(1-u).\]

Since \(U\) and \(1-U\) are both uniform on \([0,1]\), we can also use \(-\ln(u)/\lambda\) directly.

We now implement the inverse CDF and draw 10,000 samples with \(\lambda = 1\).

\tutorialbarrier
\begin{codecell}
\begin{Verbatim}[commandchars=\\\{\},fontsize=\small]
\PYG{k}{def}\PYG{+w}{ }\PYG{n+nf}{sample\PYGZus{}exponential}\PYG{p}{(}\FancyVerbBreakAfterBreak{}\PYG{n}{lambda\PYGZus{}param}\PYG{p}{,}\FancyVerbBreakAfterBreak{} \PYG{n}{size}\PYG{o}{=}\FancyVerbBreakAfterBreak{}\PYG{l+m+mi}{1}\PYG{p}{)}\PYG{p}{:}
\PYG{+w}{    }\PYG{l+s+sd}{\PYGZdq{}\PYGZdq{}\PYGZdq{}}
\PYG{l+s+sd}{    Draw samples from an exponential distribution with the inverse CDF method.}

\PYG{l+s+sd}{    lambda\PYGZus{}param is the parameter lambda, and size is the number of samples.}
\PYG{l+s+sd}{    \PYGZdq{}\PYGZdq{}\PYGZdq{}}

    \PYG{c+c1}{\PYGZsh{} Uniform random numbers between 0 and 1}
    \PYG{n}{u} \PYG{o}{=}\FancyVerbBreakAfterBreak{} \PYG{n}{np}\PYG{o}{.}\FancyVerbBreakAfterBreak{}\PYG{n}{random}\PYG{o}{.}\FancyVerbBreakAfterBreak{}\PYG{n}{uniform}\PYG{p}{(}\FancyVerbBreakAfterBreak{}\PYG{n}{size}\PYG{o}{=}\FancyVerbBreakAfterBreak{}\PYG{n}{size}\PYG{p}{)}

    \PYG{c+c1}{\PYGZsh{} Apply the inverse CDF transformation}
    \PYG{n}{samples} \PYG{o}{=}\FancyVerbBreakAfterBreak{} \PYG{o}{\PYGZhy{}}\PYG{p}{(}\FancyVerbBreakAfterBreak{}\PYG{l+m+mi}{1}\PYG{o}{/}\PYG{n}{lambda\PYGZus{}param}\PYG{p}{)} \PYG{o}{*} \PYG{n}{np}\PYG{o}{.}\FancyVerbBreakAfterBreak{}\PYG{n}{log}\PYG{p}{(}\FancyVerbBreakAfterBreak{}\PYG{l+m+mi}{1} \PYG{o}{\PYGZhy{}} \PYG{n}{u}\PYG{p}{)}

    \PYG{k}{return} \PYG{n}{samples}

\PYG{c+c1}{\PYGZsh{} Generate 10,000 samples from an exponential distribution with lambda = 1}
\PYG{n}{lambda\PYGZus{}param} \PYG{o}{=}\FancyVerbBreakAfterBreak{} \PYG{l+m+mi}{1}
\PYG{n}{samples} \PYG{o}{=}\FancyVerbBreakAfterBreak{} \PYG{n}{sample\PYGZus{}exponential}\PYG{p}{(}\FancyVerbBreakAfterBreak{}\PYG{n}{lambda\PYGZus{}param}\PYG{p}{,}\FancyVerbBreakAfterBreak{} \PYG{n}{size}\PYG{o}{=}\FancyVerbBreakAfterBreak{}\PYG{l+m+mi}{10000}\PYG{p}{)}
\end{Verbatim}
\end{codecell}

We now plot the histogram of the samples against the PDF. We also compare the sample mean and variance with the mean \(1/\lambda\) and the variance \(1/\lambda^2\) of the distribution.

\tutorialbarrier
\begin{codecell}

\end{codecell}
\tutorialplot{\includegraphics[width=4.167in,height=0.75\textheight,keepaspectratio]{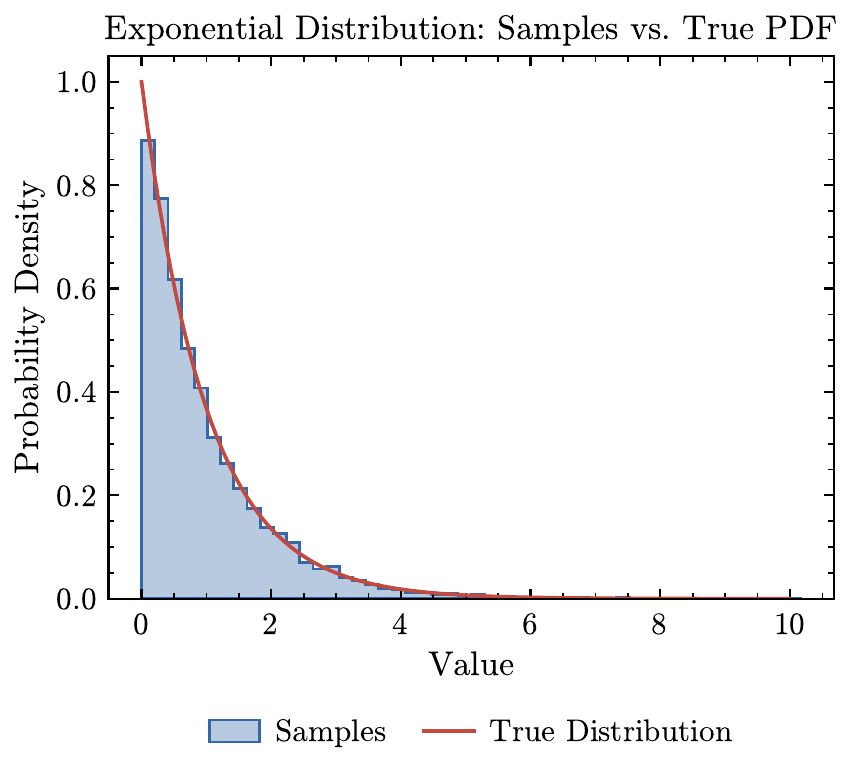}}%
  {\includegraphics[height=\tutorialheight]{tutorials/12/fig01.pdf}}
\begin{Verbatim}[fontsize=\footnotesize, xleftmargin=5mm]
Theoretical mean = 1.0000, sample mean = 0.9941
Theoretical variance = 1.0000, sample variance = 1.0090
\end{Verbatim}

The histogram follows the PDF, and the sample mean and variance are close to \(1/\lambda\) and \(1/\lambda^2\).

\subsection*{Astronomical Application: The Kroupa Initial Mass Function}

We now apply inverse transform sampling to a more complex problem, the generation of stellar masses with the Kroupa IMF.

\subsubsection*{Mathematical Formulation}

The Kroupa IMF is a piecewise power law. We use its three pieces from 0.01 to 1 solar masses and set the density to zero outside this range. The full function continues to higher masses. The probability density of the mass \(m\) is

\[ p(m) =
    \begin{cases}
      K_1 m^{-0.3} & 0.01 \leq m < 0.08 \\
      K_2 m^{-1.3} & 0.08 \leq m < 0.5 \\
      K_3 m^{-2.3} & 0.5 \leq m \leq 1,
   \end{cases}
\]

where the mass is in solar masses (\(\mathrm{M}_\odot\)). The three pieces describe three mass regimes.

\begin{itemize}
\tightlist
\item
  \textbf{Sub-stellar regime} (\(m < 0.08\,\mathrm{M}_\odot\)). The power law has the exponent \(\alpha = 0.3\). Objects in this range do not burn hydrogen and are brown dwarfs.
\item
  \textbf{Low-mass stars} (\(0.08 \leq m < 0.5\,\mathrm{M}_\odot\)). The steeper exponent \(\alpha = 1.3\) applies to low-mass stars that burn hydrogen.
\item
  \textbf{Higher-mass stars} (\(0.5 \leq m \leq 1\,\mathrm{M}_\odot\)). The steepest exponent \(\alpha = 2.3\) means that stars become rarer as the mass grows.
\end{itemize}

The constants \(K_1\), \(K_2\) and \(K_3\) make the IMF continuous at the transition masses and normalize it to unit total probability.

\subsubsection*{Determining the Normalization Constants}

To apply inverse transform sampling to the Kroupa IMF, we must first normalize it. The constants must satisfy two conditions, continuity at the mass boundaries and unit total probability.

The continuity at the boundary between brown dwarfs and hydrogen-burning stars (\(m = 0.08\,\mathrm{M}_\odot\)) gives

\[K_1 \times 0.08^{-0.3} = K_2 \times 0.08^{-1.3},\]

which yields \(K_2 = K_1 \times 0.08\). Similarly, the continuity at the transition to the higher-mass stars (\(m = 0.5\,\mathrm{M}_\odot\)) gives

\[K_2 \times 0.5^{-1.3} = K_3 \times 0.5^{-2.3},\]

which yields \(K_3 = K_2 \times 0.5 = K_1 \times 0.08 \times 0.5\).

To determine \(K_1\), we require that the total probability is one. We integrate each power law over its mass range with numerical integration, and then solve for \(K_1\).

\tutorialbarrier
\begin{codecell}

\end{codecell}
\begin{Verbatim}[fontsize=\footnotesize, xleftmargin=5mm]
K1 = 2.1162, K2 = 0.1693, K3 = 0.0846
\end{Verbatim}

\subsubsection*{Deriving the Inverse CDF}

With the normalization constants known, we can derive the inverse CDF that we need for sampling. We compute the CDF of the Kroupa IMF segment by segment, with the contribution of each mass range.

For the first segment (\(0.01 \leq m < 0.08\,\mathrm{M}_\odot\)) the CDF is

\[ F_1(m) = K_1 \int_{0.01}^{m} t^{-0.3}\,\mathrm{d}t = \frac{K_1}{0.7} \left( m^{0.7} - 0.01^{0.7} \right). \]

For the second segment (\(0.08 \leq m < 0.5\,\mathrm{M}_\odot\)) we include the probability of the first segment

\[ F_2(m) = F_1(0.08) + K_2 \int_{0.08}^{m} t^{-1.3}\,\mathrm{d}t = F_1(0.08) + \frac{K_2}{-0.3} \left( m^{-0.3} - 0.08^{-0.3} \right). \]

Similarly, for the third segment (\(0.5 \leq m \leq 1\,\mathrm{M}_\odot\))

\[ F_3(m) = F_2(0.5) + K_3 \int_{0.5}^{m} t^{-2.3}\,\mathrm{d}t = F_2(0.5) + \frac{K_3}{-1.3} \left( m^{-1.3} - 0.5^{-1.3} \right). \]

The total probability of each segment sets how we partition the uniform random variable.

\begin{itemize}
\tightlist
\item
  \(A = F_1(0.08)\) is the probability of a brown dwarf.
\item
  \(B = F_2(0.5) - F_1(0.08)\) is the probability of a low-mass star.
\item
  \(C = 1 - F_2(0.5)\) is the probability of a mass above \(0.5\,\mathrm{M}_\odot\).
\end{itemize}

To generate samples, we invert the CDF in each segment. For a uniform random variable \(u \in [0,1]\) we proceed as follows.

If \(0 \leq u < A\) (brown dwarf regime), then

\[ m = \left( \frac{0.7u}{K_1} + 0.01^{0.7} \right)^{1/0.7}. \]

If \(A \leq u < A+B\) (low-mass regime), then

\[ m = \left( \frac{-0.3(u-A)}{K_2} + 0.08^{-0.3} \right)^{-1/0.3}. \]

If \(A+B \leq u \leq 1\) (higher-mass regime), then

\[ m = \left( \frac{-1.3(u-A-B)}{K_3} + 0.5^{-1.3} \right)^{-1/1.3}. \]

A uniform random number first selects the segment by its probability and then gives the mass inside that segment.

We now implement the three inverse CDFs in a function.

\tutorialbarrier
\begin{codecell}

\end{codecell}

We now draw 100,000 masses. We also define the Kroupa IMF as a function, so that we can compare the histogram of the samples with the exact density. We print the mean mass, the median mass and the share of stars above 0.5 solar masses.

\tutorialbarrier
\begin{codecell}

\end{codecell}
\begin{Verbatim}[fontsize=\footnotesize, xleftmargin=5mm]
Mean mass = 0.1893 M☉, median mass = 0.1084 M☉
Fraction of stars above 0.5 M☉ = 0.0950
\end{Verbatim}

We now plot the histogram of the masses against the exact density, on a linear scale and on a logarithmic scale.

\tutorialbarrier
\begin{codecell}

\end{codecell}
\tutorialplot{\includegraphics[width=6.173in,height=0.75\textheight,keepaspectratio]{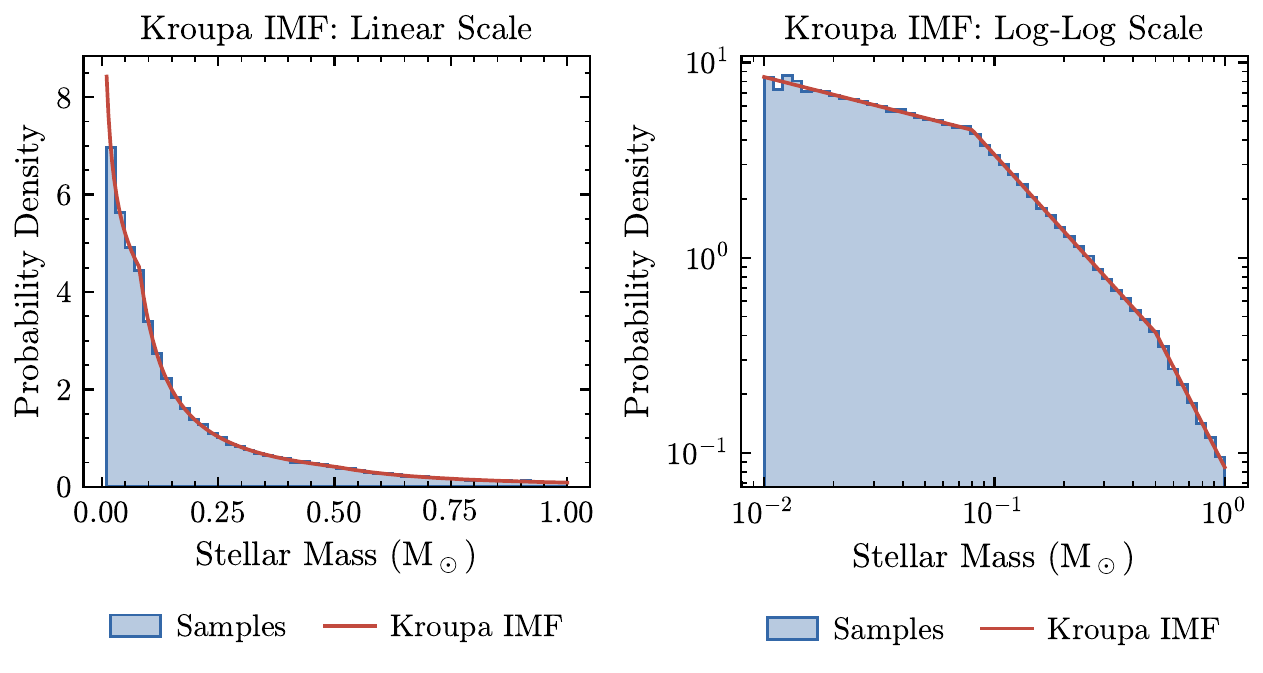}}%
  {\includegraphics[height=\tutorialheight]{tutorials/12/fig02.pdf}}

The histograms follow the Kroupa IMF on both scales, so the inverse CDF method gives masses with the right distribution. The median mass lies below the mean mass, because the stars above 0.5 solar masses, fewer than one in ten, pull the mean up.

\section*{Rejection Sampling: Complex Distributions}

Inverse transform sampling is efficient, but it needs a CDF that we can write down and invert. For many astronomical distributions, in particular those from complex physical models or empirical fits, this is not possible. Rejection sampling is an alternative (Section 12.4). It can handle a distribution when we can evaluate its PDF and cover it with a scaled proposal.

\subsection*{The Mathematical Principle}

Rejection sampling turns the problem of sampling from a complex target distribution \(p(x)\) into a simpler problem. We sample from a proposal distribution \(q(x)\) and apply a probabilistic acceptance criterion. The method is general, because it needs only the value of the target density at a point. It does not need an integral or an inverse.

The algorithm needs three components.

\begin{enumerate}
\def\labelenumi{\arabic{enumi}.}
\tightlist
\item
  A target distribution \(p(x)\) that we can evaluate, up to a normalization constant.
\item
  A proposal distribution \(q(x)\) from which we can easily generate samples.
\item
  A scaling constant \(M\) such that \(M\,q(x) \geq p(x)\) for all \(x\) in the domain. This is the covering requirement.
\end{enumerate}

The procedure uses the relation between probability densities and geometric areas. We generate points uniformly under the scaled proposal density \(M\,q(x)\) and accept only the points that fall under \(p(x)\). The accepted points are distributed according to \(p(x)\).

\subsection*{The Rejection Algorithm}

The rejection sampling algorithm has three steps.

\begin{enumerate}
\def\labelenumi{\arabic{enumi}.}
\tightlist
\item
  \textbf{Generate a proposal.} Draw \(x_*\) from the proposal distribution \(q(x)\).
\item
  \textbf{Compute the acceptance probability.} \(\alpha = p(x_*)/(M\,q(x_*))\).
\item
  \textbf{Accept or reject.} Draw \(u \sim \mathcal{U}(0,1)\). If \(u \leq \alpha\), accept \(x_*\) as a sample. Otherwise reject it and repeat.
\end{enumerate}

The pair \((x_*, u\,M\,q(x_*))\) is a point under the scaled proposal, and \(u \leq \alpha\) means that it lies under \(p(x)\). The proposal distribution and the scaling constant decide the efficiency. An ideal proposal follows the shape of the target distribution closely, so the area between \(M\,q(x)\) and \(p(x)\) is small. The rejections occur in this area.

\subsection*{Astronomical Application: Sampling from the Kroupa IMF}

We now implement rejection sampling for the Kroupa IMF. We already sampled this distribution with inverse transform sampling. Sampling it again lets us check the samples and measure the efficiency of the method.

For the proposal we use an exponential distribution with \(\lambda = 4\). It is easy to sample, as inverse transform sampling showed, and its decay roughly follows the fall of the IMF. The smallest constant that covers the IMF is the largest ratio \(p(x)/q(x)\) on 0.01 to 1. For \(\lambda = 4\) it is 2.19 at \(x = 0.01\), so we use \(M = 2.2\).

We now implement the three steps of the algorithm in a function.

\tutorialbarrier
\begin{codecell}
\begin{Verbatim}[commandchars=\\\{\},fontsize=\small]
\PYG{k}{def}\PYG{+w}{ }\PYG{n+nf}{rejection\PYGZus{}sample\PYGZus{}kroupa}\PYG{p}{(}\FancyVerbBreakAfterBreak{}\PYG{n}{num\PYGZus{}samples}\PYG{p}{,}\FancyVerbBreakAfterBreak{} \PYG{n}{lambda\PYGZus{}q}\PYG{p}{,}\FancyVerbBreakAfterBreak{} \PYG{n}{M}\PYG{p}{)}\PYG{p}{:}
\PYG{+w}{    }\PYG{l+s+sd}{\PYGZdq{}\PYGZdq{}\PYGZdq{}}
\PYG{l+s+sd}{    Draw samples from the Kroupa IMF with rejection sampling.}

\PYG{l+s+sd}{    The proposal is an exponential distribution with the parameter lambda\PYGZus{}q, and M scales it.}
\PYG{l+s+sd}{    \PYGZdq{}\PYGZdq{}\PYGZdq{}}

    \PYG{c+c1}{\PYGZsh{} Storage for the samples}
    \PYG{n}{accepted\PYGZus{}samples} \PYG{o}{=}\FancyVerbBreakAfterBreak{} \PYG{p}{[}\FancyVerbBreakAfterBreak{}\PYG{p}{]}
    \PYG{n}{accepted\PYGZus{}heights} \PYG{o}{=}\FancyVerbBreakAfterBreak{} \PYG{p}{[}\FancyVerbBreakAfterBreak{}\PYG{p}{]}
    \PYG{n}{rejected\PYGZus{}samples} \PYG{o}{=}\FancyVerbBreakAfterBreak{} \PYG{p}{[}\FancyVerbBreakAfterBreak{}\PYG{p}{]}
    \PYG{n}{rejected\PYGZus{}heights} \PYG{o}{=}\FancyVerbBreakAfterBreak{} \PYG{p}{[}\FancyVerbBreakAfterBreak{}\PYG{p}{]}

    \PYG{c+c1}{\PYGZsh{} Counter for the efficiency}
    \PYG{n}{total\PYGZus{}trials} \PYG{o}{=}\FancyVerbBreakAfterBreak{} \PYG{l+m+mi}{0}

    \PYG{c+c1}{\PYGZsh{} Density of the proposal distribution q(x)}
    \PYG{k}{def}\PYG{+w}{ }\PYG{n+nf}{q}\PYG{p}{(}\FancyVerbBreakAfterBreak{}\PYG{n}{x}\PYG{p}{)}\PYG{p}{:}
        \PYG{k}{return} \PYG{n}{lambda\PYGZus{}q} \PYG{o}{*} \PYG{n}{np}\PYG{o}{.}\FancyVerbBreakAfterBreak{}\PYG{n}{exp}\PYG{p}{(}\FancyVerbBreakAfterBreak{}\PYG{o}{\PYGZhy{}}\PYG{n}{lambda\PYGZus{}q} \PYG{o}{*} \PYG{n}{x}\PYG{p}{)}

    \PYG{c+c1}{\PYGZsh{} Continue until we have enough samples}
    \PYG{k}{while} \PYG{n+nb}{len}\PYG{p}{(}\FancyVerbBreakAfterBreak{}\PYG{n}{accepted\PYGZus{}samples}\PYG{p}{)} \PYG{o}{\PYGZlt{}} \PYG{n}{num\PYGZus{}samples}\PYG{p}{:}
        \PYG{n}{total\PYGZus{}trials} \PYG{o}{+}\FancyVerbBreakAfterBreak{}\PYG{o}{=}\FancyVerbBreakAfterBreak{} \PYG{l+m+mi}{1}

        \PYG{c+c1}{\PYGZsh{} Step 1. Draw a candidate from the proposal distribution q(x)}
        \PYG{n}{x} \PYG{o}{=}\FancyVerbBreakAfterBreak{} \PYG{n}{sample\PYGZus{}exponential}\PYG{p}{(}\FancyVerbBreakAfterBreak{}\PYG{n}{lambda\PYGZus{}q}\PYG{p}{,}\FancyVerbBreakAfterBreak{} \PYG{l+m+mi}{1}\PYG{p}{)}\PYG{p}{[}\FancyVerbBreakAfterBreak{}\PYG{l+m+mi}{0}\PYG{p}{]}

        \PYG{c+c1}{\PYGZsh{} Only consider candidates in the valid range 0.01 to 1}
        \PYG{k}{if} \PYG{n}{x} \PYG{o}{\PYGZlt{}} \PYG{l+m+mf}{0.01} \PYG{o+ow}{or} \PYG{n}{x} \PYG{o}{\PYGZgt{}} \PYG{l+m+mf}{1.0}\PYG{p}{:}
            \PYG{k}{continue}

        \PYG{c+c1}{\PYGZsh{} Evaluate the proposal density at the candidate}
        \PYG{n}{qx} \PYG{o}{=}\FancyVerbBreakAfterBreak{} \PYG{n}{q}\PYG{p}{(}\FancyVerbBreakAfterBreak{}\PYG{n}{x}\PYG{p}{)}

        \PYG{c+c1}{\PYGZsh{} Step 2. Acceptance probability of the candidate}
        \PYG{n}{px} \PYG{o}{=}\FancyVerbBreakAfterBreak{} \PYG{n}{kroupa\PYGZus{}imf}\PYG{p}{(}\FancyVerbBreakAfterBreak{}\PYG{n}{x}\PYG{p}{)}
        \PYG{n}{alpha} \PYG{o}{=}\FancyVerbBreakAfterBreak{} \PYG{n}{px} \PYG{o}{/} \PYG{p}{(}\FancyVerbBreakAfterBreak{}\PYG{n}{M} \PYG{o}{*} \PYG{n}{qx}\PYG{p}{)}

        \PYG{c+c1}{\PYGZsh{} Step 3. Accept the candidate if a uniform number is below the acceptance probability}
        \PYG{n}{u} \PYG{o}{=}\FancyVerbBreakAfterBreak{} \PYG{n}{np}\PYG{o}{.}\FancyVerbBreakAfterBreak{}\PYG{n}{random}\PYG{o}{.}\FancyVerbBreakAfterBreak{}\PYG{n}{uniform}\PYG{p}{(}\FancyVerbBreakAfterBreak{}\PYG{p}{)}
        \PYG{n}{height} \PYG{o}{=}\FancyVerbBreakAfterBreak{} \PYG{n}{u} \PYG{o}{*} \PYG{n}{M} \PYG{o}{*} \PYG{n}{qx}  \PYG{c+c1}{\PYGZsh{} the height of the point under the scaled proposal, for the plot}
        \PYG{k}{if} \PYG{n}{u} \PYG{o}{\PYGZlt{}}\PYG{o}{=}\FancyVerbBreakAfterBreak{} \PYG{n}{alpha}\PYG{p}{:}
            \PYG{n}{accepted\PYGZus{}samples}\PYG{o}{.}\FancyVerbBreakAfterBreak{}\PYG{n}{append}\PYG{p}{(}\FancyVerbBreakAfterBreak{}\PYG{n}{x}\PYG{p}{)}
            \PYG{n}{accepted\PYGZus{}heights}\PYG{o}{.}\FancyVerbBreakAfterBreak{}\PYG{n}{append}\PYG{p}{(}\FancyVerbBreakAfterBreak{}\PYG{n}{height}\PYG{p}{)}
        \PYG{k}{else}\PYG{p}{:}
            \PYG{n}{rejected\PYGZus{}samples}\PYG{o}{.}\FancyVerbBreakAfterBreak{}\PYG{n}{append}\PYG{p}{(}\FancyVerbBreakAfterBreak{}\PYG{n}{x}\PYG{p}{)}
            \PYG{n}{rejected\PYGZus{}heights}\PYG{o}{.}\FancyVerbBreakAfterBreak{}\PYG{n}{append}\PYG{p}{(}\FancyVerbBreakAfterBreak{}\PYG{n}{height}\PYG{p}{)}

    \PYG{c+c1}{\PYGZsh{} The accepted and rejected samples with their heights, and the number of candidates}
    \PYG{k}{return} \PYG{n}{np}\PYG{o}{.}\FancyVerbBreakAfterBreak{}\PYG{n}{array}\PYG{p}{(}\FancyVerbBreakAfterBreak{}\PYG{n}{accepted\PYGZus{}samples}\PYG{p}{)}\PYG{p}{,}\FancyVerbBreakAfterBreak{} \PYG{n}{accepted\PYGZus{}heights}\PYG{p}{,}\FancyVerbBreakAfterBreak{} \PYG{n}{rejected\PYGZus{}samples}\PYG{p}{,}\FancyVerbBreakAfterBreak{} \PYG{n}{rejected\PYGZus{}heights}\PYG{p}{,}\FancyVerbBreakAfterBreak{} \PYG{n}{total\PYGZus{}trials}
\end{Verbatim}
\end{codecell}

We now draw 5,000 samples. The function counts every candidate that it draws from the proposal, also those outside 0.01 to 1. We print the acceptance rate, the fraction of the candidates that we accept.

\tutorialbarrier
\begin{codecell}
\begin{Verbatim}[commandchars=\\\{\},fontsize=\small]
\PYG{c+c1}{\PYGZsh{} Parameter of the exponential proposal and scaling constant}
\PYG{n}{lambda\PYGZus{}q} \PYG{o}{=}\FancyVerbBreakAfterBreak{} \PYG{l+m+mf}{4.0}
\PYG{n}{M} \PYG{o}{=}\FancyVerbBreakAfterBreak{} \PYG{l+m+mf}{2.2}

\PYG{c+c1}{\PYGZsh{} Generate samples with rejection sampling}
\PYG{n}{kroupa\PYGZus{}samples}\PYG{p}{,}\FancyVerbBreakAfterBreak{} \PYG{n}{accepted\PYGZus{}heights}\PYG{p}{,}\FancyVerbBreakAfterBreak{} \PYG{n}{rejected\PYGZus{}samples}\PYG{p}{,}\FancyVerbBreakAfterBreak{} \PYG{n}{rejected\PYGZus{}heights}\PYG{p}{,}\FancyVerbBreakAfterBreak{} \PYG{n}{total\PYGZus{}trials} \PYG{o}{=}\FancyVerbBreakAfterBreak{} \PYGZbs{}
    \PYG{n}{rejection\PYGZus{}sample\PYGZus{}kroupa}\PYG{p}{(}\FancyVerbBreakAfterBreak{}\PYG{l+m+mi}{5000}\PYG{p}{,}\FancyVerbBreakAfterBreak{} \PYG{n}{lambda\PYGZus{}q}\PYG{p}{,}\FancyVerbBreakAfterBreak{} \PYG{n}{M}\PYG{p}{)}

\PYG{c+c1}{\PYGZsh{} The acceptance rate is the fraction of the candidates that we accept}
\PYG{n}{acceptance\PYGZus{}rate} \PYG{o}{=}\FancyVerbBreakAfterBreak{} \PYG{n+nb}{len}\PYG{p}{(}\FancyVerbBreakAfterBreak{}\PYG{n}{kroupa\PYGZus{}samples}\PYG{p}{)} \PYG{o}{/} \PYG{n}{total\PYGZus{}trials}
\PYG{n+nb}{print}\PYG{p}{(}\FancyVerbBreakAfterBreak{}\PYG{l+s+sa}{f}\PYG{l+s+s2}{\PYGZdq{}}\PYG{l+s+s2}{Acceptance rate = }\PYG{l+s+si}{\PYGZob{}}\PYG{n}{acceptance\PYGZus{}rate}\PYG{l+s+si}{:}\PYG{l+s+s2}{.4f}\PYG{l+s+si}{\PYGZcb{}}\FancyVerbBreakAfterBreak{}\PYG{l+s+s2}{ (}\PYG{l+s+si}{\PYGZob{}}\PYG{n+nb}{len}\PYG{p}{(}\FancyVerbBreakAfterBreak{}\PYG{n}{kroupa\PYGZus{}samples}\PYG{p}{)}\PYG{l+s+si}{\PYGZcb{}}\FancyVerbBreakAfterBreak{}\PYG{l+s+s2}{ of }\PYG{l+s+si}{\PYGZob{}}\PYG{n}{total\PYGZus{}trials}\PYG{l+s+si}{\PYGZcb{}}\FancyVerbBreakAfterBreak{}\PYG{l+s+s2}{ candidates)}\PYG{l+s+s2}{\PYGZdq{}}\PYG{p}{)}
\end{Verbatim}
\end{codecell}
\begin{Verbatim}[fontsize=\footnotesize, xleftmargin=5mm]
Acceptance rate = 0.4570 (5000 of 10940 candidates)
\end{Verbatim}

We now plot the target, the scaled proposal and the candidates under it. The figure shows the first 500 accepted and the first 500 rejected candidates.

\tutorialbarrier
\begin{codecell}

\end{codecell}
\tutorialplot{\includegraphics[width=5.514in,height=0.75\textheight,keepaspectratio]{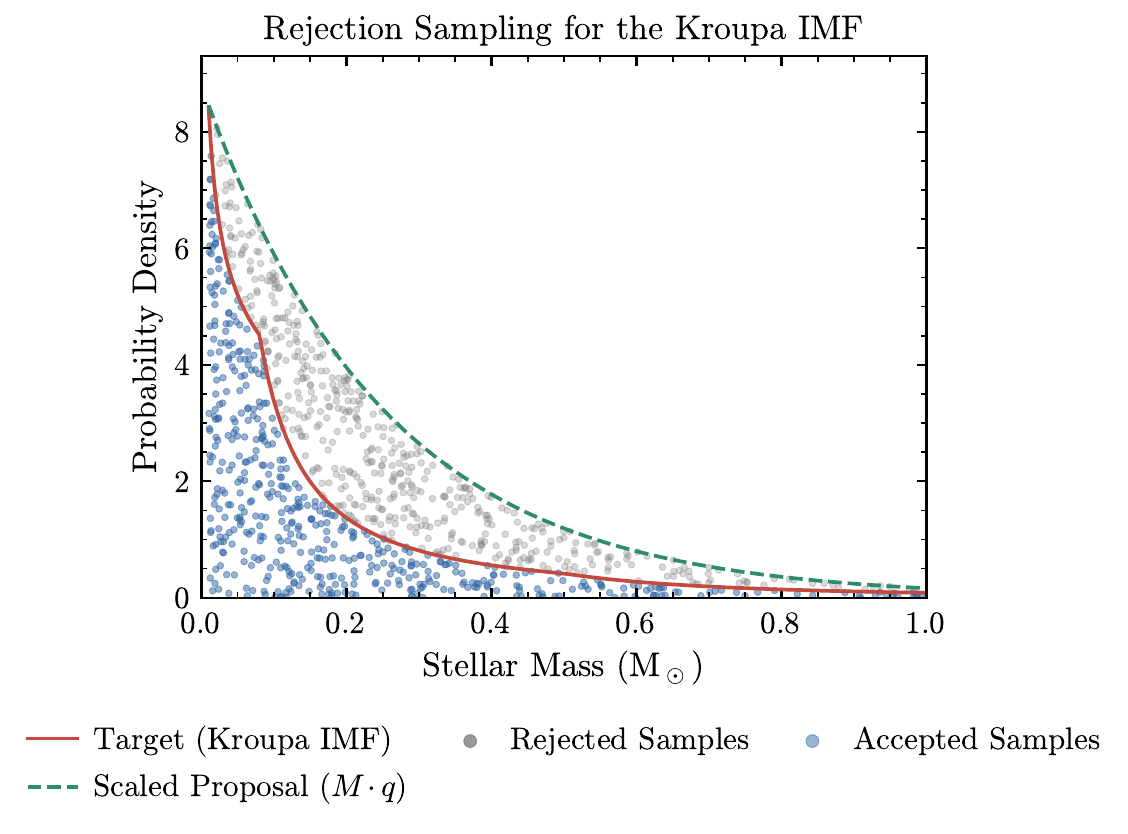}}%
  {\includegraphics[height=\tutorialheight]{tutorials/12/fig03.pdf}}

The figure shows the geometry of rejection sampling. We generate points uniformly under the scaled proposal curve \(M\,q(x)\). We accept the points that fall below the target distribution \(p(x)\) (blue) and reject those above it (gray). The area under \(M\,q(x)\) is \(M\) and the area under \(p(x)\) is 1, so we accept a share \(1/M\) of all candidates. This includes the candidates outside 0.01 to 1. The acceptance rate of 0.457 is close to \(1/M = 0.455\) for \(M = 2.2\) (Section 12.4).

As a check, we now compare the histogram of all accepted samples with the Kroupa IMF.

\tutorialbarrier
\begin{codecell}

\end{codecell}
\tutorialplot{\includegraphics[width=4.214in,height=0.75\textheight,keepaspectratio]{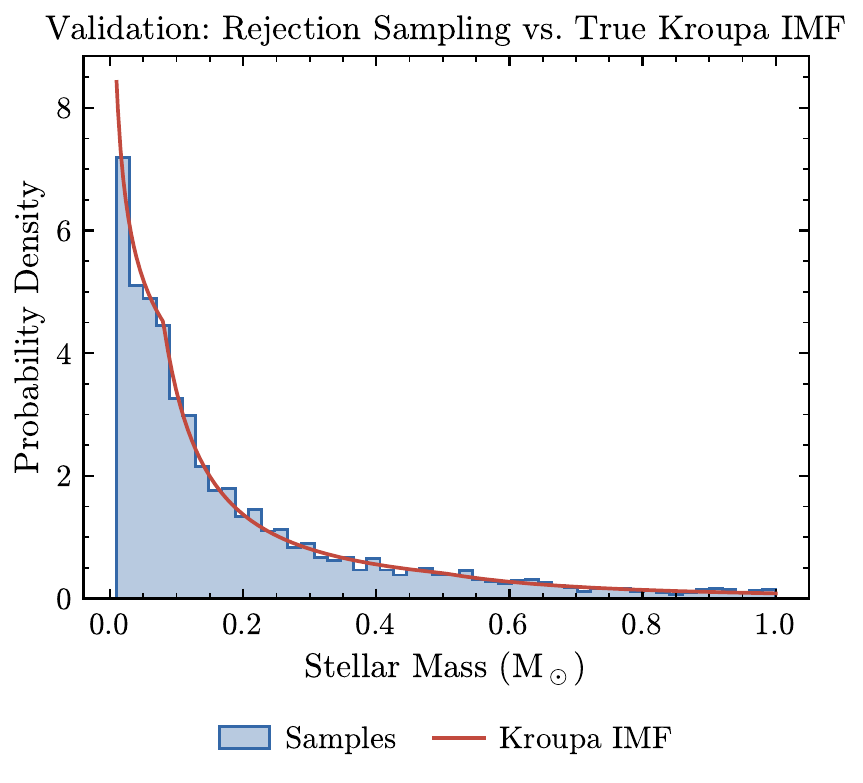}}%
  {\includegraphics[height=\tutorialheight]{tutorials/12/fig04.pdf}}

The histogram of the accepted samples follows the IMF, so rejection sampling gives masses with the right distribution.

\subsection*{Understanding Rejection Sampling Efficiency}

Several factors influence the efficiency of rejection sampling.

\begin{enumerate}
\def\labelenumi{\arabic{enumi}.}
\tightlist
\item
  \textbf{Proposal shape matching.} The exponential proposal follows the general decreasing trend of the IMF, but it cannot match the piecewise power-law structure. The two shapes differ most at masses of about 0.1 to 0.5 \(\mathrm{M}_\odot\), where \(M\,q(x)\) lies well above the IMF.
\item
  \textbf{Scaling constant selection.} The constant \(M = 2.2\) makes \(M\,q(x) \geq p(x)\) everywhere. It is already close to the smallest value that does this, because \(M\,q(x)\) almost touches the IMF at the lowest masses. A smaller \(M\) would break the covering requirement. The difference of the shapes limits the efficiency, and not the choice of \(M\).
\item
  \textbf{Domain restrictions.} The Kroupa IMF is defined on 0.01 to 1, while the exponential distribution covers all masses above 0. We must reject the samples outside the valid range. They take computing time and do not give a sample. A shifted exponential, with \(x - 0.01\) in place of \(x\), can be more efficient because it wastes fewer candidates. We keep the simple exponential here.
\end{enumerate}

In higher dimensions the volume between the scaled proposal and the target can grow quickly, so rejection sampling can become much less efficient (Section 12.4). This is the curse of dimensionality.

\section*{Importance Sampling: Efficiently Estimating Properties}

So far we generated representative samples from probability distributions. Many astronomical applications need only specific properties of the distribution, and not the samples. Examples are the mean stellar mass, the total luminosity and the expected chemical yields. Importance sampling can estimate these properties without generating samples that follow the target distribution (Section 12.5).

\subsection*{The Importance Sampling Estimator}

We consider the basic problem of calculating an expectation. For a function \(f(x)\) and a distribution with density \(p(x)\) we want

\[\mathbb{E}_{p}[f] = \int f(x)\,p(x)\,\mathrm{d}x.\]

Direct Monte Carlo estimation uses samples \(x^{(1)}, \dots, x^{(N)}\) from \(p(x)\) and approximates the expectation as

\[\mathbb{E}_{p}[f] \approx \frac{1}{N}\sum_{i=1}^{N} f(x^{(i)}).\]

This can be inefficient when sampling from \(p(x)\) is difficult, or when \(f(x)\) varies widely over the domain.

Importance sampling rewrites the problem with a proposal distribution \(q(x)\)

\[\mathbb{E}_{p}[f] = \int f(x) \frac{p(x)}{q(x)}\, q(x)\, \mathrm{d}x = \mathbb{E}_{q}\left[f(x) \frac{p(x)}{q(x)}\right].\]

We can estimate expected values under \(p(x)\) with samples from a different distribution \(q(x)\), if we weight each sample by the ratio \(w(x) = p(x)/q(x)\). With samples \(x^{(i)}\) from \(q(x)\) the estimator is

\[\mathbb{E}_{p}[f] \approx \frac{1}{N}\sum_{i=1}^{N} w(x^{(i)})\, f(x^{(i)}).\]

\subsection*{Importance Sampling with Unnormalized Distributions}

In astronomical applications we often know distributions only up to a normalization constant. A Bayesian posterior \(p(\boldsymbol{\theta}|\mathcal{D}) \propto p(\mathcal{D}|\boldsymbol{\theta})\,p(\boldsymbol{\theta})\) is an example, because its normalization constant needs an intractable integral. Theoretical models can also give the functional form of a distribution without a normalization.

For unnormalized densities \(\tilde{p}(x) = Z_p\,p(x)\) and \(\tilde{q}(x) = Z_q\,q(x)\), where \(Z_p\) and \(Z_q\) are unknown normalization constants, the estimator becomes (Section 12.5)

\[\mathbb{E}_{p}[f] \approx \frac{\sum_{i=1}^{N} \tilde{w}(x^{(i)})\, f(x^{(i)})}{\sum_{i=1}^{N} \tilde{w}(x^{(i)})},\]

where \(\tilde{w}(x^{(i)}) = \tilde{p}(x^{(i)})/\tilde{q}(x^{(i)})\) are the unnormalized importance weights. This self-normalized estimator cancels the unknown constants. It needs only the ratio of the unnormalized densities.

The choice of the proposal \(q(x)\) decides how well the estimator works. An ideal proposal would be proportional to \(|f(x)|\,p(x)\) and would concentrate the samples where the integrand is large (Section 12.6). A poor proposal gives highly variable weights. A few samples then dominate the estimate, and the statistical uncertainty stays high even for many samples.

\subsection*{The Effective Sample Size}

The effective sample size (ESS, Section 12.6) measures this problem. It is approximately the number of direct samples from \(p(x)\) that give the same precision as the \(N\) weighted samples,

\[\text{ESS} \approx \frac{\left(\sum_{i=1}^{N} \tilde{w}(x^{(i)})\right)^2}{\sum_{i=1}^{N} \tilde{w}(x^{(i)})^2}.\]

The ESS equals \(N\) when all weights are equal, and it falls as the weights differ more.

\subsection*{Astronomical Application: Mean Stellar Mass in the Kroupa IMF}

We now use importance sampling to calculate a basic property of a stellar population, the mean stellar mass of the Kroupa IMF. The mean mass turns a number of stars into a total mass.

We compare two proposal distributions to show how the choice affects the efficiency of the estimate.

\begin{enumerate}
\def\labelenumi{\arabic{enumi}.}
\tightlist
\item
  \textbf{Exponential proposal.} It roughly follows the decreasing trend of the IMF and concentrates the samples at low masses.
\item
  \textbf{Uniform proposal.} It covers all masses equally. It is simpler, but it may be less efficient.
\end{enumerate}

The mean stellar mass is

\[\mathbb{E}_{p}[m] = \int_{0.01}^{1} m\, p(m)\, \mathrm{d}m,\]

where \(p(m)\) is the Kroupa IMF. With importance sampling we estimate this integral with samples from a proposal \(q(m)\)

\[\mathbb{E}_{p}[m] \approx \frac{\sum_{i=1}^{N} m^{(i)}\, p(m^{(i)})/q(m^{(i)})}{\sum_{i=1}^{N} p(m^{(i)})/q(m^{(i)})}.\]

We could average the 100,000 masses of inverse transform sampling. Importance sampling is for the case that we cannot sample \(p(m)\), and here the exact mean lets us check it. We now define a function that draws samples from a proposal and returns them with their importance weights. Importance sampling does not need a covering constant, so \(\lambda\) of the exponential proposal is free, and we use \(\lambda = 5\). We keep only its samples between 0.01 and 1, and their density is then known up to a constant, which the self-normalized estimator cancels. The same holds for the uniform proposal, whose density we set to 1.

\tutorialbarrier
\begin{codecell}
\begin{Verbatim}[commandchars=\\\{\},fontsize=\small]
\PYG{c+c1}{\PYGZsh{} Fix the random seed so that the outputs repeat}
\PYG{n}{np}\PYG{o}{.}\FancyVerbBreakAfterBreak{}\PYG{n}{random}\PYG{o}{.}\FancyVerbBreakAfterBreak{}\PYG{n}{seed}\PYG{p}{(}\FancyVerbBreakAfterBreak{}\PYG{l+m+mi}{52}\PYG{p}{)}

\PYG{k}{def}\PYG{+w}{ }\PYG{n+nf}{importance\PYGZus{}weights}\PYG{p}{(}\FancyVerbBreakAfterBreak{}\PYG{n}{num\PYGZus{}samples}\PYG{p}{,}\FancyVerbBreakAfterBreak{} \PYG{n}{proposal}\PYG{o}{=}\FancyVerbBreakAfterBreak{}\PYG{l+s+s1}{\PYGZsq{}}\PYG{l+s+s1}{exponential}\PYG{l+s+s1}{\PYGZsq{}}\PYG{p}{)}\PYG{p}{:}
\PYG{+w}{    }\PYG{l+s+sd}{\PYGZdq{}\PYGZdq{}\PYGZdq{}}
\PYG{l+s+sd}{    Draw samples from a proposal and return them with their importance weights.}

\PYG{l+s+sd}{    The proposal is \PYGZsq{}exponential\PYGZsq{} (lambda = 5) or \PYGZsq{}uniform\PYGZsq{} between 0.01 and 1.}
\PYG{l+s+sd}{    \PYGZdq{}\PYGZdq{}\PYGZdq{}}
    \PYG{k}{if} \PYG{n}{proposal} \PYG{o}{==} \PYG{l+s+s1}{\PYGZsq{}}\PYG{l+s+s1}{exponential}\PYG{l+s+s1}{\PYGZsq{}}\PYG{p}{:}
        \PYG{c+c1}{\PYGZsh{} Use an exponential distribution as the proposal}
        \PYG{n}{lambda\PYGZus{}val} \PYG{o}{=}\FancyVerbBreakAfterBreak{} \PYG{l+m+mf}{5.0}

        \PYG{c+c1}{\PYGZsh{} Draw samples from the proposal distribution}
        \PYG{n}{samples} \PYG{o}{=}\FancyVerbBreakAfterBreak{} \PYG{n}{sample\PYGZus{}exponential}\PYG{p}{(}\FancyVerbBreakAfterBreak{}\PYG{n}{lambda\PYGZus{}val}\PYG{p}{,}\FancyVerbBreakAfterBreak{} \PYG{n}{num\PYGZus{}samples}\PYG{p}{)}

        \PYG{c+c1}{\PYGZsh{} Keep only the samples in the valid range 0.01 to 1}
        \PYG{n}{valid\PYGZus{}mask} \PYG{o}{=}\FancyVerbBreakAfterBreak{} \PYG{p}{(}\FancyVerbBreakAfterBreak{}\PYG{n}{samples} \PYG{o}{\PYGZgt{}}\PYG{o}{=}\FancyVerbBreakAfterBreak{} \PYG{l+m+mf}{0.01}\PYG{p}{)} \PYG{o}{\PYGZam{}} \PYG{p}{(}\FancyVerbBreakAfterBreak{}\PYG{n}{samples} \PYG{o}{\PYGZlt{}}\PYG{o}{=}\FancyVerbBreakAfterBreak{} \PYG{l+m+mf}{1.0}\PYG{p}{)}
        \PYG{n}{samples} \PYG{o}{=}\FancyVerbBreakAfterBreak{} \PYG{n}{samples}\PYG{p}{[}\FancyVerbBreakAfterBreak{}\PYG{n}{valid\PYGZus{}mask}\PYG{p}{]}

        \PYG{c+c1}{\PYGZsh{} Compute the proposal density at each sample}
        \PYG{n}{q\PYGZus{}values} \PYG{o}{=}\FancyVerbBreakAfterBreak{} \PYG{n}{lambda\PYGZus{}val} \PYG{o}{*} \PYG{n}{np}\PYG{o}{.}\FancyVerbBreakAfterBreak{}\PYG{n}{exp}\PYG{p}{(}\FancyVerbBreakAfterBreak{}\PYG{o}{\PYGZhy{}}\PYG{n}{lambda\PYGZus{}val} \PYG{o}{*} \PYG{n}{samples}\PYG{p}{)}

    \PYG{k}{elif} \PYG{n}{proposal} \PYG{o}{==} \PYG{l+s+s1}{\PYGZsq{}}\PYG{l+s+s1}{uniform}\PYG{l+s+s1}{\PYGZsq{}}\PYG{p}{:}
        \PYG{c+c1}{\PYGZsh{} Use a uniform distribution as the proposal}
        \PYG{n}{samples} \PYG{o}{=}\FancyVerbBreakAfterBreak{} \PYG{n}{np}\PYG{o}{.}\FancyVerbBreakAfterBreak{}\PYG{n}{random}\PYG{o}{.}\FancyVerbBreakAfterBreak{}\PYG{n}{uniform}\PYG{p}{(}\FancyVerbBreakAfterBreak{}\PYG{l+m+mf}{0.01}\PYG{p}{,}\FancyVerbBreakAfterBreak{} \PYG{l+m+mf}{1.0}\PYG{p}{,}\FancyVerbBreakAfterBreak{} \PYG{n}{num\PYGZus{}samples}\PYG{p}{)}

        \PYG{c+c1}{\PYGZsh{} The density of the uniform proposal is constant, and we set it to 1}
        \PYG{n}{q\PYGZus{}values} \PYG{o}{=}\FancyVerbBreakAfterBreak{} \PYG{n}{np}\PYG{o}{.}\FancyVerbBreakAfterBreak{}\PYG{n}{ones\PYGZus{}like}\PYG{p}{(}\FancyVerbBreakAfterBreak{}\PYG{n}{samples}\PYG{p}{)}

    \PYG{k}{else}\PYG{p}{:}
        \PYG{k}{raise} \PYG{n+ne}{ValueError}\PYG{p}{(}\FancyVerbBreakAfterBreak{}\PYG{l+s+s2}{\PYGZdq{}}\PYG{l+s+s2}{Proposal must be }\PYG{l+s+s2}{\PYGZsq{}}\PYG{l+s+s2}{exponential}\PYG{l+s+s2}{\PYGZsq{}}\PYG{l+s+s2}{ or }\PYG{l+s+s2}{\PYGZsq{}}\PYG{l+s+s2}{uniform}\PYG{l+s+s2}{\PYGZsq{}}\PYG{l+s+s2}{\PYGZdq{}}\PYG{p}{)}

    \PYG{c+c1}{\PYGZsh{} The importance weight is the target density (Kroupa IMF) divided by the proposal density}
    \PYG{n}{weights} \PYG{o}{=}\FancyVerbBreakAfterBreak{} \PYG{n}{np}\PYG{o}{.}\FancyVerbBreakAfterBreak{}\PYG{n}{vectorize}\PYG{p}{(}\FancyVerbBreakAfterBreak{}\PYG{n}{kroupa\PYGZus{}imf}\PYG{p}{)}\PYG{p}{(}\FancyVerbBreakAfterBreak{}\PYG{n}{samples}\PYG{p}{)} \PYG{o}{/} \PYG{n}{q\PYGZus{}values}

    \PYG{k}{return} \PYG{n}{samples}\PYG{p}{,}\FancyVerbBreakAfterBreak{} \PYG{n}{weights}
\end{Verbatim}
\end{codecell}

We now estimate the mean mass with both proposals for numbers of samples from 100 to 100,000. We compare the estimates with the exact mean, which numerical integration gives.

\tutorialbarrier
\begin{codecell}

\end{codecell}
\begin{Verbatim}[fontsize=\footnotesize, xleftmargin=5mm]
Exact mean stellar mass = 0.189322 M☉
\end{Verbatim}

We now plot the estimates against the number of samples.

\tutorialbarrier
\begin{codecell}

\end{codecell}
\tutorialplot{\includegraphics[width=5.041in,height=0.75\textheight,keepaspectratio]{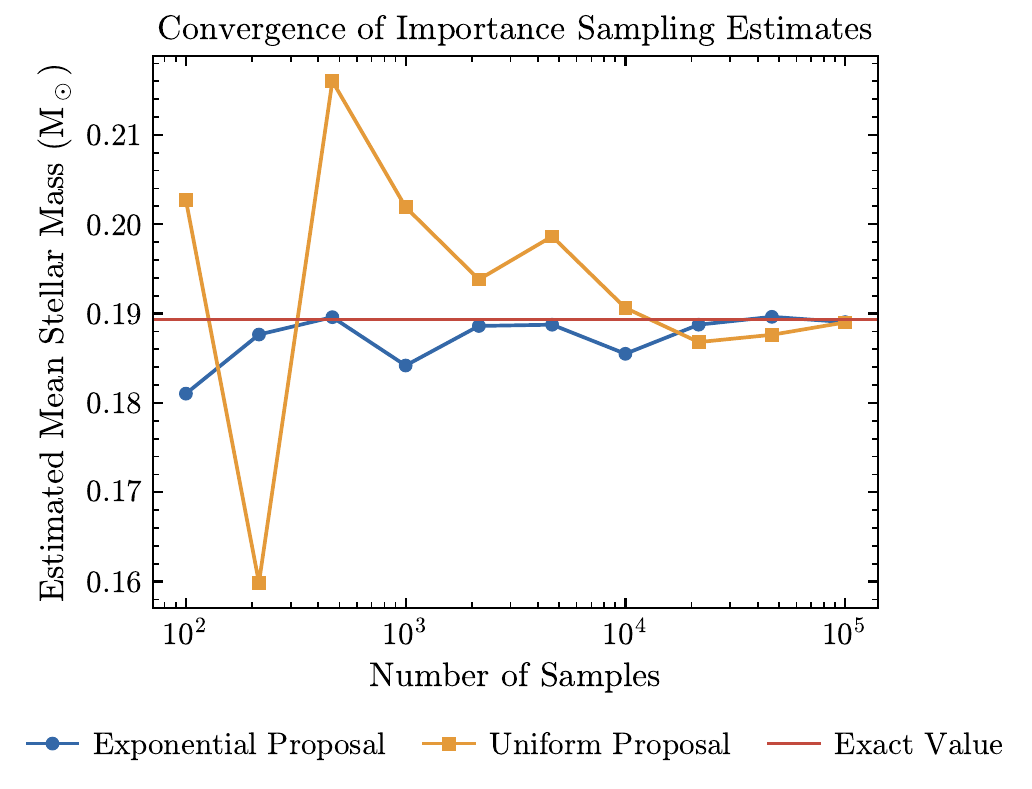}}%
  {\includegraphics[height=\tutorialheight]{tutorials/12/fig05.pdf}}

Both proposals approach the exact mean as \(N\) grows. The exponential estimate is closer for most \(N\).

The advantage of the exponential proposal can also be understood from its importance weights, through the ESS. We draw 10,000 samples from each proposal and compute the importance weights and the ESS. The exponential proposal keeps about 9,400 of its 10,000 draws, those between 0.01 and 1.

\tutorialbarrier
\begin{codecell}

\end{codecell}
\begin{Verbatim}[fontsize=\footnotesize, xleftmargin=5mm]
ESS (exponential) = 8716.9 (92.6% of 9417 samples)
ESS (uniform) = 3120.7 (31.2% of 10000 samples)
\end{Verbatim}

We now plot the distribution of the weights.

\tutorialbarrier
\begin{codecell}

\end{codecell}
\tutorialplot{\includegraphics[width=6.189in,height=0.75\textheight,keepaspectratio]{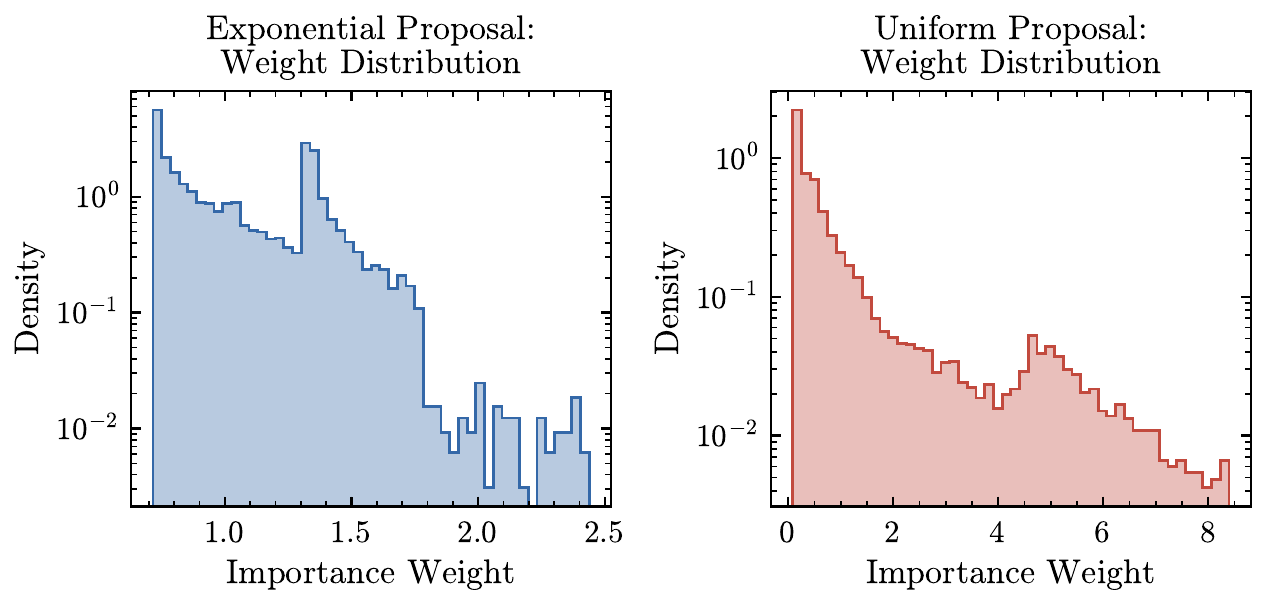}}%
  {\includegraphics[height=\tutorialheight]{tutorials/12/fig06.pdf}}

The weights depend on the mass. We plot the target and both proposals, and the weight \(p(m)/q(m)\) of every sample against its mass. We print the range of the weights of each proposal.

\tutorialbarrier
\begin{codecell}

\end{codecell}
\tutorialplot{\includegraphics[width=6.174in,height=0.75\textheight,keepaspectratio]{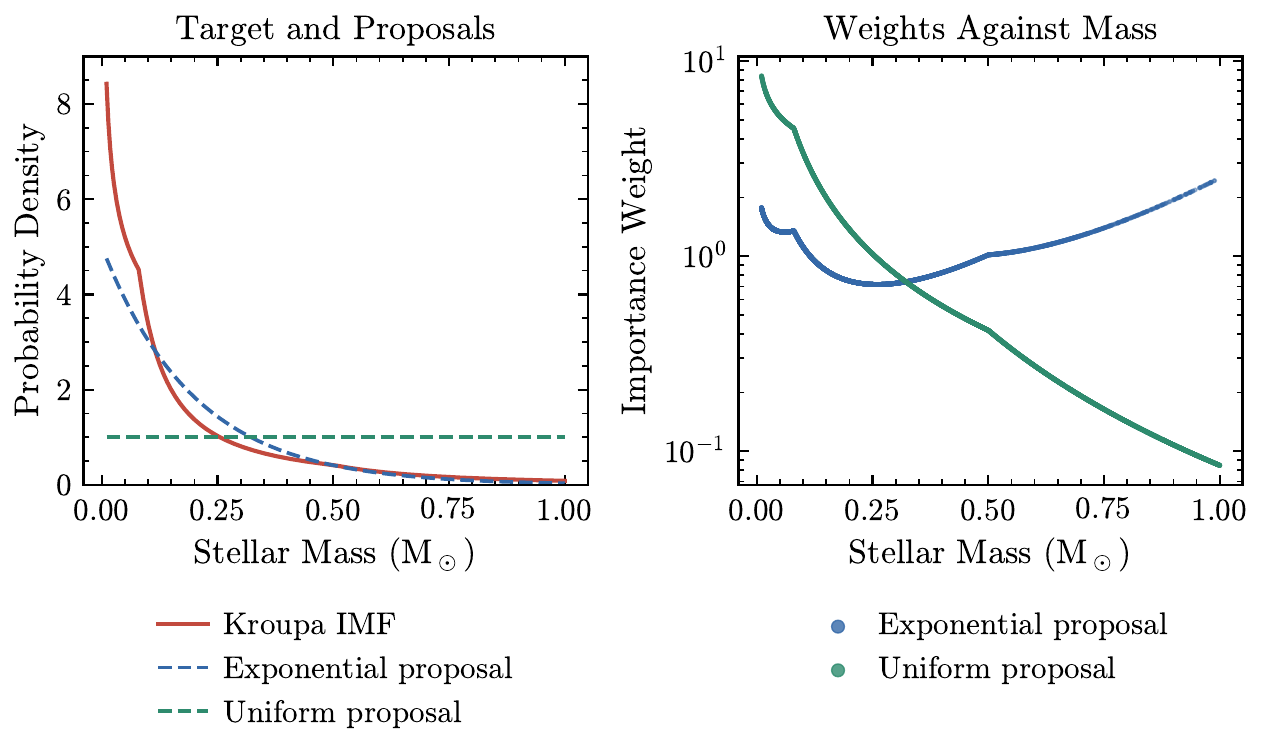}}%
  {\includegraphics[height=\tutorialheight]{tutorials/12/fig07.pdf}}
\begin{Verbatim}[fontsize=\footnotesize, xleftmargin=5mm]
Weights of the exponential proposal: 0.72 to 2.44
Weights of the uniform proposal: 0.08 to 8.39
\end{Verbatim}

The exponential proposal follows the IMF at low masses, where most stars lie, so its weights stay in a narrow range. The uniform proposal lies far below the IMF at low masses and above it at high masses, so its weights span a large range. The weights of the exponential proposal rise toward the highest masses, where it has few samples. The size of the weights has no meaning, because a constant rescales it. What matters is how much the weights vary, and the ESS is largest when they are all equal.

The exponential proposal gives a concentrated weight distribution. The uniform proposal gives a broader weight distribution with a longer tail, so its weights vary more, and the ESS measures this difference. The uniform proposal puts a large share of its samples at high masses, where the IMF has a low density. The figure of the weights shows that these samples get small weights, so fewer samples carry the estimate.

The exponential proposal has the larger ESS, so it is the better proposal for the mean mass.

\subsection*{When Efficiency Differences Matter Most}

The choice of the proposal distribution matters more in harder problems.

\begin{itemize}
\tightlist
\item
  \textbf{High-dimensional spaces.} Our stellar mass example is one-dimensional, but many astronomical problems have dozens or hundreds of parameters. Importance sampling is difficult in many dimensions, because a proposal that we can sample and that matches the target is hard to find (Section 12.6).
\item
  \textbf{Quantities set by rare stars.} Some properties depend mainly on the rare high-mass stars, for example the total luminosity, which rises steeply with the mass. The proposal must put enough samples at these masses. The ESS is a useful first gauge, and its formula treats the function as unrelated to the weights (Section 12.6). For \(m^{\alpha}\) with a large \(\alpha\), \(f\) is largest where the weights of the exponential proposal are large. The error can then exceed what the ESS suggests. Repeating the estimate or bootstrapping the samples shows the error empirically.
\item
  \textbf{Expensive functions.} Consider the expected chemical yield of a stellar population, where each calculation of \(f\) needs a stellar evolution model. Each sample then needs much computing time, so we can afford only a few samples. A poor proposal can then give misleading estimates.
\end{itemize}

In our example both proposals work. In harder problems a poor proposal can make importance sampling too inefficient to use.

\section*{Summary}

This tutorial applied three sampling methods to the distribution of stellar masses. Our main lessons follow.

\begin{enumerate}
\def\labelenumi{\arabic{enumi}.}
\tightlist
\item
  \textbf{Inverse transform sampling.} Each uniform random number gives one exact sample through the inverse CDF, so no computation is wasted. The method needs a CDF that we can invert in closed form, which limits it to distributions that we can treat mathematically.
\item
  \textbf{Rejection sampling.} The method needs only the value of the density at a point and a scaled proposal that covers it. The acceptance rate is the inverse of the scaling constant, so the efficiency depends on how well the shape of the proposal matches the target. The acceptance rate can fall quickly as the dimension grows.
\item
  \textbf{Importance sampling.} The method estimates an expectation from weighted samples of a proposal, so it does not require sampling from the target. The self-normalized estimator also handles a density that we know only up to a constant.
\item
  \textbf{Judging a proposal.} The weights show how well a proposal matches the target, and the effective sample size summarizes them. The proposal must put samples where the quantity of interest is large. The ESS is a useful first gauge of a proposal, and repeated estimates or a bootstrap give the error for one function empirically.
\end{enumerate}

With these three methods we generated a stellar population from its IMF and estimated its mean mass. The same ideas, a proposal that we can sample and a check of its fit, carry over to the posteriors of the next tutorial.

\tutorialchapter{13a}{Markov Chain Monte Carlo for Posterior Sampling}
\noindent\textit{Companion material for Chapter 13}

\section*{Introduction}

Many astronomical posteriors are too complex to sample with standard methods. We infer stellar parameters from spectra, dark matter halo properties from lensing data and redshifts from photometry. The likelihood surface is often complex, and the parameter space can have many dimensions. Rejection sampling and importance sampling (Chapter 12) work poorly when a distribution has a complex shape, several modes or many dimensions.

Markov chain Monte Carlo (MCMC) does not sample the distribution directly. We construct a Markov chain, a sequence of samples in which each sample depends only on the previous one. We design the chain so that the target distribution is its equilibrium state. A long simulation of the chain then gives samples of the target, however complex it is. MCMC is widely used for Bayesian inference in astronomy. It samples posteriors that standard methods cannot handle, so it gives the uncertainties of their parameters.

In many dimensions, rejection sampling can accept far fewer samples, and importance sampling needs a good proposal distribution. MCMC avoids these problems with local steps that follow the shape of the distribution. Its price is that successive samples are correlated. An effective analysis measures this correlation and keeps its effect small.

This tutorial uses two basic algorithms, the Metropolis algorithm and Gibbs sampling. We first estimate the mean of simulated measurements with the Metropolis algorithm and judge its chains with the tools of the chapter. We then use Gibbs sampling to see how the correlation between parameters affects the sampling efficiency.

\section*{Prerequisites and Learning Objectives}

\subsection*{Prerequisites}

This tutorial assumes the following.

\begin{itemize}
\tightlist
\item
  Probability distributions, Bayes' theorem and conditional probability (Chapter 2)
\item
  Bayesian inference, with prior distributions, likelihood functions and posterior inference (Chapters 2 and 5)
\item
  The sampling methods, including rejection and importance sampling (Chapter 12)
\item
  Python programming with NumPy and matplotlib
\item
  The multivariate Gaussian distribution and its covariance matrix (Chapters 3 and 5)
\end{itemize}

\subsection*{Learning Objectives}

After this tutorial you can do the following.

\begin{itemize}
\tightlist
\item
  Implement the Metropolis algorithm and explain how the proposal width affects the sampling efficiency
\item
  Calculate the autocorrelation of a chain and its effective sample size
\item
  Apply the Geweke and Gelman-Rubin tests to check whether chains have reached their stationary distribution
\item
  Recognize poor mixing and slow convergence, remove the burn-in, and check how thinning changes the effective sample size
\item
  Implement Gibbs sampling for a problem with tractable conditional distributions
\end{itemize}

We begin by importing the libraries and setting the plot style.

\tutorialbarrier
\begin{codecell}

\end{codecell}

\section*{The Metropolis Algorithm}

The Metropolis algorithm samples a probability distribution that we can evaluate only up to a constant. It needs only the ratio of two probabilities. We therefore do not need the normalization constant that makes many inference problems in astronomy hard.

\subsection*{The Intuition}

Recall the explorer of Chapter 13 who wants to map the population distribution of Columbus, Ohio. Each day a guidebook suggests a nearby neighborhood, and the explorer compares its population density with that of the current one. A denser neighborhood is always chosen, and a less dense one is chosen with a probability equal to the ratio of the densities. Otherwise the explorer stays for another day. The record of the visits then follows the population density. In the Metropolis algorithm the population density is the probability density, and the record of the visits is a sample of the distribution.

\subsection*{Mathematical Basis}

Two conditions make the chain converge to our target distribution \(p\). The first is detailed balance (Section 13.2),

\[p(\mathbf{x}) T(\mathbf{x}, \mathbf{x}') = p(\mathbf{x}') T(\mathbf{x}', \mathbf{x}),\]

where \(T(\mathbf{x}, \mathbf{x}')\) is the transition probability from state \(\mathbf{x}\) to state \(\mathbf{x}'\). The left side is the flow of probability from \(\mathbf{x}\) to \(\mathbf{x}'\), and the right side is the flow in the reverse direction. When the two flows are equal, one step of the chain leaves \(p\) unchanged, so \(p\) is the stationary distribution of the chain.

The second condition is ergodicity (Section 13.5). The chain must be irreducible, so that every state can reach every other state. It must be positive recurrent, so that it returns to each state in a finite expected time. It must also be aperiodic, so that it has no cycles. The chain then does not stay in disconnected regions. We typically have ergodicity in practice, and detailed balance and ergodicity together make the chain converge to \(p\).

\subsection*{The Metropolis Algorithm: Step by Step}

The Metropolis algorithm uses a symmetric proposal distribution, \(q(\mathbf{x}'|\mathbf{x}) = q(\mathbf{x}|\mathbf{x}')\), and a transition probability that satisfies detailed balance. The transition probability is \(T(\mathbf{x}, \mathbf{x}') = q(\mathbf{x}'|\mathbf{x})A(\mathbf{x}, \mathbf{x}')\), where \(A\) is the acceptance probability,

\[A(\mathbf{x}, \mathbf{x}') = \min\left(1, \frac{p(\mathbf{x}'|\mathcal{D})}{p(\mathbf{x}|\mathcal{D})}\right),\]

and \(\mathcal{D}\) is the data. More generally, the Metropolis-Hastings algorithm allows an asymmetric proposal and multiplies the ratio by \(q(\mathbf{x}|\mathbf{x}')/q(\mathbf{x}'|\mathbf{x})\) (Section 13.14). We always accept a move to a region of higher probability. We accept a move to a region of lower probability with a probability equal to the ratio of the two posteriors. Section 13.3 checks that this rule satisfies detailed balance.

The algorithm depends only on probability ratios. The posterior is proportional to the likelihood times the prior, \(p(\mathbf{x}|\mathcal{D}) \propto p(\mathcal{D}|\mathbf{x})p(\mathbf{x})\). We can therefore compute the ratio without the normalization constant,

\[\frac{p(\mathbf{x}'|\mathcal{D})}{p(\mathbf{x}|\mathcal{D})} = \frac{p(\mathcal{D}|\mathbf{x}')p(\mathbf{x}')}{p(\mathcal{D}|\mathbf{x})p(\mathbf{x})}.\]

The implementation follows these steps.

\begin{enumerate}
\def\labelenumi{\arabic{enumi}.}
\item
  \textbf{Initialize.} Choose a starting point \(\mathbf{x}^{(0)}\) in parameter space.
\item
  \textbf{Iterate.} For \(t = 1, 2, \ldots, N\), where \(N\) is the number of iterations, do the following.

  \begin{itemize}
  \tightlist
  \item
    \textbf{Propose.} Generate \(\mathbf{x}' \sim q(\mathbf{x}'|\mathbf{x}^{(t-1)})\).
  \item
    \textbf{Accept or reject.} With probability \(A(\mathbf{x}^{(t-1)}, \mathbf{x}')\), set \(\mathbf{x}^{(t)} = \mathbf{x}'\). Otherwise set \(\mathbf{x}^{(t)} = \mathbf{x}^{(t-1)}\).
  \end{itemize}
\item
  \textbf{Return.} The sequence \(\{\mathbf{x}^{(0)}, \mathbf{x}^{(1)}, \ldots, \mathbf{x}^{(N)}\}\).
\end{enumerate}

\section*{Estimating the Sample Mean}

We apply the Metropolis algorithm to a concrete problem, the estimation of the mean of a population from noisy measurements. This problem has one parameter, so it is a good test for MCMC before we turn to harder problems.

\subsection*{The Statistical Model}

We observe 1000 measurements \(\{x_i\}\) from a normal distribution with unknown mean \(\mu\) and known standard deviation \(\sigma = 1\). We simulate them with a true mean of 1. Examples of this setup are the mean radial velocity of a star cluster and the systematic offset in photometric measurements.

In the notation of the previous section, the state \(\mathbf{x}\) is the parameter \(\mu\) and the data \(\mathcal{D}\) are the measurements \(\{x_i\}\). In Bayesian terms, we seek the posterior distribution

\[p(\mu | \{x_i\}) \propto p(\{x_i\} | \mu) \times p(\mu).\]

For independent observations from a normal distribution, the likelihood is

\[p(\{x_i\} | \mu) = \prod_{i=1}^{1000} \frac{1}{\sqrt{2\pi}} \exp\left(-\frac{(x_i - \mu)^2}{2}\right).\]

We use a Cauchy prior for \(\mu\), which has heavier tails than a normal distribution,

\[p(\mu) = \frac{1}{\pi\gamma} \frac{1}{1 + \left(\frac{\mu-\mu_0}{\gamma}\right)^2},\]

where \(\mu_0 = 0\) is the location parameter and \(\gamma = 3\) is the scale parameter. The heavy tails allow large values of \(\mu\). The Cauchy prior is therefore useful when we know little about the range of the parameter.

The product of 1000 likelihood factors is too small for a floating-point number. We therefore work with the logarithm of the posterior,

\[\ln p(\mu | \{x_i\}) = \sum_{i=1}^{1000} \ln p(x_i | \mu) + \ln p(\mu) + \text{const}.\]

We now simulate the 1000 measurements and print their sample mean.

\tutorialbarrier
\begin{codecell}
\begin{Verbatim}[commandchars=\\\{\},fontsize=\small]
\PYG{c+c1}{\PYGZsh{} Define the number of measurements and the true mean}
\PYG{n}{n} \PYG{o}{=}\FancyVerbBreakAfterBreak{} \PYG{l+m+mi}{1000}
\PYG{n}{loc\PYGZus{}true} \PYG{o}{=}\FancyVerbBreakAfterBreak{} \PYG{l+m+mi}{1}

\PYG{c+c1}{\PYGZsh{} Simulate measurements from a normal distribution with a standard deviation of one}
\PYG{n}{x\PYGZus{}sample} \PYG{o}{=}\FancyVerbBreakAfterBreak{} \PYG{n}{np}\PYG{o}{.}\FancyVerbBreakAfterBreak{}\PYG{n}{random}\PYG{o}{.}\FancyVerbBreakAfterBreak{}\PYG{n}{normal}\PYG{p}{(}\FancyVerbBreakAfterBreak{}\PYG{n}{loc}\PYG{o}{=}\FancyVerbBreakAfterBreak{}\PYG{n}{loc\PYGZus{}true}\PYG{p}{,}\FancyVerbBreakAfterBreak{} \PYG{n}{scale}\PYG{o}{=}\FancyVerbBreakAfterBreak{}\PYG{l+m+mi}{1}\PYG{p}{,}\FancyVerbBreakAfterBreak{} \PYG{n}{size}\PYG{o}{=}\FancyVerbBreakAfterBreak{}\PYG{n}{n}\PYG{p}{)}

\PYG{c+c1}{\PYGZsh{} Print the sample mean}
\PYG{n+nb}{print}\PYG{p}{(}\FancyVerbBreakAfterBreak{}\PYG{l+s+sa}{f}\PYG{l+s+s2}{\PYGZdq{}}\PYG{l+s+s2}{Sample mean = }\PYG{l+s+si}{\PYGZob{}}\PYG{n}{np}\PYG{o}{.}\FancyVerbBreakAfterBreak{}\PYG{n}{mean}\PYG{p}{(}\FancyVerbBreakAfterBreak{}\PYG{n}{x\PYGZus{}sample}\PYG{p}{)}\PYG{l+s+si}{:}\PYG{l+s+s2}{.4f}\PYG{l+s+si}{\PYGZcb{}}\FancyVerbBreakAfterBreak{}\PYG{l+s+s2}{\PYGZdq{}}\PYG{p}{)}
\end{Verbatim}
\end{codecell}
\begin{Verbatim}[fontsize=\footnotesize, xleftmargin=5mm]
Sample mean = 1.0193
\end{Verbatim}

The histogram shows the simulated sample with the true mean and the sample mean.

\tutorialbarrier
\begin{codecell}

\end{codecell}
\tutorialplot{\includegraphics[width=4.449in,height=0.75\textheight,keepaspectratio]{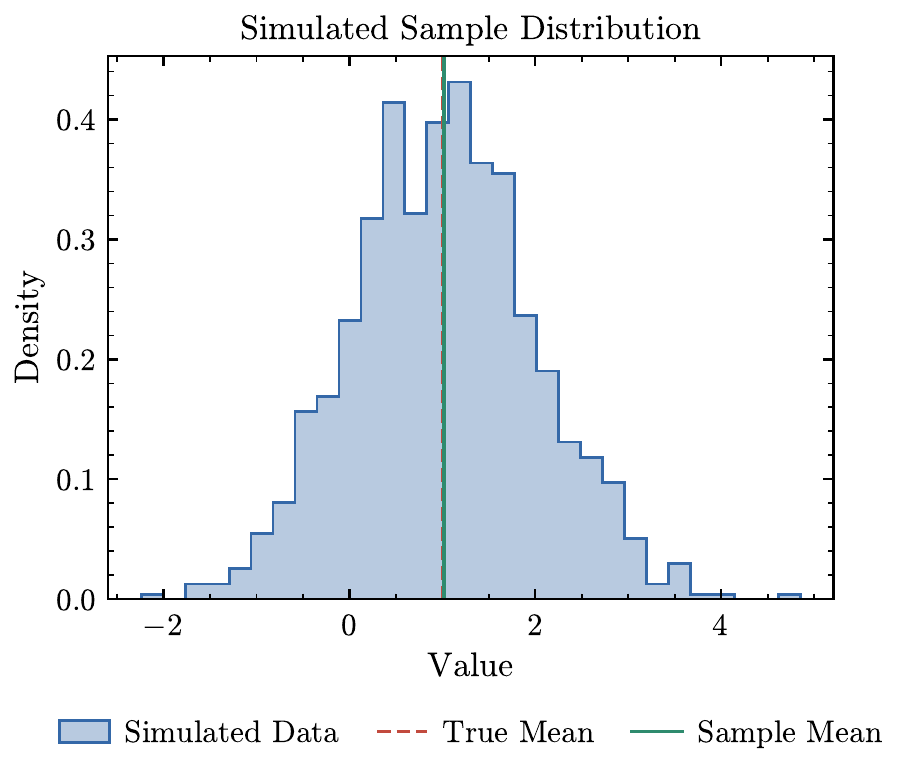}}%
  {\includegraphics[height=\tutorialheight]{tutorials/13a/fig01.pdf}}

The sample mean is a little above the true mean, because the measurements have noise. The posterior should therefore center near the sample mean and not exactly at the true mean.

\subsection*{Implementing the Log-Posterior Function}

The first step of the implementation is the log-posterior function. It sums two terms. The log likelihood measures how well a proposed mean \(\mu\) explains the data. The log prior encodes our prior beliefs about plausible values of \(\mu\).

\tutorialbarrier
\begin{codecell}
\begin{Verbatim}[commandchars=\\\{\},fontsize=\small]
\PYG{k}{def}\PYG{+w}{ }\PYG{n+nf}{log\PYGZus{}posterior}\PYG{p}{(}\FancyVerbBreakAfterBreak{}\PYG{n}{mu}\PYG{p}{,}\FancyVerbBreakAfterBreak{} \PYG{n}{x\PYGZus{}sample}\PYG{p}{)}\PYG{p}{:}
\PYG{+w}{    }\PYG{l+s+sd}{\PYGZdq{}\PYGZdq{}\PYGZdq{}Return the log posterior of the mean mu, up to a constant.\PYGZdq{}\PYGZdq{}\PYGZdq{}}

    \PYG{c+c1}{\PYGZsh{} Log likelihood (normal distribution with known standard deviation 1)}
    \PYG{n}{log\PYGZus{}likelihood} \PYG{o}{=}\FancyVerbBreakAfterBreak{} \PYG{o}{\PYGZhy{}}\PYG{l+m+mf}{0.5} \PYG{o}{*} \PYG{n}{np}\PYG{o}{.}\FancyVerbBreakAfterBreak{}\PYG{n}{sum}\PYG{p}{(}\FancyVerbBreakAfterBreak{}\PYG{p}{(}\FancyVerbBreakAfterBreak{}\PYG{n}{x\PYGZus{}sample} \PYG{o}{\PYGZhy{}} \PYG{n}{mu}\PYG{p}{)} \PYG{o}{*}\PYG{o}{*} \PYG{l+m+mi}{2}\PYG{p}{)} \PYG{o}{\PYGZhy{}} \PYG{l+m+mf}{0.5} \PYG{o}{*} \PYG{n+nb}{len}\PYG{p}{(}\FancyVerbBreakAfterBreak{}\PYG{n}{x\PYGZus{}sample}\PYG{p}{)} \PYG{o}{*} \PYG{n}{np}\PYG{o}{.}\FancyVerbBreakAfterBreak{}\PYG{n}{log}\PYG{p}{(}\FancyVerbBreakAfterBreak{}\PYG{l+m+mi}{2} \PYG{o}{*} \PYG{n}{np}\PYG{o}{.}\FancyVerbBreakAfterBreak{}\PYG{n}{pi}\PYG{p}{)}

    \PYG{c+c1}{\PYGZsh{} Log prior (Cauchy distribution)}
    \PYG{n}{log\PYGZus{}prior} \PYG{o}{=}\FancyVerbBreakAfterBreak{} \PYG{o}{\PYGZhy{}}\PYG{n}{np}\PYG{o}{.}\FancyVerbBreakAfterBreak{}\PYG{n}{log}\PYG{p}{(}\FancyVerbBreakAfterBreak{}\PYG{n}{np}\PYG{o}{.}\FancyVerbBreakAfterBreak{}\PYG{n}{pi} \PYG{o}{*} \PYG{l+m+mi}{3} \PYG{o}{*} \PYG{p}{(}\FancyVerbBreakAfterBreak{}\PYG{l+m+mi}{1} \PYG{o}{+}\FancyVerbBreakAfterBreak{} \PYG{p}{(}\FancyVerbBreakAfterBreak{}\PYG{n}{mu} \PYG{o}{/} \PYG{l+m+mi}{3}\PYG{p}{)} \PYG{o}{*}\PYG{o}{*} \PYG{l+m+mi}{2}\PYG{p}{)}\PYG{p}{)}

    \PYG{c+c1}{\PYGZsh{} Log posterior (up to a constant)}
    \PYG{k}{return} \PYG{n}{log\PYGZus{}likelihood} \PYG{o}{+}\FancyVerbBreakAfterBreak{} \PYG{n}{log\PYGZus{}prior}
\end{Verbatim}
\end{codecell}

The function takes a value of \(\mu\) and the data and returns the log posterior up to a constant. The Metropolis algorithm calls it at every step.

\subsection*{The Proposal Mechanism}

The proposal distribution decides how the chain moves through parameter space, so it sets the sampling efficiency. We use a normal distribution centered at the current value of \(\mu\). Its standard deviation \(\nu\), the proposal width, controls the step size,

\[q(\mu' | \mu) = \mathcal{N}(\mu' | \mu, \nu^2).\]

This is the Gaussian proposal of the chapter, with the covariance matrix \(\boldsymbol{\Sigma}\) reduced to the number \(\nu^2\). This proposal is symmetric, \(q(\mu' | \mu) = q(\mu | \mu')\), so the Metropolis algorithm applies. The width \(\nu\) is a tuning choice. If \(\nu\) is too small, the chain accepts most steps but moves slowly. If \(\nu\) is too large, most proposals land in regions of low probability and are rejected.

\subsection*{Implementing the Metropolis Algorithm}

The function below runs the algorithm for a given number of iterations and counts the accepted proposals. The acceptance rate gives feedback on the width \(\nu\). A very high rate suggests that the steps are too small, and a very low rate suggests that they are too large. For random walk proposals, rates of about 20 to 40 percent tend to be the most efficient (Section 13.9).

\tutorialbarrier
\begin{codecell}
\begin{Verbatim}[commandchars=\\\{\},fontsize=\small]
\PYG{k}{def}\PYG{+w}{ }\PYG{n+nf}{metropolis}\PYG{p}{(}\FancyVerbBreakAfterBreak{}\PYG{n}{mu\PYGZus{}init}\PYG{p}{,}\FancyVerbBreakAfterBreak{} \PYG{n}{nu}\PYG{p}{,}\FancyVerbBreakAfterBreak{} \PYG{n}{n\PYGZus{}iterations}\PYG{p}{)}\PYG{p}{:}
\PYG{+w}{    }\PYG{l+s+sd}{\PYGZdq{}\PYGZdq{}\PYGZdq{}Run the Metropolis algorithm and return the chain and the acceptance rate in percent.\PYGZdq{}\PYGZdq{}\PYGZdq{}}

    \PYG{c+c1}{\PYGZsh{} Initialize the chain with the starting value}
    \PYG{n}{chain} \PYG{o}{=}\FancyVerbBreakAfterBreak{} \PYG{p}{[}\FancyVerbBreakAfterBreak{}\PYG{n}{mu\PYGZus{}init}\PYG{p}{]}
    \PYG{n}{mu} \PYG{o}{=}\FancyVerbBreakAfterBreak{} \PYG{n}{mu\PYGZus{}init}
    \PYG{n}{accept\PYGZus{}count} \PYG{o}{=}\FancyVerbBreakAfterBreak{} \PYG{l+m+mi}{0}

    \PYG{c+c1}{\PYGZsh{} Run the chain}
    \PYG{k}{for} \PYG{n}{i} \PYG{o+ow}{in} \PYG{n+nb}{range}\PYG{p}{(}\FancyVerbBreakAfterBreak{}\PYG{n}{n\PYGZus{}iterations}\PYG{p}{)}\PYG{p}{:}
        \PYG{c+c1}{\PYGZsh{} Propose a new value from a normal distribution centered at mu}
        \PYG{n}{mu\PYGZus{}star} \PYG{o}{=}\FancyVerbBreakAfterBreak{} \PYG{n}{np}\PYG{o}{.}\FancyVerbBreakAfterBreak{}\PYG{n}{random}\PYG{o}{.}\FancyVerbBreakAfterBreak{}\PYG{n}{normal}\PYG{p}{(}\FancyVerbBreakAfterBreak{}\PYG{n}{mu}\PYG{p}{,}\FancyVerbBreakAfterBreak{} \PYG{n}{nu}\PYG{p}{)}

        \PYG{c+c1}{\PYGZsh{} Compute the acceptance probability (the ratio in log space)}
        \PYG{n}{log\PYGZus{}ratio} \PYG{o}{=}\FancyVerbBreakAfterBreak{} \PYG{n}{log\PYGZus{}posterior}\PYG{p}{(}\FancyVerbBreakAfterBreak{}\PYG{n}{mu\PYGZus{}star}\PYG{p}{,}\FancyVerbBreakAfterBreak{} \PYG{n}{x\PYGZus{}sample}\PYG{p}{)} \PYG{o}{\PYGZhy{}} \PYG{n}{log\PYGZus{}posterior}\PYG{p}{(}\FancyVerbBreakAfterBreak{}\PYG{n}{mu}\PYG{p}{,}\FancyVerbBreakAfterBreak{} \PYG{n}{x\PYGZus{}sample}\PYG{p}{)}
        \PYG{n}{alpha} \PYG{o}{=}\FancyVerbBreakAfterBreak{} \PYG{n+nb}{min}\PYG{p}{(}\FancyVerbBreakAfterBreak{}\PYG{l+m+mi}{1}\PYG{p}{,}\FancyVerbBreakAfterBreak{} \PYG{n}{np}\PYG{o}{.}\FancyVerbBreakAfterBreak{}\PYG{n}{exp}\PYG{p}{(}\FancyVerbBreakAfterBreak{}\PYG{n}{log\PYGZus{}ratio}\PYG{p}{)}\PYG{p}{)}

        \PYG{c+c1}{\PYGZsh{} Accept or reject}
        \PYG{n}{u} \PYG{o}{=}\FancyVerbBreakAfterBreak{} \PYG{n}{np}\PYG{o}{.}\FancyVerbBreakAfterBreak{}\PYG{n}{random}\PYG{o}{.}\FancyVerbBreakAfterBreak{}\PYG{n}{uniform}\PYG{p}{(}\FancyVerbBreakAfterBreak{}\PYG{p}{)}
        \PYG{k}{if} \PYG{n}{u} \PYG{o}{\PYGZlt{}} \PYG{n}{alpha}\PYG{p}{:}
            \PYG{n}{mu} \PYG{o}{=}\FancyVerbBreakAfterBreak{} \PYG{n}{mu\PYGZus{}star}
            \PYG{n}{accept\PYGZus{}count} \PYG{o}{+}\FancyVerbBreakAfterBreak{}\PYG{o}{=}\FancyVerbBreakAfterBreak{} \PYG{l+m+mi}{1}

        \PYG{c+c1}{\PYGZsh{} Add the current state to the chain}
        \PYG{n}{chain}\PYG{o}{.}\FancyVerbBreakAfterBreak{}\PYG{n}{append}\PYG{p}{(}\FancyVerbBreakAfterBreak{}\PYG{n}{mu}\PYG{p}{)}

    \PYG{c+c1}{\PYGZsh{} Acceptance rate in percent}
    \PYG{n}{accept\PYGZus{}rate} \PYG{o}{=}\FancyVerbBreakAfterBreak{} \PYG{n}{accept\PYGZus{}count} \PYG{o}{/} \PYG{n}{n\PYGZus{}iterations} \PYG{o}{*} \PYG{l+m+mi}{100}

    \PYG{k}{return} \PYG{n}{chain}\PYG{p}{,}\FancyVerbBreakAfterBreak{} \PYG{n}{accept\PYGZus{}rate}
\end{Verbatim}
\end{codecell}

The function returns the whole chain and its acceptance rate in percent. We now run it for three proposal widths.

The posterior of \(\mu\) is narrow, because 1000 measurements fix the mean well. Its standard deviation is close to the standard error of the mean, \(\sigma/\sqrt{N} = 1/\sqrt{1000} \approx 0.03\) (Section 3.6). This is the sampling noise of an average. The mean of \(N\) noisy measurements scatters by \(\sigma/\sqrt{N}\) around the true mean, because the noise partly cancels.

We run the algorithm with three proposal widths, \(\nu = 0.01\), 0.1 and 0.5. They are one third, 3 times and 16 times this scale. Each chain has 10,000 iterations and starts at \(\mu_{\text{init}} = 0\), about 30 standard errors away, so its first samples do not follow the posterior. We call this early stretch the burn-in.

\tutorialbarrier
\begin{codecell}
\begin{Verbatim}[commandchars=\\\{\},fontsize=\small]
\PYG{c+c1}{\PYGZsh{} Define the parameters of the experiment}
\PYG{n}{nu\PYGZus{}values} \PYG{o}{=}\FancyVerbBreakAfterBreak{} \PYG{p}{[}\FancyVerbBreakAfterBreak{}\PYG{l+m+mf}{0.01}\PYG{p}{,}\FancyVerbBreakAfterBreak{} \PYG{l+m+mf}{0.1}\PYG{p}{,}\FancyVerbBreakAfterBreak{} \PYG{l+m+mf}{0.5}\PYG{p}{]}
\PYG{n}{mu\PYGZus{}init} \PYG{o}{=}\FancyVerbBreakAfterBreak{} \PYG{l+m+mi}{0}
\PYG{n}{n\PYGZus{}iterations} \PYG{o}{=}\FancyVerbBreakAfterBreak{} \PYG{l+m+mi}{10000}
\PYG{n}{results} \PYG{o}{=}\FancyVerbBreakAfterBreak{} \PYG{p}{\PYGZob{}}\PYG{p}{\PYGZcb{}}\FancyVerbBreakAfterBreak{}

\PYG{c+c1}{\PYGZsh{} Run the Metropolis algorithm for each proposal width}
\PYG{k}{for} \PYG{n}{nu} \PYG{o+ow}{in} \PYG{n}{nu\PYGZus{}values}\PYG{p}{:}
    \PYG{n}{chain}\PYG{p}{,}\FancyVerbBreakAfterBreak{} \PYG{n}{accept\PYGZus{}rate} \PYG{o}{=}\FancyVerbBreakAfterBreak{} \PYG{n}{metropolis}\PYG{p}{(}\FancyVerbBreakAfterBreak{}\PYG{n}{mu\PYGZus{}init}\PYG{p}{,}\FancyVerbBreakAfterBreak{} \PYG{n}{nu}\PYG{p}{,}\FancyVerbBreakAfterBreak{} \PYG{n}{n\PYGZus{}iterations}\PYG{p}{)}
    \PYG{n}{results}\PYG{p}{[}\FancyVerbBreakAfterBreak{}\PYG{n}{nu}\PYG{p}{]} \PYG{o}{=}\FancyVerbBreakAfterBreak{} \PYG{p}{(}\FancyVerbBreakAfterBreak{}\PYG{n}{chain}\PYG{p}{,}\FancyVerbBreakAfterBreak{} \PYG{n}{accept\PYGZus{}rate}\PYG{p}{)}
\end{Verbatim}
\end{codecell}

\subsection*{Visualizing the Chains}

We plot the first 1000 samples of each chain. The trace plots show the value of each chain against the iteration, so they show how quickly the chain reaches the posterior. They also show how well it mixes, which means how quickly it moves through the region of high probability. Each panel gives the acceptance rate of its chain.

\tutorialbarrier
\begin{codecell}

\end{codecell}
\tutorialplot{\includegraphics[width=6.500in,height=0.75\textheight,keepaspectratio]{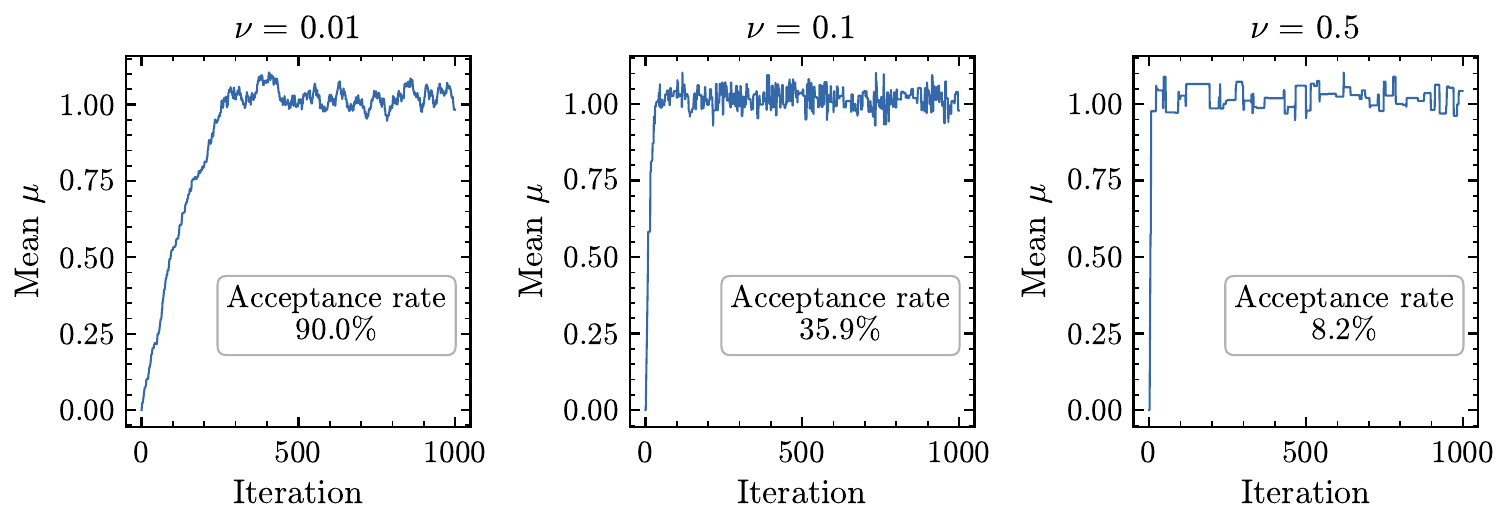}}%
  {\includegraphics[height=\tutorialheight]{tutorials/13a/fig02.pdf}}

The three widths give different chains.

\begin{itemize}
\tightlist
\item
  \textbf{Small \(\nu = 0.01\).} The chain climbs slowly from its starting value and reaches the posterior after about 250 iterations. Almost every proposal is accepted, because a tiny step stays close to the current state. Consecutive samples are highly correlated and give little new information.
\item
  \textbf{Medium \(\nu = 0.1\).} The chain reaches the posterior quickly. It then moves between states with moderate jumps and covers the range of plausible values.
\item
  \textbf{Large \(\nu = 0.5\).} The chain reaches the posterior within a few iterations. It then stays at the same value for long stretches and only occasionally jumps to a new value, because most proposals are rejected.
\end{itemize}

We discard the burn-in (Section 13.13). The traces show that all three chains have reached the posterior by iteration 500, so we discard the first 500 iterations of each chain. The convergence tests below check this choice.

\subsection*{Autocorrelation and Effective Sample Size}

We now quantify these observations with the autocorrelation of the chains. Successive samples of an MCMC chain are correlated, because each new state depends on the previous one. For a sequence of samples \(\{X_0, X_1, X_2, \ldots\}\), the autocorrelation at lag \(k\) measures how correlated two samples are that are \(k\) steps apart (Section 13.10),

\[R(k) = \frac{\mathbb{E}[(X_t - \mu_X)(X_{t+k} - \mu_X)]}{\sigma_X^2},\]

where \(\mu_X\) and \(\sigma_X^2\) are the mean and the variance of the stationary distribution. Independent samples have \(R(k) = 0\) for all \(k > 0\). For MCMC samples, \(R(k)\) typically decays as the lag increases, and the decay rate decides how many samples we need for reliable estimates.

The effective sample size (ESS) is the number of independent samples that carry the same information as our correlated chain (Section 13.11),

\[\text{ESS} = \frac{N}{1 + 2\sum_{k=1}^{\infty}R(k)},\]

where \(N\) is the number of samples in the chain. A chain with a high autocorrelation has an ESS far below \(N\). In practice we estimate \(R(k)\) from the chain and sum it only up to lag 100, because the estimate becomes noisy at large lags.

Computing \(R(k)\) at every lag directly takes \(O(N^2)\) operations. The chapter therefore uses the fast Fourier transform (FFT), but it only states the method (Section 13.10). We therefore introduce the Fourier transform first.

\subsubsection*{The Fourier Transform and the Autocorrelation}

\textbf{The Fourier transform.} A chain of \(M\) numbers \(x_0, \ldots, x_{M-1}\) can be written as a sum of sinusoids. The discrete Fourier transform gives the amount of each sinusoid,

\tutorialdisplay{\tilde{x}_f = \sum_{t=0}^{M-1} x_t\, e^{-2\pi i f t/M},}{f = 0, 1, \ldots, M-1.}

Here \(i\) is the imaginary unit, and \(e^{-2\pi i f t/M} = \cos(2\pi f t/M) - i \sin(2\pi f t/M)\) is a sinusoid that completes \(f\) cycles over the \(M\) samples. The number \(\tilde{x}_f\) is large when the chain oscillates with \(f\) cycles and small when it does not. The inverse transform recovers the chain, so no information is lost,

\[x_t = \frac{1}{M}\sum_{f=0}^{M-1} \tilde{x}_f\, e^{2\pi i f t/M}.\]

\textbf{The power spectrum.} The squared magnitude \(|\tilde{x}_f|^2 = \tilde{x}_f \tilde{x}_f^*\) is the power spectrum, where the star is the complex conjugate. It shows how much of the variation of the chain sits at each frequency.

\textbf{The Wiener-Khinchin theorem.} Let \(x_t\) be the centered chain of \(N\) samples, padded with \(N\) zeros so that \(M = 2N\). Define \(C(k)\) as the sum of the products of samples \(k\) steps apart, with \(C(-k) = C(k)\) and \(x_t = 0\) beyond the chain,

\[C(k) = \sum_{t} x_t\, x_{t+k}.\]

The theorem says that the power spectrum is the Fourier transform of \(C(k)\),

\[|\tilde{x}_f|^2 = \sum_{k=-(N-1)}^{N-1} C(k)\, e^{-2\pi i f k/M}.\]

We do not prove it here. The squared magnitude is a double sum over pairs of samples, and the phase of each pair depends only on the lag between them. Collecting the pairs by lag gives \(C(k)\). The inverse transform of the power spectrum therefore returns \(C(k)\) at every lag at once.

The padding matters, because the transform treats the chain as periodic. With \(2N\) points, the lags from \(-(N-1)\) to \(N-1\) do not overlap, so the end of the chain does not wrap around to its start.

\textbf{From \(C(k)\) to \(R(k)\).} The autocovariance \(C(k)/N\) is the average product of samples \(k\) steps apart. At lag 0 it is the variance of the chain. Dividing by it gives \(R(k) = C(k)/C(0)\), so that \(R(0) = 1\).

\textbf{The fast Fourier transform.} Each value \(\tilde{x}_f\) is a sum of \(M\) terms, so the definition needs \(O(M^2)\) operations for all frequencies. This is as slow as the direct sums of products. The FFT gives the same values in \(O(M \log M)\) operations by reusing partial sums. The route through the power spectrum needs one transform and one inverse transform, so it costs \(O(N \log N)\).

\textbf{The steps.} The function below follows four steps.

\begin{enumerate}
\def\labelenumi{\arabic{enumi}.}
\tightlist
\item
  \textbf{Center.} Subtract the mean of the chain, so that the products measure deviations from the mean.
\item
  \textbf{Transform.} Pad the chain with \(N\) zeros and take its Fourier transform with the FFT.
\item
  \textbf{Power spectrum.} Multiply the transform by its complex conjugate.
\item
  \textbf{Invert.} The inverse transform gives \(C(k)\). We divide by \(N\) for the autocovariance and by its value at lag 0 for \(R(k)\).
\end{enumerate}

We write one function for these steps, one that plots \(R(k)\) with a 95 percent band, and one that returns the ESS.

\tutorialbarrier
\begin{codecell}

\end{codecell}

The functions give \(R(k)\) of a chain, plot it, and return the ESS from the sum of \(R(k)\) up to lag 100.

We now plot the autocorrelation function of each chain after the burn-in. Each panel gives the acceptance rate and the ESS of its chain.

The red band marks the scatter of the estimated autocorrelation around zero for independent samples. At a lag \(k > 0\), the estimate is the average of \(N\) products of two independent samples, each with mean 0 and variance 1. Each product has mean 0 and variance 1, so the average has a standard error of \(1/\sqrt{N}\) (Section 3.6). By the central limit theorem (Section 2.3), the average is also nearly normal. A normal distribution has 95 percent of its probability within 1.96 standard deviations, close to \(2\sigma\), so the band is \(\pm 1.96/\sqrt{N}\).

We want the autocorrelation to fall into the band quickly. Inside the band, samples \(k\) steps apart cannot be told from independent samples, so the chain has forgotten where it was \(k\) steps earlier. The lag where the autocorrelation enters the band is roughly the number of steps that the chain needs for a new, nearly independent sample. The band is narrow because \(N\) is large, so only a small autocorrelation counts as zero.

\tutorialbarrier
\begin{codecell}

\end{codecell}
\tutorialplot{\includegraphics[width=6.500in,height=0.75\textheight,keepaspectratio]{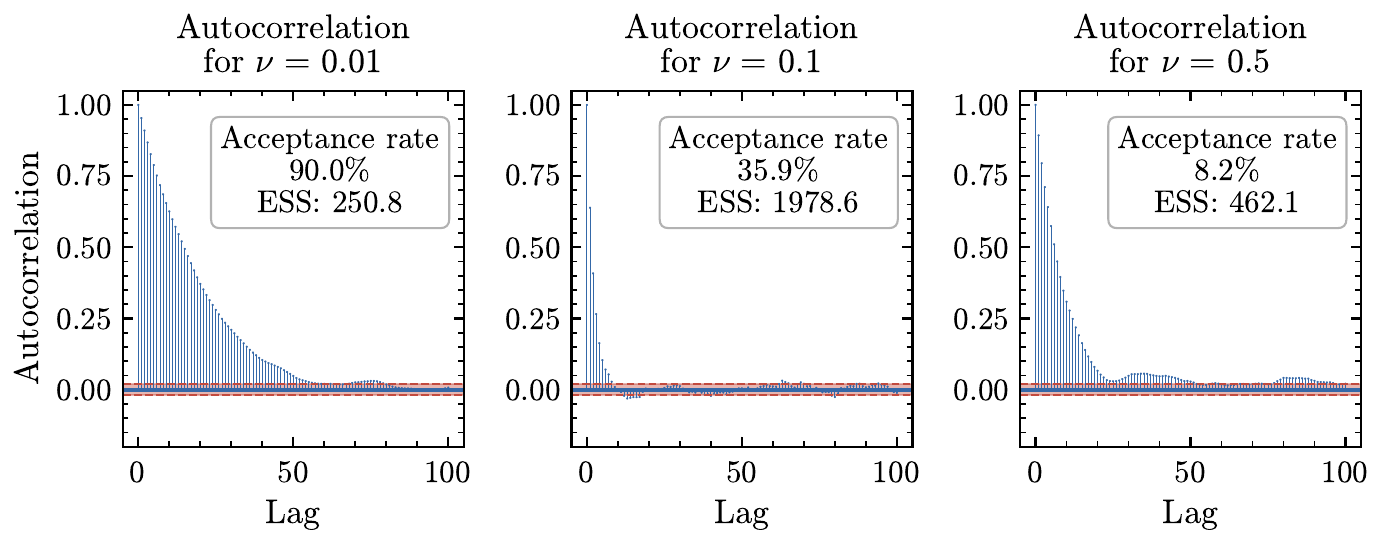}}%
  {\includegraphics[height=\tutorialheight]{tutorials/13a/fig03.pdf}}

The autocorrelation functions differ as the trace plots suggest.

\begin{itemize}
\tightlist
\item
  \textbf{Small \(\nu = 0.01\).} The autocorrelation decays slowly and stays outside the band for tens of lags. Samples that are many iterations apart still hold redundant information, so the ESS is far below the number of samples.
\item
  \textbf{Medium \(\nu = 0.1\).} The autocorrelation falls close to the band after a few lags, so the chain forgets its past quickly. The ESS is the largest of the three, so this is our most efficient sampler.
\item
  \textbf{Large \(\nu = 0.5\).} The rejected proposals leave the chain in place, so the autocorrelation starts high. The ESS lies between those of the small and the medium width.
\end{itemize}

As a rule of thumb, an ESS above 100 is enough to estimate a mean and a standard deviation (Section 13.11). All three chains meet this rule, but the ESS of the medium chain is several times that of the narrow chain.

The narrow chain accepts the most proposals but has the lowest ESS, so a high acceptance rate alone does not make an efficient chain. We must tune the proposal width to balance the acceptance rate and the mixing. The medium width has an acceptance rate between 20 and 40 percent and a rapid decay of the autocorrelation, and it gives the largest ESS.

\subsection*{Convergence Diagnostics}

The ESS assumes that a chain follows the posterior. Before we use the samples for inference, we must therefore check that the chains have converged to their stationary distribution. We judged the traces by eye. This is easy for one parameter and hard for many parameters, so we also use statistical tests (Section 13.13).

\subsubsection*{The Geweke Test}

The Geweke test compares the means of two segments of one chain. Segment A is the first 10 percent and segment B is the last 50 percent. If the chain has converged, both segments sample the same distribution, so their means agree. The function below has four steps.

\begin{enumerate}
\def\labelenumi{\arabic{enumi}.}
\tightlist
\item
  \textbf{Segments and means.} Take the first and the last part of the chain, and calculate the mean \(\bar{x}\) of each.
\item
  \textbf{Autocorrelation.} Correlated samples make a mean less certain than independent samples do. We therefore calculate \(R(k)\) of each segment with our autocorrelation function.
\item
  \textbf{Variance of the mean.} For \(N\) independent samples, the variance of the mean is \(\hat{\sigma}^2/N\) (Section 3.6). A correlated segment holds only ESS independent samples, so the variance is larger, \(\hat{\sigma}^2/\text{ESS}\). The ESS formula turns this into the expression below. A window lowers the weight of the noisy high lags in the sum, which stops at a maximum lag \(l\).
\item
  \textbf{Z-score.} The difference of the two means divided by its standard error gives \(Z\).
\end{enumerate}

For a segment of \(N\) samples with sample variance \(\hat{\sigma}^2\), the variance of its mean is

\[\hat{S} = \frac{\hat{\sigma}^2}{N}\left(1 + 2\sum_{k=1}^{l} w(k)\, R(k)\right), \quad w(k) = 1 - \frac{k}{l}.\]

The Z-score of the two segments A and B is then

\[Z = \frac{\bar{x}_A - \bar{x}_B}{\sqrt{\hat{S}_A + \hat{S}_B}}.\]

We consider the chain converged if \(|Z| < 2\), which contains about 95 percent of a standard normal distribution.

\tutorialbarrier
\begin{codecell}

\end{codecell}

The function returns the Z-score of one chain.

We first apply the test to the whole chain of each proposal width, before we remove the burn-in. A chain that still contains the climb from its starting value should fail the test, so the test shows whether a chain needs a burn-in.

\tutorialbarrier
\begin{codecell}
\begin{Verbatim}[commandchars=\\\{\},fontsize=\small]
\PYG{c+c1}{\PYGZsh{} Test convergence for each chain}
\PYG{k}{for} \PYG{n}{nu} \PYG{o+ow}{in} \PYG{n}{nu\PYGZus{}values}\PYG{p}{:}
    \PYG{n}{chain}\PYG{p}{,}\FancyVerbBreakAfterBreak{} \PYG{n}{\PYGZus{}} \PYG{o}{=}\FancyVerbBreakAfterBreak{} \PYG{n}{results}\PYG{p}{[}\FancyVerbBreakAfterBreak{}\PYG{n}{nu}\PYG{p}{]}
    \PYG{n+nb}{print}\PYG{p}{(}\FancyVerbBreakAfterBreak{}\PYG{l+s+sa}{f}\PYG{l+s+s2}{\PYGZdq{}}\PYG{l+s+s2}{ν = }\PYG{l+s+si}{\PYGZob{}}\PYG{n}{nu}\PYG{l+s+si}{\PYGZcb{}}\FancyVerbBreakAfterBreak{}\PYG{l+s+s2}{, Z\PYGZhy{}score = }\PYG{l+s+si}{\PYGZob{}}\PYG{n}{geweke\PYGZus{}test}\PYG{p}{(}\FancyVerbBreakAfterBreak{}\PYG{n}{chain}\PYG{p}{)}\PYG{l+s+si}{:}\PYG{l+s+s2}{.3f}\PYG{l+s+si}{\PYGZcb{}}\FancyVerbBreakAfterBreak{}\PYG{l+s+s2}{\PYGZdq{}}\PYG{p}{)}
\end{Verbatim}
\end{codecell}
\begin{Verbatim}[fontsize=\footnotesize, xleftmargin=5mm]
ν = 0.01, Z-score = -3.396
ν = 0.1, Z-score = -1.247
ν = 0.5, Z-score = -0.611
\end{Verbatim}

The test flags the chain with the small proposal width, because its first 10 percent still contains the climb from the starting value. The mean of that segment then differs from the mean of the last 50 percent. The chains with the medium and the large width reach the posterior within a few iterations, so they pass.

We now apply the test again after we remove the first 500 iterations of each chain.

\tutorialbarrier
\begin{codecell}
\begin{Verbatim}[commandchars=\\\{\},fontsize=\small]
\PYG{c+c1}{\PYGZsh{} Test convergence again after the burn\PYGZhy{}in}
\PYG{k}{for} \PYG{n}{nu} \PYG{o+ow}{in} \PYG{n}{nu\PYGZus{}values}\PYG{p}{:}
    \PYG{n}{chain}\PYG{p}{,}\FancyVerbBreakAfterBreak{} \PYG{n}{\PYGZus{}} \PYG{o}{=}\FancyVerbBreakAfterBreak{} \PYG{n}{results}\PYG{p}{[}\FancyVerbBreakAfterBreak{}\PYG{n}{nu}\PYG{p}{]}
    \PYG{n+nb}{print}\PYG{p}{(}\FancyVerbBreakAfterBreak{}\PYG{l+s+sa}{f}\PYG{l+s+s2}{\PYGZdq{}}\PYG{l+s+s2}{ν = }\PYG{l+s+si}{\PYGZob{}}\PYG{n}{nu}\PYG{l+s+si}{\PYGZcb{}}\FancyVerbBreakAfterBreak{}\PYG{l+s+s2}{, Z\PYGZhy{}score = }\PYG{l+s+si}{\PYGZob{}}\PYG{n}{geweke\PYGZus{}test}\PYG{p}{(}\FancyVerbBreakAfterBreak{}\PYG{n}{chain}\PYG{p}{[}\FancyVerbBreakAfterBreak{}\PYG{l+m+mi}{500}\PYG{p}{:}\PYG{p}{]}\PYG{p}{)}\PYG{l+s+si}{:}\PYG{l+s+s2}{.3f}\PYG{l+s+si}{\PYGZcb{}}\FancyVerbBreakAfterBreak{}\PYG{l+s+s2}{\PYGZdq{}}\PYG{p}{)}
\end{Verbatim}
\end{codecell}
\begin{Verbatim}[fontsize=\footnotesize, xleftmargin=5mm]
ν = 0.01, Z-score = -1.808
ν = 0.1, Z-score = -1.232
ν = 0.5, Z-score = -0.173
\end{Verbatim}

After the burn-in, all three chains pass the test, so 500 iterations are enough for this test. The narrow chain is closest to the limit. It moves the most slowly, so it is the chain most likely to need a longer burn-in.

\subsubsection*{The Gelman-Rubin Diagnostic}

The Geweke test looks at a single chain, and a single chain can pass it while it covers only part of the distribution. A second check therefore runs several chains from different starting points and compares them. If all chains converge to the same distribution, they become statistically indistinguishable after the burn-in. The Gelman-Rubin diagnostic compares the variance within the chains with the variance between them (Section 13.13).

For our parameter \(\mu\), we calculate two variances from the chains after the burn-in.

\begin{enumerate}
\def\labelenumi{\arabic{enumi}.}
\tightlist
\item
  \textbf{Within-chain variance \(W\).} The average of the variances of the individual chains.
\item
  \textbf{Between-chain variance \(B\).} The variance of the chain means, multiplied by the chain length \(N\) (n in the code). The mean of \(N\) samples has a variance about \(N\) times smaller than one sample, so the factor \(N\) makes \(B\) comparable with \(W\).
\end{enumerate}

The Gelman-Rubin statistic \(\hat{R}\) is then

\[\hat{R} = \sqrt{1 + \frac{B-W}{NW}}.\]

For converged chains, \(B\) is approximately equal to \(W\), so \(\hat{R}\) approaches 1. A value below 1.1, or below 1.05 for a stricter criterion, is generally taken as evidence of convergence. A value well above 1 shows that the chains have not converged to the same distribution.

\tutorialbarrier
\begin{codecell}

\end{codecell}

The function returns \(\hat{R}\) for a list of chains of equal length.

We run four chains with the medium proposal width \(\nu = 0.1\). Each chain has 5000 iterations and starts at a random value between \(-3\) and 3. We spread the starting points widely, because chains that start close together can agree before they have converged. We discard the first 500 samples of each chain.

\tutorialbarrier
\begin{codecell}

\end{codecell}
\begin{Verbatim}[fontsize=\footnotesize, xleftmargin=5mm]
Starting points: 2.34, -1.14, 2.11, -0.51
R-hat = 1.0000
\end{Verbatim}

We plot the first 1000 iterations of the four chains. The dashed line marks the end of the burn-in.

\tutorialbarrier
\begin{codecell}

\end{codecell}
\tutorialplot{\includegraphics[width=4.275in,height=0.75\textheight,keepaspectratio]{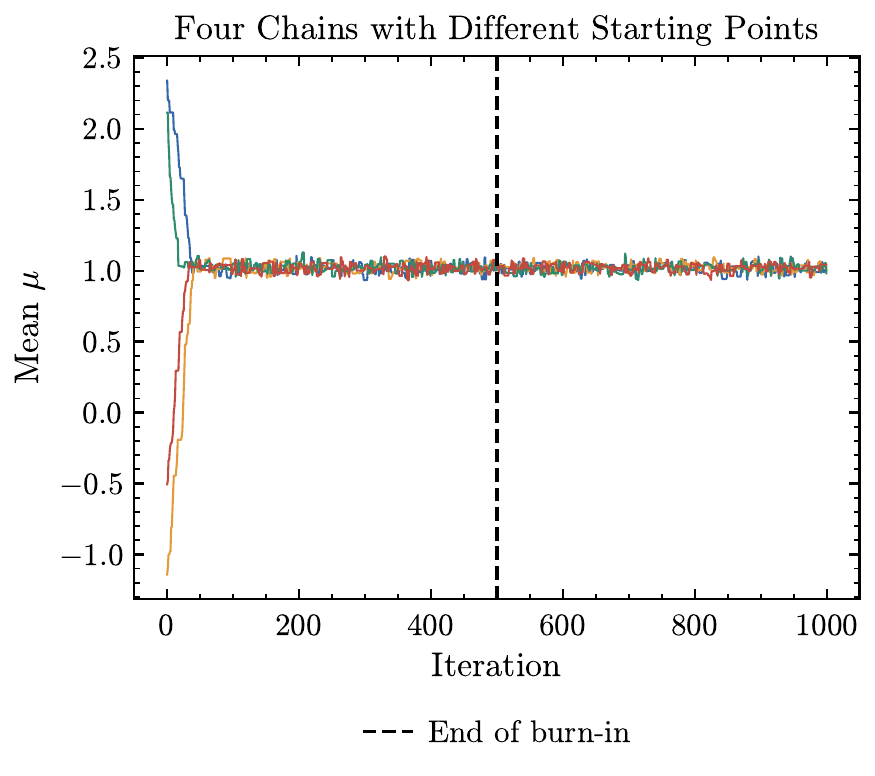}}%
  {\includegraphics[height=\tutorialheight]{tutorials/13a/fig04.pdf}}

The four chains start from different values and reach the same region within about 40 iterations, long before the end of the burn-in. The Gelman-Rubin statistic is close to 1, so they agree after the burn-in.

The Geweke test looks at parts of one chain, and the Gelman-Rubin diagnostic compares several chains. Together they indicate that the chain with the medium proposal width (\(\nu = 0.1\)) has converged. Both tests ask whether the chain has reached the posterior. They do not ask how many independent samples the chain holds, which is the question of the ESS. The chain with \(\nu = 0.5\) shows the difference, because it passes the Geweke test but has a much lower ESS than the medium chain. We therefore use both kinds of check.

\subsection*{Posterior Analysis}

We have now checked the convergence and the efficiency of the chains. We continue with the chain of the medium proposal width, which has the largest ESS. We remove the burn-in and summarize the remaining samples with the posterior mean, the standard deviation and the 95 percent interval.

\tutorialbarrier
\begin{codecell}

\end{codecell}
\begin{Verbatim}[fontsize=\footnotesize, xleftmargin=5mm]
Posterior mean = 1.019
Posterior standard deviation = 0.032
95% interval = [0.956, 1.078]
\end{Verbatim}

The histogram shows the posterior samples. A bar marks the 95 percent interval, and two lines mark the posterior mean and the true mean. The black curve is the normal distribution with the sample mean as its mean and the variance \(1/1000\). It is the posterior that the likelihood alone gives, which is the posterior for a flat prior. Our Cauchy prior is broad and nearly flat over the range of the histogram, so it changes this curve very little.

\tutorialbarrier
\begin{codecell}

\end{codecell}
\tutorialplot{\includegraphics[width=4.682in,height=0.75\textheight,keepaspectratio]{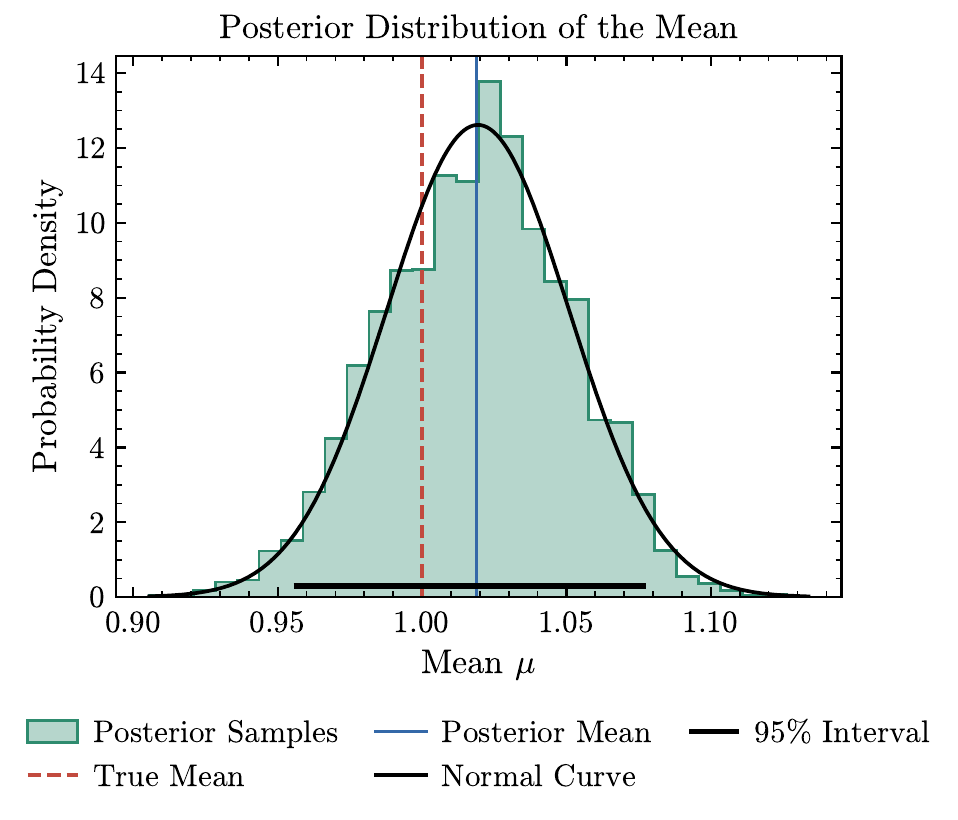}}%
  {\includegraphics[height=\tutorialheight]{tutorials/13a/fig05.pdf}}

\subsection*{Reading the Posterior Distribution}

We read three results from the posterior distribution.

\begin{enumerate}
\def\labelenumi{\arabic{enumi}.}
\tightlist
\item
  \textbf{True mean.} The true mean lies inside the 95 percent interval, so the sampler recovers the parameter. The posterior mean is close to the sample mean, as we expected.
\item
  \textbf{Uncertainty.} The posterior standard deviation agrees with the standard error of the mean, \(1/\sqrt{1000}\) (Section 3.6). This is the sampling noise of the average of 1000 measurements. The histogram follows the normal curve, so the 1000 measurements dominate the broad Cauchy prior.
\item
  \textbf{Posterior interval.} The 95 percent interval has a direct probabilistic meaning. Given our data and prior, we believe that \(\mu\) lies in this range with a probability of 95 percent.
\end{enumerate}

The posterior has no standard form, because the Cauchy prior is not conjugate to the normal likelihood (Section 5.2). MCMC does not need one, since it needs the posterior only up to a constant. It therefore also works for models whose posterior has no closed form. For one parameter, a grid would work as well (Chapter 12). The cost of a grid grows exponentially with the number of parameters, so MCMC becomes the practical choice.

\subsection*{Thinning the Chain}

The posterior summary used every sample of the chain. Many practitioners instead thin a chain, to reduce its autocorrelation and its storage, so we check what thinning does to the ESS. Thinning keeps only every \(m\)-th sample of a chain, so that the kept samples are less correlated (Section 13.12). We test it on the medium chain with \(m = 10\). We compare the number of samples and the ESS of the full chain after the burn-in with those of the thinned chain.

\tutorialbarrier
\begin{codecell}
\begin{Verbatim}[commandchars=\\\{\},fontsize=\small]
\PYG{c+c1}{\PYGZsh{} Keep every 10th sample of the medium chain after the burn\PYGZhy{}in}
\PYG{n}{thinned\PYGZus{}chain} \PYG{o}{=}\FancyVerbBreakAfterBreak{} \PYG{n}{chain\PYGZus{}post\PYGZus{}burnin}\PYG{p}{[}\FancyVerbBreakAfterBreak{}\PYG{p}{:}\PYG{p}{:}\PYG{l+m+mi}{10}\PYG{p}{]}

\PYG{c+c1}{\PYGZsh{} Compare the number of samples and the ESS of the full and the thinned chain}
\PYG{n+nb}{print}\PYG{p}{(}\FancyVerbBreakAfterBreak{}\PYG{l+s+sa}{f}\PYG{l+s+s2}{\PYGZdq{}}\PYG{l+s+s2}{Full chain: }\PYG{l+s+si}{\PYGZob{}}\PYG{n+nb}{len}\PYG{p}{(}\FancyVerbBreakAfterBreak{}\PYG{n}{chain\PYGZus{}post\PYGZus{}burnin}\PYG{p}{)}\PYG{l+s+si}{\PYGZcb{}}\FancyVerbBreakAfterBreak{}\PYG{l+s+s2}{ samples, ESS = }\PYG{l+s+si}{\PYGZob{}}\PYG{n}{calculate\PYGZus{}ess}\PYG{p}{(}\FancyVerbBreakAfterBreak{}\PYG{n}{chain\PYGZus{}post\PYGZus{}burnin}\PYG{p}{)}\PYG{l+s+si}{:}\PYG{l+s+s2}{.1f}\PYG{l+s+si}{\PYGZcb{}}\FancyVerbBreakAfterBreak{}\PYG{l+s+s2}{\PYGZdq{}}\PYG{p}{)}
\PYG{n+nb}{print}\PYG{p}{(}\FancyVerbBreakAfterBreak{}\PYG{l+s+sa}{f}\PYG{l+s+s2}{\PYGZdq{}}\PYG{l+s+s2}{Thinned chain: }\PYG{l+s+si}{\PYGZob{}}\PYG{n+nb}{len}\PYG{p}{(}\FancyVerbBreakAfterBreak{}\PYG{n}{thinned\PYGZus{}chain}\PYG{p}{)}\PYG{l+s+si}{\PYGZcb{}}\FancyVerbBreakAfterBreak{}\PYG{l+s+s2}{ samples, ESS = }\PYG{l+s+si}{\PYGZob{}}\PYG{n}{calculate\PYGZus{}ess}\PYG{p}{(}\FancyVerbBreakAfterBreak{}\PYG{n}{thinned\PYGZus{}chain}\PYG{p}{)}\PYG{l+s+si}{:}\PYG{l+s+s2}{.1f}\PYG{l+s+si}{\PYGZcb{}}\FancyVerbBreakAfterBreak{}\PYG{l+s+s2}{\PYGZdq{}}\PYG{p}{)}
\end{Verbatim}
\end{codecell}
\begin{Verbatim}[fontsize=\footnotesize, xleftmargin=5mm]
Full chain: 9501 samples, ESS = 1978.6
Thinned chain: 951 samples, ESS = 830.9
\end{Verbatim}

The thinned chain needs 10 times less storage, but its ESS is about 2.4 times lower. This is the trade-off of thinning. Each kept sample carries more information, because it is less correlated with its neighbors, but we keep far fewer samples. The removed samples still carried information, so thinning lowers the ESS. We therefore keep every sample for the posterior summary and thin a chain only when storage requires it (Section 13.12).

\section*{Gibbs Sampling}

The Metropolis algorithm treats the parameters as one unit and needs a proposal distribution, whose width we had to tune. Gibbs sampling instead updates one parameter at a time, so it needs two or more parameters and no proposal width. It applies when we can draw each parameter from its conditional distribution given the others (Section 13.15).

\subsection*{The Intuition Behind Gibbs Sampling}

The idea is to replace one hard draw from the joint distribution by several easy draws from conditional distributions. Gibbs sampling works in three steps.

\begin{enumerate}
\def\labelenumi{\arabic{enumi}.}
\tightlist
\item
  Update one parameter at a time and hold the others fixed.
\item
  Sample the new value directly from the conditional distribution of this parameter.
\item
  Cycle through all parameters in turn.
\end{enumerate}

It is useful when the conditional distributions have known forms that are easy to sample. Gibbs sampling accepts every new value, because it samples from the exact conditional distributions. It is a special case of the Metropolis-Hastings algorithm with an acceptance probability of 1 (Section 13.15).

\subsection*{Sampling from a Bivariate Gaussian Distribution}

We illustrate Gibbs sampling with a bivariate Gaussian distribution. We can write down the conditional distributions of this example in closed form. The example shows how the correlation of the parameters affects the sampling efficiency. We sample from a bivariate Gaussian distribution with zero mean and correlation coefficient \(\rho\),

\[\begin{pmatrix} x_1 \\ x_2 \end{pmatrix} \sim \mathcal{N}\left(\mathbf{0}, \begin{bmatrix} 1 & \rho \\ \rho & 1 \end{bmatrix}\right).\]

Gibbs sampling needs the conditional distributions \(p(x_1 | x_2)\) and \(p(x_2 | x_1)\). For a bivariate Gaussian distribution, they are also Gaussian (Section 5.9),

\[p(x_1 | x_2) = \mathcal{N}(\rho x_2, 1-\rho^2),\]

\[p(x_2 | x_1) = \mathcal{N}(\rho x_1, 1-\rho^2).\]

The conditional mean depends linearly on the conditioning variable and scales with \(\rho\). The conditional variance \(1-\rho^2\) decreases as the correlation increases. For \(\rho = 0\), both conditional distributions are \(\mathcal{N}(0, 1)\).

The Gibbs sampler alternates between two steps.

\begin{enumerate}
\def\labelenumi{\arabic{enumi}.}
\tightlist
\item
  Sample \(x_1^{(t+1)} \sim \mathcal{N}(\rho x_2^{(t)}, 1-\rho^2)\).
\item
  Sample \(x_2^{(t+1)} \sim \mathcal{N}(\rho x_1^{(t+1)}, 1-\rho^2)\).
\end{enumerate}

\tutorialbarrier
\begin{codecell}
\begin{Verbatim}[commandchars=\\\{\},fontsize=\small]
\PYG{k}{def}\PYG{+w}{ }\PYG{n+nf}{gibbs\PYGZus{}sampler}\PYG{p}{(}\FancyVerbBreakAfterBreak{}\PYG{n}{n\PYGZus{}samples}\PYG{p}{,}\FancyVerbBreakAfterBreak{} \PYG{n}{rho}\PYG{p}{)}\PYG{p}{:}
\PYG{+w}{    }\PYG{l+s+sd}{\PYGZdq{}\PYGZdq{}\PYGZdq{}Draw samples of a bivariate Gaussian with unit variances and correlation rho by Gibbs sampling.\PYGZdq{}\PYGZdq{}\PYGZdq{}}

    \PYG{c+c1}{\PYGZsh{} Initialize the storage for the samples}
    \PYG{n}{samples} \PYG{o}{=}\FancyVerbBreakAfterBreak{} \PYG{n}{np}\PYG{o}{.}\FancyVerbBreakAfterBreak{}\PYG{n}{zeros}\PYG{p}{(}\FancyVerbBreakAfterBreak{}\PYG{p}{(}\FancyVerbBreakAfterBreak{}\PYG{n}{n\PYGZus{}samples}\PYG{p}{,}\FancyVerbBreakAfterBreak{} \PYG{l+m+mi}{2}\PYG{p}{)}\PYG{p}{)}

    \PYG{c+c1}{\PYGZsh{} Start far from the center of the distribution to show the burn\PYGZhy{}in}
    \PYG{n}{x1} \PYG{o}{=}\FancyVerbBreakAfterBreak{} \PYG{l+m+mi}{100}
    \PYG{n}{x2} \PYG{o}{=}\FancyVerbBreakAfterBreak{} \PYG{l+m+mi}{100}

    \PYG{c+c1}{\PYGZsh{} Run the Gibbs sampler}
    \PYG{k}{for} \PYG{n}{i} \PYG{o+ow}{in} \PYG{n+nb}{range}\PYG{p}{(}\FancyVerbBreakAfterBreak{}\PYG{n}{n\PYGZus{}samples}\PYG{p}{)}\PYG{p}{:}
        \PYG{c+c1}{\PYGZsh{} Sample x1 from its conditional distribution given x2, N(rho*x2, 1\PYGZhy{}rho\PYGZca{}2)}
        \PYG{n}{x1} \PYG{o}{=}\FancyVerbBreakAfterBreak{} \PYG{n}{np}\PYG{o}{.}\FancyVerbBreakAfterBreak{}\PYG{n}{random}\PYG{o}{.}\FancyVerbBreakAfterBreak{}\PYG{n}{normal}\PYG{p}{(}\FancyVerbBreakAfterBreak{}\PYG{n}{loc}\PYG{o}{=}\FancyVerbBreakAfterBreak{}\PYG{n}{rho}\PYG{o}{*}\PYG{n}{x2}\PYG{p}{,}\FancyVerbBreakAfterBreak{} \PYG{n}{scale}\PYG{o}{=}\FancyVerbBreakAfterBreak{}\PYG{n}{np}\PYG{o}{.}\FancyVerbBreakAfterBreak{}\PYG{n}{sqrt}\PYG{p}{(}\FancyVerbBreakAfterBreak{}\PYG{l+m+mi}{1}\PYG{o}{\PYGZhy{}}\PYG{n}{rho}\PYG{o}{*}\PYG{o}{*}\PYG{l+m+mi}{2}\PYG{p}{)}\PYG{p}{)}

        \PYG{c+c1}{\PYGZsh{} Sample x2 from its conditional distribution given x1, N(rho*x1, 1\PYGZhy{}rho\PYGZca{}2)}
        \PYG{n}{x2} \PYG{o}{=}\FancyVerbBreakAfterBreak{} \PYG{n}{np}\PYG{o}{.}\FancyVerbBreakAfterBreak{}\PYG{n}{random}\PYG{o}{.}\FancyVerbBreakAfterBreak{}\PYG{n}{normal}\PYG{p}{(}\FancyVerbBreakAfterBreak{}\PYG{n}{loc}\PYG{o}{=}\FancyVerbBreakAfterBreak{}\PYG{n}{rho}\PYG{o}{*}\PYG{n}{x1}\PYG{p}{,}\FancyVerbBreakAfterBreak{} \PYG{n}{scale}\PYG{o}{=}\FancyVerbBreakAfterBreak{}\PYG{n}{np}\PYG{o}{.}\FancyVerbBreakAfterBreak{}\PYG{n}{sqrt}\PYG{p}{(}\FancyVerbBreakAfterBreak{}\PYG{l+m+mi}{1}\PYG{o}{\PYGZhy{}}\PYG{n}{rho}\PYG{o}{*}\PYG{o}{*}\PYG{l+m+mi}{2}\PYG{p}{)}\PYG{p}{)}

        \PYG{c+c1}{\PYGZsh{} Store the sample}
        \PYG{n}{samples}\PYG{p}{[}\FancyVerbBreakAfterBreak{}\PYG{n}{i}\PYG{p}{,}\FancyVerbBreakAfterBreak{} \PYG{p}{:}\PYG{p}{]} \PYG{o}{=}\FancyVerbBreakAfterBreak{} \PYG{p}{[}\FancyVerbBreakAfterBreak{}\PYG{n}{x1}\PYG{p}{,}\FancyVerbBreakAfterBreak{} \PYG{n}{x2}\PYG{p}{]}

    \PYG{k}{return} \PYG{n}{samples}
\end{Verbatim}
\end{codecell}

The function returns one row per iteration with the values of \(x_1\) and \(x_2\).

\subsection*{Impact of Correlation on Gibbs Sampling Efficiency}

We now study how the correlation coefficient affects the efficiency of Gibbs sampling. Strong correlations are common in astronomical models, so we compare a strong correlation, \(\rho = 0.95\), with a weak correlation, \(\rho = 0.1\). The conditional distributions of the two runs differ only through \(\rho\), so any difference in the efficiency comes from the correlation. Each run has 10,000 iterations and starts at \(x_1 = x_2 = 100\), far from the distribution. We start there and discard the first 100 iterations as burn-in.

\tutorialbarrier
\begin{codecell}
\begin{Verbatim}[commandchars=\\\{\},fontsize=\small]
\PYG{c+c1}{\PYGZsh{} Run Gibbs sampling for the strong and the weak correlation}
\PYG{n}{n\PYGZus{}samples} \PYG{o}{=}\FancyVerbBreakAfterBreak{} \PYG{l+m+mi}{10000}
\PYG{n}{rhos} \PYG{o}{=}\FancyVerbBreakAfterBreak{} \PYG{p}{[}\FancyVerbBreakAfterBreak{}\PYG{l+m+mf}{0.95}\PYG{p}{,}\FancyVerbBreakAfterBreak{} \PYG{l+m+mf}{0.1}\PYG{p}{]}
\PYG{n}{samples} \PYG{o}{=}\FancyVerbBreakAfterBreak{} \PYG{p}{\PYGZob{}}\PYG{n}{rho}\PYG{p}{:} \PYG{n}{gibbs\PYGZus{}sampler}\PYG{p}{(}\FancyVerbBreakAfterBreak{}\PYG{n}{n\PYGZus{}samples}\PYG{p}{,}\FancyVerbBreakAfterBreak{} \PYG{n}{rho}\PYG{p}{)} \PYG{k}{for} \PYG{n}{rho} \PYG{o+ow}{in} \PYG{n}{rhos}\PYG{p}{\PYGZcb{}}\FancyVerbBreakAfterBreak{}
\end{Verbatim}
\end{codecell}

We first plot the samples after the burn-in as dots, with the contours of the true distribution. The dots overlap where the samples are dense.

\tutorialbarrier
\begin{codecell}

\end{codecell}
\tutorialplot{\includegraphics[width=6.178in,height=0.75\textheight,keepaspectratio]{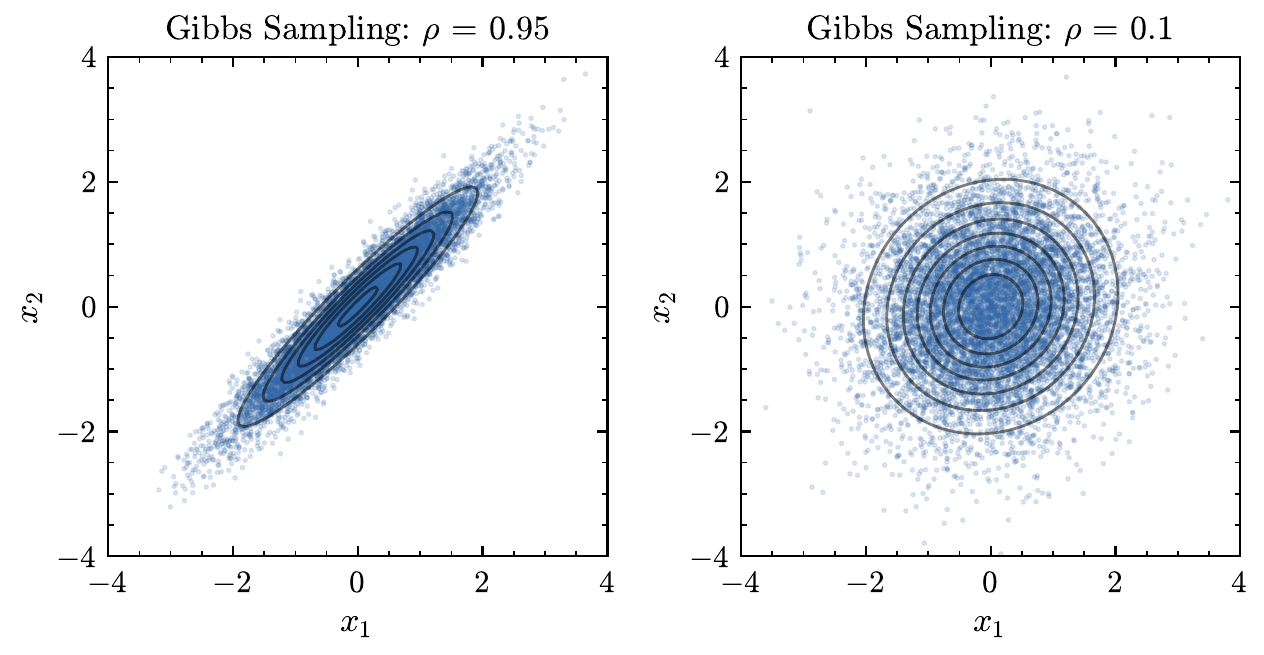}}%
  {\includegraphics[height=\tutorialheight]{tutorials/13a/fig06.pdf}}

The dots are the samples and the black contours show the true distribution. In both panels the samples fill the contours, so the sampler reproduces the true distribution for both correlations. For \(\rho = 0.95\), the distribution is highly stretched along the diagonal, which creates a ridge. Gibbs sampling moves along one axis at a time, like a climber on a narrow diagonal ridge who can move only north-south or east-west. It must therefore follow the ridge in small steps. For \(\rho = 0.1\), the distribution is nearly circular, so the updates along the axes fit its shape.

A scatter plot hides the order of the samples. We therefore plot the variables against the iteration for the first 100 iterations, which show the burn-in, and for the 100 iterations after them.

\tutorialbarrier
\begin{codecell}

\end{codecell}
\tutorialplot{\includegraphics[width=6.186in,height=0.75\textheight,keepaspectratio]{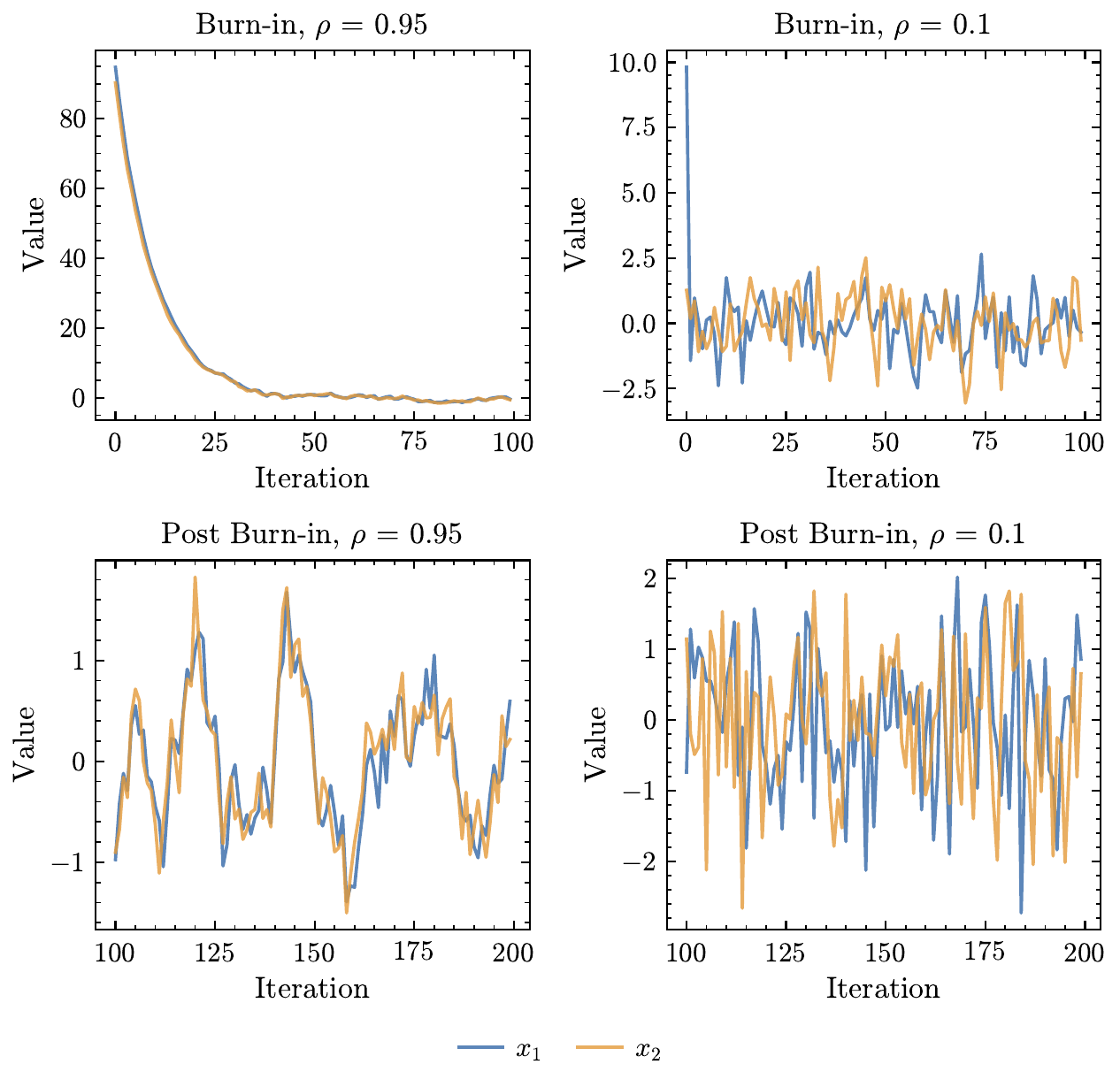}}%
  {\includegraphics[height=\tutorialheight]{tutorials/13a/fig07.pdf}}

For \(\rho = 0.95\), the chain needs tens of iterations to leave its starting point. Afterwards both variables change slowly, with long excursions in one direction before they reverse. The conditional variance \(1 - \rho^2\) is small, so each step is short and the chain needs many iterations to cross the distribution.

For \(\rho = 0.1\), the chain leaves its starting point in one iteration. Both variables then fluctuate quickly between positive and negative values. The conditional variance is close to 1, so the steps are large. In both cases the first 100 iterations cover the burn-in.

\subsection*{Autocorrelation Analysis: Measuring Sampling Efficiency}

The traces suggest that the high correlation lowers the efficiency. We measure this with the autocorrelation of \(x_1\) after the burn-in and its ESS. We show \(x_1\) only, because \(x_2\) behaves in the same way by symmetry.

For this sampler we can also derive the autocorrelation exactly, which lets us check our estimate. One iteration draws \(x_2\) with the mean \(\rho x_1\) and then draws the new \(x_1\) with the mean \(\rho x_2\). The new \(x_1\) therefore keeps the fraction \(\rho^2\) of the old \(x_1\) and adds independent noise. After \(k\) iterations the fraction is \(\rho^{2k}\), so \(R(k) = \rho^{2k}\). In the ESS formula, the geometric series \(1 + 2\sum_{k=1}^{\infty}R(k)\) then equals \((1+\rho^2)/(1-\rho^2)\), so the ESS is

\[\text{ESS} = \frac{N(1-\rho^2)}{1+\rho^2}.\]

We plot the estimated and the exact autocorrelation, with the same red band as before, and give both ESS values.

\tutorialbarrier
\begin{codecell}

\end{codecell}
\tutorialplot{\includegraphics[width=6.175in,height=0.75\textheight,keepaspectratio]{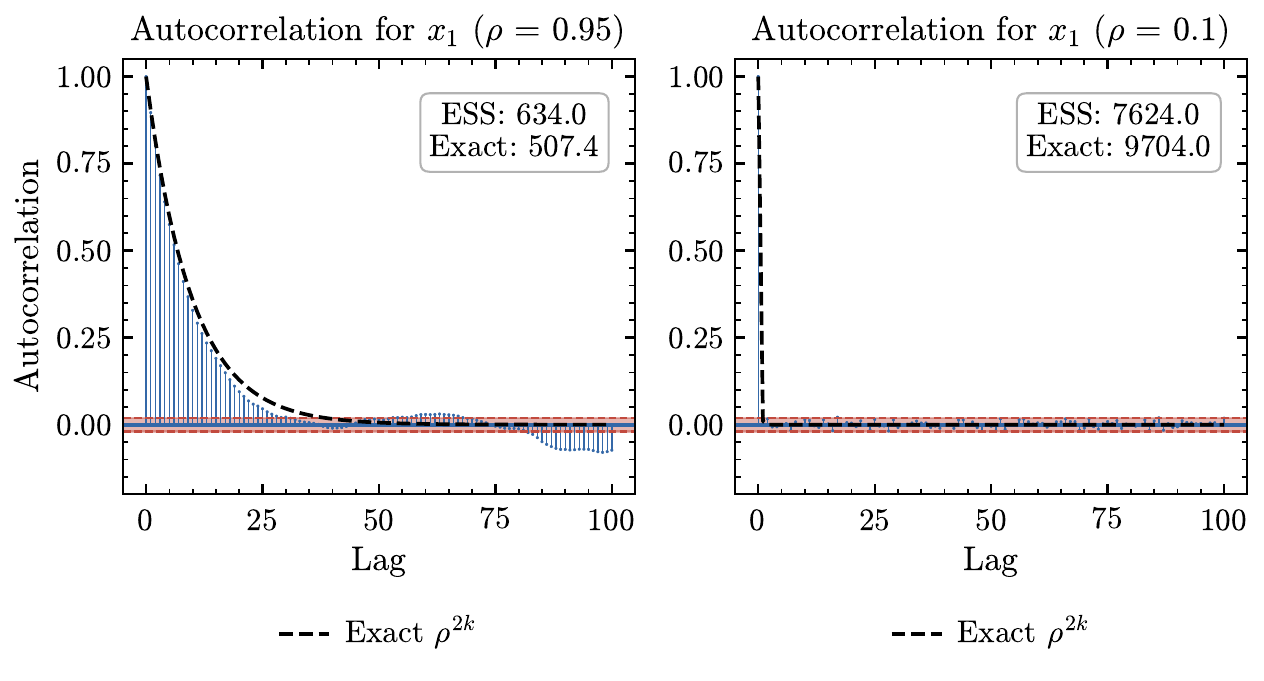}}%
  {\includegraphics[height=\tutorialheight]{tutorials/13a/fig08.pdf}}

In both panels, the estimated autocorrelation follows the dashed curve \(R(k) = \rho^{2k}\) that we derived above, so the estimate agrees with the exact result.

For \(\rho = 0.95\), the autocorrelation decays slowly. Samples many iterations apart still share information, so the ESS is far below the number of samples. At large lags the estimate scatters around the curve and dips below zero. This is sampling noise.

For \(\rho = 0.1\), the curve is already close to zero at the first lag, so the chain mixes efficiently. The ESS is then of the order of the number of samples. In both panels, the estimated ESS differs somewhat from the exact value, because it sums noisy estimates of the autocorrelation up to lag 100.

In the Metropolis runs, the acceptance rate warned us about a badly tuned width. Here it is 1 in both runs, because Gibbs sampling accepts every draw, so only the ESS shows the difference. The inefficiency comes from updating one variable at a time while the variables have a high correlation.

The remedy is block Gibbs sampling (Section 13.15), a divide and conquer strategy. We group the highly correlated parameters into blocks, so that different blocks are only weakly correlated with each other. We then update one block at a time given the others. Within a block, we draw from the joint conditional distribution when we can. Otherwise we use another sampler for that block, such as the Metropolis algorithm with a proposal that follows the correlation inside the block.

\section*{Summary}

In this tutorial we sampled a posterior with two basic MCMC algorithms, the Metropolis algorithm and Gibbs sampling. We also judged the quality of the chains. Our main lessons follow.

\begin{enumerate}
\def\labelenumi{\arabic{enumi}.}
\tightlist
\item
  \textbf{The Metropolis algorithm.} It samples a posterior that we know only up to a constant, because it needs only ratios of posterior values. A step to a higher posterior is always accepted, and a step to a lower posterior is accepted with the ratio as its probability.
\item
  \textbf{The proposal width.} A narrow proposal gives a high acceptance rate but slow motion. A wide proposal gives many rejections and repeated values. The acceptance rate is a first guide for tuning, and the ESS measures the result.
\item
  \textbf{Autocorrelation and the ESS.} Successive samples of a chain are correlated. The ESS counts the independent samples that carry the same information, so we judge a chain by its ESS and not by its length.
\item
  \textbf{Thinning.} Keeping every \(m\)-th sample saves storage but lowers the ESS, because the removed samples still carry information. We thin a chain only when storage requires it.
\item
  \textbf{Burn-in and convergence.} A chain starts where we choose, so we discard the burn-in. The Geweke test compares two parts of one chain, and the Gelman-Rubin diagnostic compares several chains from different starting points. Neither counts independent samples.
\item
  \textbf{Gibbs sampling.} It samples each parameter from its conditional distribution given the others, accepts every draw and does not need a proposal width. It is useful when the conditional distributions are simple, for example in hierarchical models. A high correlation between parameters makes one-at-a-time updates inefficient. We then divide the parameters into blocks of correlated parameters and update each block with Gibbs sampling or another sampler.
\end{enumerate}

MCMC lets us sample distributions that we cannot sample directly, in particular the posteriors of inference problems in astronomy. These posteriors are known only up to a constant and often have no closed form. They may also have many parameters, where rejection and importance sampling become inefficient (Chapter 12). A chain whose stationary distribution is the posterior replaces direct sampling with a sequence of local steps.

\tutorialchapter{13b}{Deconvolution with MCMC and Importance Sampling}
\noindent\textit{Companion material for Chapter 13}

\section*{Introduction}

Astronomical measurements carry uncertainties, and the uncertainty differs from object to object. A faint star has a larger error than a bright one, and a distant galaxy has a larger error than a nearby one. A histogram of the measurements is therefore not the distribution that we want. It is the intrinsic distribution blurred by the errors.

The blur changes what we infer. It widens a distribution, and it moves objects out of a range or into it. A fit that ignores the uncertainties gives wrong parameters, and the error grows with the size of the uncertainties. We want the parameters of the intrinsic distribution. Two examples are the dispersion of stellar velocities and the slope of the stellar mass function, the distribution of stellar masses.

The likelihood of a measurement then contains an integral over the unknown true value. For a Gaussian intrinsic distribution we can do this integral exactly. For most other distributions we cannot, so we estimate it with importance sampling (Chapter 12). We then sample the posterior of the parameters with the Metropolis algorithm (Chapter 13).

The tutorial works through three examples of increasing difficulty.

\begin{enumerate}
\def\labelenumi{\arabic{enumi}.}
\tightlist
\item
  \textbf{Velocities.} The velocity dispersion of a stellar system measures its gravitational pull, and so its mass. The velocities are Gaussian, so the likelihood has a closed form, and we learn the sampler.
\item
  \textbf{Masses.} The slope of the mass function sets how many massive stars a population holds. The masses follow a power law, so the likelihood has no closed form, and importance sampling estimates it.
\item
  \textbf{Binaries.} Unresolved binaries contaminate the masses, and a single power law gives a wrong slope. A mixture model repairs it, and it combines the closed form of the first example with the importance sampling of the second.
\end{enumerate}

\section*{Prerequisites and Learning Objectives}

\subsection*{Prerequisites}

This tutorial assumes the following.

\begin{itemize}
\tightlist
\item
  Bayesian inference with likelihoods, priors and the intrinsic scatter of a model (Chapters 4 and 5)
\item
  The variance of a sum of independent random variables (Chapter 3)
\item
  Mixture models, with their components and weights (Chapter 11)
\item
  Importance sampling, the inverse CDF method and the effective sample size of weights (Chapter 12, Tutorial 12)
\item
  The Metropolis algorithm, the acceptance rate and the burn-in of a chain (Chapter 13, Tutorial 13a)
\item
  Python programming with NumPy, matplotlib and SciPy
\end{itemize}

\subsection*{Learning Objectives}

After this tutorial you can do the following.

\begin{itemize}
\tightlist
\item
  Write the likelihood of noisy data with known, different uncertainties as an integral over the true values
\item
  Estimate such an integral with importance sampling and check it against a trapezoid integral
\item
  Sample a posterior with several parameters, and tune the proposal covariance with a pilot chain
\item
  Check a chain with its acceptance rate and its trace, and read a joint posterior
\item
  Find the failure of a model with a posterior predictive check and repair it with a mixture model
\end{itemize}

We begin by importing the libraries and setting the plot style.

\tutorialbarrier
\begin{codecell}

\end{codecell}

\section*{Noisy Measurements and the Intrinsic Distribution}

The Introduction said that a histogram of measurements is the intrinsic distribution blurred by the errors. We now write this down. Consider a star with a true mass or a true velocity that we cannot observe. We measure a value that differs from the true one by an error. We know the size of the error for each star, and the sizes differ between stars.

In symbols, object \(i\) has the following.

\begin{itemize}
\tightlist
\item
  \textbf{True value.} \(x_{\text{true},i}\), which we cannot observe. The true values follow an intrinsic distribution with the density \(p(x_{\text{true}}|\boldsymbol{\theta})\), where \(\boldsymbol{\theta}\) holds its parameters.
\item
  \textbf{Measurement.} \(x_i = x_{\text{true},i} + \delta_i\), as in Chapter 6. The error \(\delta_i\) is Gaussian with mean 0 and the known standard deviation \(\sigma_i\). Because the \(\sigma_i\) differ, the errors are heteroscedastic.
\end{itemize}

We want \(\boldsymbol{\theta}\) and not the true values.

A measurement depends on \(\boldsymbol{\theta}\) only through its unknown true value, so we average over this value. Given the parameters, the density of the measurement \(x_i\) weights the Gaussian of the error by the intrinsic density of the true value,

\[p(x_i|\boldsymbol{\theta}) = \int \mathcal{N}(x_i|x_{\text{true}}, \sigma_i^2)\, p(x_{\text{true}}|\boldsymbol{\theta})\,\mathrm{d}x_{\text{true}}.\]

For equal \(\sigma_i\) this integral is a convolution of the intrinsic density with the Gaussian, so finding the intrinsic density is called deconvolution. For independent objects the likelihood is the product \(p(\mathcal{D}|\boldsymbol{\theta}) = \prod_i p(x_i|\boldsymbol{\theta})\), where \(\mathcal{D}\) is the data. With a prior \(p(\boldsymbol{\theta})\) it gives the posterior \(p(\boldsymbol{\theta}|\mathcal{D}) \propto p(\mathcal{D}|\boldsymbol{\theta})\,p(\boldsymbol{\theta})\). We say that we marginalize over the true values. The rest of the tutorial computes this integral and samples this posterior.

\section*{An Intrinsic Gaussian Distribution}

We start with the line-of-sight velocities of stars in a stellar system, in km/s. Their distribution is a Gaussian with the mean \(\mu\) and the standard deviation \(\sigma_{\text{int}}\), the intrinsic scatter of Section 4.5. This velocity dispersion measures the gravitational pull of the system, and so its mass.

The measurement errors add to the observed width. A fit that ignores them overestimates the dispersion, and so the mass. The likelihood has a closed form, so that we can concentrate on the sampler. We simulate 1000 stars with \(\mu = 1.5\) and \(\sigma_{\text{int}} = 1\). The uncertainties \(\sigma_i\) are uniform between 0.1 and 1.5, because faint stars have larger errors than bright ones.

\tutorialbarrier
\begin{codecell}

\end{codecell}
\begin{Verbatim}[fontsize=\footnotesize, xleftmargin=5mm]
Standard deviation of intrinsic values = 1.02, of measurements = 1.36
\end{Verbatim}

We plot the intrinsic velocities and the measurements. A real survey gives us only the measurements.

\tutorialbarrier
\begin{codecell}

\end{codecell}
\tutorialplot{\includegraphics[width=4.097in,height=0.75\textheight,keepaspectratio]{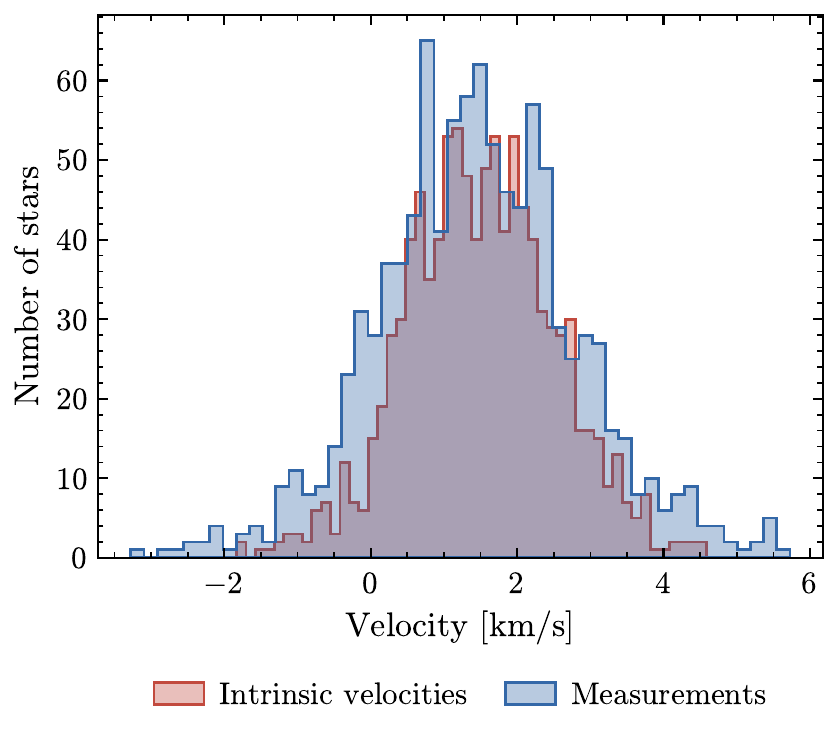}}%
  {\includegraphics[height=\tutorialheight]{tutorials/13b/fig01.pdf}}

The measurements are wider than the intrinsic velocities, because each one carries its error. A fit that ignores the uncertainties would give the standard deviation of the measurements as the intrinsic scatter, which is too large. We therefore need a likelihood that contains the errors.

\subsection*{The Likelihood and the Posterior}

We need the likelihood to sample the posterior. For a Gaussian intrinsic distribution we can do the integral above exactly, because the sum of two independent Gaussian variables is Gaussian. The variance of a sum of independent variables is the sum of their variances (Section 3.5). The intrinsic value and the error are independent, so the measurement \(x_i\) is Gaussian with the mean \(\mu\) and the variance \(\sigma_{\text{int}}^2 + \sigma_i^2\). The integral then has a closed form,

\[p(x_i|\mu, \sigma_{\text{int}}) = \mathcal{N}(x_i|\mu, \sigma_{\text{int}}^2 + \sigma_i^2).\]

This is the heterogeneous likelihood with intrinsic scatter of Section 4.5. If all \(\sigma_i\) were equal to \(\sigma\), the maximum of the likelihood would have a closed form, \(\sigma_{\text{int}}^2 = s^2 - \sigma^2\). Here \(s^2\) is the variance of the measurements. For different \(\sigma_i\) it has none (Section 4.5), so we sample the posterior.

A posterior also needs priors. We assume the following.

\begin{itemize}
\tightlist
\item
  \textbf{Mean \(\mu\).} We know little about it before we see the data. The prior is the Cauchy distribution of Tutorial 13a, \(p(\mu) = 1/\big(\pi\gamma\,[1 + (\mu/\gamma)^2]\big)\) with location 0 and scale \(\gamma = 3\). It is bell shaped like a Gaussian but has much heavier tails, so it allows large values.
\item
  \textbf{Scatter \(\sigma_{\text{int}}\).} It cannot be negative. The prior is exponential, \(p(\sigma_{\text{int}}) = \lambda e^{-\lambda \sigma_{\text{int}}}\) with the rate \(\lambda = 1\). It is zero for a negative value and favors small scatters without ruling out large ones.
\end{itemize}

Both priors vary slowly over the range that the data allow, so they hardly change the posterior.

The function below returns the logarithm of the posterior. It drops every factor that does not depend on the parameters, such as the normalization of the posterior. These factors cancel in the ratio of two posteriors, which is all that the acceptance probability needs. For \(\sigma_{\text{int}} \leq 0\) the prior is zero, so the logarithm of the posterior is \(-\infty\) and the chain rejects the proposal.

\tutorialbarrier
\begin{codecell}
\begin{Verbatim}[commandchars=\\\{\},fontsize=\small]
\PYG{k}{def}\PYG{+w}{ }\PYG{n+nf}{log\PYGZus{}posterior\PYGZus{}gauss}\PYG{p}{(}\FancyVerbBreakAfterBreak{}\PYG{n}{theta}\PYG{p}{)}\PYG{p}{:}
\PYG{+w}{    }\PYG{l+s+sd}{\PYGZdq{}\PYGZdq{}\PYGZdq{}Return the log posterior of the mean and the intrinsic scatter, up to a constant.\PYGZdq{}\PYGZdq{}\PYGZdq{}}
    \PYG{n}{mu}\PYG{p}{,}\FancyVerbBreakAfterBreak{} \PYG{n}{sigma\PYGZus{}int} \PYG{o}{=}\FancyVerbBreakAfterBreak{} \PYG{n}{theta}

    \PYG{c+c1}{\PYGZsh{} The prior is zero for a negative scatter, so the Metropolis algorithm rejects such a proposal}
    \PYG{k}{if} \PYG{n}{sigma\PYGZus{}int} \PYG{o}{\PYGZlt{}}\PYG{o}{=}\FancyVerbBreakAfterBreak{} \PYG{l+m+mi}{0}\PYG{p}{:}
        \PYG{k}{return} \PYG{o}{\PYGZhy{}}\PYG{n}{np}\PYG{o}{.}\FancyVerbBreakAfterBreak{}\PYG{n}{inf}

    \PYG{c+c1}{\PYGZsh{} Log likelihood, where each measurement is Gaussian with the variance sigma\PYGZus{}int\PYGZca{}2 + sigma\PYGZus{}i\PYGZca{}2}
    \PYG{n}{variance} \PYG{o}{=}\FancyVerbBreakAfterBreak{} \PYG{n}{sigma\PYGZus{}int}\PYG{o}{*}\PYG{o}{*}\PYG{l+m+mi}{2} \PYG{o}{+}\FancyVerbBreakAfterBreak{} \PYG{n}{sigma\PYGZus{}gauss}\PYG{o}{*}\PYG{o}{*}\PYG{l+m+mi}{2}
    \PYG{n}{log\PYGZus{}likelihood} \PYG{o}{=}\FancyVerbBreakAfterBreak{} \PYG{o}{\PYGZhy{}}\PYG{l+m+mf}{0.5} \PYG{o}{*} \PYG{n}{np}\PYG{o}{.}\FancyVerbBreakAfterBreak{}\PYG{n}{sum}\PYG{p}{(}\FancyVerbBreakAfterBreak{}\PYG{p}{(}\FancyVerbBreakAfterBreak{}\PYG{n}{x\PYGZus{}gauss} \PYG{o}{\PYGZhy{}} \PYG{n}{mu}\PYG{p}{)}\PYG{o}{*}\PYG{o}{*}\PYG{l+m+mi}{2} \PYG{o}{/} \PYG{n}{variance} \PYG{o}{+}\FancyVerbBreakAfterBreak{} \PYG{n}{np}\PYG{o}{.}\FancyVerbBreakAfterBreak{}\PYG{n}{log}\PYG{p}{(}\FancyVerbBreakAfterBreak{}\PYG{n}{variance}\PYG{p}{)}\PYG{p}{)}

    \PYG{c+c1}{\PYGZsh{} Log prior with a Cauchy distribution for the mean (location 0, scale 3) and an exponential one for the scatter (rate 1)}
    \PYG{n}{log\PYGZus{}prior} \PYG{o}{=}\FancyVerbBreakAfterBreak{} \PYG{o}{\PYGZhy{}}\PYG{n}{np}\PYG{o}{.}\FancyVerbBreakAfterBreak{}\PYG{n}{log}\PYG{p}{(}\FancyVerbBreakAfterBreak{}\PYG{l+m+mi}{1} \PYG{o}{+}\FancyVerbBreakAfterBreak{} \PYG{p}{(}\FancyVerbBreakAfterBreak{}\PYG{n}{mu} \PYG{o}{/} \PYG{l+m+mi}{3}\PYG{p}{)}\PYG{o}{*}\PYG{o}{*}\PYG{l+m+mi}{2}\PYG{p}{)} \PYG{o}{\PYGZhy{}} \PYG{n}{sigma\PYGZus{}int}

    \PYG{k}{return} \PYG{n}{log\PYGZus{}likelihood} \PYG{o}{+}\FancyVerbBreakAfterBreak{} \PYG{n}{log\PYGZus{}prior}
\end{Verbatim}
\end{codecell}

\subsection*{Sampling Several Parameters}

We now sample the posterior of \(\mu\) and \(\sigma_{\text{int}}\) with the Metropolis algorithm. Tutorial 13a used it for one parameter. Here the chain moves through the space of both parameters, so its state is the vector of parameters \(\boldsymbol{\theta} = (\mu, \sigma_{\text{int}})\).

The algorithm proposes a new state from a Gaussian centered at the current state, with a covariance matrix \(\boldsymbol{\Sigma}\), \(q(\boldsymbol{\theta}'|\boldsymbol{\theta}) = \mathcal{N}(\boldsymbol{\theta}'|\boldsymbol{\theta}, \boldsymbol{\Sigma})\) (Section 13.3). In Tutorial 13a the covariance was the square of one width \(\nu\). The proposal is symmetric, so the acceptance probability is \(A = \min(1, p(\boldsymbol{\theta}'|\mathcal{D})/p(\boldsymbol{\theta}|\mathcal{D}))\) as before. The function below draws the proposal with \texttt{np.\allowbreak{}random.\allowbreak{}multivariate\_\allowbreak{}normal}. It takes the log posterior as an argument.

\tutorialbarrier
\begin{codecell}

\end{codecell}

The covariance \(\boldsymbol{\Sigma}\) decides how efficiently the chain moves (Section 13.9). We want its shape to follow the posterior, so that the steps respect the scales and the correlations of the parameters. We do not know the posterior before we sample it, so a pilot chain gives a first estimate.

\begin{enumerate}
\def\labelenumi{\arabic{enumi}.}
\tightlist
\item
  \textbf{Pilot chain.} Run a short chain of 2000 iterations with a guessed width for every parameter. The guess only needs the right order of magnitude.
\item
  \textbf{Covariance.} Take the covariance of the pilot samples after the first 500 iterations. These early samples start away from the posterior.
\item
  \textbf{Scale.} Multiply the covariance by a factor \(c\) that sets the size of the steps.
\end{enumerate}

A larger \(c\) gives larger steps and a lower acceptance rate (Section 13.9). We choose \(c\) so that the acceptance rate is about 40 percent. With more parameters, steps of the same relative size are rejected more often, so \(c\) falls as their number grows. We use \(c = 8\) for one parameter, 3 for two and 1.5 for four. The function \texttt{tune\_\allowbreak{}proposal} follows these steps.

\tutorialbarrier
\begin{codecell}

\end{codecell}

We run the pilot chain from a guess of the parameters. We then run one chain of 5000 iterations that starts far from the posterior, so that we can see its burn-in.

\tutorialbarrier
\begin{codecell}
\begin{Verbatim}[commandchars=\\\{\},fontsize=\small]
\PYG{c+c1}{\PYGZsh{} Proposal covariance from a pilot chain with the guessed width 0.1 for both parameters}
\PYG{n}{cov\PYGZus{}gauss} \PYG{o}{=}\FancyVerbBreakAfterBreak{} \PYG{n}{tune\PYGZus{}proposal}\PYG{p}{(}\FancyVerbBreakAfterBreak{}\PYG{n}{log\PYGZus{}posterior\PYGZus{}gauss}\PYG{p}{,}\FancyVerbBreakAfterBreak{} \PYG{p}{[}\FancyVerbBreakAfterBreak{}\PYG{l+m+mf}{0.0}\PYG{p}{,}\FancyVerbBreakAfterBreak{} \PYG{l+m+mf}{1.0}\PYG{p}{]}\PYG{p}{,}\FancyVerbBreakAfterBreak{} \PYG{p}{[}\FancyVerbBreakAfterBreak{}\PYG{l+m+mf}{0.1}\PYG{p}{,}\FancyVerbBreakAfterBreak{} \PYG{l+m+mf}{0.1}\PYG{p}{]}\PYG{p}{,}\FancyVerbBreakAfterBreak{} \PYG{n}{scale}\PYG{o}{=}\FancyVerbBreakAfterBreak{}\PYG{l+m+mi}{3}\PYG{p}{)}

\PYG{c+c1}{\PYGZsh{} One chain of 5000 iterations from a starting point far from the posterior}
\PYG{n}{chain\PYGZus{}gauss}\PYG{p}{,}\FancyVerbBreakAfterBreak{} \PYG{n}{rate\PYGZus{}gauss} \PYG{o}{=}\FancyVerbBreakAfterBreak{} \PYG{n}{metropolis}\PYG{p}{(}\FancyVerbBreakAfterBreak{}\PYG{n}{log\PYGZus{}posterior\PYGZus{}gauss}\PYG{p}{,}\FancyVerbBreakAfterBreak{} \PYG{p}{[}\FancyVerbBreakAfterBreak{}\PYG{l+m+mf}{0.0}\PYG{p}{,}\FancyVerbBreakAfterBreak{} \PYG{l+m+mf}{0.5}\PYG{p}{]}\PYG{p}{,}\FancyVerbBreakAfterBreak{} \PYG{n}{cov\PYGZus{}gauss}\PYG{p}{,}\FancyVerbBreakAfterBreak{} \PYG{l+m+mi}{5000}\PYG{p}{)}
\end{Verbatim}
\end{codecell}

We plot the first 1000 iterations of the chain for both parameters, to choose the burn-in.

\tutorialbarrier
\begin{codecell}

\end{codecell}
\tutorialplot{\includegraphics[width=6.177in,height=0.75\textheight,keepaspectratio]{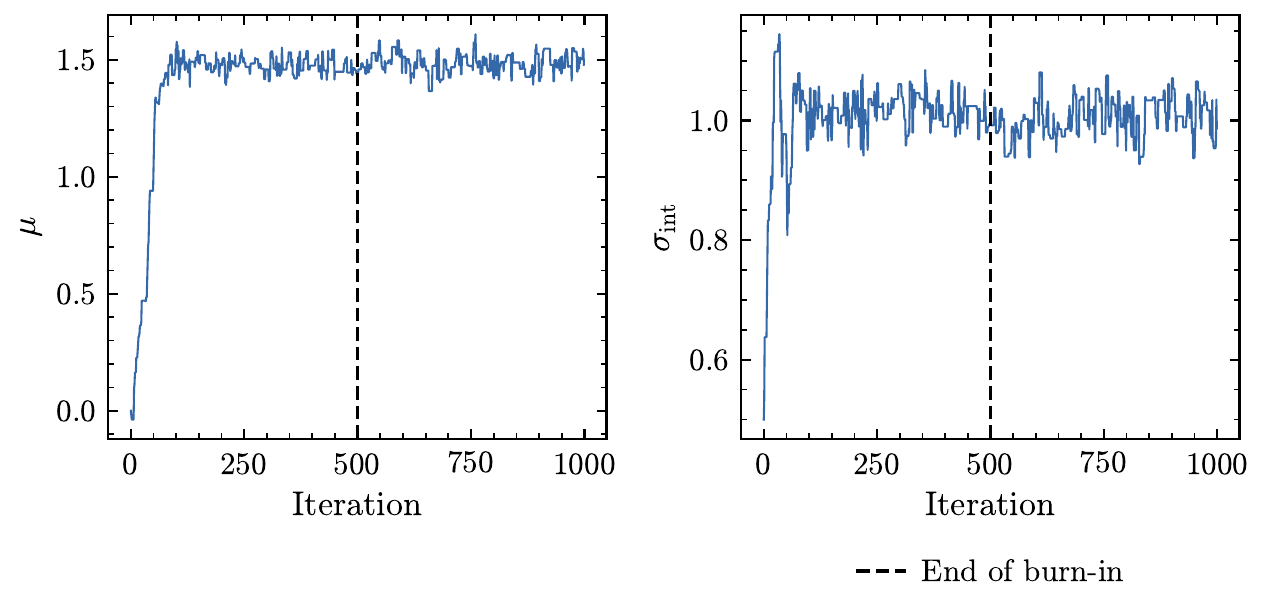}}%
  {\includegraphics[height=\tutorialheight]{tutorials/13b/fig02.pdf}}

The chain leaves its starting point and settles within about 100 iterations. The first 500 iterations therefore leave a margin, and we discard them as the burn-in. The burn-in is the part of a chain that still remembers its starting point (Section 13.13).

Tutorial 13a judged chains with the effective sample size and the Gelman-Rubin statistic. Here two simple checks are enough. The first is the acceptance rate, the share of proposals that the chain accepts. The second is the trace. The function \texttt{summarize} prints the mean and the standard deviation of every parameter after the burn-in.

\tutorialbarrier
\begin{codecell}
\begin{Verbatim}[commandchars=\\\{\},fontsize=\small]
\PYG{k}{def}\PYG{+w}{ }\PYG{n+nf}{summarize}\PYG{p}{(}\FancyVerbBreakAfterBreak{}\PYG{n}{chain}\PYG{p}{,}\FancyVerbBreakAfterBreak{} \PYG{n}{names}\PYG{p}{,}\FancyVerbBreakAfterBreak{} \PYG{n}{burn\PYGZus{}in}\PYG{p}{)}\PYG{p}{:}
\PYG{+w}{    }\PYG{l+s+sd}{\PYGZdq{}\PYGZdq{}\PYGZdq{}Print the posterior mean and standard deviation of every parameter after the burn\PYGZhy{}in.\PYGZdq{}\PYGZdq{}\PYGZdq{}}
    \PYG{k}{for} \PYG{n}{j}\PYG{p}{,}\FancyVerbBreakAfterBreak{} \PYG{n}{name} \PYG{o+ow}{in} \PYG{n+nb}{enumerate}\PYG{p}{(}\FancyVerbBreakAfterBreak{}\PYG{n}{names}\PYG{p}{)}\PYG{p}{:}
        \PYG{n+nb}{print}\PYG{p}{(}\FancyVerbBreakAfterBreak{}\PYG{l+s+sa}{f}\PYG{l+s+s2}{\PYGZdq{}}\PYG{l+s+si}{\PYGZob{}}\PYG{n}{name}\PYG{l+s+si}{:}\PYG{l+s+s2}{\PYGZlt{}6}\PYG{l+s+si}{\PYGZcb{}}\FancyVerbBreakAfterBreak{}\PYG{l+s+s2}{ mean }\PYG{l+s+si}{\PYGZob{}}\PYG{n}{chain}\PYG{p}{[}\FancyVerbBreakAfterBreak{}\PYG{n}{burn\PYGZus{}in}\PYG{p}{:}\PYG{p}{,}\FancyVerbBreakAfterBreak{}\PYG{+w}{ }\PYG{n}{j}\PYG{p}{]}\PYG{o}{.}\FancyVerbBreakAfterBreak{}\PYG{n}{mean}\PYG{p}{(}\FancyVerbBreakAfterBreak{}\PYG{p}{)}\PYG{l+s+si}{:}\PYG{l+s+s2}{.3f}\PYG{l+s+si}{\PYGZcb{}}\FancyVerbBreakAfterBreak{}\PYG{l+s+s2}{, sd }\PYG{l+s+si}{\PYGZob{}}\PYG{n}{chain}\PYG{p}{[}\FancyVerbBreakAfterBreak{}\PYG{n}{burn\PYGZus{}in}\PYG{p}{:}\PYG{p}{,}\FancyVerbBreakAfterBreak{}\PYG{+w}{ }\PYG{n}{j}\PYG{p}{]}\PYG{o}{.}\FancyVerbBreakAfterBreak{}\PYG{n}{std}\PYG{p}{(}\FancyVerbBreakAfterBreak{}\PYG{p}{)}\PYG{l+s+si}{:}\PYG{l+s+s2}{.3f}\PYG{l+s+si}{\PYGZcb{}}\FancyVerbBreakAfterBreak{}\PYG{l+s+s2}{\PYGZdq{}}\PYG{p}{)}
\end{Verbatim}
\end{codecell}

We print the acceptance rate and the summary of both parameters.

\tutorialbarrier
\begin{codecell}
\begin{Verbatim}[commandchars=\\\{\},fontsize=\small]
\PYG{c+c1}{\PYGZsh{} Acceptance rate of the chain, and the summary of both parameters after the burn\PYGZhy{}in}
\PYG{n+nb}{print}\PYG{p}{(}\FancyVerbBreakAfterBreak{}\PYG{l+s+sa}{f}\PYG{l+s+s2}{\PYGZdq{}}\PYG{l+s+s2}{Acceptance rate = }\PYG{l+s+si}{\PYGZob{}}\PYG{n}{rate\PYGZus{}gauss}\PYG{l+s+si}{:}\PYG{l+s+s2}{.0f}\PYG{l+s+si}{\PYGZcb{}}\FancyVerbBreakAfterBreak{}\PYG{l+s+s2}{ percent}\PYG{l+s+s2}{\PYGZdq{}}\PYG{p}{)}
\PYG{n}{summarize}\PYG{p}{(}\FancyVerbBreakAfterBreak{}\PYG{n}{chain\PYGZus{}gauss}\PYG{p}{,}\FancyVerbBreakAfterBreak{} \PYG{p}{[}\FancyVerbBreakAfterBreak{}\PYG{l+s+s2}{\PYGZdq{}}\PYG{l+s+s2}{μ}\PYG{l+s+s2}{\PYGZdq{}}\PYG{p}{,}\FancyVerbBreakAfterBreak{} \PYG{l+s+s2}{\PYGZdq{}}\PYG{l+s+s2}{σ\PYGZus{}int}\PYG{l+s+s2}{\PYGZdq{}}\PYG{p}{]}\PYG{p}{,}\FancyVerbBreakAfterBreak{} \PYG{n}{burn\PYGZus{}in}\PYG{o}{=}\FancyVerbBreakAfterBreak{}\PYG{l+m+mi}{500}\PYG{p}{)}
\end{Verbatim}
\end{codecell}
\begin{Verbatim}[fontsize=\footnotesize, xleftmargin=5mm]
Acceptance rate = 35 percent
μ      mean 1.481, sd 0.040
σ_int  mean 1.012, sd 0.034
\end{Verbatim}

The acceptance rate is 35 percent, inside the range of Section 13.9, and the trace has settled. The posteriors are centered near the true values, within one standard deviation for both parameters. The measurements alone would have given the scatter 1.36, which is too large. The deconvolved scatter is close to the true value 1.

\subsection*{The Joint Posterior}

The summary describes each parameter alone. The chain also gives the joint posterior of \(\mu\) and \(\sigma_{\text{int}}\), which shows how the parameters depend on each other.

A corner plot shows the posterior of each parameter on the diagonal and the joint posterior of each pair below it. The red lines mark the true values, which we know only because we simulated the data.

\tutorialbarrier
\begin{codecell}

\end{codecell}

We use the chain after the burn-in and plot its joint posterior.

\tutorialbarrier
\begin{codecell}
\begin{Verbatim}[commandchars=\\\{\},fontsize=\small]
\PYG{c+c1}{\PYGZsh{} Samples after the burn\PYGZhy{}in and their joint posterior}
\PYG{n}{samples\PYGZus{}gauss} \PYG{o}{=}\FancyVerbBreakAfterBreak{} \PYG{n}{chain\PYGZus{}gauss}\PYG{p}{[}\FancyVerbBreakAfterBreak{}\PYG{l+m+mi}{500}\PYG{p}{:}\PYG{p}{]}
\PYG{n}{corner\PYGZus{}plot}\PYG{p}{(}\FancyVerbBreakAfterBreak{}\PYG{n}{samples\PYGZus{}gauss}\PYG{p}{,}\FancyVerbBreakAfterBreak{} \PYG{p}{[}\FancyVerbBreakAfterBreak{}\PYG{l+s+sa}{r}\PYG{l+s+s1}{\PYGZsq{}}\PYG{l+s+s1}{\PYGZdl{}}\PYG{l+s+s1}{\PYGZbs{}}\PYG{l+s+s1}{mu\PYGZdl{}}\PYG{l+s+s1}{\PYGZsq{}}\PYG{p}{,}\FancyVerbBreakAfterBreak{} \PYG{l+s+sa}{r}\PYG{l+s+s1}{\PYGZsq{}}\PYG{l+s+s1}{\PYGZdl{}}\PYG{l+s+s1}{\PYGZbs{}}\PYG{l+s+s1}{sigma\PYGZus{}}\PYG{l+s+s1}{\PYGZob{}}\PYG{l+s+s1}{\PYGZbs{}}\PYG{l+s+s1}{rm int\PYGZcb{}\PYGZdl{}}\PYG{l+s+s1}{\PYGZsq{}}\PYG{p}{]}\PYG{p}{,}\FancyVerbBreakAfterBreak{} \PYG{p}{[}\FancyVerbBreakAfterBreak{}\PYG{n}{mu\PYGZus{}true}\PYG{p}{,}\FancyVerbBreakAfterBreak{} \PYG{n}{sigma\PYGZus{}int\PYGZus{}true}\PYG{p}{]}\PYG{p}{,}\FancyVerbBreakAfterBreak{} \PYG{n}{figsize}\PYG{o}{=}\FancyVerbBreakAfterBreak{}\PYG{p}{(}\FancyVerbBreakAfterBreak{}\PYG{l+m+mi}{12}\PYG{p}{,}\FancyVerbBreakAfterBreak{} \PYG{l+m+mf}{4.3}\PYG{p}{)}\PYG{p}{)}
\end{Verbatim}
\end{codecell}
\tutorialplot{\includegraphics[width=6.173in,height=0.75\textheight,keepaspectratio]{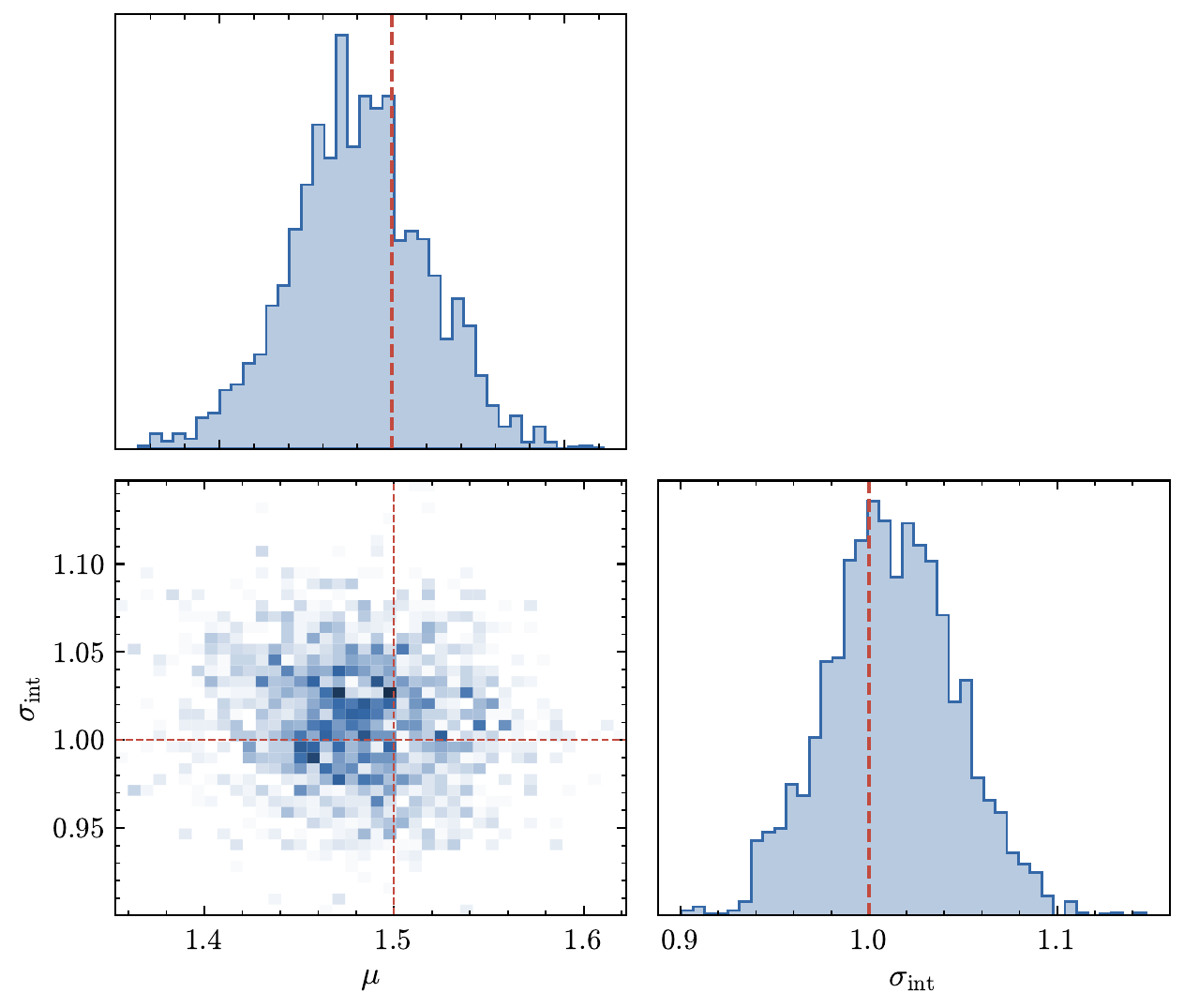}}%
  {\includegraphics[height=\tutorialheight]{tutorials/13b/fig03.pdf}}

The two parameters are nearly independent, so the covariance of the proposal is nearly diagonal. The true values lie inside both posteriors. The deconvolution recovers the intrinsic scatter that the measurements alone hide. This was easy, because the likelihood has a closed form.

\section*{An Intrinsic Power Law}

The velocities were the easy case. Stellar masses are not Gaussian, and most intrinsic distributions have no closed form. Above half a solar mass we model the mass function as a power law with sharp limits.

Its slope sets how many massive stars a population holds for each low-mass star, and so much of its light. A bias in the slope changes these conclusions. The intrinsic density \(p(x_{\text{true}}|\boldsymbol{\theta})\) of the model section is now this power law, with the single parameter \(\alpha\). We take the normalized density of Tutorial 12 on the range from \(x_{\min} = 0.5\) to \(x_{\max} = 1\) solar mass,

\[p(x_{\text{true}}|\alpha) = \frac{1-\alpha}{x_{\max}^{1-\alpha} - x_{\min}^{1-\alpha}}\, x_{\text{true}}^{-\alpha}.\]

For the simulation we take \(\alpha = 2.3\), the slope of the Kroupa initial mass function in Tutorial 12, and the density is zero outside the range. We simulate the data in two steps.

\begin{enumerate}
\def\labelenumi{\arabic{enumi}.}
\tightlist
\item
  \textbf{True masses.} We simulate 3000 stars. The inverse CDF method (Section 12.3) turns a uniform number \(u\) between 0 and 1 into a mass. It solves \(u = F(x_{\text{true}})\) for the mass.
\item
  \textbf{Errors.} We add errors with \(\sigma_i\) uniform between 0.05 and 0.1 solar masses, which is 5 to 20 percent of a mass in the range.
\end{enumerate}

In step 1, \(F\) is the cumulative distribution function (CDF) of the density,

\[F(x_{\text{true}}) = \frac{x_{\text{true}}^{1-\alpha} - x_{\min}^{1-\alpha}}{x_{\max}^{1-\alpha} - x_{\min}^{1-\alpha}}.\]

\tutorialbarrier
\begin{codecell}

\end{codecell}
\begin{Verbatim}[fontsize=\footnotesize, xleftmargin=5mm]
Measurements outside the range 0.5 to 1.0: 408 of 3000
\end{Verbatim}

We plot the true masses and the measurements, with the range of the power law.

\tutorialbarrier
\begin{codecell}

\end{codecell}
\tutorialplot{\includegraphics[width=4.178in,height=0.75\textheight,keepaspectratio]{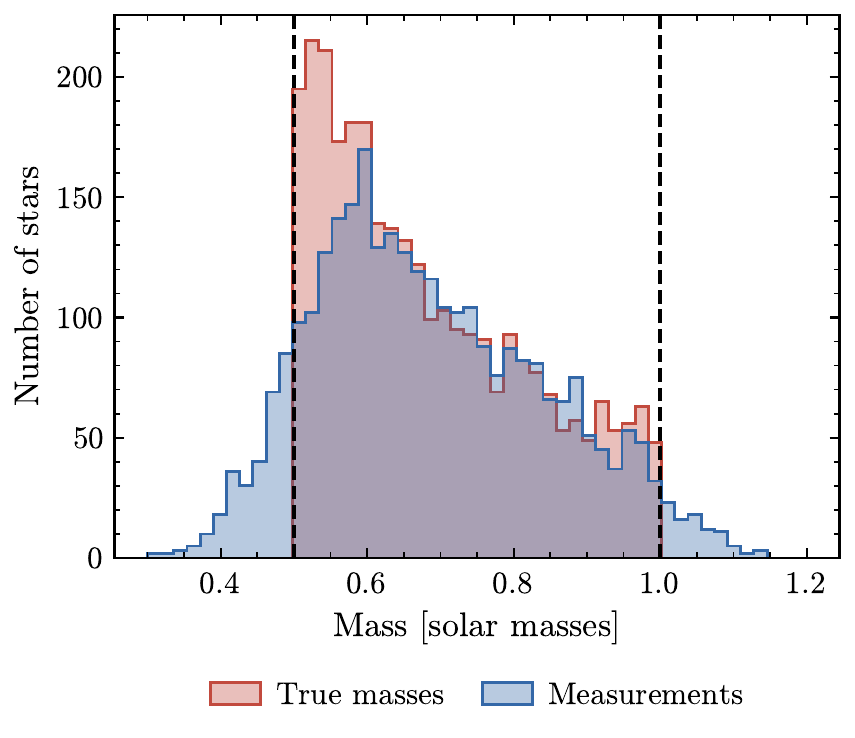}}%
  {\includegraphics[height=\tutorialheight]{tutorials/13b/fig04.pdf}}

The errors blur the edges of the range. Of the 3000 measurements, 408 lie outside the range, although every true mass lies inside it. Near the lower edge, where the density is highest, the measurements are fewer than the true masses, because many have moved below 0.5. A model of these data must account for this blur.

\subsection*{The Effect of Ignoring the Uncertainties}

Before we model the errors, we see what happens when we ignore them. We fit the power law as if the measurements were the true masses. A star outside the range has zero density under this model, so we must remove it. We maximize the likelihood of the remaining stars on a grid of slopes. The grid starts above 1, because the normalization of the density is \(0/0\) at \(\alpha = 1\).

\tutorialbarrier
\begin{codecell}

\end{codecell}
\begin{Verbatim}[fontsize=\footnotesize, xleftmargin=5mm]
Slope that ignores the uncertainties = 1.61, from 2592 stars
\end{Verbatim}

The slope is 1.61, much lower than the true value 2.3. The errors blur the steep fall of the density, so the measurements form a flatter distribution. Removing the stars outside the range also removes the stars that the errors moved out of it. A correct fit must put the errors into the likelihood.

\subsection*{Importance Sampling of the Likelihood}

Remember that \(x_i\) is the measurement of star \(i\), \(\sigma_i\) its known uncertainty and \(x_{\text{true}}\) its unknown true mass. The correct likelihood of \(x_i\) is the integral at the start of this tutorial, now over the range of the power law,

\[p(x_i|\alpha) = \int_{x_{\min}}^{x_{\max}} \mathcal{N}(x_i|x_{\text{true}}, \sigma_i^2)\, p(x_{\text{true}}|\alpha)\,\mathrm{d}x_{\text{true}}.\]

Unlike the Gaussian case, this integral has no simple closed form. We estimate it with importance sampling (Section 12.5).

The integral is the average of the Gaussian \(f(x_{\text{true}}) = \mathcal{N}(x_i|x_{\text{true}}, \sigma_i^2)\) over the intrinsic density \(p\). If we draw \(S\) masses \(x_{\text{true},i}^{(s)}\) from \(p\), we can estimate it by their plain average,

\[p(x_i|\alpha) = \int f(x_{\text{true}})\, p(x_{\text{true}}|\alpha)\,\mathrm{d}x_{\text{true}} \approx \frac{1}{S} \sum_{s=1}^{S} f\big(x_{\text{true},i}^{(s)}\big).\]

This wastes draws. Most land far from the measurement, where \(f\) is almost zero, so they carry little information. Importance sampling draws instead from a proposal density \(q_i\) that puts the draws where the product \(f\,p\) is large. We multiply and divide the integrand by \(q_i\),

\tutorialdisplay{p(x_i|\alpha) = \int f(x_{\text{true}})\, \frac{p(x_{\text{true}}|\alpha)}{q_i(x_{\text{true}})}\, q_i(x_{\text{true}})\,\mathrm{d}x_{\text{true}} \approx \frac{1}{S} \sum_{s=1}^{S} f\big(x_{\text{true},i}^{(s)}\big)\, w^{(s)},}{w^{(s)} = \frac{p\big(x_{\text{true},i}^{(s)}\big|\alpha\big)}{q_i\big(x_{\text{true},i}^{(s)}\big)},}

where the draws now come from \(q_i\). The weight \(w^{(s)}\) corrects each draw for coming from \(q_i\) and not from \(p\). The proposal can differ for every star.

\subsection*{The Proposal}

The best proposal follows the integrand \(f\,p\), the product of the Gaussian and the power law (Section 12.6). The power law varies slowly over the width of the Gaussian, so the Gaussian itself is a good choice. As a function of \(x_{\text{true}}\) it is centered at the measurement \(x_i\) with the width \(\sigma_i\).

One change is needed. The power law is zero outside the range, so a draw outside the range carries no weight. A star far outside the range would have no useful draw at all, and its estimate would be zero. We therefore cut the Gaussian to the range and renormalize it. The normalization is \(Z_i\), the probability that the Gaussian puts inside the range,

\[Z_i = \Phi\left(\frac{x_{\max} - x_i}{\sigma_i}\right) - \Phi\left(\frac{x_{\min} - x_i}{\sigma_i}\right),\]

where \(\Phi\) is the CDF of the standard Gaussian. The proposal \(q_i\) is the Gaussian divided by \(Z_i\) on the range. We plot the Gaussian, the proposal and the integrand for one star that lies at the lower edge of the range. The integrand is the Gaussian times the power law, scaled to the area 1 so that we can compare its shape with the proposal.

\tutorialbarrier
\begin{codecell}

\end{codecell}
\tutorialplot{\includegraphics[width=5.252in,height=0.75\textheight,keepaspectratio]{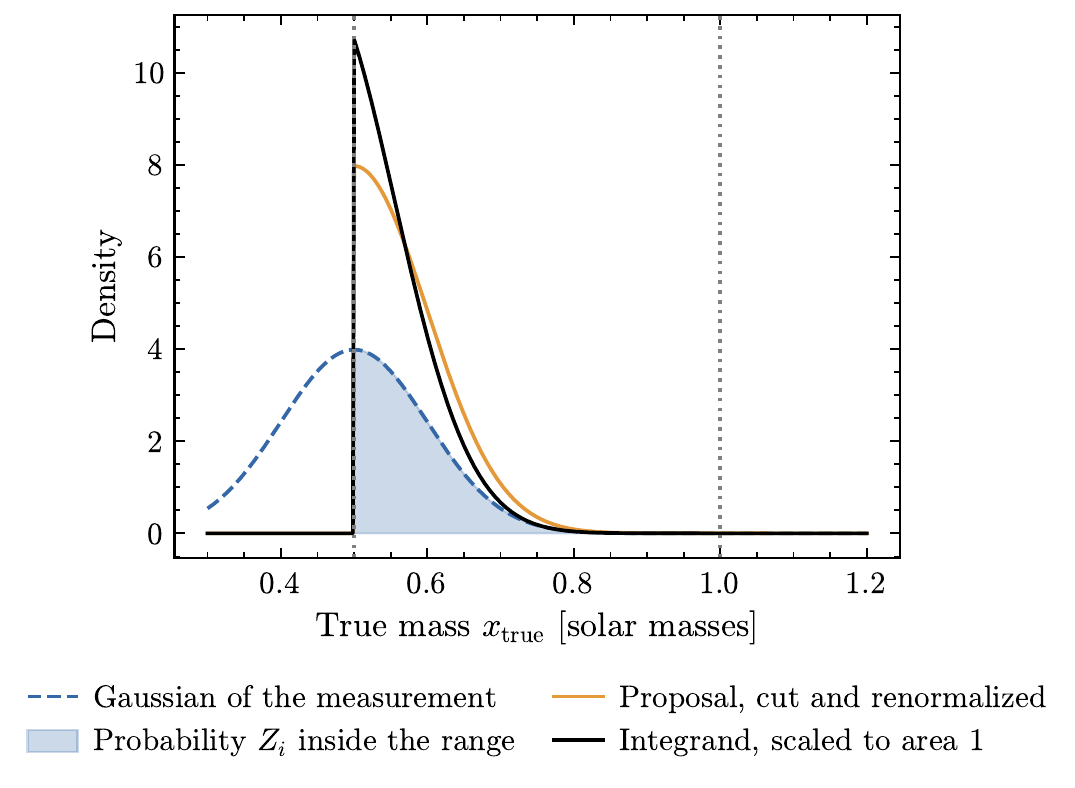}}%
  {\includegraphics[height=\tutorialheight]{tutorials/13b/fig05.pdf}}

The Gaussian puts about half of its probability outside the range, where the power law is zero. The proposal keeps only the part inside the range and scales it up by \(1/Z_i\), so that its area is 1. It follows the integrand closely, because the power law changes slowly across the width of the Gaussian.

Because \(q_i = f/Z_i\) on the range, the Gaussian \(f\) times the weight \(w^{(s)} = p/q_i\) equals \(Z_i\, p(x_{\text{true}}|\alpha)\). With \(S\) draws from \(q_i\), the estimate is

\[p(x_i|\alpha) \approx \frac{Z_i}{S} \sum_{s=1}^{S} p\big(x_{\text{true},i}^{(s)}\big|\alpha\big).\]

We draw the values with the inverse CDF method that we used to simulate the masses. The CDF of the cut Gaussian is \((\Phi((x - x_i)/\sigma_i) - \Phi(a_i))/Z_i\), where \(a_i = (x_{\min} - x_i)/\sigma_i\). Solving \(u = \text{CDF}(x)\) for \(x\) gives \(x_{\text{true},i}^{(s)} = x_i + \sigma_i\,\Phi^{-1}\big(\Phi(a_i) + u\, Z_i\big)\), where \(u\) is uniform between 0 and 1.

The draws do not depend on \(\alpha\), so we draw them once and reuse them in every evaluation. The likelihood is then a smooth function of \(\alpha\). New draws at every step would add noise to the ratio of the Metropolis algorithm. The function \texttt{make\_\allowbreak{}draws} returns \(Z_i\) and the logarithm of the draws, and \texttt{power\_\allowbreak{}law\_\allowbreak{}likelihood} returns the estimate for every star.

\tutorialbarrier
\begin{codecell}

\end{codecell}

We start with \(S = 100\) draws for every star. The estimate is \(Z_i\) times the mean of the values \(v^{(s)} = p\big(x_{\text{true},i}^{(s)}\big|\alpha\big)\), so these values play the role of the weights in the ESS (Section 12.6). The ESS tells us how many of the draws carry information,

\[\text{ESS} = \frac{\left(\sum_{s=1}^{S} v^{(s)}\right)^2}{\sum_{s=1}^{S} \left(v^{(s)}\right)^2}.\]

It is the number of equal values that carry the same information. It equals \(S\) when all values are equal and falls as they differ. The constant \(Z_i\) does not change it. Each star has its own ESS. We compute it at the true slope, which we know here, and print the mean over the stars.

\tutorialbarrier
\begin{codecell}
\begin{Verbatim}[commandchars=\\\{\},fontsize=\small]
\PYG{c+c1}{\PYGZsh{} Importance samples for every star}
\PYG{n}{Z\PYGZus{}pl}\PYG{p}{,}\FancyVerbBreakAfterBreak{} \PYG{n}{ln\PYGZus{}draws\PYGZus{}pl} \PYG{o}{=}\FancyVerbBreakAfterBreak{} \PYG{n}{make\PYGZus{}draws}\PYG{p}{(}\FancyVerbBreakAfterBreak{}\PYG{n}{x\PYGZus{}pl}\PYG{p}{,}\FancyVerbBreakAfterBreak{} \PYG{n}{sigma\PYGZus{}pl}\PYG{p}{,}\FancyVerbBreakAfterBreak{} \PYG{l+m+mi}{100}\PYG{p}{)}

\PYG{c+c1}{\PYGZsh{} ESS of every star at the true slope, one value per star}
\PYG{n}{weights} \PYG{o}{=}\FancyVerbBreakAfterBreak{} \PYG{n}{np}\PYG{o}{.}\FancyVerbBreakAfterBreak{}\PYG{n}{exp}\PYG{p}{(}\FancyVerbBreakAfterBreak{}\PYG{o}{\PYGZhy{}}\PYG{n}{alpha\PYGZus{}true} \PYG{o}{*} \PYG{n}{ln\PYGZus{}draws\PYGZus{}pl}\PYG{p}{)}
\PYG{n}{ess\PYGZus{}weights} \PYG{o}{=}\FancyVerbBreakAfterBreak{} \PYG{n}{weights}\PYG{o}{.}\FancyVerbBreakAfterBreak{}\PYG{n}{sum}\PYG{p}{(}\FancyVerbBreakAfterBreak{}\PYG{n}{axis}\PYG{o}{=}\FancyVerbBreakAfterBreak{}\PYG{l+m+mi}{1}\PYG{p}{)}\PYG{o}{*}\PYG{o}{*}\PYG{l+m+mi}{2} \PYG{o}{/} \PYG{p}{(}\FancyVerbBreakAfterBreak{}\PYG{n}{weights}\PYG{o}{*}\PYG{o}{*}\PYG{l+m+mi}{2}\PYG{p}{)}\PYG{o}{.}\FancyVerbBreakAfterBreak{}\PYG{n}{sum}\PYG{p}{(}\FancyVerbBreakAfterBreak{}\PYG{n}{axis}\PYG{o}{=}\FancyVerbBreakAfterBreak{}\PYG{l+m+mi}{1}\PYG{p}{)}

\PYG{n+nb}{print}\PYG{p}{(}\FancyVerbBreakAfterBreak{}\PYG{l+s+sa}{f}\PYG{l+s+s2}{\PYGZdq{}}\PYG{l+s+s2}{Mean ESS of one star = }\PYG{l+s+si}{\PYGZob{}}\PYG{n}{ess\PYGZus{}weights}\PYG{o}{.}\FancyVerbBreakAfterBreak{}\PYG{n}{mean}\PYG{p}{(}\FancyVerbBreakAfterBreak{}\PYG{p}{)}\PYG{l+s+si}{:}\PYG{l+s+s2}{.0f}\PYG{l+s+si}{\PYGZcb{}}\FancyVerbBreakAfterBreak{}\PYG{l+s+s2}{ of 100 draws}\PYG{l+s+s2}{\PYGZdq{}}\PYG{p}{)}
\end{Verbatim}
\end{codecell}
\begin{Verbatim}[fontsize=\footnotesize, xleftmargin=5mm]
Mean ESS of one star = 96 of 100 draws
\end{Verbatim}

A typical star keeps an ESS of 96 of 100 draws, so the proposal follows the integrand well. The values \(v^{(s)}\) of every star are nearly equal. The estimate should therefore be accurate, and we now test this.

\subsection*{Checking the Estimate against a Trapezoid Integral}

The estimate uses random numbers, so we check it against a reference that uses none. For one parameter we have one.

\begin{enumerate}
\def\labelenumi{\arabic{enumi}.}
\tightlist
\item
  \textbf{Trapezoid integral.} For every star, we place 2000 points in the range of \(x_{\text{true}}\) and apply the trapezoid rule. The rule approximates \(\int g\,\mathrm{d}x\) by \(\sum_k (g_k + g_{k+1})\,\Delta x/2\), the areas of thin trapezoids under the curve (Section 12.2). The spacing is far smaller than the width of the Gaussian, which is at least 0.05.
\item
  \textbf{Comparison.} We compare the log likelihood of the trapezoid integral with that of the estimate for \(S = 10\), 100 and 1000, on a grid of slopes. We subtract the maximum of each curve, because only the shape of the log likelihood matters for the posterior.
\end{enumerate}

\tutorialbarrier
\begin{codecell}

\end{codecell}

We plot the log likelihood of the trapezoid integral and of the three estimates.

\tutorialbarrier
\begin{codecell}

\end{codecell}
\tutorialplot{\includegraphics[width=5.236in,height=0.75\textheight,keepaspectratio]{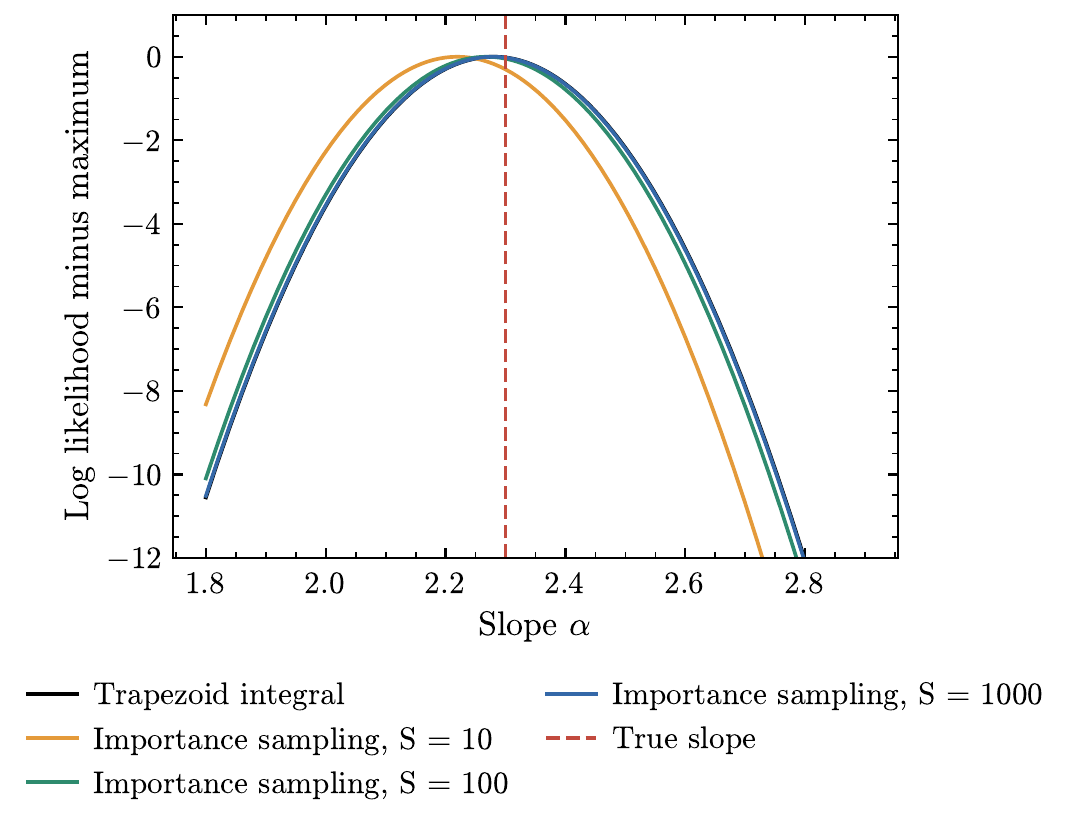}}%
  {\includegraphics[height=\tutorialheight]{tutorials/13b/fig06.pdf}}

The estimate with \(S = 10\) is far off, and its maximum lies at a lower slope. The estimates with \(S = 100\) and \(S = 1000\) lie close to the trapezoid integral. We use \(S = 100\), which is ten times cheaper than \(S = 1000\).

\subsection*{Sampling the Slope}

With the checked likelihood we sample the slope \(\alpha\), the only parameter. We assume a uniform prior between 0 and 5. This range holds every slope that we consider and favors none, so the log posterior is the log likelihood inside the range. We follow the steps of the first part, a pilot chain and one chain.

\tutorialbarrier
\begin{codecell}

\end{codecell}
\begin{Verbatim}[fontsize=\footnotesize, xleftmargin=5mm]
Acceptance rate = 40 percent
α      mean 2.270, sd 0.104
\end{Verbatim}

We compare the posterior samples with the posterior from the trapezoid integral. Because the prior is uniform, we get it by normalizing its likelihood over the slopes. The plot also marks the slope of the fit that ignores the uncertainties.

\tutorialbarrier
\begin{codecell}

\end{codecell}
\begin{Verbatim}[fontsize=\footnotesize, xleftmargin=5mm]
Trapezoid integral posterior: mean 2.280, sd 0.105
\end{Verbatim}
\tutorialplot{\includegraphics[width=5.261in,height=0.75\textheight,keepaspectratio]{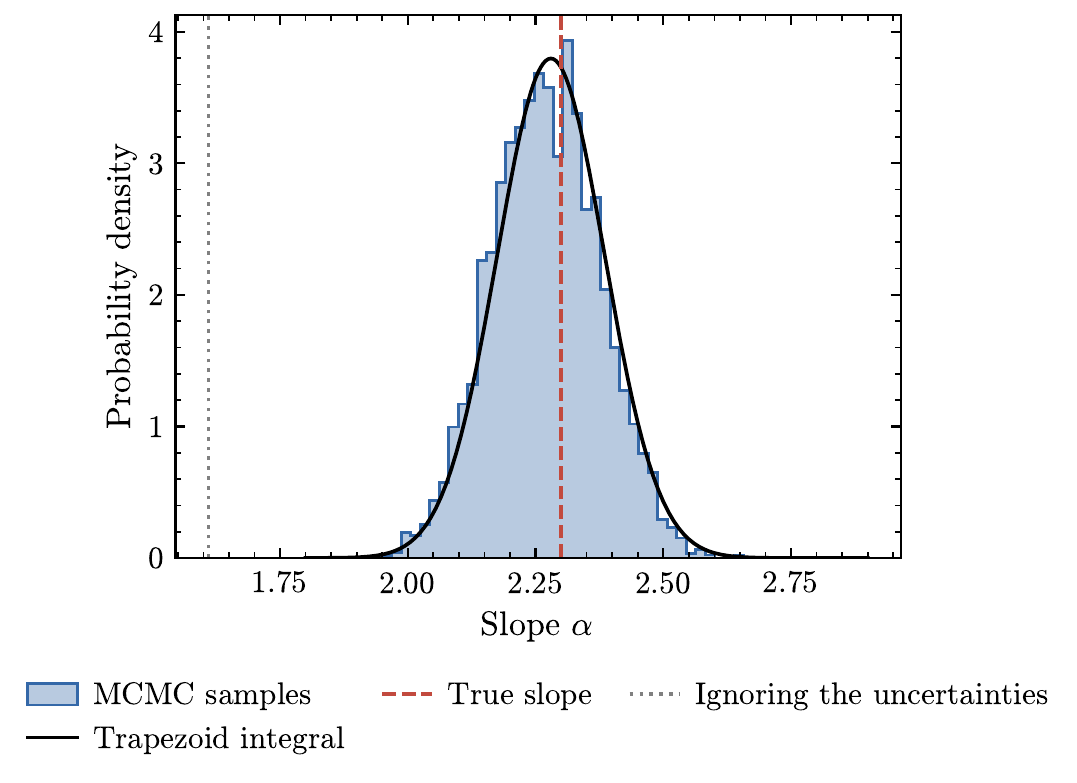}}%
  {\includegraphics[height=\tutorialheight]{tutorials/13b/fig07.pdf}}

The acceptance rate is near the 40 percent that we aimed for. The histogram of the samples follows the posterior of the trapezoid integral, and the printed means and standard deviations agree. So 100 fixed draws per star are enough. The true slope lies inside the posterior. The dotted line, the slope of the fit that ignores the uncertainties, lies far outside it. The method recovers the slope when the model describes the stars.

\section*{Contaminated Data}

So far the model describes every star. A real sample of masses can hold stars that do not follow it. Many stars have a companion that the telescope does not resolve. An unresolved binary shows the light of two stars as one star, so its inferred mass is too high. These binaries distort the mass function if we do not model them.

We assume that 10 percent of our 3000 stars are binaries. The combined mass of two stars from the range lies between 1 and 2 solar masses. We assume that their apparent masses follow a Gaussian with the mean 1.2 and the standard deviation 0.3 solar masses. The analysis does not know these values. The other 90 percent are the power-law stars of the previous part.

\tutorialbarrier
\begin{codecell}

\end{codecell}

\subsection*{A Single Power Law Fails}

An analyst who does not know about the binaries first fits the single power law of the previous part. The likelihood, the importance samples and the chain are the same.

\tutorialbarrier
\begin{codecell}

\end{codecell}
\begin{Verbatim}[fontsize=\footnotesize, xleftmargin=5mm]
Acceptance rate = 39 percent
α      mean 1.280, sd 0.103
\end{Verbatim}

The posterior is narrow, and it puts the slope at 1.28, far below the true value 2.3. The stars of the highest masses lie far above the upper edge compared with their uncertainties. A shallower power law has more density at this edge, which raises their tiny likelihood. The sampler works, but the model is wrong.

In a real survey we do not know the true slope, so we need a check that does not use the truth. A posterior predictive check compares the model with the data in three steps.

\begin{enumerate}
\def\labelenumi{\arabic{enumi}.}
\tightlist
\item
  \textbf{Draw.} Draw parameters \(\boldsymbol{\theta}\) from the posterior.
\item
  \textbf{Simulate.} Simulate a new sample of measurements for each draw, with the known uncertainties of our stars. The samples follow the posterior predictive distribution of Section 5.7, \(p(x_{\text{new}}|\mathcal{D}) = \int p(x_{\text{new}}|\boldsymbol{\theta})\, p(\boldsymbol{\theta}|\mathcal{D})\,\mathrm{d}\boldsymbol{\theta}\).
\item
  \textbf{Compare.} Compare the simulated samples with the data. If the model describes the data, the data look like one of the samples.
\end{enumerate}

The function \texttt{simulate\_\allowbreak{}measurements} simulates one sample. It draws the power law with the inverse CDF method and the binaries from a Gaussian. The function \texttt{plot\_\allowbreak{}predictive\_\allowbreak{}check} compares 200 simulated samples with the data.

\tutorialbarrier
\begin{codecell}

\end{codecell}

We run the posterior predictive check for the single power law.

\tutorialbarrier
\begin{codecell}
\begin{Verbatim}[commandchars=\\\{\},fontsize=\small]
\PYG{c+c1}{\PYGZsh{} Posterior predictive check of the single power law}
\PYG{n}{plt}\PYG{o}{.}\FancyVerbBreakAfterBreak{}\PYG{n}{figure}\PYG{p}{(}\FancyVerbBreakAfterBreak{}\PYG{n}{figsize}\PYG{o}{=}\FancyVerbBreakAfterBreak{}\PYG{p}{(}\FancyVerbBreakAfterBreak{}\PYG{l+m+mf}{13.6}\PYG{p}{,}\FancyVerbBreakAfterBreak{} \PYG{l+m+mf}{4.8}\PYG{p}{)}\PYG{p}{)}
\PYG{n}{plot\PYGZus{}predictive\PYGZus{}check}\PYG{p}{(}\FancyVerbBreakAfterBreak{}\PYG{n}{plt}\PYG{o}{.}\FancyVerbBreakAfterBreak{}\PYG{n}{gca}\PYG{p}{(}\FancyVerbBreakAfterBreak{}\PYG{p}{)}\PYG{p}{,}\FancyVerbBreakAfterBreak{} \PYG{n}{x\PYGZus{}mix}\PYG{p}{,}\FancyVerbBreakAfterBreak{} \PYG{n}{sigma\PYGZus{}mix}\PYG{p}{,}\FancyVerbBreakAfterBreak{} \PYG{n}{chain\PYGZus{}single}\PYG{p}{[}\FancyVerbBreakAfterBreak{}\PYG{l+m+mi}{500}\PYG{p}{:}\PYG{p}{]}\PYG{p}{)}
\PYG{n}{plt}\PYG{o}{.}\FancyVerbBreakAfterBreak{}\PYG{n}{legend}\PYG{p}{(}\FancyVerbBreakAfterBreak{}\PYG{n}{ncol}\PYG{o}{=}\FancyVerbBreakAfterBreak{}\PYG{l+m+mi}{3}\PYG{p}{,}\FancyVerbBreakAfterBreak{} \PYG{n}{loc}\PYG{o}{=}\FancyVerbBreakAfterBreak{}\PYG{l+s+s1}{\PYGZsq{}}\PYG{l+s+s1}{upper center}\PYG{l+s+s1}{\PYGZsq{}}\PYG{p}{,}\FancyVerbBreakAfterBreak{} \PYG{n}{bbox\PYGZus{}to\PYGZus{}anchor}\PYG{o}{=}\FancyVerbBreakAfterBreak{}\PYG{p}{(}\FancyVerbBreakAfterBreak{}\PYG{l+m+mf}{0.5}\PYG{p}{,}\FancyVerbBreakAfterBreak{} \PYG{o}{\PYGZhy{}}\PYG{l+m+mf}{0.15}\PYG{p}{)}\PYG{p}{)}
\PYG{n}{plt}\PYG{o}{.}\FancyVerbBreakAfterBreak{}\PYG{n}{show}\PYG{p}{(}\FancyVerbBreakAfterBreak{}\PYG{p}{)}
\end{Verbatim}
\end{codecell}
\tutorialplot{\includegraphics[width=5.709in,height=0.75\textheight,keepaspectratio]{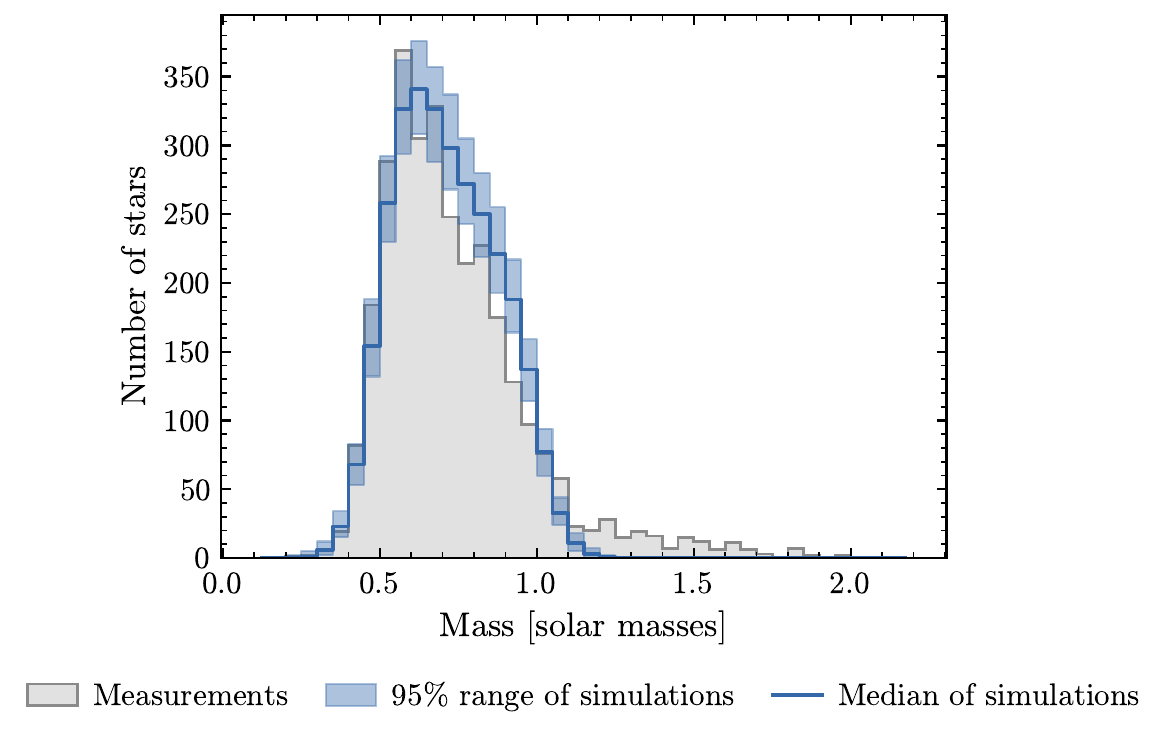}}%
  {\includegraphics[height=\tutorialheight]{tutorials/13b/fig08.pdf}}

The simulations have almost no stars above 1.2 solar masses, but the data hold a long tail of them. The model also puts all 3000 stars near the range, so it predicts too many stars between about 0.6 and 0.9 solar masses. The single power law fails the check, and we must extend the model.

\subsection*{A Mixture Model}

The single power law cannot describe the binaries, so we add them to the model. The mixture combines the Gaussian of the first part with the power law of the second. The intrinsic distribution is now a mixture of the power law and a Gaussian for the binaries. Section 11.5 treats a mixture of Gaussians, and here one component is a power law. A mixture draws each star from one component with a probability that is the weight of the component. The weights are \(1-f\) and \(f\),

\[p(x_{\text{true}}|\boldsymbol{\theta}) = (1-f)\, p(x_{\text{true}}|\alpha) + f\, \mathcal{N}(x_{\text{true}}|\mu_b, \sigma_b^2),\]

where the parameters are \(\boldsymbol{\theta} = (\alpha, \mu_b, \sigma_b, f)\). The integral is linear in the intrinsic density, so the likelihood of a measurement mixes the likelihoods of the two components.

\begin{itemize}
\tightlist
\item
  \textbf{Power law.} The importance sampling likelihood \(p(x_i|\alpha)\) of the previous part.
\item
  \textbf{Gaussian.} The closed form of the first part, with the variances of the Gaussian and of the error added.
\end{itemize}

The likelihood of the mixture is then

\[p(x_i|\boldsymbol{\theta}) = (1-f)\, p(x_i|\alpha) + f\, \mathcal{N}(x_i|\mu_b, \sigma_b^2 + \sigma_i^2).\]

We assume uniform priors for the four parameters.

\begin{itemize}
\tightlist
\item
  \textbf{Slope \(\alpha\).} Between 0 and 5, as before.
\item
  \textbf{Binary mean \(\mu_b\) and width \(\sigma_b\).} Between 0 and 2 and between 0 and 1, which are broad compared with the data.
\item
  \textbf{Fraction \(f\).} Between 0 and 0.5, so that the power law stays the main population.
\end{itemize}

\tutorialbarrier
\begin{codecell}

\end{codecell}

\subsection*{Sampling the Mixture}

The pilot chain starts from a guess of the four parameters with the width 0.05 for each. The final chain has 20000 iterations, and we discard the first 1000 to be safe.

\tutorialbarrier
\begin{codecell}

\end{codecell}
\begin{Verbatim}[fontsize=\footnotesize, xleftmargin=5mm]
Acceptance rate = 31 percent
α      mean 2.342, sd 0.127
μ_b    mean 1.177, sd 0.039
σ_b    mean 0.300, sd 0.019
f      mean 0.108, sd 0.012
\end{Verbatim}

The acceptance rate is inside the range of Section 13.9. The posterior means lie close to the true values of the four parameters, within their uncertainties. The slope is less certain than in the previous part, because the binaries replace 10 percent of the power-law stars and their parameters are unknown.

The corner plot shows the joint posterior of the four parameters, which the single numbers do not show.

\tutorialbarrier
\begin{codecell}
\begin{Verbatim}[commandchars=\\\{\},fontsize=\small]
\PYG{c+c1}{\PYGZsh{} Samples after the burn\PYGZhy{}in and their joint posterior}
\PYG{n}{samples\PYGZus{}mix} \PYG{o}{=}\FancyVerbBreakAfterBreak{} \PYG{n}{chain\PYGZus{}mix}\PYG{p}{[}\FancyVerbBreakAfterBreak{}\PYG{l+m+mi}{1000}\PYG{p}{:}\PYG{p}{]}
\PYG{n}{corner\PYGZus{}plot}\PYG{p}{(}\FancyVerbBreakAfterBreak{}\PYG{n}{samples\PYGZus{}mix}\PYG{p}{,}\FancyVerbBreakAfterBreak{} \PYG{p}{[}\FancyVerbBreakAfterBreak{}\PYG{l+s+sa}{r}\PYG{l+s+s1}{\PYGZsq{}}\PYG{l+s+s1}{\PYGZdl{}}\PYG{l+s+s1}{\PYGZbs{}}\PYG{l+s+s1}{alpha\PYGZdl{}}\PYG{l+s+s1}{\PYGZsq{}}\PYG{p}{,}\FancyVerbBreakAfterBreak{} \PYG{l+s+sa}{r}\PYG{l+s+s1}{\PYGZsq{}}\PYG{l+s+s1}{\PYGZdl{}}\PYG{l+s+s1}{\PYGZbs{}}\PYG{l+s+s1}{mu\PYGZus{}b\PYGZdl{}}\PYG{l+s+s1}{\PYGZsq{}}\PYG{p}{,}\FancyVerbBreakAfterBreak{} \PYG{l+s+sa}{r}\PYG{l+s+s1}{\PYGZsq{}}\PYG{l+s+s1}{\PYGZdl{}}\PYG{l+s+s1}{\PYGZbs{}}\PYG{l+s+s1}{sigma\PYGZus{}b\PYGZdl{}}\PYG{l+s+s1}{\PYGZsq{}}\PYG{p}{,}\FancyVerbBreakAfterBreak{} \PYG{l+s+sa}{r}\PYG{l+s+s1}{\PYGZsq{}}\PYG{l+s+s1}{\PYGZdl{}f\PYGZdl{}}\PYG{l+s+s1}{\PYGZsq{}}\PYG{p}{]}\PYG{p}{,}\FancyVerbBreakAfterBreak{} \PYG{p}{[}\FancyVerbBreakAfterBreak{}\PYG{n}{alpha\PYGZus{}true}\PYG{p}{,}\FancyVerbBreakAfterBreak{} \PYG{n}{mu\PYGZus{}b\PYGZus{}true}\PYG{p}{,}\FancyVerbBreakAfterBreak{} \PYG{n}{sigma\PYGZus{}b\PYGZus{}true}\PYG{p}{,}\FancyVerbBreakAfterBreak{} \PYG{n}{f\PYGZus{}true}\PYG{p}{]}\PYG{p}{,}\FancyVerbBreakAfterBreak{} \PYG{n}{figsize}\PYG{o}{=}\FancyVerbBreakAfterBreak{}\PYG{p}{(}\FancyVerbBreakAfterBreak{}\PYG{l+m+mi}{12}\PYG{p}{,}\FancyVerbBreakAfterBreak{} \PYG{l+m+mf}{9.5}\PYG{p}{)}\PYG{p}{)}
\end{Verbatim}
\end{codecell}
\tutorialplot{\includegraphics[width=6.500in,height=0.75\textheight,keepaspectratio]{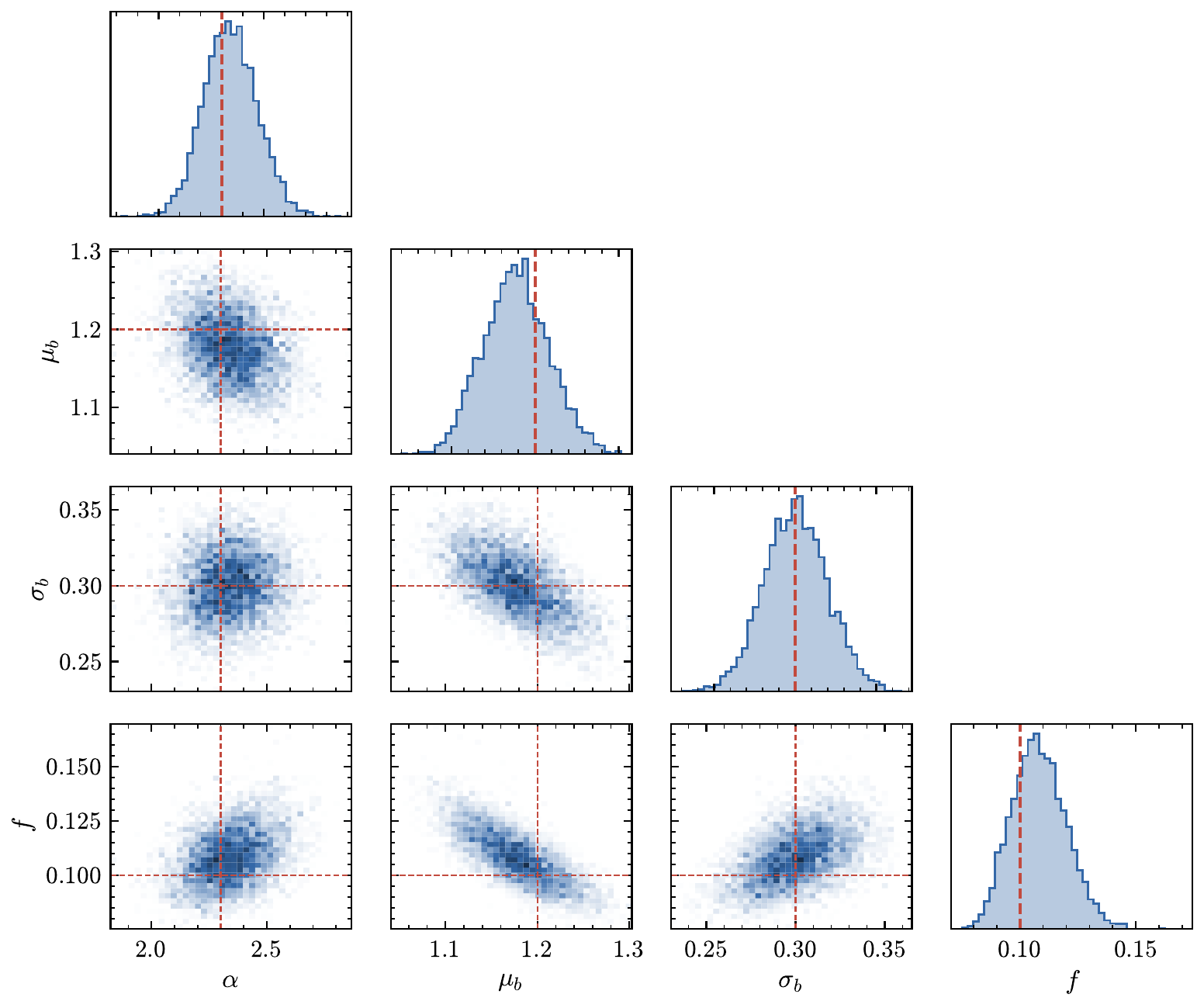}}%
  {\includegraphics[height=\tutorialheight]{tutorials/13b/fig09.pdf}}

The slope is only weakly correlated with the binary parameters, but \(\mu_b\), \(\sigma_b\) and \(f\) are highly correlated. The stars above 1 solar mass fix the upper part of the Gaussian, and the lower part hides under the power law. A lower mean therefore needs a larger width and a larger fraction to cover the same stars. The proposal covariance from the pilot chain follows these ridges (Section 13.9).

\subsection*{Checking the Mixture}

The parameters are recovered, but in a real survey we cannot compare them with the truth. We therefore repeat the posterior predictive check for the mixture, with the same plot as for the single power law.

\tutorialbarrier
\begin{codecell}
\begin{Verbatim}[commandchars=\\\{\},fontsize=\small]
\PYG{c+c1}{\PYGZsh{} Posterior predictive check of the mixture}
\PYG{n}{plt}\PYG{o}{.}\FancyVerbBreakAfterBreak{}\PYG{n}{figure}\PYG{p}{(}\FancyVerbBreakAfterBreak{}\PYG{n}{figsize}\PYG{o}{=}\FancyVerbBreakAfterBreak{}\PYG{p}{(}\FancyVerbBreakAfterBreak{}\PYG{l+m+mf}{13.6}\PYG{p}{,}\FancyVerbBreakAfterBreak{} \PYG{l+m+mf}{4.8}\PYG{p}{)}\PYG{p}{)}
\PYG{n}{plot\PYGZus{}predictive\PYGZus{}check}\PYG{p}{(}\FancyVerbBreakAfterBreak{}\PYG{n}{plt}\PYG{o}{.}\FancyVerbBreakAfterBreak{}\PYG{n}{gca}\PYG{p}{(}\FancyVerbBreakAfterBreak{}\PYG{p}{)}\PYG{p}{,}\FancyVerbBreakAfterBreak{} \PYG{n}{x\PYGZus{}mix}\PYG{p}{,}\FancyVerbBreakAfterBreak{} \PYG{n}{sigma\PYGZus{}mix}\PYG{p}{,}\FancyVerbBreakAfterBreak{} \PYG{n}{samples\PYGZus{}mix}\PYG{p}{)}
\PYG{n}{plt}\PYG{o}{.}\FancyVerbBreakAfterBreak{}\PYG{n}{legend}\PYG{p}{(}\FancyVerbBreakAfterBreak{}\PYG{n}{ncol}\PYG{o}{=}\FancyVerbBreakAfterBreak{}\PYG{l+m+mi}{3}\PYG{p}{,}\FancyVerbBreakAfterBreak{} \PYG{n}{loc}\PYG{o}{=}\FancyVerbBreakAfterBreak{}\PYG{l+s+s1}{\PYGZsq{}}\PYG{l+s+s1}{upper center}\PYG{l+s+s1}{\PYGZsq{}}\PYG{p}{,}\FancyVerbBreakAfterBreak{} \PYG{n}{bbox\PYGZus{}to\PYGZus{}anchor}\PYG{o}{=}\FancyVerbBreakAfterBreak{}\PYG{p}{(}\FancyVerbBreakAfterBreak{}\PYG{l+m+mf}{0.5}\PYG{p}{,}\FancyVerbBreakAfterBreak{} \PYG{o}{\PYGZhy{}}\PYG{l+m+mf}{0.15}\PYG{p}{)}\PYG{p}{)}
\PYG{n}{plt}\PYG{o}{.}\FancyVerbBreakAfterBreak{}\PYG{n}{show}\PYG{p}{(}\FancyVerbBreakAfterBreak{}\PYG{p}{)}
\end{Verbatim}
\end{codecell}
\tutorialplot{\includegraphics[width=5.709in,height=0.75\textheight,keepaspectratio]{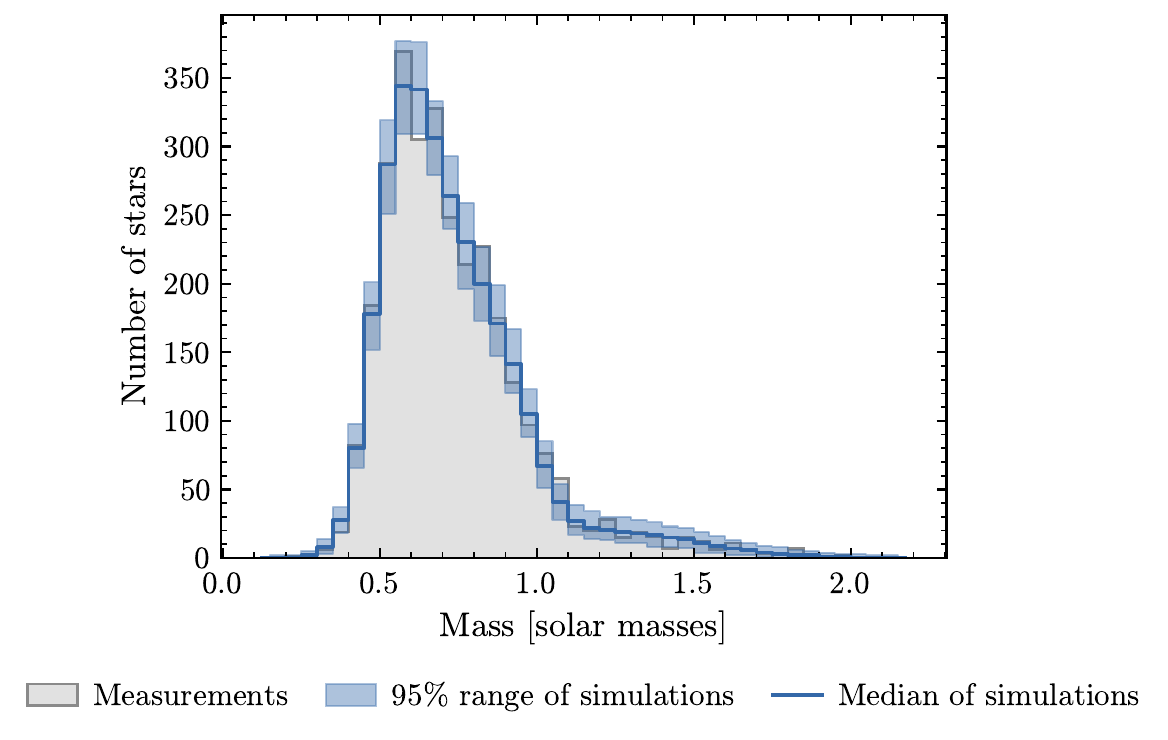}}%
  {\includegraphics[height=\tutorialheight]{tutorials/13b/fig10.pdf}}

The 95 percent range of the simulations now contains the data in most bins, including the tail above 1.2 solar masses. The mixture model passes the check, and the slope is recovered.

\section*{Limits}

The results above rest on assumptions. We note the following limits of the method.

\begin{itemize}
\tightlist
\item
  \textbf{Known Gaussian errors.} We assumed that the uncertainties \(\sigma_i\) are correct and the errors are Gaussian. If the uncertainties are too small, the intrinsic scatter absorbs the difference.
\item
  \textbf{Complete samples.} We assumed that every object enters the sample. A cut on the measured value, such as a magnitude limit, changes the likelihood of the objects that remain, and the model must include it.
\item
  \textbf{The form of the contamination.} The Gaussian shape of the binaries is an assumption. A model that passes the posterior predictive check describes the data well, but another shape for the binaries might describe them equally well.
\item
  \textbf{Random walk steps.} The random walk proposal slows down when the number of parameters grows. Hamiltonian Monte Carlo and ensemble samplers scale better (Chapter 13).
\end{itemize}

\section*{Summary}

This tutorial inferred the parameters of an intrinsic distribution from measurements with known, different uncertainties. Our main lessons follow.

\begin{enumerate}
\def\labelenumi{\arabic{enumi}.}
\tightlist
\item
  \textbf{Deconvolution as marginalization.} The likelihood of a measurement is an integral over the unknown true value. A fit that ignores the uncertainties gives biased parameters, because the errors widen the distribution and move objects over its edges.
\item
  \textbf{Importance sampling with MCMC.} When the integral has no closed form, importance sampling estimates it. A proposal that follows the integrand keeps the weights even, and draws that we make once give a smooth likelihood for the chain. A trapezoid integral and the ESS of the weights check the estimate.
\item
  \textbf{Tuning the proposal of a random walk.} A pilot chain gives a proposal covariance that follows the posterior. The acceptance rate sets the step size. A trace plot shows the burn-in, and a corner plot shows the joint posterior and its correlations.
\item
  \textbf{Posterior predictive checks and mixture models.} A posterior predictive check compares simulated samples with the data and finds a contaminating population. A mixture model repairs the failure, and its likelihood is the mixture of the likelihoods of its components.
\end{enumerate}

Deconvolution is one use of these tools. The same idea carries to richer models. A mixture describes the populations in the data, importance sampling handles the integrals over hidden quantities, and MCMC samples the posterior. Used together, they turn a statistical model of the hidden truth and the noise into inference from imperfect measurements.

\tutorialchapter{14a}{Gaussian Process Regression for Stellar Light Curves}
\noindent\textit{Companion material for Chapter 14}

\section*{Introduction}

A Gaussian process (GP) infers a function from data without a fixed parametric form, and it states how well the data fix the function. This tutorial applies GP regression to stellar light curves, the brightness of a star measured over time.

Convection and oscillations make the brightness of a star vary. In the outer layers, hot gas rises and cool gas sinks, and this granulation changes the brightness over time. Sound waves trapped inside the star make it pulsate as well. Both processes become slower as the star expands, because the convective cells are larger and a sound wave needs more time to cross the star. An evolved star has expanded, so it has a lower surface gravity, \(\log g\), than a compact one. The timescale of the variations could therefore tell us how evolved a star is, even for a star without a spectrum.

The task is to measure this timescale from light curves of the Kepler space telescope. The signals are quasi-periodic, which means that they repeat only roughly, and they contain measurement noise.

The usual methods take the Fourier transform of the light curve and fit its peaks. They need careful tuning, they have difficulty with irregular sampling, and they give limited uncertainties. They also impose a sum of sinusoids, which may not describe the stochastic variations of a star.

A GP uses no fixed basis functions. We state the statistical properties of the variations in a kernel function, and the data select the function. The length scale of the kernel is the time over which the brightness stays correlated, so it measures the timescale directly. We learn it for each star by maximizing the marginal likelihood, and we then compare it with log g.

\section*{Prerequisites and Learning Objectives}

\subsection*{Prerequisites}

This tutorial assumes the following.

\begin{itemize}
\tightlist
\item
  Multivariate Gaussian distributions, with the covariance matrix and conditioning (Chapters 3 and 5)
\item
  Maximum likelihood estimation (Chapter 4)
\item
  Gaussian processes, with kernel functions, conditioning, the marginal likelihood and the Cholesky decomposition (Chapter 14)
\item
  Python programming with NumPy arrays, matplotlib and SciPy
\end{itemize}

\subsection*{Learning Objectives}

After this tutorial you can do the following.

\begin{itemize}
\tightlist
\item
  Explain what an RBF kernel assumes about a light curve and where it falls short
\item
  Compute the prediction of a GP and read the uncertainty band of a new observation
\item
  Learn the length scale and the noise level by maximizing the marginal likelihood
\item
  Relate a GP hyperparameter to a stellar property
\item
  Recognize how the sampling interval limits what the data can resolve
\end{itemize}

We begin by importing the libraries and setting the plot style.

\tutorialbarrier
\begin{codecell}

\end{codecell}

\section*{The Kepler Stellar Light Curve Dataset}

We use 100 light curves from the Kepler space telescope, taken from the Kepler archive. Each light curve has 1000 normalized brightness values, one every 30 minutes over about 20 days. This stretch covers many cycles of the oscillations that we study. The stars span a range of evolutionary states, and the surface gravity \(\log g\) of each star is its label. We load the data and print its shape.

\tutorialbarrier
\begin{codecell}
\begin{Verbatim}[commandchars=\\\{\},fontsize=\small]
\PYG{c+c1}{\PYGZsh{} Load the dataset}
\PYG{n}{temp} \PYG{o}{=}\FancyVerbBreakAfterBreak{} \PYG{n}{np}\PYG{o}{.}\FancyVerbBreakAfterBreak{}\PYG{n}{load}\PYG{p}{(}\FancyVerbBreakAfterBreak{}\PYG{l+s+s1}{\PYGZsq{}}\PYG{l+s+s1}{../data/dataset\PYGZus{}stellar\PYGZus{}light\PYGZus{}curves.npz}\PYG{l+s+s1}{\PYGZsq{}}\PYG{p}{)}
\PYG{n}{light\PYGZus{}curve} \PYG{o}{=}\FancyVerbBreakAfterBreak{} \PYG{n}{temp}\PYG{p}{[}\FancyVerbBreakAfterBreak{}\PYG{l+s+s1}{\PYGZsq{}}\PYG{l+s+s1}{data}\PYG{l+s+s1}{\PYGZsq{}}\PYG{p}{]}
\PYG{n}{label} \PYG{o}{=}\FancyVerbBreakAfterBreak{} \PYG{n}{temp}\PYG{p}{[}\FancyVerbBreakAfterBreak{}\PYG{l+s+s1}{\PYGZsq{}}\PYG{l+s+s1}{label}\PYG{l+s+s1}{\PYGZsq{}}\PYG{p}{]}

\PYG{c+c1}{\PYGZsh{} Time in minutes, with 30 minutes between adjacent values}
\PYG{n}{time} \PYG{o}{=}\FancyVerbBreakAfterBreak{} \PYG{n}{np}\PYG{o}{.}\FancyVerbBreakAfterBreak{}\PYG{n}{arange}\PYG{p}{(}\FancyVerbBreakAfterBreak{}\PYG{l+m+mi}{0}\PYG{p}{,}\FancyVerbBreakAfterBreak{} \PYG{l+m+mi}{30}\PYG{o}{*}\PYG{n}{light\PYGZus{}curve}\PYG{o}{.}\FancyVerbBreakAfterBreak{}\PYG{n}{shape}\PYG{p}{[}\FancyVerbBreakAfterBreak{}\PYG{l+m+mi}{1}\PYG{p}{]}\PYG{p}{,}\FancyVerbBreakAfterBreak{} \PYG{l+m+mi}{30}\PYG{p}{)}

\PYG{c+c1}{\PYGZsh{} Print the shape of the dataset}
\PYG{n+nb}{print}\PYG{p}{(}\FancyVerbBreakAfterBreak{}\PYG{l+s+s2}{\PYGZdq{}}\PYG{l+s+s2}{Shape of the dataset =}\PYG{l+s+s2}{\PYGZdq{}}\PYG{p}{,}\FancyVerbBreakAfterBreak{} \PYG{n}{light\PYGZus{}curve}\PYG{o}{.}\FancyVerbBreakAfterBreak{}\PYG{n}{shape}\PYG{p}{)}
\end{Verbatim}
\end{codecell}
\begin{Verbatim}[fontsize=\footnotesize, xleftmargin=5mm]
Shape of the dataset = (100, 1000)
\end{Verbatim}

\section*{Visualizing Stellar Oscillations}

We plot ten light curves, one of every ten stars in order of surface gravity. The plot shows how the variations change with stellar evolution.

\tutorialbarrier
\begin{codecell}

\end{codecell}
\tutorialplot{\includegraphics[width=6.174in,height=0.75\textheight,keepaspectratio]{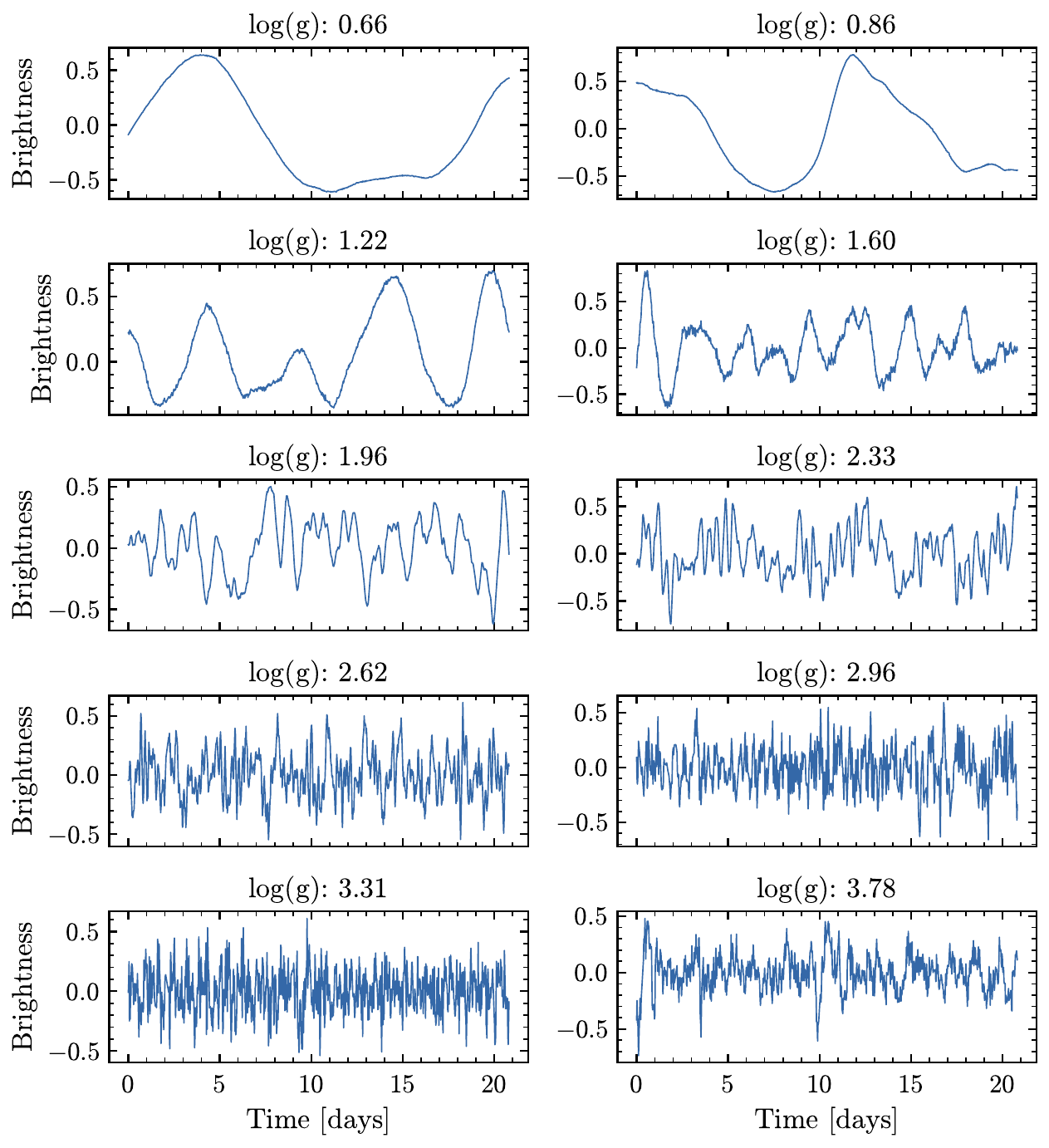}}%
  {\includegraphics[height=\tutorialheight]{tutorials/14a/fig01.pdf}}

Stars of low surface gravity (top panels) vary slowly and smoothly. Stars of high surface gravity (bottom panels) vary rapidly and look like noise. The amplitudes and phases change with time, so the variations are stochastic and suit a GP. This is the pattern of the introduction, in which the variations become faster as the star becomes more compact.

For the first fit, we focus on one light curve, that of star 50. We use it to develop and validate the method before we scale it to all 100 stars.

\section*{Gaussian Process Modeling of Stellar Light Curves}

A GP can model the underlying signal and the noise at the same time. For stellar variability, the radial basis function (RBF) kernel (Section 14.3) is a good starting point. It encodes our expectation of smooth and continuous variations of the brightness. The input \(x\) is the time and the target \(t\) is the brightness. The kernel is

\[k(x, x') = \sigma_f^2 \exp\left(-\frac{(x - x')^2}{2\ell^2}\right),\]

where \(\sigma_f^2\) sets the amplitude of the variations and \(\ell\) is the length scale, the time over which the brightness stays correlated. We fix \(\sigma_f = 1\) and learn \(\ell\), because we are interested in the timescale. The light curves have standard deviations below 1, so this choice is a simplification.

The model of the observations is

\[t_i = f(x_i) + \epsilon_i, \quad \epsilon_i \sim \mathcal{N}(0, \sigma_n^2),\]

where \(f(x)\) is the latent stellar variability and \(\sigma_n^2\) is the variance of the observation noise, so \(\sigma_n\) is the noise level. Both \(\ell\) and \(\sigma_n\) are hyperparameters, the parameters of the kernel and of the noise model. The data tell us both the timescale and the noise level.

\section*{Developing Intuition: From 2D Gaussians to GP Regression}

Before we use the GP on the light curves, we build intuition with two simpler examples. A 2D Gaussian shows how conditioning on one variable predicts another. The regression of a 1D function then shows the GP itself.

In the 2D example, \(x\) and \(y\) are two random variables, not the time. Consider points \((x, y)\) that follow a bivariate Gaussian distribution with the mean \(\boldsymbol{\mu} = [\mu_x, \mu_y]^T\) and the covariance matrix

\[\boldsymbol{\Sigma} = \begin{bmatrix} \sigma_x^2 & \sigma_{xy} \\ \sigma_{xy} & \sigma_y^2 \end{bmatrix}.\]

The conditional distribution of \(y\) given \(x\) is again Gaussian (Section 5.9),

\[p(y|x) = \mathcal{N}\left(\mu_y + \frac{\sigma_{xy}}{\sigma_x^2}(x - \mu_x), \sigma_y^2 - \frac{\sigma_{xy}^2}{\sigma_x^2}\right).\]

The mean changes linearly with \(x - \mu_x\). The variance is smaller than the marginal variance \(\sigma_y^2\) when \(x\) and \(y\) are correlated, so observing \(x\) reduces our uncertainty about \(y\). We simulate 2000 points from a 2D Gaussian with a known mean and covariance, and we implement the Gaussian density and the conditional distribution.

\tutorialbarrier
\begin{codecell}

\end{codecell}

We condition on three values of \(x\). The left panel shows the data, the density of the true distribution and the conditional mean with a range of two standard deviations. The right panel shows the three conditional distributions.

\tutorialbarrier
\begin{codecell}

\end{codecell}
\tutorialplot{\includegraphics[width=6.500in,height=0.75\textheight,keepaspectratio]{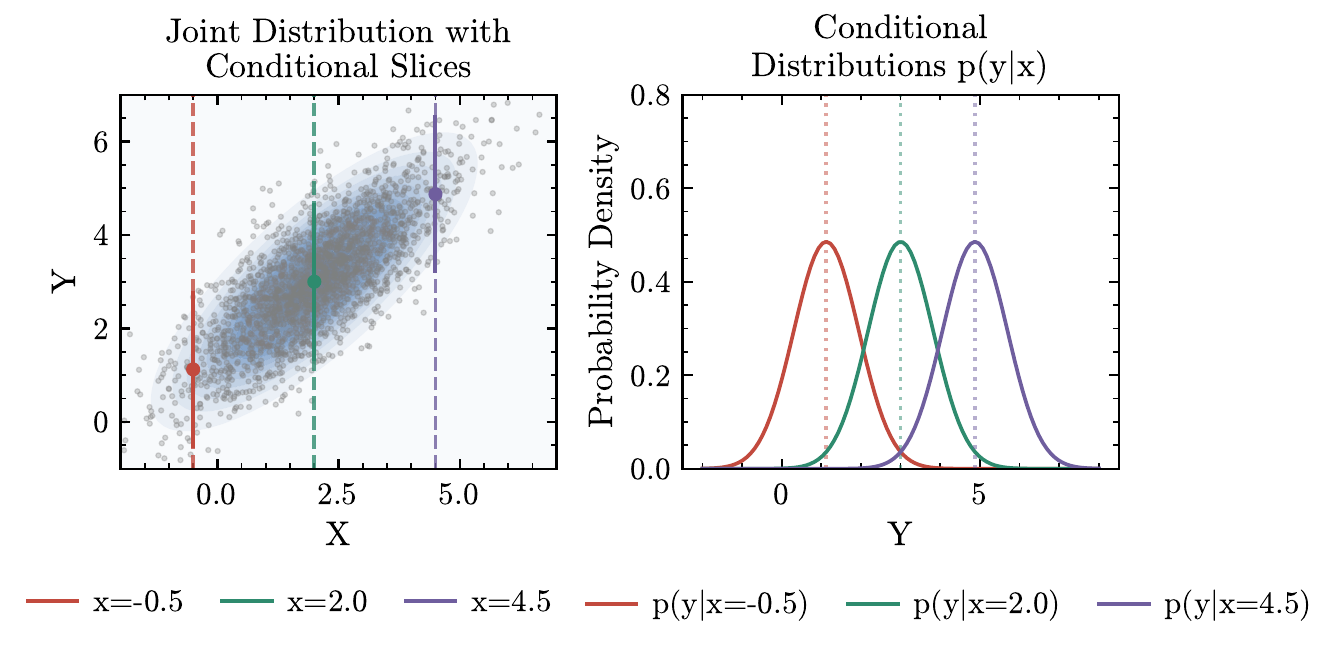}}%
  {\includegraphics[height=\tutorialheight]{tutorials/14a/fig02.pdf}}

The three conditional distributions have the same width, because the variance does not depend on \(x\), and their means rise with \(x\). Each is narrower than the marginal distribution of \(y\), because observing \(x\) reduces our uncertainty about \(y\).

\subsection*{Maximum Likelihood for the 2D Gaussian}

We simulated the points from known parameters. Suppose that we did not know them. We can estimate them by maximum likelihood (Section 4.3), which chooses the parameters that make the observed points most probable. For \(n\) points \(\mathbf{z}_i = (x_i, y_i)\), the log-likelihood is

\[\ln \mathcal{L}(\boldsymbol{\mu}, \boldsymbol{\Sigma}) = -\frac{n}{2}\ln(2\pi) - \frac{n}{2}\ln\det\boldsymbol{\Sigma} - \frac{1}{2}\sum_{i=1}^{n}(\mathbf{z}_i - \boldsymbol{\mu})^T\boldsymbol{\Sigma}^{-1}(\mathbf{z}_i - \boldsymbol{\mu}).\]

We maximize it by setting its derivatives to zero. The derivative with respect to the mean is

\[\frac{\partial \ln \mathcal{L}}{\partial \boldsymbol{\mu}} = \boldsymbol{\Sigma}^{-1}\sum_{i=1}^{n}(\mathbf{z}_i - \boldsymbol{\mu}) = \mathbf{0} \quad \Rightarrow \quad \hat{\boldsymbol{\mu}} = \frac{1}{n}\sum_{i=1}^{n}\mathbf{z}_i.\]

The maximum likelihood estimate of the mean is the sample mean. The derivative with respect to the precision matrix \(\boldsymbol{\Sigma}^{-1}\) uses the matrix identities of Section 11.6,

\[\frac{\partial \ln \mathcal{L}}{\partial \boldsymbol{\Sigma}^{-1}} = \frac{n}{2}\boldsymbol{\Sigma} - \frac{1}{2}\sum_{i=1}^{n}(\mathbf{z}_i - \boldsymbol{\mu})(\mathbf{z}_i - \boldsymbol{\mu})^T = \mathbf{0} \quad \Rightarrow \quad \hat{\boldsymbol{\Sigma}} = \frac{1}{n}\sum_{i=1}^{n}(\mathbf{z}_i - \hat{\boldsymbol{\mu}})(\mathbf{z}_i - \hat{\boldsymbol{\mu}})^T.\]

The maximum likelihood estimate of the covariance matrix is the sample covariance matrix. These are the updates of the mixture model in Chapter 11 for one component. We compute the estimates and compare them with the true values.

\tutorialbarrier
\begin{codecell}

\end{codecell}
\begin{Verbatim}[fontsize=\footnotesize, xleftmargin=5mm]
True mean: [2. 3.]
Estimated mean: [2. 3.]
True covariance:
[[2.  1.5]
 [1.5 1.8]]
Estimated covariance:
[[2.02 1.49]
 [1.49 1.78]]
\end{Verbatim}

The estimates are close to the true values, and they approach them as the number of points grows. A GP chooses its hyperparameters in the same way, with the marginal likelihood in place of this likelihood. This likelihood has a closed-form maximum, but the marginal likelihood has none, so we search for its maximum.

\subsection*{Connecting 2D Gaussians to Gaussian Processes}

A GP works in the same way with many variables. It has a joint Gaussian distribution over the function values at all times. The observed function values play the role of \(x\) in the 2D example, and the function values at new times play the role of \(y\). Conditioning on the observed values then gives the distribution of the new function values. Three ideas of the 2D example carry over.

\begin{enumerate}
\def\labelenumi{\arabic{enumi}.}
\tightlist
\item
  \textbf{Conditioning reduces uncertainty.} The conditional variance is smaller than the marginal variance, so observations constrain the prediction.
\item
  \textbf{Correlation enables prediction.} The covariance decides how much an observation at one point tells us about another. For a GP, the kernel sets it.
\item
  \textbf{Maximum likelihood chooses parameters.} The sample mean and covariance maximize the likelihood of the 2D data. For a GP, the marginal likelihood will choose the hyperparameters of the kernel.
\end{enumerate}

Next we draw eight noisy observations of a smooth function and let a GP recover it. Because we know the true function, we can check the result.

\tutorialbarrier
\begin{codecell}

\end{codecell}

\subsection*{Gaussian Process Regression: From Prior to Posterior}

In GP regression we place a prior distribution over functions, \(f \sim \mathcal{GP}(0, k)\). The mean function is zero, and the covariance function is the RBF kernel above, which says that nearby inputs have similar function values. The zero mean describes the function before we see data. It does not say that the targets have mean zero.

As in the 2D example, we need a joint Gaussian distribution. It covers the function values \(\mathbf{f}\) at the \(N\) training inputs and the function value \(f_*\) at a test input,

\[\begin{bmatrix} \mathbf{f} \\ f_* \end{bmatrix} \sim \mathcal{N}\left(\mathbf{0}, \begin{bmatrix} \mathbf{K} & \mathbf{k}_* \\ \mathbf{k}_*^T & k_{**} \end{bmatrix}\right),\]

where the blocks are as follows.

\begin{itemize}
\tightlist
\item
  \(\mathbf{K}\) is the \(N \times N\) kernel matrix at the training inputs.
\item
  \(\mathbf{k}_*\) is the vector of kernel values between the training inputs and the test input.
\item
  \(k_{**}\) is the kernel at the test input, which is the prior variance there.
\end{itemize}

We do not observe \(\mathbf{f}\) itself. We observe noisy values \(\mathbf{t} = \mathbf{f} + \boldsymbol{\epsilon}\) with \(\boldsymbol{\epsilon} \sim \mathcal{N}(\mathbf{0}, \sigma_n^2\mathrm{Id})\). The noise adds \(\sigma_n^2\) to the diagonal of the covariance of the observations, so \(\mathbf{K}\) becomes \(\mathbf{K} + \sigma_n^2\mathrm{Id}\) in the joint distribution of \(\mathbf{t}\) and \(f_*\). Conditioning on \(\mathbf{t}\), as in the 2D case, gives a Gaussian posterior for \(f_*\) with the mean and the variance (Section 14.7)

\[\mu_{f_*} = \mathbf{k}_*^T (\mathbf{K} + \sigma_n^2\mathrm{Id})^{-1} \mathbf{t},\]

\[\sigma_{f_*}^2 = k_{**} - \mathbf{k}_*^T (\mathbf{K} + \sigma_n^2\mathrm{Id})^{-1} \mathbf{k}_*.\]

The mean is a weighted sum of the observations. The weights combine the similarity of the test input to the training inputs with a correction for the correlations among them. The variance starts at the prior variance \(k_{**}\) and decreases with the information in the observations.

A new observation is \(y_* = f_* + \epsilon_*\) with \(\epsilon_* \sim \mathcal{N}(0, \sigma_n^2)\), the same noise level. The sum of two independent Gaussians is Gaussian again, so \(y_*\) has the same mean and the variance \(\sigma_{f_*}^2 + \sigma_n^2\) (Section 14.7). The bands in the plots show this range of a new observation.

Both formulas contain the inverse of \(\mathbf{K} + \sigma_n^2\mathrm{Id}\). A direct inverse costs \(\mathcal{O}(N^3)\) operations and is unstable when the matrix is poorly conditioned, for example when data points lie close together. The code therefore avoids the inverse and uses the Cholesky decomposition (Section 14.8).

The matrix \(\mathbf{K} + \sigma_n^2\mathrm{Id}\) is symmetric and positive definite, so we can write it as \(\mathbf{L}\mathbf{L}^T\), with \(\mathbf{L}\) lower triangular. The code builds this matrix in one step, by adding the noise variance to the diagonal of the kernel matrix. The factorization needs about a sixth of the operations of an inverse, and it is more stable. The cost still grows as \(\mathcal{O}(N^3)\). We compute \(\mathbf{L}\) once, and each new test input then needs only \(\mathbf{k}_*\).

With \(\mathbf{L}\) we replace the inverse by triangular solves, which are cheap and accurate. The triangular solver of SciPy does them in two steps.

\begin{enumerate}
\def\labelenumi{\arabic{enumi}.}
\tightlist
\item
  \textbf{Mean.} The mean needs \(\boldsymbol{\alpha} = (\mathbf{K} + \sigma_n^2\mathrm{Id})^{-1}\mathbf{t}\). Since \(\mathbf{L}\mathbf{L}^T\boldsymbol{\alpha} = \mathbf{t}\), we solve \(\mathbf{L}\mathbf{v} = \mathbf{t}\) for \(\mathbf{v}\) by forward substitution and then \(\mathbf{L}^T\boldsymbol{\alpha} = \mathbf{v}\) for \(\boldsymbol{\alpha}\) by backward substitution. The mean is the dot product \(\mathbf{k}_*^T\boldsymbol{\alpha}\).
\item
  \textbf{Variance.} We solve \(\mathbf{L}\mathbf{q} = \mathbf{k}_*\) by forward substitution. The variance is then \(k_{**} - \mathbf{q}^T\mathbf{q}\).
\end{enumerate}

This works because

\[\mathbf{q}^T\mathbf{q} = \mathbf{k}_*^T(\mathbf{L}\mathbf{L}^T)^{-1}\mathbf{k}_* = \mathbf{k}_*^T(\mathbf{K} + \sigma_n^2\mathrm{Id})^{-1}\mathbf{k}_*.\]

\tutorialbarrier
\begin{codecell}

\end{codecell}

The length scale \(\ell\) of the RBF kernel controls how quickly the correlation between function values decays with distance. A small length scale lets the function vary rapidly, and a large length scale enforces smoother variations. We plot the fit for four length scales.

\tutorialbarrier
\begin{codecell}

\end{codecell}
\tutorialplot{\includegraphics[width=6.180in,height=0.75\textheight,keepaspectratio]{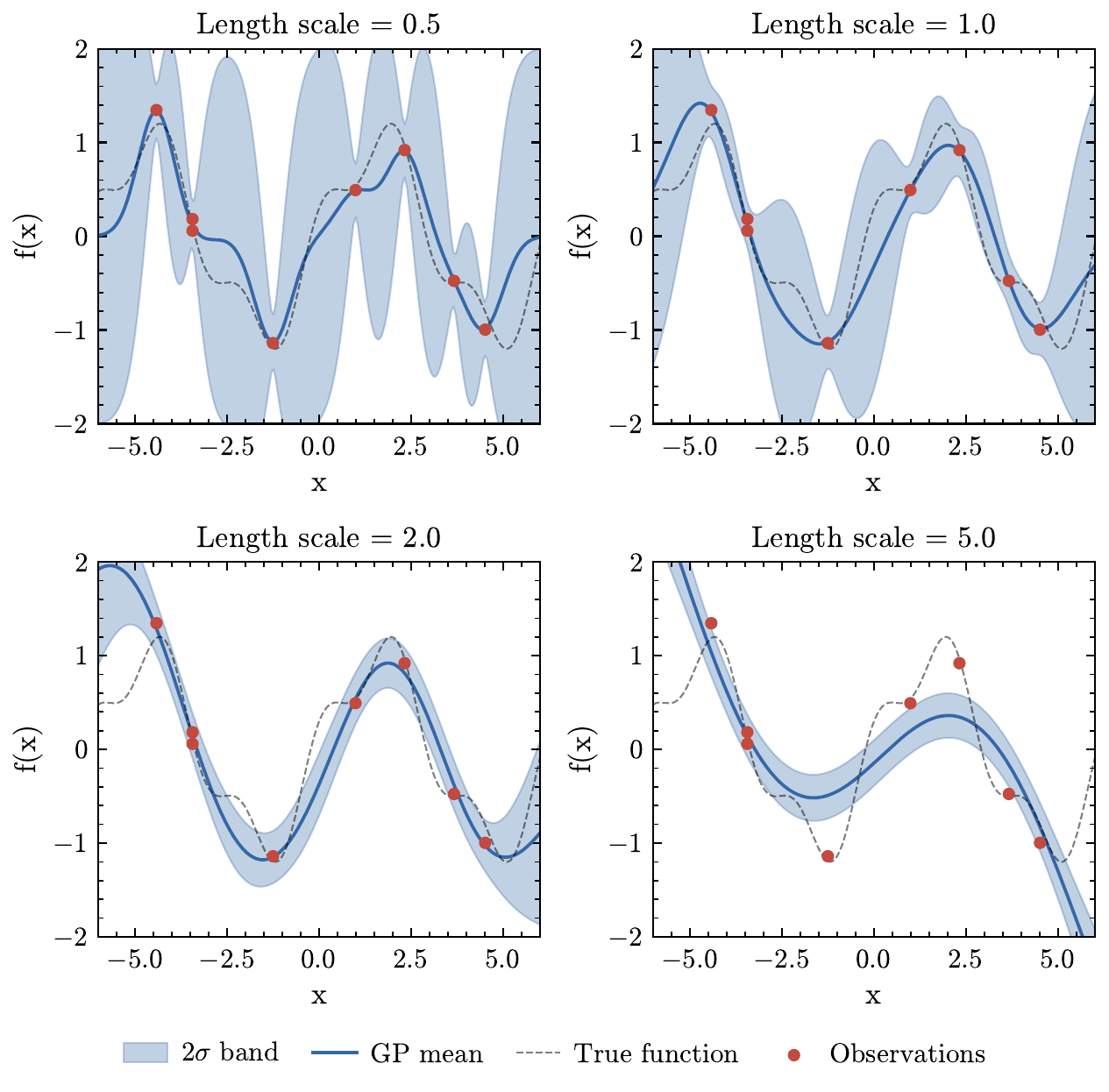}}%
  {\includegraphics[height=\tutorialheight]{tutorials/14a/fig03.pdf}}

The length scale controls how closely the GP follows the observations.

\begin{itemize}
\tightlist
\item
  \textbf{Small length scale (\(\ell = 0.5\)).} The mean wiggles between the observations and follows them closely. The band is narrow near the observations and widens quickly between them.
\item
  \textbf{Moderate length scale (\(\ell = 1\) and \(2\)).} The mean varies more smoothly and follows the trend of the observations.
\item
  \textbf{Large length scale (\(\ell = 5\)).} The mean is very smooth and misses several observations, so the fit underfits the data.
\end{itemize}

The band is narrowest at the observations, at the level of the observation noise, and it grows away from them.

\subsection*{Optimizing the Length Scale via Maximum Likelihood}

The four fits above used length scales that we picked ourselves, and they differ. We need a principled way to choose the length scale. A good length scale makes the data likely. A length scale that allows only smooth functions fits a rapidly varying signal poorly. A length scale that is too flexible makes the GP fit the noise.

The maximum likelihood principle (Section 4.3) gives the choice, as it did for the 2D Gaussian. It takes the length scale and the noise level that make the targets most probable. For any choice of \(\ell\) and \(\sigma_n\), we ask how probable the targets are. This probability is the marginal likelihood, in which we have integrated out the function values (Section 14.9). It is the distribution of \(\mathbf{t}\) in the joint Gaussian of the previous section, with the hyperparameters written out,

\[p(\mathbf{t}|\mathbf{X}, \ell, \sigma_n) = \mathcal{N}(\mathbf{t}|\mathbf{0}, \mathbf{K} + \sigma_n^2\mathrm{Id}),\]

where the kernel matrix \(\mathbf{K}\) depends on \(\ell\). The targets are one draw of this Gaussian vector. The marginal likelihood is therefore a Gaussian likelihood of the kind in Section 4.3, with a covariance that \(\ell\) and \(\sigma_n\) set. Taking the logarithm gives

\[\ln p(\mathbf{t}|\mathbf{X}, \ell, \sigma_n) = -\frac{1}{2}\mathbf{t}^T(\mathbf{K} + \sigma_n^2\mathrm{Id})^{-1}\mathbf{t} - \frac{1}{2}\ln\det(\mathbf{K} + \sigma_n^2\mathrm{Id}) - \frac{N}{2}\ln(2\pi).\]

This objective has two competing terms (Section 14.9).

\begin{enumerate}
\def\labelenumi{\arabic{enumi}.}
\tightlist
\item
  \textbf{Data fit term.} The first term, \(-\mathbf{t}^T(\mathbf{K} + \sigma_n^2\mathrm{Id})^{-1}\mathbf{t}/2\), rewards models that explain the observations well.
\item
  \textbf{Complexity penalty.} The second term, \(-\ln\det(\mathbf{K} + \sigma_n^2\mathrm{Id})/2\), penalizes models that are too flexible. The determinant sets the volume of data space that the model finds probable. A flexible model, such as one with a short length scale, spreads its probability over a larger volume. The observed data then receive a lower density.
\end{enumerate}

The balance of the two terms is Occam's razor. We find the simplest model that explains the data adequately.

The likelihood needs an inverse and a determinant, so we use the Cholesky factor \(\mathbf{L}\) of \(\mathbf{K} + \sigma_n^2\mathrm{Id}\) again.

\begin{enumerate}
\def\labelenumi{\arabic{enumi}.}
\tightlist
\item
  \textbf{Fit term.} Solve for \(\boldsymbol{\alpha}\) with the two triangular solves as before. The first term is then \(-\mathbf{t}^T\boldsymbol{\alpha}/2\).
\item
  \textbf{Log determinant.} It is \(2\sum_i \ln L_{ii}\) (Section 14.9), so the second term is \(-\sum_i \ln L_{ii}\).
\end{enumerate}

Together the terms give the form that the code evaluates,

\[\ln p(\mathbf{t}|\mathbf{X}, \ell, \sigma_n) = -\frac{1}{2}\mathbf{t}^T\boldsymbol{\alpha} - \sum_{i=1}^{N} \ln L_{ii} - \frac{N}{2}\ln(2\pi).\]

Unlike the 2D case, this likelihood has no closed-form maximum, so we search for it. We evaluate the negative log marginal likelihood on a grid of \(\log_{10}\ell\) and \(\log_{10}\sigma_n\) and mark its minimum.

\tutorialbarrier
\begin{codecell}

\end{codecell}

We find the minimum of the grid, mark it, and draw the GP with these parameters.

\tutorialbarrier
\begin{codecell}

\end{codecell}
\tutorialplot{\includegraphics[width=6.380in,height=0.75\textheight,keepaspectratio]{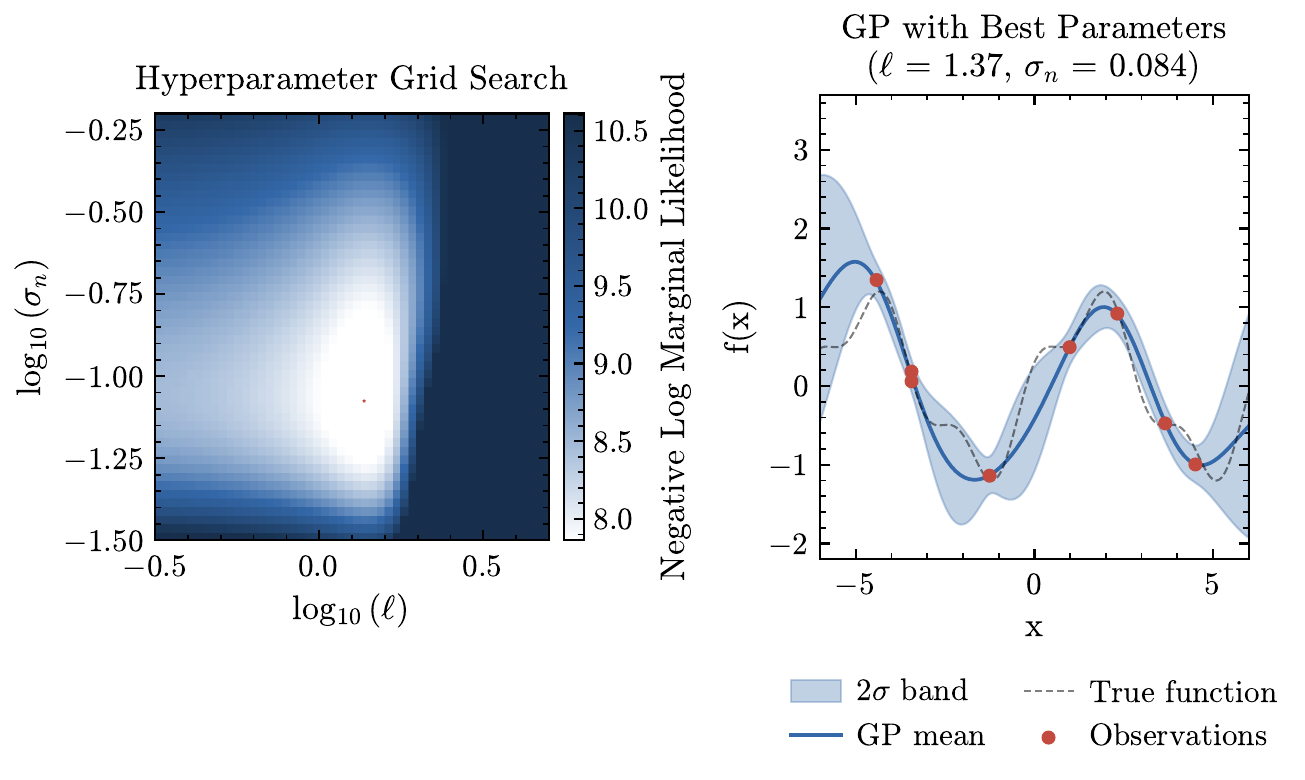}}%
  {\includegraphics[height=\tutorialheight]{tutorials/14a/fig04.pdf}}

The grid has a well-defined minimum, and the best noise level is close to the true value of 0.1 that we used in the simulation. The best length scale balances overfitting for small values and underfitting for large values. It gives a smooth interpolation that follows the main oscillation of the true function but not its small wiggle.

With this intuition we turn to the light curves. We learn the hyperparameters of each star with the same objective, the marginal likelihood. The length scale then gives a timescale that we can compare with the properties of the star. We proceed in three steps.

\begin{enumerate}
\def\labelenumi{\arabic{enumi}.}
\tightlist
\item
  \textbf{One star.} We fit one light curve as an example and validate the fit with values that the GP did not use.
\item
  \textbf{All stars.} We scale the method to all 100 stars.
\item
  \textbf{Timescale and gravity.} We compare the length scales with the surface gravity and see how the sampling limits them.
\end{enumerate}

\section*{Fitting One Light Curve}

A GP on a light curve costs \(\mathcal{O}(N^3)\) operations for \(N\) values (Section 14.8). Our light curves have \(N = 1000\) values, so one Cholesky decomposition is fast, but a grid search needs many of them.

We start with one example, star 50, which has a low surface gravity (\(\log g = 1.38\)). We hold out every tenth of its 1000 values as a test set and fit the GP to the other 900 values.

For the toy data we found the best hyperparameters on a grid, and we use a grid again. The grid has 40 length scales from about 30 to 2500 minutes and 16 noise levels from 0.003 to 0.05, as a choice. Every grid point needs one decomposition. The function below loops over the grid and returns the best pair.

\tutorialbarrier
\begin{codecell}
\begin{Verbatim}[commandchars=\\\{\},fontsize=\small]
\PYG{c+c1}{\PYGZsh{} Grid of the hyperparameters, with the length scale in minutes and the noise level as a standard deviation}
\PYG{n}{ell\PYGZus{}grid} \PYG{o}{=}\FancyVerbBreakAfterBreak{} \PYG{l+m+mi}{10}\PYG{o}{*}\PYG{o}{*}\PYG{n}{np}\PYG{o}{.}\FancyVerbBreakAfterBreak{}\PYG{n}{linspace}\PYG{p}{(}\FancyVerbBreakAfterBreak{}\PYG{l+m+mf}{1.5}\PYG{p}{,}\FancyVerbBreakAfterBreak{} \PYG{l+m+mf}{3.4}\PYG{p}{,}\FancyVerbBreakAfterBreak{} \PYG{l+m+mi}{40}\PYG{p}{)}
\PYG{n}{noise\PYGZus{}grid} \PYG{o}{=}\FancyVerbBreakAfterBreak{} \PYG{l+m+mi}{10}\PYG{o}{*}\PYG{o}{*}\PYG{n}{np}\PYG{o}{.}\FancyVerbBreakAfterBreak{}\PYG{n}{linspace}\PYG{p}{(}\FancyVerbBreakAfterBreak{}\PYG{o}{\PYGZhy{}}\PYG{l+m+mf}{2.5}\PYG{p}{,}\FancyVerbBreakAfterBreak{} \PYG{n}{np}\PYG{o}{.}\FancyVerbBreakAfterBreak{}\PYG{n}{log10}\PYG{p}{(}\FancyVerbBreakAfterBreak{}\PYG{l+m+mf}{0.05}\PYG{p}{)}\PYG{p}{,}\FancyVerbBreakAfterBreak{} \PYG{l+m+mi}{16}\PYG{p}{)}

\PYG{c+c1}{\PYGZsh{} Find the best length scale and noise level of one light curve on the grid}
\PYG{k}{def}\PYG{+w}{ }\PYG{n+nf}{fit\PYGZus{}light\PYGZus{}curve}\PYG{p}{(}\FancyVerbBreakAfterBreak{}\PYG{n}{t}\PYG{p}{,}\FancyVerbBreakAfterBreak{} \PYG{n}{lc}\PYG{p}{)}\PYG{p}{:}
\PYG{+w}{    }\PYG{l+s+sd}{\PYGZdq{}\PYGZdq{}\PYGZdq{}Return the length scale and the noise level with the highest log marginal likelihood on the grid.\PYGZdq{}\PYGZdq{}\PYGZdq{}}
    \PYG{n}{best\PYGZus{}loglik} \PYG{o}{=}\FancyVerbBreakAfterBreak{} \PYG{o}{\PYGZhy{}}\PYG{n}{np}\PYG{o}{.}\FancyVerbBreakAfterBreak{}\PYG{n}{inf}
    \PYG{k}{for} \PYG{n}{ell} \PYG{o+ow}{in} \PYG{n}{ell\PYGZus{}grid}\PYG{p}{:}
        \PYG{k}{for} \PYG{n}{noise} \PYG{o+ow}{in} \PYG{n}{noise\PYGZus{}grid}\PYG{p}{:}
            \PYG{n}{loglik} \PYG{o}{=}\FancyVerbBreakAfterBreak{} \PYG{n}{log\PYGZus{}marginal\PYGZus{}likelihood}\PYG{p}{(}\FancyVerbBreakAfterBreak{}\PYG{n}{t}\PYG{p}{,}\FancyVerbBreakAfterBreak{} \PYG{n}{lc}\PYG{p}{,}\FancyVerbBreakAfterBreak{} \PYG{n}{ell}\PYG{p}{,}\FancyVerbBreakAfterBreak{} \PYG{n}{noise}\PYG{p}{)}
            \PYG{k}{if} \PYG{n}{loglik} \PYG{o}{\PYGZgt{}} \PYG{n}{best\PYGZus{}loglik}\PYG{p}{:}
                \PYG{n}{best\PYGZus{}loglik}\PYG{p}{,}\FancyVerbBreakAfterBreak{} \PYG{n}{best\PYGZus{}ell}\PYG{p}{,}\FancyVerbBreakAfterBreak{} \PYG{n}{best\PYGZus{}noise} \PYG{o}{=}\FancyVerbBreakAfterBreak{} \PYG{n}{loglik}\PYG{p}{,}\FancyVerbBreakAfterBreak{} \PYG{n}{ell}\PYG{p}{,}\FancyVerbBreakAfterBreak{} \PYG{n}{noise}
    \PYG{k}{return} \PYG{n}{best\PYGZus{}ell}\PYG{p}{,}\FancyVerbBreakAfterBreak{} \PYG{n}{best\PYGZus{}noise}
\end{Verbatim}
\end{codecell}

The function takes the times and the brightness values of one light curve. It returns the length scale and the noise level of the grid point with the highest log marginal likelihood. We now hold out every tenth value of star 50, which gives 100 test values spread over the whole stretch. We fit the other values and print the result.

\tutorialbarrier
\begin{codecell}
\begin{Verbatim}[commandchars=\\\{\},fontsize=\small]
\PYG{c+c1}{\PYGZsh{} Hold out every tenth value of star 50 as a test set}
\PYG{n}{selected\PYGZus{}light\PYGZus{}curve} \PYG{o}{=}\FancyVerbBreakAfterBreak{} \PYG{n}{light\PYGZus{}curve}\PYG{p}{[}\FancyVerbBreakAfterBreak{}\PYG{l+m+mi}{50}\PYG{p}{]}
\PYG{n}{held\PYGZus{}out} \PYG{o}{=}\FancyVerbBreakAfterBreak{} \PYG{n}{np}\PYG{o}{.}\FancyVerbBreakAfterBreak{}\PYG{n}{zeros}\PYG{p}{(}\FancyVerbBreakAfterBreak{}\PYG{n+nb}{len}\PYG{p}{(}\FancyVerbBreakAfterBreak{}\PYG{n}{time}\PYG{p}{)}\PYG{p}{,}\FancyVerbBreakAfterBreak{} \PYG{n}{dtype}\PYG{o}{=}\FancyVerbBreakAfterBreak{}\PYG{n+nb}{bool}\PYG{p}{)}

\PYG{c+c1}{\PYGZsh{} Every tenth value is a test value}
\PYG{n}{held\PYGZus{}out}\PYG{p}{[}\FancyVerbBreakAfterBreak{}\PYG{p}{:}\PYG{p}{:}\PYG{l+m+mi}{10}\PYG{p}{]} \PYG{o}{=}\FancyVerbBreakAfterBreak{} \PYG{k+kc}{True}

\PYG{c+c1}{\PYGZsh{} Fit the GP to the other values}
\PYG{n}{best\PYGZus{}ell}\PYG{p}{,}\FancyVerbBreakAfterBreak{} \PYG{n}{best\PYGZus{}noise} \PYG{o}{=}\FancyVerbBreakAfterBreak{} \PYG{n}{fit\PYGZus{}light\PYGZus{}curve}\PYG{p}{(}\FancyVerbBreakAfterBreak{}\PYG{n}{time}\PYG{p}{[}\FancyVerbBreakAfterBreak{}\PYG{o}{\PYGZti{}}\PYG{n}{held\PYGZus{}out}\PYG{p}{]}\PYG{p}{,}\FancyVerbBreakAfterBreak{} \PYG{n}{selected\PYGZus{}light\PYGZus{}curve}\PYG{p}{[}\FancyVerbBreakAfterBreak{}\PYG{o}{\PYGZti{}}\PYG{n}{held\PYGZus{}out}\PYG{p}{]}\PYG{p}{)}
\PYG{n+nb}{print}\PYG{p}{(}\FancyVerbBreakAfterBreak{}\PYG{l+s+sa}{f}\PYG{l+s+s2}{\PYGZdq{}}\PYG{l+s+s2}{Best: length scale }\PYG{l+s+si}{\PYGZob{}}\PYG{n}{best\PYGZus{}ell}\PYG{l+s+si}{:}\PYG{l+s+s2}{.0f}\PYG{l+s+si}{\PYGZcb{}}\FancyVerbBreakAfterBreak{}\PYG{l+s+s2}{ minutes, noise level }\PYG{l+s+si}{\PYGZob{}}\PYG{n}{best\PYGZus{}noise}\PYG{l+s+si}{:}\PYG{l+s+s2}{.4f}\PYG{l+s+si}{\PYGZcb{}}\FancyVerbBreakAfterBreak{}\PYG{l+s+s2}{\PYGZdq{}}\PYG{p}{)}
\end{Verbatim}
\end{codecell}
\begin{Verbatim}[fontsize=\footnotesize, xleftmargin=5mm]
Best: length scale 584 minutes, noise level 0.0066
\end{Verbatim}

The best length scale is about 10 hours, the timescale that the GP assigns to star 50. The fitted noise level is much smaller than the variations of the light curve, which the next plot shows.

\section*{Validating the Gaussian Process Model}

A maximum of the likelihood does not show that the model describes the data. We therefore test the fitted GP of star 50 with the values that it did not use. We fit the GP once, to the 900 values. We then use it to predict the brightness at all 1000 times, which include the 100 test times, and plot the prediction. Each test value lies between fitted values that are 30 minutes apart, so this tests the interpolation of the GP.

\tutorialbarrier
\begin{codecell}

\end{codecell}
\tutorialplot{\includegraphics[width=5.061in,height=0.75\textheight,keepaspectratio]{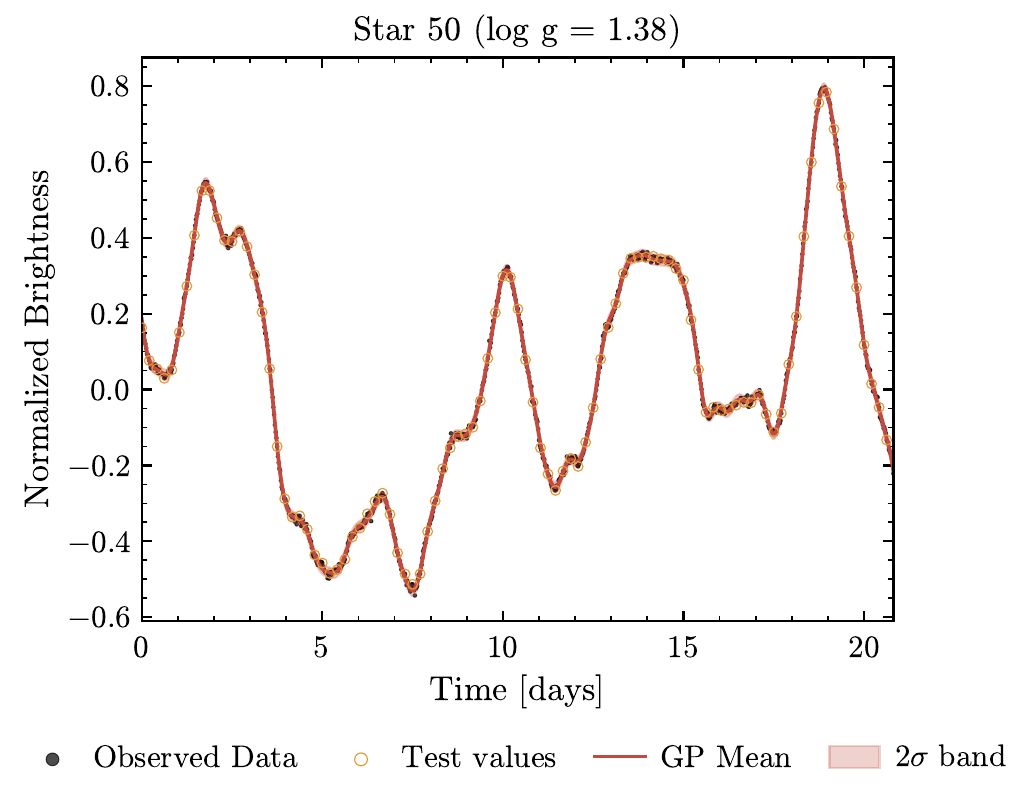}}%
  {\includegraphics[height=\tutorialheight]{tutorials/14a/fig05.pdf}}

The dark points are the observed data, and the orange points are the test values. The red line is the GP mean, and the shaded region is the 2σ band of a new observation. The GP mean follows the dominant oscillation of the light curve, also at the test values. If the band of a new observation has the right width, it holds about 95 percent of them. A much lower share would mean that the band is too narrow, and a much higher share that it is too wide.

\tutorialbarrier
\begin{codecell}
\begin{Verbatim}[commandchars=\\\{\},fontsize=\small]
\PYG{c+c1}{\PYGZsh{} Share of the test values inside the 2σ band}
\PYG{n}{residual} \PYG{o}{=}\FancyVerbBreakAfterBreak{} \PYG{n}{selected\PYGZus{}light\PYGZus{}curve} \PYG{o}{\PYGZhy{}} \PYG{n}{mean}
\PYG{n}{inside} \PYG{o}{=}\FancyVerbBreakAfterBreak{} \PYG{n}{np}\PYG{o}{.}\FancyVerbBreakAfterBreak{}\PYG{n}{abs}\PYG{p}{(}\FancyVerbBreakAfterBreak{}\PYG{n}{residual}\PYG{p}{[}\FancyVerbBreakAfterBreak{}\PYG{n}{held\PYGZus{}out}\PYG{p}{]}\PYG{p}{)} \PYG{o}{\PYGZlt{}} \PYG{l+m+mi}{2} \PYG{o}{*} \PYG{n}{std}\PYG{p}{[}\FancyVerbBreakAfterBreak{}\PYG{n}{held\PYGZus{}out}\PYG{p}{]}
\PYG{n+nb}{print}\PYG{p}{(}\FancyVerbBreakAfterBreak{}\PYG{l+s+sa}{f}\PYG{l+s+s2}{\PYGZdq{}}\PYG{l+s+s2}{Test values inside the 2 sigma band = }\PYG{l+s+si}{\PYGZob{}}\PYG{l+m+mi}{100}\PYG{+w}{ }\PYG{o}{*}\PYG{+w}{ }\PYG{n}{inside}\PYG{o}{.}\FancyVerbBreakAfterBreak{}\PYG{n}{mean}\PYG{p}{(}\FancyVerbBreakAfterBreak{}\PYG{p}{)}\PYG{l+s+si}{:}\PYG{l+s+s2}{.0f}\PYG{l+s+si}{\PYGZcb{}}\FancyVerbBreakAfterBreak{}\PYG{l+s+s2}{\PYGZpc{}}\PYG{l+s+s2}{\PYGZdq{}}\PYG{p}{)}
\end{Verbatim}
\end{codecell}
\begin{Verbatim}[fontsize=\footnotesize, xleftmargin=5mm]
Test values inside the 2 sigma band = 96%
\end{Verbatim}

The band holds about the 95 percent that we expect, so its width is right. The GP therefore predicts the test values within its stated uncertainty.

\section*{Scaling to Multiple Light Curves}

The fit of star 50 gives a useful first estimate of its timescale. We now apply the same fit to all 100 stars. The goal is one timescale for each star. The test above has shown that the fit predicts test values, so we use all values for the fit. This gives one length scale per star.

\tutorialbarrier
\begin{codecell}
\begin{Verbatim}[commandchars=\\\{\},fontsize=\small]
\PYG{c+c1}{\PYGZsh{} Fit every light curve}
\PYG{n}{fits} \PYG{o}{=}\FancyVerbBreakAfterBreak{} \PYG{p}{[}\FancyVerbBreakAfterBreak{}\PYG{n}{fit\PYGZus{}light\PYGZus{}curve}\PYG{p}{(}\FancyVerbBreakAfterBreak{}\PYG{n}{time}\PYG{p}{,}\FancyVerbBreakAfterBreak{} \PYG{n}{lc}\PYG{p}{)} \PYG{k}{for} \PYG{n}{lc} \PYG{o+ow}{in} \PYG{n}{light\PYGZus{}curve}\PYG{p}{]}

\PYG{c+c1}{\PYGZsh{} Length scale and noise level of each light curve}
\PYG{n}{best\PYGZus{}ells}\PYG{p}{,}\FancyVerbBreakAfterBreak{} \PYG{n}{best\PYGZus{}noises} \PYG{o}{=}\FancyVerbBreakAfterBreak{} \PYG{n+nb}{zip}\PYG{p}{(}\FancyVerbBreakAfterBreak{}\PYG{o}{*}\PYG{n}{fits}\PYG{p}{)}
\end{Verbatim}
\end{codecell}

With the best hyperparameters of each star, we ask how the GP timescale relates to the surface gravity. The length scale \(\ell\) is the timescale over which the brightness varies smoothly, so we expect it to follow the oscillation period. We also look at the noise level \(\sigma_n\). The left panel plots the best length scale of each star against its surface gravity, and the right panel plots the best noise level.

\tutorialbarrier
\begin{codecell}

\end{codecell}
\tutorialplot{\includegraphics[width=6.174in,height=0.75\textheight,keepaspectratio]{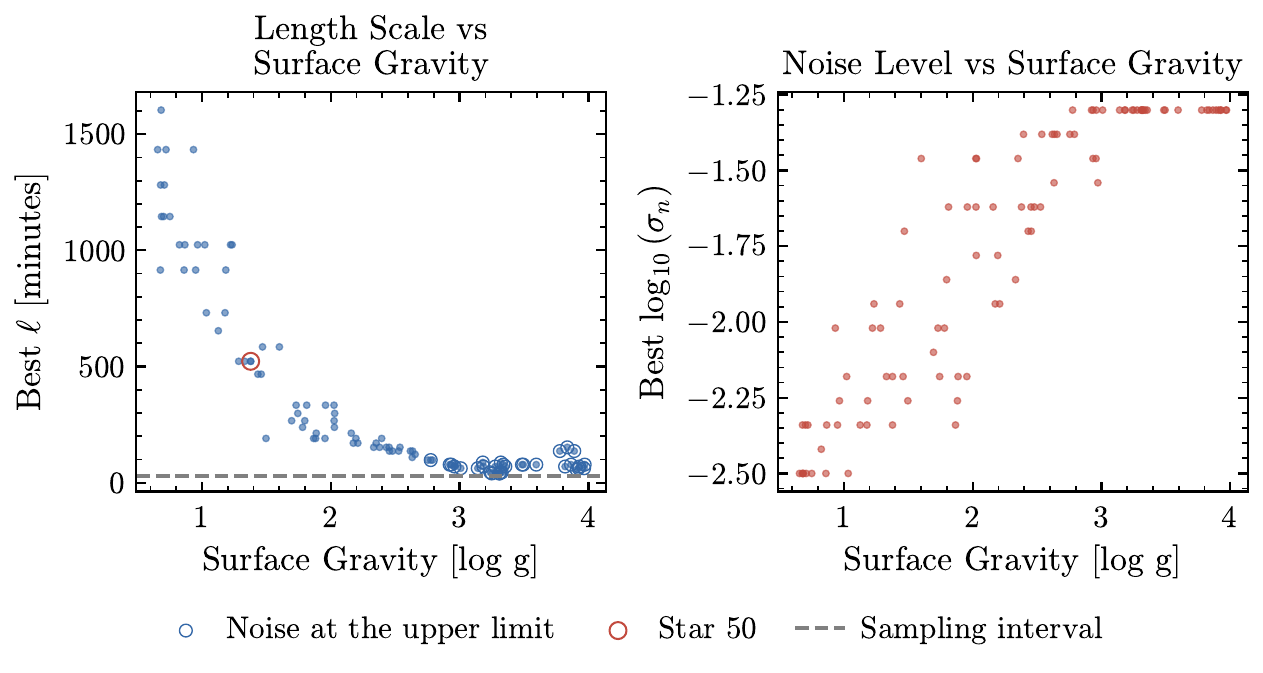}}%
  {\includegraphics[height=\tutorialheight]{tutorials/14a/fig06.pdf}}

\textbf{Length scale and gravity.} The left panel shows an inverse relation between the surface gravity and the length scale. As the stars become more compact, the length scale decreases. This is the direction that the slower variations of evolved stars predict, because sound waves take longer to cross a larger star. The red ring marks star 50, fitted here to all 1000 values. The length scale therefore ranks the stars by their evolutionary state, as the introduction proposed.

\textbf{The sampling limit.} For \(\log g\) above 3 the length scales flatten at about an hour. The data resolve a variation only if they sample it at least twice per cycle. The sampling interval of 30 minutes (dashed line) therefore sets the Nyquist limit of 60 minutes. This is the shortest period that the data resolve. These stars vary on timescales near this limit.

\textbf{The noise level.} The right panel shows that the fitted noise level rises with the surface gravity. For the stars of high surface gravity it piles up at the upper limit of 0.05 that we set for the grid. The open circles in the left panel mark these fits. When the variations are as fast as the sampling, the GP cannot tell them from the observation noise and fits part of them as noise. The flattening of the length scale depends on this limit, so we do not trust the length scales of these fits.

To improve these fits, we can fix \(\sigma_n\) to the noise level that the instrument gives. The search then has one parameter, and it may give a more stable timescale.

The length scales without an open circle are timescales that we can relate to a stellar property. Some stars of the lowest surface gravity end at the lower limit of the noise level, so we read their length scales with care.

\section*{Limitations}

\textbf{One length scale.} The RBF kernel has a single length scale, so it describes one timescale of a light curve. It smooths over variations on shorter timescales, and the fitted noise level can absorb part of them. A kernel with several terms, such as a periodic term and an RBF term (Section 14.3), could describe more than one timescale. For our goal, a characteristic timescale to compare with the properties of a star, the RBF kernel gives a useful first estimate.

\textbf{Sampling.} The sampling interval sets the shortest timescale that the data resolve. A fit near this limit treats variations as noise, as the stars of high surface gravity show.

\textbf{Grid search.} A grid is simple, and it shows the likelihood surface, but its cost grows exponentially with the number of hyperparameters. It also returns only values on the grid. A gradient-based optimizer finds the maximum with far fewer evaluations and scales to more hyperparameters (Section 14.9).

\textbf{Cost.} The cost of GP inference grows as \(\mathcal{O}(N^3)\). We used light curves of 1000 values. The full Kepler light curves have 4365 values, which would make one decomposition about 80 times slower. A large data set needs approximations of the exact computation.

\section*{Summary}

This tutorial used a GP to measure the timescale of the brightness variations of stars. We built intuition with a 2D Gaussian and a 1D function before we turned to real light curves.

Our main lessons follow.

\begin{enumerate}
\def\labelenumi{\arabic{enumi}.}
\tightlist
\item
  \textbf{Learning without features.} A GP states the character of a signal in a kernel, so we construct no features. The marginal likelihood chooses the kernel parameters and the noise level, and it balances the complexity of the model against the fit to the data.
\item
  \textbf{Prediction with uncertainty.} Conditioning a Gaussian on some of its variables gives a Gaussian again. A prediction at a new time therefore comes with a mean and a variance, which is statistically elegant and practically useful.
\item
  \textbf{A hyperparameter with a physical meaning.} The length scale of the kernel is a timescale of the variations. It follows a stellar property, as the physics of the stars predicts.
\item
  \textbf{Limitations.} The cost of GP inference grows as \(\mathcal{O}(N^3)\), so a GP is hard to use on large data sets. The sampling interval limits the shortest timescale that the data resolve, and an RBF kernel with one length scale describes only one timescale.
\item
  \textbf{Optimization.} A grid search is simple and shows the likelihood surface, but its cost grows exponentially with the number of hyperparameters. A gradient-based optimizer scales to more hyperparameters.
\end{enumerate}

In astronomy a GP models light curves and spectra, interpolates irregular or missing data, and emulates expensive simulators. Its statistical elegance comes with a cost in computing time, and this cost decides where we can use it.

\tutorialchapter{14b}{Gaussian Process Classification for Stars and Galaxies}
\noindent\textit{Companion material for Chapter 14}

\section*{Introduction}

Tutorial 14a used Gaussian processes (GPs) to fit the brightness of stars, which is a continuous quantity. Classification asks for a discrete answer, the class of an object. A continuous Gaussian distribution cannot model a class label directly.

Many astronomical problems need a class label. One example is the separation of stars from galaxies. Surveys detect far more objects than spectroscopy can classify, so we must separate them by photometry.

Photometry measures the brightness of an object in a few filters, and the difference between two filters is a color. A galaxy is extended, and a faint star in a crowded field is blended with its neighbors. The wide BP and RP windows of Gaia collect more of this extra light than the G window, so the colors of such objects shift. In this tutorial we classify stars and galaxies from two Gaia colors in this difficult regime, and we compare a GP classifier with logistic regression.

A GP classifier follows the strategy of logistic regression (Chapter 7) with one change. Inside the sigmoid function we place a function drawn from a GP, in place of a linear function. The boundary between the classes can then bend, and the uncertainty grows where objects are few.

We place the GP prior on a latent function, a hidden score that the sigmoid turns into a class probability. Discrete labels break the conjugacy of GP regression, so the posterior of the latent function is not Gaussian. Tutorial 9 met the same problem for the weights of Bayesian logistic regression. We use its two tools here, the Laplace approximation for the posterior and the probit approximation for the class probability. Cross-validation then chooses the length scale of the kernel. Finally we ask which objects need follow-up spectra.

\section*{Prerequisites and Learning Objectives}

\subsection*{Prerequisites}

This tutorial assumes the following.

\begin{itemize}
\tightlist
\item
  Gaussian processes, with kernel functions and GP regression (Chapter 14)
\item
  Logistic regression, with the sigmoid function (Chapter 7)
\item
  Bayesian logistic regression, with the Laplace and probit approximations (Chapter 9)
\item
  Cross-validation for choosing hyperparameters (Chapter 4)
\item
  Linear algebra, with matrix inverses and linear systems
\end{itemize}

\subsection*{Learning Objectives}

After this tutorial you can do the following.

\begin{itemize}
\tightlist
\item
  Explain how a latent function and the sigmoid connect a GP to class labels
\item
  Find the mode of the latent function by gradient ascent and approximate its posterior by a Gaussian
\item
  Compute the predictive probability of an object from the mean and the variance of its latent value
\item
  Compare the curved boundary of a GP classifier with the straight boundary of logistic regression
\item
  Read the class probability and the error bar of the latent score to find the objects that need follow-up observations
\item
  Choose the length scale of the kernel with cross-validation
\end{itemize}

We begin by importing the libraries and setting the plot style.

\tutorialbarrier
\begin{codecell}

\end{codecell}

\section*{The Star-Galaxy Data}

We use 600 objects, 300 stars and 300 galaxies. Each object has two colors from the Gaia photometry, BP-RP and G-RP, and a label that says star or galaxy. The classes follow \href{https://ui.adsabs.harvard.edu/abs/2022A&A...668A..99H/abstract}{Hughes et al.~(2022)}, who classify Gaia sources with labels from the Sloan Digital Sky Survey (SDSS). The galaxies have SDSS spectra that classify them as galaxies without an active nucleus. The stars are random Gaia sources, after we remove the known galaxies and quasars.

We keep faint objects whose colors are shifted by the light of neighbors or of an extended source. Gaia flags them with the BP/RP flux excess factor, the ratio of the summed BP and RP fluxes to the G flux. We keep the objects with a large factor. The shift differs from object to object, so the colors of stars and galaxies overlap. This overlap is the difficult part of the problem.

We keep 20 percent of the stars and 20 percent of the galaxies as test objects, chosen at random. Their share of stars is then the same as in the rest. The test objects take no part in the fit or in the choice of the length scale. We use them only to measure the accuracy, the share of the test objects whose predicted class agrees with the label.

\tutorialbarrier
\begin{codecell}

\end{codecell}
\begin{Verbatim}[fontsize=\footnotesize, xleftmargin=5mm]
Objects = 600 (300 stars, 300 galaxies)
Training objects = 480, test objects = 120
\end{Verbatim}

We plot the objects in the plane of the two colors.

\tutorialbarrier
\begin{codecell}

\end{codecell}
\tutorialplot{\includegraphics[width=4.148in,height=0.75\textheight,keepaspectratio]{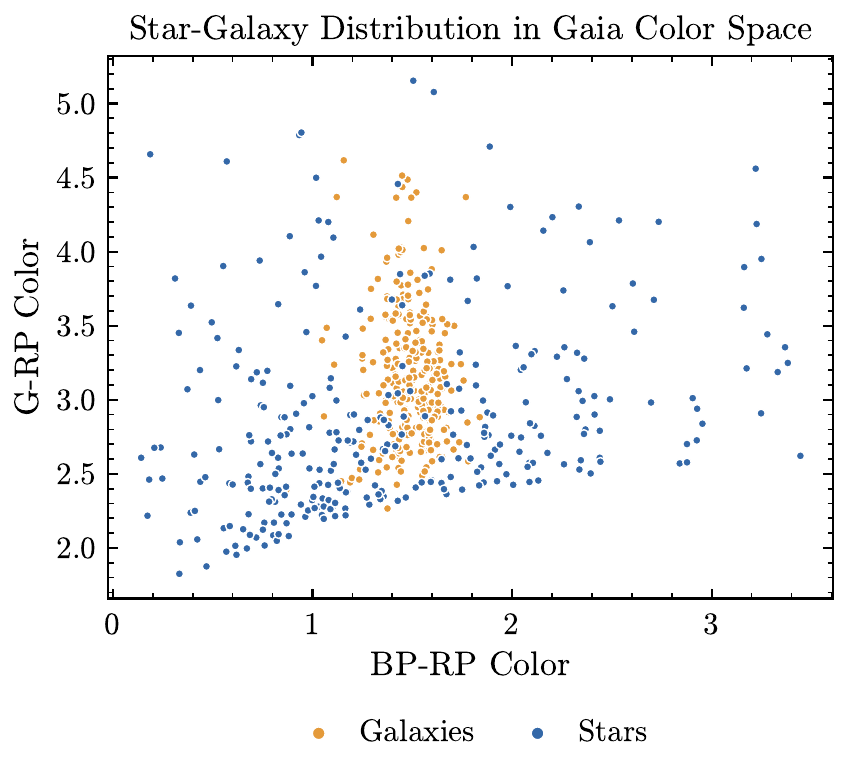}}%
  {\includegraphics[height=\tutorialheight]{tutorials/14b/fig01.pdf}}

The galaxies form one compact group. The stars spread over a wider range of colors, in two groups, one to the lower left and one to the right of the galaxies. Some stars lie among the galaxies, so the two classes overlap there. The galaxies lie between the two groups of stars, so a straight line should separate the classes poorly.

\section*{The Limits of Linear Classification}

We first fit logistic regression as a baseline (Chapter 7). It models the probability of a star as

\[p(t=1|\mathbf{x}, \mathbf{w}) = \sigma(\mathbf{w}^T\mathbf{x} + w_0),\]

where \(\mathbf{x}\) holds the two colors of an object, \(\mathbf{w}\) are the weights, \(w_0\) is the bias term and \(\sigma\) is the sigmoid function. The decision boundary lies where the probability is one half, which is where \(\mathbf{w}^T\mathbf{x} + w_0 = 0\). In the plane of the two colors this is a straight line. We fit the weights by gradient descent on the average negative log-likelihood, as in Tutorial 7, and we measure the accuracy on the test objects.

\tutorialbarrier
\begin{codecell}

\end{codecell}
\begin{Verbatim}[fontsize=\footnotesize, xleftmargin=5mm]
Logistic regression test accuracy = 0.642
\end{Verbatim}

We plot the probability of a star from logistic regression on a grid of colors. We add the decision boundary, the training objects and the misclassified test objects.

\tutorialbarrier
\begin{codecell}

\end{codecell}
\tutorialplot{\includegraphics[width=6.184in,height=0.75\textheight,keepaspectratio]{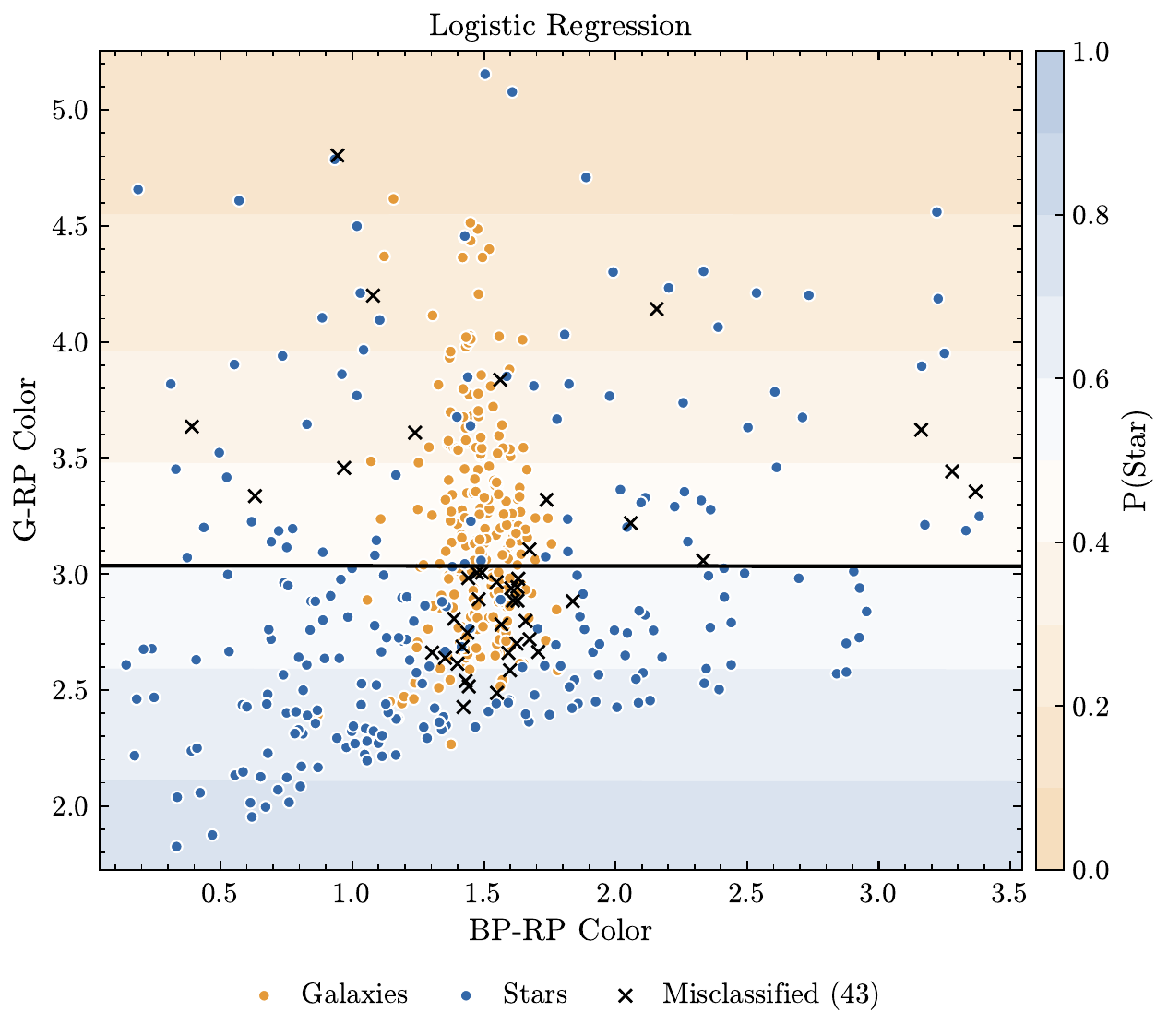}}%
  {\includegraphics[height=\tutorialheight]{tutorials/14b/fig02.pdf}}

Logistic regression classifies about 64 percent of the test objects correctly. Its decision boundary is a nearly horizontal line. The galaxies lie in a narrow band of BP-RP, with stars on both sides of the band. No straight line separates them from the stars on the left and on the right. The line can only use G-RP, and the stars and the galaxies overlap in G-RP. The misclassified objects lie among the galaxies below the line and among the stars on both sides above it.

The GP classifier replaces the linear function \(\mathbf{w}^T\mathbf{x} + w_0\) with a function drawn from a GP, so that the boundary can bend around the galaxies.

\section*{The Latent Function Approach}

Logistic regression passes a linear score through the sigmoid. We keep the sigmoid and replace the linear score by a latent function (Section 14.11). The model then passes information from the colors \(\mathbf{x}\) of an object to its label \(t\) in three steps.

\begin{enumerate}
\def\labelenumi{\arabic{enumi}.}
\tightlist
\item
  A latent function value \(a(\mathbf{x})\), which can be any real number, is the score of the object for the class star.
\item
  The sigmoid turns the score into a class probability \(p(t=1|\mathbf{x}) = \sigma(a(\mathbf{x}))\), which lies between 0 and 1.
\item
  The label is 1 with this probability and 0 otherwise.
\end{enumerate}

The sigmoid is the link function between the latent value and the class probability. We place a GP prior on the latent function,

\[a(\mathbf{x}) \sim \mathcal{GP}(0, k(\mathbf{x}, \mathbf{x}')),\]

where \(k\) is the kernel, defined next. The decision boundary is the set where \(a(\mathbf{x}) = 0\), because the probability is then one half. The kernel sets how flexible this boundary is, and the GP also gives an error bar for each prediction.

\section*{Implementing the Core Components}

We build the components step by step, and we start with the kernel. Tutorial 14a used the RBF kernel (Section 14.3) for one input, the time. Here the input has two components, and the kernel depends on the distance between two objects in the plane of the colors,

\[k(\mathbf{x}, \mathbf{x}') = \sigma_f^2 \exp\left(-\frac{\|\mathbf{x} - \mathbf{x}'\|^2}{2\ell^2}\right),\]

where \(\sigma_f^2\) is the amplitude (amplitude in the code) and \(\ell\) is the length scale. The amplitude sets the prior size of the latent score. The kernel assumes that objects with similar colors have similar latent values. A smaller length scale allows more complex boundaries. We implement the kernel and the sigmoid function below.

\tutorialbarrier
\begin{codecell}

\end{codecell}

\section*{The Laplace Approximation}

With the kernel we can fit the model to the training objects. The first task is to infer their latent values. In GP regression we observe noisy function values directly. Here we observe only class labels, and we need the latent values that best explain them.

The posterior distribution of the latent values \(\mathbf{a}\) is proportional to the GP prior times the likelihood of the observed labels,

\[p(\mathbf{a}|\mathbf{X}, \mathbf{t}) \propto p(\mathbf{t}|\mathbf{a}) p(\mathbf{a}|\mathbf{X}),\]

where

\begin{itemize}
\tightlist
\item
  \(p(\mathbf{a}|\mathbf{X}) = \mathcal{N}(\mathbf{0}, \mathbf{K})\) is the GP prior, with the kernel matrix \(\mathbf{K}\) of the training objects \(\mathbf{X}\),
\item
  \(p(\mathbf{t}|\mathbf{a}) = \prod_{i=1}^N \sigma(a_i)^{t_i}(1-\sigma(a_i))^{1-t_i}\) is the likelihood of the labels \(\mathbf{t}\) of the \(N\) training objects.
\end{itemize}

The sigmoid function is nonlinear, so this posterior is not Gaussian. As in Tutorial 9, the Laplace approximation (Section 14.12) finds the mode of the posterior, the MAP estimate. It then approximates the posterior by a Gaussian around the mode. We find the mode in this section and build the Gaussian in the section on the predictive distribution. To find the mode, we maximize the log posterior,

\[\mathbf{a}_{\text{MAP}} = \arg\max_{\mathbf{a}} \left[ \sum_{i=1}^N \left(t_i \ln \sigma(a_i) + (1-t_i)\ln(1-\sigma(a_i))\right) - \frac{1}{2}\mathbf{a}^T\mathbf{K}^{-1}\mathbf{a} \right].\]

\section*{Finding the Mode by Gradient Ascent}

To find the maximum we need the gradient of the log posterior. The log-likelihood term of object \(i\) depends on \(a_i\) only, and the derivative of the sigmoid is \(\sigma'(a) = \sigma(a)(1-\sigma(a))\) (Section 7.7). The derivative of this term is

\[\frac{\partial}{\partial a_i}\left[t_i \ln\sigma(a_i) + (1-t_i)\ln(1-\sigma(a_i))\right] = t_i(1-\sigma(a_i)) - (1-t_i)\sigma(a_i) = t_i - \sigma(a_i).\]

The gradient of the Gaussian prior term \(-\mathbf{a}^T\mathbf{K}^{-1}\mathbf{a}/2\) is \(-\mathbf{K}^{-1}\mathbf{a}\). Setting the full gradient to zero gives the condition (Section 14.12)

\[\mathbf{t} - \boldsymbol{\sigma}(\mathbf{a}) - \mathbf{K}^{-1}\mathbf{a} = \mathbf{0}.\]

The sigmoid is nonlinear, so this equation has no closed-form solution. We solve it iteratively by gradient ascent with the learning rate \(\eta\), as in Tutorial 7 and Section 14.12,

\[\mathbf{a}^{(\tau+1)} = \mathbf{a}^{(\tau)} + \eta\left(\mathbf{t} - \boldsymbol{\sigma}(\mathbf{a}^{(\tau)}) - \mathbf{K}^{-1}\mathbf{a}^{(\tau)}\right).\]

At each step we compute the residuals \(\mathbf{t} - \boldsymbol{\sigma}(\mathbf{a})\) between the labels and the current probabilities. The term \(-\mathbf{K}^{-1}\mathbf{a}\) is the gradient of the log prior. It pulls the latent values toward zero, and it pulls hardest on values that are not smooth. The log posterior is strictly concave (Section 14.12), so it has one mode, and the iteration reaches it for a small enough \(\eta\).

The kernel matrix of an RBF kernel is nearly singular, because objects with similar colors have nearly equal rows. We add a small term to its diagonal before the inversion, so that \(\mathbf{K}\) is positive definite and \(\mathbf{K}^{-1}\) exists. The function below implements the gradient ascent.

\tutorialbarrier
\begin{codecell}
\begin{Verbatim}[commandchars=\\\{\},fontsize=\small]
\PYG{c+c1}{\PYGZsh{} Small term on the diagonal of the kernel matrix, so that the matrix is positive definite}
\PYG{n}{jitter} \PYG{o}{=}\FancyVerbBreakAfterBreak{} \PYG{l+m+mf}{0.01}

\PYG{k}{def}\PYG{+w}{ }\PYG{n+nf}{find\PYGZus{}map\PYGZus{}estimate}\PYG{p}{(}\FancyVerbBreakAfterBreak{}\PYG{n}{K}\PYG{p}{,}\FancyVerbBreakAfterBreak{} \PYG{n}{t}\PYG{p}{,}\FancyVerbBreakAfterBreak{} \PYG{n}{learning\PYGZus{}rate}\PYG{o}{=}\FancyVerbBreakAfterBreak{}\PYG{l+m+mf}{0.01}\PYG{p}{,}\FancyVerbBreakAfterBreak{} \PYG{n}{max\PYGZus{}iter}\PYG{o}{=}\FancyVerbBreakAfterBreak{}\PYG{l+m+mi}{100000}\PYG{p}{,}\FancyVerbBreakAfterBreak{} \PYG{n}{tol}\PYG{o}{=}\FancyVerbBreakAfterBreak{}\PYG{l+m+mf}{1e\PYGZhy{}3}\PYG{p}{)}\PYG{p}{:}
\PYG{+w}{    }\PYG{l+s+sd}{\PYGZdq{}\PYGZdq{}\PYGZdq{}}
\PYG{l+s+sd}{    Find the MAP latent values by gradient ascent on the log posterior.}

\PYG{l+s+sd}{    The function returns the values and the number of iterations, and it raises an error without convergence.}
\PYG{l+s+sd}{    \PYGZdq{}\PYGZdq{}\PYGZdq{}}
    \PYG{n}{n} \PYG{o}{=}\FancyVerbBreakAfterBreak{} \PYG{n+nb}{len}\PYG{p}{(}\FancyVerbBreakAfterBreak{}\PYG{n}{t}\PYG{p}{)}
    \PYG{n}{a} \PYG{o}{=}\FancyVerbBreakAfterBreak{} \PYG{n}{np}\PYG{o}{.}\FancyVerbBreakAfterBreak{}\PYG{n}{zeros}\PYG{p}{(}\FancyVerbBreakAfterBreak{}\PYG{n}{n}\PYG{p}{)}  \PYG{c+c1}{\PYGZsh{} Initialize at zero}

    \PYG{c+c1}{\PYGZsh{} The inverse of the kernel matrix with the small diagonal term}
    \PYG{n}{K\PYGZus{}inv} \PYG{o}{=}\FancyVerbBreakAfterBreak{} \PYG{n}{np}\PYG{o}{.}\FancyVerbBreakAfterBreak{}\PYG{n}{linalg}\PYG{o}{.}\FancyVerbBreakAfterBreak{}\PYG{n}{inv}\PYG{p}{(}\FancyVerbBreakAfterBreak{}\PYG{n}{K} \PYG{o}{+}\FancyVerbBreakAfterBreak{} \PYG{n}{jitter} \PYG{o}{*} \PYG{n}{np}\PYG{o}{.}\FancyVerbBreakAfterBreak{}\PYG{n}{eye}\PYG{p}{(}\FancyVerbBreakAfterBreak{}\PYG{n}{n}\PYG{p}{)}\PYG{p}{)}

    \PYG{k}{for} \PYG{n}{iteration} \PYG{o+ow}{in} \PYG{n+nb}{range}\PYG{p}{(}\FancyVerbBreakAfterBreak{}\PYG{n}{max\PYGZus{}iter}\PYG{p}{)}\PYG{p}{:}
        \PYG{c+c1}{\PYGZsh{} Gradient of the log posterior}
        \PYG{n}{gradient} \PYG{o}{=}\FancyVerbBreakAfterBreak{} \PYG{n}{t} \PYG{o}{\PYGZhy{}} \PYG{n}{sigmoid}\PYG{p}{(}\FancyVerbBreakAfterBreak{}\PYG{n}{a}\PYG{p}{)} \PYG{o}{\PYGZhy{}} \PYG{n}{K\PYGZus{}inv} \PYG{o}{@} \PYG{n}{a}

        \PYG{c+c1}{\PYGZsh{} Stop when the gradient is small}
        \PYG{k}{if} \PYG{n}{np}\PYG{o}{.}\FancyVerbBreakAfterBreak{}\PYG{n}{linalg}\PYG{o}{.}\FancyVerbBreakAfterBreak{}\PYG{n}{norm}\PYG{p}{(}\FancyVerbBreakAfterBreak{}\PYG{n}{gradient}\PYG{p}{)} \PYG{o}{\PYGZlt{}} \PYG{n}{tol}\PYG{p}{:}
            \PYG{k}{return} \PYG{n}{a}\PYG{p}{,}\FancyVerbBreakAfterBreak{} \PYG{n}{iteration}

        \PYG{c+c1}{\PYGZsh{} Gradient ascent step, and a stop when the values run away}
        \PYG{n}{a} \PYG{o}{=}\FancyVerbBreakAfterBreak{} \PYG{n}{a} \PYG{o}{+}\FancyVerbBreakAfterBreak{} \PYG{n}{learning\PYGZus{}rate} \PYG{o}{*} \PYG{n}{gradient}
        \PYG{k}{if} \PYG{n}{np}\PYG{o}{.}\FancyVerbBreakAfterBreak{}\PYG{n}{abs}\PYG{p}{(}\FancyVerbBreakAfterBreak{}\PYG{n}{a}\PYG{p}{)}\PYG{o}{.}\FancyVerbBreakAfterBreak{}\PYG{n}{max}\PYG{p}{(}\FancyVerbBreakAfterBreak{}\PYG{p}{)} \PYG{o}{\PYGZgt{}} \PYG{l+m+mi}{100}\PYG{p}{:}
            \PYG{k}{break}

    \PYG{k}{raise} \PYG{n+ne}{RuntimeError}\PYG{p}{(}\FancyVerbBreakAfterBreak{}\PYG{l+s+s2}{\PYGZdq{}}\PYG{l+s+s2}{The gradient ascent did not converge}\PYG{l+s+s2}{\PYGZdq{}}\PYG{p}{)}
\end{Verbatim}
\end{codecell}

We now standardize the colors with the mean and the standard deviation of the training objects. One length scale then acts on both colors alike, and it is in units of the standard deviations of the colors. We start with the length scale \(\ell = 0.3\) and the amplitude 1. The iteration stops when the gradient is small. We find the MAP estimate for the training objects.

\tutorialbarrier
\begin{codecell}

\end{codecell}
\begin{Verbatim}[fontsize=\footnotesize, xleftmargin=5mm]
The gradient ascent converged after 8029 iterations
\end{Verbatim}

\section*{Computing the Predictive Distribution}

With the MAP estimate we can make predictions for new objects. We want the probability that a new object with the colors \(\mathbf{x}_*\) is a star. The unknown latent value \(a_*\) of the new object connects the training objects to this probability, so we average over it (Section 14.12),

\[p(t_*=1|\mathbf{x}_*, \mathbf{X}, \mathbf{t}) = \int \sigma(a_*) p(a_*|\mathbf{x}_*, \mathbf{X}, \mathbf{t})\mathrm{d}a_*.\]

The distribution of \(a_*\) depends on the latent values \(\mathbf{a}\) of the training objects, and we do not know them exactly. We average the GP prediction of \(a_*\) over their posterior,

\[p(a_*|\mathbf{x}_*, \mathbf{X}, \mathbf{t}) = \int p(a_*|\mathbf{a}, \mathbf{x}_*, \mathbf{X}) p(\mathbf{a}|\mathbf{X}, \mathbf{t})\mathrm{d}\mathbf{a}.\]

If we knew \(\mathbf{a}\), the GP would predict \(a_*\) as in Tutorial 14a, without noise,

\[p(a_*|\mathbf{a}, \mathbf{x}_*, \mathbf{X}) = \mathcal{N}\left(a_*|\mathbf{k}_*^T\mathbf{K}^{-1}\mathbf{a}, k_{**} - \mathbf{k}_*^T\mathbf{K}^{-1}\mathbf{k}_*\right).\]

Here \(\mathbf{k}_* = k(\mathbf{x}_*, \mathbf{X})\) is the vector of kernel values between the new object and the training objects. The value \(k_{**} = k(\mathbf{x}_*, \mathbf{x}_*)\) is the prior variance at the new object.

The posterior of the latent values is not Gaussian. The Laplace approximation (Section 14.12) replaces it by a Gaussian around the mode,

\[p(\mathbf{a}|\mathbf{X}, \mathbf{t}) \approx \mathcal{N}(\mathbf{a}|\mathbf{a}_{\text{MAP}}, \mathbf{H}^{-1}),\]

where \(\mathbf{H}\) is the Hessian matrix of the negative log posterior at the mode. The negative log-likelihood term of object \(i\) has the second derivative \(\sigma'(a_i) = \sigma(a_i)(1-\sigma(a_i))\) with respect to \(a_i\). It does not depend on the other latent values, so these terms form a diagonal matrix. The prior term adds \(\mathbf{K}^{-1}\). The Hessian matrix is therefore

\[\mathbf{H} = \mathbf{W} + \mathbf{K}^{-1},\]

with \(\mathbf{W} = \text{diag}(\sigma(a_1)(1-\sigma(a_1)), \ldots, \sigma(a_N)(1-\sigma(a_N)))\). The matrix \(\mathbf{W}\) contains the second derivatives of the negative log-likelihood at \(\mathbf{a}_{\text{MAP}}\).

With this Gaussian approximation we can carry out the average over \(\mathbf{a}\). Both factors are Gaussian, and the mean of the first one is linear in \(\mathbf{a}\), so the average is again a Gaussian (Section 14.12),

\[p(a_*|\mathbf{x}_*, \mathbf{X}, \mathbf{t}) \approx \mathcal{N}(a_*|\mu_*, \sigma_*^2),\]

with

\tutorialdisplay{\mu_* = \mathbf{k}_*^T\mathbf{K}^{-1}\mathbf{a}_{\text{MAP}},}{\sigma_*^2 = k_{**} - \mathbf{k}_*^T\mathbf{K}^{-1}\mathbf{k}_* + \mathbf{k}_*^T\mathbf{K}^{-1}\mathbf{H}^{-1}\mathbf{K}^{-1}\mathbf{k}_*.}

The variance has two parts. The first part is the variance of the GP prediction for known latent values. The last term comes from our uncertainty about the latent values. At the mode the gradient condition gives \(\mathbf{K}^{-1}\mathbf{a}_{\text{MAP}} = \mathbf{t} - \boldsymbol{\sigma}(\mathbf{a}_{\text{MAP}})\), which simplifies the mean. The Woodbury identity (Section 14.4), with \(\mathbf{A} = \mathbf{K}\), \(\mathbf{U} = \mathbf{V} = \mathrm{Id}\) and \(\mathbf{C} = \mathbf{W}^{-1}\), simplifies the variance. This gives

\tutorialdisplay{\mu_* = \mathbf{k}_*^T (\mathbf{t} - \boldsymbol{\sigma}(\mathbf{a}_{\text{MAP}})),}{\sigma_*^2 = k_{**} - \mathbf{k}_*^T (\mathbf{W}^{-1} + \mathbf{K})^{-1} \mathbf{k}_*.}

In words, \(\mu_*\) is our best estimate of the latent score of the new object, and \(\sigma_*\) is the error bar of this estimate. The mean is a weighted sum of the residuals at the training objects, with the kernel values as weights. The variance measures how well the training objects fix the latent value, and it grows with the distance from them.

The simplified variance has the form of the GP regression variance of Tutorial 14a, which contains \((\mathbf{K} + \sigma_n^2\mathrm{Id})^{-1}\) (Section 14.7). Here \(\mathbf{W}^{-1}\) takes the place of the noise covariance \(\sigma_n^2\mathrm{Id}\), so each label acts like a noisy measurement of its latent value. The element \(W_{ii} = \sigma(a_i)(1-\sigma(a_i))\) is largest when the probability of object \(i\) is one half, and then the measurement is the least noisy. A label near the decision boundary therefore tells us the most about the latent function, and it lowers the variance the most.

We now have the distribution of the latent score \(a_*\). The class probability averages the sigmoid over it, as in the first equation of this section. This average has no closed form for the sigmoid, so we use the probit approximation, as in Tutorial 9 (Section 9.6). It replaces the sigmoid by \(\Phi(\lambda a)\), where \(\Phi\) is the standard normal CDF and \(\lambda = \sqrt{\pi/8}\). The average of this function over a Gaussian has a closed form, \(\Phi(\lambda\mu_*/\sqrt{1 + \lambda^2\sigma_*^2})\). Writing \(\Phi(\lambda x)\) as \(\sigma(x)\) again gives

\[p(t_*=1|\mathbf{x}_*, \mathbf{X}, \mathbf{t}) \approx \sigma\left(\frac{\mu_*}{\sqrt{1 + \pi \sigma_*^2/8}}\right).\]

The mean \(\mu_*\) and the variance \(\sigma_*^2\) take the place of \(\mu_a\) and \(\sigma_a^2\) of Tutorial 9. We write \(\kappa = 1/\sqrt{1 + \pi \sigma_*^2/8}\) for the factor that scales \(\mu_*\). It is smaller than 1, and it shrinks as \(\sigma_*^2\) grows. A large error bar means that we are less sure of the score, so the probability moves toward one half.

We now implement these predictions in a function. It takes the training objects and their MAP latent values, and it returns \(\mu_*\), \(\sigma_*^2\) and the probability of a star for new objects.

\tutorialbarrier
\begin{codecell}

\end{codecell}

\section*{Visualizing GP Classification Results}

We now see how the classifier separates stars and galaxies in the plane of the colors. We compute the predictive probability of a star on a grid of colors. We also compute the predictions and the accuracy for the test objects.

\tutorialbarrier
\begin{codecell}

\end{codecell}
\begin{Verbatim}[fontsize=\footnotesize, xleftmargin=5mm]
GP classifier test accuracy = 0.883
\end{Verbatim}

We plot the probability of a star for the GP classifier, with the training objects and the misclassified test objects. The black line is the decision boundary, where the probability is one half.

\tutorialbarrier
\begin{codecell}

\end{codecell}
\tutorialplot{\includegraphics[width=6.184in,height=0.75\textheight,keepaspectratio]{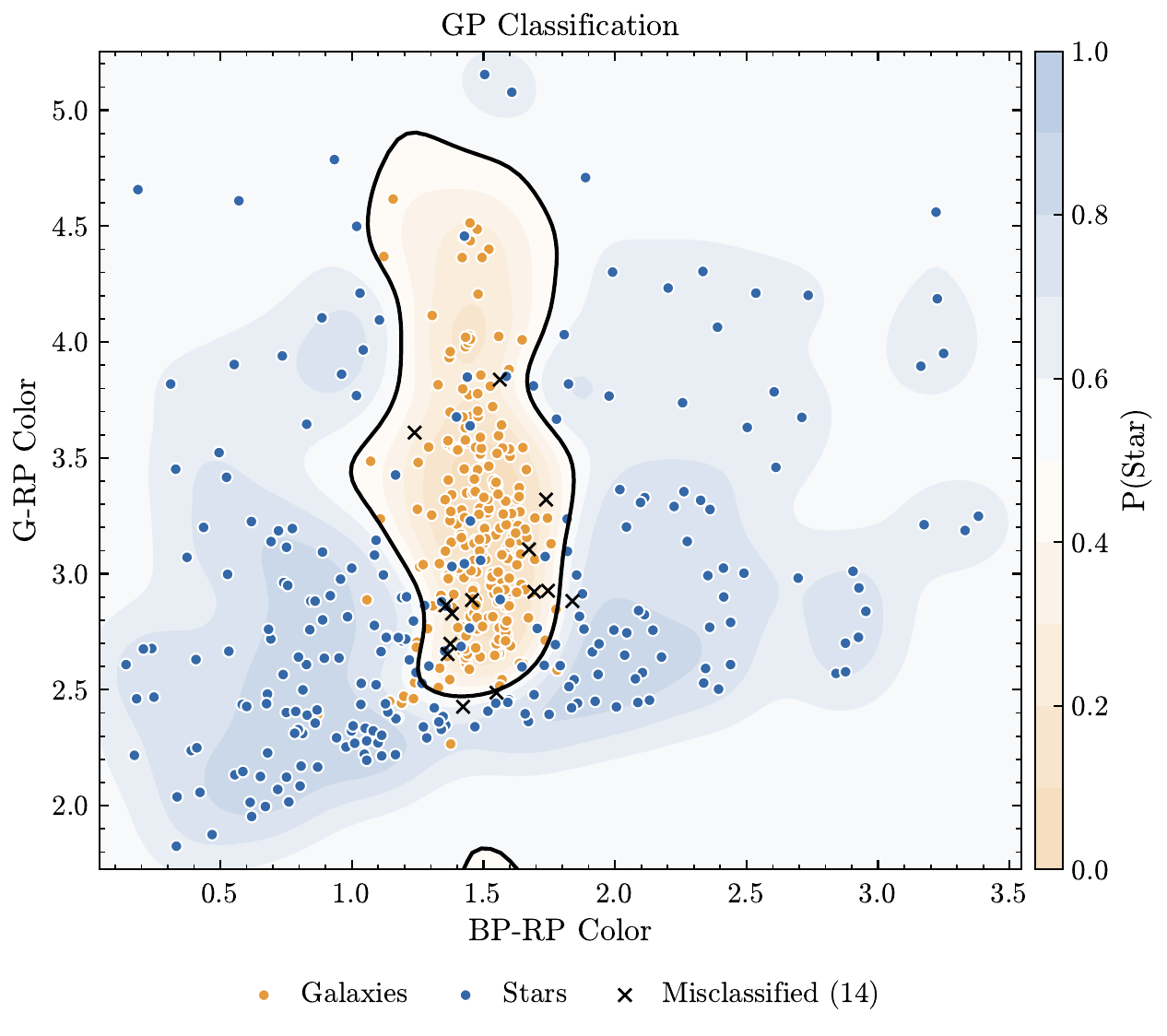}}%
  {\includegraphics[height=\tutorialheight]{tutorials/14b/fig03.pdf}}

The GP classifier misclassifies 14 of the 120 test objects. We notice three features of its probability map.

\begin{enumerate}
\def\labelenumi{\arabic{enumi}.}
\tightlist
\item
  \textbf{Curved boundary.} The boundary closes around the galaxies in a tall, narrow region. The stars lie below it and on both sides of it.
\item
  \textbf{Smooth transitions.} The probability changes smoothly from low values among the galaxies and high values among the stars to one half at the boundary.
\item
  \textbf{Far from the objects.} In the upper corners, far from the objects, the probability is about one half.
\end{enumerate}

To see why the probability returns to one half far from the objects, we plot the error bar \(\sigma_*\) of the latent score. It is the standard deviation of the latent score of a new object, and it measures how well the training objects fix that score. The right panel shows the same uncertainty as an error bar of the probability, the half-width of the range from \(\sigma(\mu_* - \sigma_*)\) to \(\sigma(\mu_* + \sigma_*)\).

\tutorialbarrier
\begin{codecell}

\end{codecell}
\tutorialplot{\includegraphics[width=6.175in,height=0.75\textheight,keepaspectratio]{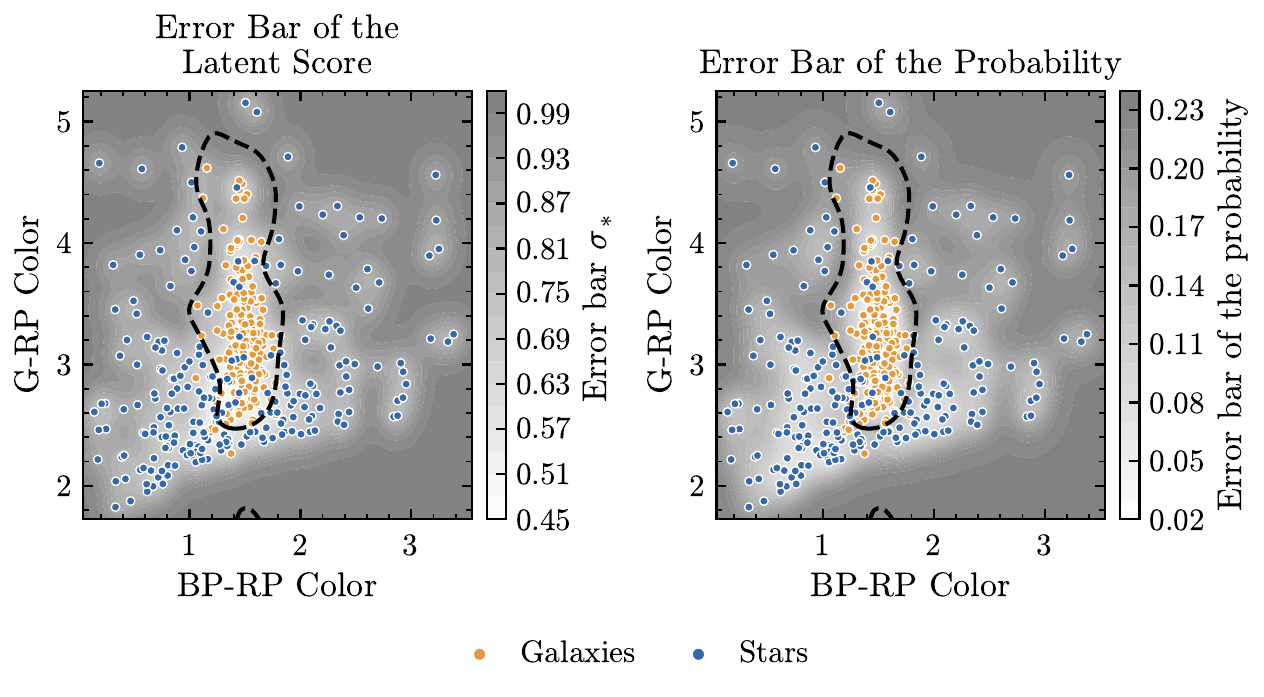}}%
  {\includegraphics[height=\tutorialheight]{tutorials/14b/fig04.pdf}}

Both error bars depend on how many training objects lie nearby.

\begin{enumerate}
\def\labelenumi{\arabic{enumi}.}
\tightlist
\item
  \textbf{Many training objects (light).} The error bar of the score is smallest here, about half the prior value. Where the class is clear, the score is far from zero and the sigmoid is flat, so the error bar of the probability is small. Near the decision boundary the score is close to zero. The sigmoid is steep there, so the same error bar of the score gives a larger error bar of the probability.
\item
  \textbf{Few training objects (grey).} The training objects thin out at the outskirts of the groups, and both error bars grow.
\item
  \textbf{No training object close (dark).} The error bar of the score approaches the prior value 1. This is the error bar that the GP prior gives before it sees any data.
\end{enumerate}

The error bar measures how much data supports the prediction. It does not measure how close the probability is to one half. Far from the objects the probability is one half with a large error bar. There \(\mathbf{k}_*\) is close to zero, so the score \(\mu_*\) is close to zero. The large error bar adds only a small pull through \(\kappa\).

\section*{The Effect of Kernel Length Scale}

So far we used the length scale \(\ell = 0.3\). The length scale controls how much the decision boundary can bend. We now fit the classifier with three length scales, and we compute its accuracy on the training objects and on the test objects. The test accuracies only illustrate the effect. They play no part in the choice of the length scale.

\tutorialbarrier
\begin{codecell}

\end{codecell}

We plot the probability of a star for each length scale, with all objects. The galaxies are orange and the stars are blue, as before.

\tutorialbarrier
\begin{codecell}

\end{codecell}
\tutorialplot{\includegraphics[width=6.305in,height=0.75\textheight,keepaspectratio]{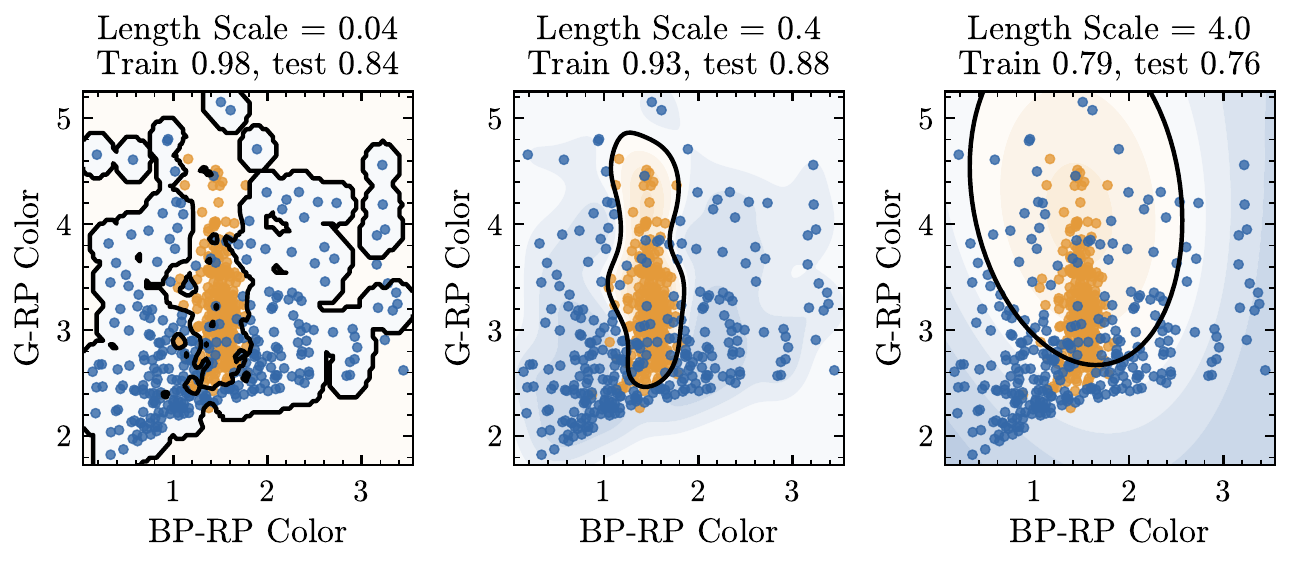}}%
  {\includegraphics[height=\tutorialheight]{tutorials/14b/fig05.pdf}}

The three length scales show how the boundary changes.

\begin{enumerate}
\def\labelenumi{\arabic{enumi}.}
\tightlist
\item
  \textbf{Length scale 0.04.} The boundary has many small islands and narrow curves around single training objects, which is overfitting. Two objects share a latent value only when they lie very close, so each object affects only its neighborhood. The accuracy on the training objects is almost perfect, and the accuracy on the test objects is lower.
\item
  \textbf{Length scale 0.4.} The boundary is smooth, and it still follows the nonlinear separation of stars and galaxies. The accuracies on the training and the test objects are close.
\item
  \textbf{Length scale 4.0.} The boundary is too smooth to follow the groups, which is underfitting. Distant objects share almost one latent value, so the boundary cannot bend around the groups. Both accuracies are lower than for 0.4.
\end{enumerate}

\section*{Selecting the Optimal Length Scale}

The three plots show that the length scale shapes the boundary, so we need a systematic way to choose it. In GP regression we maximized the marginal likelihood (Tutorial 14a). With the sigmoid likelihood the marginal likelihood has no closed form, so we use cross-validation (Sections 4.7 and 14.10).

We split the training objects into five parts, the folds, at random. For each length scale, we train on four folds and measure the accuracy on the fifth, and we repeat this for each fold. The mean accuracy over the folds estimates the accuracy on new objects. The test objects take no part in this choice. We search ten length scales on a logarithmic grid. We keep the amplitude at 1 and tune only the length scale, which shapes the boundary.

\tutorialbarrier
\begin{codecell}

\end{codecell}

We plot the mean accuracy with its standard deviation over the folds, and we mark the best length scale.

\tutorialbarrier
\begin{codecell}

\end{codecell}
\tutorialplot{\includegraphics[width=4.224in,height=0.75\textheight,keepaspectratio]{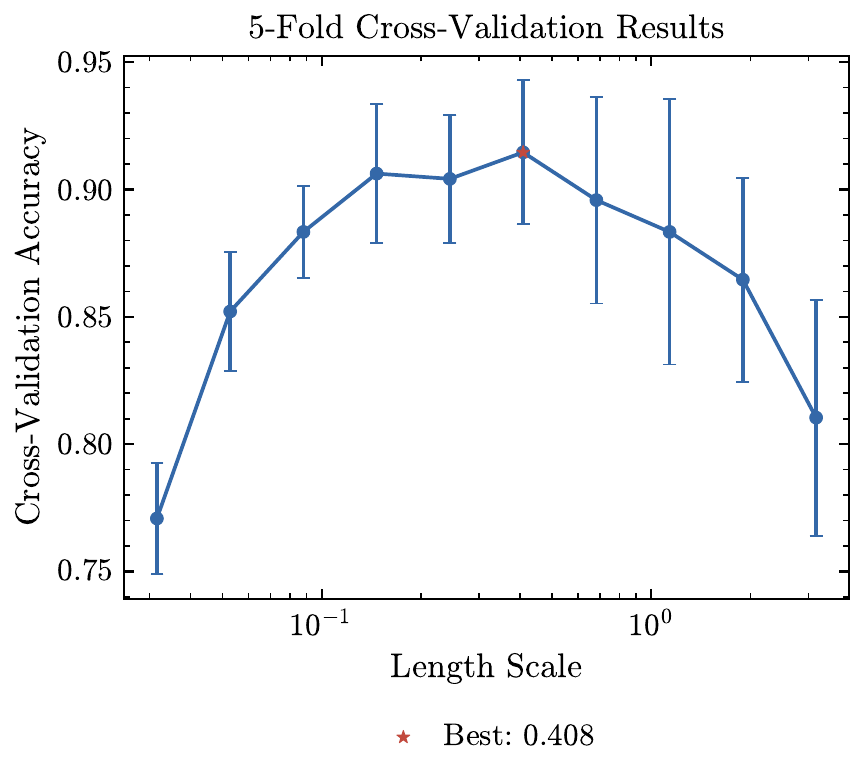}}%
  {\includegraphics[height=\tutorialheight]{tutorials/14b/fig06.pdf}}
\begin{Verbatim}[fontsize=\footnotesize, xleftmargin=5mm]
Best length scale = 0.408, accuracy = 0.915 ± 0.028
\end{Verbatim}

The curve agrees with the three panels. The accuracy is lowest for the smallest length scales, reaches a plateau, and falls for larger values. On the plateau the accuracies differ by less than the scatter over the folds, so several length scales are about equally good. We take the one with the highest mean accuracy.

\section*{The Final Model and Its Uncertainty}

We now train the final model with this length scale on the training objects and evaluate it on the test objects.

\tutorialbarrier
\begin{codecell}

\end{codecell}
\begin{Verbatim}[fontsize=\footnotesize, xleftmargin=5mm]
Test accuracy = 0.883
Stars correct = 49 of 60
Galaxies correct = 57 of 60
\end{Verbatim}

The GP classifier classifies both classes well. It also tells us which predictions rest on little data. We compute the probability map and select the quarter of the test objects with the largest error bar of the latent score.

\tutorialbarrier
\begin{codecell}
\begin{Verbatim}[commandchars=\\\{\},fontsize=\small]
\PYG{c+c1}{\PYGZsh{} Probability of a star on the grid for the final model}
\PYG{n}{\PYGZus{}}\PYG{p}{,}\FancyVerbBreakAfterBreak{} \PYG{n}{\PYGZus{}}\PYG{p}{,}\FancyVerbBreakAfterBreak{} \PYG{n}{p\PYGZus{}grid\PYGZus{}final} \PYG{o}{=}\FancyVerbBreakAfterBreak{} \PYG{n}{predict\PYGZus{}gpc}\PYG{p}{(}\FancyVerbBreakAfterBreak{}\PYG{n}{X\PYGZus{}train\PYGZus{}norm}\PYG{p}{,}\FancyVerbBreakAfterBreak{} \PYG{n}{t\PYGZus{}train}\PYG{p}{,}\FancyVerbBreakAfterBreak{} \PYG{n}{X\PYGZus{}grid\PYGZus{}norm}\PYG{p}{,}\FancyVerbBreakAfterBreak{} \PYG{n}{a\PYGZus{}map\PYGZus{}final}\PYG{p}{,}\FancyVerbBreakAfterBreak{} \PYG{n}{best\PYGZus{}ls}\PYG{p}{)}
\PYG{n}{p\PYGZus{}grid\PYGZus{}final} \PYG{o}{=}\FancyVerbBreakAfterBreak{} \PYG{n}{p\PYGZus{}grid\PYGZus{}final}\PYG{o}{.}\FancyVerbBreakAfterBreak{}\PYG{n}{reshape}\PYG{p}{(}\FancyVerbBreakAfterBreak{}\PYG{n}{n\PYGZus{}grid}\PYG{p}{,}\FancyVerbBreakAfterBreak{} \PYG{n}{n\PYGZus{}grid}\PYG{p}{)}

\PYG{c+c1}{\PYGZsh{} Misclassified test objects}
\PYG{n}{misclassified\PYGZus{}final} \PYG{o}{=}\FancyVerbBreakAfterBreak{} \PYG{n}{t\PYGZus{}pred\PYGZus{}final} \PYG{o}{!=} \PYG{n}{t\PYGZus{}test}

\PYG{c+c1}{\PYGZsh{} Test objects with the largest error bar (above the 75th percentile)}
\PYG{n}{test\PYGZus{}std\PYGZus{}final} \PYG{o}{=}\FancyVerbBreakAfterBreak{} \PYG{n}{np}\PYG{o}{.}\FancyVerbBreakAfterBreak{}\PYG{n}{sqrt}\PYG{p}{(}\FancyVerbBreakAfterBreak{}\PYG{n}{var\PYGZus{}test\PYGZus{}final}\PYG{p}{)}
\PYG{n}{high\PYGZus{}uncertainty} \PYG{o}{=}\FancyVerbBreakAfterBreak{} \PYG{n}{test\PYGZus{}std\PYGZus{}final} \PYG{o}{\PYGZgt{}} \PYG{n}{np}\PYG{o}{.}\FancyVerbBreakAfterBreak{}\PYG{n}{percentile}\PYG{p}{(}\FancyVerbBreakAfterBreak{}\PYG{n}{test\PYGZus{}std\PYGZus{}final}\PYG{p}{,}\FancyVerbBreakAfterBreak{} \PYG{l+m+mi}{75}\PYG{p}{)}
\end{Verbatim}
\end{codecell}

We plot the final model in two panels. The left panel shows the probability map with all objects. The right panel shows the test objects, colored by their probability of a star. Black crosses mark the misclassified objects, and green squares mark the objects with the largest error bar. The dashed line is the decision boundary.

\tutorialbarrier
\begin{codecell}

\end{codecell}
\tutorialplot{\includegraphics[width=6.338in,height=0.75\textheight,keepaspectratio]{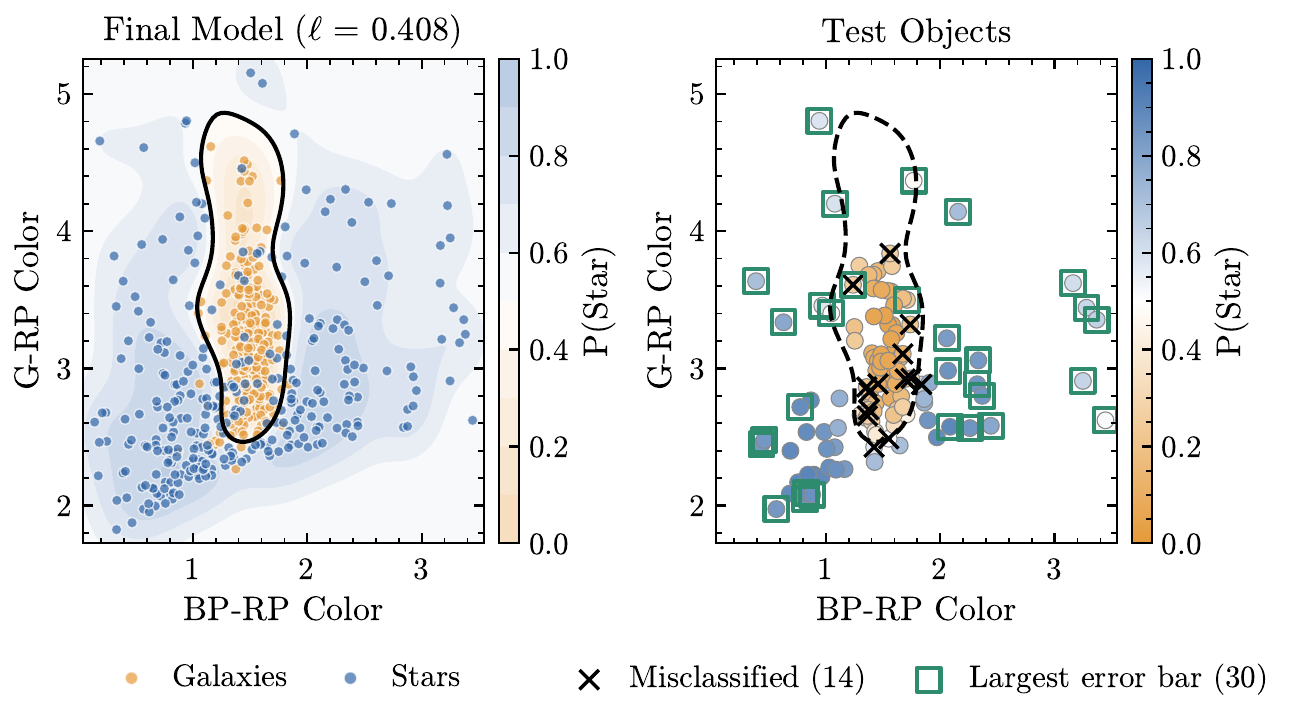}}%
  {\includegraphics[height=\tutorialheight]{tutorials/14b/fig07.pdf}}

The green squares lie at the outskirts of the groups and along the upper part of the decision boundary. The training objects are sparse there, so a large error bar is expected and correct. By eye many of these objects look like clear stars, but the model has seen little data near them. The misclassified objects lie at the edge of the galaxy group, where the classes overlap and the probability is close to one half. Objects with a probability close to one half and a large error bar are the most worth new data. The model is unsure of their class and has few training objects to learn from, so a spectrum there would change the prediction most.

\section*{Summary}

This tutorial turned a GP, a model of continuous functions, into a classifier of discrete labels. Our main lessons follow.

\begin{enumerate}
\def\labelenumi{\arabic{enumi}.}
\tightlist
\item
  \textbf{A latent function.} A GP prior on a latent function, passed through the sigmoid, gives a class probability. The boundary can have a flexible shape, while a linear model gives a straight boundary.
\item
  \textbf{Laplace and probit approximations.} Discrete labels break the conjugacy of GP regression, so the posterior of the latent function is not Gaussian. Gradient ascent finds its mode. The Laplace approximation gives a Gaussian around the mode, and the probit approximation turns the Gaussian into a class probability.
\item
  \textbf{Uncertainty.} The error bar of the latent score grows with the distance from the training objects. Far from them the probability returns to one half. A large error bar marks a prediction that rests on little data, and a probability close to one half marks objects where the classes overlap. Objects with both are the most worth new data.
\item
  \textbf{The length scale.} The length scale sets how much the boundary can bend. A value that is too small overfits, and one that is too large underfits. The marginal likelihood has no closed form here, so cross-validation on the training objects chooses the length scale.
\end{enumerate}

A GP classifier gives a nonlinear boundary and a Bayesian measure of uncertainty for each object, which is what difficult classification cases need. Where the classes overlap, it gives probabilities near one half. Where the training objects are sparse, it gives a large error bar. The cost of a GP grows with the cube of the number of objects, so this classifier suits samples of moderate size.

\tutorialchapter{15a}{Backpropagation and PyTorch}
\noindent\textit{Companion material for Chapter 15}

\section*{Introduction}

Modern deep learning models are trained with gradient descent (Section 15.3). Each step needs the gradient of the loss with respect to every weight. A network can have very many weights, so we need an efficient way to compute all of these gradients.

Backpropagation gives all gradients from one forward pass and one backward pass through the network (Section 15.4). The backward pass applies the chain rule layer by layer and reuses the numbers of the forward pass. Symbolic differentiation returns a formula that grows with every layer. Finite differences estimate each derivative from a small change of one weight, so they need one extra forward pass for each weight. Automatic differentiation libraries such as PyTorch apply the same idea to any model that we write with differentiable operations.

In this tutorial we write backpropagation in NumPy for a network with one hidden layer. The network fits a noisy cubic polynomial. We check its gradients against those of PyTorch and then train the same network with PyTorch. We also compare two optimizers, stochastic gradient descent and Adam (Section 15.3), and a learning rate schedule. Last, we meet the ReLU activation and the He initialization of the weights.

\section*{Prerequisites and Learning Objectives}

\subsection*{Prerequisites}

This tutorial assumes the following.

\begin{itemize}
\tightlist
\item
  Python programming, with NumPy arrays, matplotlib plots and classes
\item
  Neural networks, with the forward pass, activation functions and loss functions (Chapter 15)
\item
  Gradient descent, with the loss and the learning rate (Chapter 7)
\item
  Matrix multiplication and the chain rule from multivariable calculus
\end{itemize}

\subsection*{Learning Objectives}

After this tutorial you can do the following.

\begin{itemize}
\tightlist
\item
  Implement the forward and backward passes of a network with one hidden layer in NumPy
\item
  Check the gradients of your own backward pass against automatic differentiation
\item
  Explain how PyTorch records a computational graph, and how automatic differentiation differs from symbolic differentiation and finite differences
\item
  Train a network with PyTorch, use mini-batches, compare gradient descent with Adam and use a cosine learning rate schedule
\item
  Use ReLU and GELU in PyTorch and explain the He initialization
\item
  Describe how to run PyTorch code on a GPU
\end{itemize}

\section*{Understanding Backpropagation Through Implementation}

We first write backpropagation in NumPy, so that we see each step of the algorithm before we use a library.

Gradient descent needs the gradient of the loss with respect to every weight (Section 15.3). A neural network is a nested function, a chain of simple steps, so the chain rule gives these gradients (Section 15.4). Consider a tiny chain, \(z = 3x + 1\) followed by \(L = z^2\).

In the forward pass we compute \(z\) and then \(L\), and we keep the value of \(z\). In the backward pass we start from the loss and work back. The derivative of \(L\) with respect to \(z\) is \(2z\), and the derivative of \(z\) with respect to \(x\) is 3. The chain rule multiplies them, so \(\partial L/\partial x = 2z \times 3\). Each step needs only its own derivative and the value that the forward pass kept.

Backpropagation does the same for a whole network. The forward pass computes the output and stores the intermediate values. The backward pass multiplies the local derivatives with the chain rule, from the loss back to each weight.

We build the network from three kinds of components. They are a fully connected layer, an activation function and a loss function. Each component has three methods.

\begin{enumerate}
\def\labelenumi{\arabic{enumi}.}
\tightlist
\item
  The forward pass computes the output and caches the values that the backward pass needs.
\item
  The backward pass receives the gradient of the loss with respect to the output of the component. It returns the gradient with respect to its input and computes the gradients of its own parameters.
\item
  The update changes the parameters of the component (if it has any) with their gradients.
\end{enumerate}

Gradients flow backward through the components, and each component multiplies by its local derivative.

We begin by importing the libraries and setting the plot style.

\tutorialbarrier
\begin{codecell}

\end{codecell}

\subsection*{Constructing the Fully Connected Layer}

We begin with the fully connected layer, also called the linear layer. It mixes all inputs into each output with the operation \(\mathbf{z} = \mathbf{W}\mathbf{x} + \mathbf{b}\) (Section 15.1). Here \(\mathbf{W}\) is the weight matrix of shape (\texttt{out\_\allowbreak{}features}, \texttt{in\_\allowbreak{}features}) and \(\mathbf{b}\) is the bias vector. The code processes a batch of points at once, which is the set of points that we handle together. The batch array has one row for each point, so the code computes the same operation for all points as \(\mathbf{X}\mathbf{W}^T + \mathbf{b}^T\).

The method \texttt{init\_\allowbreak{}weights} sets the weights to random numbers from a standard normal distribution and the biases to zero. The weights start at random values so that the hidden units differ, because equal weights would give units that stay equal.

\tutorialbarrier
\begin{codecell}

\end{codecell}

\subsection*{Understanding the Fully Connected Layer Operations}

The forward pass computes \(\mathbf{z}\) and caches the input \(\mathbf{x}\). We evaluate a gradient at a specific point. The backward pass therefore needs the values of the forward pass (Section 15.4). The cache saves us from computing them again.

The backward pass receives the row \(\partial L/\partial \mathbf{z}\) from the next layer, as in Section 15.4. The local derivatives of \(\mathbf{z} = \mathbf{W}\mathbf{x} + \mathbf{b}\) are \(\mathbf{x}^T\) for the weights of one unit, 1 for the bias and \(\mathbf{W}\) for the input. The chain rule multiplies each of them by \(\partial L/\partial \mathbf{z}\) and gives three gradients.

\begin{enumerate}
\def\labelenumi{\arabic{enumi}.}
\tightlist
\item
  The gradient with respect to the weights is \(\partial L/\partial \mathbf{W} = (\partial L/\partial \mathbf{z})^T \mathbf{x}^T\), the outer product of a column and a row, summed over the batch.
\item
  The gradient with respect to the bias is \(\partial L/\partial \mathbf{b} = \partial L/\partial \mathbf{z}\), summed over the batch. The code stores it as a column, like \(\mathbf{b}\).
\item
  The gradient with respect to the input is \(\partial L/\partial \mathbf{x} = (\partial L/\partial \mathbf{z})\,\mathbf{W}\).
\end{enumerate}

The weights and the bias act on every point of the batch, so their gradients add the contributions of all points. The layer passes the third gradient to the previous layer. This is how gradients flow backward through the whole network.

In the code, each row of the array \texttt{g\_\allowbreak{}next\_\allowbreak{}layer} is the row \(\partial L/\partial \mathbf{z}\) of one point. The product \((\partial L/\partial \mathbf{Z})^T \mathbf{X}\) gives the weight gradient and sums over the batch, and \((\partial L/\partial \mathbf{Z})\,\mathbf{W}\) gives the input gradient of every point.

\subsection*{Implementing the Sigmoid Activation Function}

We implement the sigmoid activation function \(\sigma(z) = 1/(1 + e^{-z})\) (Section 15.5). Without activation functions, a stack of layers would collapse into one linear map. Activation functions make the network nonlinear, so that it can learn complex functions. The derivative of the sigmoid is \(\sigma'(z) = \sigma(z)(1 - \sigma(z))\), so it depends only on the output. The layer therefore caches its output and computes the derivative from it, without the original input.

\tutorialbarrier
\begin{codecell}

\end{codecell}

In the backward pass the sigmoid receives the gradient \(\partial L/\partial a\) with respect to its output \(a = \sigma(z)\) from the next layer. The chain rule gives the gradient with respect to its input, \(\partial L/\partial z = \sigma'(z)\,\partial L/\partial a\), so the sigmoid multiplies the gradient by \(\sigma'(z)\). This factor is at most 0.25, which it reaches when the output is 0.5. It is close to zero when the output is near 0 or 1, where the sigmoid saturates. Each sigmoid layer therefore shrinks the gradient that flows back to the earlier layers, and the shrinking compounds over many layers. This is the vanishing gradient problem (Section 15.5).

\subsection*{Implementing the Mean Squared Error Loss}

We need a loss function to complete the components. We use the mean squared error (MSE) of Section 15.3, which follows from maximum likelihood under Gaussian noise.

\[
L = \frac{1}{N} \sum_{i=1}^{N} (\hat{y}_i - y_i)^2.
\]

Here \(\hat{y}_i\) is the prediction of the network and \(y_i\) is the target of point \(i\).

\tutorialbarrier
\begin{codecell}

\end{codecell}

\subsection*{Understanding the Loss Function's Role in Backpropagation}

The loss function starts the backward pass. Differentiating the MSE gives

\[
\frac{\partial L}{\partial \hat{y}_i} = \frac{2}{N} (\hat{y}_i - y_i).
\]

Section 15.4 writes the loss of one example as \((1/2)(y - \hat{y})^2\). Its gradient has the factor 1 where ours has the factor \(2/N\). The loss is a scalar, and its derivative with respect to itself is 1. The argument \texttt{g\_\allowbreak{}next\_\allowbreak{}layer} of the loss therefore defaults to 1. The factor \(1/N\) keeps the size of the gradient from growing with the number of points.

\subsection*{Assembling a Complete Neural Network}

We now chain the components into a network with one hidden layer of configurable size. The forward pass runs the components in order, and the backward pass runs them in reverse order.

\tutorialbarrier
\begin{codecell}

\end{codecell}

\subsection*{Understanding the Network Architecture}

The network maps the input through a hidden layer with a sigmoid and a linear output layer to one output value.

\begin{enumerate}
\def\labelenumi{\arabic{enumi}.}
\tightlist
\item
  \textbf{Hidden layer.} Each hidden unit applies a sigmoid to a linear function of the input. It adds one smooth step to the output function (Section 15.2).
\item
  \textbf{Linear output layer.} The output is a linear combination of the hidden units, so it can take any real value, which regression needs.
\end{enumerate}

\subsection*{Training on Simulated Data}

We now train the network on simulated data. The 200 points \(x\) are uniform on \((-2, 2)\). The target is \(y = 0.5x^3 - 0.3x^2 + 0.8x + 0.2\) plus Gaussian noise of standard deviation 0.1. We split the points at random into 160 training points and 40 test points. The test points take no part in the training.

\tutorialbarrier
\begin{codecell}

\end{codecell}

\subsection*{Training and Monitoring the Network}

We train a network with 20 hidden units by gradient descent with the learning rate \(\eta = 0.03\) (Section 15.3). One epoch is one pass through the training points. Here it is one gradient step, because we use all points at once. Plain gradient descent with a small step needs many epochs, so we train for 100,000 epochs.

Each epoch has four steps.

\begin{enumerate}
\def\labelenumi{\arabic{enumi}.}
\tightlist
\item
  Run the forward pass on the training points and compute their loss.
\item
  Run the backward pass to get the gradients.
\item
  Compute the loss of the test points with the same weights, those before the update.
\item
  Update the weights with the gradients.
\end{enumerate}

The test loss only monitors the training. The update of each layer is

\[
\mathbf{W} \leftarrow \mathbf{W} - \eta \frac{\partial L}{\partial \mathbf{W}},
\]

and the same for the biases.

\tutorialbarrier
\begin{codecell}

\end{codecell}
\begin{Verbatim}[fontsize=\footnotesize, xleftmargin=5mm]
Final losses: training 0.0105, test 0.0093
\end{Verbatim}

\subsection*{Visualizing the Training Results}

We plot the two losses against the epoch. We also plot the fitted function on a fine grid of inputs, together with the data and the true cubic.

\tutorialbarrier
\begin{codecell}

\end{codecell}
\tutorialplot{\includegraphics[width=6.500in,height=0.75\textheight,keepaspectratio]{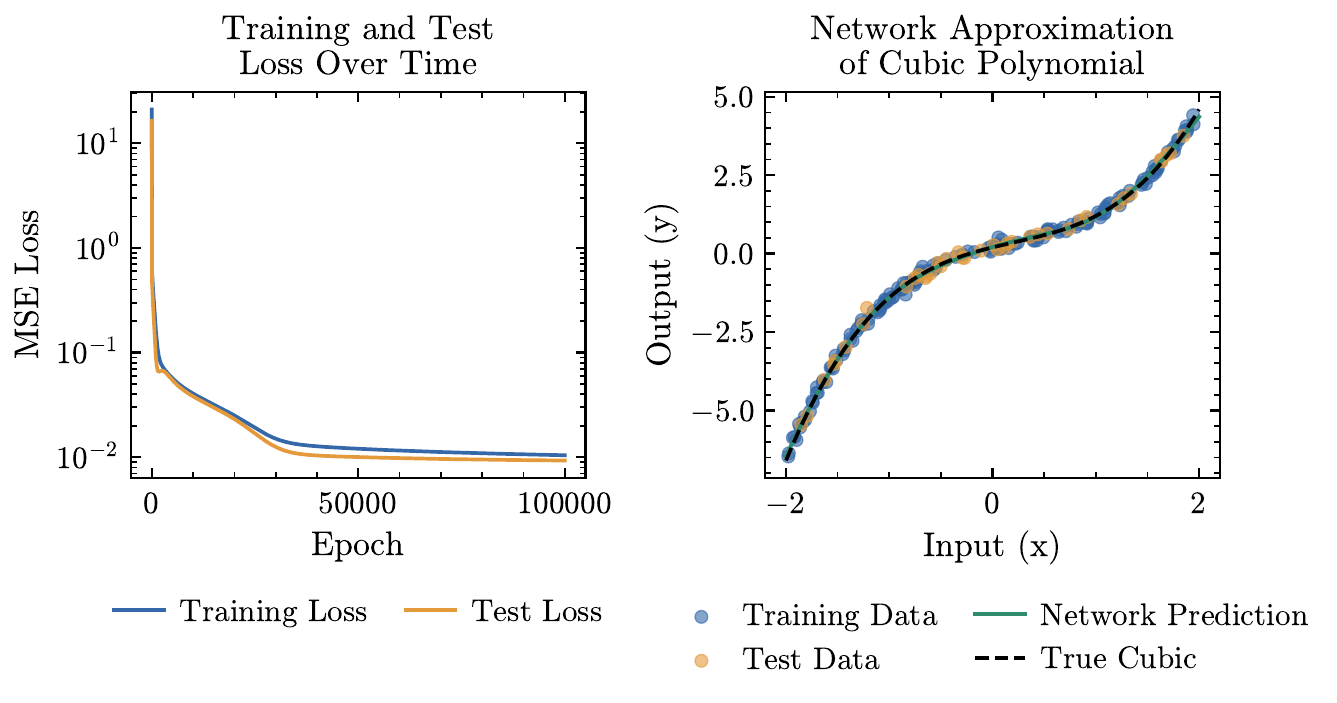}}%
  {\includegraphics[height=\tutorialheight]{tutorials/15a/fig01.pdf}}

The loss falls quickly in the first epochs and then slowly. The test loss is a little below the training loss, so the network does not overfit. An overfit network fits the training points better than new points. The noise has the variance \(0.1^2 = 0.01\), which is the loss that the true cubic gives on average. Both final losses are close to this value. The fitted function follows the cubic over the range of the data.

Our task was regression, one of the two main kinds of supervised task. The other is classification, which assigns an object to a class, such as a star or a galaxy. The output layer depends on the task (Section 15.5), and regression needs outputs that can be any real number, so ours is linear. For two classes, the output is a probability between 0 and 1, so we apply the sigmoid. For several classes, we apply the softmax, so that the probabilities sum to one. The loss changes too, from the mean squared error for regression to the cross-entropy for classification (Section 15.3).

Our fit is also an example of a general result. One hidden layer can approximate any continuous function on a closed, bounded interval to arbitrary accuracy (Section 15.2). It needs enough hidden units and an activation that is not a polynomial.

Every gradient in this network came from the local derivatives of its components, multiplied in reverse order, which is backpropagation. A backward method for every component of every model is a lot of work, so libraries automate it. PyTorch is such a library.

\section*{Automatic Differentiation with PyTorch}

PyTorch builds the backward pass from the operations of the forward pass, so we write only the forward pass. We start with tensors, build our network from PyTorch modules and check its gradients against the NumPy ones. We then train it with optimizers and a learning rate schedule, and we end with activation functions, the initial weights and the GPU. We first import PyTorch and set its random seed.

\tutorialbarrier
\begin{codecell}
\begin{Verbatim}[commandchars=\\\{\},fontsize=\small]
\PYG{c+c1}{\PYGZsh{} Import PyTorch and its modules for networks and optimizers}
\PYG{k+kn}{import}\PYG{+w}{ }\PYG{n+nn}{torch}
\PYG{k+kn}{import}\PYG{+w}{ }\PYG{n+nn}{torch}\PYG{n+nn}{.}\FancyVerbBreakAfterBreak{}\PYG{n+nn}{nn}\PYG{+w}{ }\PYG{k}{as}\PYG{+w}{ }\PYG{n+nn}{nn}
\PYG{k+kn}{import}\PYG{+w}{ }\PYG{n+nn}{torch}\PYG{n+nn}{.}\FancyVerbBreakAfterBreak{}\PYG{n+nn}{optim}\PYG{+w}{ }\PYG{k}{as}\PYG{+w}{ }\PYG{n+nn}{optim}
\PYG{k+kn}{from}\PYG{+w}{ }\PYG{n+nn}{torch}\PYG{n+nn}{.}\FancyVerbBreakAfterBreak{}\PYG{n+nn}{utils}\PYG{n+nn}{.}\FancyVerbBreakAfterBreak{}\PYG{n+nn}{data}\PYG{+w}{ }\PYG{k+kn}{import} \PYG{n}{DataLoader}\PYG{p}{,}\FancyVerbBreakAfterBreak{} \PYG{n}{TensorDataset}  \PYG{c+c1}{\PYGZsh{} Mini\PYGZhy{}batches of the data}

\PYG{c+c1}{\PYGZsh{} Set random seed for reproducibility}
\PYG{n}{\PYGZus{}} \PYG{o}{=}\FancyVerbBreakAfterBreak{} \PYG{n}{torch}\PYG{o}{.}\FancyVerbBreakAfterBreak{}\PYG{n}{manual\PYGZus{}seed}\PYG{p}{(}\FancyVerbBreakAfterBreak{}\PYG{l+m+mi}{42}\PYG{p}{)}
\end{Verbatim}
\end{codecell}

\subsection*{PyTorch Tensors and Automatic Differentiation}

PyTorch tensors are the basic data structure. They are similar to NumPy arrays, but they support automatic differentiation. Each operation on a tensor stores two things, its value and the means to compute its gradient. A tensor that we create with \texttt{requires\_\allowbreak{}grad=\allowbreak{}True} starts to record the computational graph that the differentiation needs. It then carries four pieces of information.

\begin{enumerate}
\def\labelenumi{\arabic{enumi}.}
\tightlist
\item
  \textbf{The data values}, as in a NumPy array.
\item
  \textbf{The gradient}, which \texttt{backward()} stores in the \texttt{.\allowbreak{}grad} attribute of the tensors that we create.
\item
  \textbf{The graph connectivity}, which means the references to the operations that created the tensor.
\item
  \textbf{The backward function}, which computes the local derivative, like our \texttt{backward\_\allowbreak{}pass} methods.
\end{enumerate}

We start with the gradient of \(f(x) = x^2\) at \(x = 3\) and compare it with the exact derivative \(2x\).

\tutorialbarrier
\begin{codecell}
\begin{Verbatim}[commandchars=\\\{\},fontsize=\small]
\PYG{c+c1}{\PYGZsh{} Compute the gradient of f(x) = x\PYGZca{}2 at x = 3}

\PYG{c+c1}{\PYGZsh{} Create a tensor that records its operations}
\PYG{n}{x} \PYG{o}{=}\FancyVerbBreakAfterBreak{} \PYG{n}{torch}\PYG{o}{.}\FancyVerbBreakAfterBreak{}\PYG{n}{tensor}\PYG{p}{(}\FancyVerbBreakAfterBreak{}\PYG{p}{[}\FancyVerbBreakAfterBreak{}\PYG{l+m+mf}{3.0}\PYG{p}{]}\PYG{p}{,}\FancyVerbBreakAfterBreak{} \PYG{n}{requires\PYGZus{}grad}\PYG{o}{=}\FancyVerbBreakAfterBreak{}\PYG{k+kc}{True}\PYG{p}{)}

\PYG{c+c1}{\PYGZsh{} Forward computation, then backpropagation through the recorded graph}
\PYG{n}{y} \PYG{o}{=}\FancyVerbBreakAfterBreak{} \PYG{n}{x}\PYG{o}{*}\PYG{o}{*}\PYG{l+m+mi}{2}
\PYG{n}{y}\PYG{o}{.}\FancyVerbBreakAfterBreak{}\PYG{n}{backward}\PYG{p}{(}\FancyVerbBreakAfterBreak{}\PYG{p}{)}

\PYG{c+c1}{\PYGZsh{} The gradient is stored in x.grad}
\PYG{n+nb}{print}\PYG{p}{(}\FancyVerbBreakAfterBreak{}\PYG{l+s+sa}{f}\PYG{l+s+s2}{\PYGZdq{}}\PYG{l+s+s2}{y = }\PYG{l+s+si}{\PYGZob{}}\PYG{n}{y}\PYG{l+s+si}{\PYGZcb{}}\FancyVerbBreakAfterBreak{}\PYG{l+s+s2}{\PYGZdq{}}\PYG{p}{)}
\PYG{n+nb}{print}\PYG{p}{(}\FancyVerbBreakAfterBreak{}\PYG{l+s+sa}{f}\PYG{l+s+s2}{\PYGZdq{}}\PYG{l+s+s2}{dy/dx = }\PYG{l+s+si}{\PYGZob{}}\PYG{n}{x}\PYG{o}{.}\FancyVerbBreakAfterBreak{}\PYG{n}{grad}\PYG{o}{.}\FancyVerbBreakAfterBreak{}\PYG{n}{item}\PYG{p}{(}\FancyVerbBreakAfterBreak{}\PYG{p}{)}\PYG{l+s+si}{\PYGZcb{}}\FancyVerbBreakAfterBreak{}\PYG{l+s+s2}{, exact derivative 2x = }\PYG{l+s+si}{\PYGZob{}}\PYG{l+m+mi}{2}\PYG{o}{*}\PYG{n}{x}\PYG{o}{.}\FancyVerbBreakAfterBreak{}\PYG{n}{item}\PYG{p}{(}\FancyVerbBreakAfterBreak{}\PYG{p}{)}\PYG{l+s+si}{\PYGZcb{}}\FancyVerbBreakAfterBreak{}\PYG{l+s+s2}{\PYGZdq{}}\PYG{p}{)}
\end{Verbatim}
\end{codecell}
\begin{Verbatim}[fontsize=\footnotesize, xleftmargin=5mm]
y = tensor([9.], grad_fn=<PowBackward0>)
dy/dx = 6.0, exact derivative 2x = 6.0
\end{Verbatim}

\subsection*{Understanding PyTorch's Implementation}

When we computed \texttt{y\ =\ x**2}, PyTorch did three things.

\begin{enumerate}
\def\labelenumi{\arabic{enumi}.}
\tightlist
\item
  It performed the forward computation, the squaring of \(x\).
\item
  It recorded the operation in a computational graph. The printed \texttt{grad\_\allowbreak{}fn=\allowbreak{}{\char60}PowBackward0{\char62}} names this record.
\item
  It stored the values that the backward pass needs.
\end{enumerate}

The call \texttt{y.\allowbreak{}backward()} then went through the graph in reverse and stored the gradient in \texttt{x.\allowbreak{}grad}.

We now treat a computation like one neuron of our network. It has a linear map, a sigmoid and a squared error for one example.

\tutorialbarrier
\begin{codecell}

\end{codecell}
\begin{Verbatim}[fontsize=\footnotesize, xleftmargin=5mm]
z = 2.000, y = 0.881, loss = 0.006528
dL/dW = 0.033933, dL/db = 0.016966, dL/dx = 0.025450
\end{Verbatim}

The chain rule of Section 15.4 explains the gradients. Each one is the gradient with respect to \(z\) times a local factor. We have \(\partial L/\partial b = \partial L/\partial z\), \(\partial L/\partial W = x \, \partial L/\partial z\) and \(\partial L/\partial x = W \, \partial L/\partial z\). The printed gradients agree. The printed \(\partial L/\partial b\) is \(\partial L/\partial z = 0.0170\). The value 0.0339 is \(x = 2\) times 0.0170, and 0.0255 is \(W = 1.5\) times 0.0170.

\subsection*{Automatic, Symbolic and Numerical Differentiation}

There are three ways to compute a derivative with a computer (Section 15.4). Automatic differentiation, which PyTorch uses, differs from the other two in what it keeps.

\begin{itemize}
\tightlist
\item
  \textbf{Symbolic differentiation} applies the rules of calculus to an expression and returns a formula, as computer algebra systems do. For \(f(x) = \sin(x^2)\) the formula is \(\cos(x^2)\cdot 2x\). For a deep network the formula becomes impractical.
\item
  \textbf{Finite differences} approximate the derivative by \([f(x + \Delta) - f(x)]/\Delta\) for a small step \(\Delta\). They give only an approximation, and a network with \(P\) parameters needs at least \(P + 1\) forward passes for its gradient.
\item
  \textbf{Automatic differentiation} returns no formula. It stores the derivative rule of each elementary operation and the values of the forward pass, so the chain rule becomes a product of numbers. The derivatives are exact up to rounding, and one backward pass gives the gradient with respect to all parameters.
\end{itemize}

We compute the derivative of \(f(x) = \sin(x^2)\) at \(x = 1.5\) in the three ways.

\tutorialbarrier
\begin{codecell}

\end{codecell}
\begin{Verbatim}[fontsize=\footnotesize, xleftmargin=5mm]
Symbolic: -1.8845208682
Finite differences: -1.8845249974
Automatic: -1.8845208682
\end{Verbatim}

The automatic result agrees with the formula to the digits shown, although we never wrote the formula. PyTorch kept the value of \(x^2\) and the derivative rule of each operation, and the backward pass multiplied the numbers. The finite difference differs in the sixth decimal place, because it is an approximation with the step \(\Delta\).

For training a network, this is what we need. Gradient descent uses the numerical value of the gradient at the current weights, for every weight and at every step, and not a formula. Symbolic differentiation derives a formula that we do not need, and finite differences need one extra forward pass for each parameter. Automatic differentiation needs only the forward computation, so we write the forward pass and PyTorch supplies the backward pass.

\subsection*{Inspecting the Computational Graph}

PyTorch records each operation as a node of a computational graph while the forward pass runs. The attribute \texttt{grad\_\allowbreak{}fn} of a tensor names the backward function of the operation that created it. A leaf is a tensor that we create ourselves, and not the result of an operation, so it has no \texttt{grad\_\allowbreak{}fn}.

We follow the graph from the loss back toward the inputs. The attribute \texttt{next\_\allowbreak{}functions} of a \texttt{grad\_\allowbreak{}fn} lists the backward functions of the operations that produced its inputs. We print this chain and, for each tensor, whether it is a leaf and which \texttt{grad\_\allowbreak{}fn} it has.

\tutorialbarrier
\begin{codecell}

\end{codecell}
\begin{Verbatim}[fontsize=\footnotesize, xleftmargin=5mm]
PowBackward0
    SubBackward0
        SigmoidBackward0
            AddBackward0
x: is_leaf=True, grad_fn=None
W: is_leaf=True, grad_fn=None
b: is_leaf=True, grad_fn=None
y: is_leaf=False, grad_fn=SigmoidBackward0
loss: is_leaf=False, grad_fn=PowBackward0
\end{Verbatim}

\subsection*{Understanding PyTorch's Computational Graph}

The output shows the chain of backward operations from the loss to the sum \(Wx + b\). It is the chain of Section 15.4. Each \texttt{grad\_\allowbreak{}fn} object is one backward operation.

These are backward passes. \texttt{PowBackward0} is the square in the loss, \texttt{SubBackward0} the subtraction \texttt{y\ -\ target}, \texttt{SigmoidBackward0} the sigmoid and \texttt{AddBackward0} the addition \(Wx + b\).

PyTorch distinguishes leaf tensors from non-leaf tensors.

\begin{itemize}
\tightlist
\item
  A \textbf{leaf tensor} (\(x\), \(W\), \(b\)) is one that we created with \texttt{requires\_\allowbreak{}grad=\allowbreak{}True}. Its \texttt{.\allowbreak{}grad} attribute stores the gradient after \texttt{backward()}.
\item
  A \textbf{non-leaf tensor} (\(y\), loss) is the result of an operation. Its \texttt{grad\_\allowbreak{}fn} records how PyTorch computed it.
\end{itemize}

\subsection*{Recreating Our Neural Network in PyTorch}

The tensors above modeled one neuron. A whole network has many neurons, and PyTorch groups them in modules. We now build the architecture of our NumPy network with these modules. The module \texttt{nn.\allowbreak{}Linear} holds the weights and the biases of a fully connected layer, and \texttt{nn.\allowbreak{}Sigmoid} applies the sigmoid.

A PyTorch network is a class that inherits from \texttt{nn.\allowbreak{}Module}, which collects the parameters of its layers. Its method \texttt{\_\_\allowbreak{}init\_\_} creates the layers, and its method \texttt{forward} says how an input passes through them.

The first layer has 20 weights and 20 biases. The second has 20 weights and one bias. The network therefore has 61 parameters. PyTorch draws the initial weights of \texttt{nn.\allowbreak{}Linear} at random with its own default rule. This network therefore starts from different weights than the NumPy network.

When a network applies its layers one after the other, \texttt{nn.\allowbreak{}Sequential} is a shorter way to write it. It takes the layers as arguments and runs them in order. We also build our network in this way, and the later tutorials use this form.

\tutorialbarrier
\begin{codecell}

\end{codecell}
\begin{Verbatim}[fontsize=\footnotesize, xleftmargin=5mm]
Total parameters: 61
Sequential(
  (0): Linear(in_features=1, out_features=20, bias=True)
  (1): Sigmoid()
  (2): Linear(in_features=20, out_features=1, bias=True)
)
\end{Verbatim}

\subsection*{Preparing Data for PyTorch}

We convert the NumPy arrays of the same cubic data to PyTorch tensors. The method \texttt{float()} gives 32-bit numbers, the type of the network weights. Each tensor keeps one column, as the NumPy arrays do.

\tutorialbarrier
\begin{codecell}
\begin{Verbatim}[commandchars=\\\{\},fontsize=\small]
\PYG{c+c1}{\PYGZsh{} Convert our NumPy data to PyTorch tensors}
\PYG{c+c1}{\PYGZsh{} The method float() gives 32\PYGZhy{}bit numbers, the type of the network weights}
\PYG{n}{X\PYGZus{}train\PYGZus{}torch} \PYG{o}{=}\FancyVerbBreakAfterBreak{} \PYG{n}{torch}\PYG{o}{.}\FancyVerbBreakAfterBreak{}\PYG{n}{from\PYGZus{}numpy}\PYG{p}{(}\FancyVerbBreakAfterBreak{}\PYG{n}{X\PYGZus{}train}\PYG{p}{)}\PYG{o}{.}\FancyVerbBreakAfterBreak{}\PYG{n}{float}\PYG{p}{(}\FancyVerbBreakAfterBreak{}\PYG{p}{)}
\PYG{n}{y\PYGZus{}train\PYGZus{}torch} \PYG{o}{=}\FancyVerbBreakAfterBreak{} \PYG{n}{torch}\PYG{o}{.}\FancyVerbBreakAfterBreak{}\PYG{n}{from\PYGZus{}numpy}\PYG{p}{(}\FancyVerbBreakAfterBreak{}\PYG{n}{y\PYGZus{}train}\PYG{p}{)}\PYG{o}{.}\FancyVerbBreakAfterBreak{}\PYG{n}{float}\PYG{p}{(}\FancyVerbBreakAfterBreak{}\PYG{p}{)}
\PYG{n}{X\PYGZus{}test\PYGZus{}torch} \PYG{o}{=}\FancyVerbBreakAfterBreak{} \PYG{n}{torch}\PYG{o}{.}\FancyVerbBreakAfterBreak{}\PYG{n}{from\PYGZus{}numpy}\PYG{p}{(}\FancyVerbBreakAfterBreak{}\PYG{n}{X\PYGZus{}test}\PYG{p}{)}\PYG{o}{.}\FancyVerbBreakAfterBreak{}\PYG{n}{float}\PYG{p}{(}\FancyVerbBreakAfterBreak{}\PYG{p}{)}
\PYG{n}{y\PYGZus{}test\PYGZus{}torch} \PYG{o}{=}\FancyVerbBreakAfterBreak{} \PYG{n}{torch}\PYG{o}{.}\FancyVerbBreakAfterBreak{}\PYG{n}{from\PYGZus{}numpy}\PYG{p}{(}\FancyVerbBreakAfterBreak{}\PYG{n}{y\PYGZus{}test}\PYG{p}{)}\PYG{o}{.}\FancyVerbBreakAfterBreak{}\PYG{n}{float}\PYG{p}{(}\FancyVerbBreakAfterBreak{}\PYG{p}{)}
\end{Verbatim}
\end{codecell}

\subsection*{Checking the Gradients}

Before we train the PyTorch network, we check the backward pass that we wrote in NumPy. A backward pass that we write ourselves is easy to get wrong, and automatic differentiation gives a reference. The check has three steps.

\begin{enumerate}
\def\labelenumi{\arabic{enumi}.}
\tightlist
\item
  Make a new network with 61 parameters, and keep its initial weights.
\item
  Compute the gradient of the loss with respect to all 61 parameters with our NumPy backward pass.
\item
  Compute the same gradient with PyTorch for the same weights and data, in double precision, which means 64-bit numbers as in NumPy.
\end{enumerate}

We print the largest relative difference between the two gradients.

\tutorialbarrier
\begin{codecell}

\end{codecell}
\begin{Verbatim}[fontsize=\footnotesize, xleftmargin=5mm]
Largest relative difference: 1.0e-15
\end{Verbatim}

Double precision rounds at about 1e-16, so the printed difference is rounding error. Our backward pass computes the same gradients as automatic differentiation.

\subsection*{Training with PyTorch}

We now train the PyTorch network with the same learning rate, 0.03, and the same 100,000 epochs. The loss is \texttt{nn.\allowbreak{}MSELoss}, the same mean squared error as before. The optimizer \texttt{optim.\allowbreak{}SGD} applies the gradient descent update (Section 15.3). Each of our steps uses all training points, so it is plain gradient descent as in NumPy, so the two networks are comparable.

A PyTorch training step has three calls.

\begin{enumerate}
\def\labelenumi{\arabic{enumi}.}
\tightlist
\item
  \texttt{optimizer.\allowbreak{}zero\_\allowbreak{}grad()} clears the stored gradients, because PyTorch adds each new gradient to them. This lets us add the gradients of several batches before one update, which helps when a batch does not fit in memory. Our NumPy layers overwrite their gradients at each step instead.
\item
  \texttt{loss.\allowbreak{}backward()} runs the backward pass.
\item
  \texttt{optimizer.\allowbreak{}step()} updates the weights.
\end{enumerate}

As before, we record both losses for the same weights, before the update. We compute the test loss inside \texttt{torch.\allowbreak{}no\_\allowbreak{}grad()}. PyTorch then records no graph for it, which saves memory and time.

\tutorialbarrier
\begin{codecell}

\end{codecell}
\begin{Verbatim}[fontsize=\footnotesize, xleftmargin=5mm]
Final losses: training 0.0090, test 0.0092
\end{Verbatim}

We plot the training and test losses of the two networks in the top row. In the bottom panel we plot their fitted functions, the data and the true cubic.

\tutorialbarrier
\begin{codecell}

\end{codecell}
\tutorialplot{\includegraphics[width=6.170in,height=0.75\textheight,keepaspectratio]{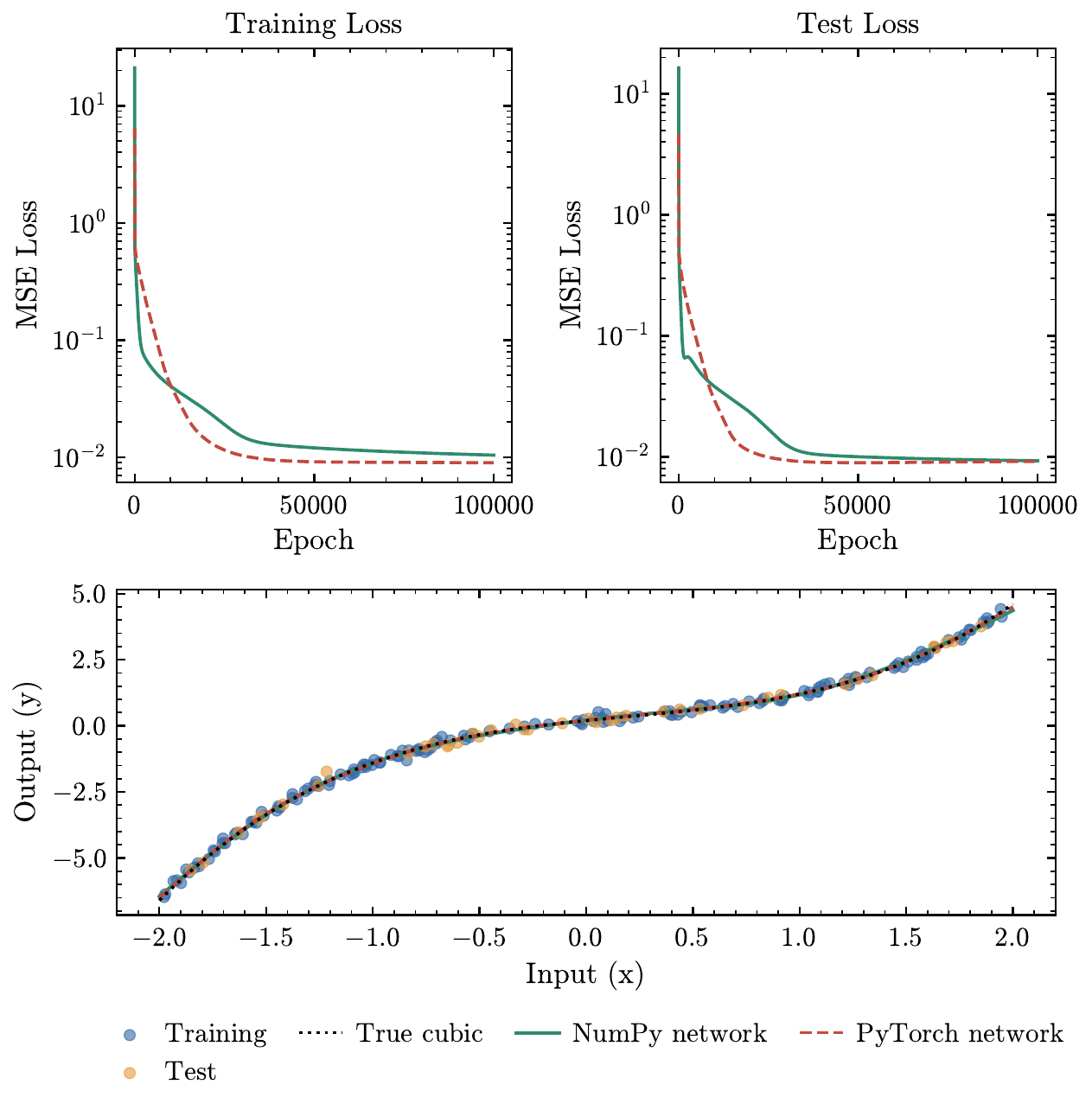}}%
  {\includegraphics[height=\tutorialheight]{tutorials/15a/fig02.pdf}}

Both networks fit the cubic, and both losses fall quickly at first and then slowly. The PyTorch training loss ends a little lower than the NumPy one, and the two test losses end almost equal. All final losses are close to the noise variance 0.01.

PyTorch differs from our NumPy network in four ways. It computes the backward pass for us, from the forward pass that we write. It provides ready-made layers, loss functions and optimizers. The same code runs on a GPU, a graphics processing unit, which is a processor built for matrix operations.

\subsection*{Comparing Different Optimizers}

PyTorch provides several optimizers, and we change one line to switch between them. We compare stochastic gradient descent with Adam (Section 15.3).

Stochastic gradient descent approximates the gradient with a small random subset of the training points, called a mini-batch. It shuffles the training points and splits them into mini-batches of a fixed size, such as 32. Each step uses the same update as before, with the loss \(L_B\) of one mini-batch \(B\) in place of the full loss \(L\). The gradient of \(L_B\) is a noisy estimate of the gradient of \(L\).

A step costs less, so a large data set gives many steps in the time of one full pass. The noise can also help the weights to leave a shallow minimum. One epoch is a pass through all mini-batches. In PyTorch, \texttt{TensorDataset} pairs the inputs with their targets, and \texttt{DataLoader} shuffles them and delivers the mini-batches.

Adam adapts the step of each parameter in two ways.

\begin{itemize}
\tightlist
\item
  It keeps a moving average of the gradient. This is momentum, because the updates continue in the direction of the recent gradients and oscillations are damped.
\item
  It keeps a moving average of the squared gradient, and it divides the step of each parameter by its square root. A parameter with consistent gradients then takes larger steps, and a parameter with highly variable gradients takes smaller steps.
\end{itemize}

We train two networks from the same initial weights for 2000 epochs with mini-batches of 32 points, so an epoch has five updates. After each epoch we compute two losses for all points. The training loss shows how fast each optimizer reduces the loss. The test loss shows how well the network predicts new points.

Gradient descent uses the learning rate 0.03 of the earlier runs, and Adam uses 0.01.

\tutorialbarrier
\begin{codecell}

\end{codecell}
\tutorialplot{\includegraphics[width=6.188in,height=0.75\textheight,keepaspectratio]{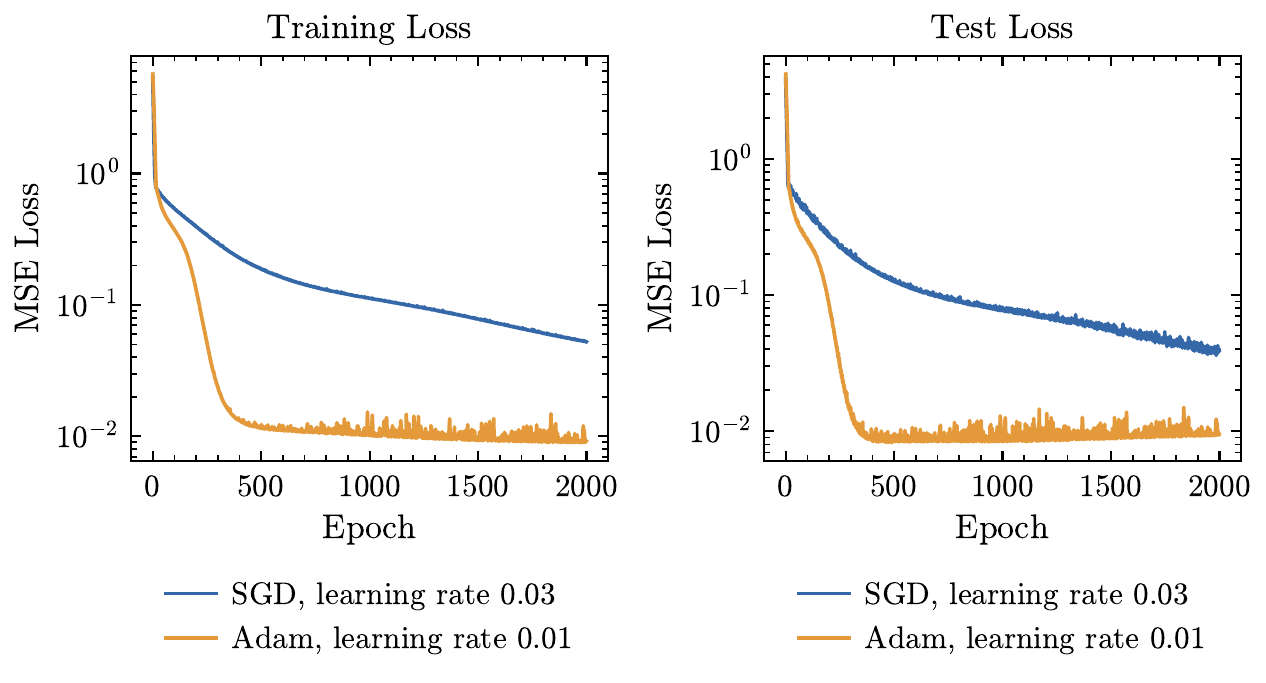}}%
  {\includegraphics[height=\tutorialheight]{tutorials/15a/fig03.pdf}}
\begin{Verbatim}[fontsize=\footnotesize, xleftmargin=5mm]
SGD, last 100 epochs: training 0.0545, test 0.0399
Adam, last 100 epochs: training 0.0094, test 0.0097
\end{Verbatim}

\subsection*{Understanding the Optimizer Comparison}

With these settings, Adam reaches a far lower training and test loss than gradient descent in the same number of epochs. In the plot, both losses fall together until about epoch 60. Then the loss of gradient descent falls slowly and steadily, because its step is proportional to the gradient and the gradient is small. Adam divides the step by the root mean square of the recent gradients. Its step does not shrink in the same way, and its loss reaches the noise level of the data after about 500 epochs.

Both optimizers use a different mini-batch for each update, so both losses fluctuate. The fluctuation of Adam is larger, because its steps stay large.

After 2000 epochs the test loss of gradient descent is still about four times the noise variance. The test loss of Adam is close to the noise variance.

\subsection*{Learning Rate Schedules}

Each optimizer above used one fixed learning rate. A fixed rate is a compromise. A large rate makes fast progress at the start, but it makes the weights jump around the minimum at the end. A small rate is precise, but it is slow. A learning rate schedule changes the rate during the training (Section 7.8).

Cosine annealing lowers the learning rate from \(\eta_0\) to a minimum \(\eta_{\rm min}\) along a cosine curve over a cycle of \(T\) steps,

\[\eta^{(\tau)} = \eta_{\rm min} + \frac{1}{2}(\eta_0 - \eta_{\rm min})\left(1 + \cos\frac{\tau\pi}{T}\right),\]

where \(\tau\) counts the steps since the start of the cycle. The curve falls slowly at first, fastest in the middle and slowly again at the end. At the end of a cycle the rate returns to \(\eta_0\), and a new cycle starts. The name comes from annealing in metallurgy, where a metal that is cooled slowly settles into an ordered structure. The high rates explore the parameter space, and the low rates improve the solution.

PyTorch provides this schedule as \texttt{optim.\allowbreak{}lr\_\allowbreak{}scheduler.\allowbreak{}CosineAnnealingWarmRestarts}. The argument \texttt{T\_0} is the length of the first cycle in epochs, and \texttt{eta\_\allowbreak{}min} is the minimum rate. The scheduler takes \(\eta_0\) from the learning rate of the optimizer. It changes the rate each time that we call \texttt{scheduler.\allowbreak{}step()}, which we do once after each epoch.

We train two networks from the same initial weights with Adam for 2000 epochs, with the mini-batches of the last section. One keeps the learning rate 0.01. The other uses cycles of 500 epochs, from 0.01 down to 0.0001. The left panel shows the two learning rates. The right panel shows the training losses as solid lines and the test losses as dashed lines. We also print the final losses of each network.

\tutorialbarrier
\begin{codecell}

\end{codecell}
\tutorialplot{\includegraphics[width=6.500in,height=0.75\textheight,keepaspectratio]{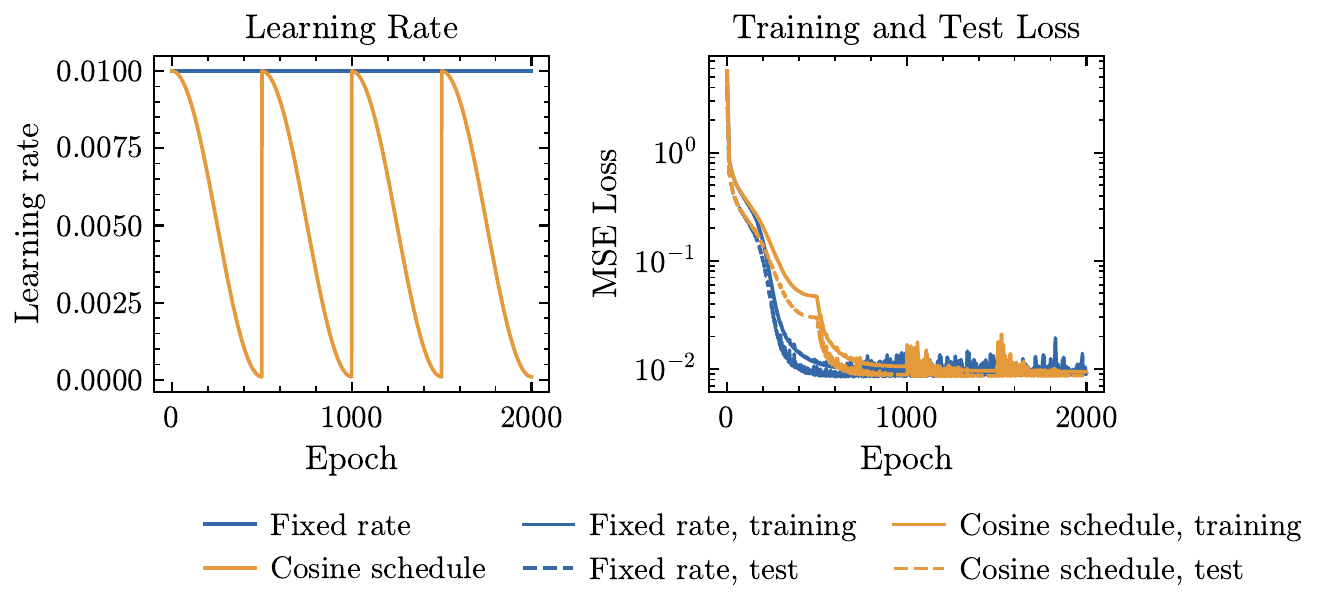}}%
  {\includegraphics[height=\tutorialheight]{tutorials/15a/fig04.pdf}}
\begin{Verbatim}[fontsize=\footnotesize, xleftmargin=5mm]
Fixed rate, last 100 epochs: training 0.0097, test 0.0094
Cosine schedule, last 100 epochs: training 0.0094, test 0.0087
\end{Verbatim}

The learning rate of the cosine run falls in four cycles and jumps back to 0.01 at each restart. In the first cycle its loss falls more slowly than that of the fixed rate, because its rate is lower. At the end of each cycle, when the rate is small, its loss is smooth. The loss of the fixed rate keeps fluctuating because of the mini-batches, and the loss of the cosine run fluctuates again after a restart.

The two runs end at about the same training loss. The test loss of the cosine run is a little lower, 0.0087 against 0.0094 in the means of the last 100 epochs. The gain is small for this simple problem. Annealing removes the noise of the mini-batches at the end of a cycle, and this matters more for problems with many minima. Tutorial 15d uses one cycle of a cosine decay for a larger network.

\subsection*{Activation Functions and Initialization in PyTorch}

The hidden units of our networks used the sigmoid. The sigmoid saturates, so its gradient becomes small for large inputs, which is the vanishing gradient problem above. Most networks therefore use the Rectified Linear Unit (ReLU) or the Gaussian Error Linear Unit (GELU) in their hidden layers (Section 15.5).

ReLU is \(\max(0, x)\). GELU is \(x\,\Phi(x)\), where \(\Phi\) is the cumulative distribution function of the standard normal distribution. PyTorch provides them as \texttt{nn.\allowbreak{}ReLU()} and \texttt{nn.\allowbreak{}GELU()}, and we switch between them by changing one line of the network. We print the three functions at five inputs.

\tutorialbarrier
\begin{codecell}

\end{codecell}
\begin{Verbatim}[fontsize=\footnotesize, xleftmargin=5mm]
Sigmoid  [0.119 0.269 0.5   0.731 0.881]
ReLU     [0. 0. 0. 1. 2.]
GELU     [-0.046 -0.159  0.     0.841  1.954]
\end{Verbatim}

ReLU is zero for negative inputs and equal to the input for positive inputs. Its gradient is 1 for positive inputs, so it does not shrink as the gradient of the sigmoid does. GELU is smooth near zero and approaches ReLU for large inputs.

The initial weights matter too, although the textbook does not cover them. If they are too small, the signal shrinks from layer to layer, and if they are too large, it grows. The He initialization, also called Kaiming initialization, draws the weights of a layer with the standard deviation \(\sqrt{2/n_{\rm in}}\). Here \(n_{\rm in}\) is the number of inputs of the layer.

ReLU sets about half of the values to zero, which halves the variance of the signal, and the factor 2 compensates for this. In PyTorch, \texttt{nn.\allowbreak{}init.\allowbreak{}kaiming\_\allowbreak{}normal\_} applies it to the weights of a layer. The cell passes random inputs through a ReLU network of 20 layers with 100 units each. It repeats this for the weight variances \(1/n_{\rm in}\), \(2/n_{\rm in}\) and \(4/n_{\rm in}\), and it prints the standard deviation of the output.

\tutorialbarrier
\begin{codecell}

\end{codecell}
\begin{Verbatim}[fontsize=\footnotesize, xleftmargin=5mm]
Variance 1/n_in: standard deviation of the output = 0.000391
He, 2/n_in: standard deviation of the output = 0.776
Variance 4/n_in: standard deviation of the output = 519
\end{Verbatim}

With the variance \(1/n_{\rm in}\) the output shrinks to almost zero after 20 layers, and with \(4/n_{\rm in}\) it grows by orders of magnitude. The He scale keeps the output at a size similar to the input. The factor 2 belongs to ReLU. A sigmoid layer keeps its output between 0 and 1 for any scale of the weights. The sigmoid networks above therefore did not need it, and they used the default of \texttt{nn.\allowbreak{}Linear}. The He initialization matters for deep ReLU networks.

\subsection*{GPU Acceleration in PyTorch}

So far all computations ran on the CPU. A GPU runs the matrix operations of a network much faster (Section 15.4), which helps large models. To use a GPU we move the model and the data to the device with \texttt{.\allowbreak{}to(device)}, where \texttt{device} names the GPU or the CPU.

The rest of the code stays the same, and we move results back to the CPU before NumPy or matplotlib use them. We move the model and the data to the device, run a forward pass and move the result back. This runs on the CPU when no GPU is available.

\tutorialbarrier
\begin{codecell}

\end{codecell}
\begin{Verbatim}[fontsize=\footnotesize, xleftmargin=5mm]
Device: cpu, predictions on the CPU with shape (160, 1)
\end{Verbatim}

\section*{Summary}

In this tutorial we went from backpropagation in NumPy to the automatic differentiation of PyTorch. Our main lessons follow.

\begin{enumerate}
\def\labelenumi{\arabic{enumi}.}
\tightlist
\item
  \textbf{Backpropagation.} The forward pass stores the intermediate values. The backward pass multiplies the local derivatives with the chain rule, from the loss back to each parameter. One backward pass gives the gradient with respect to all parameters.
\item
  \textbf{Automatic differentiation.} It differs from symbolic differentiation, which builds a formula, and from finite differences, which need one extra forward pass for each parameter. PyTorch records each operation on tensors that require gradients in a computational graph. We write the forward pass, and the library supplies the backward pass. The same code runs on a GPU after the model and the data move there.
\item
  \textbf{Optimizers.} Gradient descent steps along the scaled gradient with one learning rate. Adam sets the step of each parameter from the history of its gradients. Stochastic gradient descent uses the gradient of a random mini-batch in place of all points. A schedule such as cosine annealing changes the rate during the training.
\item
  \textbf{Activations and initialization.} ReLU and GELU replace the sigmoid in the hidden layers of most networks. The He initialization scales the weights of a ReLU layer by \(\sqrt{2/n_{\rm in}}\), which keeps the size of the signal the same from layer to layer.
\end{enumerate}

The chain rule, the forward and backward passes and a gradient-based optimizer are most of what even very complex models need. PyTorch supplies the backward pass and modules, which lowers the barrier to building them.

\tutorialchapter{15b}{Autoencoders for Galaxy Images}
\noindent\textit{Companion material for Chapter 15}

\section*{Introduction}

A survey takes images of more galaxies than we can inspect, and each image has thousands of pixels. In Tutorial 10 we reduced such images with principal component analysis (PCA, Chapter 10). PCA rebuilds each image from a few components. It assumes that an image is a linear combination of them, so it can miss subtle structure (Section 15.10). Because galaxies have different orientations, elongation can spread over several components (Section 10.9).

An autoencoder (Section 15.10) keeps the aim of PCA, a short code that rebuilds the image with a small error. It replaces the linear maps by two neural networks. The encoder \(e(\mathbf{x})\) maps an image \(\mathbf{x}\) to a latent vector \(\mathbf{z}\), which is the short code. The decoder \(d(\mathbf{z})\) rebuilds the image from \(\mathbf{z}\). The bottleneck of a few numbers makes the network keep the structure that the reconstruction needs. We train both networks together to minimize the reconstruction error.

A network with many weights can memorize a small sample (Section 15.3). We therefore keep validation images to monitor the training and test images to score the result.

In this tutorial we train an autoencoder with 32 latent values on the galaxy images of Tutorial 10. We first ask whether its nonlinear code keeps structure that PCA with 32 components misses, and we look at test galaxies for this. We then probe the latent space by decoding single latent values and by interpolating between two galaxies. Finally we use the reconstruction error to find unusual galaxies.

\section*{Prerequisites and Learning Objectives}

\subsection*{Prerequisites}

This tutorial assumes the following.

\begin{itemize}
\tightlist
\item
  Neural networks, with fully connected layers, activation functions, backpropagation and autoencoders (Chapter 15)
\item
  Basic PyTorch, with tensors, modules and a training loop (Tutorial 15a)
\item
  Principal component analysis and the singular value decomposition (Chapter 10)
\item
  Training, validation and test sets (Chapter 4)
\end{itemize}

\subsection*{Learning Objectives}

After this tutorial you can do the following.

\begin{itemize}
\tightlist
\item
  Build an autoencoder in PyTorch with an encoder, a decoder and a bottleneck
\item
  Train it with mini-batches and use early stopping on the validation loss
\item
  Recognize overfitting from the losses and from the reconstructions of training and validation images
\item
  Compare its reconstructions with those of PCA on test images
\item
  Probe the latent space by decoding single latent values and by interpolating between two images
\item
  Use the reconstruction error to find unusual images
\end{itemize}

We begin by importing the libraries and setting the plot style.

\tutorialbarrier
\begin{codecell}

\end{codecell}

\section*{The Galaxy Images}

We use the galaxy images of Tutorial 10. They are cutouts from the SDSS images of Galaxy Zoo 2, as in \href{https://ui.adsabs.harvard.edu/abs/2026ApJS..282...25M/abstract}{Ma et al.~2026}. We subsample each image to 64 × 64 pixels with values between 0 and 1. Each galaxy is therefore a point in a space of 4,096 dimensions. We load the images and print their shape and pixel values.

\tutorialbarrier
\begin{codecell}
\begin{Verbatim}[commandchars=\\\{\},fontsize=\small]
\PYG{c+c1}{\PYGZsh{} Load the galaxy images, one row for each galaxy}
\PYG{n}{galaxy\PYGZus{}images} \PYG{o}{=}\FancyVerbBreakAfterBreak{} \PYG{n}{np}\PYG{o}{.}\FancyVerbBreakAfterBreak{}\PYG{n}{load}\PYG{p}{(}\FancyVerbBreakAfterBreak{}\PYG{l+s+s1}{\PYGZsq{}}\PYG{l+s+s1}{../data/dataset\PYGZus{}images\PYGZus{}galaxy.npy}\PYG{l+s+s1}{\PYGZsq{}}\PYG{p}{)}

\PYG{c+c1}{\PYGZsh{} Print the shape (galaxies, pixels) and the pixel value statistics}
\PYG{n+nb}{print}\PYG{p}{(}\FancyVerbBreakAfterBreak{}\PYG{l+s+s2}{\PYGZdq{}}\PYG{l+s+s2}{Shape of the galaxy images =}\PYG{l+s+s2}{\PYGZdq{}}\PYG{p}{,}\FancyVerbBreakAfterBreak{} \PYG{n}{galaxy\PYGZus{}images}\PYG{o}{.}\FancyVerbBreakAfterBreak{}\PYG{n}{shape}\PYG{p}{)}
\PYG{n+nb}{print}\PYG{p}{(}\FancyVerbBreakAfterBreak{}\PYG{l+s+sa}{f}\PYG{l+s+s2}{\PYGZdq{}}\PYG{l+s+s2}{Pixel values: minimum = }\PYG{l+s+si}{\PYGZob{}}\PYG{n}{np}\PYG{o}{.}\FancyVerbBreakAfterBreak{}\PYG{n}{min}\PYG{p}{(}\FancyVerbBreakAfterBreak{}\PYG{n}{galaxy\PYGZus{}images}\PYG{p}{)}\PYG{l+s+si}{:}\PYG{l+s+s2}{.3f}\PYG{l+s+si}{\PYGZcb{}}\FancyVerbBreakAfterBreak{}\PYG{l+s+s2}{, maximum = }\PYG{l+s+si}{\PYGZob{}}\PYG{n}{np}\PYG{o}{.}\FancyVerbBreakAfterBreak{}\PYG{n}{max}\PYG{p}{(}\FancyVerbBreakAfterBreak{}\PYG{n}{galaxy\PYGZus{}images}\PYG{p}{)}\PYG{l+s+si}{:}\PYG{l+s+s2}{.3f}\PYG{l+s+si}{\PYGZcb{}}\FancyVerbBreakAfterBreak{}\PYG{l+s+s2}{, }\PYG{l+s+s2}{\PYGZdq{}}
      \PYG{l+s+sa}{f}\PYG{l+s+s2}{\PYGZdq{}}\PYG{l+s+s2}{mean = }\PYG{l+s+si}{\PYGZob{}}\PYG{n}{np}\PYG{o}{.}\FancyVerbBreakAfterBreak{}\PYG{n}{mean}\PYG{p}{(}\FancyVerbBreakAfterBreak{}\PYG{n}{galaxy\PYGZus{}images}\PYG{p}{)}\PYG{l+s+si}{:}\PYG{l+s+s2}{.3f}\PYG{l+s+si}{\PYGZcb{}}\FancyVerbBreakAfterBreak{}\PYG{l+s+s2}{\PYGZdq{}}\PYG{p}{)}
\end{Verbatim}
\end{codecell}
\begin{Verbatim}[fontsize=\footnotesize, xleftmargin=5mm]
Shape of the galaxy images = (1500, 4096)
Pixel values: minimum = 0.000, maximum = 0.899, mean = 0.065
\end{Verbatim}

We show eight galaxies in a grid. Each tile has its own brightness scale.

\tutorialbarrier
\begin{codecell}

\end{codecell}
\tutorialplot{\includegraphics[width=6.500in,height=0.75\textheight,keepaspectratio]{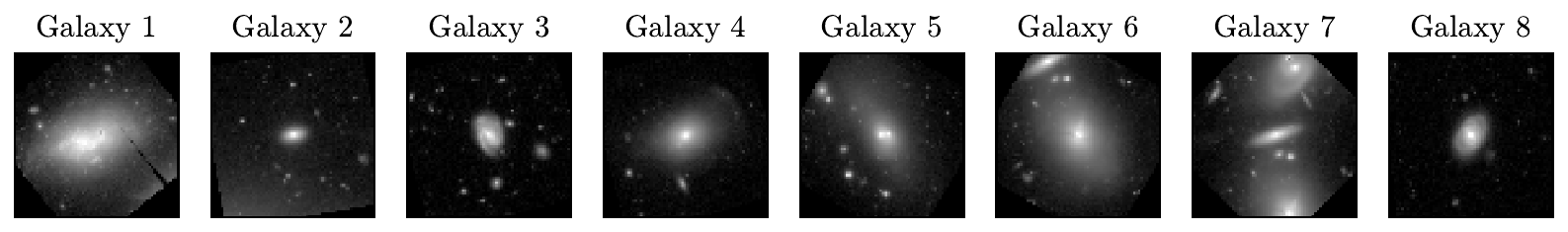}}%
  {\includegraphics[height=\tutorialheight]{tutorials/15b/fig01.pdf}}

The galaxies differ in size, shape and brightness. Some cutouts hold bright neighbors, and some have black corners from the edge of the image frame.

\section*{The Autoencoder Architecture}

We have 1,500 images of 4,096 pixels each, and we want a code of 32 numbers for each of them. PCA describes the variation of the images with linear combinations of the pixel values. An autoencoder can learn nonlinear transformations instead.

\subsection*{Mathematical Formulation}

We follow the notation of Chapter 15. The encoder \(e(\mathbf{x})\) maps an image \(\mathbf{x} \in \mathbb{R}^{D}\) to a latent vector \(\mathbf{z} \in \mathbb{R}^{M}\). The decoder \(d(\mathbf{z})\) maps \(\mathbf{z}\) back to \(\mathbb{R}^{D}\). Here \(D = 4096\) is the number of pixels and \(M = 32\) is the number of latent dimensions. The reconstruction of an image is

\[\hat{\mathbf{x}} = d(e(\mathbf{x})).\]

We minimize the mean squared error (MSE) between the input and its reconstruction,

\[\mathcal{L} = \frac{1}{ND}\sum_{i=1}^{N}\|\mathbf{x}_i - d(e(\mathbf{x}_i))\|^2,\]

where \(N\) is the number of images. This is the loss of Section 15.10 divided by \(D\), which is what \texttt{nn.\allowbreak{}MSELoss} computes.

\subsection*{Network Design}

Our architecture is symmetric. The encoder narrows the image from 4,096 values to 512, 128 and 32 values, and the decoder widens the code again. The bottleneck is the constraint that drives the learning. Without it the network could copy the input to the output (Section 15.10).

Each hidden layer applies a linear map followed by an activation function (Section 15.5). We use the GELU, a smooth variant of the ReLU, as in Tutorial 15a. Without the activations both networks would be linear, and the autoencoder would be equivalent to PCA (Section 15.10). The latent vector passes through a ReLU, so its values are not negative. The output of the decoder is a linear map without activation, so the reconstructed pixels can take any value.

We choose \(M = 32\) latent dimensions, so 4,096 pixel values become 32 numbers, a compression of 128 to 1. Tutorial 10 found that 23 components keep 80 percent of the variance of these galaxies, so 32 is a little more.

We now define the network as a PyTorch module and count its parameters. The class \texttt{nn.\allowbreak{}Sequential} applies its layers in order.

\tutorialbarrier
\begin{codecell}

\end{codecell}
\begin{Verbatim}[fontsize=\footnotesize, xleftmargin=5mm]
Parameters = 4,338,976 (encoder 2,167,456, decoder 2,171,520)
\end{Verbatim}

\section*{Training the Autoencoder}

PCA has a closed-form solution through the eigendecomposition (Chapter 10). An autoencoder needs iterative optimization with gradient descent, which takes more computing time. We first split the images, and we then choose the optimizer.

\subsection*{Data Preparation and Splits}

A network with many parameters can overfit. It then memorizes the training images and learns features that do not generalize (Section 15.3). We split the 1,500 galaxy images at random into 60 percent training images, 20 percent validation images and 20 percent test images (Section 4.7). This gives 900, 300 and 300 images. Our network has 4.3 million parameters, and the 900 training images have 3.7 million pixel values.

The split has two steps. The first sets aside 20 percent test images. The second divides the remaining 80 percent into 75 percent training images and 25 percent validation images, so that the three sets have different roles.

\begin{itemize}
\tightlist
\item
  The training images update the weights, and the network learns from them only.
\item
  The validation images monitor the overfitting and choose the epoch whose weights we keep. The network never updates on them, but our choice depends on them.
\item
  The test images take no part in the training or in any choice, so they show how the final network does on new images.
\end{itemize}

We use the pixel values as they are, so no statistic of the test images enters the preparation.

We train with mini-batches (Section 15.3). They reduce the cost of an update, and their noise can help the optimization to escape shallow minima. We use 60 images in a batch, so the training images make 15 equal batches. One epoch is one pass through all training images, which is 15 updates. A \texttt{DataLoader} splits the training images into mini-batches and shuffles them in each epoch, as in Tutorial 15a.

\tutorialbarrier
\begin{codecell}

\end{codecell}
\begin{Verbatim}[fontsize=\footnotesize, xleftmargin=5mm]
Training 900, validation 300, test 300 images, 15 batches per epoch
\end{Verbatim}

\subsection*{Optimization Strategy}

The split fixes which images the network sees, and the optimizer decides how it learns from them. We use the Adam optimizer of Tutorial 15a (Section 15.3). It rescales the step of each parameter by the size of its recent gradients. The learning rate starts at 0.001 and falls to zero along a cosine curve over 200 epochs (Tutorial 15a). A high rate at the start trains fast, and a low rate at the end lets the loss settle.

Each epoch has four steps.

\begin{enumerate}
\def\labelenumi{\arabic{enumi}.}
\tightlist
\item
  Update the weights once for each of the 15 mini-batches of training images.
\item
  Lower the learning rate by one step along the cosine curve.
\item
  Compute the MSE of all training images and of all validation images for the same weights. The losses of the mini-batches are not enough, because the weights change during the epoch.
\item
  Keep a copy of the weights if the validation loss is the lowest so far.
\end{enumerate}

The fourth step is early stopping (Section 15.3), except that we always train all 200 epochs. We also keep the reconstructions of four training images and four validation images at the epochs 1, 5, 20, 75 and 200.

\tutorialbarrier
\begin{codecell}

\end{codecell}

We plot the training loss and the validation loss against the epoch. A dashed line marks the epoch with the lowest validation loss. The first epochs lie above the top of the plot.

\tutorialbarrier
\begin{codecell}

\end{codecell}
\tutorialplot{\includegraphics[width=4.392in,height=0.75\textheight,keepaspectratio]{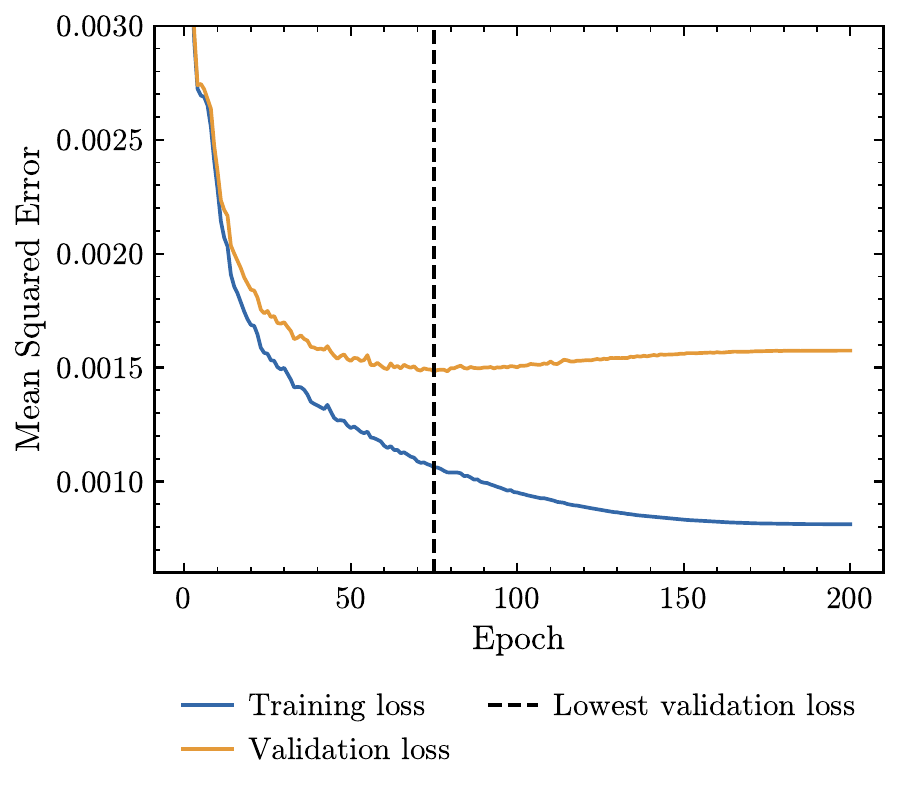}}%
  {\includegraphics[height=\tutorialheight]{tutorials/15b/fig02.pdf}}
\begin{Verbatim}[fontsize=\footnotesize, xleftmargin=5mm]
Lowest validation loss = 0.001482 at epoch 75
Final training loss = 0.000812, validation loss = 0.001574
\end{Verbatim}

Both losses fall quickly in the first epochs, and the training loss keeps falling. The validation loss reaches its lowest value at epoch 75 and then rises slowly. The gap between the two losses grows because the network fits the training images better than new images, which is overfitting. Both curves flatten at the end, because the cosine curve brings the learning rate to zero. We keep the weights of epoch 75 and not those of epoch 200, which is early stopping (Section 15.3).

\subsection*{Reconstructions During the Training}

The losses are numbers. We now look at what they mean for the images. The figure shows four training images and four validation images. The top row holds the originals, and the rows below hold the reconstructions at the epochs that we kept. The network never updates on the validation images. We also print the loss of all training and all validation images at these epochs.

\tutorialbarrier
\begin{codecell}

\end{codecell}
\begin{Verbatim}[fontsize=\footnotesize, xleftmargin=5mm]
Epoch 1: training images 0.00429, validation images 0.00443
Epoch 5: training images 0.00269, validation images 0.00275
Epoch 20: training images 0.00169, validation images 0.00184
Epoch 75: training images 0.00106, validation images 0.00148
Epoch 200: training images 0.00081, validation images 0.00157
\end{Verbatim}
\tutorialplot{\includegraphics[width=6.500in,height=0.75\textheight,keepaspectratio]{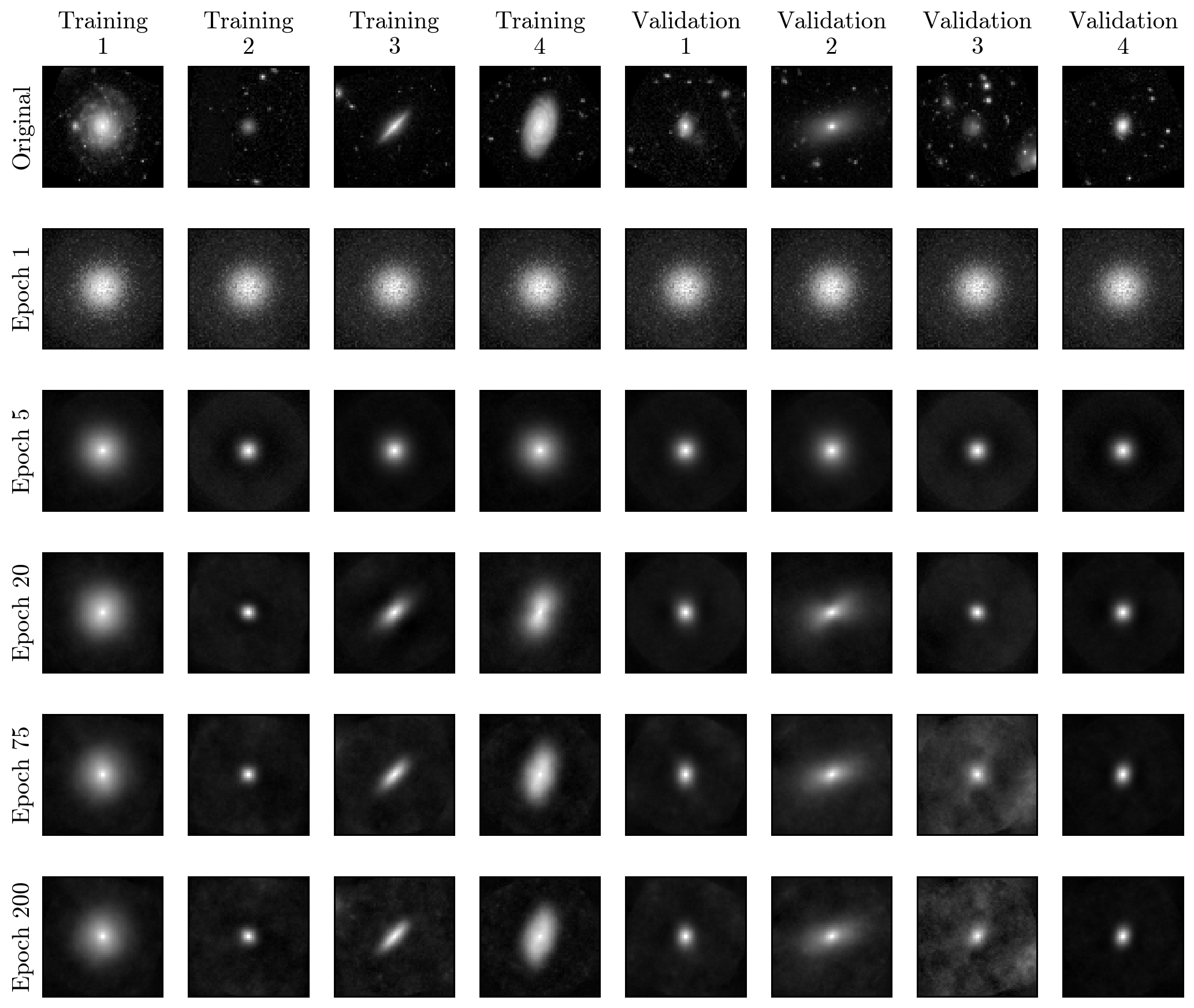}}%
  {\includegraphics[height=\tutorialheight]{tutorials/15b/fig03.png}}

At epoch 1 the output is noise around a blob. At epoch 5 it is a round source. At epoch 20 it has the elongation and the orientation of the galaxies, and at epoch 75 it is close to the final result. From epoch 75 to 200 the training loss falls and the validation loss rises. The training images gain faint specks in the background and near the edges. The network starts to store details of the training images, which is overfitting that we can now see.

\section*{Reconstruction Quality}

The training is done, and we keep the weights of the best epoch. We now compare the autoencoder with PCA on test images, to see whether its nonlinear code keeps structure that the linear PCA misses. PCA with 32 components is the baseline with the same number of latent values. In Tutorial 10 we fitted PCA to all galaxies and rebuilt the same galaxies. Here we fit PCA to the training images only, and we rebuild the test images as for the autoencoder.

We first restore the weights of the epoch with the lowest validation loss. We then reconstruct the test images.

\tutorialbarrier
\begin{codecell}
\begin{Verbatim}[commandchars=\\\{\},fontsize=\small]
\PYG{c+c1}{\PYGZsh{} Restore the weights of the epoch with the lowest validation loss}
\PYG{n}{model}\PYG{o}{.}\FancyVerbBreakAfterBreak{}\PYG{n}{load\PYGZus{}state\PYGZus{}dict}\PYG{p}{(}\FancyVerbBreakAfterBreak{}\PYG{n}{best\PYGZus{}state}\PYG{p}{)}
\PYG{n}{model}\PYG{o}{.}\FancyVerbBreakAfterBreak{}\PYG{n}{eval}\PYG{p}{(}\FancyVerbBreakAfterBreak{}\PYG{p}{)}

\PYG{c+c1}{\PYGZsh{} Reconstruct the test images}
\PYG{k}{with} \PYG{n}{torch}\PYG{o}{.}\FancyVerbBreakAfterBreak{}\PYG{n}{no\PYGZus{}grad}\PYG{p}{(}\FancyVerbBreakAfterBreak{}\PYG{p}{)}\PYG{p}{:}
    \PYG{n}{test\PYGZus{}ae} \PYG{o}{=}\FancyVerbBreakAfterBreak{} \PYG{n}{model}\PYG{p}{(}\FancyVerbBreakAfterBreak{}\PYG{n}{x\PYGZus{}test}\PYG{p}{)}\PYG{o}{.}\FancyVerbBreakAfterBreak{}\PYG{n}{cpu}\PYG{p}{(}\FancyVerbBreakAfterBreak{}\PYG{p}{)}\PYG{o}{.}\FancyVerbBreakAfterBreak{}\PYG{n}{numpy}\PYG{p}{(}\FancyVerbBreakAfterBreak{}\PYG{p}{)}
\end{Verbatim}
\end{codecell}

We fit PCA to the training images with the SVD (Section 10.8), as in Tutorial 10. The fit has four steps.

\begin{enumerate}
\def\labelenumi{\arabic{enumi}.}
\tightlist
\item
  Compute the mean image \(\overline{\mathbf{x}}\) of the training images, and subtract it from every image. The rows of \(\mathbf{X}\) are now the centered images.
\item
  Compute the SVD of the centered images, \(\mathbf{X} = \mathbf{U}\boldsymbol{\Sigma}\mathbf{V}^T\). The columns of \(\mathbf{V}\) are the principal components, the directions in pixel space along which the images vary most.
\item
  Keep the first \(M = 32\) columns of \(\mathbf{V}\), which form the matrix \(\mathbf{V}_M\).
\item
  Project an image onto these components, \(\mathbf{z} = (\mathbf{x} - \overline{\mathbf{x}})\mathbf{V}_M\), and rebuild it from its 32 coefficients (Section 10.1),
\end{enumerate}

\[\hat{\mathbf{x}} = \mathbf{z}\mathbf{V}_M^T + \overline{\mathbf{x}}.\]

The mean image and the components come from the training images only. We apply them unchanged to the test images, so no test image enters the fit. In the code, the rows of \texttt{Vt} are the columns of \(\mathbf{V}\).

\tutorialbarrier
\begin{codecell}

\end{codecell}

We show the first six test galaxies, their PCA reconstructions and their autoencoder reconstructions. Each tile has its own brightness scale.

\tutorialbarrier
\begin{codecell}

\end{codecell}
\tutorialplot{\includegraphics[width=6.500in,height=0.75\textheight,keepaspectratio]{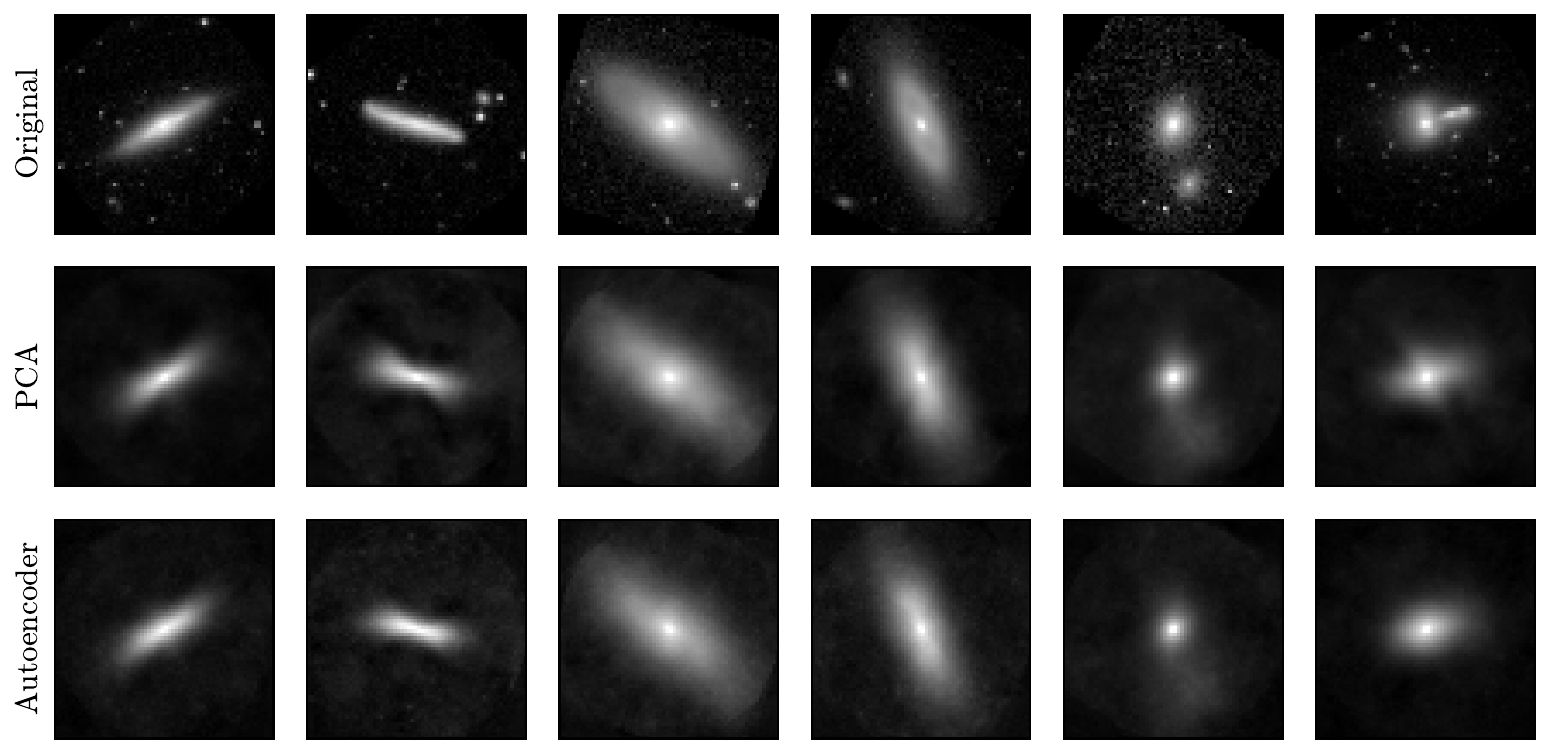}}%
  {\includegraphics[height=\tutorialheight]{tutorials/15b/fig04.pdf}}

Both methods keep the size, elongation and orientation of the main galaxy, and both drop the background noise. The autoencoder follows the structure more closely. Its thin disks are narrower and brighter than those of PCA, which spreads them out. The thin disk of the second galaxy becomes a bow tie in PCA and a bar in the autoencoder. The bow tie is the shape of the quadrupole eigengalaxies of Tutorial 10. In the last two galaxies PCA smears the light of a neighbor into the main galaxy, and the autoencoder keeps a compact core.

The autoencoder looks sharper in these six galaxies, but six are a small sample. We therefore measure the detail that each method keeps for all 300 test images. Both sets of reconstructions are smoother than the originals, because 32 numbers cannot hold the fine structure. We measure the sharpness of an image \(I\) with its gradient magnitude at each pixel,

\[g = \sqrt{\left(\frac{\partial I}{\partial x}\right)^2 + \left(\frac{\partial I}{\partial y}\right)^2},\]

where the derivatives are the differences between neighboring pixels. The value \(g\) is large at a sharp edge, where the brightness changes quickly from one pixel to the next. It is small in a smooth region. The sharpness of an image is the mean of \(g\) over its pixels.

We compute the sharpness of every test image for the original, the PCA reconstruction and the autoencoder reconstruction. We print the mean over the test images for each of the three. We also count the test images for which the autoencoder is sharper than PCA.

\tutorialbarrier
\begin{codecell}

\end{codecell}
\begin{Verbatim}[fontsize=\footnotesize, xleftmargin=5mm]
Mean gradient magnitude: original 0.0205, PCA 0.0106, autoencoder 0.0118
The autoencoder is sharper than PCA for 263 of 300 test images
\end{Verbatim}

The originals are the sharpest, because their fine structure and their noise give many sharp edges. Both reconstructions are about half as sharp, because they keep the large structure and drop most of the fine detail. The autoencoder is slightly sharper than PCA on average, and it is sharper for most of the test images. A larger sharpness means that a reconstruction keeps more detail.

\section*{The Latent Space}

The reconstructions show that the 32 numbers of the code hold the main structure of a galaxy. We now ask what these numbers mean. The latent dimensions of an autoencoder come from the optimization. We probe the code through the decoder in two ways. We first decode codes with a single non-zero value, and we then change single values in the code of a real galaxy.

\subsection*{Decoding One Latent Value}

To see what one latent dimension does, we decode a latent vector that has the value \(s\) in dimension \(k\) and zeros elsewhere (Section 15.10),

\tutorialdisplay{\mathbf{z}_k(s) = s\,\mathbf{e}_k = (0, \dots, 0, s, 0, \dots, 0),}{\hat{\mathbf{x}}_k(s) = d\left(\mathbf{z}_k(s)\right).}

Here \(\mathbf{e}_k\) is the unit vector with the value 1 in dimension \(k\). The image \(\hat{\mathbf{x}}_k(s)\) shows what dimension \(k\) adds to the output of the decoder. We use the two scales \(s = 1\) and \(s = 10\) to see how the output changes when the value grows. The encoder ends with a ReLU, so the latent values are not negative, and we use \(s > 0\). The cell decodes one vector for each of the 32 dimensions.

\tutorialbarrier
\begin{codecell}

\end{codecell}
\tutorialplot{\includegraphics[width=6.500in,height=0.75\textheight,keepaspectratio]{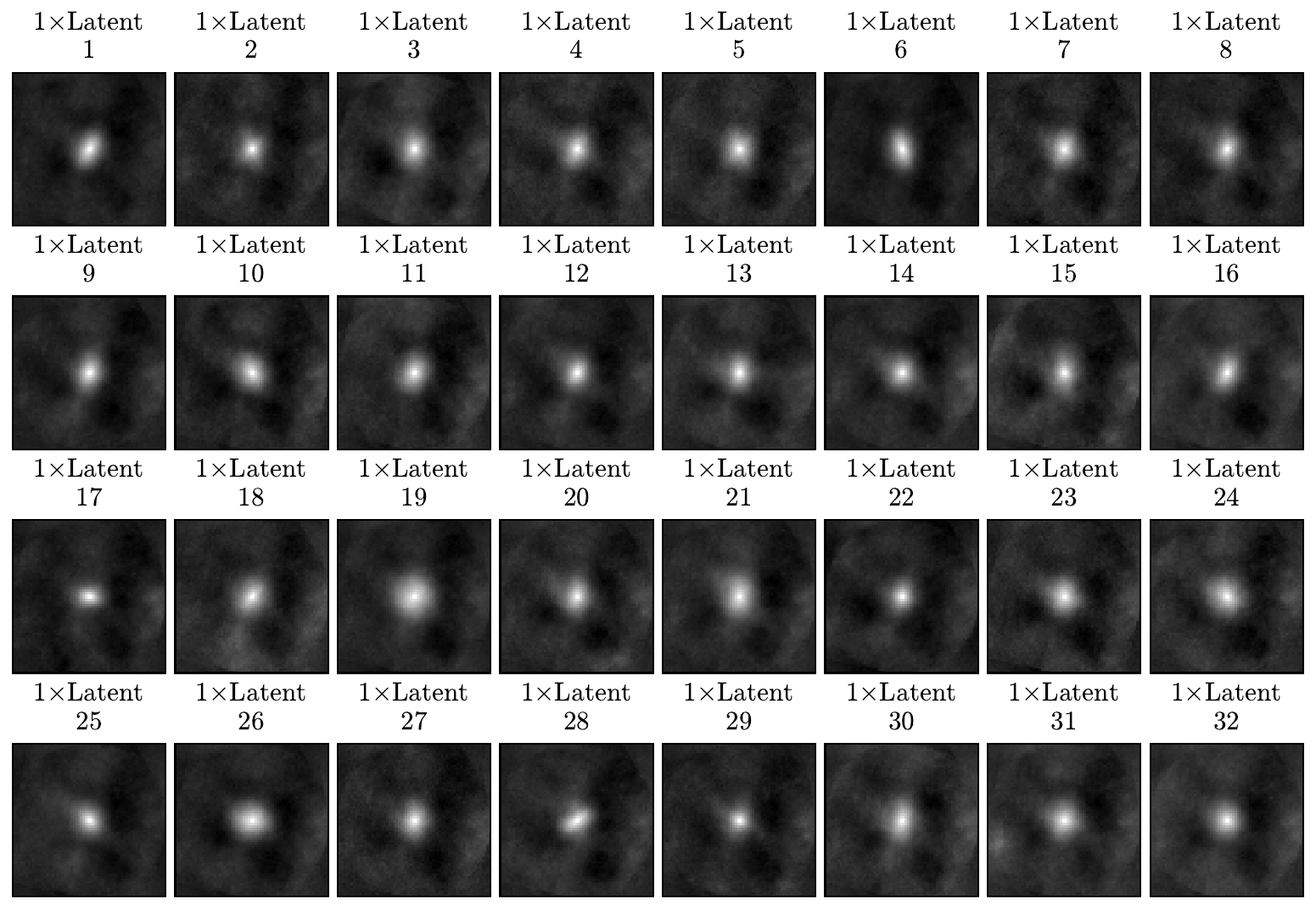}}%
  {\includegraphics[height=\tutorialheight]{tutorials/15b/fig05.pdf}}
\tutorialplot{\includegraphics[width=6.500in,height=0.75\textheight,keepaspectratio]{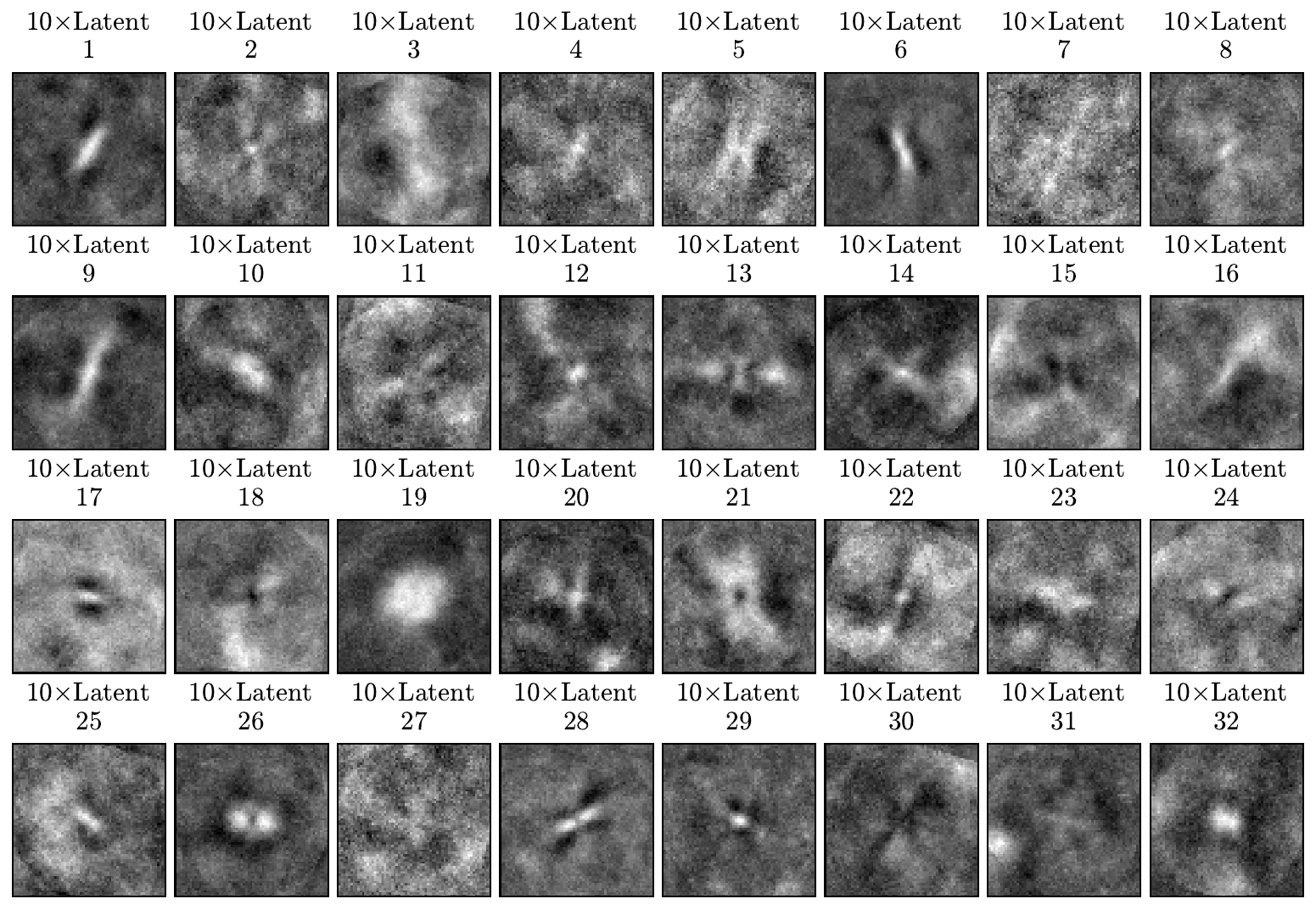}}%
  {\includegraphics[height=\tutorialheight]{tutorials/15b/fig06.pdf}}

The tiles show the decoder output for one latent dimension at a time. Different dimensions give different images. Some show elongated sources with different orientations, and others show rounder sources. The tiles at \(s = 1\) show a compact source on a mottled background. The tiles at \(s = 10\) show more structure, especially in the background.

A latent dimension comes from the optimization, so it need not match a physical property of a galaxy (Section 15.10).

\subsection*{Scaling the Code of a Real Galaxy}

The tiles above come from codes with one non-zero value, and a real galaxy has many non-zero values. We therefore start from the code \(\mathbf{z} = e(\mathbf{x})\) of a real galaxy and change one value at a time. We take galaxy 151. The encoder turns it into a code of 32 numbers. We pick the two latent dimensions with the largest values, because a large value means that the dimension is active for this galaxy.

For a dimension \(k\) and a scale \(c\) we multiply the value \(z_k\) by \(c\) and keep the other 31 values,

\tutorialdisplay{\mathbf{z}'_k(c) = (z_1, \dots, c\,z_k, \dots, z_{32}),}{\hat{\mathbf{x}}_k(c) = d\left(\mathbf{z}'_k(c)\right).}

We decode the changed codes for four scales. A title such as 2×Latent 1 means that \(c = 2\) and \(k = 1\).

\begin{itemize}
\tightlist
\item
  The scale \(c = 0\) removes the dimension.
\item
  The scale \(c = 0.5\) halves it.
\item
  The scale \(c = 1\) leaves the code unchanged, so it gives the reconstruction \(d(e(\mathbf{x}))\) of the galaxy.
\item
  The scale \(c = 2\) doubles it.
\end{itemize}

The first column of the figure shows the original galaxy, and each row follows one dimension.

\tutorialbarrier
\begin{codecell}

\end{codecell}
\begin{Verbatim}[fontsize=\footnotesize, xleftmargin=5mm]
Latent dimensions 1 and 16, with values 2.9 and 2.0
\end{Verbatim}
\tutorialplot{\includegraphics[width=6.500in,height=0.75\textheight,keepaspectratio]{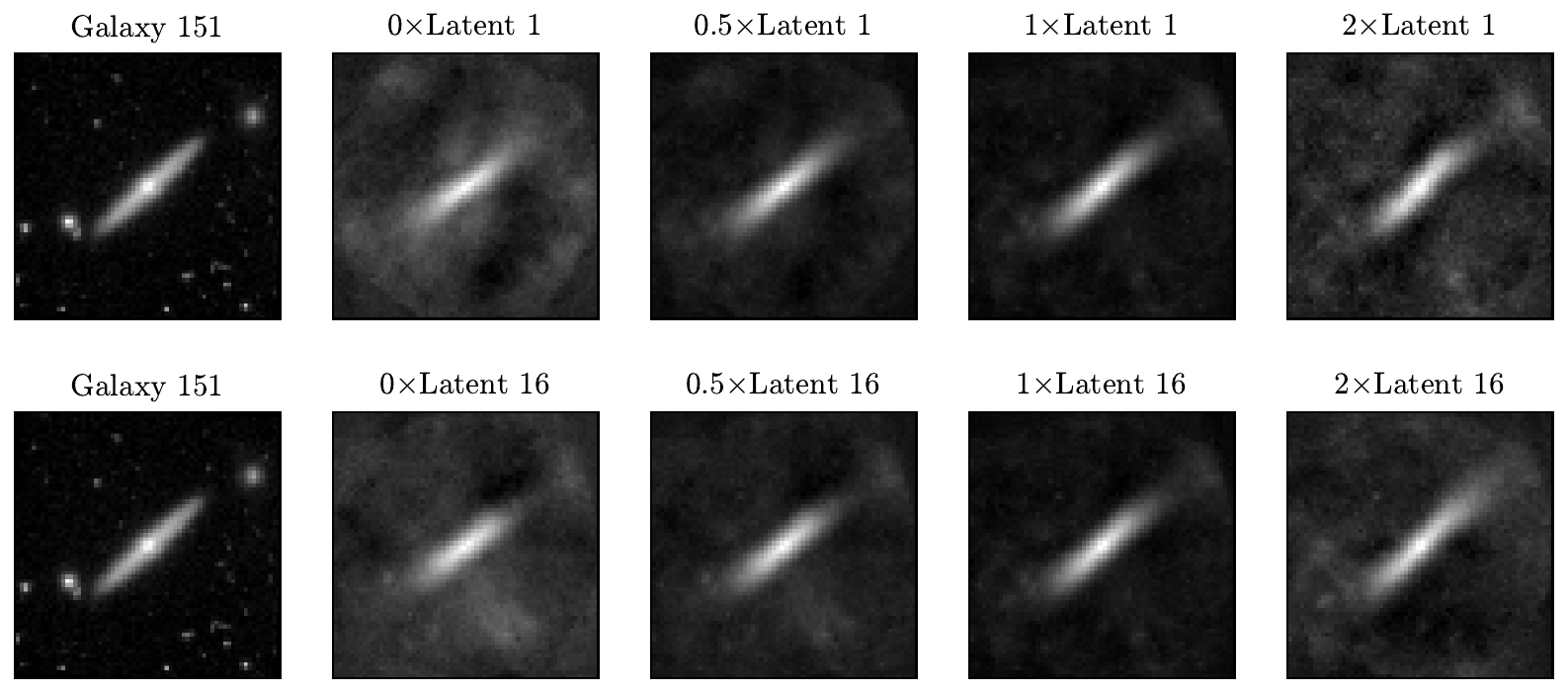}}%
  {\includegraphics[height=\tutorialheight]{tutorials/15b/fig07.pdf}}

Scaling one value changes the length, the width and the brightness of the galaxy, and the background level. The galaxy stays recognizable at every scale, because the other 31 values still describe it. The change is smoother than in the tiles above.

\section*{Two Uses of the Learned Representation}

Both uses treat the code as a description of a galaxy. We look at the interpolation between two galaxies and at the search for unusual galaxies.

\subsection*{Interpolation in Latent Space}

Scaling one value changes a galaxy in one direction. Interpolation instead moves the code from one galaxy to another and asks what lies between them. We encode two galaxies into the codes \(\mathbf{z}_{51} = e(\mathbf{x}_{51})\) and \(\mathbf{z}_{151} = e(\mathbf{x}_{151})\). Galaxy 51 is compact, and galaxy 151 is elongated.

We then take points on the straight line between the two codes and decode each of them,

\tutorialdisplay{\mathbf{z}(\alpha) = (1-\alpha)\,\mathbf{z}_{51} + \alpha\,\mathbf{z}_{151},}{\hat{\mathbf{x}}(\alpha) = d\left(\mathbf{z}(\alpha)\right).}

The number \(\alpha\) runs from 0 at galaxy 51 to 1 at galaxy 151. The first and the last tile of the figure show the two original images. The six tiles in between show \(\hat{\mathbf{x}}(\alpha)\) for six values of \(\alpha\), which the titles give.

\tutorialbarrier
\begin{codecell}

\end{codecell}
\tutorialplot{\includegraphics[width=6.500in,height=0.75\textheight,keepaspectratio]{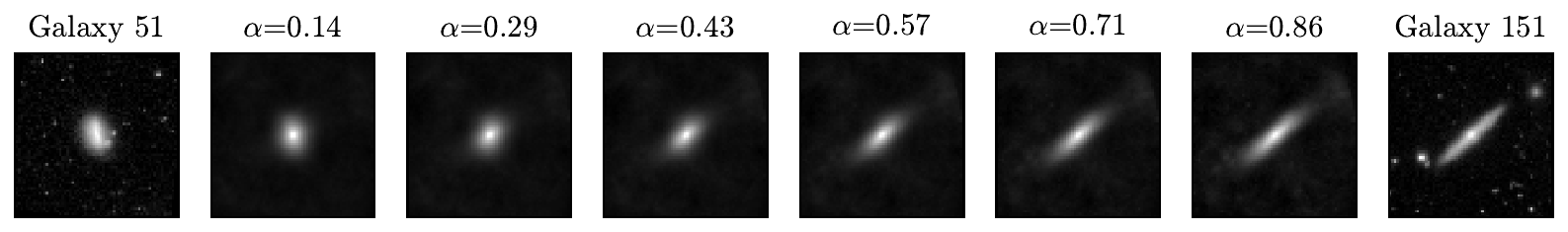}}%
  {\includegraphics[height=\tutorialheight]{tutorials/15b/fig08.pdf}}

The decoded galaxy changes gradually from a round source to an elongated and tilted one. Nearby points on the line give similar galaxies, so the latent space is smooth between these two galaxies.

\subsection*{Anomaly Detection Through Reconstruction Error}

The second use looks at how well the network rebuilds a galaxy. The autoencoder learns from the training galaxies, so it learns to rebuild the typical ones. An image that differs from them is harder to rebuild, and its reconstruction error tends to be larger (Section 15.10). The error is therefore a score for how unusual an image is.

We score the 300 validation images and the 300 test images. We leave out the training images, because the network trained on them and rebuilds them better than new images. The error of an image is its MSE over the \(D = 4096\) pixels, as in the loss,

\[E(\mathbf{x}) = \frac{1}{D}\left\|\mathbf{x} - d(e(\mathbf{x}))\right\|^2.\]

We compute \(E\) for every scored image and print the median. We then pick the eight images with the largest \(E\).

\tutorialbarrier
\begin{codecell}

\end{codecell}
\begin{Verbatim}[fontsize=\footnotesize, xleftmargin=5mm]
Median error = 0.0012
\end{Verbatim}

The figure shows the eight galaxies with the largest errors. Each column has the original image, its reconstruction and the absolute difference between them, where a dark pixel marks a large difference. The title gives the error.

\tutorialbarrier
\begin{codecell}

\end{codecell}
\tutorialplot{\includegraphics[width=6.500in,height=0.75\textheight,keepaspectratio]{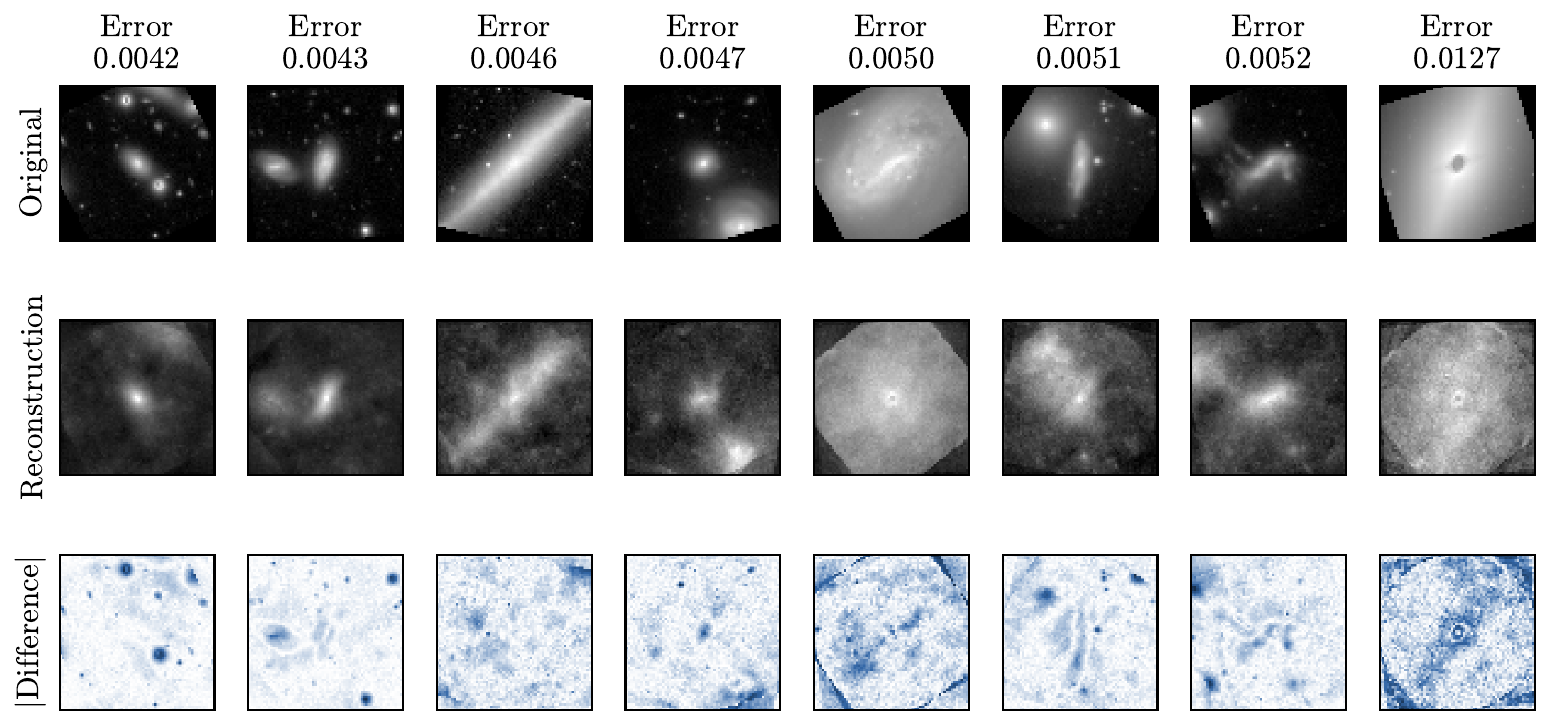}}%
  {\includegraphics[height=\tutorialheight]{tutorials/15b/fig09.pdf}}

The eight largest errors are several times the median error, and the largest is about ten times the median. In five of the eight images, two or more galaxies lie close together, and some show arcs. The reconstruction blurs or misses the fainter sources, and the difference maps show them as dark spots and arcs.

In two images a single large galaxy fills the cutout. In the eighth image a bright, smooth background fills the cutout, and the galaxy at the center is small. The tilted frame edges of three images give dark lines in the difference maps.

Whether such an image is of scientific interest depends on the application. A study of galaxy interactions looks for the pairs and the tidal arcs.

\section*{Limits and Considerations}

Our network has fully connected layers. They treat the pixels as an unordered list and ignore the grid of an image. A convolutional network applies the same small filters at every position. It finds a pattern wherever it appears and builds larger structures from local ones (Section 15.6). In principle it suits images better than our network, and a Transformer is another option (Section 15.6).

\section*{Summary}

In this tutorial we trained an autoencoder on galaxy images. Our main lessons follow.

\begin{enumerate}
\def\labelenumi{\arabic{enumi}.}
\tightlist
\item
  \textbf{Nonlinear dimension reduction.} An autoencoder keeps the aim of PCA, a short code with a small reconstruction error. It replaces the linear maps by an encoder and a decoder. The bottleneck makes the network keep the structure that the reconstruction needs. Without activation functions the autoencoder would be equivalent to PCA.
\item
  \textbf{Architecture and training choices.} The architecture and the training settings shape what the network learns. The size of the bottleneck sets the compression, and the activation, the optimizer and the learning rate schedule affect how well the training works. Validation images are the right data to compare such choices.
\item
  \textbf{Overfitting and early stopping.} A network with many parameters can fit a small sample. The training loss then keeps falling while the validation loss stops falling, and the reconstructions of training images gain details that validation images do not. Early stopping keeps the weights with the lowest validation loss, and the test images show how the network does on new images.
\item
  \textbf{The latent space.} Decoding single latent values and scaling values in the code of a galaxy show what each dimension adds to the image. A latent dimension need not match a physical property.
\item
  \textbf{Interpolation.} The points on the line between two codes decode to galaxies that change shape gradually, so the latent space is smooth between them.
\item
  \textbf{Anomaly detection.} An autoencoder rebuilds what is typical in its training sample. An object that it rebuilds poorly differs from the sample, and what this means depends on the application.
\end{enumerate}

An autoencoder learns a compact nonlinear representation of high-dimensional data from the data itself. Surveys deliver many high-dimensional images and spectra, and a learned representation helps us to compress, explore and screen them.

\tutorialchapter{15c}{Mixture Density Networks for Stellar Lithium Abundances}
\noindent\textit{Companion material for Chapter 15}

\section*{Introduction}

Two stars with the same temperature and age can differ in their lithium abundance. A prediction of the lithium abundance from these two properties therefore needs a range of values and not one number. A standard neural network predicts one number. With the squared error, this number is the mean of the possible outputs (Section 15.12). This mean gives the center of the scatter but not its width.

Protons destroy lithium at temperatures above about 2.5 million kelvin. A star therefore keeps lithium only in its cool outer convection zone, and it loses lithium when convection carries it to hotter layers. We usually measure the lithium of a star by A(Li), which is 12 plus the base-10 logarithm of the lithium to hydrogen number ratio. The unit is the dex, and one dex is a factor of 10.

A(Li) depends on the effective temperature \(T_{\rm eff}\), because cooler stars have deeper convection zones and destroy more lithium. It also falls with the age \(\tau\) of the star. Other properties change the depletion, such as the rotation of the star, its magnetic activity, its accretion history and its metallicity. Stars with similar temperatures and ages can therefore differ in lithium by a factor of several.

A Mixture Density Network (MDN) predicts a probability density and not one number (Section 15.12). A neural network computes the weights, means and widths of a Gaussian mixture from the inputs (Section 11.5). We train it by maximum likelihood (Chapters 4 and 7). The network then models the conditional density \(p(\text{A(Li)}|T_{\rm eff}, \tau)\). The hidden properties of the stars appear as its width and its shape.

We first train an MDN on the two moons, a simulated relation in which some inputs have two outputs. We then train it on the temperature, age and lithium abundance of stars in open clusters, where all stars of a cluster share one age. We compare the conditional densities with the stars, and we plot its predictions for many ages.

\section*{Prerequisites and Learning Objectives}

\subsection*{Prerequisites}

This tutorial assumes the following.

\begin{itemize}
\tightlist
\item
  Python programming with NumPy, matplotlib and PyTorch (Tutorial 15a)
\item
  Neural networks, including feedforward networks, activation functions and gradient-based training (Chapter 15)
\item
  Gaussian mixture models (Chapter 11)
\item
  Maximum likelihood estimation (Chapters 4 and 7)
\end{itemize}

\subsection*{Learning Objectives}

After this tutorial you can do the following.

\begin{itemize}
\tightlist
\item
  Build a neural network whose output is a Gaussian mixture and not a single value
\item
  Explain how the network sets the weights, means and widths of the mixture and keeps them valid
\item
  Train the network with a numerically stable negative log-likelihood
\item
  Explain why one predicted value fails for a one-to-many relation
\item
  Read a conditional density and compare it with the data
\item
  Read the predictions of the network at many ages
\end{itemize}

We begin by importing the libraries and setting the plot style.

\tutorialbarrier
\begin{codecell}

\end{codecell}

\section*{The Two-Moons Dataset}

We first test the method on the two moons, a simulated relation that shows the problem in its simplest form. The data are two interleaved half circles with noise. Each \(x\) between 0 and 1 has two values of \(y\), one on the upper moon and one on the lower moon.

No function \(y = f(x)\) describes this one-to-many relation, however complex it is. A standard neural network gives one \(y\) for each \(x\). Trained with the squared error, it predicts the mean of the two values, which lies between the moons (Section 15.12). This mean minimizes the squared error and describes neither moon.

\tutorialbarrier
\begin{codecell}

\end{codecell}
\tutorialplot{\includegraphics[width=4.260in,height=0.75\textheight,keepaspectratio]{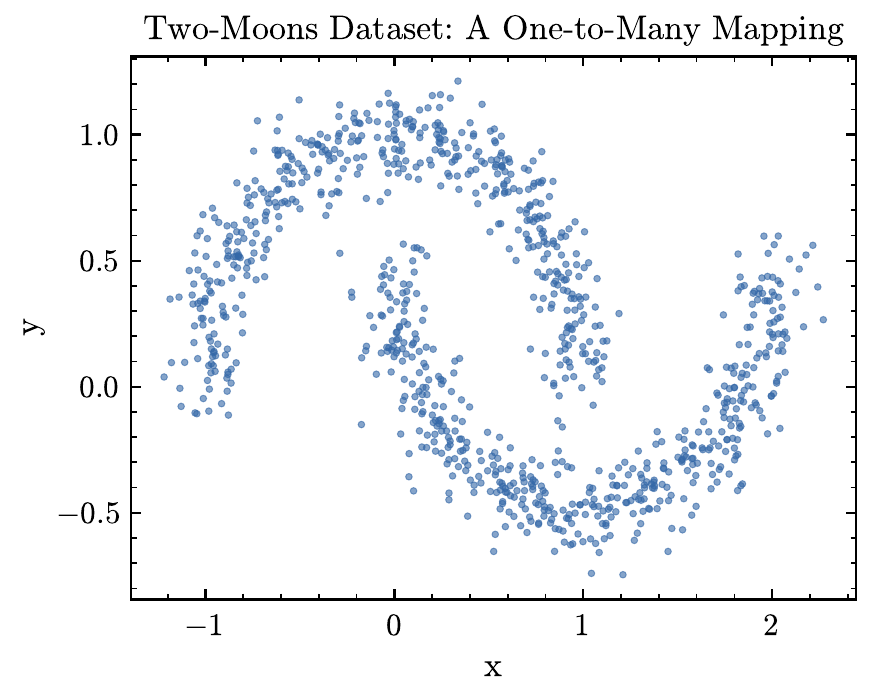}}%
  {\includegraphics[height=\tutorialheight]{tutorials/15c/fig01.pdf}}

We train a standard network with the squared error on the two moons. It has two hidden layers of 20 neurons and one output. We plot its prediction and print the prediction at \(x = 0.5\) next to the mean of the points of each moon near this \(x\).

\tutorialbarrier
\begin{codecell}

\end{codecell}
\begin{Verbatim}[fontsize=\footnotesize, xleftmargin=5mm]
Prediction at x = 0.5: 0.25
Mean of the points of the upper moon: 0.88, of the lower moon: -0.36
\end{Verbatim}
\tutorialplot{\includegraphics[width=4.260in,height=0.75\textheight,keepaspectratio]{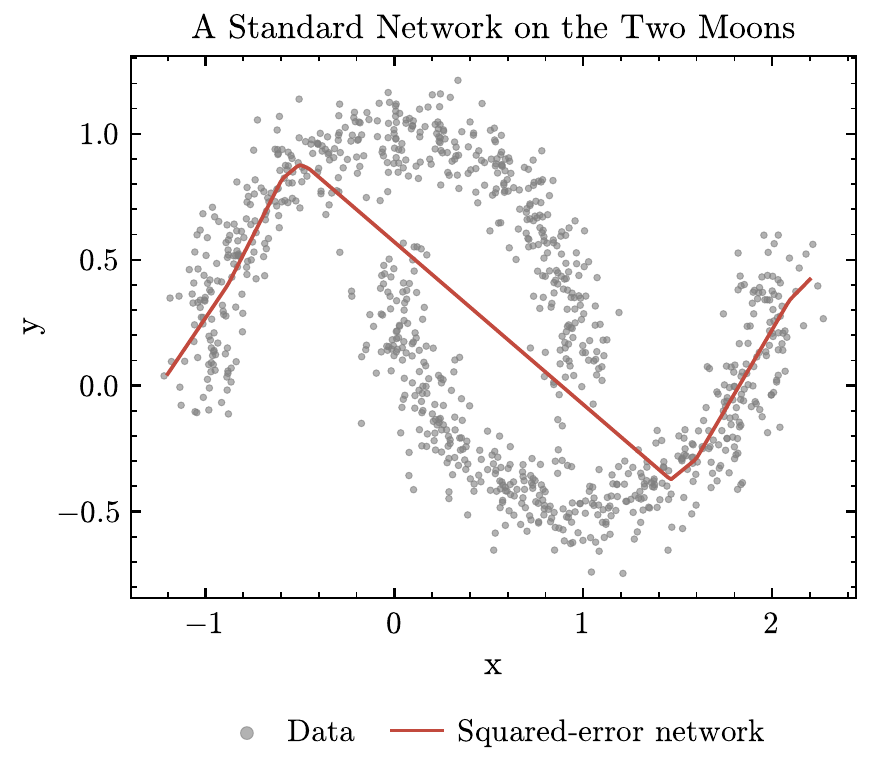}}%
  {\includegraphics[height=\tutorialheight]{tutorials/15c/fig02.pdf}}

The prediction at \(x = 0.5\) is close to the average of the two moons, so it describes neither moon. It lies in the empty space between them. The network gives one value, without a width. A model of this relation must give a density of \(y\) for each \(x\).

\section*{The Mathematical Formulation of Mixture Density Networks}

An MDN replaces a point prediction by a probability density. It does not learn a function \(\mathbf{y} = f(\mathbf{x})\). It learns the conditional density \(p(\mathbf{y}|\mathbf{x})\) (Section 15.12). The density can then show both the scatter and several groups of values in the data.

\subsection*{Gaussian Mixture Models as Output Densities}

An MDN models the conditional density as a mixture of Gaussians,

\[p(\mathbf{y}|\mathbf{x}) = \sum_{k=1}^K \pi_k(\mathbf{x}) \mathcal{N}(\mathbf{y}|\boldsymbol{\mu}_k(\mathbf{x}), \boldsymbol{\Sigma}_k(\mathbf{x})).\]

Each component of this mixture has three parameters that depend on the input \(\mathbf{x}\).

\begin{enumerate}
\def\labelenumi{\arabic{enumi}.}
\tightlist
\item
  The mixture weights \(\pi_k(\mathbf{x})\) are the relative weights of the components, with \(\sum_k \pi_k(\mathbf{x}) = 1\).
\item
  The means \(\boldsymbol{\mu}_k(\mathbf{x})\) are the centers of the components.
\item
  The covariances \(\boldsymbol{\Sigma}_k(\mathbf{x})\) set the spread and the shape of the components. We often simplify them to diagonal or spherical covariances.
\end{enumerate}

A neural network computes these parameters as functions of the input. The density then changes smoothly over the input space and follows the local structure of the data.

\subsection*{Network Architecture and Output Constraints}

The network must output the parameters of the mixture, and these parameters have to be valid. The weights must form a probability distribution, and the widths must be positive. The output layer therefore transforms the raw numbers of the network. The hidden layers use the ReLU activation (Section 15.5).

For a mixture with \(K\) components and \(D\)-dimensional outputs, the network must produce the following.

\begin{itemize}
\tightlist
\item
  \(K\) mixture weights, which form a valid probability distribution
\item
  \(K \times D\) mean values, which are unconstrained
\item
  The covariance parameters, which must give positive definite covariances
\end{itemize}

The architecture is a standard feedforward network until the output layer, where we must handle the constraints with care. A softmax transformation makes the mixture weights positive and sum to 1 (Section 15.12),

\[\pi_k(\mathbf{x}) = \frac{\exp(z_k(\mathbf{x}))}{\sum_{j=1}^K \exp(z_j(\mathbf{x}))},\]

where \(z_k(\mathbf{x})\) are the raw network outputs for the mixture weights, called logits. For simplicity we use diagonal covariances. In this tutorial the output \(y\) is one number, so each covariance is a variance \(\sigma_k^2(\mathbf{x})\). The network outputs \(\ln \sigma_k\), and the exponential function makes \(\sigma_k\) positive. The output layer therefore has \(3K\) numbers, the \(K\) means, the \(K\) values of \(\ln \sigma_k\) and the \(K\) logits.

\tutorialbarrier
\begin{codecell}

\end{codecell}

\section*{Training Mixture Density Networks}

We train an MDN by maximum likelihood (Chapters 4 and 7). The output \(y_i\) is one number here. For \(N\) pairs \((\mathbf{x}_i, y_i)\) we seek the network parameters \(\mathbf{w}\) that maximize the likelihood of the data under the conditional densities of the network,

\[L(\mathbf{w}) = \prod_{i=1}^N p(y_i|\mathbf{x}_i, \mathbf{w}).\]

The logarithm turns this product into a sum, which avoids underflow, and gives the log-likelihood (Section 15.12),

\[\ell(\mathbf{w}) = \ln L(\mathbf{w}) = \sum_{i=1}^N \ln \left[ \sum_{k=1}^K \pi_k(\mathbf{x}_i) \mathcal{N}(y_i|\mu_k(\mathbf{x}_i), \sigma_k^2(\mathbf{x}_i)) \right].\]

We minimize the negative log-likelihood, which is our loss function. In the code we take the mean over the points and not the sum. The size of the gradient then does not depend on the number of points (Section 15.3).

The inner sum over the components can underflow when every term is small. The code therefore adds the terms with the log-sum-exp function. Let \(a_k = \ln\pi_k + \ln\mathcal{N}(y|\mu_k, \sigma_k^2)\) be the logarithm of the term of component \(k\), and let \(m\) be the largest of the \(a_k\). Then

\[\ln \sum_{k=1}^K e^{a_k} = m + \ln \sum_{k=1}^K e^{a_k - m}.\]

The exponentials on the right are at most 1, and one of them is exactly 1, so the sum cannot underflow. The method \texttt{log\_\allowbreak{}prob} of \texttt{torch.\allowbreak{}distributions.\allowbreak{}Normal} gives the log of the Gaussian density of each component.

The optimization is the same as for a standard neural network. Backpropagation computes the gradients through the mixture parameters and through the weights of the network (Section 15.4). The difference lies in the loss function. It measures how well the conditional densities explain the data, and not the distance to point predictions.

\tutorialbarrier
\begin{codecell}
\begin{Verbatim}[commandchars=\\\{\},fontsize=\small]
\PYG{k}{def}\PYG{+w}{ }\PYG{n+nf}{mixture\PYGZus{}log\PYGZus{}prob}\PYG{p}{(}\FancyVerbBreakAfterBreak{}\PYG{n}{y}\PYG{p}{,}\FancyVerbBreakAfterBreak{} \PYG{n}{mu}\PYG{p}{,}\FancyVerbBreakAfterBreak{} \PYG{n}{sigma}\PYG{p}{,}\FancyVerbBreakAfterBreak{} \PYG{n}{pi}\PYG{p}{)}\PYG{p}{:}
\PYG{+w}{    }\PYG{l+s+sd}{\PYGZdq{}\PYGZdq{}\PYGZdq{}Logarithm of the mixture density at y.\PYGZdq{}\PYGZdq{}\PYGZdq{}}

    \PYG{c+c1}{\PYGZsh{} Create a Gaussian distribution for each component}
    \PYG{n}{gaussians} \PYG{o}{=}\FancyVerbBreakAfterBreak{} \PYG{n}{torch}\PYG{o}{.}\FancyVerbBreakAfterBreak{}\PYG{n}{distributions}\PYG{o}{.}\FancyVerbBreakAfterBreak{}\PYG{n}{Normal}\PYG{p}{(}\FancyVerbBreakAfterBreak{}\PYG{n}{mu}\PYG{p}{,}\FancyVerbBreakAfterBreak{} \PYG{n}{sigma}\PYG{p}{)}

    \PYG{c+c1}{\PYGZsh{} Compute the log probability of y under each component}
    \PYG{n}{log\PYGZus{}prob} \PYG{o}{=}\FancyVerbBreakAfterBreak{} \PYG{n}{gaussians}\PYG{o}{.}\FancyVerbBreakAfterBreak{}\PYG{n}{log\PYGZus{}prob}\PYG{p}{(}\FancyVerbBreakAfterBreak{}\PYG{n}{y}\PYG{p}{)}

    \PYG{c+c1}{\PYGZsh{} Weight each component by its mixture weight}
    \PYG{c+c1}{\PYGZsh{} In the logarithm the product of weight and density is the sum log(pi) + log(N)}
    \PYG{n}{weighted\PYGZus{}log\PYGZus{}prob} \PYG{o}{=}\FancyVerbBreakAfterBreak{} \PYG{n}{log\PYGZus{}prob} \PYG{o}{+}\FancyVerbBreakAfterBreak{} \PYG{n}{torch}\PYG{o}{.}\FancyVerbBreakAfterBreak{}\PYG{n}{log}\PYG{p}{(}\FancyVerbBreakAfterBreak{}\PYG{n}{pi}\PYG{p}{)}

    \PYG{c+c1}{\PYGZsh{} Sum over the components with the log\PYGZhy{}sum\PYGZhy{}exp, which stays numerically stable}
    \PYG{k}{return} \PYG{n}{torch}\PYG{o}{.}\FancyVerbBreakAfterBreak{}\PYG{n}{logsumexp}\PYG{p}{(}\FancyVerbBreakAfterBreak{}\PYG{n}{weighted\PYGZus{}log\PYGZus{}prob}\PYG{p}{,}\FancyVerbBreakAfterBreak{} \PYG{n}{dim}\PYG{o}{=}\FancyVerbBreakAfterBreak{}\PYG{l+m+mi}{1}\PYG{p}{)}

\PYG{k}{def}\PYG{+w}{ }\PYG{n+nf}{mdn\PYGZus{}loss\PYGZus{}fn}\PYG{p}{(}\FancyVerbBreakAfterBreak{}\PYG{n}{y\PYGZus{}true}\PYG{p}{,}\FancyVerbBreakAfterBreak{} \PYG{n}{mu}\PYG{p}{,}\FancyVerbBreakAfterBreak{} \PYG{n}{sigma}\PYG{p}{,}\FancyVerbBreakAfterBreak{} \PYG{n}{pi}\PYG{p}{)}\PYG{p}{:}
\PYG{+w}{    }\PYG{l+s+sd}{\PYGZdq{}\PYGZdq{}\PYGZdq{}Negative log\PYGZhy{}likelihood of the mixture, averaged over the batch.\PYGZdq{}\PYGZdq{}\PYGZdq{}}
    \PYG{k}{return} \PYG{o}{\PYGZhy{}}\PYG{n}{torch}\PYG{o}{.}\FancyVerbBreakAfterBreak{}\PYG{n}{mean}\PYG{p}{(}\FancyVerbBreakAfterBreak{}\PYG{n}{mixture\PYGZus{}log\PYGZus{}prob}\PYG{p}{(}\FancyVerbBreakAfterBreak{}\PYG{n}{y\PYGZus{}true}\PYG{p}{,}\FancyVerbBreakAfterBreak{} \PYG{n}{mu}\PYG{p}{,}\FancyVerbBreakAfterBreak{} \PYG{n}{sigma}\PYG{p}{,}\FancyVerbBreakAfterBreak{} \PYG{n}{pi}\PYG{p}{)}\PYG{p}{)}
\end{Verbatim}
\end{codecell}

We train the model with Adam on mini-batches of 64 points, as in Tutorial 15a (Section 15.3). We print the loss of all points at a few epochs. One epoch is one pass through the points, and the printed loss belongs to the weights at the end of that epoch.

\tutorialbarrier
\begin{codecell}

\end{codecell}

\section*{Application to the Two-Moons Dataset}

With the MDN and its loss function defined, we train the model on the two moons. We use two components, one for each moon. The network has two hidden layers of 20 neurons. We train for 1000 epochs with the learning rate 0.001.

\tutorialbarrier
\begin{codecell}

\end{codecell}
\begin{Verbatim}[fontsize=\footnotesize, xleftmargin=5mm]
Epoch    1, loss 1.0025
Epoch  100, loss -0.0333
Epoch 1000, loss -0.1361
\end{Verbatim}

\section*{Visualizing the Learned Densities}

We look at how the MDN represents the two moons. The network gives the weights, means and standard deviations of the two components for every \(x\). We plot the means with a band of one standard deviation, over a range of \(x\) wider than the range of the data. The lines outside the range of the data are extrapolations.

\tutorialbarrier
\begin{codecell}

\end{codecell}
\tutorialplot{\includegraphics[width=4.260in,height=0.75\textheight,keepaspectratio]{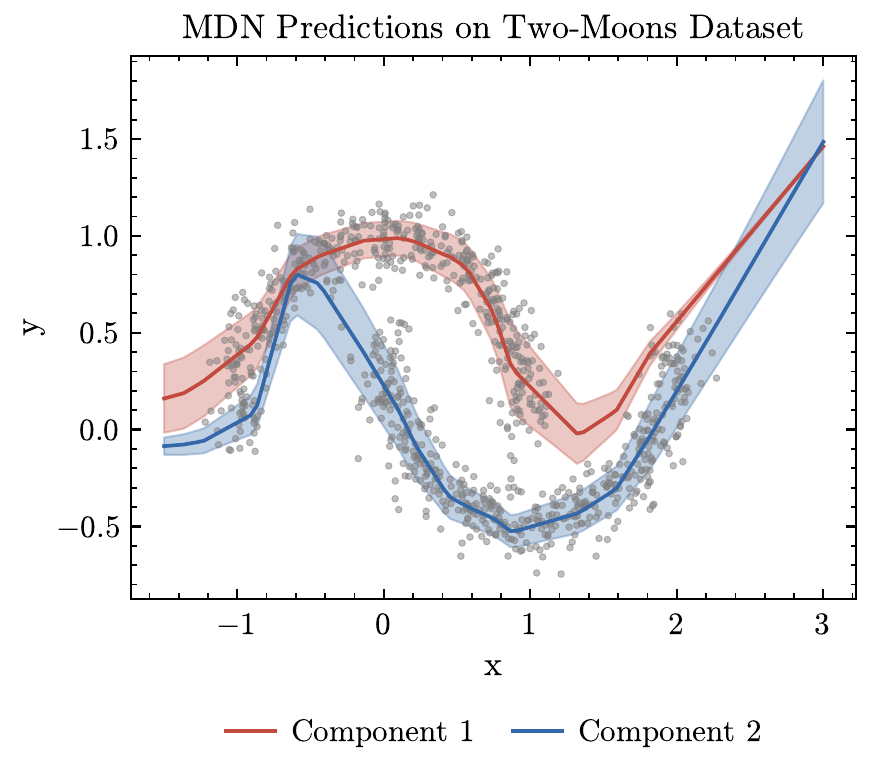}}%
  {\includegraphics[height=\tutorialheight]{tutorials/15c/fig03.pdf}}

Between \(x = 0\) and \(x = 1\), where both moons exist, the two component means follow the upper and the lower moon. The width of each band is about the width of the moon that it follows. We plot the mixture weights next, which say how much each component contributes at each \(x\).

\tutorialbarrier
\begin{codecell}

\end{codecell}
\tutorialplot{\includegraphics[width=4.150in,height=0.75\textheight,keepaspectratio]{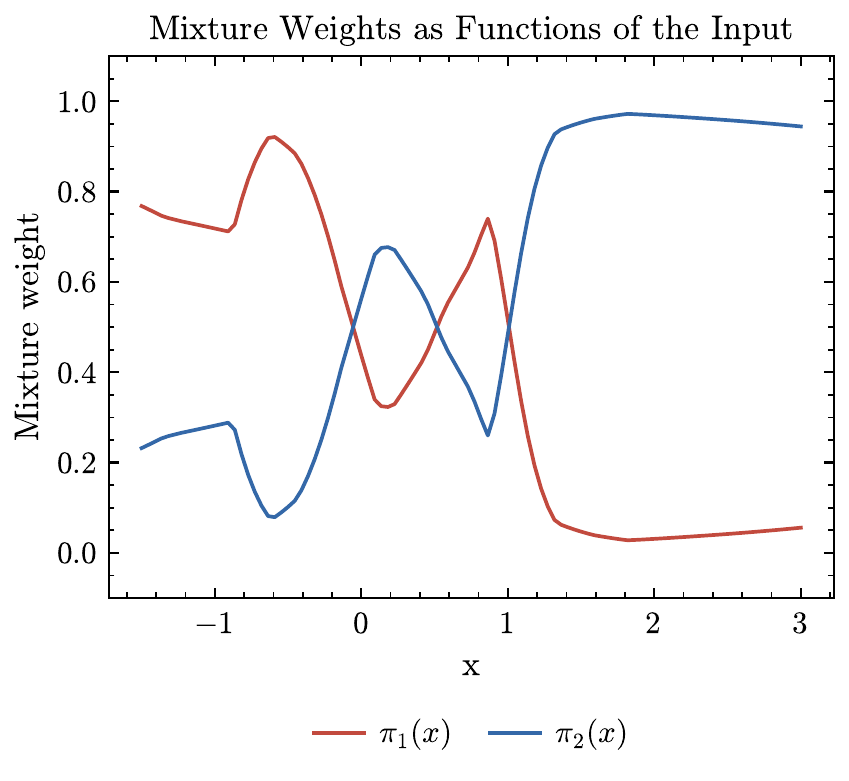}}%
  {\includegraphics[height=\tutorialheight]{tutorials/15c/fig04.pdf}}

Where both moons exist, both weights are large. Where only one moon exists, one component carries most of the weight. The weights change with \(x\), as the means do, so the whole density changes with the input.

The means, widths and weights together give the density \(p(y|x)\), which is the output of the MDN. We plot it for the moons at \(x = -0.5\), 0.5 and 1.5. The thin black curves are the two weighted components, and the short vertical lines at the bottom mark the points within 0.1 of each \(x\).

\tutorialbarrier
\begin{codecell}

\end{codecell}
\tutorialplot{\includegraphics[width=6.309in,height=0.75\textheight,keepaspectratio]{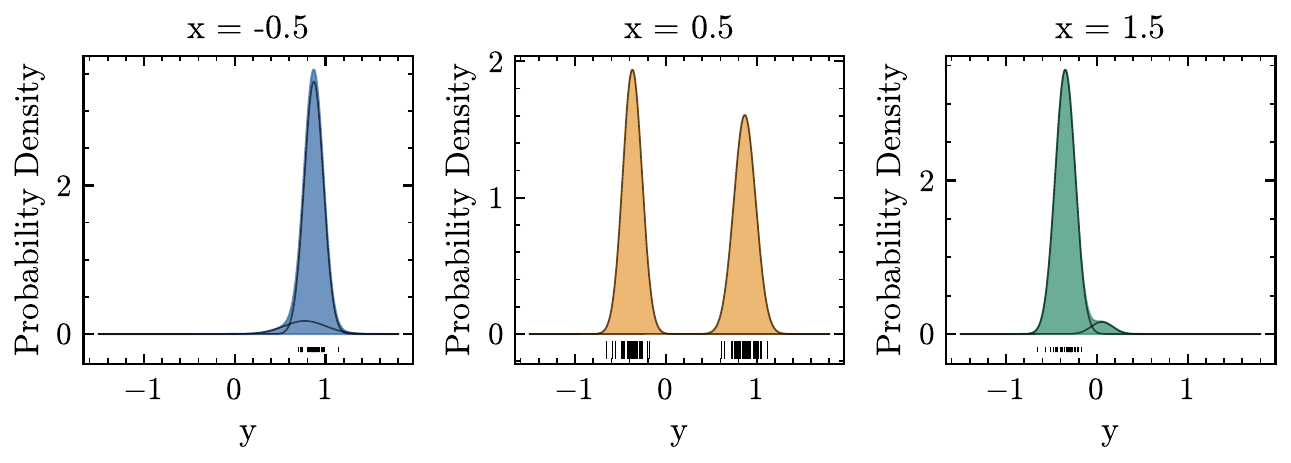}}%
  {\includegraphics[height=\tutorialheight]{tutorials/15c/fig05.pdf}}

At \(x = 0.5\) both moons exist, and each component forms one peak. At \(x = 1.5\) only the lower moon exists, and one component carries almost all the weight. At \(x = -0.5\) only the upper moon exists, and both components lie on it, one with a small weight.

\section*{The Lithium Data}

We now turn to the lithium abundance of stars. Two stars with the same temperature and age can differ in A(Li). Here we do not expect two branches as for the moons. We ask for the mean, the width and the shape of the density.

The sample is part of the open-cluster sample of \href{https://ui.adsabs.harvard.edu/abs/2025ApJ...978..107S/abstract}{Sun et al.~(2025)}, who also train a mixture density network on it. It holds the stars of five clusters, the Pleiades (0.10 Gyr), M48 (0.42 Gyr), the Hyades and Praesepe (0.65 Gyr) and NGC 3960 (1.00 Gyr). The stars of an open cluster formed from one cloud at nearly the same time, so they share an age. The age comes from fitting stellar evolution models to the colors and brightnesses of the stars of the cluster. It is usually more precise than the age of an isolated star.

The Hyades and Praesepe share one age, so the data hold only four ages. The ages are in Gyr, and 1 Gyr is one billion years.

\tutorialbarrier
\begin{codecell}

\end{codecell}
\begin{Verbatim}[fontsize=\footnotesize, xleftmargin=5mm]
358 stars at 4 ages
Age 0.10 Gyr: 112 stars, 4199 to 7700 K
Age 0.42 Gyr: 126 stars, 4350 to 7669 K
Age 0.65 Gyr: 91 stars, 5143 to 7623 K
Age 1.00 Gyr: 29 stars, 5274 to 6733 K
\end{Verbatim}

We plot A(Li) against the temperature, colored by the age.

\tutorialbarrier
\begin{codecell}

\end{codecell}
\tutorialplot{\includegraphics[width=4.641in,height=0.75\textheight,keepaspectratio]{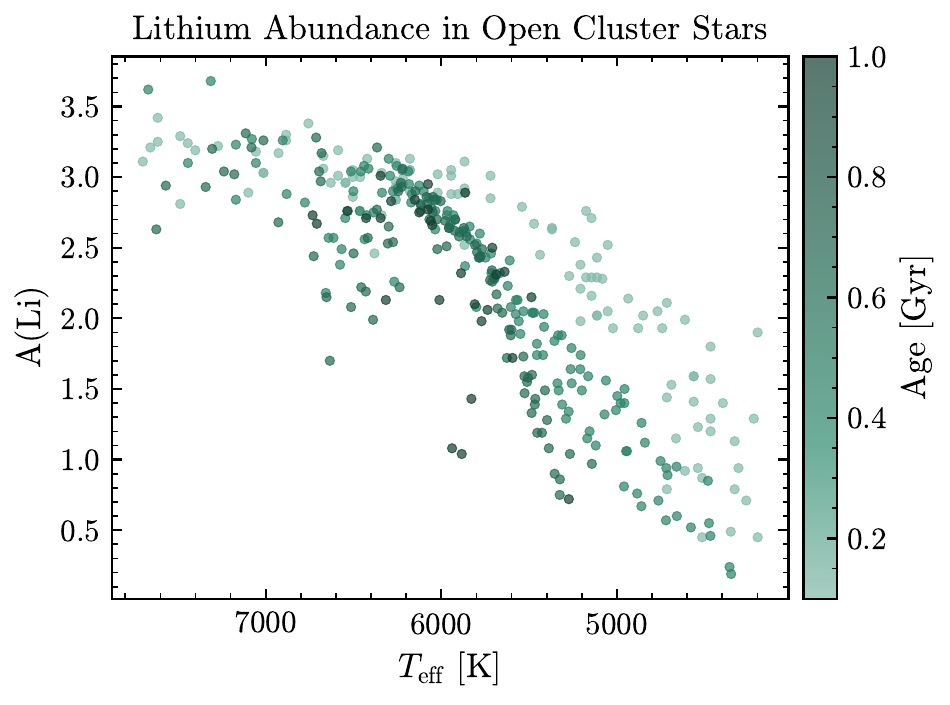}}%
  {\includegraphics[height=\tutorialheight]{tutorials/15c/fig06.pdf}}

The plot shows three features.

\begin{enumerate}
\def\labelenumi{\arabic{enumi}.}
\tightlist
\item
  \textbf{Temperature dependence.} A(Li) decreases toward cooler stars.
\item
  \textbf{Age dependence.} At fixed temperature, older clusters (darker points) tend to have lower abundances than younger clusters (lighter points).
\item
  \textbf{Scatter.} Even at fixed temperature and age, a scatter remains, often of a factor of several. Other properties of the stars cause it, such as the rotation, the magnetic activity, the accretion history and the metallicity.
\end{enumerate}

\section*{Modeling Lithium Abundance with MDNs}

We want a model of the conditional density \(p(\text{A(Li)}|T_{\rm eff}, \tau)\). An MDN predicts a mixture of Gaussians for each pair of temperature and age. The mixture gives the mean depletion and the scatter at each point, and it can have more than one peak.

The inputs are \(T_{\rm eff}\) in units of 1000 K, which is closer to the scale of the age, and the age \(\tau\) in Gyr. We use two components, as for the moons, so that the mixture can have a main peak and a tail. The network has two hidden layers of 50 neurons. We train it on all stars with Adam, the learning rate 0.0001 and 5000 epochs. In our tests, the GELU activation and a cosine schedule gave no better fit, so we keep ReLU and a constant rate.

\tutorialbarrier
\begin{codecell}

\end{codecell}
\begin{Verbatim}[fontsize=\footnotesize, xleftmargin=5mm]
Epoch    1, loss 2.7158
Epoch 1000, loss 0.2764
Epoch 5000, loss -0.0058
\end{Verbatim}

\section*{Visualizing Lithium Evolution in Parameter Space}

With the trained MDN we can explore how the conditional density changes with temperature and age. We want the mean and the standard deviation of the mixture at each point. A mixture generates a value in two steps. It picks the component \(k\) with the probability \(\pi_k(\mathbf{x})\), and it then draws \(y\) from the Gaussian of that component. With this view, the laws of total expectation and total variance of Chapter 3 give both quantities.

The mean of the mixture is the mean of the component means, weighted by the probability of each component (Section 3.4),

\[\mathbb{E}[y|\mathbf{x}] = \sum_{k=1}^K \pi_k(\mathbf{x}) \mu_k(\mathbf{x}).\]

The variance has two sources (Section 3.8). The first is the spread of \(y\) inside a component, \(\sigma_k^2\). The second is the spread of the component means around the mean of the mixture, \((\mu_k - \mathbb{E}[y|\mathbf{x}])^2\). Each source is weighted by \(\pi_k\),

\[\text{Var}[y|\mathbf{x}] = \sum_{k=1}^K \pi_k(\mathbf{x}) \left[\sigma_k^2(\mathbf{x}) + \left(\mu_k(\mathbf{x}) - \mathbb{E}[y|\mathbf{x}]\right)^2\right].\]

The first term is the width that each component has on its own. The second term is large when the means of the components lie far apart, even if each component is narrow. If all components have the same mean, it is 0, and the variance is the weighted mean of the component variances. We write \(\sigma_{\rm A(Li)}\) for the square root of the variance, the total scatter of the mixture at a point.

We first predict at the four ages of the data. For each age we use the temperatures from the coolest to the hottest star of the cluster, so the predictions stay inside the data.

\tutorialbarrier
\begin{codecell}

\end{codecell}

We plot the predicted mean over the stars, and \(\sigma_{\rm A(Li)}\), for each age of a cluster.

\tutorialbarrier
\begin{codecell}

\end{codecell}
\tutorialplot{\includegraphics[width=6.175in,height=0.75\textheight,keepaspectratio]{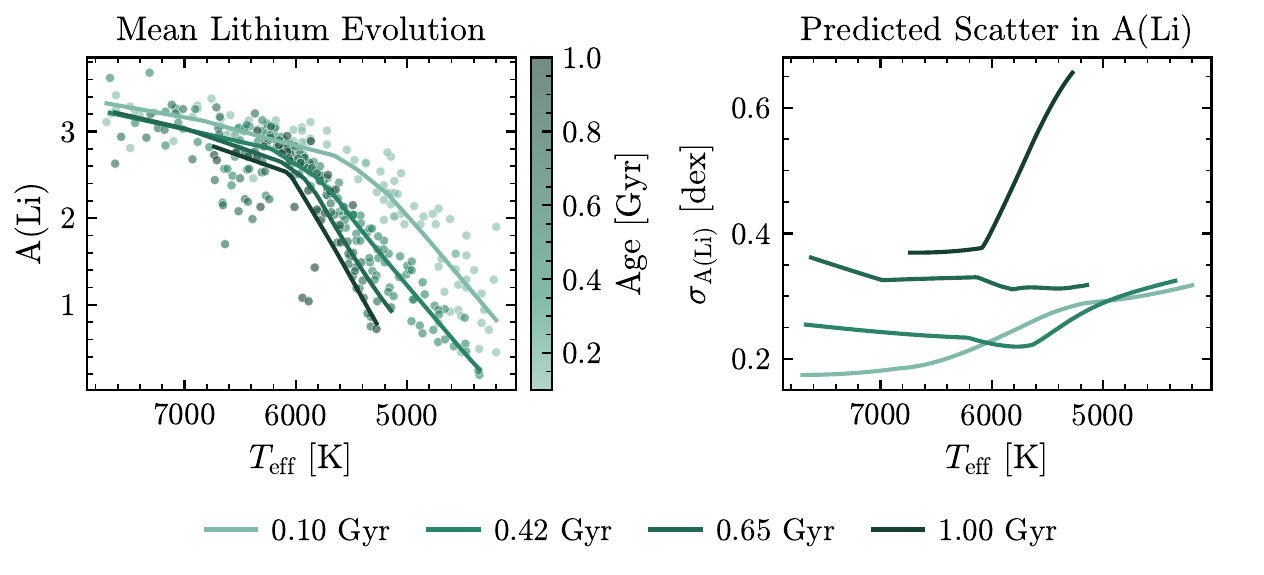}}%
  {\includegraphics[height=\tutorialheight]{tutorials/15c/fig07.pdf}}

\section*{Predictions at Many Ages}

The four lines above show four separate ages. The age is an input of the network, so the MDN gives a prediction at any age. We plot the mean and the scatter for many ages from 0.10 to 1.00 Gyr, to see how they change with the age. All lines use the temperatures where all four clusters have stars, so that every line lies inside the data.

\tutorialbarrier
\begin{codecell}

\end{codecell}

We plot the lines with the color of their age. The thick lines are the four cluster ages.

\tutorialbarrier
\begin{codecell}

\end{codecell}
\tutorialplot{\includegraphics[width=6.178in,height=0.75\textheight,keepaspectratio]{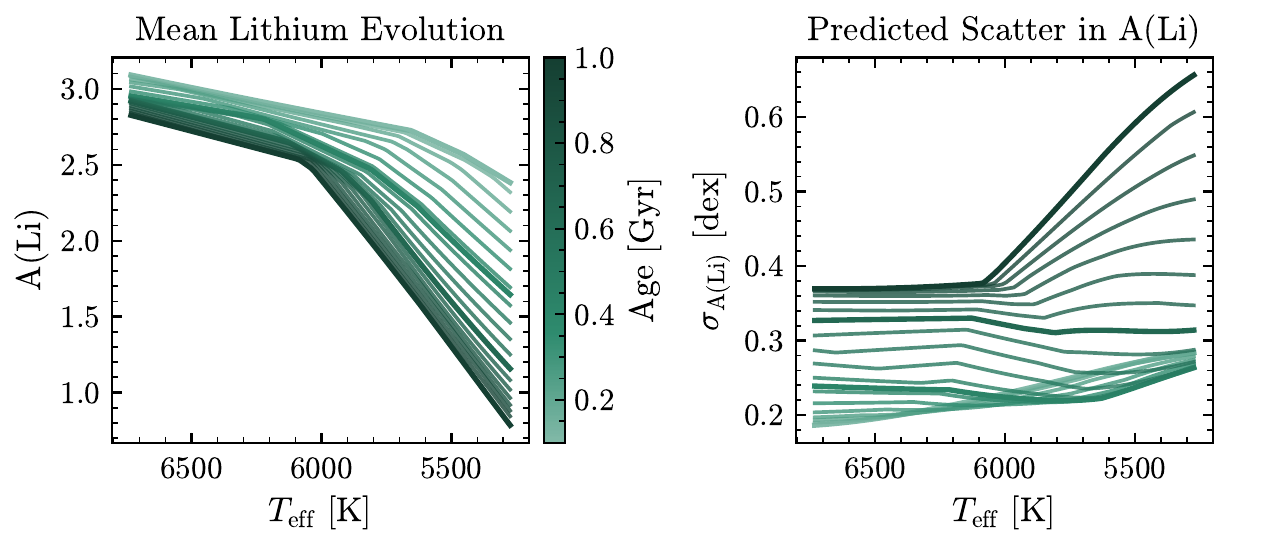}}%
  {\includegraphics[height=\tutorialheight]{tutorials/15c/fig08.pdf}}

The lines change smoothly with the age, and their colors stay in order. The band is also a check of overfitting. Nothing in the data ties the lines between the four ages, so an overfit network would wiggle there. We see no wiggles. If wiggles appeared, we would train for fewer epochs or use a smaller network.

The band shows the physics of lithium depletion that we described at the start. The mean of A(Li) falls toward cooler stars, because the convection zone is deeper there. It also falls with age, because the destruction continues. At 6000 K, the mean falls from about 2.8 at 0.10 Gyr to about 2.4 at 1.00 Gyr. The fall is larger for cooler stars.

The right panel shows the scatter \(\sigma_{\rm A(Li)}\). At 6000 K it grows from about 0.2 dex for the young clusters to about 0.4 dex at 1.00 Gyr. For the oldest ages it also grows toward cooler stars. The measurement errors, typically 0.02 to 0.08 dex in A(Li), are small compared with these widths.

Several processes can cause the scatter. A star that rotates faster at the beginning is thought to deplete more lithium by mixing, but we know only its present rotation. Other causes are the magnetic activity, the accretion history and the metallicity. The network cannot separate these causes, so the hidden properties appear as the width and the shape of the mixture.

\section*{Conditional Densities at Three Points}

The mean and the standard deviation do not show the shape of the conditional density. We therefore plot the whole conditional density \(p(\text{A(Li)}|T_{\rm eff}, \tau)\) at three points.

We choose a hot young point, a solar-type middle-aged point and a cool older point. Each point lies inside the temperature range of the cluster with that age. Each panel shows the following.

\begin{itemize}
\tightlist
\item
  The colored curve is the density of the mixture, the sum of the weighted Gaussians.
\item
  The thin black curves are the weighted components.
\item
  The dashed line marks the mean of the mixture.
\item
  The short vertical lines mark the A(Li) of the stars of that cluster near the temperature.
\end{itemize}

\tutorialbarrier
\begin{codecell}

\end{codecell}
\begin{Verbatim}[fontsize=\footnotesize, xleftmargin=5mm]
Hot, young               mixture 3.01 ± 0.20, 12 stars 2.94 ± 0.18
Solar-type, middle-aged  mixture 2.32 ± 0.22, 18 stars 2.28 ± 0.22
Cool, older              mixture 1.60 ± 0.31, 19 stars 1.49 ± 0.28
\end{Verbatim}
\tutorialplot{\includegraphics[width=6.500in,height=0.75\textheight,keepaspectratio]{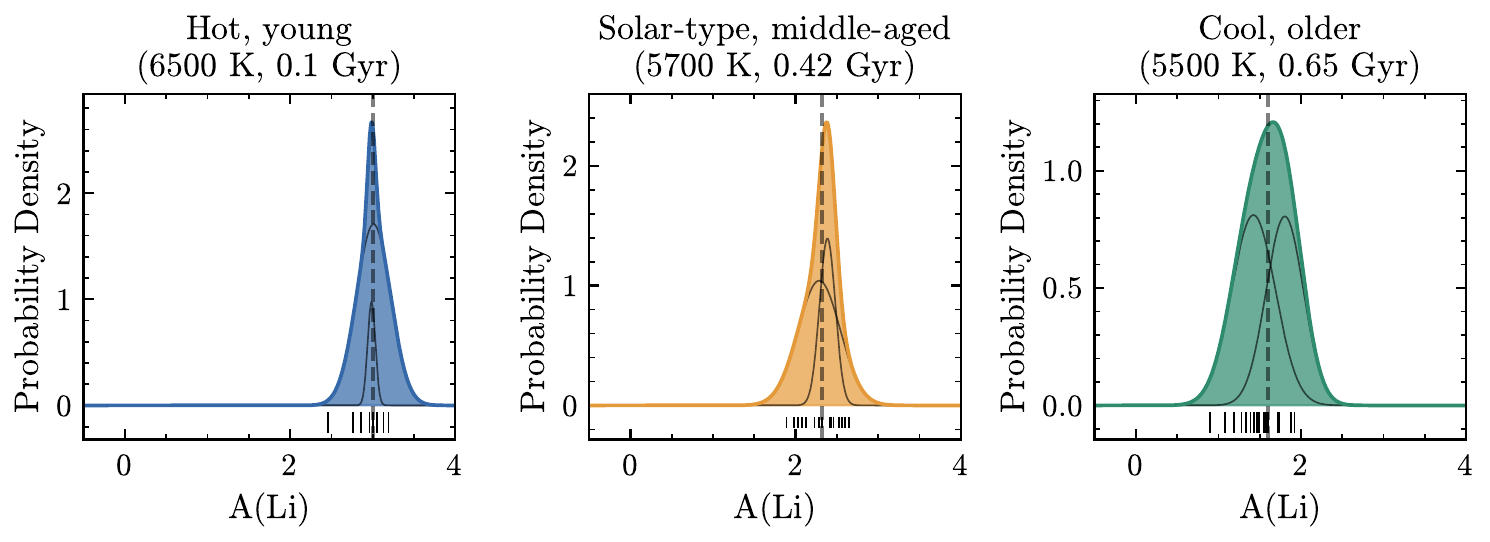}}%
  {\includegraphics[height=\tutorialheight]{tutorials/15c/fig09.pdf}}

At the hot young point the density is one narrow peak near 3.0. At the solar-type point the density has a narrow peak near 2.4 and a broad component toward lower A(Li). At the cool older point the density is broad. The width grows from the first point to the last, as the right panel above shows. Each density has one peak, so the stars at a point form one group, and the mixture describes its width and its tail. The mean and the standard deviation of the mixture are close to those of the stars at each point.

\section*{Summary}

A standard neural network predicts one value for each input. A Mixture Density Network predicts a density instead. Our main lessons follow.

\begin{enumerate}
\def\labelenumi{\arabic{enumi}.}
\tightlist
\item
  \textbf{Conditional density estimation.} An MDN models the conditional density \(p(\mathbf{y}|\mathbf{x})\) and not a function \(\mathbf{y} = f(\mathbf{x})\). A standard neural network predicts the mean of the outputs, which can lie between two groups of values. An MDN can describe the scatter and several groups of values at one input.
\item
  \textbf{Parameterization by a network.} The network outputs the weights, means and widths of a Gaussian mixture. A softmax keeps the weights positive and summing to 1, and an exponential keeps the widths positive.
\item
  \textbf{Maximum likelihood training.} We minimize the negative log-likelihood by gradient descent, as for other networks. The log-sum-exp function keeps the sum over the components stable.
\item
  \textbf{A probabilistic output.} The mixture gives the mean, the width and the shape of the density at each input. The prediction is a probability density, so it describes the scatter of the data and the mean.
\item
  \textbf{Lithium in open clusters.} The predicted mean of A(Li) falls toward cooler stars and is lower for older clusters. The predicted scatter is larger for older clusters. The predictions change smoothly with the age.
\end{enumerate}

Many quantities in astronomy have a scatter that changes with the inputs or more than one possible value, so a prediction needs a density. The MDN is the basic network model for this case. It shows that a neural network is more than a deterministic mapping from \(\mathbf{x}\) to \(\mathbf{y}\), because its outputs can define a probability density.

\tutorialchapter{15d}{Normalizing Flows for Stellar Abundances}
\noindent\textit{Companion material for Chapter 15}

\section*{Introduction}

The elemental abundances of stars in the Milky Way form sequences. The ratios {[}Fe/H{]} and {[}Mg/Fe{]} are abundance ratios relative to the Sun. In their plane, the stars follow two curved sequences with a lower density between them.

The two sequences come from two kinds of supernovae. Massive stars explode soon after they form and make magnesium. The upper sequence holds stars enriched mainly by these explosions. Type Ia supernovae are explosions of white dwarfs in binary systems. The white dwarf gains mass from a companion, and the explosion occurs later and makes iron. The lower sequence holds stars whose gas also received this iron.

Surveys measure the abundances of many elements, and the abundances of one star form a point in a space with one dimension for each element. A model of the density in this space can generate mock stars and tell how typical a star is. The space has many dimensions, so the model needs many parameters, and we cannot check it by eye. The structure is also not simple. Already the plane of {[}Fe/H{]} and {[}Mg/Fe{]} shows two curved sequences, which a density with a fixed shape, such as a Gaussian, cannot follow.

A normalizing flow (Section 15.13) takes another route. An invertible neural network transforms a standard Gaussian into the density of the stars. The change of variables (Section 3.2) then gives the exact likelihood, so we can train the flow by maximum likelihood (Section 4.3).

We build the most basic flow, RealNVP (Real-valued Non-Volume Preserving), from scratch in PyTorch. A one-dimensional map shows how an invertible map changes a density. We then build coupling layers, which keep the map invertible and its Jacobian cheap to compute. We train the flow on the abundances of stars from APOGEE, a spectroscopic survey of the Milky Way. The sample follows \href{https://ui.adsabs.harvard.edu/abs/2022ApJ...927..209T/abstract}{Ting \& Weinberg (2022)}. We check the samples of the flow against stars that we did not use for training.

\section*{Prerequisites and Learning Objectives}

\subsection*{Prerequisites}

This tutorial assumes the following.

\begin{itemize}
\tightlist
\item
  Neural networks, backpropagation and the basics of PyTorch (Chapter 15, Tutorial 15a)
\item
  Training a network on a negative log-likelihood with Adam (Tutorial 15c)
\item
  The transformation of random variables and the Jacobian determinant (Chapter 3)
\item
  Maximum likelihood estimation, and training and test data (Chapter 4)
\item
  The multivariate Gaussian (Chapter 2)
\end{itemize}

\subsection*{Learning Objectives}

After this tutorial you can do the following.

\begin{itemize}
\tightlist
\item
  Explain how a probability density changes under an invertible map, and the role of the Jacobian determinant
\item
  Explain why a coupling layer is invertible and has a cheap Jacobian
\item
  Build an invertible neural network from coupling layers
\item
  Train a normalizing flow by maximum likelihood, stop at the lowest validation loss and generate samples with its inverse
\item
  Compare the samples of a flow with test stars in several projections
\end{itemize}

We begin by importing the libraries and setting the plot style.

\tutorialbarrier
\begin{codecell}

\end{codecell}

\section*{The Abundance Data}

The data hold 13 abundances for each of 23,986 stars. The elements are Mg, O, Si, S, Ca, Na, Al, K, Cr, Fe, Ni, V and Co.~The notation {[}X/H{]} is the logarithmic abundance ratio of the element X and hydrogen, relative to the Sun,

\[[\text{X/H}] = \log_{10}\left(\frac{N_{\text{X}}}{N_{\text{H}}}\right)_{\star} - \log_{10}\left(\frac{N_{\text{X}}}{N_{\text{H}}}\right)_{\odot},\]

where \(N\) is the number density of atoms. A star with {[}Fe/H{]} = -1 has an iron to hydrogen ratio ten times lower than the Sun. Because the scale is logarithmic, the ratio {[}Mg/Fe{]} is {[}Mg/H{]} minus {[}Fe/H{]}.

We split the stars at random into 60 percent training stars, 20 percent validation stars and 20 percent test stars (Section 4.7). The training stars fit the flow, the validation stars tell us when to stop the training, and the test stars check the result. The abundances have standard deviations of about 0.2 to 0.3, while the base Gaussian of a flow has unit width. We therefore standardize each abundance with the mean and the standard deviation of the training stars.

\tutorialbarrier
\begin{codecell}

\end{codecell}
\begin{Verbatim}[fontsize=\footnotesize, xleftmargin=5mm]
14391 training stars, 4797 validation stars, 4798 test stars
13 abundances for each star
\end{Verbatim}

We look at the plane of {[}Fe/H{]} and {[}Mg/Fe{]}. Darker hexagons hold more points, on a logarithmic scale.

\tutorialbarrier
\begin{codecell}

\end{codecell}
\tutorialplot{\includegraphics[width=4.652in,height=0.75\textheight,keepaspectratio]{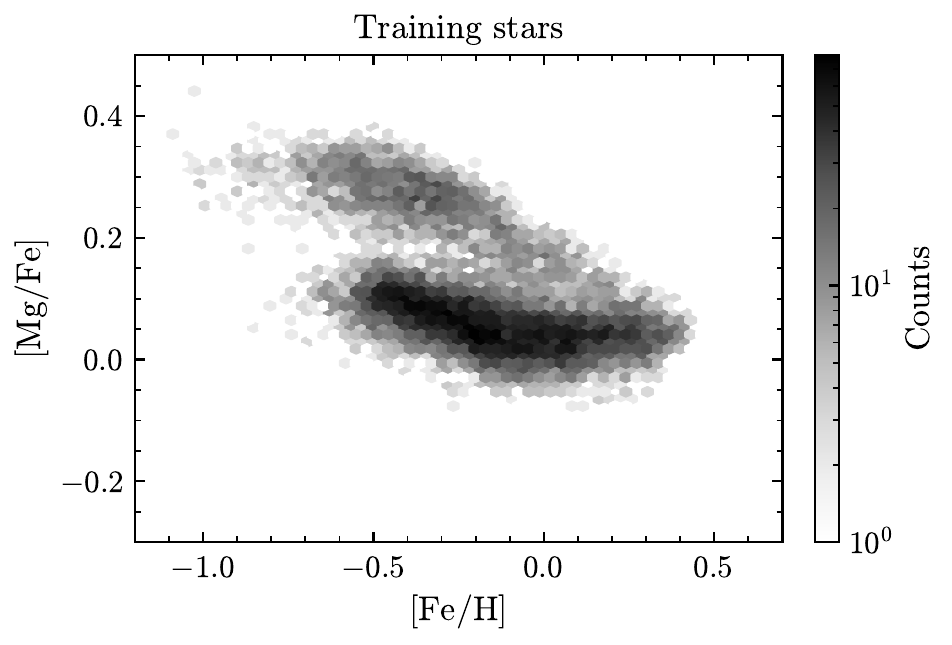}}%
  {\includegraphics[height=\tutorialheight]{tutorials/15d/fig01.pdf}}

The stars form a dense lower sequence and a fainter upper sequence, with a lower density between them.

\section*{The Basic Ideas of Normalizing Flows}

A normalizing flow learns the density without a chosen parametric family. The idea has three parts. We choose a simple base density, a standard Gaussian. We learn an invertible map \(f\) that carries a variable \(\mathbf{z}\) with the base density to the data \(\mathbf{x}\). The inverse map \(f^{-1}\) carries the data back to \(\mathbf{z}\).

Invertible means that every \(\mathbf{z}\) gives one \(\mathbf{x}\), and every \(\mathbf{x}\) gives one \(\mathbf{z}\). The two maps give us two abilities. To generate a new star, we draw \(\mathbf{z}\) from the standard Gaussian and compute \(\mathbf{x} = f(\mathbf{z})\). To evaluate the density of a star, we compute \(\mathbf{z} = f^{-1}(\mathbf{x})\) and use the change of variables (Section 3.2),

\[p(\mathbf{x}) = p(\mathbf{z})\left|\det\left(\frac{\partial \mathbf{z}}{\partial \mathbf{x}}\right)\right| = p(f^{-1}(\mathbf{x}))\left|\det\left(\frac{\partial f^{-1}}{\partial \mathbf{x}}\right)\right|.\]

The determinant of the Jacobian matrix measures how \(f^{-1}\) stretches space. If \(f^{-1}\) maps a region of \(\mathbf{x}\) to a larger region of \(\mathbf{z}\), \(p(\mathbf{x})\) is higher than \(p(\mathbf{z})\). Both regions hold the same probability.

The logarithm of \(p(\mathbf{x})\) is the log-likelihood of a star. We train the map by maximizing the mean log-likelihood of the training stars, so the loss is minus this mean. The loss needs \(f^{-1}\) and the determinant of its Jacobian for every star at every step.

\section*{A Flow in One Dimension}

In one dimension we know such a map already, from the inverse CDF method (Section 12.3). Let \(F\) be the cumulative distribution function (CDF) of \(x\), and let \(\Phi\) be the CDF of the standard Gaussian. For \(x\) drawn from the density, the number \(u = F(x)\) is uniform between 0 and 1, and \(z = \Phi^{-1}(u)\) is a standard Gaussian. The map

\[z = f^{-1}(x) = \Phi^{-1}(F(x))\]

therefore carries any density to the standard Gaussian. It is invertible, because the CDFs increase everywhere and so each \(x\) gives one \(z\). Its inverse, \(x = f(z) = F^{-1}(\Phi(z))\), generates samples of \(x\). In many dimensions this method has difficulties, so a flow learns the map with a neural network. Before that, we check the density formula in one dimension with a map that we can write down, the sigmoid function.

We start from a standard Gaussian \(z\) and map it with the sigmoid function

\[x = f(z) = \frac{1}{1 + e^{-2z}}.\]

This map is invertible, because it increases at every \(z\). Its derivative \(\mathrm{d}x/\mathrm{d}z = 2x(1-x)\) is positive for every \(x\) between 0 and 1, so each \(x\) comes from exactly one \(z\). Solving for \(z\) gives the inverse map, the logit function divided by 2,

\[z = f^{-1}(x) = \frac{1}{2}\ln\frac{x}{1-x}.\]

In one dimension the Jacobian determinant is the derivative of \(f^{-1}\),

\[\frac{\mathrm{d}f^{-1}}{\mathrm{d}x} = \frac{1}{2x(1-x)}.\]

We draw 100,000 samples of \(z\), map them to \(x\) with the sigmoid function and map them back with its inverse, the logit function. The curve in the middle panel is the density \(p(f^{-1}(x))\,|\mathrm{d}f^{-1}/\mathrm{d}x|\) that the change of variables gives. The curves in the other two panels are the standard Gaussian density.

\tutorialbarrier
\begin{codecell}

\end{codecell}
\tutorialplot{\includegraphics[width=6.500in,height=0.75\textheight,keepaspectratio]{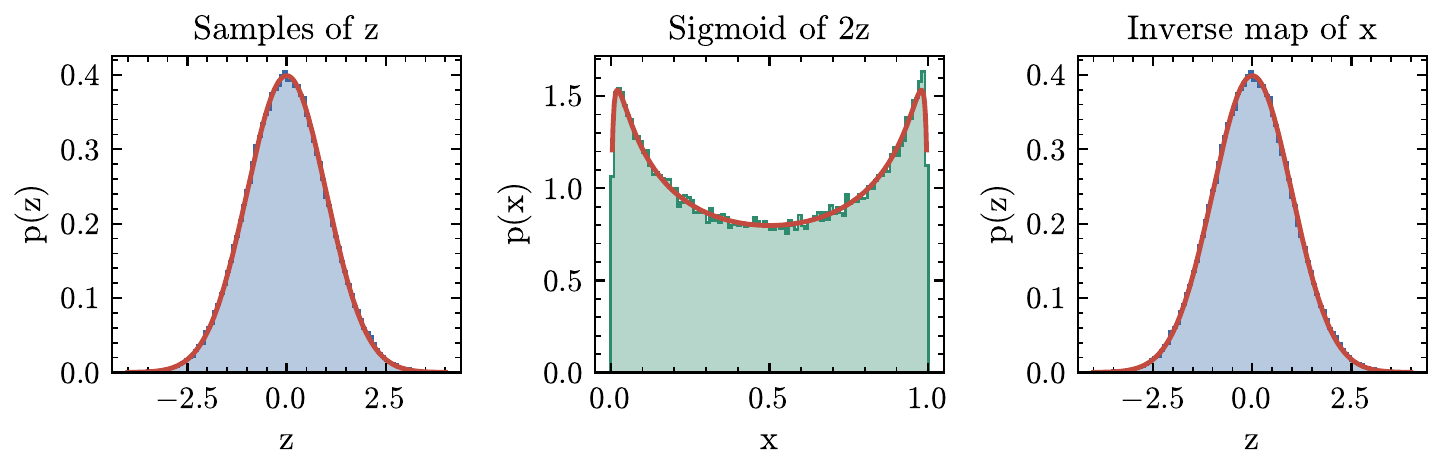}}%
  {\includegraphics[height=\tutorialheight]{tutorials/15d/fig02.pdf}}

The left panel shows the standard Gaussian. The sigmoid map turns it into a density with two peaks, one near 0 and one near 1. The curve from the change of variables follows the histogram. The inverse map recovers the standard Gaussian in the right panel. We can therefore describe a complex density through its relation to a simple one.

\section*{The Two Requirements}

In this example we chose the map. For the abundances we do not know a map to a Gaussian, so we learn it with a neural network. The network must meet two requirements.

It must be invertible, so that we can map \(\mathbf{x}\) to \(\mathbf{z}\) for the density and \(\mathbf{z}\) to \(\mathbf{x}\) for sampling. A general ReLU network does not meet this. A ReLU sets all negative inputs to 0, so two different stars can give the same output.

The Jacobian determinant must also be cheap to compute. The loss needs it for every star at every step, and the determinant of a general \(D \times D\) matrix needs \(O(D^3)\) operations. We therefore need a design in which both requirements hold by construction (Section 15.13).

\section*{Introducing RealNVP}

RealNVP (\href{https://ui.adsabs.harvard.edu/abs/2016arXiv160508803D/abstract}{Dinh et al.~2016}) meets both requirements with coupling layers. A layer keeps some dimensions fixed and transforms the others with a map that depends on the fixed ones. The fixed dimensions pass through unchanged, so one layer cannot transform every dimension. We therefore stack several layers, and each layer fixes different dimensions.

We write a layer in the direction from the data \(\mathbf{x}\) to the base \(\mathbf{z}\), which is the direction of \(f^{-1}\). Take two abundances, \(x_1\) for {[}Fe/H{]} and \(x_2\) for {[}Mg/Fe{]}. We leave \(x_1\) unchanged, which we call freezing it, and we transform \(x_2\) by a one-dimensional map, a stretch and a shift,

\tutorialdisplay{z_1 = x_1,}{z_2 = x_2\, e^{s(x_1)} + t(x_1).}

The numbers \(s\) and \(t\) depend on \(x_1\), and they come from two neural networks. For each fixed \(x_1\), the map from \(x_2\) to \(z_2\) is a straight line with the positive slope \(e^{s}\). A line with a positive slope is a simple invertible map in one dimension. The exponential keeps the slope positive for any output of the network. Because the networks see \(x_1\), the line changes with \(x_1\). The map can then stretch and shift {[}Mg/Fe{]} differently at low and at high {[}Fe/H{]}, as the two sequences require.

Both requirements hold by construction. For the inverse, the first equation gives \(x_1 = z_1\). With \(x_1\) known, we run the networks forward to get \(s(x_1)\) and \(t(x_1)\), and we solve the line for \(x_2\),

\[x_2 = \big(z_2 - t(x_1)\big)\, e^{-s(x_1)}.\]

We never invert a neural network. For the Jacobian, \(z_1\) does not depend on \(x_2\), so the Jacobian matrix is lower triangular,

\tutorialdisplay{\mathbf{J} = \begin{bmatrix} 1 & 0 \\ \partial z_2/\partial x_1 & e^{s(x_1)} \end{bmatrix},}{\ln|\det\mathbf{J}| = s(x_1).}

The determinant is the product of the diagonal, so we never need the entry \(\partial z_2/\partial x_1\), which holds the derivatives of the networks. This is why the layer fixes one dimension. The networks can be as complicated as we like.

A coupling layer does the same in \(D\) dimensions. It freezes \(d\) of the \(D\) dimensions and transforms the other \(D-d\) dimensions, based on the frozen ones. For simple notation we write the frozen dimensions first,

\[\mathbf{z}_{1:d} = \mathbf{x}_{1:d},\]

\[\mathbf{z}_{d+1:D} = \mathbf{x}_{d+1:D} \odot \exp(s(\mathbf{x}_{1:d})) + t(\mathbf{x}_{1:d}).\]

The subscripts select dimensions, and \(\odot\) is the element-wise product. The scale \(s\) and the shift \(t\) are neural networks that map \(\mathbb{R}^d\) to \(\mathbb{R}^{D-d}\). The name Non-Volume Preserving refers to \(\exp(s)\), which changes volumes, while a shift alone preserves them.

The same two properties hold in \(D\) dimensions. The layer is invertible for three reasons.

\begin{enumerate}
\def\labelenumi{\arabic{enumi}.}
\tightlist
\item
  \textbf{Coupling structure.} One group of dimensions stays unchanged, and the other group is transformed.
\item
  \textbf{Algebraic invertibility.} The transformation of the second group is a scale and a shift, which we invert algebraically.
\item
  \textbf{No inversion of the networks.} We only evaluate the networks \(s\) and \(t\), and we never invert them.
\end{enumerate}

The Jacobian is cheap for three reasons.

\begin{enumerate}
\def\labelenumi{\arabic{enumi}.}
\tightlist
\item
  \textbf{Triangular structure.} The frozen dimensions do not change, so the Jacobian matrix is triangular.
\item
  \textbf{Diagonal determinant.} The determinant of a triangular matrix is the product of its diagonal elements.
\item
  \textbf{Efficient computation.} This reduces an \(O(D^3)\) operation to an \(O(D)\) operation.
\end{enumerate}

In the code, a binary mask picks the frozen dimensions, so that we need not reorder the data. The networks \(s\) and \(t\) then map \(\mathbb{R}^D\) to \(\mathbb{R}^D\). We multiply the input by the mask, so that the networks see only the frozen dimensions. We multiply their output by one minus the mask, so that only the other dimensions change.

\section*{Constructing a RealNVP from Scratch}

The equations are short, but a layer has several steps. We build one layer for \(D = 5\) inputs, with the first \(d = 2\) dimensions frozen, and we print each step.

The networks \(s\) and \(t\) are multilayer perceptrons (MLPs) with three hidden layers and ReLU activations (Section 15.5). The class \texttt{MLP} builds the network with \texttt{nn.\allowbreak{}Sequential}, which applies its layers one after the other. It maps \(D\) values to \(D\) values. This is more general than the equations, where \(s\) and \(t\) map \(d\) values to \(D-d\) values. We create one network for \(s\) and one for \(t\), with 64 neurons in each hidden layer. We fix the seed, so that this example does not depend on the earlier cells.

\tutorialbarrier
\begin{codecell}

\end{codecell}

A mask selects the frozen dimensions. Here the mask is \((1, 1, 0, 0, 0)\), and the inverse mask, one minus the mask, is \((0, 0, 1, 1, 1)\). We multiply a random input by the mask and feed the result to the network \(s\). We then multiply the output of the network by the inverse mask.

\tutorialbarrier
\begin{codecell}

\end{codecell}
\begin{Verbatim}[fontsize=\footnotesize, xleftmargin=5mm]
x: [-0.605 -0.149 -0.785 -0.048 -0.884]
x times the mask: [-0.605 -0.149  0.     0.     0.   ]
s: [ 0.     0.    -0.132  0.104  0.111]
\end{Verbatim}

The mask sets the last three numbers of the input to 0 and keeps the first two. The network \(s\) therefore sees only the first two dimensions, because the other inputs are always 0. The inverse mask keeps the output of the network for the last three dimensions and sets the output for the first two to 0. An MLP that maps \(\mathbb{R}^D\) to \(\mathbb{R}^D\) then acts as the map from \(\mathbb{R}^d\) to \(\mathbb{R}^{D-d}\) of the equations.

We compute the shift \(t\) in the same way. Then we scale and shift the last three dimensions of the input.

\tutorialbarrier
\begin{codecell}
\begin{Verbatim}[commandchars=\\\{\},fontsize=\small]
\PYG{c+c1}{\PYGZsh{} The shift, and the scaled and shifted transformed dimensions}
\PYG{n}{t} \PYG{o}{=}\FancyVerbBreakAfterBreak{} \PYG{n}{t\PYGZus{}network}\PYG{p}{(}\FancyVerbBreakAfterBreak{}\PYG{n}{x\PYGZus{}frozen}\PYG{p}{)} \PYG{o}{*} \PYG{n}{inverse\PYGZus{}mask}
\PYG{n}{z\PYGZus{}transformed} \PYG{o}{=}\FancyVerbBreakAfterBreak{} \PYG{n}{x} \PYG{o}{*} \PYG{n}{inverse\PYGZus{}mask} \PYG{o}{*} \PYG{n}{torch}\PYG{o}{.}\FancyVerbBreakAfterBreak{}\PYG{n}{exp}\PYG{p}{(}\FancyVerbBreakAfterBreak{}\PYG{n}{s}\PYG{p}{)} \PYG{o}{+}\FancyVerbBreakAfterBreak{} \PYG{n}{t}
\PYG{n+nb}{print}\PYG{p}{(}\FancyVerbBreakAfterBreak{}\PYG{l+s+s2}{\PYGZdq{}}\PYG{l+s+s2}{t:}\PYG{l+s+s2}{\PYGZdq{}}\PYG{p}{,}\FancyVerbBreakAfterBreak{} \PYG{n}{t}\PYG{p}{[}\FancyVerbBreakAfterBreak{}\PYG{l+m+mi}{0}\PYG{p}{]}\PYG{o}{.}\FancyVerbBreakAfterBreak{}\PYG{n}{detach}\PYG{p}{(}\FancyVerbBreakAfterBreak{}\PYG{p}{)}\PYG{o}{.}\FancyVerbBreakAfterBreak{}\PYG{n}{numpy}\PYG{p}{(}\FancyVerbBreakAfterBreak{}\PYG{p}{)}\PYG{o}{.}\FancyVerbBreakAfterBreak{}\PYG{n}{round}\PYG{p}{(}\FancyVerbBreakAfterBreak{}\PYG{l+m+mi}{3}\PYG{p}{)} \PYG{o}{+}\FancyVerbBreakAfterBreak{} \PYG{l+m+mf}{0.0}\PYG{p}{)}
\PYG{n+nb}{print}\PYG{p}{(}\FancyVerbBreakAfterBreak{}\PYG{l+s+s2}{\PYGZdq{}}\PYG{l+s+s2}{Transformed dimensions:}\PYG{l+s+s2}{\PYGZdq{}}\PYG{p}{,}\FancyVerbBreakAfterBreak{} \PYG{n}{z\PYGZus{}transformed}\PYG{p}{[}\FancyVerbBreakAfterBreak{}\PYG{l+m+mi}{0}\PYG{p}{]}\PYG{o}{.}\FancyVerbBreakAfterBreak{}\PYG{n}{detach}\PYG{p}{(}\FancyVerbBreakAfterBreak{}\PYG{p}{)}\PYG{o}{.}\FancyVerbBreakAfterBreak{}\PYG{n}{numpy}\PYG{p}{(}\FancyVerbBreakAfterBreak{}\PYG{p}{)}\PYG{o}{.}\FancyVerbBreakAfterBreak{}\PYG{n}{round}\PYG{p}{(}\FancyVerbBreakAfterBreak{}\PYG{l+m+mi}{3}\PYG{p}{)} \PYG{o}{+}\FancyVerbBreakAfterBreak{} \PYG{l+m+mf}{0.0}\PYG{p}{)}
\end{Verbatim}
\end{codecell}
\begin{Verbatim}[fontsize=\footnotesize, xleftmargin=5mm]
t: [ 0.     0.    -0.08  -0.067 -0.093]
Transformed dimensions: [ 0.     0.    -0.768 -0.121 -1.081]
\end{Verbatim}

The frozen dimensions pass through unchanged. The two masks do not overlap, so the output \(\mathbf{z}\) is the sum of the frozen part and the transformed part.

\tutorialbarrier
\begin{codecell}
\begin{Verbatim}[commandchars=\\\{\},fontsize=\small]
\PYG{c+c1}{\PYGZsh{} The output of the layer, the frozen dimensions plus the transformed dimensions}
\PYG{n}{z} \PYG{o}{=}\FancyVerbBreakAfterBreak{} \PYG{n}{x\PYGZus{}frozen} \PYG{o}{+}\FancyVerbBreakAfterBreak{} \PYG{n}{z\PYGZus{}transformed}
\PYG{n+nb}{print}\PYG{p}{(}\FancyVerbBreakAfterBreak{}\PYG{l+s+s2}{\PYGZdq{}}\PYG{l+s+s2}{z:}\PYG{l+s+s2}{\PYGZdq{}}\PYG{p}{,}\FancyVerbBreakAfterBreak{} \PYG{n}{z}\PYG{p}{[}\FancyVerbBreakAfterBreak{}\PYG{l+m+mi}{0}\PYG{p}{]}\PYG{o}{.}\FancyVerbBreakAfterBreak{}\PYG{n}{detach}\PYG{p}{(}\FancyVerbBreakAfterBreak{}\PYG{p}{)}\PYG{o}{.}\FancyVerbBreakAfterBreak{}\PYG{n}{numpy}\PYG{p}{(}\FancyVerbBreakAfterBreak{}\PYG{p}{)}\PYG{o}{.}\FancyVerbBreakAfterBreak{}\PYG{n}{round}\PYG{p}{(}\FancyVerbBreakAfterBreak{}\PYG{l+m+mi}{3}\PYG{p}{)}\PYG{p}{)}
\end{Verbatim}
\end{codecell}
\begin{Verbatim}[fontsize=\footnotesize, xleftmargin=5mm]
z: [-0.605 -0.149 -0.768 -0.121 -1.081]
\end{Verbatim}

The first two numbers of \(\mathbf{z}\) equal those of \(\mathbf{x}\), and the other three change. We use these steps again for every layer, so we write them once as a function, which maps \(\mathbf{x}\) to \(\mathbf{z}\) for one layer.

\tutorialbarrier
\begin{codecell}
\begin{Verbatim}[commandchars=\\\{\},fontsize=\small]
\PYG{c+c1}{\PYGZsh{} Map of one coupling layer from the data x to the base z, with the networks s\PYGZus{}network and t\PYGZus{}network}
\PYG{k}{def}\PYG{+w}{ }\PYG{n+nf}{to\PYGZus{}base}\PYG{p}{(}\FancyVerbBreakAfterBreak{}\PYG{n}{x}\PYG{p}{,}\FancyVerbBreakAfterBreak{} \PYG{n}{mask}\PYG{p}{)}\PYG{p}{:}
    \PYG{n}{x\PYGZus{}frozen} \PYG{o}{=}\FancyVerbBreakAfterBreak{} \PYG{n}{x} \PYG{o}{*} \PYG{n}{mask}
    \PYG{n}{s} \PYG{o}{=}\FancyVerbBreakAfterBreak{} \PYG{n}{s\PYGZus{}network}\PYG{p}{(}\FancyVerbBreakAfterBreak{}\PYG{n}{x\PYGZus{}frozen}\PYG{p}{)} \PYG{o}{*} \PYG{p}{(}\FancyVerbBreakAfterBreak{}\PYG{l+m+mi}{1} \PYG{o}{\PYGZhy{}} \PYG{n}{mask}\PYG{p}{)}
    \PYG{n}{t} \PYG{o}{=}\FancyVerbBreakAfterBreak{} \PYG{n}{t\PYGZus{}network}\PYG{p}{(}\FancyVerbBreakAfterBreak{}\PYG{n}{x\PYGZus{}frozen}\PYG{p}{)} \PYG{o}{*} \PYG{p}{(}\FancyVerbBreakAfterBreak{}\PYG{l+m+mi}{1} \PYG{o}{\PYGZhy{}} \PYG{n}{mask}\PYG{p}{)}
    \PYG{k}{return} \PYG{n}{x\PYGZus{}frozen} \PYG{o}{+}\FancyVerbBreakAfterBreak{} \PYG{p}{(}\FancyVerbBreakAfterBreak{}\PYG{n}{x} \PYG{o}{*} \PYG{p}{(}\FancyVerbBreakAfterBreak{}\PYG{l+m+mi}{1} \PYG{o}{\PYGZhy{}} \PYG{n}{mask}\PYG{p}{)} \PYG{o}{*} \PYG{n}{torch}\PYG{o}{.}\FancyVerbBreakAfterBreak{}\PYG{n}{exp}\PYG{p}{(}\FancyVerbBreakAfterBreak{}\PYG{n}{s}\PYG{p}{)} \PYG{o}{+}\FancyVerbBreakAfterBreak{} \PYG{n}{t}\PYG{p}{)}
\end{Verbatim}
\end{codecell}

\section*{Inverting RealNVP}

Sampling needs the inverse map from \(\mathbf{z}\) to \(\mathbf{x}\). It follows the two-dimensional case. The frozen dimensions are unchanged, so \(\mathbf{x}_{1:d} = \mathbf{z}_{1:d}\). We solve the second equation for the other dimensions,

\[\mathbf{x}_{d+1:D} = (\mathbf{z}_{d+1:D} - t(\mathbf{z}_{1:d})) \odot \exp(-s(\mathbf{z}_{1:d})).\]

The networks take the frozen dimensions as input, and these are the same in \(\mathbf{x}\) and \(\mathbf{z}\). We therefore run \(s\) and \(t\) forward and invert the scale and the shift algebraically. We train \(s\) and \(t\) with standard forward passes and backpropagation, and the inverse needs no extra training. The networks can have any architecture.

We write the inverse of one layer as a function, like the map to the base.

\tutorialbarrier
\begin{codecell}
\begin{Verbatim}[commandchars=\\\{\},fontsize=\small]
\PYG{c+c1}{\PYGZsh{} Map of one coupling layer from the base z to the data x, with s and t evaluated on the frozen dimensions of z}
\PYG{k}{def}\PYG{+w}{ }\PYG{n+nf}{to\PYGZus{}data}\PYG{p}{(}\FancyVerbBreakAfterBreak{}\PYG{n}{z}\PYG{p}{,}\FancyVerbBreakAfterBreak{} \PYG{n}{mask}\PYG{p}{)}\PYG{p}{:}
    \PYG{n}{z\PYGZus{}frozen} \PYG{o}{=}\FancyVerbBreakAfterBreak{} \PYG{n}{z} \PYG{o}{*} \PYG{n}{mask}
    \PYG{n}{s} \PYG{o}{=}\FancyVerbBreakAfterBreak{} \PYG{n}{s\PYGZus{}network}\PYG{p}{(}\FancyVerbBreakAfterBreak{}\PYG{n}{z\PYGZus{}frozen}\PYG{p}{)} \PYG{o}{*} \PYG{p}{(}\FancyVerbBreakAfterBreak{}\PYG{l+m+mi}{1} \PYG{o}{\PYGZhy{}} \PYG{n}{mask}\PYG{p}{)}
    \PYG{n}{t} \PYG{o}{=}\FancyVerbBreakAfterBreak{} \PYG{n}{t\PYGZus{}network}\PYG{p}{(}\FancyVerbBreakAfterBreak{}\PYG{n}{z\PYGZus{}frozen}\PYG{p}{)} \PYG{o}{*} \PYG{p}{(}\FancyVerbBreakAfterBreak{}\PYG{l+m+mi}{1} \PYG{o}{\PYGZhy{}} \PYG{n}{mask}\PYG{p}{)}
    \PYG{k}{return} \PYG{n}{z\PYGZus{}frozen} \PYG{o}{+}\FancyVerbBreakAfterBreak{} \PYG{p}{(}\FancyVerbBreakAfterBreak{}\PYG{n}{z} \PYG{o}{*} \PYG{p}{(}\FancyVerbBreakAfterBreak{}\PYG{l+m+mi}{1} \PYG{o}{\PYGZhy{}} \PYG{n}{mask}\PYG{p}{)} \PYG{o}{\PYGZhy{}} \PYG{n}{t}\PYG{p}{)} \PYG{o}{*} \PYG{n}{torch}\PYG{o}{.}\FancyVerbBreakAfterBreak{}\PYG{n}{exp}\PYG{p}{(}\FancyVerbBreakAfterBreak{}\PYG{o}{\PYGZhy{}}\PYG{n}{s}\PYG{p}{)}
\end{Verbatim}
\end{codecell}

\section*{The Jacobian of RealNVP}

The second requirement is a cheap Jacobian determinant. The change of variables needs its logarithm (Section 3.2),

\[\ln p(\mathbf{x}) = \ln p(\mathbf{z}) + \ln\left|\det\left(\frac{\partial \mathbf{z}}{\partial \mathbf{x}}\right)\right|.\]

As in the two-dimensional case, the Jacobian matrix of a coupling layer is

\[\mathbf{J} = \frac{\partial \mathbf{z}}{\partial \mathbf{x}} = \begin{bmatrix} \mathrm{Id}_{d} & \mathbf{0}_{d \times (D-d)} \\ \frac{\partial \mathbf{z}_{d+1:D}}{\partial \mathbf{x}_{1:d}} & \text{diag}(\exp(s(\mathbf{x}_{1:d}))) \end{bmatrix}.\]

This is a lower-triangular block matrix, so its determinant is the product of its diagonal elements. The lower-left block does not enter,

\[\det(\mathbf{J}) = \prod_{j=d+1}^{D} \exp(s(\mathbf{x}_{1:d}))_j = \exp\left(\sum_{j=d+1}^{D} s(\mathbf{x}_{1:d})_j\right).\]

We work with the logarithm, which is stable numerically and is a sum,

\[\ln|\det(\mathbf{J})| = \sum_{j=d+1}^{D} s(\mathbf{x}_{1:d})_j.\]

We do not compute a \(D \times D\) determinant. We only sum the outputs of the network \(s\), an \(O(D)\) operation.

We extend the map to the base so that it also returns the sum of \(s\).

\tutorialbarrier
\begin{codecell}
\begin{Verbatim}[commandchars=\\\{\},fontsize=\small]
\PYG{c+c1}{\PYGZsh{} The map of one coupling layer to the base that also returns the log\PYGZhy{}determinant of the Jacobian, the sum of s}
\PYG{k}{def}\PYG{+w}{ }\PYG{n+nf}{to\PYGZus{}base\PYGZus{}log\PYGZus{}det}\PYG{p}{(}\FancyVerbBreakAfterBreak{}\PYG{n}{x}\PYG{p}{,}\FancyVerbBreakAfterBreak{} \PYG{n}{mask}\PYG{p}{)}\PYG{p}{:}
    \PYG{n}{x\PYGZus{}frozen} \PYG{o}{=}\FancyVerbBreakAfterBreak{} \PYG{n}{x} \PYG{o}{*} \PYG{n}{mask}
    \PYG{n}{s} \PYG{o}{=}\FancyVerbBreakAfterBreak{} \PYG{n}{s\PYGZus{}network}\PYG{p}{(}\FancyVerbBreakAfterBreak{}\PYG{n}{x\PYGZus{}frozen}\PYG{p}{)} \PYG{o}{*} \PYG{p}{(}\FancyVerbBreakAfterBreak{}\PYG{l+m+mi}{1} \PYG{o}{\PYGZhy{}} \PYG{n}{mask}\PYG{p}{)}
    \PYG{n}{t} \PYG{o}{=}\FancyVerbBreakAfterBreak{} \PYG{n}{t\PYGZus{}network}\PYG{p}{(}\FancyVerbBreakAfterBreak{}\PYG{n}{x\PYGZus{}frozen}\PYG{p}{)} \PYG{o}{*} \PYG{p}{(}\FancyVerbBreakAfterBreak{}\PYG{l+m+mi}{1} \PYG{o}{\PYGZhy{}} \PYG{n}{mask}\PYG{p}{)}
    \PYG{n}{z} \PYG{o}{=}\FancyVerbBreakAfterBreak{} \PYG{n}{x\PYGZus{}frozen} \PYG{o}{+}\FancyVerbBreakAfterBreak{} \PYG{p}{(}\FancyVerbBreakAfterBreak{}\PYG{n}{x} \PYG{o}{*} \PYG{p}{(}\FancyVerbBreakAfterBreak{}\PYG{l+m+mi}{1} \PYG{o}{\PYGZhy{}} \PYG{n}{mask}\PYG{p}{)} \PYG{o}{*} \PYG{n}{torch}\PYG{o}{.}\FancyVerbBreakAfterBreak{}\PYG{n}{exp}\PYG{p}{(}\FancyVerbBreakAfterBreak{}\PYG{n}{s}\PYG{p}{)} \PYG{o}{+}\FancyVerbBreakAfterBreak{} \PYG{n}{t}\PYG{p}{)}
    \PYG{k}{return} \PYG{n}{z}\PYG{p}{,}\FancyVerbBreakAfterBreak{} \PYG{n}{s}\PYG{o}{.}\FancyVerbBreakAfterBreak{}\PYG{n}{sum}\PYG{p}{(}\FancyVerbBreakAfterBreak{}\PYG{n}{dim}\PYG{o}{=}\FancyVerbBreakAfterBreak{}\PYG{l+m+mi}{1}\PYG{p}{)}
\end{Verbatim}
\end{codecell}

\section*{The Coupling Layer in PyTorch}

A flow needs many layers, each with its own mask and its own networks, so we pack the steps into a class. The class holds the mask and the two networks. It computes \(s\) and \(t\), maps \(\mathbf{x}\) to \(\mathbf{z}\) and returns the sum of \(s\), and maps \(\mathbf{z}\) to \(\mathbf{x}\). The sum of \(s\) is the log-determinant, because \(s\) is zero for the frozen dimensions. The method \texttt{register\_\allowbreak{}buffer} stores the mask with the layer, and training does not change it. The parameters of the flow are then the weights of the networks only.

\tutorialbarrier
\begin{codecell}

\end{codecell}

\section*{The Coupling Strategy}

In one layer, the frozen dimensions never change, so one layer cannot reshape the whole density. We therefore stack several layers and change which dimensions are frozen. Let \(h_1\) freeze the dimensions 1 to \(d\), and let \(h_2\) freeze the other dimensions. Both layers map the data toward the base. The map \(f^{-1}\) and its inverse \(f\) are

\tutorialdisplay{f^{-1} = h_2 \circ h_1,}{f = h_1^{-1} \circ h_2^{-1},}

so \(f\) applies the inverse layers in reverse order. The Jacobian matrices multiply by the chain rule, so the determinants multiply and the log-determinants add,

\[\ln|\det(\mathbf{J}_{f^{-1}})| = \ln|\det(\mathbf{J}_{h_1})| + \ln|\det(\mathbf{J}_{h_2})|.\]

Each layer adds one cheap term.

Our flow stacks 8 layers in 4 pairs, and each pair is like \(h_1\) and \(h_2\). In each pair, the first layer freezes a random set of 6 of the 13 abundances, and the second layer freezes the other 7. The two masks of a pair are opposite, so each pair transforms every abundance exactly once. Every abundance is then transformed by exactly 4 of the 8 layers.

The flow class keeps the layers in an \texttt{nn.\allowbreak{}ModuleList}, which registers them so that their parameters train. It maps \(\mathbf{x}\) to \(\mathbf{z}\) by running the layers in order, and it maps \(\mathbf{z}\) to \(\mathbf{x}\) by running the inverse layers in reverse order. It also computes the log-density,

\[\ln p(\mathbf{x}) = \ln p(\mathbf{z}) + \ln|\det(\mathbf{J})|.\]

The base density \(p(\mathbf{z})\) is a standard Gaussian in \(D\) dimensions, the product of \(D\) standard Gaussians,

\[\ln p(\mathbf{z}) = -\frac{1}{2}\sum_{j=1}^{D} z_j^2 - \frac{D}{2}\ln 2\pi.\]

\tutorialbarrier
\begin{codecell}

\end{codecell}

We create the flow with 8 layers and 128 neurons in each hidden layer of \(s\) and \(t\). The masks come in pairs. We fix the seed again, so that the masks and the initial weights do not depend on the earlier cells.

\tutorialbarrier
\begin{codecell}
\begin{Verbatim}[commandchars=\\\{\},fontsize=\small]
\PYG{c+c1}{\PYGZsh{} Opposite masks in pairs of layers, with a random choice of the 6 frozen abundances}
\PYG{n}{mask\PYGZus{}rng} \PYG{o}{=}\FancyVerbBreakAfterBreak{} \PYG{n}{np}\PYG{o}{.}\FancyVerbBreakAfterBreak{}\PYG{n}{random}\PYG{o}{.}\FancyVerbBreakAfterBreak{}\PYG{n}{default\PYGZus{}rng}\PYG{p}{(}\FancyVerbBreakAfterBreak{}\PYG{l+m+mi}{42}\PYG{p}{)}
\PYG{n}{masks} \PYG{o}{=}\FancyVerbBreakAfterBreak{} \PYG{p}{[}\FancyVerbBreakAfterBreak{}\PYG{p}{]}
\PYG{k}{for} \PYG{n}{\PYGZus{}} \PYG{o+ow}{in} \PYG{n+nb}{range}\PYG{p}{(}\FancyVerbBreakAfterBreak{}\PYG{l+m+mi}{4}\PYG{p}{)}\PYG{p}{:}
    \PYG{n}{pair\PYGZus{}mask} \PYG{o}{=}\FancyVerbBreakAfterBreak{} \PYG{n}{torch}\PYG{o}{.}\FancyVerbBreakAfterBreak{}\PYG{n}{zeros}\PYG{p}{(}\FancyVerbBreakAfterBreak{}\PYG{n}{n\PYGZus{}dim}\PYG{p}{)}

    \PYG{c+c1}{\PYGZsh{} A random set of 6 abundances is frozen in the first layer of the pair}
    \PYG{n}{pair\PYGZus{}mask}\PYG{p}{[}\FancyVerbBreakAfterBreak{}\PYG{n}{torch}\PYG{o}{.}\FancyVerbBreakAfterBreak{}\PYG{n}{tensor}\PYG{p}{(}\FancyVerbBreakAfterBreak{}\PYG{n}{mask\PYGZus{}rng}\PYG{o}{.}\FancyVerbBreakAfterBreak{}\PYG{n}{permutation}\PYG{p}{(}\FancyVerbBreakAfterBreak{}\PYG{n}{n\PYGZus{}dim}\PYG{p}{)}\PYG{p}{[}\FancyVerbBreakAfterBreak{}\PYG{p}{:}\PYG{n}{n\PYGZus{}dim} \PYG{o}{/}\PYG{o}{/} \PYG{l+m+mi}{2}\PYG{p}{]}\PYG{p}{)}\PYG{p}{]} \PYG{o}{=}\FancyVerbBreakAfterBreak{} \PYG{l+m+mi}{1}
    \PYG{n}{masks} \PYG{o}{+}\FancyVerbBreakAfterBreak{}\PYG{o}{=}\FancyVerbBreakAfterBreak{} \PYG{p}{[}\FancyVerbBreakAfterBreak{}\PYG{n}{pair\PYGZus{}mask}\PYG{p}{,}\FancyVerbBreakAfterBreak{} \PYG{l+m+mi}{1} \PYG{o}{\PYGZhy{}} \PYG{n}{pair\PYGZus{}mask}\PYG{p}{]}

\PYG{c+c1}{\PYGZsh{} Each row is the mask of one layer, with 1 for a frozen abundance}
\PYG{n+nb}{print}\PYG{p}{(}\FancyVerbBreakAfterBreak{}\PYG{n}{torch}\PYG{o}{.}\FancyVerbBreakAfterBreak{}\PYG{n}{stack}\PYG{p}{(}\FancyVerbBreakAfterBreak{}\PYG{n}{masks}\PYG{p}{)}\PYG{o}{.}\FancyVerbBreakAfterBreak{}\PYG{n}{int}\PYG{p}{(}\FancyVerbBreakAfterBreak{}\PYG{p}{)}\PYG{o}{.}\FancyVerbBreakAfterBreak{}\PYG{n}{numpy}\PYG{p}{(}\FancyVerbBreakAfterBreak{}\PYG{p}{)}\PYG{p}{)}  \PYG{c+c1}{\PYGZsh{} torch.stack joins the masks into one matrix}

\PYG{c+c1}{\PYGZsh{} Create the flow, with the seed fixed so that the weights do not depend on the earlier cells}
\PYG{n}{\PYGZus{}} \PYG{o}{=}\FancyVerbBreakAfterBreak{} \PYG{n}{torch}\PYG{o}{.}\FancyVerbBreakAfterBreak{}\PYG{n}{manual\PYGZus{}seed}\PYG{p}{(}\FancyVerbBreakAfterBreak{}\PYG{l+m+mi}{42}\PYG{p}{)}
\PYG{n}{flow} \PYG{o}{=}\FancyVerbBreakAfterBreak{} \PYG{n}{RealNVP}\PYG{p}{(}\FancyVerbBreakAfterBreak{}\PYG{n}{masks}\PYG{p}{,}\FancyVerbBreakAfterBreak{} \PYG{n}{n\PYGZus{}hidden}\PYG{o}{=}\FancyVerbBreakAfterBreak{}\PYG{l+m+mi}{128}\PYG{p}{)}
\PYG{n+nb}{print}\PYG{p}{(}\FancyVerbBreakAfterBreak{}\PYG{l+s+sa}{f}\PYG{l+s+s2}{\PYGZdq{}}\PYG{l+s+s2}{The flow has }\PYG{l+s+si}{\PYGZob{}}\PYG{n+nb}{sum}\PYG{p}{(}\FancyVerbBreakAfterBreak{}\PYG{n}{p}\PYG{o}{.}\FancyVerbBreakAfterBreak{}\PYG{n}{numel}\PYG{p}{(}\FancyVerbBreakAfterBreak{}\PYG{p}{)}\PYG{+w}{ }\PYG{k}{for}\PYG{+w}{ }\PYG{n}{p}\PYG{+w}{ }\PYG{o+ow}{in}\PYG{+w}{ }\PYG{n}{flow}\PYG{o}{.}\FancyVerbBreakAfterBreak{}\PYG{n}{parameters}\PYG{p}{(}\FancyVerbBreakAfterBreak{}\PYG{p}{)}\PYG{p}{)}\PYG{l+s+si}{\PYGZcb{}}\FancyVerbBreakAfterBreak{}\PYG{l+s+s2}{ parameters.}\PYG{l+s+s2}{\PYGZdq{}}\PYG{p}{)}
\end{Verbatim}
\end{codecell}
\begin{Verbatim}[fontsize=\footnotesize, xleftmargin=5mm]
[[1 0 0 1 0 0 0 1 0 1 1 0 1]
 [0 1 1 0 1 1 1 0 1 0 0 1 0]
 [0 0 1 0 1 1 0 1 0 0 1 0 1]
 [1 1 0 1 0 0 1 0 1 1 0 1 0]
 [0 1 0 0 1 0 1 1 1 0 0 0 1]
 [1 0 1 1 0 1 0 0 0 1 1 1 0]
 [0 1 0 0 0 1 1 0 1 1 0 1 0]
 [1 0 1 1 1 0 0 1 0 0 1 0 1]]
The flow has 583888 parameters.
\end{Verbatim}

Each row is the mask of one layer. The rows come in pairs of opposite masks. Every column has four zeros, so each abundance is transformed by four layers.

\section*{Training the Flow}

We train the flow by maximum likelihood (Section 4.3). The loss is the negative log-likelihood, averaged over the training stars. We minimize it with the Adam optimizer (Section 15.3) on mini-batches of 512 stars. An epoch is one pass through the training stars. The learning rate decays along a cosine curve from \(3 \times 10^{-4}\) to \(10^{-5}\) over 150 epochs.

A flow can fit the noise of the training stars. We therefore train all 150 epochs, compute the loss of the validation stars after every epoch, and select the epoch with the lowest validation loss. This is early stopping (Section 15.3), and we keep the weights of the selected epoch. The test stars take no part in this choice.

\tutorialbarrier
\begin{codecell}

\end{codecell}
\begin{Verbatim}[fontsize=\footnotesize, xleftmargin=5mm]
Lowest validation loss -4.767 at epoch 54, test loss -4.751
\end{Verbatim}

We plot the loss of the training stars and of the validation stars at every epoch.

\tutorialbarrier
\begin{codecell}

\end{codecell}
\tutorialplot{\includegraphics[width=4.392in,height=0.75\textheight,keepaspectratio]{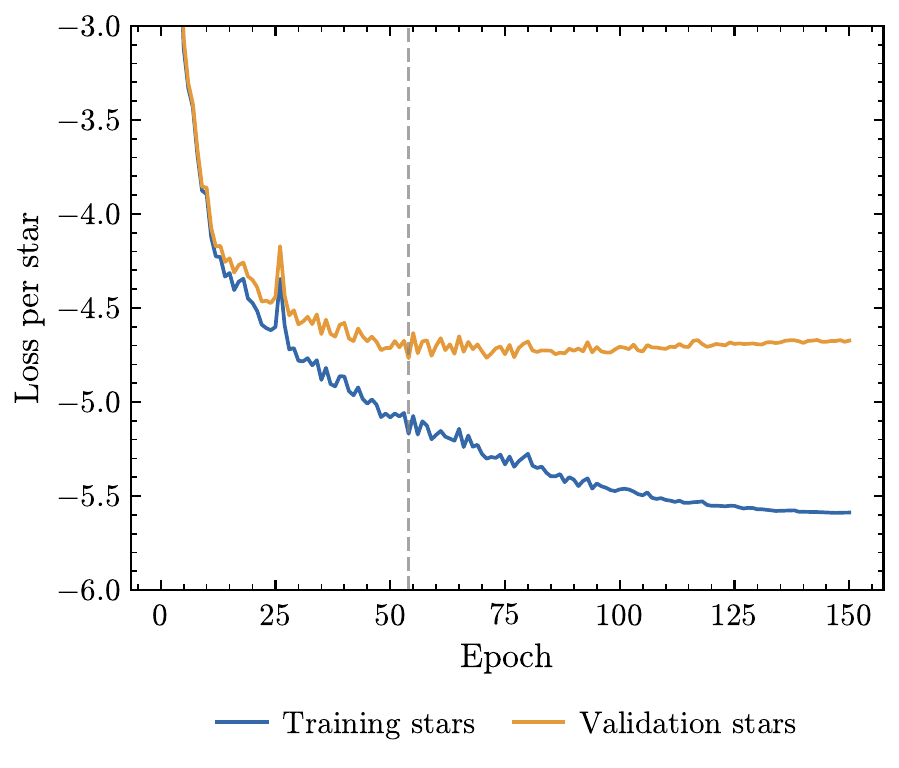}}%
  {\includegraphics[height=\tutorialheight]{tutorials/15d/fig03.pdf}}

The training loss keeps falling to the end, while the validation loss stops falling at about epoch 50 and stays near \(-4.7\). Its lowest value is at epoch 54, and it rises slowly to \(-4.67\) at epoch 150. The gap between the two losses grows to about 0.9 per star. The flow fits the training stars better than new stars, which is overfitting. More epochs do not help, so we keep the weights of epoch 54, which is early stopping (Section 15.3). The test loss at these weights is \(-4.751\), so the test log-likelihood of the flow is 4.751 per star.

\section*{Samples from the Flow}

So far we used the flow to evaluate densities. It also generates stars. To generate stars, we draw \(\mathbf{z}\) from the standard Gaussian, apply the map \(f\) and undo the standardization. The flow describes all 13 abundances. We look first at the plane of {[}Fe/H{]} and {[}Mg/Fe{]}, and then at other projections. We generate as many samples as there are test stars and compare them with the test stars.

\tutorialbarrier
\begin{codecell}

\end{codecell}
\tutorialplot{\includegraphics[width=6.426in,height=0.75\textheight,keepaspectratio]{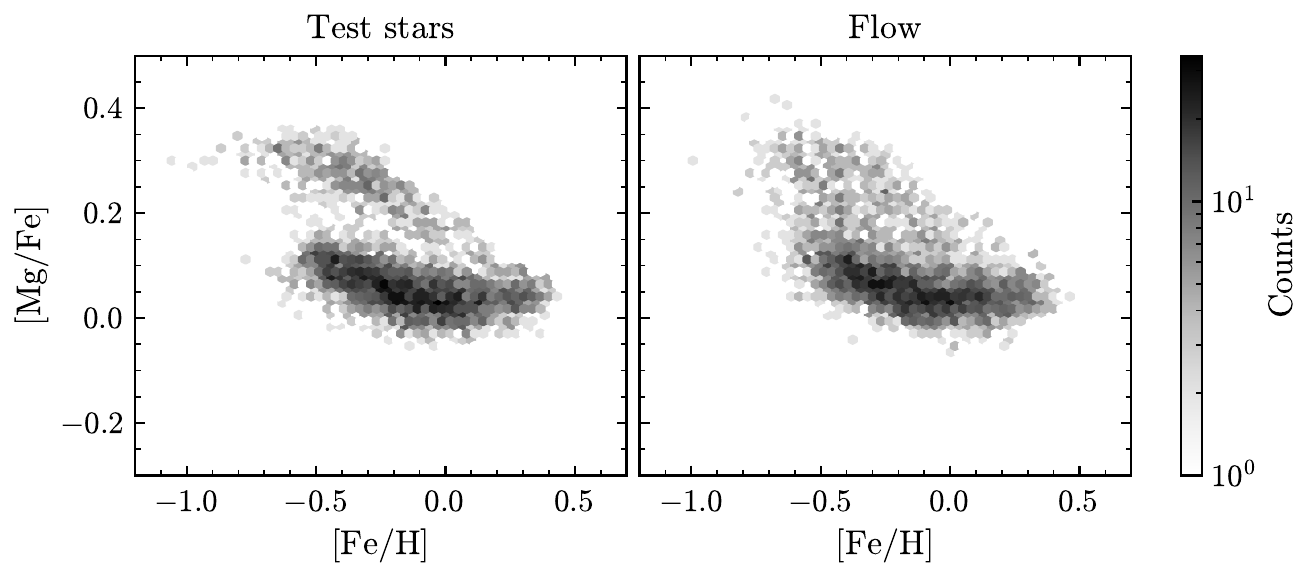}}%
  {\includegraphics[height=\tutorialheight]{tutorials/15d/fig04.pdf}}

The samples of the flow follow the structure of the stars in this plane, the two sequences with a lower density between them. They also fill part of the gap between the sequences, so the density is not yet as sharp as that of the stars. A normalizing flow must keep every layer invertible with a cheap Jacobian. RealNVP is the most basic design that meets this, and it limits how sharp the density can become. More expressive flows exist, and methods such as flow matching free the network from the invertibility constraint (Section 15.13).

The plane shows only two of the 13 dimensions, so the test log-likelihood is the more complete comparison.

\section*{The Other Abundances}

The flow models the 13 abundances together, so we look at other projections. We plot {[}X/Fe{]} against {[}Fe/H{]} for the other 12 elements. The grey density shows the flow samples of the previous section, without the hexagons that hold fewer than 3 samples. The red contours enclose 50, 80 and 95 percent of the test stars.

\tutorialbarrier
\begin{codecell}

\end{codecell}
\tutorialplot{\includegraphics[width=6.500in,height=0.75\textheight,keepaspectratio]{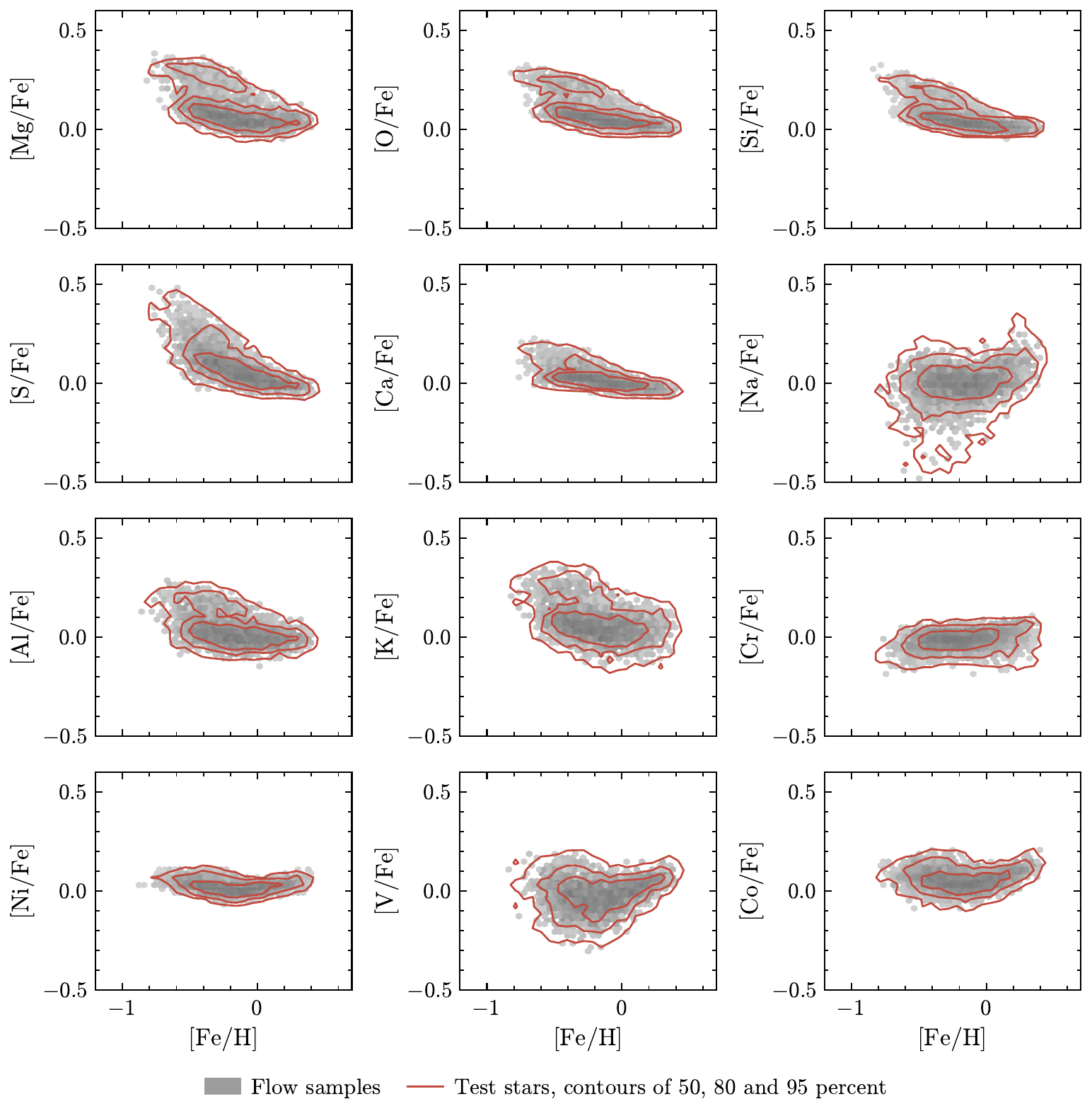}}%
  {\includegraphics[height=\tutorialheight]{tutorials/15d/fig05.pdf}}

The density of the flow and the contours of the stars overlap in all 12 projections. For {[}Mg/Fe{]}, {[}O/Fe{]} and {[}Si/Fe{]}, the stars form two sequences, and the density follows both. For the other elements, the stars form one band or one cloud, and the density covers it.

\section*{Summary}

A normalizing flow models a density with an invertible neural network. Our main lessons follow.

\begin{enumerate}
\def\labelenumi{\arabic{enumi}.}
\tightlist
\item
  \textbf{Exact density and sampling.} An invertible map between a standard Gaussian and the data gives the density of the data through the change of variables. The map itself generates new data.
\item
  \textbf{Invertibility and a cheap Jacobian.} The loss needs the density, so the map must be invertible. Its Jacobian determinant must also be cheap for every star at every step. A general neural network meets neither, so a flow restricts the architecture.
\item
  \textbf{Coupling layers.} A coupling layer freezes some dimensions and scales and shifts the others with networks of the frozen ones. The inverse is algebraic, and the Jacobian is triangular, so its log-determinant is a sum. One layer leaves the frozen dimensions unchanged, so we stack layers with different frozen dimensions. The log-determinants of the layers add.
\item
  \textbf{Limited expressiveness.} A coupling layer is a restricted map, and the samples of our flow still fill part of the gap between the sequences. More expressive flows, and flow-based methods such as flow matching, relax the restriction (Section 15.13).
\item
  \textbf{Overfitting and early stopping.} The training loss keeps falling while the validation loss stops falling and then rises slightly. Early stopping keeps the weights with the lowest validation loss, and the test stars give the final log-likelihood.
\end{enumerate}

An invertible neural network extends the use of neural networks from point predictions to density estimation. It gives the exact likelihood of each object and generates mock objects. It is also a basis for simulation-based inference, where neural density estimators learn likelihoods and posteriors from simulations (Section 15.14).

\cleardoublepage


\begin{thebibliography}{226}
\providecommand{\natexlab}[1]{#1}
\providecommand{\url}[1]{\texttt{#1}}
\expandafter\ifx\csname urlstyle\endcsname\relax
  \providecommand{\doi}[1]{doi: #1}\else
  \providecommand{\doi}{doi: \begingroup \urlstyle{rm}\Url}\fi

\bibitem[Abril-Pla et~al.(2023)Abril-Pla, Andreani, Carroll, Dong, Fonnesbeck,
  Kochurov, et~al.]{Abril-Pla2023}
O.~Abril-Pla, V.~Andreani, C.~Carroll, L.~Dong, C.~J. Fonnesbeck, M.~Kochurov,
  et~al.
\newblock {PyMC: a modern, and comprehensive probabilistic programming
  framework in Python}.
\newblock \emph{PeerJ Computer Science}, 9:\penalty0 e1516, 2023.

\bibitem[Adcock(1878)]{Adcock1878}
R.~J. Adcock.
\newblock {A problem in least squares}.
\newblock \emph{The Analyst}, 5\penalty0 (2):\penalty0 53--54, 1878.

\bibitem[Adlam and Pennington(2020)]{Adlam2020}
B.~Adlam and J.~Pennington.
\newblock {Understanding double descent requires a fine-grained bias-variance
  decomposition}.
\newblock \emph{arXiv preprint arXiv:2011.03321}, 2020.

\bibitem[Aigrain and Foreman-Mackey(2023)]{Aigrain2023}
S.~Aigrain and D.~Foreman-Mackey.
\newblock {Gaussian process regression for astronomical time series}.
\newblock \emph{Annual Review of Astronomy and Astrophysics}, 61\penalty0
  (1):\penalty0 329--371, 2023.

\bibitem[Akaike(1974)]{Akaike1974}
H.~Akaike.
\newblock {A new look at the statistical model identification}.
\newblock \emph{IEEE Transactions on Automatic Control}, 19\penalty0
  (6):\penalty0 716--723, 1974.

\bibitem[Akritas and Bershady(1996)]{Akritas1996}
M.~G. Akritas and M.~A. Bershady.
\newblock {Linear regression for astronomical data with measurement errors and
  intrinsic scatter}.
\newblock \emph{The Astrophysical Journal}, 470:\penalty0 706, 1996.

\bibitem[Albert and Chib(1993)]{AlbertChib1993}
J.~H. Albert and S.~Chib.
\newblock {Bayesian analysis of binary and polychotomous response data}.
\newblock \emph{Journal of the American Statistical Association}, 88\penalty0
  (422):\penalty0 669--679, 1993.

\bibitem[Allwein et~al.(2000)Allwein, Schapire, and Singer]{Allwein2001}
E.~L. Allwein, R.~E. Schapire, and Y.~Singer.
\newblock {Reducing multiclass to binary: A unifying approach for margin
  classifiers}.
\newblock \emph{Journal of Machine Learning Research}, 1:\penalty0 113--141,
  2000.

\bibitem[Anderson(1958)]{Anderson1958}
T.~W. Anderson.
\newblock \emph{{An introduction to multivariate statistical analysis}}.
\newblock Wiley, 1958.

\bibitem[Aronszajn(1950)]{Aronszajn1950}
N.~Aronszajn.
\newblock {Theory of reproducing kernels}.
\newblock \emph{Transactions of the American Mathematical Society}, 68\penalty0
  (3):\penalty0 337--404, 1950.

\bibitem[Arthur and Vassilvitskii(2007)]{ArthurVassilvitskii2007}
D.~Arthur and S.~Vassilvitskii.
\newblock {k-means++: the advantages of careful seeding}.
\newblock In \emph{Proceedings of the Eighteenth Annual ACM-SIAM Symposium on
  Discrete Algorithms}, pages 1027--1035. Society for Industrial and Applied
  Mathematics, 2007.

\bibitem[Bailer-Jones et~al.(1998)Bailer-Jones, Irwin, and von
  Hippel]{BailerJones1998}
C.~A.~L. Bailer-Jones, M.~Irwin, and T.~von Hippel.
\newblock {Automated classification of stellar spectra – II. Two-dimensional
  classification with neural networks and principal components analysis}.
\newblock \emph{Monthly Notices of the Royal Astronomical Society},
  298\penalty0 (2):\penalty0 361--377, 1998.

\bibitem[Bayes(1763)]{Bayes1763}
T.~Bayes.
\newblock {LII. An essay towards solving a problem in the doctrine of chances.
  By the late Rev. Mr. Bayes, F. R. S. communicated by Mr. Price, in a letter
  to John Canton, A. M. F. R. S}.
\newblock \emph{Philosophical Transactions of the Royal Society of London},
  53:\penalty0 370--418, 1763.

\bibitem[Belkin et~al.(2019)Belkin, Hsu, Ma, and Mandal]{Belkin2019}
M.~Belkin, D.~Hsu, S.~Ma, and S.~Mandal.
\newblock {Reconciling modern machine-learning practice and the classical
  bias–variance trade-off}.
\newblock \emph{Proceedings of the National Academy of Sciences}, 116\penalty0
  (32):\penalty0 15849--15854, 2019.

\bibitem[Bentler and Weeks(1980)]{BentlerWeeks1980}
P.~M. Bentler and D.~G. Weeks.
\newblock {Linear structural equations with latent variables}.
\newblock \emph{Psychometrika}, 45\penalty0 (3):\penalty0 289--308, 1980.

\bibitem[Berkson(1944)]{Berkson1944}
J.~Berkson.
\newblock {Application of the logistic function to bio-assay}.
\newblock \emph{Journal of the American Statistical Association}, 39\penalty0
  (227):\penalty0 357--365, 1944.

\bibitem[Bevington(1969)]{Bevington1969}
P.~R. Bevington.
\newblock \emph{{Data reduction and error analysis for the physical sciences}}.
\newblock McGraw-Hill, 1969.

\bibitem[Bezdek(1981)]{Bezdek1981}
J.~C. Bezdek.
\newblock \emph{{Pattern recognition with fuzzy objective function
  algorithms}}.
\newblock Plenum Press, New York, 1981.

\bibitem[Bishop(1994)]{Bishop1994}
C.~M. Bishop.
\newblock {Mixture density networks}.
\newblock Technical Report NCRG/94/004, Neural Computing Research Group, Aston
  University, Birmingham, UK, 1994.

\bibitem[Bishop(2006)]{Bishop2006}
C.~M. Bishop.
\newblock \emph{{Pattern recognition and machine learning}}.
\newblock Springer, 2006.

\bibitem[Bliss(1934)]{Bliss1934}
C.~I. Bliss.
\newblock {The method of probits}.
\newblock \emph{Science}, 79\penalty0 (2037):\penalty0 38--39, 1934.

\bibitem[Bommasani et~al.(2021)Bommasani, Hudson, Adeli, Altman, Arora, von
  Arx, et~al.]{Bommasani2022}
R.~Bommasani, D.~A. Hudson, E.~Adeli, R.~Altman, S.~Arora, S.~von Arx, et~al.
\newblock {On the opportunities and risks of foundation models}.
\newblock \emph{arXiv preprint arXiv:2108.07258}, 2021.

\bibitem[Box and Muller(1958)]{BoxMuller1958}
G.~E.~P. Box and M.~E. Muller.
\newblock {A note on the generation of random normal deviates}.
\newblock \emph{The Annals of Mathematical Statistics}, 29:\penalty0 610--611,
  1958.

\bibitem[Box and Tiao(1973)]{Box1973}
G.~E.~P. Box and G.~C. Tiao.
\newblock \emph{{Bayesian inference in statistical analysis}}.
\newblock Addison-Wesley, Reading, MA, 1973.

\bibitem[Bozdogan(1987)]{Bozdogan1987}
H.~Bozdogan.
\newblock {Model selection and Akaike's information criterion (AIC): the
  general theory and its analytical extensions}.
\newblock \emph{Psychometrika}, 52\penalty0 (3):\penalty0 345--370, 1987.

\bibitem[Bridle(1990)]{Bridle1990}
J.~S. Bridle.
\newblock {Training stochastic model recognition algorithms as networks can
  lead to maximum mutual information estimation of parameters}.
\newblock In D.~S. Touretzky, editor, \emph{Advances in Neural Information
  Processing Systems}, volume~2, pages 211--217. Morgan Kaufmann, 1990.

\bibitem[Brown et~al.(2020)Brown, Mann, Ryder, Subbiah, Kaplan, Dhariwal,
  et~al.]{Brown2020}
T.~B. Brown, B.~Mann, N.~Ryder, M.~Subbiah, J.~Kaplan, P.~Dhariwal, et~al.
\newblock {Language models are few-shot learners}.
\newblock \emph{arXiv preprint arXiv:2005.14165}, 2020.

\bibitem[Carroll and Ruppert(1996)]{CarrollRuppert1996}
R.~J. Carroll and D.~Ruppert.
\newblock {The use and misuse of orthogonal regression in linear
  errors-in-variables models}.
\newblock \emph{The American Statistician}, 50\penalty0 (1):\penalty0 1--6,
  1996.

\bibitem[Carroll et~al.(2006)Carroll, Ruppert, Stefanski, and
  Crainiceanu]{Carroll2006}
R.~J. Carroll, D.~Ruppert, L.~A. Stefanski, and C.~M. Crainiceanu.
\newblock \emph{{Measurement error in nonlinear models: A modern perspective,
  Second edition}}.
\newblock Chapman and Hall/CRC, 2006.

\bibitem[Casella and Berger(2002)]{Casella2002}
G.~Casella and R.~L. Berger.
\newblock \emph{{Statistical inference}}.
\newblock Duxbury Press, 2002.

\bibitem[Cauchy(1847)]{Cauchy1847}
A.~L. Cauchy.
\newblock {Méthode générale pour la résolution des systèmes d'équations
  simultanées}.
\newblock \emph{Comptes Rendus de l'Académie des Sciences de Paris},
  25:\penalty0 536--538, 1847.

\bibitem[Christensen et~al.(2001)Christensen, Meyer, Knox, and
  Luey]{Christensen2001}
N.~Christensen, R.~Meyer, L.~Knox, and B.~Luey.
\newblock {Bayesian methods for cosmological parameter estimation from cosmic
  microwave background measurements}.
\newblock \emph{Classical and Quantum Gravity}, 18\penalty0 (14):\penalty0
  2677--2688, 2001.

\bibitem[Comon(1994)]{Comon1994}
P.~Comon.
\newblock {Independent component analysis, a new concept?}
\newblock \emph{Signal Processing}, 36\penalty0 (3):\penalty0 287--314, 1994.

\bibitem[Cowles and Carlin(1996)]{Cowles1996}
M.~K. Cowles and B.~P. Carlin.
\newblock {Markov chain Monte Carlo convergence diagnostics: a comparative
  review}.
\newblock \emph{Journal of the American Statistical Association}, 91\penalty0
  (434):\penalty0 883--904, 1996.

\bibitem[Cox(1958)]{Cox1958}
D.~R. Cox.
\newblock {The regression analysis of binary sequences}.
\newblock \emph{Journal of the Royal Statistical Society: Series B
  (Methodological)}, 20\penalty0 (2):\penalty0 215--232, 1958.

\bibitem[Cox and Hinkley(1974)]{Cox1974}
D.~R. Cox and D.~V. Hinkley.
\newblock \emph{{Theoretical statistics}}.
\newblock Chapman and Hall, 1974.

\bibitem[Cram{\'e}r(1946)]{Cramer1946}
H.~Cram{\'e}r.
\newblock \emph{{Mathematical methods of statistics}}.
\newblock Princeton University Press, 1946.

\bibitem[Cranmer et~al.(2020)Cranmer, Brehmer, and Louppe]{Cranmer2020}
K.~Cranmer, J.~Brehmer, and G.~Louppe.
\newblock {The frontier of simulation-based inference}.
\newblock \emph{Proceedings of the National Academy of Sciences}, 117\penalty0
  (48):\penalty0 30055--30062, 2020.

\bibitem[Cressie(1993)]{Cressie1993}
N.~Cressie.
\newblock \emph{{Statistics for spatial data}}.
\newblock Wiley, 1993.

\bibitem[Csat{\'o} and Opper(2002)]{CsatoOpper2002}
L.~Csat{\'o} and M.~Opper.
\newblock {Sparse on-line Gaussian processes}.
\newblock \emph{Neural Computation}, 14\penalty0 (3):\penalty0 641--668, 2002.

\bibitem[Curry(1944)]{Curry1944}
H.~B. Curry.
\newblock {The method of steepest descent for non-linear minimization
  problems}.
\newblock \emph{Quarterly of Applied Mathematics}, 2\penalty0 (3):\penalty0
  258--261, 1944.

\bibitem[Cybenko(1989)]{Cybenko1989}
G.~Cybenko.
\newblock {Approximation by superpositions of a sigmoidal function}.
\newblock \emph{Math. Control Signal Systems}, 2:\penalty0 303--314, 1989.

\bibitem[Davison and Hinkley(1997)]{Davison1997}
A.~C. Davison and D.~V. Hinkley.
\newblock \emph{{Bootstrap methods and their application}}.
\newblock Cambridge University Press, 1997.

\bibitem[Day(1969)]{Day1969}
N.~E. Day.
\newblock {Estimating the components of a mixture of normal distributions}.
\newblock \emph{Biometrika}, 56\penalty0 (3):\penalty0 463--474, 1969.

\bibitem[DeGroot(1970)]{DeGroot1970}
M.~H. DeGroot.
\newblock \emph{{Optimal statistical decisions}}.
\newblock McGraw-Hill, 1970.

\bibitem[Deming(1943)]{Deming1943}
W.~E. Deming.
\newblock \emph{{Statistical adjustment of data}}.
\newblock Wiley, 1943.

\bibitem[Dempster et~al.(1977)Dempster, Laird, and Rubin]{Dempster1977}
A.~P. Dempster, N.~M. Laird, and D.~B. Rubin.
\newblock {Maximum likelihood from incomplete data via the EM algorithm}.
\newblock \emph{Journal of the Royal Statistical Society. Series B
  (Methodological)}, 39\penalty0 (1):\penalty0 1--38, 1977.

\bibitem[Dennis and Schnabel(1983)]{DennisSchnabel1983}
J.~E. Dennis, Jr. and R.~B. Schnabel.
\newblock \emph{{Numerical methods for unconstrained optimization and nonlinear
  equations}}.
\newblock Prentice-Hall, 1983.

\bibitem[Devlin et~al.(2018)Devlin, Chang, Lee, and Toutanova]{Devlin2019}
J.~Devlin, M.-W. Chang, K.~Lee, and K.~Toutanova.
\newblock {BERT: pre-training of deep bidirectional transformers for language
  understanding}.
\newblock \emph{arXiv preprint arXiv:1810.04805}, 2018.

\bibitem[Devroye(1986)]{Devroye1986}
L.~Devroye.
\newblock \emph{{Non-uniform random variate generation}}.
\newblock Springer, 1986.

\bibitem[Dietterich and Bakiri(1991)]{DietterichBakiri1991}
T.~G. Dietterich and G.~Bakiri.
\newblock {Error-correcting output codes: A general method for improving
  multiclass inductive learning programs}.
\newblock In \emph{Proceedings of the Ninth National Conference on Artificial
  Intelligence}, pages 572--577. AAAI Press, 1991.

\bibitem[Dietterich and Bakiri(1995)]{DietterichBakiri1995}
T.~G. Dietterich and G.~Bakiri.
\newblock {Solving multiclass learning problems via error-correcting output
  codes}.
\newblock \emph{Journal of Artificial Intelligence Research}, 2\penalty0
  (1):\penalty0 263--286, 1995.

\bibitem[Dinh et~al.(2016)Dinh, Sohl-Dickstein, and Bengio]{Dinh2017}
L.~Dinh, J.~Sohl-Dickstein, and S.~Bengio.
\newblock {Density estimation using Real NVP}.
\newblock \emph{arXiv preprint arXiv:1605.08803}, 2016.

\bibitem[Duda et~al.(2001)Duda, Hart, and Stork]{Duda2001}
R.~O. Duda, P.~E. Hart, and D.~G. Stork.
\newblock \emph{{Pattern classification (2nd ed.)}}.
\newblock Wiley, 2001.

\bibitem[Dunn(1973)]{Dunn1973}
J.~C. Dunn.
\newblock {A fuzzy relative of the ISODATA process and its use in detecting
  compact well-separated clusters}.
\newblock \emph{Journal of Cybernetics}, 3\penalty0 (3):\penalty0 32--57, 1973.

\bibitem[Durkan et~al.(2019)Durkan, Bekasov, Murray, and
  Papamakarios]{Durkan2019}
C.~Durkan, A.~Bekasov, I.~Murray, and G.~Papamakarios.
\newblock {Neural spline flows}.
\newblock \emph{arXiv preprint arXiv:1906.04032}, 2019.

\bibitem[Eadie et~al.(2023)Eadie, Speagle, Cisewski-Kehe, Foreman-Mackey,
  Huppenkothen, Jones, Springford, and Tak]{Eadie2023}
G.~M. Eadie, J.~S. Speagle, J.~Cisewski-Kehe, D.~Foreman-Mackey,
  D.~Huppenkothen, D.~E. Jones, A.~Springford, and H.~Tak.
\newblock {Practical guidance for Bayesian inference in astronomy}.
\newblock \emph{arXiv preprint arXiv:2302.04703}, 2023.

\bibitem[Eckart and Young(1936)]{EckartYoung1936}
C.~Eckart and G.~Young.
\newblock {The approximation of one matrix by another of lower rank}.
\newblock \emph{Psychometrika}, 1\penalty0 (3):\penalty0 211--218, 1936.

\bibitem[Efron(1975)]{Efron1975}
B.~Efron.
\newblock {The efficiency of logistic regression compared to normal
  discriminant analysis}.
\newblock \emph{Journal of the American Statistical Association}, 70\penalty0
  (352):\penalty0 892--898, 1975.

\bibitem[Efron(1979)]{Efron1979}
B.~Efron.
\newblock {Bootstrap methods: Another look at the jackknife}.
\newblock \emph{The Annals of Statistics}, 7\penalty0 (1):\penalty0 1--26,
  1979.

\bibitem[Elman(1990)]{Elman1990}
J.~L. Elman.
\newblock {Finding structure in time}.
\newblock \emph{Cognitive Science}, 14\penalty0 (2):\penalty0 179--211, 1990.

\bibitem[Faber and Jackson(1976)]{FaberJackson1976}
S.~M. Faber and R.~E. Jackson.
\newblock {Velocity dispersions and mass-to-light ratios for elliptical
  galaxies}.
\newblock \emph{The Astrophysical Journal}, 204:\penalty0 668--683, 1976.

\bibitem[Feroz and Hobson(2008)]{Feroz2008}
F.~Feroz and M.~P. Hobson.
\newblock {Multimodal nested sampling: an efficient and robust alternative to
  Markov chain Monte Carlo methods for astronomical data analyses}.
\newblock \emph{Monthly Notices of the Royal Astronomical Society},
  384\penalty0 (2):\penalty0 449--463, 2008.

\bibitem[Ferrarese and Merritt(2000)]{Ferrarese2000}
L.~Ferrarese and D.~Merritt.
\newblock {A fundamental relation between supermassive black holes and their
  host galaxies}.
\newblock \emph{The Astrophysical Journal Letters}, 539\penalty0 (1):\penalty0
  L9--L12, 2000.

\bibitem[Finney(1947)]{Finney1947}
D.~J. Finney.
\newblock \emph{{Probit analysis: A statistical treatment of the sigmoid
  response curve}}.
\newblock Cambridge University Press, Cambridge, 1947.

\bibitem[Fisher(1922)]{Fisher1922}
R.~A. Fisher.
\newblock {On the mathematical foundations of theoretical statistics}.
\newblock \emph{Philosophical Transactions of the Royal Society of London.
  Series A, Containing Papers of a Mathematical or Physical Character},
  222:\penalty0 309--368, 1922.

\bibitem[Fisher(1936)]{Fisher1936}
R.~A. Fisher.
\newblock {The use of multiple measurements in taxonomic problems}.
\newblock \emph{Annals of Eugenics}, 7\penalty0 (2):\penalty0 179--188, 1936.

\bibitem[Fishman(1996)]{Fishman1996}
G.~S. Fishman.
\newblock \emph{{Monte Carlo: concepts, algorithms, and applications}}.
\newblock Springer, 1996.

\bibitem[Ford(2005)]{Ford2005}
E.~B. Ford.
\newblock {Quantifying the uncertainty in the orbits of extrasolar planets}.
\newblock \emph{The Astronomical Journal}, 129\penalty0 (3):\penalty0
  1706--1717, 2005.

\bibitem[Foreman-Mackey et~al.(2013)Foreman-Mackey, Hogg, Lang, and
  Goodman]{Foreman-Mackey2013}
D.~Foreman-Mackey, D.~W. Hogg, D.~Lang, and J.~Goodman.
\newblock {emcee: the MCMC hammer}.
\newblock \emph{Publications of the Astronomical Society of the Pacific},
  125\penalty0 (925):\penalty0 306--312, 2013.

\bibitem[Foreman-Mackey et~al.(2017)Foreman-Mackey, Agol, Ambikasaran, and
  Angus]{Foreman-Mackey2017}
D.~Foreman-Mackey, E.~Agol, S.~Ambikasaran, and R.~Angus.
\newblock {Fast and scalable Gaussian process modeling with applications to
  astronomical time series}.
\newblock \emph{The Astronomical Journal}, 154\penalty0 (6):\penalty0 220,
  2017.

\bibitem[Forsythe(1972)]{Forsythe1972}
G.~E. Forsythe.
\newblock {Von Neumann's comparison method for random sampling from the normal
  and other distributions}.
\newblock \emph{Mathematics of Computation}, 26:\penalty0 817--826, 1972.

\bibitem[Fuller(1987)]{Fuller1987}
W.~A. Fuller.
\newblock \emph{{Measurement error models}}.
\newblock Wiley, 1987.

\bibitem[Fürnkranz(2002)]{Fuerkranz2002}
J.~Fürnkranz.
\newblock {Round robin classification}.
\newblock \emph{Journal of Machine Learning Research}, 2:\penalty0 721--747,
  2002.

\bibitem[Gauss(1823)]{Gauss1823}
C.~F. Gauss.
\newblock \emph{{Theoria combinationis observationum erroribus minimis
  obnoxiae}}.
\newblock H. Dieterich, 1823.

\bibitem[Gebhardt et~al.(2000)Gebhardt, Bender, Bower, Dressler, Faber,
  Filippenko, et~al.]{Gebhardt2000}
K.~Gebhardt, R.~Bender, G.~Bower, A.~Dressler, S.~M. Faber, A.~V. Filippenko,
  et~al.
\newblock {A relationship between nuclear black hole mass and galaxy velocity
  dispersion}.
\newblock \emph{The Astrophysical Journal Letters}, 539\penalty0 (1):\penalty0
  L13--L16, 2000.

\bibitem[Gelfand and Smith(1990)]{Gelfand1990}
A.~E. Gelfand and A.~F.~M. Smith.
\newblock {Sampling-based approaches to calculating marginal densities}.
\newblock \emph{Journal of the American Statistical Association}, 85\penalty0
  (410):\penalty0 398--409, 1990.

\bibitem[Gelman and Rubin(1992)]{Gelman1992}
A.~Gelman and D.~B. Rubin.
\newblock {Inference from iterative simulation using multiple sequences}.
\newblock \emph{Statistical Science}, 7:\penalty0 457--472, 1992.

\bibitem[Gelman et~al.(2013)Gelman, Carlin, Stern, Dunson, Vehtari, and
  Rubin]{Gelman2013}
A.~Gelman, J.~B. Carlin, H.~S. Stern, D.~B. Dunson, A.~Vehtari, and D.~B.
  Rubin.
\newblock \emph{{Bayesian data analysis (3rd ed.)}}.
\newblock Chapman and Hall/CRC, 2013.

\bibitem[Geman and Geman(1984)]{Geman1984}
S.~Geman and D.~Geman.
\newblock {Stochastic relaxation, Gibbs distributions, and the Bayesian
  restoration of images}.
\newblock \emph{IEEE Transactions on Pattern Analysis and Machine
  Intelligence}, 6\penalty0 (6):\penalty0 721--741, 1984.

\bibitem[Geweke(1992)]{Geweke1992}
J.~Geweke.
\newblock {Evaluating the accuracy of sampling-based approaches to the
  calculation of posterior moments}.
\newblock In J.~M. Bernardo, J.~O. Berger, A.~P. Dawid, and A.~F.~M. Smith,
  editors, \emph{Bayesian Statistics 4}, pages 169--193. Oxford University
  Press, Oxford, 1992.

\bibitem[Gilks et~al.(1996)Gilks, Richardson, and Spiegelhalter]{Gilks1996}
W.~R. Gilks, S.~Richardson, and D.~J. Spiegelhalter, editors.
\newblock \emph{{Markov chain Monte Carlo in practice}}.
\newblock Chapman and Hall, 1996.

\bibitem[Goertzel(1949)]{Goertzel1949}
G.~Goertzel.
\newblock {Quota sampling and importance functions in stochastic solution of
  particle problems}.
\newblock Technical Report ORNL-434, Oak Ridge National Laboratory, Oak Ridge,
  TN, 1949.
\newblock Also issued as AECD-2793.

\bibitem[Golub and Kahan(1965)]{GolubKahan1965}
G.~Golub and W.~Kahan.
\newblock {Calculating the singular values and pseudo-inverse of a matrix}.
\newblock \emph{Journal of the Society for Industrial and Applied Mathematics
  Series B Numerical Analysis}, 2\penalty0 (2):\penalty0 205--224, 1965.

\bibitem[Golub and Reinsch(1970)]{GolubReinsch1970}
G.~H. Golub and C.~Reinsch.
\newblock {Singular value decomposition and least squares solutions}.
\newblock \emph{Numerische Mathematik}, 14:\penalty0 403--420, 1970.

\bibitem[Golub and Van~Loan(1980)]{GolubVanLoan1980}
G.~H. Golub and C.~F. Van~Loan.
\newblock {An analysis of the total least squares problem}.
\newblock \emph{SIAM Journal on Numerical Analysis}, 17\penalty0 (6):\penalty0
  883--893, 1980.

\bibitem[Gonçalves et~al.(2020)Gonçalves, Lueckmann, Deistler, Nonnenmacher,
  Öcal, Bassetto, et~al.]{Goncalves2020}
P.~J. Gonçalves, J.-M. Lueckmann, M.~Deistler, M.~Nonnenmacher, K.~Öcal,
  G.~Bassetto, et~al.
\newblock {Training deep neural density estimators to identify mechanistic
  models of neural dynamics}.
\newblock \emph{eLife}, 9:\penalty0 e56261, 2020.

\bibitem[Goodfellow et~al.(2014)Goodfellow, Pouget-Abadie, Mirza, Xu,
  Warde-Farley, Ozair, Courville, and Bengio]{Goodfellow2014}
I.~J. Goodfellow, J.~Pouget-Abadie, M.~Mirza, B.~Xu, D.~Warde-Farley, S.~Ozair,
  A.~Courville, and Y.~Bengio.
\newblock {Generative adversarial networks}.
\newblock \emph{arXiv preprint arXiv:1406.2661}, 2014.

\bibitem[Greenberg et~al.(2019)Greenberg, Nonnenmacher, and
  Macke]{Greenberg2019}
D.~S. Greenberg, M.~Nonnenmacher, and J.~H. Macke.
\newblock {Automatic posterior transformation for likelihood-free inference}.
\newblock \emph{arXiv preprint arXiv:1905.07488}, 2019.

\bibitem[Greene et~al.(2021)Greene, Lancaster, Ting, Koposov, Danieli, Huang,
  Jiang, Greco, and Strader]{Greene2021}
J.~E. Greene, L.~Lancaster, Y.-S. Ting, S.~E. Koposov, S.~Danieli, S.~Huang,
  F.~Jiang, J.~P. Greco, and J.~Strader.
\newblock {A search for wandering black holes in the {Milky Way} with {Gaia}
  and {DECaLS}}.
\newblock \emph{The Astrophysical Journal}, 917:\penalty0 17, 2021.

\bibitem[Gregory(2005{\natexlab{a}})]{Gregory2005}
P.~C. Gregory.
\newblock {A Bayesian analysis of extrasolar planet data for HD 73526}.
\newblock \emph{The Astrophysical Journal}, 631\penalty0 (2):\penalty0
  1198--1214, 2005{\natexlab{a}}.

\bibitem[Gregory(2005{\natexlab{b}})]{Gregory2005book}
P.~C. Gregory.
\newblock \emph{{Bayesian logical data analysis for the physical sciences: A
  comparative approach with {Mathematica} support}}.
\newblock Cambridge University Press, Cambridge, 2005{\natexlab{b}}.

\bibitem[Hartigan(1975)]{Hartigan1975}
J.~A. Hartigan.
\newblock \emph{{Clustering algorithms}}.
\newblock Wiley, 1975.

\bibitem[Hartigan and Wong(1979)]{HartiganWong1979}
J.~A. Hartigan and M.~A. Wong.
\newblock {Algorithm AS 136: a k-means clustering algorithm}.
\newblock \emph{Journal of the Royal Statistical Society. Series C (Applied
  Statistics)}, 28\penalty0 (1):\penalty0 100--108, 1979.

\bibitem[Hastings(1970)]{Hastings1970}
W.~K. Hastings.
\newblock {Monte Carlo sampling methods using Markov chains and their
  applications}.
\newblock \emph{Biometrika}, 57\penalty0 (1):\penalty0 97--109, 1970.

\bibitem[He et~al.(2016)He, Zhang, Ren, and Sun]{He2016}
K.~He, X.~Zhang, S.~Ren, and J.~Sun.
\newblock {Deep residual learning for image recognition}.
\newblock In \emph{2016 IEEE Conference on Computer Vision and Pattern
  Recognition (CVPR)}, pages 770--778. IEEE, 2016.

\bibitem[Hinton and Salakhutdinov(2006)]{HintonSalakhutdinov2006}
G.~E. Hinton and R.~R. Salakhutdinov.
\newblock {Reducing the dimensionality of data with neural networks}.
\newblock \emph{Science}, 313\penalty0 (5786):\penalty0 504--507, 2006.

\bibitem[Hinton et~al.(2012)Hinton, Srivastava, Krizhevsky, Sutskever, and
  Salakhutdinov]{Hinton2012}
G.~E. Hinton, N.~Srivastava, A.~Krizhevsky, I.~Sutskever, and R.~R.
  Salakhutdinov.
\newblock {Improving neural networks by preventing co-adaptation of feature
  detectors}.
\newblock \emph{arXiv preprint arXiv:1207.0580}, 2012.

\bibitem[Ho et~al.(2020)Ho, Jain, and Abbeel]{Ho2020}
J.~Ho, A.~Jain, and P.~Abbeel.
\newblock {Denoising diffusion probabilistic models}.
\newblock \emph{arXiv preprint arXiv:2006.11239}, 2020.

\bibitem[Hochreiter and Schmidhuber(1997)]{HochreiterSchmidhuber1997}
S.~Hochreiter and J.~Schmidhuber.
\newblock {Long short-term memory}.
\newblock \emph{Neural Computation}, 9\penalty0 (8):\penalty0 1735--1780, 1997.

\bibitem[Hoerl and Kennard(1970)]{HoerlKennard1970}
A.~E. Hoerl and R.~W. Kennard.
\newblock {Ridge regression: Biased estimation for nonorthogonal problems}.
\newblock \emph{Technometrics}, 12\penalty0 (1):\penalty0 55--67, 1970.

\bibitem[Hogg and Foreman-Mackey(2018)]{Hogg2018}
D.~W. Hogg and D.~Foreman-Mackey.
\newblock {Data analysis recipes: using Markov chain Monte Carlo}.
\newblock \emph{Astrophysical Journal Supplement Series}, 236\penalty0
  (1):\penalty0 11, 2018.

\bibitem[Hornik et~al.(1989)Hornik, Stinchcombe, and
  White]{HornikStinchcombeWhite1989}
K.~Hornik, M.~Stinchcombe, and H.~White.
\newblock {Multilayer feedforward networks are universal approximators}.
\newblock \emph{Neural Networks}, 2\penalty0 (5):\penalty0 359--366, 1989.

\bibitem[Hotelling(1933)]{Hotelling1933}
H.~Hotelling.
\newblock {Analysis of a complex of statistical variables into principal
  components}.
\newblock \emph{Journal of Educational Psychology}, 24\penalty0 (6):\penalty0
  417--441, 1933.

\bibitem[Hughes et~al.(2022)Hughes, Bailer-Jones, and Jamal]{Hughes2022}
A.~C.~N. Hughes, C.~A.~L. Bailer-Jones, and S.~Jamal.
\newblock {Quasar and galaxy classification using {Gaia} {EDR3} and
  {CatWise2020}}.
\newblock \emph{Astronomy \& Astrophysics}, 668:\penalty0 A99, 2022.

\bibitem[Hwang et~al.(2022)Hwang, Ting, and Zakamska]{Hwang2022}
H.-C. Hwang, Y.-S. Ting, and N.~L. Zakamska.
\newblock {The eccentricity distribution of wide binaries and their individual
  measurements}.
\newblock \emph{Monthly Notices of the Royal Astronomical Society},
  512:\penalty0 3383, 2022.

\bibitem[Isobe et~al.(1990)Isobe, Feigelson, Akritas, and Babu]{Isobe1990}
T.~Isobe, E.~D. Feigelson, M.~G. Akritas, and G.~J. Babu.
\newblock {Linear regression in astronomy. I.}
\newblock \emph{The Astrophysical Journal}, 364:\penalty0 104, 1990.

\bibitem[Jacot et~al.(2018)Jacot, Gabriel, and Hongler]{Jacot2018}
A.~Jacot, F.~Gabriel, and C.~Hongler.
\newblock {Neural tangent kernel: convergence and generalization in neural
  networks}.
\newblock \emph{arXiv preprint arXiv:1806.07572}, 2018.

\bibitem[Jaynes(2003)]{Jaynes2003}
E.~T. Jaynes.
\newblock \emph{{Probability theory: The logic of science}}.
\newblock Cambridge University Press, Cambridge, 2003.

\bibitem[Jolliffe(2002)]{Jolliffe2002}
I.~T. Jolliffe.
\newblock \emph{{Principal component analysis (2nd ed.)}}.
\newblock Springer, 2002.

\bibitem[Kahn(1950)]{Kahn1950}
H.~Kahn.
\newblock {Random sampling (Monte Carlo) techniques in neutron attenuation
  problems, I \& II}.
\newblock \emph{Nucleonics}, 6\penalty0 (5--6):\penalty0 27--33, 37, 60--65,
  1950.

\bibitem[Kahn(1955)]{Kahn1955}
H.~Kahn.
\newblock {Use of different Monte Carlo sampling techniques}.
\newblock RAND Paper P-766, RAND Corporation, Santa Monica, CA, 1955.

\bibitem[Kahn and Harris(1951)]{KahnHarris1951}
H.~Kahn and T.~E. Harris.
\newblock {Estimation of particle transmission by random sampling}.
\newblock In A.~S. Householder, G.~E. Forsythe, and H.~H. Germond, editors,
  \emph{Monte Carlo Method}, volume~12 of \emph{National Bureau of Standards
  Applied Mathematics Series}, pages 27--30. National Bureau of Standards,
  Washington, DC, 1951.

\bibitem[Kahn and Marshall(1953)]{KahnMarshall1953}
H.~Kahn and A.~W. Marshall.
\newblock {Methods of reducing sample size in Monte Carlo computations}.
\newblock \emph{Journal of the Operations Research Society of America},
  1\penalty0 (5):\penalty0 263--278, 1953.

\bibitem[Kalos and Whitlock(1986)]{KalosWhitlock1986}
M.~H. Kalos and P.~A. Whitlock.
\newblock \emph{{Monte Carlo methods: Volume I: basics}}.
\newblock Wiley, 1986.

\bibitem[Kass and Raftery(1995)]{KassRaftery1995}
R.~E. Kass and A.~E. Raftery.
\newblock {Bayes factors}.
\newblock \emph{Journal of the American Statistical Association}, 90\penalty0
  (430):\penalty0 773--795, 1995.

\bibitem[Kaufman and Rousseeuw(1990)]{KaufmanRousseeuw1990}
L.~Kaufman and P.~J. Rousseeuw.
\newblock \emph{{Finding groups in data: an introduction to cluster analysis}}.
\newblock Wiley, 1990.

\bibitem[Kelly(2007)]{Kelly2007}
B.~C. Kelly.
\newblock {Some aspects of measurement error in linear regression of
  astronomical data}.
\newblock \emph{The Astrophysical Journal}, 665\penalty0 (2):\penalty0
  1489--1506, 2007.

\bibitem[Kendall and Stuart(1977)]{Kendall1977}
M.~G. Kendall and A.~Stuart.
\newblock \emph{{The advanced theory of statistics, Volume 1: Distribution
  theory}}.
\newblock C. Griffin, 1977.

\bibitem[Kennedy and O'Hagan(2001)]{KennedyOHagan2001}
M.~C. Kennedy and A.~O'Hagan.
\newblock {Bayesian calibration of computer models (with discussion)}.
\newblock \emph{Journal of the Royal Statistical Society: Series B (Statistical
  Methodology)}, 63\penalty0 (3):\penalty0 425--464, 2001.

\bibitem[Kennicutt(1998)]{Kennicutt1998}
R.~C. Kennicutt.
\newblock {The global Schmidt law in star-forming galaxies}.
\newblock \emph{The Astrophysical Journal}, 498\penalty0 (2):\penalty0
  541--552, 1998.

\bibitem[Kimeldorf and Wahba(1970)]{KimeldorfWahba1970}
G.~S. Kimeldorf and G.~Wahba.
\newblock {A correspondence between Bayesian estimation on stochastic processes
  and smoothing by splines}.
\newblock \emph{Annals of Mathematical Statistics}, 41\penalty0 (2):\penalty0
  495--502, 1970.

\bibitem[Kinderman and Monahan(1977)]{Kinderman1977}
A.~J. Kinderman and J.~F. Monahan.
\newblock {Computer generation of random variables using the ratio of uniform
  deviates}.
\newblock \emph{ACM Transactions on Mathematical Software}, 3:\penalty0
  257--260, 1977.

\bibitem[Kingma and Ba(2015)]{Kingma2015}
D.~P. Kingma and J.~Ba.
\newblock {Adam: a method for stochastic optimization}.
\newblock In \emph{3rd International Conference on Learning Representations
  (ICLR 2015)}, San Diego, CA, 2015.

\bibitem[Kingma and Welling(2013)]{KingmaWelling2013}
D.~P. Kingma and M.~Welling.
\newblock {Auto-encoding variational Bayes}.
\newblock \emph{arXiv preprint arXiv:1312.6114}, 2013.

\bibitem[Kroupa(2001)]{Kroupa2001}
P.~Kroupa.
\newblock {On the variation of the initial mass function}.
\newblock \emph{Monthly Notices of the Royal Astronomical Society},
  322\penalty0 (2):\penalty0 231--246, 2001.

\bibitem[Kullback and Leibler(1951)]{Kullback1951}
S.~Kullback and R.~A. Leibler.
\newblock {On information and sufficiency}.
\newblock \emph{Annals of Mathematical Statistics}, 22\penalty0 (1):\penalty0
  79--86, 1951.

\bibitem[Kuss and Rasmussen(2005)]{KussRasmussen2005}
M.~Kuss and C.~E. Rasmussen.
\newblock {Assessing approximate inference for binary Gaussian process
  classification}.
\newblock \emph{Journal of Machine Learning Research}, 6\penalty0
  (57):\penalty0 1679--1704, 2005.

\bibitem[Lahav et~al.(1995)Lahav, Naim, Buta, Corwin, de~Vaucouleurs, Dressler,
  et~al.]{Lahav1995}
O.~Lahav, A.~Naim, R.~J. Buta, H.~G. Corwin, G.~de~Vaucouleurs, A.~Dressler,
  et~al.
\newblock {Galaxies, human eyes, and artificial neural networks}.
\newblock \emph{Science}, 267\penalty0 (5199):\penalty0 859--862, 1995.

\bibitem[Laplace(1820)]{Laplace1820}
P.~S. Laplace.
\newblock \emph{{Théorie analytique des probabilités}}.
\newblock Courcier (Paris), 1820.

\bibitem[Le~Cam(1986)]{LeCam1986}
L.~Le~Cam.
\newblock \emph{{Asymptotic methods in statistical decision theory}}.
\newblock Springer, 1986.

\bibitem[LeCun et~al.(1998)LeCun, Bottou, Bengio, and Haffner]{Lecun1998}
Y.~LeCun, L.~Bottou, Y.~Bengio, and P.~Haffner.
\newblock {Gradient-based learning applied to document recognition}.
\newblock \emph{Proceedings of the IEEE}, 86\penalty0 (11):\penalty0
  2278--2324, 1998.

\bibitem[Lee et~al.(2017)Lee, Bahri, Novak, Schoenholz, Pennington, and
  Sohl-Dickstein]{Lee2018}
J.~Lee, Y.~Bahri, R.~Novak, S.~S. Schoenholz, J.~Pennington, and
  J.~Sohl-Dickstein.
\newblock {Deep neural networks as Gaussian processes}.
\newblock \emph{arXiv preprint arXiv:1711.00165}, 2017.

\bibitem[Lelli et~al.(2016)Lelli, McGaugh, and Schombert]{Lelli2016}
F.~Lelli, S.~S. McGaugh, and J.~M. Schombert.
\newblock {SPARC: mass models for 175 disk galaxies with Spitzer photometry and
  accurate rotation curves}.
\newblock \emph{The Astronomical Journal}, 152\penalty0 (6):\penalty0 157,
  2016.

\bibitem[Lewis and Bridle(2002)]{Lewis2002}
A.~Lewis and S.~Bridle.
\newblock {Cosmological parameters from CMB and other data: a Monte Carlo
  approach}.
\newblock \emph{Phys. Rev. D}, 66\penalty0 (10):\penalty0 103511, 2002.

\bibitem[Lindley(1965)]{Lindley1965}
D.~V. Lindley.
\newblock \emph{{Introduction to probability and statistics from a Bayesian
  viewpoint}}.
\newblock Cambridge University Press, 1965.

\bibitem[Liu(2001)]{Liu2001}
J.~S. Liu.
\newblock \emph{{Monte Carlo strategies in scientific computing}}.
\newblock Springer, 2001.

\bibitem[Lloyd(1982)]{Lloyd1982}
S.~Lloyd.
\newblock {Least squares quantization in PCM}.
\newblock \emph{IEEE Transactions on Information Theory}, 28\penalty0
  (2):\penalty0 129--137, 1982.

\bibitem[Loredo(1992)]{Loredo1992}
T.~J. Loredo.
\newblock {The promise of Bayesian inference for astrophysics}.
\newblock In E.~D. Feigelson and G.~J. Babu, editors, \emph{Statistical
  Challenges in Modern Astronomy}, pages 275--297. Springer-Verlag, New York,
  1992.

\bibitem[Loshchilov and Hutter(2017)]{Loshchilov2017}
I.~Loshchilov and F.~Hutter.
\newblock {{SGDR}: Stochastic gradient descent with warm restarts}.
\newblock In \emph{International Conference on Learning Representations}, 2017.

\bibitem[Luce(1959)]{Luce1959}
R.~D. Luce.
\newblock \emph{{Individual choice behavior: A theoretical analysis}}.
\newblock Wiley, 1959.

\bibitem[Lupton(1993)]{Lupton1993}
R.~Lupton.
\newblock \emph{{Statistics in theory and practice}}.
\newblock Princeton University Press, 1993.

\bibitem[Ma et~al.(2026)Ma, Sun, Jing, et~al.]{Ma2026}
C.~Ma, Z.~Sun, T.~Jing, et~al.
\newblock {Can {AI} dream of unseen galaxies? {C}onditional diffusion model for
  galaxy morphology augmentation}.
\newblock \emph{The Astrophysical Journal Supplement Series}, 282:\penalty0 25,
  2026.

\bibitem[MacKay(1992)]{MacKay1992}
D.~J.~C. MacKay.
\newblock {Bayesian interpolation}.
\newblock \emph{Neural Computation}, 4\penalty0 (3):\penalty0 415--447, 1992.

\bibitem[MacKay(2003)]{MacKay2003}
D.~J.~C. MacKay.
\newblock \emph{{Information theory, inference, and learning algorithms}}.
\newblock Cambridge University Press, 2003.

\bibitem[MacQueen(1967)]{MacQueen1967}
J.~MacQueen.
\newblock {Some methods for classification and analysis of multivariate
  observations}.
\newblock In L.~M. Le~Cam and J.~Neyman, editors, \emph{Proceedings of the
  Fifth Berkeley Symposium on Mathematical Statistics and Probability},
  volume~1, pages 281--297. University of California Press, 1967.

\bibitem[Marsaglia(1977)]{Marsaglia1977}
G.~Marsaglia.
\newblock {The squeeze method for generating gamma variates}.
\newblock \emph{Computers \& Mathematics with Applications}, 3\penalty0
  (4):\penalty0 321--325, 1977.

\bibitem[Matheron(1963)]{Matheron1963}
G.~Matheron.
\newblock {Principles of geostatistics}.
\newblock \emph{Economic Geology}, 58\penalty0 (8):\penalty0 1246--1266, 1963.

\bibitem[Matthews et~al.(2018)Matthews, Rowland, Hron, Turner, and
  Ghahramani]{Matthews2018}
A.~G. de~G. Matthews, M.~Rowland, J.~Hron, R.~E. Turner, and Z.~Ghahramani.
\newblock {Gaussian process behaviour in wide deep neural networks}.
\newblock \emph{arXiv preprint arXiv:1804.11271}, 2018.

\bibitem[McCullagh and Nelder(1989)]{McCullaghNelder1989}
P.~McCullagh and J.~A. Nelder.
\newblock \emph{{Generalized linear models (2nd ed.)}}.
\newblock Chapman and Hall, 1989.

\bibitem[McCulloch and Pitts(1943)]{McCullochPitts1943}
W.~S. McCulloch and W.~Pitts.
\newblock {A logical calculus of the ideas immanent in nervous activity}.
\newblock \emph{Bulletin of Mathematical Biophysics}, 5:\penalty0 115--133,
  1943.

\bibitem[McFadden(1974)]{McFadden1973}
D.~McFadden.
\newblock {Conditional logit analysis of qualitative choice behavior}.
\newblock In P.~Zarembka, editor, \emph{Frontiers in Econometrics}, pages
  105--142. Academic Press, 1974.

\bibitem[Metropolis et~al.(1953)Metropolis, Rosenbluth, Rosenbluth, Teller, and
  Teller]{Metropolis1953}
N.~Metropolis, A.~W. Rosenbluth, M.~N. Rosenbluth, A.~H. Teller, and E.~Teller.
\newblock {Equation of state calculations by fast computing machines}.
\newblock \emph{Journal of Chemical Physics}, 21\penalty0 (6):\penalty0
  1087--1092, 1953.

\bibitem[Minka(2001)]{Minka2001}
T.~P. Minka.
\newblock {Expectation propagation for approximate Bayesian inference}.
\newblock In \emph{Proceedings of the Seventeenth Conference on Uncertainty in
  Artificial Intelligence}, pages 362--369, San Francisco, CA, USA, 2001.
  Morgan Kaufmann Publishers Inc.

\bibitem[Minsky and Papert(1969)]{MinskyPapert1969}
M.~Minsky and S.~Papert.
\newblock \emph{{Perceptrons: an introduction to computational geometry}}.
\newblock MIT Press, 1969.

\bibitem[Murtagh and Heck(1987)]{Murtagh1987}
F.~Murtagh and A.~Heck.
\newblock \emph{{Multivariate data analysis}}, volume 131 of \emph{Astrophysics
  and Space Science Library}.
\newblock D. Reidel, Dordrecht, 1987.

\bibitem[Nakkiran et~al.(2019)Nakkiran, Kaplun, Bansal, Yang, Barak, and
  Sutskever]{Nakkiran2019}
P.~Nakkiran, G.~Kaplun, Y.~Bansal, T.~Yang, B.~Barak, and I.~Sutskever.
\newblock {Deep double descent: where bigger models and more data hurt}.
\newblock \emph{arXiv preprint arXiv:1912.02292}, 2019.

\bibitem[Neal(1996)]{Neal1996}
R.~M. Neal.
\newblock \emph{{Bayesian learning for neural networks}}.
\newblock Springer, 1996.

\bibitem[Ng and Jordan(2001)]{Ng2001}
A.~Y. Ng and M.~I. Jordan.
\newblock {On discriminative vs. generative classifiers: A comparison of
  logistic regression and naive Bayes}.
\newblock In T.~G. Dietterich, S.~Becker, and Z.~Ghahramani, editors,
  \emph{Advances in Neural Information Processing Systems 14 (NIPS 2001)},
  pages 841--848. MIT Press, 2001.

\bibitem[O'Hagan(1978)]{OHagan1978}
A.~O'Hagan.
\newblock {Curve fitting and optimal design for prediction}.
\newblock \emph{Journal of the Royal Statistical Society: Series B
  (Methodological)}, 40\penalty0 (1):\penalty0 1--24, 1978.

\bibitem[Papamakarios et~al.(2018)Papamakarios, Sterratt, and
  Murray]{Papamakarios2019}
G.~Papamakarios, D.~C. Sterratt, and I.~Murray.
\newblock {Sequential neural likelihood: fast likelihood-free inference with
  autoregressive flows}.
\newblock \emph{arXiv preprint arXiv:1805.07226}, 2018.

\bibitem[Pearson(1894)]{Pearson1894}
K.~Pearson.
\newblock {III. Contributions to the mathematical theory of evolution}.
\newblock \emph{Philosophical Transactions of the Royal Society of London. A},
  185:\penalty0 71--110, 1894.

\bibitem[Pearson(1901)]{Pearson1901}
K.~Pearson.
\newblock {LIII. On lines and planes of closest fit to systems of points in
  space}.
\newblock \emph{The London, Edinburgh, and Dublin Philosophical Magazine and
  Journal of Science}, 2\penalty0 (11):\penalty0 559--572, 1901.

\bibitem[Polyak and Juditsky(1992)]{PolyakJuditsky1992}
B.~T. Polyak and A.~B. Juditsky.
\newblock {Acceleration of stochastic approximation by averaging}.
\newblock \emph{SIAM Journal on Control and Optimization}, 30\penalty0
  (4):\penalty0 838--855, 1992.

\bibitem[Quiñonero-Candela and Rasmussen(2005)]{QuinoneroCandela2005}
J.~Quiñonero-Candela and C.~E. Rasmussen.
\newblock {A unifying view of sparse approximate Gaussian process regression}.
\newblock \emph{Journal of Machine Learning Research}, 6\penalty0
  (65):\penalty0 1939--1959, 2005.

\bibitem[Raffel et~al.(2020)Raffel, Shazeer, Roberts, Lee, Narang, Matena,
  Zhou, Li, and Liu]{Raffel2020}
C.~Raffel, N.~Shazeer, A.~Roberts, K.~Lee, S.~Narang, M.~Matena, Y.~Zhou,
  W.~Li, and P.~J. Liu.
\newblock {Exploring the limits of transfer learning with a unified
  text-to-text transformer}.
\newblock \emph{Journal of Machine Learning Research}, 21\penalty0
  (140):\penalty0 1--67, 2020.

\bibitem[Raiffa and Schlaifer(1961)]{Raiffa1961}
H.~Raiffa and R.~Schlaifer.
\newblock \emph{{Applied statistical decision theory}}.
\newblock Division of Research, Graduate School of Business Administration,
  Harvard University, Boston, 1961.

\bibitem[Rasmussen and Williams(2006)]{RasmussenWilliams2006}
C.~E. Rasmussen and C.~K.~I. Williams.
\newblock \emph{{Gaussian processes for machine learning}}.
\newblock MIT Press, 2006.

\bibitem[Redner and Walker(1984)]{RednerWalker1984}
R.~A. Redner and H.~F. Walker.
\newblock {Mixture densities, maximum likelihood and the EM algorithm}.
\newblock \emph{SIAM Review}, 26\penalty0 (2):\penalty0 195--239, 1984.

\bibitem[Rezende and Mohamed(2015)]{Rezende2016}
D.~J. Rezende and S.~Mohamed.
\newblock {Variational inference with normalizing flows}.
\newblock \emph{arXiv preprint arXiv:1505.05770}, 2015.

\bibitem[Robbins and Monro(1951)]{Robbins1951}
H.~Robbins and S.~Monro.
\newblock {A stochastic approximation method}.
\newblock \emph{Annals of Mathematical Statistics}, 22\penalty0 (3):\penalty0
  400--407, 1951.

\bibitem[Robert and Casella(1999)]{Robert1999}
C.~P. Robert and G.~Casella.
\newblock \emph{{Monte Carlo statistical methods}}.
\newblock Springer, 1999.

\bibitem[Roberts and Smith(1994)]{Roberts1994}
G.~O. Roberts and A.~F.~M. Smith.
\newblock {Simple conditions for the convergence of the Gibbs sampler and
  Metropolis-Hastings algorithms}.
\newblock \emph{Stochastic Processes and their Applications}, 49\penalty0
  (2):\penalty0 207--216, 1994.

\bibitem[Rosenblatt(1958)]{Rosenblatt1958}
F.~Rosenblatt.
\newblock {The perceptron: a probabilistic model for information storage and
  organization in the brain}.
\newblock \emph{Psychological Review}, 65\penalty0 (6):\penalty0 386--408,
  1958.

\bibitem[Rubinstein(1981)]{Rubinstein1981}
R.~Y. Rubinstein.
\newblock \emph{{Simulation and the Monte Carlo method}}.
\newblock Wiley, 1981.

\bibitem[Rubinstein and Hastie(1997)]{Rubinstein1997}
Y.~D. Rubinstein and T.~Hastie.
\newblock {Discriminative vs informative learning}.
\newblock In \emph{KDD'97}, pages 49--53. AAAI Press, 1997.

\bibitem[Rumelhart et~al.(1986)Rumelhart, Hinton, and
  Williams]{RumelhartHintonWilliams1986}
D.~E. Rumelhart, G.~E. Hinton, and R.~J. Williams.
\newblock {Learning representations by back-propagating errors}.
\newblock \emph{Nature}, 323:\penalty0 533--536, 1986.

\bibitem[Schwarz(1978)]{Schwarz1978}
G.~Schwarz.
\newblock {Estimating the dimension of a model}.
\newblock \emph{The Annals of Statistics}, 6\penalty0 (2):\penalty0 461--464,
  1978.

\bibitem[Selim and Ismail(1984)]{SelimIsmail1984}
S.~Z. Selim and M.~A. Ismail.
\newblock {K-means-type algorithms: a generalized convergence theorem and
  characterization of local optimality}.
\newblock \emph{IEEE Transactions on Pattern Analysis and Machine
  Intelligence}, PAMI-6\penalty0 (1):\penalty0 81--87, 1984.

\bibitem[Shannon(1948)]{Shannon1948}
C.~E. Shannon.
\newblock {A mathematical theory of communication}.
\newblock \emph{The Bell System Technical Journal}, 27:\penalty0 379--423,
  623--656, 1948.

\bibitem[Sharma(2017)]{Sharma2017}
S.~Sharma.
\newblock {Markov chain Monte Carlo methods for Bayesian data analysis in
  astronomy}.
\newblock \emph{Annual Review of Astronomy and Astrophysics}, 55\penalty0
  (1):\penalty0 213--259, 2017.

\bibitem[Sivia and Skilling(2006)]{Sivia2006}
D.~S. Sivia and J.~Skilling.
\newblock \emph{{Data analysis: A Bayesian tutorial}}.
\newblock Oxford University Press, 2006.

\bibitem[Skilling(2006)]{Skilling2006}
J.~Skilling.
\newblock {Nested sampling for general Bayesian computation}.
\newblock \emph{Bayesian Analysis}, 1\penalty0 (4):\penalty0 833--859, 2006.

\bibitem[Snelson and Ghahramani(2005)]{SnelsonGhahramani2006}
E.~Snelson and Z.~Ghahramani.
\newblock {Sparse Gaussian processes using pseudo-inputs}.
\newblock In \emph{Advances in Neural Information Processing Systems 18 (NIPS
  2005)}, pages 1257--1264. MIT Press, 2005.

\bibitem[Song et~al.(2020)Song, Sohl-Dickstein, Kingma, Kumar, Ermon, and
  Poole]{Song2021}
Y.~Song, J.~Sohl-Dickstein, D.~P. Kingma, A.~Kumar, S.~Ermon, and B.~Poole.
\newblock {Score-based generative modeling through stochastic differential
  equations}.
\newblock \emph{arXiv preprint arXiv:2011.13456}, 2020.

\bibitem[Srivastava et~al.(2014)Srivastava, Hinton, Krizhevsky, Sutskever, and
  Salakhutdinov]{Srivastava2014}
N.~Srivastava, G.~E. Hinton, A.~Krizhevsky, I.~Sutskever, and R.~R.
  Salakhutdinov.
\newblock {Dropout: a simple way to prevent neural networks from overfitting}.
\newblock \emph{Journal of Machine Learning Research}, 15\penalty0
  (56):\penalty0 1929--1958, 2014.

\bibitem[Stein(1999)]{Stein1999}
M.~L. Stein.
\newblock \emph{{Interpolation of spatial data: Some theory for kriging}}.
\newblock Springer, 1999.

\bibitem[Steinhaus(1956)]{Steinhaus1956}
H.~Steinhaus.
\newblock {Sur la division des corps mat\'{e}riels en parties}.
\newblock \emph{Bulletin de l'Acad\'{e}mie Polonaise des Sciences, Classe III},
  4\penalty0 (12):\penalty0 801--804, 1956.

\bibitem[Stone(1974)]{Stone1974}
M.~Stone.
\newblock {Cross-validatory choice and assessment of statistical predictions}.
\newblock \emph{Journal of the Royal Statistical Society. Series B
  (Methodological)}, 36:\penalty0 111--147, 1974.

\bibitem[Storrie-Lombardi et~al.(1992)Storrie-Lombardi, Lahav, Sodr{\'e}, and
  Storrie-Lombardi]{StorrieLombardi1992}
M.~C. Storrie-Lombardi, O.~Lahav, L.~Sodr{\'e}, Jr., and L.~J.
  Storrie-Lombardi.
\newblock {Morphological classification of galaxies by artificial neural
  networks}.
\newblock \emph{Monthly Notices of the Royal Astronomical Society},
  259\penalty0 (1):\penalty0 8P--12P, 1992.

\bibitem[Stuart and Ord(1994)]{Stuart1994}
A.~Stuart and J.~K. Ord.
\newblock \emph{{Kendall's advanced theory of statistics, Volume 1:
  Distribution theory}}.
\newblock Edward Arnold, 1994.

\bibitem[Sun et~al.(2025)Sun, Ting, Liu, et~al.]{SunC3PO3}
Q.~Sun, Y.-S. Ting, F.~Liu, et~al.
\newblock {{C3PO}. {III}. On the lithium signatures following planet engulfment
  by stars}.
\newblock \emph{The Astrophysical Journal}, 978:\penalty0 107, 2025.

\bibitem[Sun et~al.(2023)Sun, Ting, and Cai]{Sun2023}
Z.~Sun, Y.-S. Ting, and Z.~Cai.
\newblock {Quasar factor analysis -- an unsupervised and probabilistic quasar
  continuum prediction algorithm with latent factor analysis}.
\newblock \emph{The Astrophysical Journal Supplement Series}, 269\penalty0
  (1):\penalty0 4, 2023.

\bibitem[Taylor(1982)]{Taylor1982}
J.~R. Taylor.
\newblock \emph{{An introduction to error analysis: The study of uncertainties
  in physical measurements}}.
\newblock University Science Books, 1982.

\bibitem[Tegmark et~al.(2004)Tegmark, Strauss, Blanton, Abazajian, Dodelson,
  Sandvik, et~al.]{Tegmark2004}
M.~Tegmark, M.~A. Strauss, M.~R. Blanton, K.~Abazajian, S.~Dodelson,
  H.~Sandvik, et~al.
\newblock {Cosmological parameters from SDSS and WMAP}.
\newblock \emph{Physical Review D}, 69\penalty0 (10):\penalty0 103501, 2004.

\bibitem[Theil(1969)]{Theil1969}
H.~Theil.
\newblock {A multinomial extension of the linear logit model}.
\newblock \emph{International Economic Review}, 10\penalty0 (3):\penalty0
  251--259, 1969.

\bibitem[Thrane and Talbot(2019)]{Thrane2019}
E.~Thrane and C.~Talbot.
\newblock {An introduction to Bayesian inference in gravitational-wave
  astronomy: Parameter estimation, model selection, and hierarchical models}.
\newblock \emph{Publications of the Astronomical Society of Australia},
  36:\penalty0 e010, 2019.

\bibitem[Tibshirani(1996)]{Tibshirani1996}
R.~Tibshirani.
\newblock {Regression shrinkage and selection via the lasso}.
\newblock \emph{Journal of the Royal Statistical Society. Series B
  (Methodological)}, 58\penalty0 (1):\penalty0 267--288, 1996.

\bibitem[Tierney(1994)]{Tierney1994}
L.~Tierney.
\newblock {Markov chains for exploring posterior distributions}.
\newblock \emph{The Annals of Statistics}, 22\penalty0 (4):\penalty0
  1701--1728, 1994.

\bibitem[Tierney and Kadane(1986)]{TierneyKadane1986}
L.~Tierney and J.~B. Kadane.
\newblock {Accurate approximations for posterior moments and marginal
  densities}.
\newblock \emph{Journal of the American Statistical Association}, 81\penalty0
  (393):\penalty0 82--86, 1986.

\bibitem[Tikhonov(1963)]{Tikhonov1963}
A.~N. Tikhonov.
\newblock {Solution of incorrectly formulated problems and the regularization
  method}.
\newblock \emph{Soviet Mathematics Doklady}, 4:\penalty0 1035--1038, 1963.

\bibitem[Ting(2025)]{Ting2025}
Y.-S. Ting.
\newblock {Why machine learning models systematically underestimate extreme
  values}.
\newblock \emph{The Open Journal of Astrophysics}, 8:\penalty0 95, 2025.

\bibitem[Ting(2026)]{Ting2026}
Y.-S. Ting.
\newblock {Why machine learning models systematically underestimate extreme
  values {II}: How to fix it with {LatentNN}}.
\newblock \emph{The Open Journal of Astrophysics}, 9:\penalty0 59559, 2026.

\bibitem[Ting and Ji(2025)]{TingJi2025}
Y.-S. Ting and A.~Ji.
\newblock {Quantifying bursty star formation in dwarf galaxies}.
\newblock \emph{The Open Journal of Astrophysics}, 8:\penalty0 7, 2025.

\bibitem[Ting and Weinberg(2022)]{TingWeinberg2022}
Y.-S. Ting and D.~H. Weinberg.
\newblock {How many elements matter?}
\newblock \emph{The Astrophysical Journal}, 927:\penalty0 209, 2022.

\bibitem[Ting et~al.(2018)Ting, Hawkins, and Rix]{TingHawkinsRix2018}
Y.-S. Ting, K.~Hawkins, and H.-W. Rix.
\newblock {A large and pristine sample of standard candles across the {Milky
  Way}: {$\sim$}100,000 red clump stars with 3\% contamination}.
\newblock \emph{The Astrophysical Journal Letters}, 858:\penalty0 L7, 2018.

\bibitem[Titsias(2009)]{Titsias2009}
M.~K. Titsias.
\newblock {Variational learning of inducing variables in sparse Gaussian
  processes}.
\newblock In \emph{Proceedings of the Twelfth International Conference on
  Artificial Intelligence and Statistics}, pages 567--574, 2009.

\bibitem[Titterington et~al.(1985)Titterington, Smith, and
  Makov]{Titterington1985}
D.~M. Titterington, A.~F.~M. Smith, and U.~E. Makov.
\newblock \emph{{Statistical analysis of finite mixture distributions}}.
\newblock Wiley, 1985.

\bibitem[Trotta(2008)]{Trotta2008}
R.~Trotta.
\newblock {Bayes in the sky: Bayesian inference and model selection in
  cosmology}.
\newblock \emph{Contemporary Physics}, 49\penalty0 (2):\penalty0 71--104, 2008.

\bibitem[Tully and Fisher(1977)]{TullyFisher1977}
R.~B. Tully and J.~R. Fisher.
\newblock {A new method of determining distances to galaxies}.
\newblock \emph{Astronomy and Astrophysics}, 54:\penalty0 661--673, 1977.

\bibitem[Van~Huffel and Vandewalle(1991)]{VanHuffelVandewalle1991}
S.~Van~Huffel and J.~Vandewalle.
\newblock \emph{{The total least squares problem: Computational aspects and
  analysis}}.
\newblock Society for Industrial and Applied Mathematics, 1991.

\bibitem[Vapnik(1995)]{Vapnik1995}
V.~N. Vapnik.
\newblock \emph{{The nature of statistical learning theory}}.
\newblock Springer, 1995.

\bibitem[Vaswani et~al.(2017)Vaswani, Shazeer, Parmar, Uszkoreit, Jones, Gomez,
  Kaiser, and Polosukhin]{Vaswani2017}
A.~Vaswani, N.~Shazeer, N.~Parmar, J.~Uszkoreit, L.~Jones, A.~N. Gomez,
  {\L}.~Kaiser, and I.~Polosukhin.
\newblock {Attention is all you need}.
\newblock In \emph{Advances in Neural Information Processing Systems},
  volume~30, pages 5998--6008. Curran Associates, 2017.

\bibitem[Verde et~al.(2003)Verde, Peiris, Spergel, Nolta, Bennett, Halpern,
  et~al.]{Verde2003}
L.~Verde, H.~V. Peiris, D.~N. Spergel, M.~R. Nolta, C.~L. Bennett, M.~Halpern,
  et~al.
\newblock {First-year Wilkinson Microwave Anisotropy Probe (WMAP) observations:
  parameter estimation methodology}.
\newblock \emph{Astrophysical Journal Supplement}, 148:\penalty0 195--211,
  2003.

\bibitem[von Hippel et~al.(1994)von Hippel, Storrie-Lombardi, Storrie-Lombardi,
  and Irwin]{vonHippel1994}
T.~von Hippel, L.~J. Storrie-Lombardi, M.~C. Storrie-Lombardi, and M.~J. Irwin.
\newblock {Automated classification of stellar spectra -- {I. Initial results
  with artificial neural networks}}.
\newblock \emph{Monthly Notices of the Royal Astronomical Society},
  269\penalty0 (1):\penalty0 97--104, 1994.

\bibitem[von Neumann(1951)]{VonNeumann1951}
J.~von Neumann.
\newblock {Various techniques used in connection with random digits}.
\newblock In A.~S. Householder, G.~E. Forsythe, and H.~H. Germond, editors,
  \emph{Monte Carlo Method}, volume~12 of \emph{National Bureau of Standards
  Applied Mathematics Series}, pages 36--38. National Bureau of Standards,
  Washington, DC, 1951.

\bibitem[Vrard et~al.(2016)Vrard, Mosser, and Samadi]{Vrard2016}
M.~Vrard, B.~Mosser, and R.~Samadi.
\newblock {Period spacings in red giants. {II}. Automated measurement}.
\newblock \emph{Astronomy and Astrophysics}, 588:\penalty0 A87, 2016.

\bibitem[Wahba(1990)]{Wahba1990}
G.~Wahba.
\newblock \emph{{Spline models for observational data}}.
\newblock Society for Industrial and Applied Mathematics, 1990.

\bibitem[Walker(1969)]{Walker1969}
A.~M. Walker.
\newblock {On the asymptotic behaviour of posterior distributions}.
\newblock \emph{Journal of the Royal Statistical Society. Series B
  (Methodological)}, 31\penalty0 (1):\penalty0 80--88, 1969.

\bibitem[Wall and Jenkins(2003)]{Wall2003}
J.~V. Wall and C.~R. Jenkins.
\newblock \emph{{Practical statistics for astronomers}}.
\newblock Cambridge University Press, 2003.

\bibitem[Wei et~al.(2022)Wei, Tay, Bommasani, Raffel, Zoph, Borgeaud,
  et~al.]{Wei2022}
J.~Wei, Y.~Tay, R.~Bommasani, C.~Raffel, B.~Zoph, S.~Borgeaud, et~al.
\newblock {Emergent abilities of large language models}.
\newblock \emph{arXiv preprint arXiv:2206.07682}, 2022.

\bibitem[Williams(1996)]{Williams1996}
C.~K.~I. Williams.
\newblock {Computing with infinite networks}.
\newblock In \emph{Advances in Neural Information Processing Systems 9 (NIPS
  1996)}, pages 295--301. MIT Press, 1996.

\bibitem[Williams and Barber(1998)]{WilliamsBarber1998}
C.~K.~I. Williams and D.~Barber.
\newblock {Bayesian classification with Gaussian processes}.
\newblock \emph{IEEE Transactions on Pattern Analysis and Machine
  Intelligence}, 20\penalty0 (12):\penalty0 1342--1351, 1998.

\bibitem[Wolfe(1970)]{Wolfe1970}
J.~H. Wolfe.
\newblock {Pattern clustering by multivariate mixture analysis}.
\newblock \emph{Multivariate Behavioral Research}, 5\penalty0 (3):\penalty0
  329--350, 1970.

\bibitem[Zellner(1971)]{Zellner1971}
A.~Zellner.
\newblock \emph{{An introduction to Bayesian inference in econometrics}}.
\newblock Wiley, 1971.

\bibitem[Zellner and Rossi(1984)]{ZellnerRossi1984}
A.~Zellner and P.~E. Rossi.
\newblock {Bayesian analysis of dichotomous quantal response models}.
\newblock \emph{Journal of Econometrics}, 25\penalty0 (3):\penalty0 365--393,
  1984.

\end{thebibliography}
\end{document}